\pdfoutput=1
\documentclass[3p,times,book,onecolumn]{elsarticle}

\usepackage{titlesec}
\titleclass{\subsubsubsection}{straight}[\subsection]

\newcounter{subsubsubsection}[subsubsection]
\renewcommand\thesubsubsubsection{\thesubsubsection.\arabic{subsubsubsection}}
\titleformat{\subsubsubsection}
  {\normalfont\normalsize\itshape}{\thesubsubsubsection}{1em}{}
\titlespacing*{\subsubsubsection}
{0pt}{3.25ex plus 1ex minus .2ex}{.5ex plus .2ex}

\makeatletter
\renewcommand\paragraph{\@startsection{paragraph}{5}{\z@}%
  {3.25ex \@plus1ex \@minus.2ex}%
  {-1em}%
  {\normalfont\normalsize\bfseries}}
\renewcommand\subparagraph{\@startsection{subparagraph}{6}{\parindent}%
  {3.25ex \@plus1ex \@minus .2ex}%
  {-1em}%
  {\normalfont\normalsize\bfseries}}

\titleclass{\subsubsubsubsection}{straight}[\subsection]

\newcounter{subsubsubsubsection}[subsubsubsection]
\renewcommand\thesubsubsubsubsection{\thesubsubsubsection.\arabic{subsubsubsubsection}}
\titleformat{\subsubsubsubsection}
  {\normalfont\normalsize\itshape}{\thesubsubsubsubsection}{1em}{}
\titlespacing*{\subsubsubsubsection}
{0pt}{3.25ex plus 0ex minus .2ex}{.5ex plus .2ex}

\makeatletter
\renewcommand\paragraph{\@startsection{paragraph}{6}{\z@}%
  {3.25ex \@plus1ex \@minus.2ex}%
  {-1em}%
  {\normalfont\normalsize\bfseries}}
\renewcommand\subparagraph{\@startsection{subparagraph}{7}{\parindent}%
  {3.25ex \@plus1ex \@minus .2ex}%
  {-1em}%
  {\normalfont\normalsize\bfseries}}
\def\toclevel@subsubsubsection{4}
\def\toclevel@subsubsubsubsection{5}
\def\toclevel@paragraph{6}
\def\toclevel@paragraph{7}
\def\l@subsubsubsection{\@dottedtocline{4}{7em}{4em}}
\def\l@subsubsubsubsection{\@dottedtocline{5}{10em}{5em}}
\def\l@paragraph{\@dottedtocline{6}{14em}{6em}}
\def\l@subparagraph{\@dottedtocline{7}{17em}{7em}}
\makeatother

\usepackage[utf8]{inputenc}
\usepackage[english]{babel}

\usepackage{graphicx,epsfig,amsmath,color}

\usepackage{multirow}
\usepackage{cuted}
\usepackage[version=4]{mhchem}
\usepackage{siunitx}
\usepackage{caption}
\usepackage{subcaption}
\usepackage{mathtools}

\usepackage{hyperref}
\usepackage{booktabs}
\usepackage{float}
\usepackage[super]{nth}

\usepackage{tikz}
\usepackage{pgfplots}
\pgfplotsset{compat=1.5}

\usepackage{lineno}
\definecolor{red}{rgb}{1,0,0}
\definecolor{blue}{rgb}{0,0,1}

\newcommand\barparenb[1]{\overset{(-)}{#1}}

\newcommand{\tit}[1]{\text{\textit{#1}}}

\DeclareSIUnit\year{yr}
\DeclareSIUnit\min{min}
\DeclareSIUnit\h{h}
\DeclareSIUnit\d{d}
\DeclareSIUnit\PeV{PeV}
\DeclareSIUnit\MB{MB}
\DeclareSIUnit\GB{GB}
\DeclareSIUnit\TB{TB}
\DeclareSIUnit\PE{PE}
\DeclareSIUnit\inch{inch}
\DeclareSIUnit\rad{rad}
\DeclareSIUnit\speedoflight{\tit{c}}

\makeatletter
\def\ps@pprintTitle{%
  \let\@oddhead\@empty
  \let\@evenhead\@empty
  \def\@oddfoot{\reset@font\hfil\thepage\hfil}
  \let\@evenfoot\@oddfoot
}
\makeatother

\begin{document}
%%%%%\linenumbers \pagewiselinenumbers

\begin{frontmatter}
\begin{titlepage}

\title{The European Spallation Source neutrino Super Beam Conceptual Design Report}

%\cortext[contact]{Corresponding author, marcos.dracos@in2p3.fr}
\newcommand{\authorlist}{

\author[cern,uu]{A.~Alekou}
\author[iphc]{E.~Baussan\corref{coeditor}}
\author[ess]{A.K.~Bhattacharyya}
\author[ess]{N.~Blaskovic Kraljevic}
\author[kth,okc]{M.~Blennow}
\author[unisof]{M.~Bogomilov}
\author[ess]{B.~Bolling}
\author[iphc]{E.~Bouquerel}
\author[ess]{O.~Buchan}
\author[ulund1]{A.~Burgman\corref{coeditor}}
\author[uu]{C.J.~Carlile}
\author[ulund1]{J.~Cederkall}
\author[ulund1]{P.~Christiansen}
\author[ulund2,ess]{M.~Collins}
\author[infn2]{E.~Cristaldo Morales}
\author[agh]{P.~Cupiał}
\author[iphc]{L.~D'Alessi}
\author[ess]{H.~Danared}
\author[uu]{D.~Dancila}
\author[iphc]{J.~P.~A.~M.~de~Andr\'{e}}
\author[cern]{J.P.~Delahaye}
\author[iphc]{M.~Dracos\corref{contact}}
\author[cern]{I.~Efthymiopoulos}
\author[uu]{T.~Ekel\"{o}f\corref{contact}}
\author[ess]{M.~Eshraqi\corref{coeditor}}
\author[ncsr]{G.~Fanourakis}
\author[ul]{A.~Farricker}
\author[uam]{E.~Fernandez-Martinez\corref{coeditor}}
\author[ess]{B.~Folsom\corref{coeditor}}
\author[nu]{T.~Fukuda}
\author[ess]{N.~Gazis\corref{coeditor}}
\author[ess]{B.~G\r{a}lnander}
\author[ncsr]{Th.~Geralis}
\author[rbi,uoh]{M.~Ghosh\corref{coeditor}}
\author[cu]{G.~Gokbulut}
\author[rbi]{L.~Halić}
\author[ess]{M.~Jenssen}
\author[cu]{A.~Kayis Topaksu}
\author[ess]{B.~Kildetoft}
\author[rbi]{B.~Kliček\corref{coeditor}}
\author[agh]{M.~Kozioł}
\author[rbi]{K.~Krhač}
\author[agh]{Ł.~Łacny}
\author[ess]{M.~Lindroos}
\author[ess]{C.~Maiano}
\author[ess]{C.~Marrelli}
\author[ess]{C.~Martins}
\author[infn1]{M.~Mezzetto}
\author[ess]{N.~Milas}
\author[cu]{M.~Oglakci}
\author[kth,okc]{T.~Ohlsson}
\author[uu]{M.~Olveg\r{a}rd\corref{coeditor}}
\author[uam]{T.~Ota}
\author[ulund1]{J.~Park\fnref{ibs}}
\author[ess]{D.~Patrzalek}
\author[unisof]{G.~Petkov}
\author[iphc]{P.~Poussot}
\author[ess]{R.~Johansson}
\author[ijclab]{S.~Rosauro-Alcaraz}
\author[lu]{D.~Saiang}
\author[agh]{B.~Szybiński}
\author[agh]{J.~Snamina}
\author[ncsr]{G.~Stavropoulos}
\author[rbi]{M.~Stipčević}
%\author[ihep]{J.Y.~Tang}
\author[ess]{R.~Tarkeshian}
\author[infn2]{F.~Terranova}
\author[iphc]{J.~Thomas}
\author[uhh]{T.~Tolba\corref{coeditor}}
\author[ess]{E.~Trachanas}
\author[unisof]{R.~Tsenov}
\author[unisof]{G.~Vankova-Kirilova}
\author[csns]{N.~Vassilopoulos} 
\author[cern]{E.~Wildner}
\author[iphc]{J.~Wurtz}
\author[ncsr]{O.~Zormpa}
\author[uu]{Y.~Zou}

\address[agh]{AGH University of Science and Technology, al. Mickiewicza 30, 30-059 Krakow, Poland}

\address[cu]{University of Cukurova, Faculty of Science and Letters, Department of Physics, 01330 Adana, Turkey}

\address[lu]{$Lule\aa~University~of~Technology$}

\address[ncsr]{Institute of Nuclear and Particle Physics, NCSR Demokritos, Neapoleos 27, 15341 Agia Paraskevi, Greece}

%\address[ihep]{Institute of High Energy Physics, Chinese Academy of Sciences, Yuquan Road 19B, Shijingshan District, Beijing 100049, China}

\address[csns]{Spallation Neutron Science Center, Dongguan 523803, China} 

%\address[cern]{CERN, Esplanade des Particules 1, 1217 Meyrin, Suisse}

\address[cern]{CERN, 1211 Geneva 23, Switzerland}

\address[uhh]{Institute for Experimental Physics, Hamburg University, 22761 Hamburg, Germany}

\address[ulund1]{Department of Physics, Lund University, P.O Box 118, 221 00 Lund, Sweden}

\address[ulund2]{Faculty of Engineering, Lund University, P.O Box 118, 221 00 Lund, Sweden}

\address[ess]{European Spallation Source, Box 176, SE-221 00 Lund, Sweden}

\address[uam]{Departamento de Fisica Teorica and Instituto de Fisica Teorica, IFT-UAM/CSIC, Universidad Autonoma de Madrid, Cantoblanco, 28049, Madrid, Spain}

\address[infn2]{University of Milano-Bicocca and INFN sez. di Milano-Bicocca, Milano, Italy}

\address[infn1]{INFN sez. di Padova, Padova, Italy}

%\address[uniri]{University of Rijeka, Department of Physics, 51000 Rijeka, Croatia}

\address[unisof]{Sofia University St. Kliment Ohridski, Faculty of Physics, 1164 Sofia, Bulgaria}

\address[iphc]{IPHC, Universit\'{e} de Strasbourg, CNRS/IN2P3, Strasbourg, France}

\address[kth]{Department of Physics, School of Engineering Sciences, KTH Royal Institute of Technology, Roslagstullsbacken 21, 106 91 Stockholm, Sweden}

\address[okc]{The Oskar Klein Centre, AlbaNova University Center, Roslagstullsbacken 21, 106 91 Stockholm, Sweden}

\address[uu]{Department of Physics and Astronomy, FREIA Division, Uppsala University, P.O. Box 516, 751 20 Uppsala, Sweden}

\address[ul]{Cockroft Institute (A36), Liverpool University, Warrington WA4 4AD, UK}

\address[rbi]{Center of Excellence for Advanced Materials and Sensing Devices, Ruđer Bo\v{s}kovi\'c Institute, 10000 Zagreb, Croatia}

\address[uoh]{School of Physics, University of Hyderabad, Hyderabad 500046, India} 

\address[ijclab]{P\^ole Th\'eorie, Laboratoire de Physique des 2 Infinis Ir\'ene Joliot Curie (UMR 9012) CNRS/IN2P3, 15 rue Georges Clemenceau, 91400 Orsay, France}

\address[nu]{Department of Physics, Nagoya University, Nagoya 464–8602, Japan}

\cortext[contact]{Corresponding authors: tord.ekelof@physics.uu.se, marcos.dracos@in2p3.fr}
\cortext[coeditor]{Co--editor}

\fntext[ibs]{Now at The center for Exotic Nuclear Studies, Institute for Basic Science, 34126 Daejeon, Korea}

}

\authorlist

\begin{keyword}
ESS, ESSnuSB, Super Beam, neutrino, oscillations, long baseline
\end{keyword}

\end{titlepage}

\end{frontmatter}

\clearpage

%\begin{strip}
\tableofcontents
%\end{strip}

\pagebreak

%test biblio \cite{Nunokawa:2007qh} and \cite{GerigkMontesinos} 
\setcounter{footnote}{0}

\setcounter{figure}{0}
\numberwithin{figure}{section}
\setcounter{equation}{0}
\numberwithin{equation}{section}
\setcounter{table}{0}
\numberwithin{table}{section}

\section{Executive Summary} 
\label{executivesummary}

An EU-supported Design Study has been carried out during the years 2018-2022 of how the \SI{5}{\mega\watt} linear accelerator (linac) of the European Spallation Source under construction in Lund, Sweden, can be used to produce the world’s most intense long-baseline neutrino beam for CP violation discovery in the leptonic sector and, in particular, precision measurement of the CP violating phase $\delta_{CP}$. The project is called the European Spallation Source neutrino Super Beam (ESS$\nu$SB). This Conceptual Design Report describes the design of:

\begin{itemize}
  \item	the required upgrade of the ESS linac,
  \item	the accumulator ring used to compress the linac pulses from \SI{2.86}{\milli\second} to \SI{1.2}{\micro\second},  
  \item the target station where the \SI{5}{\mega\watt} proton beam is used to produce the intense neutrino beam,  
  \item	the near detector which is used to monitor the neutrino beam as well as measure neutrino cross-sections, and  
  \item	the large underground far detector where the magnitude of the oscillation appearance of $\nu_{e}$ from $\nu_{\mu}$ is measured.
  \end{itemize}
  
The report also provides the results of the physics performance study of the neutrino research facility that is proposed. 

About a decade ago, results were published of the measurement of the neutrino mass-state mixing angle $\theta_{13}$ which was found to be $8.6^{\circ}$, a value that was much higher than that which had until then been presumed. This result made the discovery of leptonic CP violation significantly more viable in practice than previously foreseen, as well as shifting the optimal place for such a measurement from the first neutrino oscillation maximum to the second oscillation maximum where, given the relatively high $\theta_{13}$ value, the CP violation signal is close to 3 times larger than at the first. As the second maximum is located further away from the neutrino source, a higher intensity is required when measuring at the second maximum as compared to the first maximum, in order to obtain the same event statistics. 

At about the same time as the results of the measurement of the high $\theta_{13}$ value became known, the decision was taken to build the European Spallation Source ESS at Lund in southern Sweden. With its \SI{5}{\mega\watt} proton linear accelerator, ESS will produce the world’s highest flux of slow neutrons for materials science studies, generated by the spallation process. The neutrons will be produced in \SI{2.86}{\milli\second} pulses at \SI{14}{\hertz}. The ESS$\nu$SB conceptual design study has demonstrated that, given the high power and inherent upgrade capacity of the ESS linear accelerator, the linac can be used to produce, in addition to the high intensity spallation-neutron flux, and concurrently with it, a neutrino beam sufficiently intense to provide a statistically significant number of events at the second oscillation maximum. Thanks to these factors the evaluated performance of the proposed ESS$\nu$SB research facility is considerably higher than that of the other proposed neutrino super beam facilities, which are constrained to make measurements at the first oscillation maximum.

In order to reach large enough event statistics, an additional requirement is that the far water Cherenkov detector should have an appropriately large mass. For ESS$\nu$SB the designed mass is 540 kton. In order to ensure a negligibly small background of neutrinos from cosmic rays, this detector will be installed in a mine \SI{1000}{\meter} underground and the data-collection time-gate for each of the 14 neutrino pulses per second generated by ESS will only be \SI{1.2}{\micro\second}. The use of this short time-gate will require that the pulses from the ESS linac be compressed from \SI{2.86}{\milli\second} to \SI{1.2}{\micro\second}. In order to accomplish such a compression, each \SI{2.86}{\milli\second} linac pulse will be fed into a \SI{380}{\meter} circumference accumulator ring, located underground on the ESS site, and be extracted in just one turn. This pulse compression is also necessary in order to be able to operate the target magnetic horns with sufficiently short current pulses. The current required for an efficient focussing of the pions emitted from the target in the forward direction into the decay tunnel is \SI{350}{\kilo\ampere}. In order to keep the heating of the horns to a manageable level, the current must be delivered in very short pulses with a flat top that still need to be at least as long as the proton pulses from the accumulator, i.e. \SI{1.2}{\micro\second}.

The design of the ESS$\nu$SB research facility is largely conditioned by the above considerations. The doubling of the power of the linac from \SI{5}{\mega\watt} to\SI{10}{\mega\watt} is facilitated by the fact that adequate space has already been foreseen in the ESS modulators feeding the accelerating cavities, that can be used to install twice the amount of charging capacitors. In order to be able to feed the very high charge of 2.5$\times10^{14}$ protons per pulse into the accumulator, H$^{-}$ ions will be accelerated in the linac and stripped of their two electrons at the injection point into the accumulator. For this, an H$^{-}$ source has to be furnished at the side of the ESS linac proton source. There are several mechanisms that will result in a fraction of the H$^{-}$ ions being stripped of their electrons and therefore lost in the accelerator or the transfer line to the accumulator, causing activation of the accelerator and transfer line components. In order for this activation to be kept at an acceptable level, the beam losses must not exceed \SI{1}{\watt\per\meter}. The simulations of the operation of the linac and the transfer line with the selected design have demonstrated that all of these requirements can be fulfilled and that the required \SI{5}{\mega\watt} H$^{-}$ beam for neutrino production can therefore be produced, concurrently with the \SI{5}{\mega\watt} proton beam for spallation neutron generation.

The design of the underground accumulator ring and its beam optics must satisfy a number of different conditions. Among these are that the emittance must not exceed $60\pi$\ mm\ mrad, that the temperature of the carbon electron-stripping foils must not exceed \SI{2000}{\kelvin}, and that the beam losses in the accumulator must be kept sufficiently low by sequential collimation to keep the irradiation of the beam line components below \SI{1}{\watt\per\meter}. Furthermore, the edges of the \SI{100}{\nano\second} gap in the circulating beam, needed for the extraction of the circulating beam and generated by chopping in the low energy part of the accelerator, must be kept sufficiently sharp using radiofrequency cavities in order to limit the irradiation of the extraction-region components in the ESS linac. The rise time of the current in the extraction magnets must not be longer than \SI{100}{\nano\second} in order for the beam extraction to fit within the time gap made available, which is achieved by having the inductance of these magnets sufficiently low. The design of the accumulator ring has been optimised through many iterations and the results of the simulation of the final design show that it will be feasible to deliver the required short and intense proton pulses to the target station.

The underground target station design is conditioned by the requirement that the target must stand the formidable shocks, heat dissipation and radiation damage of a \SI{5}{\mega\watt} proton beam delivered in \SI{1.2}{\micro\second} short pulses 14 times a second. In order to manage these severe conditions, the design adopted foresees the chopping of the \SI{2.86}{\milli\second} pulse into four sub-pulses in the low energy part of the linac, the compression in sequence of these four sub-pulses in the accumulator ring, and their extraction into a magnet switchyard, which will direct each of the four sub-pulses to one of four separate targets, each of which will thus receive a \SI{1.25}{\mega\watt} beam. Each of the four targets is designed as a \SI{78}{\cm} long tube, \SI{3}{\cm} in diameter and filled with \SI{3}{\milli\meter} titanium balls that are cooled by a high-pressure transverse helium-gas flow. The extreme heating, high irradiation and radiation damage of these targets, the surrounding focusing magnetic horns, the walls of the \SI{50}{\meter} long pion decay tunnel and the water-cooled beam stop at the end of the tunnel have been simulated and found to be within tolerable limits.

The neutrino beam resulting from the decay of the pions produced in the four targets will be directed towards the 1 kton underground near detector located on the ESS site \SI{250}{\meter} from the target station and the 540 kton underground far detector, located at a distance of \SI{360}{\kilo\meter} from ESS. The two detectors are based on the water Cherenkov technique. The near detector is in addition equipped with a tracking detector mounted inside a dipole magnet and consisting of one million \SI{1}{\cubic\cm} scintillator cubes read out with wavelength-shifting fibres, and with an emulsion stack detector immersed in water. The excavation of the two cylindrical far detector caverns, each \SI{78}{\meter} high and \SI{78}{\meter} in diameter, at a depth of \SI{1000}{\meter} in the Zinkgruvan mine, represents a unique geotechnical design-challenge, requiring the rock strength and rock pressure to be measured using core drillings before the final design can be certified. About 92,000 20$''$ single-photon sensitive photomultipliers will be mounted on the walls of these caverns, providing a 30\% photocathode coverage.

The physics performance of the ESS$\nu$SB research facility has been evaluated considering two baselines corresponding to the positions of two active mines, Zinkgruvan at \SI{360}{\kilo\meter} and Garpenberg at \SI{540}{\kilo\meter} from ESS in Lund. The result of the evaluation is that, although the CP violation discovery capability is comparable for the two baselines, the accuracy with which $\delta_{CP}$ can be measured is higher with the far detector being installed at Zinkgruvan. After 10 years of data-taking with the detector located in Zinkgruvan, leptonic CP violation can be detected with more than 5 standard deviation significance over 70\% of the range of values that the CP violation phase angle $\delta_{CP}$ can take, and $\delta_{CP}$ can be measured with a standard error less than $8^{\circ}$ irrespective of the measured value of $\delta_{CP}$. These results demonstrate the uniquely high physics performance of the proposed ESS$\nu$SB research facility.

The geological conditions of the ESS site allow for that all installations needed for the operation of ESS$\nu$SB will be located underground at a depth between 10 and 30 meters, thereby eliminating all risk for radiation hazard above ground level. It has also been possible to located the ESS$\nu$SB installations in such a way that there will be no interference with the existing ESS installations.

This CDR contains a preliminary estimate of the construction cost of ESS$\nu$SB which is on the level of 1.4 B\texteuro. This estimate does not include the cost of the associated civil engineering work, which has not yet been evaluated in detail. The plan is to achieve, after another period of design work, a Technical Design Report, including a more accurate cost estimate, and to seek the necessary support for approval and financing of the ESS$\nu$SB construction project. The plan for the following period of build-up and commissioning of the facility is such that ESS$\nu$SB will be ready for start of data-taking operation around year 2035. The operation of the facility is foreseen to continue for several decades, possibly including intermediate upgrades. 

Throughout, the Design Study leading up to the present CDR has had the strong support of the ESS management. The international nature of the ESS$\nu$SB project is demonstrated by the list of the authors’ institutions of the present CDR, which all have, together with the EU Horizon 2020 Frameworks Programme and the European Cooperation in Science and Technology Programme COST, provided the personnel, technical and financial resources for the four years of Design Study. 

Significant effort has been spent on different outreach activities with the aim of informing the scientific and general public of the goals and achievement of the ESS$\nu$SB Design Study. One part of this has been the production of two video films, each ca \SI{6}{\min} long, one intended for the scientifically literate public and the other for the general public and both of which are available at the ESS$\nu$SB home page https://essnusb.eu/.

\clearpage

\setcounter{figure}{0}
\numberwithin{figure}{section}
\setcounter{equation}{0}
\numberwithin{equation}{section}
\setcounter{table}{0}
\numberwithin{table}{section}

\section{Introduction} 
\label{introduction}
%{\bfseries TODO: Tord}

This Conceptual Design Report (CDR) gives an account of the results of a 4 years’ design study of a European neutrino Super Beam facility ESS$\nu$SB to be based on the use of the ESS linear accelerator (linac), currently under construction in Lund, Sweden. When this linac will have reached its full design performance it will be the world’s most powerful accelerator, delivering a proton beam of \SI{5}{\mega\watt} average power. The original purpose of creating such a powerful proton beam is to provide the world’s most brilliant neutron spallation source. The linac will deliver the \SI{5}{\mega\watt} by accelerating ca 10$^{15}$ protons in each of 14 pulses/second of length \SI{2.86}{\milli\second}, to \SI{2}{\giga\electronvolt} energy, implying a duty cycle of just 4\%. By accelerating an additional 14 pulses, interleaved with these 14 pulses, the duty cycle can be raised to 8\% and the extra \SI{5}{\mega\watt} used to produce a neutrino Super Beam of world-unique intensity. 

The motivation for this experiment stems from the discovery and publication in 2012 of the measurement of the neutrino mixing angle $\theta_{13}$ which was found to be about $8.6^\circ$, a value much higher than that which had earlier been presumed. Were $\theta_{13}$ to have been small, leptonic CP violation could only have been discovered from neutrino-oscillation measurements using non-conventional neutrino beams of very high flux such as one provided by a Neutrino Factory operated at the first neutrino oscillation maximum. The relatively high value of $\theta_{13}$ thus revealed made CP violation discovery significantly more viable using conventional neutrino beams. On top of that, this allows the far neutrino detector to be located at the second oscillation maximum, where the CP violation signal is almost 3 times larger than at the first maximum, implying less sensitivity to systematic errors.

On the other hand, for the same neutrino energy the second oscillation maximum is located further away from the neutrino source, and thus a higher intensity is required when measuring at the second oscillation maximum as compared to the first, in order to provide the same event statistics. The decision to build the high power ESS linac - rendering a high-intensity neutrino Super Beam also possible - was taken at about the same time as the high $\theta_{13}$ value was made public. These two facts combined to change, at that point in time, the optimal strategy for CP violation searches to have the neutrino detector placed, not at the first oscillation maximum, but at the second oscillation maximum.

ESS$\nu$SB, being the only neutrino long-baseline experiment in the world that is based on this new strategy, will therefore have a sensitivity significantly higher to CP violation as compared to other long baseline neutrino experiments, the baselines of which were fixed before the discovery of the high $\theta_{13}$ value. 

To discover leptonic CP violation and, in particular, to measure the value of the leptonic CP violation phase $\delta_{CP}$, is of paramount scientific interest. The observed value of $\delta_{CP}$ will provide clues to the Big Bang mechanism, as to why there is matter in the Universe and not only radiation, on what dark matter is made of and on why leptons and quarks appear in a variety of flavours. The key to making conclusive measurements of $\delta_{CP}$ is \textit{high precision}, which, provided that a sufficiently high event statistics can be reached, translates into achieving a sufficiently low ratio of the size of the irreducible systematic errors to the size of the signal. There are definite limits to how much the current systematic errors, of which irreducible neutrino-nucleus-interactions modelling-uncertainties represent an important part, can be reduced. However, these limits can thus effectively be lowered by using the highest available accelerator power, thereby allowing the matter-antimatter asymmetry to be accurately measured at the second neutrino oscillation maximum rather than at the first maximum. 

\begin{figure}[H]
  \centering
\includegraphics[width=0.8\textwidth,trim={0 0.2cm 0 0},clip]{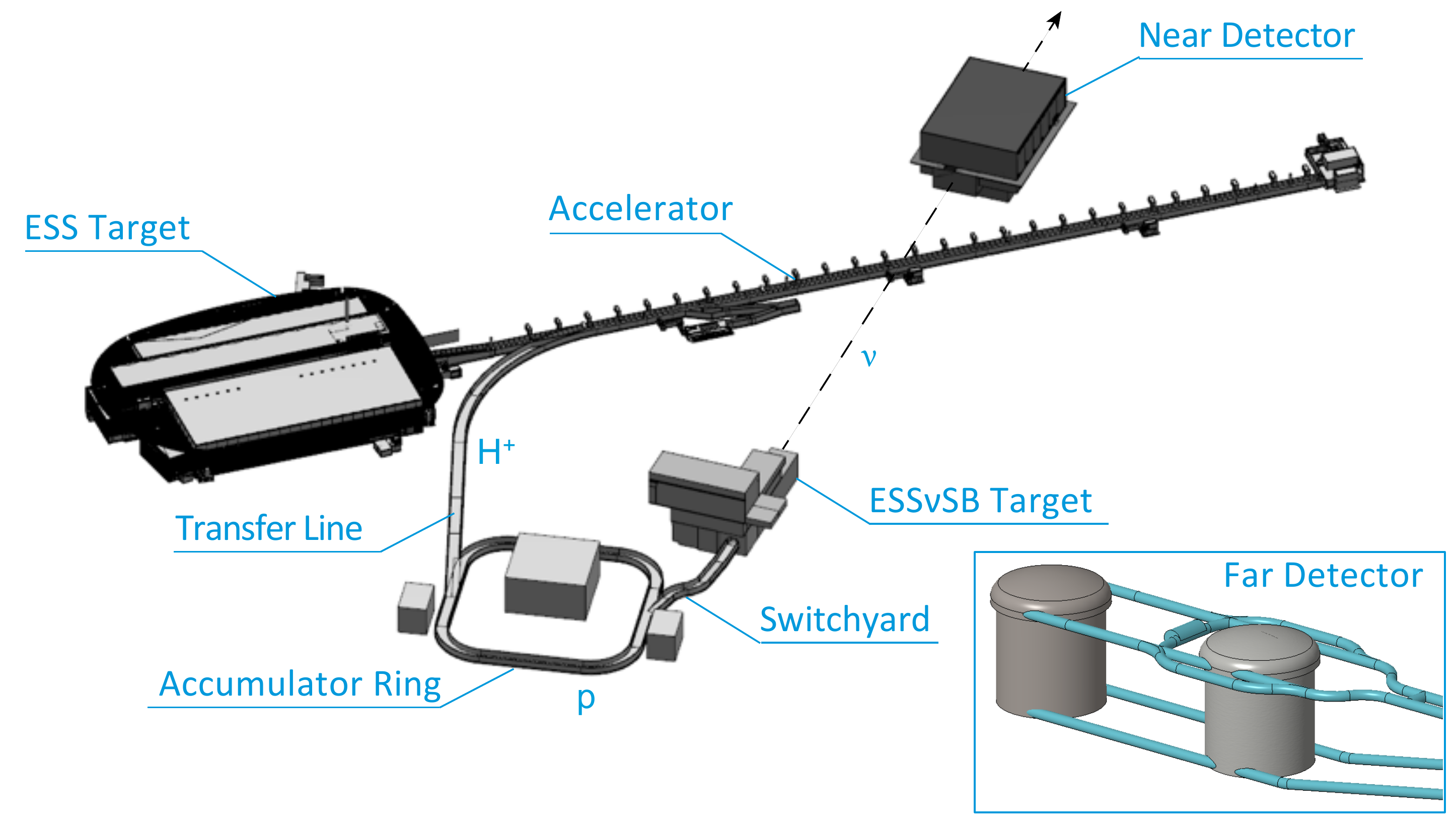}
  \caption{The layout of the ESS$\nu$SB components on the ESS and the far detector sites.}
    \label{fig:layoutint}
\end{figure}

The results of the work made on the design of the main components of the ESS$\nu$SB facility, which are the ESS linac upgraded to \SI{10}{\mega\watt}, the pulse accumulator ring, the target station and the near and far neutrino detectors, will be described in the following sections of this report. The general layout of the proposed facility is shown in Fig.~\ref{fig:layoutint}. A cost estimation of each major component of the proposed facility is provided in 2022 prices. The results of the evaluation of the physics performance for leptonic CP violation discovery and, in particular, the precision with which it will be possible to measure the CP violation phase $\delta_{CP}$ are also included in this report.

\clearpage

\setcounter{figure}{0}
\numberwithin{figure}{section}
\setcounter{equation}{0}
\numberwithin{equation}{section}
\setcounter{table}{0}
\numberwithin{table}{section}

\section{Proton Driver}
\label{protondriver}

\subsection{Introduction}
The ESS$\nu$SB linac upgrade consists of the modifications of the ESS proton linac needed in order to be able to produce and accelerate interleaved pulses of H$^-$ for the generation of the proposed neutrino super beam, while the acceleration of protons for the production of spallation neutrons continues uninterrupted. 

\begin{figure*}[ht!]
\begin{center}
\includegraphics[width=0.85\textwidth]{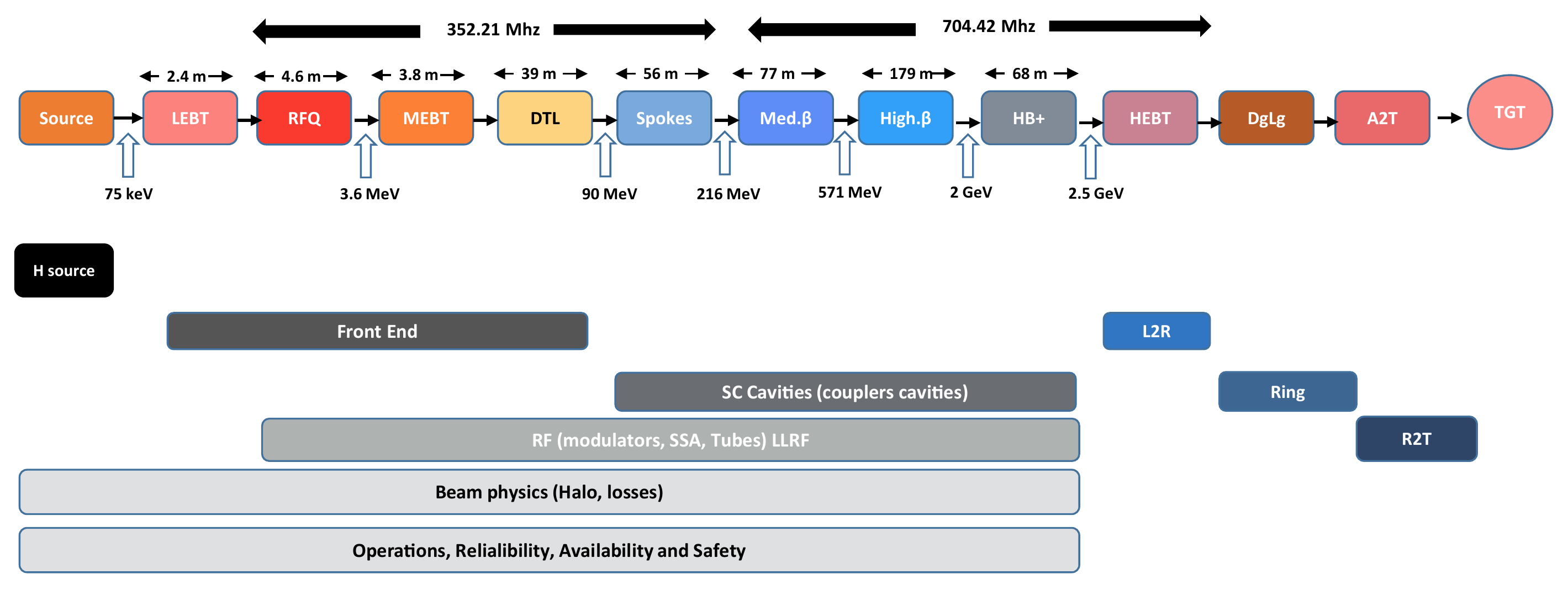}
\caption{\small Proton driver layout.}
\label{fig:protondriver}
\end{center}
\end{figure*}

This section presents the upgrade requirements and cost estimates for the linac, its auxiliary systems, as well as design considerations for the linac-to-ring (L2R) transfer line. The schematic in Fig.~\ref{fig:protondriver} details each of the portions of the linac which must be reviewed for the present upgrade study. Detailed cost estimates of subsystems such as cryogenics and electrical power will be presented in the later subsections, beginning with a review of the RF subsystems' modifications in Section~\ref{sect:RF_systems}.

The low energy beam transport (LEBT) and medium energy beam transport (MEBT) lines have been modelled and simulated in terms of the necessary modifications for transporting both proton and H$^-$ bunches.  

\begin{figure*}[ht!]
\centering
\includegraphics[width=0.75\textwidth]{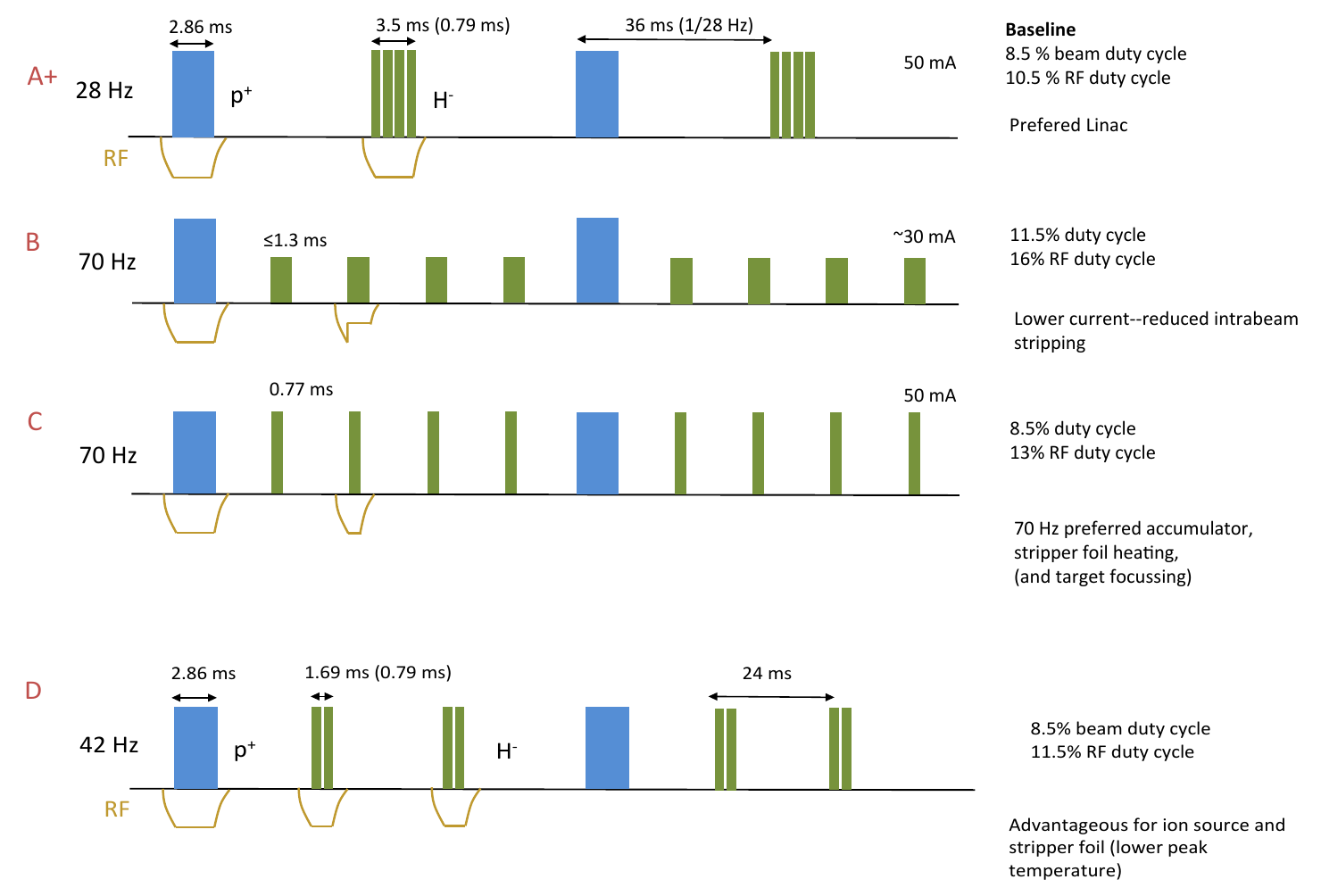}
\caption{Pulse structure alternatives. Option A+ is the baseline design, with minimal powering requirements for the superconducting RF cavities.}
\label{fig:pulse_struct_options}
\end{figure*}

\begin{table}[ht!]%[H]
  \begin{center}
    \caption{Nominal ESS design parameters versus ESS$\nu$SB upgrade$^*$}
    %\label{tab:linac_comparison}
    \label{table:nomESSvsupgrade}
    %\vspace{-0.35 cm}
    \begin{tabular}{lccl}
    &\textbf{Nominal Linac} & \textbf{ESS$\nu$SB Linac} &  \textbf{Units} \\
\hline
Species & p & p and H$^-$\\
Energy & 2.0 & 2.5 &GeV\\
Current & 62.5 & 62.5  50 (p) / (H$^-$)  & mA \\
Average beam power & 5 & 10 & MW \\
Linac length & 352.2 & ${\sim}$423 & m
\\
Macro pulse length & 2.86 & $>$ 2.86 (p) / 2.9 (H$^-$) & ms
\\
Protons per pulse & $10^{15}$ & $8.3\cdot 10^{14}$(p) / $8.9\cdot 10^{14}$(H$^-$) & 
\\
Sub-pulse length & N/A & ${\sim}$0.65 & ms
\\
Repetition rate & 14 & 28$^\dagger$ & Hz \\
Beam duty cycle & 4 & 8 & \% \\

Total losses & $<$~1 & $<$~1 & W/m \\
\hline
     \end{tabular}
\\~\\
\footnotesize{$^*$The baseline pulsing scheme is assumed here, see Fig.~\ref{fig:pulse_struct_options}, Option A+.} \\
$~~~~~~~~~~~~$\footnotesize{$^\dagger$14\,Hz proton $+$ 14\,Hz H$^-$, with the H$^-$ pulses consisting of four subpulses.}
   \end{center}
\end{table}

A comprehensive study of H$^-$ beam stripping was performed, with simulations of the beam transport and evaluation of the loss magnitude for the baseline beam parameters; this included end-to-end simulations of the upgraded linac and trajectory analysis of ionisation energy deposited from stripped particles (see Section~\ref{sect:beam_losses}). The upgrade requirements for the superconducting (SC) linac sectors and the linac-to-ring (L2R) transfer line have also been studied both in terms of beam dynamics and stripping.

A detailed study of the modifications needed on the modulators was also performed, showing the possible upgrade paths for the modulators, with an evaluation of the energy efficiency of each option, their cost and their added footprint on the klystron gallery. This analysis is summarised in Section~\ref{sect:RF_systems}.

\subsection{\label{subsect:pulse_struct}Pulse Structure}
Different pulsing schemes of the H$^-$ beam have been considered, see Fig.~\ref{fig:pulse_struct_options}. To accommodate the rise and fall times of the extraction magnet in the accumulator ring, and also owing to space charge and beam instabilities, the full 2.86\,ms beam pulses cannot be injected in one filling of the ring. Each beam pulse must instead be split into several sub-pulses or batches. The batch length is limited by the storage time in the accumulator ring -- about 1000 turns, corresponding to 1.3 ms -- before instabilities are likely to develop \cite{zou_challenges_2019}. 

        \begin{figure*}[ht!]
        \centering
        \includegraphics[width=0.75\paperwidth]{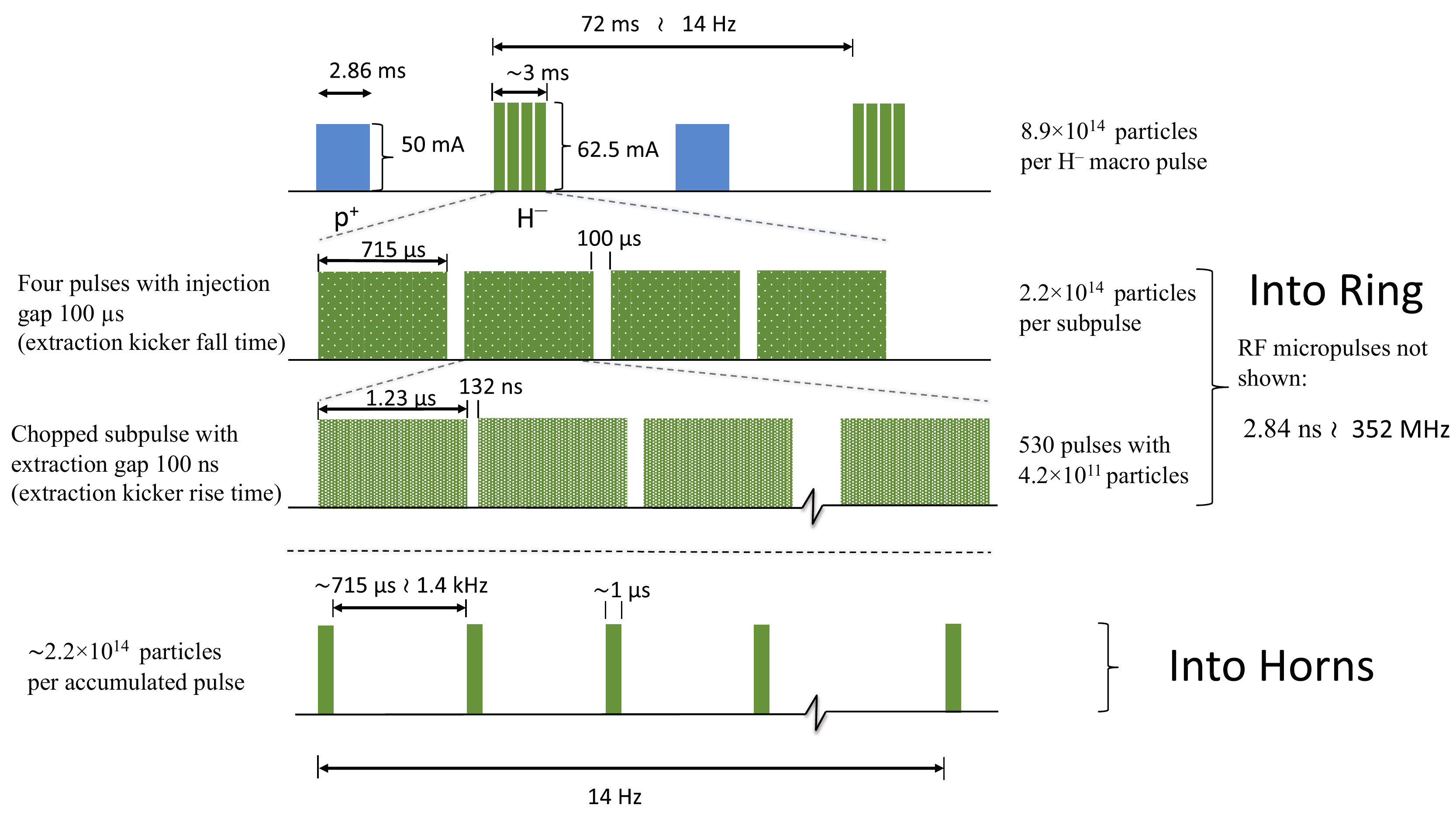}
        \caption{Detailed schematic of the nominal ESS$\nu$SB pulsing scheme, Option A+ from Fig.~\ref{fig:pulse_struct_options}.}
        \label{fig:pulse_struct_detailed_A}
    \end{figure*}
    
A pulsing scheme with an overall 28\,Hz macro-pulse structure has been selected as the baseline design, corresponding to Option A+ in Fig.~\ref{fig:pulse_struct_options}. The pulse length in this case is limited by the flat-top length of the present RF modulators (3.35\,ms). Since the filling time of the superconducting cavities is about 0.3\,ms, and the time to stabilise the radio frequency (RF) regulation is about 0.1\,ms, the effective pulse length is limited to about 2.9\,ms. If the pulse length from the modulators could be extended, this would decrease the demand of the current. Other pulsing schemes have also been reviewed (Options B and C) where the H$^-$ beam is pulsed at 70\,Hz, (4 out of 5 pulses are H$^-$). Compared with Option A+, these alternatives relax the performance demands on the accumulator ring (in particular, the stripping foil) and the target focusing system, but come with substantial costs due primarily to RF system upgrades and annual electrical running costs. 

The total number of particles delivered to the accumulator ring will be $8.9\times 10^{14}$ per pulse cycle (macro-pulse), divided into four batches of $2.3\times 10^{14}$, as shown in Fig.~\ref{fig:pulse_struct_detailed_A}. Further beam parameters are summarised in Table~\ref{table:nomESSvsupgrade}. Each batch is stacked in the accumulator ring, compressing the pulses to 1.2\,\SI{}{\micro\second}, which are subsequently extracted to the target. By splitting the macro pulse into four batches, the power on each target is limited to 1.25 MW, and the space charge tune shift \cite{wolski_2014_ch12} in the accumulator ring is limited to an acceptable level \cite{zou_challenges_2019,zou_2019_accum_design_status}.

As mentioned above, the disadvantage to a 70\,Hz pulsing is a higher total load on the RF system (along with uncertainty on the duty-cycle rating for the nominal ESS superconducting cavity couplers, see Section~\ref{sect:RF_systems}); this is due to the filling time of the superconducting (SC) cavities of 0.3\,ms. Option B, with a pulse length of up to 1.3\,ms, has the advantage of allowing for a lower beam current of about 30\,mA, relaxing the demands on the ion source requirements and reducing beam stripping losses, particularly intra-beam stripping (IBSt). The impact on the RF modulator systems of the different pulsing options are also discussed in Section~\ref{sect:RF_systems}, with a detailed analysis available in~\cite{galnander:2019_D2.2}. 

\subsection{Front End Baseline and Overview \label{sect:front_end_overview}}
\begin{figure*}[ht!]
    \centering
    \includegraphics[width=0.85\textwidth]{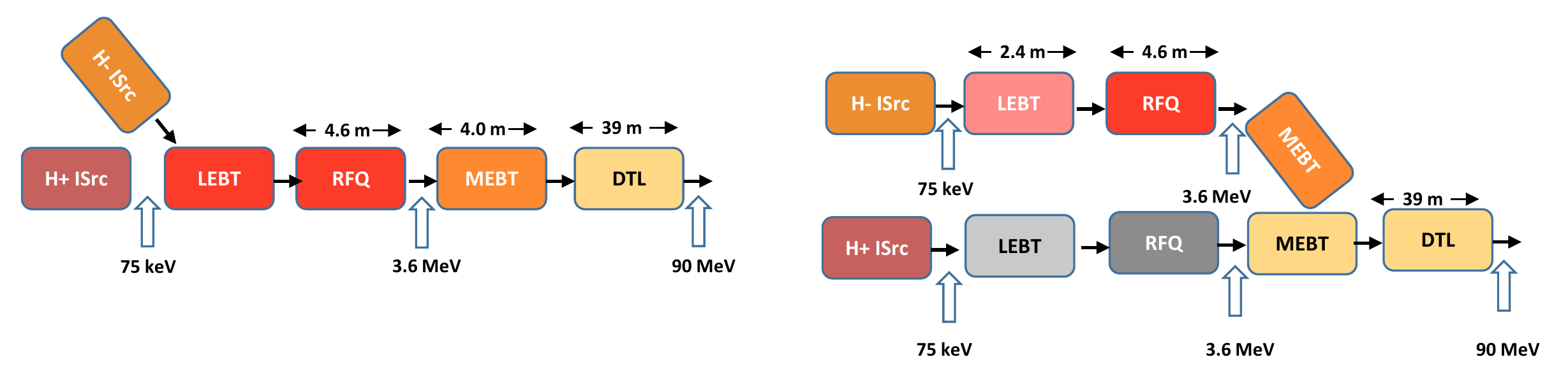}
    \caption{Two front-end layouts for a merged proton/H$^-$ beamline. The left-hand layout uses a common LEBT and RFQ, while the right-hand layout merges the two species in the MEBT.}
    \label{fig:frontend_options}
\end{figure*}

The upgrade of the ESS linac for ESS$\nu$SB requires an additional ion source for delivering H$^-$ ions at the correct energy, and a low energy beam transport (LEBT) to merge this beam and the nominal proton beam into the radiofrequency quadrupole (RFQ) for the first stage of acceleration. Alternatively, the two beams could be merged in the medium energy beam transport (MEBT), see Fig.~\ref{fig:frontend_options}. This option with separate RFQ units and merging MEBT sections could more straightforward to realise in terms of beam dynamics, since the transport of the two beams is independent.

\begin{figure*}[ht!]
    \centering
    \includegraphics[width=0.85\textwidth]{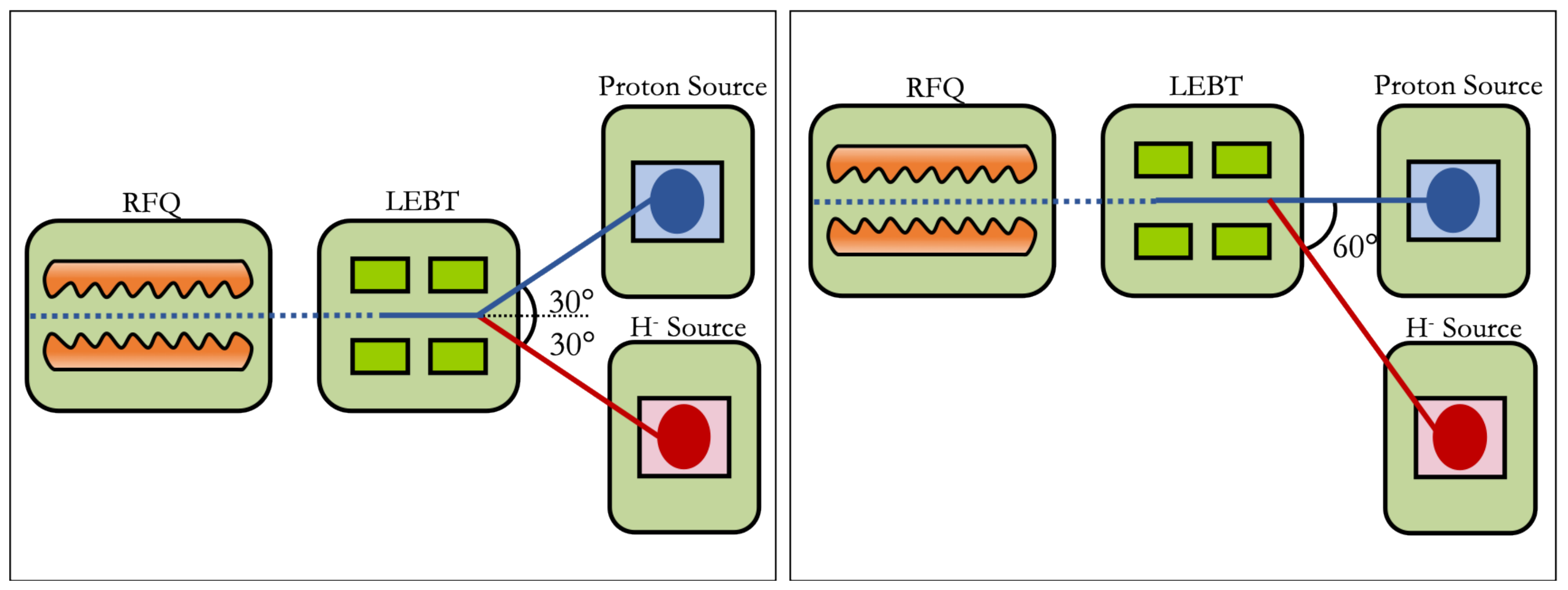}
    \caption{Two layouts for merging the proton/H$^-$ sources. The left-hand layout requires moving the current proton source and bending both beams at 30$^\circ$, the right-hand side layout leaves the proton source unchanged and bends the H$-$ beam by 60$^\circ$.}
    \label{fig:source_options}
\end{figure*}

However, the option with shared RFQ and MEBT would be less expensive and require less downtime for nominal ESS operation. Extensive beam dynamics simulations with this option have shown no prohibitive drawbacks, making it the favoured baseline design. A few key considerations for this design are as follows:

\begin{itemize}
    \item The MEBT chopper will need to be redesigned to deflect along both transverse axes due to the opposite charges of the accelerating ion species. Otherwise, the entire linac design is charge-agnostic in terms of transverse dynamics, despite having vertical and horizontal envelope profiles reversed for protons and H$^-$. The chopper will also need to be redesigned to accommodate the fast beam-chopping requirements for creating extraction gaps in the beam.
    \item The use of switching magnets (changing sign / magnitude between proton and H$^-$ pulses) is optional, though it may be beneficial in the MEBT for better matching with the drift-tube linac (DTL). Without switching magnets, compromising the matching of both ion species at the MEBT/DTL interface incurs a small overall emittance growth downstream.
    \item The RFQ is designed for compatibility with the nominal ESS proton source, and H$^-$ transmission is poorer (especially for the case of injection at 60$^\circ$, see Fig.~\ref{fig:source_options}, right-hand side). This means requiring a substantially greater current from the H$^-$ source. However, as the presently installed RFQ is designed for a maximum 5\% duty cycle and the upgraded RF duty cycle would be 10--15\%, a redesign of the RFQ would be necessary regardless, and may be more forgiving in terms of H$^-$ losses. The maximum 5\% duty cycle of the present RFQ is imposed by the cooling needed for resistive losses. Thus, the main design issue is allowing for a greater heat load.
\end{itemize}

For this baseline design with beams merged in the LEBT, two layouts have been studied in terms of source position: a symmetric layout (the ${\pm}$30$^\circ$) and an asymmetric one ($+$60$^\circ$ layout) as shown in Fig.~\ref{fig:source_options}. In other words, merging the sources requires moving the current proton source and bending both beams at 30$^\circ$, or leaving the proton source and bending the H$-$ beam by 60$^\circ$. This second layout is less costly in terms of equipment and may require less downtime for nominal ESS operation, but increases dispersion dramatically (thus reducing RFQ transmission). Space limitations in the tunnel may also be unavoidable and prohibitive for the second layout, especially in terms of structural requirements, personnel access issues, or wiring and grounding. Specifically, the right-hand solution's H$^-$ source position makes it difficult to pass and that the drop hatch area has to be used for logistics and included in each search of the tunnel. Meanwhile, left-hand solution means that a realignment of the beam is also required for the proton beam by adjusting the steering magnets.

Both layouts have been simulated in terms of losses and RFQ--DTL transmission. At present, the symmetric layout is taken as the baseline and has been used for end-to-end simulations.

As mentioned above, the total RF duty cycle with both proton and H$^-$ beams will be between 10\% and 16\%. The maximum 5\% duty cycle of the present RFQ is imposed by the cooling needed for resistive losses. Thus, for the nominal case of merging the beams into a single redesigned RFQ, the main design issue is allowing for a greater heat load.

For a more in-depth discussion of the options for the LEBT and MEBT design, see Sections~\ref{sect:lebt_design} and \ref{sect:mebt_design}.

\subsection{Ion Source Design \label{subsect:ion_src}}
The requirements of the H$^-$ source are closely related to the pulsing structure of the beam shown in Fig.~\ref{fig:pulse_struct_options}. For the baseline solution (Option A+), which pulses the H$^-$ source at 14\,Hz, a long pulse of 3\,ms is needed. For the alternative schemes with 70\,Hz, a pulse length of around 1\,ms is needed.

There will inevitably be losses along the linac and the cumulative radiative activation from power loss along the linac for the proton and H$^-$ beams must not exceed 1\,W/m for safe, hands-on maintenance (or 0.1\,W/dm, in terms of machine protection limits~\cite{Mohkov:2000ue}). In the high-energy sections of the linac, stripping losses due to intrabeam collisions, which has been identified as a major loss mechanism at the Spallation Neutron Source (SNS) are expected to dominate, along with stripping due to Lorentz-shifted electromagnetic forces~\cite{shishlo_2012,folsom:2021strp}. Although though these losses are of great concern in terms of activation, they only involve a small fraction of the acceleratedbeam. 

However, losses in the low-energy sectors can involve significant fractions of the beam being lost; with, for example, the residual gas stripping, intrabeam stripping, and other losses in the LEBT and RFQ estimated to comprise about 25\% of the initial bunch population exiting the source. 

A large fraction of beam losses are expected to occur in the LEBT and the transmission through the RFQ, where the beam is bunched. Simulations show that roughly 10--15\% losses are expected in transport through the LEBT and RFQ, depending on ion source emittance. For the H$^-$ beam, there are also losses expected due to stripping, since the extra electron has a binding energy of only 0.75~eV. The stripping loss mechanisms have been summarised in \cite{folsom:2021strp}, and will be discussed in detail in Section~\ref{sect:beam_losses}.

Assuming 25\% losses in total, for a 60\,mA beam to reach the accumulator ring, a 80\,mA beam current is needed from the ion source (required for Option A+ in Fig.~\ref{fig:pulse_struct_options}). The limiting factor for the current is the pulse length of the flat-top of the RF modulators being 3.35 ms in the present design. With a superconducting cavity filling time of about 0.3 ms and the time to stabilise the regulation at 0.1 ms, the effective supported pulse duration is about 2.9 ms. For Options B and C, the beam current requirements from the ion source are more relaxed since the overall beam current requirements are lower.

The beam from the ion source has to match the acceptance of the RFQ, and therefore the emittance should ideally be less than 0.25\,$\pi$\,mm\,mrad; higher emittance will lead to higher loss and downstream emittance growth. However, since a new RFQ design will needed, and since the 80\,mA required current is at the upper end of conventional H$^-$ source design~\cite{han:2021_sns_srce,katsuhiro:2018_jparcsrce}, it can be expected that this emittance requirement will relax. Simulations completed for this report using the present RFQ design show that for source emittances of up to ${\sim}$0.38\,$\pi$\,mm\,mrad, the required current can be accelerated through the upgraded 2.5\,GeV linac, with emittance growth limited to ${<}\thinspace5\%$ of the ideal case.

%Another aspect to consider here is the extraction voltage, which should be the same for both beams at 75\,kV for the option where  assuming the baseline where a common LEBT and RFQ are used.
The pulse-to-pulse variations in beam current, and the flat-top stability, are also important parameters. These must be managed in order to avoid field and phase variation in the cavities, which would increase losses in the linac. The variations of the beam current are assumed to be acceptable within $\pm 3\%$. 
The requirements for the H$^-$ ion source are summarised in Table~\ref{tab:srce_upg_params}. 
\begin{table}[ht!]
\begin{center}
\caption{Required H$^-$ source parameters for the ESS$\nu$SB upgrade}
\begin{tabular}{lccc}
\label{tab:srce_upg_params}
\textbf{Parameter} & \textbf{Option A+ Baseline} & \textbf{Option B} & \textbf{Option C} \\
\hline Current & $80 \mathrm{\,mA}$ & $30 \mathrm{\,mA}$ & $70 \mathrm{\,mA}$ \\
Repetition frequency & $14 \mathrm{\,Hz}$ & $70 \mathrm{\,Hz}$ & $70 \mathrm{\,Hz}$ \\
Pulse length & $3 \mathrm{\,ms}$ & $\leq 1.3 \mathrm{\,ms}$ & $0.7 \mathrm{\,ms}$ \\
Duty cycle & $4 \%$ & $\sim 6 \%$ & $4 \%$ \\
Extraction voltage & $75~\mathrm{kV}$ & $75~\mathrm{kV}$ & $75 \mathrm{kV}$ \\
Emittance RMS norm & $0.25 \ \pi \ \mathrm{mm} \  \mathrm{mrad}$ & $0.25 \ \pi \ \mathrm{mm} \ \mathrm{mrad}$ & $0.25 \ \pi  \ \mathrm{mm} \ \mathrm{mrad}$\\
\hline
\end{tabular}
\end{center}
\end{table}

\subsubsection{\label{subsubsect:ion_src_baseline} Ion Source Baseline}
A variety of H$^-$ ion sources have been studied to meet the linac requirements and there is expertise at leading facilities with ion sources having similar performance as that needed for ESS$\nu$SB. The strongest limitation here is that the ESS$\nu$SB ion source will require an output current, emittance, and repetition rate roughly matching the limits of technological performance of those currently in operation. 
For selection of the H$^-$ ion source, recent reviews of various ion source types can be found in~\cite{Faircloth_2018,Stockli:2018kze,dudnikov:2012}; further reviews are also available~\cite{peters_2000_isrc_rvw,moehs:2005,welton:2002}. Additionally, a workshop on the subject was held in 1994, which focused on surveying possible ion sources for SNS ~\cite{alonso:1996}.

At present, the favoured source design is that of SNS at Oak Ridge National Lab in Tennessee, USA. It is an RF-antenna multicusp (i.e. multipole magnet) volume and surface source; this source operates via inductive excitation of the plasma using a porcelain-coated copper antenna. Caesium is added for enhancing the surface ionisation rate~\cite{Faircloth_2018,stockli_SNS_injector_2017} and the beam is injected at an angle to improve electron extraction~\cite{stockli:2018sns_src}.

Recent experience at SNS shows that such an ion source can be operated routinely, delivering 50 to 60\,mA H$^-$ beams into the RFQ at a 6\% duty cycle, with availability of ${\sim}$99.5\%~\cite{stockli_SNS_injector_2017}. Additional labs in the US have adopted this design for its performance and ease of installation~\cite{stockli_lansce_source_upg_2020}; SNS is also now testing the performance of an external-antenna type source (not pictured) with improvements in efficiency (output current vs. input power) and a smoother beam pulse profile.

The lifetime remains limited for these ion sources before their caesium supply must be replenished, despite improvements in the last few years; it is now on the order of ten weeks. At Japan’s Particle Accelerator Research Complex (J-PARC) in Tokai, Japan, there has recently been a development of a similar source with ${\sim}$0.25\,$\pi$\,mm\,mrad emittance and ${\sim}$65\,mA beam current~\cite{Ueno_2017}. 

The characterisation of multicusp magnets for plasma confinement may also be a worthy avenue of study, with the influence of pole count~\cite{HOSSEINZADEH2014416} and the use of virtual cusps~\cite{Sharma_2020} strongly affecting plasma characteristics. The use of pulsed, switching multicusp magnets (e.g. from virtual-cusp to non-virtual-cusp modes) may also be worth investigating, with the goal of leveraging the benefits of various modes. Such technology may be more feasible from an engineering standpoint thanks to recent developments with compact switching multipole magnets~\cite{Mitsuda:IPAC2019-THPTS027}. 

In the following section, more details are provided onthis favoured ion-source type, along with other available technologies.

\subsubsection{\label{subsect:ion_src_overview} Available Ion-Source Technologies}
The production of H$^-$ ion beams is more complex than providing proton beams; and an H$^-$ ion, once formed,  can easily be stripped to neutral hydrogen atom, H$^0$, since the binding energy (also termed electron affinity) of the outer electron is only 0.75~eV. This can be compared with, for example, the electron binding energy of neutral hydrogen at 13.6~eV.

There are essentially two types of ion sources in use at accelerator facilities, one is the surface production type -- Penning or magnetron sources -- where the plasma discharge is generated by an applied DC voltage. The other main type is the RF volume source, where the plasma discharge is driven by an applied electromagnetic field with a high frequency~(MHz~range). 

In a volume source, the production of ions takes place by first creating highly excited ro-vibrational hydrogen molecules, H$^2$*, by collision with fast electrons. In a second step, a slow electron (${\sim}$1~eV) is attached to the H$^2$* which dissociates into H$^-$ and H$^0$. In order for this process to be successful, the fast and slow electrons need to be separated. This is done by a magnetic filter field which separates the plasma into two distinct regions: one with fast electrons and one with slow electrons, where the H$^-$ can be produced~\cite{Faircloth_2018,bacal:2005}. 

In high-current H$^-$ sources, caesium (Cs) is commonly used for increasing the production rate of H$^-$ in the surface process, since Cs has the lowest work function of all elements at 2.1~eV (and by surface adsorption, it can also reduce the work function of other metals). Molybdenum is also commonly used in ion sources, owing to its low sputtering rate, but has a relatively high work function of 4.2~eV.

Moreover, a variety of metals can be coated with a sub-monolayer of Cs, to reduce the work function further than possible with solid Cs; this yields a theoretical optimum coverage of around 0.6\,monolayers, having a work function of about 1.5 eV~\cite{Chou_2012}. The low work function is important for an electron to be easily transferred from the cathode surface toward a hydrogen molecule to form an H$^-$ ion. Maintaining an optimal coverage of Cs throughout the discharge is thus very important. 

The most promising ion sources to meet the requirements of the future ESS$\nu$SB ion source have been identified as follows: the Penning ion source (surface) in use at ISIS, Rutherford Appleton Laboratory (RAL), UK; and the RF volume sources in use at SNS and J-PARC. From this point, the text remains focused on these ion sources and the on-going development at RAL and SNS, with a few comments on other source types.

\subsubsubsection{Penning Ion Sources}
The ion source in use at ISIS, RAL, is a surface plasma ion source of Penning type, see Fig.~\ref{fig:penning_source}~\cite{Faircloth_2018}.
The Penning source was first developed by Dudnikov~\cite{dudnikov:2012} and has also been used at Los Alamos National Laboratory (LANL)~\cite{allison:1977_hminsrcs}. The Penning ion source can produce high currents, up to 100\,mA, provide high-emission current densities {{$>$}1\,A$\thinspace$cm$^{-2}$ and have a fairly low energy spread of ${<}1$~eV. In routine operation, the Penning 1X source at ISIS produces 55\,mA in 0.25\,ms pulses at 50\,Hz (1.5\% duty cycle).

Studies pursuing longer pulse lengths are ongoing at RAL. With the Penning 1X source, H$^-$ ion pulses of 6\,mA at 1\,ms and 50\,Hz can be reached with a stable flat-top current. However, for longer pulse-lengths at a pulse frequency of 50\,Hz there is a droop in the beam current~\cite{faircloth:2012_fets}. In order to reach a more stable flat-top current for longer pulses, a research program is underway at the Vessel for Extraction and Source Plasma Analyses (VESPA) test stand~\cite{lawrie:2016_vespa} and at the Front End Test Stand (FETS). One mitigation that is explored is to have a controlled power supply that can counteract the droop by increasing power in the discharge. Additionally, by improving the stability in the beginning of the discharge, more of the pulse can be used in the extracted beam. Today, about 0.3 ms is required for the discharge to stabilise and noise to reduce before beam is extracted.

Another area of study is to scale up the dimensions of the ion source, which is done in a Penning 2X source~\cite{faircloth:2018_penning2x}. The flat-top droop is believed to be related to thermal variations during the discharge, which causes the Cs coverage of the electrode surfaces to deviate from ideal conditions for H$^-$ production ($\sim$0.6 monolayers). By scaling the size of the electrode surfaces, and thereby the plasma volume, the thermal variation of the components is reduced, which will increase the pulse stability and the potential ion current~\cite{smith:1992_penningscaling,faircloth:2015_penning)}. The aim of the Penning~2X source is to deliver a 60\,mA beam at 2\,ms and 50\,Hz (10\% duty cycle). 

One disadvantage with the Penning source is its relatively short lifetime of about four weeks, limited by sputtering of the anode and cathode. The refurbishing process is also tedious, including mounting new parts to high mechanical precision. The sources routinely in use at ISIS are exchanged every two weeks as a preventive maintenance. 

Using Cs is an effective way of enhancing H$^-$ ion production, and also reduces co-extracted electrons. However, one needs to ensure that Cs is not transported further through the LEBT to reach the RFQ, or other accelerating structures, where it may cause unwanted electron emission. In the Penning~1X type source, this is done by a cold trap, operated at roughly $-$5 degrees C, along with a 90-degree bending magnet placed directly after the ion source. In the Penning~2X type source, initial focusing is provided by an Einzel lens, and the Cs is trapped by a carbon gettering system~\cite{lawrie:2017_einzel,sarmento:2018_Cscapt}. The extraction voltage used is 18\,kV, which is lower than the requirements for the ESS$\nu$SB ion source. However, the energy can be increased with a post-acceleration gap, which is used at FETS to accelerate the beam to 65 keV~\cite{faircloth:2012_fets}. 
The performance of the Penning ion source comes close to meeting the requirements of the ESS$\nu$SB ion source (see Table~\ref{tab:ion_sources}). With the development program taking place at RAL, the Penning ion source could potentially reach the long pulse requirements of 3\,ms at 14\,Hz and 80\,mA.
 
 \begin{figure*}[ht!]
    \centering
    \includegraphics[width=0.85\textwidth]{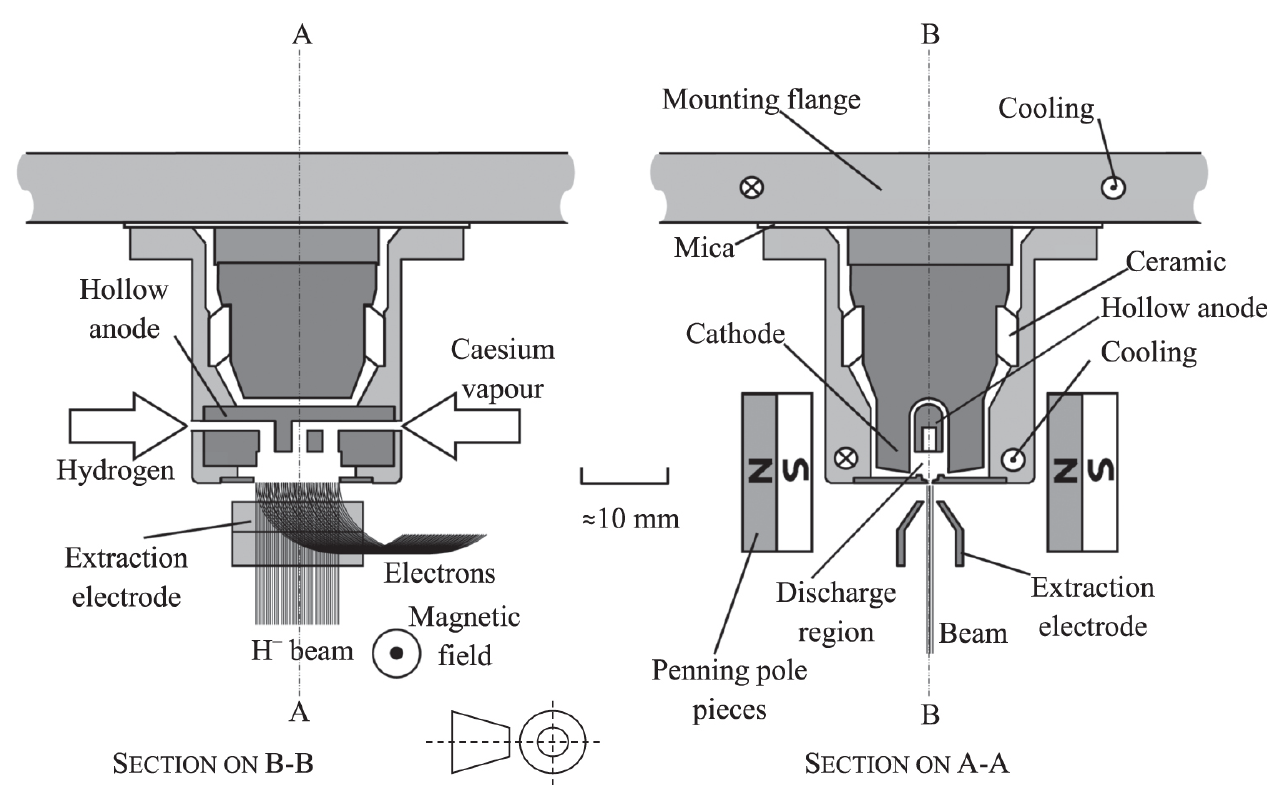}
    \caption{Schematic cross section of a Penning ion source, from \cite{Faircloth_2018}.}
    \label{fig:penning_source}
\end{figure*}

\subsubsubsection{RF Volume and Surface H$^-$ Sources}
The SNS ion source has an internal RF porcelain-coated copper antenna for inductive excitation of the plasma, and permanent cusp magnets surrounding the plasma chamber to create a magnetic field which confines the plasma, see Fig.~\ref{fig:SNS_source_STOCKLI}~\cite{stockli:2018sns_src}. The SNS ion source is based on a design from Lawrence Berkley National Laboratory LBNL~\cite{keller:2002_snssource}, with a filter magnetic field separating the fast- and slow-electron regions. This ion source is of volume type; however, it has an additional Cs collar near the outlet, which enhances the ionization rate. This source therefore combines the phenomena of volume and surface ion production. The plasma is excited using 55--65\,kW of 2\,mHz~RF, the plasma is sustained using a low power 13.56\,mHz~RF at 200\,W. In this way, the plasma is quickly ignited when the high power RF is turned on.

The SNS ion source is operated routinely with 50 to 60\,mA H$^-$ beams directed into the RFQ at a 6\% duty cycle (1\,mA, 50\,Hz). Caesium is added using caesium-chromate cartridges at the outlet collar; the Cs release is adjusted by controlling the temperature. The amount of Cs used is in the order of 10\,mg, for one operation cycle, without the need for continuous Cs injection~\cite{stockli:2018_RFsrc}. An updated design places the RF antenna external to the plasma chamber (not pictured); although this technology is much less mature than the internal-antenna design, early results show improvements in efficiency~\cite{sarmento_sns_ext_int_2020}.

The LEBT used at SNS is a short electrostatic type, of about 12 cm, without any diagnostic capability (on the test stand, sources are fitted to a false RFQ aperture, beam-current toroid and Faraday cup for RFQ input-current estimates). The extraction voltage is 65\,kV and could potentially be increased to 75\,kV to meet the requirements of the ESS$\nu$SB ion source~\cite{stockli:private_comm}. Recently, the development of a magnetic LEBT has been studied, which would allow for diagnostics and prevent problems with electric discharges \cite{han2011physics}. 

\begin{figure*}[ht!]
\centering
\includegraphics[width=0.46\textwidth]{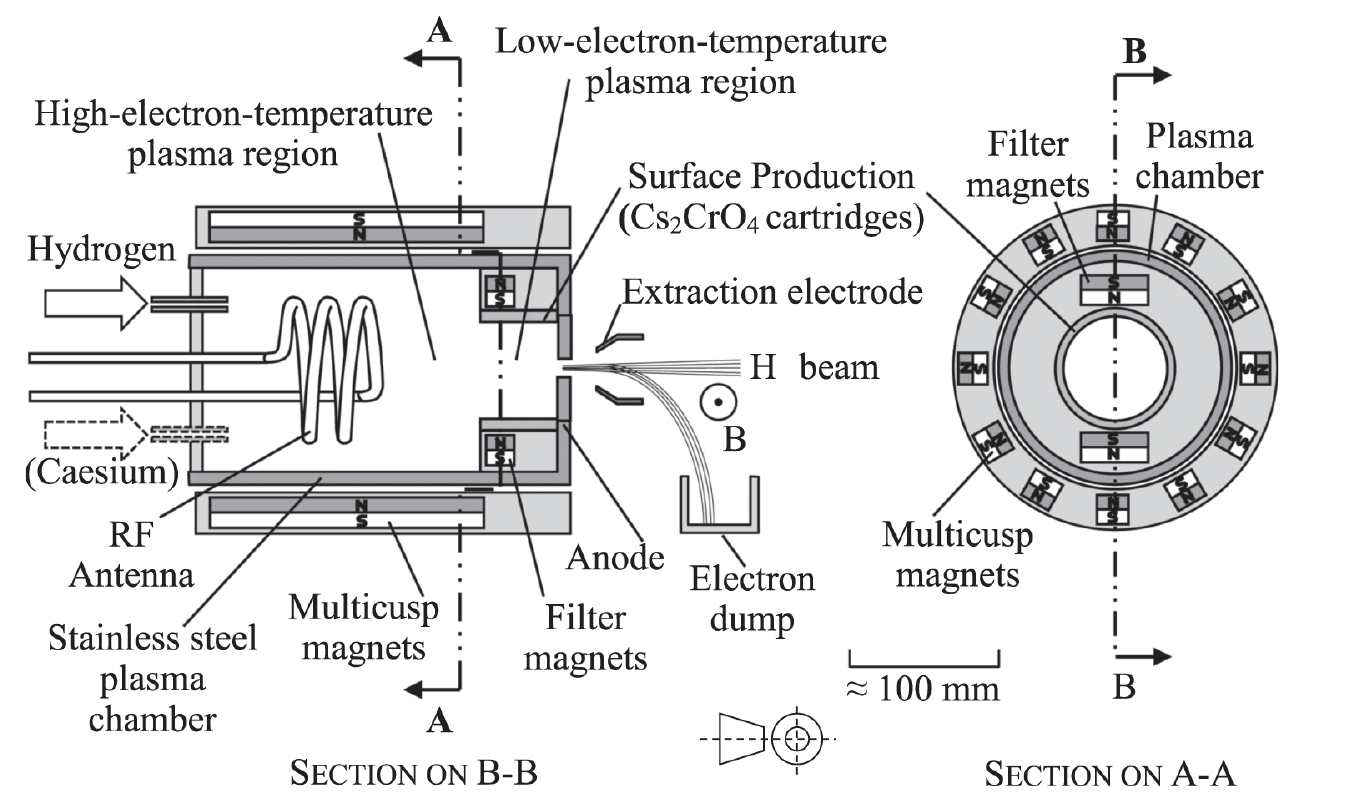}
\includegraphics[width=0.46\textwidth]{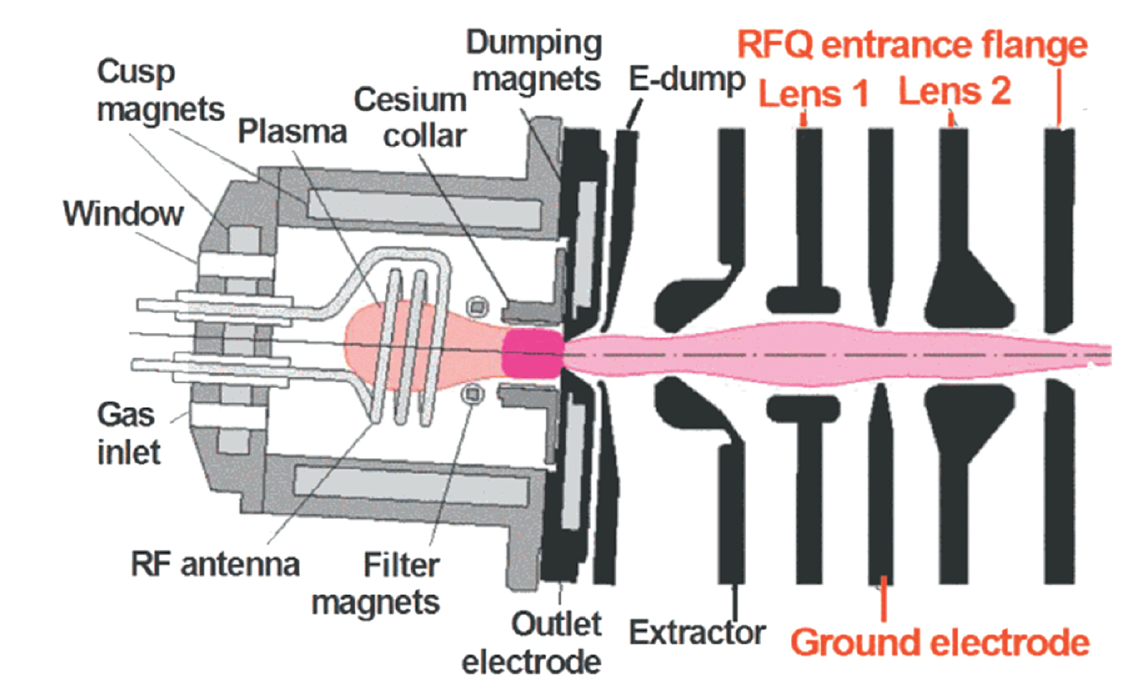}
\caption{General schematic of an internal RF negative ion source (left) and a cross section of the SNS RF ion source (right), both from \cite{Faircloth_2018}.}
\label{fig:SNS_source_STOCKLI}
\end{figure*}

At SNS, there has been success in addressing issues limiting the source lifetime and availability. Problems with the insulation of the porcelain-coated antenna were solved by improving the coating procedure and careful selection of the antennas. More recent improvements include increasing the electron dump efficiency and minimizing electrical discharge problems in the electrostatic LEBT. These developments have led to a high power, high duty-cycle RF H$^-$ ion source with a long lifetimes of up to 14 weeks and availabilities of ${\sim}$99.5\%~\cite{stockli:2018_RFsrc,stockli_SNS_injector_2017}. 

The performance of the SNS ion source is summarised in Table~\ref{tab:ion_sources}. Its regular operation is with 1\,ms pulse length and 60\,Hz. It should also be possible to extend the pulse length to 3\,ms at 14\,Hz to reach ESS$\nu$SB requirements~\cite{stockli:private_comm}. Assuming tests with longer pulses prove successful, the SNS RF ion source will stand as the most feasible alternative for the ESS$\nu$SB ion source. At J-PARC there has recently been a development of a similar source, with reports of 66\,mA ion beam currents~\cite{Ueno_2017}. 

\begin{table}[ht!]
\begin{center}
 \caption{Summary of parameters for the RAL penning source and the SNS RF source~\cite{Faircloth_2018,stockli:private_comm,lawrie:private_comm}}
    \label{tab:ion_sources}

\begin{tabular}{lccc} 
\textbf{Parameter} & \textbf{RAL Penning 1X ISIS} & \textbf{RAL Penning 2X FETS} & $\begin{matrix} \textbf{SNS , Oak Ridge} \\ \textbf
{RF surface enhanced} \\ \textbf{volume source} \end{matrix}$\\
\hline Beam pulse length $(\mathrm{ms})$ & $0.25 \mathrm{\,ms}$ & $2 \mathrm{\,ms}$ & $1 \mathrm{\,ms}$ \\
Repetition frequency & $50 \mathrm{\,Hz}$ & $50 \mathrm{\,Hz}$ & $60 \mathrm{\,Hz}$ \\
Beam current & $55 \mathrm{\,mA}$ & $100 \mathrm{\,mA}$ & $>60 \mathrm{\,mA}$ \\
Duty cycle & $1.25 \%$ & $10 \%$ & $6 \%$ \\
Lifetime & 5 weeks & 2 weeks & 14 weeks \\
Cs consumption & $\sim 0.7 \mathrm{~g} /$ week & $\sim 3.5 \mathrm{~g} /$ week & $\sim 2 \mathrm{mg} /$ week \\
Emittance RMS norm & $0.25 \pi \mathrm{mm}$ mrad & $0.3 \pi \mathrm{mm} \mathrm{mrad}$ & $0.25 \pi \mathrm{mm}$ mrad \\
LEBT & $\begin{matrix}  \text{Sector magnet 90 degree} \\ \text{plus Cs cold trap} \\ \text{Magnetic LEBT} \end{matrix}$    & 
$\begin{matrix}  \text{Einzel lens,} \\ \text{carbon Cs trap} \\ \text{Magnetic LEBT} \end{matrix}$  & Electrostatic LEBT \\
$\begin{matrix}  \text{Emittance RMS norm} \\ \text{after inital beam } \\ \text{transport stage} \end{matrix}$    
 & $0.7 \pi \mathrm{mm}$ mrad & $0.3 \pi \mathrm{mm}$ mrad & N/A \\
Extraction voltage & $18(35) \mathrm{kV}$ & $18(65) \mathrm{kV}$ & $65 \mathrm{kV}$\\
\hline
\end{tabular}
\end{center}    
\end{table}

\subsubsubsection{Other Ion Sources}
A few other available sources are mentioned here, with some commentary on their performance:

The magnetron surface-plasma ion sources used at Fermi National Laboratory (FNAL)~\cite{bollinger:2017_fnal_mgntrn} and Brookhaven National Laboratry (BNL)~\cite{alessi:2002_bnl_mgntrn} are similar to the Penning source, but use a different geometry. Such ion sources can produce high currents, on the order of 100\,mA, with a long lifetime of {$>$}9 months but have a higher noise level, about 10\%, and higher energy spread than Penning sources; and have typically been used at relatively low beam duty cycle of about 0.5\%~\cite{Faircloth_2018}. However, recent improvements to the BNL source and its LEBT transport line have resulted in a 120--130\,mA current at a 7\,Hz repetition rate and 600-1000\,\SI{}{\micro\second} pulse length, with an 80\,mA beam transmitted through the RFQ~\cite{zelenski:2021_BNL_src}.

In addition to SNS, RF sources using external antennas have been used at the Deutsches Elektronen-Synchrotron (DESY), reaching 80\,mA at a duty cycle of 0.8\%~\cite{peters_2000_isrc_rvw} and the ion source for Linac~4 at the European Council for Nuclear Research (CERN) is of a similar design, producing 45\,mA at a low duty cycle of 0.04\%~\cite{Faircloth_2018}.

There is also a development at RAL of a Cs-free RF source. This ion source will have a 30\,mA current at a 5\% duty cycle. This source is designed to meet the demands of the regular user beam at ISIS, which will be sufficient if the transmission is increased by introducing a MEBT into the linac~\cite{tarvainen:2018_ral_src}.

\subsubsubsection{Summary}
Studies have been performed by the authors of this report on the Penning source at RAL and the RF source at SNS closer, since these were identified as being the most promising ion sources to meet the requirements for ESS$\nu$SB. For the pulsing Option A+, with long pulses of 3 ms, the closest to match is the Penning source at RAL. However this source type is quite service-demanding and requires relatively high amount of Cs. The RF source at SNS has a higher lifetime and uses less Cs. For both ion sources it remains to be seen whether they can deliver long pulses which can maintain a stable flat-top current of roughly 80\,mA over 3\,ms at 14\,Hz. 

For the pulsing Options B and C, with 50--80\,mA, {$\sim$}1 ms at 70\,Hz, both types of source are feasible. The RF source seems the most promising from the lifetime point of view, and the requirements are nearly met by the state-of-the-art ion sources at SNS and J-PARC. 

There are ongoing discussions between ESS$\nu$SB and both RAL and SNS, in order to follow their developments and determine if there are any areas of research needing attention in order to meet the upgrade requirements for ESS$\nu$SB (since neither of these sources do so as of today). However, steady improvements in performance of the ion sources lead to the expectation that they will meet the requirements within the schedule of the ESS$\nu$SB project. Moreover, it is worth recalling from the introductory remarks to this section that preliminary simulations with a relaxed emittance limit from the source of 0.38\,$\pi$\,mm\,mrad showed acceptable transmission through the RFQ as well as through the DTL, with total emittance growth at the end of the linac less than 10\% versus the nominal 0.25\,$\pi$\,mm\,mrad source emittance.

\subsection{Low Energy Beam Transport (LEBT) Design\label{sect:lebt_design}}
In studies of the front end, the primary focus has been on finding solutions where a common LEBT, RFQ, and MEBT are used for the proton and H$^-$ beams, as in the left panel in Fig.~\ref{fig:frontend_options}. The alternative, which uses separate front ends, would be more straightforward, but more expensive.

One challenge with the LEBT is that the beam is highly space-charge dominated; this implies the beam tends to ``blow up'' spatially, has increased emittance, and that the beam transport generally should be kept as short as possible. Space-charge compensation is therefore critical; this involves an injected inert gas or mixture of gases~\cite{valerio:2014_linac4lebt}. The process for the space charge compensation is different for protons and H$^-$. For protons, space-charge compensation is largely imparted by electrons, while for H$^-$ it is induced by positively charged ions, which are heavier and alter the underlying dynamics. With an excessive gas pressure, there is also a risk of introducing stripping losses with the H$^-$ ions (see Section~\ref{sect:beam_losses}), although the required gas concentration is likely to be at least an order of magnitude below the point of problematic stripping, it can cause a non-negligible portion of total losses, and should be simulated carefully in the design stages~\cite{valerio:2014_linac4lebt,chou:2005_hmins_strp,folsom:2021_hmin_strp}.

As discussed above, there are two options for how to integrate the ion source with the LEBT, with one option keeping the proton ion source and the beam aligned, and installing the H$^-$ ion source at an angle of 60 degrees, Fig.~\ref{fig:source_options} (left) or with both ion sources displaced by an angle of ${\pm}$30 degrees, Fig.~\ref{fig:source_options} (right). It should be noted that the first option requires a switching dipole to be introduced; the second option can use a fixed-field dipole magnet, since the proton and H$^-$ beams have different charge. 

%\subsubsection{Mechanical Considerations}
Moreover, when merging the beam in the LEBT and using the same RFQ, the energy of the different species should be equal. The platform of the proton ion source is +75\,kV, and so the platform voltage of the H$^-$ source will be -75\,kV relative to ground. The ion sources must therefore lie on two different platforms, with a grounded cage between for shielding. The shortest distance in air at the present proton ion source is about 185\,mm, without any discharge problems, so both options in Fig.~\ref{fig:source_options} seem feasible from a high-voltage perspective. 

There are physical limitations in the linac tunnel and how accelerator equipment can be added in different directions. From a building design perspective, it is easier to add components in the area to the left of the beam line (eastward, facing the direction of the beam). In the front-end building, the wall to the right in the tunnel is solid and several components are installed in the intervening space, which makes it easier to install equipment on the left-hand side. This makes the 60-degree option easier to accommodate. 

\subsubsection{Simulations of Beam Transport}
Simulations of the beam transport through the LEBT and RFQ have been carried out utilising the \textsc{TraceWin} software~\cite{TraceWin}. This effort began with 60-degree layout, since in this case the proton beam line would be more or less unchanged. The bending magnet has a bending radius of 0.4\,m and edge angles of 20.5 degrees, which balances the focusing effects in horizontal and vertical directions. The first solenoid is kept as close as possible to the ion source.

The dipole bend introduces dispersion, and in principle the bend could be made achromatic. However, this would require a longer beam line with greater decoherence from space charge forces; this has therefore been avoided. The dispersion introduced is relatively small, about 0.5\,m. and considering the limited momentum spread of the beam, in the order of 0.1\% determined mostly by the ripple of the power supply for the extraction voltage, this dispersion will not affect the beam significantly. 

Simulations with the 60 degree layout, assuming a beam from the ion source of emittance 0.14\,$\pi$\,mm\,mrad, and assuming a space charge compensation of 95\%, give a transmission of about 93\% from the ion source to the end of the RFQ. However, simulations with emittance from the ion source of 0.2\,$\pi$\,mm\,mrad and assuming 85\% space charge, give a transmission of only 60\% from the ion source to the end of the RFQ. This is because the beam envelopes in this case become large and particles at large amplitudes are affected by non-linear fields in the solenoids and fringe fields of the dipole. This result indicates that the beam transport in a LEBT with a 60 degree bend is sensitive to both space charge and initial beam distribution from the ion source.  

Simulations with a 30 degree magnet show that the beam transport becomes much less sensitive to input emittance and space charge compensation (since the envelopes are smaller) and is not influenced by aberration and fringe fields to the same degree. Simulations with a distribution of 0.2\,$\pi$\,mm\,mrad and space charge compensation of 95\% show a transmission of about 95\%. Using the same input emittance, but space charge compensation of 85\%, also gives a transmission of about 95\%. A distribution of 0.25\,$\pi$\,mm\,mrad, which is the nominal requirement for the ion source, and space charge compensation of 85\%, which is a reasonable assumption for the LEBT~\cite{lawrie:private_comm} gives a transmission of 84\%. The beam distribution after the RFQ for this case is shown in Figure 9, tracking 200 k macro particles. The output emittance from the RFQ is about 0.25 $\pi$ mm mrad. A beam density plot in the LEBT is shown in Figure 10 for this case.

As mentioned in Section~\ref{subsect:ion_src_overview}, further simulations at an upper-limit emittance of 0.38\,$\pi$\,mm\,mrad were also carried out for the 30 degree case assuming a source capable of delivering ${\sim}85$\,mA. Here the transmission through the RFQ is limited to ${\sim}70\%$, but the beam is delivered to the end of the linac with a modest emittance growth of ${<}10\%$ versus the nominal case.

These simulations demonstrate that the case with a source placement of ${\pm}$30 degrees is much less sensitive to the initial ion source distribution and the degree of space charge compensation than the 60 degree design. However, the impact on ESS operations due to installation is also a significant factor; this is covered in depth in Section~\ref{sect:operations}. Although both options have serious merits and drawbacks, considering all these aspects discussed here and those discussed in Section~\ref{sect:front_end_overview}, the present baseline is the 30-degree option.

\subsection{Medium Energy Beam Transport (MEBT) Design\label{sect:mebt_design}}

The function of the MEBT is to match the RFQ output beam to the DTL; to characterise the beam with different diagnostics; to clean the head of pulse using a fast chopper (2.86\,ms long for the proton beam); and to clean the transverse halo using scrapers, see Fig.~\ref{fig:mebt_schematic}. The chopper is also part of the machine protection system, which shuts down the linac on detection of any serious faults. Details of the ESS MEBT design and matching can be found in~\cite{Miyamoto:2016_matching_linac,Miyamoto:2014_ess_mebt}.

For the H$^-$ beam, the MEBT will also be used to chop the beam in order to create extraction gaps for the accumulator ring. The beam envelopes from the RFQ are similar for the proton and H$^-$ beams, but since the DTL has permanent quadrupole magnets, the orientation of the beam envelopes are opposite for H$^-$ and proton beams, see Fig.~\ref{fig:proton_hminus_beamshapes}. 

\begin{figure*}[ht!]
    \centering
    \includegraphics[width=0.85\textwidth]{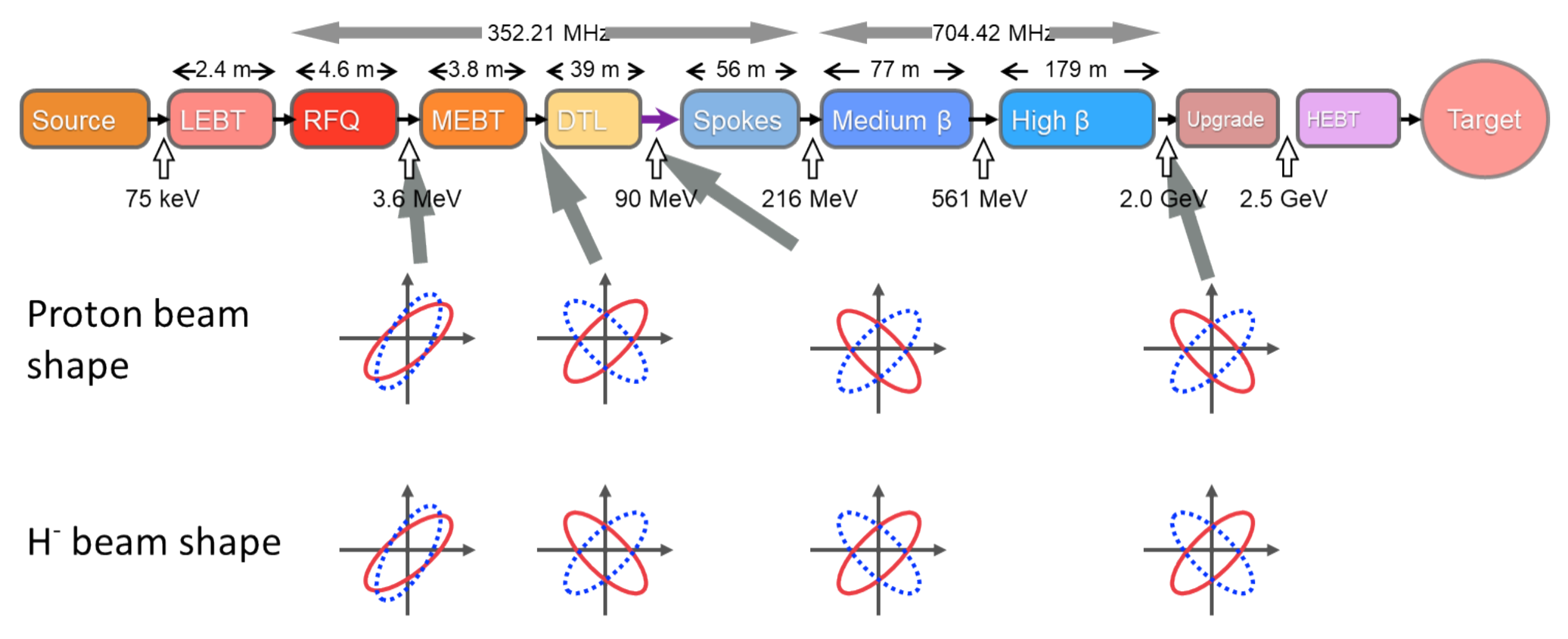}
    \caption{Beam envelopes for different ion species at different stages in the linac.}
    \label{fig:proton_hminus_beamshapes}
\end{figure*}

\begin{figure*}[ht!]
    \centering
    \includegraphics[width=0.70\textwidth,trim={0 0.7cm 0 0},clip]{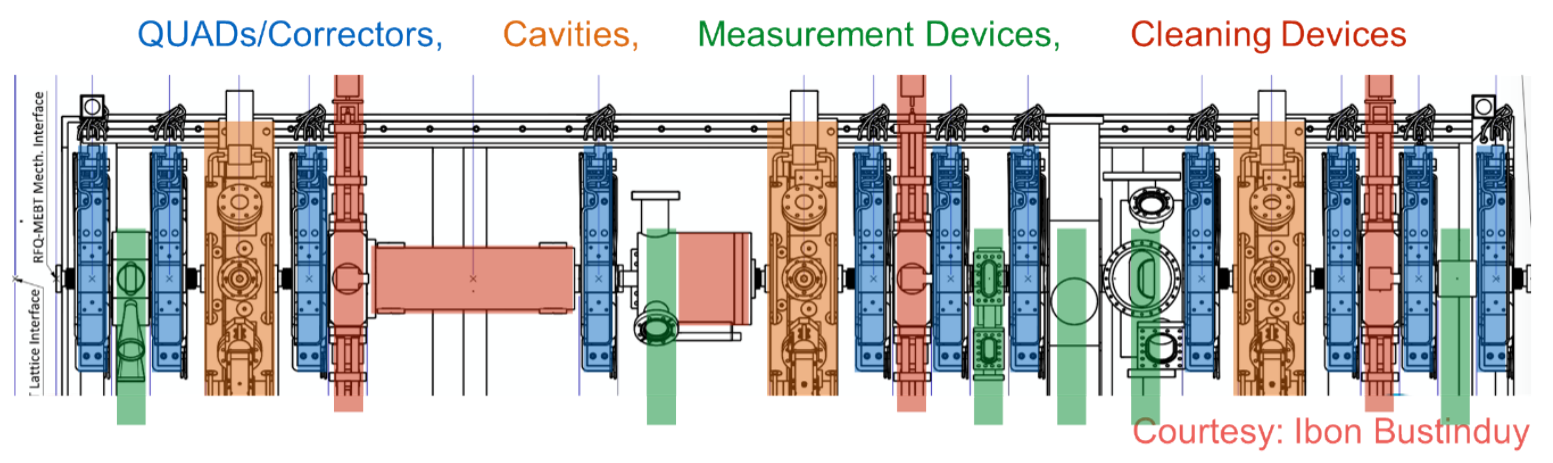}
    \caption{Schematic MEBT drawing with different functions indicated.}
    \label{fig:mebt_schematic}
\end{figure*}

As a starting point for the MEBT design, the proton MEBT is used (see Fig.~\ref{fig:mebt_schematic}). The ESS MEBT contains 11 quadrupoles for transverse focusing and 3 buncher cavities for longitudinal focusing, opposing the space charge forces in the beam. Figure~\ref{fig:mebt_sigs_eps} shows the beam envelopes and expected emittance development taken from simulations. 

The MEBT must be redesigned in order to meet the requirements of the H$^-$ beam. The chopper for the standard ESS MEBT is designed to remove the head and tail of the 2.86\,ms pulse. It consists of electrostatic plates, and a dump in one plane, with a quadrupole that is used to deflect the beam. This does not work, however, for a beam with the opposite charge (without switching the polarity of the quadrupole). Alternative methods for chopping will be discussed shortly.

\begin{figure*}[ht!]
    \centering
    \includegraphics[width=0.70\textwidth]{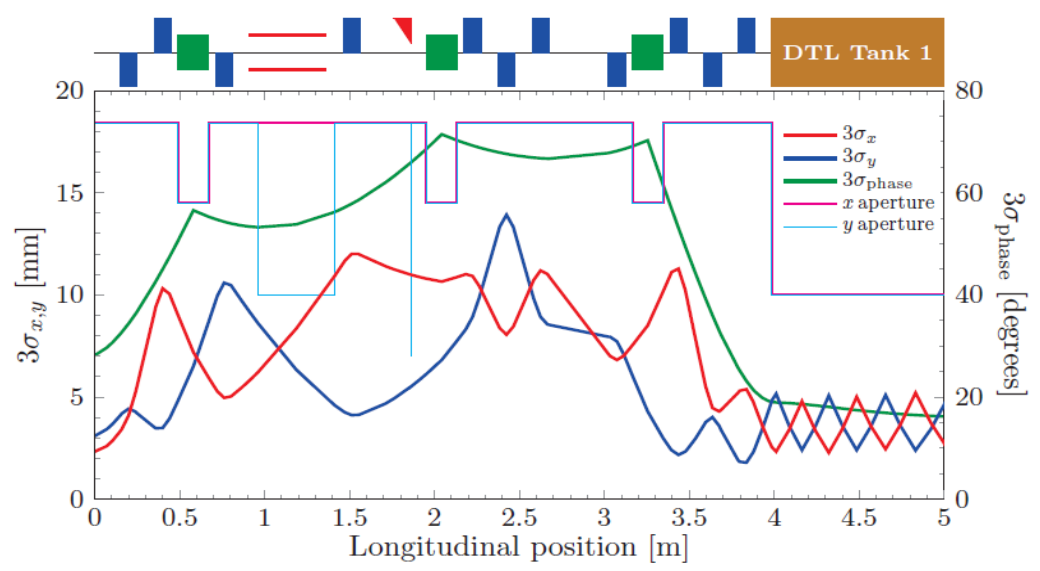}
    \\
    \includegraphics[width=0.70\textwidth]{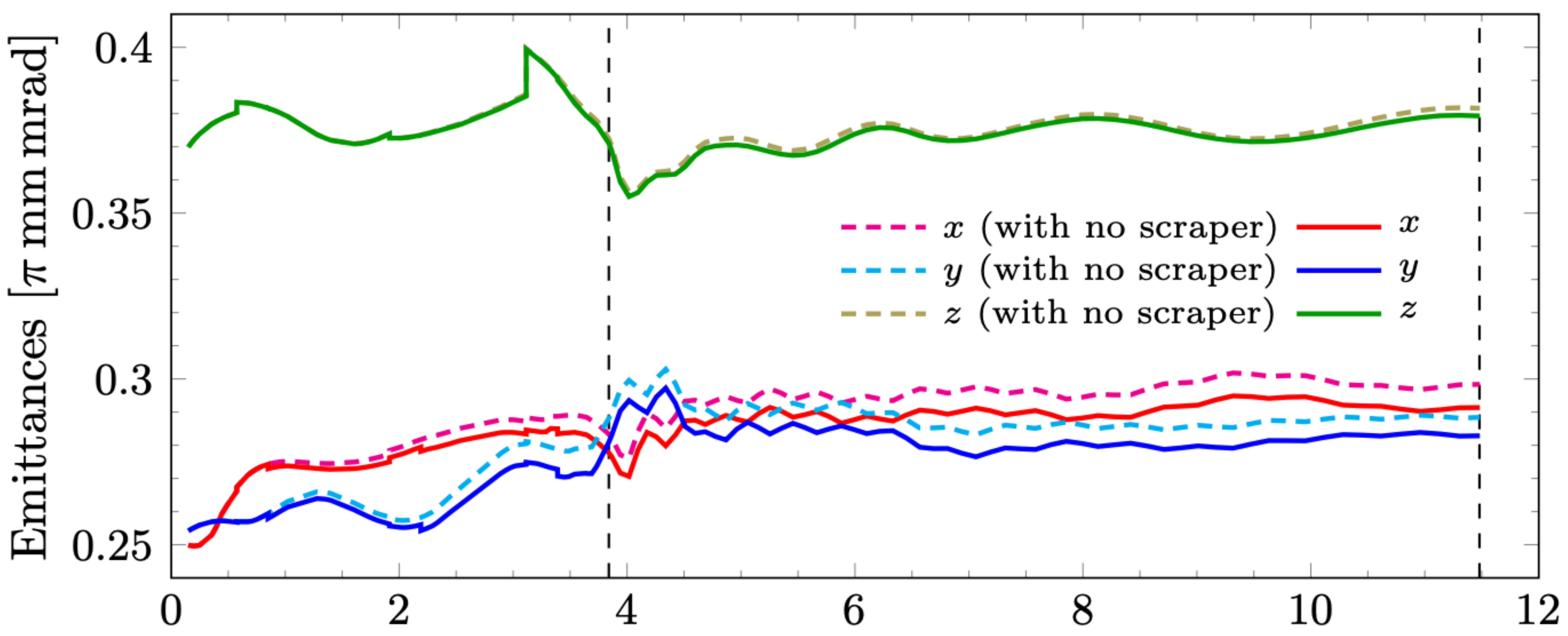}
    \caption{Standard proton MEBT, simulations of envelopes (top) and emittance growth for both the MEBT and first DTL tank (bottom), from~\cite{Miyamoto:2014_ess_mebt}.}
    \label{fig:mebt_sigs_eps}
\end{figure*}

The first elements in the MEBT are used to form a beam which is circular at the chopper, both for the proton and H$^-$ beams. The focusing elements after the dump will be used to match the beam to the DTL. Some elements may have to be pulsed between the proton and H$^-$ beams, 28\,Hz in the baseline design, but it is desirable to find a MEBT design that switches as few elements as possible. Simulations performed in \textsc{TraceWin} show an emittance growth of ${~3\%}$ for the case having no switching magnets versus having both beams matched to optimal Twiss parameters at the MEBT-to-DTL interface via switching. The exact design of the quadrupoles will need to be studied further, especially the choice of iron or ferrite-based design.

The critical design criteria for the MEBT are summarised here:
\begin{itemize}
    \item Use as few switching magnets and/or focusing elements as possible; these switch the value between p and H$^-$. 
    \item The longitudinal phase spread should be limited to less than ${\pm}$25 degrees (rms) in order to avoid non-linear fields in the buncher cavities, which would lead to halo and/or emittance growth longitudinally. 
    \item In case one or more cavities fail in the superconducting (SC) section of the linac, parts of the SC linac need to be retuned to the order of 10 MeV. This means that no dispersion can be allowed downstream of the MEBT, and that an achromatic solution should be sought as a translation stage. 
    \item For the option of merging beams in the MEBT, the new RFQ and ion source need to be displaced from the proton RFQ by about three meters in order to allow access, and space for installations. It also requires further investigation on whether it is feasible to install a second H$^-$ front end, considering the required waveguides and other auxiliary equipment.
\end{itemize}

\subsubsection{Option for a Separate H$^-$ MEBT with a 45-degree Translation Stage} 

In an alternative design for the H$^-$ MEBT, the standard MEBT design was used, then split after the diagnostics unit and before the last four quadrupoles, where the translation stage with two 45-degree magnets is introduced, see Fig.~\ref{fig:mebt_2x_schematic} and Fig.~\ref{fig:mebt_merge_lattice}. The translation section includes three buncher cavities to focus the bunch longitudinally. The quadrupoles and the buncher cavities are matched to create an achromatic section, so that the dispersion and its gradient remains zero at the MEBT exit. As mentioned above, the new RFQ and ion source must be displaced from the proton RFQ by about 3 meters in order to allow access and space for installations; the length of the 45-degree translation stage thus becomes 4.2\,meters.

The proton and H$^-$ beams merge at a second dipole magnet. Thus, for optimal beam conditions, the last four quads and the last buncher cavity need to act in a switching mode, since they require different settings for each of the two beams to provide matching to the DTL. In the DTL, permanent quadrupoles are used for focusing, so no switching can be done between the two beams.

The first section of the MEBT is essentially identical to the nominal proton MEBT, and includes the beam chopper and diagnostics. The chopper cavity is rotated 90 degrees compared to the proton one, to deflect the beam in $x$ instead of $y$, and the fifth quadrupole from the MEBT entrance acts as a kick-magnifying element. The quadrupoles in the first section have approximately the same values those in the nominal MEBT.

Table~\ref{tab:mebt_merge_procon} lists the major drawbacks and benefits of this scheme. One additional advantage with merging the beams in the MEBT is that the proton MEBT can be kept more or less intact except for the last section of the MEBT, consisting of four quads and one buncher cavity. 

However, the space restrictions, emittance growth, and other listed drawbacks have left this option as a backup, with the design of merging further upstream into a common RFQ being favoured as a baseline, as discussed in Sections~\ref{sect:front_end_overview} and \ref{sect:lebt_design}. Further details on this study can found in~\cite{blaskovic:2020_MS10}.

\begin{figure*}[ht!]
    \centering
    \includegraphics[width=0.85\textwidth]{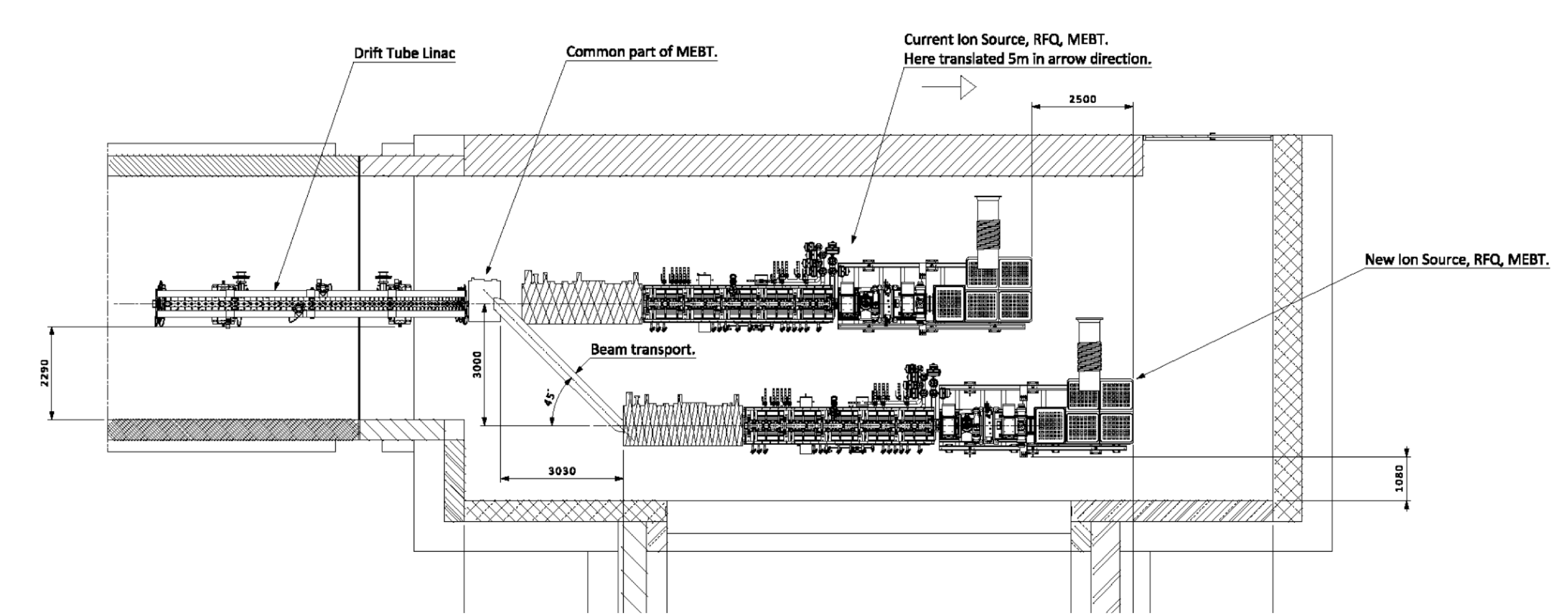}
    \caption{Detailed Schematic for the 45-degree merge-in-MEBT option.}
    \label{fig:mebt_2x_schematic}
\end{figure*}

\begin{figure*}[ht!]
    \centering
    \includegraphics[width=0.70\textwidth]{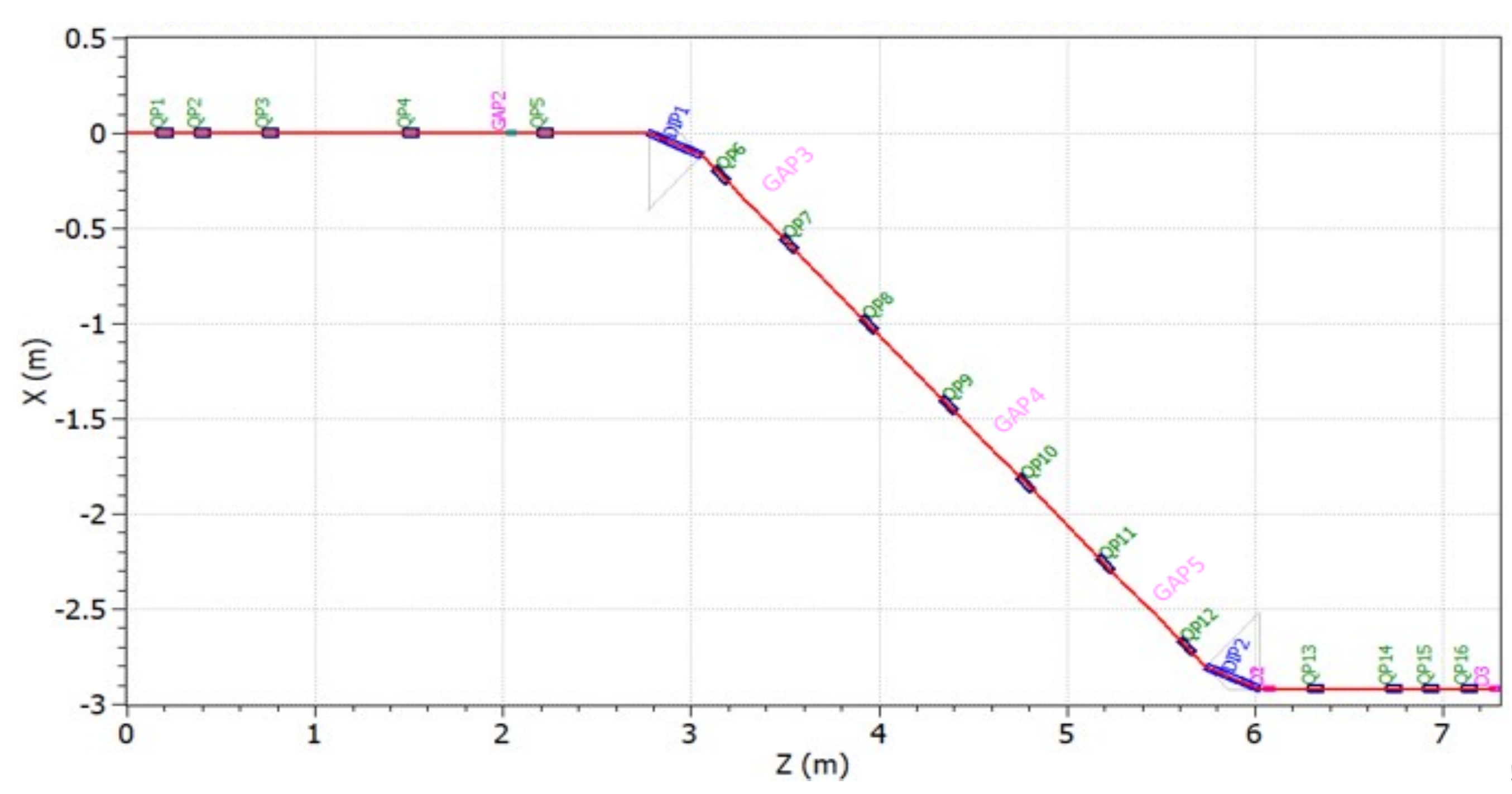}
    \caption{MEBT lattice including the translation stage, indicating the quadrupoles and buncher cavities.}
    \label{fig:mebt_merge_lattice}
\end{figure*}

\begin{table}[ht!]
\begin{center}
\caption{Pros and cons of merging the beams in the MEBT.}
\label{tab:mebt_merge_procon}
\begin{tabular}{p{6cm} p{6cm}}
%\hline
\multicolumn{1}{c}{\textbf{Pro}} &
  \multicolumn{1}{c}{\textbf{Con}} \\ \hline
%\rowcolor[HTML]{F2F2F2} 
Independence  of the proton and H$^-$ beamlines up to the DTL. & Space   restrictions for two parallel front ends are severe, probably untenable.\\ \hline
The   proton and H$^-$ ion sources can be commissioned and tested separately,   including the RFQs. & Dispersion   diagnostics need to be introduced. \\ \hline
Avoids commissioning a new common RFQ, which will be a lengthy interference with the proton beam operation. & The dipole and four quads have to be switched, at about 70\,Hz, with zero field for the proton beam (The magnets could be made iron free or low-remanence soft steel).\\ \hline
Avoids requiring the chopper to function both for the proton and H$^-$ beams. & Significantly higher emittance growth compared to a single MEBT. \\ \hline
The second RFQ can be identical to the present RFQ design. (Merging in the LEBT requires an RFQ with improved cooling.) & \\ \hline
\end{tabular}
\end{center}
\end{table}

\subsubsection{Matching the MEBT to the DTL}
Tracking calculations were performed in \textsc{TraceWin}, which does multi-particle tracking with the PICNIC 3D space-charge routine. The beam was tracked in the MEBT and DTL with the input beam distribution provided by tracking an H$^-$ beam through the LEBT and RFQ.

% For good matching to the DTL, we find that the Twiss parameters for the H$^-$ beam out of the MEBT should have the following values:

% \begin{align*}
% &\beta_{x-x'} = 0.733, \alpha_{x-x'} = -4.005,\\
% &\beta_{y-y'} = 0.196, \alpha_{y-y'} = 1.228,\\
% &\beta_{z-z'} = 0.370, \alpha_{z-z'} = 0.098 
% \end{align*}

A precision matching of the MEBT was performed using the four final quadrupoles, along with its second and fifth buncher cavities, using multi-particle tracking, to a precision of 2.4e-4 by \textsc{TraceWin}'s optimisation routine. Some oscillations (beat) in the emittances in the DTL can be seen due to imperfect matching; this can probably be improved with further refinement.

\begin{figure*}[ht!]
    \centering
    \includegraphics[width=0.42\textwidth]{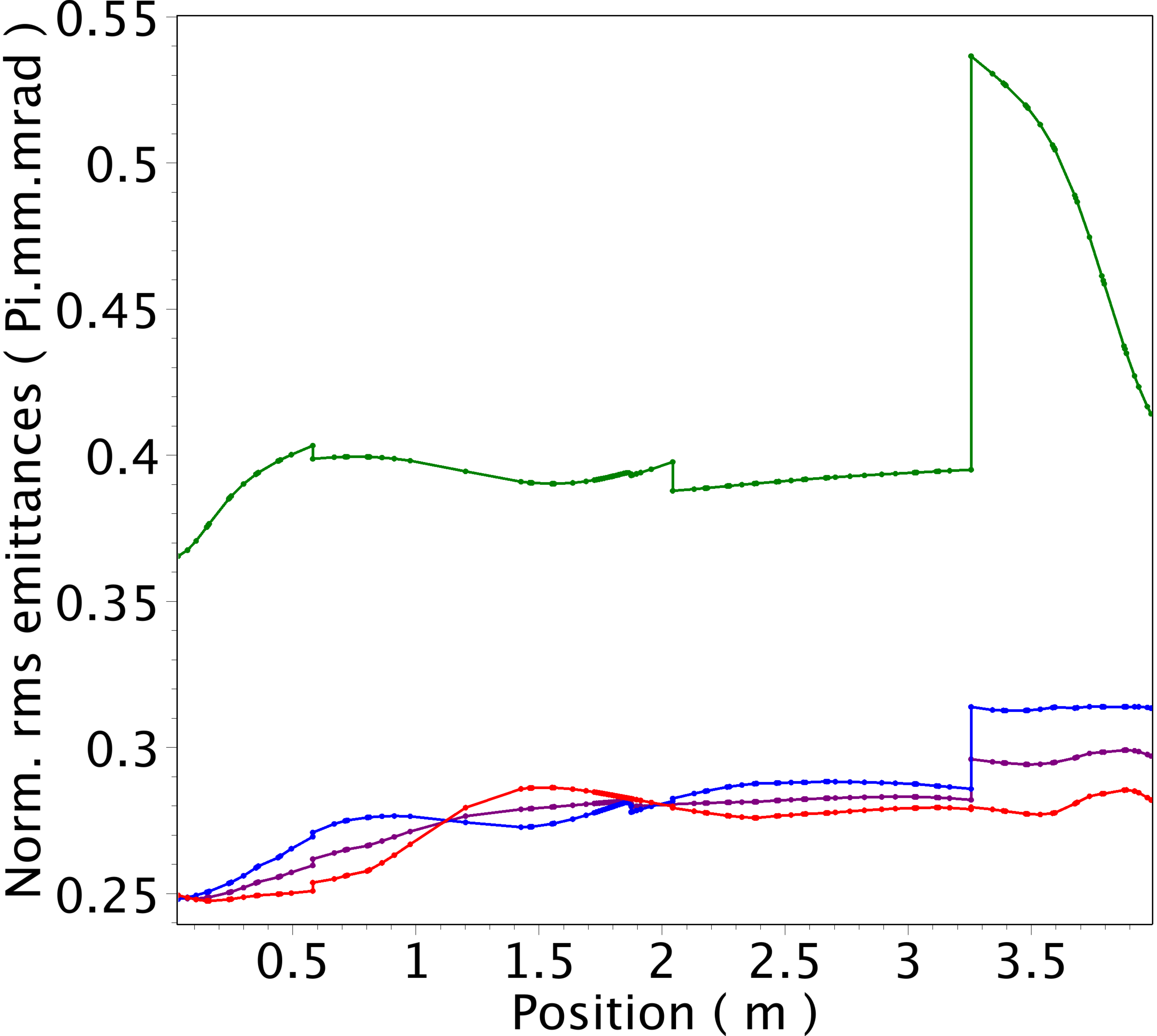} 
    \includegraphics[width=0.42\textwidth]{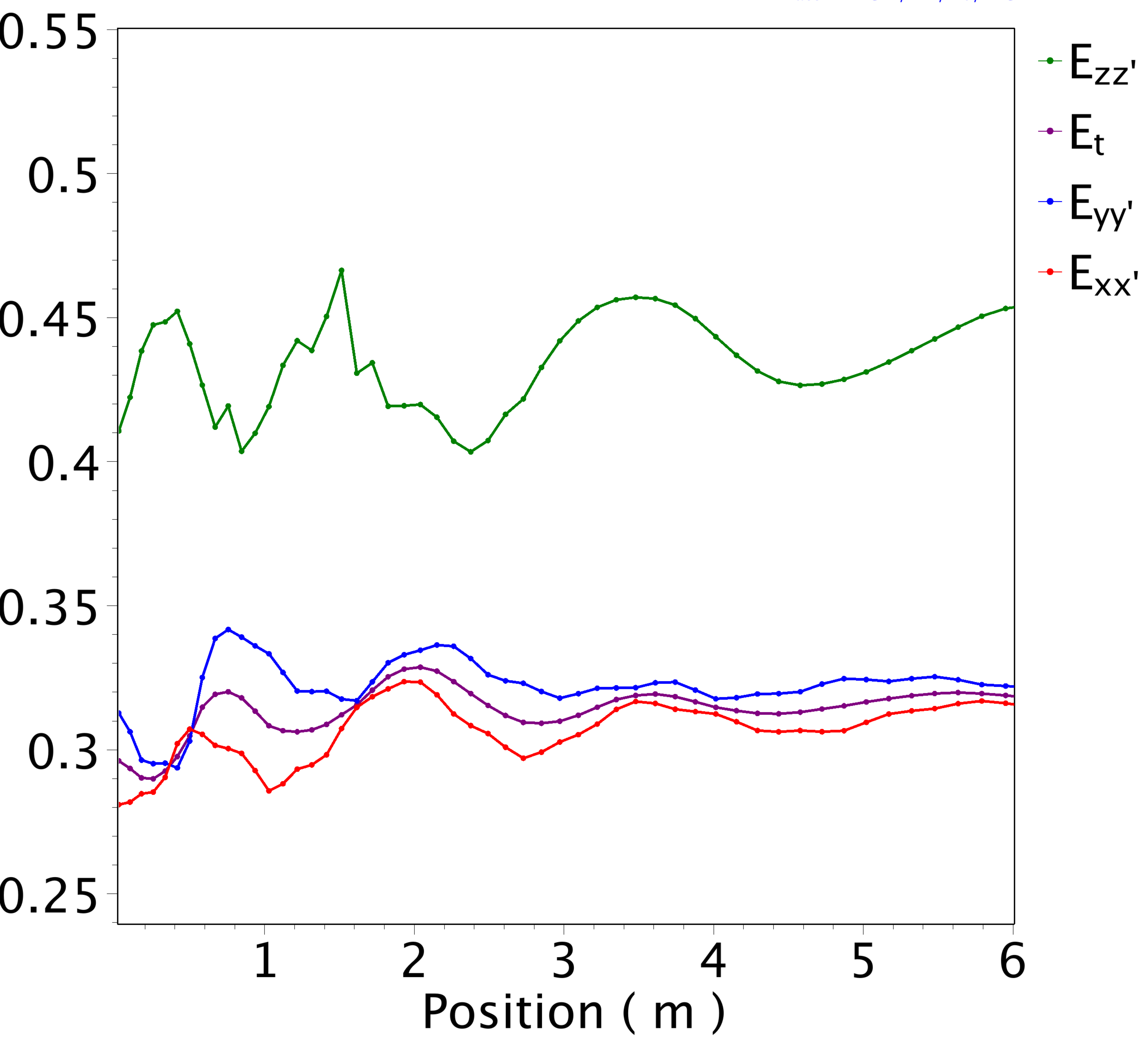}
    \caption{Emittance in the MEBT and the first DTL tank for a nominal H$^-$ beam.}
    \label{fig:mebt_eps_Hminus}
\end{figure*}

It is worth noting in Fig.~\ref{fig:mebt_eps_Hminus}, that there is a noticeable emittance growth. This can be compared to the standard proton MEBT where the emittance growth is about 10\% transverse, see Fig.~\ref{fig:mebt_sigs_eps}. The emittance growth is, to a large extent, taking place in the dipoles, where there is a coupling between x, z and dp/p. In an achromat with two dipoles, this emittance growth will cancel after the second dipole; but this does not take place in this lattice which has a 3$\pi$ phase advance. This issue is discussed further below. %The optics in the translation stage may require better optimization: there are large envelope oscillations in x, with rather narrow waists, which will drive emittance growth due to non-linear space charge forces. 

Additional matching studies were performed using the technique described in~\cite{Miyamoto:2016_matching_linac}, with a modified solver to accommodate the interleaved pulses of H$^-$ and protons without the use of switching magnets. This results in a modest emittance growth versus the single-species simulations, reaching the end of the linac at roughly 3${\sim}$10\% above the nominal case for both H$^-$ and protons (depending on emittance and current parameters of the ion source). Further optimisation of the matching should be performed in order to finalise the MEBT chopper design.

\subsubsection{Chopping} 
% FROM MS2
% One alternative for chopping is to use the middle buncher cavity for increasing the deflection of the beam. This buncher cavity focuses the beam in the longitudinal direction, and defocuses the beam in horizontal and vertical directions -- this scheme will work for both beam species. The design of chopper has to be adapted to chop the beam at the frequency required to create the extraction gaps (around 740 kHz, 1/1.35 \SI{}{\micro\second}). The increased beam heat load needs to be taken into account, as well as the increased pulsing frequency demands of the chopper including its power supply requirements. Both approaches to chopping
The chopper in the standard proton MEBT is used for removing the head and tail of the macro pulse of 2.86\,ms. For the H$^-$ beam, the chopper will also be used to create extraction gaps for the accumulator ring~\cite{galnander:2018_D2.1}.

This means chopping off about 0.13\,\SI{}{\micro\second} every 1.35\,\SI{}{\micro\second}, or 670~kHz, about 2000 times for every macro pulse of 3\,ms. This is a considerably higher frequency than the standard chopper. It also means that about 10\% of the H$^-$ beam will be dumped in the chopper beam dump. With a beam current of 62\,mA, this corresponds to a power of 22\,kW at 3.62\,meV, and with a beam duty cycle of 5\%, an average power of 1.1\,kW. The chopper and chopper dump will thus have to be redesigned to meet these requirements. 

Assuming the same design of the RFQ as for the standard proton MEBT, envelopes are exchanged in $x$ and $y$ between proton and H$^-$ if using the same gradient strengths for the quadrupoles. The chopper must be designed to deflect the beam in $x$, with the fourth quadrupole used to amplify the kick in $x$. (For protons, the chopper cavity deflects in the $y$ direction.) The chopping is efficient with a full voltage applied of ${\pm}$2.5\,kV (the total voltage difference is 5\,kV), see Fig.~\ref{fig:mebt_chopper}.

\begin{figure*}[ht!]
    \centering
    \includegraphics[width=0.85\textwidth]{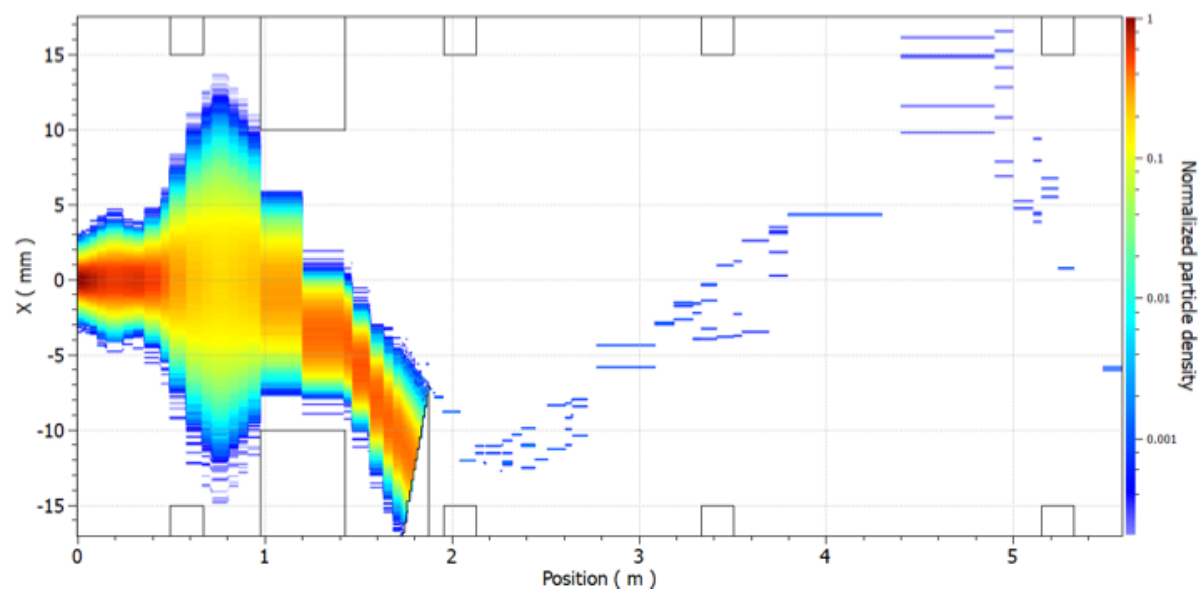} 
    \caption{Chopper at full deflection; H$^-$ beam deflected to the intended beam dump.}
    \label{fig:mebt_chopper}
\end{figure*}

One of the challenges with 5\,kV chopper is to achieve a rise-time as fast as the bunch spacing (2.84\,ns). The present requirements for the chopper are 4\,kV and 10~ns for the ESS MEBT. There will thus be partially-chopped bunches that are deflected but do not feel the full chopper voltage, and thus will be lost in the MEBT or the DTL. For the standard MEBT, such losses are on an acceptable level~\cite{Miyamoto:2016_matching_linac}; but these losses will be roughly 2000 times greater for the H$^-$ case. For the merge-in-MEBT option, these lost particles can probably be handled within the MEBT and not reach higher energies, since the translation stage is situated after the chopper. However, for the baseline option of merging in the front end, this may be more difficult and should be studied further.

\subsection{Superconducting Linac}
The medium-to-high-beta acceleration process is also being studied with the \textsc{TraceWin} program. It is of primary interest that the beam exiting the linac is matched to the accumulator ring in order to have efficient injection and to avoid losses. The simulations presented in the above sections focused on the first part of the accelerator, from the ion source to the DTL. Simulations of the beam acceleration and transport from the entrance of the DTL to the injection point to the accumulator ring -- with and without errors -- were performed as well for the H$^-$ beam. The summary of the losses from the machine errors and different H$^-$ stripping sources are reported in~\cite{folsom:2020_D2.3}, with results of a supplementary study also available in \cite{folsom:2021_hmin_strp}; this issue is also discussed in detail in Section~\ref{sect:beam_losses}.

From the DTL onwards, the settings of the quadrupoles can be identical for the H$^-$ and proton beams, although the trajectory correction for the two species will most likely need to be different. This requires that the steerer magnets operate in pulsed mode, which will mean modifying their corresponding power converters.

The RF power needed for the combined ESS$\nu$SB and ESS beams is nearly double the current design (although there most subsystems, as well as overall grid power requirements have significant percentages of overhead reserved for contingency and upgrade). This issue has been studied in detail and is reported in~\cite{galnander:2019_D2.2} and \cite{folsom:2021_D2.4}. Such modifications to the RF system were evaluated for different pulsing schemes of the linac; these are discussed further in Section~\ref{sect:RF_systems}.

In the superconducting cavities, it is a concern that the RF couplers feeding from the waveguides may experience electrical breakdown due to the increased duty cycle. Based on conditioning data, these couplers are rated to handle the required 10\% duty cycle for accelerating both the proton and H$^-$ beams. Thus, the risk of their needing to be retrofitted or replaced is considered very low. However, since the scenario of their needing replacement presents a significant disruption risk to ESS operations, a detailed replacement plan is presented in Section~\ref{sect:operations}.

\subsection{Linac-to-Ring (L2R) Transfer Line \label{sect:L2R}}

The L2R transfer line has been designed to transport the fully accelerated 2.5\,GeV H$^-$ beam from the upgraded high-$\beta$ line (HBL) at the end of the linac to the accumulator ring~(AR) \cite{alekou:2020_D3.2}. The L2R line bends the beam both horizontally away from the linac and vertically down towards the underground AR as shown in the L2R tunnel engineering drawing in Fig.~\ref{fig:L2R}. The beam is horizontal both entering and leaving the L2R line, but at a height difference of 7.864\,m.

The transfer line lattice is based on a quadrupole-doublet cell structure, taken from the ESS high energy beam transport (HEBT) line design. The cell length is 8.52\,m long, each quadrupole is 0.35\,m long, and the separation of the mid-points of the quadrupoles in each doublet is 1.08\,m. The remaining part of a cell constitutes a drift space which is 7.09\,m long. It is in these long drifts where the dipole magnets are placed for the transfer line.

In designing the lattice, it is important to stay below the ESS beam-loss limit of 1\,W/m \cite{Tchelidze2019} which restricts the radioactive activation of the accelerator elements. To this end, the dipole strengths are set to 0.15\,T, which limits H$^-$ losses from Lorentz stripping (see Section~\ref{sect:beam_losses}) to a fractional loss of $5.7 \times 10^{-8}/$m \cite{Keating:1995zz}, corresponding to a stripping loss in the dipole magnets of 0.3\,W/m for a 5\,MW beam. The 0.15\,T field corresponds to a dipole bending radius of 73.5\,m, which is used for all horizontal and vertical dipole magnets in the transfer line. All dipole magnets in the L2R line have been taken to be sector dipoles.

Following the tunnel engineering design \cite{JohanssonDrawing} in Fig.~\ref{fig:L2R}, the beam is bent horizontally by $\theta = 68.75^\circ$, starting from the ESS linac tunnel (point A). For the L2R lattice design, the horizontal bending is performed over 16 lattice cells each of 8.52 meters. Each cell therefore bends the beam horizontally by 4.3$^\circ$, and this can be achieved using a single 0.15\,T horizontally bending dipole of length 5.512\,meters per lattice cell.

\begin{figure*}[ht!]
\vspace*{-0.5\baselineskip}
\begin{center}
\includegraphics[width=0.4\columnwidth]{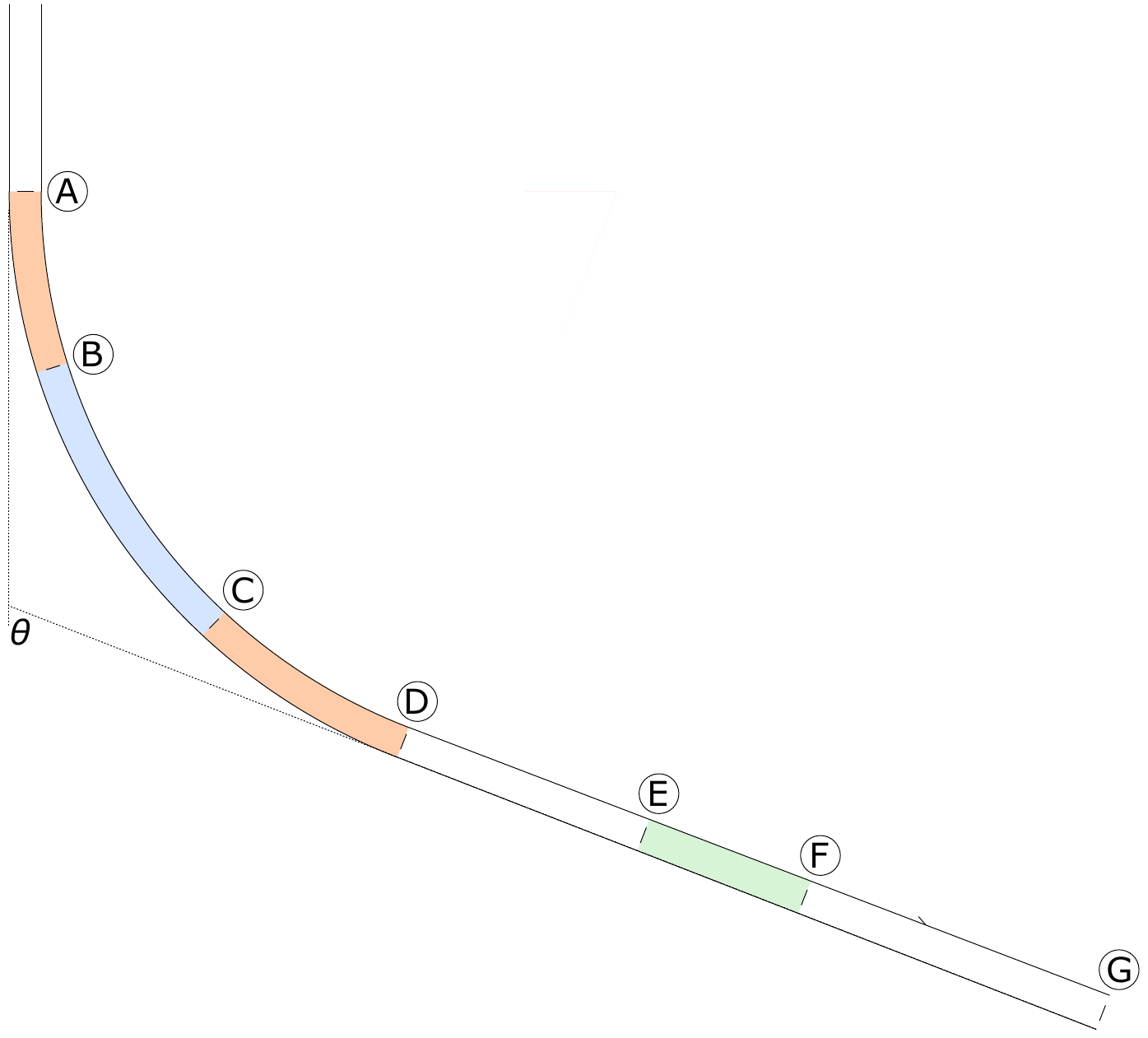}
\\
\includegraphics[width=0.7\columnwidth]{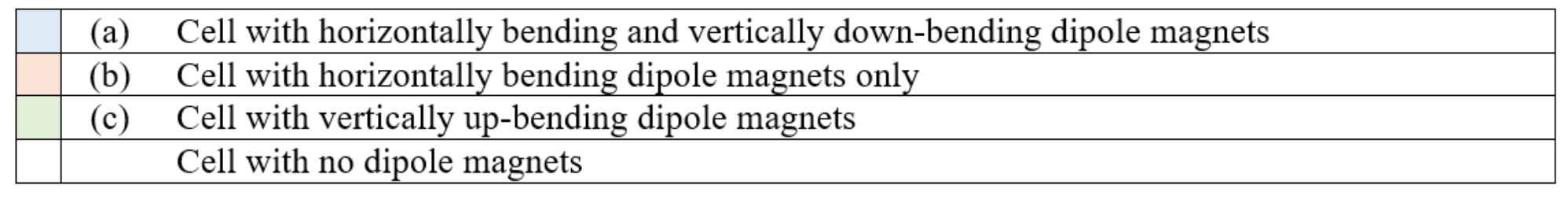}
\caption{Layout of the L2R transfer line \cite{JohanssonDrawing}. Sections are colour-coded according to the presence of horizontal and/or vertical bending}
\label{fig:L2R}
\end{center}
\end{figure*}

Regarding the vertical bending, the tunnel engineering design calls for a maximum vertical tunnel incline of 3.38$^\circ$. The tunnel incline is progressively increased in section B--C in Fig.~\ref{fig:L2R}, from horizontal to 3.38$^\circ$ \cite{JohanssonDrawing}. After the section of maximum incline (section C--E), the tunnel incline is progressively decreased (section E--F) to return to fully horizontal upon reaching the AR (section F--G). Injection into the AR occurs at, or immediately after, point G in Fig.~\ref{fig:L2R}.

Such a design can be achieved by assigning 8 lattice cells in section B--C and 4 lattice cells in section E--F, with an incline change of 0.42$^\circ$ per cell and 0.84$^\circ$ per cell, respectively. This calls for single 0.15\,T vertically bending dipole magnets per cell of length 0.542\,m and 1.084\,m, respectively. It is noted that the longer magnets are not problematic in section E--F as these magnets do not need to share the drift space with horizontally bending magnets.

The dipole and quadrupole distributions within the different 8.52\,m lattice cells is shown in Fig.~\ref{fig:CellOptions}. In the section with both horizontal and vertical bending (section B--C in Fig.~\ref{fig:L2R}), both the horizontally and vertically bending magnets are distributed to maintain an equal separation of 0.345\,m between adjacent dipoles -- and between the dipoles and adjacent quadrupoles. In this way, the magnet separations are maximised in order to minimise cross-talk between magnetic field edge effects.

In the sections with horizontal bending only (sections A--B and C--D in Fig.~\ref{fig:L2R}), the location of the horizontally bending dipole magnet is preserved within the lattice cell (see Fig.~\ref{fig:CellOptions}(b)); this ensures that the beam optics in the horizontal plane is identical throughout sections A--D. For section E--F with vertical bending only, given the freedom of the vertical dipole magnet location, it is placed in the centre of the long drift between quadrupoles, as shown in Fig.~\ref{fig:CellOptions}(c).

\begin{figure*}[ht!]
\vspace*{-0.5\baselineskip}
\begin{center}
\includegraphics[width=\columnwidth]{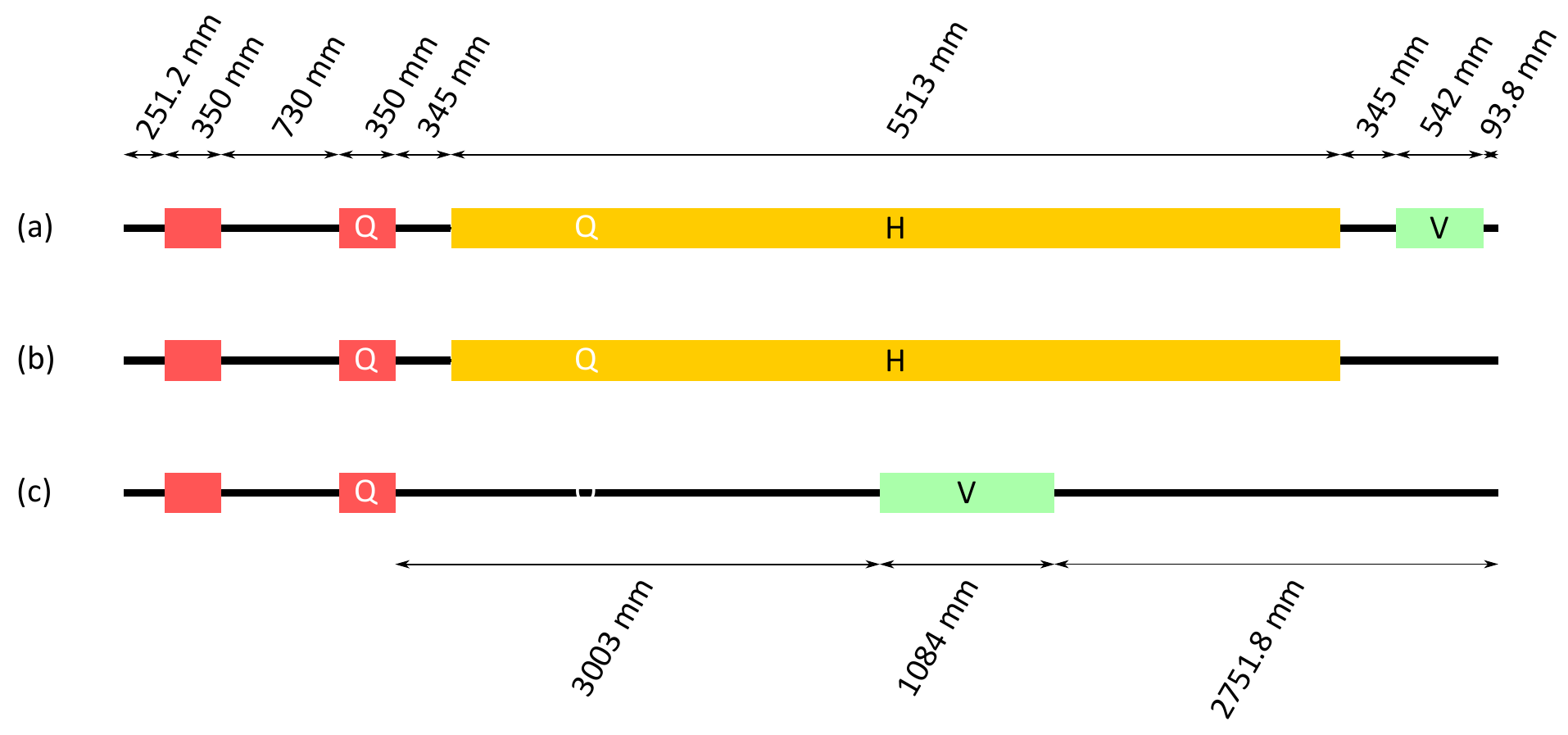}
\caption{Quadrupole (Q), horizontally bending dipole (H) and vertically bending dipole (V) distributions within 8.52\,m lattice cells for (a)--(c) sections listed in the legend of Fig.~\ref{fig:L2R}.}
\label{fig:CellOptions}
\end{center}
\end{figure*}

Using 16 horizontally bending, 8 vertically down-bending and 4 vertically up-bending dipole magnets, an AR depth of 7.864 m can be achieved. The assignment of the 32 lattice cells is shown in Table~\ref{tab:CellDistribution}, and the locations at which the sections of horizontal and vertical bending start and end\footnote{Note that the locations in Table~\ref{tab:SectionStart} are at the start of the first or the end of the last magnet of the magnet type in question, e.g. location B is at the start of the first vertical dipole magnet.} are shown in Table~\ref{tab:SectionStart}.

\begin{table}[ht!]
\centering
\caption{Assignment of lattice cells according to the presence of horizontal and/or vertical bending; the colour key follows the legend in Fig.~\ref{fig:L2R}.}
\label{tab:CellDistribution}
\includegraphics[width=\columnwidth]{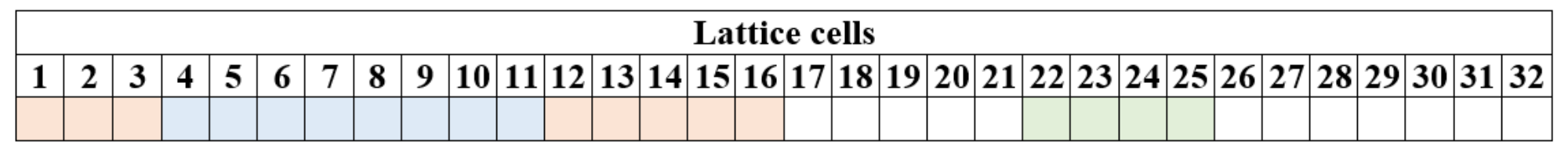}
\end{table}

\begin{table}[ht!]
\centering
\caption{Section start and end locations, as defined in Fig.~\ref{fig:L2R}, for the proposed lattice, measured from the start of the L2R line.}
\label{tab:SectionStart}
\begin{tabular}{c c}
%\hline
\textbf{Point} & \textbf{Location along beamline (m)} \\
\hline
 A     & 0.000                       \\
 B     & 31.418                      \\
 C     & 91.600                      \\
 D     & 133.313                     \\
 E     & 181.578                     \\
 F     & 208.222                     \\
 G     & 270.614                     \\
\hline
\end{tabular}
\end{table}

\subsubsection{Beam Dynamics}

Beam dynamics simulations have been performed in \textsc{TraceWin} from the start of the DTL to the end of the L2R transfer line \cite{Blaskovic2021}. The L2R quadrupole strengths are optimised to make the L2R line achromatic (that is, with no output horizontal and vertical dispersion) and to limit intra-beam stripping by avoiding beam sizes that are too small (see also Section~\ref{sect:beam_losses}). A matched beam is also ensured between accelerator sectors by matching the Twiss parameters sector-to-sector.

The optimised choice of phase advance is shown in Fig.~\ref{fig:PhaseAdvance}. The phase advance up to and including the HBL matches closely that used for protons, to allow concurrent operation of protons and H$^-$ ions. There is no step in the transverse phase advance on entering the L2R line to ensure efficient matching; beam matching is performed in \textsc{TraceWin} using the last 2 cells in the HBL section and the phase advance adjustment is minimal, as shown in Fig.~\ref{fig:PhaseAdvance}.

\begin{figure*}[ht!]
\vspace*{-0.5\baselineskip}
\begin{center}
\includegraphics[width=0.6\columnwidth]{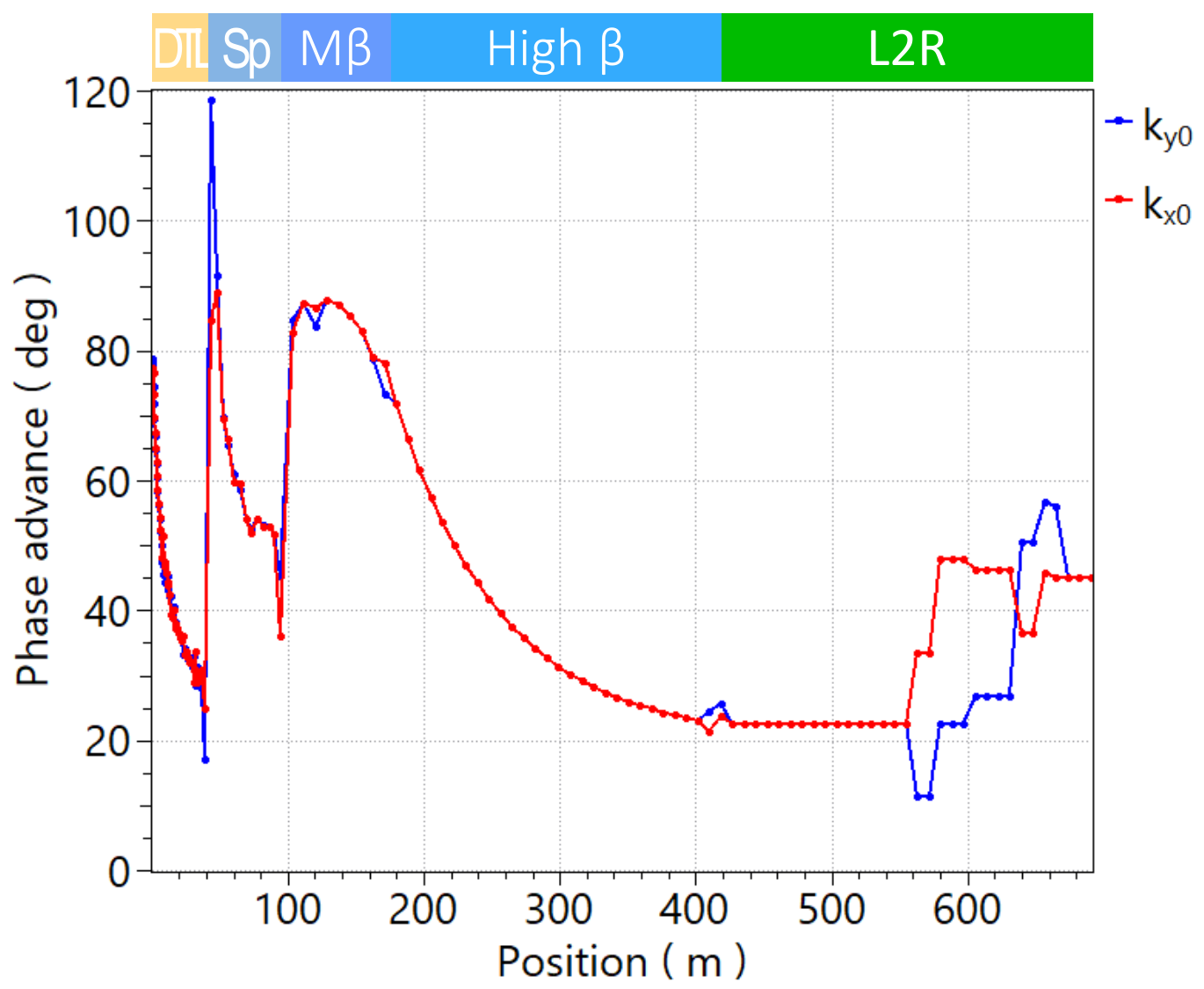}
\caption{Transverse $x$ (red) and $y$ (blue) phase advance per cell along the linac and L2R line. Positions are given relative to the start of the DTL. The lattice sections are indicated along the top of the plot.}
\label{fig:PhaseAdvance}
\end{center}
\end{figure*}

In order to make the L2R line achromatic, the total phase advance is set to a multiple of 180$^{\circ}$ \cite{Satogata2011}. A relatively low transverse phase advance of 22.5$^{\circ}$/cell is used at the beginning of the L2R line, providing relatively weak transverse beam focusing \cite{Holzer2011} and a large transverse beam size of up to $\pm 4$\,mm (root mean square) in the dispersive section. Critically, this large beam size limits the intra-beam stripping losses \cite{Maruta2016} to under 0.4\,W/m, at or below the level observed in the linac (Fig.~\ref{fig:IntraBeamStripping}).

The magnitude of intra-beam stripping losses decreases along the L2R line as the longitudinal bunch length increases in the absence of accelerating cavities for longitudinal focusing. Therefore, one can take the opportunity to ramp up the phase advance from 22.5$^{\circ}$/cell to 45$^{\circ}$/cell (Fig.~\ref{fig:PhaseAdvance}), increasing the transverse focusing and reducing the transverse beam size to below $\pm 2$\,mm (root mean square). This is beneficial in that the number of cells with vertically up-bending dipole magnets (section E--F in Fig.~\ref{fig:L2R}) can be reduced from 8 to 4; thus reducing the number of magnets required for the project, while maintaining the 180$^{\circ}$ requirement for keeping the line achromatic.

\begin{figure*}[ht!]
\vspace*{-0.5\baselineskip}
\begin{center}
\includegraphics[width=0.7\columnwidth]{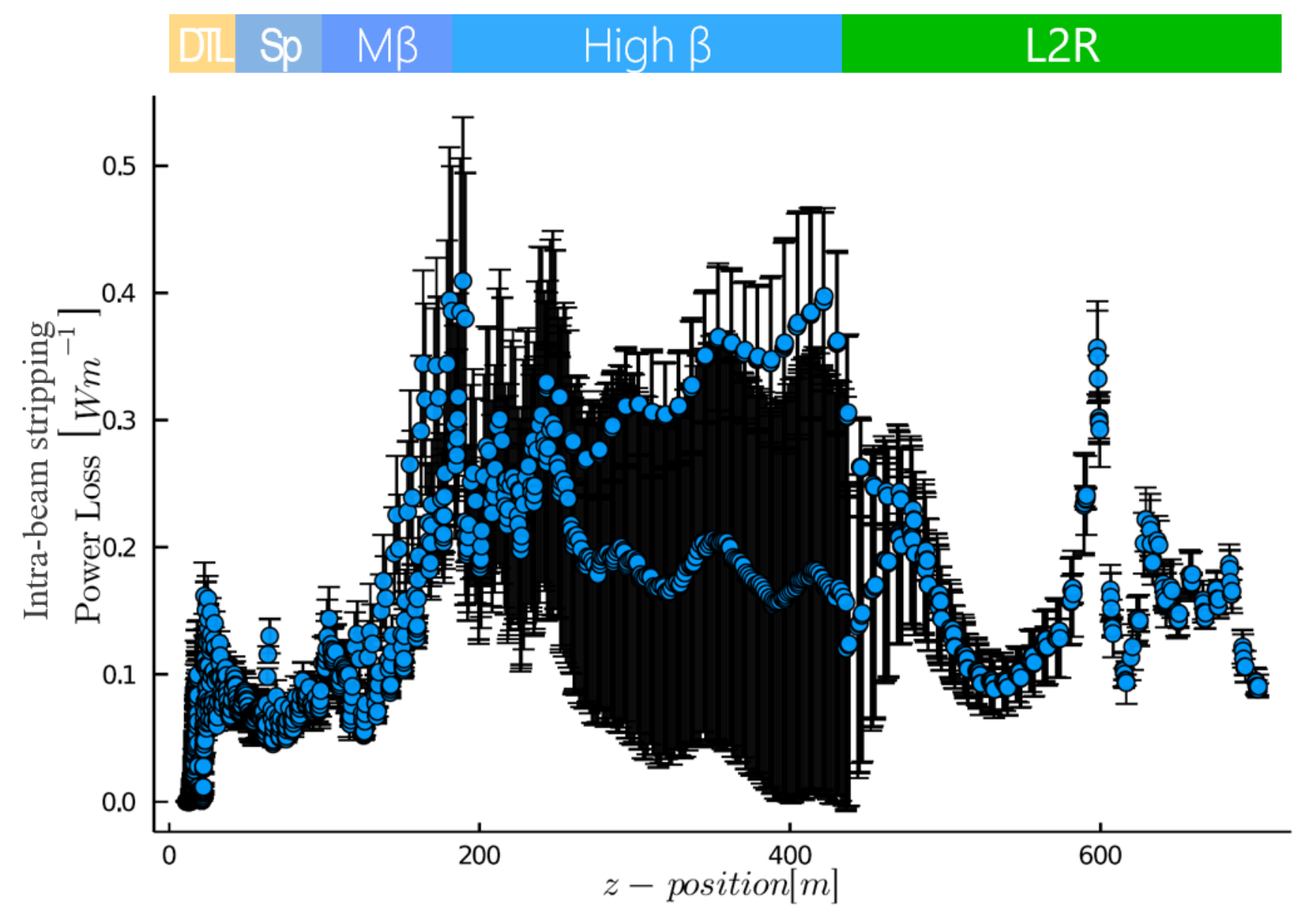}
\caption{Intra-beam stripping power loss along the linac and L2R line. Positions are given relative to the start of the DTL. The lattice sections are indicated along the top of the plot.}
\label{fig:IntraBeamStripping}
\end{center}
\end{figure*}

In the absence of accelerating cavities in the L2R line, the energy spread grows due to space-charge forces. Figure~\ref{fig:EnergyEnvelope} shows how the energy spread increases from $\pm 0.4$\,meV to $\pm 2.3$\,meV (1 sigma) along the L2R line. Introducing accelerating cavities towards the end of the L2R line, operated at the zero-crossing (i.e. at an RF phase of $-90^\circ$ from the RF peak), allows bunches to be rotated in the longitudinal phase space to significantly reduce the outgoing energy spread. Figure~\ref{fig:EnergyEnvelope} shows how using four such pi-mode structures (PIMS) with 4 MV/m \cite{Gerigk2009} towards the end of the L2R line brings the energy spread down from $\pm 2.3$\,meV to $\pm 0.6$\,meV.

\begin{figure*}[ht!]
\vspace*{-0.5\baselineskip}
\begin{center}
\includegraphics[width=0.8\columnwidth]{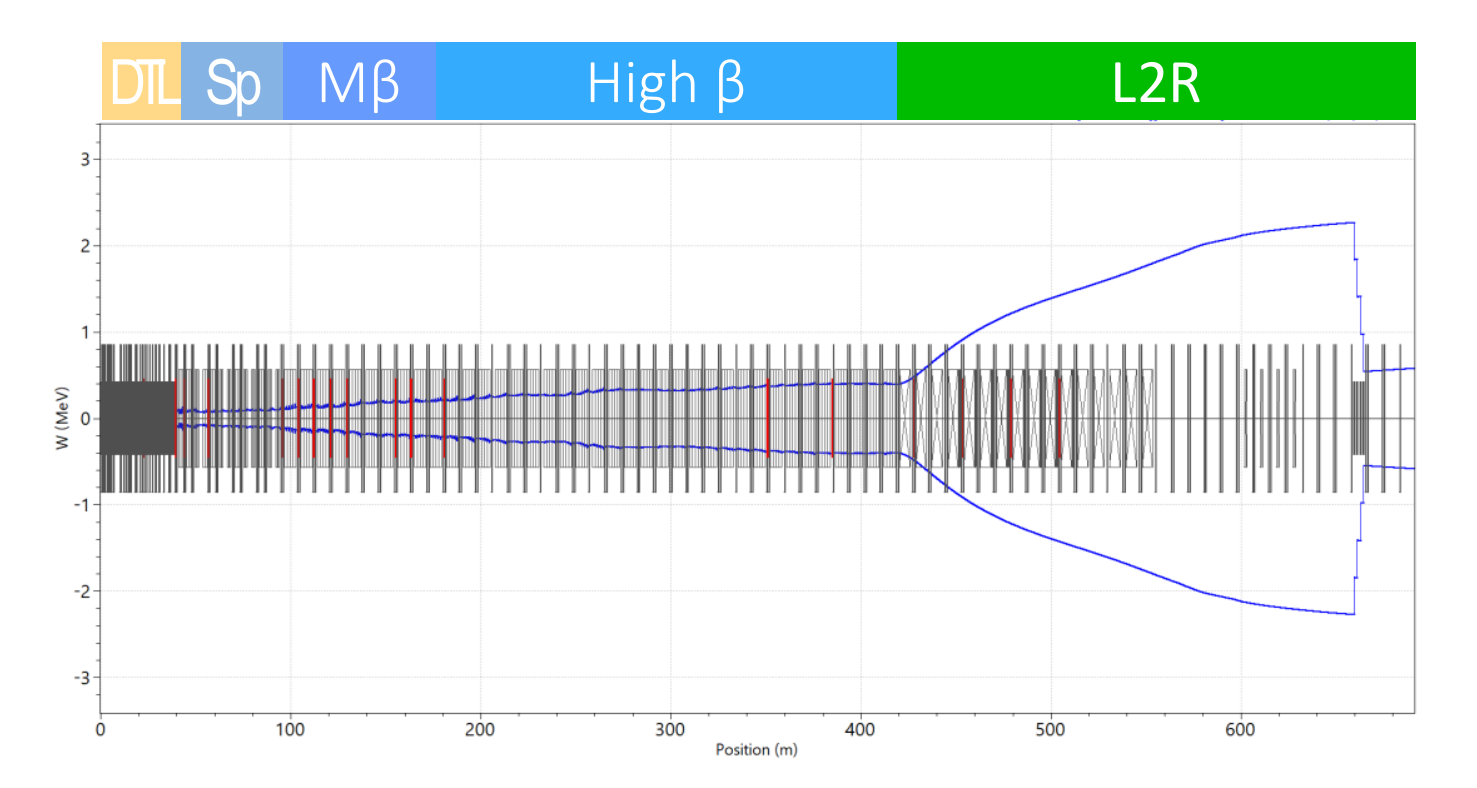}
\caption{Energy spread (1\,$\sigma$) along the linac and L2R line, with four cavities operated at the zero-crossing near the end of the L2R line. Positions are given relative to the start of the DTL. The lattice sections are indicated along the top of the plot.}
\label{fig:EnergyEnvelope}
\end{center}
\end{figure*}

\subsubsection{Energy Collimation}

Given the strong horizontal dispersion in the L2R line, its energy acceptance is $\pm 20$\,meV. When the energy error exceeds this limit, the beam is lost in the beam pipe, whose aperture is $\pm 50$\,mm. Therefore, beam energy collimation is necessary for the protection of beam line equipment.

For the results presented here, a multiparticle beam was tracked through an error-free lattice from the start of the DTL to the end of HBL line. This provided the Twiss parameters at the input to the L2R line. To test the energy acceptance of the L2R line, the longitudinal emittance was blown up from 0.446\,$\pi$\,mm\,mrad to 1000\,$\pi$\,mm\,mrad, and a beam distribution with a random longitudinal phase-plane ellipse was adopted in \textsc{TraceWin}.

The most effective beam-energy collimation locations were found to fall after the second and third horizontal dipole magnets in the L2R line, at 16.4\,m and 24.9\,m from the start of the L2R line, respectively. Given the strong correlation of particle offset on energy in the dispersive section of the L2R line, the phase advance between the collimators was found to be non-critical for the collimation efficiency. In order to distribute the energy deposition further, each collimator can be split into a triplet of closely spaced collimators. For example, three successive collimators with apertures of $\pm 12$, $\pm 10$ and $\pm 8$\,mm after the second dipole magnet have been tested; along with a further three successive collimators with apertures of $\pm 13$, $\pm 10.5$ and $\pm 8$\,mm located 24.9\,m downstream from the third dipole magnet. The collimators within each triplet are separated by 250\,mm.

The results are shown in Fig.~\ref{fig:EnergyCollimation}. The energy collimation keeps the full energy spread within the allowed $\pm 20$\,meV (Fig.~\ref{fig:EnergyCollimation}[a]), therefore maintaining the beam within the $\pm 50$\,mm beam pipe aperture (Fig.~\ref{fig:EnergyCollimation}[b]). Fig.~\ref{fig:EnergyCollimation}(c) shows how the energy deposition is shared amongst the collimators for this particular incoming energy distribution.

\begin{figure*}[ht!]
    \centering
    \begin{subfigure}[b]{0.55\textwidth}
        \centering
        \includegraphics[width=\textwidth,trim={0 0 0 0.35cm},clip]{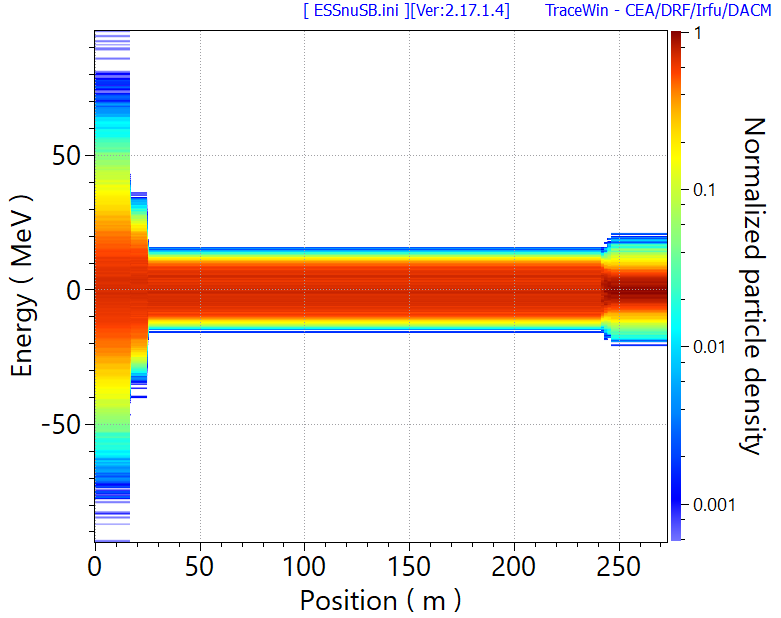}
        \caption{}
    \end{subfigure}
    \\
    \begin{subfigure}[b]{0.46\textwidth}
        \centering
        \includegraphics[width=\textwidth,trim={0 0 0 0.2cm},clip]{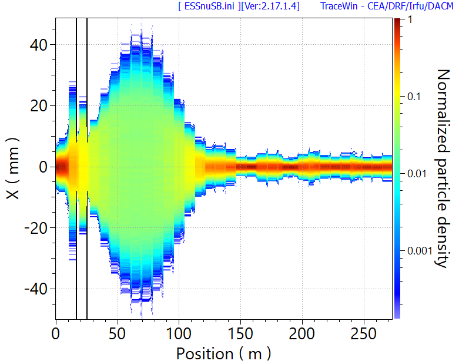}
        \caption{}
    \end{subfigure}
    \hfill
    \begin{subfigure}[b]{0.46\textwidth}
        \centering
        \includegraphics[width=\textwidth,trim={0 0 0 0.35cm},clip]{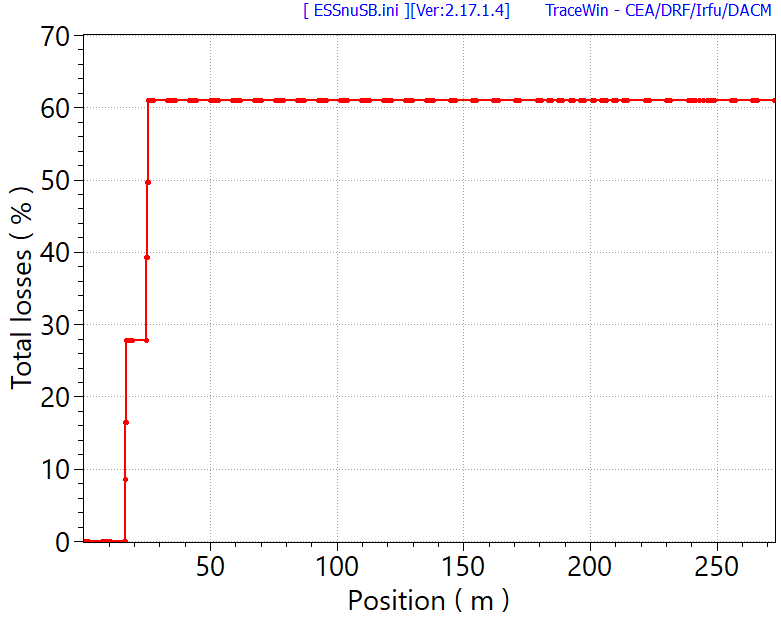}
        \caption{}
    \end{subfigure}
    \caption{(a) Energy and (b) transverse $x$ particle densities, and (c) integrated particle losses, for the L2R line with triplets of $x$-collimators after both the second and the third L2R dipoles.}
    \label{fig:EnergyCollimation}
\end{figure*}

The simulation does not include collimation for incoming--transverse position or angle errors. An additional collimation system to address these incoming transverse errors would most likely call for pairs of collimators separated by a phase advance of $90^\circ$ for both position and angle collimation \cite{Catalan-Lasheras:2001qmv}.

\subsection{Proton Driver Beam Dynamics Summary}
Simulations of the H$^-$ beam have been completed along the entirety of the linac both to the line-of-sight beam dump (not illustrated) and through the L2R transfer line (Fig.~\ref{fig:endtoend_linac}).
\begin{figure}[ht!]
    \centering
    \includegraphics[width=0.46\textwidth]{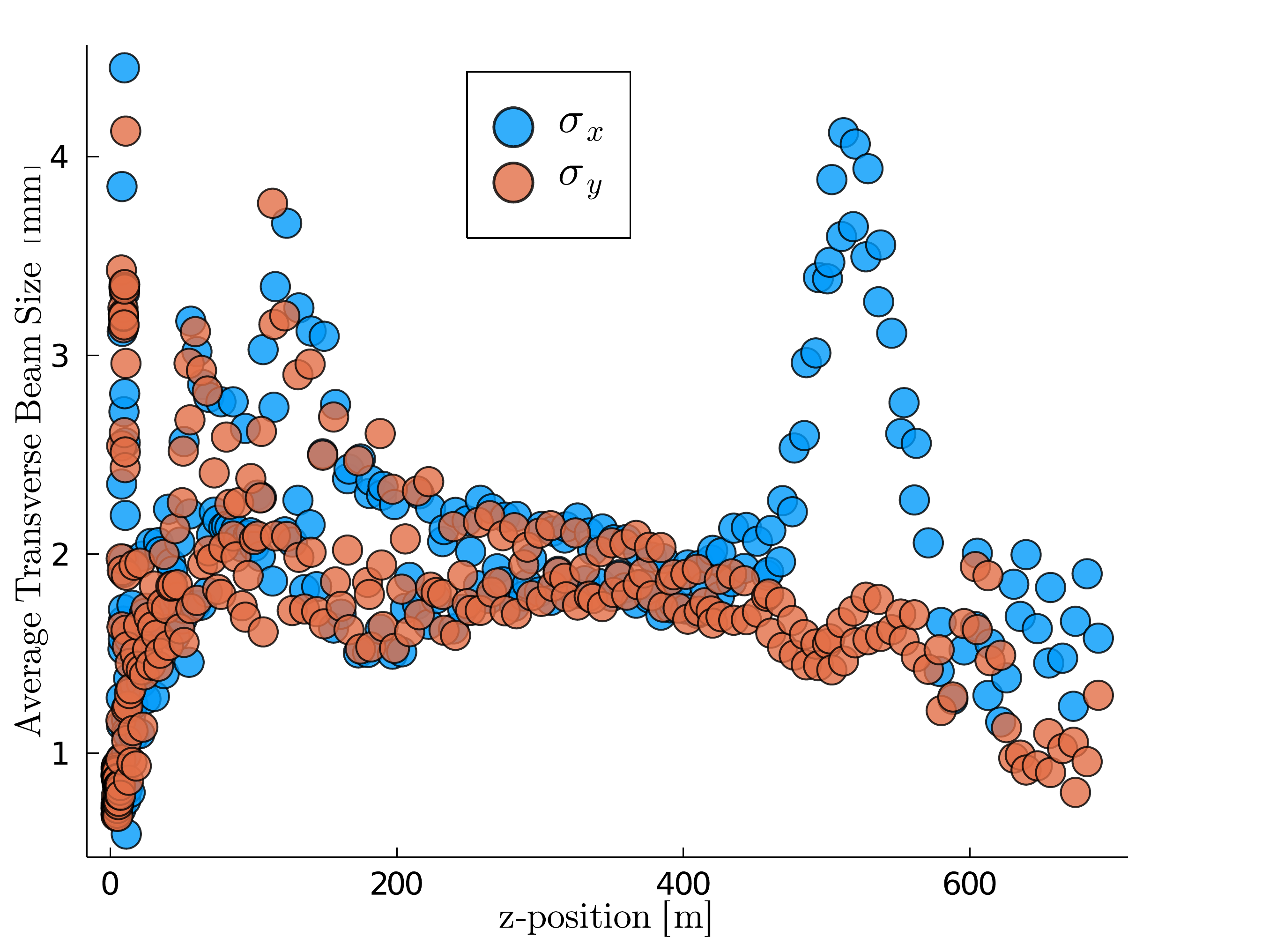}
    \includegraphics[width=0.46\textwidth]{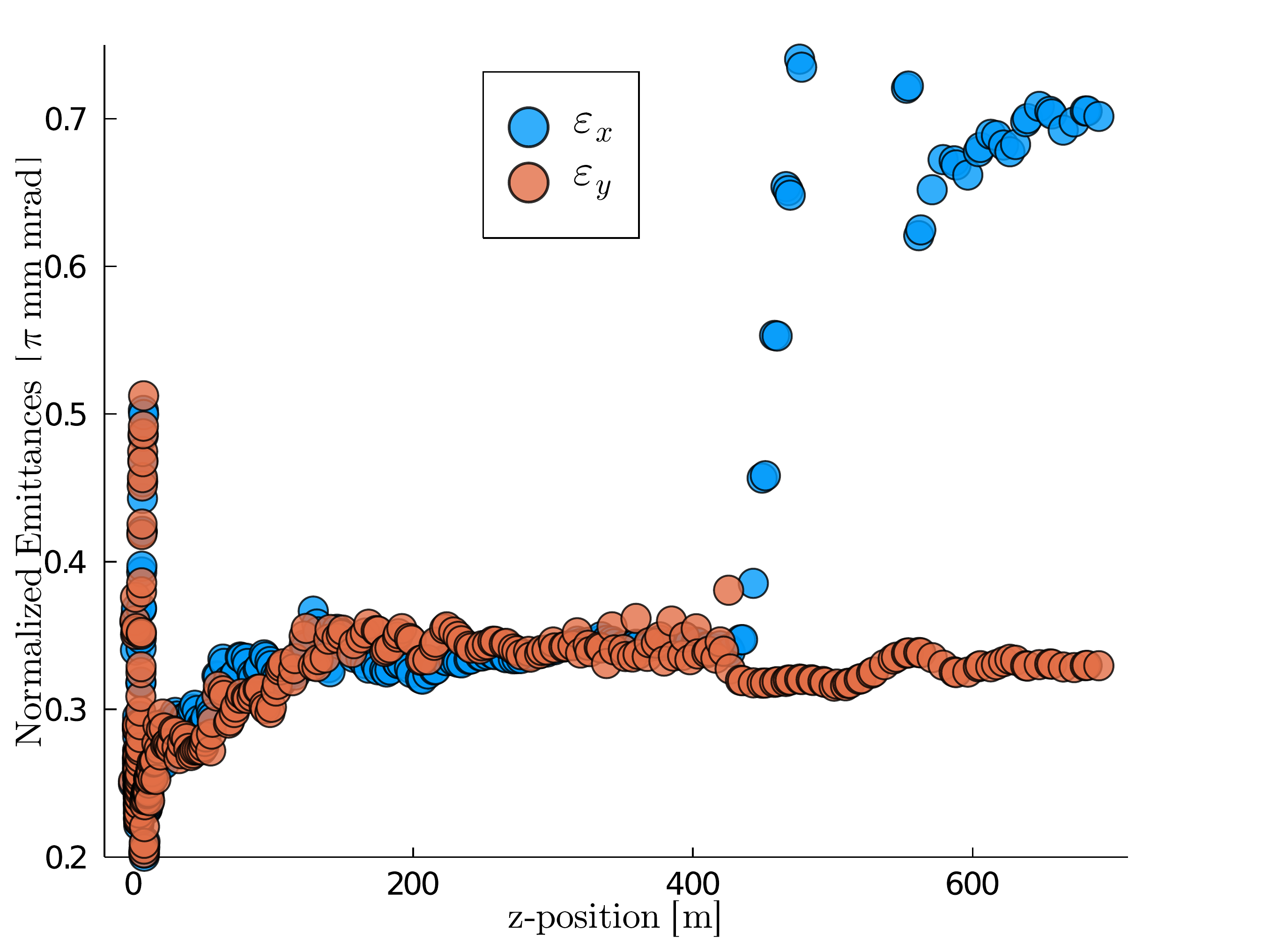}
    \caption{Average transverse beam sizes (left) and emittances (right) for the H$^-$ beam through the length of the linac and L2R transfer line.}
    \label{fig:endtoend_linac}
\end{figure}
This set of simulations was completed in \textsc{TraceWin}. Here, it is worth noting that normalized emittance levels throughout the linac are comparable with the nominal ESS proton linac at an average of $\epsilon{\sim}0.35~\mathrm{\pi\,mm\,mrad}$. These simulations demonstrate that beam transport is viable without the use of switching magnets throughout the linac. The blow-up seen for both size and emittance in the transfer line (beyond the 400-meter point) is intentional: this reduces intra-beam stripping (see Sections~\ref{sect:beam_losses} and \ref{sect:L2R}). The results here are strongly dependent on the H$^-$ source meeting design current and emittance requirements (roughly 80\,mA and 0.25\,$\pi$\,mm\,mrad, respectively). The optics for this beam line were optimised for both protons and H$^-$, such that an each species shared an emittance growth beyond nominal of approximately 2--5\% from the MEBT to the end of the linac.

\subsection{Beam Losses}
\label{sect:beam_losses}

\subsubsection{\label{sec:level1}Introduction}

The use of H$^-$ is essential in the design of the accumulator ring, where charge-exchange injection strips both electrons and puts them in orbit adiabatically with already-inserted protons -- a process which effectively circumvents Liouville's theorem by allowing the beam's phase-space density to be increased. However, at any point between the ion source and injection point, H$^-$ can be easily stripped of its outer electron a owing to the electron's low binding energy (0.75~eV~\cite{armstrong_empirical_1963}). This makes H$^-$ a particularly difficult species in terms of beamline transport.

Stripping leads to the production of electrons, which have negligible energy; and neutral H$^0$ atoms. Because of the low energy of the stripped electron, the H$^0$ carries nearly all the energy of the original H$^-$. Because it is uncharged, the H$^0$ follows a drift trajectory, eventually striking a machine-element wall or a line-of-sight beam dump. 

There are four main types of H$^-$ stripping: residual gas, blackbody radiation, Lorentz-force or field-induced, and intrabeam stripping (IBSt).

For the linac, IBSt and Lorentz stripping are of primary concern in terms of component activation, as these are prevalent at high energies, where deposited ionising radiation scales with power loss from the beam and can reach conventional limits. For the linac-to-ring (L2R) transfer line, all four types of stripping must be evaluated.

At beam energies greater than ${\sim}$100\,meV, activation of machine components could become a concern if the loss values exceed acceptable limits. The beam-loss limit at ESS has been set to 1~ W/m, from a commonly accepted standard~\cite{Mohkov:2000ue}. This ensures a maximum 1\,msv/h ambient dose rate due to activation at 30\,cm from a surface of any given accelerator component, after 100 days of irradiation and 4 hours of cool-down~\cite{Tchelidze2019}.

Since the addition of the H$^-$ beam will increase the total duty cycle of the linac to about 8\% (depending on the pulsing schemes), and with the loss for both beams in total kept below 1\,W/m, the beam loss in the linac for H$^-$ is limited to 0.5\,W/m. In the L2R, where only H$^-$ is present, this limit can be relaxed to 1\,W/m.

Simulations predict the loss from the proton beam to be well below 1\,W/m (estimated at roughly 0.1\,W/m). If this is confirmed during beam commissioning and operation of the proton beam, the loss limit for the H$^-$ beam may be relaxed provisionally.

\subsubsection{Residual Gas Stripping}

This type of stripping occurs when an H$^-$ ion collides with a neutral gas molecule in the beam pipe. It can be modelled in terms of fractional beam loss of total particle count $N$ per unit length $L$ \cite{armstrong_empirical_1963,gillespie_high-energy_1977,carneiro_numerical_2009,raparia_2016}: 

\begin{align}
    &\tau = \sum_i \frac{1}{\rho_i \sigma_i \beta c}
    \nonumber \\
    &\frac{\Delta N}{L} = \frac{1}{\tau \beta c}~~~,
\end{align}
In this formula, $\tau$ is the H$^-$ lifetime, $i$ is the residual gas species with a molecular density $\rho_i$ and a scattering cross section of $\sigma_i$, and $\beta c$ is the particle velocity.

This type of stripping is suppressed for high energies and high vacuum levels. For ESS$\nu$SB, it is then only a major concern as the beam leaves the ion source into the LEBT and neutral gas is injected as a space-charge compensation measure for avoiding emittance blow-up. Depending on the gas species used and the degree of space-charge compensation needed, the gas pressure required for well-saturated space charge compensation can be roughly an order of magnitude below where stripping causes substantial beam loss \cite{valerio-lizarraga_negative_2015}, or high enough to cause serious beam loss~\cite{raparia_2016}. It should be stressed, however, that the energies here are so low that activation and prompt radiation is not significant. 

In the L2R transfer line, ultra-high vacuum levels are no longer required. A high vacuum is still required to prevent ordinary gas scattering. \textsc{TraceWin} has built-in capability for estimating both gas scattering and gas stripping, and predicts power loss from stripping to be below 0.05\,W/m at nominal ESS vacuum levels~\cite{ravelli:2019_ESS_vacuum}.

\subsubsubsection{Double Stripping in Residual Gas}
A secondary effect that must be considered here is that a proportion of the ions may also be fully stripped to protons; this is often referred to as double stripping. Such protons can ultimately cause activation or structural damage, particularly if they are inadvertently accelerated to higher energies.

Some success has been reported in J-PARC in Tokai, Japan using a chicane after the chopper in the MEBT to divert protons produced by double stripping. Nevertheless, their recommendation is to use a quadrupole FODO lattice instead of solenoid focusing in the LEBT to avoid proton capture at the outset~\cite{ikegami_2012_beamcommiss}.

If allowed to reach higher energies, double-stripped protons can even survive a frequency jump from the normal to superconducting sections of the linac: they do not survive the doubled frequency at SNS, but they do survive tripled frequency at J{\-}PARC~\cite{maruta_beamloss_2011,plum2013challenges,plum_beam_2014,plum_beam_2016}. Thus, double stripping is not likely to lead to the production of high-energy protons at ESS$\nu$SB, where the frequency is doubled between normal and superconducting linac sections. However, long-term cumulative damage from undetected double stripping ``hot spots'' as reported in~\cite{raparia_double_proton_fixed} may still be a risk, and should be prevented.

A general preventative measure to reduce the risk of such damages could be collimation of either double-stripped protons or H0 particles (using stripping foils in the latter case). 

\subsubsection{Blackbody Radiation Stripping}

The infrared photons emitted by the beam pipe or other machine parts, when Lorentz-shifted into reference frame of the beam, can cause significant H$^-$ stripping. 
\begin{figure*}[ht!]
\centering
  \includegraphics[width=0.75\textwidth]{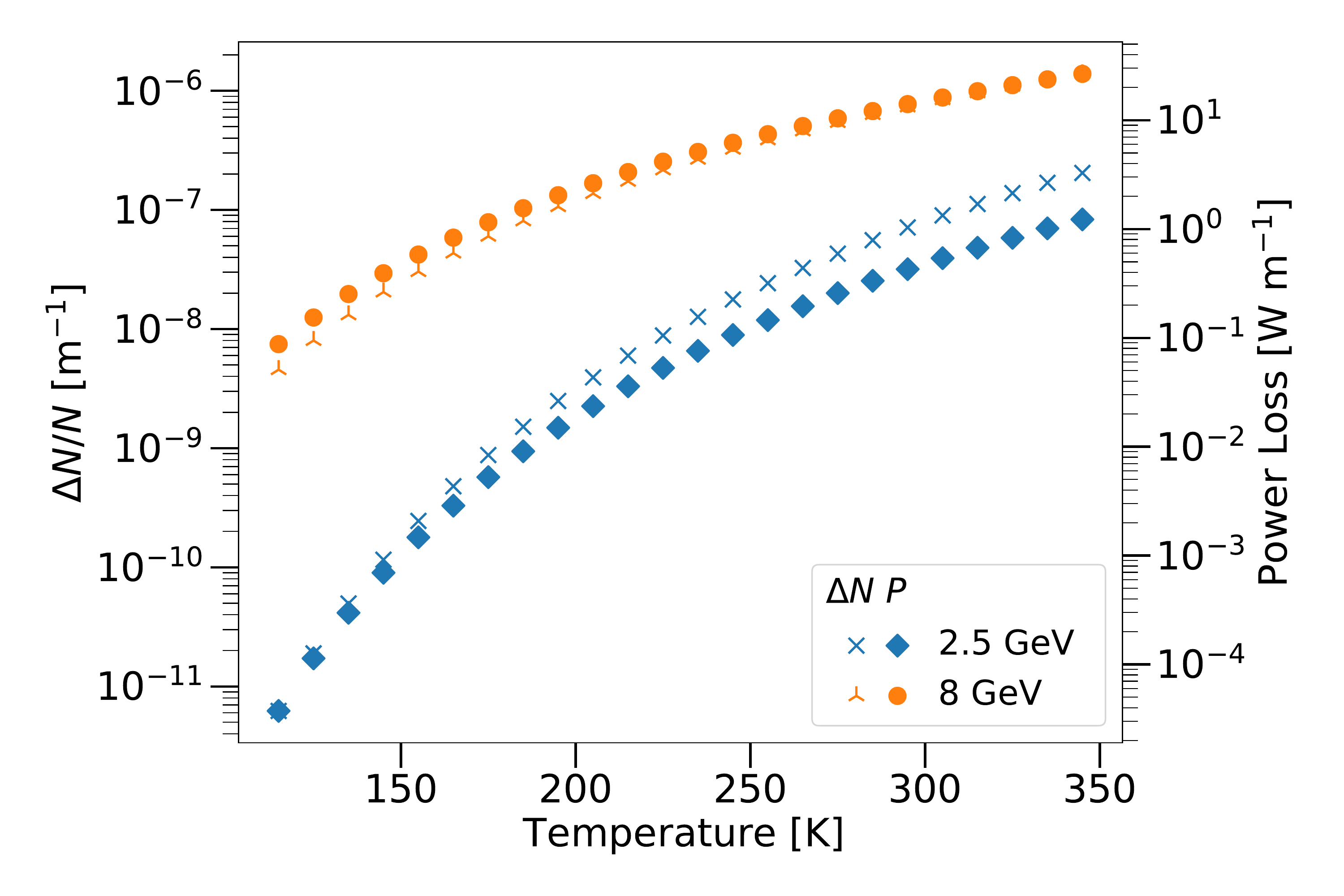}
  \vspace*{-5 mm}
  \caption{Temperature dependence of blackbody-radiation stripping for 60\,mA beams at a 4\% duty cycle for the ESS$nu$SB beam energy of 2.5\,GeV, with curves for a more restrictive 8\,GeV case shown as reference. Left and right-hand markers in the legend correspond to fractional loss rate and beam power loss scaled by the left and right-hand $y$ axes, respectively. Reproduced from~\cite{folsom:2021_hmin_strp}.}
  \label{fig:blackbody_v_T}
\end{figure*}

Here, the fractional loss per unit length is used as a figure of merit, which can be taken as~\cite{carneiro:2003_beam,carneiro_numerical_2009,bryant_atomic_2006, du_photodet_1988}

\begin{align}
\label{eq:BB_eqs}
    &\frac{\Delta N}{N}\frac{1}{L} 
    =\int_{0}^\infty d\epsilon \int_0^\pi d\alpha \frac{d^3 r}{d\Omega d\nu d l} 
    \\ \nonumber
    &\frac{d^3 r}{d\Omega d\nu d l} =\frac{\left(1+\beta cos\alpha \right)n(\nu,r)\sigma(v')}{4\pi\beta}~~~,
\end{align}

\noindent where $h$ is Planck's constant; $n(\nu,r)$ is the spectral density of thermal photons (a function of frequency $\nu$, and radius $r$); $\alpha$ is the angle between the incoming photons and the beam; and the factor $\epsilon$ is defined as $\epsilon=h\nu/E_0$, with $E_{0}=0.7543~\mathrm{eV}$ being the electron binding energy for $H^-$. The stripping cross section in the beam frame is represented by $\sigma(\nu')$. This integral can be broken into factors which can be evaluated analytically and numerically, for further detail, see~\cite{armstrong_empirical_1963,carneir_beamsdoc,bryant_atomic_2006,herling_blackbody_2009}.

In the linac, the low temperature of the superconducting cryomodules suppresses blackbody stripping. However, the L2R transfer lines has no such nominal cooling needs, and as shown in Fig.~\ref{fig:blackbody_v_T}, beam power loss from blackbody stripping is roughly 0.4\,W/m for a 2.5\,GeV beam at room temperature.  

Cooling the L2R to 100--200~K would limit blackbody stripping, and is expected to be a cost-effective solution, given the existing cooling infrastructure serving the nominal ESS linac.

However, low-emissivity beam-pipe coatings may a viable alternative; Eq.~(\ref{eq:BB_eqs}) and the results in Fig.~\ref{fig:blackbody_v_T} assume a 100\% emissivity, which is expected to scale linearly with blackbody stripping. Emissivity is dependent on surface characteristics including roughness and susceptibility to oxidation. For stainless steel, an emissivity of up to 0.4 can be observed, whereas copper, nickel, or gold can have values as low as 0.03~\cite{vollmer:2010_handbook-thermal-imaging,huang:2010,setien:2014}. 

A coating of TiN is often used in accelerators to reduce secondary electron yield and may be a strong candidate for blackbody stripping reduction, owing to its mechanical and thermal properties~\cite{klocek:2017_handbook-infrared,YUSTE:2011_1784,ZHAO:2008_1272}. Similarly, the non-evaporable getter (NEG) TiZrV coating is also effective in reducing secondary-electron emission, which, depending on surface roughness, may also indicate a low infrared emissivity~\cite{Hilleret:2000_multipacting,pimpec:2004_SecondaryElectronYield}.

It should be noted that such metal-based coatings have emissivities in the 0.2 range, but similar carbon-based coatings should be avoided, with emissivities closer to 0.8~\cite{taborelli:2014_coatings}. A variety of alternatives may also be worth further study, as discussed in \cite{chiba:2005_low-emissivity, hwang_high_2019}.

\subsubsection{\label{LZST}Lorentz Stripping}
A transverse external magnetic field either from focusing quadrupoles or from dipoles is Lorentz transformed into an electrical field as~\cite{folsom:2021_hmin_strp,jackson:1999_classical}

\begin{equation}
    \label{lor_E_B}
    |E_\perp| = \beta\gamma c |B_\perp|~~~,
\end{equation}

\noindent where $|B_\perp|$ and $|E_\perp|$ are the originating magnetic field and beam-frame electric field, respectively; and where $\beta$ and $\gamma$ are the relativistic Lorentz factors while $c$ is light speed in vacuum. This relation indicates that H$^-$ beams at GeV energies can have problematic stripping from the field gradients needed for focusing and steering. 

For the simpler case of dipole magnets (approximated as a uniform field near the beam axis) the Lorentz stripping probability per unit length can be modelled as \cite{keating:1995_electric-field,jackson:1999_classical}
 \begin{align}
    \frac{\Delta N}{N}&\frac{1}{L} = \frac{|B_{\perp}|}{A_1}\operatorname{exp}\left({-\frac{A_2}{\beta \gamma c \left|B_{\perp}\right|}}\right)
    \\ \nonumber
    A_1 &= 3.073 \times 10^{-6}~\mathrm{s\,V / m}
    \\ \nonumber 
    A_2 &= 4.414 \times 10^9~\mathrm{V / m}~~~.
    \end{align}

\noindent where $A_1$ and $A_2$ are empirical-fit parameters. With this equation, calculating Lorentz stripping for bunches traversing dipole magnet fields is straightforward. Quadrupoles, however, have a transverse field-strength dependence which makes particles in the outer halo or misaligned beams more likely to undergo stripping. This probability of Lorentz stripping through a quadrupole can be thus calculated as

\begin{equation}
P = \int^{2\pi}_{0}\int^r_0 f(r',\sigma)\frac{\Delta N}{N} \frac{1}{L} r' dr d\theta~~~,
\label{lor_quad}
\end{equation}

\noindent where $L$ is the magnet length and $f(r',\sigma)$ is a radially symmetric particle density distribution \cite{blaskovickraljevic:ipac2021-tupab175}. The general result is that a small RMS transverse beam size $\sigma_{\perp}$ advantageous. However, the opposite is true for IBSt. This converse dependence on beam size will be analysed shortly.  

\subsubsection{Intra-beam Stripping (IBSt)}

The phenomenon of IBSt arises from the collisions of H$^-$ ions within a bunch, and can be the predominant form of stripping in high-intensity H$^-$ linacs~\cite{shishlow:2012_IBSt}. The fractional loss rate per length, in the laboratory frame, is given as 

\begin{equation}
\frac{\Delta N}{N}\frac{1}{L} = 
\frac{N\sigma_{max_{IB}}
\sqrt{\gamma^2\theta_x^2+\gamma^2\theta_y^2+\theta_z^2}
}%
{8\pi^2\gamma^2\sigma_x\sigma_y\sigma_z}F\left({\gamma\theta_x,\gamma\theta_y,\theta_z}\right),
\label{ibs_lossperm}
\end{equation}
\\
\noindent  where $F\left({\gamma\theta_x,\gamma\theta_y,\theta_z}\right)$ is the shape function for momentum spreads with a weak dependence on its parameters from 1 to 1.15. (For more detail, see~\cite{lebedev:2012_IBSt,cohen:1986_stripping,folsom:2021_hmin_strp}). The factor $\sigma_{max_{IB}}$ is the maximum stripping cross-section; this is approximately

\begin{equation}
\nonumber
\sigma_{max_{IB}} \approx \frac{240 a_{0}^{2} \alpha_{f}^{2} \ln{\left(1.97\frac{\alpha_{f} + \beta}{\alpha_{f}} \right)}}{\left(\alpha_{f} + \beta\right)^{2} }~~\leq~~4\times 10^{-15}~\mathrm{cm}^{-2},
\end{equation}

\noindent where $\alpha_f$ is the fine-structure constant, $a_0$ is the Bohr radius, and $\beta=v / c$ is the velocity between ions. The bunch sizes and angular momentum spreads (RMS) are then respectively defined as
\begin{equation}
\nonumber
\sigma_{x,y} = \sqrt{\epsilon_{x,y}\beta_{x,y}}
    \quad\quad
    \theta_{x,y} = \sqrt{\frac{\epsilon_{x,y}}{\beta_{x,y}}}~~,
\end{equation}
\noindent where the factors $\epsilon_{x,y}$ and $\beta_{x,y}$ are the transverse Twiss parameters.

\begin{figure*}[ht!]
\centering
  \includegraphics[width=0.75\textwidth]{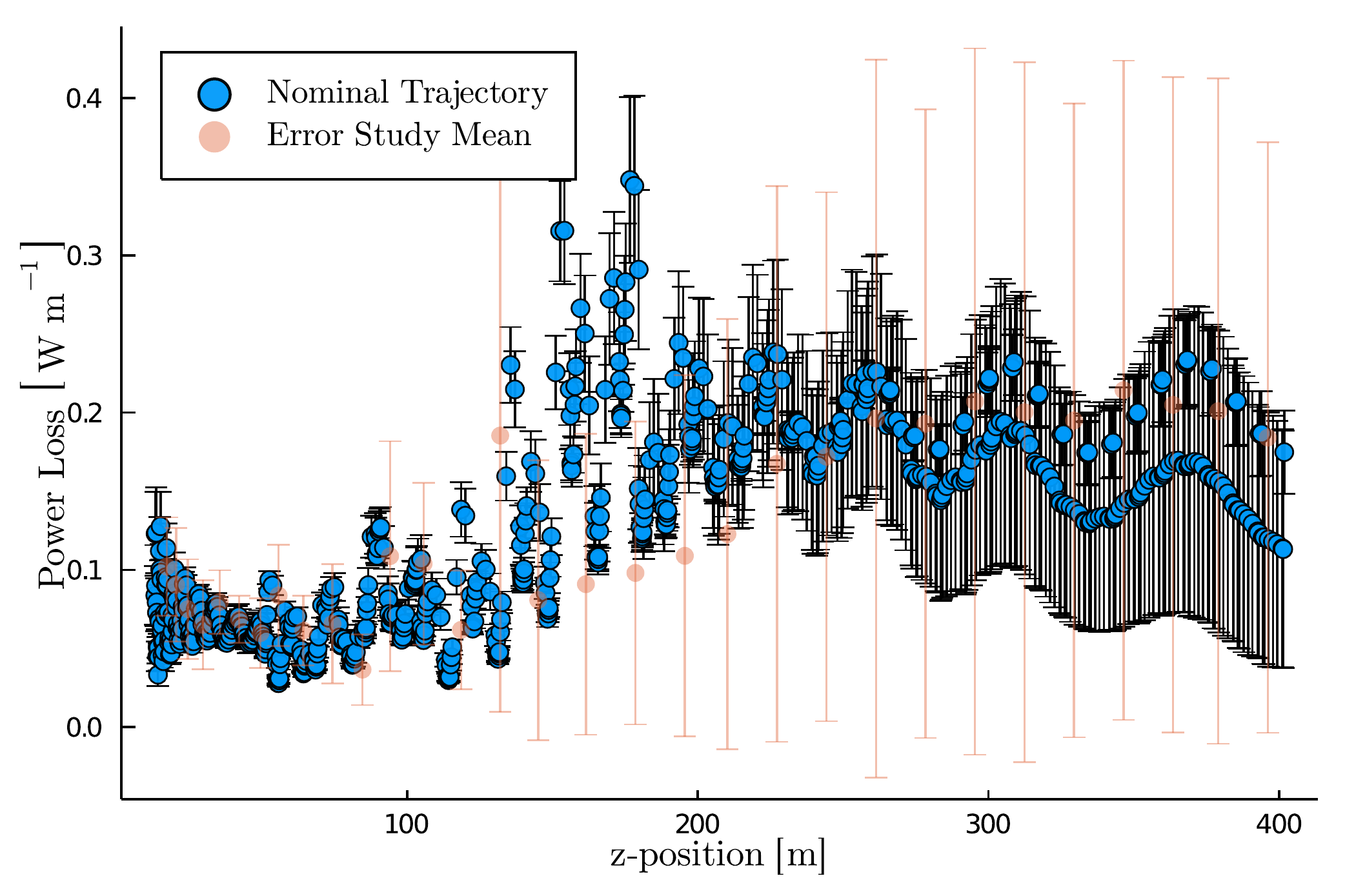}
  \caption{Intra-beam stripping (IBSt) power losses for a nominal trajectory and a corresponding error study for the ESS$\nu$SB linac (H$^-$ only). Static and dynamic machine errors are incorporated along with beam misalignment. Standard deviations are from the cumulative trajectories of 100 trials.}
  \label{fig:IBSt_power_errstud}
\end{figure*}

Since Eq.~(\ref{ibs_lossperm}) has an inverse dependence on bunch length $\sigma_z$, IBSt can be limited by maximising length -- minimising momentum spread has a similar effect. One can also infer dependencies on accelerating phase and RF frequency: both can be reduced to mitigate stripping. 

Relaxed transverse focusing also limits IBSt, and has been shown to be highly effective \cite{shishlow:2012_IBSt}. In doing so, one may use accelerating cavity defocusing strength as a limiting parameter for a minimally focused beam~\cite{lebedev:2012_IBSt}.

Figure~\ref{fig:IBSt_power_errstud} shows the power loss rate per meter derived from Eq.~(\ref{ibs_lossperm}) and corresponding power loss for a 5\,MW, 2.5\,GeV H$^-$ linac. These calculations were done as a post-processing step, with trajectories outputted from \textsc{TraceWin}. This post-processing approach is feasible since the loss rates from IBSt are orders of magnitude below that of a bunch population and do not affect the overall dynamics of the bunch (i.e. beyond the LEBT, it is \textit{activation} from stripping losses that is critical, not whether such losses affect the delivery of a desired beam current).

Figure~\ref{fig:IBSt_power_errstud} also shows error study trajectories, which account for static and dynamic dipole, quadrupole, and cavity errors, as well as beam misalignments. (Specifically: 0.1\,mm and 10\,mrad static errors for the beam position and momentum, respectively; 0.1\,mm for displacement, 1$^\circ$ rotation, and 0.5\% gradient errors for quadrupoles; and 1\% field and 0.5\% phase error for cavities.)

\subsubsection{H$^0$ Traversal and Power Deposition}
For the simulation underlying Fig.~\ref{fig:power_dep_w_err}, the neutral H$^0$ created by IBSt are tracked from the instant they are stripped to when they collide with the beam pipe or other machine element. As these particles are not affected by external fields, this calculation could be performed using an elementary trajectory integrator.

Error studies were also done for this set of power losses using the same error parameter set as above.  Comparing these results with Fig.~\ref{fig:IBSt_power_errstud}, one can see a significant improvement in power deposited versus instantaneous power lost.

\begin{figure*}[ht!]
\centering
  \includegraphics[width=0.75\textwidth]{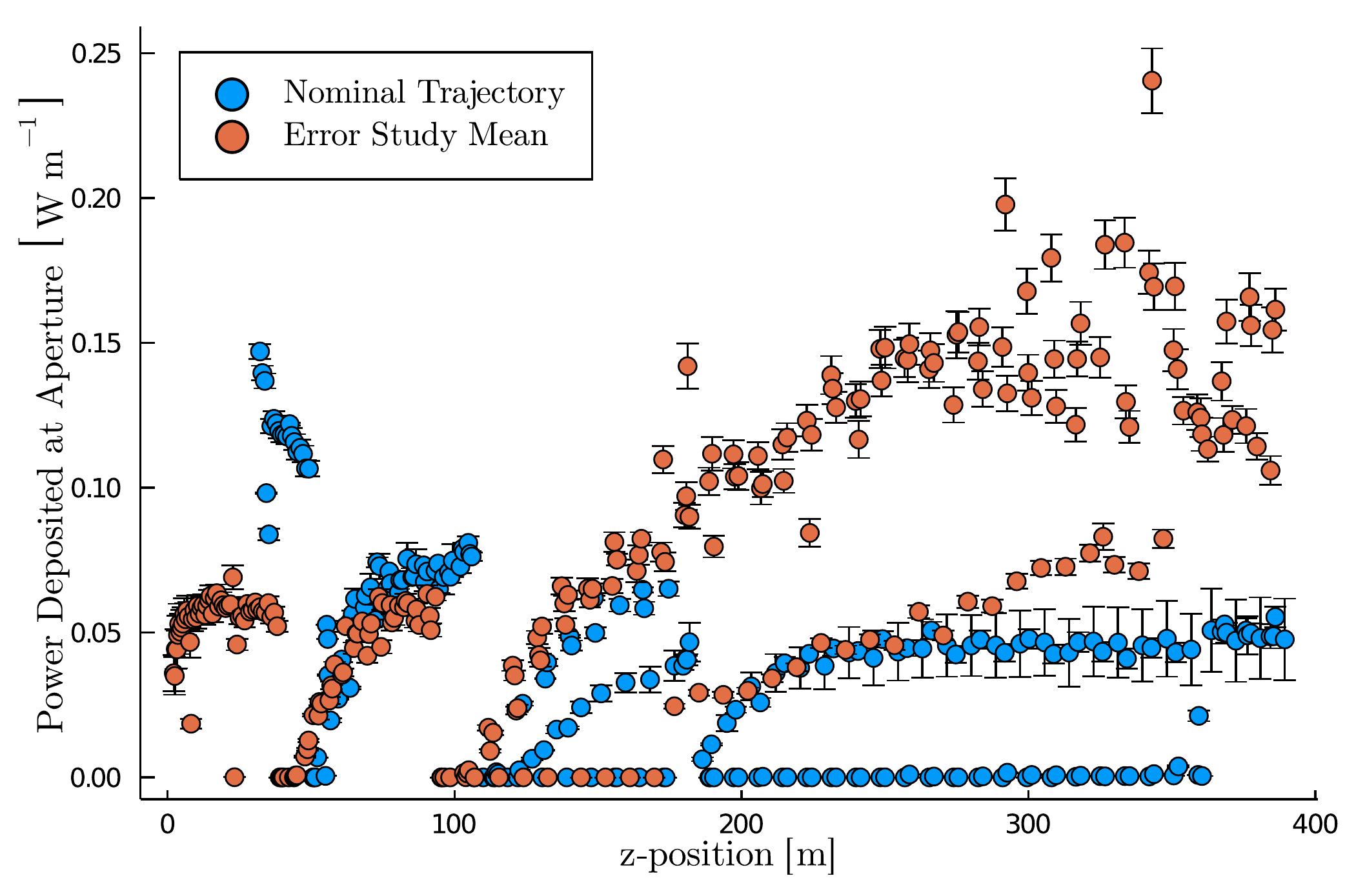}
  \caption{Power deposition of H$^0$ into beam-pipe and machine element walls for a 62.5{\thinspace}mA H$^-$ beam accelerating to 2.5\,GeV for the ESS$\nu$SB linac. Only IBSt is accounted for.}
  \label{fig:power_dep_w_err}
\end{figure*}

Figure~\ref{fig:H0distance} further illustrates this notion, showing the distances travelled by the stripped H$^0$ particles before reaching a collision point (with a remaining $\sim$50\,W deposited into the beam dump at the end of the linac). However, this advantage does not come in to play for the L2R transfer line: in this case the H$^0$ particles will collide shortly after passing the nearest bending dipole.

\begin{figure*}[ht!]
\centering
  \includegraphics[width=0.75\textwidth]{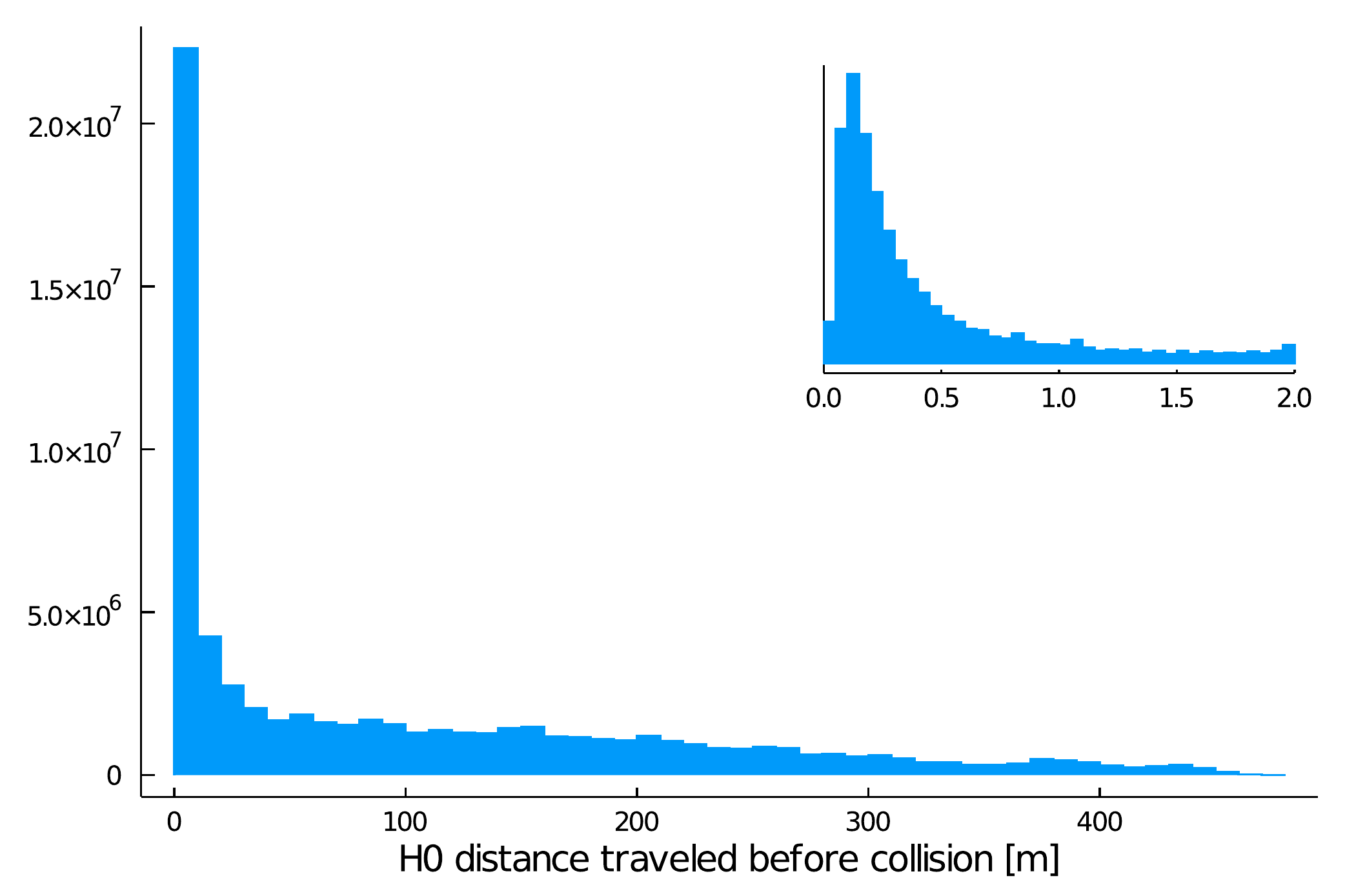}
  \caption{Traversal of H$^0$ particles from the point of IBSt to collision with the nearest machine-element wall or aperture for the ESS$\nu$SB linac. Inset shows the depositions occurring within 2\,m of stripping.
  }
  \label{fig:H0distance}
\end{figure*}

\subsubsubsection{\label{beam_params} Optics, Emittance, and Optimization}

For quadrupoles, as mentioned above, relaxed quadrupole focusing can improve the IBSt rate while being detrimental for Lorentz stripping, as particles furthest from the beam axis are closer to the magnet pole tips (specifically, the Hamiltonian describing a quadrupole's magnetic vector potential has quadratic dependence on position, extending radially outward from the magnets' transverse centre).

Figure~\ref{fig:lor_v_ibs} shows the inter-dependence for these two types of H$^-$ stripping, along with blackbody stripping, for a dummy FODO lattice of 20 cells with buncher cavities. Figure~\ref{fig:lor_v_IBS_quadgrad} extrapolates this result to quadrupole strength. An energy of 2.5\,GeV and current of 62.5\,mA was used for all test points.

This analysis follows Eq.~(\ref{ibs_lossperm}) for IBSt, and Eq.~(\ref{lor_quad}) adapted to a Gaussian distribution for Lorentz stripping; this is integrated numerically over a radius $r$ as:
\begin{equation}
    \frac{\Delta N}{N}\frac{1}{L} = 
    \int_{0}^{3 \sigma} \frac{ \sqrt{2} G r^{\prime}}{2 \sqrt{\pi} A_{1} \sigma}
    \operatorname{exp}\left[-\frac{\left(-\mu+r^{\prime}\right)^{2}}{2\sigma^{2}}\right] \operatorname{exp}\left[-\frac{A_{2}}{G \beta\gamma c}\right]
     d r~~~.
\end{equation}
A misalignment of $\mu = 1.5$\,mm is used in Fig.~\ref{fig:lor_v_ibs} for the upper error limit; and the factor $G$ is quadrupole field gradient in T/m. The blackbody-stripping results are based on Eq.~(\ref{eq:BB_eqs}) convolved to a Gaussian beam.

For this test, different trials were performed with varied quadrupole strengths and gap voltages over a phase-advance range of 1--90$^{\circ}$. Beam parameters for each run were matched to the increasing phase advance, with longitudinal-to-transverse emittance ratios also varied from 0.5 to 2 at each phase-advance step, giving a variety of bunch sizes for each test lattice.

This result is \textit{not} specific to the ESS$\nu$SB lattice as-is, since the present baseline keeps all optics identical for protons and H$^-$ where possible, and thus favours a smaller beam size. However, these results provide a benchmark for optimisation, should relaxed quadrupole strengths become necessary to reduce IBSt in the linac. Thus, at present, Lorentz stripping is expected to be negligible along the ESS linac, which has an average transverse beam size of 2.5\,mm for both particle species. For H$^-$ in the L2R transfer line, this size is increased to 4.0\,mm to further reduce IBSt as illustrated in Section~\ref{sect:L2R}; here, Lorentz stripping in quadrupoles remains low, effectively nearing the optimised crossing point for the IBSt and Lorentz stripping curves shown in Fig.~\ref{fig:lor_v_ibs}.

\begin{figure*}[ht!]
\centering
  \includegraphics[width=0.7\textwidth]{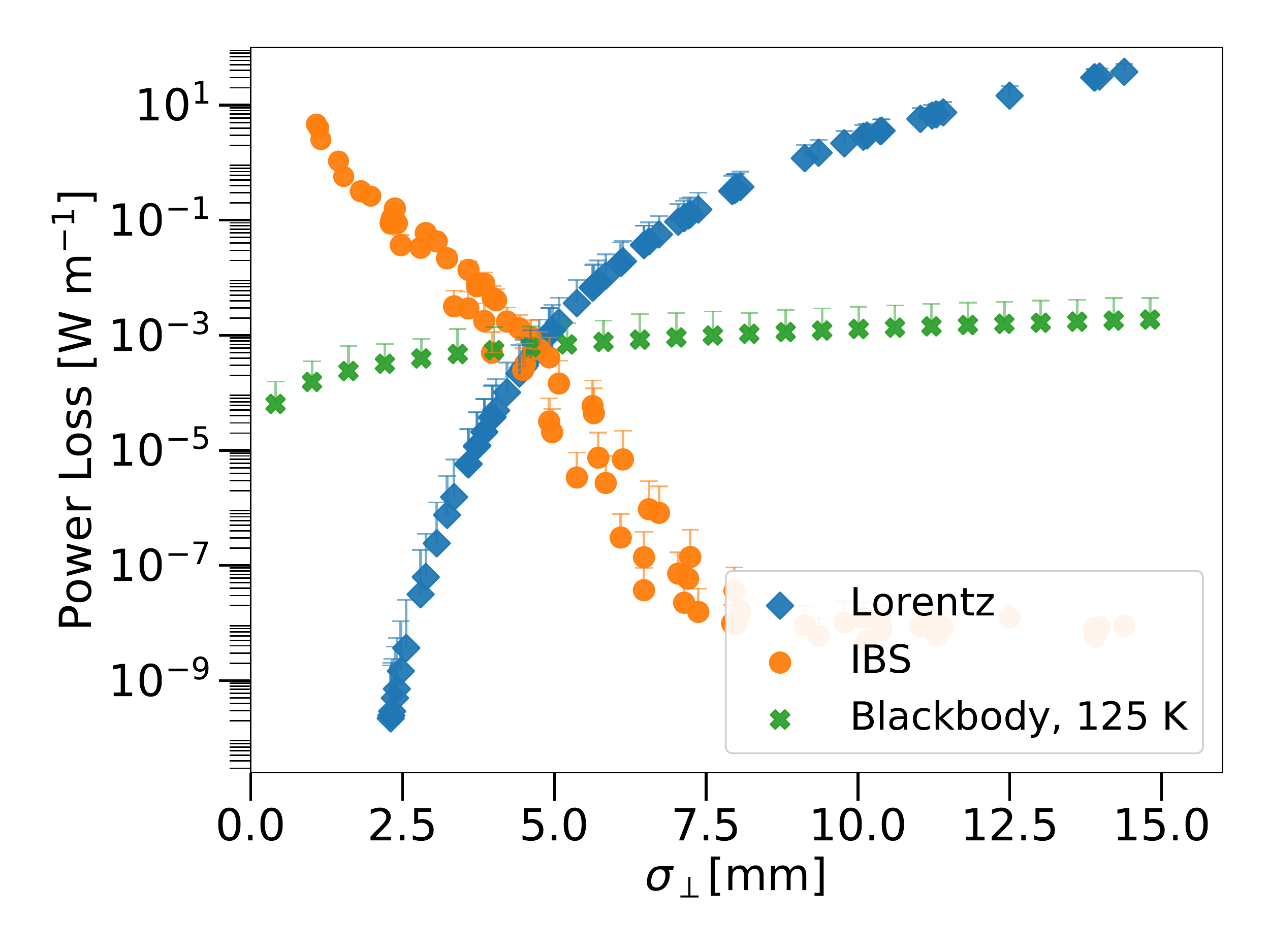}
  \vspace*{-5 mm}
  \caption{Dependence of blackbody, IBSt, and Lorentz stripping (quadrupoles only) on average transverse beam size $\sigma_\perp$ for a 2.5\,GeV, 62.5\,mA beam traversing toy FODO lattices (6${\thinspace}$m~$\times$~20 cells) of one quadrupole pair and one bunching gap per cell. Beam parameters are determined by setting phase advance and solving for optimum inputs. A range of 1--90$^\circ$ phase advance runs gives the shown range of beam sizes. Blackbody stripping is simulated separately, assuming a constant $\sigma_\perp$ for each point.}
  \label{fig:lor_v_ibs}
\end{figure*}

\begin{figure*}[ht!]
\centering
  \includegraphics[width=0.7\textwidth]{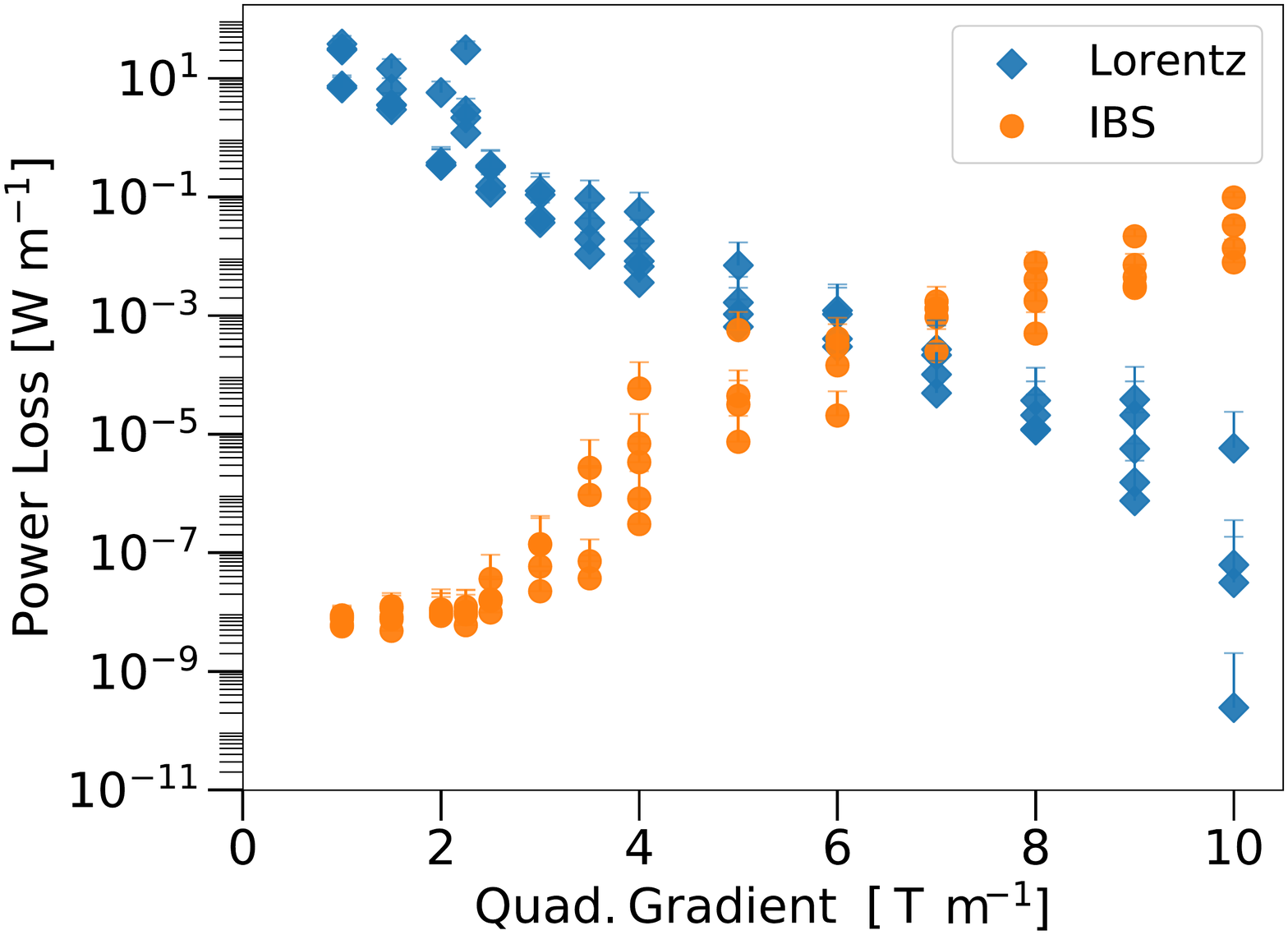}
  \caption{Dependence of IBSt and Lorentz stripping (quadrupoles only) on field gradient for the same lattice and optimisation scheme as Fig.~\ref{fig:lor_v_ibs}.  
  }
  \label{fig:lor_v_IBS_quadgrad}
\end{figure*}

For emittance, the situation is simpler: transverse emittance growth implies a larger $\sigma_{\perp}$, which means greater Lorentz stripping; it also means increased intra-bunch velocities, causing more IBSt. Thus, a linac design with limited emittance growth is particularly important for H$^-$ beams. This trait was confirmed by comparing preliminary lattice designs with high emittance growth to the a finalised, well-matched baseline design; a reduction of IBSt rates by a factor of 1.5${\sim}$2 was observed.

\subsubsection{\label{sect:L2R_Losses} Linac-to-Ring Transfer Line Losses}

In the linac, the proportion of H$^0$ power deposited into the line-of-sight beam dump affords some relief in terms of component activation. This is not the case, however, for the L2R transfer line. As mentioned earlier, since L2R is curved, the stripped H$^0$ collide with the nearest bend after passing a dipole magnet.

Activation from residual gas stripping should also be mentioned: at 2.5\,GeV, a higher vacuum level is needed to prevent problematic stripping losses than that needed to prevent losses due to scattering. However, the nominal ESS vacuum pressure required for ideal longevity of the ion pumps~\cite{ravelli:2019_ESS_vacuum} is such that gas stripping levels may be taken as negligible.

Lorentz stripping is also compounded in transfer lines. Although it is below problematic levels in the quadrupoles, in the bending dipoles it makes a significant contribution. (The Lorentz stripping scales with the bending radius, e.g $\rho=73$\,m for a field strength limited to $|B|=0.15$\,T in a 5\,MW, 2.5\,GeV beam \cite{blaskovickraljevic:ipac2021-tupab175}.) 

Additionally, blackbody-radiation stripping can become problematic, as discussed earlier. Without remediation, this leads to a problematic power loss of roughly 0.4\,W/m as shown in Fig.~\ref{fig:blackbody_v_T}.

The L2R stripping levels can then be summarised as follows:
\begin{itemize}
    \item Blackbody Radiation  $~ \sim 0.4$\,W/m (Room temperature, 100\% emmissivity)
    \item IBSt  $~~~~~~~~~~~~~~~~~~~~~~~\sim 0.3$\,W/m
    \item Lorentz $~~~~~~~~~~~~~~~~~~~\sim0.29$\,W/m (Dipoles)
    \item Residual Gas  $~~~~~~~~~~~~< 0.05$\,W/m
    \end{itemize}

This puts the activation level of L2R transfer line near the 1\,W/m limit. Since the IBSt and Lorentz stripping levels are not likely to be improved further beyond minor refinements, it is imperative that measures be taken to reduce blackbody stripping (either by cooling the beam pipe to at least approximately 200~K or reducing emissivity of the beam-pipe surface by at least a factor of two). However, as mentioned earlier, some well-polished metals have emissivities less than 0.1, and stainless steel can have emissivity as low as 0.4. Thus, blackbody radiation may be more likely to fall roughly a factor of two below the simulated 100\% emissivity, leaving the total stripping loss rate at an adequate level of 0.8\,W/m.

An additional design measure is advisable: placing local beam dumps after each dipole bend. The trajectories of stripped particles should be calculable to a scale comparable to the transverse beam size, so the placement of such dumps may be fairly simple.

\subsection{RF Systems}
\label{sect:RF_systems}
In this section, a description of the RF systems used for the proton driver of the neutrino facility is described and the upgrade on the RF power sources and stations is outlined. Some details on cost estimates are also provided throughout.

\subsubsection{RF Modulator Upgrade}

The RF modulators designed for the current ESS linac are based on the stacked multi-level (SML) topology and are each rated for peak power 11.5\,MW and average power 650\,kVA~\cite{collins:2016_SMLmodulators}. This modulator topology permits control of output pulse amplitude as well as offering variable pulse length on a pulse-to-pulse basis, thus allowing output of any one of the proposed pulsing schemes for ESS$\nu$SB.

For the ESS linac baseline design at 2.0\,GeV a total of 33 modulators are needed. For upgrading to 2.5\,GeV, an additional 32 high-beta cavities are needed, requiring 8 additional modulators. In total, 36 modulators must be upgraded to meet the increased power demands of ESS$\nu$SB, and 8 new ones must be constructed, for a total of 41 modulators. 

Upgrading the SML modulators will require no additional footprint in the accelerator gallery, but will extend the height of the modulators by 0.5\,m. The estimated time of work in order to perform the upgrade is approximately two weeks per modulator, provided that the necessary components can be prepared and pre-cabled in advance.

Additional insulated-gate bipolar transistors will be necessary in the upgrade \cite{folsom:2021_D2.4}, allowing for increased cooling capacity without altering the nominal heatsink design. 

It should be noted that the peak pulse power, and therefore also output pulse voltage and current amplitudes, are in the proposed pulsing schemes not greater than that specified for the baseline ESS RF modulators. This means that all components are already appropriately scaled for peak pulse power conditions, and that the upgrade will largely concern the capability of the modulators for handling increased average power.

The oil-tank assembly, including the high-voltage components contained within, is designed conservatively with respect to necessary isolation distance, using mineral oil as insulating medium. Since the oil is continuously circulated for a high heat-transfer capacity (and since the high-voltage components are relatively large with respect to the power they convert) it has been determined that the oil tank assembly will require no upgrades to handle the increased average power represented by the proposed scenarios.

To study the performance of the present SML modulators for the new pulsing conditions of the proposed scenarios, a simulation model was developed using MATLAB Simulink~\cite{matlab:2020}.
First, in order to maintain flicker-free operation, charging power was proportionally increased with the idea of recovering the lost energy of all pulses (1 proton pulse plus 1--4 H$^-$ pulses) in time for the next proton pulse. Consequently, in scenarios B and C, reduced capacitor-bank voltage is available for each of the four H$^-$ pulses. SML modulator pulse generation control systems were, in these cases, adjusted to compensate. The following text details the simulation results for each scenario with particular focus on 1) rise time, used in deriving modulator beam efficiency, 2) pulse flat top ripple, and 3) the dynamics of the capacitor charger waveforms.

Figure~\ref{fig:modulator_sims_optA} shows simulation results for the upgraded SML modulator with the baseline 28\,Hz pulsing scheme from Fig.~\ref{fig:pulse_struct_options}. In this scenario, pulse repetition rate is doubled from the nominal ESS value. Consequently, with doubled charging power, the following H$^-$ pulse is identical to the first pulse. Flat top ripple is seen to be within specification, and pulse rise time is verified to be just over 120~\SI{}{\micro\second}. This pulse rise time corresponds to a modulator-beam efficiency of 82\%, and to an annual electricity cost of M€14.6 for all 41 modulators (this is the lowest-cost option, representing is a 100\% increase from the nominal ESS).

Simulations for the alternative pulsing schemes using the existing modulator design with additional capacitor banks, or with pulse-transformer technology, were performed and reported in \cite{galnander:2019_D2.2}. The 28\,Hz scheme was found to outperform the 70\,Hz options considerably in terms of efficiency and annual electricity cost, which leaves it as a strongly favoured baseline.

\begin{figure*}[ht!]
\centering
\includegraphics[width=0.8\textwidth]{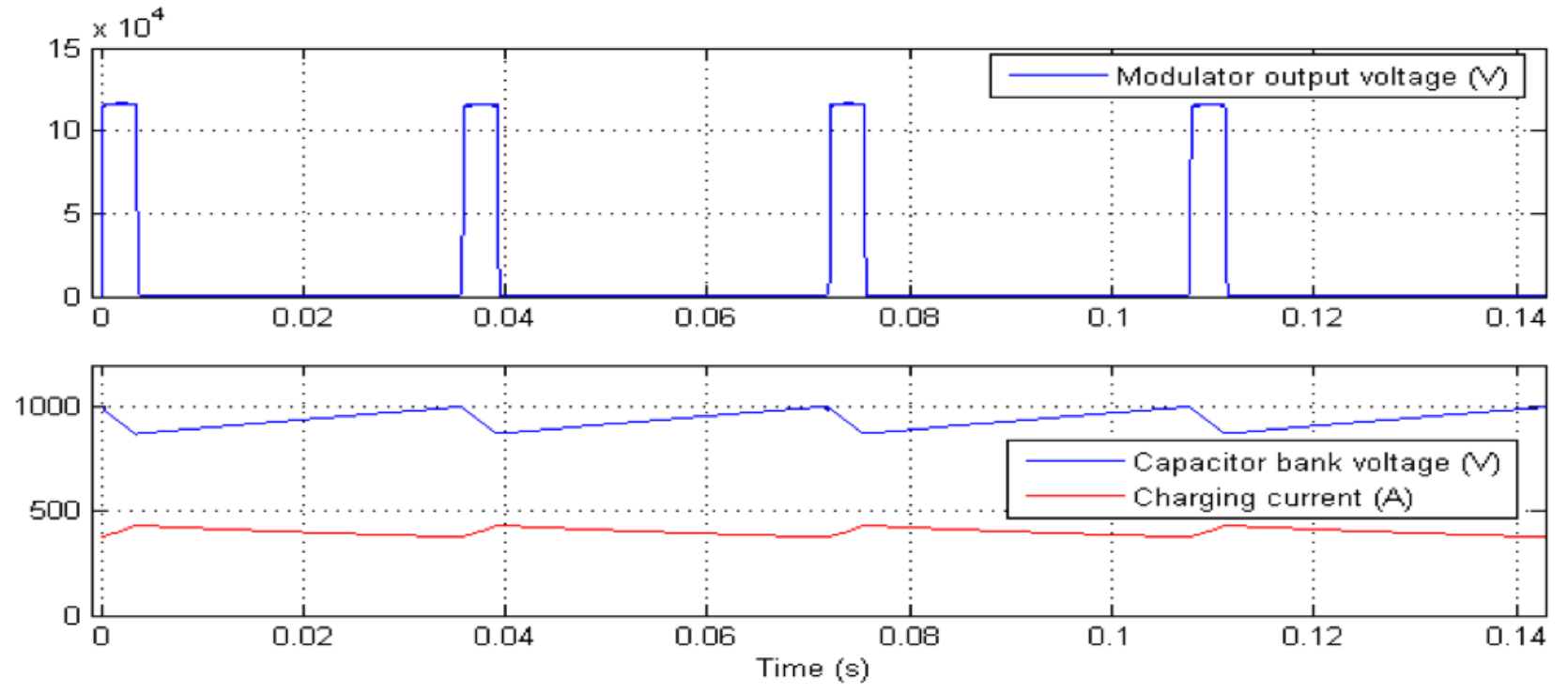}
\includegraphics[width=0.8\textwidth]{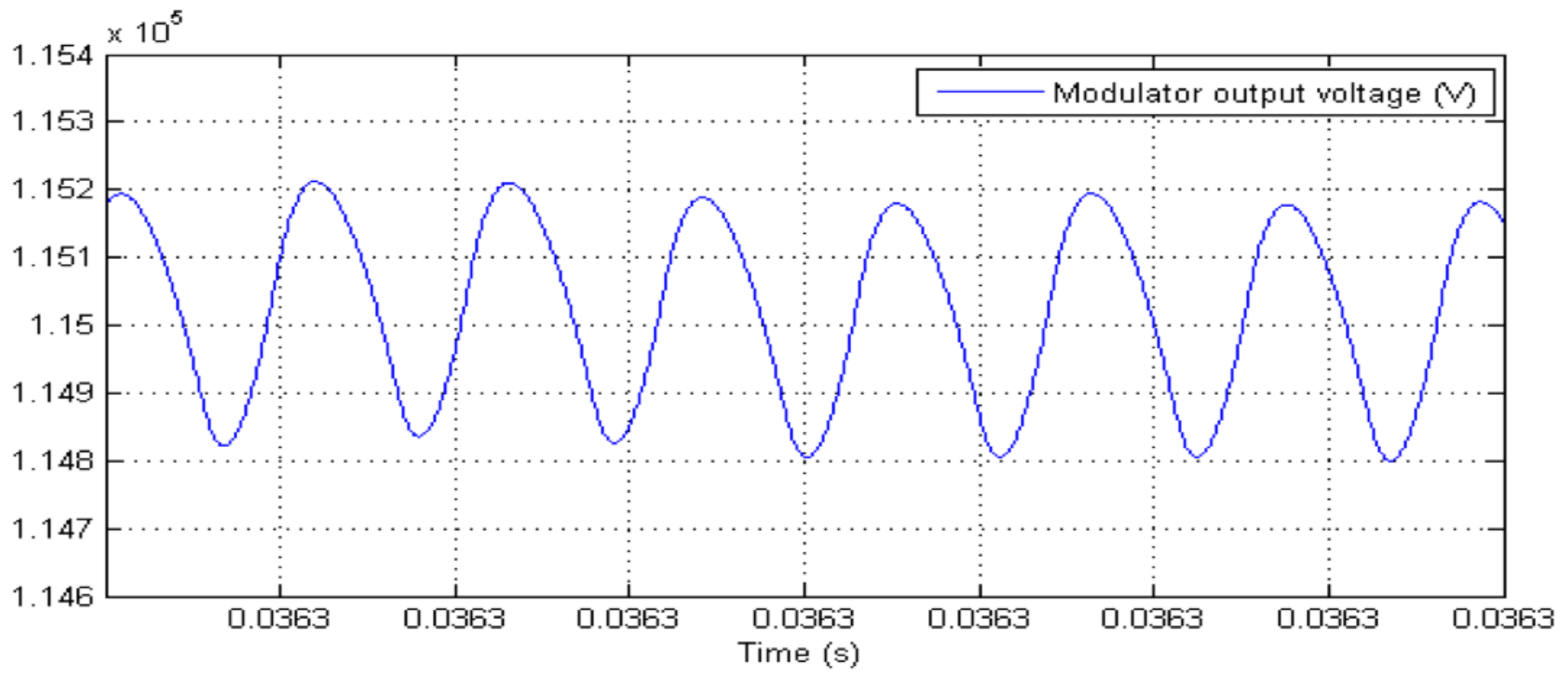}
\caption{Simulation results for pulsing Option A+ from Fig.~\ref{fig:pulse_struct_options} ; Top: modulator output voltage and capacitor bank charging waveforms, Bottom: flat top ripple, confirmed to be within acceptable limits.
}
\label{fig:modulator_sims_optA}
\end{figure*}

\subsubsection{Grid-to-Modulator Transformers}

For the nominal ESS design, there are eleven transformers feeding grid power to three modulators each. These are rated for 2000\,kVA, with each modulator requiring 650\,kVA. Taking a conservative estimate of a 100\% increase in power for the ESS$\nu$SB upgrade, these transformers can be reused -- but only supplying one modulator each instead of three each. 

Beyond this, 15 new transformers would be needed, each delivering power to two modulators (for the remaining 22 upgraded and 8 newly installed modulators). These new transformers would need ratings of 2600\,kVA, 600\,V, and 50\,Hz, with an estimated total cost for the 15 new units at M€~5.6.

\subsubsection{Klystron Upgrades}
\label{sect:klystron_upgrades}
Only Pulsing Scheme A from Fig.~\ref{fig:pulse_struct_options} is considered here, as it is presently a strongly favoured baseline. (For details on the klystron power requirements for other pulsing schemes see~\cite{galnander:2019_D2.2}.) Given the timeframe of the ESS$\nu$SB project, it can be assumed that the klystrons in operation on the ESS linac will be nearing the end of their lifetime. The requirement for klystrons in operation in the linac (352 MHz, 3 MW klystrons and 704 MHz, 1.5\,MW klystrons) was for a lifetime exceeding 60,000 hours (about 10 years). Hence, it is reasonable to consider that they will have to be replaced once ESS$\nu$SB is operational.

There are several options for the replacement:
\begin{enumerate}
    \item In the first option, the klystrons are replaced with similar devices. The klystron beam voltage, current, and the RF circuit design are left unchanged. The only modifications required would be related to the increased average power dissipation (with collector size and cooling redesigned for operation at 28 Hz), and possibly shielding. Klystron manufacturers have been involved in order to understand the implication of the possible modifications and the cost and some details will be discussed shortly.
    \item In the second option, the klystrons can be replaced with tubes that have been redesigned and optimised for the new operating conditions and increased efficiency. Again, two alternatives can be evaluated:
    \begin{enumerate}
        \item In this case a complete tube re-design can be studied, including optimisation of:
        \begin{itemize}
            \item  Micro-perveance 
            \item Electron gun
            \item Cavity circuit
            \item Cooling
            \item Collector
        \end{itemize}

    Increasing the beam voltage to about 140\,kV could have a positive effect on the klystron efficiency, as demonstrated by several studies. An 800\,mHz klystron has been designed for the Future Circular Collider (FCC) at CERN (134 kV, 12.5 A, 1.3 MW), showing an efficiency of about 80\% in 3D simulations~\cite{garoby:2018_essdesign,cai:2019high_eff_kly}. This klystron has specifications similar to the ESS requirements, and it makes use of the new bunching methods for high efficiency, with no major technological changes. Another innovative design with two HV stages has shown an efficiency of 82\% in 3D simulations~\cite{cai:2020_clic_klys}, but this technology is quite new and might be not mature enough by the time procurement must be negotiated.
    
    At the same time, increasing the klystron beam voltage has important implications from the modulator design point of view, with several components needing replacement and possibly requiring major design changes. This option should thus be evaluated with the help of the power converter experts.
    A recent report \cite{marrelli:2019_high_eff_kly} demonstrates by particle-in-cell (PIC) simulations an efficiency for this option of 73\% (about 9\% higher than the one achieved by the klystrons currently installed on the ESS linac). For the 352\,mHz klystrons (RFQ-DTL) such work has not been carried out yet. As there will be only six of these klystrons, it should be evaluated whether the efficiency gain would justify the investment in a new design.
    The cost of the different scenarios has to be evaluated against their advantages. Klystron manufacturers have to be involved in all cases, in order to assess the feasibility. Scenarios 1 and 2b will be discussed in the following paragraphs.
        \end{enumerate} 
\end{enumerate}

Scenario 2a will not be considered further at this time, as it would require major changes on the modulator design and collaboration with the power converter group in order to assess the feasibility of such a design.

\subsubsubsection{Modifications Required on ESS klystrons for the ESS$\nu$SB Upgrade}
This section will focus on the modifications required to operate the current klystrons at 28\,Hz repetition rate (Scenario 1).
The klystrons presently used for the ESS linac have been provided by three different manufacturers: Thales and CPI for the normal-conducting linac (352.21\,mHz) and Thales, CPI, and Canon for the medium and high beta part of the linac (704.42\,mHz). A first enquiry regarding the modifications required to operate the klystrons at 28 Hz has been carried out, with the following outcome~\cite{priv:2021_Canon,priv:2021_cpi,priv:2021_Thales}:

\begin{flushleft}
\textbf{352.21 MHz Klystrons}
\end{flushleft}

\begin{itemize}
 \item Thales TH2179D klystrons (352.21 MHz, 3 MW peak)
    \begin{itemize}
        \item Some modifications in the oil tank cooling circuit are required to dissipate the heat caused by the current in the limiting resistors.
        \item The flow rate of the body should be slightly increased (16 to $25~\mathrm{l} / \mathrm{min}$ ).
        \item The lead shielding will need to be locally adjusted.
        \item The presently installed collector can dissipate about $560 \mathrm{\,kW}$ (average). The margin should be enough to ensure safe operation at $28 \mathrm{\,Hz}$, but additional thermomechanical simulations are needed. The temperature of inlet water at $50^{\circ} \mathrm{C}$ should also be considered.
        \item The collector flow rate should also be increased from 300 to $600~\mathrm{l} / \mathrm{min}$.
    \end{itemize}
\item CPI VKP-8352A (352.21 MHz, 2.9 MW peak)
    \begin{itemize}
        \item CPI provided a detailed report including simulations of the thermomechanical behaviour of the collector at $28~\mathrm{\,Hz}$, and of the klystron output cavity and output window.
        \item For the collector, CPI found a limited margin, and recommends a collector upgrade. CPI is currently under contract with ESS Bilbao to repair a VKP-8352A klystron and upgrade it with a larger collector (from the VKP-8352C klystron). This collector will be capable of operating at the double duty requirement of $28~\mathrm{\,Hz}$. The klystron will have a new model number.
        \item The output window and output cavity are capable of higher duty.
        \item Increased coolant flow rate would be required at $28~\mathrm{\,Hz}$ for the collector and body circuits. The electromagnet flow rate would not change.
        \item The $X$-ray shielding should be adequate for the higher duty, but there may be some areas that need upgrading. It is difficult to estimate the impact on life of operating the existing tubes at $28~\mathrm{\,Hz}$. The only vulnerable component is the collector. CPI will review the current design for projected cycling fatigue.
    \end{itemize}
\end{itemize}

\begin{flushleft}
\textbf{704.42\,mHz Klystrons}
\end{flushleft}

\begin{itemize}
    \item Thales TH2180 klystrons (704.42\,mHz, $1.5$\,MW peak)
    \begin{itemize}
        \item The oil tank cooling seems sufficient, but this will need to be confirmed by further testing.
        \item The flow rate of the body should be slightly increased (from 14 to $16~\mathrm{l} / \mathrm{min}$ ).
        \item The lead shielding will need to be locally adjusted.
        \item A water-cooling circuit will have to be added for the output waveguide transformer and RF window.
        \item The collector should be able to dissipate the increased power, but this has to be confirmed by further simulations. However, the margin is higher in this case (compared to the $352 \mathrm{MHz}$ klystrons). Collector flow rate must be increased to $275~\mathrm{l} / \mathrm{min}$.
    \end{itemize}
\item Canon E37504 (704.42 MHz, $1.5$ MW peak)
    \begin{itemize}
        \item The collector cooling will need an increased flow rate $(365 ~\mathrm{l} / \mathrm{min})$, which will bring the pressure drop just below the upper limit of 3~barg. An alternative is to decrease the inlet temperature of the collector cooling water to $43^{\circ}\mathrm{C}$ instead of the $50^{\circ}\mathrm{C}$ currently allotted. Otherwise, no other issues are foreseen with the collector design: the average power density on the collector inner wall will increase from $122\mathrm{\,W} / \mathrm{cm}^2$ at $14\mathrm{\,Hz}$ to $243\mathrm{\,W} / \mathrm{cm}^2$ at $28\mathrm{\,Hz}$. This is within the range of performance experience of other $\mathrm{CETD}$ products.
        \item No issues for the body and window design. The power dissipation will double but will still be well within the absolute rating for this klystron. The temperature of the outlet water will increase about six degrees with the present flow rate.
        \item $\mathrm{X}$-ray shielding is expected to be sufficient with the new repetition rate, but this should be confirmed with testing, and it might need to be locally adjusted.
        \item It is difficult to estimate the impact of the higher duty cycle on the klystron lifetime. It is reasonable to expect a shorter lifetime, and if the warranty is to be kept as it is for the present klystrons, the cost will increase accordingly.
    \end{itemize}
    \item CPI VKP-8292A (704.42 MHz, $1.5$ MW peak)
    \begin{itemize}
        \item CPI provided a detailed report including simulations of the thermomechanical behavior of the collector at $28 \mathrm{\,Hz}$, and of the klystron output cavity and output window.
        \item Thermomechanical simulations of the collector show an instantaneous power density distribution for the existing design of about 3800 W/cm2, which is considered too high. A modified geometry has been studied and it is able to reduce the maximum value from $3800 \mathrm{\,W} / \mathrm{cm}^2$ to $2575\mathrm{\,W} / \mathrm{cm}^2$. The reduced power density for the proposed VKP8292A collector design results in predicted mechanical performance within guidelines for reliable performance at the specified pulse parameters. Therefore, $\mathrm{CPI}$ recommends a collector upgrade for the $28~\mathrm{Hz}$ duty cycle.
        \item According to thermomechanical simulations carried out by CPI, the output window and output cavity are capable of higher duty cycles.
        \item The X-ray shielding should be adequate for the higher duty cycle, but there may be some areas that need upgrade.
        \item Increased coolant flow rate would be required at $28~\mathrm{Hz}$ for the collector and body circuits. The electromagnet flow rate would not change.
        \end{itemize}
\end{itemize}

\subsubsubsection{Efficiency and Electricity Consumption for the Upgraded Medium and High Beta Klystrons}

\begin{flushleft}
\textbf{Scenario 1}
\end{flushleft}

As already mentioned, three different types of klystrons are being used for the ESS linac. They have been procured from three different manufacturers, Thales, Canon and CPI. The klystrons have similar specifications, with efficiencies between 63\% and 66\%, in worst and best cases. These data have not been fully confirmed in site acceptance tests yet, but have been provided by the manufacturers.
The calculations below are based on the data provided by Canon~\cite{canon:2021_klyreport}, as Canon klystrons are used in both the medium and high beta linac. Similar performance values are expected from the other klystrons. 

It is possible to increase the klystron efficiency in operation at beam voltages lower than the nominal by introducing a mismatch on the output of the klystron. This has been demonstrated and could present advantages, especially for the first medium-beta klystrons (which will operate at lower output power). However, this improvement will not be considered in the calculations to follow, since it is assumed that the same technique could also be adopted for the second Scenario (a--b), bringing the same advantage to the three different cases.

The power required by the cavities in the medium and high beta linac is shown in Fig.~\ref{fig:kly_pow_curves_msmtch}. In order to estimate the efficiency of the klystrons at the operating point, some considerations must be taken into account:
\begin{itemize}
\item	The output power required from the klystron will have to include the losses in the RF distribution system (RFDS), which are estimated to be about 5\%.
\item	In order for the low-level RF (LLRF) systems to be able to regulate the power, the amplifiers have to operate in the linear regime. This means that the klystrons cannot be operated at saturation, but at a power level which is about 25\% below saturation. 
\item	In order to increase the overall efficiency, the first klystrons of the normal-conducting linac can be operated at beam voltages lower than nominal (nominal here is defined as the beam voltage required to reach 1.5 MW output power) since they will have to provide lower power to the first cavities (the power required from the first medium beta cavities is lower than 400\,kW). However, four klystrons share the same modulator, so the chosen voltage has to accommodate four klystrons at the time.
\item	The klystron saturated efficiency when operating at lower beam voltages is lower than the one at nominal voltage. This is shown in the following figures for the Canon E37504 klystrons (no mismatch assumed).
\item	The modulator efficiency is estimated to be around 82\%~\cite{galnander:2019_D2.2}.
\end{itemize}

\begin{figure*}[ht!]
\centering
\includegraphics[width=0.6\textwidth]{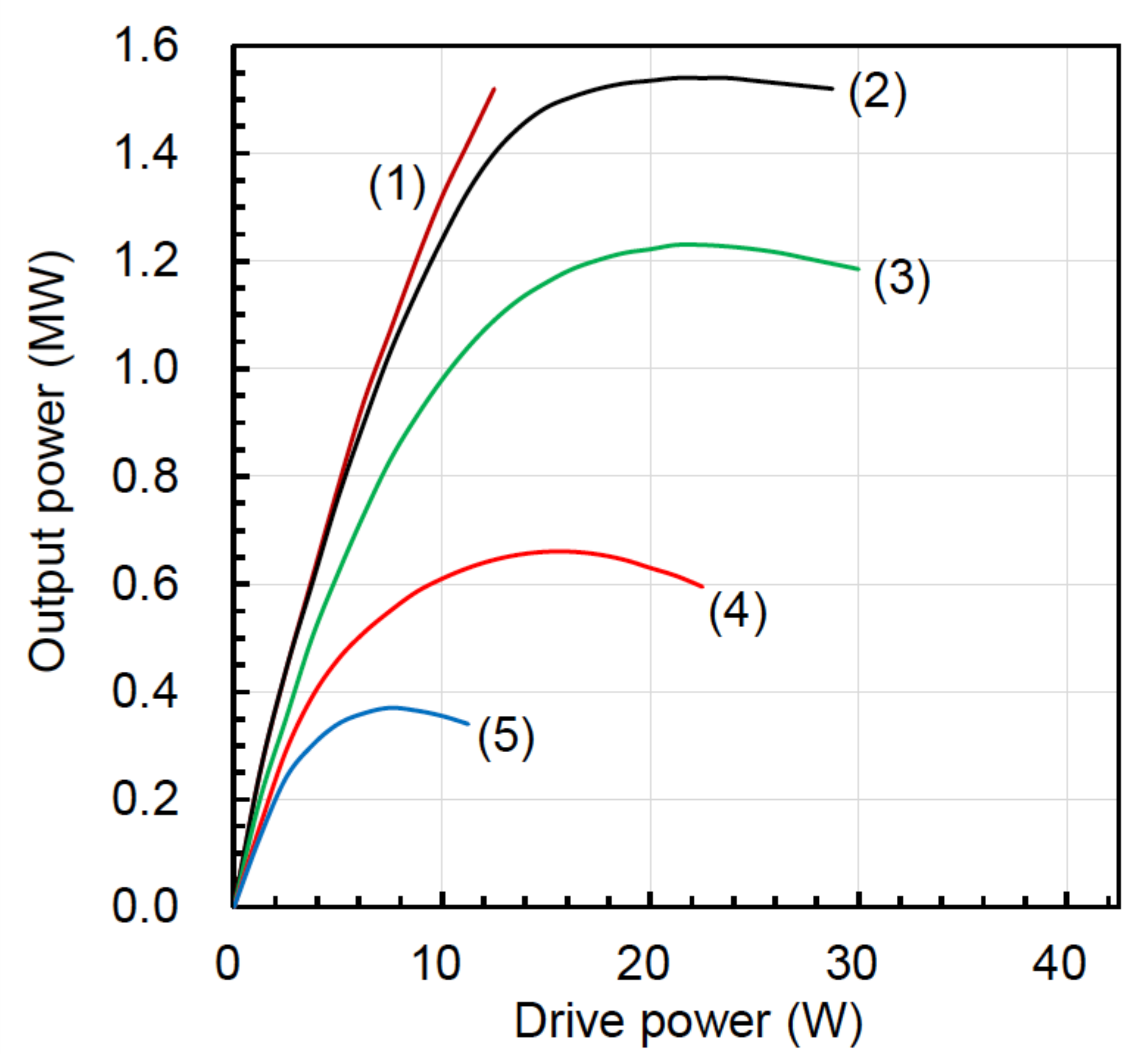}
\caption{Power curves for SN18K010 Canon klystron. Curve 4 and 5 have been obtained by introducing a mismatch at the klystron output. This case will not be considered in the following calculations, so the klystron output power is considered at these voltage levels is slightly lower.}
\label{fig:kly_pow_curves_msmtch}
\end{figure*}

\begin{table}[ht!]
\begin{center}
\caption{Output power for different klystron settings}
\label{table:kly_pow_curves_msmtch}
\begin{tabular}{lcccccc}
%\hline 
\textbf{No.}  & \textbf{Beam} &  \textbf{Beam} & \multicolumn{2}{c}{ \textbf{Solenoid current} } &  \textbf{RF power}  & \textbf{Iris} \\
  & \textbf{voltage (kV))} &  \textbf{current (A)} &  \textbf{1} & \textbf{2} &  \textbf{(MW)}  &  \\  
 \hline
 1 & $115.2$ & $23.9$ & $10.0$ & $10.0$ & $1.50$ & \\
 2 & $107.6$ & $21.4$ & $10.0$ & $10.0$ & $1.54$ & \\
 3 & $97.6$ & $18.8$ & $8.5$ & $8.5$ & $1.23$ & \\
 4 & $76.0$ & $13.2$ & $8.0$ & $8.0$ & $0.66$ & Iris 1 \\
 5 & $61.2$ & $9.6$ & $6.0$ & $7.5$ & $0.37$ & Iris 2 \\
 \hline
 \end{tabular}
\end{center}
\end{table}

Due to the limited efficiency of the systems in the chain from the input to modulators to the beam, the power goes up as the distance to the beam increases; this is illustrated in Fig.~\ref{fig:kly_peak_pow_prfl}.

\begin{figure*}[ht!]
\centering
\includegraphics[width=0.85\textwidth]{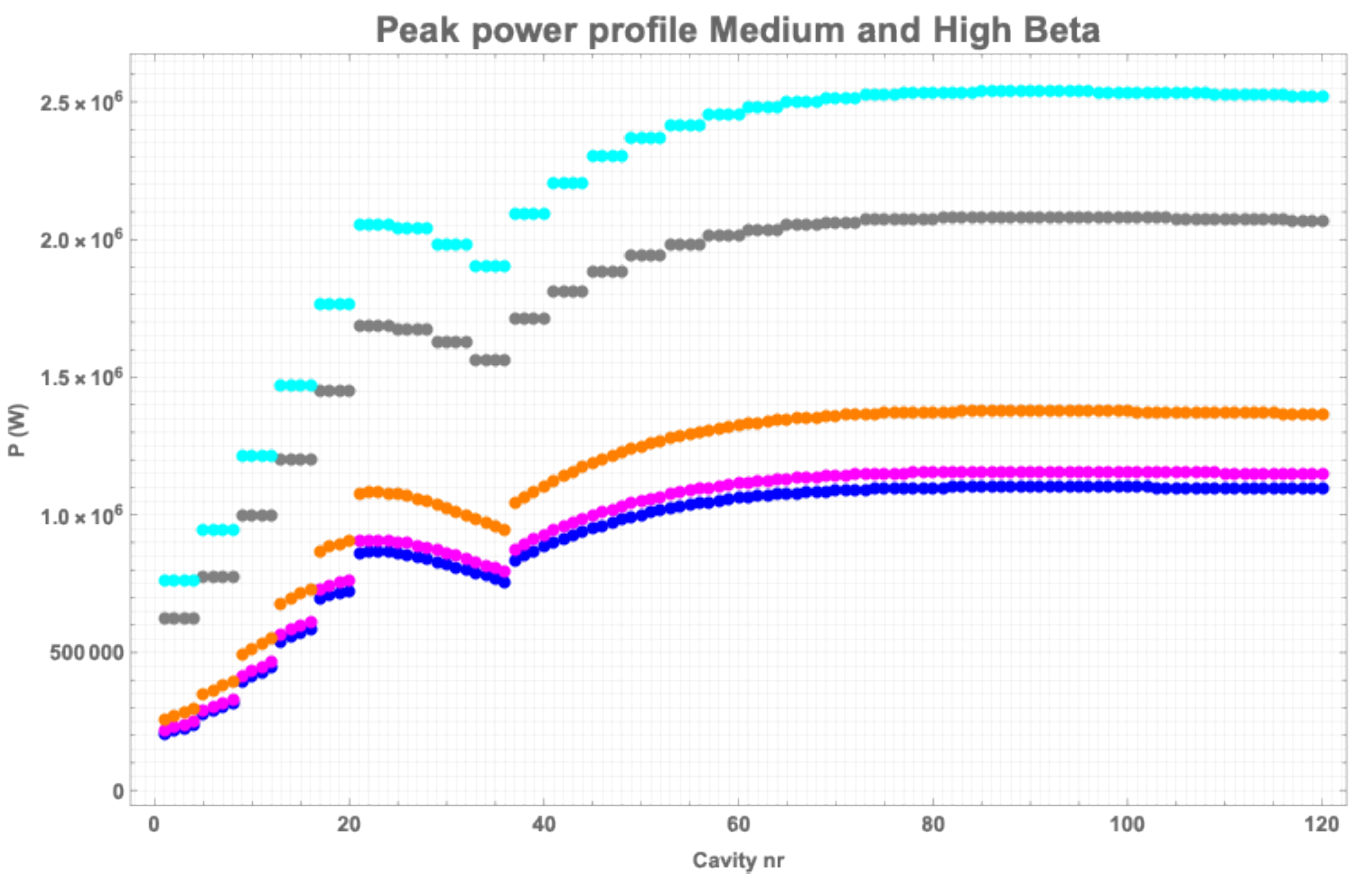}
\caption{Peak power profiles for the superconducting linac at different stages in the RF chain. Blue -- power to cavity; magenta -- klystron output power; orange -- klystron saturated power; grey -- power provided from the modulator to klystrons; cyan -- modulator input power.}
\label{fig:kly_peak_pow_prfl}
\end{figure*}

Assuming that each klystron will operate for about 5500 h/year (with a duty cycle of 9.8\%), and a cost of electricity of about 0.11 €/kWh; this means, for the cost of electricity:
\begin{itemize}
    \item Medium Beta: 3.35 M€/year
    \item High Beta: 12.26 M€/year
    \item Total: 15.6 M€/year
\end{itemize}

\begin{flushleft}
\textbf{Scenario 2b}
\end{flushleft}

In this case the maximum modulator beam voltage is not modified, while the perveance can still be changed by reducing the electron beam current.

CPI is currently studying a new high-efficiency klystron design for the medium and high beta sections of the linac. Here, the tube micro-perveance is kept unchanged at 0.6; they then attempt to increase the efficiency by using the new core-stabilisation method (CSM) bunching technique. The first simulation results show a very limited improvement (about 2\% with respect to the current design, leading to a maximum of 68\% efficiency), but more studies and optimisations are ongoing, and the company believes that there is high margin for improvement.

For the medium--high beta klystrons, a high efficiency design study has also already been carried out at ESS (HEK-ESS-8)~\cite{marrelli:2019_high_eff_kly}, demonstrating an efficiency of almost 74\% in simulations using the KlyC 1.5~D code~\cite{Cai:2018_klyc_code}, and about 73\% in MAGIC~2D~PIC simulations~\cite{woods:2011_magic_pic}. Some results from Magic~2D and KlyC simulations are shown in Fig.~\ref{fig:kly_sims}.

\begin{figure*}[ht!]
\centering
\includegraphics[width=0.95\textwidth]{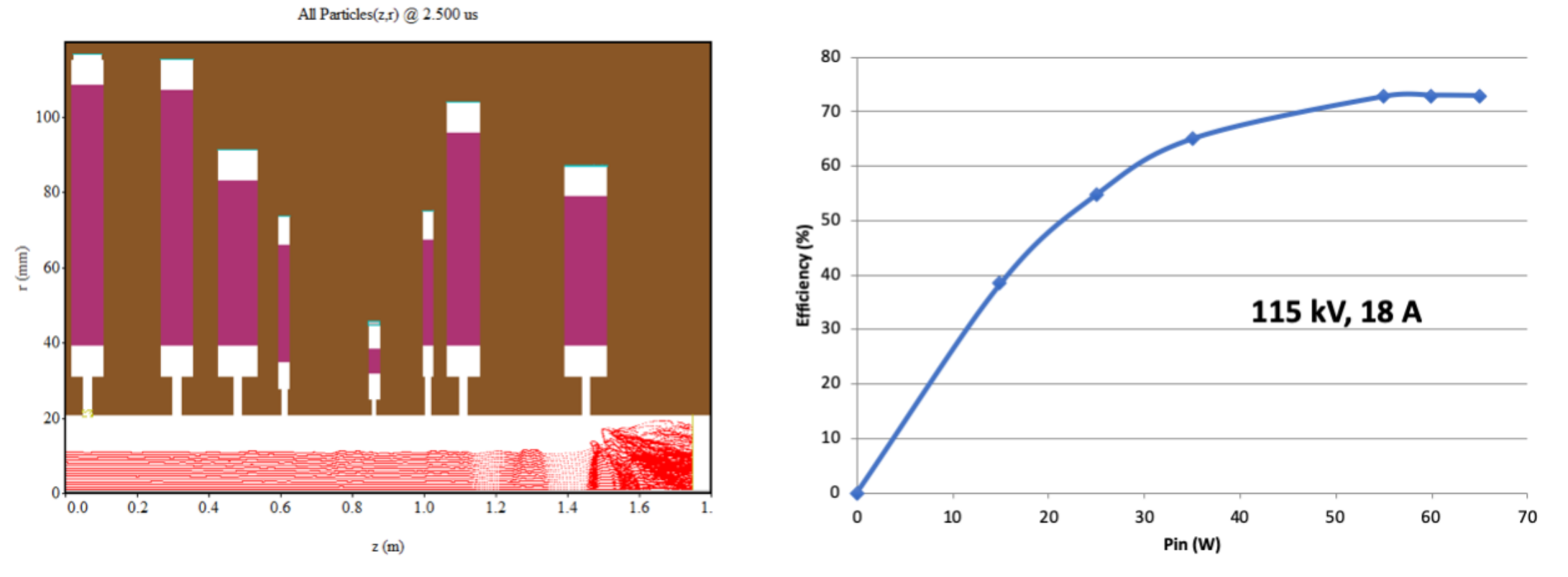}

\includegraphics[width=0.95\textwidth]{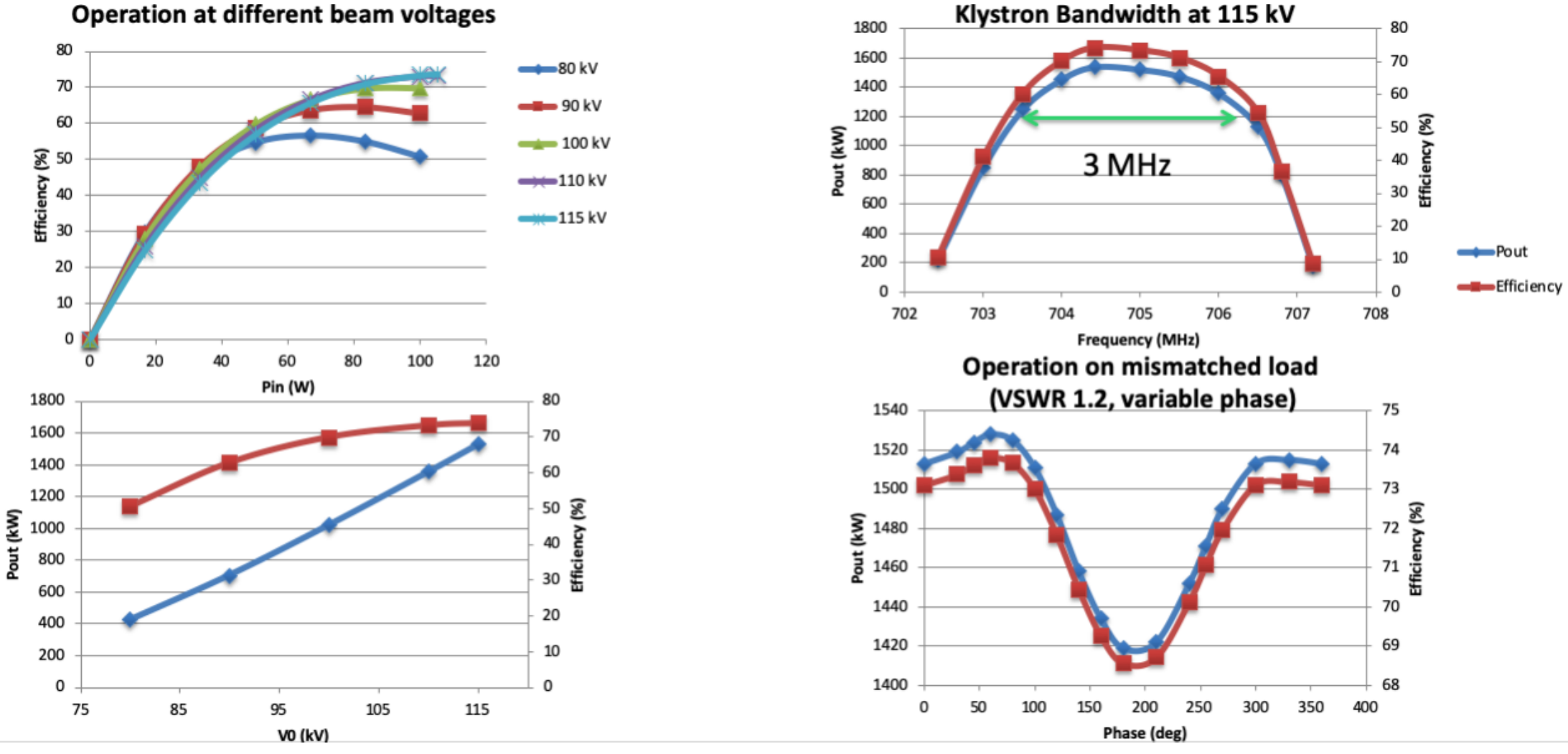}

\caption{Clockwise from top left: Electron beam trajectory simulation in a klystrons, efficiency vs. input power, klystron bandwidth, operation on a mismatched load, efficiency and output power vs. cathode voltage, and efficiency vs. input power at different cathode voltages.}
\label{fig:kly_sims}
\end{figure*}

In the following calculations, results from KlyC simulations are used. The HEK-ESS-8 klystron has a slightly lower micro-perveance compared with the klystrons currently in use at ESS (0.46 versus 0.6). The klystron efficiency at nominal high voltage (115\,kV) and output power (1.53\,MW) is 73.8\% (see Fig.~\ref{fig:kly_sat_eff}). This efficiency is lower when the klystron is operated at lower beam voltages, but it could be improved by adding a mismatch at the klystron output as discussed above (and already demonstrated for the present ESS klystrons).

\begin{figure*}[ht!]
\centering
\includegraphics[width=0.45\textwidth]{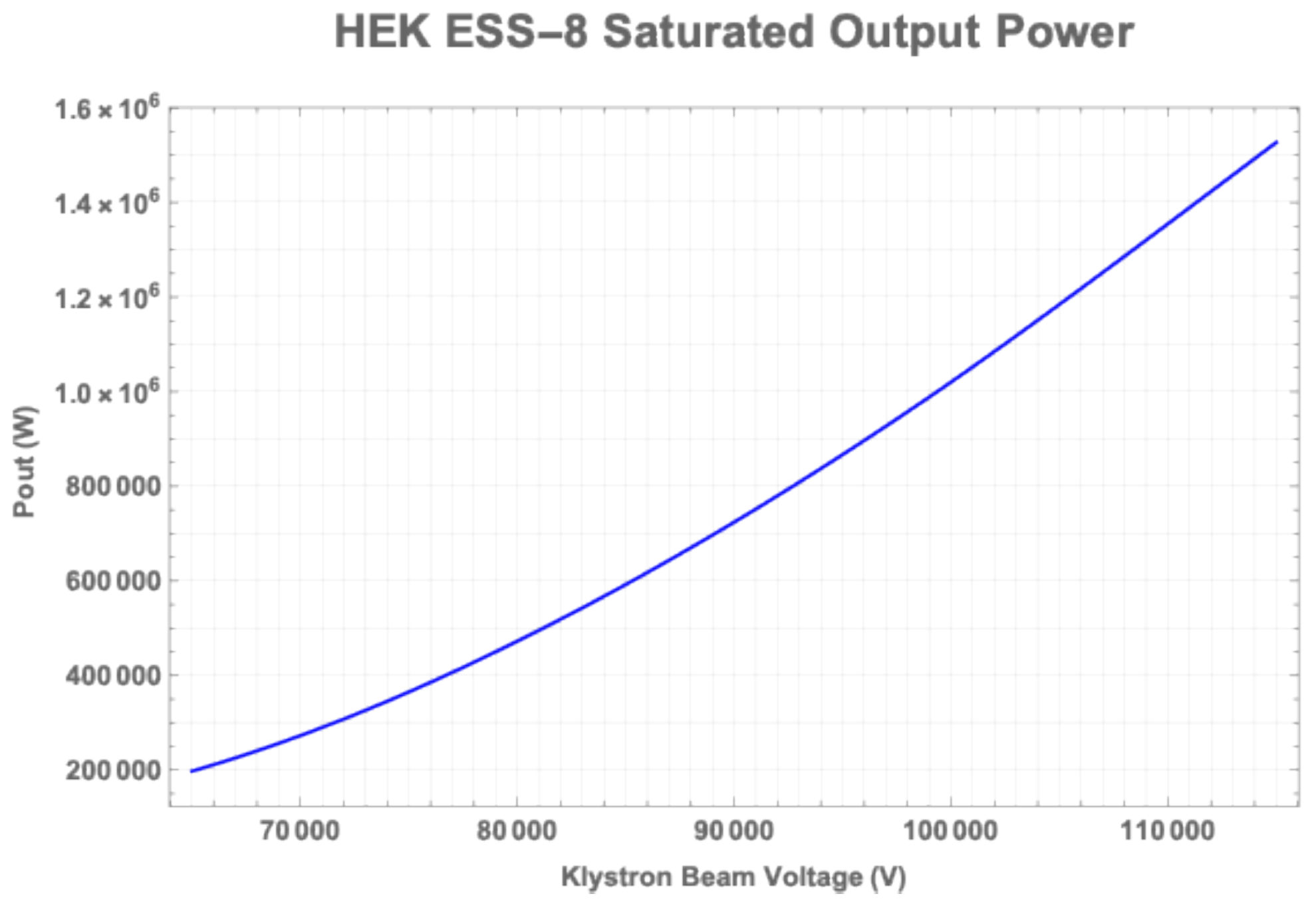}
\includegraphics[width=0.45\textwidth]{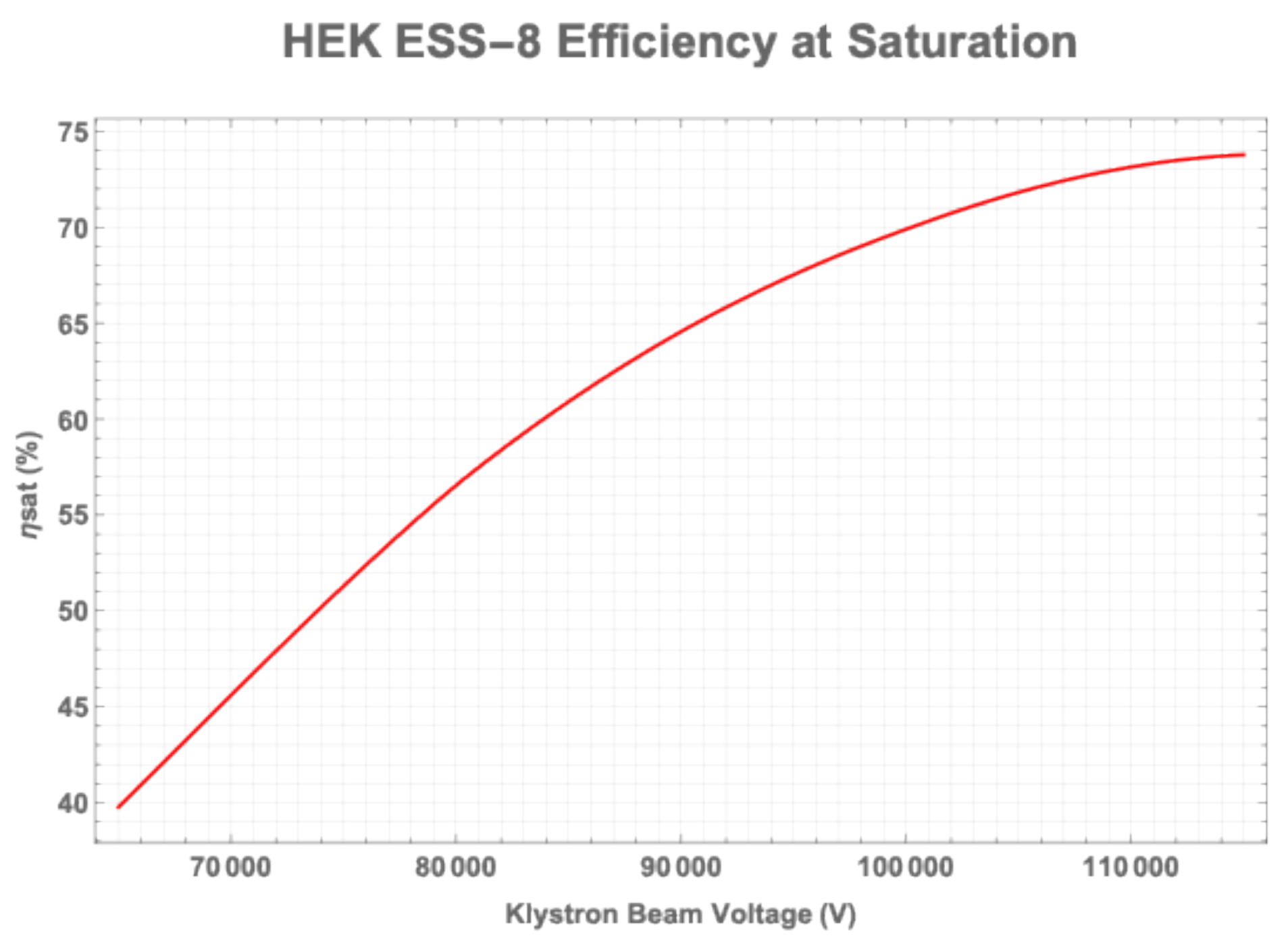}
\caption{Saturated output power (top) and efficiency at saturation (bottom) of the HEK-ESS-8 klystrons.}
\label{fig:kly_sat_eff}
\end{figure*}

The power profile along the superconducting linac at different stages of the RF chain is shown in Fig.~\ref{fig:kly_peak_pow_2}. 

\begin{figure*}[ht!]
\centering
\includegraphics[width=0.85\textwidth]{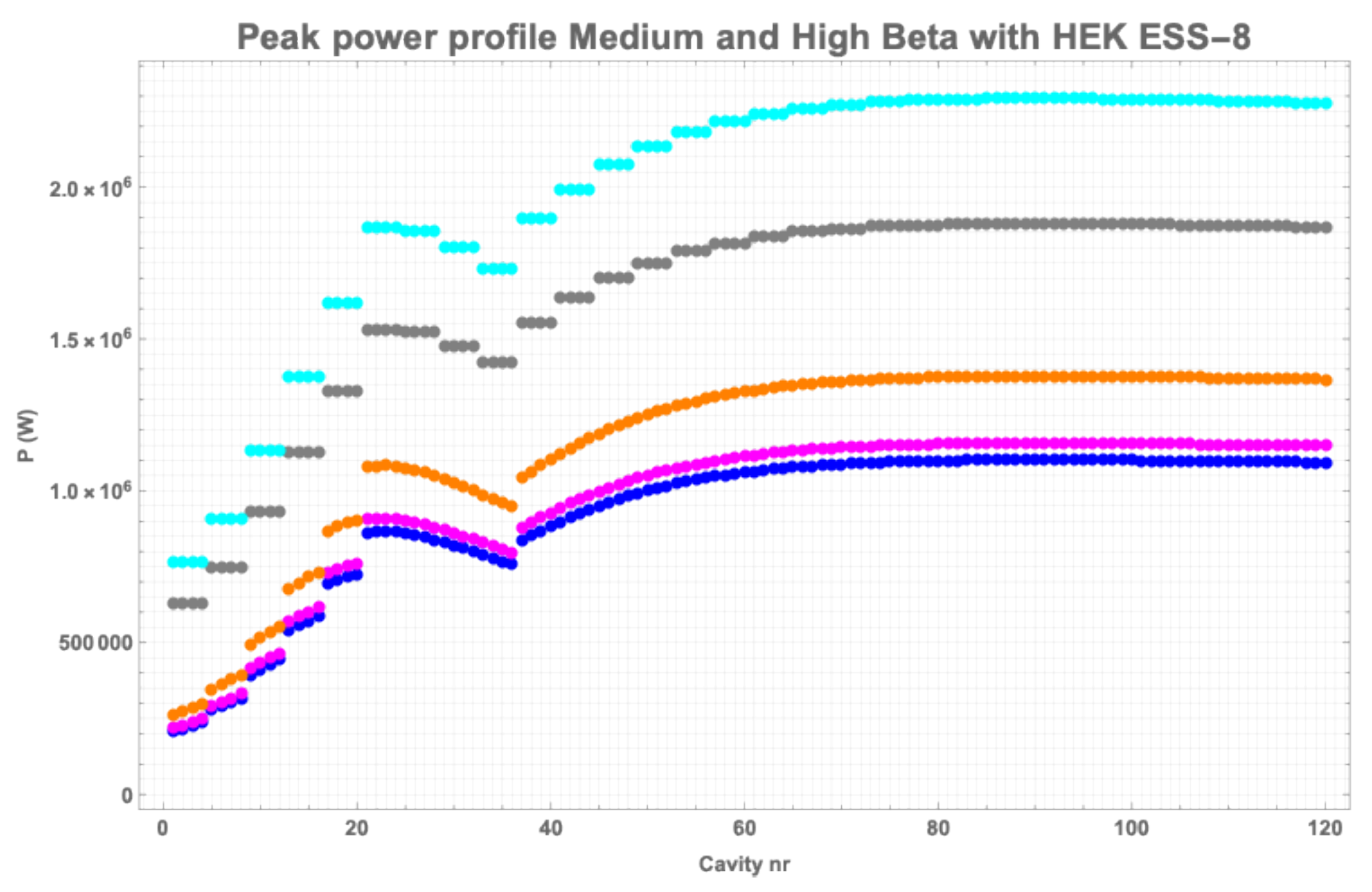}
\caption{Blue -- power to cavity; magenta -- klystron output power; orange -- klystron saturated power; grey -- power provided from the modulator to klystrons; cyan -- modulator input power.}
\label{fig:kly_peak_pow_2}
\end{figure*}

Assuming that each klystron will operate for about 5500 h/year (with a duty cycle of 9.8\%), and a cost of electricity of about 0.11 €/kWh one finds, for the cost of electricity:

\begin{itemize}
    \item Medium Beta: 3.1 M€/year
    \item High Beta: 11.08 M€/year
    \item Total: 14.17 M€/year
\end{itemize}

A comparison of the peak power provided by the modulator to the klystron in the case standard klystrons are used (blue) or HEK-ESS-8 klystrons are used is shown Figs.~\ref{fig:kly_modtokly} and \ref{fig:kly_gridtocav}.

\begin{figure*}[ht!]
\centering
\includegraphics[width=0.75\textwidth]{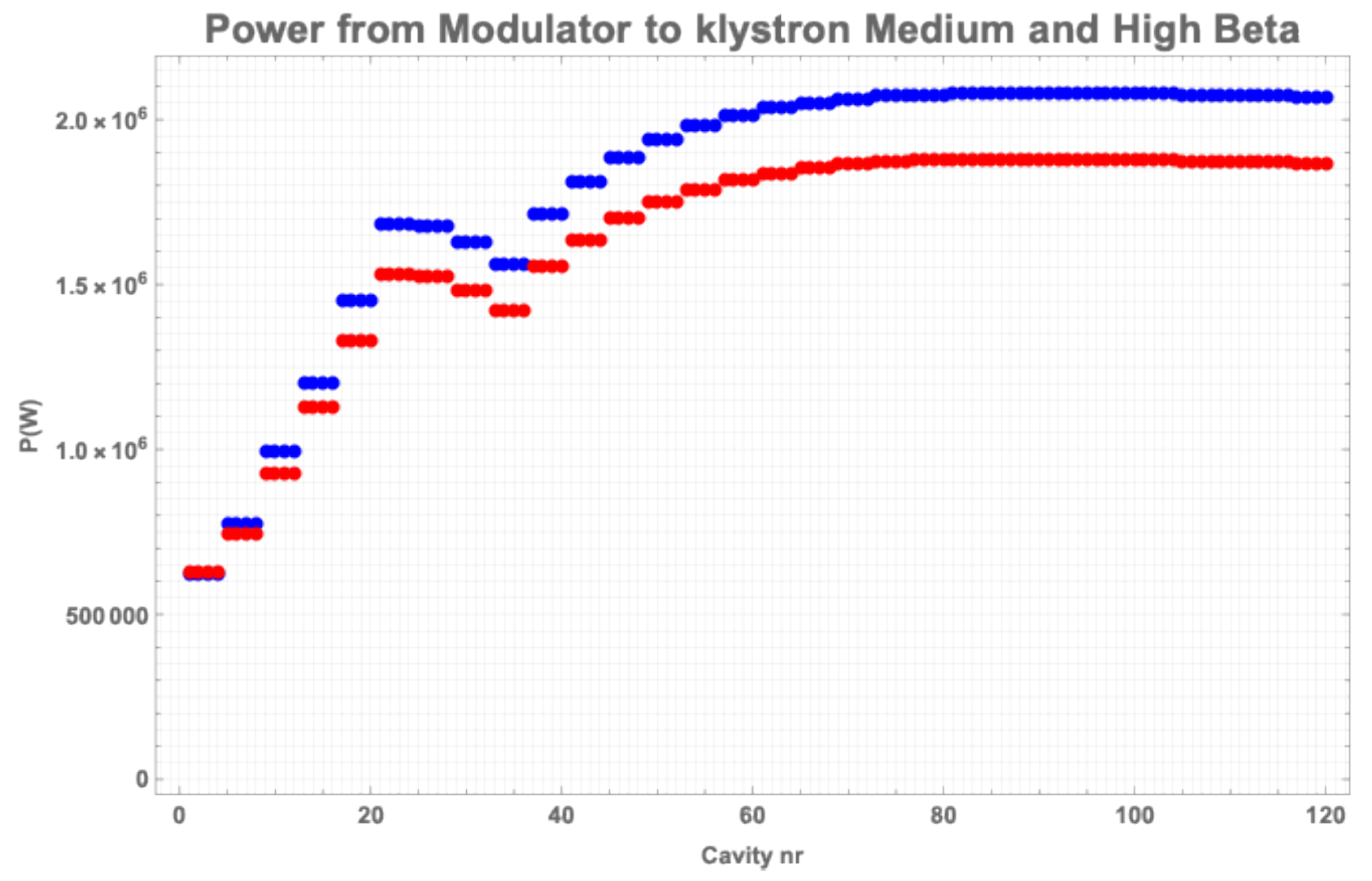}
\caption{ Peak power from modulator to klystrons in the elliptical part of the linac. Blue -- ESS baseline linac; red -- HEK ESS-8 klystrons.}
\label{fig:kly_modtokly}
\end{figure*}

\begin{figure*}[ht!]
\centering
\includegraphics[width=0.75\textwidth]{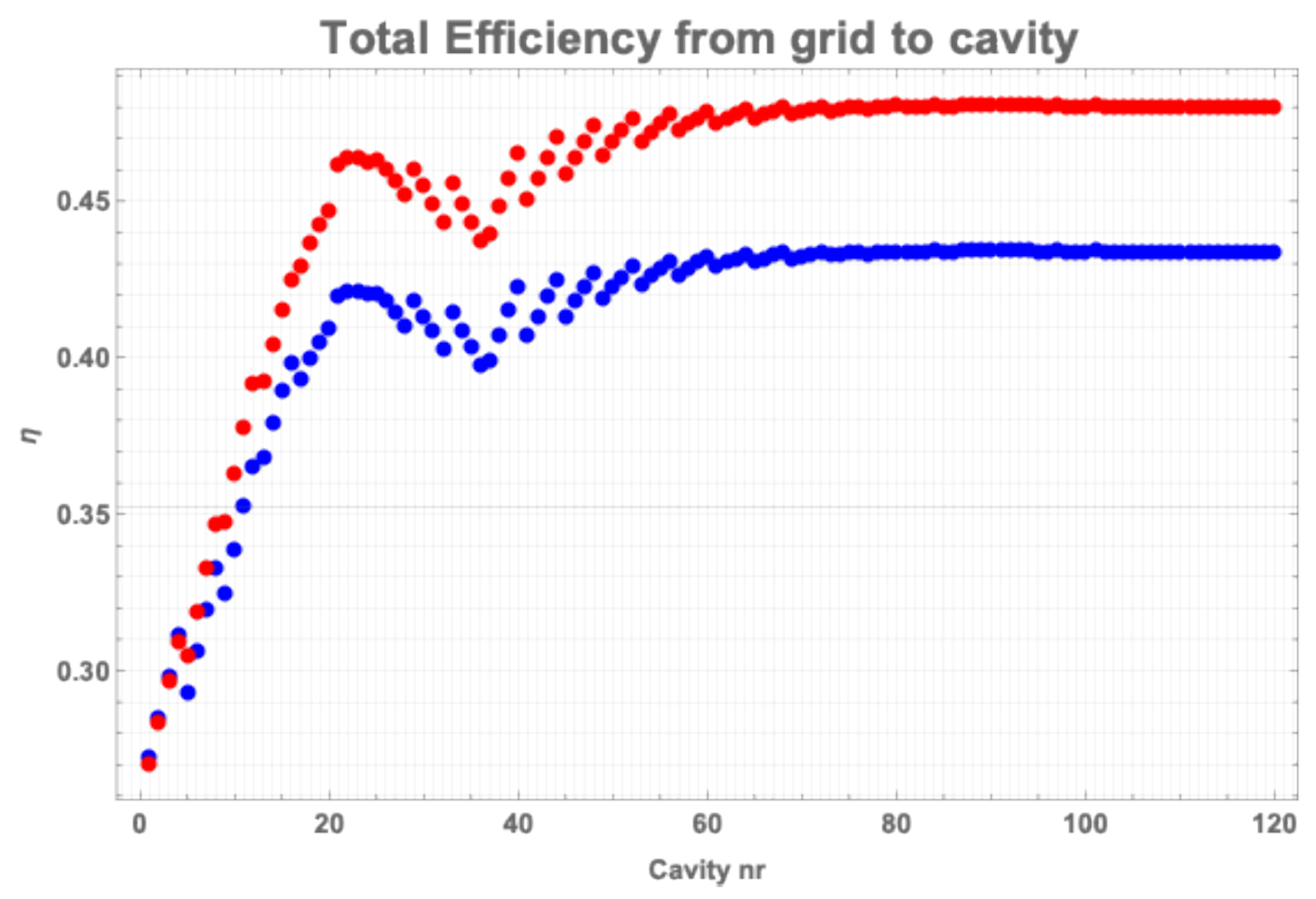}
\caption{Klystron efficiency along the elliptical linac. Blue -- ESS baseline linac; red -- HEK ESS-8 klystrons.}
\label{fig:kly_gridtocav}
\end{figure*}

It can be seen from the analysis presented herein that several options are available for the klystron upgrade in terms of cost, efficiency, and power consumption. Making the most appropriate choice may depend heavily on the end-of-lifetime plan for the existing ESS klystrons.

\subsubsection{RF Distribution System}
The standard waveguide and coaxial components are expected to be capable of operation at 28\,Hz without any critical issues. However, the thermal impact of the higher average power on some ``special'' components (magic tee, switches, phase shifter, etc.) should be evaluated. Also, the present cooling solution for the stubs connecting the klystron gallery to the accelerator tunnel should be re-evaluated.

\subsubsection{Circulators and Loads}
The high-power loads and circulators currently installed in the ESS linac are supplied by AFT Microwave. For the high-power loads, it is likely that it will be possible to upgrade the current devices by modifying the cooling circuit and possibly by introducing an additional attenuating section in a modular way~\cite{priv:2021_AVT}.

Regarding the circulators, these would require more detailed studies (according to AFT).

\subsubsection{Low-Level RF (LLRF) Subsystems}
The real-time operation of LLRF is controlled by events from the timing system and can be run at higher repetition rates. What becomes critical at higher rates is the shorter computation time available for the data processing required between pulses. However, this can be mitigated by updating only some of the algorithms at every second pulse at the highest repetition rates without sacrificing performance significantly~\cite{priv:2021_Asvens}. Therefore, no major issue is expected.

\subsubsection{Existing Solid-State Power Amplifiers (SSPAs) for Buncher Cavities}
From a first enquiry, according to the supplier~\cite{priv:2021_BTesa}:
\begin{itemize}
    \item The current design of the SSPA could withstand the new conditions but:
    \begin{itemize}
        \item A new power supply (PS) drawer would need to be added in order to maintain the desired redundancy.
        \item One load of the 5-way combiner would be replaced by a bigger one (1600\,W instead of 1250\,W) in order to keep a similar safety margin. The combiner heatsink is already prepared to host the new load. Thus, no mechanical modifications would be needed on the heatsink.
    \end{itemize}

\item The new PS drawer could be implemented in two different ways:
	\begin{itemize}
    \item With a bigger rack: 44 rectifier units (RUs) instead of 41 RUs, if possible. This option has less of an impact in terms of mechanical redesign.
    \item Using the same 41 RU rack but trying to accommodate the new PS drawer. This option has a bigger redesign impact.
\end{itemize}
\end{itemize}

\subsubsection{Tetrode Amplification for Spoke Cavities }
The spoke cavities of the ESS linac each use a pair of tetrodes as an RF power source, and an upgrade of the anode power supplies will be needed (consisting of increased capacitor chargers). Considering the lifetime of the tetrodes powering the linac's spoke section, another solution could be to use a solid-state amplifier, or a klystron-based solution, at the time when ESS$\nu$SB is built. In summation, there are 26 spoke cavities which are powered using 52 tetrodes; each pair of tetrodes is fed by an anode power supply, making a total of 26 anode power supplies in this part of the linac.

Research is ongoing at the FREIA laboratory in Uppsala, Sweden, on the reliability and feasibility of tetrode solutions for both ESS and ESS$\nu$SB. At this laboratory, two 400\,kW RF stations at 352.21\,mHz have been installed. One manufactured by Itelco-Electrosys~\cite{padamsee:2014_sc_rfcav} was commissioned in 2015; the other was manufactured by DB Science and commissioned in 2016~\cite{Jobs:2017_freia_report}. These stations were used for testing spoke cryomodule prototypes for ESS.

Unfortunately, multiple issues with both RF stations delayed or inhibited normal operations. The tetrodes have shown certain reliability issues and the new version of the vacuum tube, i.e., the TH595A (in replacement of the TH595) has yet to prove its reliability. However, the most worrying issue for the future operations of ESS is that there is only a single manufacturer worldwide supplying such vacuum tubes, which may pose serious supply issues. Recently, prices are up by 40\%, with a six month lead-time for orders. The future of these tubes is not guaranteed, as the market is shifting towards solid-state technology.

\subsubsubsection{Spoke Cavity Solid-State Power Amplifiers (SSPAs)}
A pioneering work using SSPAs for high power applications began in 2014 at the SOLEIL synchrotron in Saint-Aubin, France, with a high-power amplifier at 352.21\,mHz, combining several 330\,W MOSFET modules for the first time: 40 kW for the booster, then $2 {\times} 180~\mathrm{kW}$ for the storage ring; these demonstrated extremely high reliability~\cite{marchand:2011_high_pow_sspa}. Seven high-power amplifiers (150\,kW) were also constructed at ESRF~\cite{jacob:2011_SSPAs_ESRF} in Grenoble, France. The technology is now proposed commercially by an increasing number of industrial suppliers (e.g., Thales at 200\,mHz for CERN, and IBA at 176\,mHz for MYRRHA).

A major leap forward comes from CERN, where 200\,mHz tetrodes have been replaced by SSPAs for the upgrade of the acceleration system of the Super Proton Synchrotron (SPS). The transistors are assembled in sets of 4 per module, for a total of 2 kilowatts. A sum of 2560 modules, i.e. 10,240 transistors, will then be spread across 32 towers. The full system will be able to provide two times $2{\times}21.6~\mathrm{MW}$.

Power combination is crucial for enabling the economic competitiveness of a system of this size, and relies on a cavity combiner. A prototype was realised at ESRF with a 144:1 cavity combiner~\cite{crisp:2013_cern-liu-sps}. Since the power is distributed across hundreds of transistors, if a few transistors stop working, the RF system will not stop completely, which would be the case for a vacuum tube-based station.

The cavity combiner is designed with the magnetic field independent of the azimuthal and vertical positions, such that 100 input ports are homogeneously coupled using inductive coupling via loops. Then, feeding these ports with efficient SSPA modules allows for high RF power combination with high efficiency, within one stage of combination. Another advantage is that a variable number of inputs are allowed.  Not requiring a set number at the design stage allows for redundant implementation~\cite{langlois:2011_sspa_esrf}.

\subsubsubsection{Research and development at FREIA on SSPAs}

Work is also ongoing at FREIA laboratory on the development of SSPA technology for ESS, MAX~IV (a neighbouring fourth-generation synchrotron light source in Lund), and ESS$\nu$SB. Specifically, upgrading the ESS accelerator power from 5\,MV to 10\,MV to serve the neutrino beam ESS$\nu$SB experiment is required due to the accelerator's pulse frequency increasing from 14 Hz to 28 Hz (5\% to 10\% duty cycle); this will put even more stringent requirements on the RF amplifiers. 

A 10\,kW prototype SSPA has been built at FREIA~\cite{Dancila:2017_10kwSSPA} and will be used as a starting point to design, construct, and test a 400\,kW SSPA station. This development may be of decisive importance for as a replacement of the current vacuum-tube-based RF power amplifiers in order to guarantee a long-term stable and more energy efficient operation of the accelerators.

The ongoing research resulted in several publications on a compact, 10\,kW, solid-state RF power amplifier operating at 352\,mHz. During tests, a total output peak power of 10.5\,kW has been measured. The amplifier combines SSPA modules at 1.25\,kW each, built around a commercially available LDMOS transistor in a single-ended architecture with 71\% efficiency in pulsed operation~\cite{Haapala:2016_kWSSPA}. A feedback-controlled RF power flatness compensation was also demonstrated at a 10\,kW level~\cite{duc:2019_10kWSSPA_feedback}. 

The forthcoming work on project will involve producing a 400\,kW prototype power station based on combining $4{\times}100$\,kW units, each unit composed of $70{\times}1.5$\,kW SSPA modules. A current budget of approximately 3.5\,m Swedish Krona is available at FREIA from a Vetenskapsr{\aa}det (VR) grant to purchase transistors, manufacture a cavity combiner procured by Swedish industry and assemble a 100\,kW prototype. Commissioning of the prototype is planned for late 2022. Replacement of the spoke power amplifiers with solid-state power amplifiers is estimated to cost about 1.2\,m€ per unit, for a total of 31.2\,m€ to power the 26 Spoke cavities sections.

\subsubsection{Superconducting Cryomodule Couplers}
Given that the average current during the pulse decreases in the ESS$\nu$SB operation mode (for the H$^-$ beam) due to the high-frequency chopping of the ESS$\nu$SB (extraction gap generation), the coupler design envelope of maximum delivered power is not affected (${\sim}$1.1\,MW)~\cite{Arcambal:2021tnt}. 

Therefore, the only remaining issue is accommodating the increased duty cycle. The present coupler was designed to accommodate the 10{\%} duty cycle of the SPL~\cite{charrier2008704}, which is not a hard limit. The 28\,Hz option is thus well within its operating envelope. Increasing the duty cycle and repetition rate for the 70\,Hz scheme (5${\times}$ the nominal ESS value of 14\,Hz), the duty cycle increases further; whether or not this can be accommodated needs to be verified.

The conditioning of the couplers is done at CEA for repetition rates of up to 48\,Hz (limitations of the power source did not permit going to higher repetition rates), but at shorter pulse lengths due to the absence of beam which creates standing waves (reflection) at the coupler~\cite{Arcambal:2015set,Arcambal:2017fwi}.

The hard limit here is the electrical breakdown on peak power, and these coupler have a wide margin for average-power dissipation capabilities. The external conductor is gas cooled and the internal antenna is water cooled. There is also the possibility to cool the ceramic window, but at the ESS duty cycle this is not necessary, and the circuit is not connected to the water manifolds. However, for the higher-duty cycles, it may be necessary to bring the water cooling to the ceramic window.

\subsection{Cryogenic Systems}
The total cryogenic load for nominal operation is two-thirds static load and one-third dynamic load. If the dynamic load is doubled (from ESS$\nu$SB's H$^-$ beam) then the total load increases by one-third. The nominal ESS cryogenic plant has capacity to cover the needs of additional cryomodules in the linac's contingency space; these are the cryomodules needed to bring both the proton and H$^-$ beams to 2.5\,GeV. A 30\% overhead margin is also included in the design to account for unforeseen cryogenic losses (e.g. badly performing RF cavities)

In total, the ESS cryogenics design already includes an extra capacity of 63\% (including the contingency space and overhead margin) which is expected to adequately accommodate the extra 33\% heat load generated by doubling the dynamic load. 

Dynamic heat comes largely from heat that must be collected from the accelerating cavities and is directly related to the quality factor of those elements. If the quality factor does not degrade beyond nominal when the cavities are assembled into cryomodules -- a fact that is considered unlikely -- then the present cryogenic plant should be able to meet the cooling needs of the upgraded ESS linac. 

In terms of impact on operations: the installation of new cryogenic distribution lines in the high-energy part of the linac can only be done when the machine is stopped. The primary cost for this upgrade, then, is the new cryogenic distribution lines for the high-energy area in the accelerator tunnel. 

Despite the estimate that existing cryogenic plant can accommodate the linac upgrade, the cooling needs of the of newly constructed ESS$\nu$SB facilities (the L2R, accumulator, horns, and target) may also deplete a considerable percentage of the cooling capacity overhead. This may result in relatively tight margins, which can impact overall availability of the facility.

\subsubsection{Diagnostics}

\subsubsubsection{Beam Current Monitors (BCM)}
The beam current monotor (BCM) sensors -- that is, the AC current-transformers (ACCTs) -- including their analogue front-end / back-end electronics -- work equally well with positively and negatively charged particles. The existing BCM system has already successfully measured sub-mA proton beams with pulse lengths as short as 1~\SI{}{\micro\second}. However, with such short pulses, the ACCT bandwidth limitation (i.e. 1\,mHz) becomes visible in the rise-fall times of the measured beam pulse.

The linac upgrade will put new requirements on the BCM, primarily due to the combined proton and H$^-$ pulses, with the H$^-$ pulses being as short as 1~\SI{}{\micro\second} and their separation being as small as 0.75 ms. The effect of the BCM bandwidth limitation and droop rate on measuring a combined proton / H$^-$ beam and new timing requirements will require more detailed study. Considering the new linac layout, this may even require customised ACCT sensors with a better compromise between the bandwidth and droop rate, particularly on the sections that simultaneously accelerate proton and H$^-$ pulses. Similarly, the existing BCM firmware/software, including the machine protection functions, will need to be modified and tested and verified. This is where most of the required effort is likely to be necessary. 

\subsubsubsection{Further Scopes}
The following topics must also be re-evaluated for the ESS$\nu$SB linac upgrade:

\begin{itemize}

\item Additional systems for the new H$^-$ ion source and LEBT sections, including new LEBT chopper and upgraded MEBT chopper.
\item Laser-based instrumentation, transverse and longitudinal (mode-locked). Laser-wires typically measure a 1D beam profile and/or 2D transverse emittance from the products of photo-detached ions as a laser beam is scanned across the H$^-$ beam~\cite{Gibson:2018_laser_scan}.
\item Addition of beam-in-gap measurement if needed. This could be a function of the laser-based instrumentation.
\item Additional collimation and associated instrumentation as needed.
\item Additional instrumentation to support the new cryo-modules, particularly BLMs
\item Replacement of analogue front ends for systems that must resolve the extraction gap 
\item Upgrade of data acquisition and protection functions consistent with the increases of the repetition rate, data rate, and beam power. Two classes of upgrades: firmware/software alone and replacement/modification of electronics.
\item Power deposition assessment and signal estimation for beam-intercepting devices, particularly in cases where atomic electrons or H$^0$ are stopped. Modification to these devices as required, such as different beam-intercepting materials.
\item Upgraded instrumentation to accommodate any special studies/tuning beam modes. Ring injection and accumulation may benefit from unusual chopping configurations.

\end{itemize}

\subsection{Operations and Installation}
\label{sect:operations}
\subsubsubsection{Upgraded LEBT}
For the H$^-$ beam, all new components will be needed upstream of the second solenoid. All LEBT components upstream of this point will have to be either newly installed and commissioned or reinstalled or recommissioned. To minimise the risk of impact on nominal ESS neutron production, the elements required for accelerating protons along the nominal linac towards the neutron source must be prioritised.

\subsubsubsection{RFQ Upgrade}
As discussed previously, an upgraded RFQ must be designed and procured since the duty cycle to deliver the two beams will increase by at least a factor of two (up to a minimum of 8\%) -- the maximum duty cycle (due to cooling) is currently at 5\%. This new RFQ will have to be installed, conditioned, and commissioned. To mitigate the risks of affecting nominal ESS neutron production, the main conditioning could be done inside a test-stand bunker before the swap of the RFQs.

\subsubsubsection{MEBT Upgrade}
The current MEBT may need replacement of bunchers and new pulsing quadrupoles; and it will certainly require a higher-frequency chopper. Simulations and tests are required to prove that the rest of the components in this section will withstand the increased duty cycle. The worst case scenario would be that a more substantial number of the MEBT components have to be replaced.

\subsubsubsection{Upgraded DTL tanks}
The DTL tanks are rated for a maximum duty cycle of 10\%, which indicates that they will withstand the increased duty cycle. However, simulations and testing will have to be done to demonstrate this, and there are concerns that the considerable increase in deposited heat may be exceed the present cooling capacity. In the worst case scenario, tests and simulations results would show that the DTL tanks could not withstand the increased duty cycle and that replacements are needed for all 5 tanks.

\subsubsubsection{Upgraded Modulator Capacitors}
As detailed previously, the modulator capacitors for the nominal ESS linac are each rated for a peak power of 11.5\,MW and average power 660\,kVA. Since the peak power does not increase, only the capacitor chargers of the modulators need to be upgraded to deliver power at an increased repetition rate.

\subsubsubsection{Upgraded Cooling Systems}
The cryogenic distribution system is expected to be able to handle the additional cooling to the upgraded couplers and the additional 8 HBL cryomodules. However, these expansion activities result in some risks of affecting the neutron production program, since warm-ups and cooldowns take time, potential leaks may have to be fixed, along with other minor concerns.

The doubling in average power requires modifications of the skids (as discussed in Deliverable~2.1 \cite{galnander:2018_D2.1}), which is thus also included in the suggested timeline.

\subsubsubsection{Upgraded Superconducting Linac}
The cryomodule (CM) couplers may have to be upgraded due to insufficient cooling. Replacing the couplers at ESS would require the establishing of a clean room (preferably near Test Stand $\#$2, where site acceptance tests are currently performed on procured cryomodules). Furthermore, simulations and tests are required to prove that the spoke, medium-beta (MB), and high-beta (HB) cryomodules will withstand the increased duty cycle after the coupler upgrades. The steerer magnets also have to be upgraded to pulsed steerer magnets, and within the High Energy Beam Transport (HEBT) section, eight new HB cryomodules must be installed for the energy upgrade from 2.0 to 2.5\,GeV.

\subsubsubsection{Installation Windows} \label{installations}
As presented in \cite{ICHEP2020}, a possible ESS$\nu$SB schedule is as follows:
\begin{itemize}
  \item 2021: End of ESS$\nu$SB Conceptual Design Study, CDR, and preliminary costing 
  \item 2022-2024: Preparatory Phase, TDR 
  \item 2025-2026: Preconstruction Phase, International Agreement 
  \item 2027-2034: Construction of the facility and detectors, including commissioning
\end{itemize}

The operational challenges would begin with the construction and installations of the linac upgrade. In the report \cite{ESSops}, for 2028 and onwards, the schedule for normal operations of the neutron source includes two long shutdowns per year:
\begin{itemize}
  \item Winter (ca 8.9 weeks)
  \item Summer (ca 8.1 weeks)
\end{itemize}
This amounts to a total of 17 weeks of shutdowns per year. This condition is projected to apply for the year 2027; for the present planning study, these windows are extended through the entire construction phase.

In order to mitigate the risks with swapping the RFQ, it is highly beneficial to properly condition it within a test-stand bunker beforehand. Due to the limited time constraints, and since eight new HB cryomodules are to be ordered and installed, this (or another) test-stand bunker can and is assumed to be used for conditioning the CMs before installations in the tunnel. Following such a procedure should allow for a smoother integration into the linac.

\subsection*{Schedule Prototype}
To address the contingency of having to upgrade all CM couplers, a procedure is considered such that 3--5 CMs are swapped during multiple shutdown periods (winters and summers, \cite{ESSops}). The estimated time has been determined by constructing a prototype procedure for the CM swaps as shown in Fig.~\ref{fig:CMswap_procedure}.

\begin{figure*}[ht!]
    \centering
    \includegraphics[width=\textwidth]{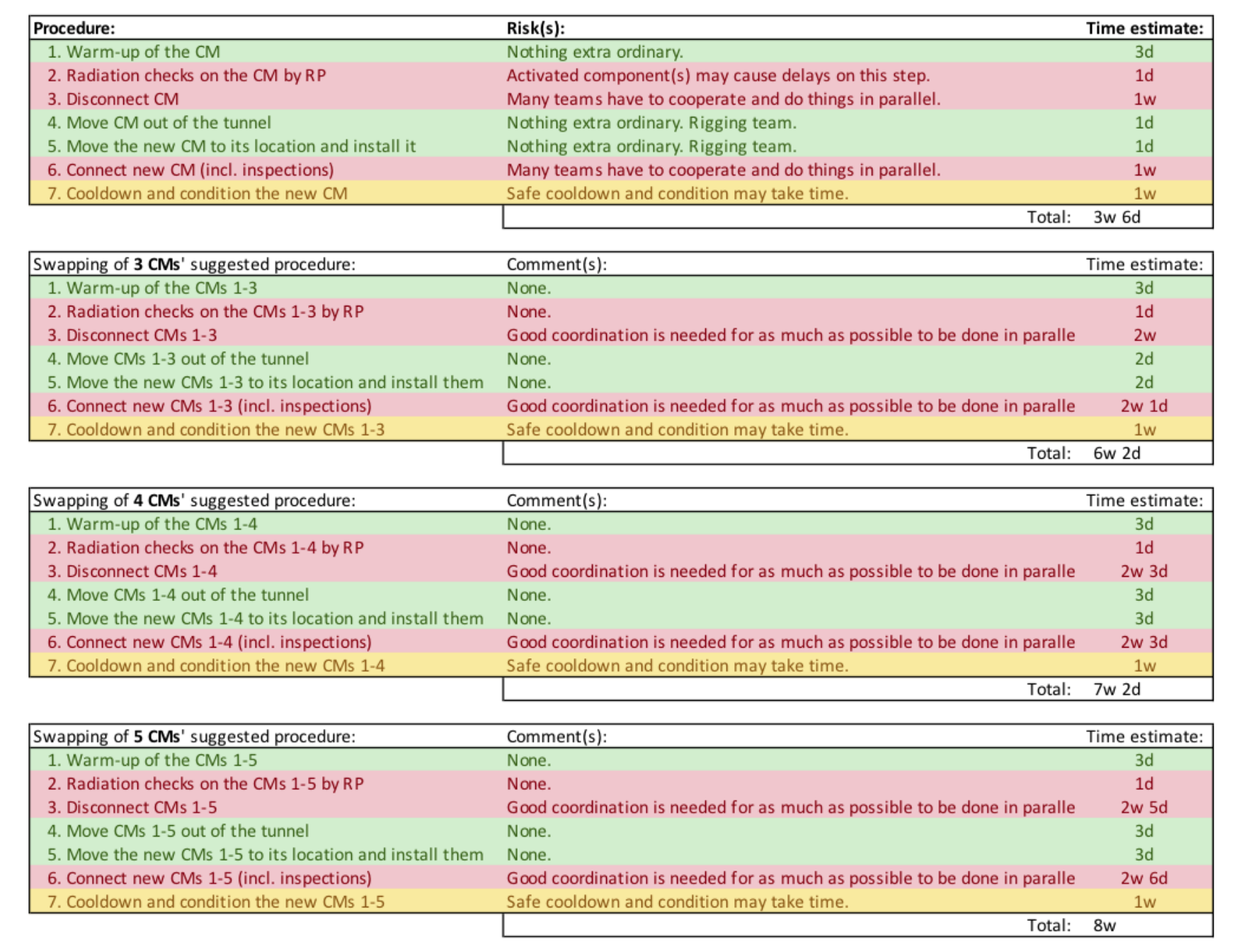}
    \caption{A swap procedure considered for each cryomodule during an ESS neutron program shutdown period (8--9 weeks each).}
    \label{fig:CMswap_procedure}
\end{figure*}

For this schedule prototype, the colour-coding of the different activities are defined in terms of risk for affecting the neutron-production program:
\begin{itemize}
  \item Red = High (or very high) risk of a serious negative impact
  \item Yellow = Medium risk of a moderate negative impact
  \item Green = Low risk
\end{itemize}

By these definitions, attempting to swap 5 CMs during a single shutdown period is a high-risk activity, while a 3--4 CM swap is a medium-risk activity. To mitigate the potential negative impact on neutron production, fewer high-risk activities are preferred over fast completion of the linac upgrade. An attempt at solving this planning challenge can be seen in Figure~\ref{fig:CMswap_timeline}.

\begin{figure*}[ht!]
    \centering
    \includegraphics[width=\textwidth]{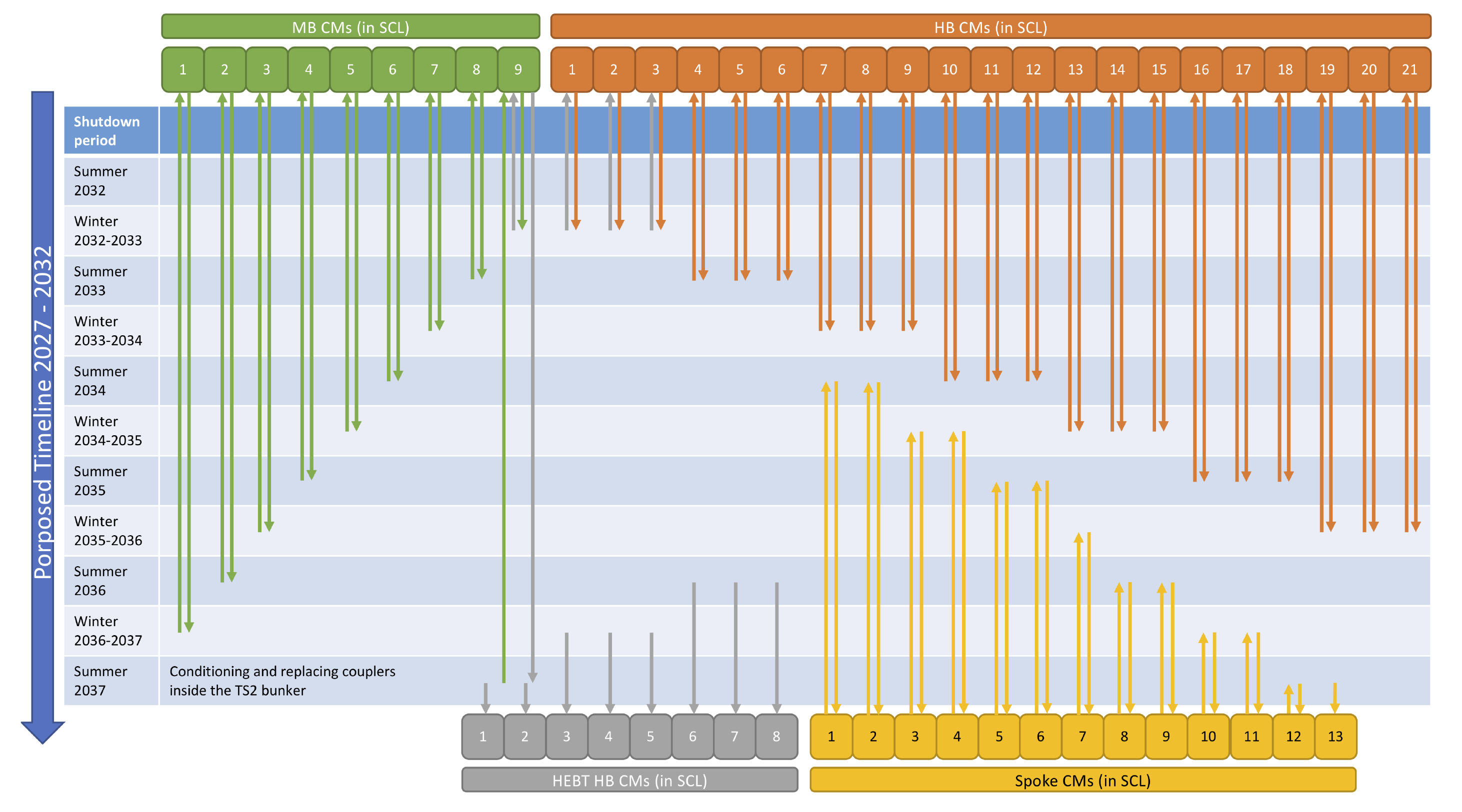}
    \caption{A visual outline for scheduling the CM coupler upgrades during the ESS neutron production program and its shutdown periods. This plan takes as few high-risk activities as possible, but remains within the ESS$\nu$SB upgrade timeline.}
    \label{fig:CMswap_timeline}
\end{figure*}

As can be seen in Fig.~\ref{fig:CMswap_timeline}, the CM coupler upgrade plan shows a possible solution and is used in the initial schedule prototype, attached in Fig.~\ref{fig:timeline}. Note that activities in the NCL/SCL areas during Autumn and Spring may also be planned during any downtime of the machine which may occur (e.g. for machine studies, or maintenance). 

It becomes evident that, during the ESS$\nu$SB upgrade of the ESS linac, seven shutdown periods involve a high risk of impacting nominal neutron production. %The full spreadsheet is attached in a separate document, which includes motivations of each activity's risk index.

\subsubsubsection{Analysis of the Timeline Prototype} \label{results}
%After studying the outcome of the schedule prototype, it is clear that some of the required linac installation work carries high risks of impacting the nominal ESS neutron program.
From further assessment, it can be concluded that the tasks with the highest risk -- and the most challenging from an operations point of view -- are those involving a vital system or component which has to be reinstalled, reconditioned, and/or recommissioned.

One such system is the RFQ. The current RFQ has to be disconnected from all electrical wiring, network cables, water cooling, etc., and the new RFQ must then be integrated into the linac (including installations, re-conditioning, and commissioning with beam); this must all be accomplished during a time window of less than nine weeks, as per Fig.~\ref{fig:CMswap_timeline}. It is assumed, as mentioned above, that the main conditioning of the RFQ has taken place inside a test-stand bunker, and is completed prior to its installation and integration into the linac.

The nominal RFQ conditioning was conducted in June 2021, reaching nominal conditions in roughly five weeks. Hence, as this section is a vital part of the ESS linac to obtain neutron production, this activity can be considered as having the highest risk of affecting the ESS neutron production program negatively.

\begin{figure*}[ht!]
    \centering
    \includegraphics[width=0.9\textwidth]{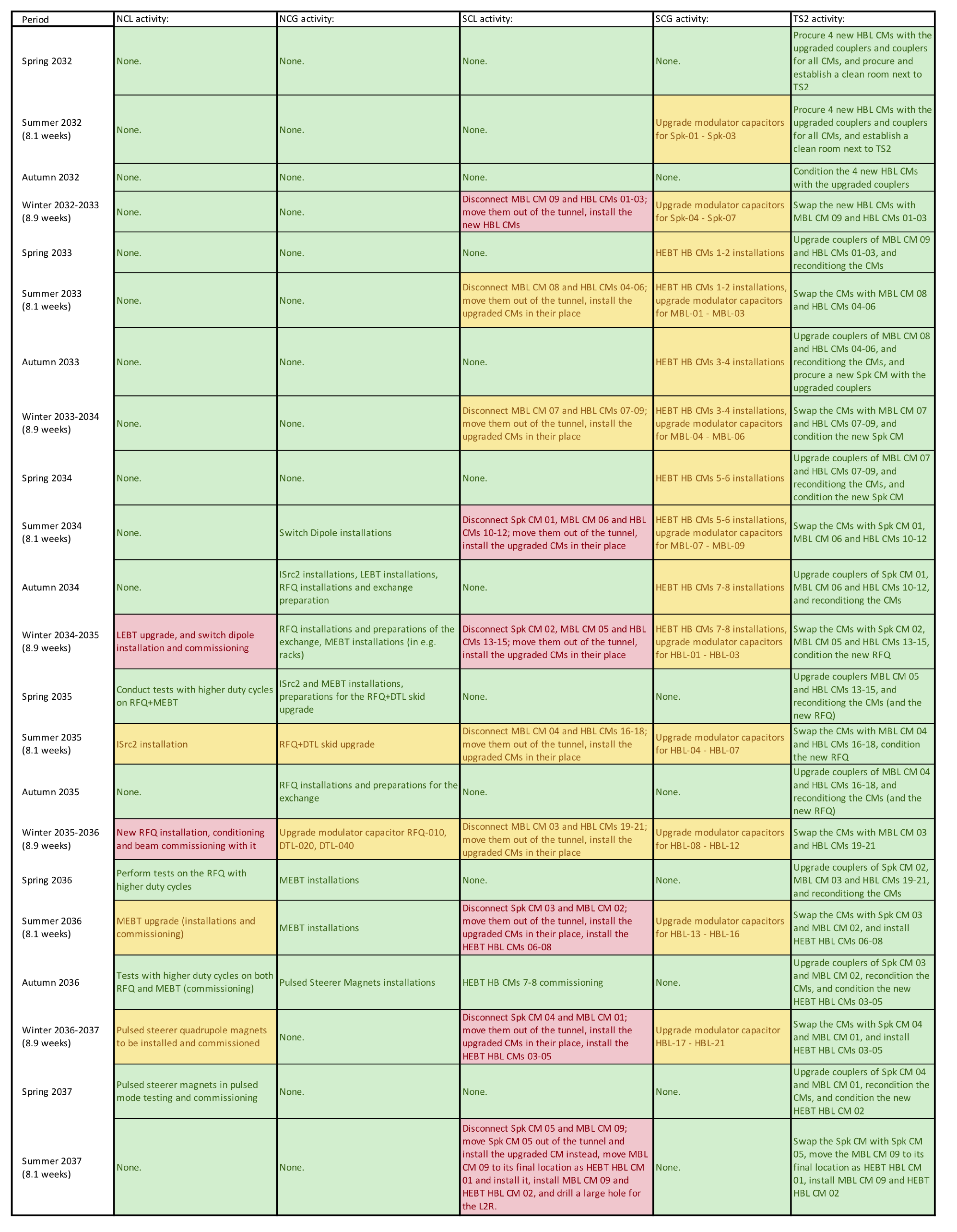}
    \caption{The first suggested timeline for the installations, aimed at mitigating the operational challenges arising from the required linac upgrade.}
    \label{fig:timeline}
\end{figure*}

Another section vital for neutron production is the LEBT, of which all components upstream of the second solenoid must be replaced, along with the installation of a new switch dipole magnet for merging the proton and H$^-$ beams from the two ion sources. This will require a new support platform. As can be seen in Figure~\ref{fig:timeline}, its suggested installation period is during the winter shutdown 2029--2030. This activity carries a high risk since this section is required by the linac for neutron production. 

Using the left-hand layout of Fig.~\ref{fig:source_options}, the proton ion source must be moved and realigned from the downstream beam axis at a 30-degree angle. For this approach, the realignment has to be done during the LEBT upgrade, which includes the installation of the switch dipole. This installation has a very high risk of causing a negative impact on the neutron production, as the LEBT section's systems are partially replaced with new systems needing to be properly integrated and recommissioned. As a mitigation effort, scheduling this high-risk action at the same time as the installation of the first 2 HEBT cryomodules can reduce the total number of high-risk periods.

%Unless a trajectory good enough for both the proton and hydrogen ion bunches are found, the steerer and focusing magnets have to be upgraded to pulsed magnets (which is in alignment with the ESS$\nu$SB baseline). Hence, for this project, the upgrade is assumed to be required, meaning that all steerer magnets must be upgraded to pulsed magnets. This means that ca 120 and ca 240 steerer and focusing magnets, respectively, have to be replaced in the Linac [TODO: Linac lattice reference]. In terms of the cooling system, no upgrade is expected to be required as the pulsed mode will result in net power going down [TODO : Insert reference here].

\subsection{Safety}
 
The ESS is a complex facility where several hazards might occur. These hazards include radioactive hazards as well as non-radioactive hazards. Although ESS is not defined as a nuclear facility according to Swedish regulations, ESS emphasises the objective of setting radiation shielding and safety as a main priority for all phases of the project: from design, through construction and operation, to decommissioning. This should be no different with the ESS$\nu$SB project.

When starting the operation of ESS, there will be no significant radioactive inventory. During operation, penetrating fast neutrons are generated in the target and by proton beam losses in the accelerator. The main inventory of nuclides will be in the target, and thus it is in the target station where most radioactivity will be generated. This means that the main hazards arise from radioactive sources. However, other hazards, here termed non-radiation hazards, must be addressed as well. Examples are hazards originating from cryogenics, high-voltage, electromagnetic fields, heavy equipment, working at high elevation, transports etc. Thus, in order to protect the ESS staff, the public, and the environment, it is necessary that ESS states and defines specific General Safety Objectives (GSO). The GSO will serve as a guiding document at ESS, giving necessary input of how to design the ESS facility. Many of the work already done for the ESS proton linac in developing the GSO will have to be revisited or done for part of the ESS$\nu$SB upgrade.

At the time when the linac upgrade for the ESS$\nu$SB project starts, many of the procedures and internal safety regulations at ESS will be well established. However, the power upgrade in the linac, which will be doubled from 5 to 10\,MW, will have to re-addressed. Radiation regarding neutrons generated from losses (coming from H$^-$ losses and stripping) must be evaluated in order to guarantee staff and public safety, but also non-radiation hazards will have to be studied. It import to stress that all cavities will operate at a doubled duty cycle (for the baseline ESS$\nu$SB H$^-$ pulsing scheme) and thus the effects from X-rays coming from the electromagnetic fields have to be reviewed. The same in-depth analysis will have to be performed for the transport lines and accumulator as well.

With extensive competence developed in this area, given the experience with the spallation source, ESS should have sufficient expertise on hand to aid the ESS$\nu$SB project to address such safety issues in due time~\cite{Jacobsson:IPAC11-WEPC166}.

\subsection{Cost}
In this section, an overview is given of costs for the linac upgrades needed for ESS$\nu$SB. A summary of the main costs is given in Table~\ref{tab:ESSvsbLinacCost}. 

For the upgrade, an $H^{-}$ source and new LEBT will be necessary -- the amount presented in Table~\ref{tab:ESSvsbLinacCost} is thus based on the cost of the original proton source and LEBT at ESS. As discussed earlier, the most promising $H^{-}$ source is the SNS design, which is used as a baseline. However, significant development is needed in order to achieve the desired output current; this uncertainty may dictate a large contingency budget needed for development.

\begin{table}[ht!]%[H]
  \begin{center}
    \caption{ESS$\nu$SB linac upgrade cost breakdown}
    \label{tab:ESSvsbLinacCost}
    \begin{tabular}{lc}
    \textbf{Item}  &\textbf{Cost [M\texteuro]} \\
\hline
Ion Source and Low-Energy Beam Transport (LEBT) & 5.0\\
Radio-Frequency Quadrupole & 6.9\\
Medium Energy Beam Transport (MEBT) Upgrade & 3.0\\
Drift-Tube Linac with BPMs, BCMs & 13.4\\
High-Beta Linac (HBL) Upgrade & 10.4\\
33 Modulator Upgrades & 3.5\\
8 New Modulators & 9.0\\
15 Grid—Modulator Transformers & 5.6\\
11 Grid—Modulator Transformers Retrofitted &  0.5\\
26 Solid State Spoke Amplifiers & 26.0\\
New Klystrons for upgraded HBL & 12.1\\
Remaining Klystron Refurbishment/Replacement & 25.2\\
Cryogenics,Water Cooling,Civil Eng. & 12.0\\
\textbf{Total} & \textbf{132.6}\\
\hline
     \end{tabular}
   \end{center}
\end{table}

In terms of civil engineering, the current baseline linac design of merging beams in the LEBT is fairly conservative, so some savings may be possible in this regard (as opposed to merging beams further downstream in the MEBT).

Currently, it is highly doubtful that the DTL can handle the necessary 10\% duty cycle. For the nominal proton linac, it disperses a heat load of roughly 4.5 MW. Doubling this is non-trivial. As a baseline, it is assumed that a new DTL is needed. This is taken as 85\% of the cost of the current DTL, owing to reduced development costs. If other novel cooling solutions can be implemented, this cost may be dramatically reduced.

Similarly, the RFQ redesign will mainly involve added cooling capacity, with an estimated 20\% reduction in overall cost from the original ESS RFQ, again due to redundant development. 

For the RF upgrade of the ESS-designed stacked modulators, the cost involves the retrofitting of the existing modulators and new modulators for the upgraded HBL and SML modulator upgrades. As an additional figure of merit, the modulators' annual electricity cost for the baseline pulsing scheme increases from 7.2\,m€ to 14.4\,m€.

The kilovolt-amperage (kVA) requirements for grid power (wall-plug) feeding the upgraded modulator transformers have also been reviewed. The estimate presented in Table~\ref{tab:ESSvsbLinacCost} reuses existing transformers to power one modulator each instead of three, then adds fifteen newly procured higher-kVA transformers which power two modulators each. The transformers are rated to last the facility's lifetime, so this reuse plan is considered low-risk. Retrofitting the existing transformers will require new control boards and busbars.

For klystrons already installed in the linac, it is estimated that collector replacement is only necessary for roughly one fourth of the procured klystrons. However, collector replacement could reach as high as 80\% of the original klystron cost. It is assumed for now that a refurbishment cost for the klystrons comes at ${\sim}$60\% the original price. These estimates are fairly conservative, the more nuanced analysis from Section~\ref{sect:klystron_upgrades} gives a variety of options which are likely to fall well under this cost point.

For the new klystrons, the current design should need no major overhauls. Moreover, the development work on water cooling for the refurbished klystrons should be applicable to the new klystrons. To estimate the new klystron cost, a mean price per unit of 330~k{\texteuro} was used for existing klystrons, plus an additional 15\% development cost .

For the SCL cavities, the couplers are rated sufficient for the 10\% duty cycle necessary for ESS$\nu$SB. Early conditioning results support this rating, so their replacement is categorised as a risk, not a baseline necessity.

Costs for the most significant items in the linac upgrade plan may depend heavily on coordination with ESS scheduled refurbishments and replacements. The solid-state spoke amplifiers (SSPAs) are a good example or this coordination. Although they are designed to be more reliable and efficient than tetrodes, the capital cost is still significant. The most optimistic scenario is having solid-state amplifiers procured to replace tetrodes at end-of-lifetime and being able to add the additional required units for ESS$\nu$SB in a modular fashion.

For this cost estimate, it was assumed that ESS$\nu$SB must retrofit or replace all klystrons and install all SSPAs with no shared costs. In the scenario where ESS pays for all but the increased capacitor banks for the SSPAs, and charges only a nominal fraction above the nominal klystron price, a savings of 20 M{\texteuro} or more could be expected.

\subsection{Summary}
The unprecedented power of the ESS linac set the impetus to investigate the feasibility of increasing its duty cycle to be the driver for a neutrino oscillation experiment. The target and horn requirements put a limit on the final pulse length, which is obtainable only by adding an accumulator between the linac and the target. For improved injection to such an accumulator, the H$^-$ beam is accelerated in the linac and through a charge-exchange system -- in the subsequent injection stage, the ring already contains circulating ions. This process entails the addition of an H$^-$ ion source to the facility; the specifics have been presented on how and where the beam from this additional ion source is merged with the proton beam of the ESS linac. The acceleration of negatively charged ions in the same lattice that is used for the acceleration and transport of positively charged protons was simulated and no prohibitive issues were found. The losses specific to H$^-$ were studied in detail; it was found that these losses are within acceptable limits (additional mitigating actions to further reduce the losses were also proposed).

The baseline pulse structure was established to be 14 $\times$ 4 batches of 650\,$\mu s$ with a gap of 100\,$\mu s$ between batches, leaving the 2.86\,ms proton pulses for the nominal ESS neutron production interleaved with 2.9\, ms H$^-$ pulses of 62.5\,mA peak current. Increasing the pulse length to 3.5\,ms would even allow for a reduction to the peak current to 50\,mA, reducing the H$^-$ specific losses even further. This depends on the possibility of increasing the RF pulses to 4\,ms. 
Generating the RF at 28\,Hz requires modifications in the power chain from the grid to the cavity, the major components in this chain were studied and solutions were presented whenever a change was needed, and the capability of the components which could operate at increased duty cycle were also discussed.
Finally, the preliminary schedule for the installation and modification of components, compatible with the ESS's neutron programme, were presented. 
\clearpage
\clearpage

\setcounter{figure}{0}
\numberwithin{figure}{section}
\setcounter{equation}{0}
\numberwithin{equation}{section}
\setcounter{table}{0}
\numberwithin{table}{section}

\section{Accumulator Ring} \label{sec:accumulator}

The purpose of the ESS$\nu$SB accumulator is to transform long pulses of H$^-$ ions, delivered by the ESS linac, into very short and intense pulses of protons through multi-turn injection and single-turn extraction. A long transfer line brings the H$^-$ ions from the end of the linac to the injection point of the ring. This transfer line is described in the previous chapter. A second transfer line brings the compressed pulses from the ring extraction point to the beam switchyard where they are  distributed over the four targets. The ring and transfer lines together constitute 900\,m of new beam line which has been designed with strict requirements on beam loss control~\cite{Baussan:2013zcy}. In addition, this chapter includes a short discussion on the possibilities of using ESSnuSB accelerator complex to produce short neutron pulses for material science.

%------------------------------------------------------------------------------------------------------
\subsection{Accumulator Ring Lattice and Optics}\label{subsec:lattice}
%------------------------------------------------------------------------------------------------------
%
The lattice for the ESS$\nu$SB accumulator ring has been developed based on the following main requirements:
\begin{enumerate}
    \item The circumference of the ring must be such that the duration of the extracted beam pulse is less than 1.5\,\SI{}{\micro\second}, in order to comply with the requirements of the neutrino horns that determine the final shape of the neutrino super-beam. For a proton beam with a kinetic energy of 2.5\,GeV, this means a ring circumference of less than 433\,m.
    \item The ring, the associated transfer lines, and the beam switchyard must fit physically within the perimeter of the ESS site.
    \item The ring must be able to store 2.23${\times}10^{14}$ particles per filling at a beam energy of 2.5\,GeV, with an uncontrolled beam loss below 1\,W/m in order to minimise activation of the accelerator equipment.
 \end{enumerate}   
 
The ring has similarities in terms of size and beam parameters to existing machines in operation, namely, the Proton Synchrotron Booster (PSB) at CERN \cite{Newborough:2020obe,Forte:2016qjp}, the Rapid Cycling Synchrotron (RCS) at the China Spallation Neutron Source (CSNS) \cite{Wang:2010bw}, the SNS accumulator ring \cite{Wei:2000wa}, and the RCS at the J-PARC which provides beam to the Material and Life Science Experimental Facility (MLF) \cite{Yamamoto:2022RCS}. Table~\ref{tab:otherrings} lists selected parameters of comparable rings as a comparison to the ESS$\nu$SB accumulator ring.

%----------------------------------------------------
\begin{table}[htbp]%[H]
  \begin{center}
    \caption{A comparison with other machines of ring and beam parameters.}
    \label{tab:otherrings}
    \vspace{0.25 cm}
    \begin{tabular}{lcccc}
 & \textbf{ESSnuSB} & \textbf{PSB} & \textbf{SNS} & \textbf{J-PARC RCS} \\
\hline
Ring circumference  & 384\,m & 157\,m & 248\,m & 348\,m \\
Beam energy at injection  & 2.5\,GeV & 160\,MeV & 1.0\,GeV & 400\,MeV \\
Beam energy at extraction & 2.5\,GeV & 2.0\,GeV & 1.0\,GeV & 3.0\,GeV \\
Number of injected particles & 2.3e14 & 1.6E13 & 1.5E14 &  8.3E13 \\
Number of stored bunches per ring & 1 & 1 & 1 & 1 \\
Repetition rate & $4\times14$\,Hz & 1\,Hz & 60\,Hz & 25\,Hz \\ 
Average beam power & 5\,MW &  & 1.4\,MW & 1.0\,MW \\
\hline
\end{tabular}
\end{center}
\end{table}

The SNS accumulator has the most similarities because it serves as an accumulator only, whereas the other rings are synchrotrons which increase the beam energy before extraction. The number of particles stored in each fill is slightly reduced for present SNS operation, but with the planned proton power upgrade~\cite{Galambos:2018koi}, the bunch intensity will be similar to that of ESS$\nu$SB. Nevertheless, the target average beam power for ESS$\nu$SB is much higher; this is possible because each linac pulse is split into four batches, as illustrated in Fig.~\ref{fig:pulse_struct_detailed_A}. In addition, the ESS$\nu$SB beam energy is higher, which results in fewer issues from space-charge forces. 
%----------------------------------------------------
\begin{figure}[htbp]
  \begin{center}
    \mbox{\epsfig{file=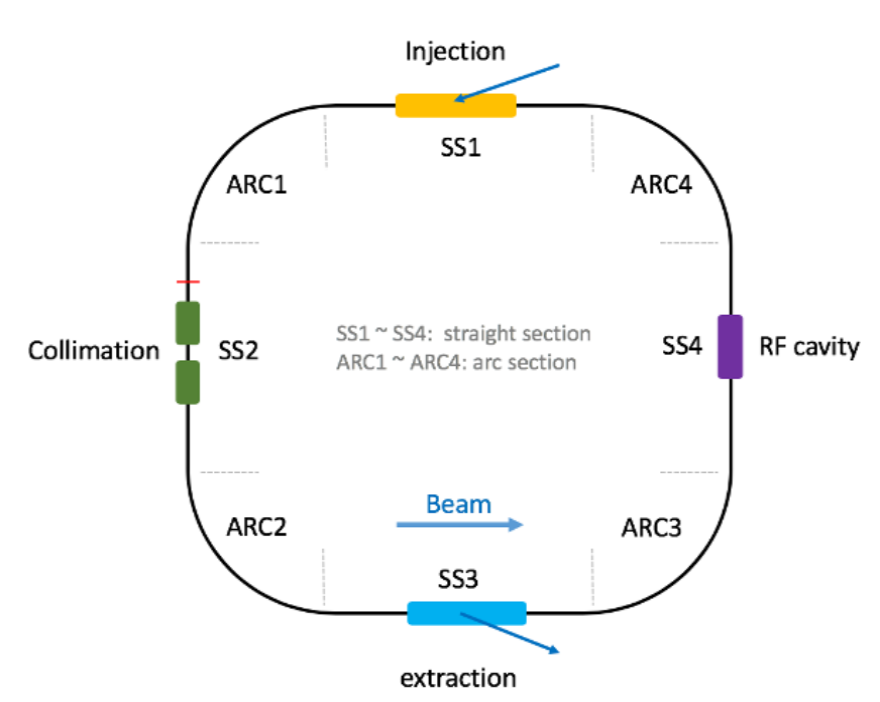,width=9cm}}
\caption{The accumulator ring layout.}
    \label{fig:accumulator_layout}
  \end{center}
\end{figure}
%----------------------------------------------------

The overall layout of the accumulator is designed with four relatively short arcs connected by longer straight sections, taking the SNS accumulator as a basis (see Fig.~\ref{fig:accumulator_layout}). This layout offers ample space in the straight sections for all necessary equipment, such as the injection and extraction system, RF systems for longitudinal beam control, and a collimation system for managing controlled beam loss. The 4-fold symmetry makes it possible to suppress systematic resonances, but not as efficiently as with the 16-fold symmetry of the PSB, where the beam is stored for a much longer duration.

The arcs contain four focusing-defocusing (FODO) cells, each with two dipole magnets and with the dipole magnet centred between two quadrupole magnets. The number and length of the dipole magnets have been chosen to reach the desired bending radius with a moderate magnetic field strength of 1.3\,T.

The advantage of a FODO structure versus, for example, a quadrupole triplet structure, or a double-bend achromat, is that it provides a smoothly varying optical $\beta$-function both in the horizontal and the vertical plane. In addition, a FODO structure allows all the quadrupole magnets to be manufactured with identical configuration, and be driven by a limited number of power supply units, which makes construction and operation less costly.
%------------------------------------------------------------------------------------------
\begin{figure}[htbp]
\begin{center}
\begin{minipage}{0.47\linewidth}
\includegraphics[width=0.99\textwidth]{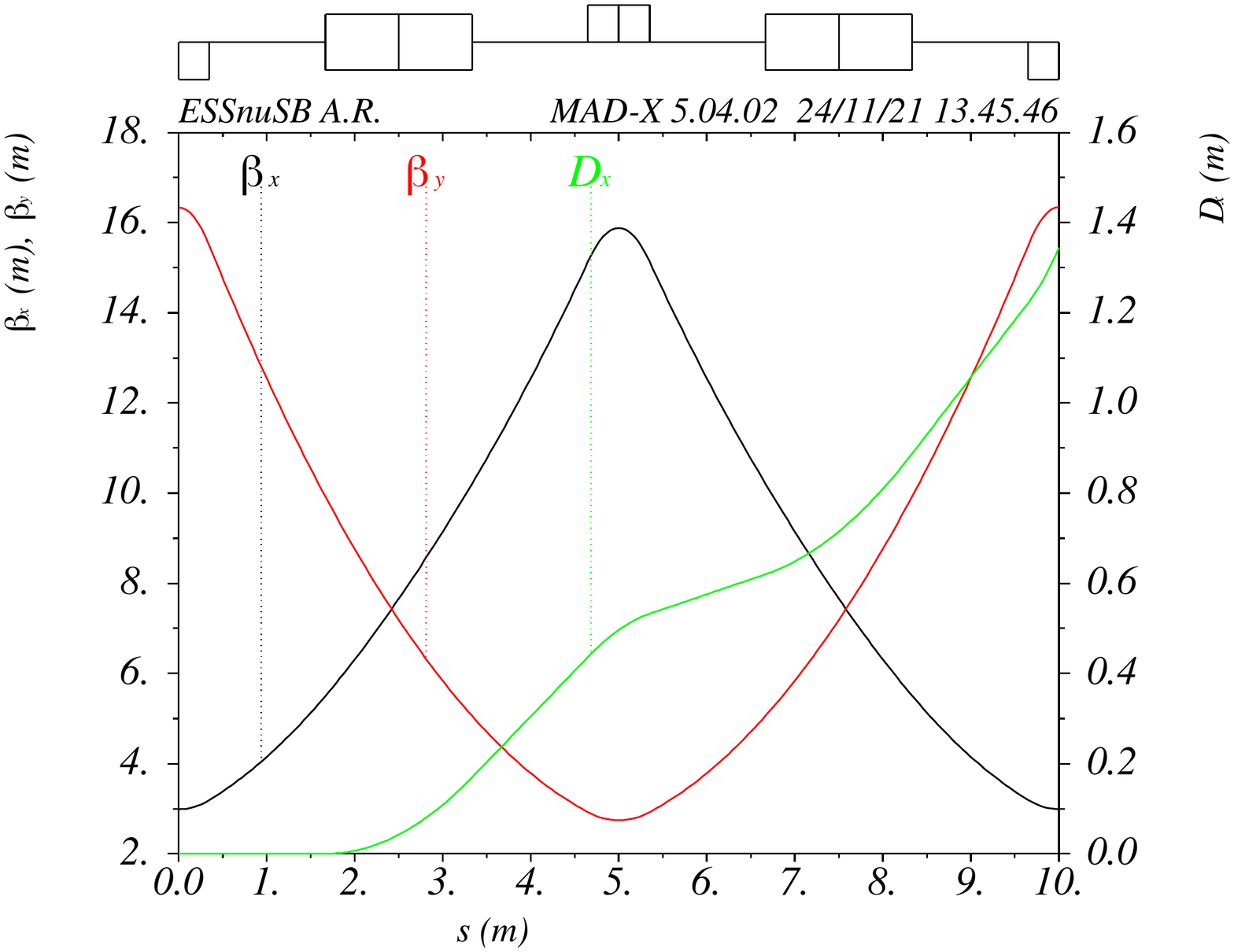}
\caption{\label{fig:FODO}Optical $\beta$-functions and the horizontal dispersion $D_x$ in one FODO cell.}
\end{minipage}\hspace{2pc}%
\begin{minipage}{0.47\linewidth}
\includegraphics[width=0.99\textwidth]{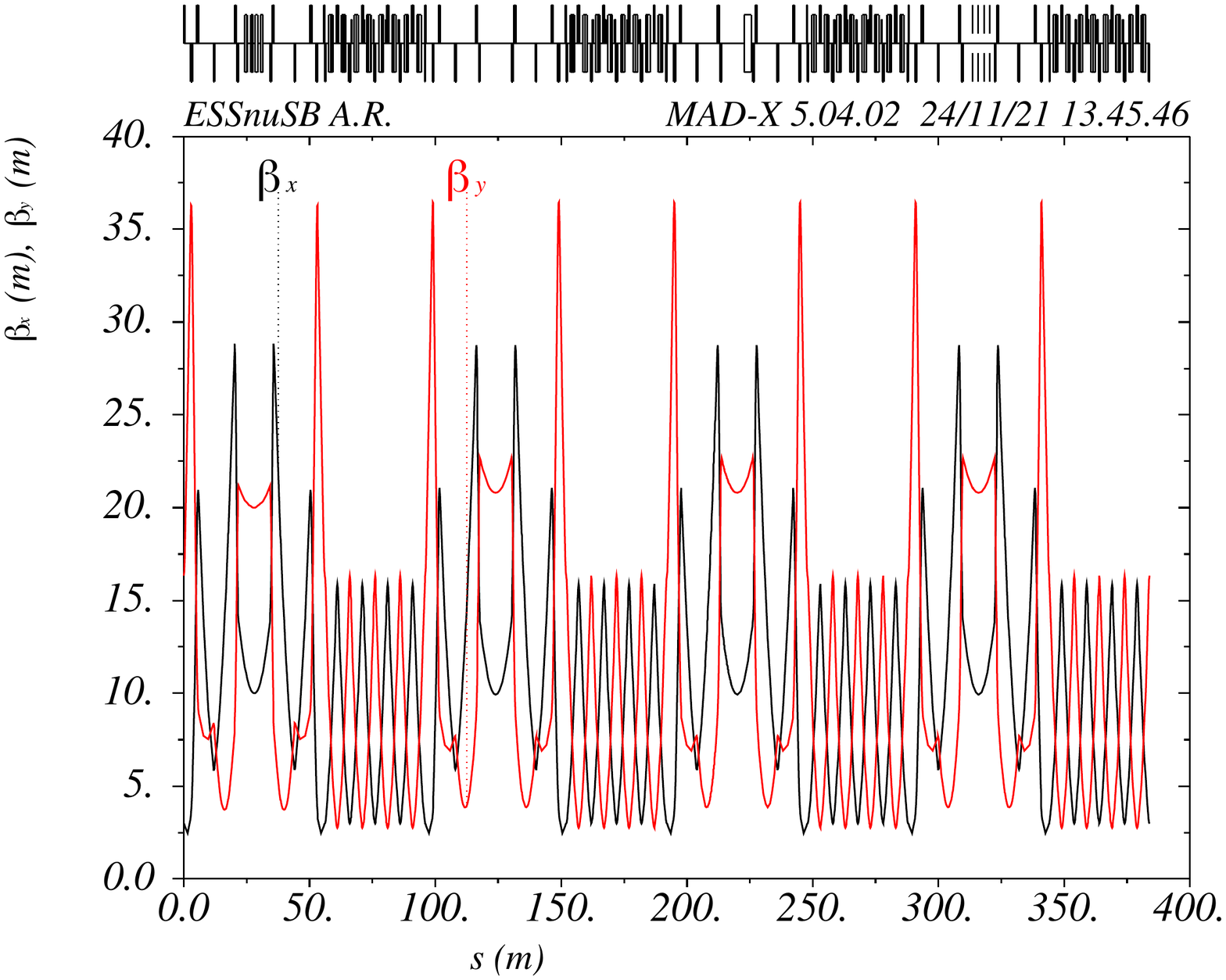}
\caption{\label{fig:ring_betas}Optical $\beta$-functions throughout the ring.}
\end{minipage} 
\end{center}
\end{figure}
%------------------------------------------------------------------------------------------

The focusing strength of the FODO quadrupoles are chosen to provide a $2\pi$ phase advance in the horizontal plane and a vertical phase advance near $2\pi$. This closes the dispersion generated in the horizontal plane by the dipole magnets, which means that the straight sections connecting the arcs are dispersion free. Figure~\ref{fig:FODO} shows the optical $\beta$-function in the horizontal and vertical planes, together with the horizontal dispersion function $D_x$, in one of the FODO cells. See Table~\ref{tab:ring} for a selection of lattice parameters in the ring.
%----------------------------------------------------
\begin{table}[htbp]%[H]
  \begin{center}
    \caption{Overall parameters of the accumulator ring lattice and beam.}
    \label{tab:ring}
    \vspace{0.25 cm}
    \begin{tabular}{lccc}
\textbf{Parameter} & \textbf{symbol} & \textbf{value} & \textbf{unit} \\
\hline
Ring circumference  & $L$   & 384 & m \\
Arc length          &       & 40 & m \\
Length of straight section &    & 56 & m\\
Relativistic Lorentz factor & $\gamma_r$ & 3.664 & \\ 
Transition Lorentz factor & $\gamma_t$ & 5.825 & \\ 
Relativistic speed & $\beta$ & 0.962 & c  \\ 
Revolution time & & 1.331 & \SI{}{\micro\second} \\
Magnetic rigidity & $|B\rho|$ & 11.03 & Tm \\
 \hline
 Tune, horizontal & $Q_x$ & 8.24 & \\
 Tune, vertical & $Q_y$ & 8.31 & \\
Chromaticity, horizontal & $\xi_x$ & -11.1 & \\
Chromaticity, vertical & $xi_y$ & -12.5 & \\
Momentum compaction factor & $\alpha$ & 0.0295 & \\
Phase slip factor & $\eta$ & -0.0450 & \\
\hline
Maximum horizontal beta & & 28.5 & m \\
Minimum horizontal beta & & 2.85 & m\\
Average horizontal beta & $<\beta_x>$ & 10.8 & m\\
Maximum vertical beta  & & 36.4 & m\\
Minimum vertical beta & & 2.74 & m\\
Average vertical beta & $<\beta_y>$ &11.4 & m \\
Maximum horizontal dispersion & & 4.879 & m \\
\hline
Number of FODO cells per arc & & 4 & \\ 
Phase advance per FODO cell, horizontal && 0.25 & $\pi$ \\
Phase advance per FODO cell, vertical && 0.267 & $\pi$ \\
Bending radius of arc dipole && 8.48 & m\\
Magnetic field strength in arc dipole & & 1.301 & T\\
Length of arc dipole && 1.666 & m\\
Number of dipoles per arc & & 8 & \\
Length of arc quadrupoles & & 0.7 & m\\
Number of quadrupoles per arc && 8 & \\
    \hline
    \end{tabular}
  \end{center}
\end{table}
%----------------------------------------------------

%------------------------------------------------------------------------------------------
%     Figure: Optical functions in one FODO cell 
%------------------------------------------------------------------------------------------
%\begin{figure}[htbp]
%\begin{center}
%\mbox{\epsfig{file=figures/accumulator/FODO.pdf,width=12cm}}
%\caption{ Optical $\beta$-functions and the horizontal dispersion $D_x$ in one FODO cell.}
%\label{fig:FODO}
%\end{center}
%\end{figure}
%------------------------------------------------------------------------------------------

Figure~\ref{fig:ring_betas} shows the beta functions for both transverse planes in the full ring. The optical functions exhibit a four-fold symmetry, with only small variations from sector to sector. This symmetry is crucial for suppressing structural resonances. 
%------------------------------------------------------------------------------------------
%     Figure: Full ring optical beta functions
%------------------------------------------------------------------------------------------
%\begin{figure}[htbp]
%\begin{center}
%\mbox{\epsfig{file=figures/accumulator/ring_betas.pdf,width=12cm}}
%\caption{ Optical $\beta$-functions throughout the ring.}
%\label{fig:ring_betas}
%\end{center}
%\end{figure}
%------------------------------------------------------------------------------------------

Additional ``trim'' coils on the quadrupole magnets can be used to restore the designed super-periodicity, in the case of manufacturing and alignment errors, etc. This scheme is implemented at SNS~\cite{Henderson:2005xd} and will be considered also for ESS$\nu$SB.

By adding five short sextupole magnets to each arc section, the natural chromaticity can be fully corrected using limited sextupole strength (\~{}2\,T/m$^2$). Correcting the chromaticity may turn out to be important if the beam energy spread is large, since an uncorrected chromaticity would yield a chromatic tune spread proportional to the chromaticity and the momentum spread. Considering the limited sextupole magnet strength, the dynamic aperture is not expected to be reduced dramatically. In the SNS accumulator, sextupole magnets are never used during normal operation despite the significantly larger beam energy spread~\cite{Cousineau:HB2016-MOAM4P40}. So far, the study has not revealed a need for using the sextupole magnets in the ESS$\nu$SB accumulator, but they are included in the final design nonetheless, in order to be adaptable to future changes of the conditions.

\subsection{Injection\label{sec:accummulator_injection}}
Also the lattice in the injection region is inspired by the SNS accumulator ring~\cite{Wei:2000wa}, which has had success in their high-power beam accumulation with phase-space injection and H$^-$ stripping using a foil. Furthermore, there are advanced experimental activities on using laser-assisted stripping currently performed at the SNS~\cite{Cousineau:2017lrm,Cousineau:2017dsn,Gorlov:2019wfl}. If the efforts are successful, it would be beneficial if also the laser stripping setup can be relatively easily adapted to the ESS$\nu$SB accumulator ring. %\red{This topic will be discussed in more detail later in this chapter.}

In the injection region there are four dipole magnets, two on each side of the centre, which are used to create a permanent orbit bump. This bump brings the ideal orbit of the circulating beam closer to the point where the incoming beam is injected. This layout facilitates the merging of the two beams. Inside the orbit bump, the dispersion is non-zero (as opposed to elsewhere in any of the straight sections). The configuration of the permanent orbit bump is illustrated in Fig.~\ref{fig:injection_bump_magnets}, as seen from the top of the accumulator. The red rectangles in the centre of the bottom figure mark the position and length of the dipoles forming the permanent orbit bump depicted in the top graph. The stripping of the H$^-$ ions will take place near the peak of the permanent orbit bump. 
%------------------------------------------------------------------------------------------
%     Figure: Injection orbit bump
%------------------------------------------------------------------------------------------
\begin{figure}[htbp]
\begin{center}
\mbox{\epsfig{file=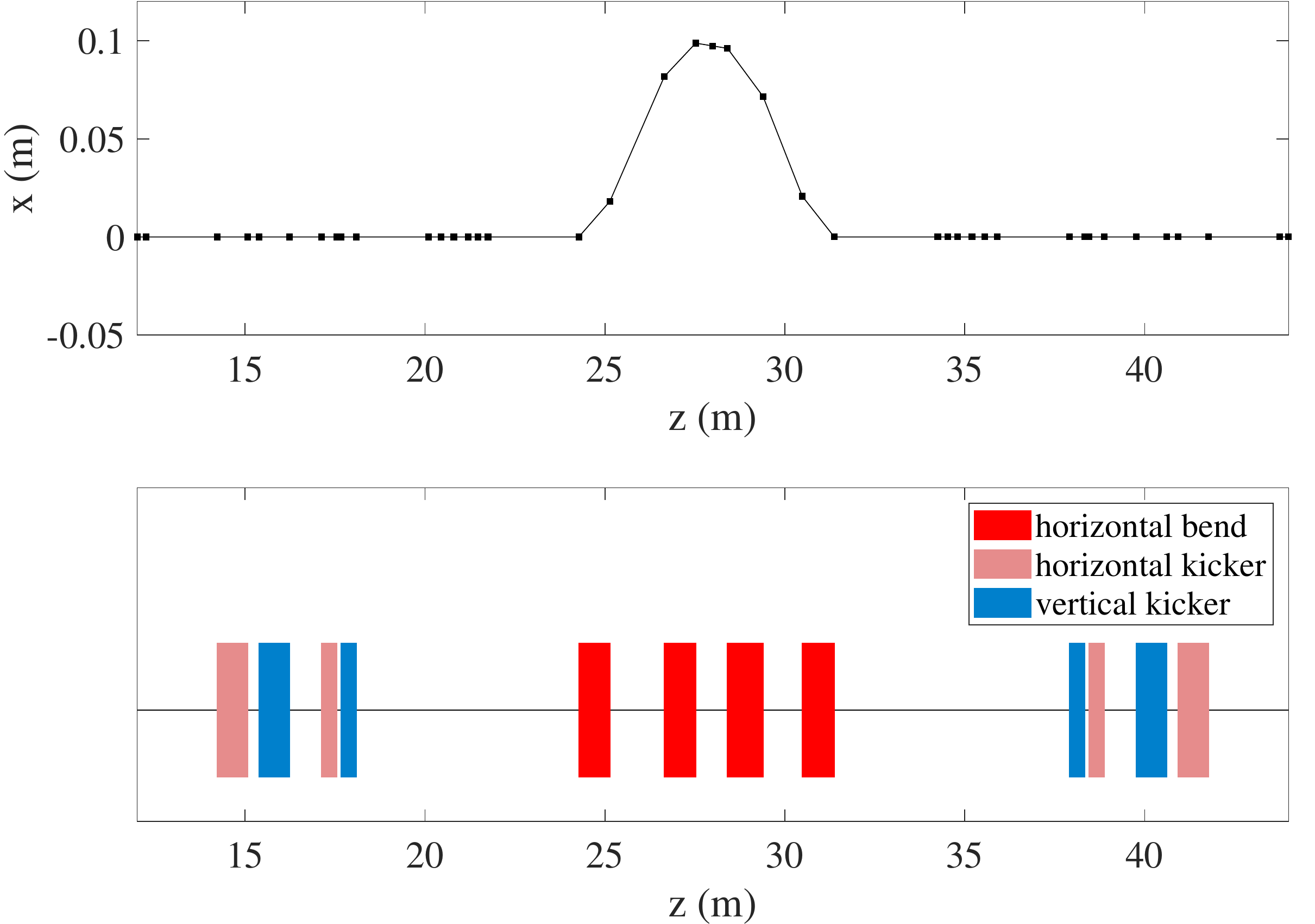,width=9cm}}
\caption{The permanent orbit bump (upper), to facilitate injection, formed by four dipole magnets marked as red boxes in the lower frame. Four fast kicker magnets in each plane are used to generate an orbit bump additional to the permanent one.}
\label{fig:injection_bump_magnets}
\end{center}
\end{figure}
%------------------------------------------------------------------------------------------

The injection region contains four fast kicker dipole magnets that generate an orbit bump in addition to the permanent orbit bump. These kickers are represented by the pink and blue rectangles before and after the constant field dipoles in Figure~\ref{fig:injection_bump_magnets}. The amplitude of this additional bump will vary during the injection so that the injected beam, which has a very small emittance, fills the larger phase space in the ring, through so called phase-space painting.

The phase-space painting has been optimised with two main goals:
\begin{itemize}
  \item To form a beam distribution which minimises the effects of space charge.
  \item To minimise the number of times that an already circulating particle traverses the stripper foil. Ideally, each particle only crosses the foil at injection and never after it has already been stripped, due to the fact that each foil crossing leads to a heat deposition, and to scattering, in the foil. Since this is not feasible in practice, it becomes important to minimise the foil hits by varying how the bump changes with time.
\end{itemize}

An extensive study of different painting procedures was made as part of the ring design. There are two classes of phase-space painting: correlated painting and anti-correlated painting. In correlated painting, the circulating beam is moved with respect to the injection point in the same direction in the horizontal plane as in the vertical plane. Normally, this means starting close to the injection point and gradually moving away, both horizontally and vertically, as the injection proceeds. This produces a beam which resembles a square in real space.

In anti-correlated painting, the beam is moved in opposite directions. In the ESS$\nu$SB case, the closed orbit and therefore the circulating beam starts close to the injection point in the horizontal plane and gradually moves away while the beam approaches the injection point in the vertical plane. This process yields a round beam. Since a round beam is desirable when the beam hits the target, anti-correlated painting was selected for the ESSnuSB injection. The disadvantage is that this scheme, as a general rule, results in twice as many stray foil hits as correlated pointing; this occurs since there are more circulating particles when the beam approaches the foil in the vertical plane.

The way the injection bump amplitude varies with time during injection will affect the final beam distribution, as well as the space charge forces during the injection. The aim has been to generate a final distribution that is as flat as possible, both because it leads to fewer space charge issues at the end of injection, but also because a flat distribution has advantages in terms of the thermal response of the target. However, since a beam will not circulate in the ring for more than a few turns after a filling, it would be interesting to, in the future, investigate the effect of having a painting procedure that instead minimises space charge effects \emph{during} injection.
%-------------------
\subsubsection{Numerical Simulations}
%------------------------------------------------------------------------------------------
%     Figure: Fast injection bump
%------------------------------------------------------------------------------------------
\begin{figure}[htbp]
    \begin{center}
    \mbox{\epsfig{file=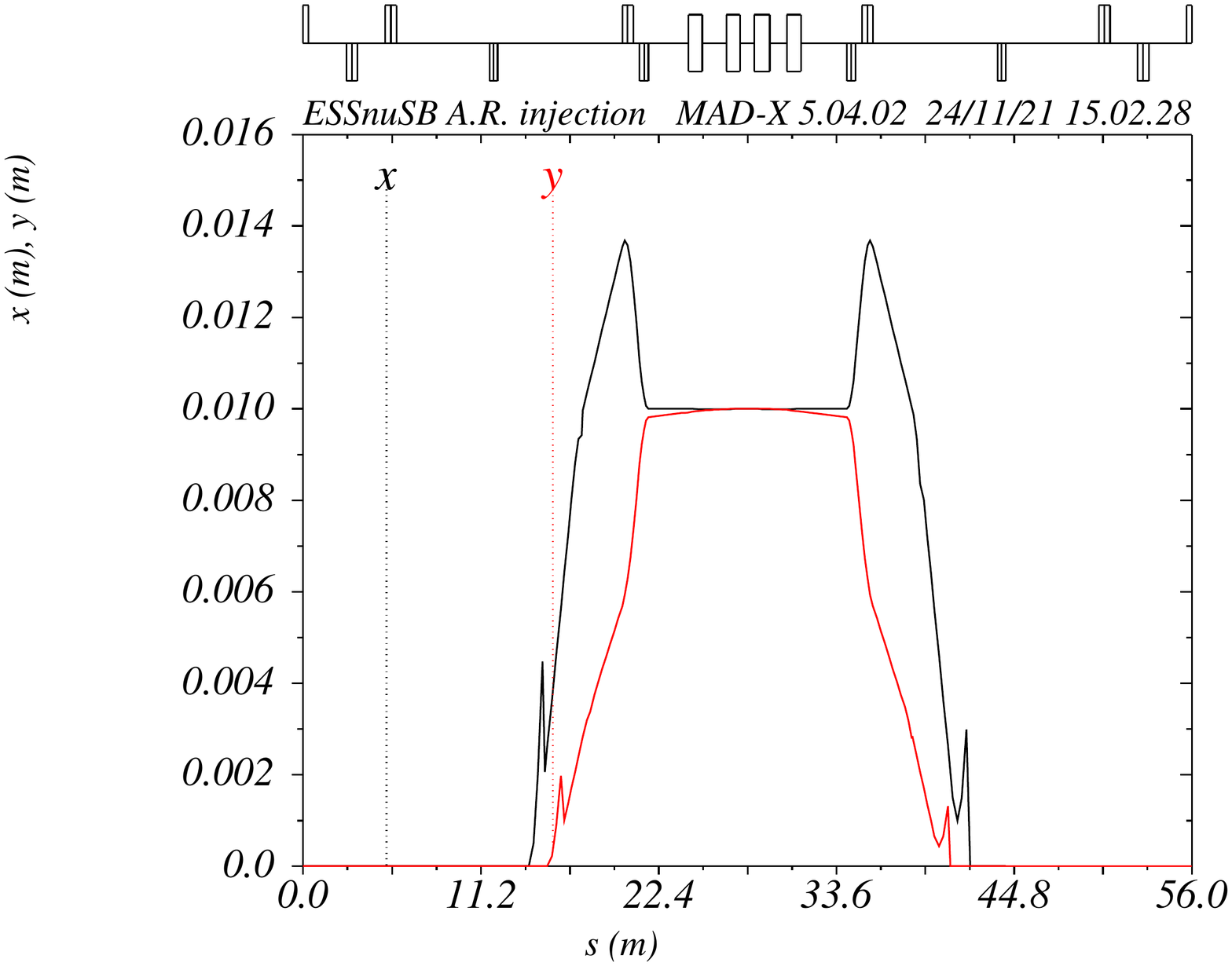,width=12cm}}
    \caption{ Example of a 10\,mm orbit bump generated by the injection kicker magnets illustrated in Fig.~\ref{fig:injection_bump_magnets}.}
    \label{fig:fast_bump}
    \end{center}
\end{figure}
%------------------------------------------------------------------------------------------

Multi-particle simulations have been performed with the Particle-In-Cell (PIC) code PyORBIT~\cite{Shishlo:2015:1272} and external PTC libraries~\cite{Schmidt:2002vp} to evaluate the injection procedure. In this simulation framework, the magnetic lattice is represented by a so-called flat file in which all elements are split into a given number of slices. The slices are associated with computation nodes where mathematical operations are applied. The beam is represented by macro-particles, and these macro-particles are transported through the lattice slice by slice. Space charge effects are computed at every node with a sliced 2D (two-dimensional) model, also referred to as 2.5D model, in which the beam is sliced in the longitudinal direction. For each slice, the projected transverse distribution is placed over a grid to calculate the space charge forces using the Fast Fourier Transform (FFT). These forces are then scaled according to the longitudinal line density of the slice. Finally, the force is translated to a kick given to each grid unit, before the beam is propagated to the next node. An overall aperture 200\,mm is applied to the ring in order to handle lost particles. This aperture is also used in the calculation of indirect space charge, i.e. space charge induced by mirror charges in the vacuum chamber.

The number of lattice nodes, macro-particles and space charge grid bins are optimised in a contingency study, to find the optimum level where the accuracy as well as the simulation CPU time are satisfactory.

The simulation of the whole injection is made by setting the strength of a given set of magnets on a turn-by-turn basis. In this way the amplitude of the closed-orbit bump can be controlled with respect to the injection point. The location and dimensions of a stripper foil is defined and every particle crossing through the foil is registered. Particle scattering from the interaction with the foil is included through a model including multi-Coulomb scattering, elastic and inelastic nuclear scattering. No significant halo formation has been observed due to foil scattering.   

Figure~\ref{fig:fast_bump} shows the displacement of the orbit trajectory in the injection region when the fast kicker magnets are powered. In this example, the amplitude of the orbit bump is set to +10\,mm at the middle point, although an amplitude of at least $\pm50$\,mm can be reached with the present design. The horizontal and vertical amplitudes can be set independently of each other, which provides excellent flexibility. The fast orbit bump is local, which implies that the beam trajectory outside of the injection region remains unaffected when the bump is activated.

Both correlated and anti-correlated painting have been successfully tested. Figure~\ref{fig:ring:orbit_bumps} shows six examples of how the amplitude of the fast orbit bump changes along the injection in both transverse planes. Both correlated and anti-correlated painting are shown. These orbit-bump functions, ranging from a slow, linear function to a more rapid exponential dependence, have been used to optimise the final beam distribution, which are depicted in Fig.~\ref{fig:ring:painting_optimisation}.
%----------------------------------------------------
\begin{figure}[ht!]
  \begin{center}
    \mbox{\epsfig{file=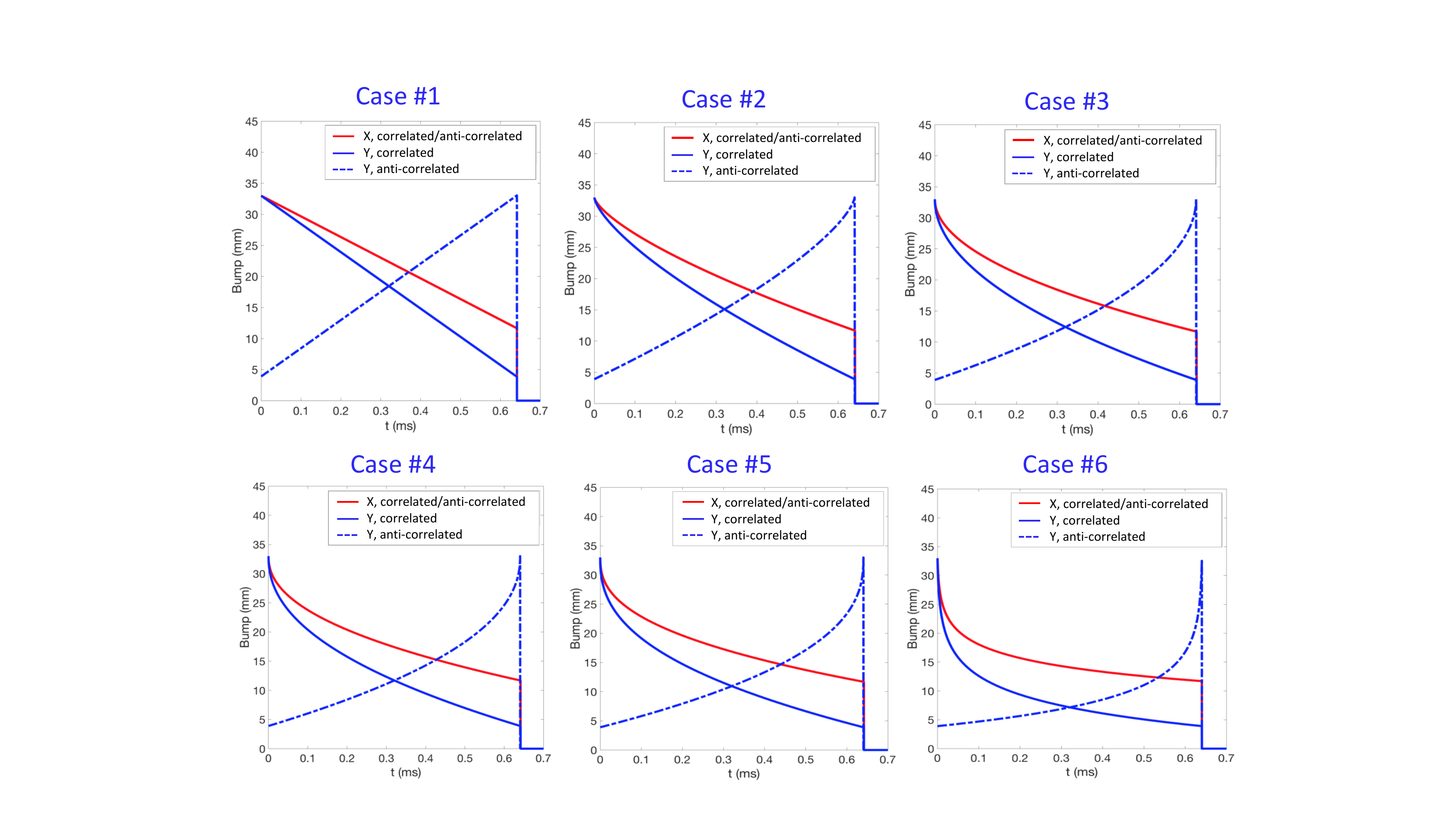,clip, trim=5cm 0cm 5cm 2cm,width=15.5cm}}
\caption{The closed-orbit bump amplitude as a function of time during the injection. Six different functions are shown, ranging from a linear slope to a more dramatic exponential slope.}
    \label{fig:ring:orbit_bumps}
  \end{center}
\end{figure}
%----------------------------------------------------
%----------------------------------------------------
\begin{figure}[htbp]
  \begin{center}
    \mbox{\epsfig{file=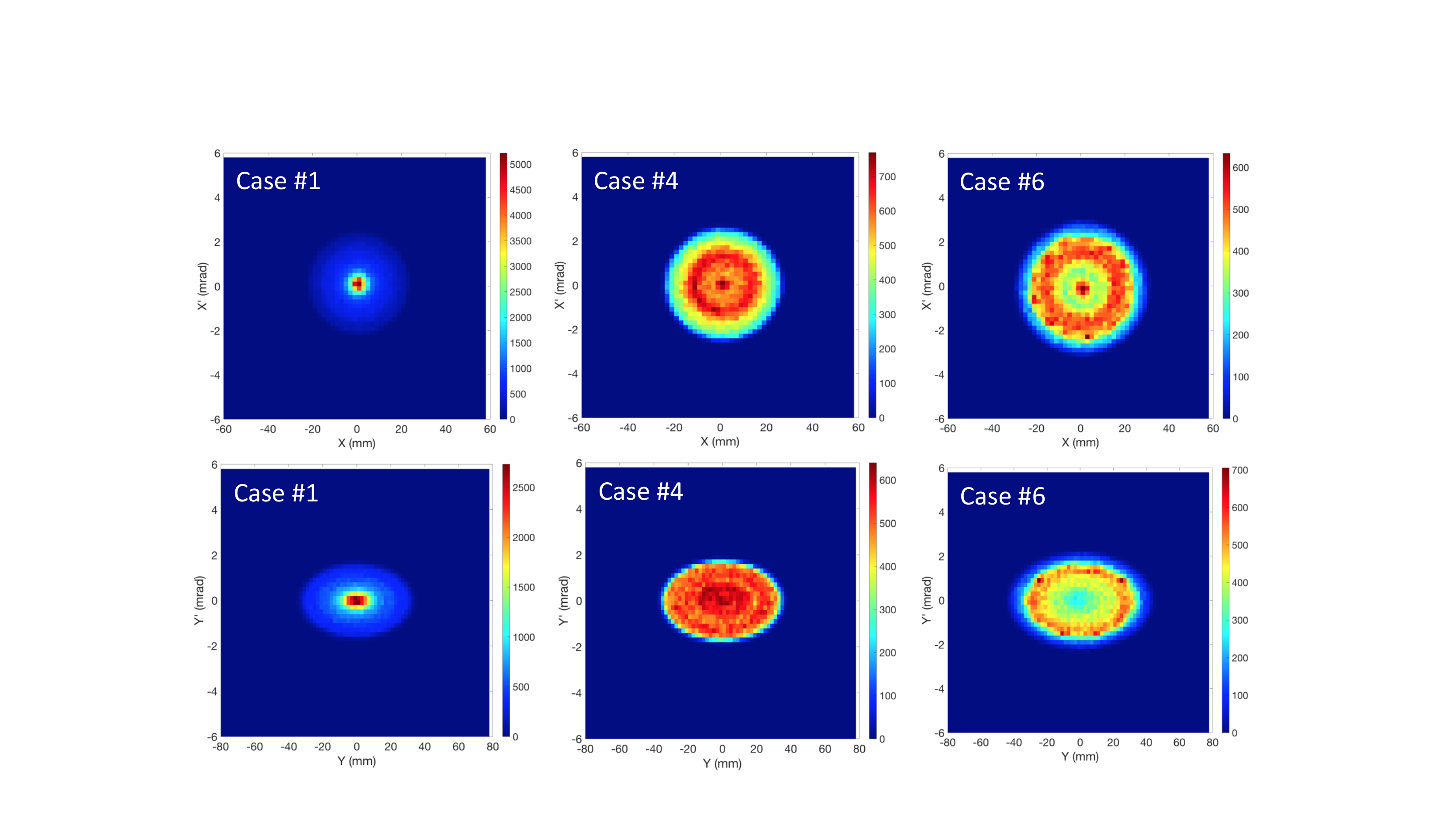,clip, trim=4cm 1cm 4cm 3cm,width=15cm}}
\caption{The particle density distribution in horizontal (top row) and vertical (bottom row) phase space resulting from anti-correlated painting with three selected cases of the orbit bump functions, shown in Fig.~\ref{fig:ring:orbit_bumps}. The distribution goes from under-painted (case \#1, left), where the particle density is very high in the centre, to over-painted (right, case \#6), with a underpopulated core. The best case (middle, case \#4) gives the most uniform beam distributions.}
    \label{fig:ring:painting_optimisation}
  \end{center}
\end{figure}
%----------------------------------------------------

An example of the rectangular beam shape resulting from correlated painting is shown in Fig.~\ref{fig:ring:correlated_painting}. This particular distribution has been produced using the orbit bump function in Case \#4. The difference in particle density when space charge is activated or ignored in the simulations is very small, which indicates that the painting process fulfils its purpose. 
%----------------------------------------------------
\begin{figure}[ht!]
  \begin{center}
    \mbox{\epsfig{file=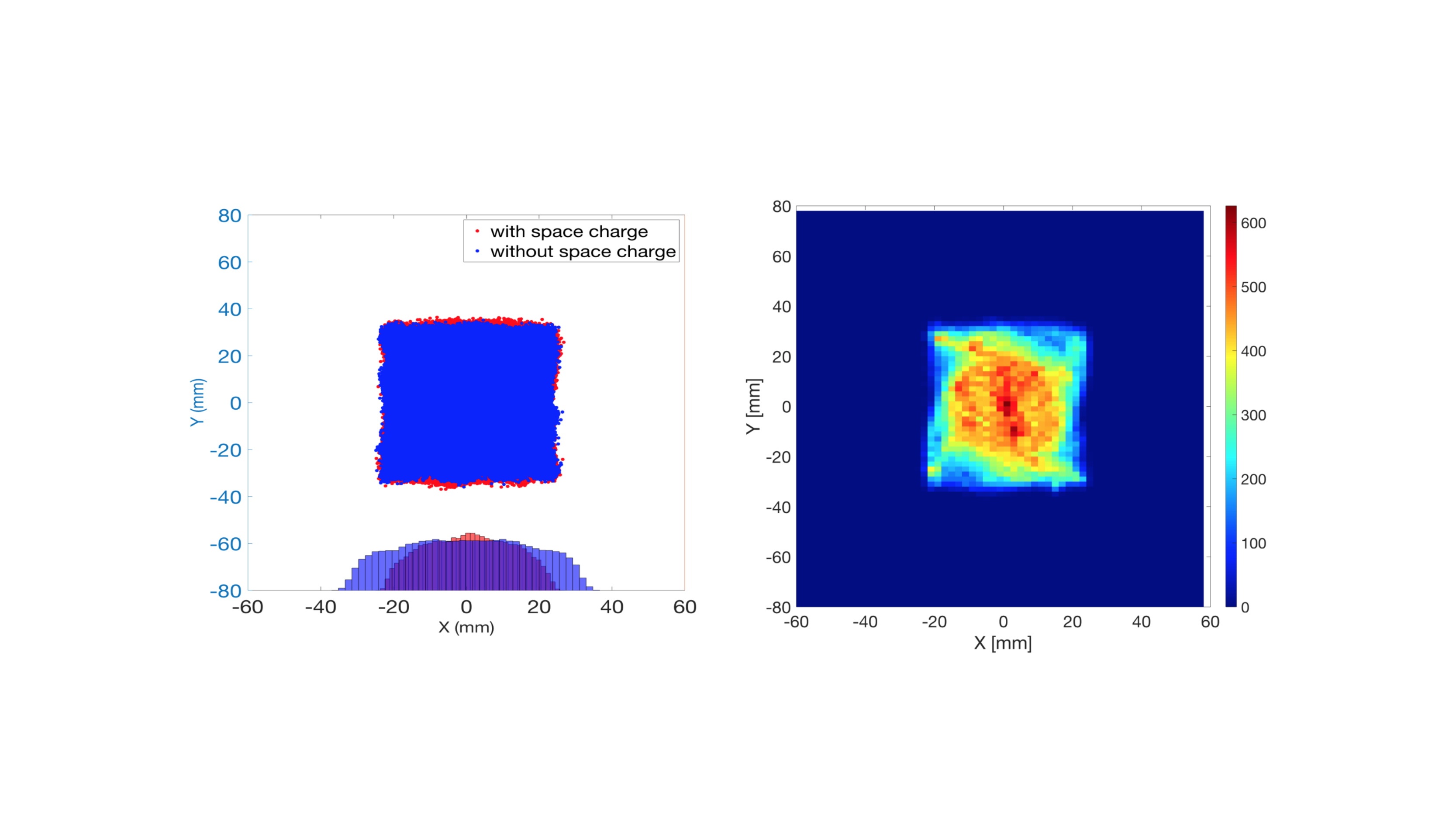,clip, trim=4cm 3cm 3cm 4cm,width=12cm}}
\caption{The particle distribution (left) and the particle density distribution (right) when correlated painting is used. }
    \label{fig:ring:correlated_painting}
  \end{center}
\end{figure}
%----------------------------------------------------

Since a round beam is optimal for the target, anti-correlated painting is the preferred option. Figure~\ref{fig:ring:final_transverse_distributions} shows the transverse distribution in horizontal and vertical phase space as well as in real space, when the orbit bump function denoted Case \#4 is used in the anti-correlated configuration. This is the procedure that best meets the requirements targeted in the optimisation of the injection process: a uniform transverse beam distribution, an elliptical beam shape and minimised number of stray foil hits.
%----------------------------------------------------
\begin{figure}[ht!]
  \begin{center}
    \mbox{\epsfig{file=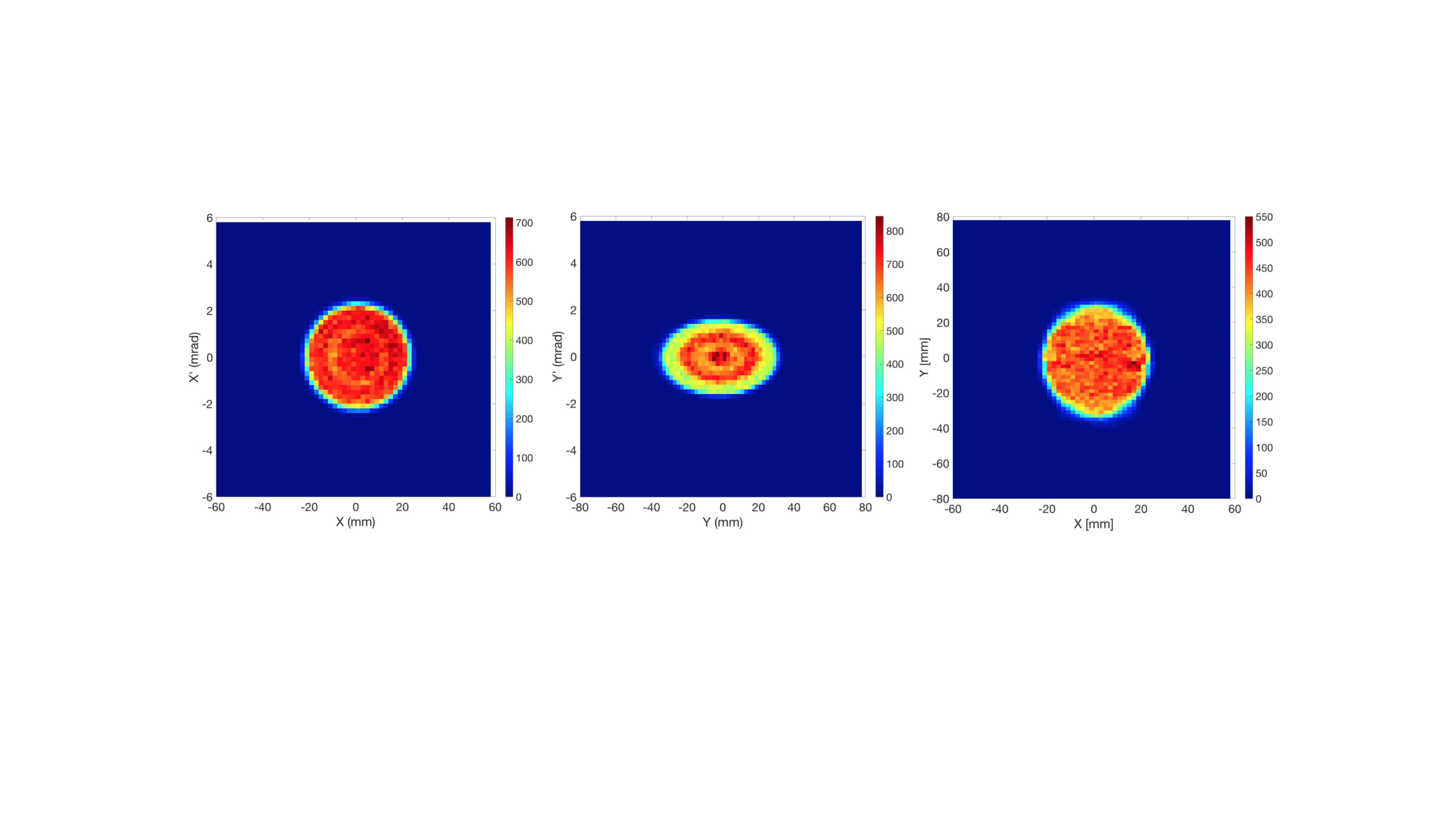,clip, trim=4cm 6cm 4cm 5cm,width=16cm}}
\caption{The particle density distribution in horizontal phase space (left), vertical phase space (middle) and real space (right) resulting from anti-correlated painting with the orbit bump function in case \#4, see Fig.~\ref{fig:ring:orbit_bumps}.}
    \label{fig:ring:final_transverse_distributions}
  \end{center}
\end{figure}
%----------------------------------------------------

The resulting transverse beam emittance is illustrated in Fig.~\ref{fig:ring_emittance_tune}, together with the corresponding tune diagram. In this case, 100\% of the simulated particles are contained in a geometrical emittance of about 70$\pi$\,mm\,mrad, and with a tune spread well below 0.05. These results assume and incoming beam energy spread of the order of 0.02\% which is the same as at the end of the linac.

%----------------------------------------------------
\begin{figure}[htbp]
  \begin{center}
    \mbox{\epsfig{file=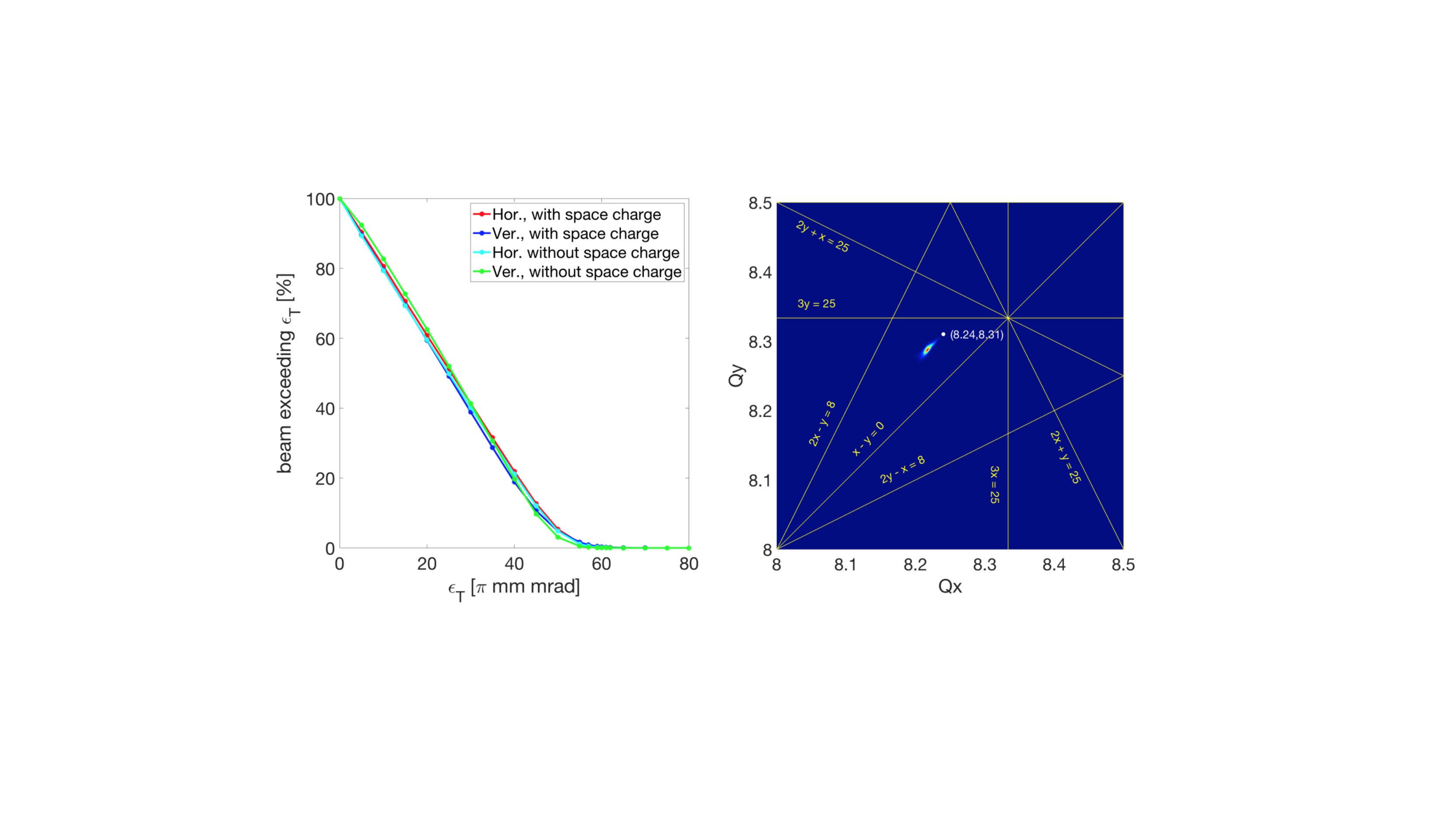,clip, trim=6cm 5cm 7cm 4cm,width=13cm}}
\caption{Transverse beam emittance $\epsilon_T$ with and without space charge (left) and the corresponding tune diagram (right). The plot to the left shows the fraction of the beam that is contained within a specific beam emittance. When space charge is considered, the 100\% beam emittance is roughly 70$\pi$\,mm\,mrad and the tune spread is less than 0.05.}
    \label{fig:ring_emittance_tune}
  \end{center}
\end{figure}
%----------------------------------------------------
%
\subsubsection{H$^-$ stripping}
%----------------------------------------------------
The painting process has been optimised for injection with foil stripping. Limiting the thermal load on the stripper foil is a focus - and one of the main challenges - of the accumulator ring design and operation~\cite{Plum:2016cfe}. The painting is strongly affected by this challenge, in the sense that the thermal load on the stripper foil sets a limit for how small the transverse emittance can be. Painting to a small emittance implies that the circulating beam travels closer to the injected beam, which results in more unwanted foil crossings and a higher peak temperature in the foil.

The thermal load on the stripper foil has been estimated by 1) registering all the foil hits in the numerical simulation of the injection; 2) translating each foil hit to an energy deposition; 3) calculating the temperature in the foil as a function of time with a temperature model which includes blackbody radiation as only cooling mechanism. The model has been bench marked with a similar model used at SNS, a model which in turn has been compared with measurements~\cite{Liaw:1999ar,Beebe-Wang:2001nwy}. A carbon foil with emissivity 0.8 has been assumed in the study.

% Add paragraph and figure on foil thickness and stripping efficiency.

Figure~\ref{fig:ring:foil_density} shows the energy density in a carbon foil of density 500\,\SI{}{\micro\gram/\centi\meter\squared}, corresponding to 99\% stripping efficiency~\cite{Gulley:1996zz,Chou:2007zz}, after the injection of one batch. In the left plot, the injected beam has the same beta function $\beta_i$ as the circulating beam, $\beta_m$, at the point of injection. In this case, the peak temperature is at the centre of the stripper foil, which implies that the peak temperature cannot be reduced by further optimising the painting. In the plot to the right the beam size of the injected beam has been increased so that $\beta_i=2\beta_m$ which results in an energy density which is spread over a larger part of the foil. Note that also the foil size has been slightly increased in order to make sure that all injected $H^-$ ions are stripped. Through this measure the peak energy density per batch is decreased by about 30\%. 
%----------------------------------------------------
\begin{figure}[htbp]
  \begin{center}
    \mbox{\epsfig{file=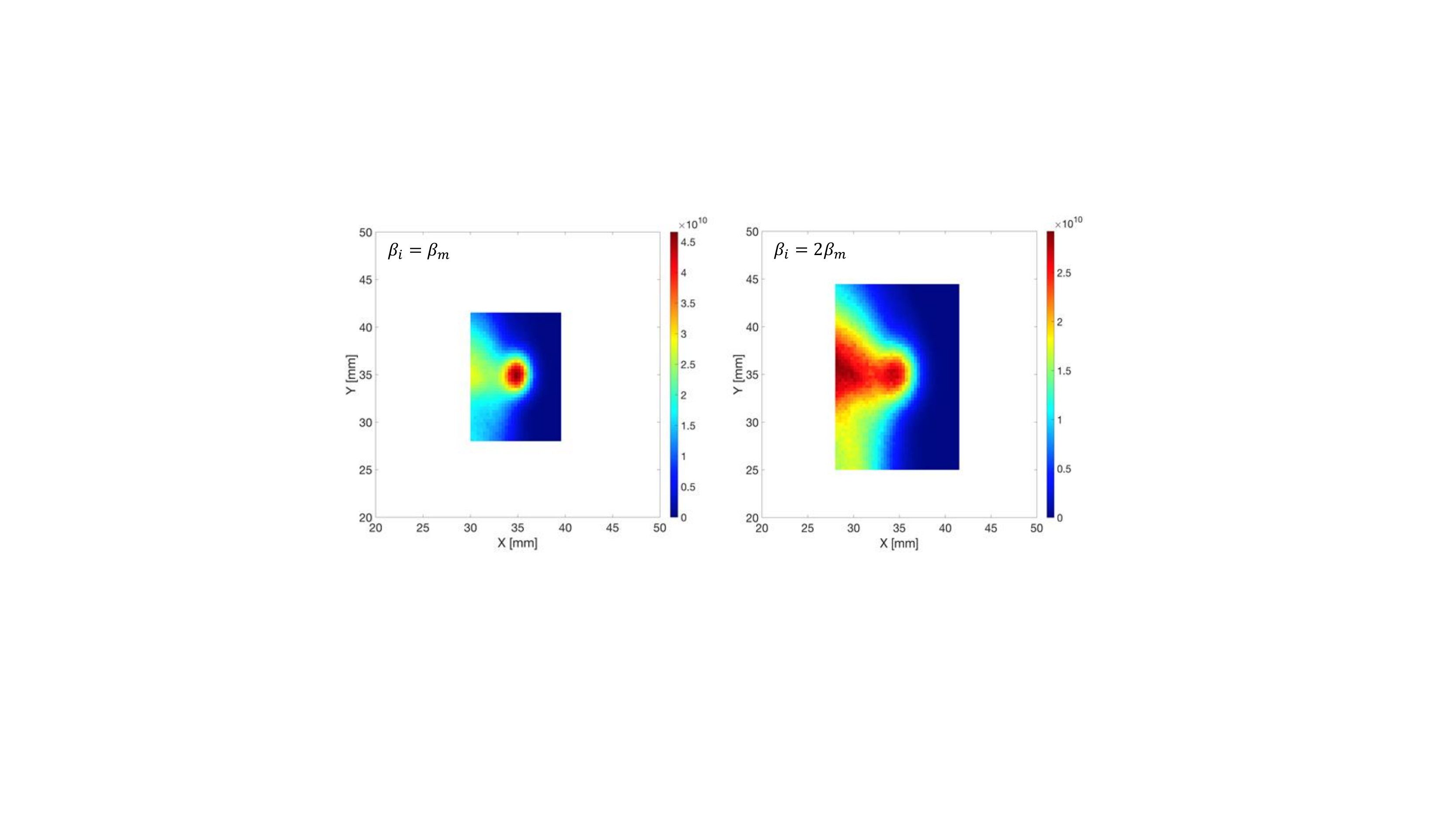,clip, trim=8cm 6cm 8cm 5cm,width=14cm}}
\caption{Particle density distribution in the stripper foil using matched injection (left) and mismatched injection (right).}
    \label{fig:ring:foil_density}
  \end{center}
\end{figure}
%----------------------------------------------------
%By calculating the peak temperature in the foil following several batches it is possible to estimate the steady-state behaviour. In Fig.~\ref{fig:ring:foil_temp_mismatch} the maximum temperature, in the steady-state mode, is shown as a function of the mismatch parameter $\beta_i/\beta_m$.
%%----------------------------------------------------
%\begin{figure}[ht!]
%  \begin{center}
%    \mbox{\epsfig{file=figures/accumulator/accumulator_foil_temp_beta.pdf,clip, trim=10cm 3cm 10cm 3cm,width=8cm}}
%\caption{\red{Peak temperature in the foil as a function of the mismatch parameter $\beta_i/\beta_m$.}}
%    \label{fig:ring:foil_temp_mismatch}
%  \end{center}
%\end{figure}
%----------------------------------------------------

The peak temperature can be further reduced by increasing the effective foil surface area. A preliminary investigation indicates that by replacing the single foil of density 500\,\SI{}{\micro\gram/\centi\meter\squared} with four foils of density 125\,\SI{}{\micro\gram/\centi\meter\squared} the peak temperature in the steady-state mode can be reduced by almost 800\,degrees. This assumes a comparable total stripping efficiency for 500 and $4\times125$\,\SI{}{\micro\gram/\centi\meter\squared}, which must be confirmed, in particular since accurate models of $H^0$ stripping have not been identified in this study. In addition, a detailed study of the placing of these sequential stripper foils, so that the blackbody radiation and/or the convoy electron emitted by the first one is not absorbed by the second, etc, will be required.

Figure~\ref{fig:ring:foil_steadystate_temp} shows the full result of the injection optimisation, from a careful choice of painting scheme, to mismatched injection and having several thin foils placed after each other. The steady-state peak temperature is reached within a couple of pulses and stays around 1900\,K, almost 100\,K below the target value of 2000\,K at which temperature the sublimation rate of carbon becomes unreasonably high~\cite{Plum:2016cfe}.

%----------------------------------------------------
\begin{figure}[ht!]
  \begin{center}
    \mbox{\epsfig{file=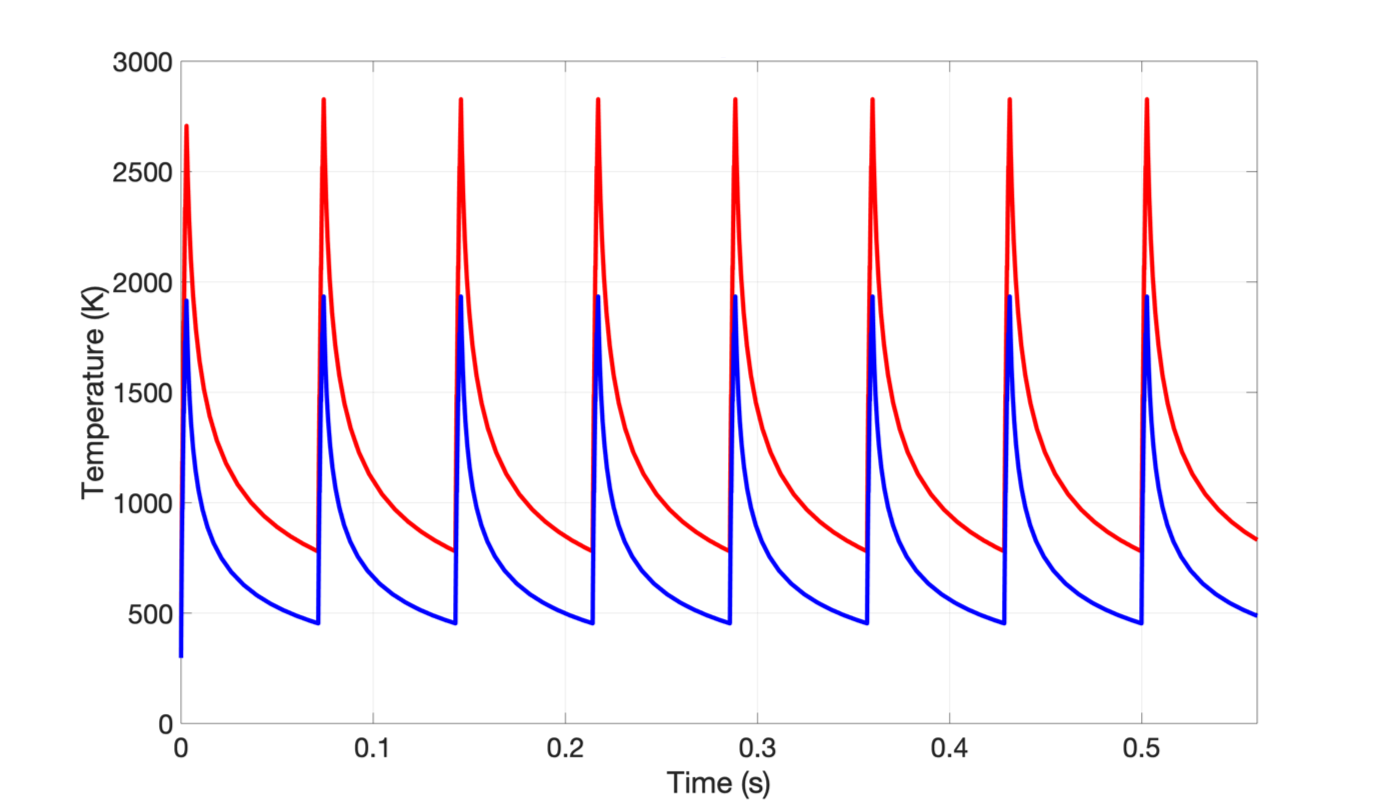,%clip, trim=5cm 1cm 6cm 3cm,
    width=12cm}}
\caption{Peak foil temperature as a function of time following several injected pulses. With mismatched injection and four sequential stripper foils the peak temperature stays below 2000\,K. }
    \label{fig:ring:foil_steadystate_temp}
  \end{center}
\end{figure}
%----------------------------------------------------

The conclusion of the painting optimisation study is that 60$\pi$\,mm\,mrad is the minimum emittance that can be tolerated with foil stripping. By carefully setting up the injection painting, choosing the adequate mismatch parameter, and by using sequential thin stripper foils instead of a single thicker foil, the steady-state peak temperature in the foil can be kept below 2000\,K. 

If, at a later stage, laser stripping can be employed in the ESS$\nu$SB accumulator ring, the situation changes. Laser stripping requires a different optical setup, where the H$^-$ beam is squeezed in one dimension in order to create sufficient overlap with the laser beam~\cite{Cousineau:2017lrm}. In this case, the circulating proton beam is not a threat to the stripping procedure, and a smaller emittance is desirable since it directly affects the required aperture, and, in turn, the cost of vacuum chambers and magnets.

To this end, the painting process was further modified to generate a beam with successively smaller emittance. At 40$\pi$\,mm\,mrad in both planes, the tune shift is of the order of 0.1, and negligible halo formation is observed. Even if the beam is painted to an even smaller phase-space area, the final vertical emittance does not decrease further; it instead remains near 40$\pi$\,mm\,mrad while the horizontal emittance can reach 25$\pi$\,mm\,mrad. Halo formation is observed in this case. Although the painting process can be further optimised for smaller emittance, it may be the case that space charge forces set a lower limit on the achievable emittance to 40$\pi$\,mm\,mrad.

%------------------------------------------------------------------------------------------------------
\subsection{Longitudinal Dynamics and Radiofrequency Cavities}\label{sec:accumulator_rf}
%------------------------------------------------------------------------------------------------------
Single-turn extraction requires a beam-free gap of at least 100\,ns. Injecting a coasting beam and adiabatically forming such an extraction gap within the ring has proven unfeasible, due to the fact that both the momentum spread and the phase slip factor are small, which makes the beam quite stiff. Such a process would then take thousands of turns~\cite{Machida:2016ESS} 
%$$\frac{\Delta t}{T}=\eta\frac{\Delta p}{p}$$
rendering it incompatible with the 100\,\SI{}{\micro\second} batch-to-batch distance. Instead, a 10\% beam gap (i.e. 133\,ns) is formed by chopping in the early stages of the linac (see Section~\ref{subsect:pulse_struct}) at a frequency corresponding to the revolution frequency in the ring. While the beam is accumulated in the ring, this gap must not shrink below 100\,ns -- something that can only be guaranteed by the use of RF cavities. These RF cavities will be accommodated in the last straight section in the ring, labeled SS3 in Fig.~\ref{fig:accumulator_layout}.

Conventionally, dual-harmonic cavities are used for this kind of application \cite{Wei:2000wa}, since they offer a large energy acceptance and effective beam-gap preservation. The cavity voltage is specified to guarantee a clean extraction gap while limiting the induced energy spread. 

As an alternative, barrier RF systems~\cite{Griffin:1983bt} have become increasingly popular for bunch manipulations in rings and synchrotrons. The barrier RF system utilises a waveform which leaves the centre of the beam unaffected. Only the head and the tail of the bunch are forced back towards the core as the beam traverses the cavity. Figure~\ref{fig:accumulator:rf_waveforms} shows an example of the RF voltage waveforms in the two cases.

\begin{figure}[htbp]
\begin{center}
\begin{minipage}{0.46\linewidth}
\includegraphics[width=0.99\textwidth]{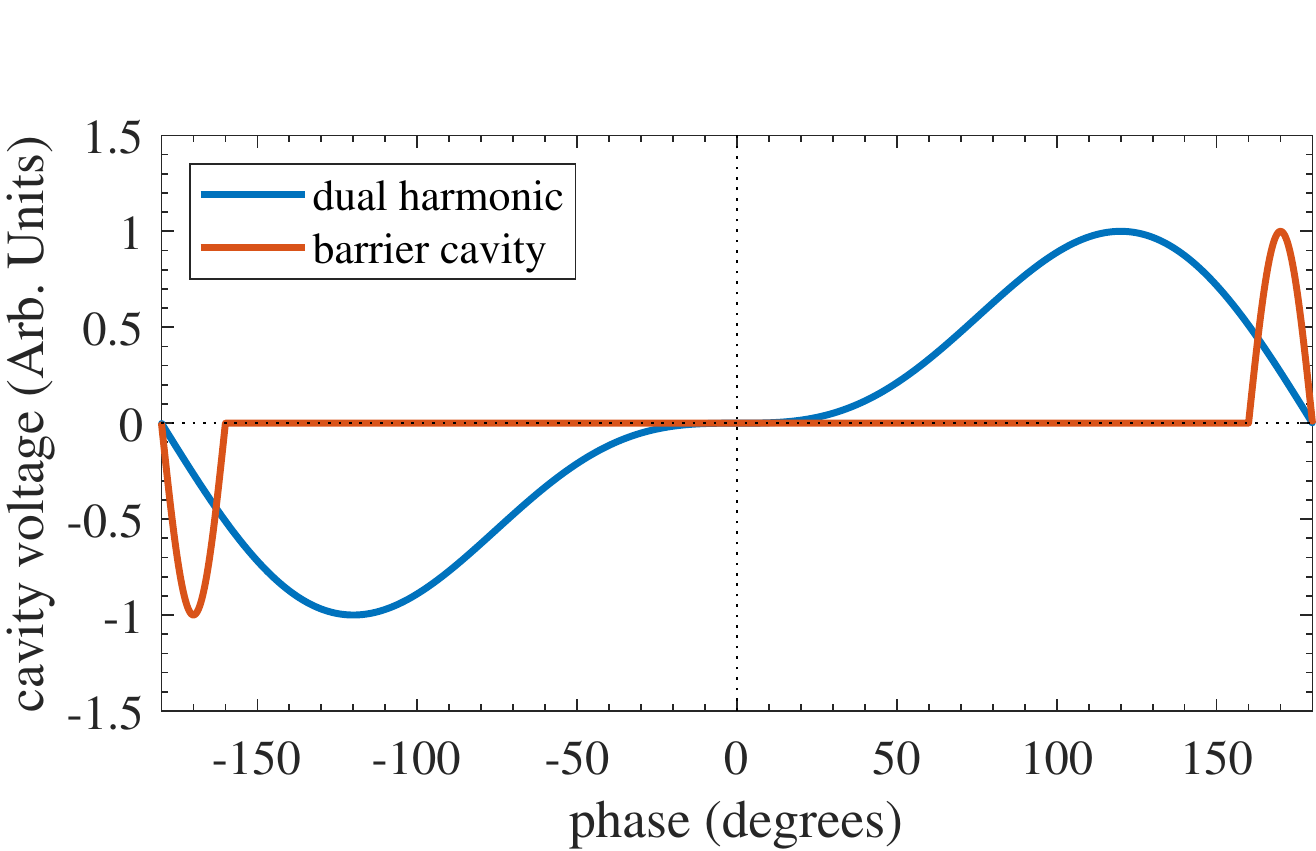}
\caption{\label{fig:accumulator:rf_waveforms}RF waveforms of a dual-harmonic cavity and a barrier RF cavity with a single sinusoidal pulse with a frequency 9 times the first harmonic, i.e the revolution frequency.}
\end{minipage}\hspace{2pc}%
\begin{minipage}{0.48\linewidth}
\includegraphics[width=0.99\textwidth]{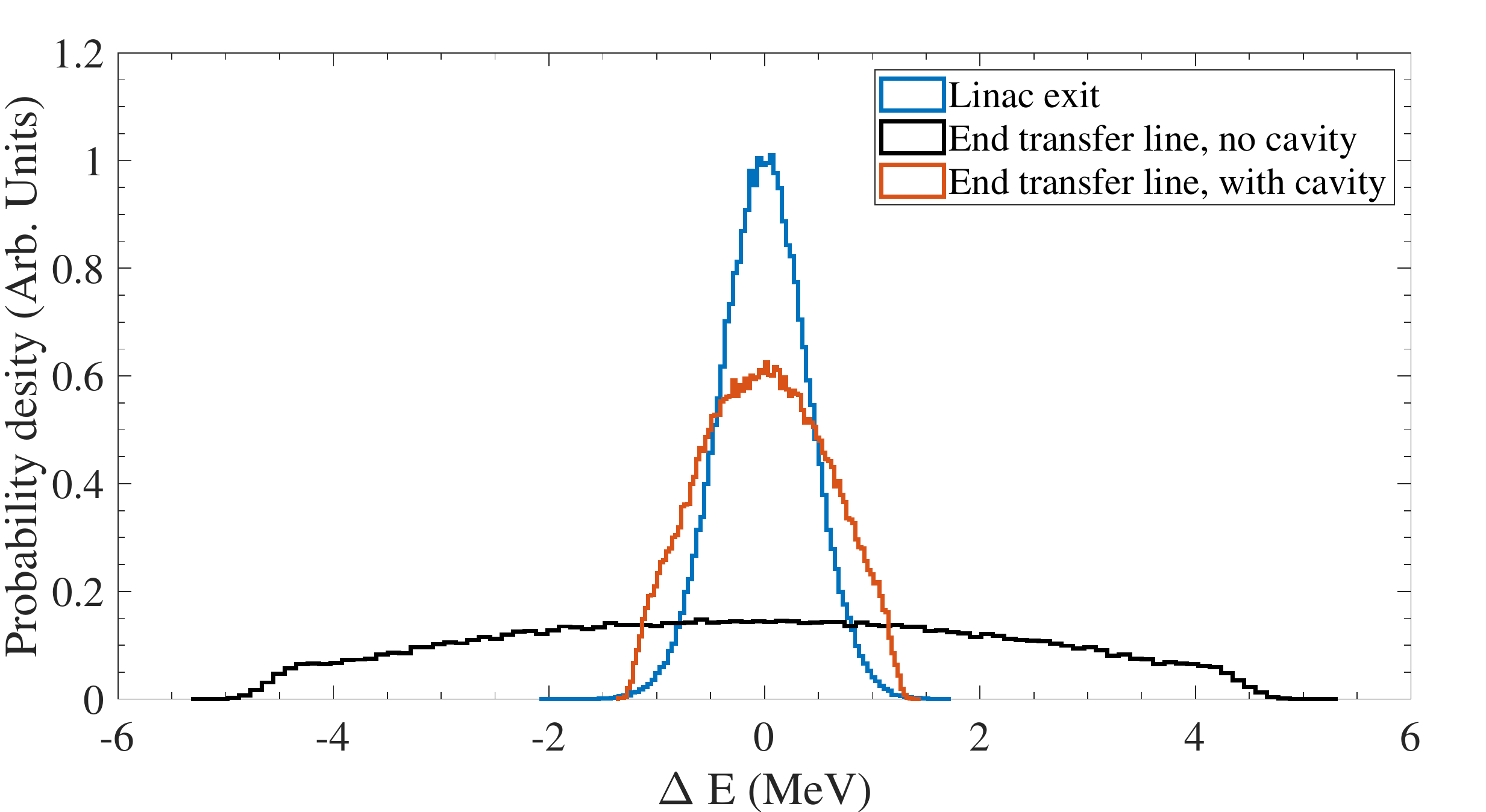}
\caption{\label{fig:accumulator:initial_energy_distribution}The beam energy distribution before accumulation.}
\end{minipage} 
\end{center}
\end{figure}

%----------------------------------------------------
%\begin{figure}[htb]
  %\begin{center}
    %\mbox{\epsfig{file=figures/accumulator/rf_waveform_dual_barrier,width=7cm}}
%\caption{RF waveforms of a dual-harmonic cavity and a barrier RF cavity with a single sinusoidal pulse with a frequency 9 times the first harmonic, i.e the revolution frequency.}
   % \label{fig:accumulator:rf_waveforms}
  %\end{center}
%\end{figure}
%----------------------------------------------------

Three cavities of 1\,m length each have been included in the ring lattice for simulation purposes, and the two different types of RF cavities described above have been studied through multi-particle simulations using PyORBIT~\cite{Shishlo:2015:1272} and external PTC libraries~\cite{Schmidt:2002vp}. Two separate initial conditions have been considered. The first is when the beam energy spread is controlled in the linac-to-ring transfer line so that the energy spread of the incoming beam is similar to that at the end of the linac. This option, requires additional RF cavities at the end of the transfer line, as described in Section~\ref{sect:L2R}. The second situation is when the energy spread is allowed to grow along the transfer line, due to space charge. The beam energy profile at the end of the transfer line is shown in Fig.~\ref{fig:accumulator:initial_energy_distribution} for these two cases, together with the energy profile at the end of the linac. 
%----------------------------------------------------
%\begin{figure}[htbp]
  %\begin{center}
    %\mbox{\epsfig{file=figures/accumulator/incoming_energy_profile,width=7cm}}
%\caption{The beam energy distribution before accumulation.}
   % \label{fig:accumulator:initial_energy_distribution}
  %\end{center}
%\end{figure}
%----------------------------------------------------

The longitudinal particle distribution at the end of injection is shown in Fig.~\ref{fig:accumulator:long_dist}. When the incoming energy spread is large (i.e. when there are no cavities in the transfer line) a voltage of at least 20\,kV is needed to maintain the gap with a barrier RF system. When the incoming energy spread is controlled through PIMS cavities in the transfer line, 10\,kV is enough to preserve the gap. Note that only the head and tail of the bunch are affected by the cavity, so that the induced energy spread is small.

%----------------------------------------------------
\begin{figure}[ht!]
  \begin{center}
    \mbox{\epsfig{file=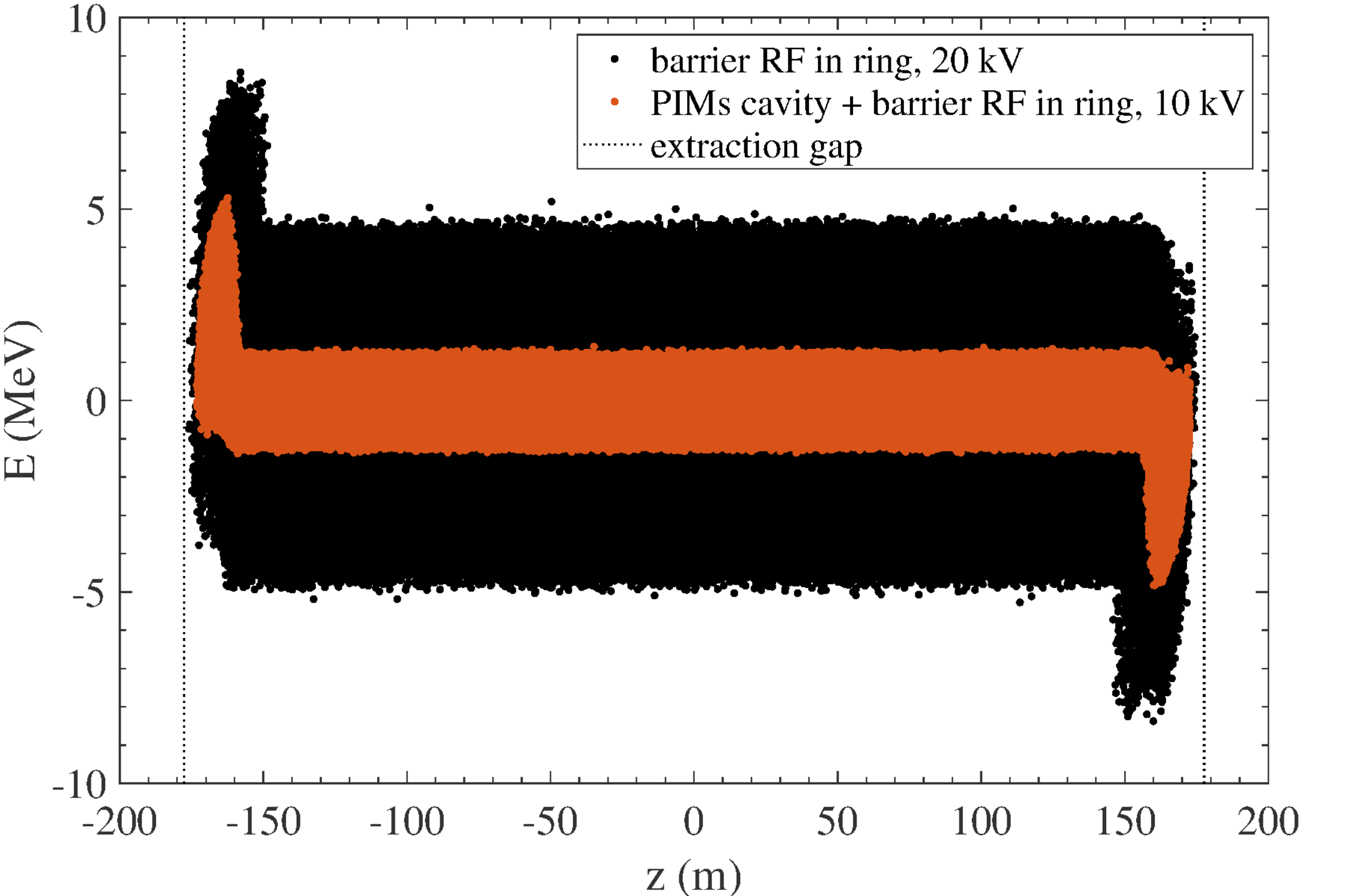,width=8cm}}
\caption{The longitudinal particle distribution at the end of injection. If the incoming energy spread is large, i.e. the case of having no cavities in the transfer line (see Fig.~\ref{fig:accumulator:initial_energy_distribution}), it takes a barrier RF cavity with a voltage of 20\,kV to preserve the extraction gap, the border of which is marked with the vertical dotted lines. With a reduced incoming spread, 5-10\,kV are enough. }
    \label{fig:accumulator:long_dist}
  \end{center}
\end{figure}
%----------------------------------------------------

Having a low spread in beam energy gives the advantage of a small chromatic tune spread since the chromatic tune shift is directly proportional to the energy deviation of a particle. If the chromatic tune spread is small enough, correcting the natural chromaticity in the ring with the use of sextupole magnets may be unnecessary. This, in turn, means that the dynamic aperture, which is generally reduced by the use of sextupole magnets, remains unaffected.

Conversely, having a low energy spread may prove problematic because of the risk of longitudinal microwave instabilities. A way of estimating this risk is through the Keil--Schnell stability criterion \cite{Keil:1969zza} with which it is possible to calculate the longitudinal impedance threshold for a set of given parameters. It reads~\cite{Wolski:2007:USPAS,Sagota:2015:instabilities}
\begin{equation}
    \label{eq:Keill-Schnell}
    \left|\frac{Z_{||}}{n}\right|\leq F\frac{\beta^2E|\eta|}{q\bar{I}}\left(\frac{\Delta p}{p}\right)^2 
\end{equation}
where $F$ is the form factor, determined by the longitudinal energy distribution, $E$ is the total beam energy, $\eta$ is the phase slip factor, $\bar{I}$ is the average beam current in the ring, and $\Delta p/p$ is the momentum spread. 
For the current design, the incoming energy spread would be about 0.2\%, at minimum. With $|\eta|=0.04$, $\bar{I}=27$\,A (at extraction) and $F=1$ the longitudinal impedance threshold is $|Z_{||}|=21\,\Omega$ for a momentum spread of 0.2\%, and 48\,$\Omega$ if the momentum spread is 0.3\%. Above these values the beam may become unstable. The extraction kicker magnets are expected to be the primary impedance contributor, and it is crucial that the impedance of the ring is estimated in the technical phase of the project, along with its resulting effects.

%%%%%%%%%%%%%%%%%%%%%%%%%%%%%%%%%%%%%%%%%%%%%%%%%%%%%%%%%%%%%%%%%%%%%%%%%%%%%%%%
%
%                  Ring collimation
%
%%%%%%%%%%%%%%%%%%%%%%%%%%%%%%%%%%%%%%%%%%%%%%%%%%%%%%%%%%%%%%%%%%%%%%%%%%%%%%%%

\subsection{Collimation}\label{subsec:collimation}
In the operation of a high-intensity proton accumulator such as this, it is of paramount importance to minimise uncontrolled beam loss in order to reduce component activation and to make hands-on maintenance possible. This is done by designing a collimation system that will be used for \emph{controlled} beam loss. In other words, the collimation system gets rid of beam particles that before they can damage sensitive equipment elsewhere in the ring.

A two-stage collimation system has been designed for the ESS$\nu$SB accumulator. It consists of a thin scraper to scatter halo particles, followed by a set of secondary collimators to absorb those scattered particles. Phase advances between scraper and secondary collimators, together with the type, the material, the thickness of collimators and the relationship with the physical acceptance, have been studied in detail and numerical simulations have been performed to evaluate the performance of the collimation system.

Given that 5\,MW of average beam power would be stored in the ring, the total fractional uncontrolled beam loss in the ring must be less than $10^{-4}$, for which a collimation system with a collimation efficiency beyond 90\% is required, according to experience from similar accelerators, e.g. SNS~\cite{Jeon:1999PAC} and J-PARC~\cite{Yamamoto:2008zzh}. 

\subsubsection{Two-stage Collimation System}
The two-stage collimation system chosen for the ESS$\nu$SB accumulator includes a thin scraper, which is the primary collimator that increases the scattering angle of the beam halo particles. This is followed by four thick collimators, two in the horizontal plane and two in the vertical plane; these are the secondary collimators which absorb the scattered beam-halo particles. The main advantage compared to the traditional single-stage collimation system is an increase in the impact parameter of particles hitting the front of the secondary collimators. This effect can dramatically improve the collimation efficiency, which is defined as the ratio of particles lost on the collimators to the total particle loss in the ring.

The straight section following the injection section will accommodate the collimation system, see Fig.~\ref{fig:accumulator_layout}. To leave enough room for downstream placement of secondary collimators, the primary collimator is located before the centre of the straight section, where $\beta_x=\beta_y$. Figure~\ref{fig:primary_collimator} shows the lattice function for one lattice super-period and the location of the primary collimator.
%-------------------------------------------------------------------------------
%     Figure: Primary collimator location
%-------------------------------------------------------------------------------
\begin{figure}[ht!]
  \begin{center}
    \mbox{\epsfig{file=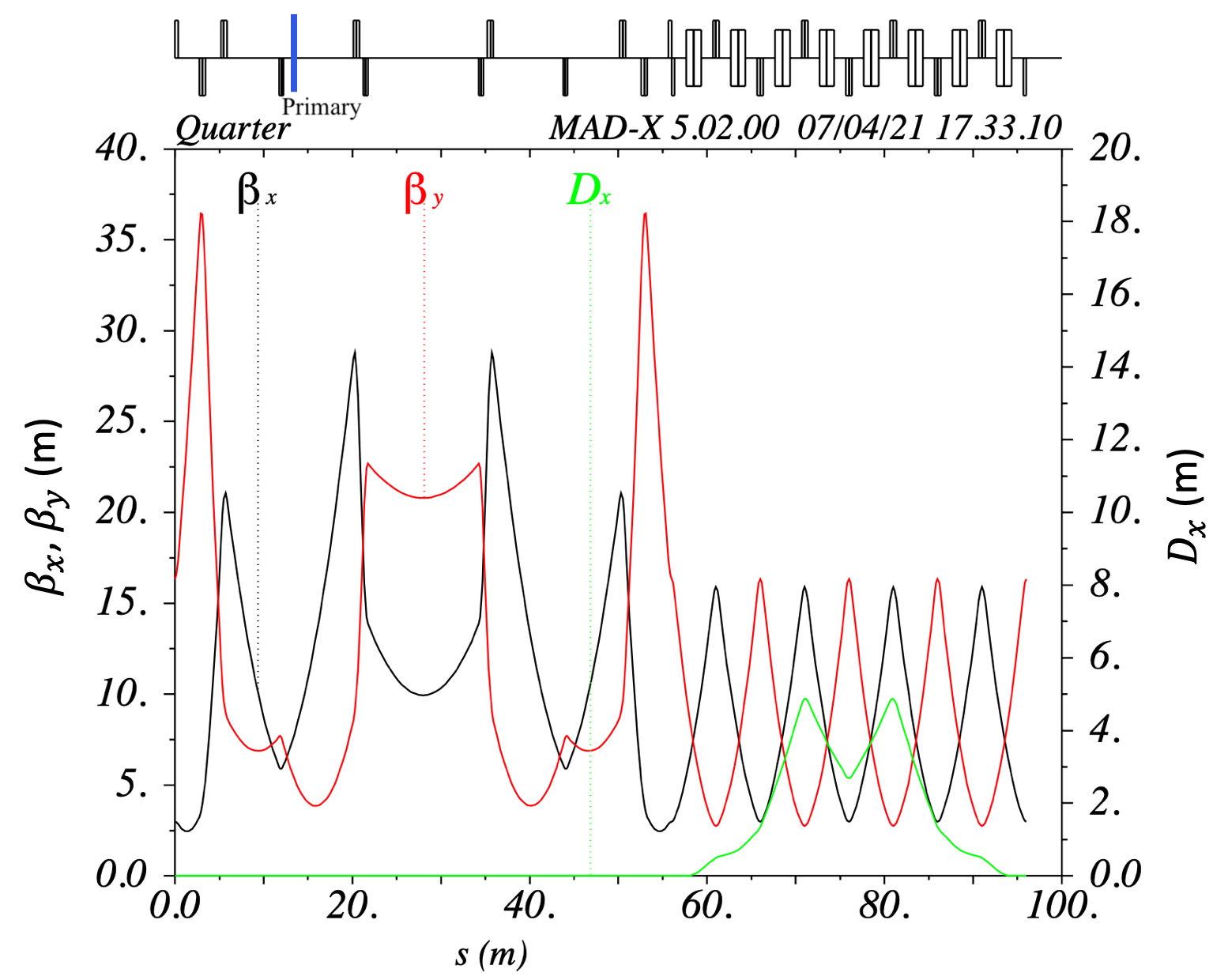,width=10cm}}
    \caption{ Optical beta functions $\beta_x$ and $\beta_y$ and horizontal dispersion $D_x$ in one lattice super-period. The blue line at the top marks the location of the primary collimator.}
    \label{fig:primary_collimator}
  \end{center}
\end{figure}
%-------------------------------------------------------------------------------

The optimal phase advance between the primary and the secondary collimator, for maximum interception efficiency, can be calculated as~\cite{Jeanneret:1998ed}:
\begin{equation}
  \label{eq:opt_phase_advance}
  \mu_{\rm opt}=\arccos\left(\frac{n_{\rm prim}}{n_{\rm sec}}\right)
  \end{equation}
where $n_{\rm prim}$ and $n_{\rm sec}$ are half apertures of primary and secondary collimators normalised to RMS beam size. For the ESS$\nu$SB accumulator, considering the acceptance of primary and secondary collimators as $70\pi$\,mm\,mrad and $120\pi$\,mm\,mrad, respectively, we have $n_{\rm prim}=2.35$, $n_{\rm sec}=3.07$, and the optimal phase advance from Eq.~(\ref{eq:opt_phase_advance}) is approximately 40$^\circ$. Thus, the secondary collimators are ideally located at 40$^\circ$ and its complementary location 140$^\circ$. Considering the real lattice situation, the phase advances of the secondary collimators are listed in Table~\ref{tab:ring_collimators}.

%-------------------------------------------------------------------------------
%     Table: Collimators and phase advances
%-------------------------------------------------------------------------------
\begin{table}[ht!]%[H]
  \begin{center}
    \caption{A list of the collimator elements and the phase advance, with respect to the primary collimator, at which they are located.}
    \label{tab:ring_collimators}
    \vspace{0.25 cm}
    \begin{tabular}{llc}
\textbf{Collimator type} & \textbf{Name} & \textbf{Horizontal/Vertical phase advance} \\
      \hline
Primary horizontal \& vertical & Prim   & $0^\circ$ / $0^\circ$ \\
Secondary vertical 1    & Sec\_V1       & $20^\circ$ / $38^\circ$ \\
Secondary horizontal 1  & Sec\_H1       & $40^\circ$ / $96^\circ$ \\
Secondary vertical 2    & Sec\_V2       & $90^\circ$ / $121^\circ$ \\
Secondary horizontal 2  & Sec\_H2       & $130^\circ$ / $154^\circ$ \\
\hline
    \end{tabular}
  \end{center}
\end{table}

\subsubsection{Material and Thickness of Primary Collimator}
The material choice of the primary collimator normally follows two rules: 1) small energy loss, and 2) large multi-Coulomb scattering angle. Figure~\ref{fig:angle_eloss_material} shows a comparison of RMS scattering angle along with energy loss for a 2.5\,GeV proton beam interacting with a variety of materials. After also considering availability, heat tolerance, melting point, and thermal conductivity, tantalum, tungsten, and platinum have been identified as acceptable choices for the primary collimator.

The choice of thickness for the primary collimator is also a balance between scattering angle and energy loss in the proton beam. A thick scraper increases the probability of large-angle scattering, which may cause particle loss downstream before reaching the first secondary collimator. Figure~\ref{fig:angle_eloss_tantalum} shows an example of the RMS scattering angle and energy loss for a 2.5\,GeV proton beam interacting with tantalum collimators of different thicknesses. With a desired RMS scattering angle larger than e.g. 4\,mrad and energy loss less than roughly 1\%, a thickness of 6--20\,g/cm$^2$ is expected to be suitable.

\begin{figure}[ht!]
\begin{center}
\begin{minipage}{0.47\linewidth}
\includegraphics[width=0.99\textwidth]{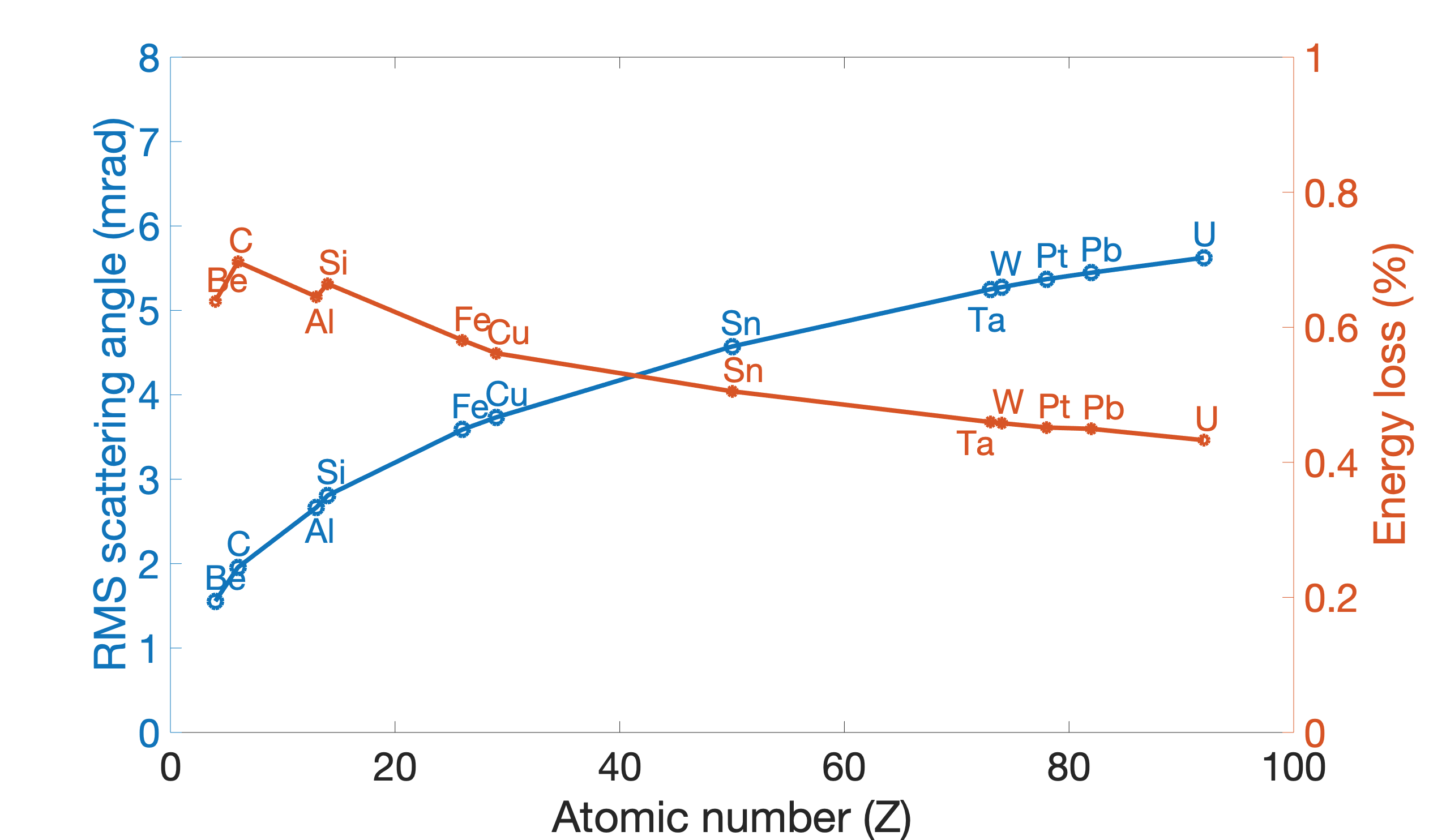}
\caption{\label{fig:angle_eloss_material}RMS scattering angle and energy loss for 2.5\,GeV protons interacting with different materials, each having a thickness of 10\,g/cm$^2$.}
\end{minipage}\hspace{2pc}%
\begin{minipage}{0.47\linewidth}
\includegraphics[width=0.99\textwidth]{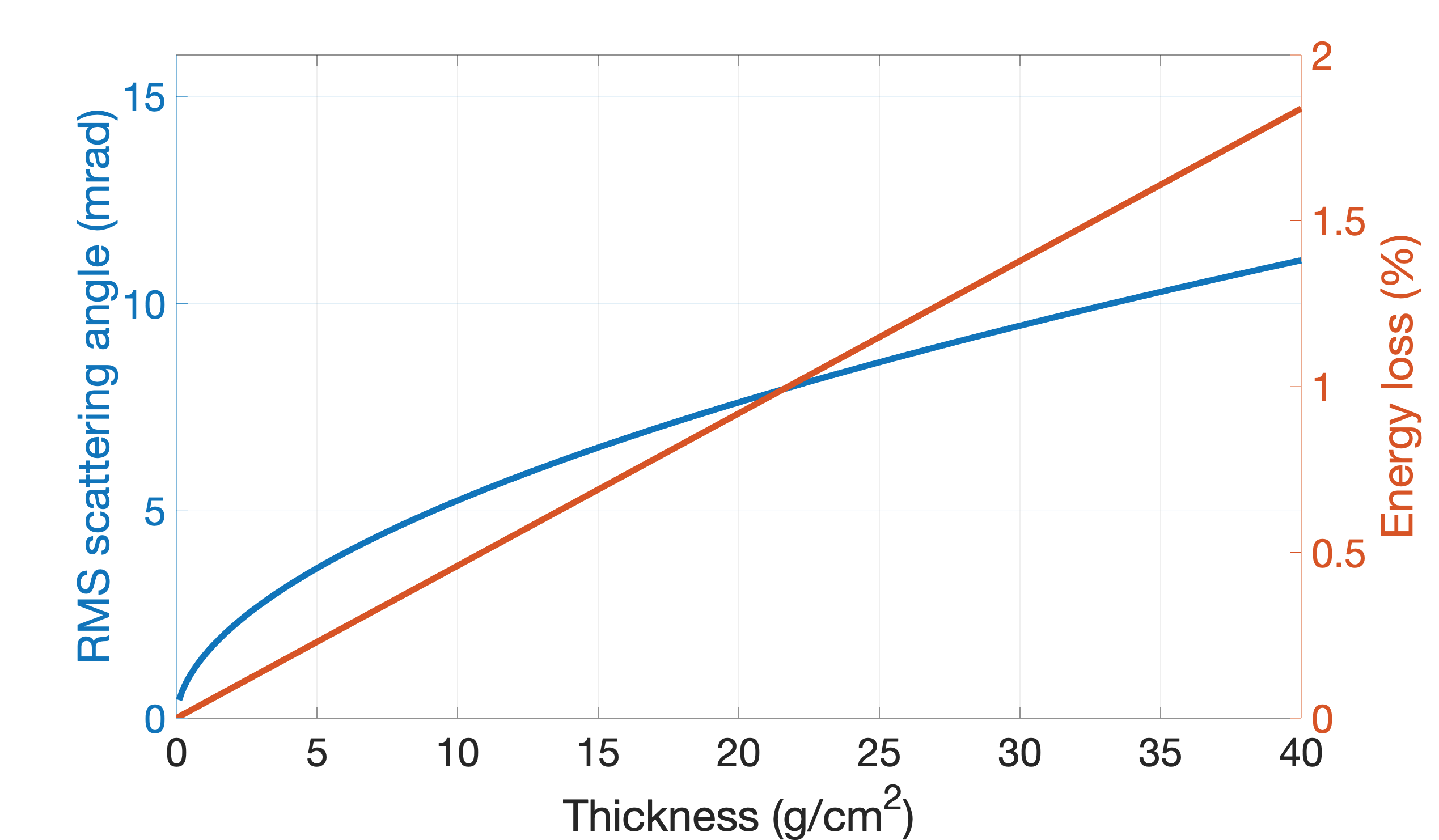}
\caption{\label{fig:angle_eloss_tantalum}RMS scattering angle and energy loss for 2.5\,GeV protons interacting with different thicknesses of tantalum.}
\end{minipage} 
\end{center}
\end{figure}

%-------------------------------------------------------------------------------
%     Figure: Scattering angle and energy loss for different materials
%-------------------------------------------------------------------------------
%\begin{figure}[ht!]
  %\begin{center}
    %\mbox{\epsfig{file=figures/accumulator/Figure_2.png,width=10cm}}
    %\caption{RMS scattering angle and energy loss for 2.5\,GeV protons interacting with different materials, each having a thickness of 10\,g/cm$^2$.}
    %\label{fig:angle_eloss_material}
  %\end{center}
%\end{figure}
%-------------------------------------------------------------------------------
%-------------------------------------------------------------------------------
%     Figure: Scattering angle and energy loss per thickness for Tantalum
%-------------------------------------------------------------------------------
%\begin{figure}[ht!]
  %\begin{center}
    %\mbox{\epsfig{file=figures/accumulator/Figure_3.png,width=10cm}}
    %\caption{RMS scattering angle and energy loss for 2.5\,GeV protons interacting with different thicknesses of tantalum.}
    %\label{fig:angle_eloss_tantalum}
  %\end{center}
%\end{figure}
%-------------------------------------------------------------------------------

\subsubsection{Numerical Simulations}
Multi-particle simulations have been performed with the PyORBIT code~\cite{Shishlo:2015:1272} to evaluate the performance of the collimation system.

The collimation system only affects the beam halo particles; thus, to simplify the simulation process, initial particles are give a larger amplitude than the primary collimator aperture, starting at the front of the primary collimator. This means that all simulated particles hit the primary collimator first. The initial particle distribution is an ``L'' type in real space, with a typical width of 10\,\SI{}{\micro\meter} and a uniform distribution, which matches the shape of the primary collimator. Space charge effects are not included in the simulation. 

Tantalum with 10\,g/cm$^2$ thickness is chosen for the primary collimator and tungsten with 1.5\,m length for the secondary collimators. In order to compare the particle loss rate for different collimator types, three configurations have been studied:
\begin{enumerate}
\item All collimators of two-sided type
\item The first collimator of rectangular type, the others of two-sided type
  \item All collimators of rectangular type
\end{enumerate}
Figure~\ref{fig:coll_loss_type} shows the particle loss rate in the collimation section for the different configurations. For Option~1, the major particle loss in the collimation section but not in the absorber blocks appears just after the first secondary collimator. Changing the first secondary collimator to a rectangular type (Option~2) reduces the beam loss after the first secondary collimator from around 7\% to less than 1\%. For Option~3, beam loss after the other secondary collimators can also be reduced. Figure~\ref{fig:coll_loss_map} shows the beam loss map for the full ring using Option~3. Here, the collimation efficiency reaches 97\%, which is 2\% higher than Option~2 and 12\% higher than Option~1.
%-------------------------------------------------------------------------------
%     Figure: Energy loss in collimators per type
%-------------------------------------------------------------------------------
\begin{figure}[ht!]
  \begin{center}
    \mbox{\epsfig{file=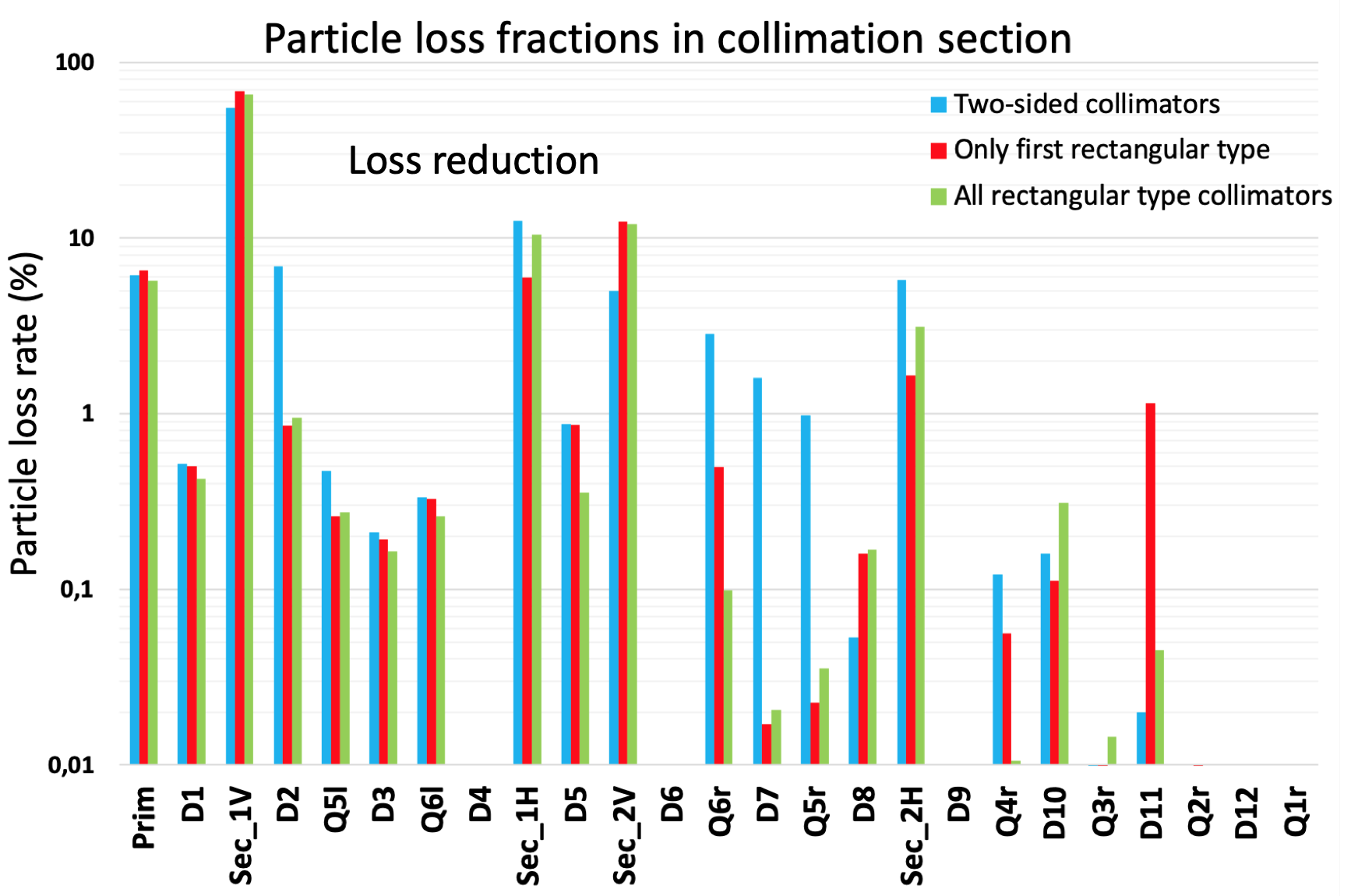,width=10cm}}
    \caption{Particle loss rate in collimation sections with different collimator types.}
    \label{fig:coll_loss_type}
  \end{center}
\end{figure}
%-------------------------------------------------------------------------------
%-------------------------------------------------------------------------------
%     Figure: Beam loss map with collimators
%-------------------------------------------------------------------------------
\begin{figure}[ht!]
  \begin{center}
    \mbox{\epsfig{file=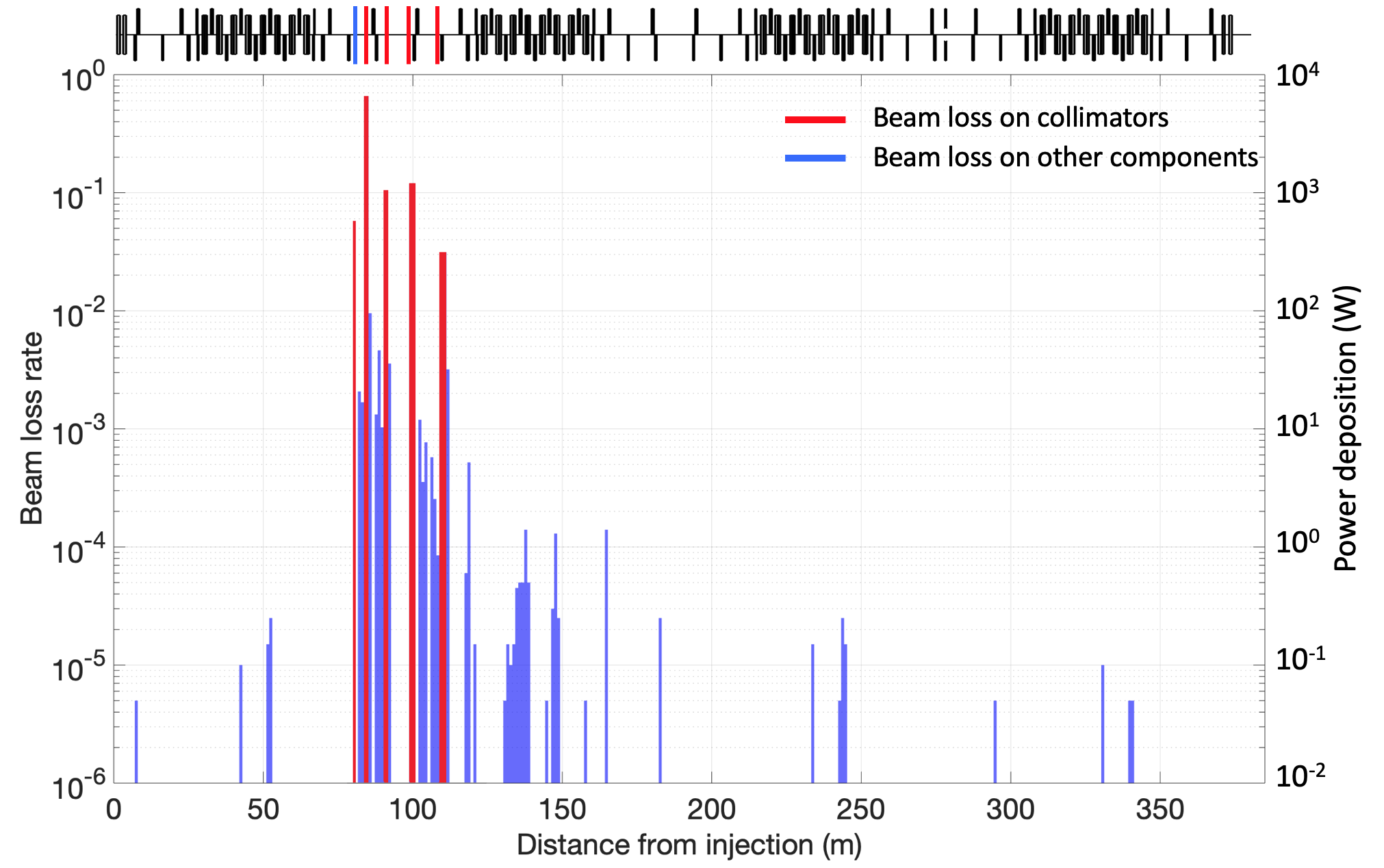,width=10cm}}
    \caption{Beam-loss map for the accumulator ring for Option~3 (all rectangular-type collimators).}
    \label{fig:coll_loss_map}
  \end{center}
\end{figure}
%-------------------------------------------------------------------------------

Studies of the relationship between collimation efficiency, ring acceptance, and primary collimator thickness have also been performed. Figure~\ref{fig:coll_efficiency_acceptance} shows the relationship between collimation efficiency and ring acceptance. The collimation efficiency increases with the ring acceptance and comes to a quasi-flat top when the acceptance reaches approximately 200$\pi$\,mm\,mrad. Figure~\ref{fig:coll_efficiency_prim_thickness} shows collimation efficiency and power loss before reaching the secondary collimators as a function of primary-collimator thickness. The collimation efficiency rises to a quasi-flat top when the thickness of primary collimation reaches roughly 6\,g/cm$^2$, while power loss continues increasing due to the large-angle scattering of the primary collimator.
%-------------------------------------------------------------------------------
%     Figure: Scattering angle and energy loss for different materials
%-------------------------------------------------------------------------------
\begin{figure}[ht!]
  \begin{center}
    \mbox{\epsfig{file=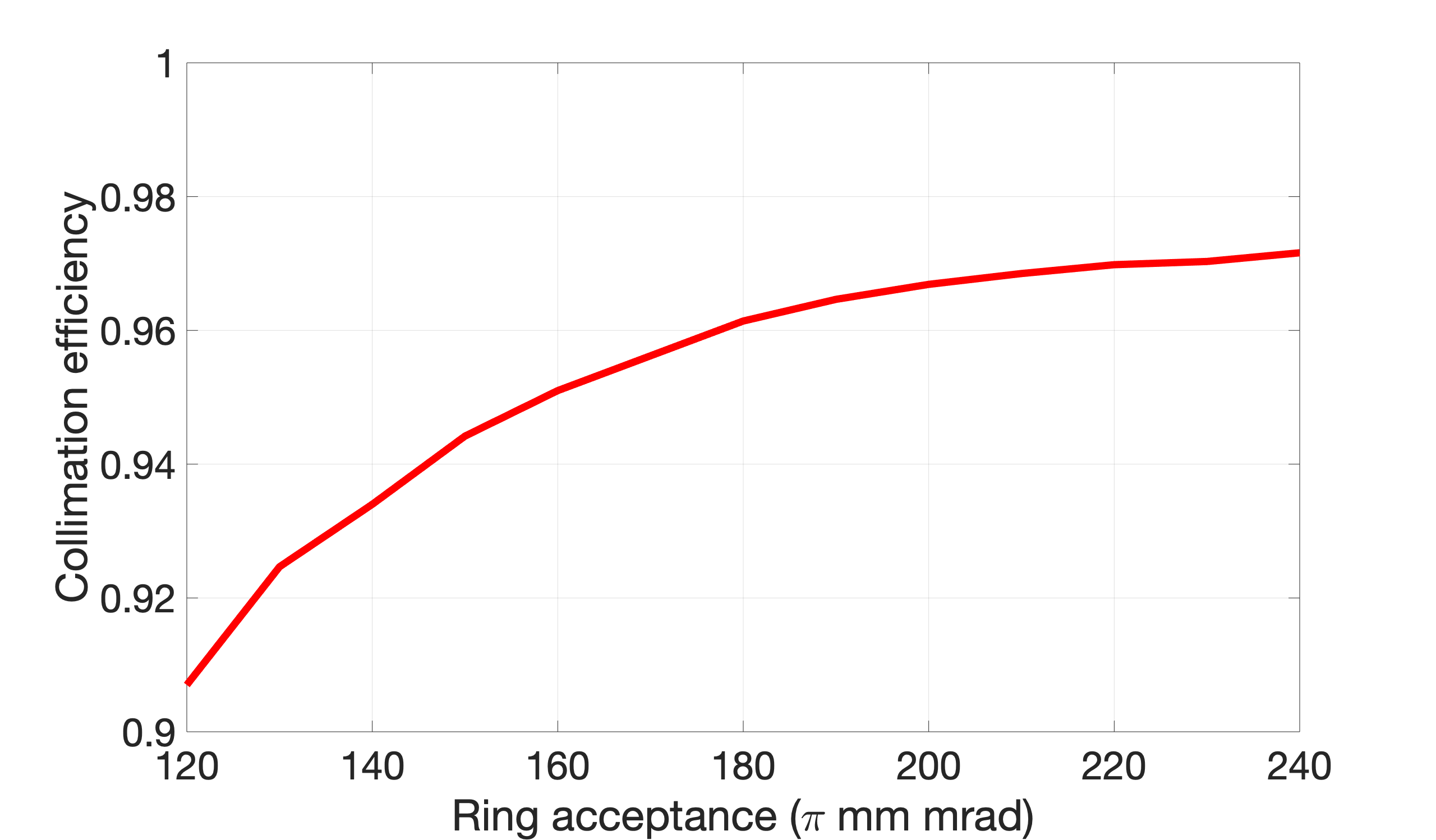,width=10cm}}
    \caption{Relationship between collimation efficiency and ring acceptance.}
    \label{fig:coll_efficiency_acceptance}
  \end{center}
\end{figure}
%-------------------------------------------------------------------------------
%-------------------------------------------------------------------------------
%     Figure: Scattering angle and energy loss per thickness for Tantalum
%-------------------------------------------------------------------------------
\begin{figure}[ht!]
  \begin{center}
    \mbox{\epsfig{file=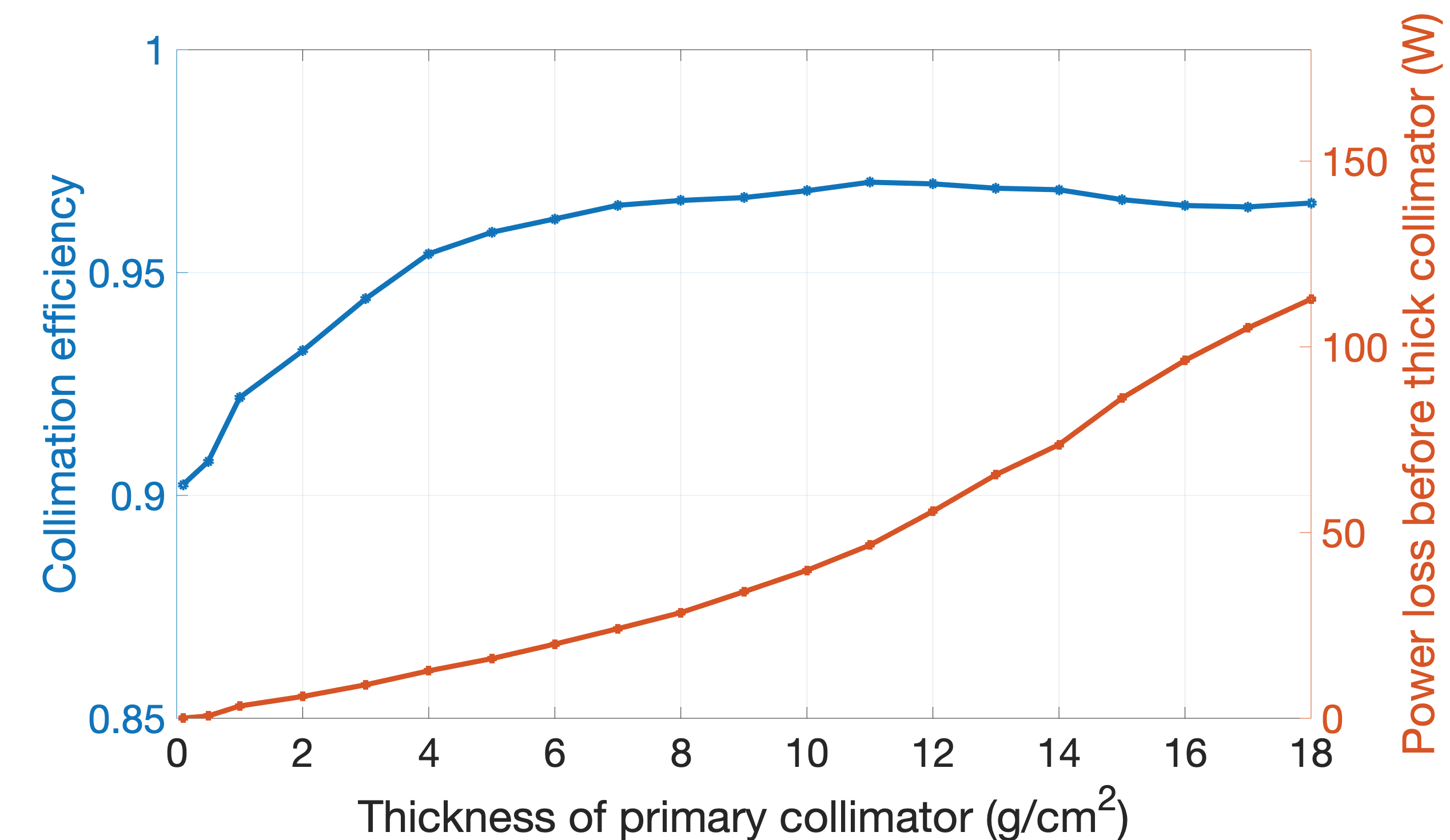,width=10cm}}
    \caption{Collimation efficiency and the power loss before reaching secondary collimators as a function of the primary-collimator thickness.}
    \label{fig:coll_efficiency_prim_thickness}
  \end{center}
\end{figure}
%-------------------------------------------------------------------------------

%%%%%%5%%%%%%%%%%%%%%%%%%%%%%%%%%%%%%%%%%%%%%%%%%%%%%%%%%%%%%%%%%%%%%%%%%%%%%%%
In conclusion, a 97\% collimation efficiency can be achieved using a two-stage collimation system where the primary collimator consists of an 10\,g/cm$^2$ thick, L-shaped scraper of tantalum, and the secondary collimator consists of four tungsten absorber blocks, 1.5\,m long each. The primary collimator acceptance is 70$\pi$\,mm\,mrad, the secondary collimator acceptance 120$\pi$\,mm\,mrad, and the machine acceptance is 200$\pi$\,mm\,mrad. Position acceptance mechanisms in the horizontal and vertical planes are foreseen for the collimator elements.   

%------------------------------------------------------------------------------
%%%%%%%%%%%%%%%%%%%%%%%%%%%%%%%%%%%%%%%%%%%%%%%%%%%%%%%%%%%%%%%%%%%%%%%%%%%%%%%%
%
%                  Single-turn extraction
%
%%%%%%%%%%%%%%%%%%%%%%%%%%%%%%%%%%%%%%%%%%%%%%%%%%%%%%%%%%%%%%%%%%%%%%%%%%%%%%%%

\subsection{Single-turn Extraction}\label{subsec:extraction}
Once the accumulator ring has been filled, the now-compressed beam pulse is transferred to the ring-to-target (R2T) transfer line through fast, single-turn extraction. The pulse at this point is 1.2\,\SI{}{\micro\second} long, in compliance with the target and horn requirements, and carries and energy of nearly 90\,kJ.

The extraction system is located in the straight section on the opposite side of the ring from the injection region (labeled SS3 in Fig.~\ref{fig:accumulator_layout}). The extraction system, illustrated in Fig.~\ref{fig:accumulator:extraction_layout} consists of four sets of fast kicker magnets, located around the central quadrupole pair in the straight section SS3, and a septum magnet. Each kicker provides a small kick in the vertical direction so that when the beam arrives at the septum magnet it encounters a horizontally deflecting field (see Fig.~\ref{fig:accumulator:septum}). Such a scheme was successfully implemented in the accumulator ring at SNS to extract the 1.3\,GeV proton beam with 24\,kJ per pulse~\cite{Wei:2000wa}. The scheme suits the ESS$\nu$SB well since the beam, when arriving at the target, must be directed $2.29^\circ$ downwards in order for the neutrino super-beam to point towards the far detector. Therefore, there is no need for cancelling the vertical angle provided by the kicker magnets, and there is no need of having a tilted septum magnet, which is the case in the SNS accumulator ring.

For simplicity, a horizontal deflection angle of 16.8$^\circ$ is used, matching that of the SNS septum magnet. A 3\,m long magnet at 1.0\,T gives a beam separation of approximately 50\,cm at the exit. This means that the septum magnet can be placed at a minimal distance to the subsequent quadrupoles, thus maximising the available space in the straight section for placing the kickers. A large distance between kicker and septum magnet means that the kick provided can be smaller, which facilitates the kicker magnet design in terms of impedance, aperture, and ramping speed. 

%-------------------------------------------------------------------------------
%     Figure: Extraction layout
%-------------------------------------------------------------------------------
\begin{figure}[htb]
  \begin{center}
    \mbox{\epsfig{file=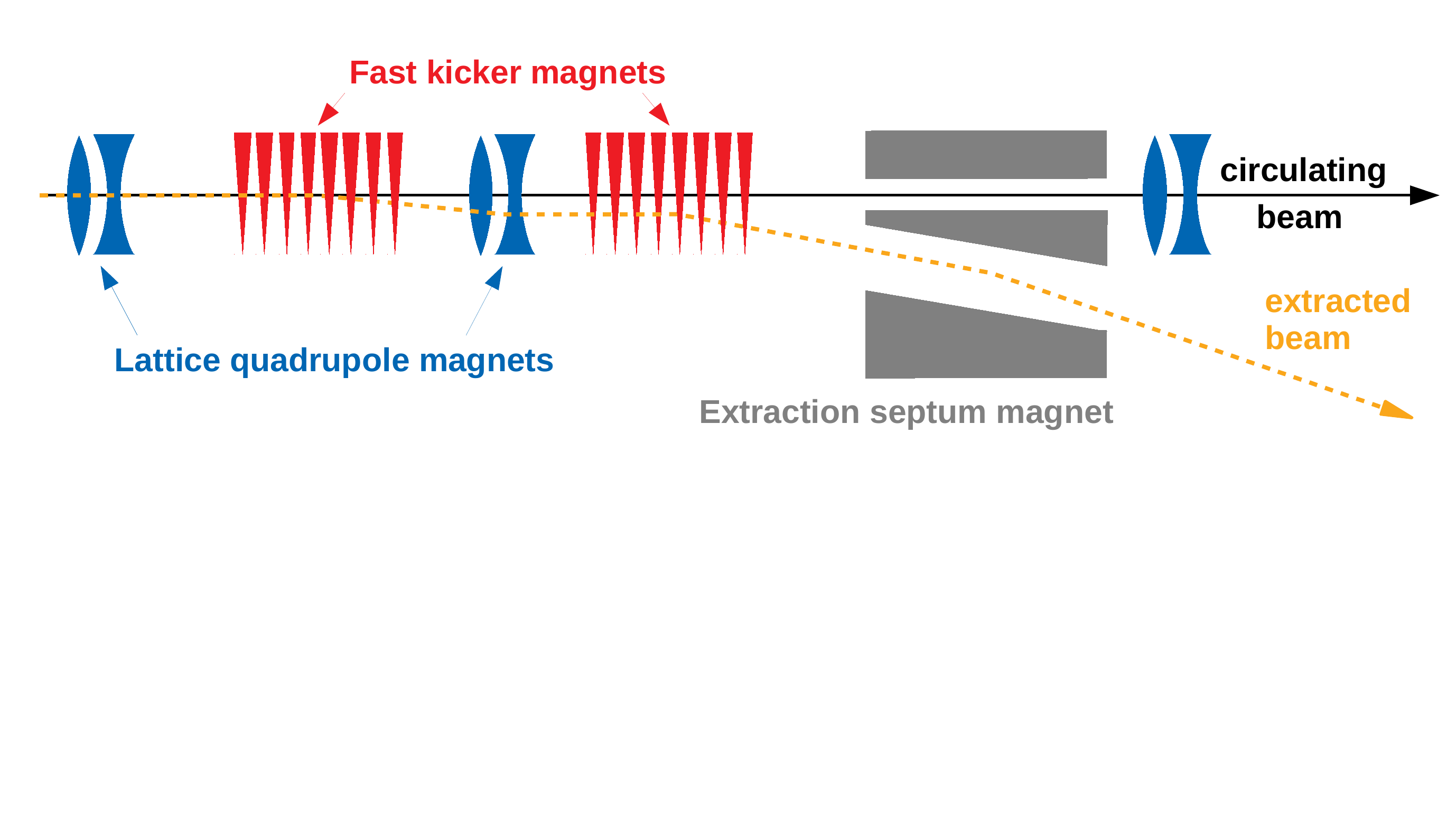,clip, trim=0cm 6cm 0cm 0cm, width=0.8\textwidth}}
    \caption{ Schematic of the extraction system.}
    \label{fig:accumulator:extraction_layout}
  \end{center}
\end{figure}
%-------------------------------------------------------------------------------
%-------------------------------------------------------------------------------
%     Figure: Septum magnet
%-------------------------------------------------------------------------------
\begin{figure}[htb]
  \begin{center}
    \mbox{\epsfig{file=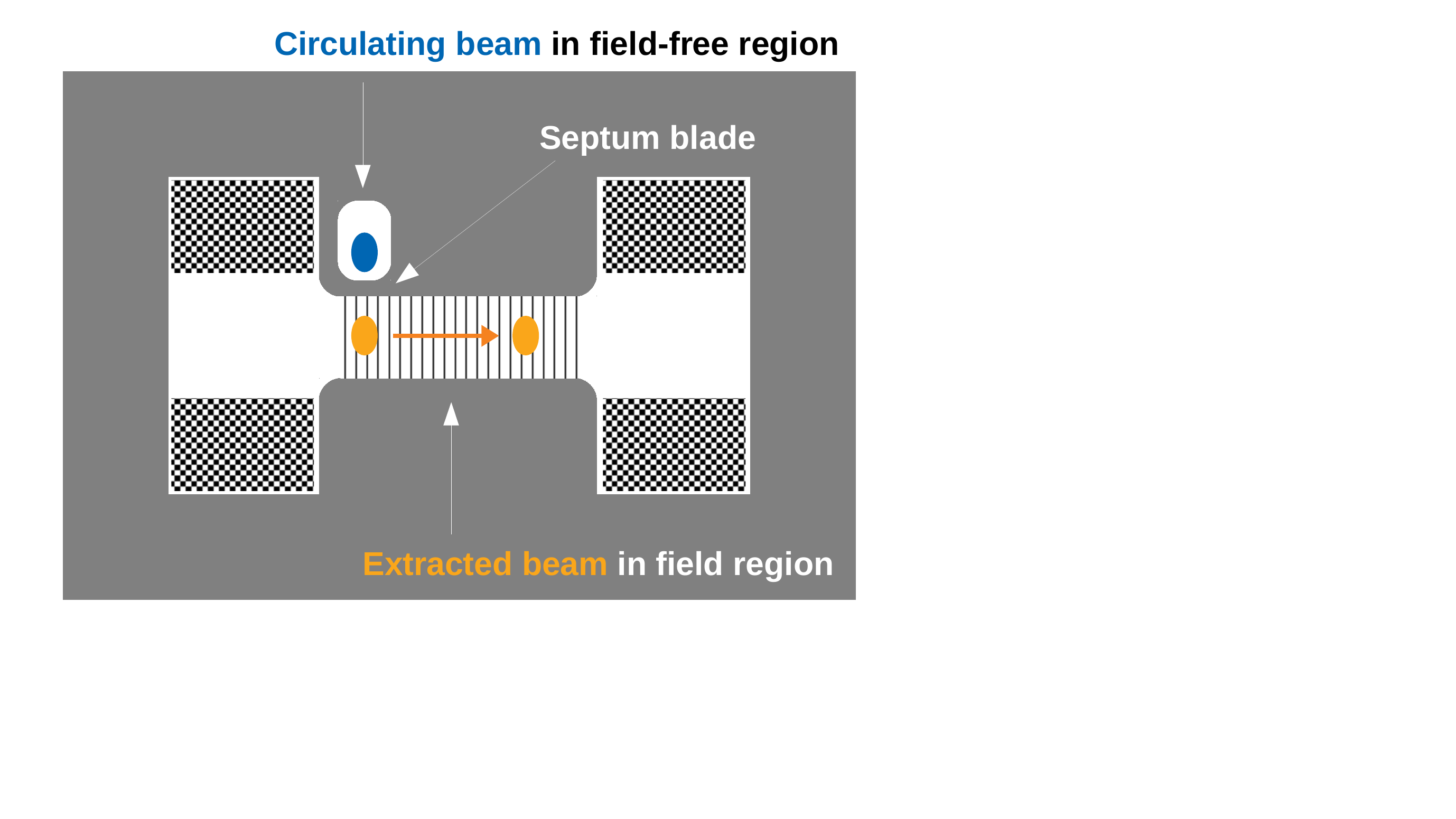,clip, trim=0cm 4cm 10cm 0cm, width=0.5\textwidth}}
    \caption{ Schematic view of the Lambertson septum magnet~\cite{Rank:2005:1591644}: The beam encounters no magnetic field while injection is ongoing. Once the injection is complete, the kicker magnets are powered and steer the beam towards the other side of the septum blade, where the magnetic dipole field deflects the beam horizontally.}
    \label{fig:accumulator:septum}
  \end{center}
\end{figure}
%-------------------------------------------------------------------------------

The total vertical deflection $\theta_y$ is determined from the desired vertical position $y$ of the closed orbit at the entrance of the septum, through
\begin{equation}
  \label{eq:accumulator:extraction_kick}
  \theta_y=\frac{y}{\sqrt{\beta_{y,k}\beta_{y,s}}\sin\mu_y},
  \end{equation}
where $\beta_{y,k}$ and $\beta_{y,s}$ are the vertical beta functions at the location of the kicker and septum respectively, and $\mu_y$ is the phase-advance between the two points. In order to minimise the kicker deflection angle, the phase-advance between the kicker and the septum should be close to $(2n+1)\pi/2$, where $n$ is an integer. In addition, the septum should be placed as far away from the kicker as possible. Furthermore, in order to reduce the individual kicker strength, and minimise the impact of an element failure, a larger number of shorter kickers is considered.

The vertical distance from the closed orbit to the extracted beam is calculated from
\begin{equation}
  \label{eq:accumulator:extraction_position}
  y_{\rm extr}=\sigma_0+\sigma_{\rm extr}+2\cdot y_m + a,
  \end{equation}
with $\sigma_0$ and $\sigma_{\rm extr}$ being the full beam width of the circulating and extracted beam, respectively, at the entrance of the septum, estimated from the optical functions:
\begin{equation}
  \label{eq:accumulator:extraction_width}
\sigma_{0, {\rm extr}}=\sqrt{\beta_sJ_s}+\delta\cdot\Delta_y
  \end{equation}
where $\beta_s$ is the value of the optical beta function at the entrance of the septum magnet and $J_s$ is the septum acceptance of the circulating and extracted beams. For a septum-blade width of $a=10$\,mm, and an aperture margin of $y_m=10$\,mm, we obtain a value $y_{\rm extr}= -161.0$\,mm.

In the proposed configuration, 16 kickers, in groups of four, are placed around the central quadrupole doublet in the SS3 straight section. Each of the 16 kickers has a length of 0.3\,m. In order to simplify their design, each element in a given group has the same aperture and strength. Since the aperture of the kickers is expected to increase towards the extraction point (to accommodate the beam deflection) their strength is scaled accordingly. Based on these conditions, using MAD-X~\cite{MADX}, the strength of the kickers was matched to the desired $y_{\rm extr}$-coordinate for the extracted beam (at the entrance of the septum) as defined above.

The apertures of all the elements in the extraction section were calculated to enclose the full beam envelope, plus a 1\,cm contingency margin. This is to account for possible variations of kicker fields during extraction or single-element failures. Table~\ref{tab:accumulator:extraction_kicker_params} includes the apertures of the extraction kickers, as well as designated kick angles and field strengths. The beam envelopes of the circulating and the extracted beam are shown in Fig.~\ref{fig:accumulator:extraction_envelopes}. The 16 kickers are shown in groups of 4 in different shades of green; the Lambertson septum is illustrated in magenta; and the lattice quadrupoles are shown in brown. A 180$\pi$\,mm\,mrad beam acceptance has been used in this calculation -- a value that has been explicitly chosen to be larger than the acceptance of the secondary collimators, which define the maximum beam envelope -- in order to have a safety margin in the design. 

A septum blade thickness of 1\,cm is assumed to be sufficient for ensuring that no field penetrates into the field-free opening. The circulating beam requires an aperture in the field-free region of about 120\,mm in the horizontal plane and 148\,mm in the vertical plane. The extracted beam will require a similar vertical aperture, but more than half a meter in the horizontal plane, where the 16.8$^\circ$ deflection takes place. 
%-------------------------------------------------------------------------------
%     Figure: Beam envelopes in extraction region
%-------------------------------------------------------------------------------
\begin{figure}[htb]
  \begin{center}
    \mbox{\epsfig{file=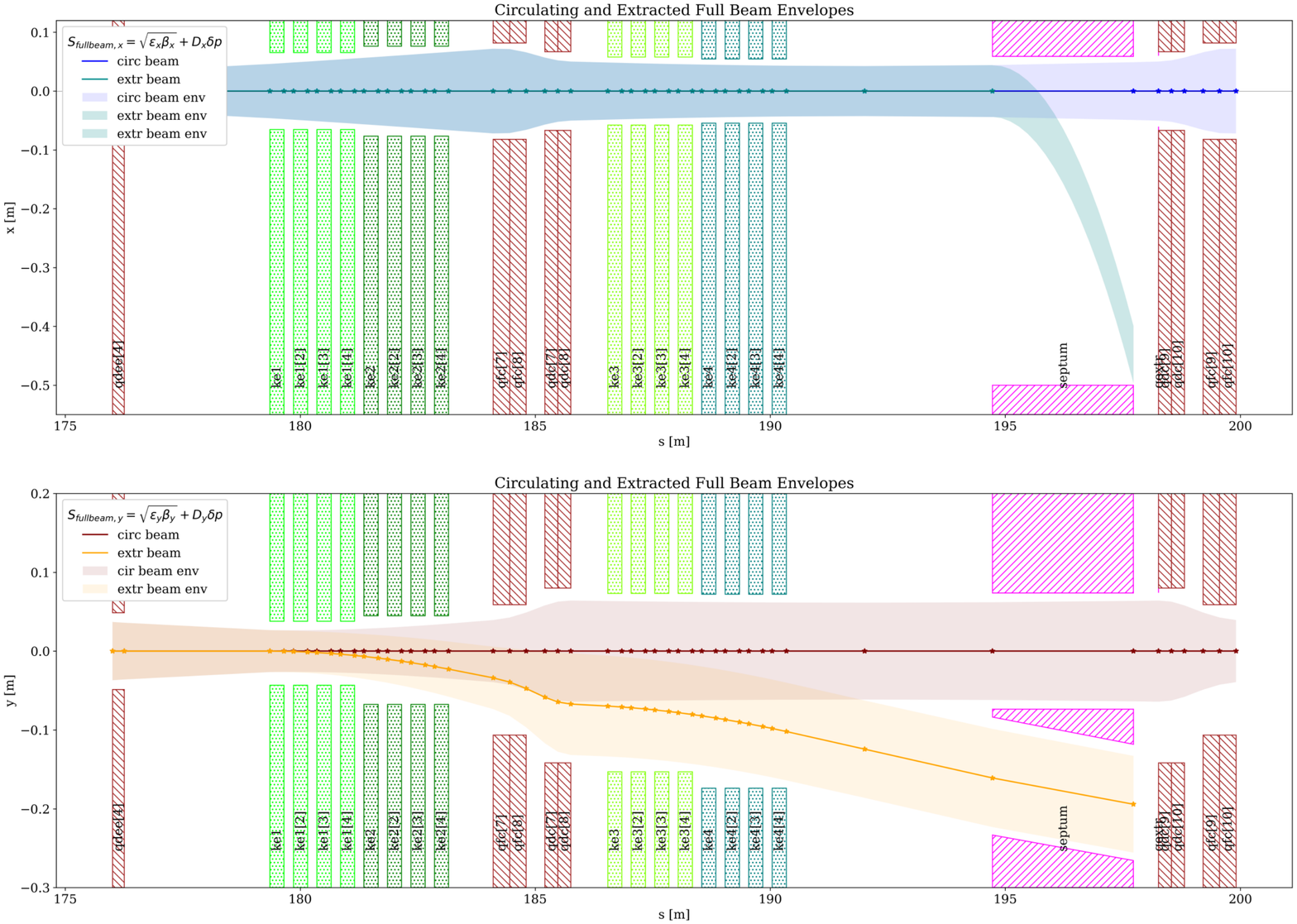,width=12cm}}
    \caption{Beam envelopes in the extraction region.}
    \label{fig:accumulator:extraction_envelopes}
  \end{center}
\end{figure}
%-------------------------------------------------------------------------------
%-------------------------------------------------------------------------------
%     Table: Extraction kicker params
%-------------------------------------------------------------------------------
\begin{table}[htbp]%[H]
  \begin{center}
    \caption{Kicker design parameters. The total deflection angle is -21.9\,mrad.}
    \label{tab:accumulator:extraction_kicker_params}
    \vspace{0.25 cm}
    \begin{tabular}{lccccc}
      %\hline
      \textbf{Kicker no.} &  \textbf{unit}    & \textbf{1-4}  & \textbf{5-8}   & \textbf{9-12}  & \textbf{13-16}\\
      \hline
      Horizontal aperture &  mm  & $\pm$65     & $\pm$76    & $\pm$58    & $\pm$55 \\
      Vertical aperture   & mm   &  [-43, 38]  &  [-68, 45]  &  [-153, 73] &  [-174, 72] \\
      Length              & m    &  0.3        & 0.3 & 0.3 & 0.3 \\
      Kick angle          & mrad & -1.54        & -1.39 & -1.34 & -1.20 \\
      Field strength      & mT   & -57         & -51 & -49 & -44 \\
      \hline
    \end{tabular}
  \end{center}
\end{table}
%-------------------------------------------------------------------------------

\subsubsection{Extraction Kickers}
As described in Section~\ref{sec:accumulator_rf}, the kicker must be able to deflect the entire circulating beam in a single turn. This means that the ramping of the magnetic field in the kicker magnets must be done within the duration of the beam-free extraction gap (100\,ns). To this end, a kicker-field rise-time is defined as the time it takes for the kicker to go from 2\% to 98\% of the nominal field strength, as depicted in Fig.~\ref{fig:accumulator:extraction_kicker_ramping}. There may be remaining ripple in the pulse but it should remain below $\pm2$\% of the nominal field strength. The fall time of the kicker pulse is defined by the interval between the extraction of a pulse and the start of injection for the next pulse (i.e. a maximum of 100\,\SI{}{\micro\second}). The pulse duration should be around 1.5\,\SI{}{\micro\second}.
%-------------------------------------------------------------------------------
%     Figure: Ramping of the extraction kickers
%-------------------------------------------------------------------------------
\begin{figure}[htb]
  \begin{center}
    \mbox{\epsfig{file=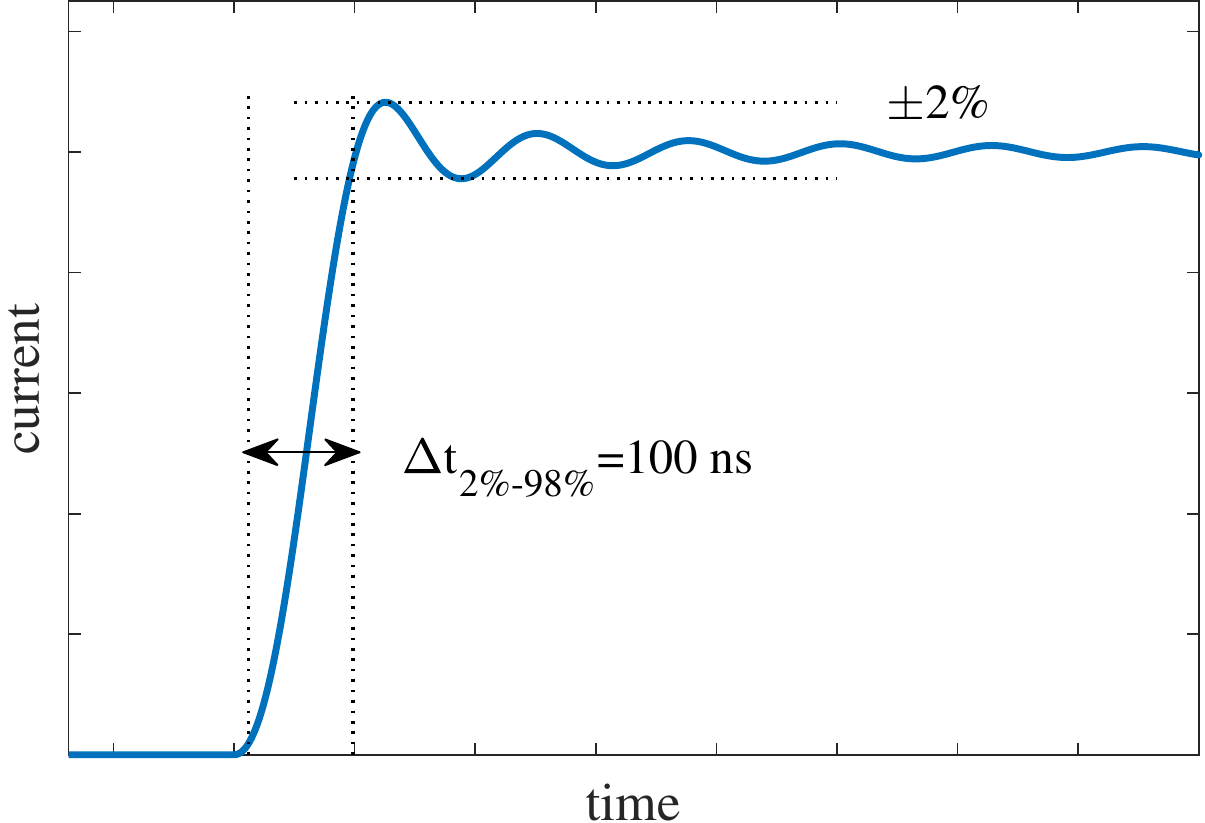,width=8cm}}
    \caption{ Illustration of the extraction kicker ramping.}
    \label{fig:accumulator:extraction_kicker_ramping}
  \end{center}
\end{figure}
%-------------------------------------------------------------------------------

The specifications described above will serve as input to an engineering study for kicker-magnet optimisation and construction. However, the required parameters are comparable to those of kickers already constructed for the SNS accumulator~\cite{Tsoupas:2000yj} and the CERN Proton Synchrotron Booster extraction region~\cite{Metzmacher:2061508}. Development of the ESS$\nu$SB extraction kickers will also include the pulse-forming network needed for powering the magnets. Since the time between pulses is short, it may be necessary to use two power supplies per magnet.

\subsubsection{Failure Scenarios}
Two different failure scenarios have been studied: a) a $\pm2\%$ variation in the kicker field, b) the failure of a single kicker element in the sixteen-kicker assembly.

The first case simulates a possible uncorrected ripple in the kicker field strength as explained above until it is stabilised to the nominal value. To take a conservative estimate, the error is assigned in all sixteen kickers (which in reality can be compensated automatically among the four power supplies). Simulations indicate that the chosen apertures are sufficient to provide clearance and avoid losses even in this pessimistic case.

The second case was simulated in MAD-X by setting the field strength to zero for one kicker at a time. Also in this case, the chosen apertures provide sufficient margin for the beam to pass without losses. For both cases, the resulting trajectory variations and resulting losses in the beam transfer line up to the target have not been studied.
%-------------------------------------------------------------------------------

%%%%%%5%%%%%%%%%%%%%%%%%%%%%%%%%%%%%%%%%%%%%%%%%%%%%%%%%%%%%%%%%%%%%%%%%%%%%%%%

%------------------------------------------------------------------------------
\subsection{Generation of Short Neutron Pulses}\label{subsec:slow_extraction}
%------------------------------------------------------------------------------

%\red{Infor from Colin to be rewritten: The idea originates from the early days (mid to late 1980s) at ISIS where the short >1\,\SI{}{\micro\second} pulse length was ideal for epithermal and thermal neutrons but inefficient for cold neutrons $>4$\,\SI{}{\angstrom} wavelength. The moderating time for the undermoderated neutrons (i.e. 0.5-2\,\SI{}{\angstrom}) that an accelerator source provides is $\sim7\lambda$. Rule of thumb. For cold neutrons the moderating time (in the Maxwell-Boltzmann distribution) is $\sim22\lambda$ even for the small moderators on accelerator sources. I was building a cold neutron spectrometer at ISIS then, swimming against the tide, but obviously 1\SI{}{\micro\second} proton pulses to generate $>100\lambda$ neutron pulses meant that the resultant peak intensity was severely compromised.}
The ESS$\nu$SB accumulator ring could, in principle, serve an important purpose also for the neutron community. The ESS facility itself provides long neutron pulses to the users, but some users wish for high-intensity short pulses of epithermal neutrons, i.e. neutrons in the energy range 0.025-0.4\,eV. Considering the typical moderating time of about 100\,\SI{}{\micro\second}~\cite{McGinnis:2013sza} for these neutrons, there is no great advantage of using the compressed 1.2\,\SI{}{\micro\second} pulse to generate the neutrons, since this would require a specialised target to handle such a pulse~\cite{Andersen:2020}. As an alternative, a compressed pulse at lower intensity could be used, but at the cost of a subsequently reduced neutron flux. The other alternative is to use the ESS$\nu$SB accumulator to produce a proton pulse of 50-100\,\SI{}{\micro\second} through slow extraction from the ring. In this way, the delivered neutron pulse duration would be of the same order but the instantaneous power delivered to the neutron target, and thus the levels of induced mechanical stress, is limited. Thus, the ESS$\nu$SB infrastructure could also be highly impactful for the neutron user community, as a complement to the long pulses of ESS. \cite{Andersen_2016,McGinnis:2013sza}.

The most promising scenario for the ESS$\nu$SB accumulator is the use of slow resonant extraction at a half-integer or integer resonance. The more commonly used third-order resonance, employed at the J-PARC Main Ring~\cite{Muto:2019sbv}, the SPS at CERN\cite{Fraser:2693913}, and at GSI~\cite{Singh:2018nek} takes place over several seconds (on the order of 1000 turns or more), and would be too slow to extract the beam in 100\,\SI{}{\micro\second}. An extraction time of 50\,\SI{}{\micro\second} corresponds to about 75 turns in the accumulator ring, since the revolution time is about 1.3\,\SI{}{\micro\second}. Resonant extraction with as few as 40 turns has been demonstrated at SPS using integer resonance \cite{Baconnier1977, Stoel:2018euk}, which indicates that this should be a feasible approach.  

Resonant extraction will be designed with a dedicated electrostatic septum, using an identical Lambertson septum as that used in fast extraction. To minimise losses it is also important to, for example, keep the so-called step size at the extraction sufficiently large. Moreover, there are a number of special techniques to reduce losses for resonant extraction such as dynamic bump, bent-crystal diffusers, and carbon wires~\cite{Stoel:2018euk}. J-PARC  has reached impressive extraction efficiencies of 99.5\% with careful optimisation employing a dynamic bump \cite{Arakaki:2012zza}. More realistic loss levels for the ESS$\nu$SB accumulator ring are estimated to be approximately 5--10\%. 

As slow, resonant, extraction is inherently lossy, the losses cannot be eliminated entirely if resonant extraction is to be employed for neutron production. Mitigation measures should include careful beam optics design, electrostatic septum design, and careful optimisation at beam operation. Most of the unavoidable losses will happen on the septum at the point of extraction. This will have the the undesired effect of activation, which limits the feasibility of handling in terms of  repair and preventive maintenance. Remote handling, including robotics, could be introduced in order to reduce the dose levels to personnel. Ultimately, the only way to limit the average power to an acceptable level is by reducing the duty cycle for resonant extraction. If, for example, every tenth pulse is extracted for short pulse neutron production, this yields an average loss of 50\,kW, at 90\% extraction efficiency. Further analysis of the activation levels and shielding requirements are needed to determine the acceptable duty cycle for neutron production.

The moderation of the neutrons produced by the medium-duration proton pulse will have the effect that any variation in the proton intensity during the pulse is smeared. Thus, the extraction spill-quality, which is usually an important parameter for slow extraction, is of less importance in this case. This gives more degrees of freedom in terms of the excitation of the extraction, and therefore more possibilities to reduce the beam loss during the spill.

Another approach that has been considered employs a target design which is capable of handling short high-power pulses on the order of 1\,\SI{}{\micro\second}, similarly to the case at SNS following the Proton Power Upgrade~\cite{Galombos:2020:PPU}. In such a scenario the same fast-kicker extraction as for neutrino production, with the same low losses, could be used for the neutron production. This is an interesting alternative, which -- with a specially adapted neutron moderator design -- could give comparable neutron-flux levels, at about the same duty cycle;  but this would require more research and development to ensure its technical feasibility.  

%------------------------------------------------------------------------------
\subsection{Injection Beam Dump}\label{subsec:injection_dump}
%------------------------------------------------------------------------------
%\subsubsection{Introduction and General Considerations}
Straight behind the ring injection, see Fig.~\ref{fig:accumulator_layout}, there will be a short beam line towards a beam dump, here referred to as the injection beam dump. The purpose of the beam dump is to safely dispose of incoming particles that are not fully stripped. A stripping efficiency of about 98\% can be expected, which leaves an average beam power of 100\,kW to be handled by the dump. The dump is designed to handle up to 10\% of the average beam power, mostly coming from $H^-$ ions which miss the stripping foil, or from foil defects, or from partially stripped $H^0$. Radiation shielding is required on top of the beam dump for radiation safety, both at operation, and during service, to reduce the radiation levels to acceptable limits. Shielding towards the ground is required to prevent activation of soil and groundwater, primarily from tritium, which could potentially be transported through the groundwater.

In the conceptual design of the ESS$\nu$SB injection dump, the design and experience of the SNS injection beam dump has been carefully studied. Comparing the parameters, SNS has an $H^-$ beam of 1\,GeV, and 1.44\,mA average current. This gives a power of 1.44\,MW, with the injection beam dump designed for losses up to 150\,kW. In comparison, the beam parameters for ESS$\nu$SB, with an upgrade to 2.5\,GeV, and an average $H^-$ beam current of 2\,mA are set to reach 5\,MW average power and a loss level of 10\%, which corresponds to 500\,kW. The design of the beam transport around the injection region will be conceptually similar to the SNS design, see Fig.~\ref{fig:sns_injdump}. Further details on the SNS injection beam dump can be found in \cite{Johnson:2000:shielding} and \cite{SNS:2010:safety}.

\begin{figure}[htb]
    \centering 
    \includegraphics[width=10cm]{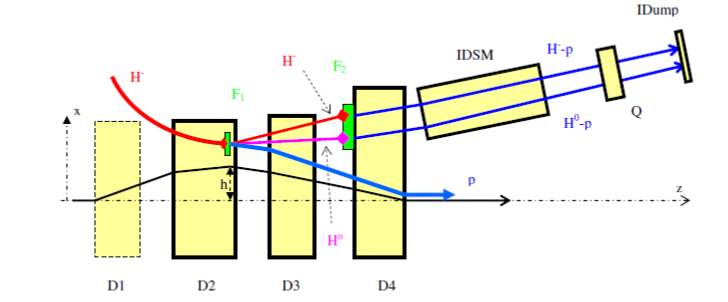}
    \caption{SNS injection section, shown schematically, where F1 is the stripper foil; F2 is a second, thick stripper foil to fully strip $H^-$ and $H^0$ to protons; IDSM is the Injection Dump Septum Magnet; and Q is a defocusing quadrupole. From: \cite{Wang:2008zzk}.}
    \label{fig:sns_injdump}
\end{figure}

\subsubsection{Energy Deposition in the Beam Dump}
The levels of heat deposition and activation have been simulated using the FLUKA code \cite{FLUKAweb, Battistoni:2015:FLUKA, Bohlen:2014:FLUKA}. The model of the beam dump in FLUKA consists of a copper cylinder of 25\,cm radius and 120\,cm length, with an effective density set to 90\% of nominal in order to model cooling water channels. In the simulation, the proton beam energy is 2.5\,GeV, and the current is 0.2\,mA corresponding to losses of 10\%. The beam is assumed to be Gaussian, with size $\sigma_x, \sigma_y = 25 mm$, (59\,mm FWHM) and an angular divergence of 1\,mrad FWHM. In the simulations, no time structure of the pulses is considered. 

%-----------------------------------
\begin{figure}[htb]
\begin{center}
\begin{tabular}{cc}
\mbox{\epsfig{file=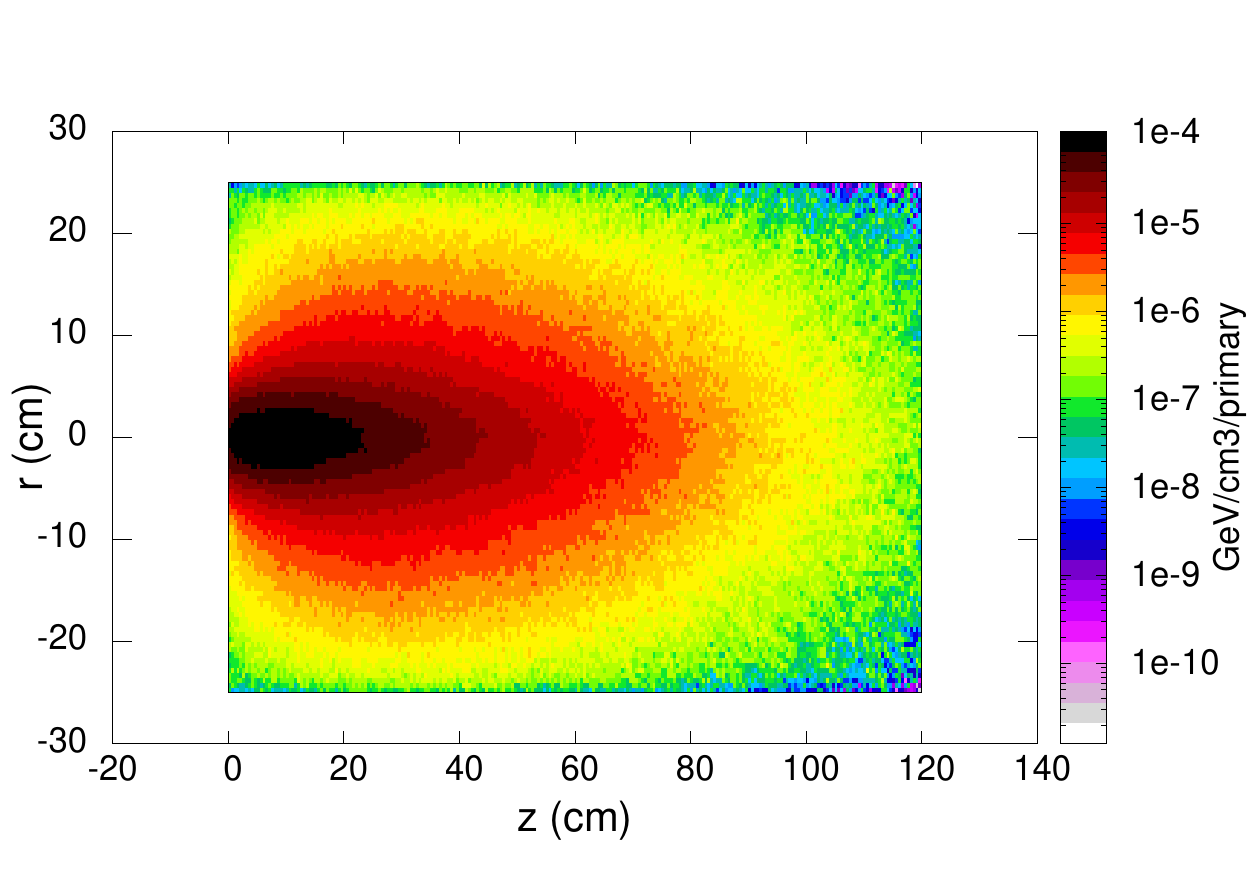,clip, trim=0cm 0cm 0cm 1cm,height=5cm}} &
\mbox{\epsfig{file=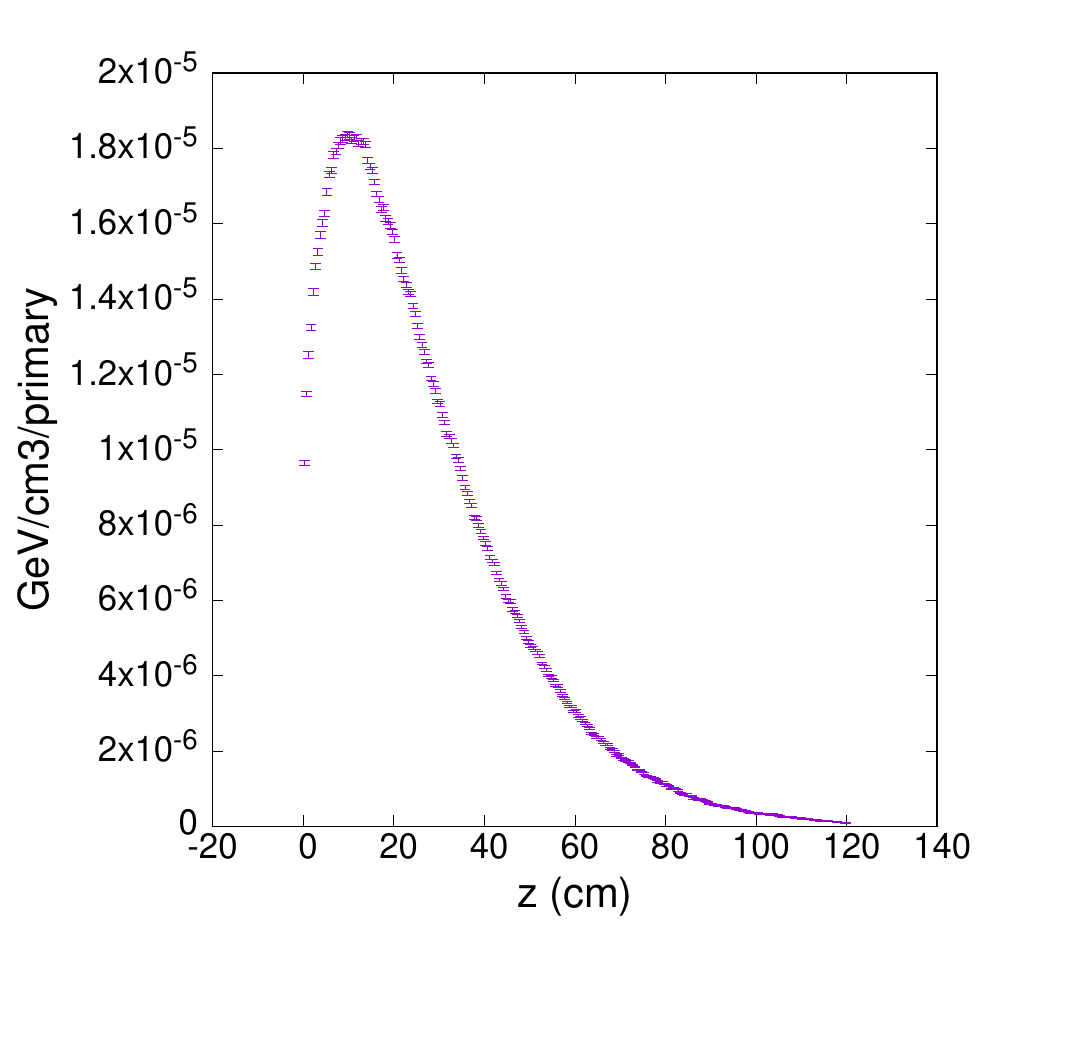,clip,trim=0cm 1cm 0cm 0cm,height=5.5cm}} \\
(a) 2D projection. & (b) 1D projection. 
\end{tabular}
\caption{Distribution of energy deposition in the copper beam dump.}
\label{fig:injdump_edep}

\end{center}
\end{figure}
%-----------------------------------

The resulting energy deposition in the beam dump is shown in Fig.~\ref{fig:injdump_edep}. Here, a 2D radial projection, integrated over the polar coordinate, is shown in (a); a one dimensional projection, integrated over the radial and polar coordinates, is shown in (b). The maximum energy deposition is about 30\,W/cm$^3$ and occurs at $z\approx9$\,cm.  

%\begin{figure}[htb]
%\begin{subfigure}[b]{0.55\linewidth}
%    \centering 
%    \includegraphics[width=\linewidth]{figures/accumulator/injection_dump_2d_edep.pdf}
%    \caption{2D projection.}
%\end{subfigure}
%\hfill
%\begin{subfigure}[b]{0.45\linewidth}
%    \centering 
%    \includegraphics[width=\linewidth]{figures/accumulator/injection_dump_1d_edep.pdf} %\begin{minipage}{.1cm}
%        \vfill
%    \end{minipage}
%    \caption{1D projection.}
%\end{subfigure}
%\caption{Distribution of energy loss in the beam dump.}
%\label{fig:injdump_edep}
%\end{figure}

For comparison, the SNS beam dump has a radius of 15\,cm and a length of 80\,cm and is built up of copper disks which are stacked into a cylinder with the thickness adapted so that each disk does not take a heat load exceeding 5\,kW. Further details are available in~\cite{Popova:2022:shielding}. Due to the higher energy at ESS$\nu$SB, the dimensions have to be increased for the primary and secondary particles to be deposited in the dump. Because of this higher energy, the energy deposition will be spread out over a larger depth, which is an advantage compared to SNS in terms of local heating.

Different dimensions have been tested and it has been concluded that a 25\,cm radius is required and a length of 120\,cm would be sufficient. A larger radius gives more margin and less heating and activation to the outer shielding, which should be considered when optimising the overall structure. 

\subsubsection{Shielding Calculations}
In the present analysis of the shielding, the focus has been on the shielding atop the injection beam dump. Here there is need for a service area -- both for the beam window to the injection dump and for the injection dump itself. 

The attenuation of the shielding can be calculated semi-empirically using the Moyer model \cite{Sullivan:1992ti, NCRP144} or a similar approach. Such models are generally valid above 1\,GeV; with, for example, a point loss in a target, and thick shielding (more than three mean free paths for inelastic nuclear interactions -- in order for the nuclear cascade to develop and average over the energy loss from secondary particles.) Iron has been chosen as the main shielding material, since it performs well in terms of shielding high-energy neutrons. However, iron has fairly low attenuation of low-energy neutrons, which dominate the particle fluence after a few attenuation lengths. Therefore, a concrete layer has been added outside the iron shielding, since concrete has a high attenuation of low-energy neutrons.  
The effective density of iron has been set to 95\% in order to allow for water cooling to compensate for heating from deposited energy due to prompt radiation and activation. A model of the shielding is shown in Fig.~\ref{fig:injdump_layout}, with the beam entering from the left. 

\begin{figure}[htb]
    \centering 
    \includegraphics[width=10cm]{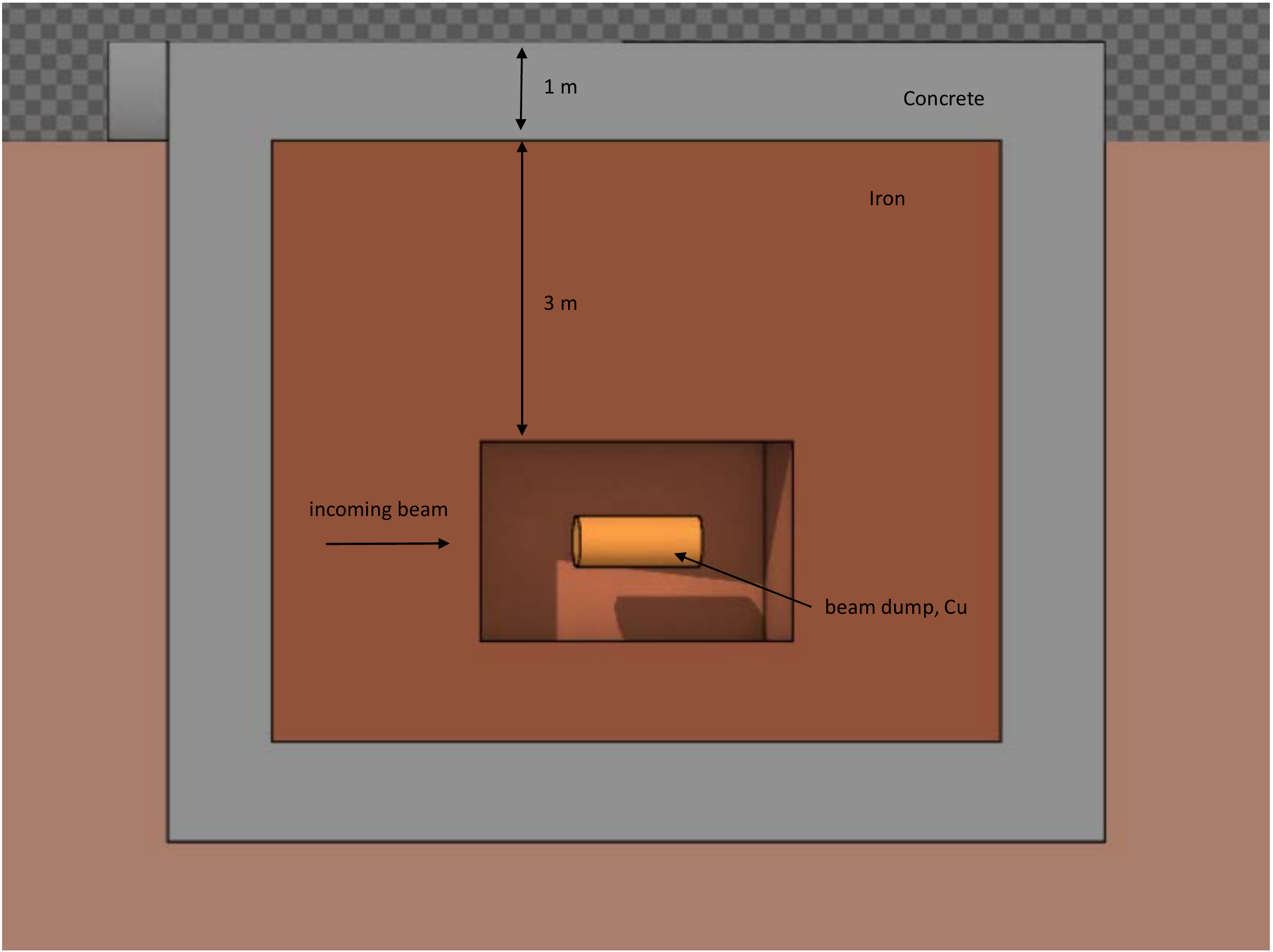}
    \caption{Schematic layout of the injection beam dump with shielding.}
    \label{fig:injdump_layout}
\end{figure}

For a point source, the dose-rate outside the shielding can be calculated from~\cite{Sullivan:1992ti}
\begin{equation}
    H=H_0\frac{e^{-d/\lambda}}{R^2}~{\mathrm Sv/h,} 
\label{eq:dose_attenuation}
\end{equation}
\noindent where $H_0$ is the source term in Sv/m$^2$, $d$ is the thickness of shielding, $\lambda$ is the attenuation length (mean free path) for nuclear inelastic interactions, and $R$ is the distance from the source. 

The data for the attenuation lengths, $\lambda$, are taken from \cite{pdg2021}. For iron, $\lambda = 132$\,g/cm$^{-2}$, corresponding to $\lambda=16.7$\,cm at a nominal density of 7.87\,g/cm$^{-2}$. Assuming a density of 95\% to compensate for cooling channels, this gives $\lambda=17.6$\,cm. Concrete has an attenuation coefficient of $\lambda=43$\,cm at a density of 2.3\,g/cm$^3$. The attenuation varies with specific concrete composition but with these numbers a first estimate can be made.

The total attenuation factor in Eq.~(\ref{eq:dose_attenuation}), $e^{-d/\lambda}/R^2$, for a shielding consisting of 3\,m iron and 1\,m concrete at 90$^\circ$, and a distance 5\,m from the source, then becomes $1.0{\times}10^{-10}$. 
 
\subsubsection{Source-Term Calculations}
The source term, $H_0$, has been calculated using FLUKA~\cite{FLUKAweb, Battistoni:2015:FLUKA, Bohlen:2014:FLUKA}, again using a copper cylinder of radius 25\,cm and 120\,cm length as a target surrounded by vacuum, with a beam of $1.25{\times}10^{15}$ particles per second (0.2\,mA) consisting of 2.5\,GeV protons; and again assuming a density of 90\% of the nominal value to compensate for cooling water channels. 

The proton beam matches the parameters used in the heat deposition calculations: Gaussian, with size $\sigma_x, \sigma_y = 25$\, mm, (59\,mm FWHM), and an angular divergence of 1\,mrad FWHM. The dose rate distribution is shown in 2D, Fig.~\ref{fig:injdump_doserate}, integrated over the polar coordinate, and in 1D, integrated over the polar and the radial coordinates, and averaged over $z=30-40$\,cm, where the dose rate has its highest values. 

As can be seen from Fig.~\ref{fig:injdump_doserate} b), the calculated dose rate at 1\,m radius, is $1.9\cdot10^5$\,Sv/h, which corresponds to a source term $H_0$ of $1.9\cdot10^5$\,Svh$^{-1}$m$^{-2}$. This can be compared to semi-empirical source term calculations following Ref.~\cite{Sullivan:1992ti}, which gives $1.6\cdot10^5$\,Svh$^{-1}$m$^{-2}$.

The dose rate outside the shielding at $R = 5$\,m is then attenuated with a factor $1.0\cdot10^{-10}$ according to Eq.~(\ref{eq:dose_attenuation}) giving a dose rate directly outside the shielding of 19\,\SI{}{\micro\sievert/\hour}. This dose rate would be acceptable for a radiation worker (corresponding to 6\,mSv per year at 300\,h occupancy) and depending on the classification of the area, and occupancy, one could possibly reduce the shielding to some extent. 

The present calculation should be confirmed with FLUKA simulations or other another Monte-Carlo code; this should be complemented with activation calculations, and radiation levels at various cool-down times, i.e. elapsed time after beam shut-off. As mentioned above, the shielding towards the ground, and possible effects of tritium production will have to be studied. 

\begin{figure}[htb]
\begin{subfigure}[b]{0.53\linewidth}
    \centering 
    \includegraphics[width=\linewidth]{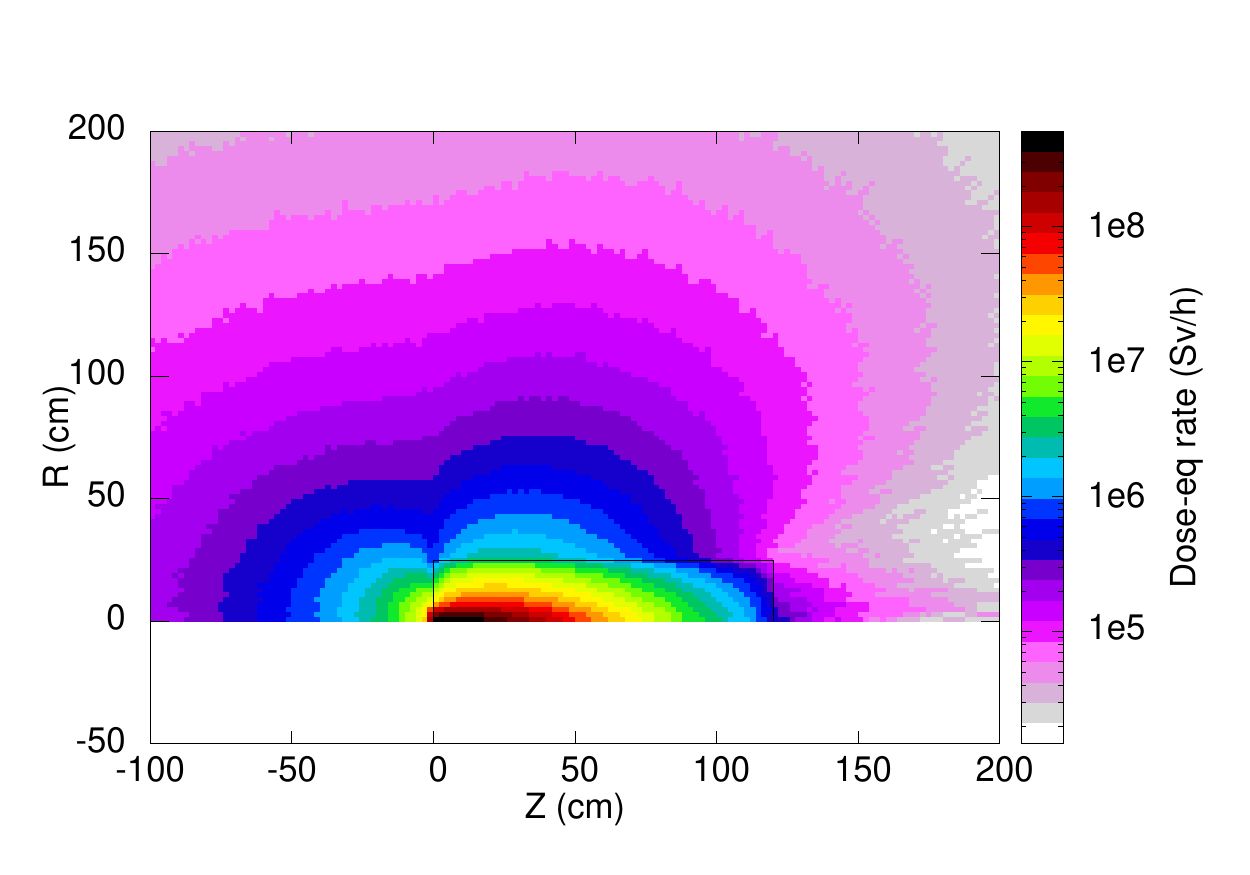}
    \caption{2D projection.}
\end{subfigure}
\begin{subfigure}[b]{0.47\linewidth}
    \centering 
    \includegraphics[width=\linewidth]{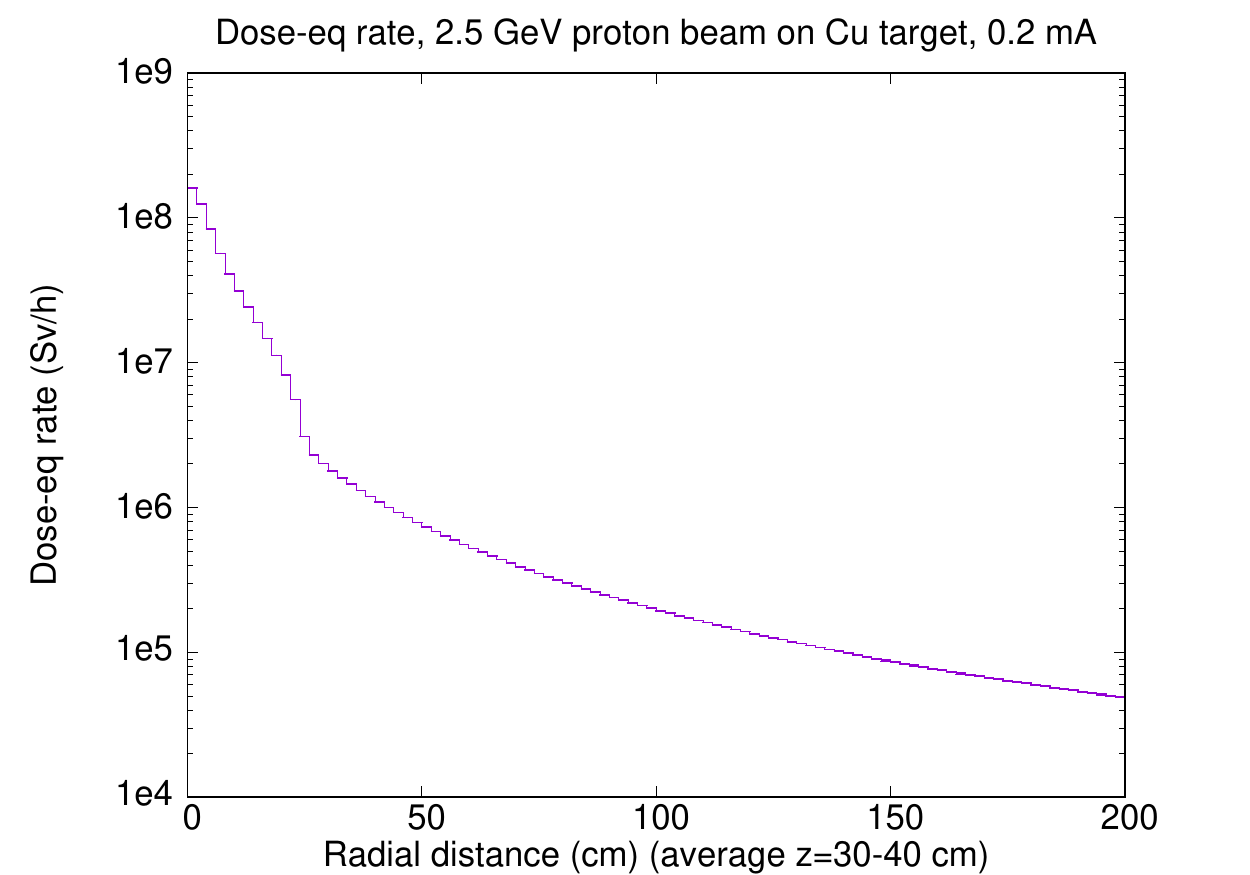}
    \caption{1D projection.}
\end{subfigure}
\caption{Distribution of dose-rate for source term calculations.}
\label{fig:injdump_doserate}
\end{figure}

\subsubsection{Summary and Outlook}
The conceptual design of the injection beam dump consists of a water-cooled copper cylinder of 25\,cm radius and 120\,cm length, surrounded by 3\,m of iron inner shielding and 1\,m of concrete outer shielding. The power loss and prompt radiation levels have been calculated for a beam loss of 0.5\,MW, 0.2\,mA average beam, 10\% beam loss. Under normal operation, a lower beam loss is expected -- on the order of 2\%. The power loss reaches 30\,W/cm$^3$ at a depth of 9\,cm, and the resulting dose rate directly atop the dump is 19\,\SI{}{\micro\sievert/\hour}, as calculated by performing source-term calculations with FLUKA along with semi-empirical calculations of the attenuation of the shielding.  

The heating resulting from the power loss can be handled reasonably well by water cooling, but the detailed thermal and mechanical response has yet to be studied in detail, in order to make sure that the temperature in the beam dump is kept below 400{$^\circ$}C to avoid softening of the copper. Such a study will include the actual pulse structure of the beam. One should also keep in mind corrosion problems for copper, owing to radiolysis of water; and possibly consider copper-alloys for mitigation. High neutron flux can also lead to material damage, such as swelling, creep, or embrittlement~\cite{Fabritsiev:1996:Cu}.

The injection beam dump has been designed to be able to handle up to 10\% losses. A sustained loss level of 10\% is not expected under long-term operation. However, as experience at SNS shows, it is important that the dump is designed with a sufficient margin to handle radiation levels and thermal stresses of up to 10\%~\cite{Evans:2020}. One of the problems experienced at SNS -- which is an additional constraint for increasing the average beam power -- is thermal stress in the concrete and consequent risk of cracking, which reduce the radiation-shielding effectiveness. Alternatives to consider for reducing the beam dump size are temperature monitoring and active cooling of the concrete. 

Careful 3D simulations of beam transport are essential, particularly for minimising losses~\cite{Wang:2007ai}, and should be carried out as part of a technical-design phase of the project. However, we do not foresee this issue to be seriously prohibitive: it is more a matter of careful magnet design with full control of the 3D fields, including fringe fields.

A strategy to manage the convoy electrons must also be made. At 5\,MW average beam power the convoy electrons carry about 5\,kW. A water-cooled electron collector was designed at SNS\cite{Plum:2016cfe}. A similar system could be envisaged for the ESSnuSB.
%------------------------------------------------------------------------------
%
%              The R2S transfer line
%
%------------------------------------------------------------------------------
\subsection{Transfer Line from the Ring to the Beam Switchyard}\label{subsec:R2S}
%------------------------------------------------------------------------------
As the compressed pulses are extracted from the accumulator they are transported through a beam line leading up to the switchyard that feeds the target station. The purpose of this beamline, called the ring-to-switchyard (R2S), %\subsubsection{Requirements}\label{subsec:R2S_requirements}
is to bring the protons, with as little losses as possible, into correct alignment with the desired neutrino beam direction.
%------------------------------------------------------
\begin{figure}[htb]
\begin{center}
\mbox{\epsfig{file=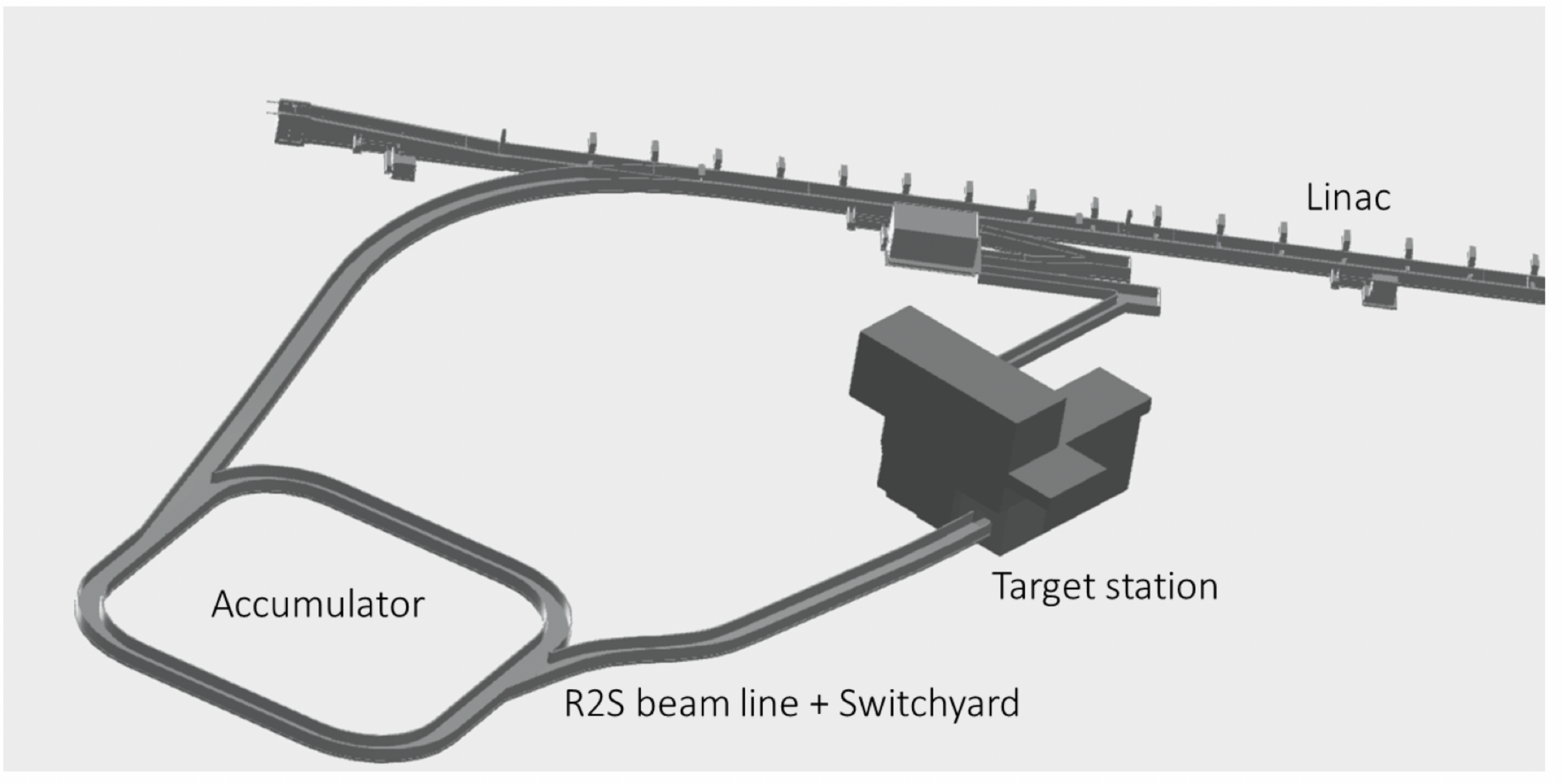,width=10cm}}
\caption{Overview drawing showing the location of the accumulator and the target station, connected by the R2S beam line \cite{Johansson:2020ESS}.}
\label{fig4.1}
\end{center}
\end{figure}
%------------------------------------------------------
In the latest ESS$\nu$SB layout~\cite{Gazis:2019ESS,Johansson:2020ESS}, a line drawn from the beam target through the decay tunnel makes an angle of 16.8$^\circ$ in the horizontal plane with the straight section of the accumulator containing the extraction. In addition, the target station should point at an angle of 2.29$^\circ$ vertically with respect to the horizon. These angles must be respected while designing the R2S beam line in order to ensure the particles produced by the target to follow their way up in the direction of the beam dump and on to the near and far detectors. From Ref.~\cite{Johansson:2020ESS}, the coordinates of the exit of the extraction septum magnet and those of the centre of the target station are found to be (101.9; -280.3; -12.5)\,m and (249.1; -127.4; -22.35)\,m respectively Fig.~\ref{fig4.1}. The distance between the target station and the accumulator is then estimated to be 212.40\,m, with the depth of the target station at 22.35\,m below the ground level (9.85\,m below the extraction point at the exit of the septum). All of these values confirm the need for a transfer line with unfolding of the horizontal and vertical planes.

The protons will exit the extraction septum magnet at an angle of 16.80$^\circ$ and 1.14$^\circ$ in the horizontal and vertical planes, respectively, seen from the accumulator reference orbit. Having a transfer line pointed in the intended neutrino-beam direction will ease the subsequent design of the switchyard. This transfer line must be as short as possible in order to minimise the cost of the equipment comprising it. Figure~\ref{fig4.2}~shows the first suggested layout of the beamline~\cite{Johansson:2020ESS} together with the direction of the neutrino beam in the $(x,y)$ plane (left) and the $(x,z)$ plane (right).
%------------------------------------------------------
\begin{figure}[htb]
\begin{center}
\mbox{\epsfig{file=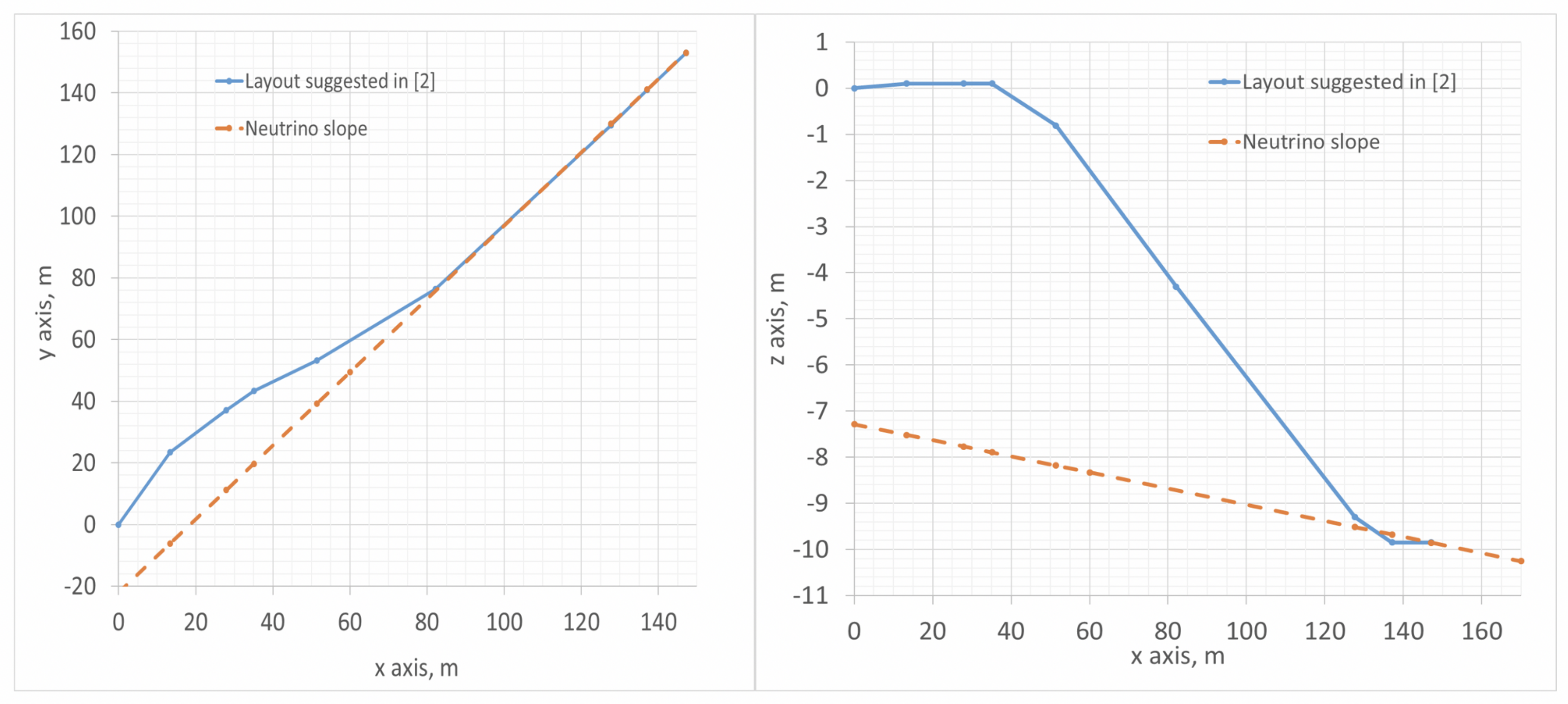,width=15cm}}
\caption{Horizontal and vertical plots of the R2S from \cite{Johansson:2020ESS} and slopes of the neutrino direction. To simplify the simulations, the extraction of the beam at the exit of the septum has the coordinates $x, y, z$ (0; 0; 0).}
\label{fig4.2}
\end{center}
\end{figure}
%------------------------------------------------------
%
%-------------------------------------------------------------------------------
\subsubsection{Design of the R2S Beamline}\label{subsec:R2S_lattice}
%-------------------------------------------------------------------------------
%\subsubsubsection{Initial Conditions}
%------------------------------------------------------------------------------------------------------
The particle distributions used in the design process of the R2S beam line are presented in Fig.~\ref{fig4.3}. These distributions come from the anti-correlated painting process and contain $240500$ macroparticles. Table~\ref{tab4.1} contains the main characteristics of the incoming beam.
%------------------------------------------------------
\begin{figure}[htb!]
\begin{center}
\mbox{\epsfig{file=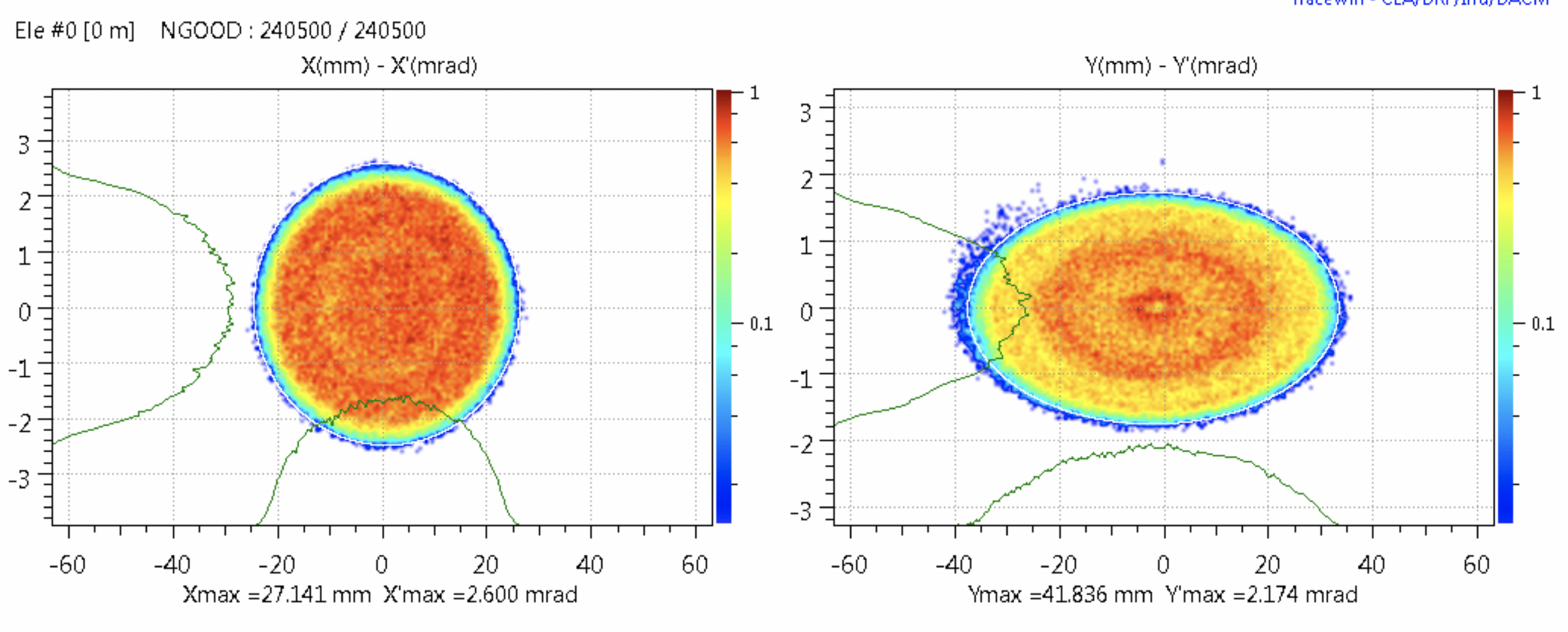,width=13cm}}
\caption{Transverse phase space of the particles used in the design of the R2S transfer line (240500 macroparticles).}
\label{fig4.3}
\end{center}
\end{figure}
%------------------------------------------------------
%------------------------------------------------------
\begin{table}[htb]%[H]
  \begin{center}
    \caption{Beam parameters of the input beam.}
    \label{tab4.1}
    \vspace{0.25 cm}
    \begin{tabular}{lccl}
\textbf{Parameter} & \textbf{symbol} & \textbf{value} & \textbf{unit} \\
\hline
Kinetic energy & $E$ & 2.5 & GeV \\
Relativistic momentum & $p$ & 3.3 & GeV/c \\
Emittances (rms, normalised) & $\epsilon_x/\epsilon_y$ & 27/26 & $\pi$\,mm\,mrad\\
Emittances (99.95\%, normalised) & $\epsilon_x/\epsilon_y$ & 244/234 & $\pi$\,mm\,mrad\\ 
Beam size (rms) & $\sigma_x/\sigma_y$ & 11.3/15.8 & mm\\ 
Beam divergence (rms) & $\sigma_x'\sigma_y'$ & 1.1/0.77 & mrad  \\ 
Twiss parameters & $\beta_x/\beta_y$ & 10/20 & m \\
Twiss parameters & $\alpha_x/\alpha_y$ & -0.0005/0.003 &  - \\
    \hline
    \end{tabular}
  \end{center}
\end{table}
%------------------------------------------------------

%\subsubsection{Defining the Lattice}

The transfer line must contain both vertical and horizontal bends. To minimise the necessary magnetic field per magnet, several dipoles were used to bend the protons in steps. Simulations show that three vertical and eight horizontal bends, 2\,m long each, are sufficient for constructing the transfer line. Table~\ref{tab4.2} presents the main characteristics of the dipole magnets.
%------------------------------------------------------
\begin{table}[htb]%[H]
  \begin{center}
    \caption{Main parameters of the dipoles composing the R2S.}
    \label{tab4.2}
    \vspace{0.25 cm}
    \begin{tabular}{lccc}
\textbf{Bending plane} & \textbf{Magnet length (m)} & \textbf{Max. Bending angles ($^\circ$)} & \textbf{Max. induced magnetic fields (mT)} \\
\hline
Horizontal & 2 & 7.2 & 690 \\
Vertical & 2 & 10 & 960 \\
    \hline
    \end{tabular}
  \end{center}
\end{table}
%------------------------------------------------------

 Quadrupole doublets are then placed in between the dipole magnets for maintaining a small beam envelope and ensure a good transmission. The lattice layout with dipoles and quadrupoles is shown in Fig.~\ref{fig4.4} whereas Figure~\ref{fig4.5} shows the geometric layout of the R2S compared to the original suggestion and the neutrino direction. 
 %
%------------------------------------------------------
\begin{figure}[htb]
\begin{center}
\mbox{\epsfig{file=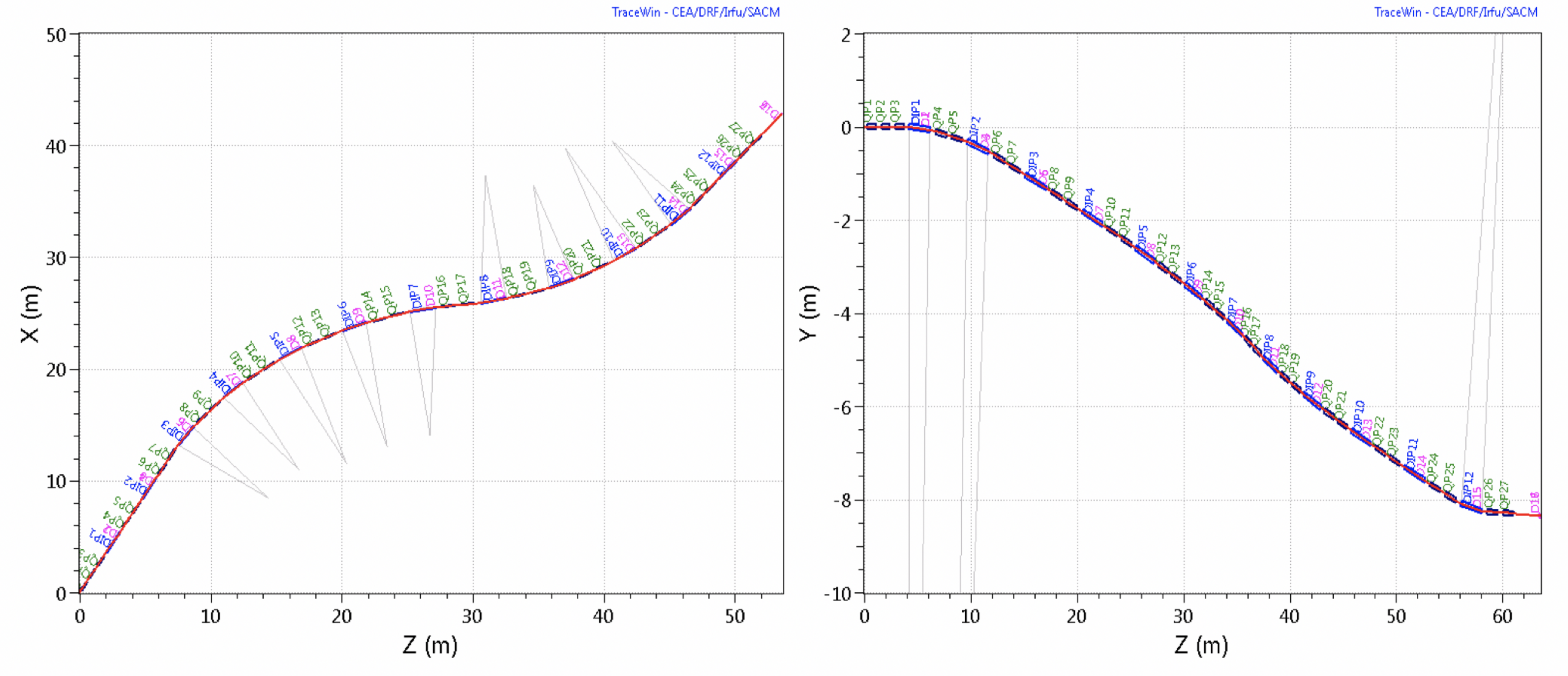,width=15cm}}
\caption{ Horizontal (left) and vertical (right) synoptics of the R2S beam line.}
\label{fig4.4}
\end{center}
\end{figure}
%------------------------------------------------------
%------------------------------------------------------
\begin{figure}[htb]
\begin{center}
\mbox{\epsfig{file=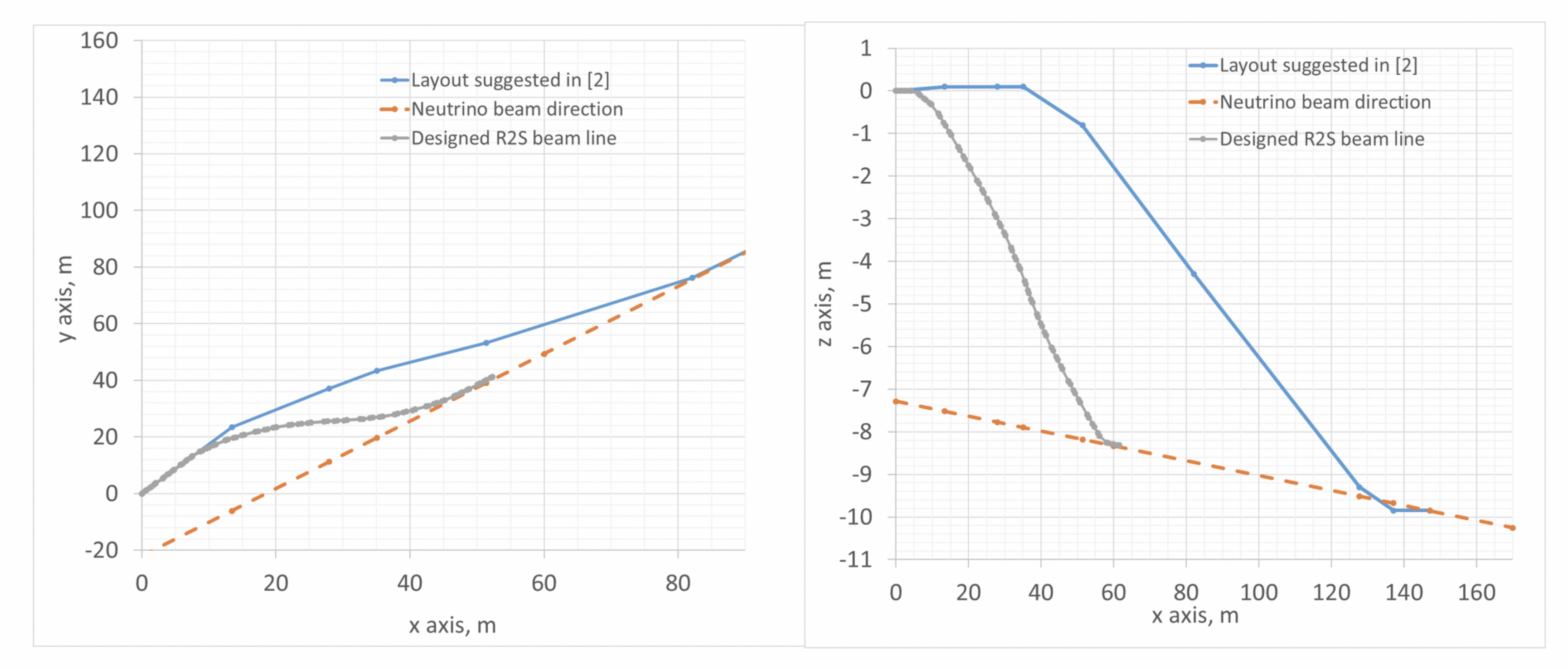,width=15cm}}
\caption{ Coincidence of the end of the R2S beam line, designed with TraceWin, and the neutrino beam direction in the horizontal (left) and vertical (right) planes.}
\label{fig4.5}
\end{center}
\end{figure}
%------------------------------------------------------

 By adjusting the values of the different quadrupoles composing the beam line, a minimum beam size is achieved and maintained all along the transfer line as displayed in Fig.~\ref{fig4.6}. In the final design the R2S measures 72\,m in length and meets all the initial requirements. 
%------------------------------------------------------
\begin{figure}[htb]
\begin{center}
\mbox{\epsfig{file=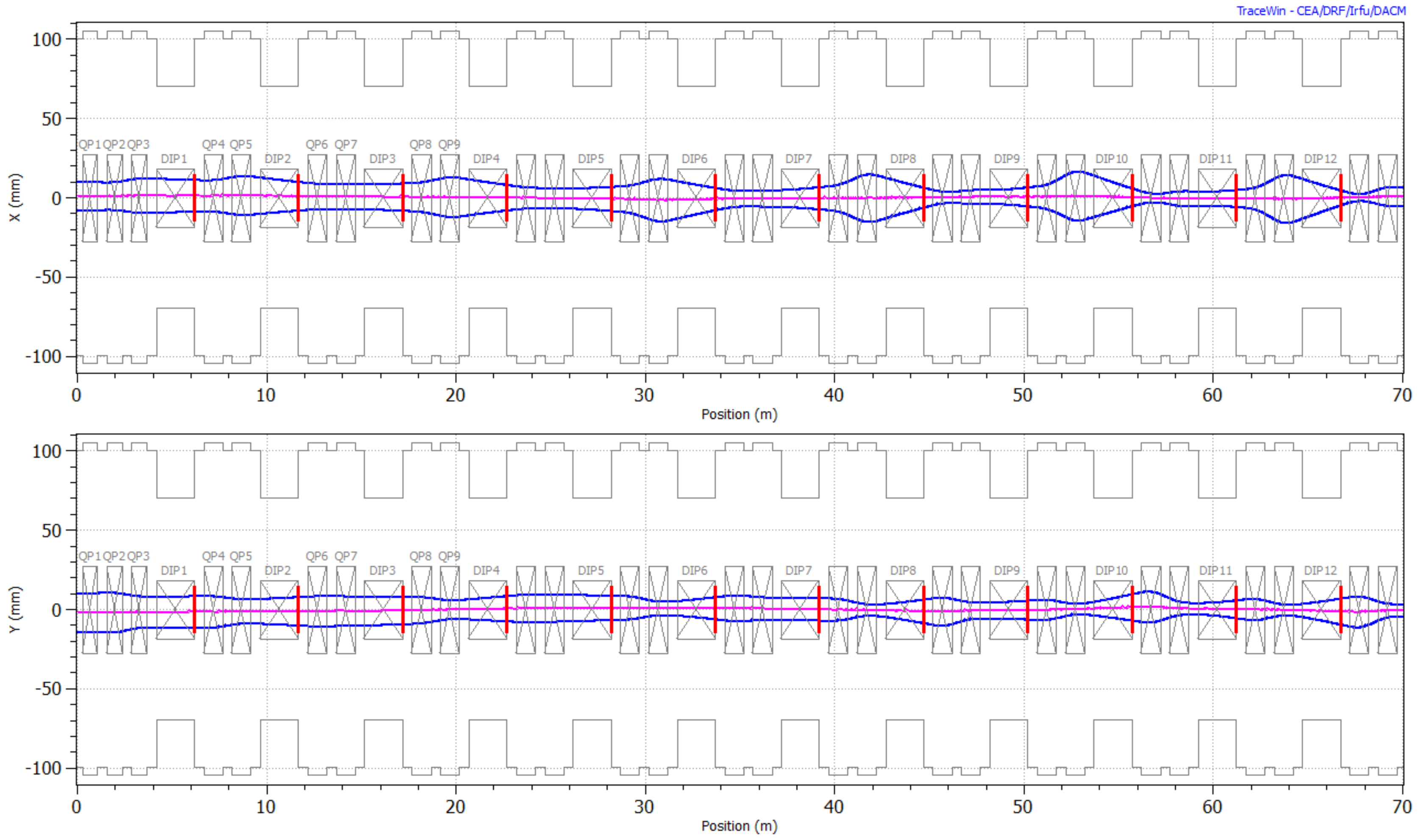,width=15cm}}
\caption{RMS transverse beam envelopes along the R2S beam line.}
\label{fig4.6}
\end{center}
\end{figure}
%------------------------------------------------------

%------------------------------------------------------------------------------
\subsection{Beam Switchyard}\label{subsec:BSY}
%------------------------------------------------------------------------------
Once the proton beam is aligned with the neutrino-beam direction by the R2S beamline, it will be distributed to the four targets by a beam switchyard (BSY). The BSY will not only transport and distribute the protons onto the targets but also focus them to the desired shape and size centred on the target, ideally a circular beam with a radius not larger than 1.5\,cm.

Figure~\ref{fig5.1} shows the layout of the target station. The distance between the centres of the targets is 3\,m. The distance was later changed to 2.5\,m, but since switchyard designs have been made for the distances 2\,m and 3\,m there is no doubt that a good design can be made for values within this range, if the baseline changes.
%------------------------------------------------------
\begin{figure}[htb]
\begin{center}
\mbox{\epsfig{file=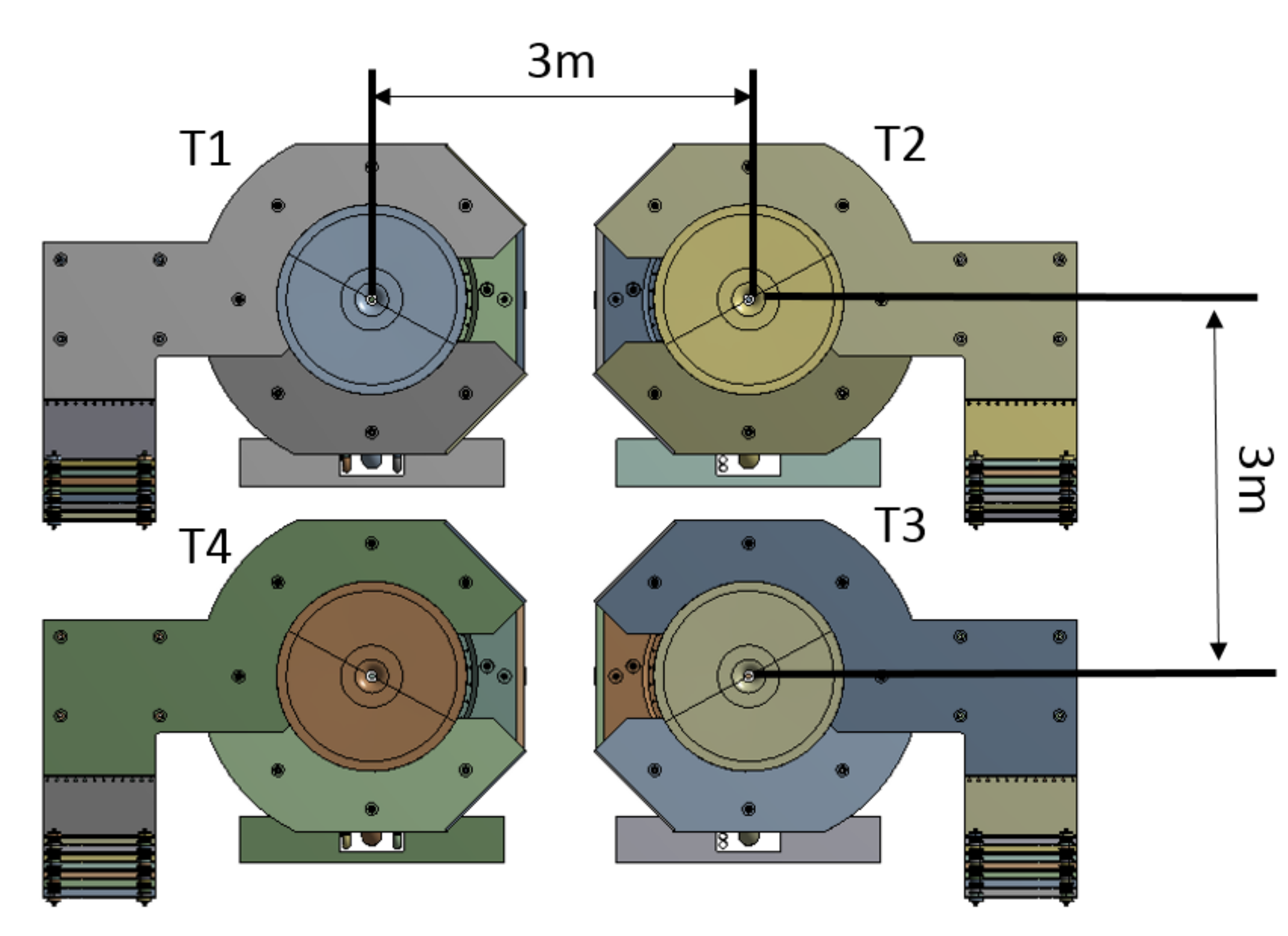,width=10cm}}
\caption{ Face view of the target station layout. The four targets are named T1, T2, T3 and T4.}
\label{fig5.1}
\end{center}
\end{figure}
%------------------------------------------------------

In the baseline pulsing scheme, the linac is pulsed at 14\,Hz. As a result the accumulator will deliver a set of compressed pulses or batches to the BSY at 14\,Hz, but where the batch-to-batch frquency is 1.1\,kHz. Each compressed pulse or batch has a 1.2\,\SI{}{\micro\second} duration and the spacing, which leaves less than 0.9\,ms for the beam switching from target to target.
%------------------------------------------------------------------------------------------------------
\subsubsection{Design of the BSY}
%------------------------------------------------------------------------------------------------------
A previous design suggested diagonally splitting the proton beam onto the target station~\cite{Bouquerel:2013:Switchyard}. However, due to a change in the baseline parameters of the beam (i.e. time structure) coming out from the accumulator, it was no longer feasible to use this configuration. In addition, this design carries a weakness in terms of safety, where a beam dump would be needed to protect the system in cases of dipole magnet failure. Therefore, a new layout was investigated. This new layout suggests driving the protons from one section to the next using conventional dipoles. 
%------------------------------------------------------
\begin{figure}[htb]
\begin{center}
\mbox{\epsfig{file=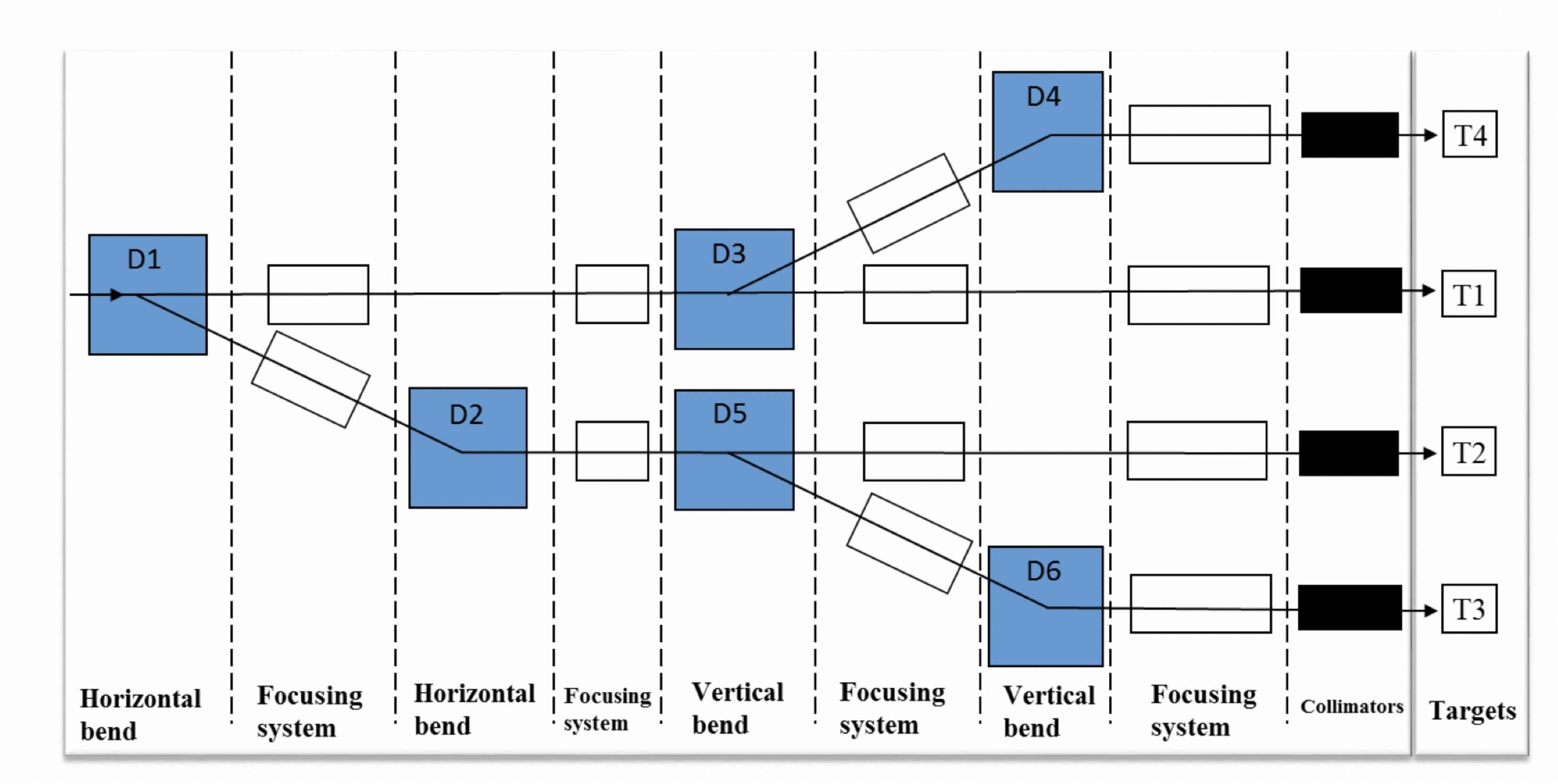,width=13cm}}
\caption{ Schematic layout of the beam switchyard. D1 and D2 are dipoles that bend the beam in the horizontal plane. D3, D4, D5, and D6 bend the beam in the vertical plane.}
\label{fig5.2}
\end{center}
\end{figure}
%------------------------------------------------------

Figure~\ref{fig5.3} shows the working principle of the switching scheme. The switchyard contains six dipoles to bend the beam in both the horizontal and the vertical planes. Sets of quadrupoles maintain minimal transverse envelops of the beam as it propagates through the beamlines. Each branch of the BSY will be equipped with a collimator at its exit.
%------------------------------------------------------
\begin{figure}[htb]
\begin{center}
\mbox{\epsfig{file=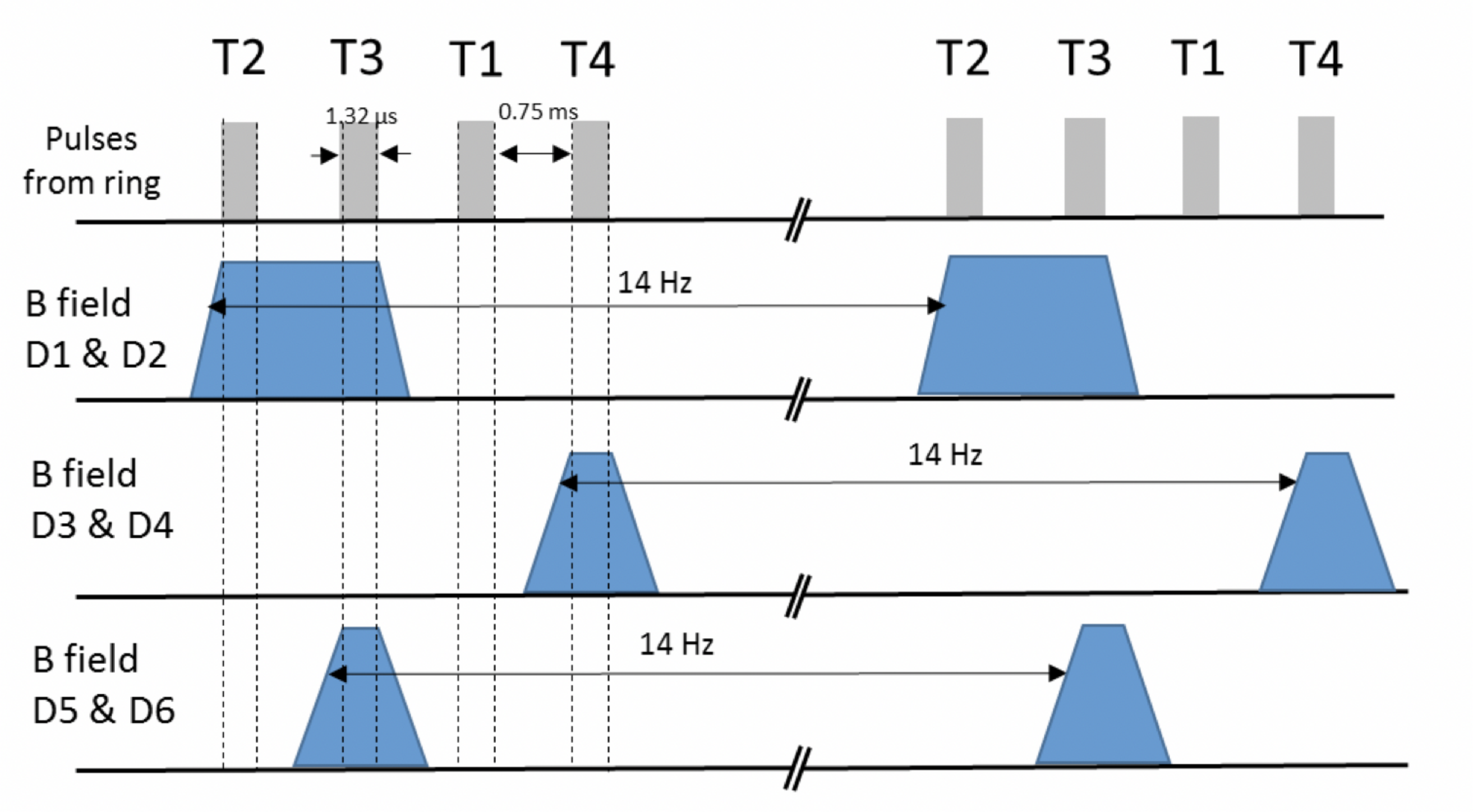,width=11cm}}
\caption{ Time structure of the incoming beam in relation to the dipoles of the BSY.}
\label{fig5.3}
\end{center}
\end{figure}
%------------------------------------------------------

%------------------------------------------------------------------------------------------------------
\subsubsection{From D1 to T1}
%------------------------------------------------------------------------------------------------------
The first branch of the BSY transports and focuses the beam onto the target named ``T1''. The T1 axis is 
the main axis of the beam upon exiting the R2S line. To reach T1, the dipoles D1 and D3 act as a drift tube. In other words, no induced magnetic field is needed from their magnets. 
Figure~\ref{fig5.4} shows the beam envelopes for the T1 branch. According to these simulations, this branch has a transmission of 100\% and is achromatic at its extremity. The radii of the beam in the middle of the target are expected to be 14.90\,mm and 14.94\,mm in $x$ and $y$, respectively. Table~\ref{tab5.1} presents the main beam parameters at the target location. 
%------------------------------------------------------
\begin{table}[htbp]
  \begin{center}
    \caption{Beam parameters at the target T1.}
    \label{tab5.1}
    \vspace{0.25 cm}
    \begin{tabular}{lccl}
\textbf{Parameter} & \textbf{symbol} & \textbf{value} & \textbf{unit} \\
\hline
Max beam size & $X/Y$ & 14.90/14.94 & mm\\ 
Max beam divergence & $X'/Y'$ & 9.60/5.94 & mrad\\ 
Twiss parameters &$\beta_x/\beta_y$ & 2.95/3.47 & m \\
Twiss parameters & $\alpha_x/\alpha_y$ & -0.21/-0.42 & - \\
    \hline
    \end{tabular}
  \end{center}
\end{table}
%------------------------------------------------------
%
%------------------------------------------------------
\begin{figure}[htb]
\begin{center}
\mbox{\epsfig{file=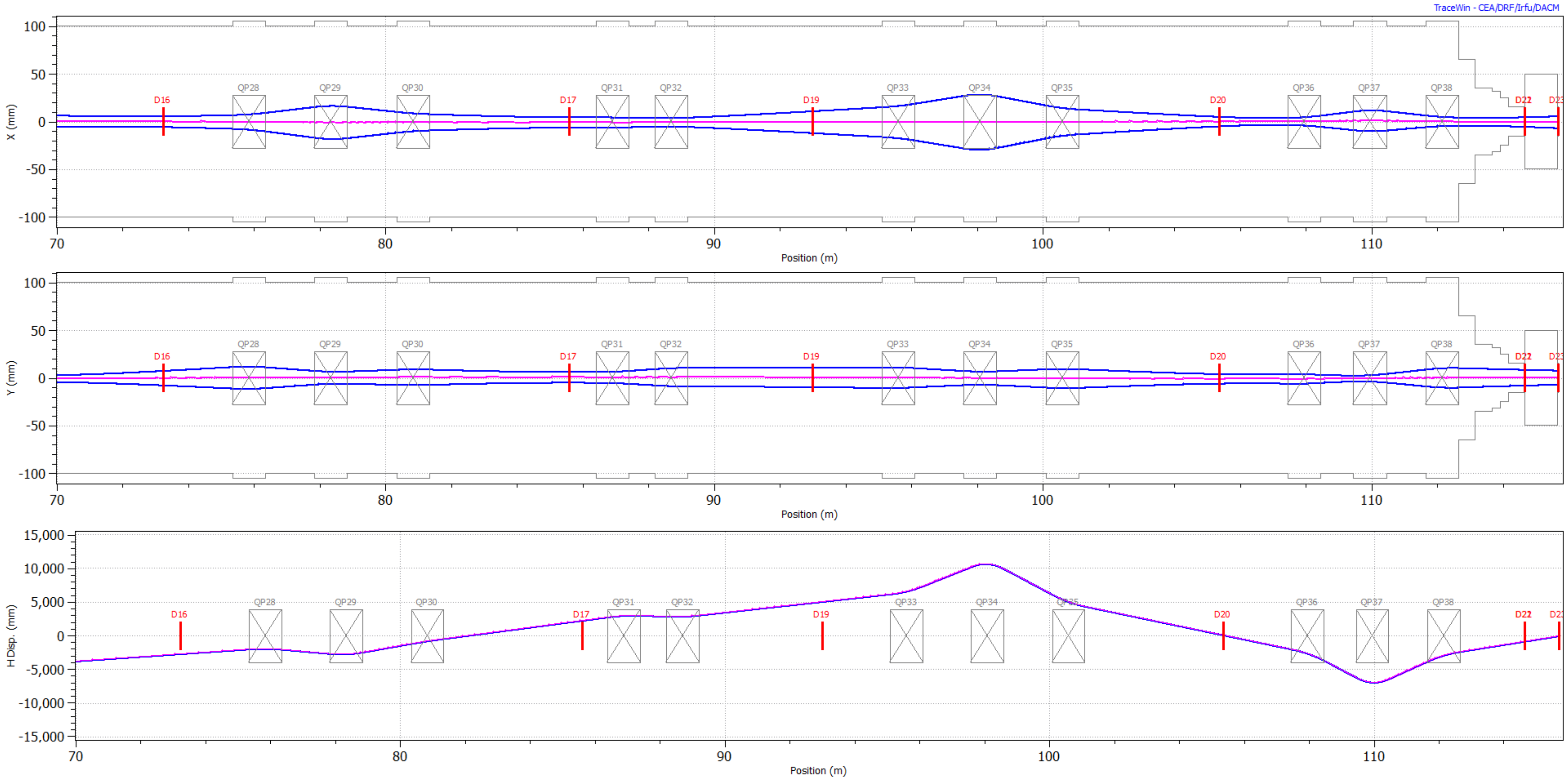,width=12cm}}
\caption{ RMS transverse beam envelopes and dispersion along the T1 branch of the BSY.}
\label{fig5.4}
\end{center}
\end{figure}
%------------------------------------------------------
%
%------------------------------------------------------------------------------------------------------
\subsubsection{From D1 to T2}
%------------------------------------------------------------------------------------------------------
The second branch of the BSY transports and focuses the beam onto the target T2. The dipoles D1 and D2 must induce a magnetic field to bend the protons horizontally. The designated angle of deflection must 
require both a moderate magnetic field in the dipole magnets (less than 1\,T, capable of switching on/off with respect to the time structure of the incoming beam) and a sufficient space downstream to allow for positioning equipment. To this end, an angle of 244\,mrad was selected. To deflect the beam with such an angle, a magnetic field of 890\,mT for a 3\,m long dipole would be necessary. Such dipoles have an inductance of around 15\,mH, an intensity consumption of 58\,kAt (kilo-Amp-turns), and a 40\,mm-radius gap. Figure~\ref{fig5.5} shows the transverse beam envelops in the $x-y$ plane through the T2 branch. The maximum radii for the beam at the centre of the target are expected to be 14.66\,mm and 14.98\,mm in $x$ and $y$, respectively. Table~\ref{tab5.2} presents the main beam parameters at the target location.
%------------------------------------------------------
\begin{figure}[htb]
\begin{center}
\mbox{\epsfig{file=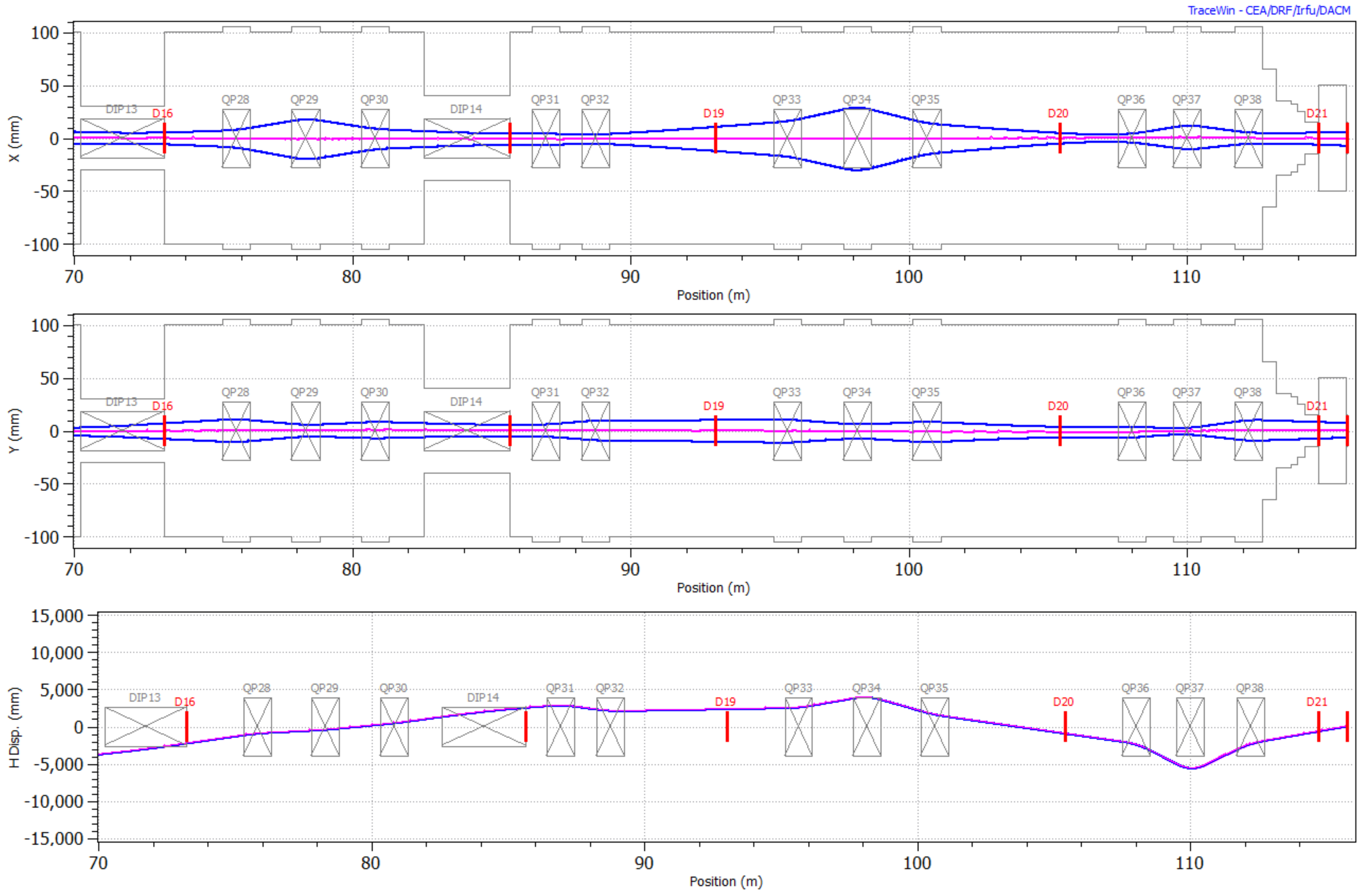,width=12cm}}
\caption{ RMS transverse beam envelopes and dispersion along the T2 branch of the BSY.}
\label{fig5.5}
\end{center}
\end{figure}
%------------------------------------------------------
%------------------------------------------------------
\begin{table}[htb]
  \begin{center}
    \caption{Beam parameters at the middle of the target T2.}
    \label{tab5.2}
    \vspace{0.25 cm}
    \begin{tabular}{lccl}
\textbf{Parameter} & \textbf{symbol} & \textbf{value} & \textbf{unit} \\
\hline
Max beam size & $X/Y$ & 14.66/14.98 & mm\\ 
Max beam divergence & $X'/Y'$ & 8.07/5.89 & mrad\\ 
Twiss parameters &$\beta_x/\beta_y$ & 3.25/3.25 & m \\
Twiss parameters & $\alpha_x/\alpha_y$ & -0.27/-0.37 & - \\
    \hline
    \end{tabular}
  \end{center}
\end{table}
%------------------------------------------------------

%------------------------------------------------------------------------------------------------------
\subsubsection{From D1 to T3}
%------------------------------------------------------------------------------------------------------
The third branch of the BSY transports and focuses the beam onto the target T3. This beam line is the most complicated of the system, since four dipoles should be active simultaneously: D1 and D2, as well as D5 and D6 to respectively bend the beam horizontally and vertically. Dipoles D5 and D6 are similar to D5 and D6. Figure~\ref{fig5.6} shows the transverse beam envelops in the $x--y$ plane through the T3 branch. The maximum radii of the beam at the middle of the target are expected to be 12.58\,mm and 14.99\,mm in $x$ and $y$, respectively. Table~\ref{tab5.3} presents the main beam parameters at the target location.%------------------------------------------------------
\begin{table}[htb]
  \begin{center}
    \caption{Beam parameters at the middle of the target T3.}
    \label{tab5.3}
    \vspace{0.25 cm}
    \begin{tabular}{lccl}
\textbf{Parameter} & \textbf{symbol} & \textbf{value} & \textbf{unit} \\
\hline
Max beam size & $X/Y$ & 12.58/14.99 & mm\\ 
Max beam divergence & $X'/Y'$ & 9.97/5.81 & mrad\\ 
Twiss parameters &$\beta_x/\beta_y$ & 3.14/3.24 & m \\
Twiss parameters & $\alpha_x/\alpha_y$ & -0.30/-0.37 & - \\
    \hline
    \end{tabular}
  \end{center}
\end{table}
%------------------------------------------------------
%
%------------------------------------------------------
\begin{figure}[htb]
\begin{center}
\mbox{\epsfig{file=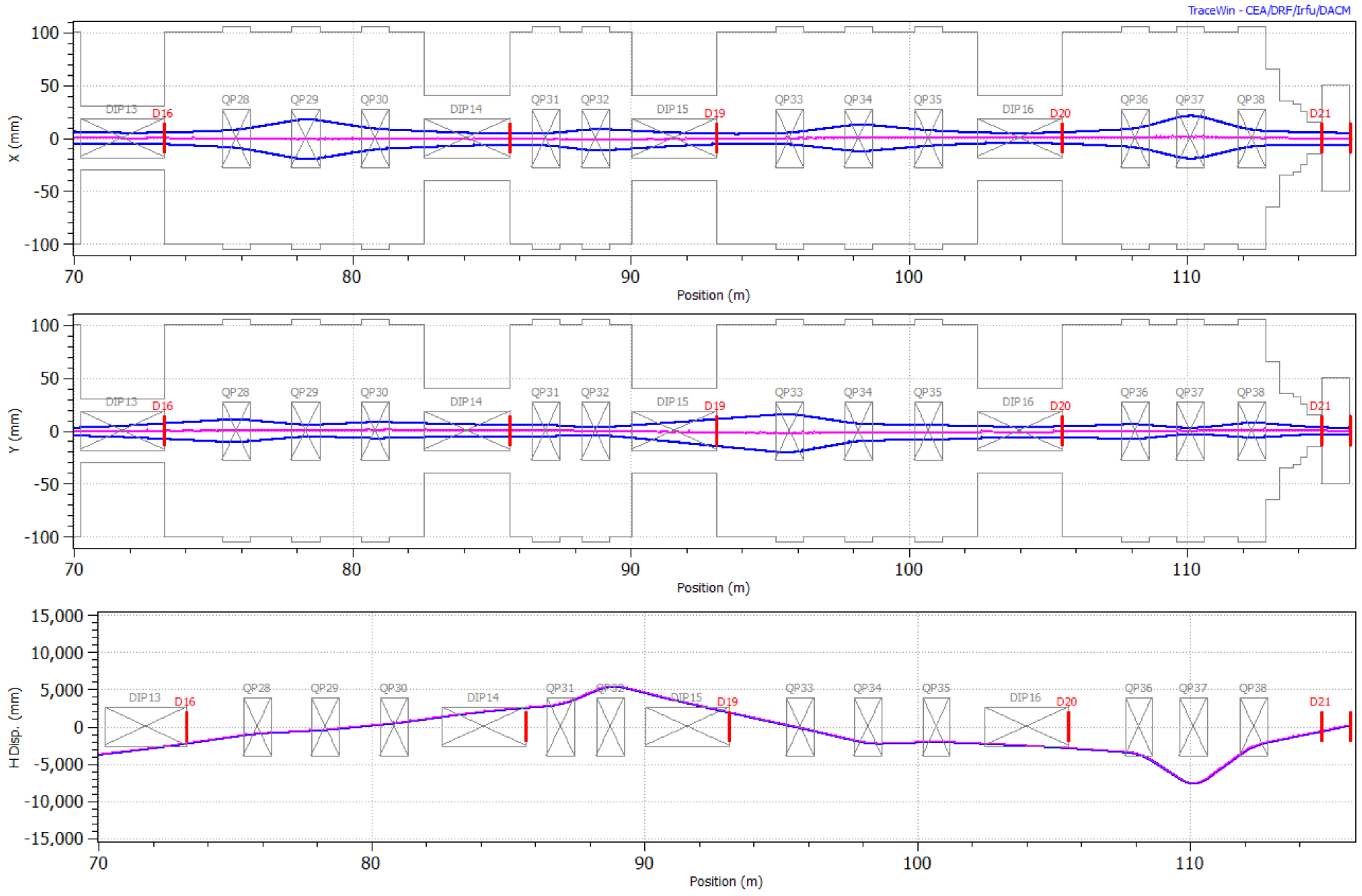,width=12cm}}
\caption{ RMS transverse beam envelopes and dispersion along the T3 branch of the BSY.}
\label{fig5.6}
\end{center}
\end{figure}
%------------------------------------------------------
%
%------------------------------------------------------------------------------------------------------
\subsubsection{From D1 to T4}
%------------------------------------------------------------------------------------------------------
The fourth branch of the BSY transports and focuses the beam onto target T4. Two dipoles are necessary to bend the beam vertically. Figure~\ref{fig5.7} shows the transverse beam envelops in the $x--y$ plane through the T4 branch. The maximum radii of the beam at the middle of the target are expected to be 14.38\,mm and 14.99\,mm in $x$ and $y$, respectively. Table~\ref{tab5.4} presents the main beam parameters at the target location. 
%------------------------------------------------------
\begin{table}[htb]
  \begin{center}
    \caption{Beam parameters at the middle of target T4.}
    \label{tab5.4}
    \vspace{0.25 cm}
    \begin{tabular}{lccl}
\textbf{Parameter} & \textbf{symbol} & \textbf{value} & \textbf{unit} \\
\hline
Max beam size & $X/Y$ & 14.38/14.99 & mm\\ 
Max beam divergence & $X'/Y'$ & 10.48/5.88 & mrad\\ 
Twiss parameters &$\beta_x/\beta_y$ & 2.81/3.36 & m \\
Twiss parameters & $\alpha_x/\alpha_y$ & -0.30/-0.40  & -\\
    \hline
    \end{tabular}
  \end{center}
\end{table}
%------------------------------------------------------
%------------------------------------------------------
\begin{figure}[htb]
\begin{center}
\mbox{\epsfig{file=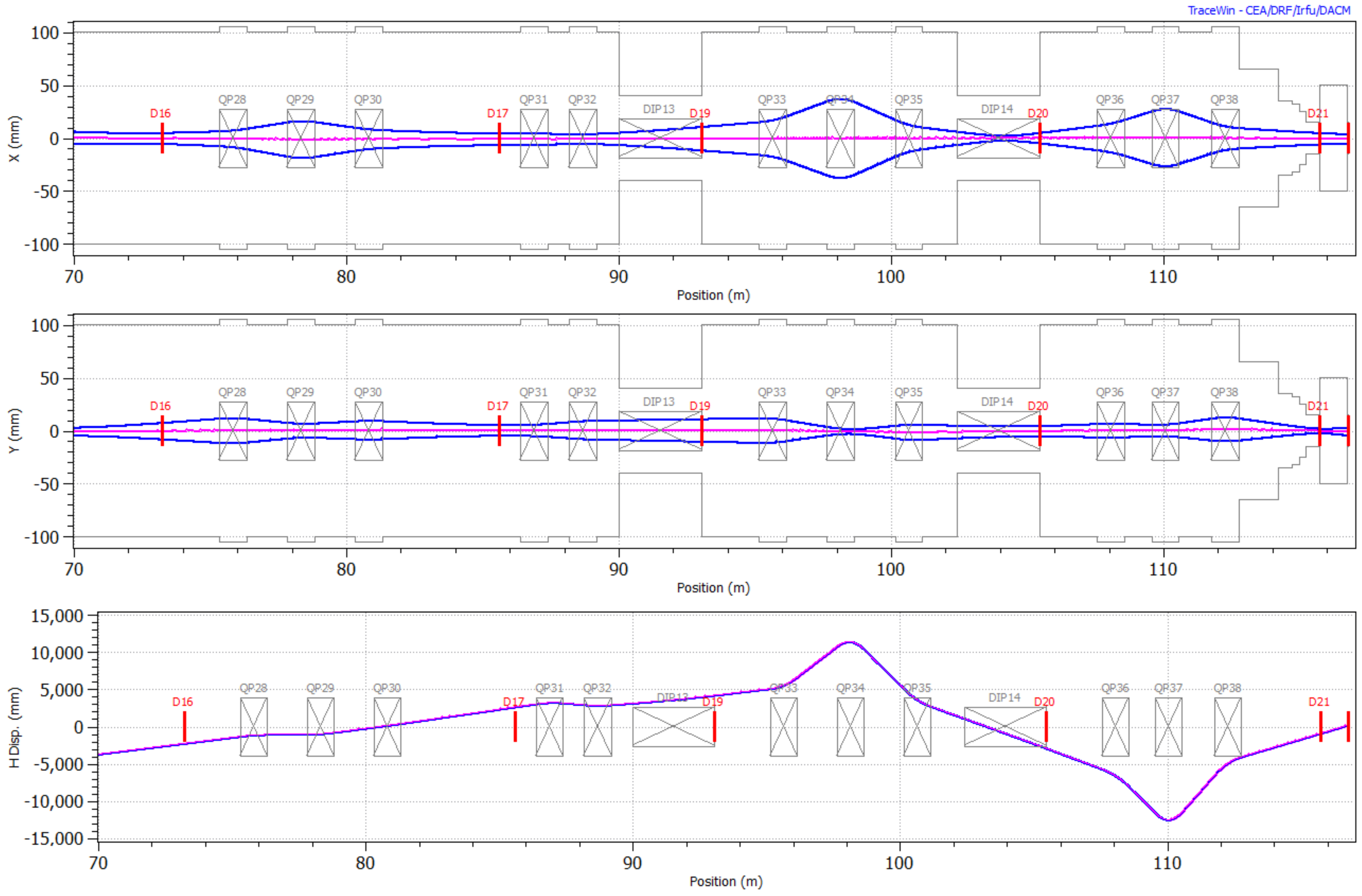,width=12cm}}
\caption{{\small RMS transverse beam envelopes and dispersion along the T4 branch of the BSY.}}
\label{fig5.7}
\end{center}
\end{figure}
%------------------------------------------------------

%------------------------------------------------------------------------------------------------------
\subsubsection{Overall Layout of the BSY}
%------------------------------------------------------------------------------------------------------
Figures~\ref{fig5.8} and \ref{fig5.9} show the overall synoptics and 3D views of the BSY, respectively. The system has a total length of 45\,m. 
%------------------------------------------------------
\begin{figure}[htb]
\begin{center}
\mbox{\epsfig{file=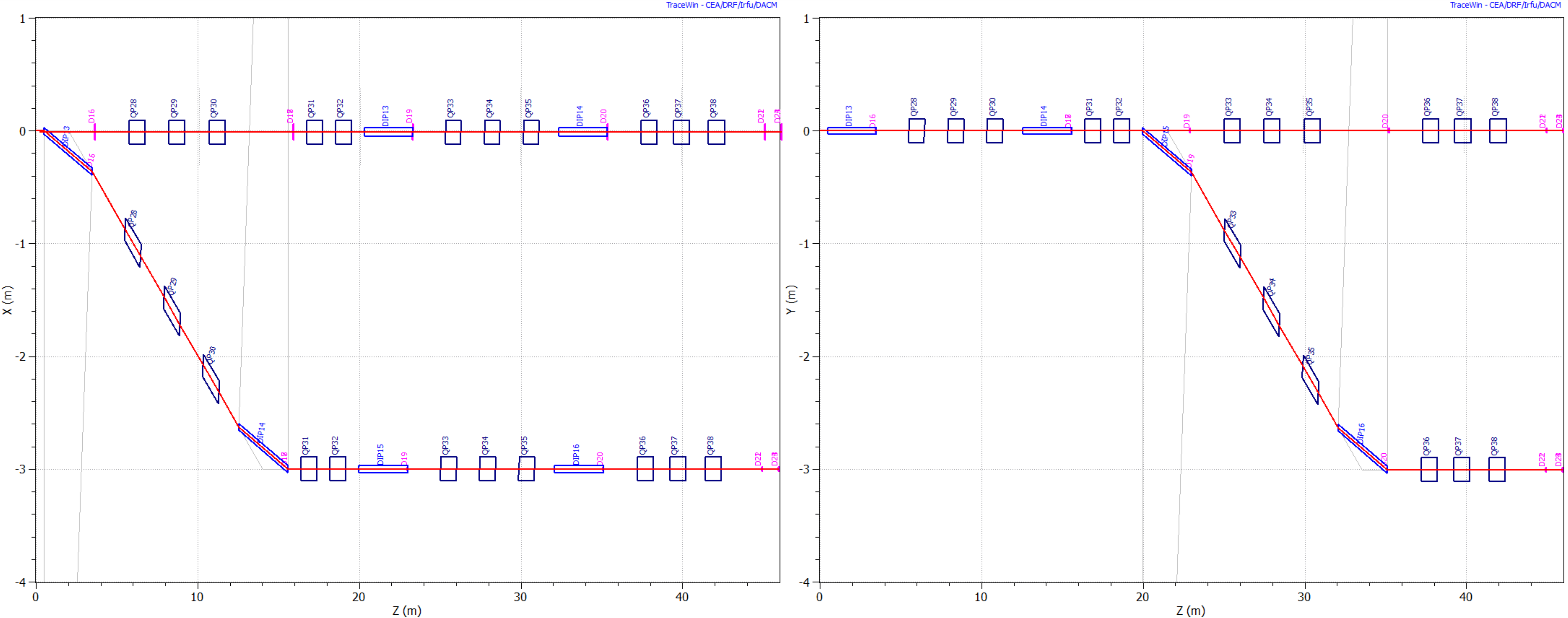,width=13cm}}
\caption{ Horizontal (left) and vertical (right) synoptics of the BSY.}
\label{fig5.8}
\end{center}
\end{figure}
%------------------------------------------------------
%------------------------------------------------------
\begin{figure}[htb]
\begin{center}
\mbox{\epsfig{file=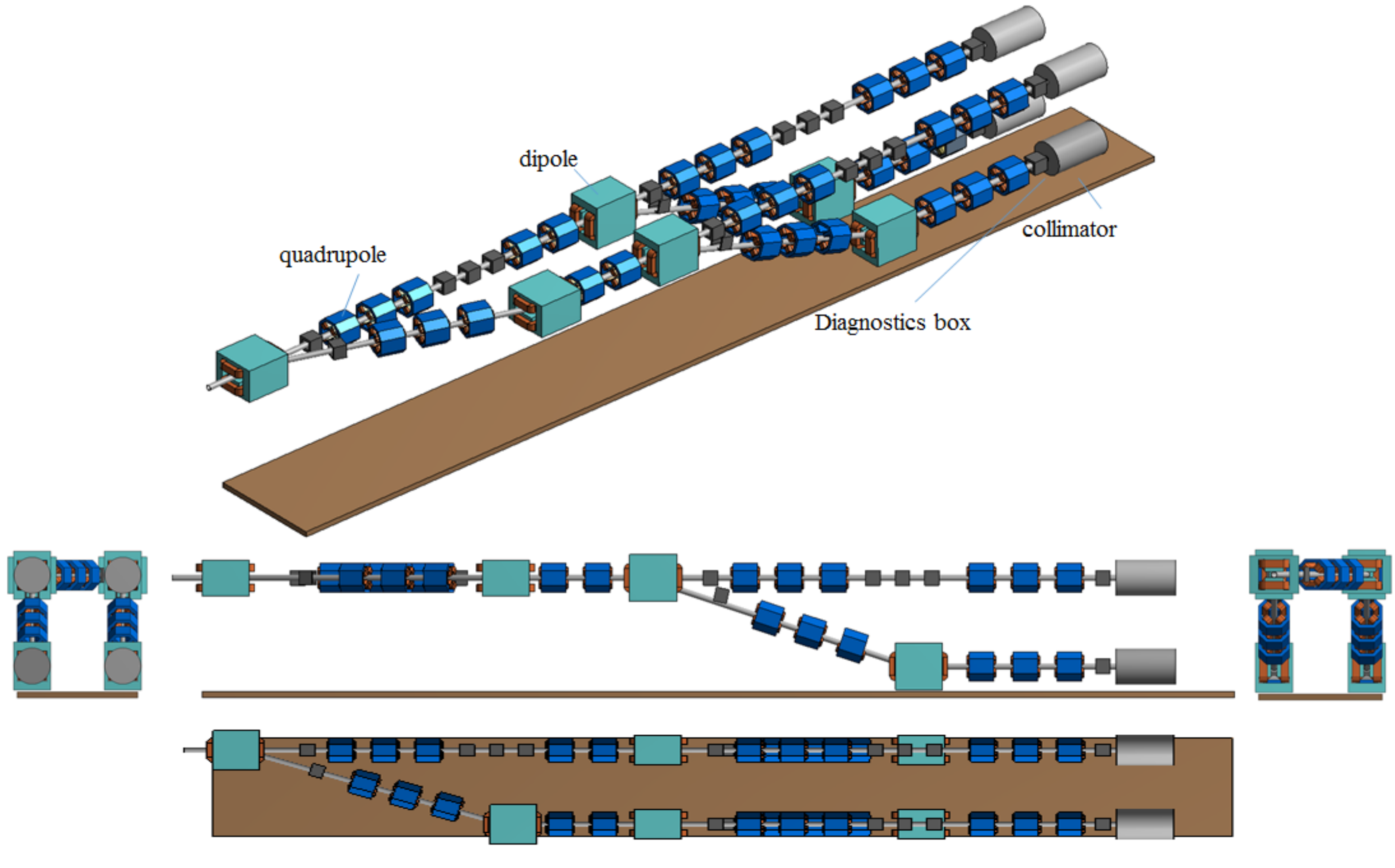,width=15cm}}
\caption{ Overall 3D views of the BSY: top, side (middle), and isometric (bottom) projections. Possible locations for adding diagnostics capable of measuring several characteristics of the beam (i.e., size, position, energy, etc.) are identified with boxes.}
\label{fig5.9}
\end{center}
\end{figure}
%------------------------------------------------------

Although simulations show 100\% transmission of the beam along the four branches of the BSY, collimators will be placed at the terminus of each beam line. These protect the horns in case of unwanted fluctuations of the beam before it reaches the targets. It must be capable of handling beam halo as well as losses due to any malfunction in the operation of the upstream system. The baffles are mainly made of graphite. Its dimensions would be 2\,m in length with an outer radius of 70\,cm. A conical aperture with a downstream radius of 1.5\,cm, as depicted in Fig.~\ref{fig5.12}, is an option to be considered. The distance between the end of the collimators and the targets is 1025\,mm. Following the amount of heat load that is expected to be deposited in the collimator, the use of He (or even air) cooling in closed circuit may be feasible here. 
%------------------------------------------------------
\begin{figure}[htb]
\begin{center}
\mbox{\epsfig{file=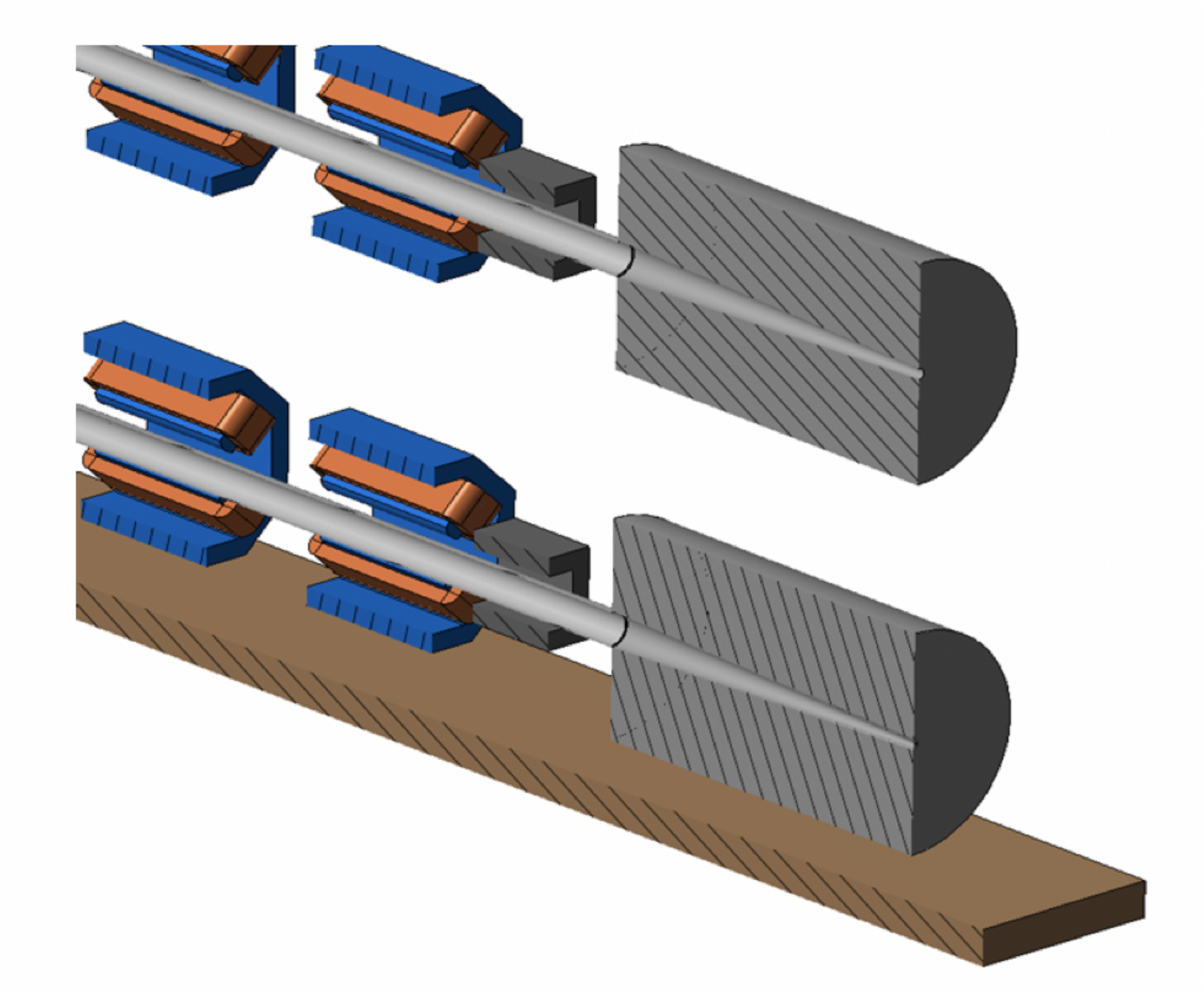,width=8cm}}
\caption{ Section plane of the collimators.}
\label{fig5.12}
\end{center}
\end{figure}
%------------------------------------------------------

%
%-------------------------------------------------------------------------------
%\subsubsection{Discussion}
%-------------------------------------------------------------------------------
%
%-------------------------------------------------------------------------------
\subsubsection{Alternatives to Dipoles}
%-------------------------------------------------------------------------------
The feasibility of having adequate rise and fall times for dipoles must be tested with respect to the operating frequency. Alternatively, dipoles D1, D3 and D5 could also be a combination of a kicker and a septum. Indeed, according to simulations, a pair of fast kickers of 0.5\,m long each could be used to kick the beam for a total angle of 1$^\circ$, then the septum would further deflect the beam. Each kicker would require a peak current of 383\,A to induce the necessary magnetic field of 193\,mT. Such kickers can have rise and fall times of less than 10\,\SI{}{\micro\second} and operate at frequencies up to 10\,kHz~\cite{Stover:2004EPAC}. Figure~\ref{fig5.10} shows the time structure of the BSY when kickers are used. Detailed investigations on such technology shall be performed during the technical design phase. 
%------------------------------------------------------
\begin{figure}[htb]
\begin{center}
\mbox{\epsfig{file=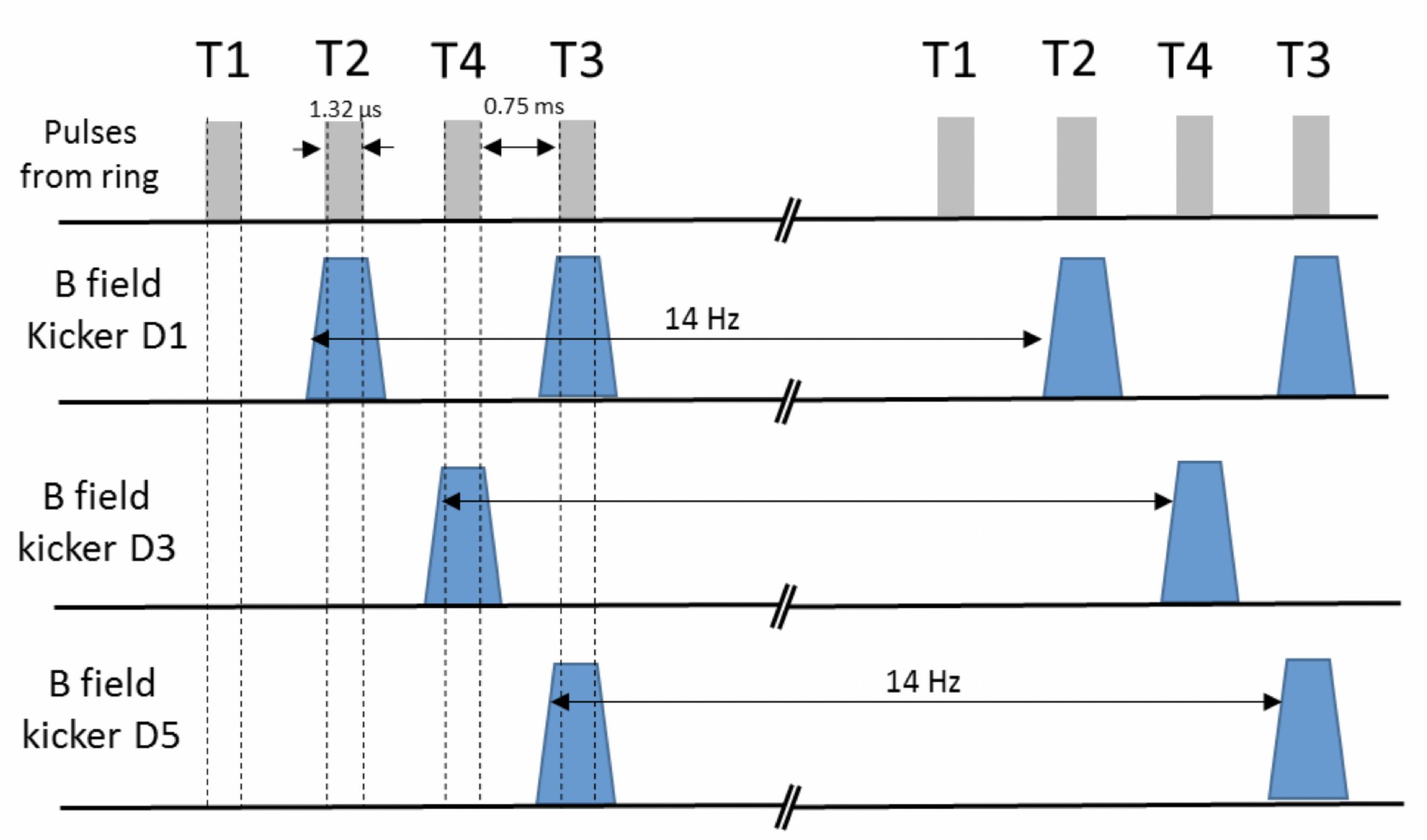,width=11cm}}
\caption{Time structure of kickers for the BSY.}
\label{fig5.10}
\end{center}
\end{figure}
%------------------------------------------------------
%
%-------------------------------------------------------------------------------
\subsubsection{Failure Scenarios}
%-------------------------------------------------------------------------------

Several failure scenarios are possible when operating the BSY. In this report, two of these are discussed. In case of failure of one of the dipoles, the protons would travel straight to the branch leading to the target T1. In addition, the BSY contains enough quadrupoles so that in case of failure of one of them, the beam envelop would be corrected automatically to match as closely as possible the required beam parameters at the target location. For example, if two quadrupoles of the T3 branch fail at the same time, preliminary calculations show that focusing the beam onto the target is still possible, and achromaticity can be achieved (Fig.~\ref{fig5.11}). In this case the beam would measure 12.48\,mm (horizontal) and 14.95\,mm (vertical) within the target (Table~\ref{tab5.5}). Nevertheless, according to these preliminary results, this specific scenario would lead to 0.2\,\% losses in the collimators.
%------------------------------------------------------
\begin{table}[htbp]
 \begin{center}
   \caption{Beam parameters at the middle of the target T2 when two quadrupoles of the branch are not functioning (failure scenario).}
    \label{tab5.5}
    \vspace{0.25 cm}
    \begin{tabular}{lccl}
\textbf{Parameter} & \textbf{symbol} & \textbf{value} & \textbf{unit} \\
\hline
Max beam size & $X/Y$ & 12.48/14.95 & mm\\ 
Max beam divergence & $X'/Y'$ & 8.31/12.81 & mrad\\ 
Twiss parameters &$\beta_x/\beta_y$ & 1.82/3.17 & m \\
Twiss parameters & $\alpha_x/\alpha_y$ & -0.08/-2.57 & - \\
    \hline
    \end{tabular}
  \end{center}
\end{table}
%------------------------------------------------------
%------------------------------------------------------
\begin{figure}[htb]
\begin{center}
\mbox{\epsfig{file=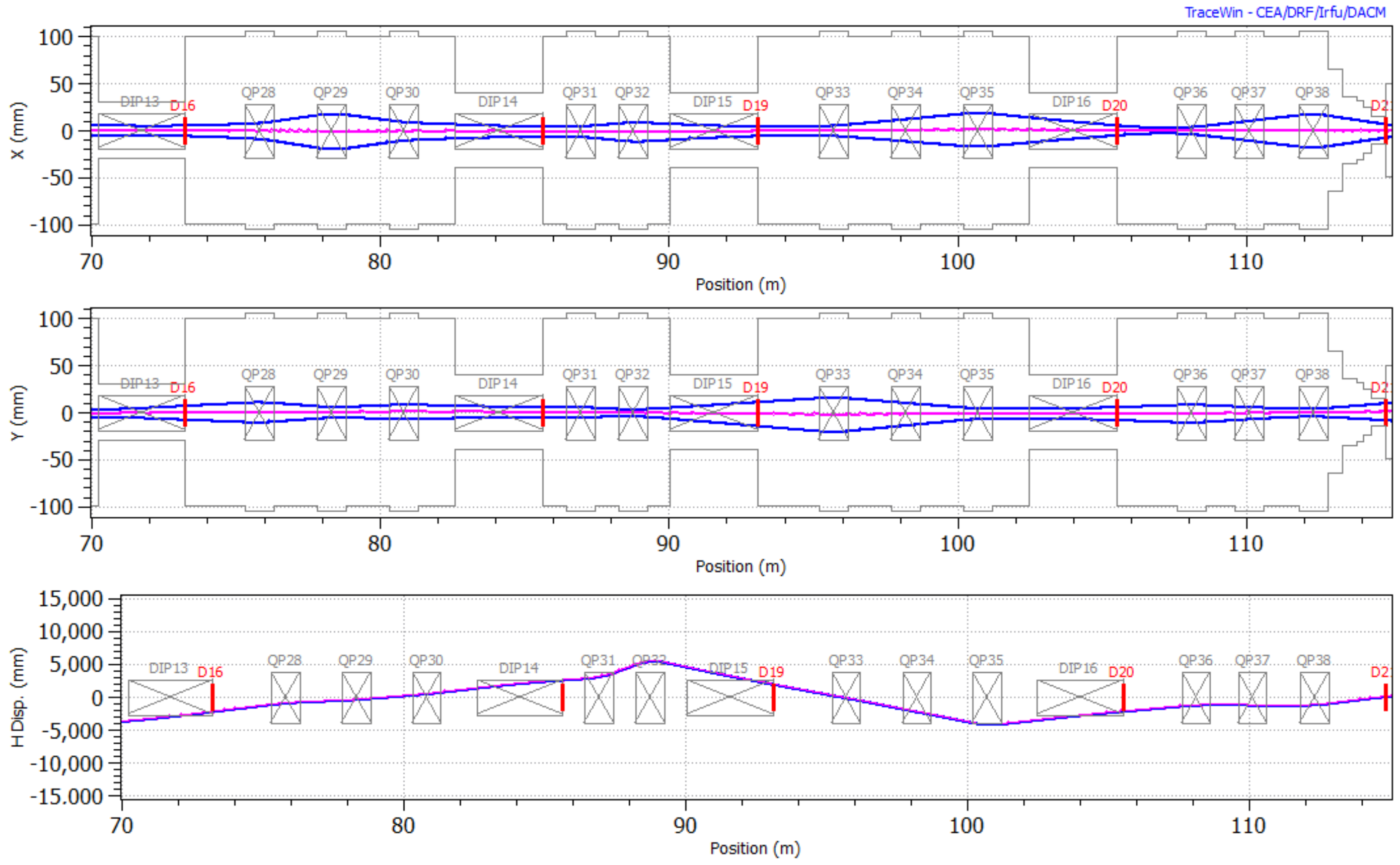,width=12cm}}
\caption{ RMS transverse beam envelopes and dispersion along the T3 branch of the BSY when two quadrupoles, Q34 and Q37, are not functioning (failure scenario).}
\label{fig5.11}
\end{center}
\end{figure}
%------------------------------------------------------

%
%-------------------------------------------------------------------------------

%%%%%%%%%%%%%%%%%%%%%%%%%%%%%%%%%%%%%%%%%%%%%%%%%%%%%%%%%%%%%%%%%%%%%%%%%%%%%%%%%%%%%%%%%%%%%%%%%%%%%%

%-------------------------
\subsection{Vacuum system\label{sec:ring_vacuum}}
The beam vacuum requirements of the ESS$\nu$SB accelerator complex is mostly expected to be in line with similar operating machines.  
Due to their single-pass nature, the vacuum level in the transfer lines, both before and after the ring, is expected to of the order of $10^{-7}$\,mbar. This means that at the point of extraction from the linac the two beam lines will have comparable vacuum levels so that there is no need for differential pumping. 

A higher vacuum level, about $10^{-8}$\,mbar, will be needed in the ring where the head of each injected batch will spend about 0.8\,ms making 600 turns of revolution. Even though the beam will be extracted immediately after the injection is complete, there is a risk of instabilities arising in the ring. At this high beam intensity, and an almost coasting beam, the biggest concern is likely to be the instability caused by the interaction of beam protons with electrons released from the inner side of the vacuum chamber \cite{Keil:333011}. The risk can be mitigated by suppressing the emission of secondary electrons by coating the inner surface with a ceramic material, titanium nitride (TiN). 

 A TiN coating forms a barrier that prevents outgassing, primarily of hydrogen from the stainless-steel vacuum chamber. This helps to maintain an acceptable vacuum level. Such a coating also suppresses the emission of secondary electrons, which is an important measure against the formation of electron clouds. It is these electron clouds that can, in combination with the high-current proton beam, give rise to electron-proton instabilities as the proton bunch interacts with the electron could coherently on a turn-by-turn basis.
 
The emission and reflection of blackbody radiation in the vacuum chamber is also suppressed by the TiN coating. Studies of H$^-$ stripping processes indicate that stripping due to blackbody radiation can be an issue in the transfer line (see Section~\ref{sect:beam_losses}), unless the emission and reflection is mitigated through coating or cooling of the vacuum chamber. It is therefore expected that the vacuum chamber in the linac-to-ring transfer line will have to be coated (although polishing of the beam-pipe may also be a feasible alternative).

\subsection{Safety}
\label{sec:accumulator:safety}
The design of the accumulator ring and transfer lines is governed by one major safety aspect: that the uncontrolled beam loss must not supersede 1\,W/m (see also Section~\ref{sect:beam_losses}). This is the empirical limit below which the activation of the accelerator equipment is low enough to allow hands-on maintenance 1\,hr after a beam stop. The large curvature of the linac-to-ring-transfer line, the two-stage collimation system, and the generous machine acceptance in the ring are a few results of this limitation. The loss limit is thus important both for the machine availability and for the safety of the personnel.

It is already apparent at this stage that certain sectors in the accelerator chain will have a heightened risk of beam loss. These sectors will suffer more activation and will therefore be sensitive areas for human intervention. The injection and extraction regions, the injection dump line, and the vicinity of the collimators are such sectors. Note that a large fraction of the losses in these regions are considered controlled, and therefor may be higher than the 1\,W/m limit.

A detailed understanding of beam-loss patterns in the beam lines will require further study, where instabilities, magnet position and field errors, correction technique, and further effects are modelled. Only then will it be possible to fully judge the activation risks, and determine the need for extra shielding near to the accelerator. As with all particle accelerators, a global radiation monitoring system will be employed to control the radiation levels. In addition, dosimetry for any personnel entering the accelerator area will be necessary. 

An extensive system of beam diagnostics and instrumentation will also be necessary. Several hundreds of beam loss monitors will be distributed along the beam line, combining two main types: fast monitors, to detect fast losses that could harm the machine; and slow varieties, to monitor losses over longer term. The former will be connected to a machine protection system with feedback to the H$^-$ source. If an error occurs which results in a rapid spike in losses, the source must be stopped on the next pulse. Roughly 200 beam position and current monitors will be employed in the beam-loss control and machine-protection scheme.

Longitudinal and transverse beam profile monitors, a diagnostic system to measure the particle tune in the ring, and a system to detect particles in the extraction gap will be crucial tools to safely operate the complex at 5\,MW (along with a variety of other instruments and methods). The commissioning of the machine will be done in stages, heavily relying on the beam diagnostics, by slowly increasing the number of injected turns and the total intensity passing through the beam lines. 

The 900 meters of new beam line is a challenge in terms of safety preparations -- in the sense that a plan must be made for a multitude of different failures. A study of such failure scenarios will be left to a later stage in the project. The existing systems at ESS, and the experience of the ESS staff will help in facilitating the planning and the implementation of all the ESS$\nu$SB safety systems. 

\subsection{Costing}
The accumulator work package includes 900~meters of new beam line with roughly 500 magnetic elements of many types (dipoles, quadrupoles, sextupoles, corrector dipoles, injection and extraction kicker magnets, septum, etc.) hundreds of beam diagnostic instruments; several RF cavities and collimation systems; and other complex equipment to ensure safe operation and reach the desired performance.

Considering that this is a new construction, and that the project has reached the end of a conceptual design phase, the cost estimate in Table~\ref{tab:accumulator:cost} is approximate and preliminary. For a more precise view of the total cost, it will be necessary to study each cost item with a degree of technical detail that is unfeasible at this stage. The values in the table are based on the experience of building comparable accelerator facilities, and on the manufacturing of similar equipment and systems. In particular, the actual costs related to the construction of the SNS facility up through 2009 have been used as a starting point.

The SNS accumulator and transfer lines have many similarities with the ESS$\nu$SB accumulator and transfer lines~\cite{Wei:2000wa}. However, the ESS$\nu$SB beam energy is roughly twice that of SNS. On the other hand, the beam lines have partially been adapted so as to limit the magnetic field strength needed to steer the beam. Furthermore, the transverse beam emittance in the SNS accumulator is 160$\pi$\,mm\,mrad (99\%, geometric) which is almost double that of ESS$\nu$SB (70$\pi$\,mm\,mrad). This means that the ESS$\nu$SB aperture can be smaller, a fact that brings the magnet cost down. A comparable magnet-packing factor (i.e. magnet length per unit length), similar number of magnets per species, and comparable magnet properties have been assumed. Through this simple comparison, the cost of DC accelerator magnets can be scaled with the length of the specific beam line. These estimates have been cross-checked with expert estimates of the cost of individual magnets types at CERN~\cite{Lopez:2021:priv} and Scanditronix Magnet~AB~\cite{SCX:2022}. 
%---------------------------------------------------------------------
%      Table: Cost of WP3
%---------------------------------------------------------------------
\begin{table}[ht!]%[H]
  \begin{center}
    \caption{ESS$\nu$SB accumulator and transfer lines cost breakdown}
    \label{tab:accumulator:cost}
    \begin{tabular}{lc}
    \textbf{Item}  &\textbf{Cost [M\texteuro]} \\
\hline
DC magnets and power supplies & 50 \\
Injection system & 11\\
Extraction system & 7\\
RF systems & 16\\
Collimation  & 8\\
Beam instrumentation & 19\\
Vacuum system & 24\\
Control system & 30\\
\textbf{Total} & \textbf{165}\\
\hline
     \end{tabular}
   \end{center}
\end{table}
%---------------------------------------------------------------------

The estimated cost of the injection system includes four DC magnets to create a permanent orbit bump, four injection kicker magnets with power supply, a stripper foil assembly, and a system to safely extract the convoy electrons. In addition, an injection beam dump is included at the end of a short beam line for the ions that have not be fully stripped.

For the extraction system, the 4$\times$4 kicker magnets, the Lambertson septum magnet, and auxiliaries are included. 

RF cavities and associated equipment will be installed in the L2R transfer line and in the ring itself. There is required R\&D work to adapt existing devices to the specific conditions of the ESS$\nu$SB, in particular the barrier RF cavities that are planned for bunch manipulations inside the ring. 

Collimation systems for transverse beam cleaning are planned for three separate locations: the end of the linac-to-ring transfer line; within the ring; and at the end of the beam switchyard, just before the targets. Similar costs have been assumed for these three systems, although slightly different designs may ultimately be used. The cost of collimators installed at CERN, primarily at the Large Hadron Collider (LHC), and in the SNS accumulator ring, have been used as a starting point for the ESS$\nu$SB cost estimate.

An extensive collection of beam instruments will be a crucial part of the facility. Due to the high beam power, over 250 beam loss monitors are needed, both fast and slow types. In addition to this, about 180 beam position monitors, 12 current monitors, 20 transverse and longitudinal profile monitors, as well as diagnostics for tune and beam-in-gap monitors will be needed. A similar instrument density and cost-per-accelerator-section as SNS have been assumed for the ESS$\nu$SB beam diagnostics. However, there is a factor of uncertainty in the cost of beam instrumentation owing to the high average beam power, which may call for the development of new types of monitors. 

The vacuum and control systems will be directly connected to the existing systems and infrastructure at ESS, thus it was logical to make the cost estimate for these systems in collaboration with specific experts at ESS. The cost of a vacuum system is in line with the general guideline of 15\,k\texteuro/m, but with an additional cost of coating the vacuum chambers with TiN -- in both the ring and the L2R transfer line -- to suppress the emission of secondary electrons as well as photons which cause beam loss. This cost is also comparable to that of the SNS vacuum system -- where TiN coating is used in the ring -- scaled to ESS$\nu$SB conditions. The fact that some vacuum support and infrastructure is already in place may reduce the final price.

\subsection{Summary}
\clearpage

% Use this to number the figures, equations and tables with section number in front
\setcounter{figure}{0}
\numberwithin{figure}{section}
\setcounter{equation}{0}
\numberwithin{equation}{section}
\setcounter{table}{0}
\numberwithin{table}{section}

\section{Target Station} \label{targetstation}

The target station facility is a key element of the ESS$\nu$SB project, which will convert the \SI{5}{\mega\watt} proton beam with a \SI{14}{\hertz} frequency into an intense neutrino beam. These particles are produced by the decay-in-flight of secondary mesons (mostly pions), which are created by the interaction of the proton beam within the target and focused into a decay tunnel by a hadronic collector. The technology chosen is based on a solution already studied in the framework of the European design study EUROnu~\cite{Edgecock:2013lga}, and it consists of four solid targets, each embedded in a magnetic horn. This hadronic collector, termed a ``four-horn system'' and shown in Fig.~\ref{fig:TargetStationOverview}, is supplied by a custom design power unit, producing a high magnetic field capable of focusing the pions inside the decay tunnel. 

\begin{figure}[!ht]
\begin{center}
\includegraphics[width=0.75\linewidth]{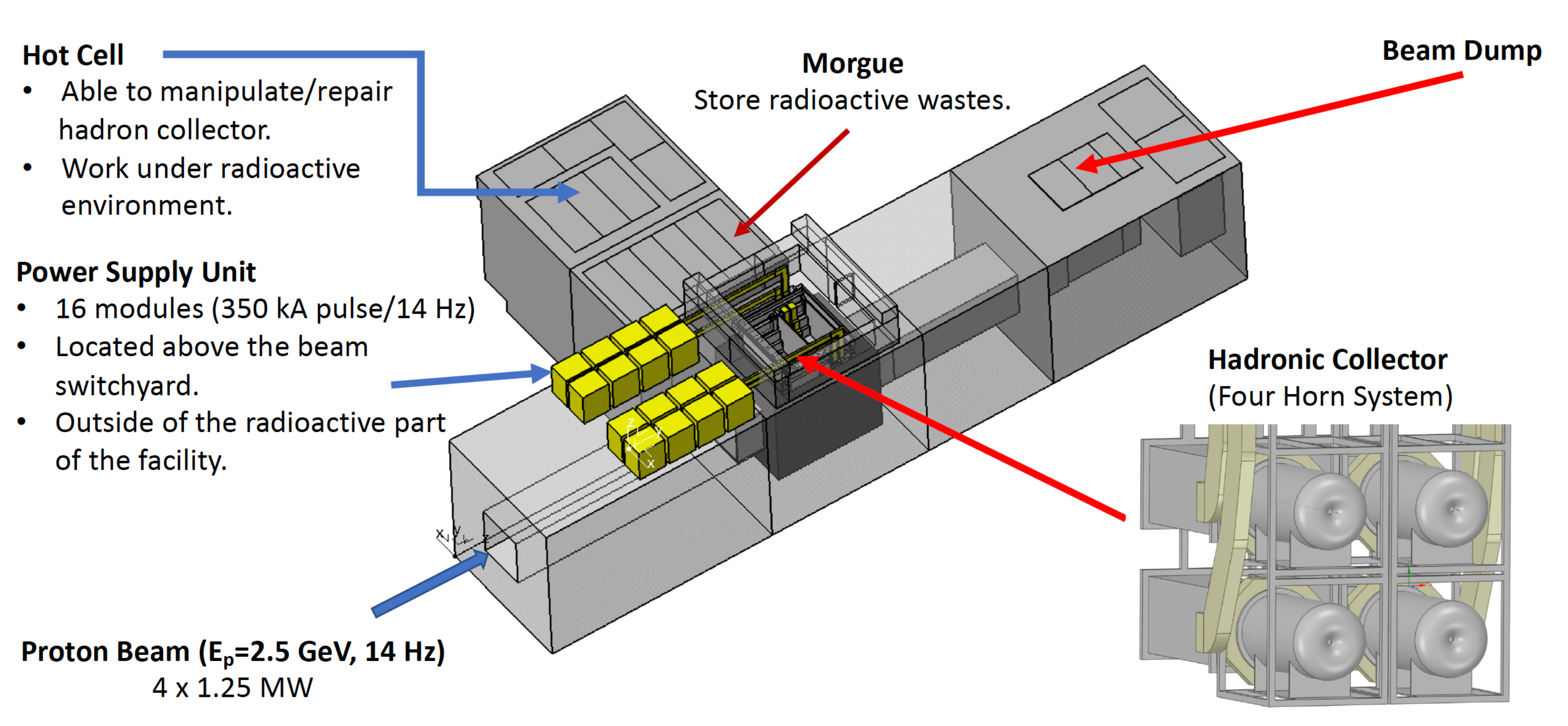}
\caption{Overview of the target station facility.}
\label{fig:TargetStationOverview}
\end{center}
\end{figure}

The revision of the baseline parameters with respect to EUROnu~\cite{Edgecock:2013lga} imposes significant changes on the present project. Most of the elements must be refined; in particular, the shape of the individual horns of the target station and the dimensions of the facility. From a technical point of view, the high power deposition inside the target imposes more stringent constraints on the granular target in terms of cooling and mechanical performance. The consequence of the change of the horn shape and the fast commutation between the four horns imposes a need for a new power-supply-unit scheme based on a modular approach.

In addition, the facility building hosting the target station will be subject to an intense flux of radiation, producing high energy deposition and high activation levels in the surrounding materials. This facility will also be equipped with essential elements including a hot cell to repair the hadronic collector, a morgue to store radioactive elements, and a beam dump to stop remaining particles. The design of such a facility represents a considerable challenge in terms of physics and engineering, and has to be compliant with Swedish safety regulations. In the following text, separate sections will address the aforementioned evolution of elements, starting with the defining factors of the target station and followed by the proposed technological solutions.

\subsection{Hadronic Collector}

In all neutrino Super Beam experiments, neutrinos are produced by the decay-in-flight of mesons inside the decay tunnel of the facility, through the following reactions:
$$
\begin{array}{lcl}
\pi^+ & \longrightarrow & \mu^+ + \nu_\mu \\  \\
\pi^- & \longrightarrow  & \mu^- + \overline{\nu}_\mu 
\end{array}
$$
The magnetic horns surrounding the target focus $\pi^+$ ($\pi^-$) and defocus $\pi^-$ ($\pi^+$) particles, thanks to the toroidal magnetic field generated by a short-pulse electric current running through the horn skin, to ultimately produce the $\nu_\mu$ ($\overline{\nu}_\mu$) beam. The polarity of the current allows for the selection between $\nu_\mu$ (positive polarity) and $\overline{\nu}_\mu$ (negative polarity) modes. Other types of neutrino flavors are also produced by the decay of other types of particles (like kaons) at lower rates. 

The four-horn system, shown in Fig.~\ref{fig:targetstation}, allows for an effective reduction of the incoming beam power of \SI{5}{\mega\watt} delivered by the linac to a more acceptable level of \SI{1.25}{\mega\watt} per target, with \SI{2.5}{\giga\eV} proton kinetic energy, \SI{1.3}{\micro\second} pulse duration, and a \SI{14}{\hertz} repetition rate. Each pulse will therefore deliver $2.23 \times 10^{14} $ protons/pulse/target. The transverse beam profile is characterised by a quasi-uniform distribution formed by the anti-correlated painting technique in the accumulator ring (see Section~\ref{sec:accummulator_injection}).

\begin{figure}[h!]
\begin{center}
\begin{subfigure}[b]{0.35\linewidth}
    \centering 
    \includegraphics[width=1.\linewidth]{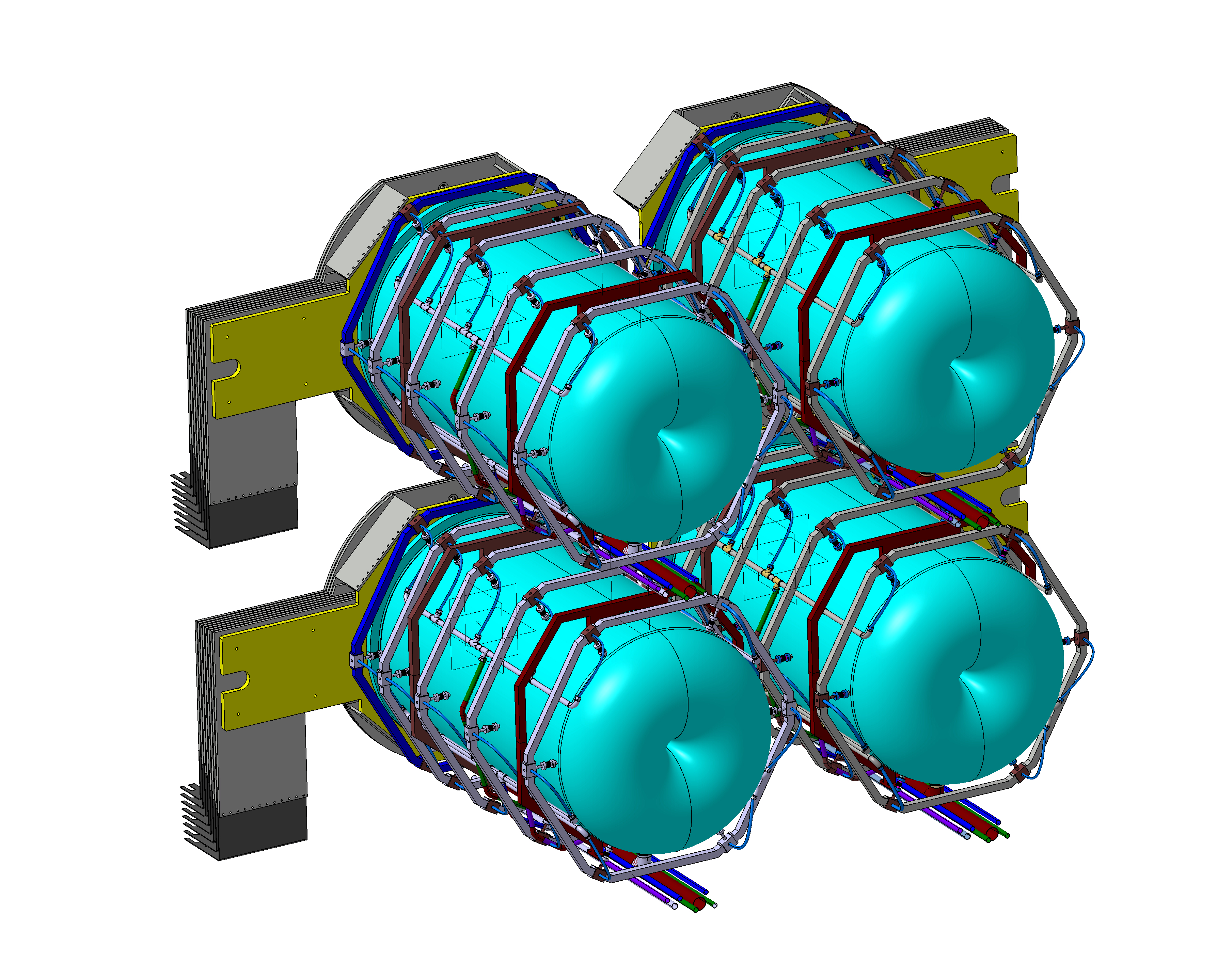}
    \caption{The four-horn system.}
\end{subfigure}
\hspace{1.cm}
\begin{subfigure}[b]{0.35\linewidth}
    \centering 
    \includegraphics[width=0.9\linewidth]{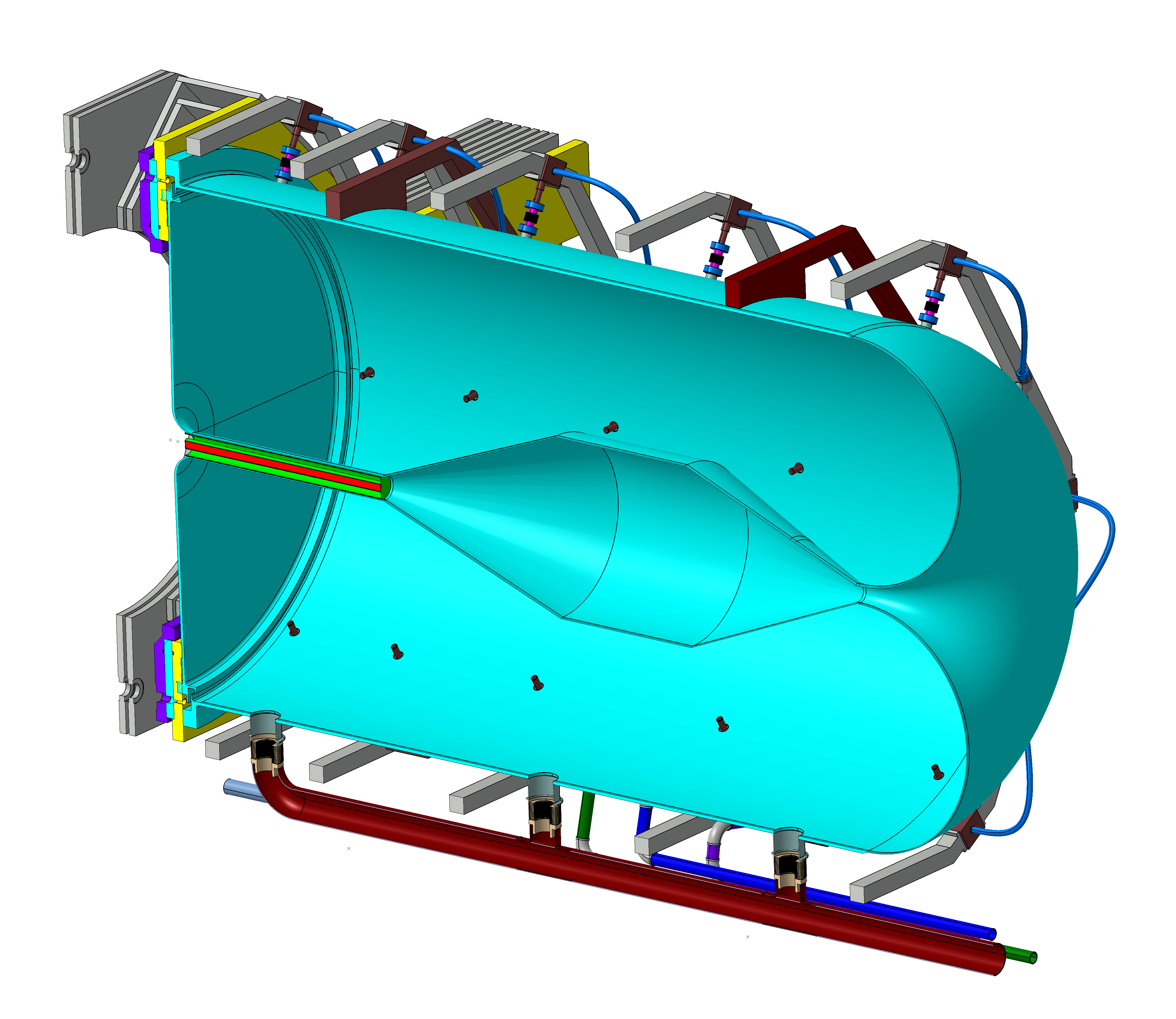}
    \caption{Transverse view of one horn.}
\end{subfigure}
\caption{The target station for ESS$\nu$SB.}
\label{fig:targetstation}
\end{center}
\end{figure}

The beam distribution on the targets is obtained after the transport of the proton beam to the switchyard, its profile in the transverse plane has a diameter of $\sim$\SI{2.8}{\cm} and a divergence of 5~mrad. Due to the high power and short pulse duration of the proton beam, the ESS$\nu$SB target will therefore be operating under severe conditions.

\subsubsection{Horn Shape Optimisation}
\label{sec:targetstation:geneticalgorithm_improvement}
Due to the relatively low energy of the protons from the ESS linac, the pions exiting the target are expected to have a large angular spread. The shape of the horns was determined by using a deep-learning method based on a genetic algorithm. This technique is an evolutionary algorithm, in which a set of different geometric configurations of the system to be optimised is allowed to evolve towards the best figure of merit (FoM), which is taken to be the fraction of $\delta_{CP}$. The configuration with the best FoM value after several iterations, known as ``generations'', is taken as the optimised configuration of the system. 

\begin{figure}[h!]
\begin{center}
\begin{subfigure}[b]{0.45\linewidth}
  \includegraphics[width=0.95\columnwidth]{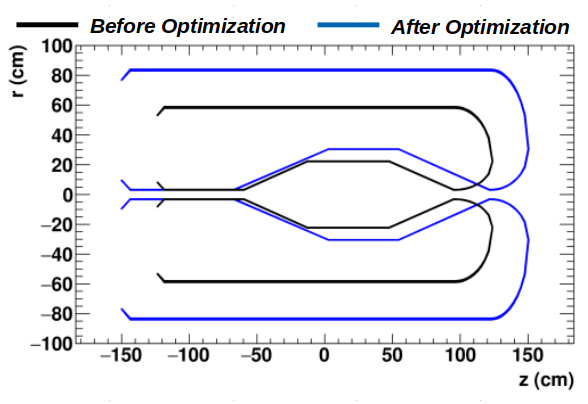}
  \caption{Evolution of the horn profile.}
\end{subfigure}
\hspace{1.cm}
\begin{subfigure}[b]{0.45\linewidth}
  \includegraphics[width=1.\columnwidth]{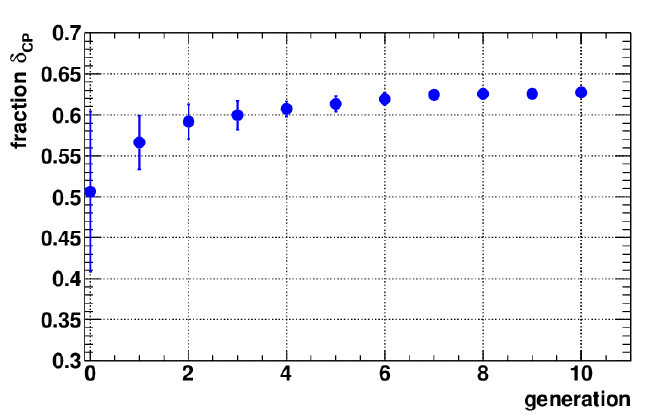}
  \caption{Genetic algorithm convergence throughout the iterative process.}
\end{subfigure}
\caption{Optimisation of the horn profile.}
\label{fig:NuFlux2}
\end{center}
\end{figure}

A simplified geometry consisting of the four targets, together with the horn assembly and the decay tunnel has been considered and simulated with \textsc{FLUKA}~\cite{Battistoni:2015:FLUKA}. The system is characterised by a parameter set defining the geometry of the horn and the length of the decay tunnel. Each parameter has been allowed to evolve within a range scaling between 0.5 and 1.5 times the initial reference values. The results of these calculations suggest an overall increase in size for the target station, as shown in  Fig.~\ref{fig:HornNewDimension}.

\begin{figure*}[!h]
\centering

  \begin{subfigure}[b]{0.48\textwidth}
  \centering
  \includegraphics[width=0.8\columnwidth]{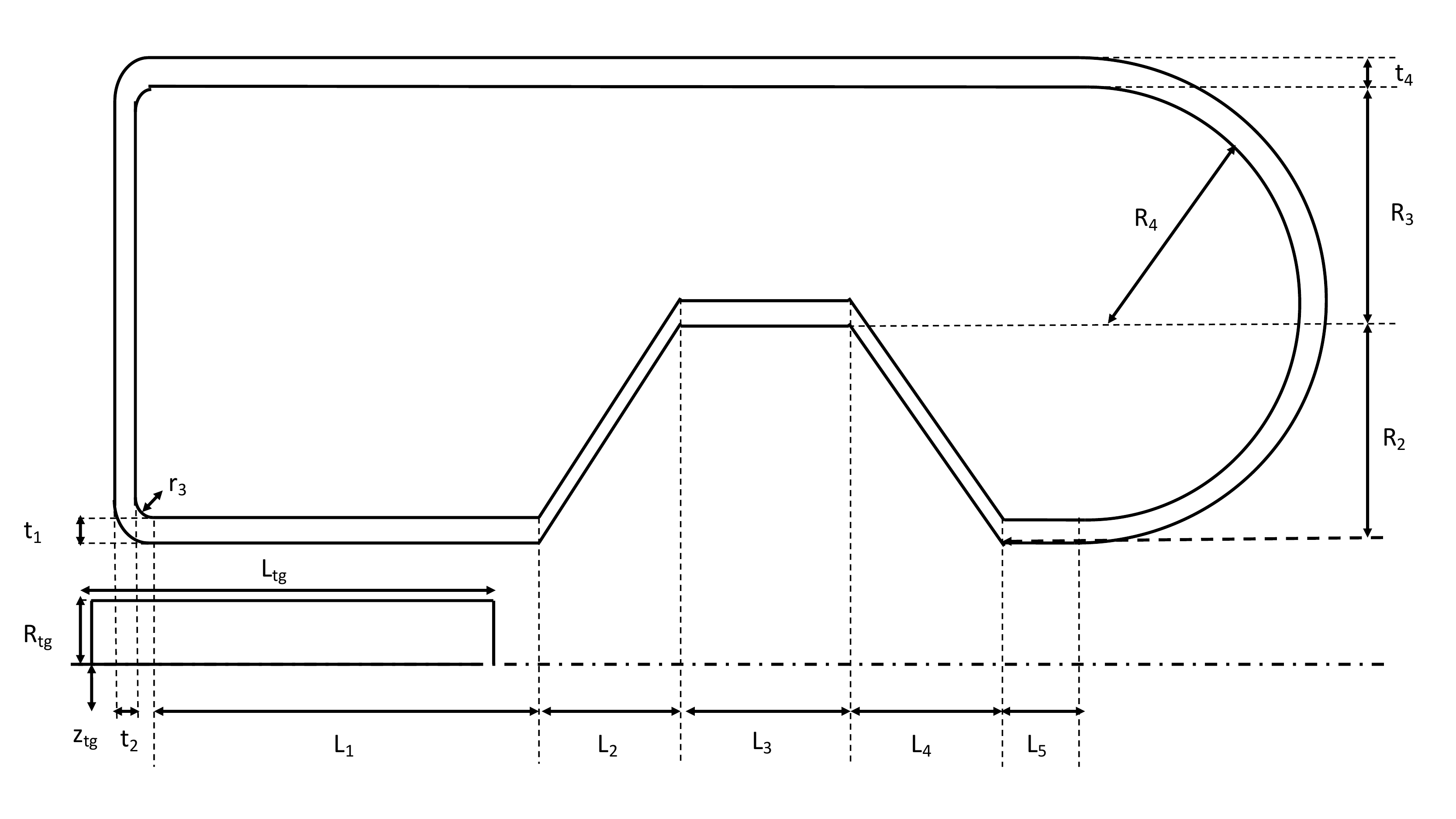}
  \end{subfigure}
  \begin{subtable}[b]{0.48\textwidth}
   \centering
   \begin{tabular}{cc} 
   \textbf{Parameters} 						& \textbf{Value (mm)}  \tabularnewline \hline
   $L_1$, $L_2$, $L_3$, $L_4$, $L_5$	& 766, 697, 519, 670, 10.8 \tabularnewline 
   $t_1$, $t_2$,  $t_3$, $t_4$			& 3, 3, 10 	 \tabularnewline
   $r_1$, $r_2$, $r_3$					& 108, 108, 66 	 \tabularnewline
   $R_{tg}$, $L_{tg}$, $z_{tg}$	 		& 15, 780, 69 	 \tabularnewline 
   $R_1$, $R_2$, $R_3$, $R_4$			& 30, 273, 558, 412.5  \tabularnewline \hline 
   \end{tabular}
   \vspace{0.7cm}
  \end{subtable}
  \caption{Definition of the horn parameters and their values obtained after optimisation.}   
  \label{fig:HornNewDimension}
\end{figure*}

This physics optimisation procedure resulted in a nearly finalised horn shape. However, some additional modifications -- the most important of which was the diameter of the horn neck (the part directly encircling the target) -- were determined by the target cooling requirements. The final horn dimensions, for which the results are discussed below, are shown in the technical drawing in Section~\ref{section:MagneticHornDesign}. The neutrino fluxes obtained for the final horn shape are represented in Fig.~\ref{fig:NuFLux1} with the flavor composition given in Table~\ref{tab:NuFlux}.

\begin{figure}[h!]
 \begin{subfigure}[b]{0.48\textwidth}
  \centering
  \includegraphics[width=0.9\linewidth]{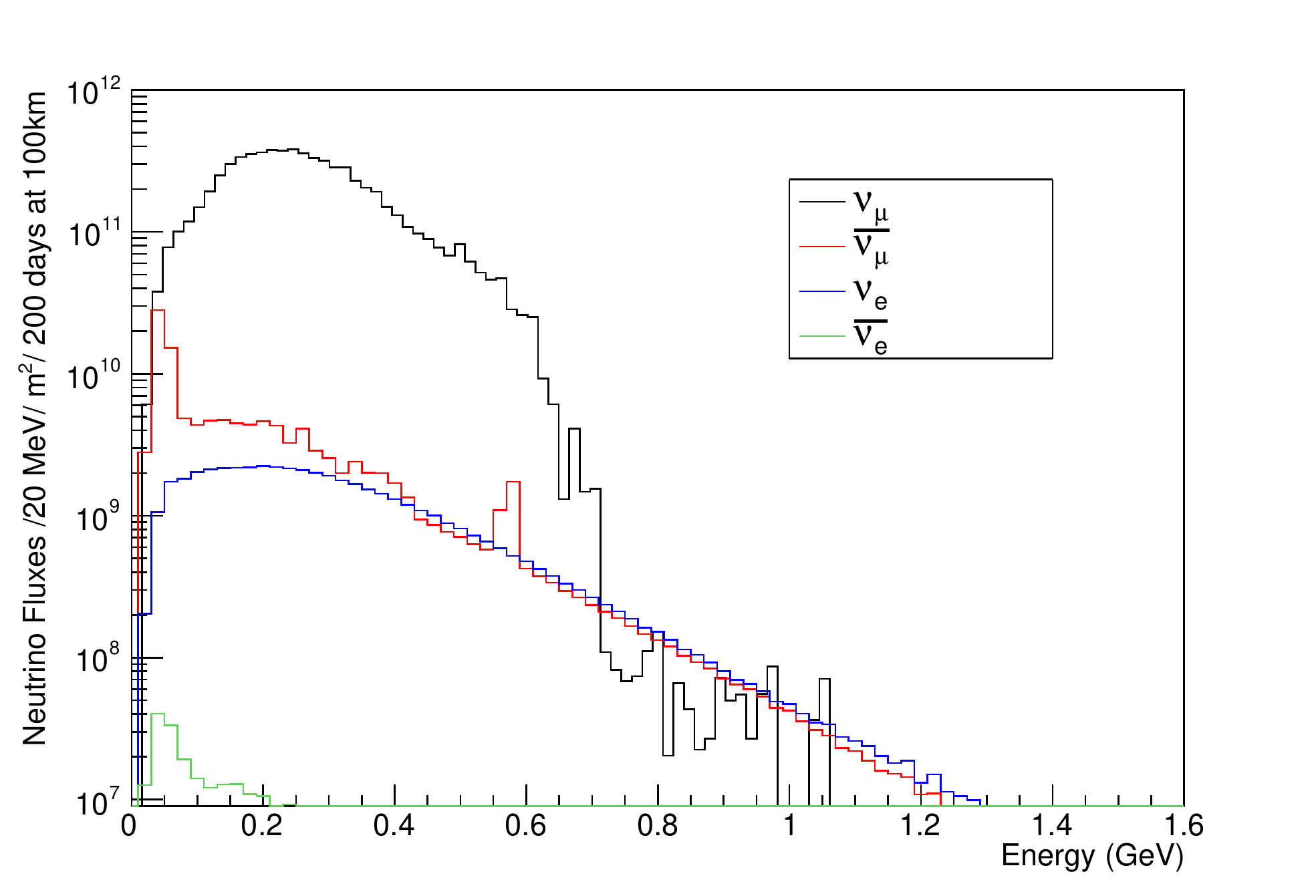}
  \caption{Positive polarity}
 \end{subfigure}
 \begin{subfigure}[b]{0.48\textwidth}
  \centering
  \includegraphics[width=0.9\linewidth]{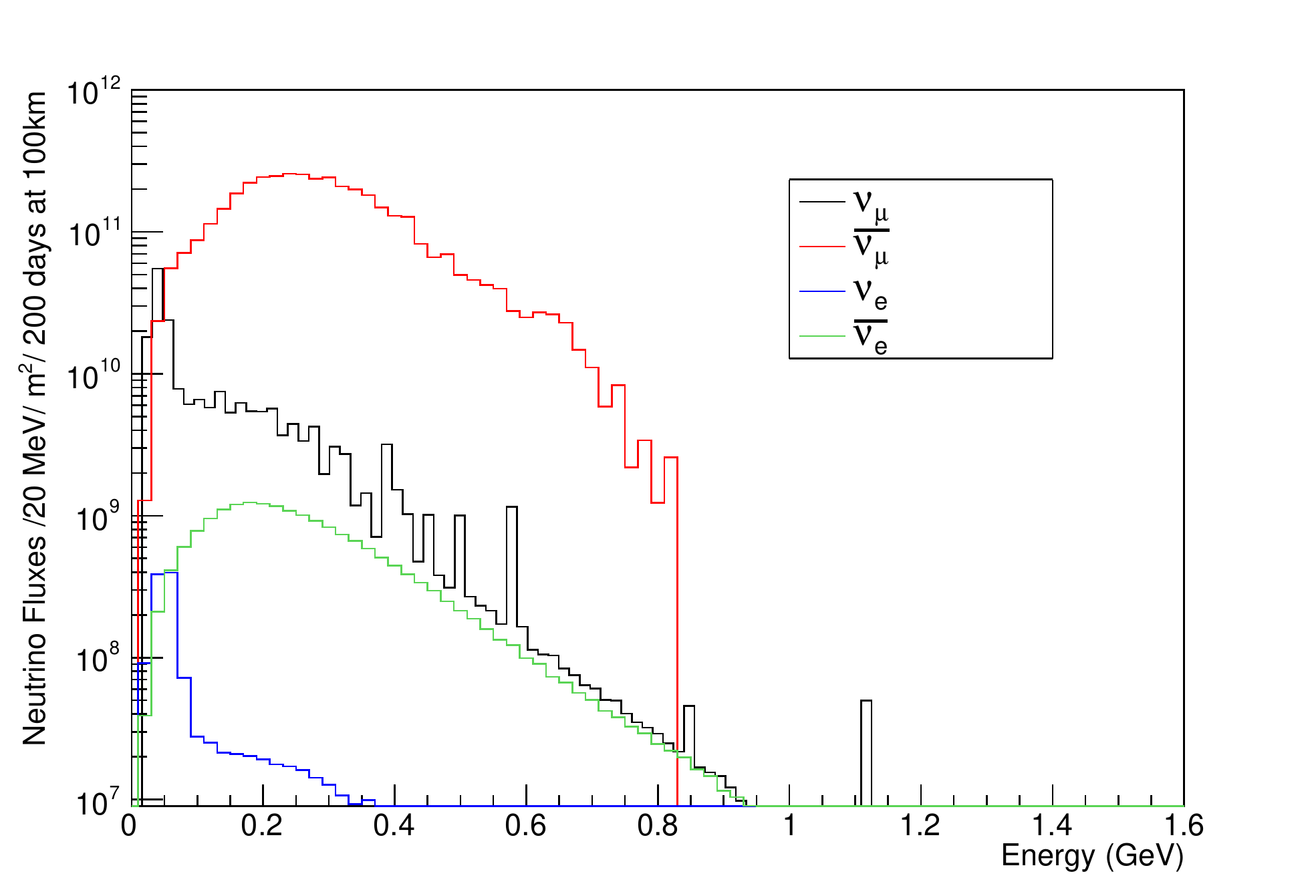}
  \caption{Negative polarity}
  \end{subfigure}
  \caption{Neutrino fluxes.}
  \label{fig:NuFLux1}
\end{figure}

\begin{table}[h!]
 \centering
  \caption{Neutrino flux composition at \SI{100}{\kilo\metre}.}
 \begin{tabular}{|c|cc|cc|} \hline
 & \multicolumn{2}{c|}{$\nu$ Mode}  &  \multicolumn{2}{c|}{$\overline{\nu}$ Mode } \tabularnewline
 \hline 
 Flavor  & $N_{\nu} (10^{10}/m^2)$ & $\%$ & $N_{\nu} (10^{10}/m^2)$ & $\%$ \tabularnewline
 \hline 
 $\nu_{\mu}$ 		    	& 674 	& 97.6 	& 20 	& 4.7   \tabularnewline 
 $\overline{\nu}_{\mu}$ 	& 11.8 	& 1.7 	& 396 	& 94.8 \tabularnewline
 $\nu_{e}$ 			& 4..76 	&  0.67 	& 0.13 	& 0.03 \tabularnewline
 $\overline{\nu}_{e}$ 		& 0.03 	& 0.03 	& 1.85 	& 0.43 \tabularnewline \hline
 \end{tabular}
 \label{tab:NuFlux}
\end{table}

%--
\subsubsection{Granular Target Design}

A granular target solution was proposed in the early 2000s by P. Sievers at CERN \cite{Sievers:2001dh,Pugnat:2002bm}, and was then adopted in the context of the EUROnu project \cite{Edgecock:2013lga}. The main advantage of such a design in comparison with a monolithic target lies in its possibility of the cooling medium flowing directly through the target, which allows for a better heat removal from the target regions with the highest power deposition (specifically, from the centre of the target co-aligned with the axis of the impinging proton beam). In addition, the stress levels are much lower in a granular target compared to a monolithic design. This target concept has been studied in the frame work of the ESS$\nu$SB project.

\begin{figure}[t]
  \centering
  \includegraphics[width=0.8\columnwidth]{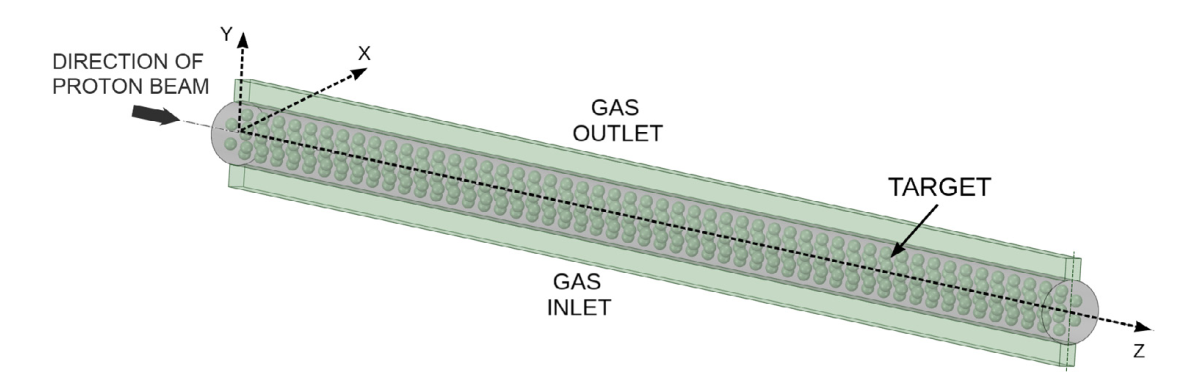}
  \caption{3D model of the target concept based on a packed bed of titanium spheres.}
  \label{fig:TargetDesign}
\end{figure}

\subsubsubsection{Target Cooling Concept}
\label{sub:target-cooling-concept}

The conceptual model of a helium-cooled target is shown in Fig.~\ref{fig:TargetDesign}. The granular target under consideration is a rod of {\SI{78}{cm}} length and {\SI{3}{\centi\meter}} diameter, consisting of titanium spheres with a mean diameter of {\SI{3}{\milli\meter}}, placed within a container made of titanium. It has been assumed for the purpose of this study that the porosity of the target is equal to 0.34, so that 66\% of the target volume includes titanium spheres, while the space between the spheres is filled with the coolant (gaseous helium).

The impinging beam will consist of {\SI{1.3}{\micro \second}} proton pulses, repeated at a frequency of {\SI{14}{\hertz}}. As a result of the interaction of the beam with the target spheres, an estimated {\SI{138}{\kilo \watt}} will be deposited in each target as heat. A map of the average power deposition inside the target is shown in Fig.~\ref{fig:PowerDeposition}.
Due to the high power deposited  (almost three times higher than in the EUROnu project), efficient heat removal is a critical issue.

\begin{figure}[h!]
\centering
  \includegraphics[width=0.7\columnwidth]{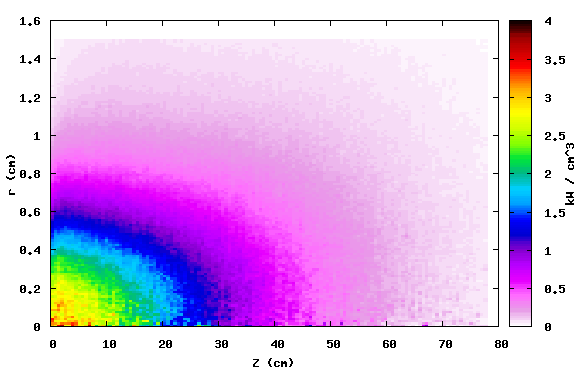}
  \caption{Map of power deposition inside a titanium granular target.}
  \label{fig:PowerDeposition}
\end{figure}

Heat energy is deposited by the beam into the spheres during a short time at the beginning of each cycle; afterwards, it is transferred to the surrounding medium over the remainder of the cycle ($t_c=1/f=\SI{0.071}{\second}$). The principal mechanism of heat transfer is the transmission of heat from the surface of the spheres to the helium flowing outside, known as forced heat convection. Because of the large amount of energy released in the titanium spheres, a high helium mass-flow rate is required. Passing such a big quantity of gas in the axial direction through the target is not feasible, not only due to the target length and its small diameter, but particularly as a result of the high sphere-packing ratio. This is the main reason for selecting transverse target cooling, as shown in Fig.~\ref{fig:TargetDesign}. 

In order to facilitate the study of granular target cooling, an analytical model has been proposed in \cite{Cupial2021}. This model allows for a rapid estimation of the influence of different parameters, such as the target geometry, the packing ratio and the flow rate on target cooling. Figure~\ref{fig:helium_parameters_first_sector} presents the average temperature $T_a$, pressure $p$, velocity $u$, and density $\rho$ of helium within the first target sector (the first \SI{8}{\centi \meter} of the target length) plotted against the transverse coordinate $y_s$, obtained using this model (full details are given in \cite{Cupial2021}). The results of the maximum sphere temperature $(T_s)_{max}$, minimum sphere temperature $(T_s)_{min}$, and $\Delta T_s$ vs. the transverse coordinate $y_s$ are shown in Fig.~{\ref{fig:sphere_temperature_first_sector}}.

\begin{figure}[htbp]
\begin{center}
\begin{minipage}{0.47\linewidth}
\includegraphics[width=0.99\textwidth]{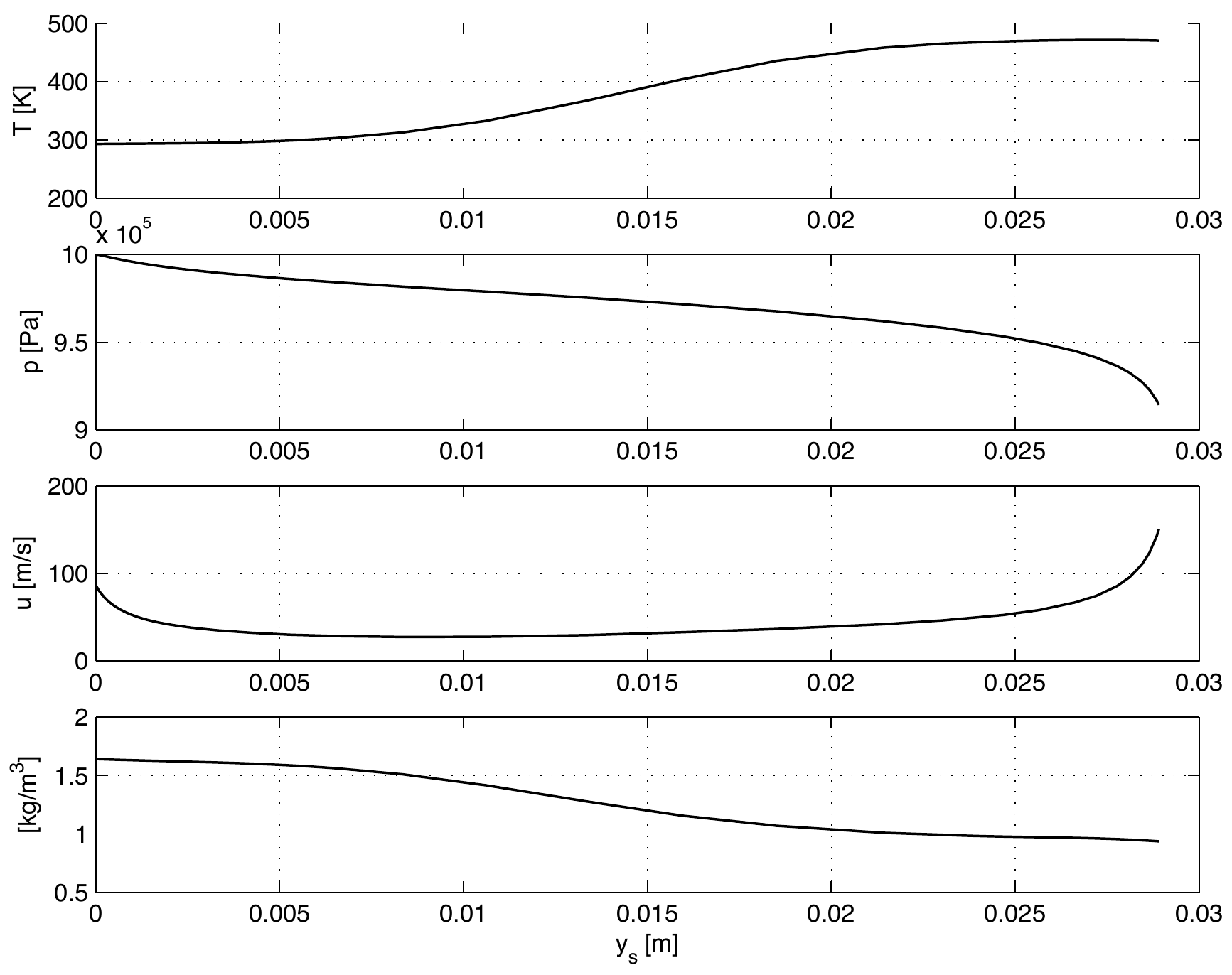}
\caption{\label{fig:helium_parameters_first_sector}Helium flow parameters obtained from the analytical approach for transverse flow (first sector).}
\end{minipage}\hspace{2pc}%
\begin{minipage}{0.47\linewidth}
\includegraphics[width=0.99\textwidth]{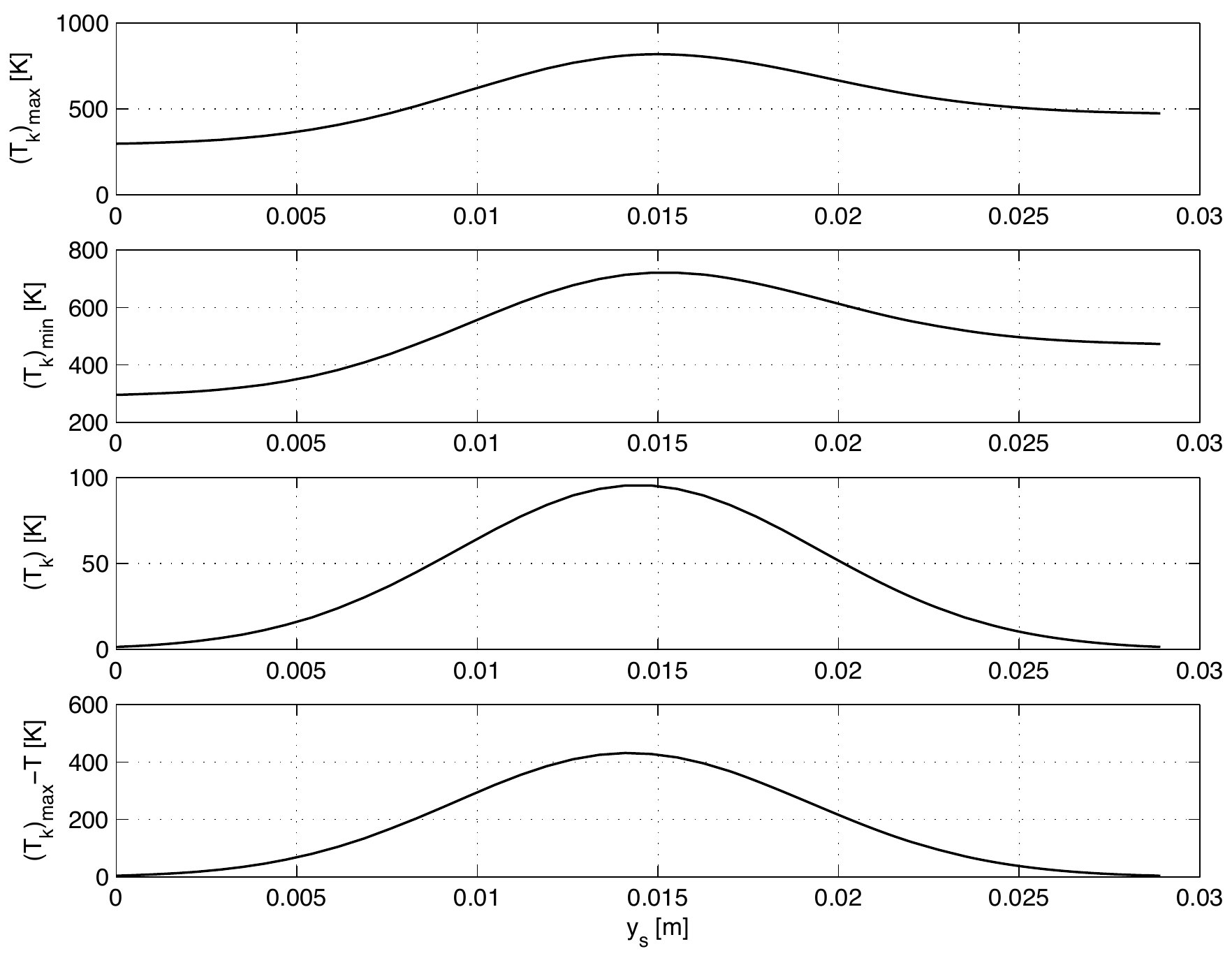}
\caption{\label{fig:sphere_temperature_first_sector}$(T_s)_{max}$, $(T_s)_{min}$ and $\Delta T_s$  of spheres in each beam cycle (first sector).}
\end{minipage} 
\end{center}
\end{figure}

%\begin{figure}[t!]
%\center
%\includegraphics[width=0.35\columnwidth]{figures/targetstation/target/Art_Figure_3_Raport.pdf}
%\caption{Helium flow parameters obtained from the analytical approach for transverse flow (first sector).}
%\label{fig:helium_parameters_first_sector}
%\end{figure}

%\begin{figure}[b!]
%\center
%\includegraphics[width=0.35\columnwidth]{figures/targetstation/target/Art_Figure_4_Raport.pdf}
%\caption{$(T_s)_{max}$, $(T_s)_{min}$ and $\Delta T_s$  of spheres in each beam cycle (first sector).}
%\label{fig:sphere_temperature_first_sector}
%\end{figure}

Figure~\ref{fig:fluent_contour} shows steady-state results from a CFD (Computational Fluid Dynamics) analysis performed using \textsc{ANSYS Fluent}~\cite{FluentUserGuide15}, for a helium global mass flow rate $\dot{m} =$ \SI[per-mode=symbol]{0.3}{\kilogram\per\second} (mass flow for the whole target) and with the power distribution inside the target given in Fig.~\ref{fig:PowerDeposition}. The numerical calculations for the packed-bed target are used a porous-medium approach, as proposed by Ergun~\cite{Ergun1949}. The superficial velocity shown in Fig.~\ref{fig:fluent_contour} is a related to the local velocity $u$ and the porosity of the porous medium. The profiles shown do not change significantly along the target length, but since the most proton beam power is released in the initial part of the target, the temperature (apart from the target initial part) decreases downstream from the point of beam impact.

\begin{figure}[h!]
\center
\includegraphics[width=0.75\columnwidth]{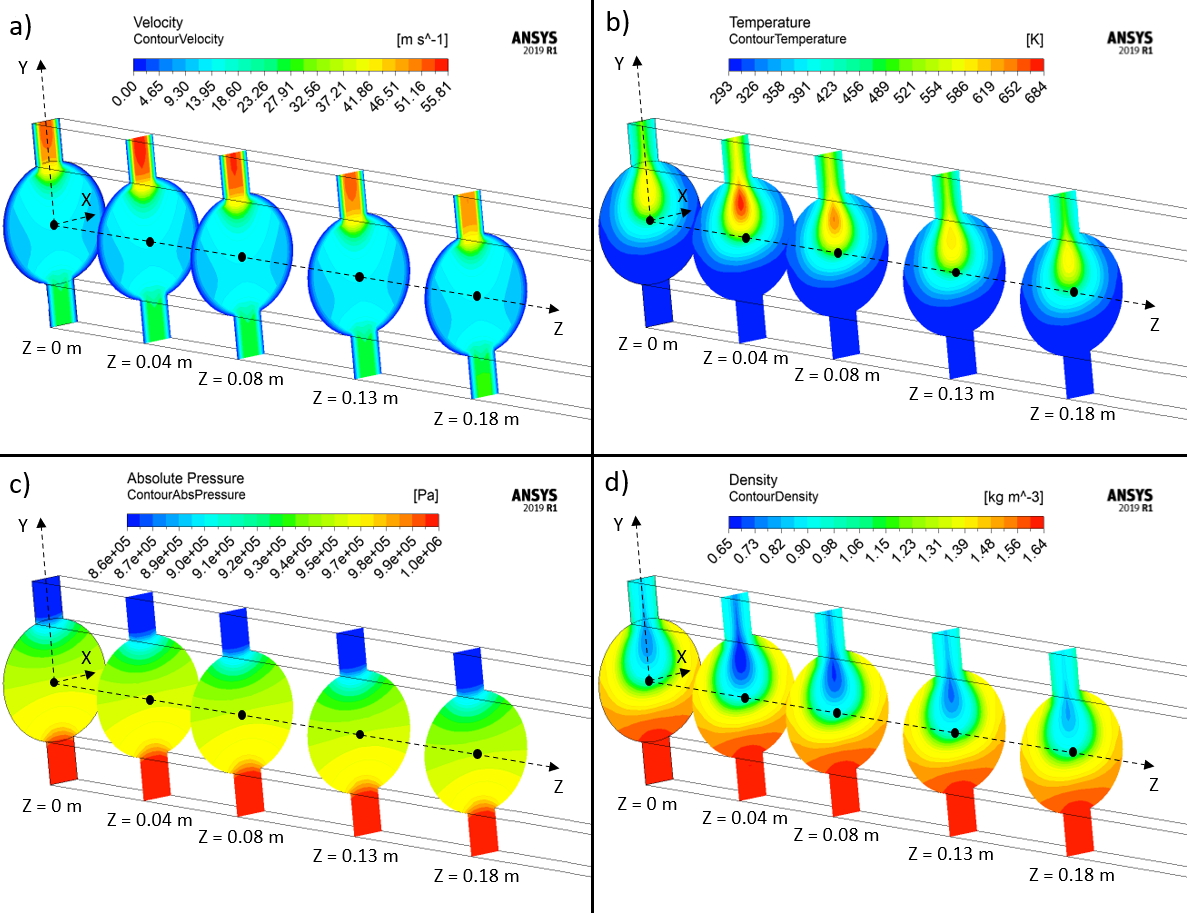}
\caption{Distribution of superficial velocity [\SI[per-mode=symbol]{}{\meter \per \second}] (a), temperature [\SI{}{\kelvin}] (b), absolute pressure [\SI{}{\pascal}] (c), and density [\SI[per-mode=symbol]{}{\kilogram \per \meter \cubed}] (d) of helium flowing upwards in the transverse direction through the first two sectors of the target under constant helium mass flow \SI[per-mode=symbol]{300}{\gram\per\second} and with the non-homogeneous power deposition inside the target having a total value of $\dot{Q} = $\SI{138.53}{\kilo\watt}}
\label{fig:fluent_contour}
\end{figure}

Both the analytical and the CFD simulations show that the velocity of the flowing gas decreases in the vicinity of the target axis, where the beam power density is the highest, and is higher at the outlet (Figs.~\ref{fig:helium_parameters_first_sector}, \ref{fig:fluent_contour}a). The decrease in the gas velocity in regions near the target axis results in a significant reduction in the heat transfer coefficient there, which in turn reduces heat exchange between the spheres and the gas. As a result, there is a considerable increase in the temperature of the spheres in this region. The maximum temperature of the spheres has been found to be approximately {\SI{750}{\kelvin}}, well below the melting point of titanium (about {\SI{2000}{\kelvin}}).

\subsubsubsection{Static and Dynamic Stress in the Target Spheres}

The power density averaged over the region of the "hottest" sphere is about \SI[per-mode=symbol]{2.8e9}{\watt \per \meter \cubed} (compare Fig.~\ref{fig:PowerDeposition}). Taking into account that the power is deposited only in the titanium spheres, which occupy 0.66\,\% of the volume, the power density used in the calculations is \SI[per-mode=symbol]{4.25e9}{\watt \per \meter \cubed}. For the material properties specified in Table~\ref{tab:sphere-mat-properties}, the maximum value of the steady-state stress components and the von~Mises stress is equal to \SI{50.6}{\mega \pascal}, for a sphere with \SI{1.5}{\milli \meter} radius. The steady-state stress is lower for smaller sphere radii. It is reduced to \SI{22.5}{\mega \pascal} and \SI{12.7}{\mega \pascal}, for sphere radii of \SI{1}{\milli \meter} and \SI{0.75}{\milli \meter}, respectively. The stress in a sphere can be shown to be independent on the heat transfer coefficient on the sphere surface.

\begin{table}[h!]
    \centering
    \caption{Material properties of titanium spheres used in stress calculations}
    \label{tab:sphere-mat-properties}
    \begin{tabular}{c|c}
        mass density & \SI{4.51e3}{kg/m^3} \\
        Young's Modulus & \SI{106}{\giga \pascal} \\
        Poisson's ratio  & 0.34 \\
        specific heat  & \SI[per-mode=symbol]{600}{\joule \per \kilogram \per \kelvin} \\
        thermal expansion coefficient & \SI{8.4e-6}{1/K} \\
        thermal conductivity& \SI[per-mode=symbol]{17}{\watt \per \meter \per \kelvin} \\
    \end{tabular}
\end{table}

However, the temperature on the spheres depends on the heat transfer coefficient. This has been calculated in the previous section, and it is equal to about \SI[per-mode=symbol]{4600}{\watt \per \meter \squared \per \kelvin}, for spheres of \SI{1.5}{\milli\meter} radius. Using this value, the maximum temperature increase above the temperature of cooling helium at the titanium sphere core is equal to: \SI{555.7}{\celsius}, \SI{349.6}{\celsius} and \SI{254.4}{\celsius}, respectively, for the sphere radii of \SI{1.5}{\milli\meter}, \SI{1}{\milli\meter} and \SI{0.75}{\milli\meter}. The last two values are expected to be even lower, since the heat transfer coefficient tends to decrease with the sphere radius. 

Thermal shock in the spheres that make up the target is an important issue. Analytical models that make use of wave propagation in solids have been studied for rods, discs, and cylinders in \cite{Sievers:1974}. These analytical solutions provide the necessary insight into the complex phenomena resulting from thermal shock.
The numerical study of thermal shock, e.g., using the finite element method, calls for much care, in terms of the size of the elements used and the integration time \cite{Davenne:2016}. As a rule, unrealistically low stress levels are obtained if the mesh is not fine enough or the integration time step is too large. In order to obtain reliable results, a code using explicit integration schemes can be used. The results discussed below have been calculated using \textsc{ANSYS LSDyna}. 

The energy released during each pulse by the proton beam in \SI{1}{\gram} of titanium that makes up the spheres is estimated, based on the above steady-state value and taking into account that the proton beam pulses are repeated every \SI{14}{\hertz}, to be about \SI[per-mode=symbol]{67}{\joule \per \gram} per cycle. For the specific heat value used ($c_p=\SI[per-mode=symbol]{600}{\joule \per \kilogram \per \kelvin}$), this results in a linear temperature increase of the "hottest" titanium spheres during the duration of a pulse by \SI{112}{\kelvin}.

\begin{figure}[h!]
\begin{center}
\begin{subfigure}[b]{0.46\linewidth}
  \includegraphics[width=\columnwidth]{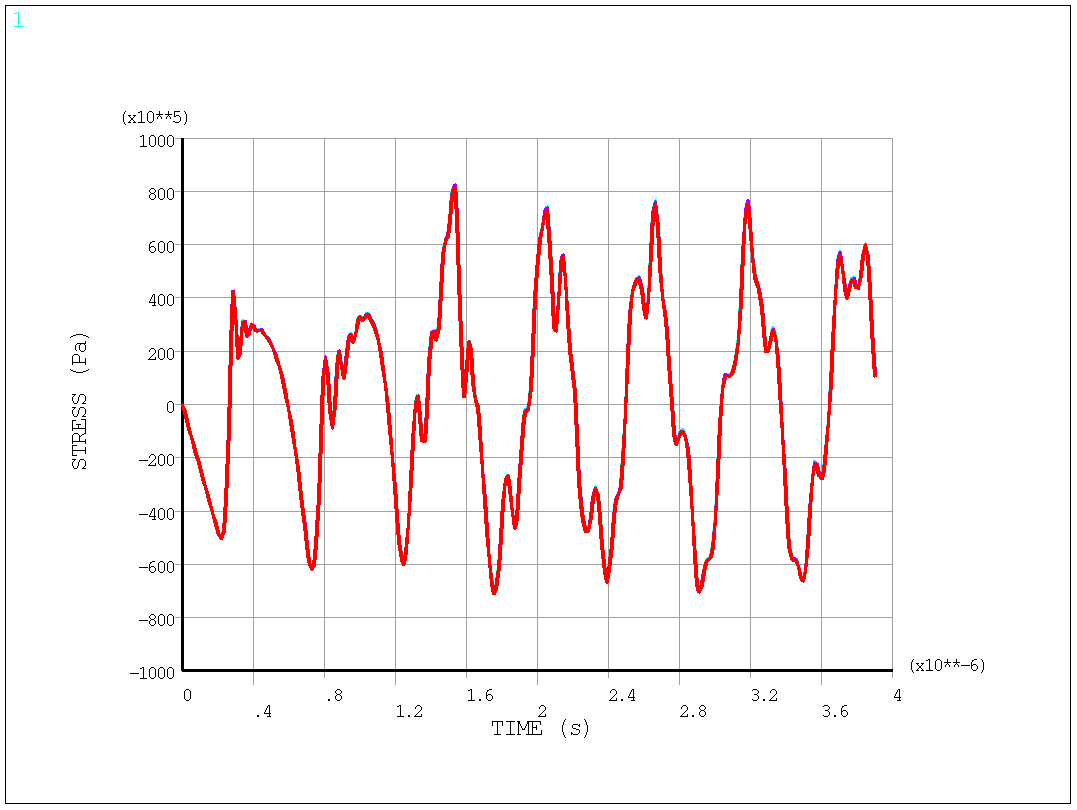}
  \caption{For a sphere with radius \SI{1.5}{\milli \meter}.}
  \label{fig:spheres-LSDyna-1.5mm}
\end{subfigure}
\hspace{1.cm}
\begin{subfigure}[b]{0.46\linewidth}
  \includegraphics[width=\columnwidth]{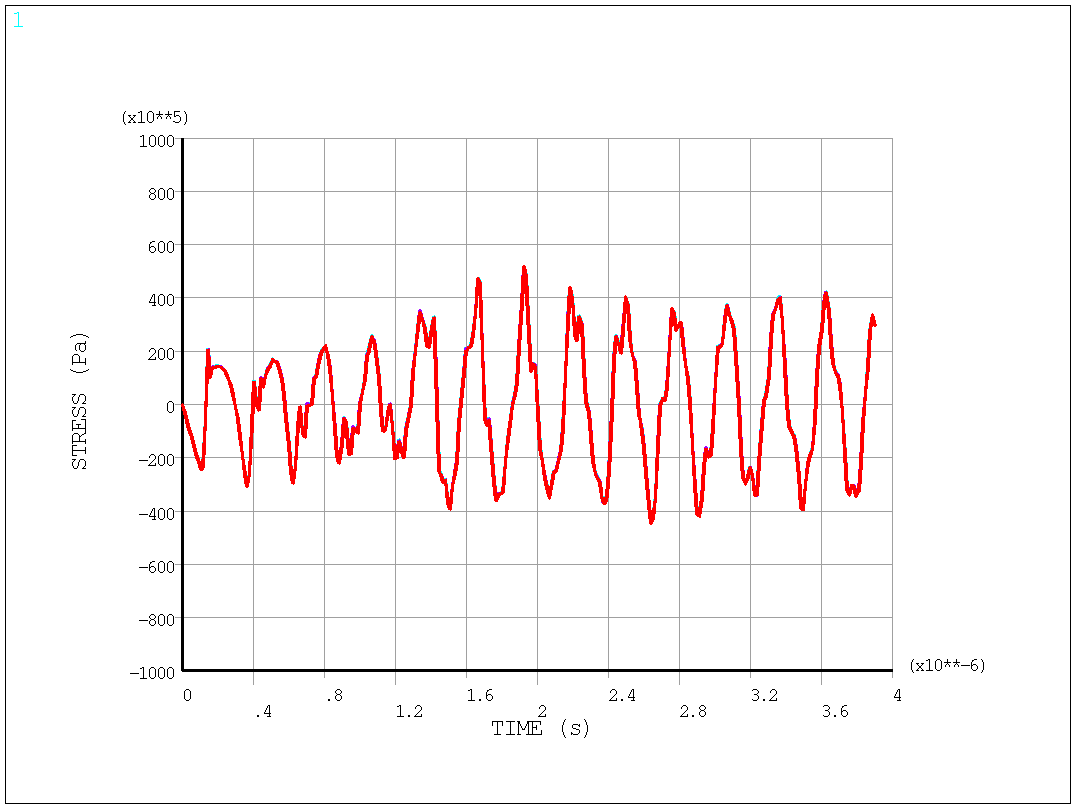}
  \caption{For a sphere with radius \SI{0.75}{\milli \meter}.}
  \label{fig:spheres-LSDyna-0.75mm}
\end{subfigure}
\caption{Dynamic stress at the sphere core.}
\label{fig:spheres-LSDyna}
\end{center}
\end{figure}

Figure~\ref{fig:spheres-LSDyna-1.5mm} shows the dynamic stress at the sphere centre, where the stress is the highest, over a time interval three times the pulse length, for a sphere with a radius of \SI{1.5}{\milli \meter}. The sphere is free to deform everywhere on its surface. The principal stress components, which are equal to the stress components in the spherical coordinate system $\sigma_r$, $\sigma_\phi$ and $\sigma_ \theta$, coincide for this point. One can see from this figure that the beam pulse length of \SI{1.3}{\micro \second} is longer than the period of oscillations of a sphere. The maximum value of dynamic stress is equal to \SI{83}{\mega \pascal}. For comparison, about half of this value would be found at a point located at a distance of one-half radius from the centre. Additive superposition of the stress wave at the end of the pulse appears in the plot, as has been discussed for discs in \cite{Sievers:1974}. Since the dynamic stress has an oscillatory character, it will combine with the static stress, irrespective of the sign of the latter. The maximum static plus dynamic stress is below the tensile strength of titanium (\SI{220}{\mega \pascal}), and especially that of some titanium alloys (the mechanical properties of which can, however, deteriorate significantly with temperature). Material and fatigue issues in the presence of high temperature and radiation will require additional study.

As a general rule, the stress due to thermal shock of a given pulse length tends to decrease for smaller radii, but the additive superposition mentioned above can also play an important role. Figure~\ref{fig:spheres-LSDyna-0.75mm} shows the stress components for a sphere with radius equal to \SI{0.75}{\milli \meter}. The maximum value of dynamic stress in this case is equal to \SI{52}{\mega \pascal}. The packing fraction is equal to 0.66, the same as for the spheres with \SI{1.5}{\milli \meter} radius.

The value of radius equal to \SI{1.5}{\milli \meter} has been used when studying target cooling to allow for an efficient flow of helium gas through a packed-bed target. 
Reducing the sphere size imposes more severe conditions on the cooling system performance, since smaller radii lead to a higher pressure drop of helium. For this reason, the radius of the spheres was kept equal to \SI{1.5}{\milli \meter} as a baseline. Concerning the stress in the spheres, it would be advantageous to reduce the sphere size. In order to make the final selection of the sphere radius, some experimental results will be required, primarily to test how the sphere size affects the flow of cooling helium through a packed-bed target.

\subsubsection{Magnetic Horn Design}
\label{section:MagneticHornDesign}

Figure~\ref{fig:magnetic-horn-sketch} shows the final optimised shape of the ESS$\nu$SB horn. 

\begin{figure}[h!]
\centering
      \includegraphics[width=0.70\columnwidth]{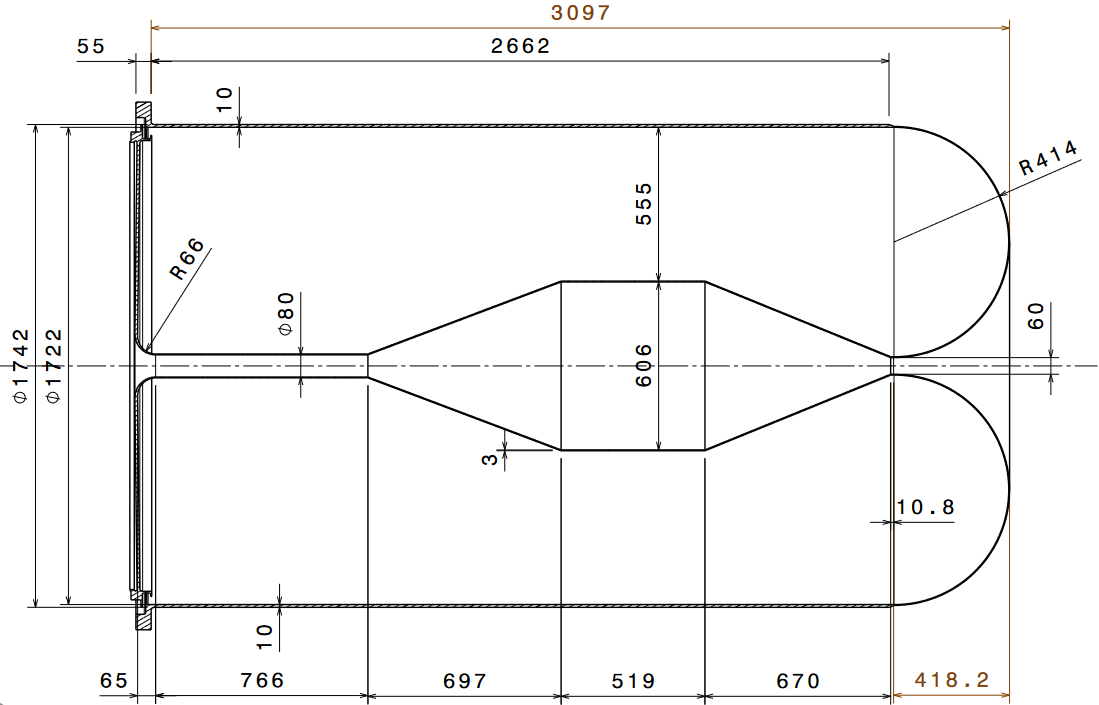}
  \caption{Sketch of the magnetic horn proposed for ESS$\nu$SB.}
  \label{fig:magnetic-horn-sketch}
\end{figure}

Different aspects must be considered in a horn design, including the magnetic field calculation, horn cooling, as well as the mechanical stresses due to magnetic, mechanical, and thermal loading. All results discussed below have been obtained using the \textsc{ANSYS} finite element code. 

%\noindent\textbf{Magnetic field due to current pulsing.} 
\paragraph{\textbf{Magnetic field due to current pulsing}.} 
The magnetic field is generated when the horn undergoes a \SI{100}{\micro \second} current pulse of magnitude \SI{350}{\kilo \ampere}. A semi-sinusoidal current pulse will be provided by a power supply unit, which is discussed in detail in section ~\ref{section:PowerSupplyUnit}. Figure~\ref{fig:horn-magnetic-flux-density} shows the magnetic flux density inside a magnetic horn at time $t=\SI{50}{\micro \second}$, when the current takes on its maximum value. The maximum calculated magnetic flux density is about~\SI{2.1}{\tesla}.

\begin{figure}[h!]
\centering
  \includegraphics[width=0.5\columnwidth]{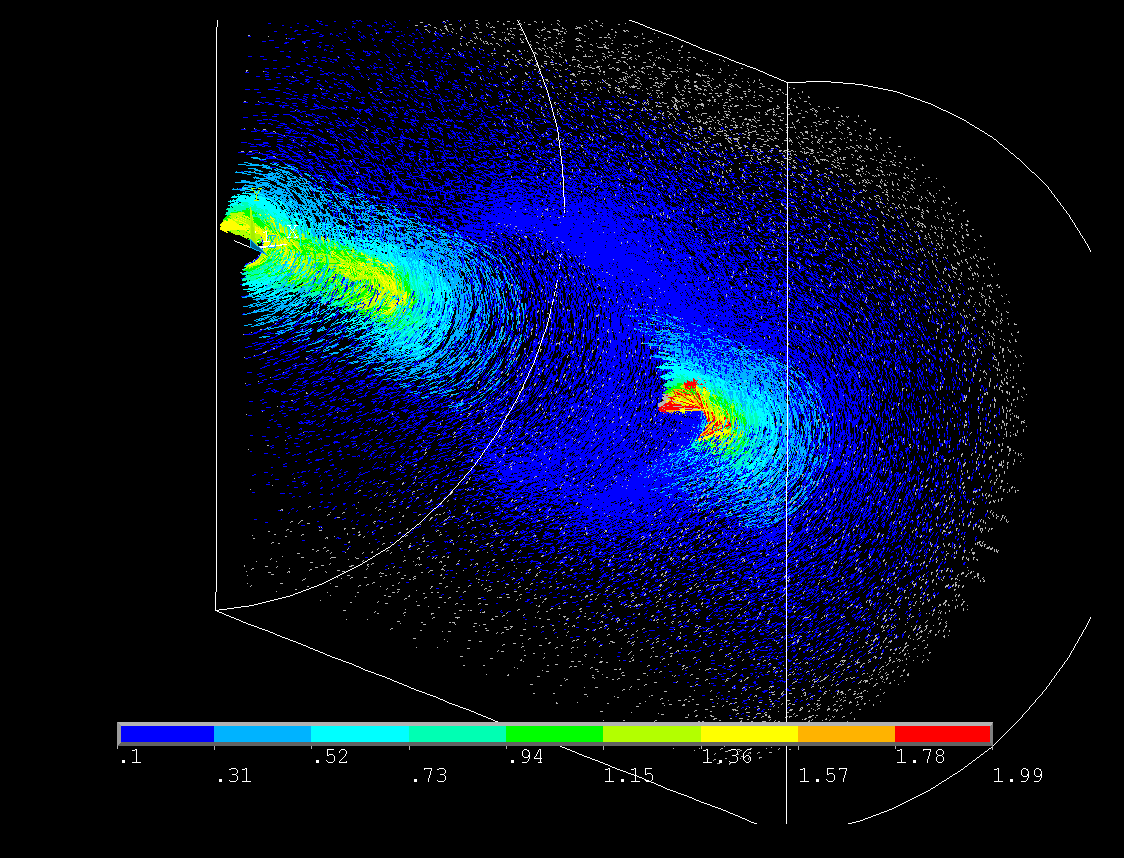}
  \caption{Magnetic flux density in tesla inside a horn and surrounding space, at time \SI{50}{\micro \second} when the value of current is equal to \SI{350}{\kilo \ampere} (after the current discharge).}
  \label{fig:horn-magnetic-flux-density}
\end{figure}

%\noindent\textbf{Power deposition inside the magnetic horn.} 
\paragraph{\textbf{Power deposition inside the magnetic horn}.} 

The power deposited inside the magnetic horn during each cycle comes primarily from two sources: the secondary particles leaving the target during a beam impact (Fig.~\ref{fig:horn-power-secondary-particles}) and the Joule heat losses caused by the high-amplitude current pulsing through the horn skin (Fig.~\ref{fig:horn-power-joule-heat}). In calculating Joule heat losses, an electromagnetic analysis has been performed, which accounts for the skin effect in the horn wall. A summary of the power deposited in different horn sections can be found in Table~\ref{tab:horn-power-deposition}.

\begin{figure}[b!]
\centering
  \includegraphics[width=0.75\columnwidth]{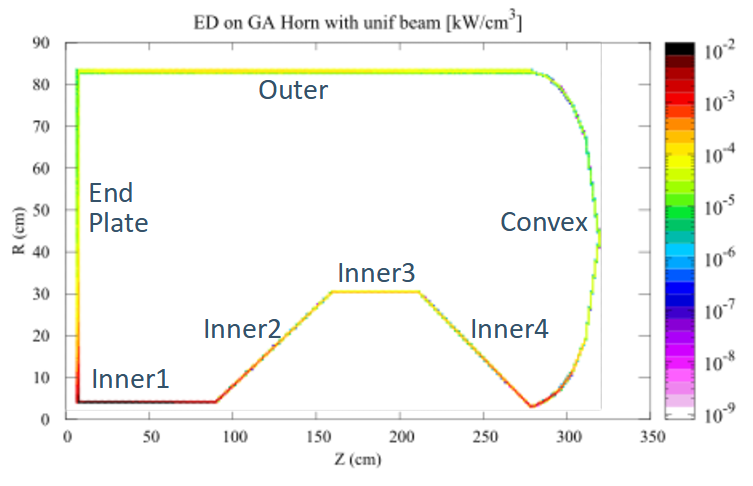}
  \caption{Power density in \SI[per-mode=symbol]{}{\kilo \watt \per \centi \meter \cubed} inside the magnetic horn skin due to secondary particles.}
  \label{fig:horn-power-secondary-particles}
\end{figure}

\begin{figure}[b!]
\centering
  \includegraphics[width=0.7\columnwidth]{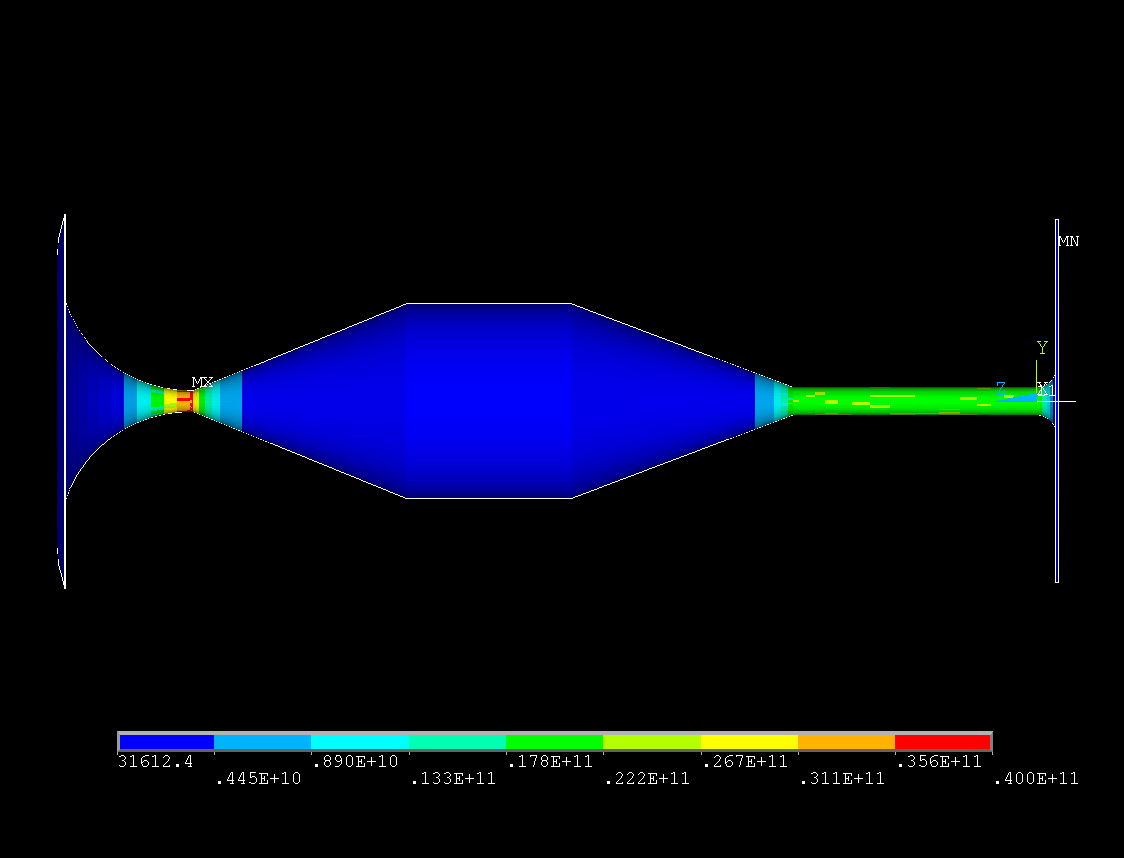}
  \caption{Joule power density per unit volume in \SI[per-mode=symbol]{}{\watt \per \meter \cubed} in the horn skin, at time \SI{50}{\micro \second} (horn outer skin is not shown).}
  \label{fig:horn-power-joule-heat}
\end{figure}

\begin{table}[h!]
    \centering
    \caption{Power deposition inside the horn by secondary particles and Joule heat losses.}
    \label{tab:horn-power-deposition}
    \begin{tabular}{c|c c | c}
        & Sec. Part. [\SI{}{\kilo \watt}] & Joule Heat [\SI{}{\kilo \watt}] & Total [\SI{}{\kilo \watt}] \\
        \hline
         End Plate & 5.28 & 1.25 & 6.53\\
         Inner1 & 9.99 & 5.80 & 15.79\\
         Inner2 & 2.37 & 1.79 & 4.16\\
         Inner3 & 1.00 & 0.54 & 1.54\\
         Inner4 & 1.50 & 1.91 & 3.41\\
         Convex & 1.74 & 2.60 & 4.34\\
         Outer & 13.97 & 1.07 & 15.04\\
         \hline
         Total & 35.85 & 14.96 & 50.81\\
    \end{tabular}

\end{table}

%\noindent\textbf{Estimation of the required water mass flow rates.} 

\paragraph{\textbf{Estimation of the required water mass flow rates}.} 
Based on the balance of energy transferred to the water from each of the separate horn sections and the increase in the internal energy of cooling water, the mass flow rate of water required for the operation of the cooling system can be calculated from the formula: 

\begin{equation}
    \dot{m} = \frac{P_{SP}+ P_{JH}}{c_w \Delta T_w}
\end{equation}

\noindent in which: $P_{SP}$ and $P_{JH}$ is the power from the secondary particles and from Joule heat losses in a given section, respectively; $c_w$ is the specific heat of water; and $\Delta T_w$ is an increase in the water temperature.

Assuming the specific heat of water $c_w=\SI[per-mode=symbol]{4193}{\joule \per \kilogram \per \kelvin}$ and the increase in water temperature due to the transmission of energy from the cooled surface equal to $\Delta T_w = \SI{10}{\celsius}$, the mass flow rate for each section of the horn can be estimated as listed in Table~\ref{tab:horn-mass-flow-rate-estimation}, and sketched in Fig.~\ref{fig:horn-cooling-estimation}.

\begin{table}[h!]
    \centering
    \caption{Mass flow rate for different horn sections, assuming a water temperature increase of $\Delta T_w = \SI{10}{\celsius}$}
    \label{tab:horn-mass-flow-rate-estimation}
    \begin{tabular}{c|c c}
       & $\dot{m}$ [\SI[per-mode=symbol]{}{\kilogram \per \second}] & Fraction [\%] \\
        \hline
         End Plate  & 0.156 & 12.82 \\
         Inner1  & 0.378 & 31.06 \\
         Inner2  & 0.100 & 8.21 \\
         Inner3  & 0.037 & 3.04 \\
         Inner4  & 0.082 & 6.74 \\
         Convex  & 0.104 & 8.55 \\
         Outer  & 0.360 & 29.58 \\
         \hline
         Total & 1.217 & 100 \\
    \end{tabular}
\end{table}

\begin{figure}[h!]
\centering
  \includegraphics[width=0.50\columnwidth]{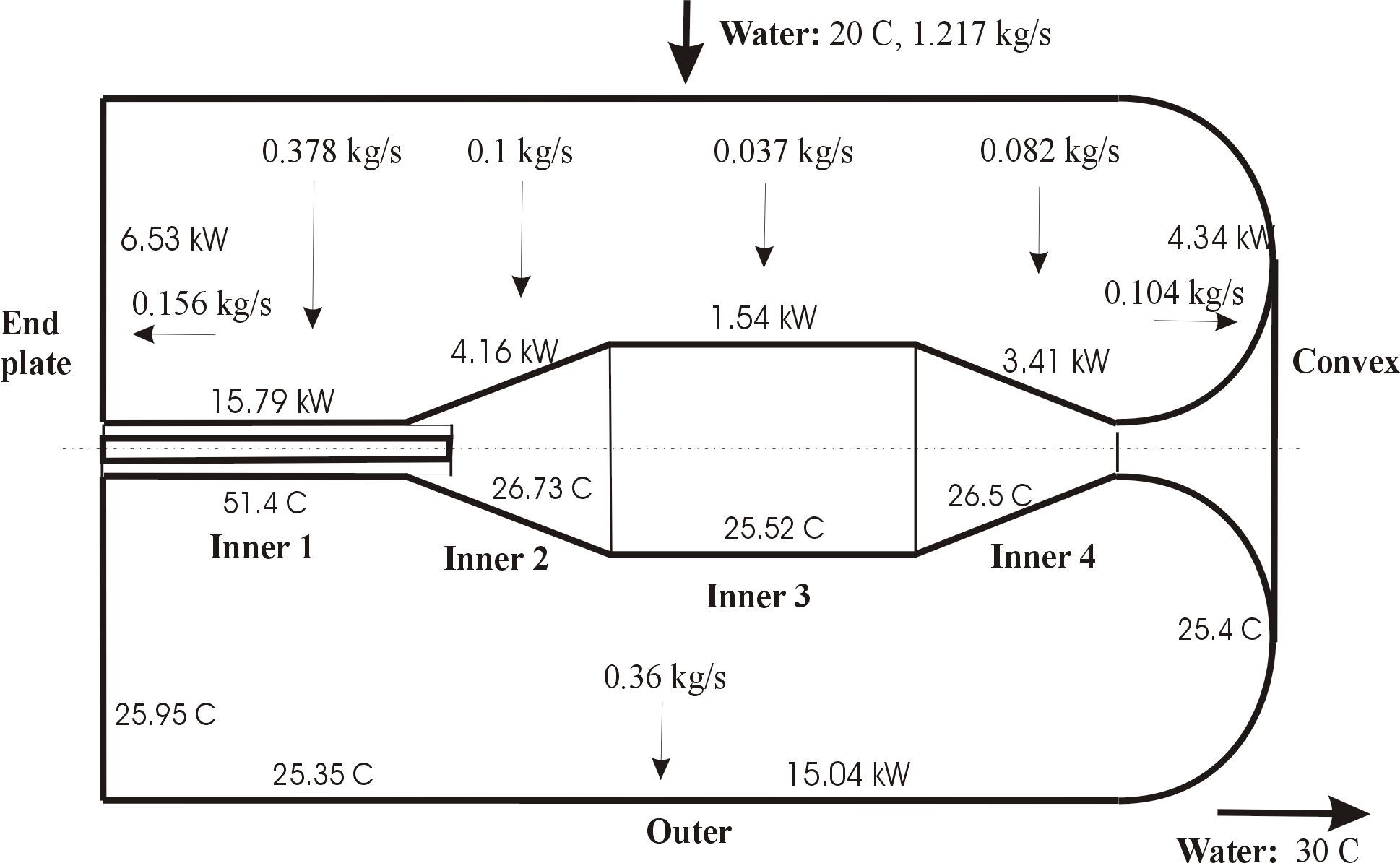}
  \caption{Model of the horn with values for each section of power deposition, required mass flow rate, and temperature.}
  \label{fig:horn-cooling-estimation}
\end{figure}

%\noindent\textbf{Mechanical stresses in the horn.} 
\paragraph{\textbf{Horn average temperature in steady-state operation}.} Given the power deposition by the secondary particles and Joule heat losses included in Table \ref{tab:horn-power-deposition}, the average temperature of each horn section has been estimated analytically. The value of the heat transfer coefficient h~=~\SI[per-mode=symbol]{3000}{\watt \per \meter \squared \per \kelvin} has been used for the entire horn surface. The results are collected in Table~\ref{tab:horn-avg-temperature-estimation}. 

\begin{table}[h!]
    \centering
    \caption{Average horn temperature for each section of the horn, assuming a heat transfer coefficient h~=~\SI[per-mode=symbol]{3000}{\watt \per \meter \squared \per \kelvin} and water temperature $T_w = \SI{25}{\celsius}$}.
    \label{tab:horn-avg-temperature-estimation}
    \begin{tabular}{c|c c}
       & $A$ [\SI{}{\meter \squared}] & $T_{max}$ [\SI[per-mode=symbol]{}{\celsius}] \\
        \hline
         End Plate & 2.294 & 25.95 \\
         Inner1 & 0.199 & 51.40 \\
         Inner2 & 0.803 & 26.73 \\
         Inner3  & 0.988 & 25.52 \\
         Inner4  & 0.757 & 26.50 \\
         Convex  & 3.63 & 25.40 \\
         Outer  & 14.52 & 25.35 \\
    \end{tabular}
\end{table}

Assuming a cooling water temperature of $T_w = \SI{25}{\celsius}$, the highest temperature (\SI{51.40}{\celsius}) is expected to occur on the horn walls in the direct vicinity of the integrated target (the ``Inner1'' section in Fig.~\ref{fig:horn-cooling-estimation}).

Numerical analyses have also been performed with respect to the cooling of the horn, using the same values of deposited power. The temperature calculated using \textsc{ANSYS} is \SI{51.26}{\celsius}, which is comparable to the results obtained using the simplified analytical approach. 

%\noindent\textbf{Mechanical stresses in the horn.} 
\paragraph{\textbf{Mechanical stresses in the horn}.} 
\label{sec:mechanical-stresses-horn}

Figure~\ref{fig:MaxStressDisp} gives the results of the structural analysis of a magnetic horn under a single current pulse. The maximum stress in the horn shell is equal to \SI{28.2}{\mega \pascal} and the maximum longitudinal displacement is equal to \SI{0.1}{\milli \meter}. The maximum values occur at different horn locations, but also at different times. These values fall entirely within the acceptable limits.

\begin{figure}[h!]
\centering
  \begin{subfigure}[b]{0.45\columnwidth}
  \includegraphics[width=\columnwidth]{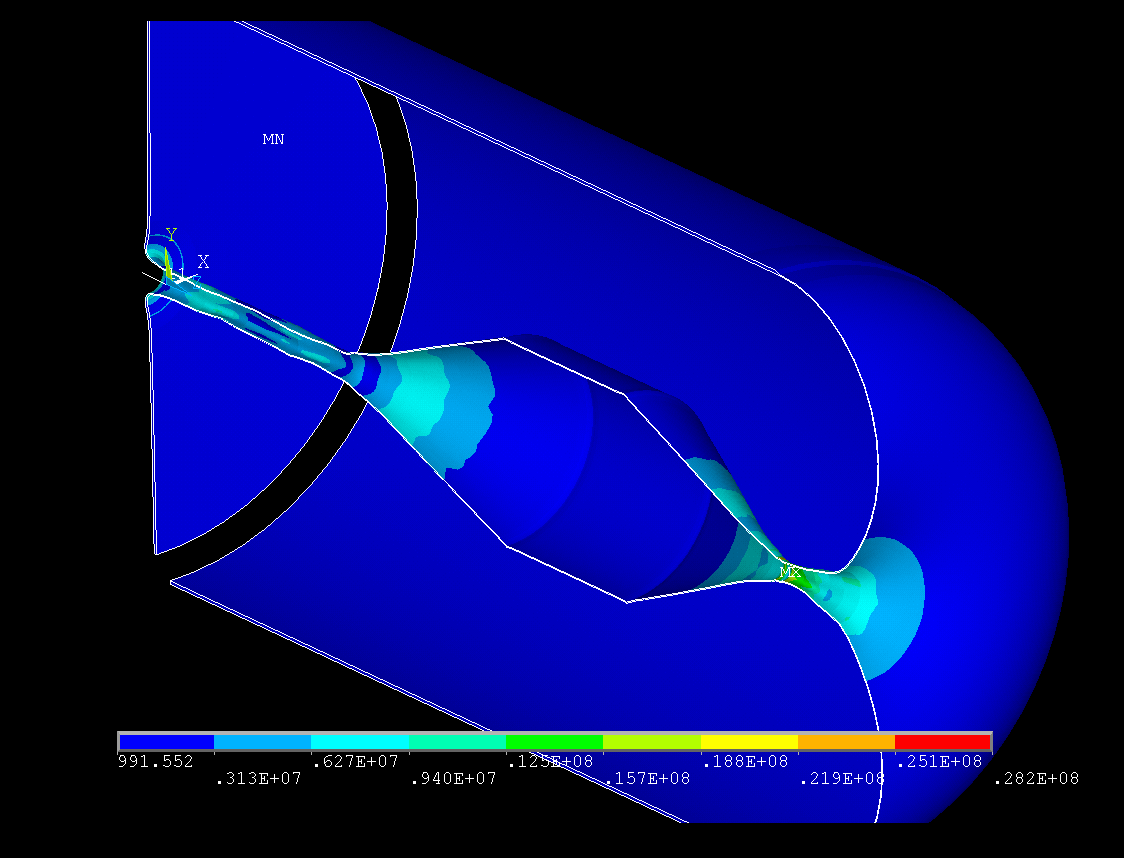}
  \caption{Maximum stress levels in the horn skin due to magnetic forces (at $t = \SI{90}{\micro \second}$).}
  \label{fig:MaxStressDispA}
  \end{subfigure}
  \hfil
  \begin{subfigure}[b]{0.45\columnwidth}
  \includegraphics[width=\columnwidth]{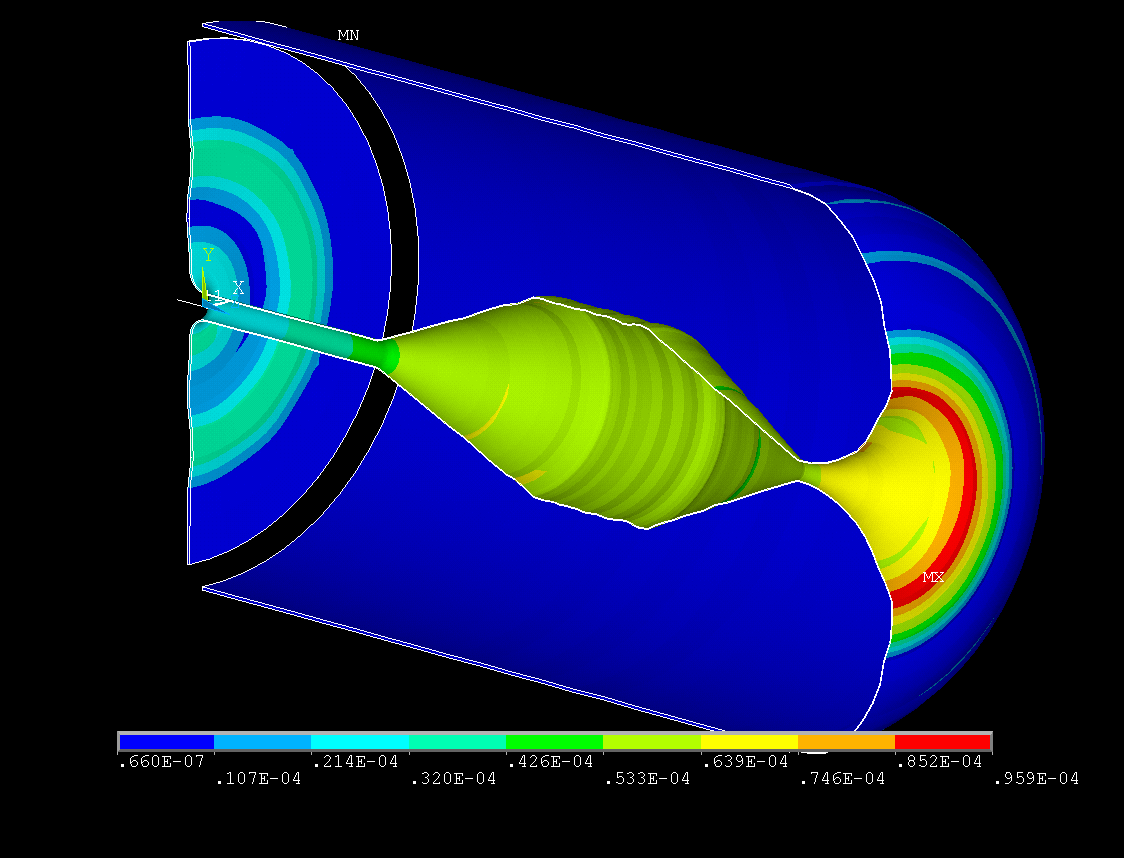}
  \caption{Maximum longitudinal displacement from magnetic forces (at $t = \SI{64}{\milli \second}$).}
  \label{fig:MaxStressDispB}
  \end{subfigure}
  \caption{Results of horn structural analysis due to magnetic forces.}
  \label{fig:MaxStressDisp}
\end{figure}

%\begin{figure}[h!]
%\centering
%  \begin{subfigure}[b]{0.45\columnwidth}
%  \includegraphics[width=\columnwidth]{figures/targetstation/stresses/Phi80Valerie_SS_2fix_2kW_temp.png}
%  \caption{heat transfer coef. h = \SI[per-mode=symbol]{2}{\kilo \watt \per \meter \squared \per \kelvin}}
  \label{fig:HornTemperature2000}
% \end{subfigure}
% \hfil
% \begin{subfigure}[b]{0.45\columnwidth}
% \includegraphics[width=\columnwidth]{figures/targets%tation/stresses/Phi80Valerie_SS_2fix_3kW_temp.png}
%\caption{heat transfer coef. h = \SI[per-mode=symbol]{3}{\kilo \watt \per \meter %\squared \per \kelvin}}
%  \label{fig:HornTemperature3000}
%  \end{subfigure}
%  \caption{Temperature map inside the horn due to the power deposited 
%by secondary particles and Joule heat losses}
%  \label{fig:HornTemperature}
%\end{figure}

%--
% To be clarified how these temperatures related to the previous paragraph
%
%Figure~\ref{fig:HornTemperature} shows the steady-state horn temperature distribution, brought about by both  the secondary particles and Joule heat losses. Two different values of heat transfer coefficient between the surface of the horn and the cooling water have been assumed in this calculation: \SI[per-mode=symbol]{2000}{} and \SI[per-mode=symbol]{3000}{\watt \per \meter \squared \per \kelvin}. The bulk temperature of water was taken to be equal to \SI{303.15}{\kelvin} (\SI{30}{\celsius}). The maximum $\Delta T$ (difference between the horn skin and the bulk temperature) is \SI{40.6}{\kelvin} in the former case and \SI{27.4}{\kelvin} in the latter.\\

The steady-state stress levels due to the calculated thermal load are shown in Fig.~\ref{fig:HornTemperatureStress}, for the heat transfer coefficient of \SI[per-mode=symbol]{3000}{\watt \per \meter \squared \per \kelvin}. The maximum steady-state stress is low (below \SI{10}{\mega \pascal}).

\begin{figure}[H]
    \centering
    \includegraphics[width=0.55\columnwidth]{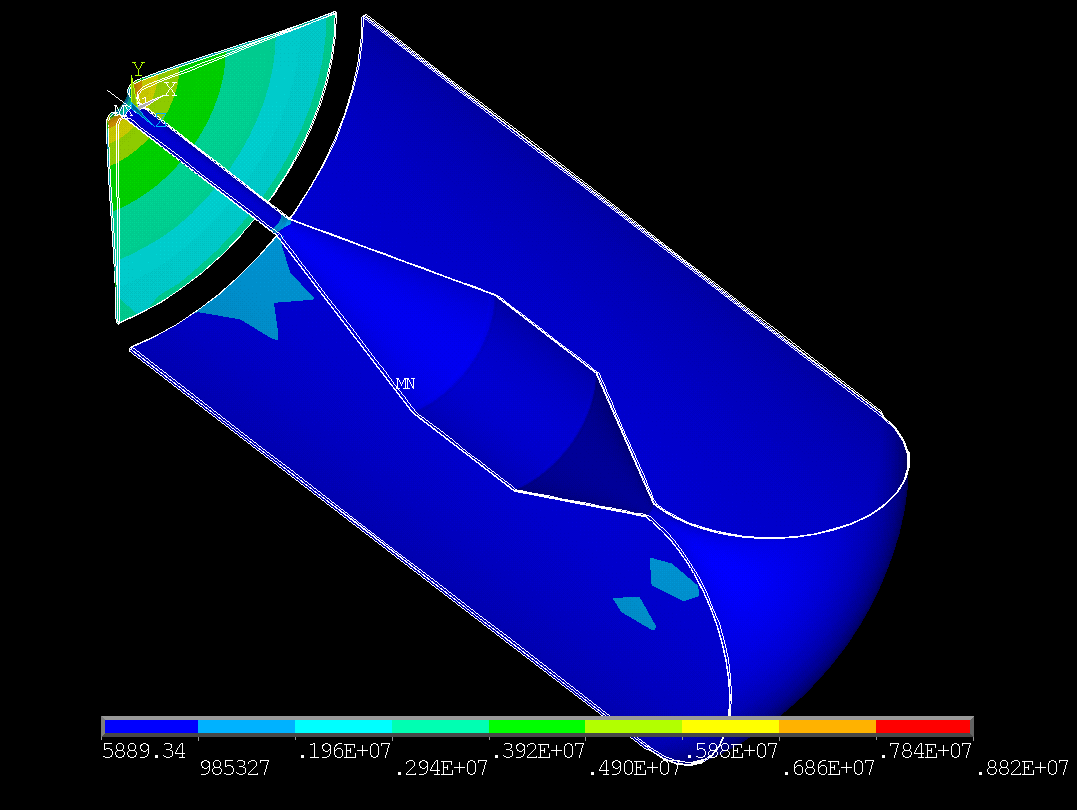}
    \caption{Thermal stress in a magnetic horn, assuming heat transfer coefficient between horn and water h = \SI[per-mode=symbol]{3000}{\watt \per \meter \squared \per \kelvin}.}
    \label{fig:HornTemperatureStress}
\end{figure}

\subsubsection{Target-horn integration issues}
\label{TargetIntegration}

Figure~\ref{fig:integration-shell-in-out} shows the direction of gas flow through the target shell, with the gas inlet located at the bottom and the gas outlet at the top of the shell. Figure~\ref{fig:integration-cross-section} shows the cross-section of the proposed integration concept of a target inside a horn. The design consists of: the granular target spheres placed inside a~\SI{2}{\milli \meter} thick titanium container (in red), the outer \SI{2}{\milli \meter} thick shell (also in red), two \SI{2}{\milli \meter} thick plates which connect the target container to the outer shell, and the horn inner conductor (in violet). There is a \SI{1}{\milli \meter} thick layer of gaseous helium (in blue) between the outer shell and the horn inner conductor, which provides thermal insulation. The location of the attachment of the plates has been chosen for two reasons: Firstly, they separate the lower channel, into which cooling helium is pumped, from the upper channel, from which hot helium will be extracted. Secondly, they act as a link between the target container, target shell and the horn wall. They are also placed in a region of a relatively low temperature, which ensures the acceptable levels of thermal stress. 

\begin{figure}[h!]
\begin{center}
\begin{minipage}{0.47\linewidth}
\includegraphics[width=0.99\textwidth]{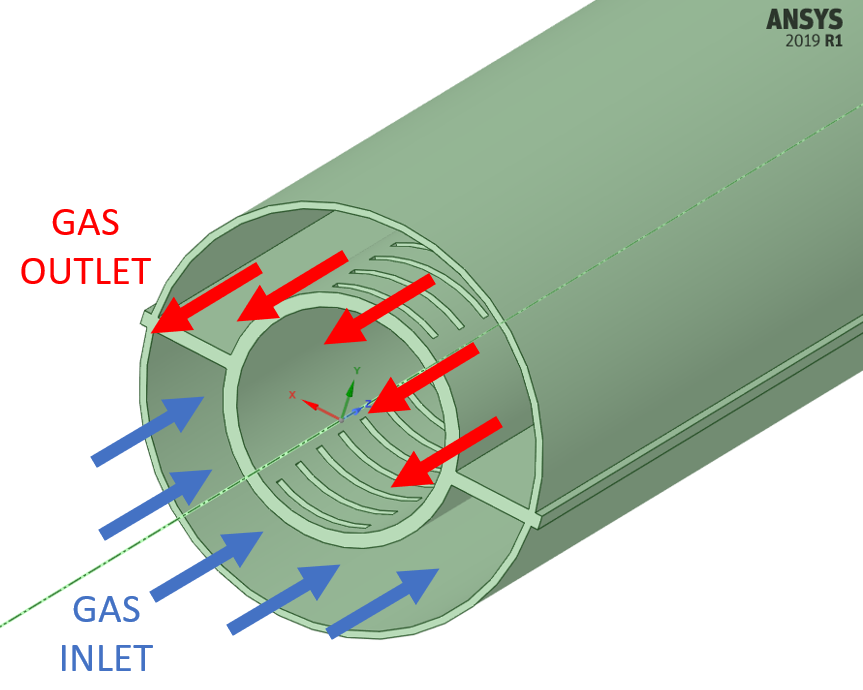}
\caption{\label{fig:integration-shell-in-out}Direction of helium inflow and outflow.}
\end{minipage}\hspace{2pc}%
\begin{minipage}{0.47\linewidth}
\includegraphics[width=0.99\textwidth]{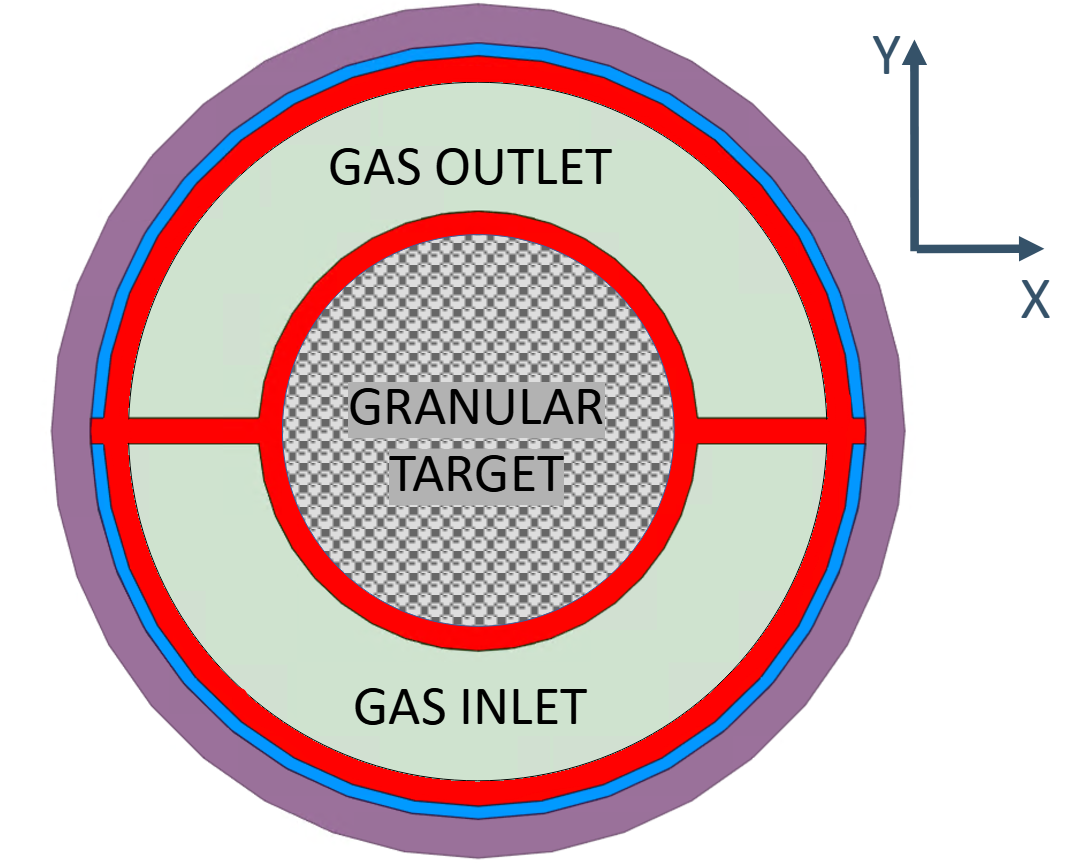}
\caption{\label{fig:integration-cross-section}Cross-section of the proposed horn-target integration (horn in purple, target container, target shell and connecting plates in red, thermal isolation in blue) used for the calculations.}
\end{minipage} 
\end{center}
\end{figure}

Initially, three gas inlet and three gas outlet channels were considered for cooling the target, as was proposed in the EUROnu project. However, the calculations have shown that for such a layout most of the gas tends to flow directly towards the outlets, omitting the central area of the target located around the z-axis, where the most power is deposited. For this reason, in order to ensure the flow of the gas through the centre of the granular target, a design has been proposed in which the gas inlet is located on the opposite side of the gas outlet, as shown in Fig.~\ref{fig:integration-shell-in-out}. 

Figure~\ref{fig:shell-holes} shows the top-down view of a layout of the openings in the target container, to allow for the transverse flow of helium through the granular target. The width of each opening is \SI{1.5}{\milli\meter}, the distance between them is \SI{4.59}{\milli\meter}(there are 128 holes along the shell length of \SI{78}{\centi\meter}), and the total surface of the holes is equal to twice the overall cross-section area of the gas inlet and outlet channels. The size and distribution of the openings at the bottom (gas inlet) and the top (gas outlet) of the target container are the same. 

\begin{figure}[h!]
\centering
  \includegraphics[width=0.9\columnwidth]{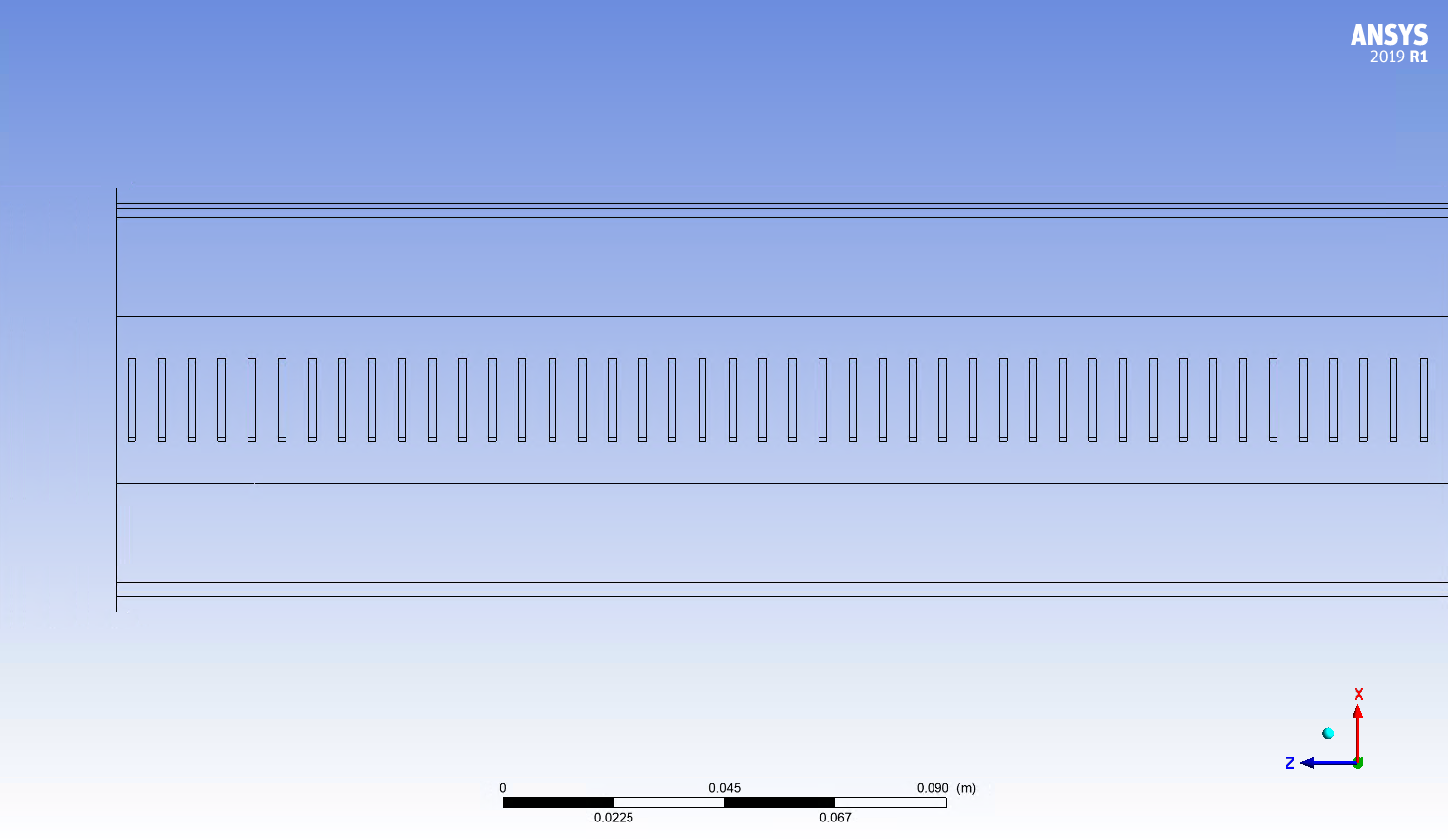}
  \caption{Pattern of helium entry and exit openings of uniform size in the target container.}
  \label{fig:shell-holes}
\end{figure}

Figure~\ref{fig:results-titanium-80mm-same} shows the numerical results obtained with inner horn diameter $d = \SI{80}{\milli \meter}$, helium mass flow rate $\dot{m} = \SI[per-mode=symbol]{0.3}{\kilogram \per \second}$, power deposition $P = \SI{138.53}{\kilo \watt}$, and the titanium granular target modelled as a porous medium. It has also been assumed in these calculations that the heat transfer coefficient at the surface of the horn sprayed by water jets is $\SI[per-mode=symbol]{3000}{\watt \per \meter \squared \per \kelvin}$, and that the space between the shell and the horn is filled with helium.

Figure~\ref{fig:shell-disp-stress-a} shows the temperature in the outer shell, under the conditions specified above. Assuming that the container and the outer shell are connected by plates along their length, and that the plates are fixed on their outer edges, the maximum deflection of 0.28\,mm occurs at the end of the outer shell, whereas the maximum von Mises stress in the outer shell is around \SI{100}{\mega \pascal}. This is the maximum value of stress, except for very localised stress concentrations at both ends of the plates, which will depend on the plate-attachment configuration. The plots of the displacement and von Mises stress are shown in Figs.~\ref{fig:shell-disp-stress-b} and \ref{fig:shell-disp-stress-c}, respectively.

As far as the electrical insulation is concerned, the horns will be insulated from both the support frame and the cooling manifold. Additionally, the target must not be grounded.

\begin{figure}[H]
\centering
  \begin{subfigure}[b]{0.45\columnwidth}
  \includegraphics[width=\columnwidth]{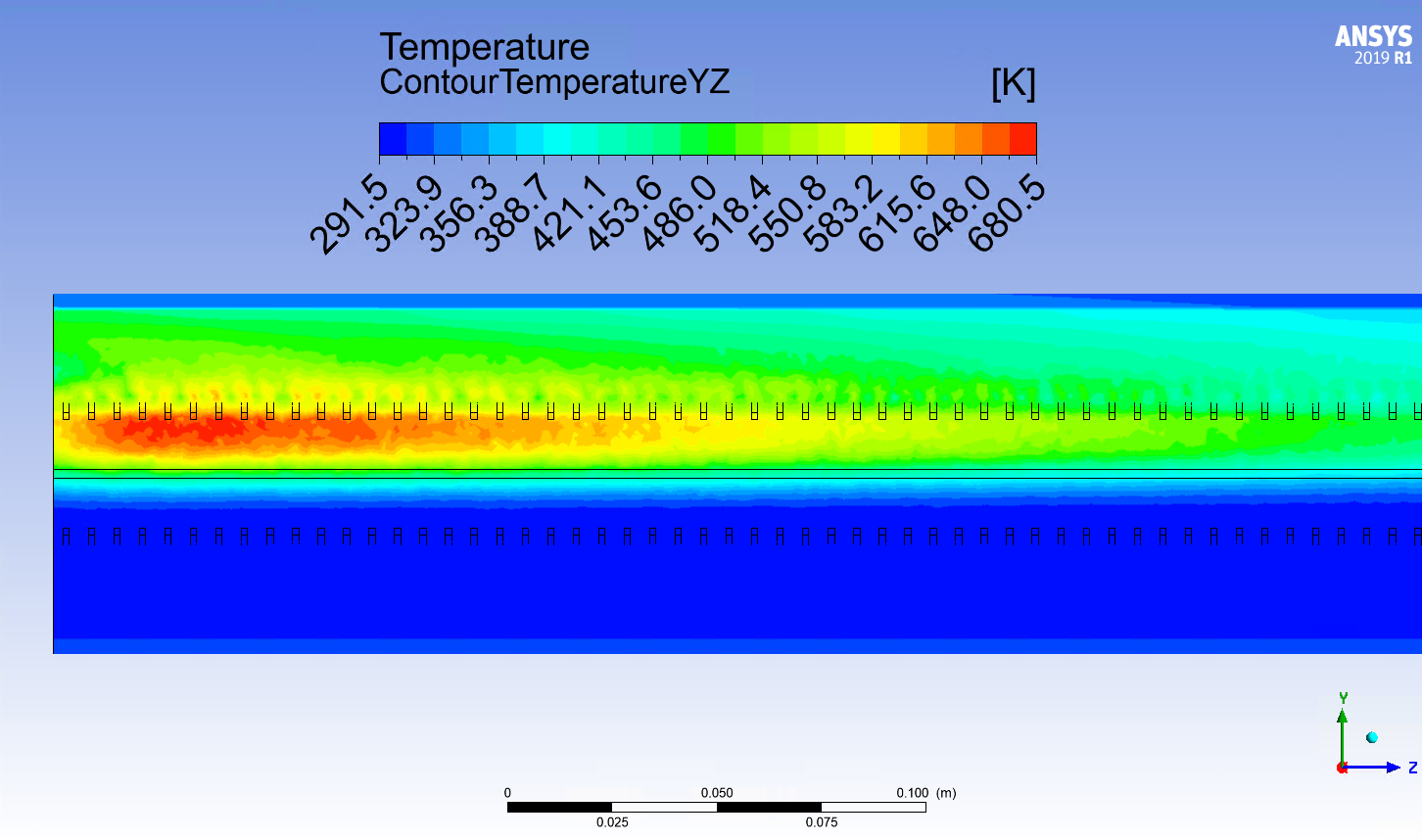}
  \caption{Helium temperature.}
  \label{fig:mass-flow-03-temperature}
  \end{subfigure}
  \hfil
  \begin{subfigure}[b]{0.45\columnwidth}
  \includegraphics[width=\columnwidth]{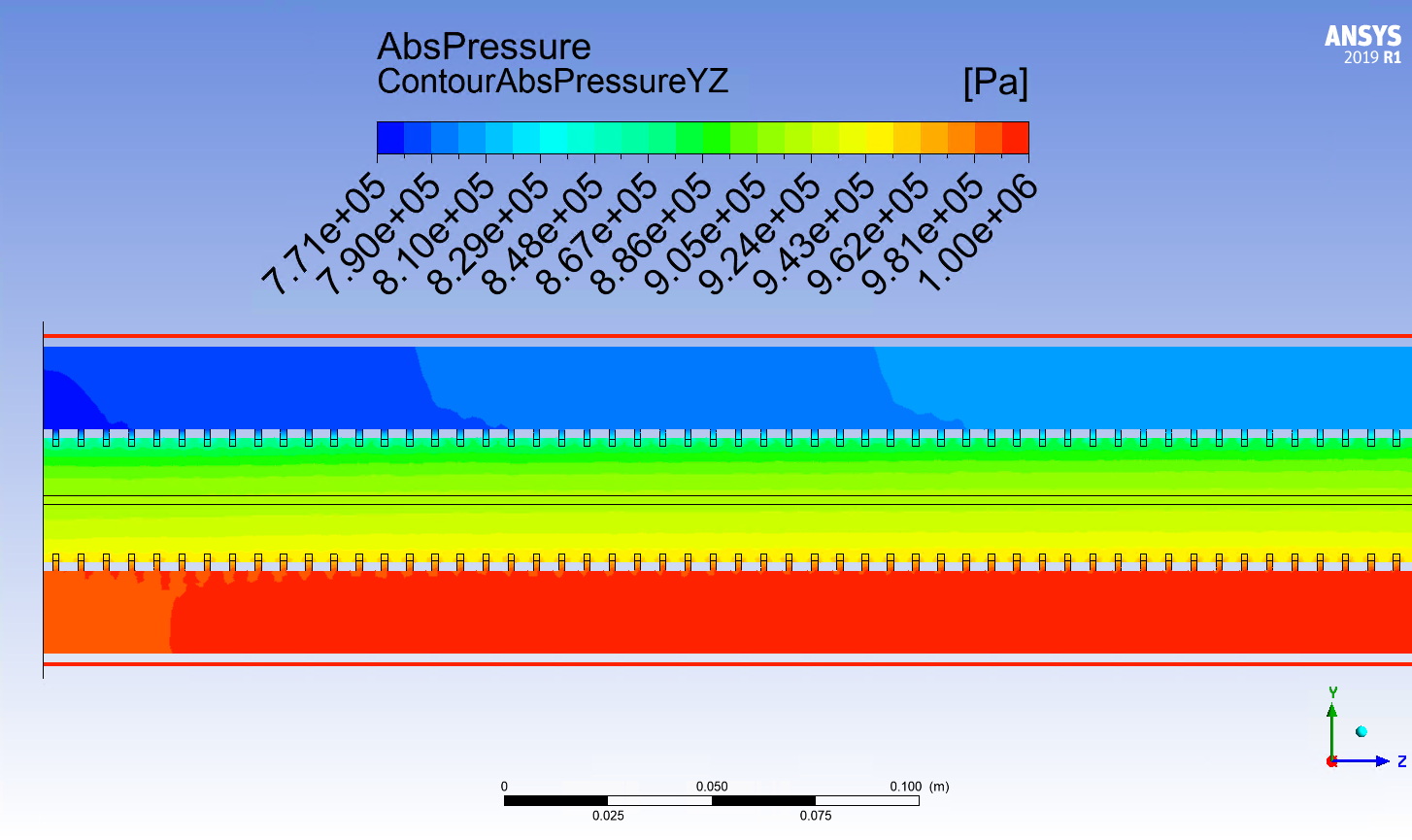} 
  \caption{Helium pressure.}
  \label{fig:mass-flow-03-pressure}
  \end{subfigure}\\
  \begin{subfigure}[b]{0.45\columnwidth}
  \includegraphics[width=\columnwidth]{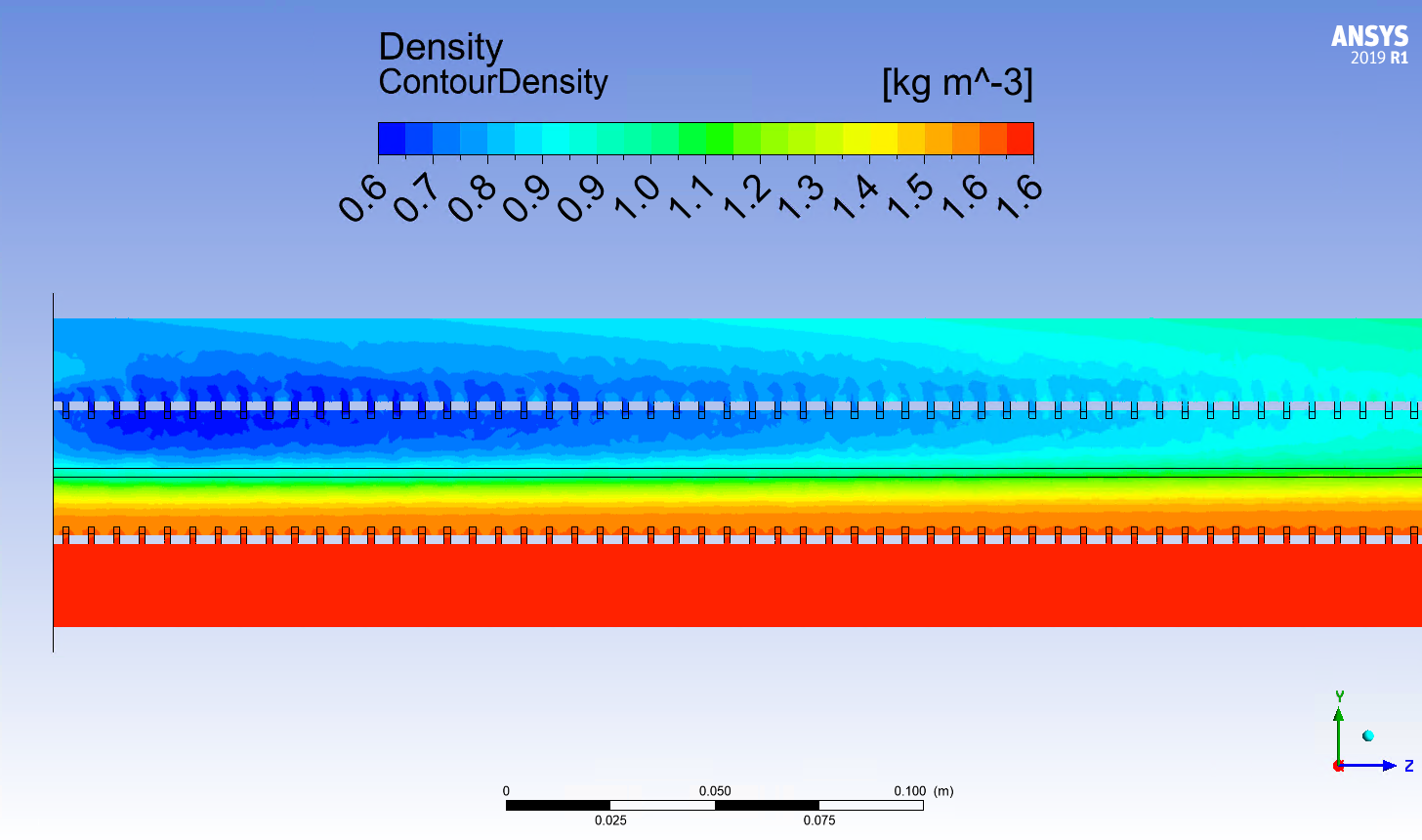}
  \caption{Helium density.}
  \label{fig:mass-flow-03-density}
  \end{subfigure}
  \hfil
  \begin{subfigure}[b]{0.45\columnwidth}
  \includegraphics[width=\columnwidth]{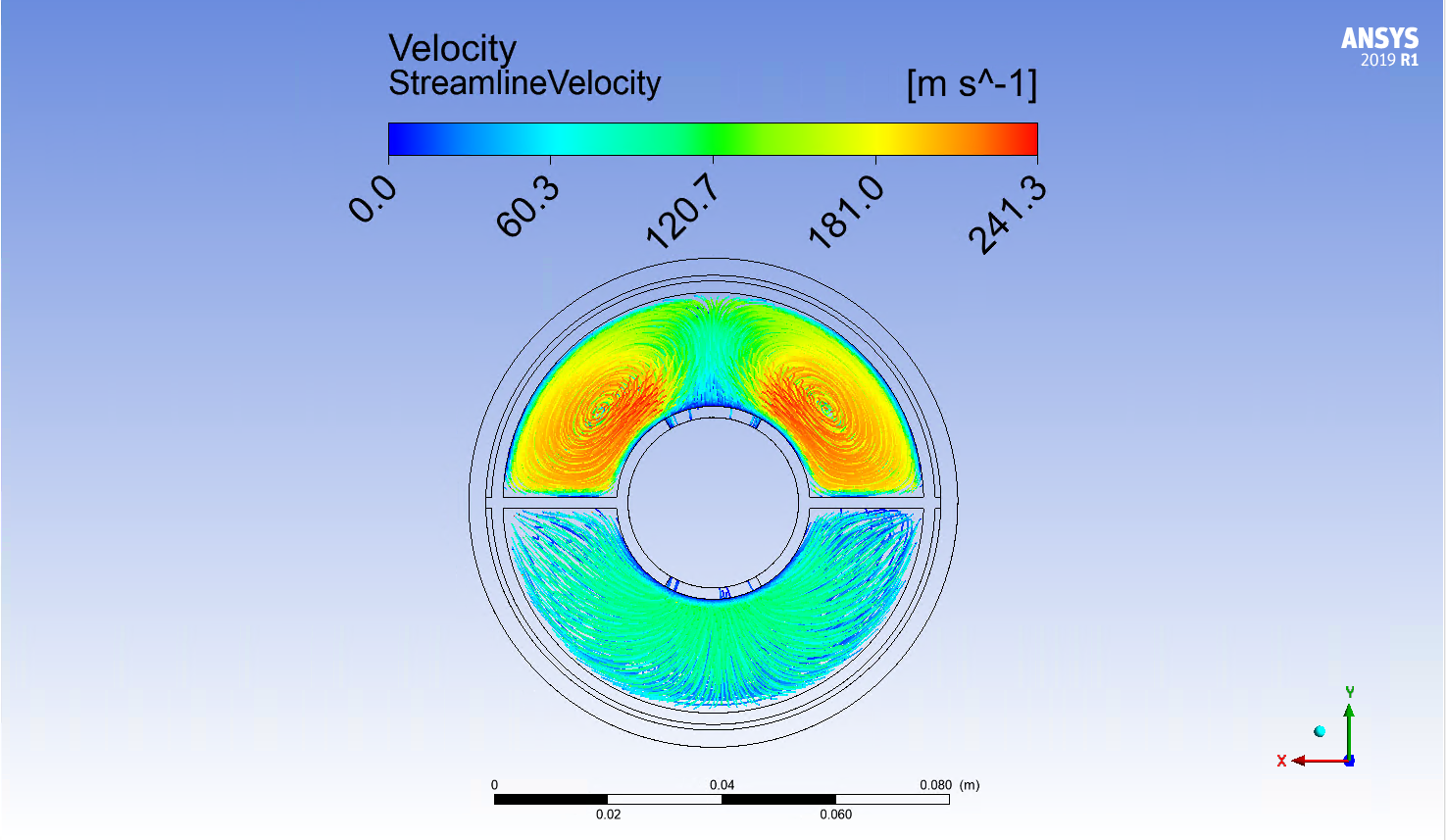}
  \caption{Helium velocity streamlines.}
  \label{fig:mass-flow-03-velocity}
  \end{subfigure}
  \caption{Numerical results for helium flow under a steady-state power deposition condition for titanium spheres, inner horn diameter $d=\SI{80}{\milli \meter}$ and the pattern of shell holes of uniform size.}
  \label{fig:results-titanium-80mm-same}
\end{figure}

\begin{figure}[h!]
     \centering
     \begin{subfigure}[b]{0.45\columnwidth}
         \centering
         \includegraphics[width=\columnwidth]{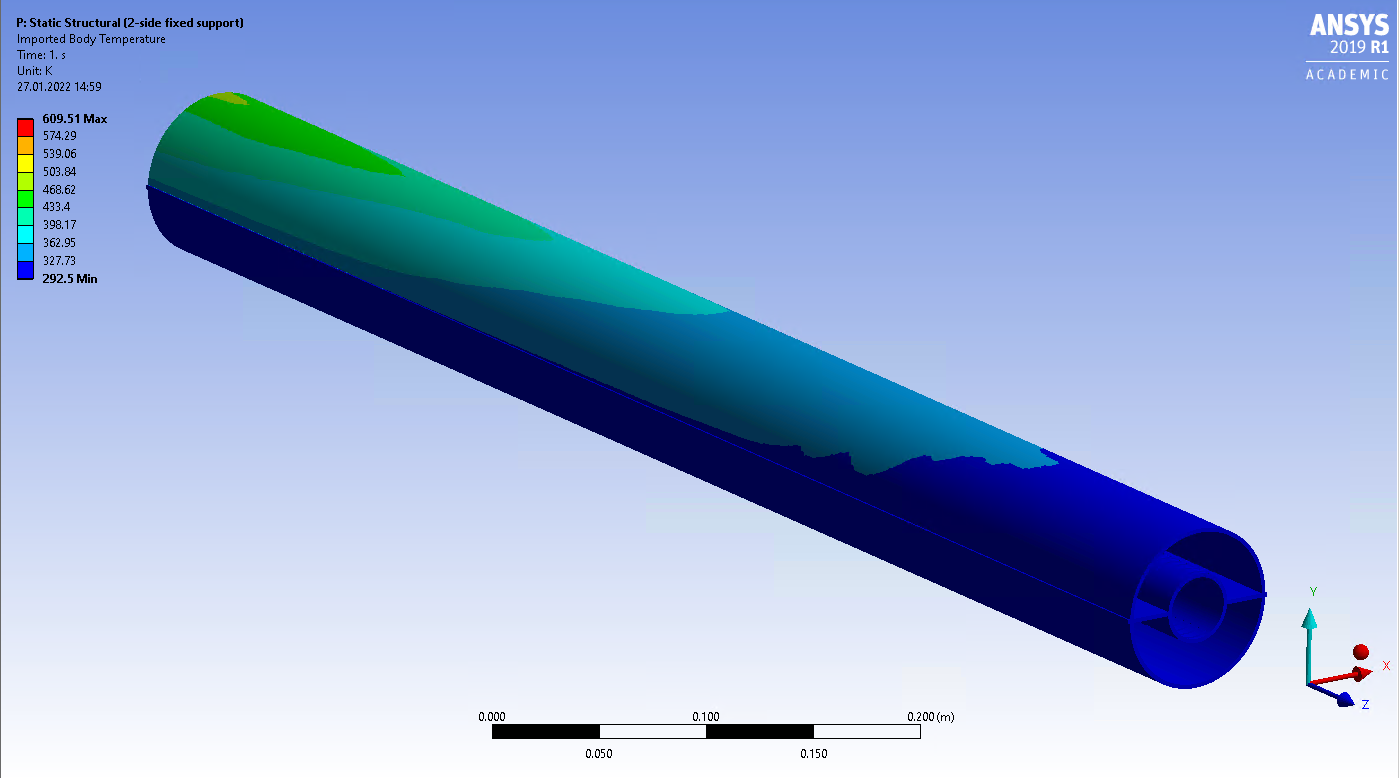}
        \caption{Temperature in the outer shell}
        \label{fig:shell-disp-stress-a}
     \end{subfigure}
     \\
     \begin{subfigure}[b]{0.45\columnwidth}
         \centering
         \includegraphics[width=\columnwidth]{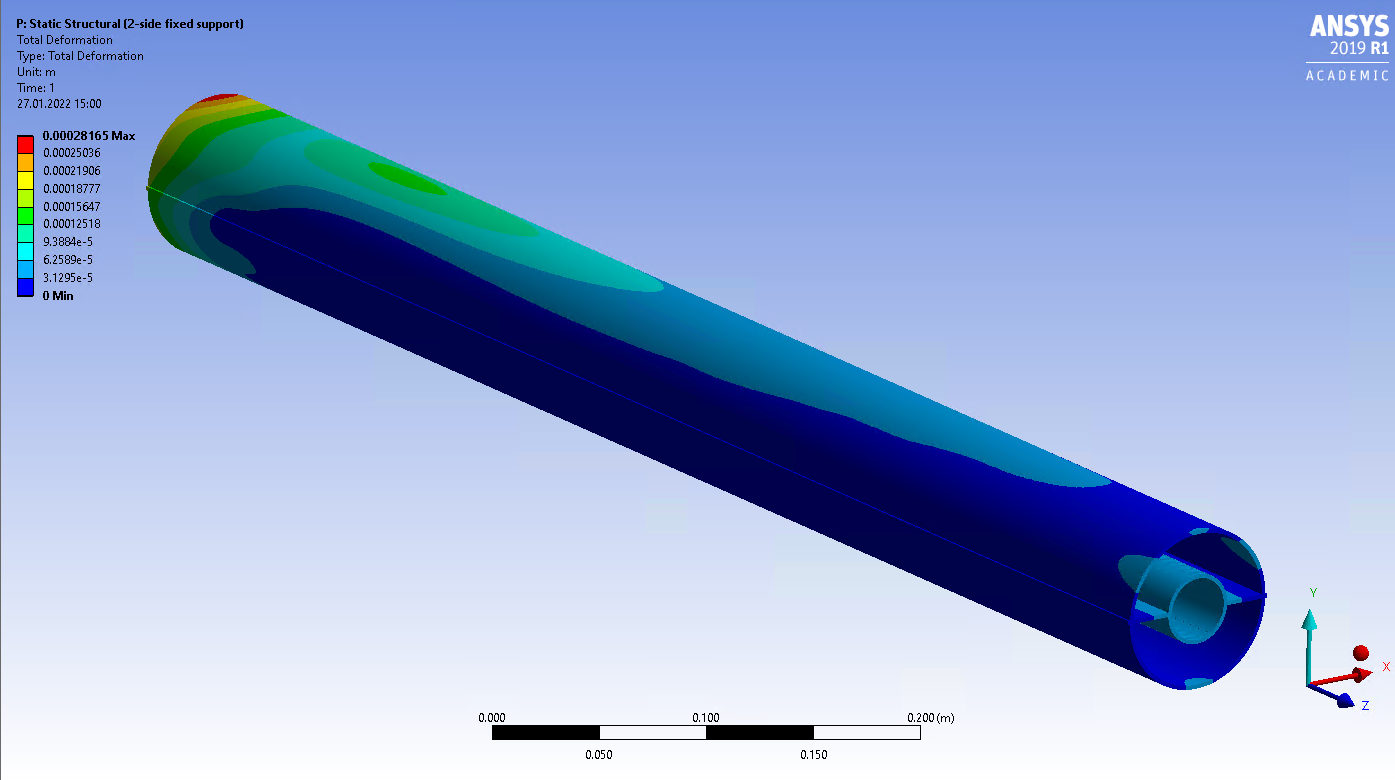}
         \caption{Shell deformation}
         \label{fig:shell-disp-stress-b}
     \end{subfigure}
     \hfill 
     \begin{subfigure}[b]{0.45\columnwidth}
         \centering
         \includegraphics[width=\columnwidth]{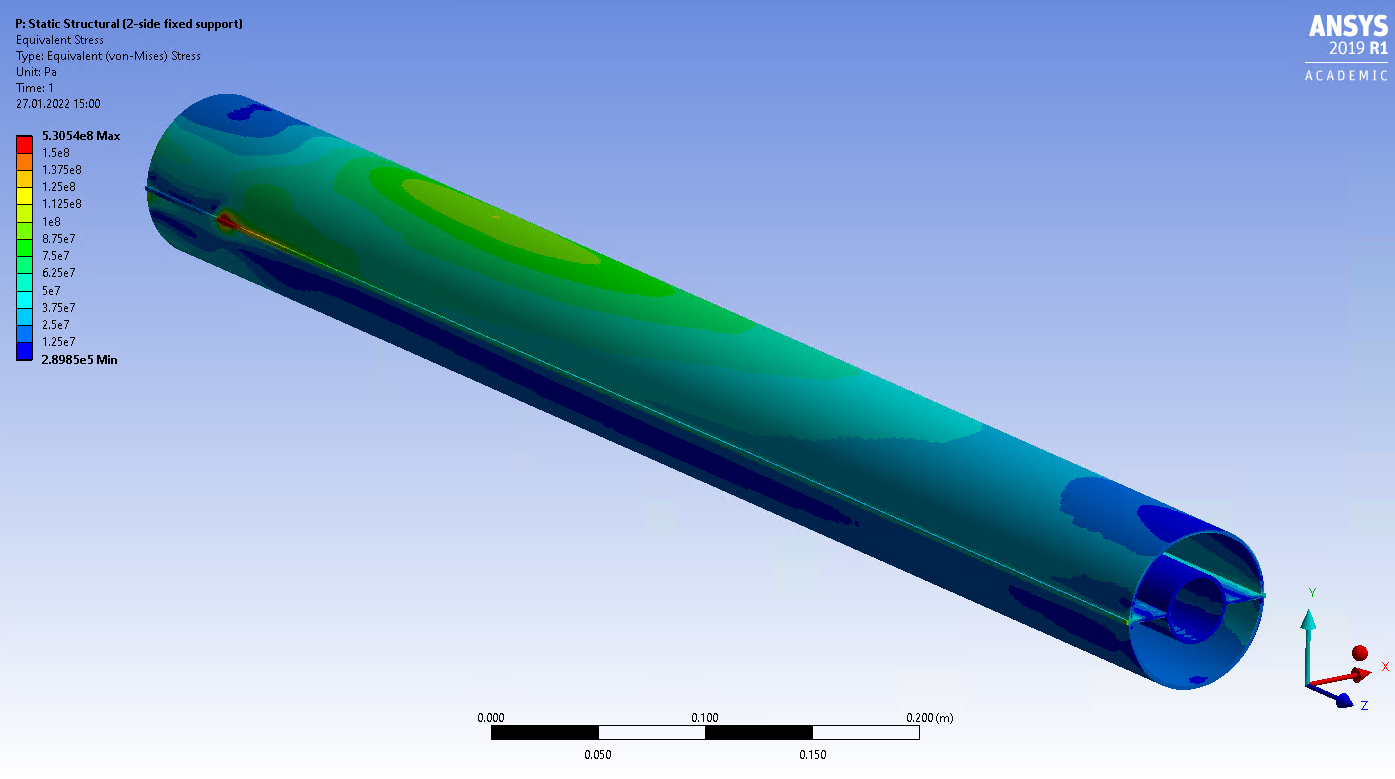}
         \caption{von Mises stress}
         \label{fig:shell-disp-stress-c}
     \end{subfigure}
        \caption{Mechanical analysis of the shell and target container}
        \label{fig:shell-disp-stress}
\end{figure}

\subsubsection{Four-Horn Support System}

\begin{figure}[b!]
  \centering
  \includegraphics[width=0.8\columnwidth]{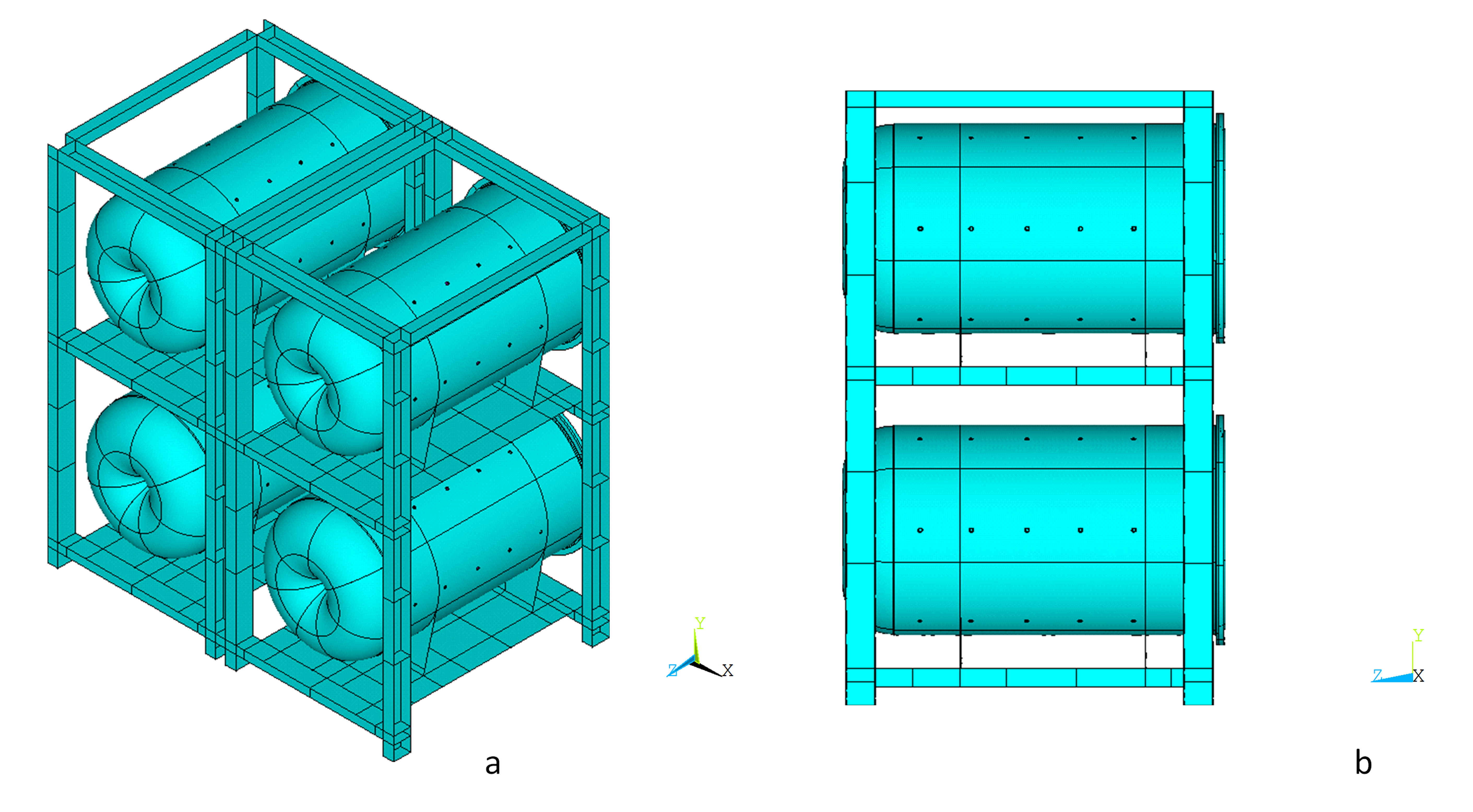}
  \caption{(a) General view of four horn support system, (b) side view}
  \label{fig:support1}
\end{figure}

The static and dynamic analyses of the four-horn support frame are discussed in the following subsections.

\subsubsubsection{Static Analysis of the Four-Horn Support System}
Figure~\ref{fig:support1} shows the general concept of a four-horn support frame. The proposed solution takes into account both the assembly and operation conditions. For this purpose, a frame system was proposed which provides a relatively easy access to the horns and their additional equipment (water jets, drain pipes, etc.), as well as sufficient stiffness resulting in minimum vertical deflections of the horns themselves (y-direction). In the proposed approach, each horn is directly supported on two saddle supports, which are located at positions in the z-direction that ensure equalisation of the maximal values of the bending moment distributed along the z-axis. These points are taken at $z_1=\SI{570}{\mm}$ and $z_2=\SI{2100}{\mm}$, measured from the inner edge of the collar of the horn, at the upstream end of the horn. All the elements of the frame will be made of aluminium. It is also assumed that the frame will be rigidly supported at several location on its top and bottom, including the corner points.

The saddle supports are placed on a horizontal supporting plate, which is attached to channel sections at both ends parallel to the x-axis. These are joined by means of welding in the four corners to the frame system. Taking into account the mass of the horn and its surrounding equipment (not modelled) the stiffness of the plate alone is too low to keep the vertical deflections within the required range. Therefore, in order to increase the plate stiffness, a set of three vertical ribs is introduced, welded to the bottom of the plate. 

For the supporting frame, a channel section has been chosen. The proposed aluminium channel section has the standard dimensions (C-cross section: $\SI{250}{\mm} \times \SI{150}{\mm} \times \SI{18}{\mm}$), and it consists of commercially accessible elements. The numerical results obtained for the designed structure are as follow:
\begin{itemize}
    \item The maximum absolute value of vertical deflection for the horn assembly: $u_y=\SI{2.06}{\mm}$; (Fig.~\ref{fig:support6})
    \item The maximum absolute value of vertical deflection for the support system: $u_y=\SI{1.85}{\mm}$; (Fig.~\ref{fig:support7})
    \item The maximum value of equivalent stress for the horn assembly: $\sigma_{eqv}=\SI{39.3}{\mega \pascal}$; 
    % (Figure \ref{fig:support8})
    \item The maximum value of equivalent stress in the supporting plate: $\sigma_{eqv}=\SI{18.8}{\mega \pascal}$; 
    % (Figure \ref{fig:support9})
\end{itemize}

\begin{figure}[H]
\begin{center}
\begin{minipage}{0.47\linewidth}
\includegraphics[width=0.99\textwidth]{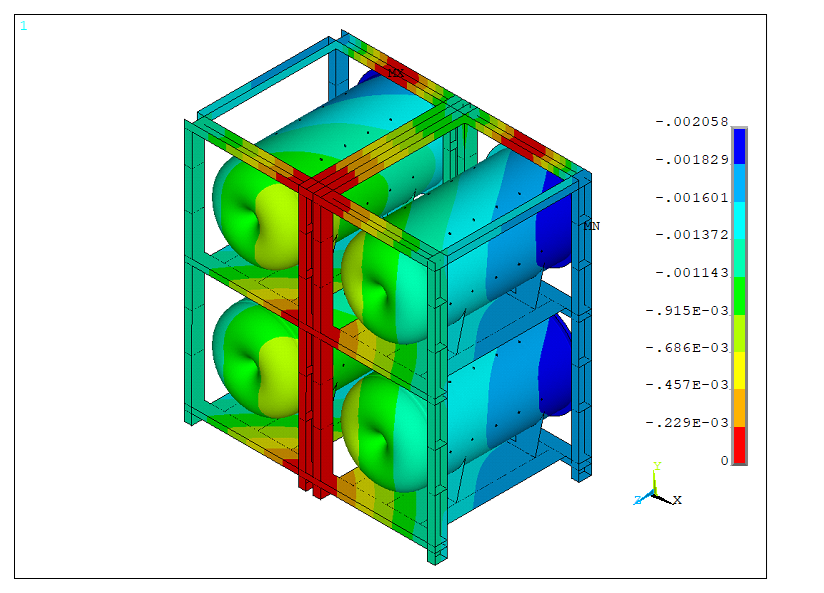}
\caption{\label{fig:support6}Distribution of vertical deflections in four horn assembly -- results in \SI{}{\mm}.}
\end{minipage}\hspace{2pc}%
\begin{minipage}{0.47\linewidth}
\includegraphics[width=0.99\textwidth]{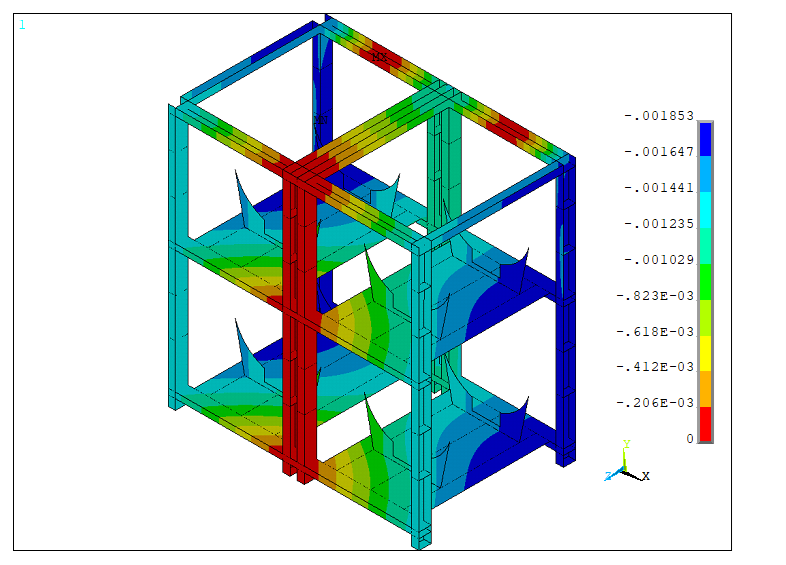}
\caption{\label{fig:support7}Distribution of vertical deflections in supporting frame -- results in mm.}
\end{minipage} 
\end{center}
\end{figure}

%\begin{figure}[H]
  %\centering
  %\includegraphics[width=0.6\columnwidth]{figures/targetstation/support/support6.png}
  %\caption{Distribution of vertical deflections in four horn assembly -- results in \SI{}{\mm}.}
  %\label{fig:support6}
%\end{figure}

%\begin{figure}[H]
  %\centering
  %\includegraphics[width=0.6\columnwidth]{figures/targetstation/support/support7.png}
  %\caption{Distribution of vertical deflections in supporting frame -- results in mm.}
  %\label{fig:support7}
%\end{figure}

% \begin{figure}[H]
%   \centering
%   \includegraphics[width=0.8\columnwidth]{figures/targetstation/support/support8.png}
%   \caption{Distribution of equivalent stress in four horn assembly -– results in \SI{}{\pascal}}
%   \label{fig:support8}
% \end{figure}

% \begin{figure}[H]
%   \centering
%   \includegraphics[width=0.8\columnwidth]{figures/targetstation/support/support9.png}
%   \caption{Distribution of equivalent stress in supporting frame -– results in \SI{}{\pascal}}
%   \label{fig:support9}
% \end{figure}

The results illustrating the discrepancies from the perfect horizontal position of the top and bottom horn axes, defined by the vertical deflections in two chosen points (see Fig.~\ref{fig:support5}) are as follows: 

\begin{itemize}
    \item Top horn: $ \Delta u_y = \left| -1.58 + 1.21 \right|\SI{}{\mm} = \SI{0.37}{mm} $
    \item Bottom horn: $ \Delta u_y = \left| -1.57 + 1.22 \right|\SI{}{\mm} = \SI{0.35}{mm} $
\end{itemize}

%\begin{itemize}
%    \item Top horn:
%    \begin{equation*}
%        \Delta u_y = \left| -1.58 + 1.21 %\right|\SI{}{\mm} = \SI{0.37}{mm}
%    \end{equation*}
%    \item Bottom horn:
%    \begin{equation*}
%        \Delta u_y = \left| -1.57 + 1.22 %\right|\SI{}{\mm} = \SI{0.35}{mm}
%    \end{equation*}
%\end{itemize}

\begin{figure}[H]
  \centering
  \includegraphics[width=0.5\columnwidth]{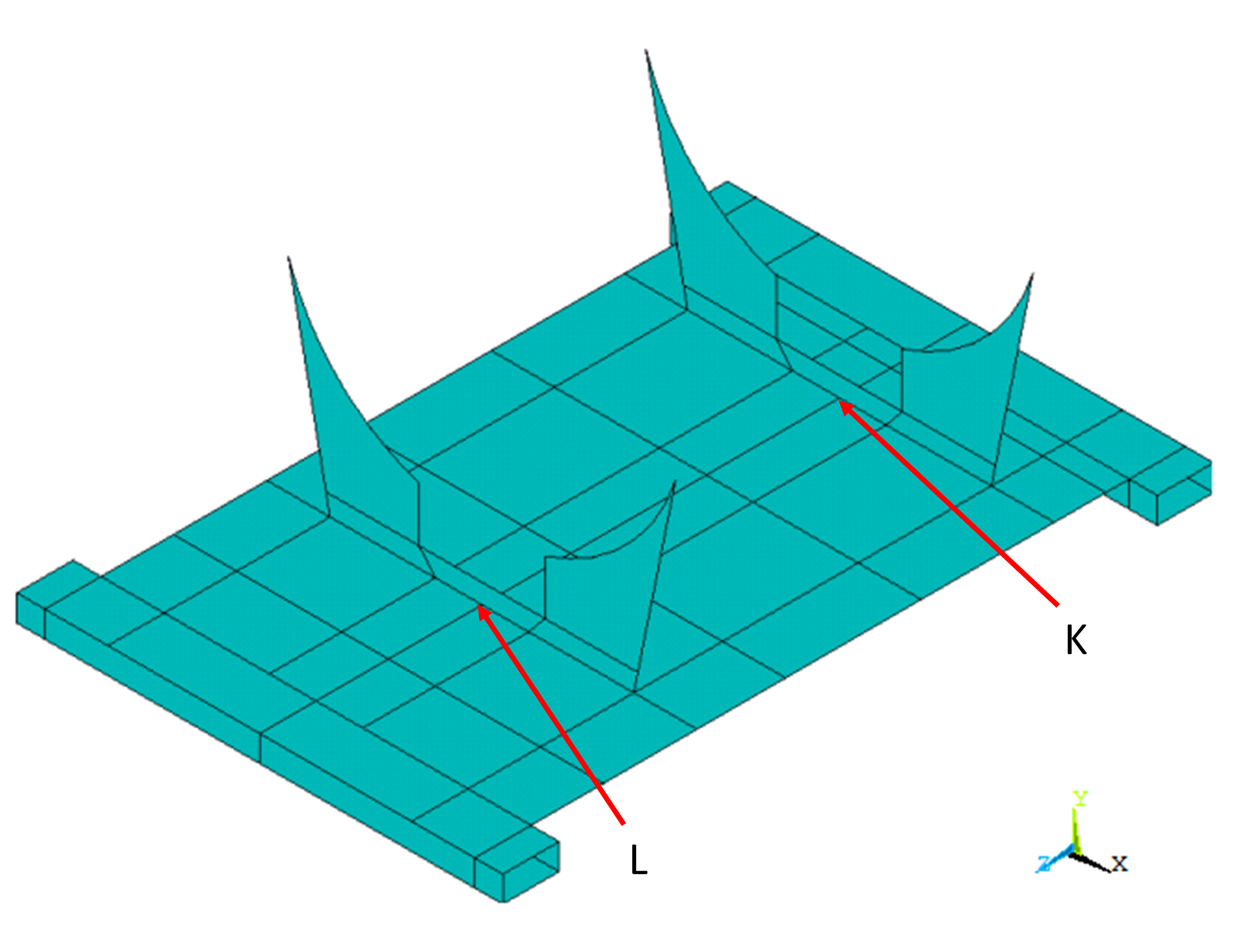}
  \caption{Control point K and L location.}
  \label{fig:support5}
\end{figure}

\subsubsubsection{Dynamic Analysis of the Frame with Four Horns}

Table~\ref{tab:frame-frequencies} gives the lowest natural frequencies of a frame with the four horns. The thickness of the plates of the saddle supports is equal to \SI{20}{mm} -- the value that was also used for the static analysis. These frequencies have been obtained under the condition that the frame is restrained at its top and bottom. It is important to ensure that this condition will be met in the final design. Should either the top or the bottom part of the frame be flexibly attached, much lower frequencies will result, which would lead to unacceptably high frame displacement due to the magnetic pulses. 

\begin{table}[H]
    \centering
    \caption{Ten lowest natural frequencies of the frame with the four horns}
    \label{tab:frame-frequencies}
    \begin{tabular}{l|c c c c c c c c c c}
        Mode number & 1 & 2 & 3 & 4 & 5 & 6 & 7 & 8 & 9 & 10 \\
        \hline
        Frequency $\left[ \SI{}{\hertz}\right]$ & \SI{10.53}{} & \SI{10.72}{} & \SI{10.86}{} & \SI{11.20}{} & \SI{11.20}{} & \SI{23.59}{} & \SI{23.93}{} & \SI{24.82}{} & \SI{24.82}{} & \SI{26.29}{}
    \end{tabular}
\end{table}

The frame has been designed so that its natural frequencies with the four horns are not near the pulse-repetition frequency of \SI{14}{\hertz}, to avoid the risk of resonance. However, the natural frequencies may change if additional elements add substantial mass or stiffness to the structure. To avoid this effect, both the striplines and the cooling water pipes need to be connected flexibly to the horns (for the pipes, a flexible connection has been proposed using bellows which surround the water jet nozzles). Figure~\ref{fig:mode-shapes} shows the mode shapes of the lowest vibration mode (Fig.~\ref{fig:mode-shapes-a}); as well as that of a higher mode, with a frequency of \SI{23.93}{\hertz}~(Fig.~\ref{fig:mode-shapes-c}), the first mode that displays a substantial deformation of the frame.

\begin{figure}[h!]
     \centering
     \begin{subfigure}[b]{0.45\columnwidth}
         \centering
         \includegraphics[width=\columnwidth]{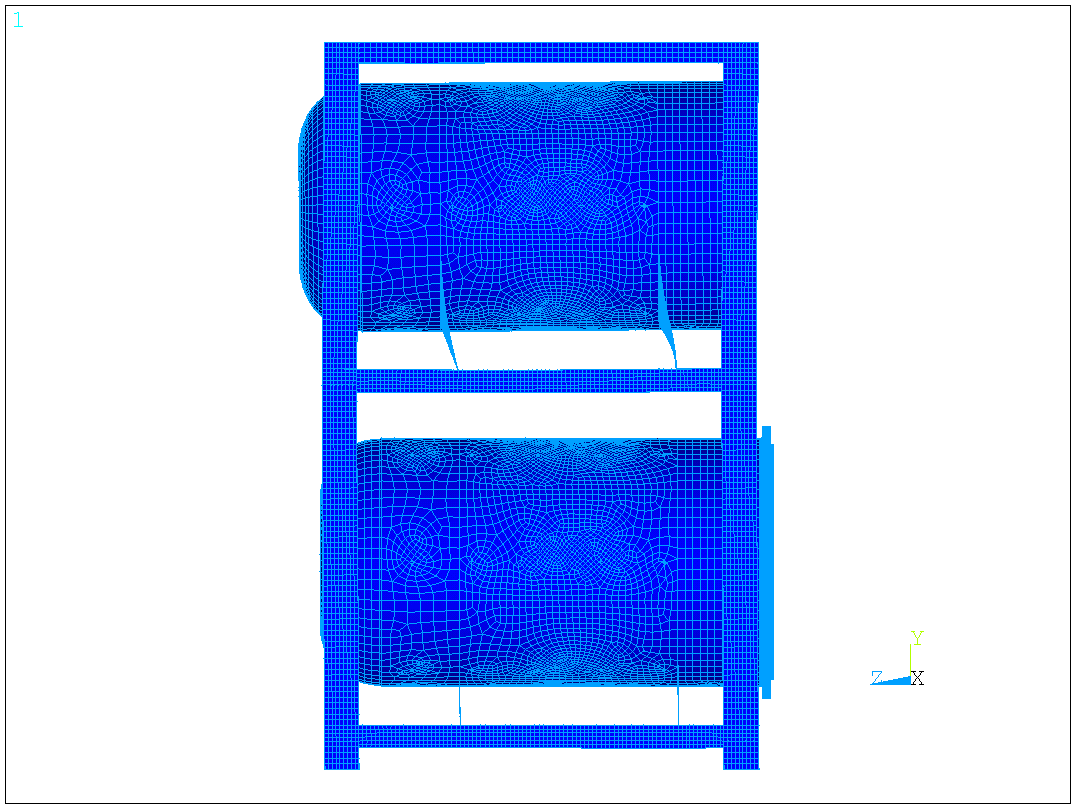}
        \caption{Mode shape with frequency \SI{10.53}{\hertz}}
        \label{fig:mode-shapes-a}
     \end{subfigure}
     \hfill 
     \begin{subfigure}[b]{0.45\columnwidth}
         \centering
         \includegraphics[width=\columnwidth]{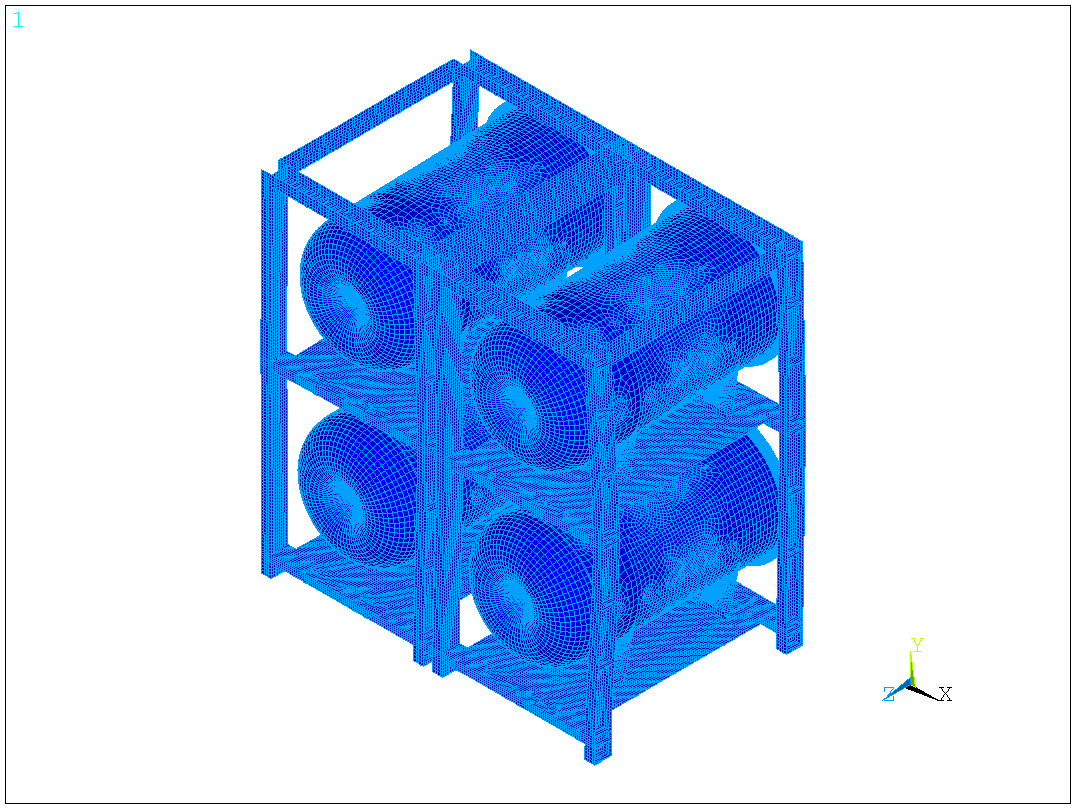}
         \caption{Mode shape with frequency \SI{23.93}{\hertz}}
         \label{fig:mode-shapes-c}
     \end{subfigure}
        \caption{Mode shapes of the frame with the four horns}
        \label{fig:mode-shapes}
\end{figure}

The analysis of the stress in the horn discussed in section~\ref{sec:mechanical-stresses-horn} used the Lorentz forces obtained from a finite-element magnetic analysis. All components of the displacement of the nodes that are in contact with the cradle supports have been set to zero in this calculation. In order to estimate the displacement and stress due to a current pulse for a horn mounted in the support frame and for the frame itself, an approximate approach is used here. This is unavoidable, due to the very large finite-element model used.

It has been assumed that the magnetic field everywhere within a horn is toroidal, so that it can be approximated by a magnetic flux $B_\varphi(r)=\mu_0I/(2\pi r)$. It has been verified, by comparing with the finite-element electromagnetic results, that this is a reasonable approximation of the field everywhere inside the horn, including the region near its end. The Maxwell stress tensor formula is used to account for the action of the magnetic forces on the horn skin. In this approach, the traction (force per unit area) acting on a surface element with the normal $\bf{n}$ is equal to:

\begin{equation}
\textbf{T}_n=\frac{1}{\mu_0}\left[(\textbf{B}\circ\textbf{n})\textbf{B}-\frac{1}{2}B^2\textbf{n}\right]
\end{equation}

Here, $\textbf{n}$ is a unit vector that points from the horn skin towards the horn interior, where $\textbf{B}$ is the magnetic induction vector as the horn skin is approached from the inside. It has been verified analytically, that with the assumed direction of the magnetic induction vector, the first component disappears everywhere on the horn surface, including the toroidal end part. As a result, the surface traction reduces to a negative pressure $-\frac{1}{2\mu_0}B^2(r)$.

\begin{figure}[ht!]
\centering
  \includegraphics[width=0.5\columnwidth]{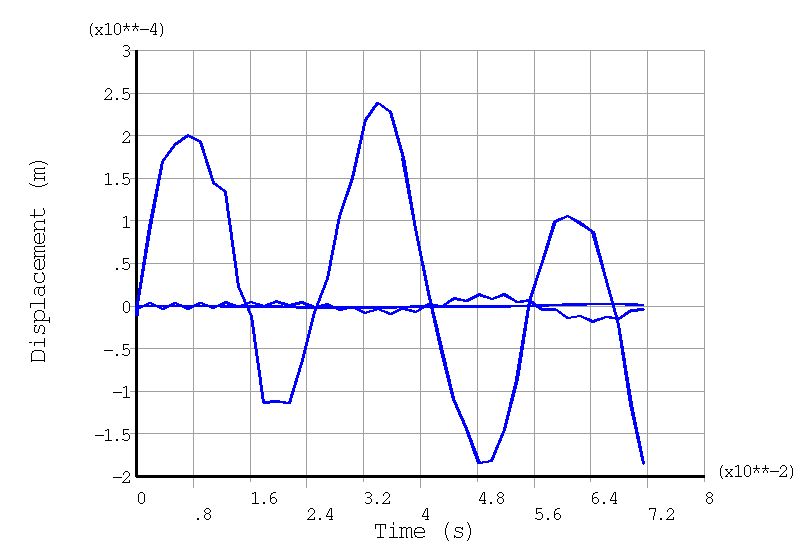}
  \caption{Components of displacements at a point on the horn neck, under a single pulse of current of amplitude \SI{350}{\kilo \ampere} and \SI{100}{\micro \second} duration.}
  \label{fig:displacement-components-horn}
\end{figure}

Figure~\ref{fig:displacement-components-horn} shows the three components of displacement at a point on the horn neck, resulting from a single pulse with a current of amplitude \SI{350}{\kilo \ampere} and a \SI{100}{\micro \second} duration. The curve with the highest-amplitude oscillations corresponds to the axial displacement, with a maximum value of about \SI{0.25}{\milli \meter}. This displacement is determined by the magnetic forces that act on the curved end part of the horn, as well as by the stiffness of the cradle supports. The transverse displacements are small, which is important from the point of view of beam alignment. It has also been found that the dynamic displacement of the frame is considerably smaller than that of the horn, so that the vibration transmitted from the horn to the frame appears to be at an acceptable level. Higher amplitudes can appear under a sequence of pulses, but these should still be acceptable in the absence of resonance.

The maximum calculated von Mises stress during a single pulse is about \SI{24.5}{\mega \pascal}, and it occurs in the horn inner conductor. This value is comparable to the more accurate result given in section \ref{sec:mechanical-stresses-horn} (the smaller value of stress and greater displacement is consistent with the flexibility of the support). Considerably lower values of stress are found on the support frame. These values should be feasible for aluminium, also when the four horns are pulsed simultaneously. Additional studies are still needed to confirm this conclusion.

%--
%\clearpage
\subsection{Power Supply Unit} \label{section:PowerSupplyUnit}
The magnetic field in each horn is produced by a power supply unit (PSU), which delivers \SI{350}{\kilo\ampere} peak-current pulses to each of the four horns, synchronised with the proton beam pulses coming from the switchyard (see Fig.~\ref{fig:pulses}).

\begin{figure}[h!]
    \centering
    \includegraphics[width=0.7\linewidth]{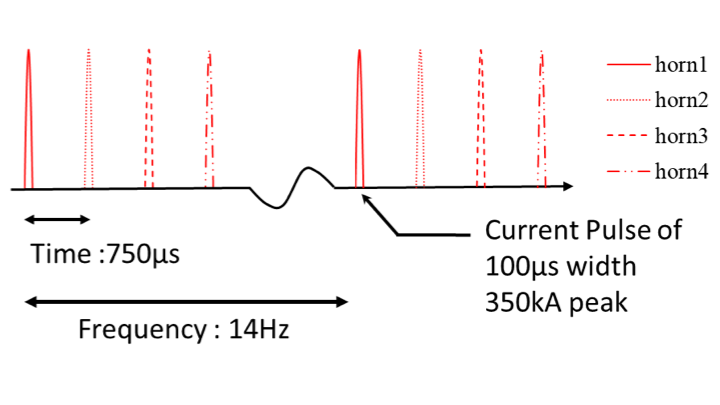} 
    \caption{Proton pulsing scheme.}
    \label{fig:pulses}
\end{figure}

The power supply unit has been developed in accordance with the following main requirements:
\begin{itemize}
\item Each horn is pulsed by a half-sinusoid current waveform of \SI{100}{\micro\second} width and \SI{350}{\kilo\ampere} peak current, with a very high RMS current of \SI{9.3}{\kilo\ampere}. The magnetic horn behaves as a ``shunt'' with a very low inductance of \SI{1.24}{\micro\henry} and a low resistance value of \SI{0.414}{\milli\ohm}.
\item The pulses will be generated at a high operating frequency of \SI{14}{\hertz} for each horn, with a \SI{750}{\micro\second} delay time between each horn.
\item The maximum operating voltage is restricted by electrical insulation and technical constraints to $<$~\SI{20}{\kilo\volt}.
\item The proposed electrical solution must reduce the thermal dissipation in the horns to a minimum. 
\item The PSU electrical consumption is minimised by developing a solution that allows for maximal energy recovery.
\item The components have been designed to accept very stringent electrical constraints during their operation.
\item A water cooling system will be implemented to improved durability.
\end{itemize}

The delivery of such high-intensity pulses and the fast commutation between the horns at the level of \SI{750}{\micro\second} imposed by the linac presents serious constraints in terms of lifetime for the electrical components. The proposed design is based on a modular approach, and it should guarantee a reasonable stability of the PSU with the minimum of maintenance for the duration of the experiment.

\subsubsection{Principle of $\mu$s Pulse Generation}
The power supply unit consists of 16 modules connected in parallel, capable of delivering \SI{350}{\kilo\ampere} to each horn, with \SI{100}{\micro\second} duration, at \SI{14}{\hertz}. The principle of generating short pulses by each module is based on an oscillating circuit, which consists of two parts, as shown in Fig.~\ref{fig:PSU_circuit}.

\begin{figure*}[!h]
\centering

  \begin{subfigure}[b]{0.48\textwidth}
  \centering
  \includegraphics[width=0.85\linewidth]{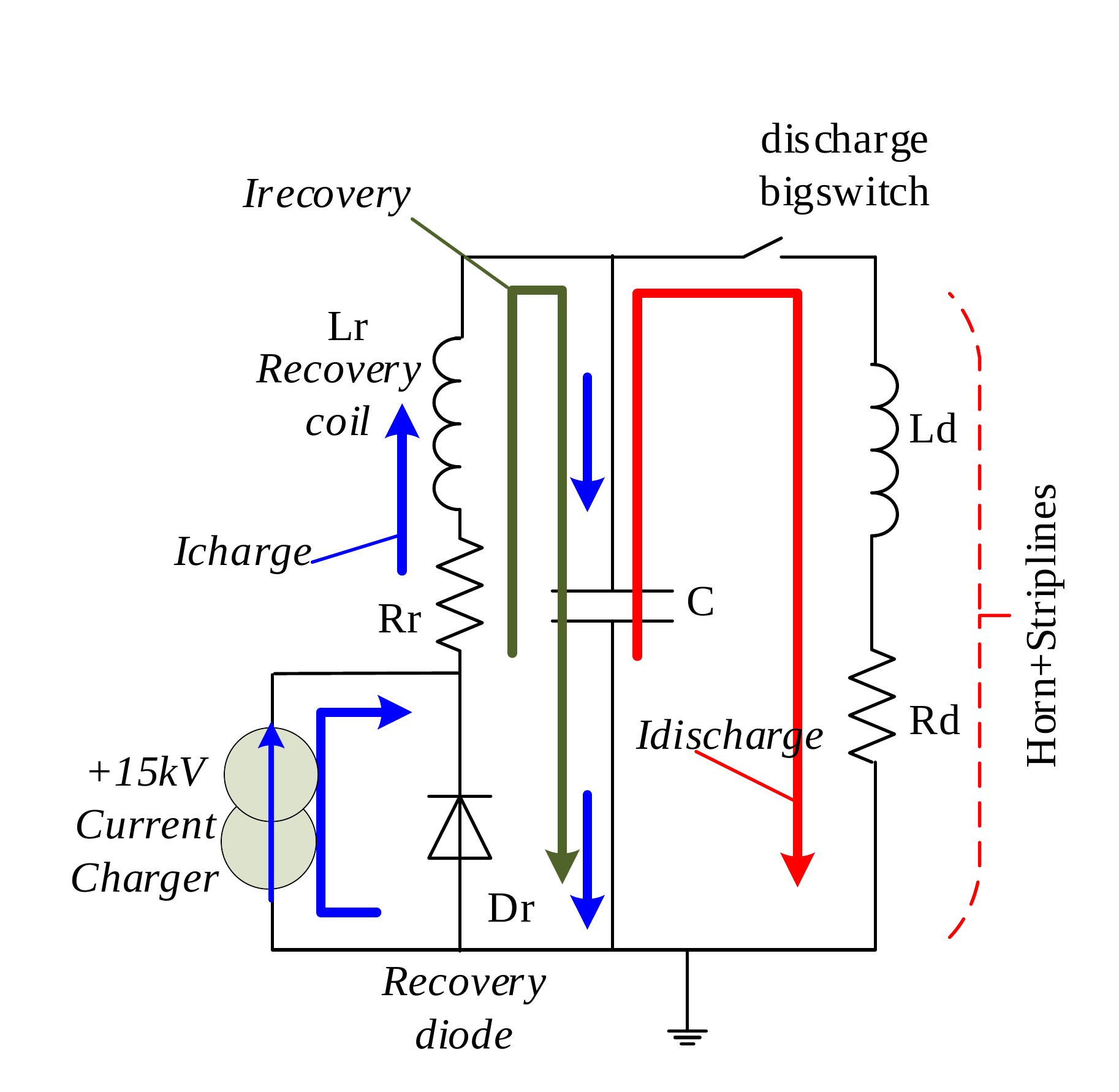}
  \end{subfigure}
  \begin{subtable}[b]{0.48\textwidth}
   \centering
    \begin{tabular}{|c|c|c|} \hline
                                    &  Discharge                    & Recovery \\
                                    & (Horn + Striplines)           & \\ \hline
      C = \SI{1000}{\micro\farad}   & $L_d$ = \SI{1.38}{\micro\henry} & $L_r$ \SI{0.25}{\milli\henry} \\ 
                                    & $R_d$ = \SI{1.07}{\micro\ohm} & $R_r$ \SI{5.94}{\micro\ohm} \\ \hline 
    \end{tabular}
  \vspace{1.5cm}
  \end{subtable}
  \caption{Electrical diagram for delivering \SI{350}{\kilo\ampere} peak current during \SI{100}{\micro\second} half period (left), Electrical values of the components (right).}
  \label{fig:PSU_circuit}
\end{figure*}

The charger circuit supplies +\SI{14}{\kilo\volt} and will charge the main capacitor $C$, which is common to the discharge circuit. It also contains a recovery circuit with a resistor $R_r$, diode $D_r$ and the coil $L_r$, ensuring good energy recovery. Once the capacitor is charged, the discharge big switch is turned on and a +\SI{44}{\kilo\ampere} pulse with \SI{100}{\micro\second} duration is delivered by the discharge circuit, which consists of a resistor $R_d$ and a coil $L_d$ representing the electrical properties of the stripline and the horn.

\begin{figure}[h!]
    \centering
    \includegraphics[width=0.65\linewidth]{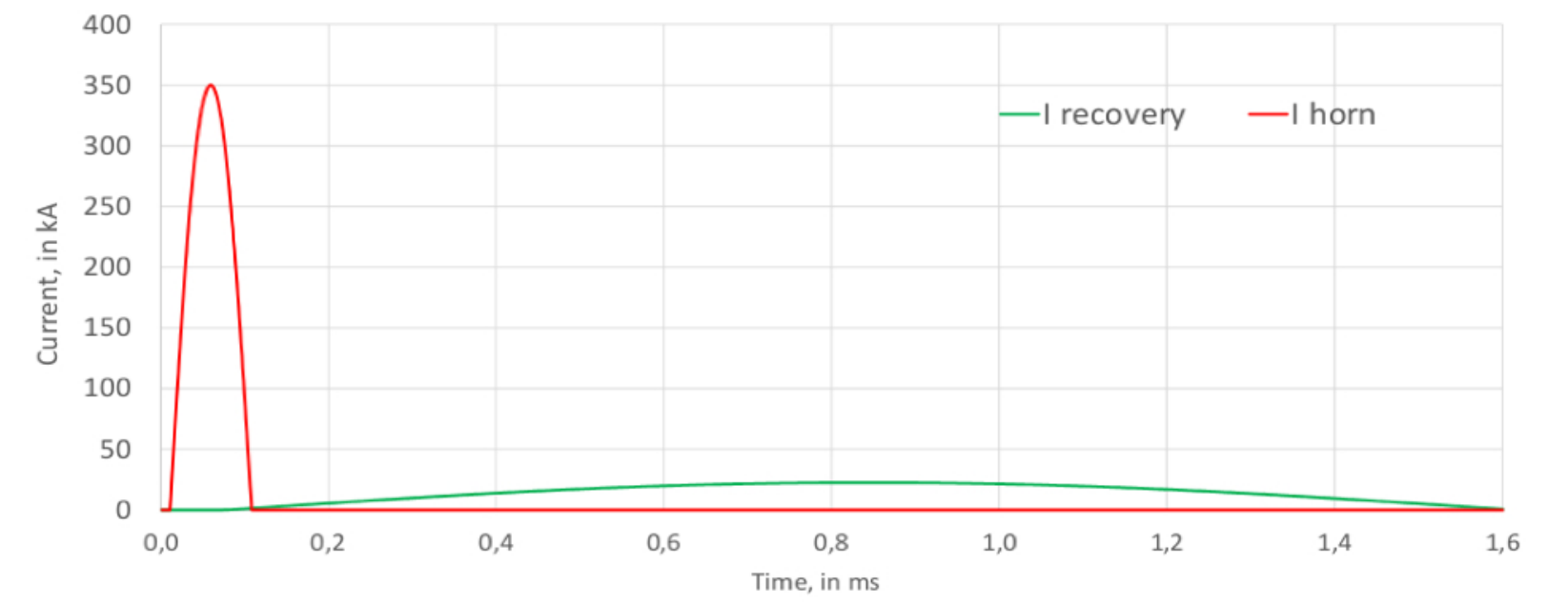}
    \includegraphics[width=0.70\linewidth]{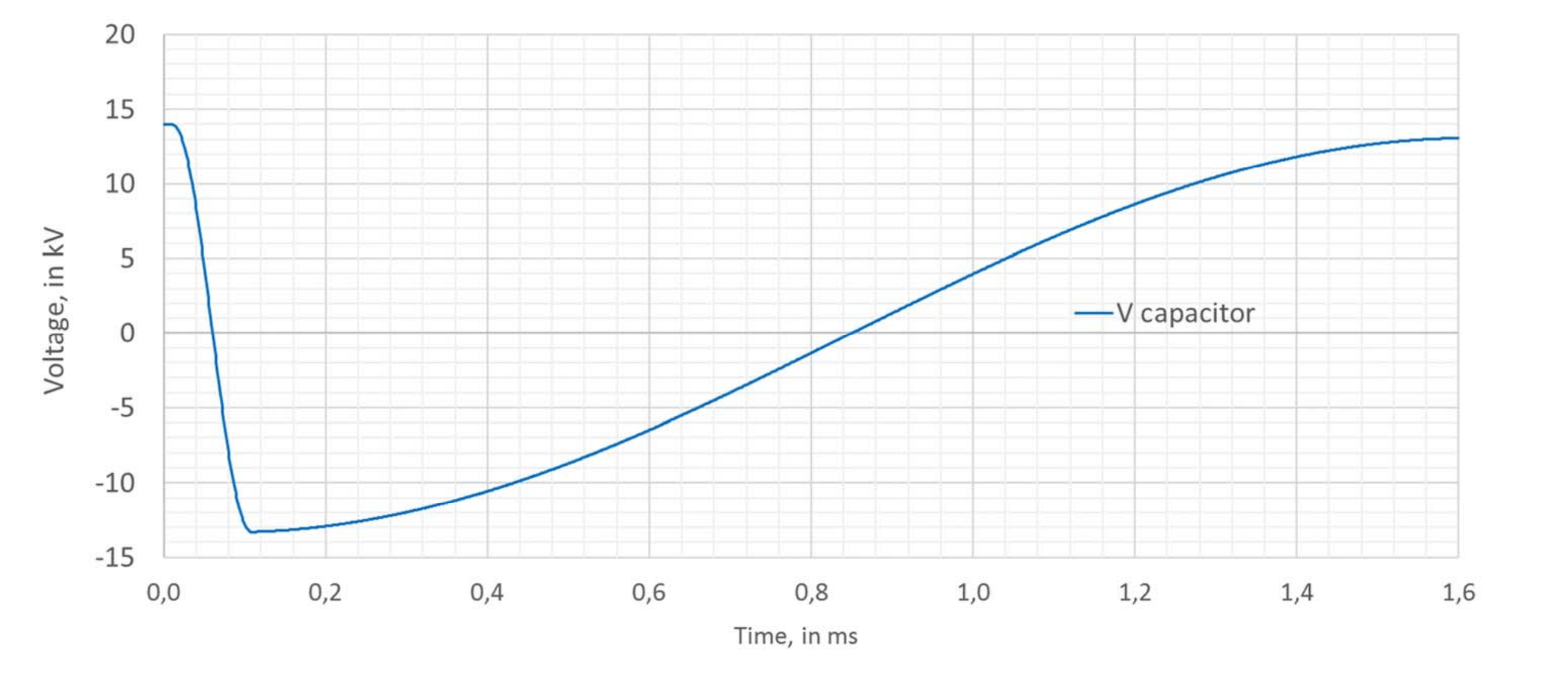}
    \caption{Principle of delivering \SI{350}{\kilo\ampere} peak current during a \SI{100}{\micro\second} half period.}
    \label{fig:PSU_Pulse}
\end{figure}

At the beginning of the sequence, all 16 PSU modules are fully charged and ready, and eight of them are simultaneously discharged in the stripline to produce the \SI{350}{\kilo\ampere} full intensity in one horn, as shown in Fig.~\ref{fig:PSU_Pulse}. After \SI{750}{\micro\second} another eight modules are simultaneously discharged to supply another horn, and the cycle continues. 

\begin{table}[h!]
\small
  \begin{center}
    \caption{Electrical constraints of the PSU for one horn pulsed at \SI{14}{\hertz}.}
    \label{tab:psu_horn_elec_constraints}
    \begin{tabular}{lcc}
         		 	 			& \textbf{Discharge} 	& \textbf{Recovery} \\ \hline
	$\frac{di(t)}{dt}$	 [\SI{}{\ampere\per\micro\second}]  				& 11000		& 40 \\
	$I_{Peak}$	[\SI{}{\kilo\ampere}]  	& 350		& 22.4 \\
	$I_{RMS}$	[\SI{}{\kilo\ampere}]  	& 9.3			& 2.4 \\ \hline
     \end{tabular}
  \end{center}
\end{table}

The values of the critical electrical constraints in the discharge and recovery stages are given in Table~\ref{tab:psu_horn_elec_constraints}. 
The high peak current and the RMS values make it impossible to find any switch components or coils capable of working under such demanding electrical constraints. Therefore, a modular approach has been developed to accommodate this limitation. 

\subsubsection{Modular Approach and PSU Implementation}

The modular approach is necessary to reduce these electrical constraints. In this case, a good compromise between electrical safety and the lifetime of components can be obtained. The PSU will use eight outputs, called units, capable of delivering \SI{44}{\kilo\ampere} each. These units
connected in parallel can deliver the nominal current of \SI{350}{\kilo\ampere} to each horn, as shown in Fig.~\ref{fig:PSU_Scheme}.

\begin{figure}[h!]
    \centering 
    \includegraphics[width=0.6\linewidth]{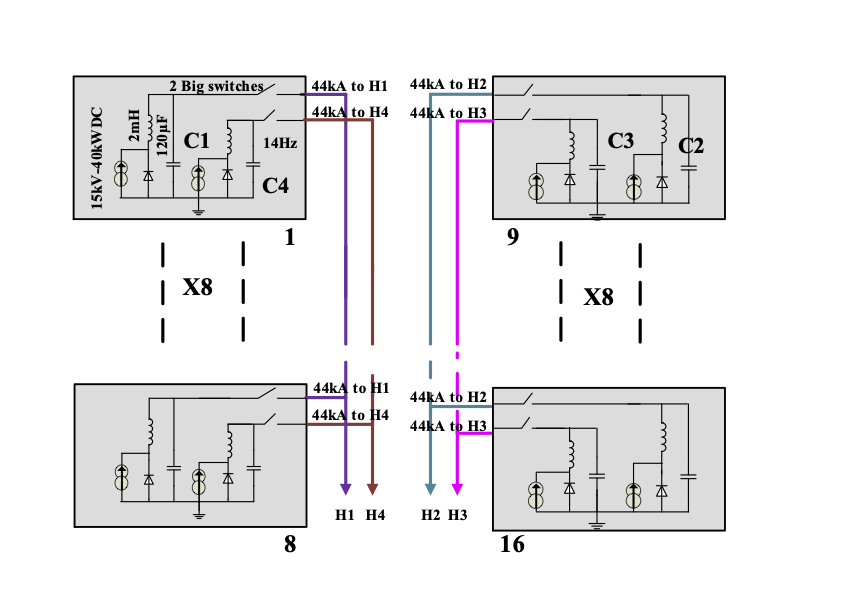}
    \caption{Principle of delivering \SI{350}{\kilo\ampere} with 16 modules.}
    \label{fig:PSU_Scheme}
\end{figure}

The electrical constraints for one \SI{44}{\kilo\ampere} unit are given in Table~\ref{tab:psu_unit_elec_constraints}. For the sake of recovery, the frequency of the charger must be higher than the output frequency. The low value of the electrical resistivity in the discharge and recovery 
circuit enables limiting the charger current to 3\% of the total current used, to provide the necessary \SI{44}{\kilo\ampere}. Therefore, a charger delivering only \SI{18}{\kilo\watt} of average power and \SI{24}{\kilo\watt} of peak power and working at \SI{15}{\hertz} is sufficient. 
\begin{table}[h!]
\small
  \begin{center}
    \caption{Electrical constraints on a unit circuit delivering \SI{44}{\kilo\ampere}.}
    \label{tab:psu_unit_elec_constraints}
    \begin{tabular}{lcccccc}
    & \textbf{Frequency  [\SI[detect-weight]{}{\hertz}]}  & \boldmath{$\frac{di(t)}{dt}$} \textbf{[\SI[detect-weight]{}{\ampere\per\micro\second}]}  & \boldmath{$I_{Peak}$}	\textbf{[\SI[detect-weight]{}{\ampere}]} & \boldmath{$I_{Average}$} \textbf{[\SI[detect-weight]{}{\ampere}]} &  \boldmath{$I_{Average}$} \textbf{[\SI[detect-weight]{}{\ampere}]} & \textbf{Q [\SI[detect-weight]{}{\coulomb}]}\\ \hline
	Charge 		& 15 & 	    & 2 	  & 4 	      & 1.35   & 0.09 \\ %\hline
	Recovery	 	& 14 & 5.9    & 2800   & 300   & 39.5   & 2.8 \\ %\hline
	Discharge 	& 14 & 1500 & 44500 & 1200 & 40.85 & 2.9 \\ \hline
     \end{tabular}
  \end{center}
\end{table}

Figure~\ref{fig:PSU_Implementation} shows the dimensions of one module and the position of all 16 modules outside of the target station facility. The length of the striplines (in yellow) from the horns to the last module has been limited to 33~m due to electrical requirements.

\begin{figure}[h!]
\begin{center}
\begin{subfigure}[b]{0.4\linewidth}
    \centering
    \includegraphics[width=\linewidth]{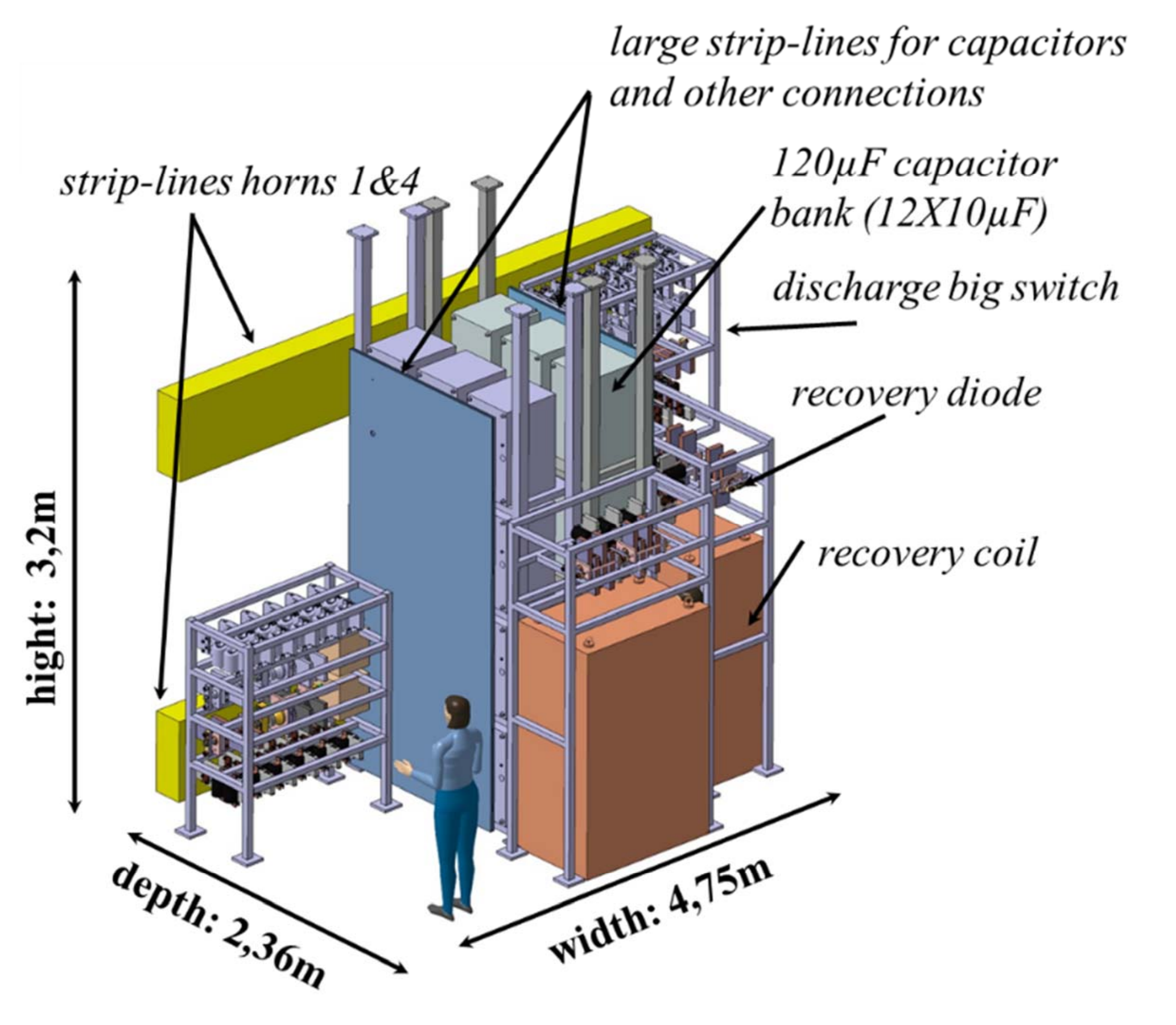}
    \caption{PSU module with 2$\times$\SI{44}{\kilo\ampere} output.}
\end{subfigure}
\hspace{1.cm}
\begin{subfigure}[b]{0.5\linewidth}
    \centering
    \includegraphics[width=\linewidth]{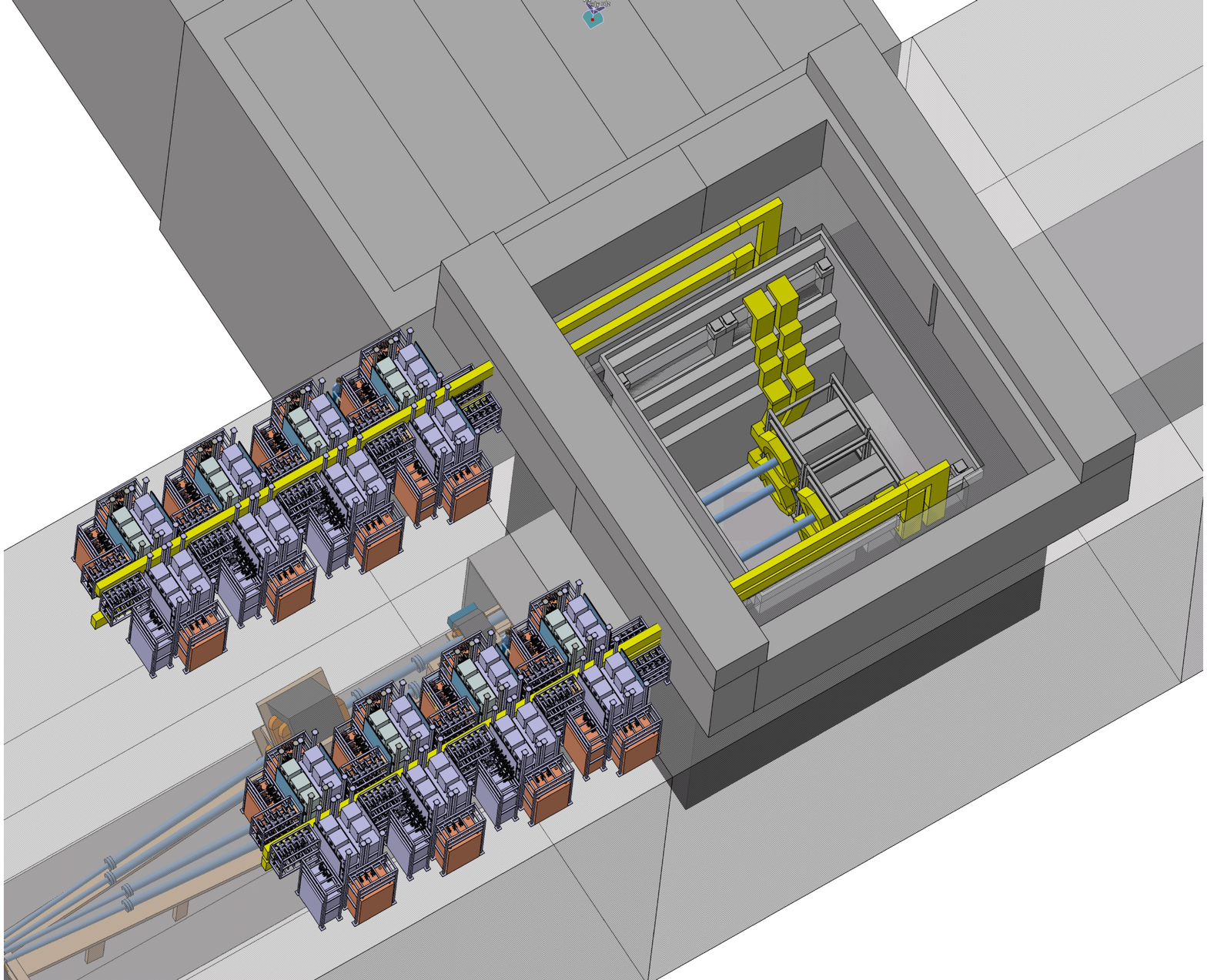}
    \caption{PSU room located above the switchyard tunnel.}
\end{subfigure}
\caption{Power supply unit.}
\label{fig:PSU_Implementation}
\end{center}
\end{figure}

%--
\subsection{Target Station Facility}

The target station facility hosting the four-horn system will receive the \SI{5}{\mega\watt} proton beam and convert it to a very intense neutrino ``Super Beam''. Because of the significant amount of power deposited in the full layout and the production of radionuclei, the main elements of the structure (shown in Fig.~\ref{fig:TargetStation3D}) have been analysed through Monte Carlo simulations, to propose a reliable safety solution in terms of radioactivation, which will be able to meet the regulation rules in Sweden.

\begin{figure}[h!]
\begin{subfigure}[b]{0.45\linewidth}
\includegraphics[width=\linewidth]{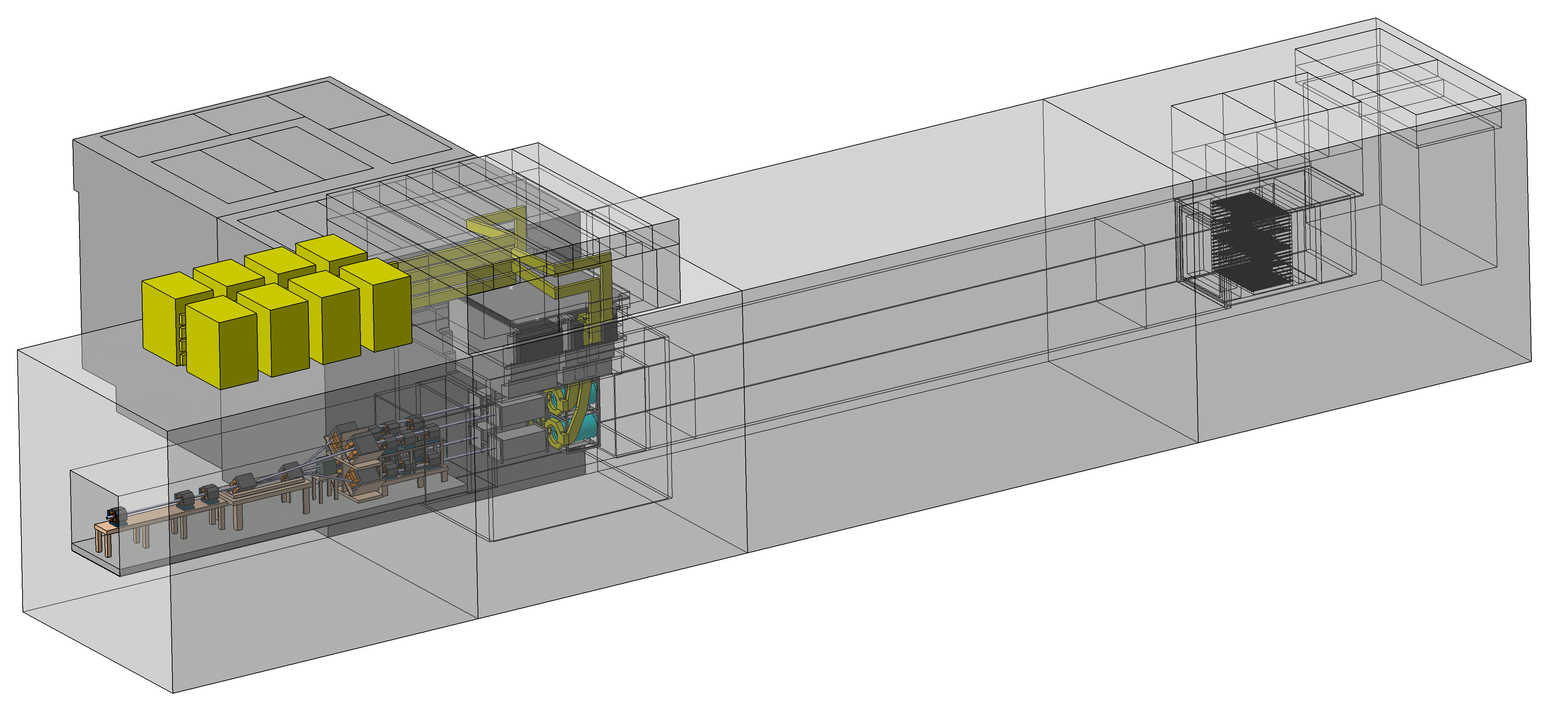}
\caption{3D schematic view.}
\end{subfigure}
\hspace{1.cm}
\begin{subfigure}[b]{0.45\linewidth}
\includegraphics[width=\linewidth]{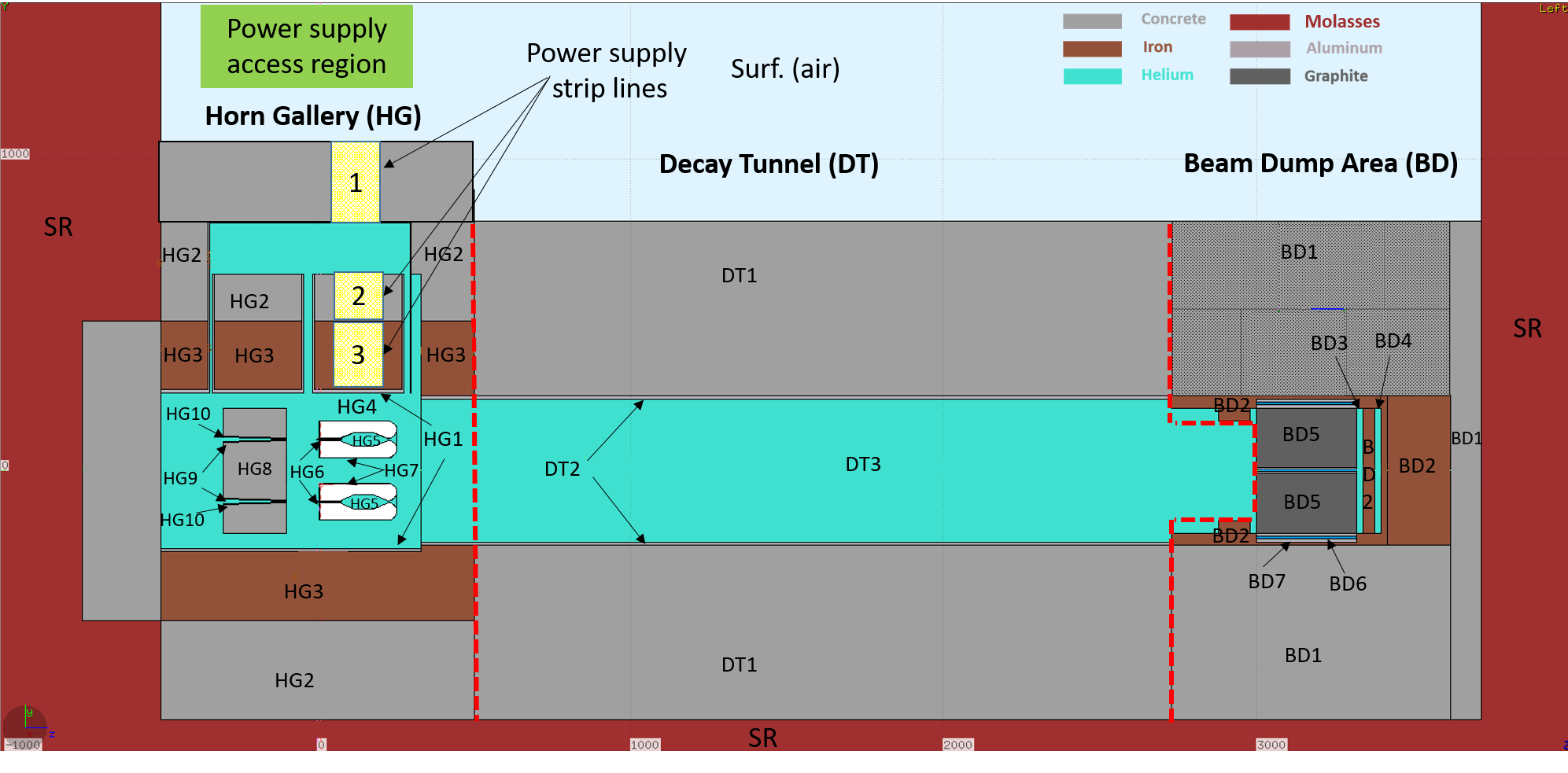}
\caption{Monte Carlo model used for calculations in \textsc{FLUKA}.}
\end{subfigure}
\caption{Target Station Facility.}
\label{fig:TargetStation3D}
\end{figure}

In the following sections, the decay tunnel and the beam dump will be discussed in detail from a technical point of view, along with the radiation safety aspects. 

%--
\subsubsection{Decay Tunnel}
\label{subsubsection:DT}
The dimensions of the decay tunnel play an important role in defining the neutrino beam composition delivered to the detectors. An optimisation procedure based on a genetic algorithm (see Section~\ref{sec:targetstation:geneticalgorithm_improvement}) fixed the width and height of the decay tunnel at \SI{5}{\meter} and the length at \SI{50}{\meter}. The decay tunnel will be filled with helium gas contained in a \SI{10}{cm} thick vessel made of aluminium, and fully surrounded by a \SI{5.5}{\meter} layer of concrete, in order to protect the underground site hosting the target station facility (see Fig.~$\ref{fig:DT_Layout}$).

\begin{figure}[h!]
    \centering
    \includegraphics[width=0.68\linewidth]{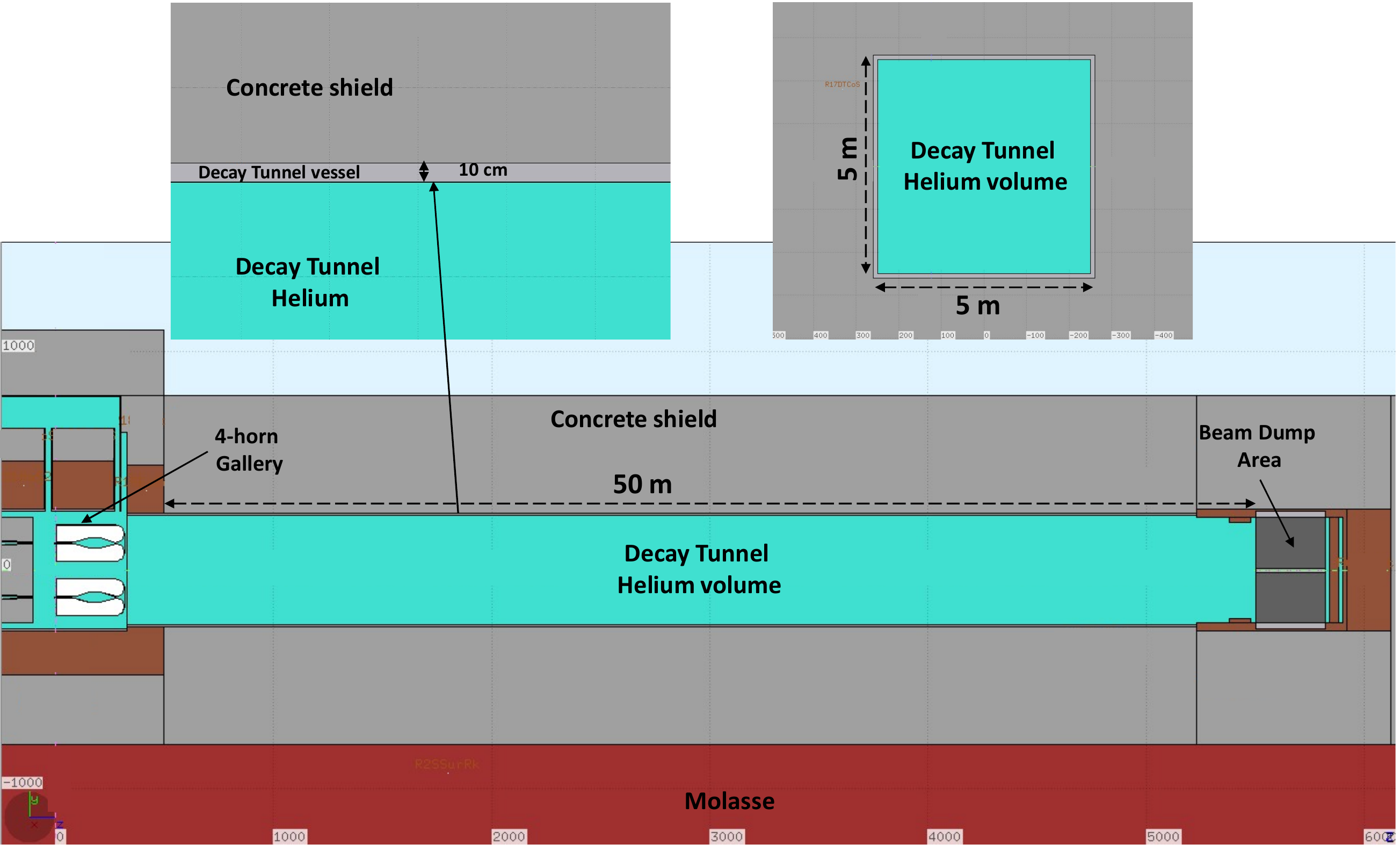} 
    \caption{ESS$\nu$SB decay tunnel layout.}
    \label{fig:DT_Layout}
\end{figure}

Dedicated \textsc{FLUKA} power deposition studies have shown that the secondary meson beam deposits \SI{840}{\kilo\watt} and \SI{612}{\kilo\watt}, respectively, in the decay tunnel vessel and in the concrete shield surrounding it. The corresponding distribution (in~\SI{}{\kilo\watt\per\cubic\centi\metre}) is shown for both parts in Fig.~\ref{fig:DT_Edep}, and it requires the design of an efficient cooling system to evacuate the deposited power.

\begin{figure}[h!]
    \centering
    \includegraphics[width=0.6\linewidth]{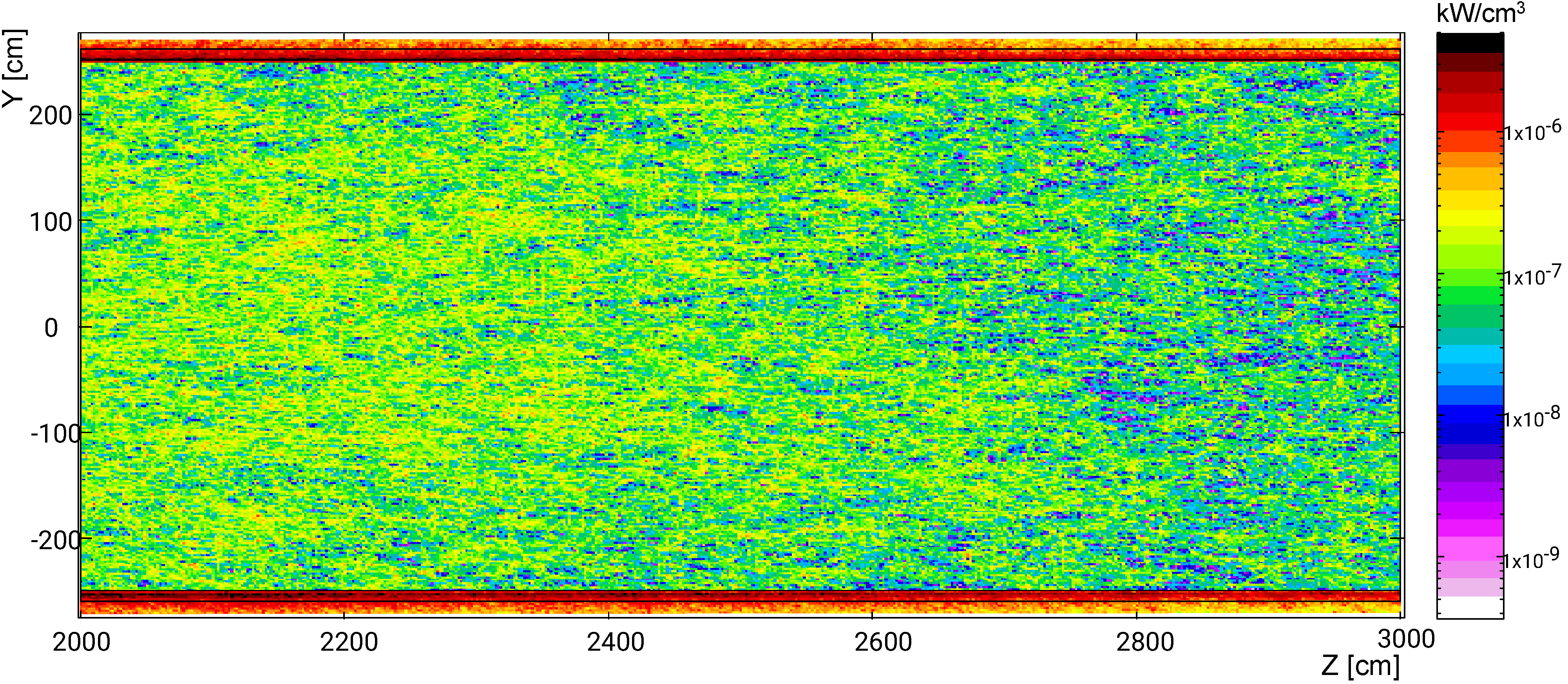} 
    \caption{Power deposition density distribution inside the decay tunnel in~\SI{}{\kilo\watt\per\cubic\centi\metre}.}
    \label{fig:DT_Edep}
\end{figure}

\subsubsection{Beam Dump}
\label{subsubsection:BD}
The beam dump of the target station facility plays an important role in shielding/protecting the underground site from being radio-activated by the remaining radiation reaching the end of the decay tunnel. The \SI{5}{\mega\watt} proton beam power and the geometry of the four-horn system requires a non-standard design compared to presently running experiments. It will have to adopt a non-trivial structure that fulfils specific needs to be able to withstand such an unprecedented radiation environment. Among these, the most important are:

\begin{itemize}
\item to be able to withstand the power deposition from remaining secondaries from the \SI{5}{\mega\watt} proton beam interaction with the four targets,
\item to offer maximum shielding for the underground site to confine all radiation,
\item the grade of the core material must be a trade-off between the material properties and the price. However, this work adopts a graphite grade between Sec Carbon Ltd PSG-11 and MSG\cite{Graphit} as the baseline material for the beam dump core.
\end{itemize}

The performance of the ESS$\nu$SB beam dump has been evaluated by a Monte-Carlo simulation based on \textsc{FLUKA} coupled to a Finite Element Method Analysis simulation based on \textsc{COMSOL}\cite{comsol}. This study focuses on the thermo-structural analysis of the beam dump cores and the cooling system. The initial design is inspired by previous project, such as EUROnu \cite{Baussan:2012wf}.

\paragraph{\textbf{Estimation of power deposition in the Beam Dump}.} 

%Fig~\ref{fig:Edep_1} shows the full beam dump core with \SI{5}{\meter}$\times$\SI{5}{\meter} cross-section (left) and the expanded view corresponding to one horn (right).

Fig~\ref{fig:Edep_1} shows an expanded view corresponding to one horn of the full beam dump core with \SI{5}{\meter}$\times$\SI{5}{\meter} dimensions. The distribution from the one-horn model shows that the secondary beam deposition profile on the surface of the dump core follows a 2D-Gaussian distribution, with $\sigma \sim$ \SI{36.17}{\cm}. Figure~$\ref{fig:Edep_2}$ shows the power density distribution in the beam dump core body, along its depth (in the direction of the beam-axis). The maximum power deposition is found at the core depth between \SI{15}{\cm} and \SI{70}{\cm} from the front face of the beam dump core block. These results demonstrate that the ESS$\nu$SB beam dump will be operating under a severe thermal load, which implies a non-trivial design of the geometry and its cooling scheme.
%Two configurations have been considered in the simulations, the first considers the full beam dump core with \SI{5}{\meter}$\times$\SI{5}{\meter} dimensions and a small design corresponding to the footprint of one horn as shown in Fig~\ref{fig:Edep_1}.

\begin{figure}[h!]
    \centering
    %\includegraphics[width=0.3\linewidth]{figures/targetstation/Fig1_BD_Left.png} 
    %\hspace{1.5cm}
    \includegraphics[width=0.4\linewidth]{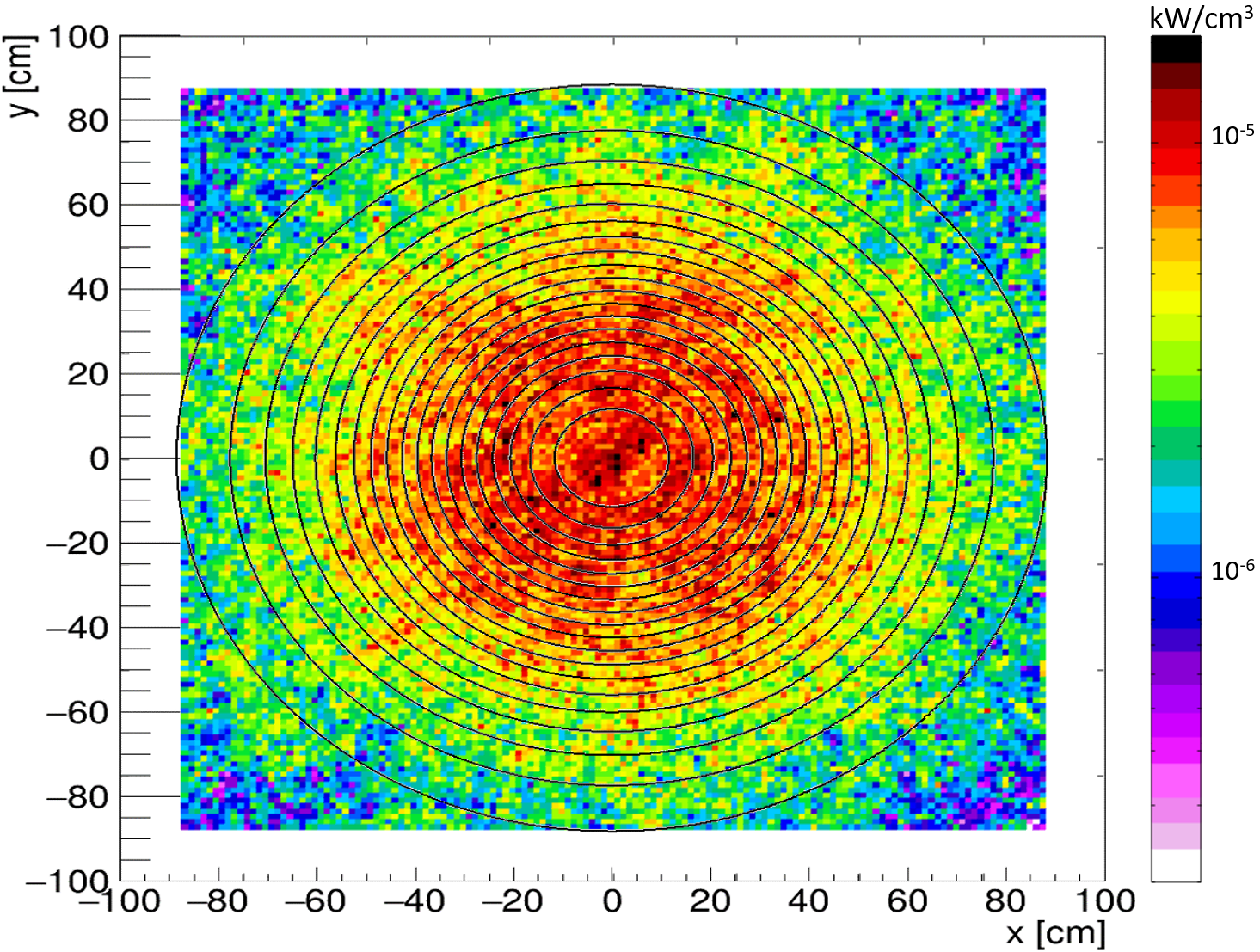} 
    \caption{%(Left) Power density on the core front surface of the four horns. (Right)
    Power surface density of one horn shows a circular Gaussian distribution with $\sigma \sim$ \SI{36.17}{\cm} (black circles).}
    \label{fig:Edep_1}
\end{figure}

%A total power deposition value of $\sim$~\SI{826}{\kilo\watt}/4-horns (or $\sim$~\SI{206.5}{\kilo\watt}/horn) was found in the beam dump core body. 

\begin{figure}[h!]
    \centering
    \includegraphics[width=0.8\linewidth]{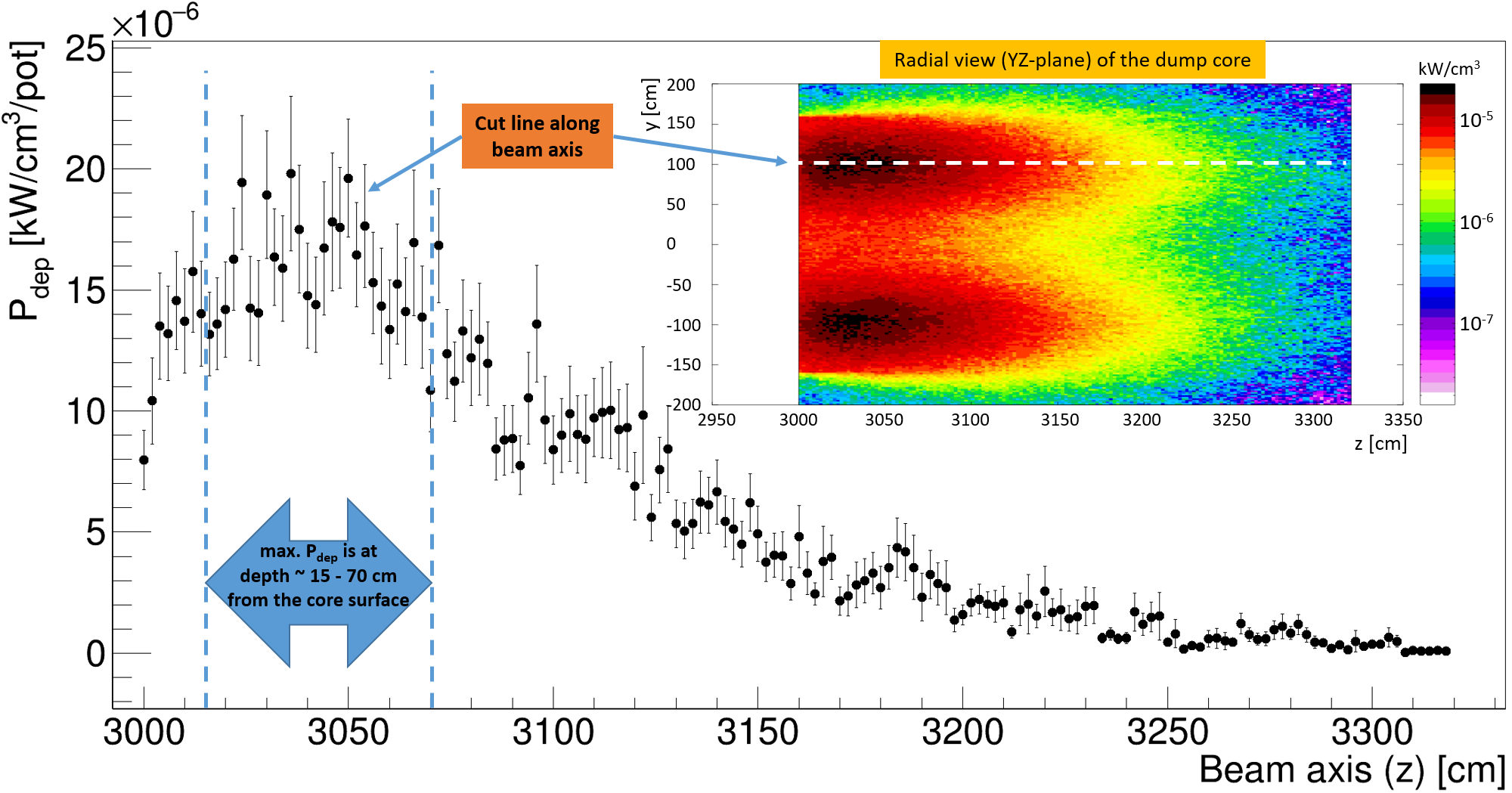} 
    \caption{E$_{dep}$ density (kW/cm$^{3}$) deposited into the one-block core (black dots). (Upper frame) shows the maximum value of the E$_{dep}$ density along the depth (beam axis) of the dump core (white dashed line).}
    \label{fig:Edep_2}
\end{figure}

The beam dump structure depends on four independent core segmented blocks (seg-blocks). Each block in position faces one of the four horns, in order to distribute the total power deposition between four independent cores. Zig-zag blocks with \SI{1}{\cm} opening between them have been considered to allow for the thermal expansion of individual blocks. The four segments are separated by a cross-shaped support structure. Figure~$\ref{fig:Seg-Block-const}$ shows the geometrical structure, dimensions and the assembly/disassembly procedure of the proposed segmented block core and its cooling system.

\begin{figure}[h!]
    \centering
    \includegraphics[width=0.6\linewidth]{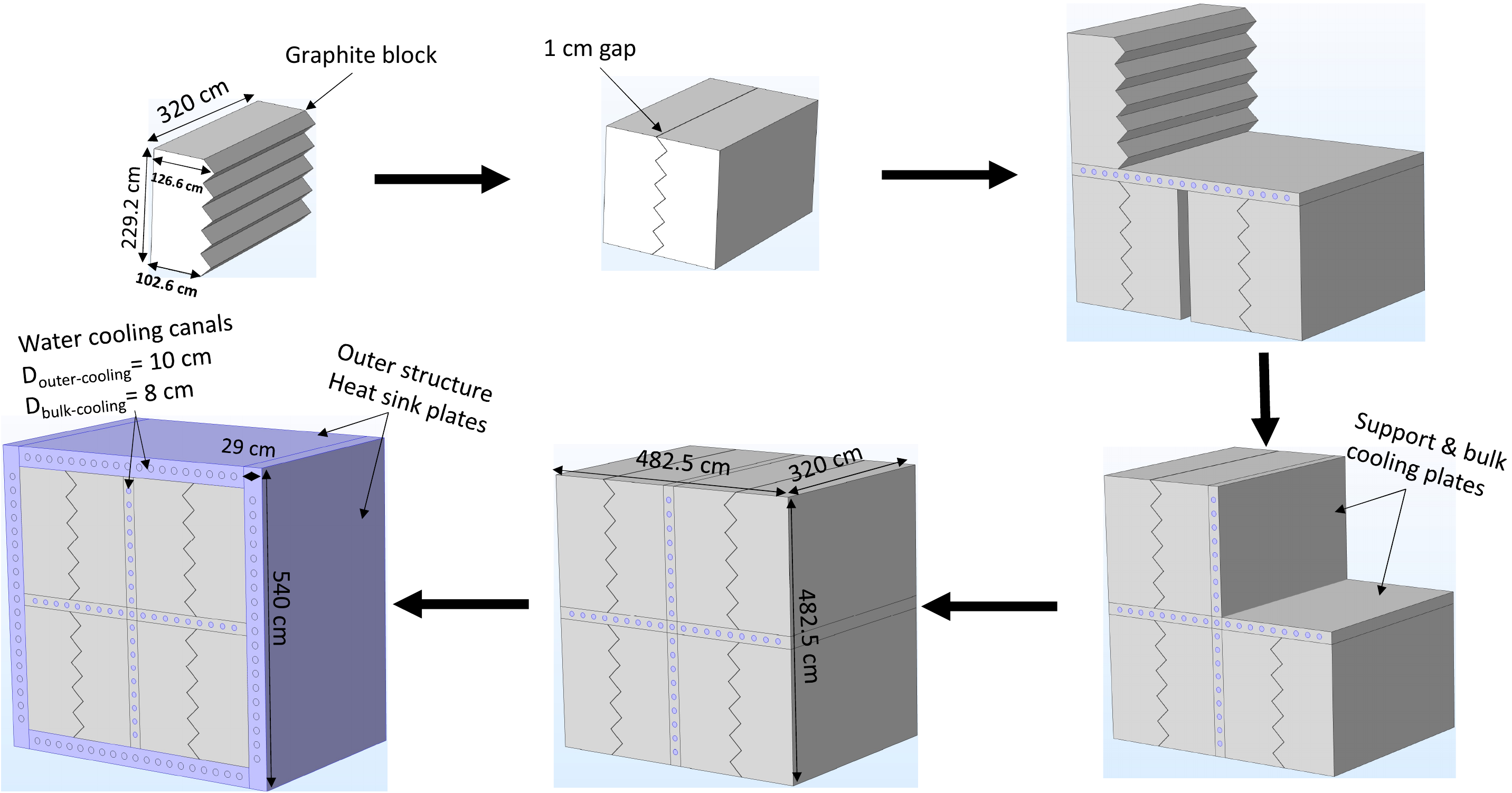}
    \caption{Segmented dump core with water cooling technique geometry and assembly process.}
    \label{fig:Seg-Block-const}
\end{figure}

The left panel in Fig.~$\ref{fig:Seg-Block}$ shows the design simulated with \textsc{COMSOL}, while the right panel shows the power deposition density distribution in the full \SI{5}{\meter}$\times$\SI{5}{\meter} core, the heat sink plates and the support structure. A total power deposition of \SI{854}{\kilo\watt}/4-horns (or \SI{213.5}{\kilo\watt}/horn) was found in the complete core body.

\begin{figure}[h!]
\begin{center}
\begin{subfigure}[b]{0.45\linewidth}
  \includegraphics[width=0.85\columnwidth]{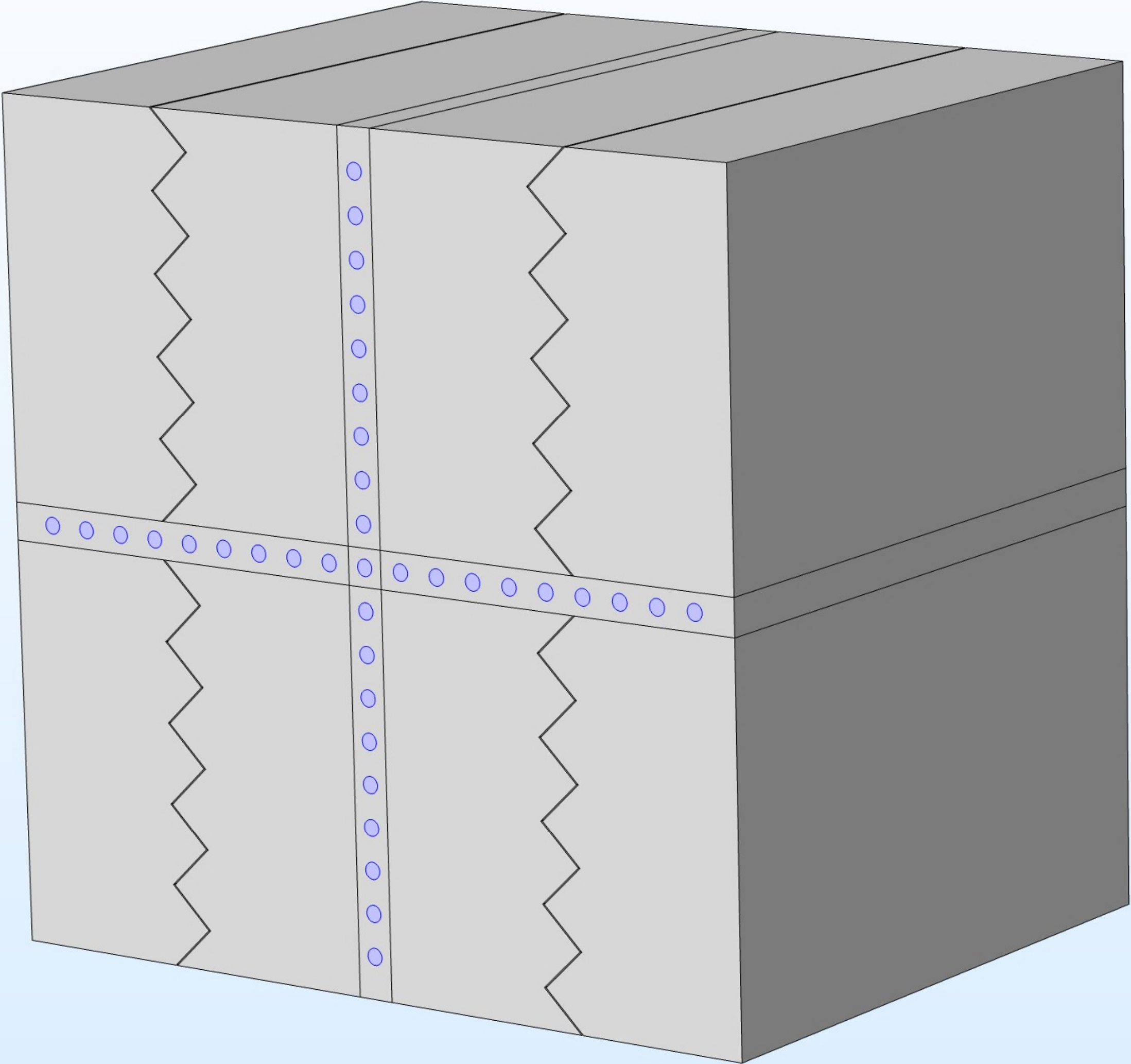}
  \caption{\textsc{COMSOL} 3D drawing of the segmented blocks core design, without cooling system.}
\end{subfigure}
\hspace{1.cm}
\begin{subfigure}[b]{0.45\linewidth}
  \includegraphics[width=1.\columnwidth]{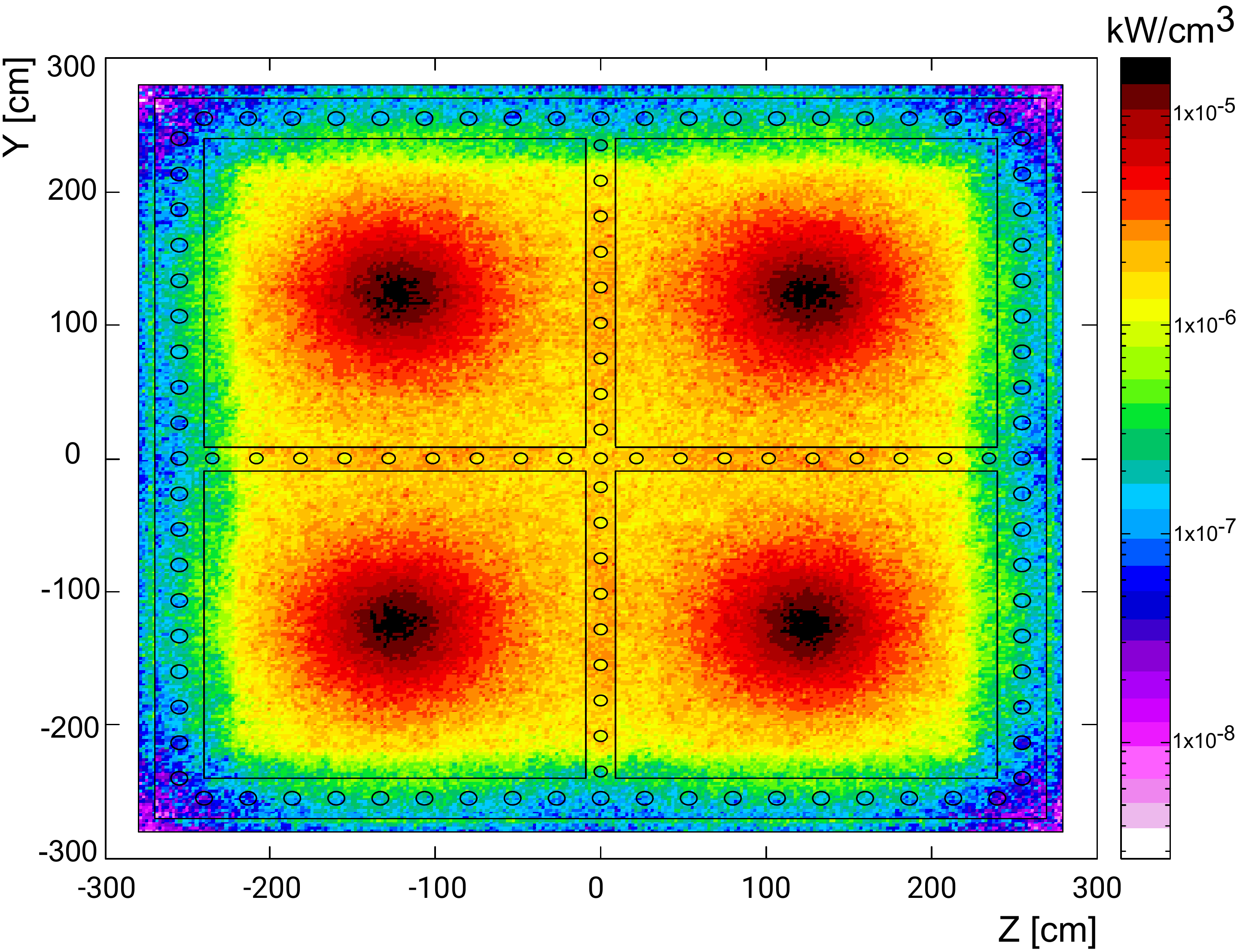}
  \caption{\textsc{FLUKA} power desposition surface density distribution from the 4-horns, on the core surface and cooling heat sink plates.}
\end{subfigure}
\caption{Beam Dump}
\label{fig:Seg-Block}
\end{center}
\end{figure}

\paragraph{\textbf{Thermodynamic Analysis}.} \textsc{COMSOL}~FEA\footnote{FEA : Finite Element Analysis} simulations were performed in order to examine the thermal performance and thermo-structural behaviour of the proposed beam dump design with respect to the applied thermal load. In order to avoid direct contact of leaked water with the graphite blocks, the use of \SI{30}{\cm} support plates as heat sinks with water channels drilled inside the plate body has been considered. The support- and the outer heat-sink plates are assumed to be made of a copper-chromium-zirconium (CuCr1Zr-UNS C18150 \cite{CuAlloy}) alloy, similar to that used for the new proton synchroton booster- (PSB~\cite{PSBBeamDump}) and the ESS beam dumps. 

A fully developed \SI{15}{\celsius} water flow in the channels was assumed. The results show that the cooling efficiency, evaluated by the maximum temperature found in the entire volume of the beam dump structure, reaches a quasi-plateau starting from \SI{40}{\liter\per\min}. A flow rate of \SI{50}{\liter\per\min} will be used in discussing the following simulation results. Moreover, some conservative hypotheses have been tested regarding the thermal transfer between the heat sink plates and the graphite blocks. A ”perfect contact” was assumed between the heat sink plates and the graphite blocks, i.e. COMSOL assumes that the thermal contact resistivity = 0 [m$^{2}\cdot$K/W] at the interfaces between these two bodies. While this assumption is not optimal, due to the surface roughness of the two surfaces in contact, calculating the joint conductance at the interface, $\it{h}_{j}$, on the other hand, is not a trivial process and depends on so many ”practical” parameters that one cannot easily guess. 

\begin{figure}[h!]
    \centering
    \includegraphics[width=0.23\linewidth]{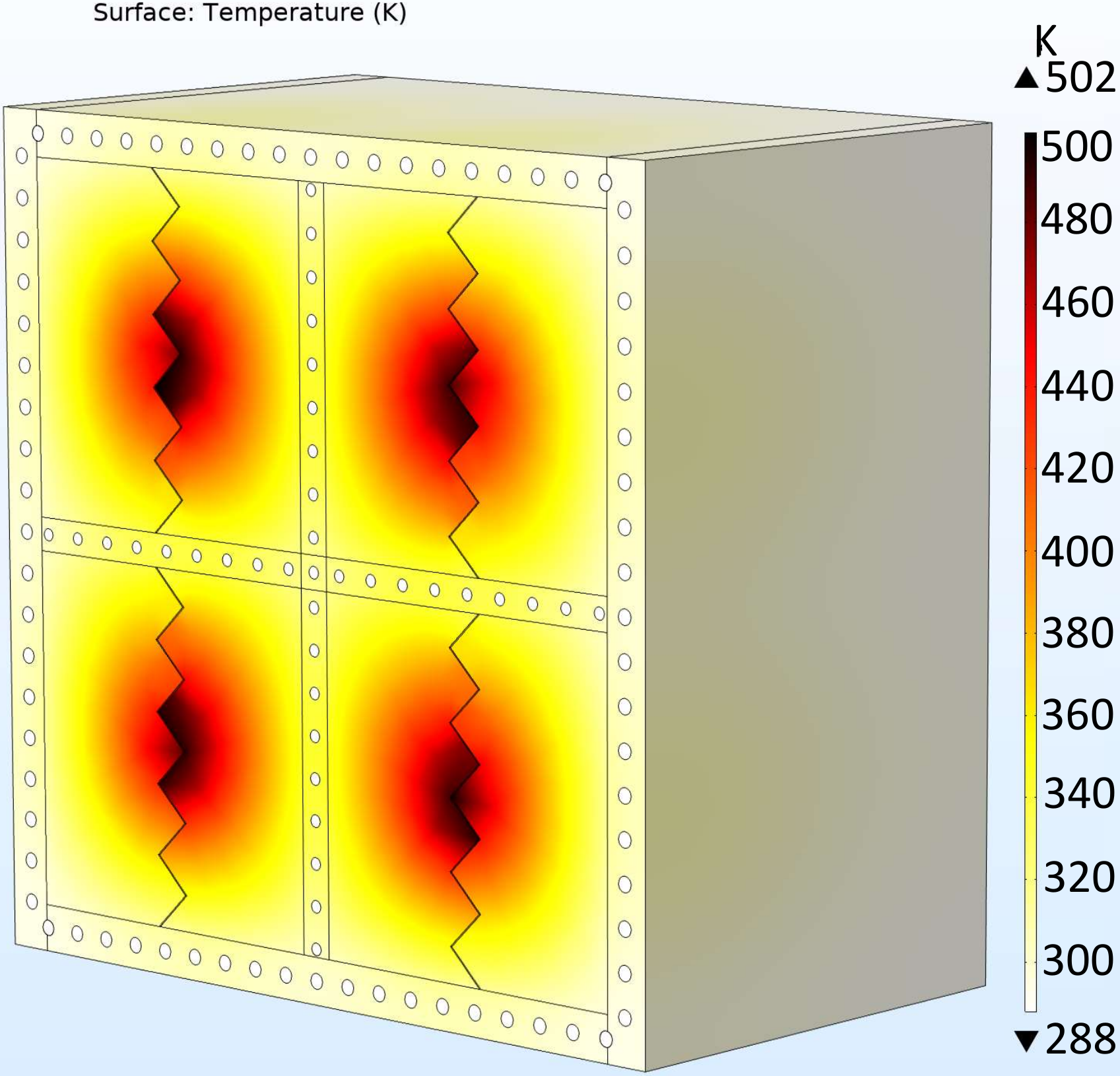}
    ~~~~\includegraphics[width=0.23\linewidth]{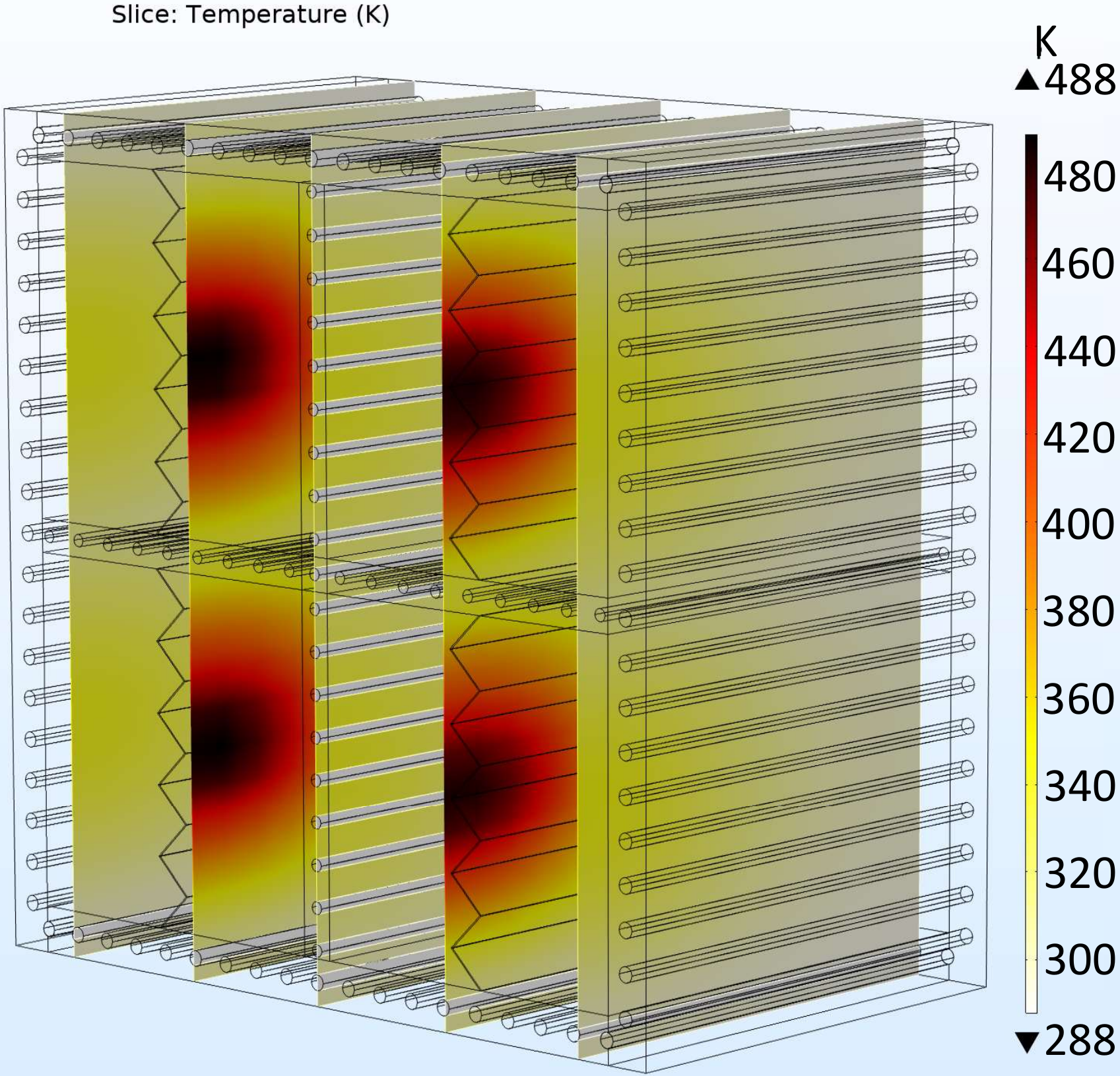}

    \includegraphics[width=0.23\linewidth]{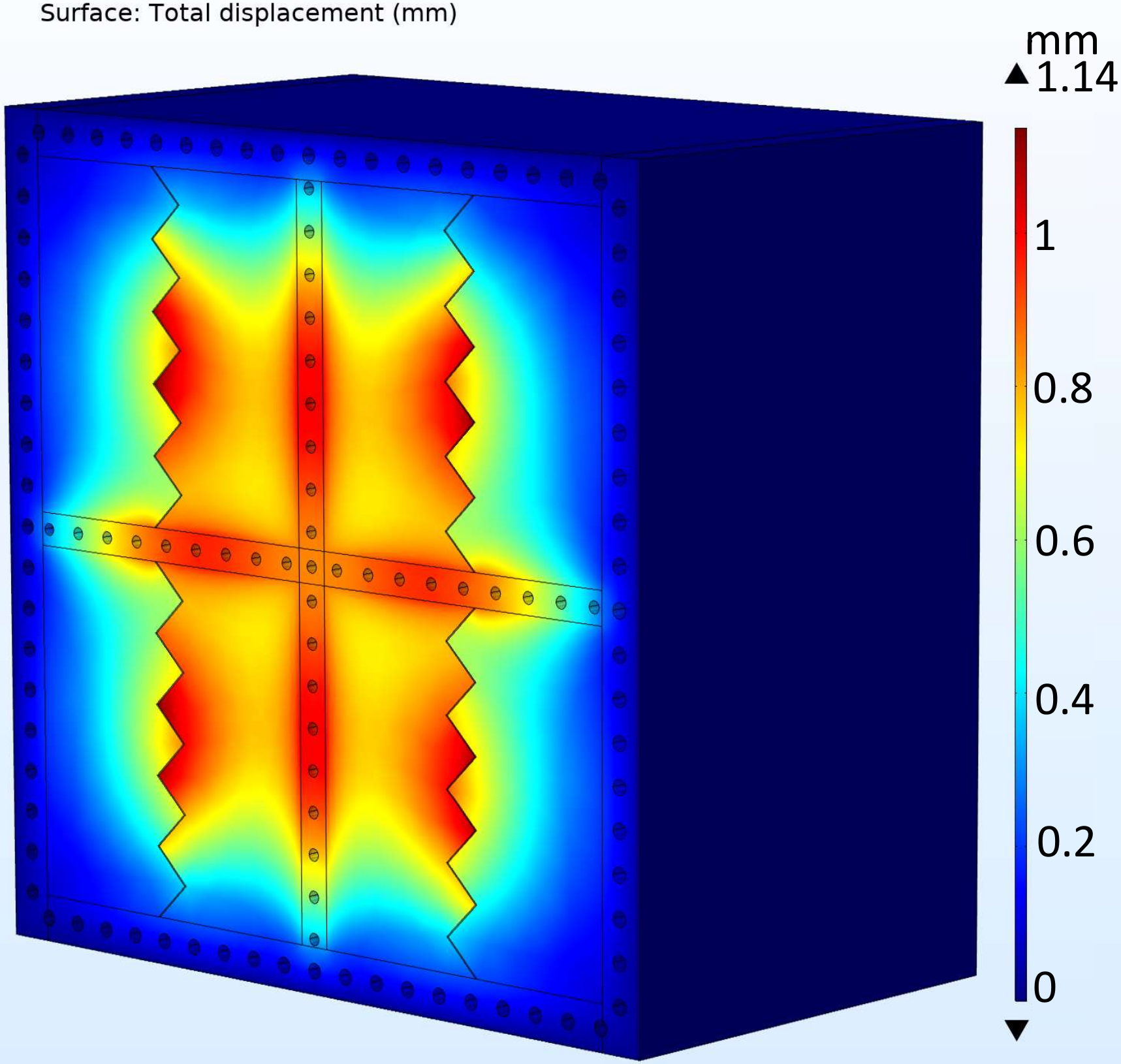}
    ~~~~\includegraphics[width=0.23\linewidth]{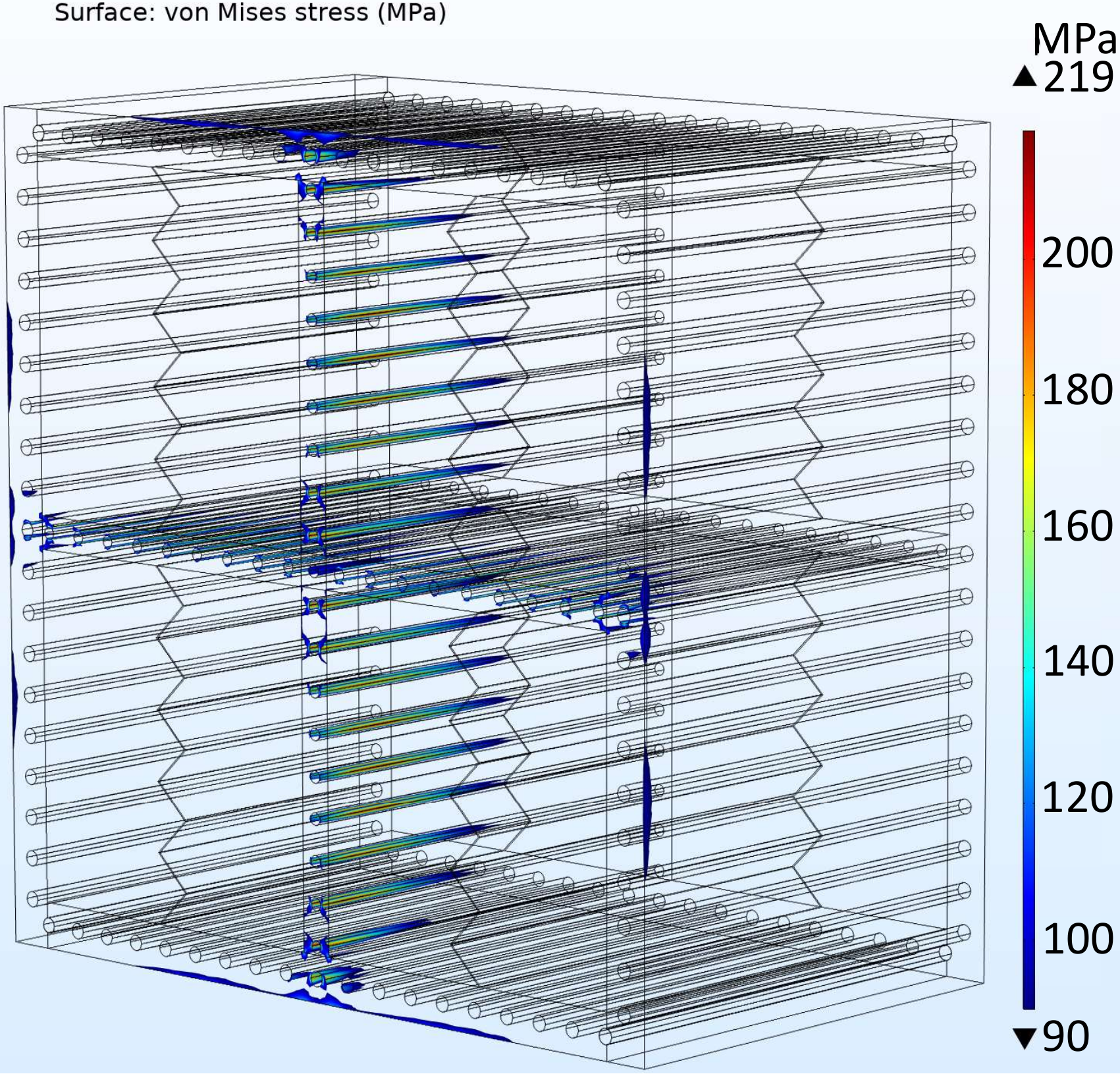}
    \caption{Temperature distribution on the surface (upper-left) and in 2D planes in the bulk (upper-right) of the dump core. Structural deformation level (lower-left) and von-Mises stress distribution (lower-right).}
    \label{fig:res}
\end{figure}

Figure~$\ref{fig:res}$ shows the temperature distribution on the surface (upper-left) and in a 2D cross-sectional planes across the bulk (upper-right) of the beam dump core. The distributions show that the highest temperature, with a value of \SI{502}{\kelvin}, is found at the interface between the two blocks of each segment, which then degrades through the core bulk. This value is compared to \SI{761}{\kelvin} and \SI{774}{\kelvin} for the one-block and EUROnu-like structures, respectively (putting in mind that these two latter structures adopt surface cooling only). 
The temperature of the cooling water increases at the exit of the channels, compared to that at the entrance, by a maximum value of \SI{10}{\kelvin}. Figure~$\ref{fig:res}$ (lower-right) shows the von Mises stress distribution in the segmented block structure. While the graphite blocks of the surface cooling structure do not show high stress values, the stress on the surfaces of the middle heat-sink channels is high, with a maximum value of $\sim$~\SI{219}{\mega\Pa}. This high value is due to the thermal expansion of the graphite blocks, which exert additional outward pressure on the heat-sink plates. However, it is still below the yield strength of the Cu alloy, $\sim$~\SI{310}{\mega\Pa} (@ \SI{20}{\celsius}). The highest von Mises stress for the one-block and the EUROnu-like structures can be found to be \SI{325}{\mega\Pa} and \SI{322}{\mega\Pa}, respectively. 

The maximum von Mises stress found in the graphite bulk of all structures is less than \SI{10}{\mega\Pa}. Moreover, the deformation in the core structure due to such a thermal load and the stresses were studied, and they are presented in Fig.~$\ref{fig:res}$ (lower-left). A very small deformation was found in the segmented block core, with a maximum displacement of \SI{1.14}{\milli\meter}, compared to \SI{3.5}{\milli\meter} and \SI{3.82}{\milli\meter} in the one-block and the EUROnu-like cores, respectively. It is important to note that that \textsc{COMSOL} accounts for the differences in the coefficient for thermal expansion (CTE) between the different materials in contact.

To evaluate the radiation hardness performance, all studies have been repeated for a potential one- and all-horn failure scenario. In the first case, the \SI{5}{\mega\watt} of beam power will be shared by the other three horns, i.e. \SI{1.66}{\mega\watt}/target, a condition that will cause higher local power deposition in the target-station parts and the beam dump core. Again, the segmented block design has a better cooling and mechanical performance with T$_{max}$ $\sim$~ \SI{650}{\kelvin}, compared to T$_{max}$ $\sim$~\SI{864}{\kelvin} and T$_{max}$~$\sim$~ \SI{957}{\kelvin} for the one-block and EUROnu-like designs, respectively. The dynamic response of the beam dump core to the beam impacts (e.g. the von Mises stress and the temperature evolution as a function of time) is a very important parameter that will still need to be studied.

\paragraph{\textbf{Beam Dump Stopping Efficiency}.} Here, the stopping efficiency at the end of the decay tunnel is studied for the proposed core structure, with respect to the unabsorbed primary protons and the remaining secondary particles. The particle yield, $\eta_{par}$ (par denotes particle type), was measured at different locations inside the beam dump: \textbf{a}) on the upstream face of the beam dump core, \textbf{b}) on the downstream (exiting) face of the beam dump, and \textbf{c}) on the downstream (exiting) face of the last concrete block of the beam dump structure. The core stopping efficiency is then calculated by taking the ratios of $\eta_{par}$ at \textbf{b} and \textbf{c} with respect to that at \textbf{a}.
The stopping efficiency at \textbf{b} was found to be $>$ 98$\%$ for all scored particles. Moreover, no particles were found exiting the boundary surface \textbf{c} (i.e. no particles passing through the beam dump to the underground site behind it) with respect to the number of generated pot (protons-on-target). This indicates that the overall protection performance complies with the safety requirements for the ESS underground site.

\begin{table*}[h!]
\small
    \caption{Summary of the results obtained from studying the essential design evaluation parameters.}
    \label{tab:BD_Det_Tab}
    \begin{tabular}{lccccccc}
      \textbf{Core} &  & \textbf{Cooling} & \textbf{Mass} & \textbf{Max.} & \textbf{Max.} & \textbf{Max. stress} & \textbf{Stopping}\\
      \textbf{geometrical} & \textbf{Material(s)} & \textbf{technique} & \textbf{(main piece)} & \textbf{Temp.} & \textbf{displacement} & \textbf{(von Mises)} & \textbf{power}\\
      \textbf{structure} & & & \textbf{[kg]} & \textbf{[K]} & \textbf{[mm]} & \textbf{[MPa]} & \boldmath{[$\%$]}\\
      \hline
      One-block & Graphite/Cu & water (side) & 99,840 & 761 & 3.5 & 325 & $>$ 99$\%$ \\
      EUROnu & Graphite/Cu & water (side) & 50,169 & 774 & 3.82 & 322 & $>$ 98$\%$ \\
      Seg-blocks & Graphite/Cu-alloy & water (side/bulk) & 16,390 & 502 & 1.14 & 219 & $>$ 98$\%$\\\hline
    \end{tabular}
\end{table*}

%--
\subsection{Radiation Safety} 
\label{sub:radiation safety}
Due to the high intensity of the primary proton beam delivered on the four targets, high interaction rates are expected between the resulting secondary beam and the different parts of the target station. The different density and Z-composition of the materials that compose these parts are expected to cause different levels of radio-activation. The main objective is to evaluate the radiation protection risks in the target station complex, in order to respect/obey the applicable ESS Radiation Safety Functions (RSF) (the IAEA standards + ESS specifications) to prevent or mitigate the radiological hazards (i.e. a dose uptake to on-site personnel and to the public). In particular, the prompt and residual dose-equivalent rates, the radio-activation and radioisotopes formation and energy deposition estimation are the most important radiation safety parameters that will be reported here. Moreover, the level of radio-activation of the different parts of the target station facility depends on the running mode of the experiment. Three main running modes: commissioning, normal, and emergency running modes have been considered. Because the design of the commissioning procedure of the individual parts of the facility is still in progress, the activation studies for the running mode are not finalised, and therefore will not be presented in this document. However, for the case of the emergency operating mode (i.e., with three active horns instead of four) the total beam power is still preserved. Thus, only local positions within the different internal structures will be affected, and the integrated values of the studied safety parameters over the whole volume of the individual parts will be the same. The proton beam profile used in the \textsc{FLUKA} simulations was taken from section~\ref{subsec:R2S}, generated at the entrance of the baffle/collimator system, $\sim$~\SI{3}{\meter} before the target canister surface. An exposure time of one operation-year (i.e.~200 days) has been assumed in the simulations, with~3.19$\times$10$^{15}$ pot/s/horn (or 5.4$\times$10$^{22}$ pot/op-year/horn) of \SI{2.5}{\giga\electronvolt} beam kinetic energy and \SI{1.25}{\mega\watt}/target beam power.

\subsubsection{Shielding}
\label{sub:Shielding}
The design of the target station shielding was optimised based on the general radiation protection guidelines, the prompt and residual dose calculations, and the radio-activation analysis of the different parts of the target station facility. The primary source of radiation will be the 4 horns and targets in the 4-horn gallery area, the decay-tunnel vessel in the decay tunnel area, as well as the support structure and the graphite cores in the beam dump area. To provide radiological shielding for the underground site surrounding the horn/target area, the helium vessel of the 4-horn gallery will be surrounded on all sides by $\sim$\,\SI{2.2}{\meter} thick iron inner-shield followed by $\sim$\,\SI{3}{\meter} thick concrete outer-shield. Although integrating concrete with sodium-acetate enhances the thermomechanical performance of concrete~\cite{NaConcrete}, a low-sodium concrete is recommended for the shielding in order to limit the formation of radioactive sodium isotopes. In the present simulations, normal concrete (i.e., with low-sodium content) is assumed. Moreover, the outer concrete shield will need to be sealed to prevent air leakage from the region immediately surrounding the 4-horn helium vessel into the target station atmosphere. This was achieved by adding an extra \SI{3}{\meter} thick concrete block on top of the 4-horn gallery area. One issue here is that there must be a way to open the shielding in order to have access to the components within the 4-horn gallery. This is achieved by constructing the top of each shield with stacked concrete blocks that can be moved aside by a crane. Preliminary simulations also indicated that
$\sim$\,\SI{2.5}{\meter} thick concrete block must be placed before (upstream to) the baffle/collimator system, in order to protect the switchyard area from back-scattered radiation coming from the horns and the targets. In order to protect the underground site beyond (downstream), the beam dump will be surrounded by $\sim$\,\SI{5.5}{\meter} concrete shielding on all sides, and with $\sim$\,\SI{3.4}{\meter} alternating layers of iron blocks, plus a final concrete block furthest downstream. The decay tunnel will be surrounded by $\sim$\,\SI{5.5}{\meter} concrete shielding on all sides, as well.

\subsubsection{Dose Equivalent Rate Calculations}
\label{sub:Dose-rate}
The prompt dose equivalent rates were calculated from the convolution of the fluence of neutrons, protons, charged pions, and muons with their respective energy-dependent fluence-to-effective dose conversion coefficients. The residual dose equivalent rates were calculated by the convolution of the fluence of photons, electrons, and positrons from $\gamma$- and $\beta$-decays with the respective energy-dependent fluence-to-effective dose conversion coefficients. The dose conversion coefficient \textit{EWT74} provided by the \textsc{FLUKA} database was used. Various cooling times: \SI{1}{\second}, \SI{1}{\hour}, 1 day, 10 days, 1 and 3 months after turning off the beam were assumed for the residual dose equivalent rates calculations. The radiation categorisation of the specific zones of the target station complex relies on the Radiation Protection Ordinance (RPO) of ESS. This was categorised as follows: a) the 4-horn gallery, the decay tunnel, and the beam dump areas are classified as \textit{prohibited-zones}, with no access permitted during operation. b) the power supply unit area and target station utility rooms are classified as \textit{controlled-zones}, with access permitted to authorised personnel only. This is the work area above the decay tunnel and beam dump concrete shields, in which the annual radiation doses may exceed 3/10~ths of the annual maximum permissible doses for exposed workers (i.e. $\sim$\,3 $\mu$Sv/h). c) The control and data acquisition rooms, at the ground level, are classified as \textit{supervised-zones}, with access permitted at all times, with low occupancy of $<$ 1.5 $\mu$Sv/h. All dose calculations in these zones are based on the dose equivalent rates in an assumed 2$\times$2$\times$2 m$^{3}$ volume of water as representative for human workers.

\begin{figure}[h!]
    \centering
    \includegraphics[width=0.8\linewidth]{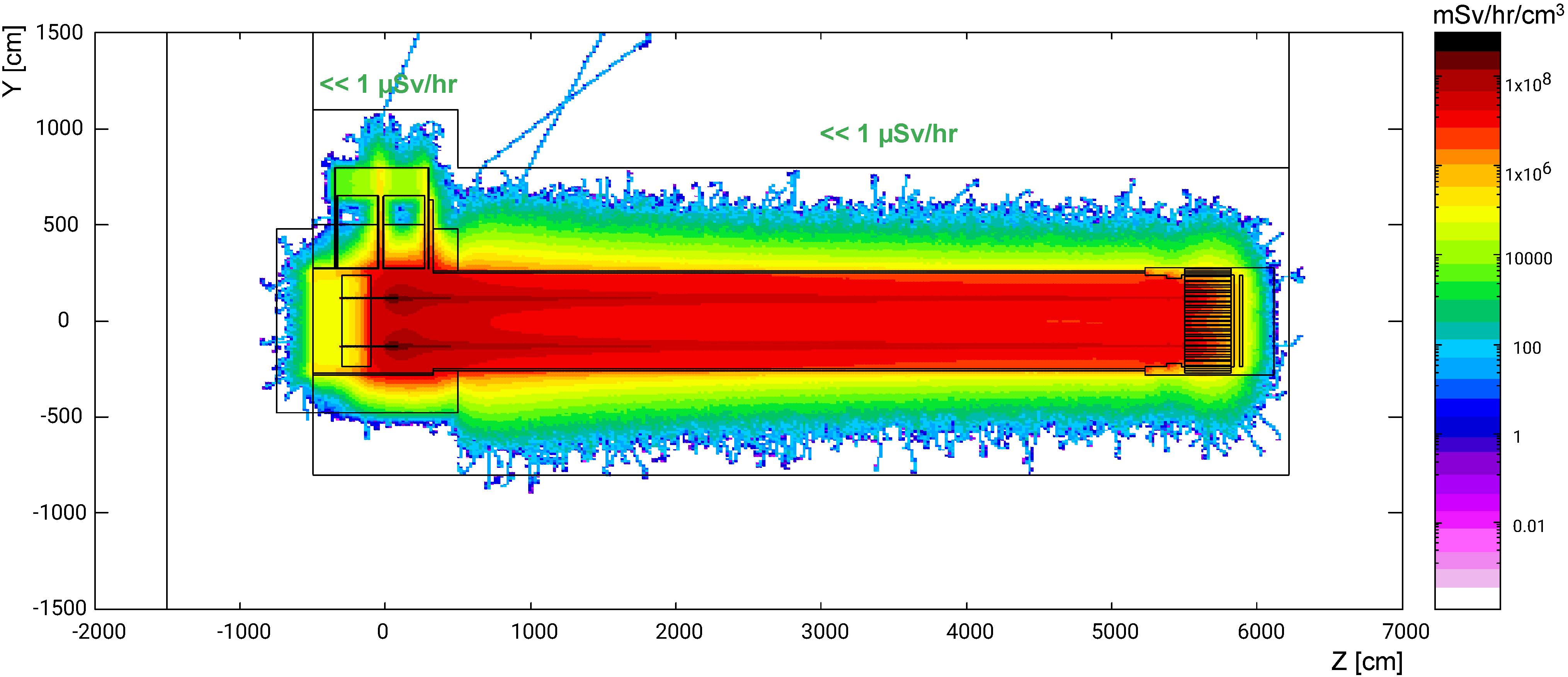}
    \includegraphics[width=0.8\linewidth]{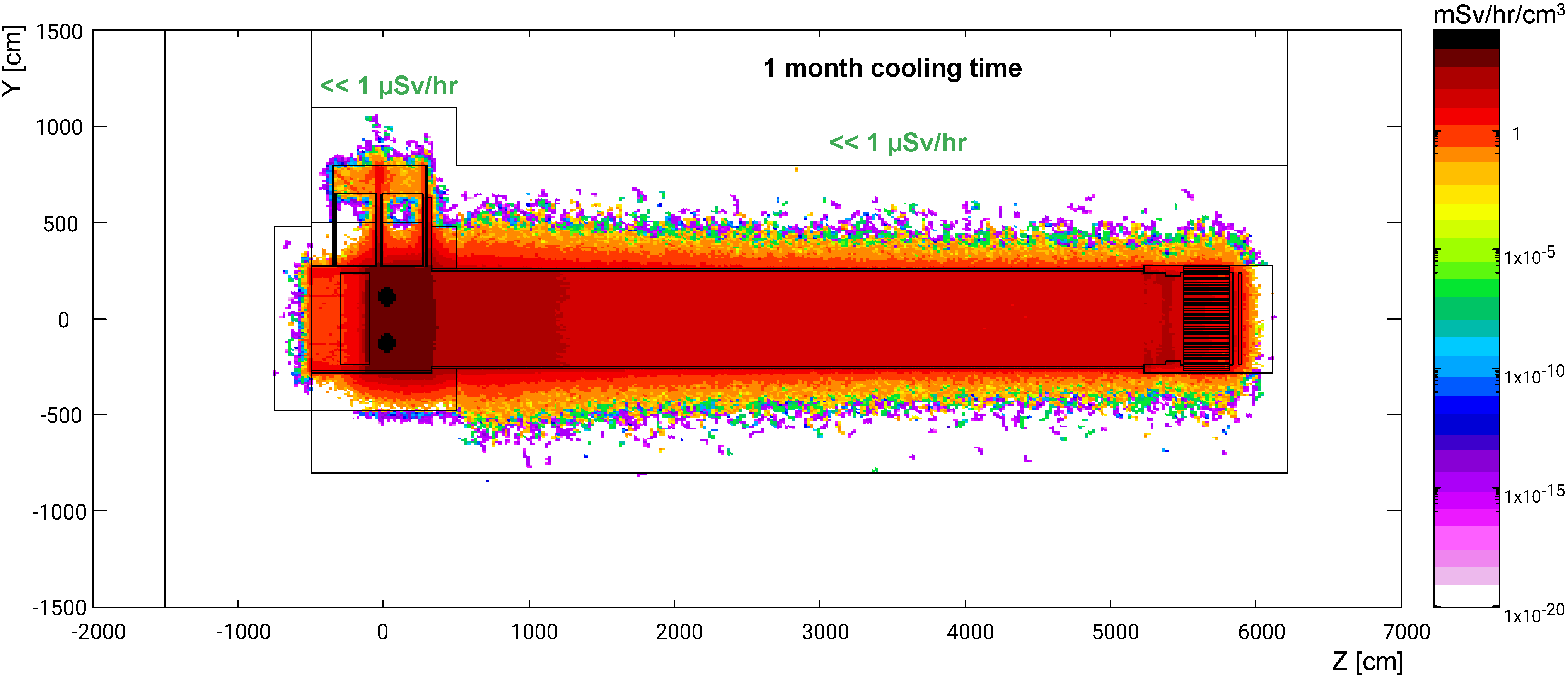}
    \caption{A radial view of the prompt (top) and residual (bottom) dose equivalent rate distributions (in [mSv/h]) in the target station facility. The residual distribution is simulated after a one-month cool down.}
    \label{fig:dose-eq}
\end{figure}

Figure~\ref{fig:dose-eq} shows the prompt (top) and the residual (bottom) dose equivalent rate distributions in the target station facility, with the residual distribution shown after a one month cool down. The prompt simulations show that during irradiation the dose equivalent rate does not exceed an acceptable level of 1\,$\mu$Sv/h in the areas above the 4-horn system, the decay tunnel, the beam dump and in the underground site/earth behind. This indicates the efficacy of the shielding system and allowing for accessibility of these zones during irradiation. The bottom frame shows that the residual dose equivalent values after one month cooling time, for all designed accessible areas, are also below 1\,$\mu$Sv/h. It is worth noting that the residual dose equivalent rates as a function of the cooling time (from 1~second to 3~months) were studied for each section of the target station facility.

\subsubsection{Radio activation and Radioisotope Formation}
\label{sub:Activation}
The radio activation in the internal structure and the helium volume of the target station facility was evaluated for a one operation-year (operation year). The total activation level, in Bq, was studied as a function of cooling time (from \SI{1}{\second} to 3 months), for each part of the target station complex. Moreover, a detailed inventory of all radioisotopes produced has been studied as well. The main objective of this detailed activation map is to provide \textit{a priori} the target station operators with the level of activation and the type of radiation expected, before entering (e.g. in emergency situations) the specific areas at different time periods up to three months after shutting off the beam.

\begin{figure}
    \centering
    \includegraphics[width=0.8\linewidth]{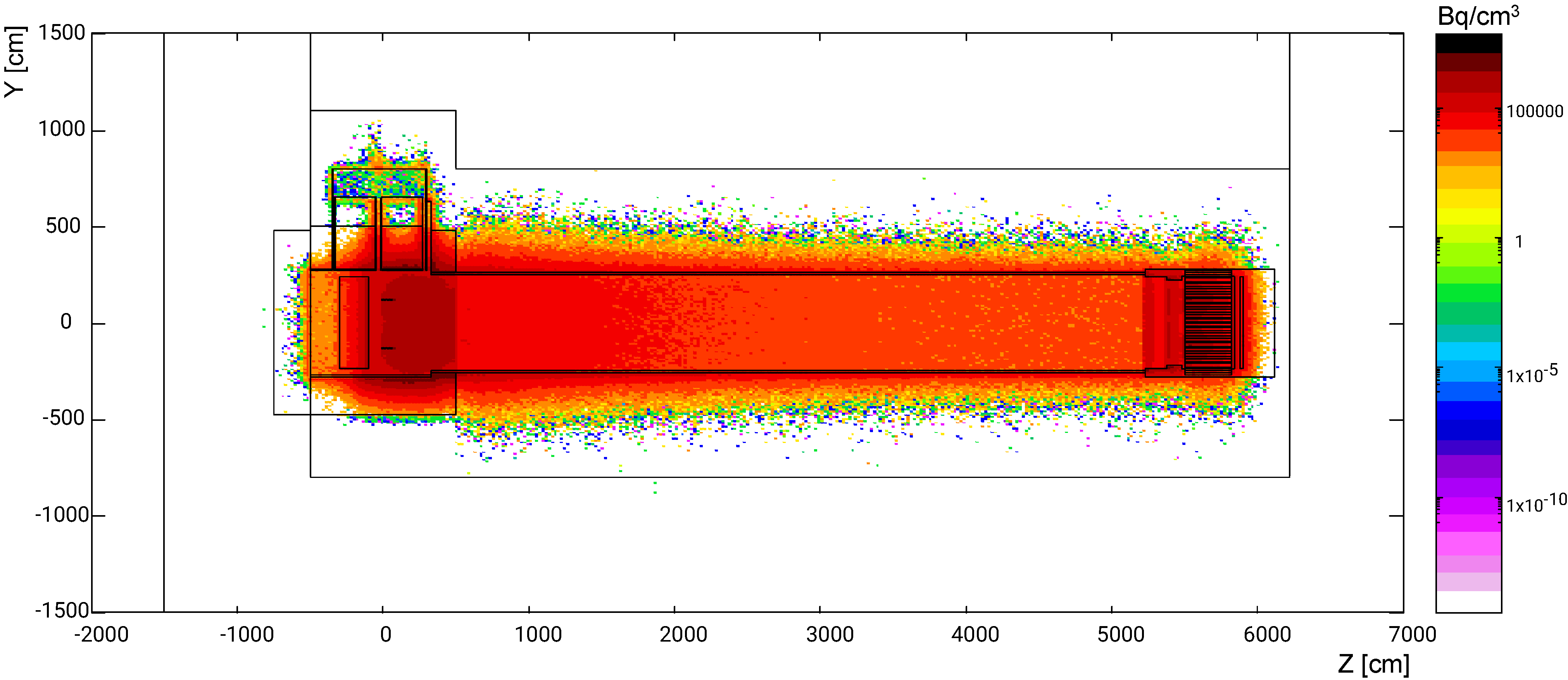}
    \caption{Radiation activation distribution in the target station facility after a one month cooling time.}
    \label{fig:RadioActivation}
\end{figure}

In the most critical helium regions, that is in the vessel surrounding the four-horn target system, a total activity of $\sim$\,5.8$\times$10$^{10}$ Bq was found, up to three months after shutting off the beam. This exhibited the formation of the tritium isotope, $\ce{^{3}_{1}H}$, which has a significantly lower radiological impact than the airborne radionuclides which would result from an alternative air-filled target station. It is worth mentioning that the current simulations do not assume He circulation in any part of the facility. A similar study has been applied to the cooling water of the beam dump, assuming two scenarios: a) a non-circulating water system, wherein the cooling water is exposed to the secondary beam for 200 days and b) a circulating water system, with a specific amount of cooling water exposed to the secondary beam for $\sim$\,\SI{33}{\second} only (assuming a flow rate of \SI{50}{\liter\per\min}, as discussed in section~\ref{subsubsection:BD}). The simulations show that the water activation level reaches $\sim$\,2$\times$10$^{11}$ Bq and $\sim$\,7$\times$10$^{6}$ after three-month cooling time, for the non-circulated and circulated scenarios, respectively. This also demonstrates the formation of tritium and beryllium $\ce{^{7}_{4}Be}$ radioisotopes in the cooling water. While the $\ce{^{7}_{4}Be}$ is expected to disappear naturally after approximately one year's cooling time, the $\ce{^{3}_{1}H}$ will remain present much longer. This makes it necessary to consider including a tritium handling system in the target station facility or, as an alternative, to connect the ESS$\nu$SB water cooling system with that of ESS, which includes a tritium handling system.

\begin{figure}[h!]
\begin{subfigure}[b]{0.45\linewidth}
  \includegraphics[width=1.05\columnwidth]{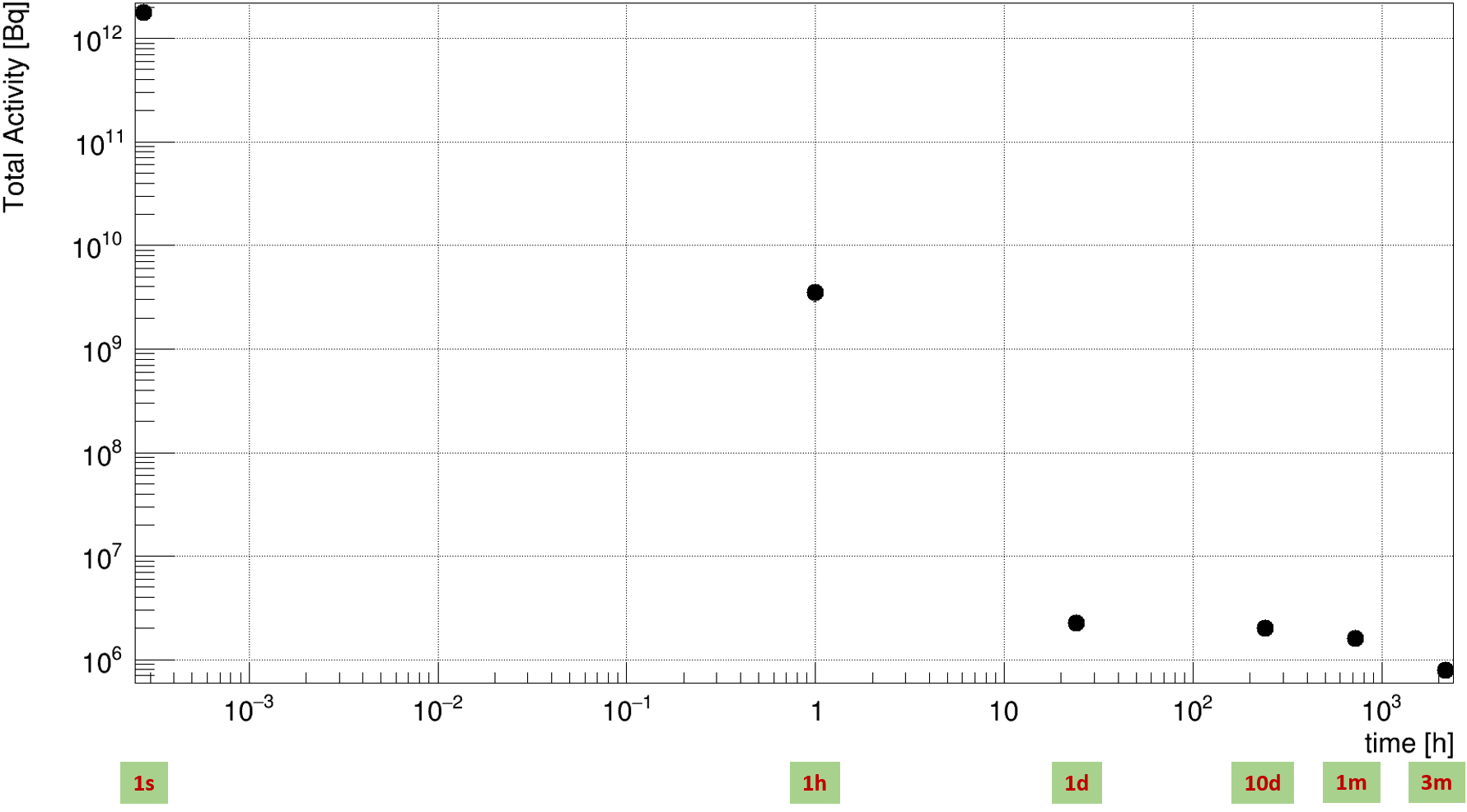}
  \caption{Total activity as a function of cooling time.}
\end{subfigure}
\hspace{1.cm}
\begin{subfigure}[b]{0.5\linewidth}
  \includegraphics[width=0.84\columnwidth]{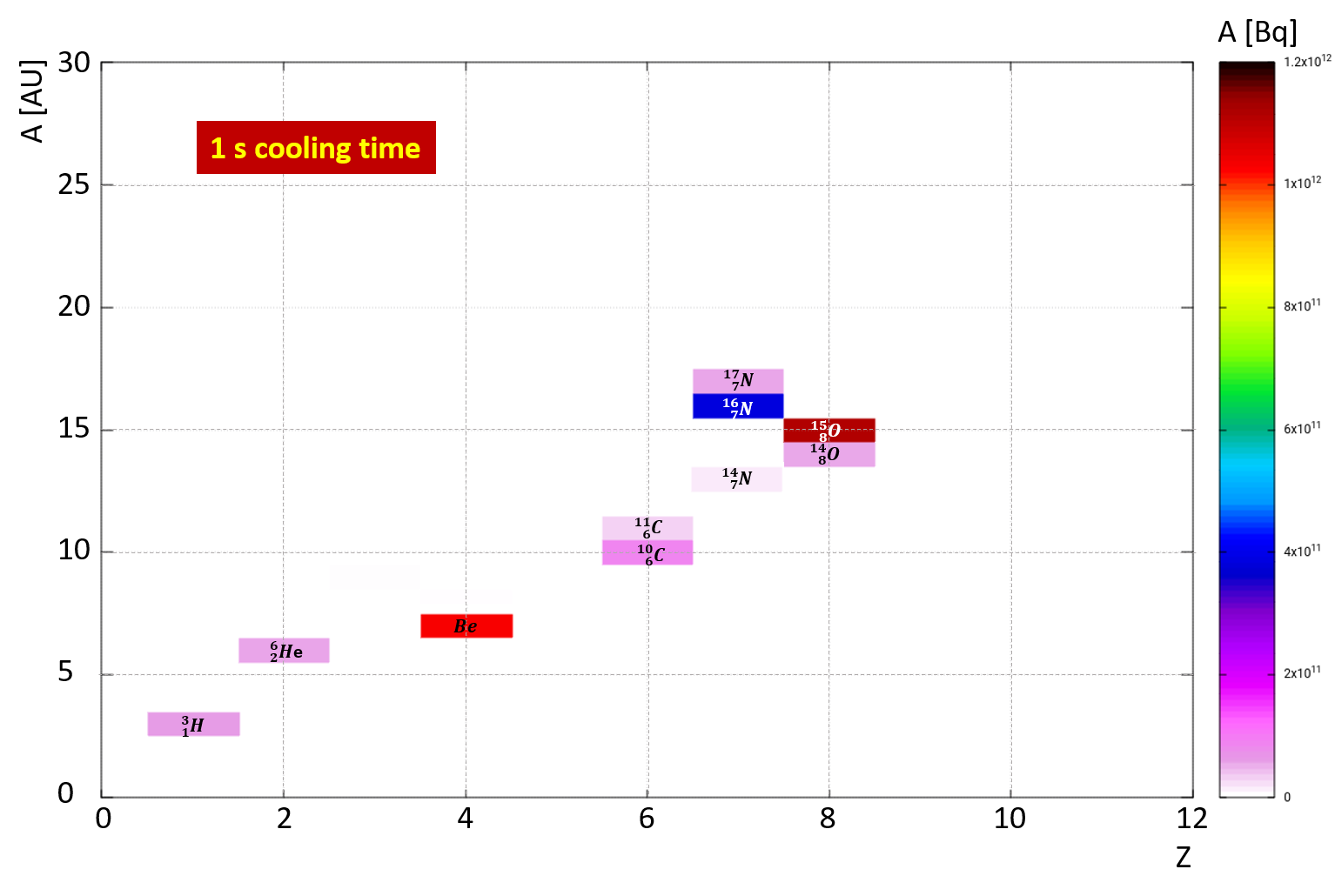}
  \caption{Radioisotopes produced after \SI{1}{\second} cooling time in the BD cooling water. 
%  The color bar in the right-hand figure represents the activity of the isotope
  }
\end{subfigure}
\caption{Time evolution of the activation}
\label{fig:TotAct_vs_time}
\end{figure}

Figure~\ref{fig:RadioActivation} shows the remaining radiation activation level in the target station facility after one month of cooling time. Figure~\ref{fig:TotAct_vs_time} (left) shows an example of the total activation in the beam-dump cooling water as a function of cooling time (the error bars represent statistical errors only), whereas the right figure shows the radioisotopes produced in the cooling water after \SI{1}{\second} of beam shutdown (the x-axis represents the element/charge number, Z, and the y-axis represents the atomic number, A).

\subsubsection{Power Deposition studies}
\label{sub:Energy-Dep}
The simulated energy-deposition distribution in the internal structures of the target station facility are presented in Fig.~$\ref{fig:Edep_all}$. These results show that the strongest power deposition, with a value of almost \SI{1}{\mega\watt}, is found in the beam-dump graphite core, as well as in the decay-tunnel vessel, the 4-horn gallery iron shield and the decay-tunnel concrete shield, with roughly \SI{840}{\kilo\watt}, \SI{640}{\kilo\watt} and \SI{612}{\kilo\watt} total power deposition values, respectively. These parts will be cooled by water pipes. The power deposition and the corresponding thermo-mechanical behaviour of the target, horns, decay tunnel and beam dump are presented in detail in sections \ref{sub:target-cooling-concept}, \ref{section:MagneticHornDesign}, \ref{subsubsection:DT} and \ref{subsubsection:BD}, respectively.

\begin{figure}
    \centering
    \includegraphics[width=0.8\linewidth]{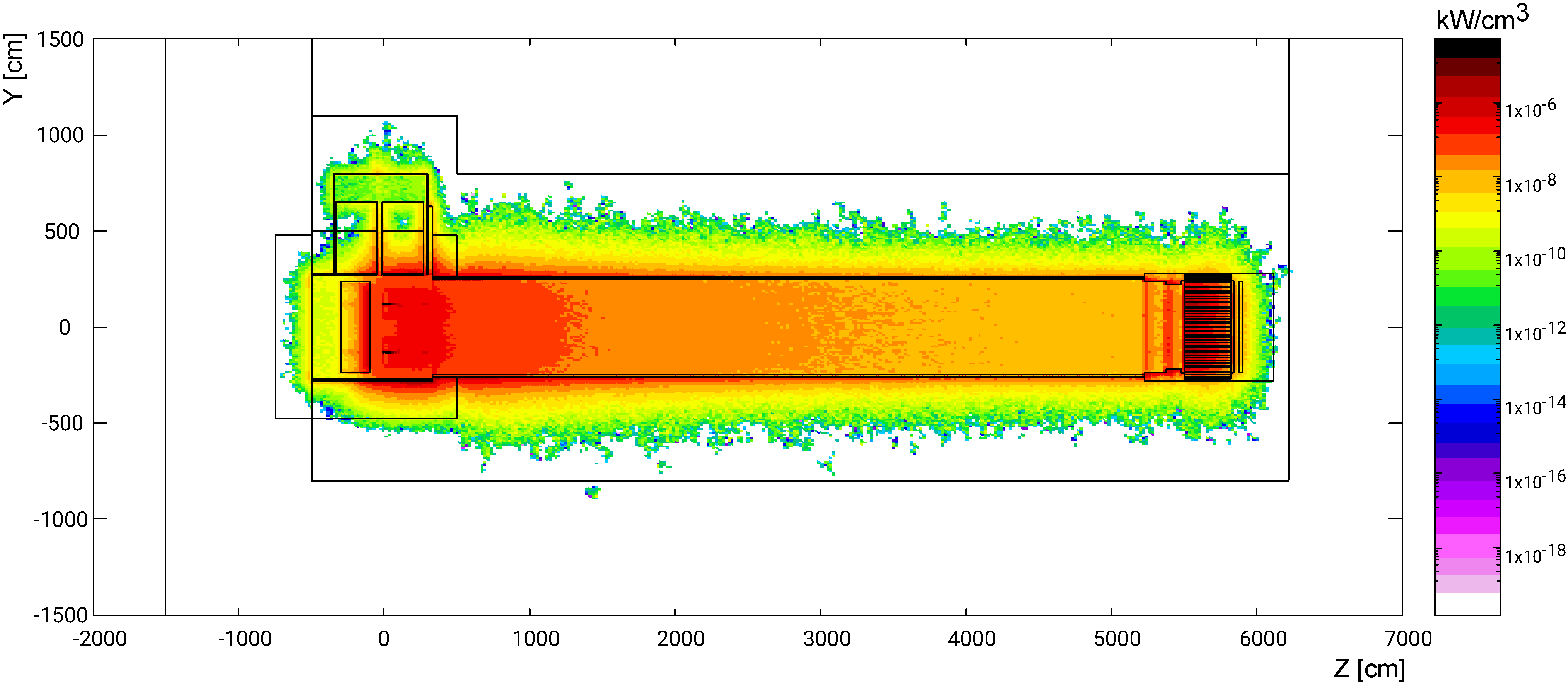}
    \caption{Energy deposition distribution in the target station facility.}
    \label{fig:Edep_all}
\end{figure}

%--
\subsection{Costing}

The ESS$\nu$SB target station follows the same breakdown structure as most other Super Beam experiments. Thus, its cost estimation is based on comparable elements from existing facilities. However, the design of the elements related to specific working conditions of the ESS proton driver are based on solutions which are at the limit of the present technology. The resulting cost of such elements is estimated with a conservative approach, and is summarised in Table~\ref{tab:ESSTargetStationCost}. 

\begin{table}[ht!]
  \begin{center}
    \caption{Summary of the ESS$\nu$SB Target Station Costs.}
    \label{tab:ESSTargetStationCost}
    \begin{tabular}{lc}
    \textbf{Target Station item}  &\textbf{Cost [M\texteuro]} \\
\hline
Target Station & 32 \\
Proton Beam Window System & 5.2 \\
PSU + Striplines & 5.4\\
Target and Horn Replacement System & 40.4\\
Facility Building Structure& 26.6\\
General System \& Services &  21.8\\ 
\textbf{Total} & \textbf{131.4} \\ 
\hline
     \end{tabular}
   \end{center}
\end{table}

\paragraph{Target Station (30~M\texteuro)} The design of this system attempts to accommodate the extreme working conditions to which the target and the hadronic collector will be subjected during operation. Many hypotheses have been made in terms of the technology chosen for the target and for the simulation of the cooling of each horn, using a conservative approach. 

\begin{itemize}
\item Target + Integration System (15~M\texteuro). The evaluation of the cost of the target system and all secondary equipment (in particular 
the target integration system, the helium cooling system and the target replacement system) has been done based on previous experiments, without feedback from industrial suppliers. The design of these elements is unique, considering the exceptionally high radiation levels to 
which the target station will be exposed. This estimate was made assuming four targets; however, their lifetime is not well-known at the present stage of the design study. Spare targets will most likely be needed, which will increase the cost.

\item Hadronic Collector System (15~M\texteuro). Again, this estimate is for four horns. There is some uncertainty regarding the behaviour 
of Aluminium Al T6061 in a 5~MW environment. In addition, the high voltage ($+12$\,kV) delivered by the power supply unit may cause electrical breakdown to the neighbouring environment. Solutions to this problem may also increase the cost. 

\item Four-horn support structure (2~M\texteuro). 

\end{itemize}
~\noindent Furthermore, an experimental prototype will be required to test many of the elements of the target station, to confirm the simulation results.

\paragraph{Power Supply Unit + Striplines (5.4~M\texteuro)} The specific structure of the proton pulses arriving from the linac imposes fast commutation between individual horns, which can be solved with the modular approach. The total cost of the elements constituting the 16~modules of the power supply unit are detailed in the following list:
\begin{itemize}
\item Big switches (1529.6~k\texteuro )
\item Bench capacitors (1494.4~k\texteuro )
\item Recovery coils (268.8~k\texteuro )
\item Recovery diodes (208.0 ~k\texteuro)
\item 24 kW chargers (691.2~k\texteuro )
\end{itemize}
~\noindent The high voltage (+\SI{12}{\kilo\volt}) between the striplines requires electrical insulation. 
The proposed solution is based on a ceramic deposited by laser technology on each side of the striplines, for such a solution the estimates are as follows:
\begin{itemize}
\item Striplines (282~k\texteuro) 
\item Ceramic Insulation (928~k\texteuro) 
\end{itemize}
~\noindent Joule heating of the striplines, the high irradiation environment and the mechanical constraints represent a technological risk and will require additional R\&D.

\paragraph{Target and Horn Replacement System (40.4~M\texteuro)} This cost is based on similar Super Beam facilities. The full target station design must specify the cost more precisely.

\paragraph{Proton Beam Window System (5.2~M\texteuro)} This element comprises the window material, using beryllium as a baseline, and a complex system for beam-window cooling. The beam power of ESS$\nu$SB is the same as that of the ESS target station. Thus, the total cost of its beam window system is presumed to be a good approximation.
 
\paragraph{General Systems and Services (21.8~M\texteuro)} These estimates, in which a conservative approach has been used, are based on similar experiments:
\begin{itemize}
\item Active cooling system (9.6~M\texteuro)
\item Vacuum and pumping system (8.9~M\texteuro)
\item Handling system (3.3~M\texteuro)
\end{itemize}
 ~\noindent One should also pay attention to the strict ecological regulatory standards in Sweden, which may increase the above cost. For example, it will be required that the heat produced by the target station be recycled to the municipal district heating system. 
 
\paragraph{Facility Building Structure (26.6~M\texteuro)}
\begin{itemize}
\item Horn Gallery (20.8~M\texteuro)
\item Decay Tunnel (2.1~M\texteuro)
\item Beam Dump (3.3~M\texteuro)
\end{itemize}
 ~\noindent 
In this estimation, the operations related to civil engineering (like excavation) are not taken into account. The plan for the treatment of radioactive waste, such as tritium, will take advantage of the nominal ESS infrastructure and capabilities. However, this is not taken into account in this part of the cost estimate.

\subsection{Summary}

The target station has been designed to provide an intense neutrino beam from a 5\,MW proton beam at \SI{2.5}{\giga\electronvolt}, delivered by the ESS linac. The sensitivity of the experiment has been optimised through deep learning techniques. This has resulted in an optimised horn shape as well as an optimised geometry of the entire target station.
A cooling method for the packed-bed target has been developed, making use of the transverse flow of gaseous helium through the target pores. The calculated temperatures of the helium cooling gas and the titanium spheres making up the target appear to fall within acceptable limits. 

Target integration is more challenging than for most comparable experiments, mainly due to the high level of power deposited in the target, which is concentrated into a very limited space within a magnetic horn. An integration concept has been proposed, which prevents heat from being transferred from the target to the horn conductor. Mechanical deformations and the stress in the target container caused by the temperature gradient appear to be feasible. Experimental verification is still required to confirm the simulation results. This applies in particular to the flow of helium gas through a porous medium, the determination of the heat transfer coefficient between the water spray and the horn skin, as well as the behaviour of the spheres inside the target container. Material issues pertaining to high-temperature and radiation exposure require further study.

The proposed scheme for the power supply unit will allow for the required 350\,kA electrical pulses to be applied to each horn at a frequency of 14\,Hz, taking into account the challenging rapid commutation between the four horns (imposed by the pulse structure of the proton beam coming from the linac). The modular approach using 44\,kA units connected in parallel, with eight striplines for each horn, reduces the critical constraints on all components. This should permit a 10-year running time for the experiment.A new concept of the beam dump has been proposed as part of the target station facility design, capable of withstanding ${\sim}$850\,kW power deposition. 
\clearpage

% ************ Style guide for the WP5 CDR ************
%
% This is the language and latex style agreed upon by the editors of the CDR, tuned to the specifics of the WP5:
% 1. British English, passive voice.
% 2. InspireHEP style bibtex labels and fields if available
% 3. Consistent labeling of figures, tables, equations, and more - fig:detectors:my_figure, table:detectors:my_table, eq:detectors:my_equation, ...
% 4. No titles inside figures, descriptions go in captions (unless there is a very good reason for it). Use .pdf instead of .png whenever possible.

% Use this to number the figures, equations and tables with section number in front
\setcounter{figure}{0}
\numberwithin{figure}{section}
\setcounter{equation}{0}
\numberwithin{equation}{section}
\setcounter{table}{0}
\numberwithin{table}{section}

%\newcommand{\tit}[1]{\text{\textit{#1}}}
%\newcommand{\ABtodo}[1]{\textbf{[#1]}}

%\DeclareSIUnit\year{yr}
%\DeclareSIUnit\min{min}
%\DeclareSIUnit\h{h}
%\DeclareSIUnit\d{d}
%\DeclareSIUnit\PeV{PeV}
%\DeclareSIUnit\MB{MB}
%\DeclareSIUnit\GB{GB}
%\DeclareSIUnit\TB{TB}
%\DeclareSIUnit\PE{PE}
%\DeclareSIUnit\inch{inch}
%\DeclareSIUnit\rad{rad}
%\DeclareSIUnit\speedoflight{\tit{c}}

\section{Detectors} 
\label{Detectors}
%{\bfseries TODO: Roumen, Klicek Budimir, Alexander Burgman (alexander.burgman$@$nuclear.lu.se)}

%\subsection {Introduction}

The principal objective of the ESS$\nu$SB project is to study the feasibility and design of a super-beam neutrino facility based on the European Spallation Source (ESS) proton linear accelerator to measure the CP-violating phase $\delta_{CP}$ in the lepton sector. %The main goal is to demonstrate that the ESS accelerator can be effectively used to generate an intense neutrino beam concurrently with the spallation neutron production by doubling the average proton beam power. 
The detector complex of the facility consists of a megaton-scale water Cherenkov (WC) far detector (FD) located at a distance of \SI{360}{\kilo\meter} from the beam source and an efficient suite of near detectors (ND).
%, it will allow for a search of the Charge-Parity Violation (CPV) in the lepton sector and a measurement of the CPV phase .
%
%We note that results from previously published ESS$\nu$SB deliverables and proceedings have in some cases been directly included in this chapter. 

We note that previously published ESS$\nu$SB results have been directly included in this chapter. 

\subsection{Software Tools}
\label{sec:detectors:software}

%The water Cherenkov detector evaluation has been done using several softwares with different purposes:
Unless otherwise noted, the following software packages have been used in the analysis of the detector performance in this section:

\begin{description}
 \item[\textsc{Genie}] The neutrino interaction vertex generator, \textsc{Genie v3.0.6}~\cite{Andreopoulos:2009rq, Andreopoulos:2015wxa,Tena-Vidal:2021rpu}, was used to simulate neutrino vertices. The G18\_10a\_00\_000 cross-section tune was used for this, and is shown in Fig.\ref{fig:detectors:genie_crossection} for neutrino interactions in water over an energy interval relevant for ESS$\nu$SB.
 
 \item[\textsc{Geant4}] Particle propagation through the detectors and surrounding material was simulated by \textsc{Geant4 v10.4.1}~\cite{Agostinelli:2002hh,Allison:2006ve,Allison:2016lfl}. It was not used directly, in favour of the dedicated simulation frameworks listed below (see \textsc{EsbRoot} and \textsc{WCSim} below).
 
 \item[\textsc{EsbRoot}] The simulation framework \textsc{EsbRoot}~\cite{detectors:EsbRoot} has been developed in-house, based on \textsc{FairRoot}~\cite{Al-Turany:2012zfk}. It was used directly for the simulation of the super-fine-grained detector (SFGD) detector and to set up the dependencies and environment for \textsc{WCSim} and \textsc{fiTQun}.
 
 \item[\textsc{WCSim}] The water Cherenkov simulation software \textsc{WCSim}~\cite{detectors:WCSim} was used for particle transport post-vertex, and simulation of the detector response of both near and far WC detectors. \textsc{WCSim} is based on \textsc{Geant4} and developed within the \textsc{Hyper-Kamiokande} collaboration. The default version was modified for needs of this project.
 
 \item[\textsc{fiTQun}] The events simulated in WC detectors are reconstructed using the \textsc{fiTQun} software~\cite{Missert:2017qdz,Jiang:2019xwn}, also native to the \textsc{Hyper-Kamiokande} collaboration. \textsc{fiTQun} fits the detector response to several particle hypotheses, including variations in, for example, particle flavor, vertex position, particle direction and momentum, as well as in the number of registered sub-events (estimating the number of final-state visible particles in the interaction). The default version was modified for needs of this project.

 \item[\textsc{EsbRootView}] A versatile event viewer for the EsbRoot data model developed in-house \cite{Barrand:2021azm}. It allows visualisation of simulated events both as still and animated scenes. Seamlessly works on multiple platforms including handheld devices and web browsers.

 %(expand).
\end{description}

\begin{figure}[!hbtp]
\centering
\begin{subfigure}[b]{0.495\textwidth}
\centering
\includegraphics[width=\textwidth]{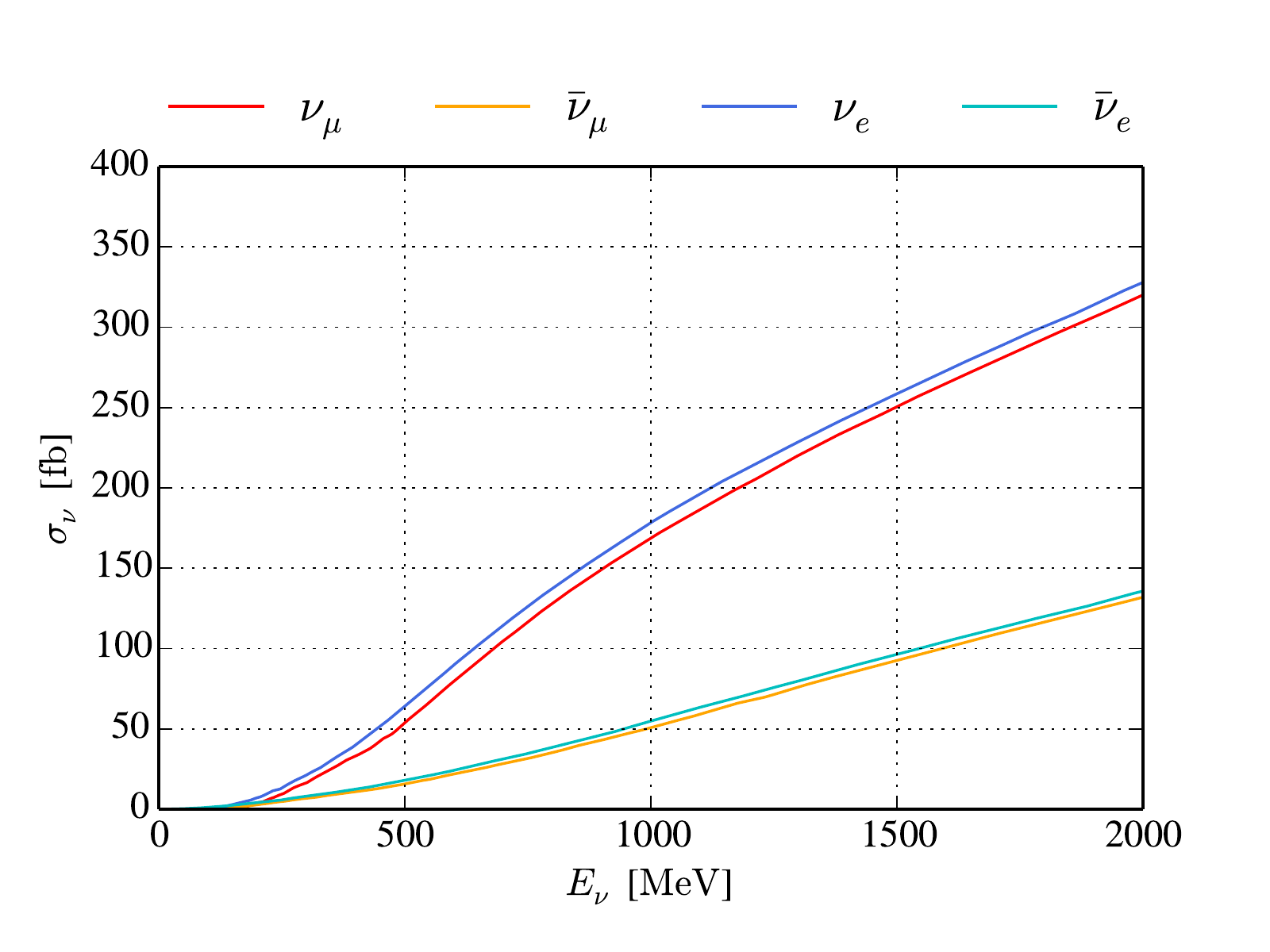}
    \caption{Charged current (CC) interaction cross-section.}
    \label{fig:detectors:genie_crossection_cc}
  \end{subfigure}
  \hfill
  \begin{subfigure}[b]{0.495\textwidth}  
    \centering 
\includegraphics[width=\textwidth]{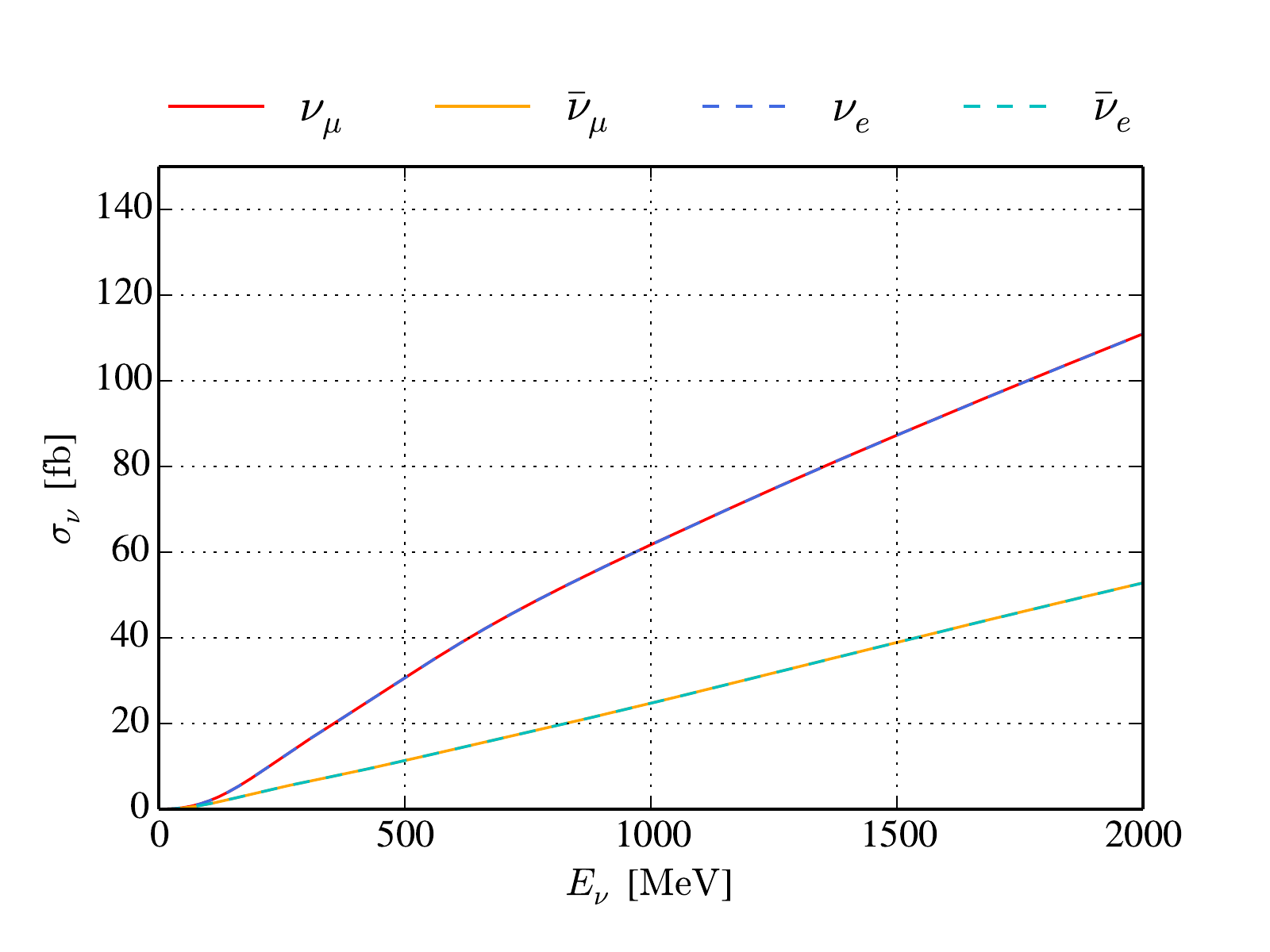}
    \caption{Neutral current (NC) interaction cross-section.}
    \label{fig:detectors:genie_crossection_nc}
  \end{subfigure}
\caption{The total neutrino-nucleus interaction cross-section with water, $(^1\text{H})_2(^{16}\text{O})$, over the incident neutrino energy for (\subref{fig:detectors:genie_crossection_cc}) charged current events and (\subref{fig:detectors:genie_crossection_nc}) neutral current.}
\label{fig:detectors:genie_crossection}
\end{figure}

\subsection {Near Detectors}

The main purpose of the near detector is to reduce the systematic uncertainties, and thus increase the physics potential of the project, especially with respect to the $\delta_{\tit{CP}}$ measurement.
%The aim of this document is to describe the design of the ND and its expected performance.
Specifically, it measures:
\begin{itemize}
 \item The unoscillated flux of neutrinos and, in particular, the electron neutrino contribution.
 \item The neutrino interaction cross section in water for all four neutrino flavors.
\end{itemize}

% Beam content:
%   Pos pol, nue:    4787055186350.0 ~= 4.787e+12
%   Pos pol, numu: 685403796525410.0 ~= 6.854e+14
%   Neg pol, nue:    1981772361640.0 ~= 1.982e+12
%   Neg pol, numu: 415528544109400.0 ~= 4.155e+14
% Flavor ratio:
%   Pos pol, nue/(nue+numu) = 6.936e-03
%   Neg pol, nue/(nue+numu) = 4.747e-03
Special attention is paid to the electron neutrino cross-section (electron neutrinos constitute $<\SI{1}{\percent}$ of the unoscillated neutrino beam) as these are required for the oscillation analysis conducted with far detector data. This requires a well-performing identification of electron-like events in the muon-rich sample~\cite{Burgman:2022h0}.

% this figure to be included in the introduction chapters
%\begin{figure}[!htbp]
%    \centering
%    \includegraphics[width=0.8\linewidth]{figures/detectors/NDlayout.png} 
%    \caption{A schematic layout of the ESS site with the proposed ESS$\nu$SB facilities (in dark blue): the accumulator ring, the target station and the near detector hall. The blue arrow indicates the central direction of the neutrino beam.}
%    \label{fig:detectors:NDlayout}
%\end{figure}

The near detector must be located close to the beam production point in order to measure the unoscillated beam, but far enough away to measure primarily the central part of the beam where the beam spectral shape approximates the shape that will be measured with the far detector.
%In this way, the near detector will be homogeneously illuminated and detect a similar neutrino flux as the far detector will observe. 
%According to the current layout of the ESS site (Figs.~\ref{fig:2} and \ref{fig:4} \ABtodo{check figure numbers when they become clearer}),
According to the current layout of the ESS site (Fig.~\ref{fig:layoutint}),
the near detector must be placed at least ${\sim}\SI{250}{\m}$ from the neutrino target.
This is taken as the baseline near-detector position. The properties of the neutrino beam at this location are described in Section~\ref{sec:detectors:NDbeamcomposition}.

%A schematic layout of the ESS site with the near detector position indicated is shown in Fig.\ref{fig:detectors:NDlayout}. 

\begin{figure}[thb]
\centering
\begin{tikzpicture}
\node at (0,0) { \includegraphics[width=0.8\textwidth]{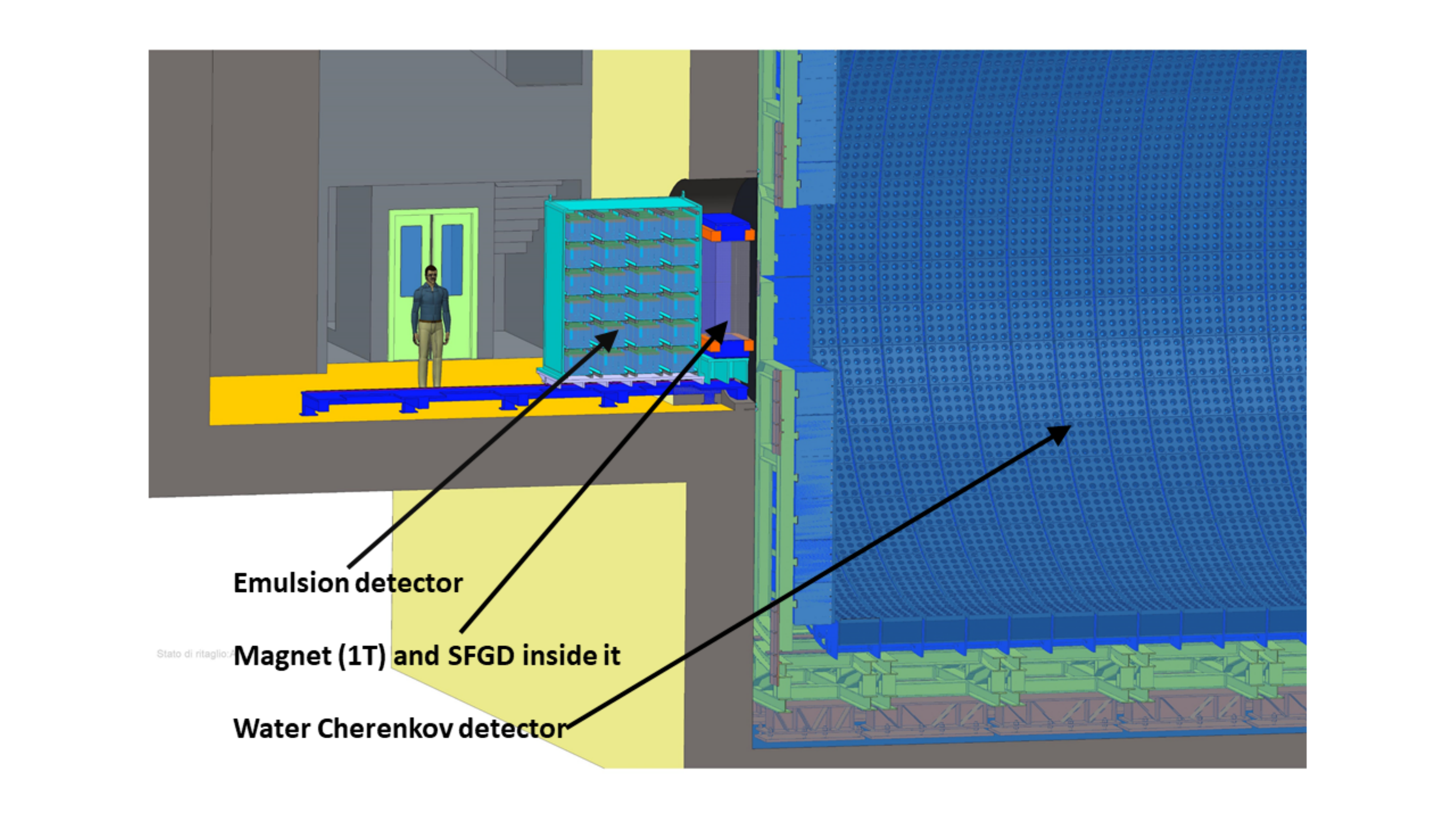} };
\end{tikzpicture}
\caption{The three detector components of the near detector complex (shown in the order they are encountered by the neutrino beam from the left-hand side of the image): The emulsion detector, the SFGD, and the water Cherenkov detector. Also shown is the window in the water Cherenkov detector which allows cross-over events from the SFGD to enter the instrumented water volume unimpeded. \label{fig:detectors:nd_cavern_close}}
\end{figure}

The near detector complex of the ESS$\nu$SB will consist of three separate, but coupled, detectors using different neutrino-detector technologies.
This design is based on a review of the near detectors at other accelerator-based neutrino oscillation experiments,
%which has partly been performed within the COST action CA15139,
as well as the specific requirements for the ESS$\nu$SB near detector.
% (see Fig.\ref{fig:detectors:NDcomplex}).
The three detectors will be:
\begin{itemize}
 \item A kiloton-scale water Cherenkov detector used for event-rate measurement and flux normalisation, neutrino interaction cross-section measurements in water, as well as event reconstruction comparison with the far detector~\cite{Burgman:2022h0}.
 \item A magnetised fine-grained tracker (super fine-grained detector, SFGD) is used for precision measurements of neutrino cross-sections in the available energy region (${\sim}\,\SI{60}{}$--$\SI{600}{\MeV}$), and is directly upstream (w.r.t.\ the neutrino beam) of the water volume.
 \item An emulsion detector setup modelled after that of the NINJA experiment~\cite{Hiramoto:2013}, to be installed directly upstream from the SFGD.
\end{itemize}

The three components are shown in Fig.~\ref{fig:detectors:nd_cavern_close}, and each component is described separately below.

\subsubsection{Near-Detector Building and Cavern}

The near detector for ESS$\nu$SB is planned to be located within the premises of the ESS campus outside Lund, Sweden. There are several reasons for this placement. Firstly, a location on the ESS site will give easier access to the services that are needed to operate the different detectors; secondly, the existing site is large enough to provide a sufficient distance between the target and the detector to avoid direct interference between the two sites during installation. The planned distance between the target and the near-detector building is \SI{250}{\m}.

A neutrino beam directed towards Zinkgruvan or Garpenberg will point at an angle below the horizon of ${\sim}\SI{2}{\degree}$. This results in a depth below ground at the near detector location of ${\sim}\SI{10}{\m}$ if the target is located at the surface. Consequently, the total depth below ground level will be determined by this value and the surface depth of the target. Given the depth of the target station in the current design, the central axis of the neutrino beam is estimated to have a depth of \SI{15}{\m} -- \SI{20}{\m} below ground level at \SI{250}{\m} from the production point. This requires the centre of the near detector complex to sit at the same depth, which in turn requires an intermediary staging area for transport of detector components to the bottom of the cavern. A conceptual model for the near detector building and cavern is shown in Fig.~\ref{fig:detectors:nd_cavern_layout}.

\begin{figure}[bth]
\centering
\begin{subfigure}[b]{0.57\textwidth}
\centering
\includegraphics[width=\textwidth]{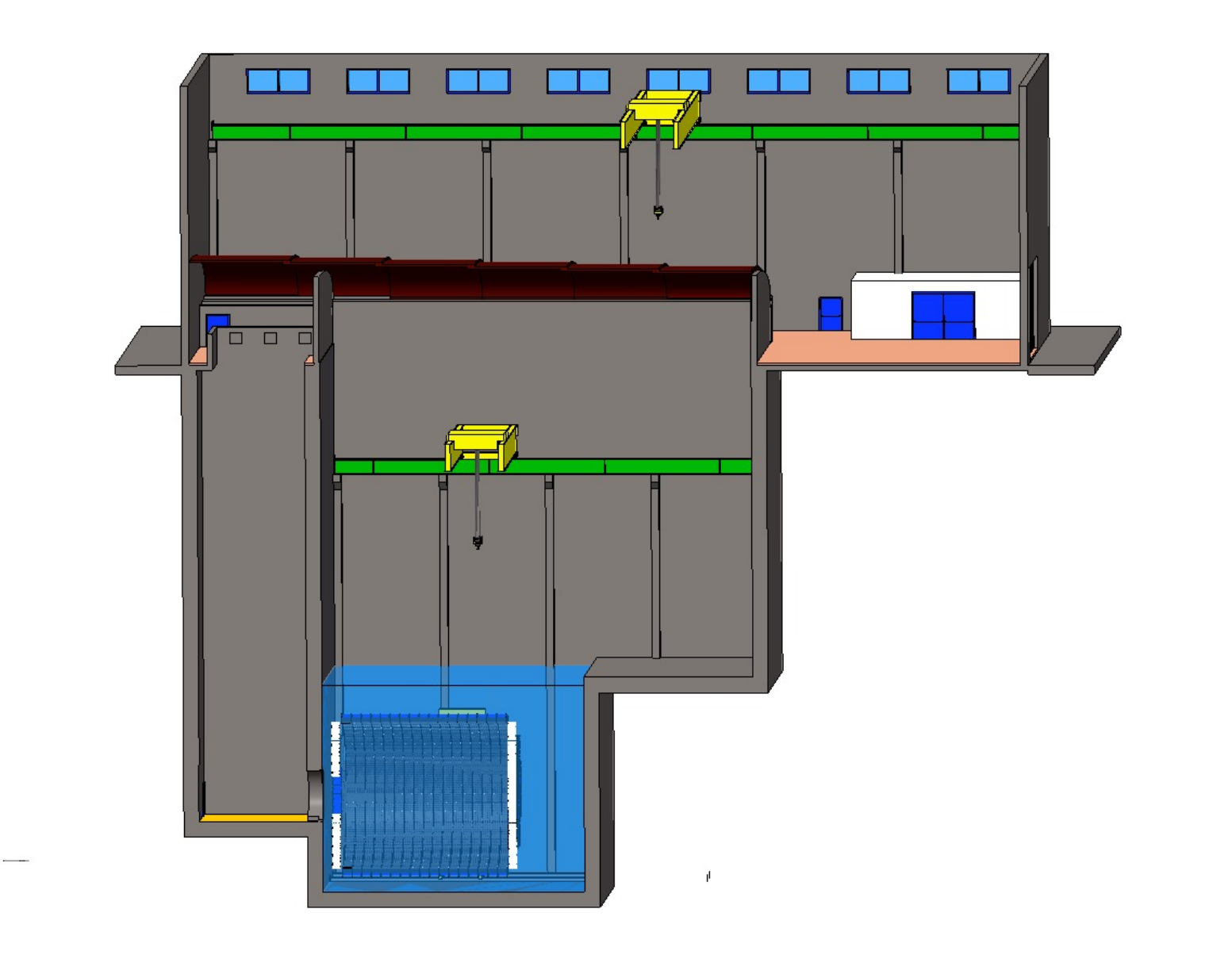}
\caption{A full rendering of the near-detector cavern and building.}
\label{fig:detectors:nd_cavern_layout_full}
\end{subfigure}
\hfill
\begin{subfigure}[b]{0.4\textwidth}  
\centering 
\includegraphics[width=\textwidth]{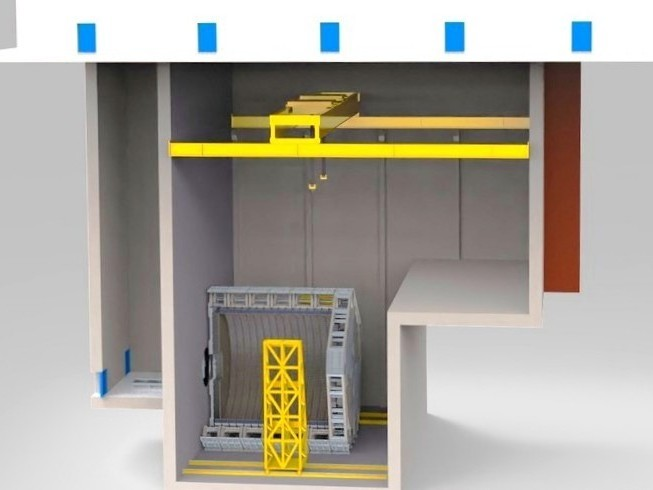}
\caption{An artistic rendering of the near-detector cavern, including the water Cherenkov component in the mounting phase.}
\label{fig:detectors:nd_cavern_layout_artistic}
\end{subfigure}
\caption{
Two views of the design for the near detector building and cavern, with the neutrino beam arriving from the left-hand side of the figure.
The left-most shaft in both figures will house the SFGD and Emulsion detectors, while the water Cherenkov component will be housed in the deep central area. An intermediary staging area is included for mounting the detector in the deep cavern.
\label{fig:detectors:nd_cavern_layout}}
\end{figure}

As previously described, the conceptual design of the near detector suite includes three different detector types: an emulsion detector, a fine-grained tracker, and a water Cherenkov detector.
A water Cherenkov detector is included in the near detector suite not only to measure the beam parameters of interest, but to do so using the same detection technique as in the far detector.
The high neutrino flux at the near detector will provide an opportunity to tune the simulation and reconstruction software in order to improve the analysis of the far detector events -- despite the geometrical differences between the detectors, as well as differences in beam composition due to flavor oscillations and different angular coverage of the two detectors.

The general concept for the water Cherenkov detector is to immerse the detector frame, which holds the photo sensors and the read-out modules, in a pool of high-purity water. This design is adopted instead of using a closed metal cylinder, which is often the choice for small water Cherenkov detectors.
The walls of the detector pool will be cladded to avoid having wall material dissolving into the water volume.
The pool will also be filled and emptied in such a manner that any flow brings water from the inner volume of the detector structure to the outer volume before the water is drained into storage or flushed out. This approach minimises migration of contaminants into the inner volume of the detector.
%Service areas for pumping and filtering systems are planned to be located in the room to the right of the detector pool indicated in the left panel of Fig.\ref{fig:detectors:nd_cavern_layout}.

In addition to the main detector pool, the near detector building also houses a detector hall for both the emulsion detector and the tracker.
This hall is shown in the left-hand side in the models in Fig.~\ref{fig:detectors:nd_cavern_layout},
with a zoomed-in view shown in Fig.~\ref{fig:detectors:nd_cavern_close}.
A $\sim\SI{3}{\m}$ diameter carbon-fiber composite window will act as the interface between the tracker and the water volume. This will have a low thickness in order to minimise energy loss for particles that traverse the interface.
Relevant safety arrangements to avoid flooding of the tracker hall, will be necessary, both in terms of operational practices and from the use of additional sliding doors that can separate the pool from the tracker hall in case of an emergency (the latter arrangement is not shown in Fig.~\ref{fig:detectors:nd_cavern_layout}.)

\subsubsection{Neutrino Beam Composition}
\label{sec:detectors:NDbeamcomposition}

The neutrino flux is discussed in Section~\ref{sec:targetstation:geneticalgorithm_improvement}, using the optimised focusing horn (see Fig.~\ref{fig:NuFlux2}) and a decay tunnel length of \SI{50}{\m}.

The resulting neutrino flux is plotted in Fig.~\ref{fig:detectors:ndwc_totalflux} and summed in Table~\ref{tbl:detectors:ndwc_totalflux}, at a distance of \SI{100}{\km} from the target station within the central \SI{100}{\m\squared} square ($\SI{10}{\m}\times\SI{10}{\m}$) around the beam axis.
The resulting expected numbers of neutrino interactions in the instrumented volume of the water Cherenkov near detector are listed in Table~\ref{tbl:detectors:ndwc_numinteractions} per flavour, interaction type, and horn polarity; these expected interactions are also shown as a function of neutrino energy in Fig.\ref{fig:detectors:ndwc_numinteractions}.

% positive, nue,numu,anue,anumu
% 4.76172341e+12,   6.73605595e+14,   2.53317754e+10, 1.17982016e+13
% negative, nue,numu,anue,anumu
% 1.26837454e+11,   1.98083171e+13,   1.85493491e+12, 3.95720227e+14
{
\begin{table}[p]
\footnotesize
%\scriptsize
\centering
\caption{Total neutrino flux \SI{100}{\km} downstream of the target station, in the central $\SI{10}{\m}\times\SI{10}{\m}$, per running-year, for each horn polarity.
\label{tbl:detectors:ndwc_totalflux}}
\begin{tabular}{ r r r r r }
\textbf{Total neutrino flux}  &                ~  &                ~  &                ~  &                ~  \\
           ~  &  \textbf{$\nu_\mu$}  &  \textbf{$\nu_e$}  &  \textbf{$\bar\nu_\mu$}  &  \textbf{$\bar\nu_e$}  \\
\hline
Positive polarity  &    \SI{6.74e14}{}  &    \SI{4.76e12}{}  &    \SI{1.18e13}{}  &    \SI{2.53e10}{}  \\
Negative polarity  &    \SI{1.98e13}{}  &    \SI{1.27e11}{}  &    \SI{3.96e14}{}  &    \SI{1.85e12}{}  \\
\hline
\end{tabular}
\end{table}
}

{
\begin{table}[p]
\footnotesize
%\scriptsize
\centering
\caption{Number of expected neutrino interactions in the detector per running year, per flavour and interaction type, and per each horn polarity.
\label{tbl:detectors:ndwc_numinteractions}}
\begin{tabular}{ r r r r r r r r r }
%          \textbf{Positive polarity}  &                ~  &                ~  &                ~  &                ~  &                ~  &                ~  &                ~  &                ~  \\
          \textbf{All interactions}  &                ~  &                ~  &                ~  &                ~  &                ~  &                ~  &                ~  &                ~  \\
                                   ~  &  \textbf{$\nu_\mu$ CC}  &  \textbf{$\nu_e$ CC}  &  \textbf{$\bar\nu_\mu$ CC}  &  \textbf{$\bar\nu_e$ CC}
                                      &  \textbf{$\nu_\mu$ NC}  &  \textbf{$\nu_e$ NC}  &  \textbf{$\bar\nu_\mu$ NC}  &  \textbf{$\bar\nu_e$ NC} \\
\hline
%                   All interactions  &    52014764.9263  &    1074587.14009  &    92467.0793719  &    1109.66754673  &    44153507.5280  &    611114.920818  &    335608.079892  &    871.563850821  \\
Positive polarity  &    \SI{5.20e7}{}  &    \SI{1.07e6}{}  &    \SI{9.25e4}{}  &    \SI{1.11e3}{}  &    \SI{4.42e7}{}  &    \SI{6.11e5}{}  &    \SI{3.36e5}{}  &    \SI{8.72e2}{}  \\
%\hline
%                                   ~  &                ~  &                ~  &                ~  &                ~  &                ~  &                ~  &                ~  &                ~  \\
%          \textbf{Negative polarity}  &                ~  &                ~  &                ~  &                ~  &                ~  &                ~  &                ~  &                ~  \\
%                                   ~  &  \textbf{$\nu_\mu$ CC}  &  \textbf{$\nu_e$ CC}  &  \textbf{$\bar\nu_\mu$ CC}  &  \textbf{$\bar\nu_e$ CC}
%                                      &  \textbf{$\nu_\mu$ NC}  &  \textbf{$\nu_e$ NC}  &  \textbf{$\bar\nu_\mu$ NC}  &  \textbf{$\bar\nu_e$ NC} \\
%\hline
%                   All interactions  &    1063386.14084  &    8898.35854805  &    9743488.59294  &    188973.330204  &    880745.878012  &    11080.8734824  &    9652849.63447  &    135527.488425  \\
Negative polarity  &    \SI{1.06e6}{}  &    \SI{8.90e3}{}  &    \SI{9.74e6}{}  &    \SI{1.89e5}{}  &    \SI{8.81e5}{}  &    \SI{1.11e4}{}  &    \SI{9.65e6}{}  &    \SI{1.36e5}{}  \\
\hline
\end{tabular}
\end{table}
}

% FIGURE
% flux
\begin{figure}[p]
\centering
\begin{subfigure}[b]{0.495\textwidth}
\centering
\includegraphics[width=\textwidth]{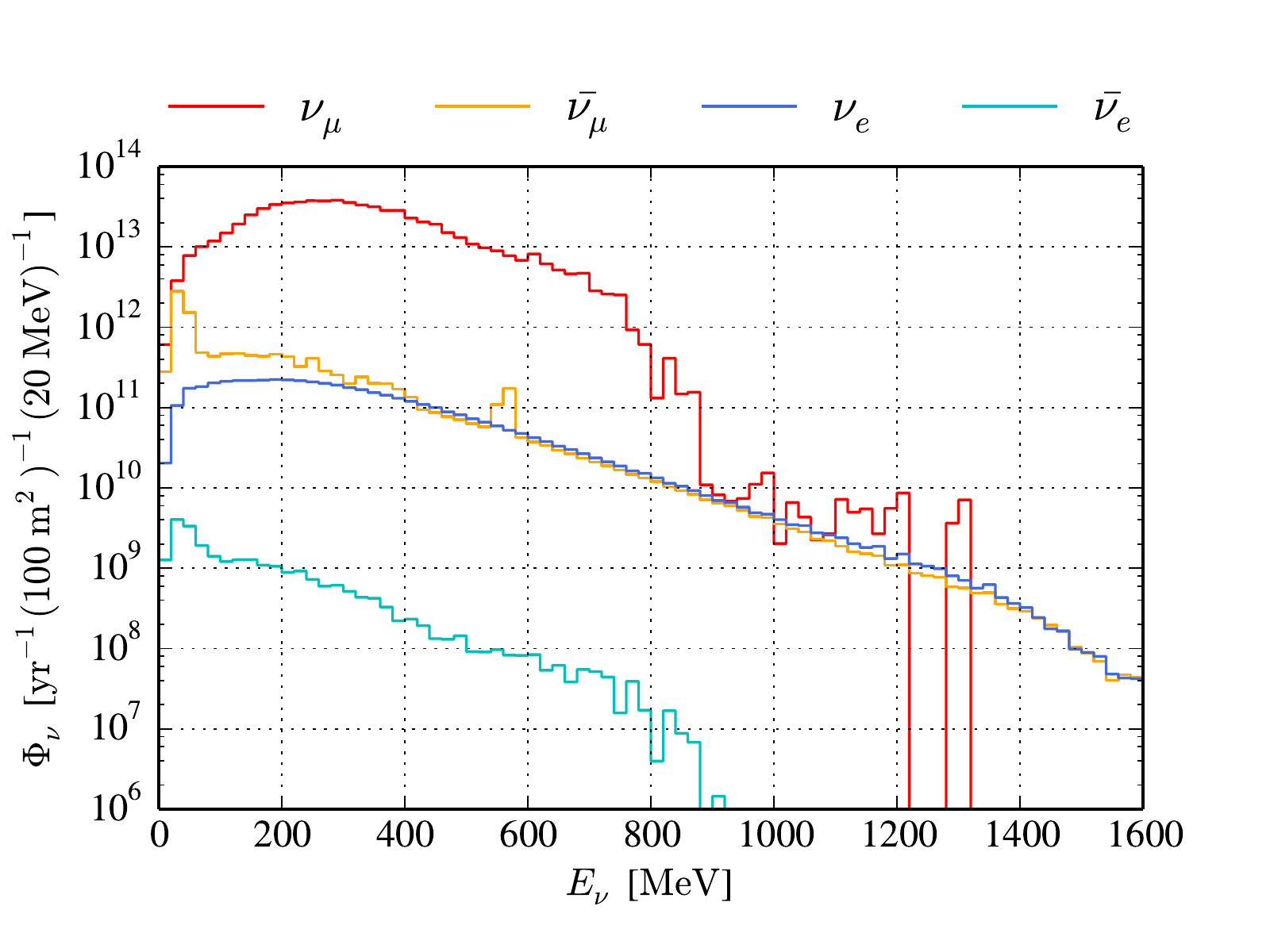}
\caption{Positive polarity.}
\label{fig:detectors:ndwc_totalflux_pospol}
\end{subfigure}
\hfill
\begin{subfigure}[b]{0.495\textwidth}  
\centering 
\includegraphics[width=\textwidth]{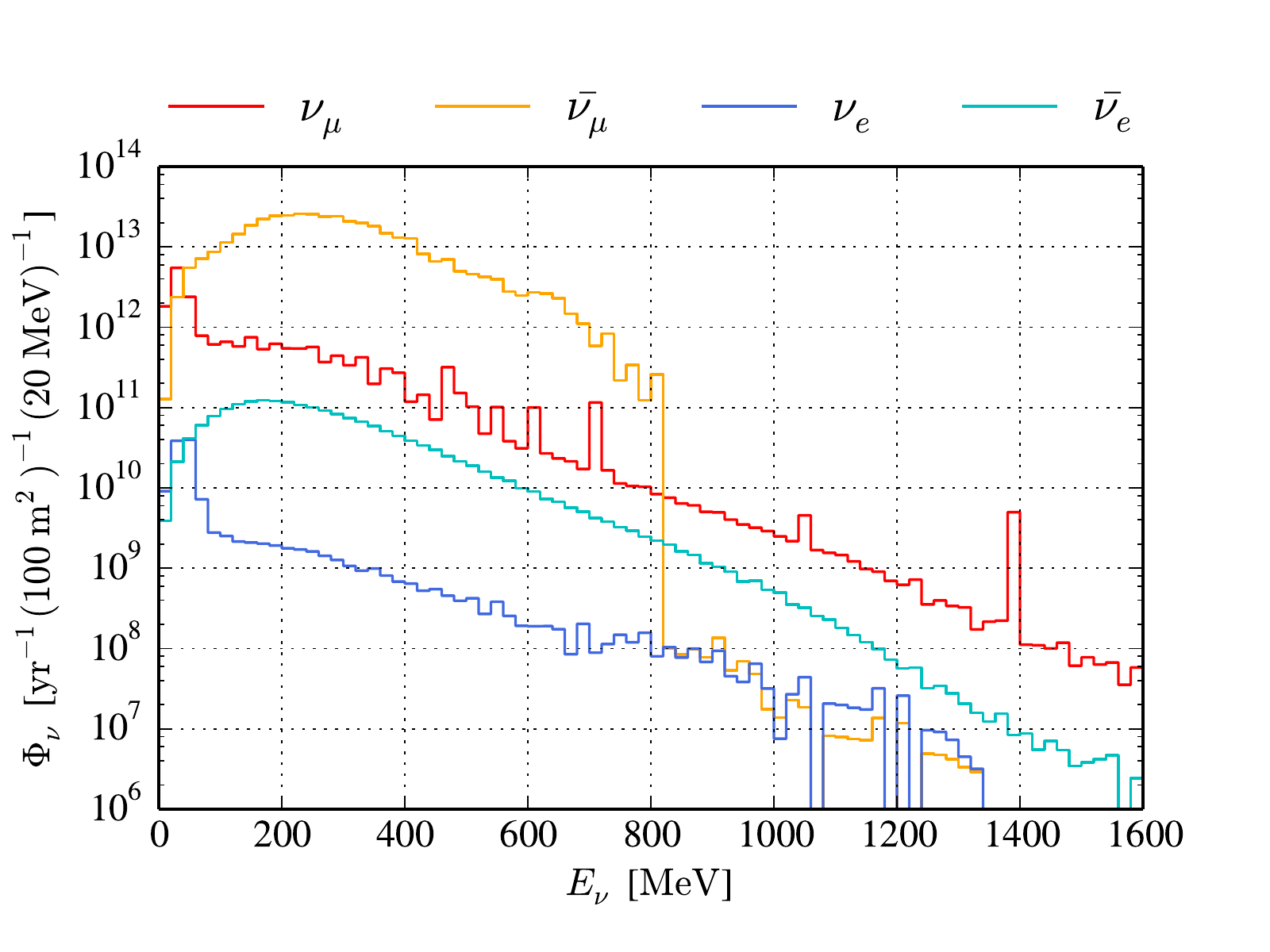}
\caption{Negative polarity.}
\label{fig:detectors:ndwc_totalflux_negpol}
\end{subfigure}
\caption{Neutrino flux \SI{100}{\km} downstream of the target station, in the central $\SI{10}{\m}\times\SI{10}{\m}$, per running-year, for each horn polarity as functions of the incident neutrino energy. The sharp spectral features (e.g.\ for the negative polarity $\nu_\mu$ spectrum) only have a statistical origin and no connection to physical processes.
\label{fig:detectors:ndwc_totalflux}}
\end{figure}

% FIGURE
% interactions
\begin{figure}[p]
\centering
\begin{subfigure}[b]{0.495\textwidth}
\centering
\includegraphics[width=\textwidth]{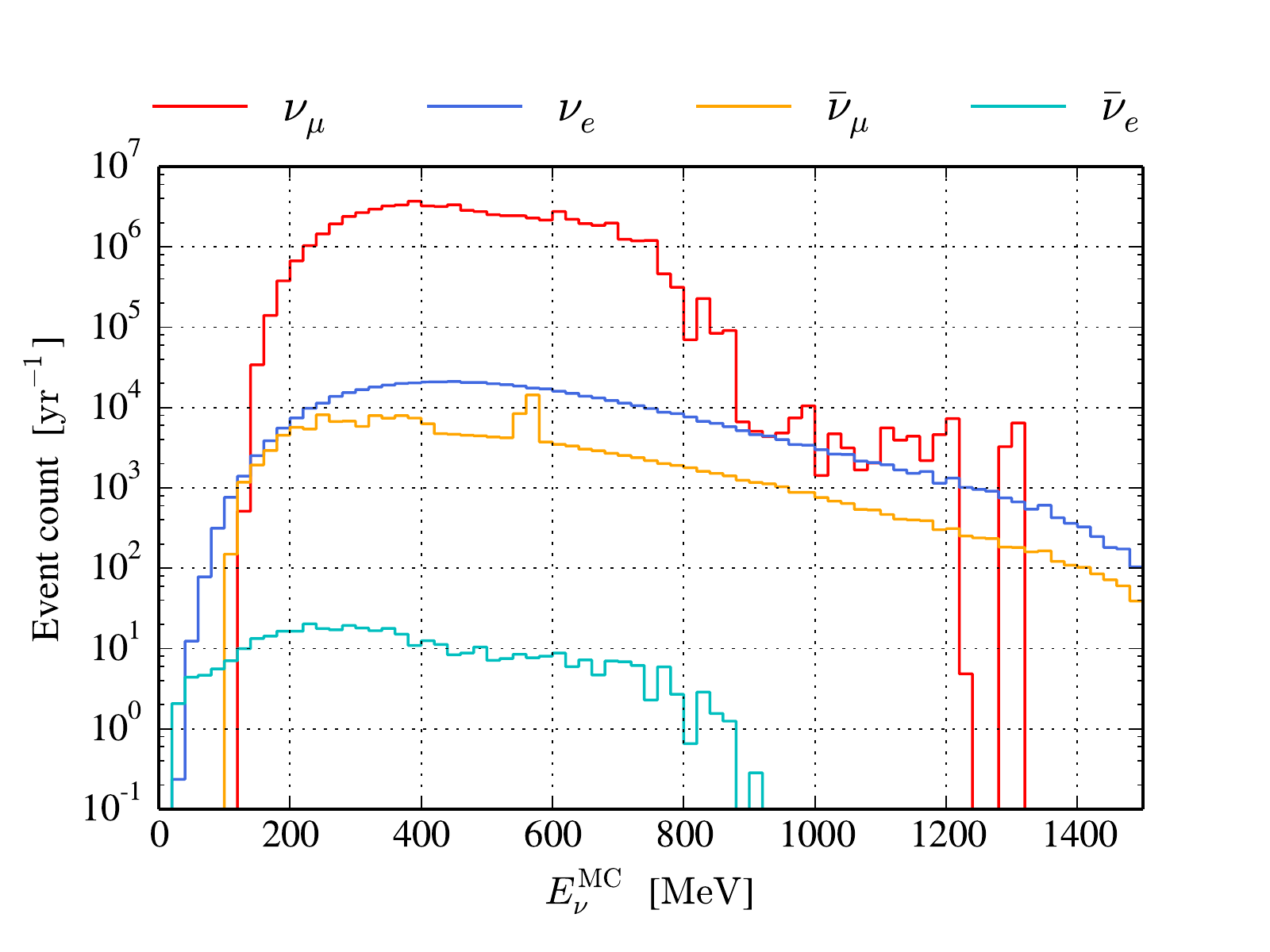}
\caption{Positive polarity, charged current interaction.}
\label{fig:detectors:ndwc_numinteractions_pospolcc}
\end{subfigure}
\hfill
\begin{subfigure}[b]{0.495\textwidth}  
\centering 
\includegraphics[width=\textwidth]{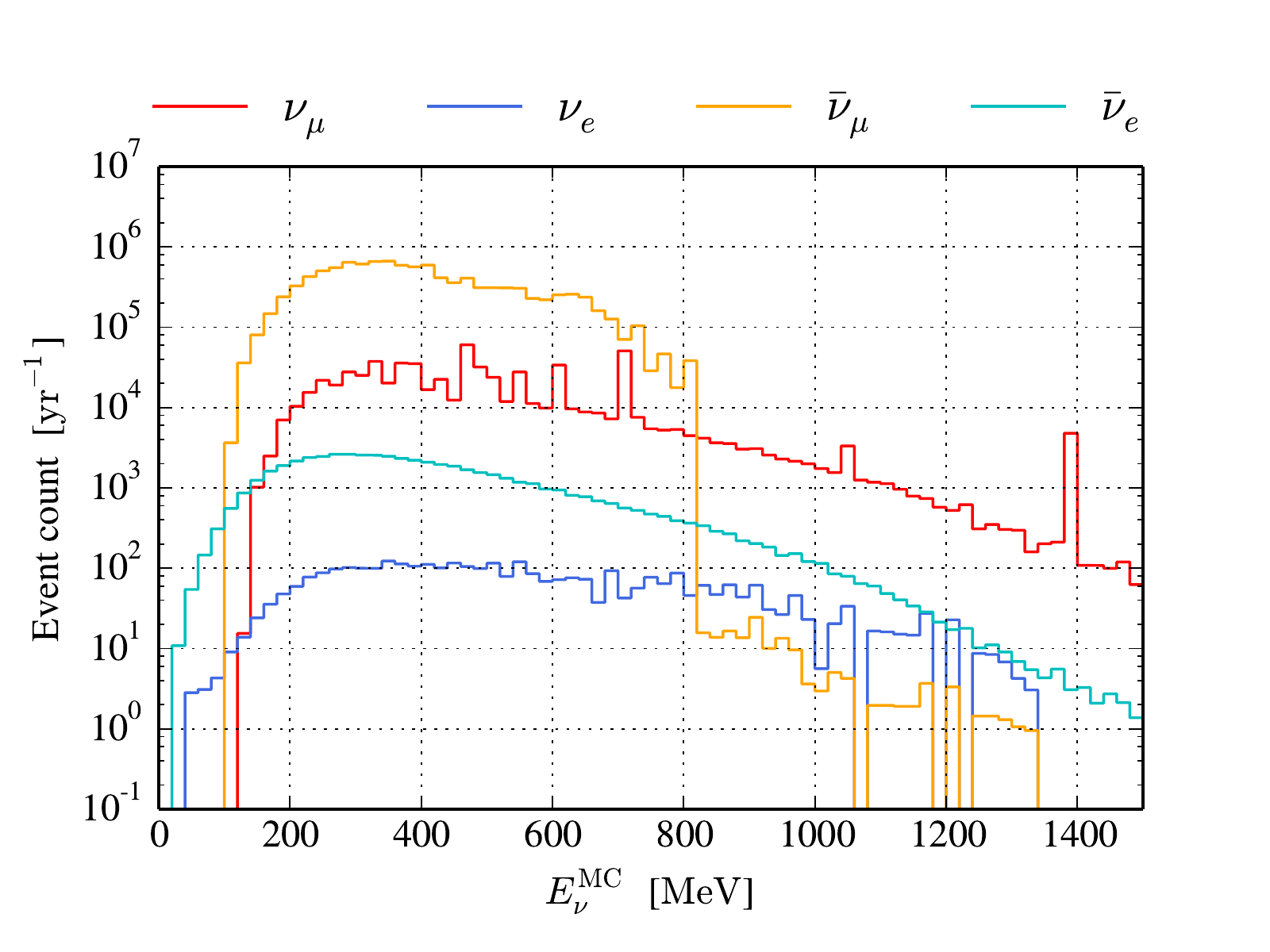}
\caption{Negative polarity, charged current interaction.}
\label{fig:detectors:ndwc_numinteractions_negpolcc}
\end{subfigure}
\hfill
\begin{subfigure}[b]{0.495\textwidth}  
\centering 
\includegraphics[width=\textwidth]{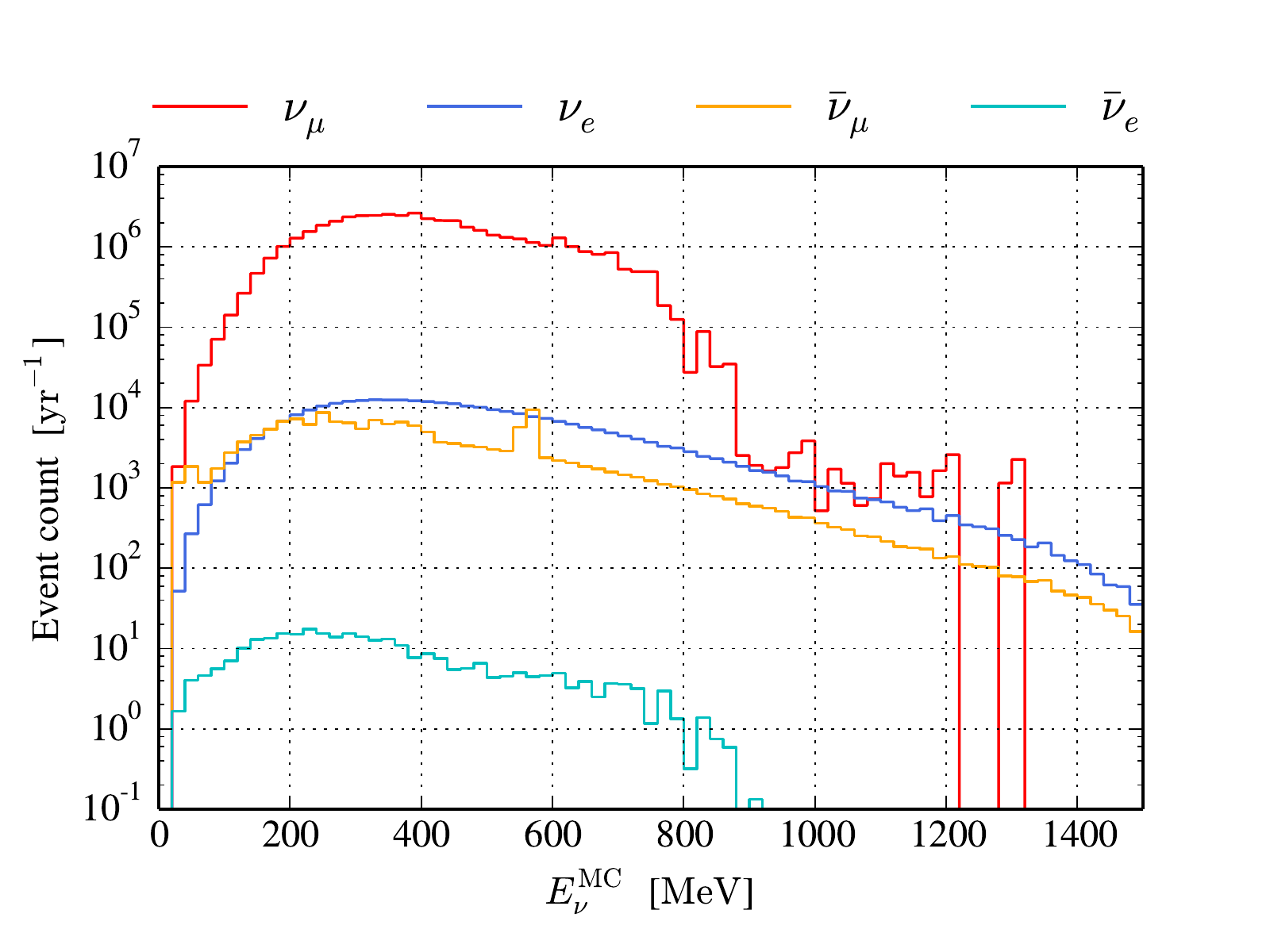}
\caption{Positive polarity, neutral current interaction.}
\label{fig:detectors:ndwc_numinteractions_pospolnc}
\end{subfigure}
\hfill
\begin{subfigure}[b]{0.495\textwidth}  
\centering 
\includegraphics[width=\textwidth]{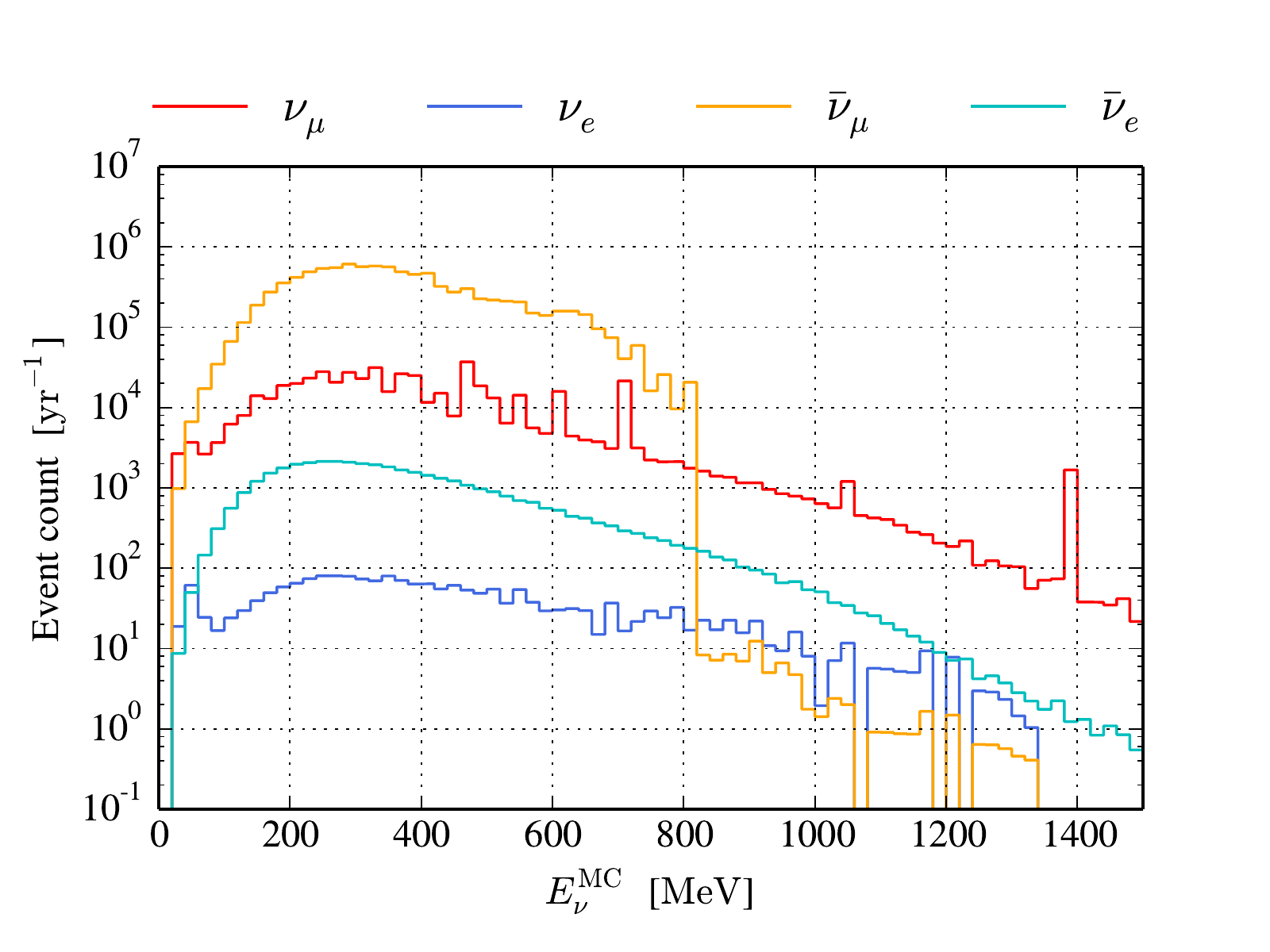}
\caption{Negative polarity, neutral current interaction.}
\label{fig:detectors:ndwc_numinteractions_negpolnc}
\end{subfigure}
\caption{Number of expected neutrino interactions in the detector per running year, per flavour and interaction type, and for different horn polarities (positive -- neutrino mode, negative -- antineutrino mode), as functions of the incident neutrino energy. Similar to the spectra shown in Fig.~\ref{fig:detectors:ndwc_totalflux}, the spectra shown here exhibit several sharp spectral features of statistical origin.
\label{fig:detectors:ndwc_numinteractions}}
\end{figure}

\subsubsection {Water Cherenkov Detector}
%\bigskip

The water Cherenkov component of the near detector complex will consist of a horizontally oriented detector tank submerged in clean water, having the cylinder axis aligned with the neutrino beam direction.
The interior will be instrumented with more than \SI{22000}{} photo-multiplier tubes (PMTs) each having a \SI{3.5}{\inch} diameter,
and resulting in a ${\sim}\SI{30}{\percent}$ instrumentation coverage of the detector cylinder.

As mentioned previously, the primary purpose of including a water Cherenkov component in the near detector complex is to
enable measurement of the non-oscillated neutrino beam using the same technology as in the far detectors.
This allows for measurements of the composition, absolute flux, and energy spectrum of the neutrino beam;
as well as a measurement of the interaction cross-section for electron-(anti)neutrinos incident on nuclei in water, $\sigma_{\nu_eN}$.
Measurements of the latter in current experiments have large uncertainties (see Fig. 51.1 in \cite{Zyla:2020zbs}), so this precision measurement is imperative
to reduce systematic uncertainties for the appearance/measurement of electron-(anti)neutrinos.

As the distance between the neutrino beam production point and the near-detector is small,
negligible oscillation will occur before the neutrino beam arrives at the ND.
Therefore, at the near detector, the beam will consist of $>\SI{98}{\percent}$ muon-neutrinos, with the remainder dominated by electron-neutrinos.
The low fraction of $\nu_e$ presents a challenge for measuring $\sigma_{\nu_eN}$,
which requires an efficient $\nu_e$ event selection strategy.
A two-step process has been developed in order to achieve this:
(1) to distinguish electron events from muon events, and
(2) to distinguish electron-neutrino events from muon-neutrino events.
These steps are described in detail in the following sections.

%The water Cherenkov detector evaluation has been done using several softwares with different purposes:
%\begin{description}
% \item[\textsc{Genie}] The neutrino interaction vertex generator, \textsc{Genie}~\cite{Andreopoulos:2009rq, Andreopoulos:2015wxa,Tena-Vidal:2021rpu}, was used to simulate neutrino vertices.
% \cite{Andreopoulos:2009rq,Andreopoulos:2015wxa,Tena-Vidal:2021rpu}
% \item[\textsc{WCSim}] The water Cherenkov simulations software \textsc{WCSim}~\cite{detectors:WCSim} was used for particle transport post-vertex and simulation of the detector response.
% \cite{wcsimgithub}
%                       \textsc{WCSim} is based on \textsc{Geant4}~\cite{Agostinelli:2002hh,Allison:2006ve,Allison:2016lfl} and developed within the \textsc{Hyper-Kamiokande} collaboration. 
% \cite{GEANT4:2002zbu,Allison:2006ve,Allison:2016lfl}
% \item[\textsc{fiTQun}] The events simulated events are reconstructed using the \textsc{fiTQun} software~\cite{Missert:2017qdz,Jiang:2019xwn},
% \cite{fitqun:2017,Super-Kamiokande:2019gzr}
%                        also native to the \textsc{Hyper-Kamiokande} collaboration.
%                        \textsc{fiTQun} fits the detector response to several particle hypotheses, including variations in
%                        e.g.\ particle flavor, vertex position, particle direction and momentum, as well as in the number of
%                        registered sub-events (estimating the number of final-state visible particles in the interaction).
%\end{description}

\subsubsubsection{Detector Geometry and Instrumentation}

The size of the water Cherenkov detector has been optimised based on two basic parameters: the event reconstruction performance, and the total cost. A main criterion for the event reconstruction efficiency is to minimise the number of charged particles that leave the detector volume without depositing their full energy. Figure~\ref{fig:detectors:lepton_range} shows the ranges for muons and electrons in water, for kinetic energies up to \SI{500}{\MeV}, which covers the majority of the expected energy range for the ESS$\nu$SB beam. One can see that the muon range is ${\sim}\SI{2.3}{\m}$ at \SI{500}{\MeV}, while the electron range is considerably shorter at \SI{0.6}{\m}. Consequently, it is expected that a detector that fully contains all events up to \SI{500}{\MeV}, and whose interaction vertices are in the centre of the detector, would have a radius of ${\sim}\SI{2.5}{\m}$. In addition to this observation, it is worth noting that the Cherenkov cone must be allowed to expand over a certain distance in order for well-defined Cherenkov rings to be projected on the inner mantle of the detector. This is necessary for making a good selection of particle type from the shape of the ring. %The exact determination of this distance for a given case requires more detailed simulations.
The optimisation of the detector tank dimensions and instrumentation density is discussed later in this section. %Efforts on this topic are discussed in other sections of this chapter.

\begin{figure}[htb]
\centering
\begin{tikzpicture}
%\node at (0,0) { \textit{\Large{}PlaceHolder} };
\node at (0,0) { \includegraphics[width=0.495\textwidth]{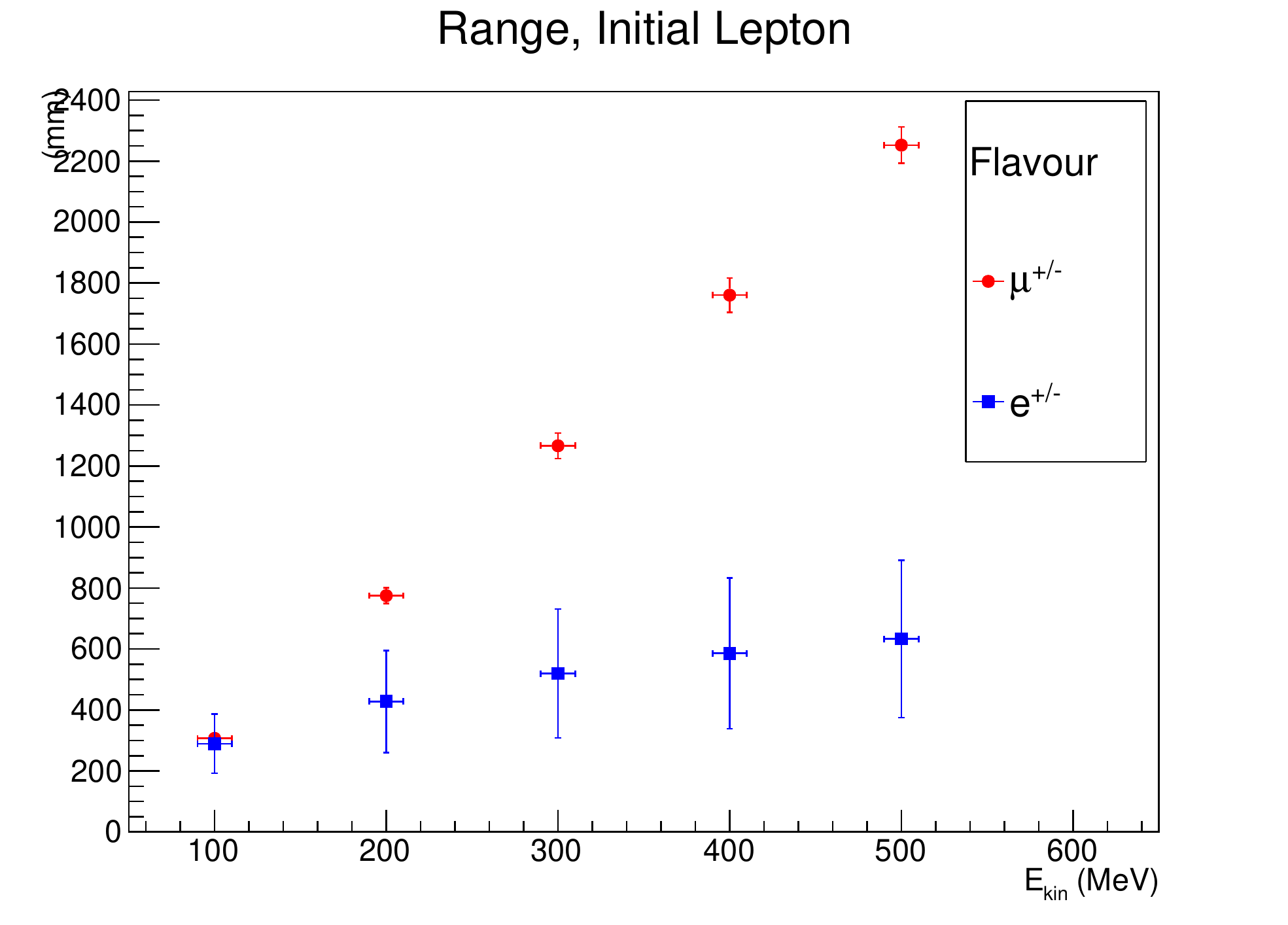} };
\end{tikzpicture}
\caption{Range in water for charged leptons with kinetic energy up to \SI{500}{\MeV}~\cite{burgman:msc2015}.
\label{fig:detectors:lepton_range}}
\end{figure}

In short, one can conclude that a good separation between muon and electron events requires a few meters of light propagation in addition to the stopping range mentioned above. Furthermore, the capability to discriminate between electron and muon events also depends on the density of the photo sensors (i.e.\ the spatial resolution of the photon impact points on the mantle of the detector). 
To investigate these points in more detail, a set of simulations were carried out using different detector volumes and sensor coverages.

\begin{figure}[phtb]
\centering
\begin{subfigure}[b]{0.35\textwidth}\centering
\includegraphics[width=\textwidth]{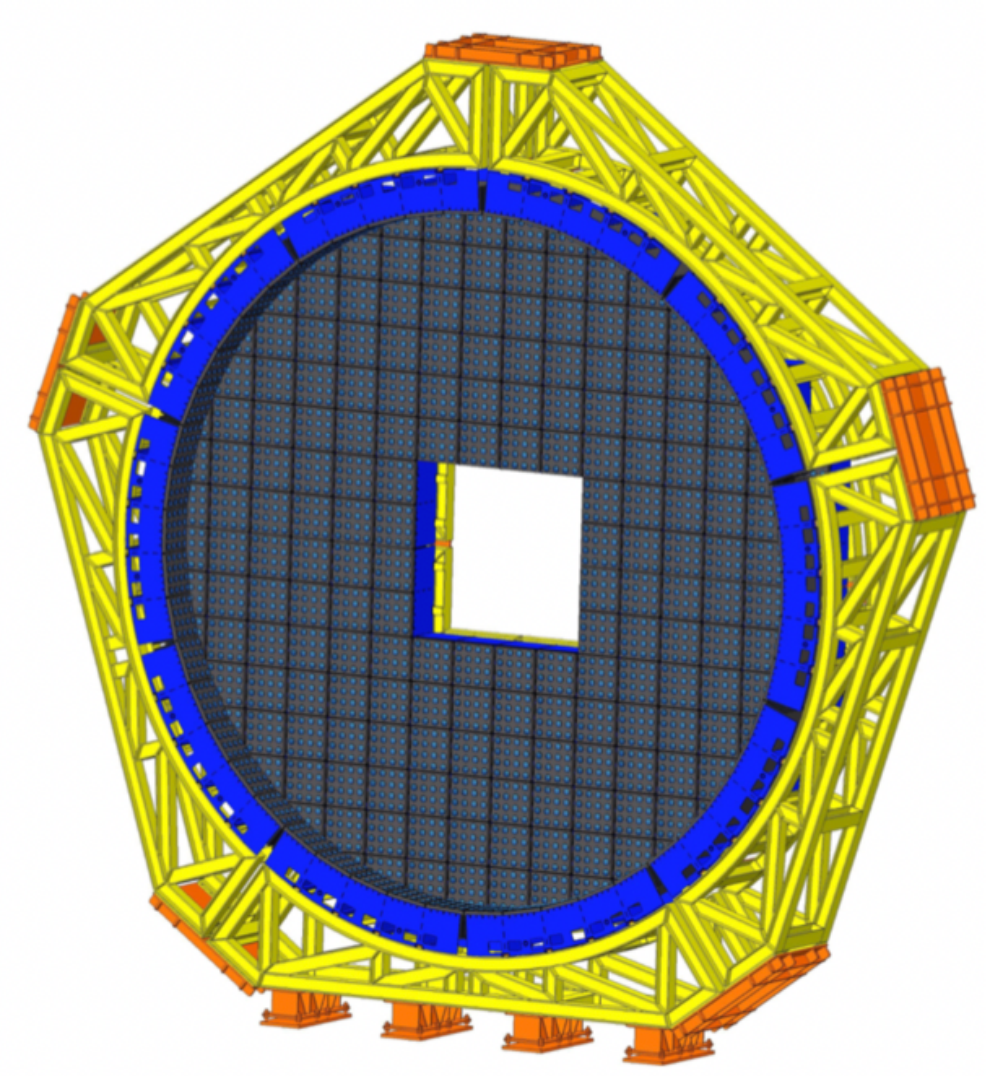}
\caption{The upstream end-cap of the detector, with the opening which provides an interface to the tracking detector located in the adjacent room.}
\label{fig:detectors:nd_tank_model_endcap}\end{subfigure}
\hfill
\begin{subfigure}[b]{0.45\textwidth}\centering 
\includegraphics[width=\textwidth]{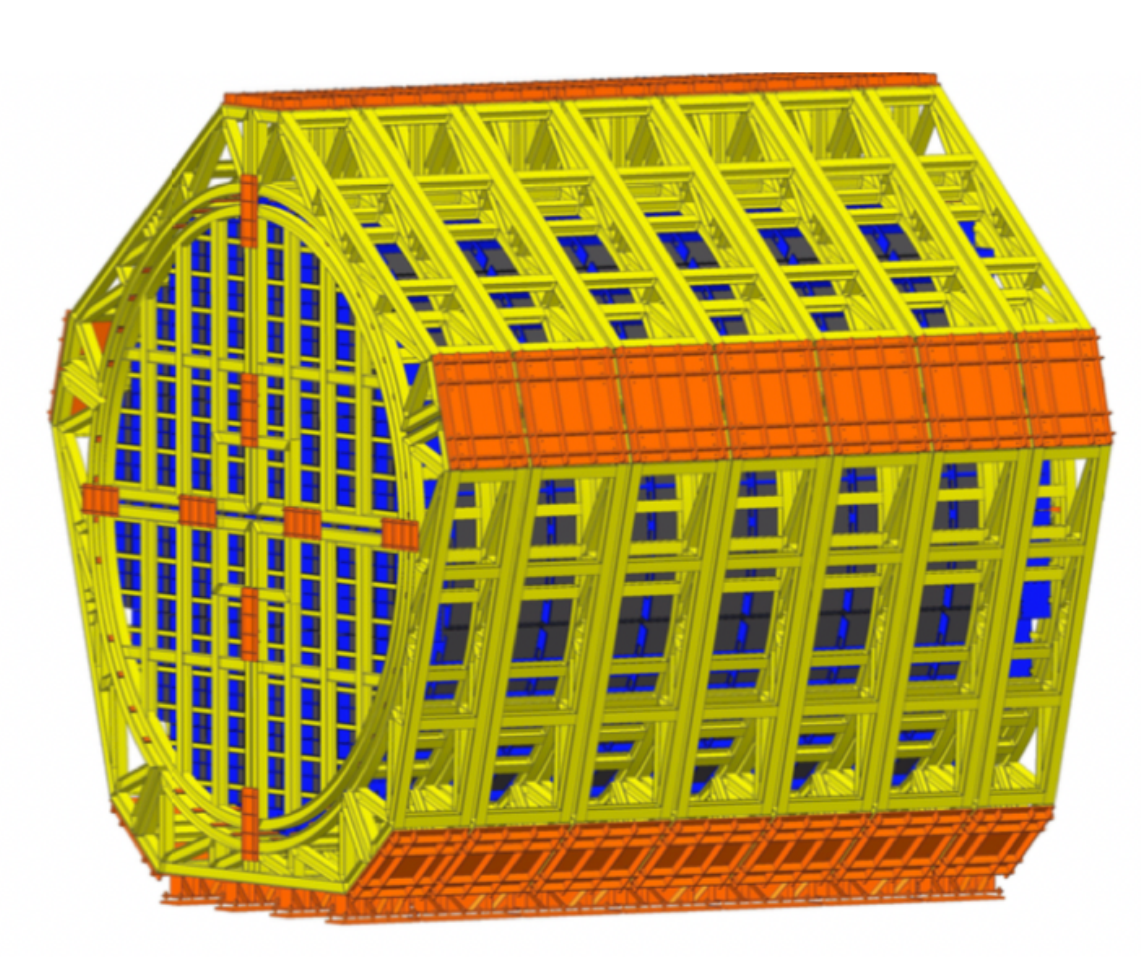}
\caption{The structural frame of the water Cherenkov detector tank.\vspace{2.0\baselineskip}}
\label{fig:detectors:nd_tank_model_structureframe}\end{subfigure}
%\hfill
%\begin{subfigure}[b]{0.47\textwidth}\centering 
%\includegraphics[width=\textwidth]{figures/detectors/nd_instr_jc_frames-fig1redone}
%\caption{A cross-sectional view of the detector structure tank along with its dimensions given in \SI{}{\mm}.}
%\label{fig:detectors:nd_tank_model_tankmeasurements}\end{subfigure}
%\hfill
%\begin{subfigure}[b]{0.5\textwidth}\centering 
%\includegraphics[width=\textwidth]{figures/detectors/nd_instr_jc_mounting-fig2}
%\caption{An illustration of the detector in a mounting stage. Here a previously considered detector tank with a larger diameter is shown. This is included as the mounting principle remains.}
%\label{fig:detectors:nd_tank_model_mounting}\end{subfigure}

\vspace{0.7cm}
%\hfill
\begin{subfigure}[b]{0.45\textwidth}\centering 
\includegraphics[width=\textwidth]{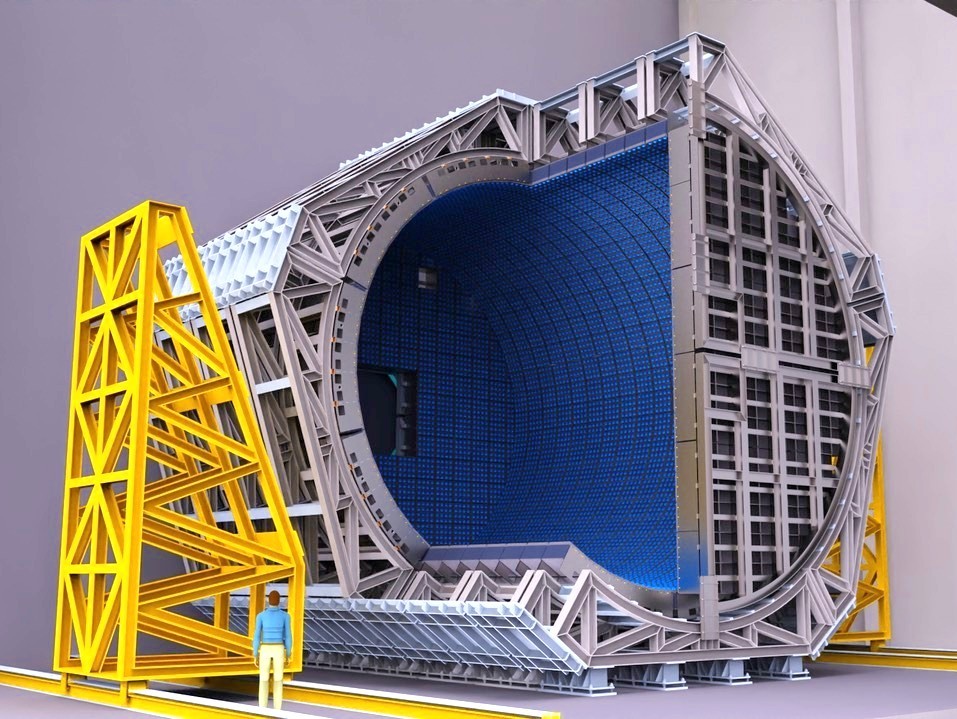}
\caption{An illustration of the detector structure in a mounting phase. Note the window with the interface to the SFGD.}
\label{fig:detectors:nd_tank_model_realisticmounting}\end{subfigure}
\hfill
\begin{subfigure}[b]{0.45\textwidth}\centering 
\includegraphics[width=\textwidth]{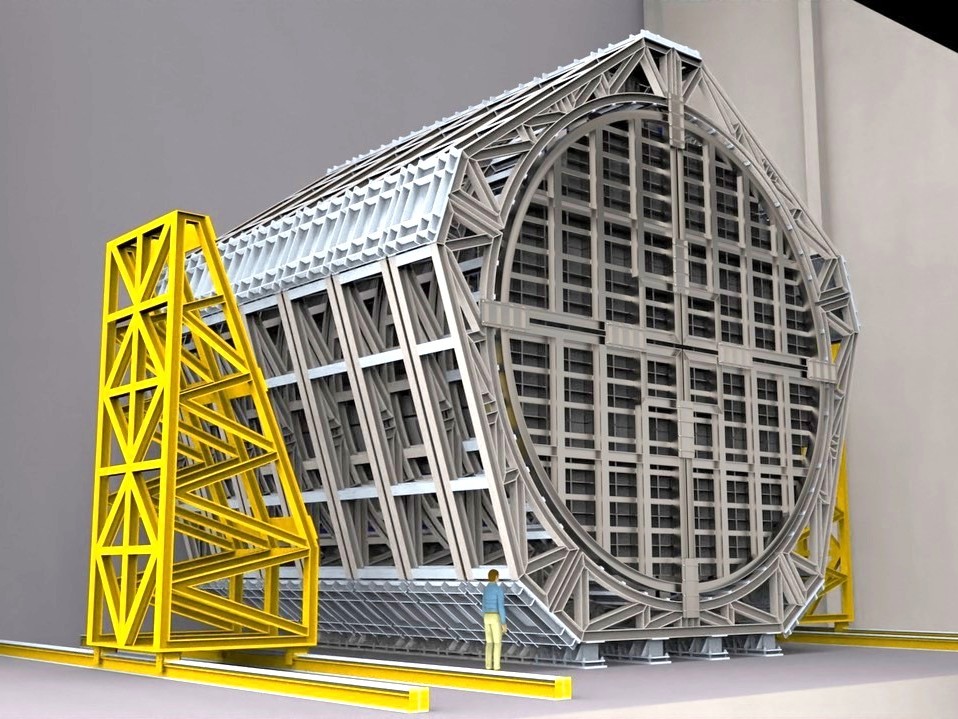}
\caption{An illustration of the detector structure with assembly complete.\vspace{1.0\baselineskip}}
\label{fig:detectors:nd_tank_model_realistictankstructure}\end{subfigure}
\caption{
Several views of the near detector water Cherenkov detector structure.
The detector frame will consist of five mantle sections and two end-cap sections, indicated in yellow in
(\subref{fig:detectors:nd_tank_model_endcap}) and (\subref{fig:detectors:nd_tank_model_structureframe}).
%and gray in
%(\subref{fig:detectors:nd_tank_model_realisticmounting}, \subref{fig:detectors:nd_tank_model_realistictankstructure}).
The mantle sections are further divided into five arches, which are fastened together using fixation structures, indicated in orange.
%in
%(\subref{fig:detectors:nd_tank_model_endcap}, \subref{fig:detectors:nd_tank_model_structureframe}).
The arches and end-caps will house the photo-sensor modules, indicated in blue and black, where each module shown houses 16 PMTs in a $4\times4$ grid (see Fig.\ref{fig:detectors:nd_pmt_housing} for further detail).
\label{fig:detectors:nd_tank_model}}
\end{figure}

Using a detector tank length of \SI{10.96}{\m}, two different detector tank radii -- \SI{9.44}{\m} and \SI{14.16}{\m} -- were combined with two respective photo-sensor coverages: \SI{30}{\percent} and \SI{40}{\percent}, to form four distinct detector configurations. For each detector configuration, \SI{6000}{} electron events and \SI{6000}{} muon events were simulated, equally distributed between positive and negative charge. The events were homogeneously distributed over the detector volume, with isotropic direction distribution and uniform kinetic energy distribution up to \SI{1}{\GeV}.

It was determined that the four detector configurations performed equally well, considering both energy reconstruction of the initial particles and particle flavor identification. The energy reconstruction average error varied between \SI{5.8}{\percent} and \SI{6.3}{\percent} for electron events, and between \SI{4.8}{\percent} and \SI{5.1}{\percent} for muon events.
Assuming the lepton-flavor identification criterion given in %Eq.~(\ref{eqn:detectors:nd_pid}),
Sec.~\ref{sct:detectors:ndwc_chlep},
the misidentification rate for electron events as muon events varied between \SI{0.3}{\percent} and \SI{0.7}{\percent} over the different detector geometries, and for muon events identified as electron events between \SI{0.2}{\percent} and \SI{1.0}{\percent}. No consistent correlation was observed for any detector configuration with the performances of the energy reconstruction and flavor identification. Since the four detector geometries gave similar performance in these measures, it was concluded that the smaller and more sparsely instrumented detector (radius \SI{9.44}{\m}, length \SI{10.96}{\m}, photo-sensor coverage \SI{30}{\percent}) is sufficient for this experiment.
%0.3 -> 0.308, 0.7 -> 0.682, 0.2 -> 0.248, 1.0 -> 0.987

In conclusion, the simulations show that a detector with a diameter and length of \SI{10}{\m}, and a photo sensor coverage of \SI{30}{\percent} provides as good performance as the larger alternatives that were investigated. Smaller volumes are not considered, as the range-out probability for events starting in the detector will increase (as indicated by the ranges given in Fig.~\ref{fig:detectors:lepton_range}). The smaller detector diameter and sparser PMT coverage, of \SI{9.44}{\m} and \SI{30}{\percent} respectively, are thus selected. Due to restrictions imposed by the PMT installations, the diameter and length of the tank are set to \SI{9.4}{\m} and \SI{10.8}{\m} respectively, yielding a total detector volume of \SI{750}{\m\cubed}.
%The exact dimensions are a consequence of the sensor coverage and the size of the photo multiplier tubes.
A model of this geometry is shown in Fig.~\ref{fig:detectors:nd_tank_model}.

%\begin{figure}[!phtb]
%    \centering
%     \includegraphics[width=0.9\linewidth]{figures/detectors/SmallPMT_one.png} 
%      \includegraphics[width=0.9\linewidth]{figures/detectors/SmallPMT_data.png} 
%    \caption{R14689 PMT data}
%    \label{fig:detectors:SmallPMT}
%\end{figure}

The detector frame consists of seven sections, five of these make up the central part of the detector, with the remaining two comprising the back and front end-caps, respectively. The central sections are divided into five arches that are mounted in the staging areas and then lowered to the floor of the detector pool for final assembly. The two end-caps differ slightly from each other; namely, there is an opening providing an interface area between the tracker and the upstream end-cap of the detector, whereas the downstream end-cap has a complete coverage of photo sensor modules. 

\begin{figure}[thb]
\centering
\begin{subfigure}[b]{0.4\textwidth}
\centering
\begin{tikzpicture}
\node[rotate=90] at (0,0) { \includegraphics[height=\textwidth]{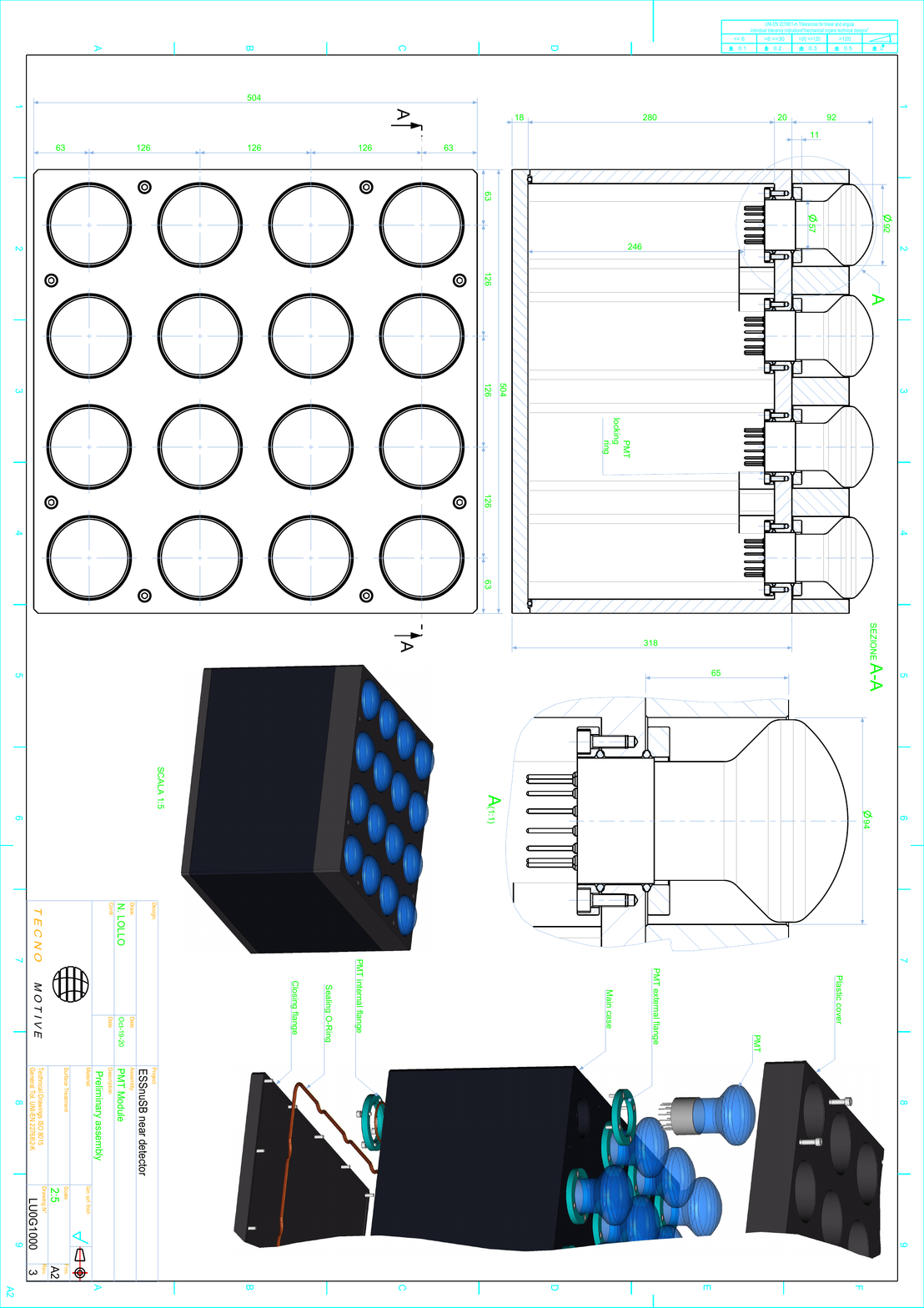} };
\end{tikzpicture}
\caption{Design of the photo-sensor housing module.\vspace{3.0\fontcharht\font`A}}
\label{fig:detectors:nd_pmt_housing_specs}
\end{subfigure}
\hfill
\begin{subfigure}[b]{0.57\textwidth}  
\centering 
\includegraphics[width=\textwidth]{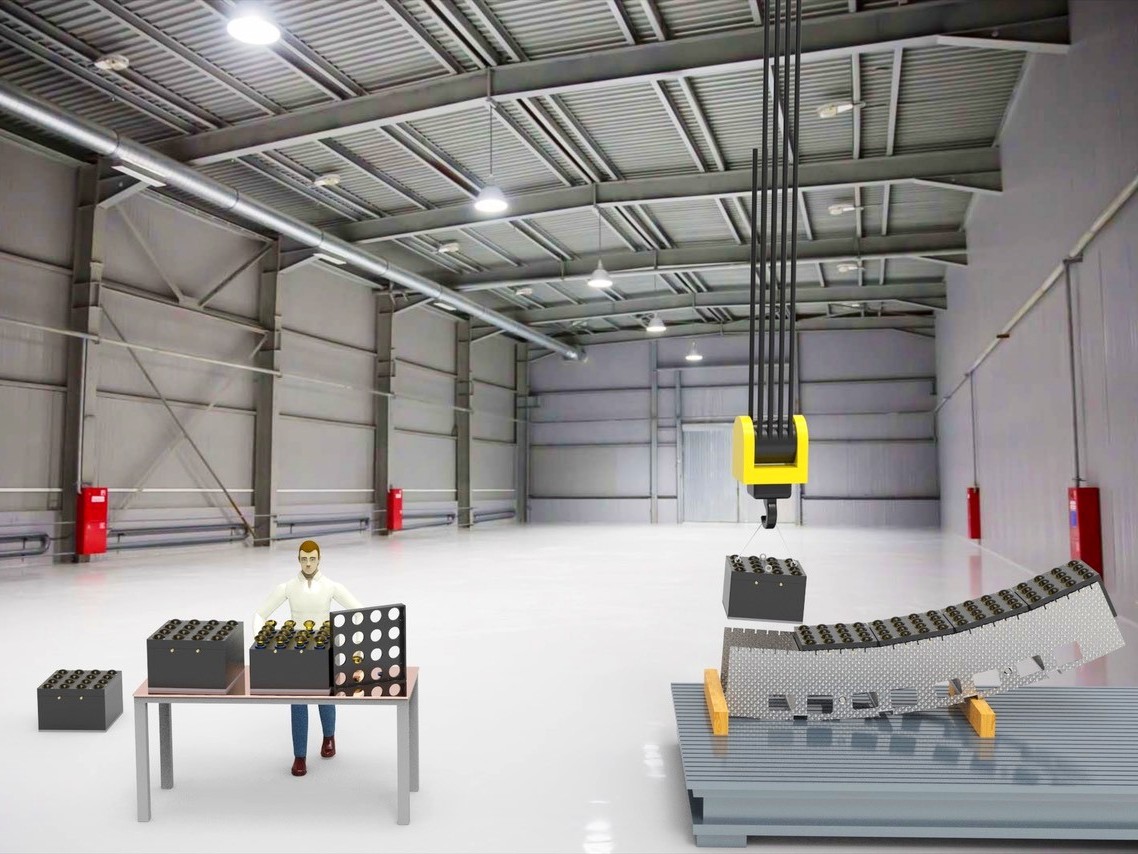}
\caption{Illustration of the assembly of two PMT mounting boxes by a technician, as well as the mounting of six photo-module boxes into an arch for subsequent mounting on the detector frame (see Fig.~\ref{fig:detectors:nd_tank_model})}
\label{fig:detectors:nd_pmt_housing_scene}
\end{subfigure}
%\begin{tikzpicture}
%\node[rotate=90] at (0,0) { \includegraphics[height=0.4\textwidth]{figures/detectors/nd_instr_jc_lu0g1000-2-201020} };
%\node at (8.3,0.9) { \includegraphics[width=0.57\textwidth]{figures/detectors/scena_montaggio.39_coladj_crop} };
%\end{tikzpicture}
\caption{Details (a) and depiction of the mounting (b) of photo-multiplier tubes.
\label{fig:detectors:nd_pmt_housing}}
\end{figure}

The internal part of the frame is covered with photo sensor modules that house \SI{16}{} photo multiplier tubes each. The preferred type is the Hamamatsu R14689 which has a transit-time spread of \SI{1.5}{\ns}~\cite{hamamatsu:R14689}. The tubes are mounted in a waterproof box that has space for readout electronics and a plastic veto detector to discard events which create charged particles outside of the detector volume. The data conversion takes place inside the waterproof module, and the data is transferred to the counting room via optical fibre. Similarly, each module will be fed with one power cable which feeds the power converters for the photo multipliers and the electronics. A drawing of the detector module is shown in Fig.~\ref{fig:detectors:nd_pmt_housing_specs}. After the photo sensors have been mounted in the module, the modules are mounted in arches as shown in Fig.~\ref{fig:detectors:nd_pmt_housing_scene}. The arches are mounted onto the frame as indicated in Fig.~\ref{fig:detectors:nd_tank_model_realisticmounting}.

The detector volume will be filled with clean water from the interior, with an inlet pipe at the bottom; to fill the main pool an outlet pipe sits at a slightly lower height near the inlet. The inlet is connected to an adjacent water purification plant along with a drainage pipe put at floor level. The detector modules must be fitted water-tight in order to prevent additional leakage between the inner and outer water volumes, apart from the flow through the outlet. Any potential water flow should occur from the inner to the outer volume to avoid contamination of the detector volume from particulate matter from the walls, floor, or the truss. The detector pool is covered with an extendable opaque roof (shown in red and black in Fig.~\ref{fig:detectors:nd_cavern_layout_full}).

\newcommand{\wcsmath}{\text{MC}}
\newcommand{\mcmath}{\text{MC}}
\newcommand{\fqmath}{\text{FQ}}
\newcommand{\fqemath}{\fqmath{e}}
\newcommand{\fqmumath}{\fqmath{\mu}}
\newcommand{\fqpizeromath}{\fqmath{\pi0}}
\newcommand{\fqnommath}{\fqmath{\text{nom}}}
\newcommand{\nllmath}{\tit{NLL}}
\newcommand{\nllemath}{\nllmath^{\fqemath}}
\newcommand{\nllmumath}{\nllmath^{\fqmumath}}
\newcommand{\nllratiomath}{\nllmumath/\nllemath}
\newcommand{\nllpizeromath}{\nllmath^{\fqpizeromath}}
\newcommand{\nllepizeroratiomath}{\nllemath/\nllpizeromath}
\newcommand{\nllmupizeroratiomath}{\nllmumath/\nllpizeromath}
\newcommand{\ekinmath}{\tit{E}_{\tit{kin}}}
\newcommand{\ekinemath}{\ekinmath^{\fqemath}}
\newcommand{\ekinmumath}{\ekinmath^{\fqmumath}}
\newcommand{\ekinnommath}{\ekinmath^{\fqnommath}}
\newcommand{\ekinwcsmath}{\ekinmath^{\wcsmath}}
\newcommand{\ekinratiomath}{\ekinmumath/\ekinemath}
\newcommand{\ekinewcsratiomath}{{\ekinemath}/{\ekinwcsmath}}
\newcommand{\ekinmuwcsratiomath}{{\ekinmumath}/{\ekinwcsmath}}
\newcommand{\ekinnomwcsratiomath}{{\ekinnommath}/{\ekinwcsmath}}
\newcommand{\qmath}{\tit{Q}^{\fqmath}}
\newcommand{\dwnommath}{\tit{d}_\tit{wall}^{\fqnommath}}
\newcommand{\dwdnommath}{\tit{d}_\tit{dir-wall}^{\fqnommath}}
\newcommand{\deltaxyzmath}{\Delta \bar{x}^{\fqmumath-\fqemath}}
\newcommand{\deltathetaphimath}{\Delta \hat{x}^{\fqmumath-\fqemath}}
\newcommand{\enumath}{\tit{E}_\nu}
\newcommand{\enuemath}{\enumath^{\fqemath}}
\newcommand{\enumumath}{\enumath^{\fqmumath}}
\newcommand{\enunommath}{\enumath^{\fqnommath}}
\newcommand{\enumcmath}{\enumath^{\mcmath}}
\newcommand{\enunommcratiomath}{\enunommath/\enumcmath}
\newcommand{\nsemath}{\tit{N}_{\tit{subev}}^{\fqmath}}
\newcommand{\costhetamath}{\cos\theta}
\newcommand{\costhetamcmath}{\cos\theta^{\wcsmath}}
\newcommand{\eidmath}{e^{\text{ID}}}
\newcommand{\muidmath}{\mu^{\text{ID}}}

\newcommand{\wcs}{$\wcsmath$}
\newcommand{\fq}{$\fqmath$}
\newcommand{\fqe}{$\fqemath$}
\newcommand{\fqmu}{$\fqmumath$}
\newcommand{\fqnom}{$\fqnommath$}
\newcommand{\nllnd}{$\nllmath$}
\newcommand{\nlle}{$\nllemath$}
\newcommand{\nllmu}{$\nllmumath$}
\newcommand{\nllratio}{$\nllratiomath$}
\newcommand{\nllpizero}{$\nllpizeromath$}
\newcommand{\nllepizeroratio}{$\nllepizeroratiomath$}
\newcommand{\nllmupizeroratio}{$\nllmupizeroratiomath$}
\newcommand{\ekin}{$\ekinmath$}
\newcommand{\ekine}{$\ekinemath$}
\newcommand{\ekinmu}{$\ekinmumath$}
\newcommand{\ekinnom}{$\ekinnommath$}
\newcommand{\ekinwcs}{$\ekinwcsmath$}
\newcommand{\ekinratio}{$\ekinratiomath$}
\newcommand{\ekinewcsratio}{$\ekinewcsratiomath$}
\newcommand{\ekinmuwcsratio}{$\ekinmuwcsratiomath$}
\newcommand{\ekinnomwcsratio}{$\ekinnomwcsratiomath$}
\newcommand{\q}{$\qmath$}
\newcommand{\dwnom}{$\dwnommath$}
\newcommand{\dwdnom}{$\dwdnommath$}
\newcommand{\deltaxyz}{$\deltaxyzmath$}
\newcommand{\deltathetaphi}{$\deltathetaphimath$}
\newcommand{\enu}{$\enumath$}
\newcommand{\enue}{$\enuemath$}
\newcommand{\enumu}{$\enumumath$}
\newcommand{\enunom}{$\enunommath$}
\newcommand{\enumc}{$\enumcmath$}
\newcommand{\enunommcratio}{$\enunommcratiomath$}
\newcommand{\nse}{$\nsemath$}
\newcommand{\costheta}{$\costhetamath$}
\newcommand{\costhetamc}{$\costhetamcmath$}
\newcommand{\eid}{$\eidmath$}
\newcommand{\muid}{$\muidmath$}

\subsubsubsection{Charged Lepton Identification and Reconstruction}
\label{sct:detectors:ndwc_chlep}

In order to properly reconstruct and identify neutrino events for muon and electron neutrinos, it is important to properly reconstruct and identify charged muons and electrons.
Samples of events, each consisting of single electrons ($e^+$, $e^-$) or muons ($\mu^+$, $\mu^-$), in the near detector were simulated using the WCSim software, each with \SI{54000}{} events.
The events were distributed homogeneously over the detector tank with an isotropic direction distribution, and uniformly distributed over kinetic energies up to \SI{1.2}{\GeV}.

The fiTQun software was then employed to reconstruct the events.
Several data selection criteria (see below) were defined for the reconstructed properties in order to reduce the number of misidentified and poorly reconstructed events in the sample.
An emphasis was made on rejecting misidentified muon events, since a main objective for the near detector is to select a high-purity sample of electron-neutrino candidate events. 
The criteria listed after \emph{the trigger} are collectively denoted as \emph{the charged lepton selection criteria}.

\emph{The trigger} ---
An event is rejected if there is no registered light in the detector.
This corresponds to real neutrino interactions in the detector that produce too little light to trigger a detector response.

\emph{The sub-Cherenkov criterion} ---
This criterion is applied in order to reject muon events where the muon has an energy below the Cherenkov threshold.
Such muon events do not themselves produce Cherenkov light, and are therefore mainly detected through their electron
decay product -- and are thus promptly misidentified as such.
It was found that these low energy muons have a substantial discrepancy between their reconstructed kinetic energies using the muon and electron hypotheses, {\ekinmu} and {\ekine} respectively. The criterion is set such that events with values of $\ekinratiomath>2.8$ are rejected, as shown in Fig.~\ref{fig:detectors:nd_chlep_cutsL2}.

\begin{figure}[htb]
\centering
\begin{tikzpicture}
%\node at (0,0) { \textit{\Large{}PlaceHolder} };
\node at (0.225\textwidth, 0.00\textwidth) { \includegraphics[width=0.495\textwidth]{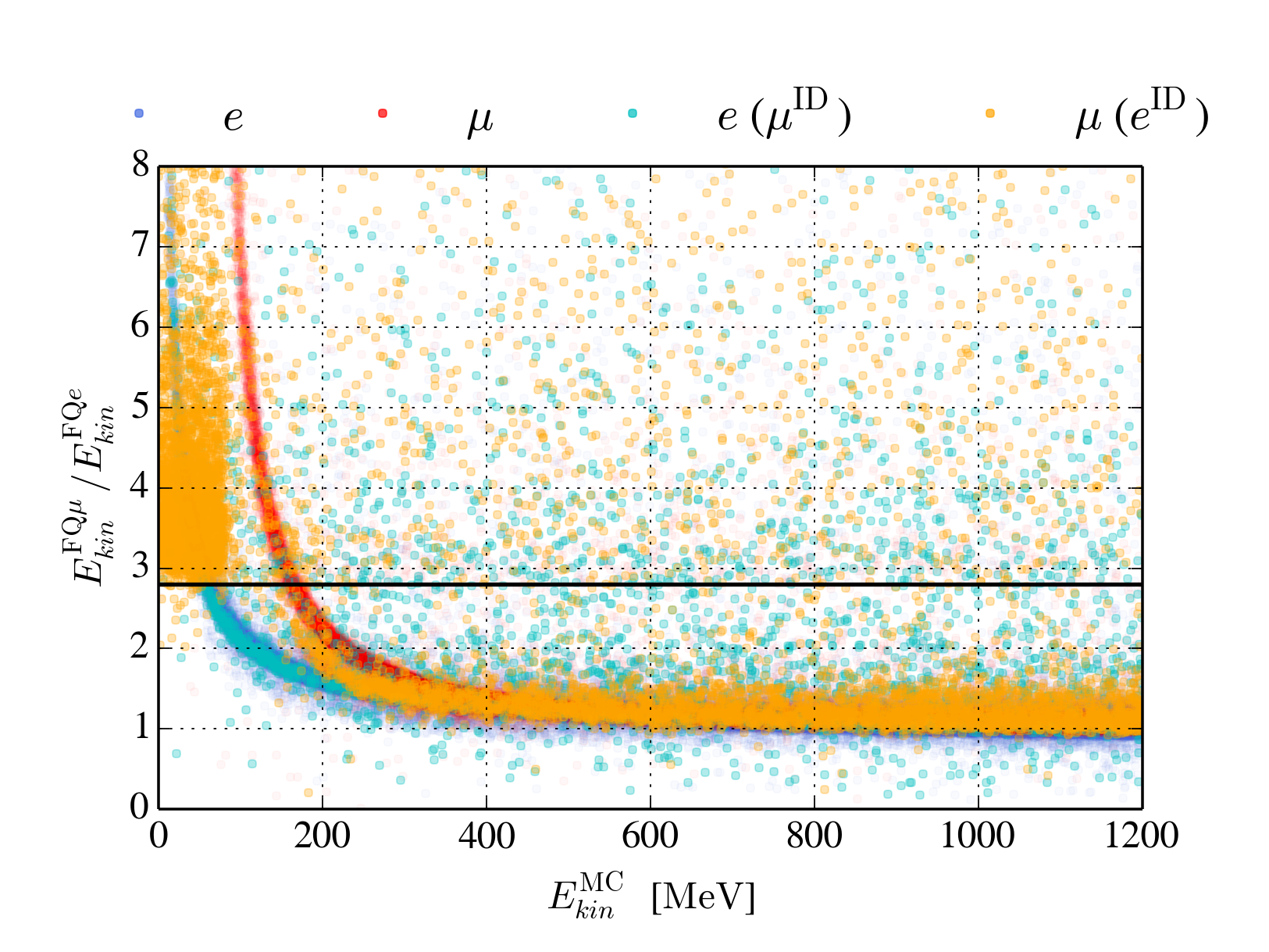} };
\end{tikzpicture}
\caption{Event distributions per flavor and reconstructed flavor over the {\ekinratio} and {\ekinwcs} variables. The \emph{sub-Cherenkov criterion} is marked with a solid black line at $\ekinratiomath=2.8$. All events above this line are rejected.
\label{fig:detectors:nd_chlep_cutsL2}}
\end{figure}

\emph{The reconstruction quality criteria} ---
Several factors contribute to the quality with which an event can be detected. Among these are the proximity of the event to the detector tank wall and the registered brightness (total registered charge) of the event. These selection criteria are implemented in order to reject events that have a poor energy reconstruction, and thus involve straight cuts on the registered charge of the event, {\q}, and on the shortest distance between the reconstructed event vertex and the detector tank wall, {\dwnom}, where the super-script represents the preferred (nominal) reconstruction flavor hypothesis. An event needs to meet the following criteria in order to be accepted:
\begin{align}
\qmath  &\geq \SI{1000}{\PE} \nonumber \\
\dwnommath &\geq \SI{30}{\cm}
\label{eqn:detectors:cutsL3}
\end{align}

\noindent where \SI{1}{\PE} is the average registered charge of a detected photon.
The event distributions over these variables, along with the ratio between the reconstructed and true kinetic energy, \ekinnomwcsratio, are shown in Fig.~\ref{fig:detectors:nd_chlep_cutsL3}.

\emph{The Cherenkov-ring resolution criterion} ---
It was observed that events with a short distance to the detector tank wall in the direction of the reconstructed particle are more likely to be mis-reconstructed and mis-identified than events with a longer distance. This distance is illustrated in Fig.~\ref{fig:detectors:nd_chlep_cutsL4_dwdnomillustration}; this criterion is defined such that events with a distance to the wall along the event direction, {\dwdnom}, shorter than
\SI{250}{\cm} are rejected. Event distributions over the {\dwdnom} and {\ekinnomwcsratio} variables are displayed in Fig.\ref{fig:detectors:nd_chlep_cutsL4_ekindwd} along with the selection criterion.

\begin{figure}[p]
\centering
\begin{subfigure}[b]{0.495\textwidth}
\centering
\includegraphics[width=\textwidth]{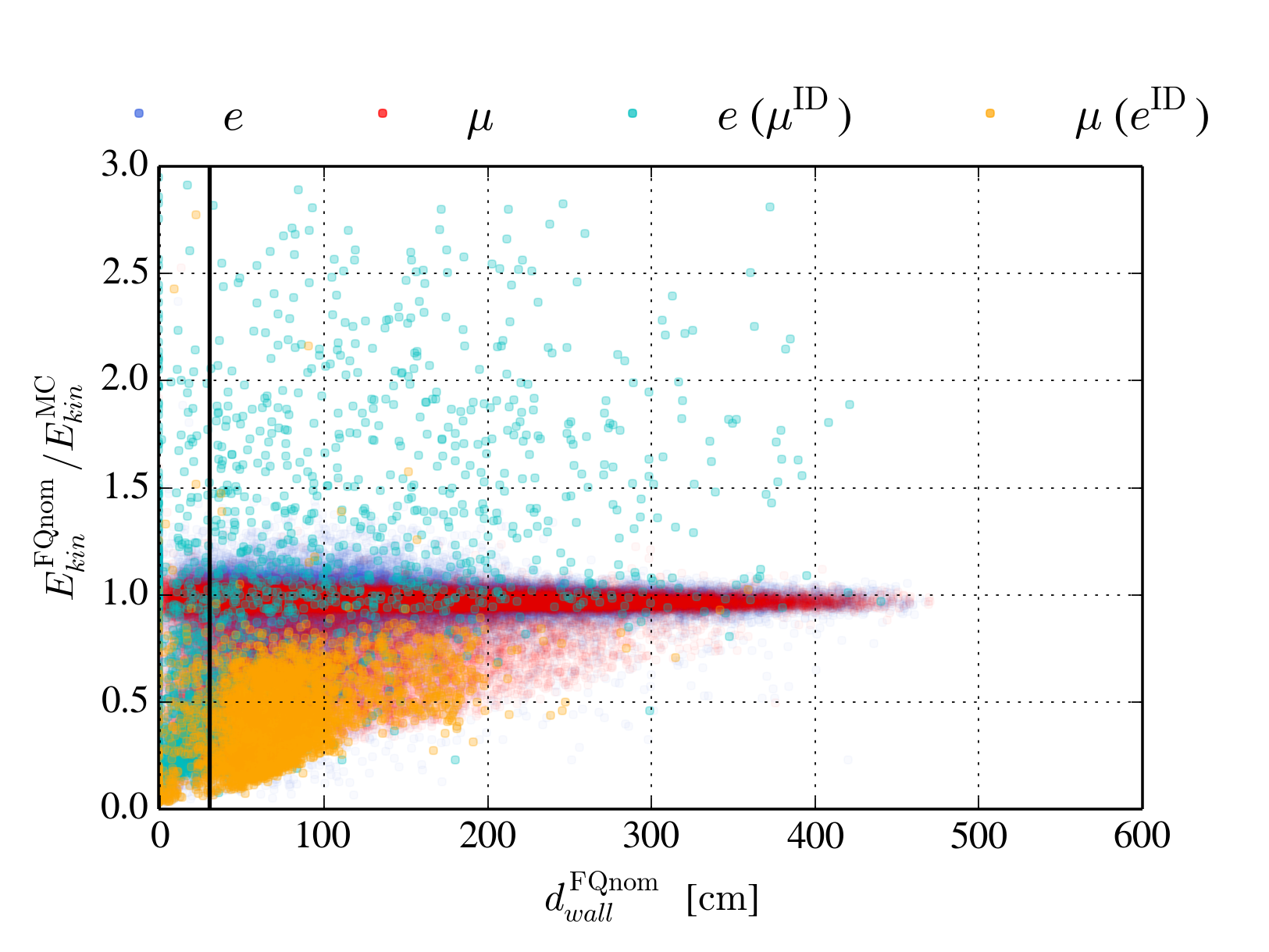}
\caption{Event distributions over the {\ekinnomwcsratio} and {\dwnom} variables.
All events to the left of the black line at $\dwnommath=\SI{30}{\cm}$ are rejected.}
\label{fig:detectors:nd_chlep_cutsL3_ekindw}
\end{subfigure}
\hfill
\begin{subfigure}[b]{0.495\textwidth}  
\centering 
\includegraphics[width=\textwidth]{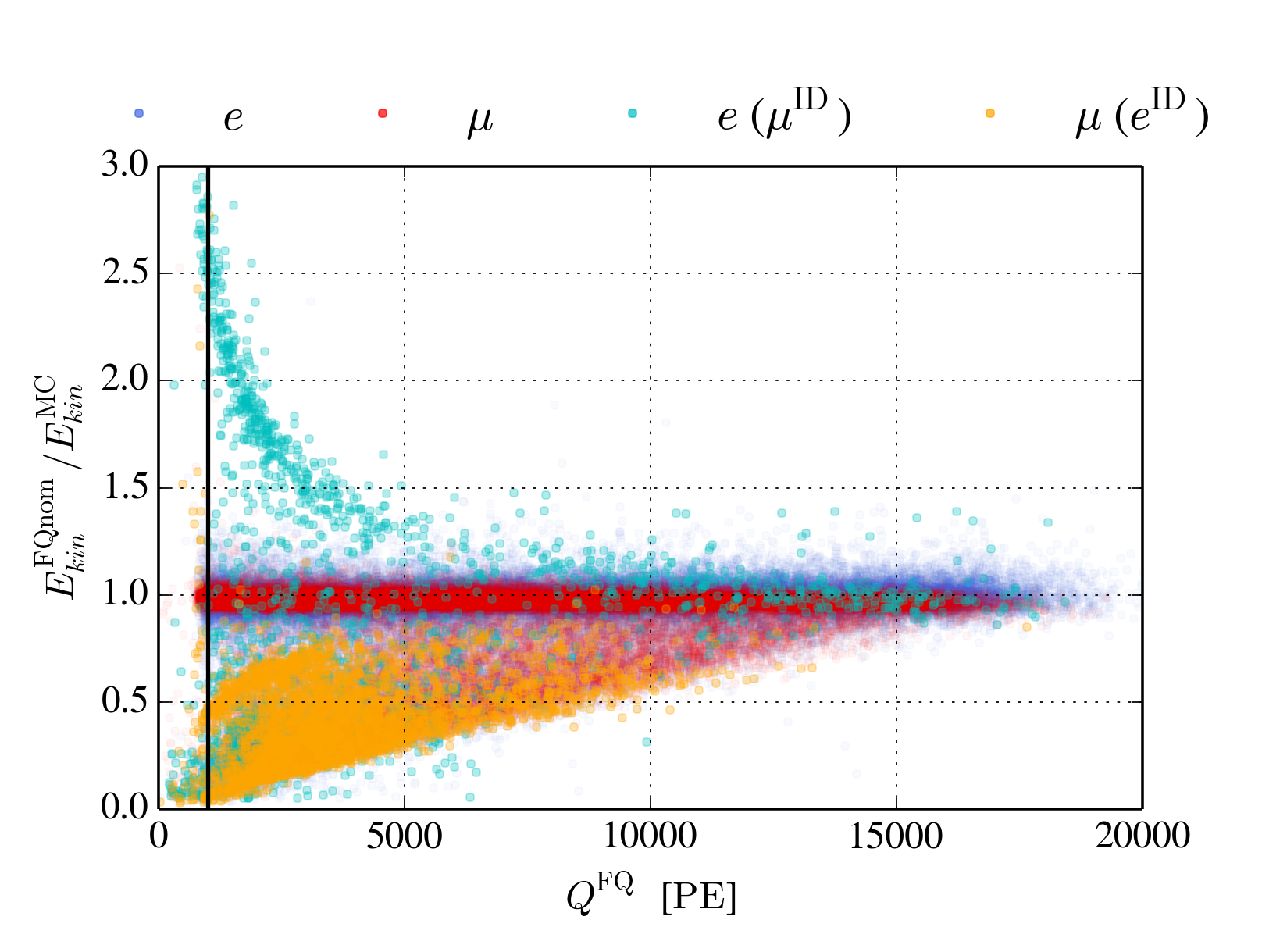}
\caption{Event distributions over the {\ekinnomwcsratio} and {\q} variables.
All events to the left of the black line at $\qmath=1000~\text{PE}$ are rejected.}
\label{fig:detectors:nd_chlep_cutsL3_ekinq}
\end{subfigure}
\caption{
Event distributions per flavor and reconstructed flavor before applying the \emph{reconstruction quality criteria}, where the selection criteria are illustrated as black lines.
\label{fig:detectors:nd_chlep_cutsL3}
}
\end{figure}

\begin{figure}[p]
\centering
\begin{subfigure}[b]{0.495\textwidth}
\centering
\includegraphics[width=\textwidth]{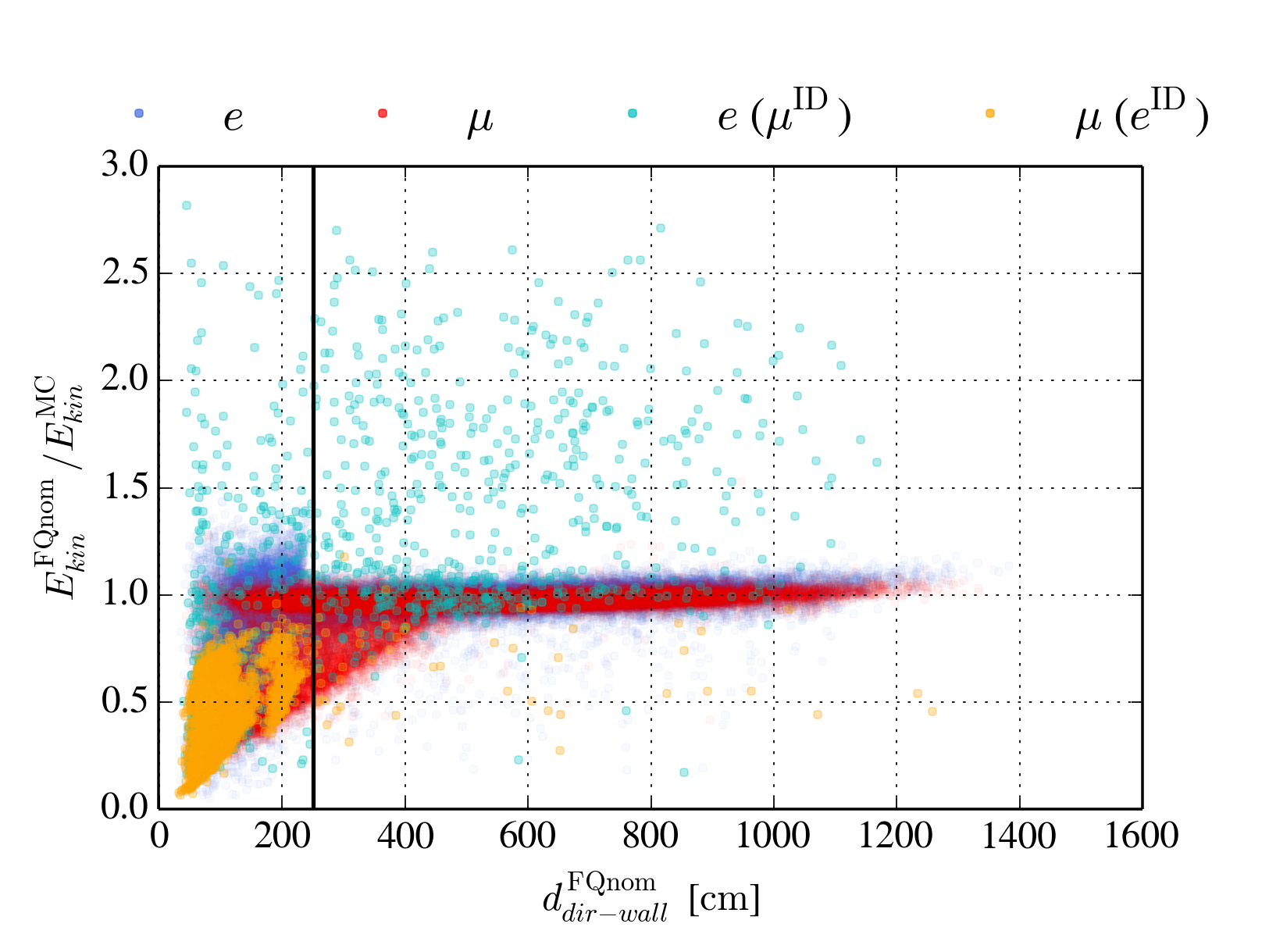}
\caption{Event distributions per flavor and reconstructed flavor over the {\ekinnomwcsratio} and {\dwdnom} variables. The \emph{Cherenkov-ring resolution criterion} is illustrated as a black line at $\dwdnommath=\SI{250}{\cm}$. All events to the left of this line are rejected.\vspace{\baselineskip}}
\label{fig:detectors:nd_chlep_cutsL4_ekindwd}
\end{subfigure}
\hfill
\begin{subfigure}[b]{0.495\textwidth}  
\centering 
\includegraphics[width=\textwidth]{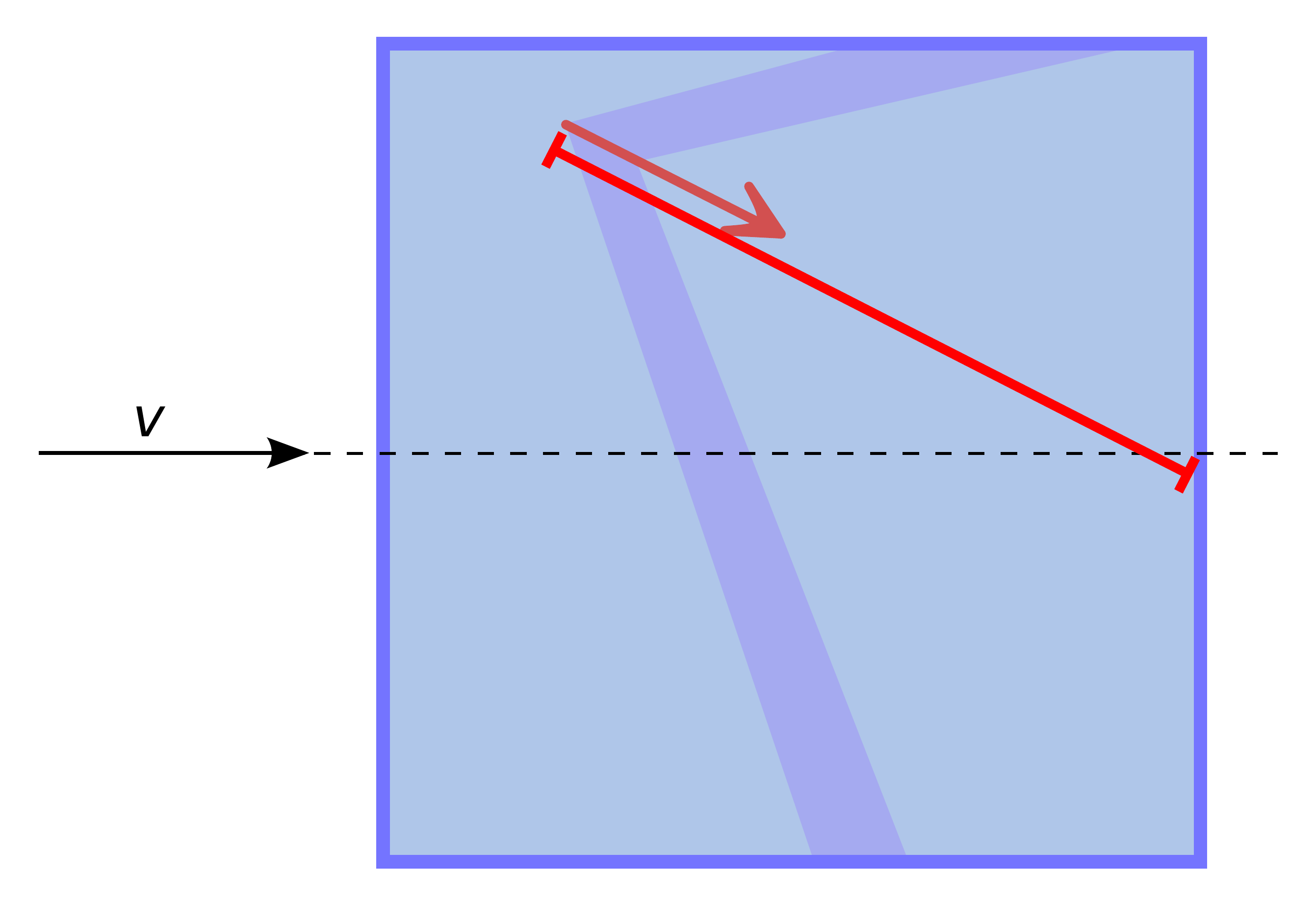}
\caption{Illustration of the {\dwdnom} variable as a red line connecting the reconstructed event vertex with the wall of the detector tank (blue volume) in the reconstructed direction of the event (dark red arrow). The Cherenkov light of the event is illustrated in purple, and the incident direction of the neutrino beam is illustrated with a black arrow.}
\label{fig:detectors:nd_chlep_cutsL4_dwdnomillustration}
\end{subfigure}
\caption{
Event distributions before the application of the \emph{Cherenkov-ring resolution criterion} along with an illustration of the {\dwdnom} variable.
\label{fig:detectors:nd_chlep_cutsL4}}
\end{figure}

The collected selection efficiency of these cuts is \SI{54.9}{\percent} for electron events and \SI{50.3}{\percent} for muon events.
This is shown as a function of kinetic energy and direction in Fig.~\ref{fig:detectors:nd_chlep_eff}.
For both electron and muon events, the efficiency plateaus at ${\sim}\SI{60}{\percent}$ above a threshold energy
around \SI{80}{\MeV} for electrons and \SI{200}{\MeV} for muons.
Events below these energies are rejected by the sub-Cherenkov criterion.
A slight bias of lower efficiency can be seen for events directed along $\costhetamcmath\approx\pm0.5$.
This direction corresponds to events directed diagonally in the detector.

\begin{figure}[thb]
\centering
\begin{subfigure}[b]{0.495\textwidth}
\centering
\includegraphics[width=\textwidth]{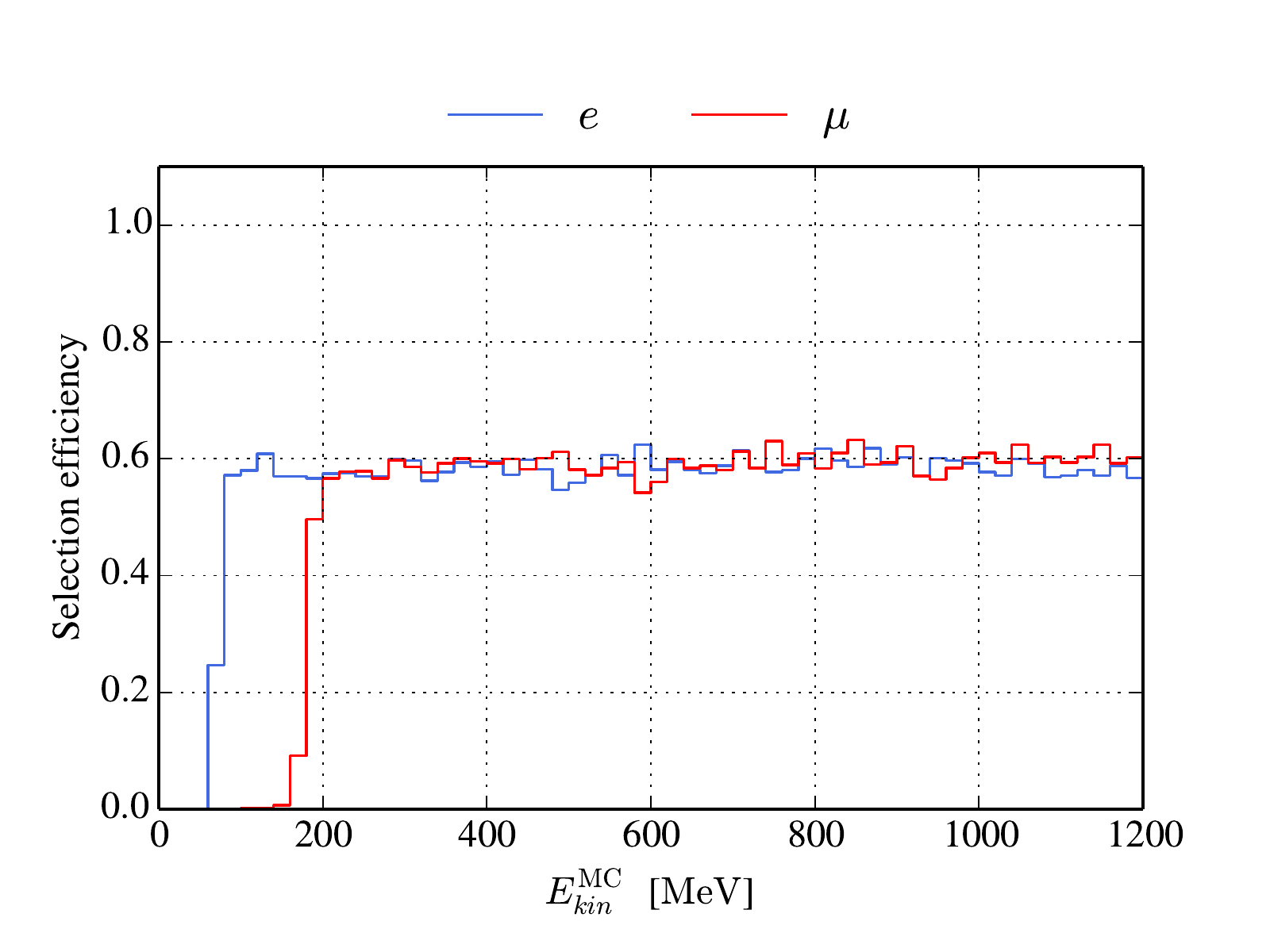}
\caption{Selection efficiency over {\ekinwcs}.}
\label{fig:detectors:nd_chlep_eff_ekin}
\end{subfigure}
\hfill
\begin{subfigure}[b]{0.495\textwidth}  
\centering 
\includegraphics[width=\textwidth]{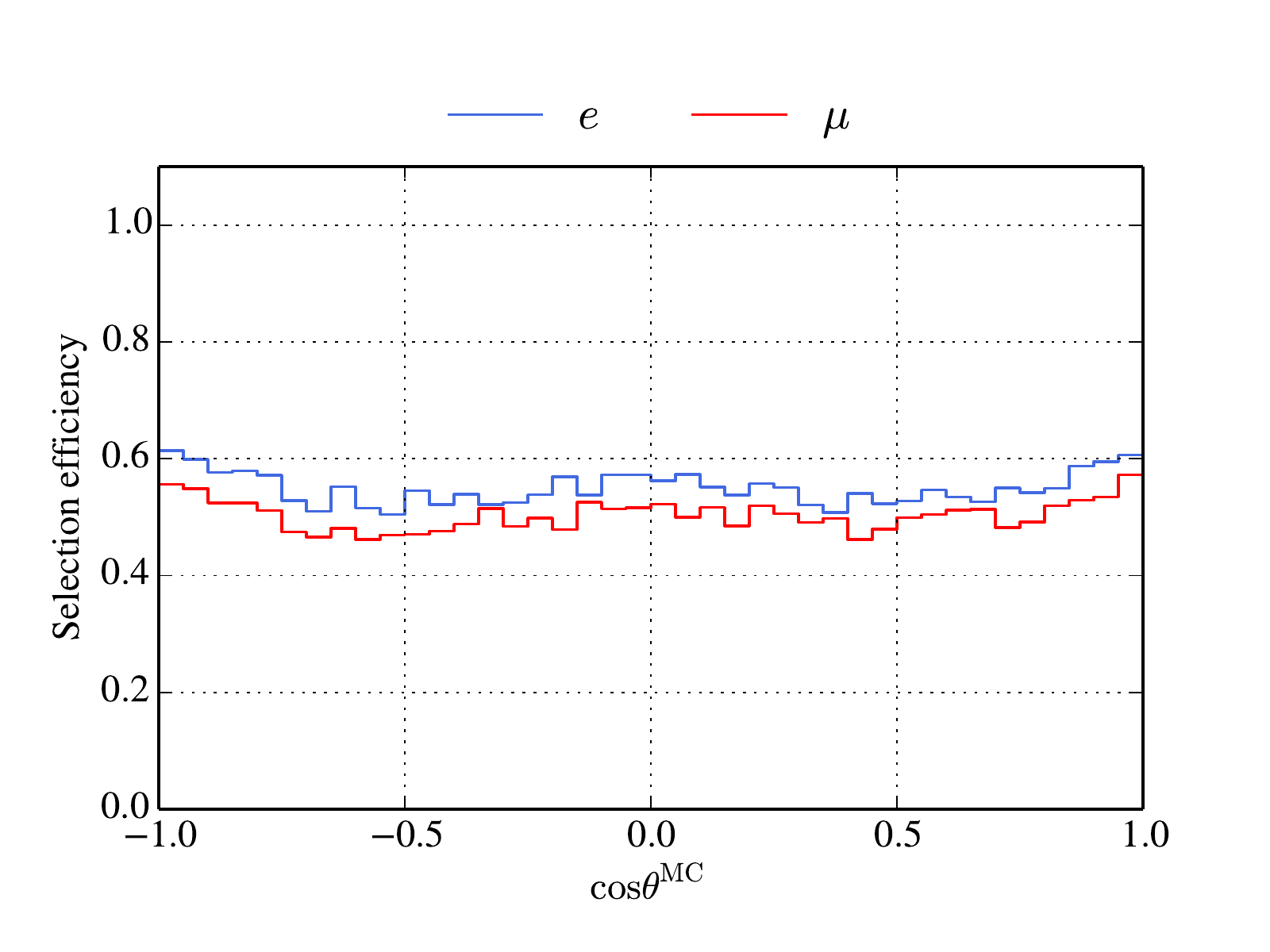}
\caption{Selection efficiency over {\costhetamc}.}
\label{fig:detectors:nd_chlep_eff_costheta}
\end{subfigure}
\caption{Selection efficiency of the charged lepton criteria for muon and electron events over the {\ekinwcs} and {\costhetamc} variables.
\label{fig:detectors:nd_chlep_eff}}
\end{figure}

Each event is identified as electron-like or muon-like through the ratio between their negative-log-likelihood ({\nllnd}) for each hypothesis, {\nllratio}.
If this {\nllnd}-ratio of a given event is above or equal to 1.01 the event is classified as electron-like, {\eid}; and as muon-like, {\muid}, if not.
\begin{align}
\nllratiomath  &\geq 1.01 ~~ \Rightarrow ~~ \eidmath \nonumber \\
\nllratiomath  &<    1.01 ~~ \Rightarrow ~~ \muidmath
\label{eqn:detectors:nd_pid}
\end{align}

The event distributions for muon and electron events are shown in Fig.~\ref{fig:detectors:nd_chlep_pid}
over the {\nllnd}-ratio and over the true kinetic energy along with the identification criterion at 1.01.
Using this identification criterion, the mis-identification rates are \SI{1.1}{\percent} and \SI{0.09}{\percent} for electrons and muons, respectively.
The higher mis-identification rate of electrons is acceptable, owing to the low electron-neutrino content of the neutrino beam.

\begin{figure}[!bht]
\centering
\begin{tikzpicture}
%\node at (0,0) { \textit{\Large{}PlaceHolder} };
\node at (0.0           ,-0.00\textwidth) { \includegraphics[width=0.495\textwidth]{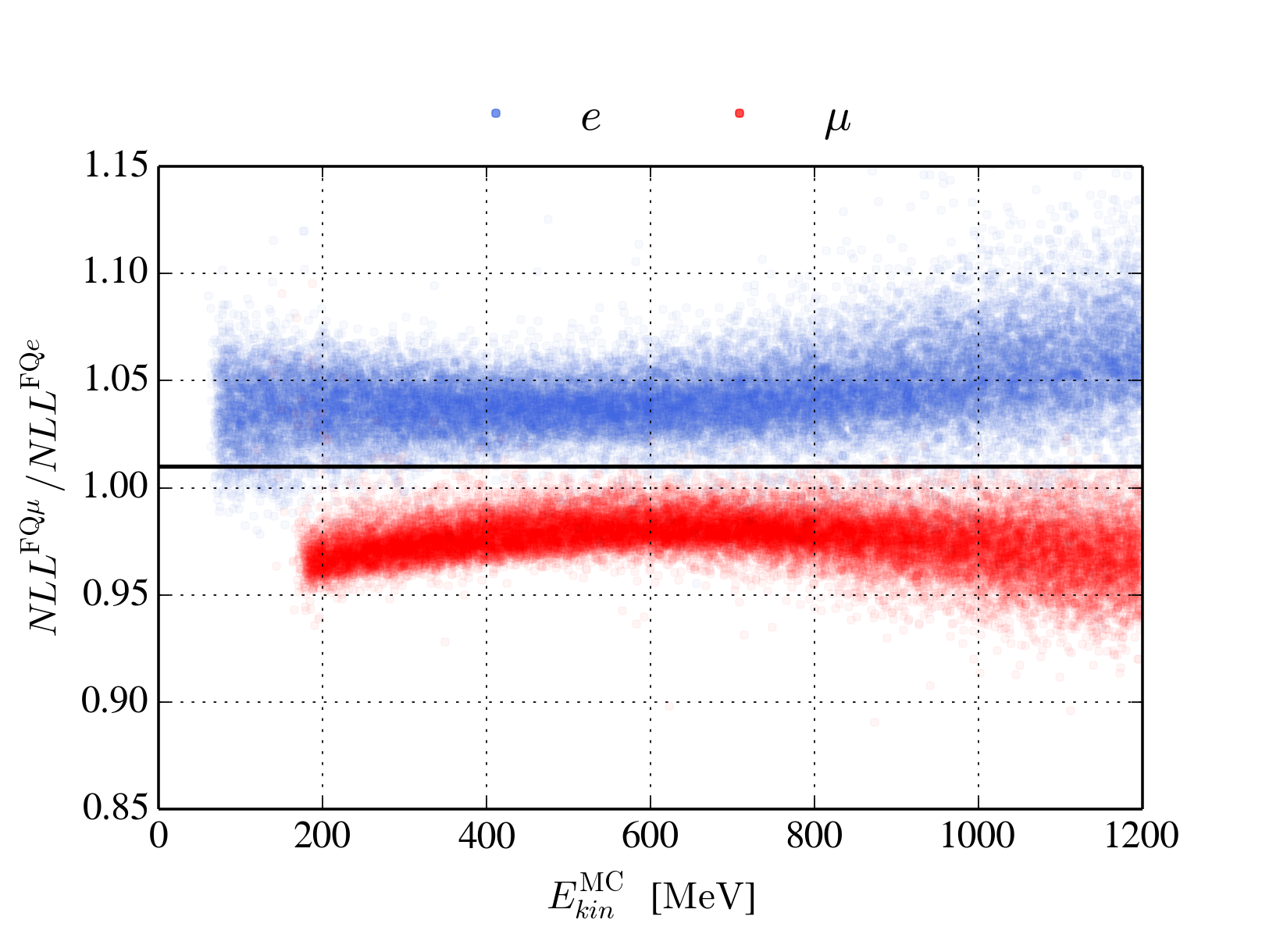} };
\end{tikzpicture}
\caption{Event distributions for muon and electron events over the {\nllratio} and {\ekinwcs} variables. The flavour identification criterion, Eq.~(\ref{eqn:detectors:nd_pid}), is shown as a black line at $\nllratiomath=1.01$. Events above this line are identified as electron-like, and events falling below are identified as muon-like.
\label{fig:detectors:nd_chlep_pid}}
\end{figure}

Event distributions after the final level of data selection are shown in Figs.~\ref{fig:detectors:nd_chlep_reco}
over the true and reconstructed energy and direction. Both muon and electron events primarily fall along the diagonal of these figures, indicating that the reconstruction performs well. A population of muon events are mis-reconstructed with their energies too low (below the diagonal in Fig.~\ref{fig:detectors:nd_chlep_reco_muekin}). These are identified as muons that exit the active detector volume before depositing their full energy.

\begin{figure}[p]
\centering
\begin{subfigure}[b]{0.495\textwidth}
\centering
\begin{tikzpicture}
\node at (0,0) { \includegraphics[width=\textwidth]{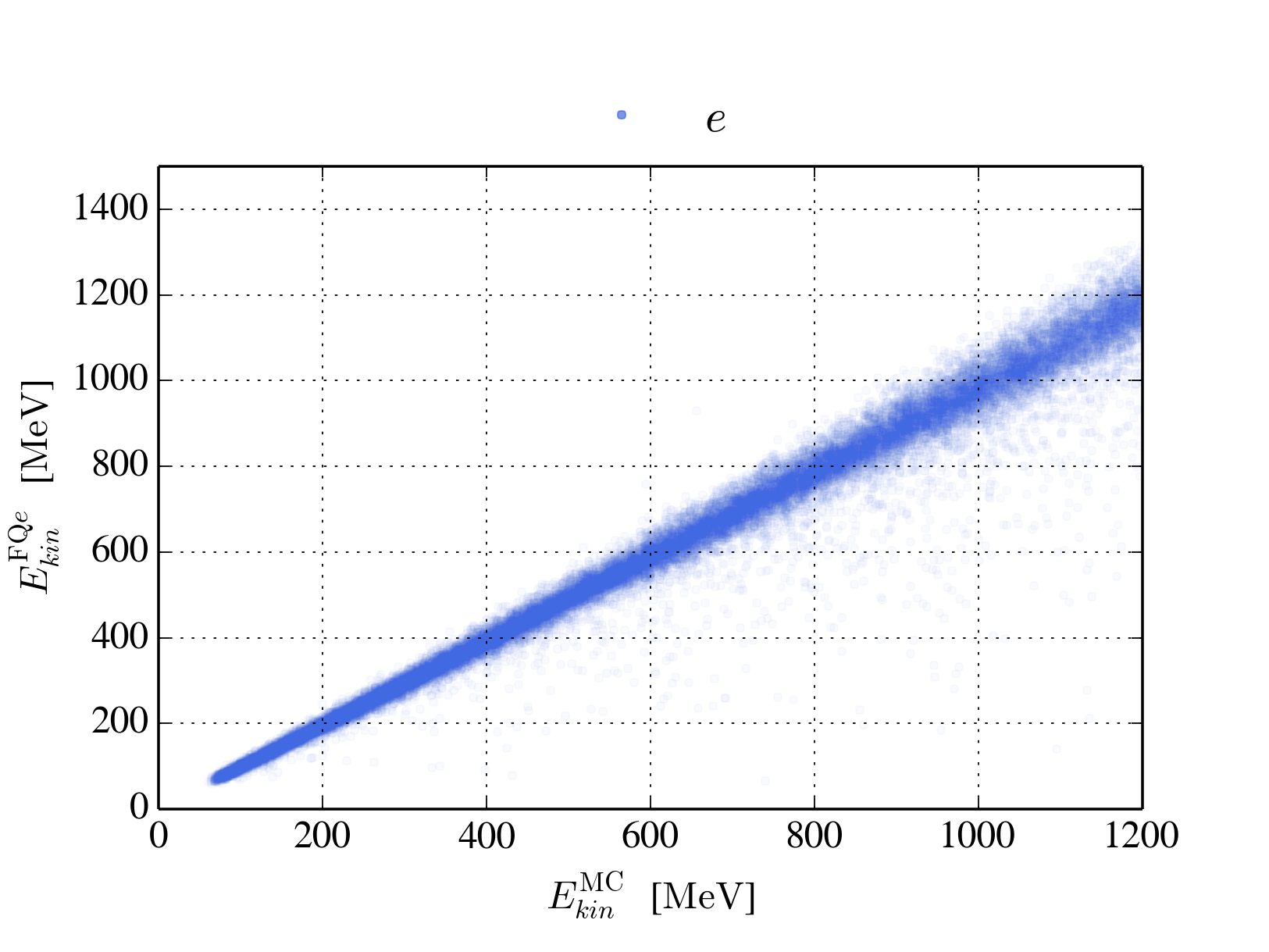} };
\fill[white] (-0.09\textwidth, 0.26\textwidth) rectangle ( 0.09\textwidth, 0.32\textwidth);
\end{tikzpicture}
\caption{}
\label{fig:detectors:nd_chlep_reco_eekin}
\end{subfigure}
\hfill
\begin{subfigure}[b]{0.495\textwidth}  
\centering 
\begin{tikzpicture}
\node at (0,0) { \includegraphics[width=\textwidth]{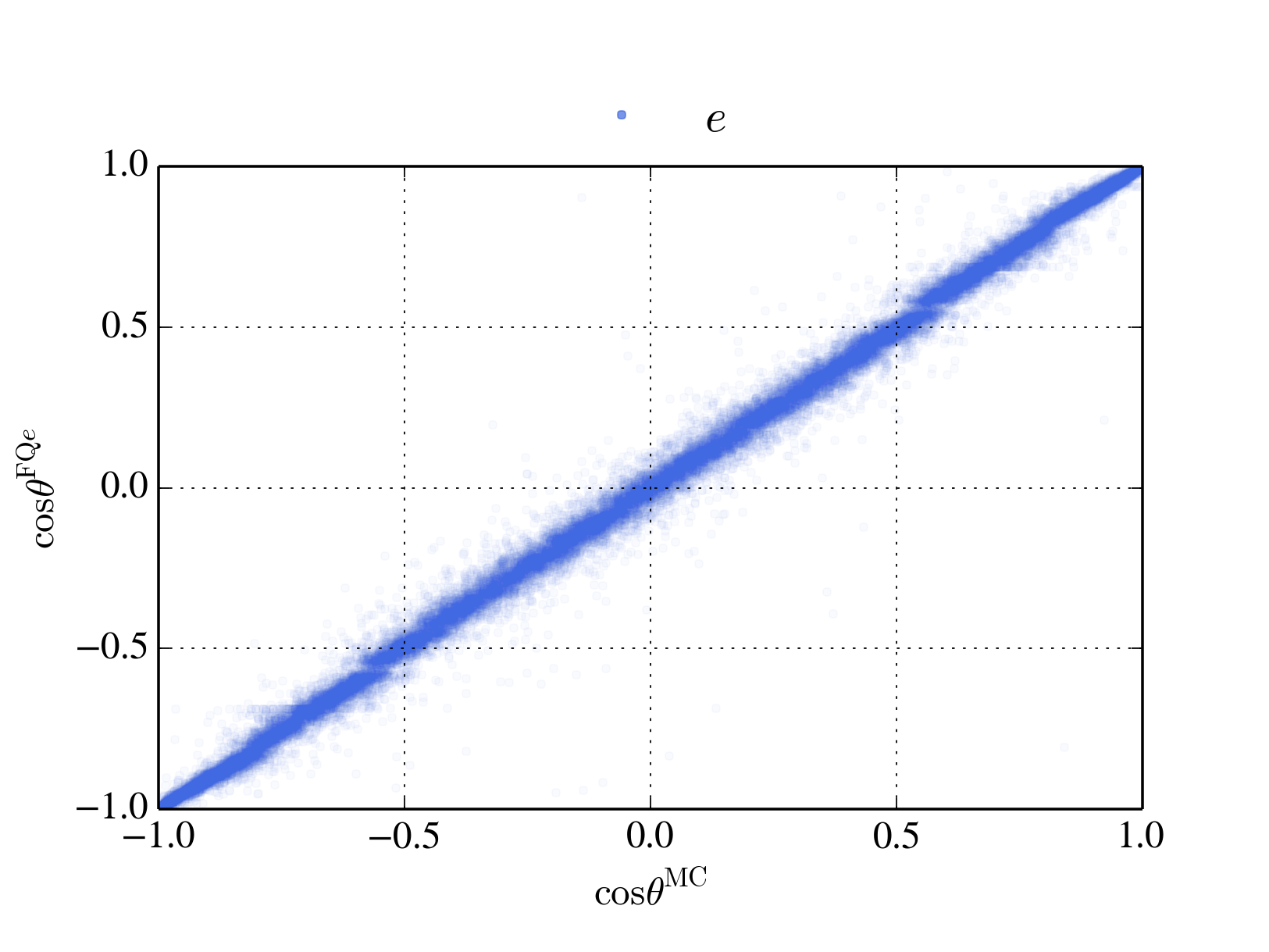} };
\fill[white] (-0.09\textwidth, 0.26\textwidth) rectangle ( 0.09\textwidth, 0.32\textwidth);
\end{tikzpicture}
\caption{}
\label{fig:detectors:nd_chlep_reco_ecostheta}
\end{subfigure}
\hfill
\begin{subfigure}[b]{0.495\textwidth}  
\centering 
\begin{tikzpicture}
\node at (0,0) { \includegraphics[width=\textwidth]{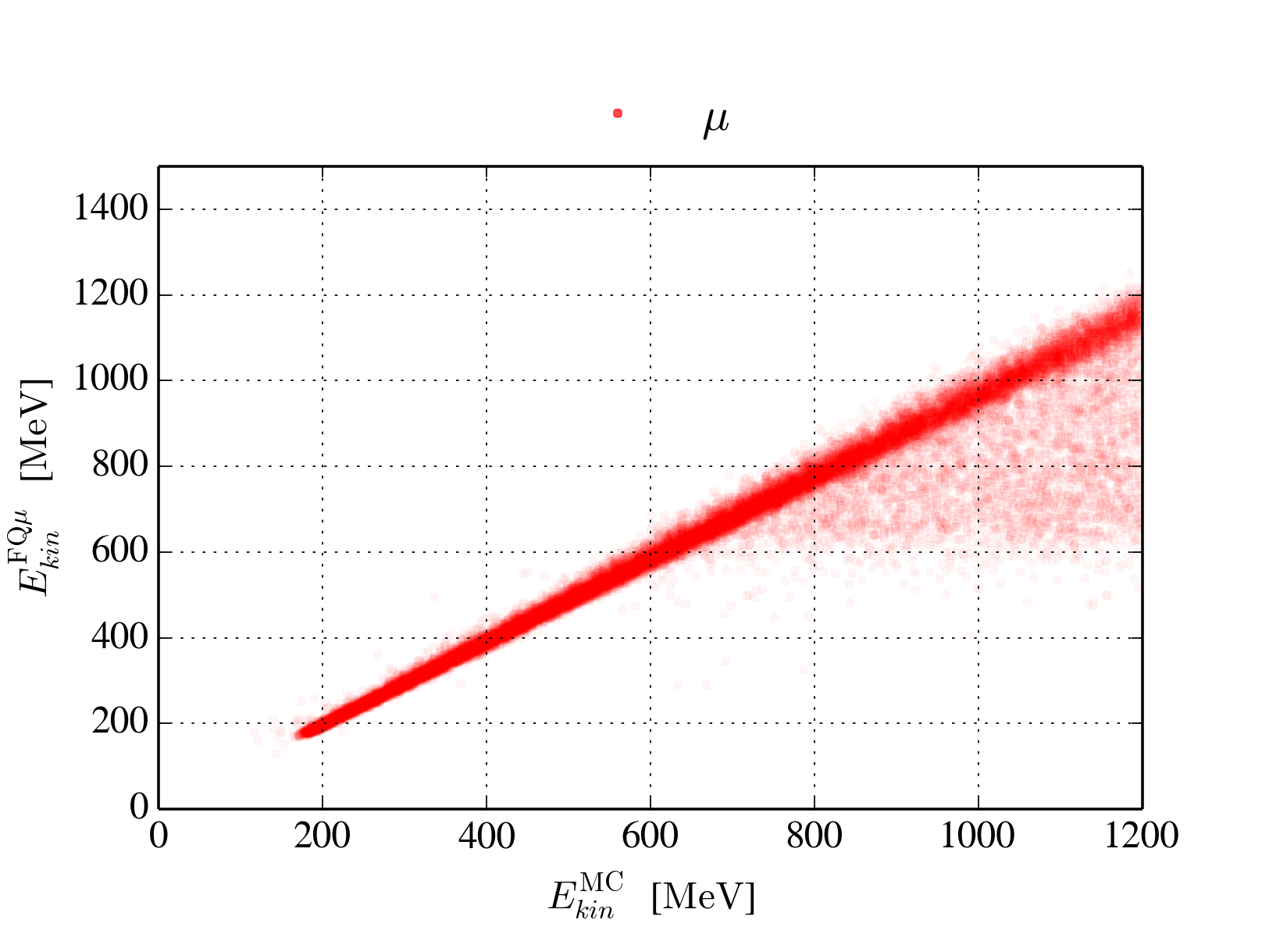} };
\fill[white] (-0.09\textwidth, 0.26\textwidth) rectangle ( 0.09\textwidth, 0.32\textwidth);
\end{tikzpicture}
\caption{}
\label{fig:detectors:nd_chlep_reco_muekin}
\end{subfigure}
\hfill
\begin{subfigure}[b]{0.495\textwidth}  
\centering 
\begin{tikzpicture}
\node at (0,0) { \includegraphics[width=\textwidth]{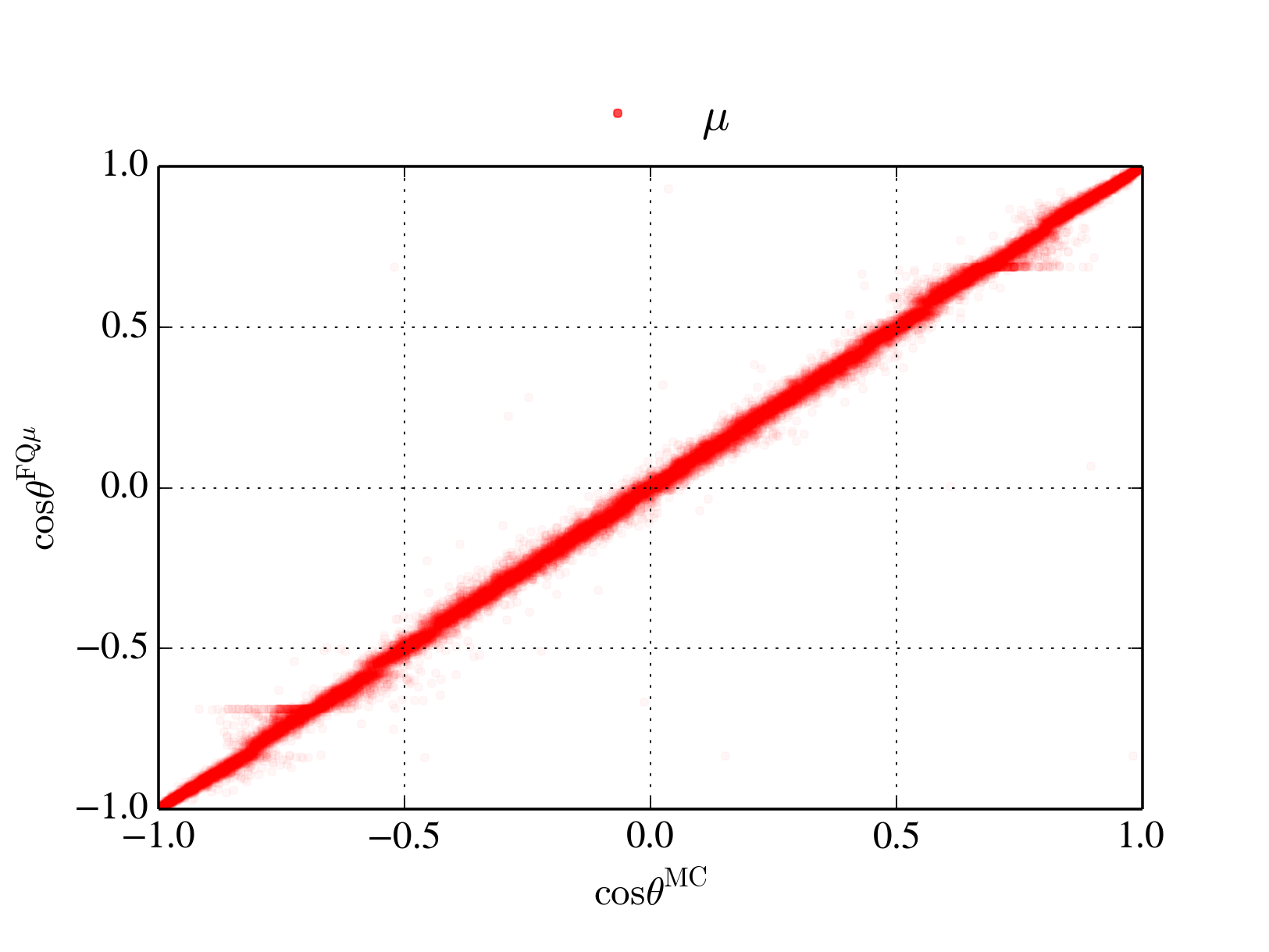} };
\fill[white] (-0.09\textwidth, 0.26\textwidth) rectangle ( 0.09\textwidth, 0.32\textwidth);
\end{tikzpicture}
\caption{}
\label{fig:detectors:nd_chlep_reco_mucostheta}
\end{subfigure}
%\begin{tikzpicture}
%\node at (0.0           ,-0.35\textwidth) { \includegraphics[width=0.5\textwidth]{figures/detectors/ERECO_210430__fqe_ekin__wcs_ekin__L5} };
%\node at (0.5\textwidth,-0.35\textwidth) { \includegraphics[width=0.5\textwidth]{figures/detectors/ERECO_210430__fqmu_ekin__wcs_ekin__L5} };
%\fill[white] (-0.045\textwidth,-0.225\textwidth) rectangle ( 0.045\textwidth,-0.19\textwidth);
%\fill[white] ( 0.455\textwidth,-0.225\textwidth) rectangle ( 0.545\textwidth,-0.19\textwidth);
%\node at (0.0           ,-0.00\textwidth) { \includegraphics[width=0.5\textwidth]{figures/detectors/ERECO_210430__fqe_costheta__wcs_costheta__L5} };
%\node at (0.5\textwidth,-0.00\textwidth) { \includegraphics[width=0.5\textwidth]{figures/detectors/ERECO_210430__fqmu_costheta__wcs_costheta__L5} };
%\fill[white] (-0.045\textwidth, 0.125\textwidth) rectangle ( 0.045\textwidth, 0.16\textwidth);
%\fill[white] ( 0.455\textwidth, 0.125\textwidth) rectangle ( 0.545\textwidth, 0.16\textwidth);
%\node[anchor=north,align=center] at (-0.10\textwidth, 0.07\textwidth) {{\Large$e$}};
%\node[anchor=north,align=center] at ( 0.40\textwidth, 0.07\textwidth) {{\Large$\mu$}};
%\node[anchor=north,align=center] at (-0.10\textwidth,-0.27\textwidth) {{\Large$e$}};
%\node[anchor=north,align=center] at ( 0.40\textwidth,-0.27\textwidth) {{\Large$\mu$}};
%\end{tikzpicture}
\caption{Event samples of electron (\subref{fig:detectors:nd_chlep_reco_eekin}, \subref{fig:detectors:nd_chlep_reco_ecostheta}) and muon (\subref{fig:detectors:nd_chlep_reco_muekin}, \subref{fig:detectors:nd_chlep_reco_mucostheta}) events, shown over true and reconstructed kinetic energy (\subref{fig:detectors:nd_chlep_reco_eekin}, \subref{fig:detectors:nd_chlep_reco_muekin}) and direction (\subref{fig:detectors:nd_chlep_reco_ecostheta}, \subref{fig:detectors:nd_chlep_reco_mucostheta}). The true angle and energy are given by the {\costhetamc} and {\ekinwcs} variables, and the reconstructed variables are calculated using the correct flavor-assumption, given as {\fqe} and {\fqmu} for electrons and muons, respectively.
\label{fig:detectors:nd_chlep_reco}}
\end{figure}

The performance of the energy and direction reconstruction can be estimated as follows. The ratio between the reconstructed and true kinetic energies is $0.95\pm0.08$ and $0.97\pm0.06$ for the electron and muon event samples, respectively, where the uncertainties represent the spread of the distributions.
% 0.9502994164990798 +/- 0.08372081018935326
% 0.9748114499275058 +/- 0.06438153487834992
Correspondingly, the difference between the reconstructed and true directions, {\costheta}, is
$0.00\pm0.03$ and $0.00\pm0.04$ for the electron and muon event samples, respectively.
% -0.00040306866462943387 +/- 0.029652096153122694
% -0.0003764524828911423 +/- 0.044661066258716006
The muon sample shows a larger error in reconstructed energy than the electron sample, considering both the average error and width of the spread. This is attributed to the population of muons falling below the diagonal in Fig.~\ref{fig:detectors:nd_chlep_reco_muekin} with kinetic energies exceeding ${\sim}\SI{700}{\MeV}$. These are identified as events where the muon travels beyond the instrumented detector volume before depositing all of its energy, which yields a lower total brightness to the event than otherwise expected.

\subsubsubsection{Neutrino identification and reconstruction}
\label{sct:detectors:ndwc_nuvtx}

In addition to having an efficient and well-performing reconstruction algorithm of single charged leptons (as demonstrated in the previous section), it is important to construct a data-selection algorithm that accounts for the full neutrino event.

% FIGURE
% L6 cuts
\begin{figure}[p]
\centering
\begin{subfigure}[b]{0.495\textwidth}
\centering
\includegraphics[width=\textwidth]{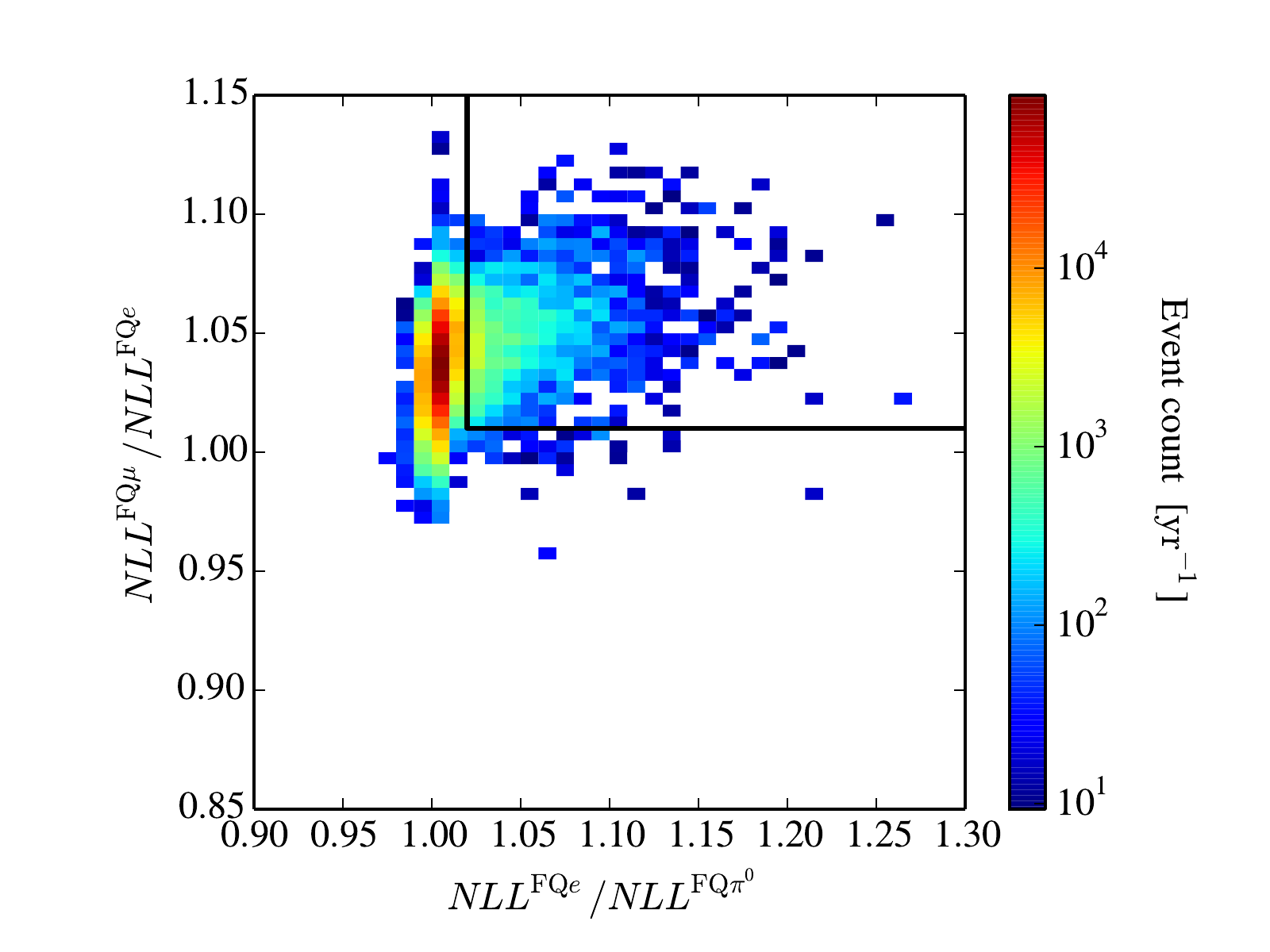}
\caption{}
\label{fig:detectors:nd_nuvtx_cutL6_nueepi}
\end{subfigure}
\hfill
\begin{subfigure}[b]{0.495\textwidth}  
\centering 
\includegraphics[width=\textwidth]{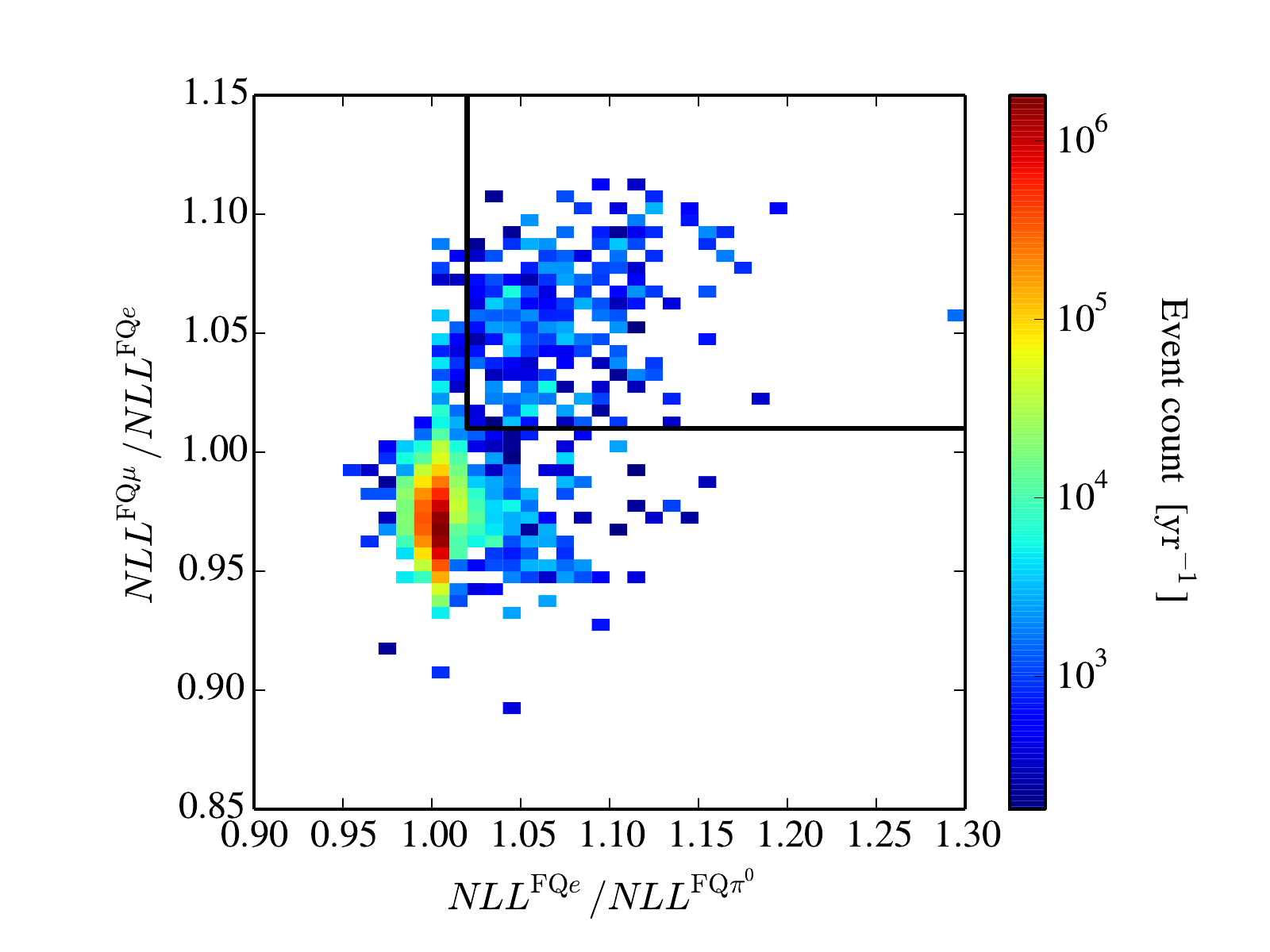}
\caption{}
\label{fig:detectors:nd_nuvtx_cutL6_numuepi}
\end{subfigure}
\hfill
\begin{subfigure}[b]{0.495\textwidth}  
\centering 
\includegraphics[width=\textwidth]{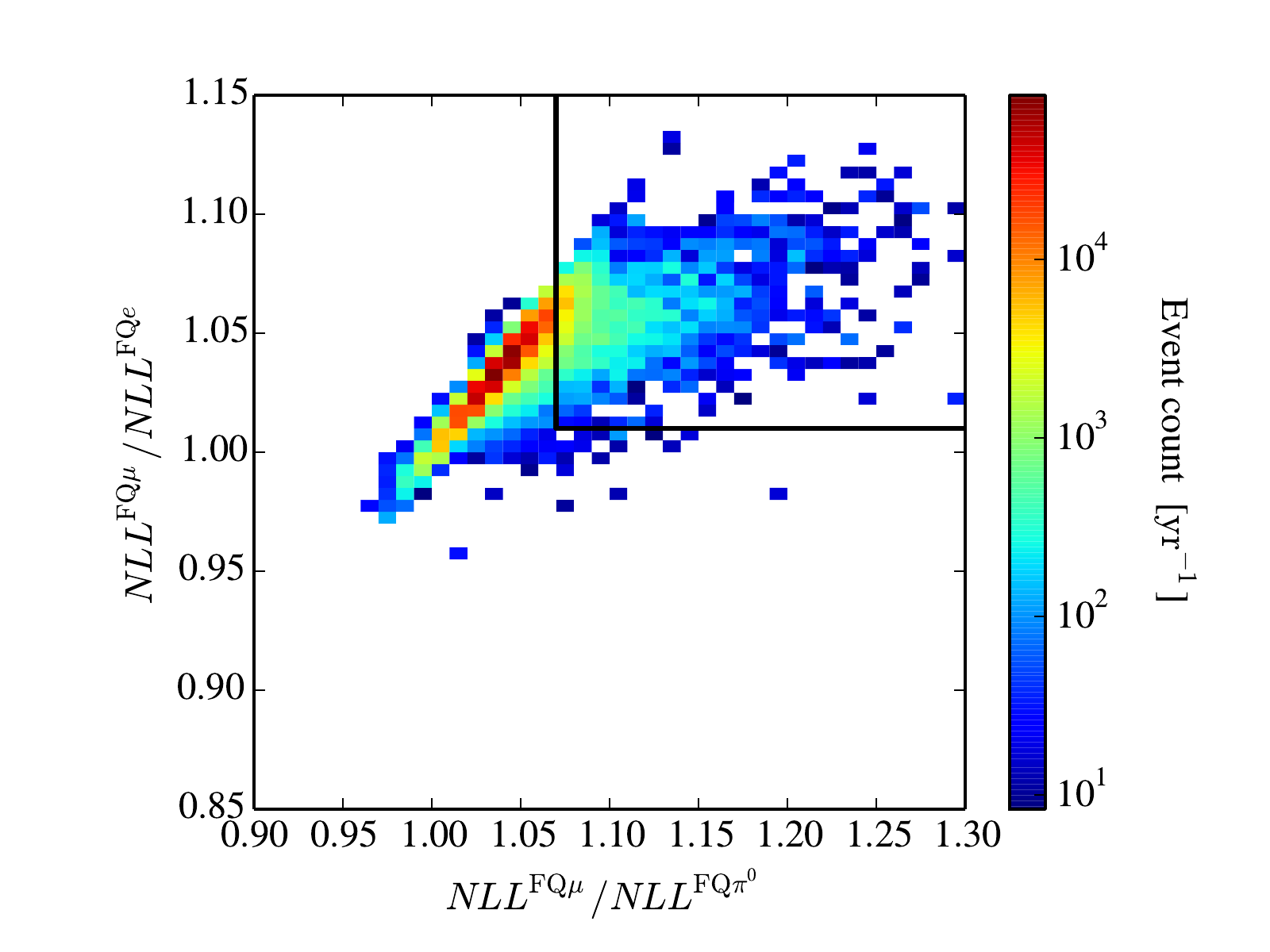}
\caption{}
\label{fig:detectors:nd_nuvtx_cutL6_nuemupi}
\end{subfigure}
\hfill
\begin{subfigure}[b]{0.495\textwidth}  
\centering 
\includegraphics[width=\textwidth]{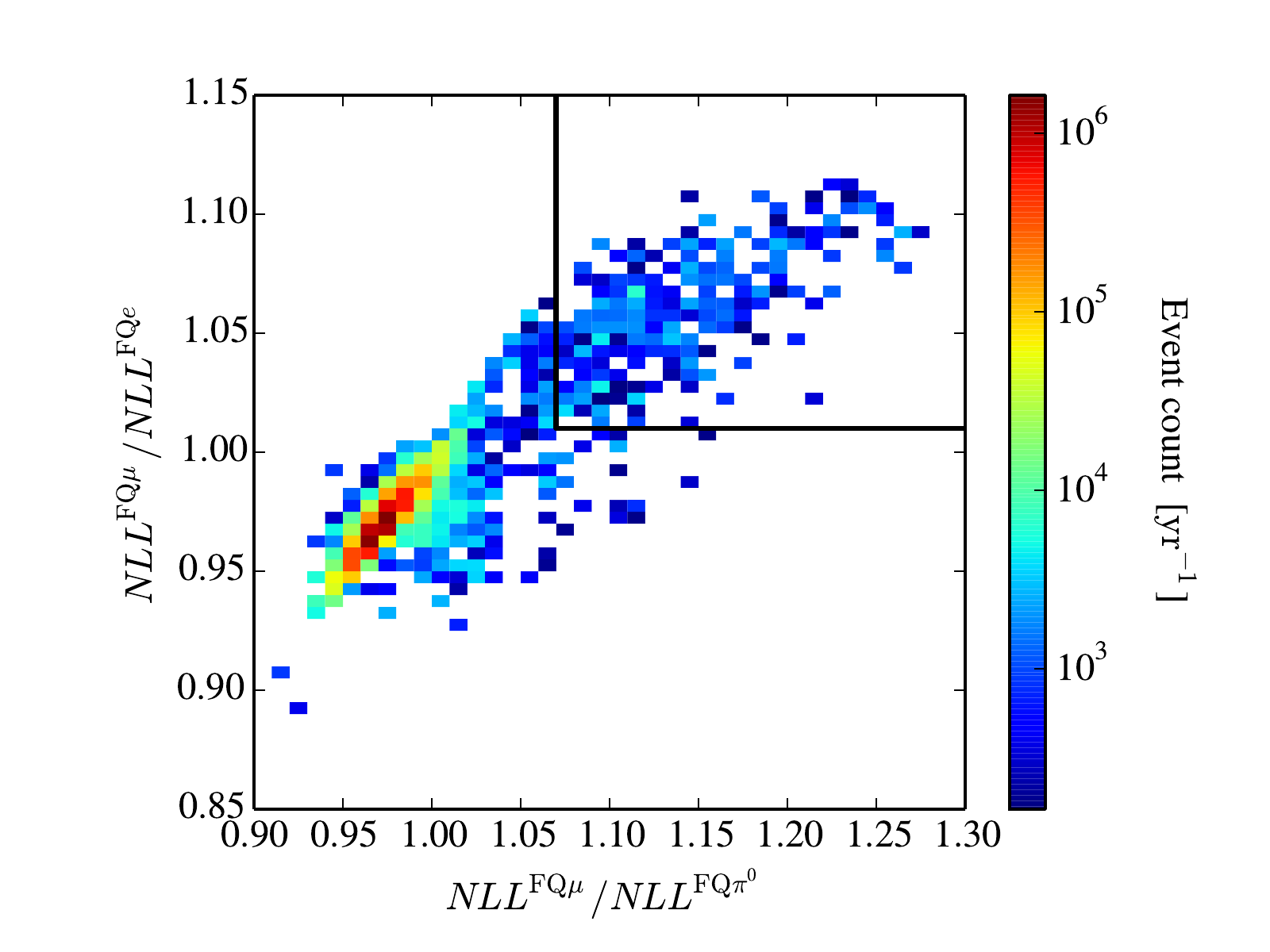}
\caption{}
\label{fig:detectors:nd_nuvtx_cutL6_numumupi}
\end{subfigure}
\hfill
\begin{subfigure}[b]{0.495\textwidth}  
\centering 
\includegraphics[width=\textwidth]{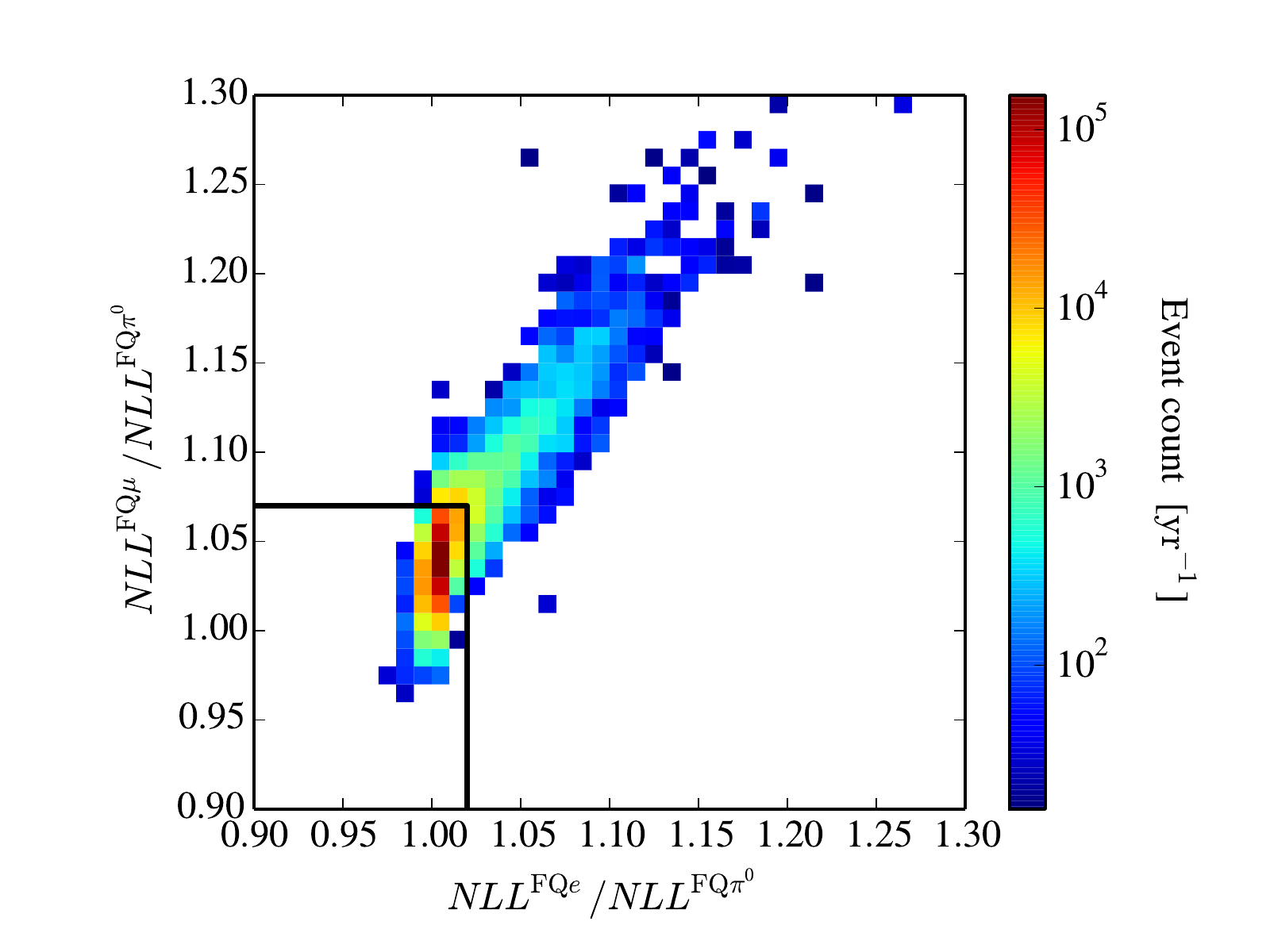}
\caption{}
\label{fig:detectors:nd_nuvtx_cutL6_nuemue}
\end{subfigure}
\hfill
\begin{subfigure}[b]{0.495\textwidth}  
\centering 
\includegraphics[width=\textwidth]{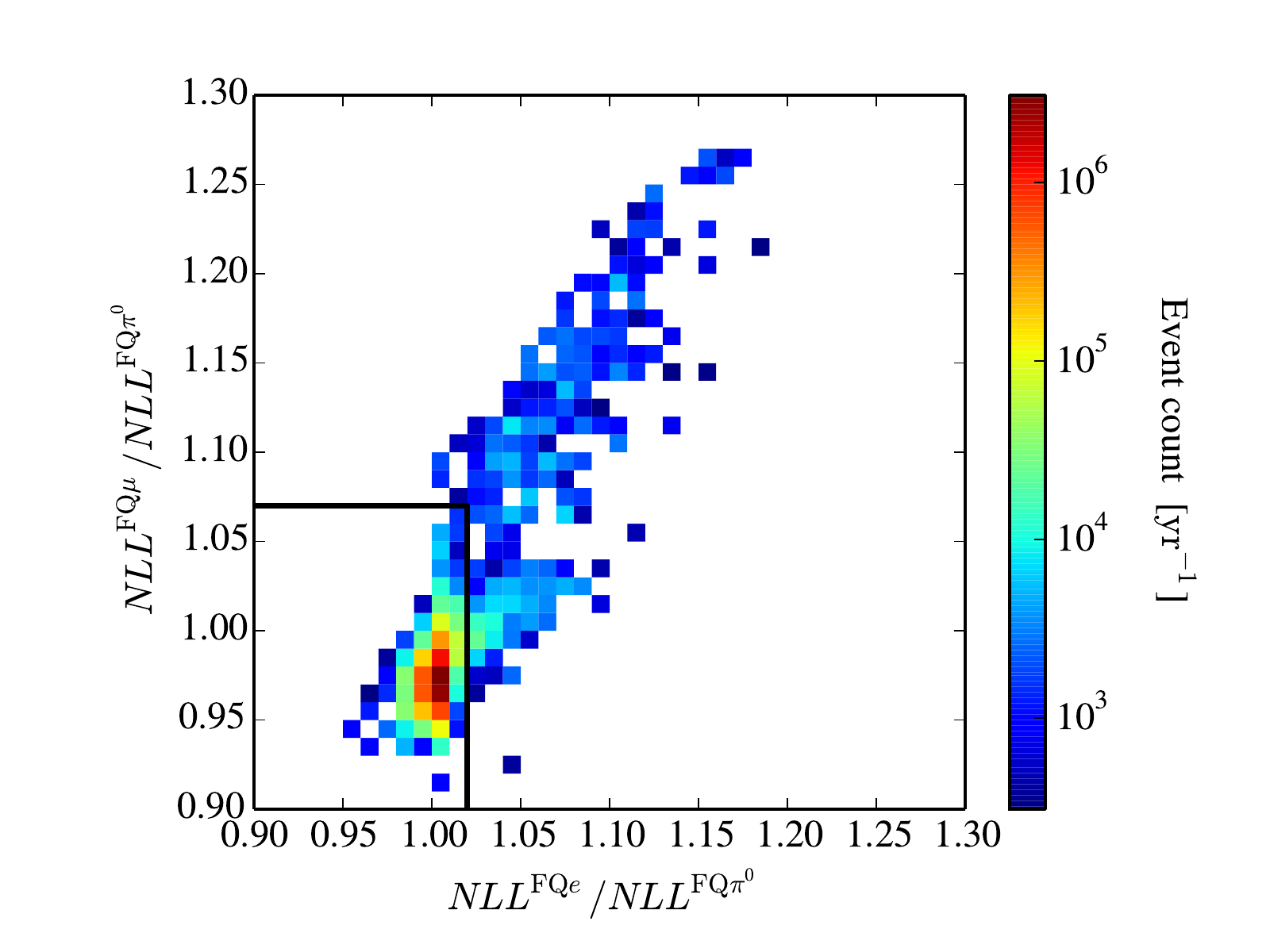}
\caption{}
\label{fig:detectors:nd_nuvtx_cutL6_numumue}
\end{subfigure}
%\begin{tikzpicture}
%\node[anchor=north,align=center] at ( 0.06\textwidth,-0.74\textwidth) {{\Large$\nu_e$}};
%\node[anchor=north,align=center] at ( 0.56\textwidth,-0.74\textwidth) {{\Large$\nu_\mu$}};
%\node[anchor=north,align=center] at ( 0.06\textwidth,-0.39\textwidth) {{\Large$\nu_e$}};
%\node[anchor=north,align=center] at ( 0.56\textwidth,-0.39\textwidth) {{\Large$\nu_\mu$}};
%\node[anchor=north,align=center] at ( 0.06\textwidth,-0.04\textwidth) {{\Large$\nu_e$}};
%\node[anchor=north,align=center] at ( 0.56\textwidth,-0.04\textwidth) {{\Large$\nu_\mu$}};
%\end{tikzpicture}
\caption{
Event distributions for electron
(\subref{fig:detectors:nd_nuvtx_cutL6_nueepi}, \subref{fig:detectors:nd_nuvtx_cutL6_nuemupi}, \subref{fig:detectors:nd_nuvtx_cutL6_nuemue}) and muon (\subref{fig:detectors:nd_nuvtx_cutL6_numuepi}, \subref{fig:detectors:nd_nuvtx_cutL6_numumupi}, \subref{fig:detectors:nd_nuvtx_cutL6_numumue}) events over the
{\nllratio} and {\nllepizeroratio} variables (\subref{fig:detectors:nd_nuvtx_cutL6_nueepi}, \subref{fig:detectors:nd_nuvtx_cutL6_numuepi}), along with the {\nllratio} and {\nllmupizeroratio} variables (\subref{fig:detectors:nd_nuvtx_cutL6_nuemupi}, \subref{fig:detectors:nd_nuvtx_cutL6_numumupi}), and the {\nllmupizeroratio} and {\nllepizeroratio} variables (\subref{fig:detectors:nd_nuvtx_cutL6_nuemue}, \subref{fig:detectors:nd_nuvtx_cutL6_numumue}).
The \emph{pion-like criteria}, described in Eq.~(\ref{eqn:detectors:nd_nuvtx_cutsL6}), are displayed as black lines in these figures. All events falling above and to the right of these lines are rejected.
\label{fig:detectors:nd_nuvtx_cutL6}}
\end{figure}

In order to achieve this, a sample of \SI{e5}{} neutrino interaction events were simulated for each combination of neutrino flavor ($\nu_e$, $\bar\nu_e$, $\nu_\mu$, $\bar\nu_\mu$) and interaction type (charged current (CC), neutral current (NC)) using the \textsc{GENIE} neutrino interaction simulation software. The events were simulated with a uniform distribution over interacting neutrino energy up to $\SI{1.6}{\GeV}$ with pure water, $(^1\text{H})_2(^{16}\text{O})$, as target material.
The resulting final-state particles were inserted into the \textsc{WCSim} simulation of the water Cherenkov detector, distributed homogeneously over the detector tank, and with the initial neutrino direction set along the neutrino beam axis. The resulting detector response was then reconstructed using the \textsc{fiTQun} software, and the selection criteria developed for charged leptons was applied to the sample.

This produced a sample of neutrino interaction events, which was weighted to represent the expected neutrino beam, and then used to develop additional data-selection criteria. These are used for rejecting events with poor particle-identification performance and poor performance in reconstructing the initial neutrino energy. These additional criteria are collectively denoted as \emph{the neutrino selection criteria}.
The initial neutrino energy, {\enu}, was calculated assuming that it interacted through quasi-elastic scattering (QES), given by the following relation~\cite{MiniBooNE:2010bsu}:
\begin{align}
E_\nu = \frac{\tit{m}_\tit{F}^2-\tit{m}_\tit{IB}^2-\tit{m}_\tit{l}^2+2\tit{m}_\tit{IB}\tit{E}_\tit{l}}
             {2\left(\tit{m}_\tit{IB}-\tit{E}_\tit{l}+\tit{p}_\tit{l}\cos\theta_\tit{l}\right)}
\label{eqn:detectors:enu_qes}
\end{align}

\noindent Here $\tit{m}_\tit{F}$ represents the final state mass of the nucleon,
$\tit{m}_\tit{IB}=\tit{m}_\tit{I}-\tit{E}_\tit{B}$ represents the bound state energy
of the target nucleon, with $\tit{m}_\tit{I}$ as the initial-state free nucleon mass and $\tit{E}_\tit{B}$ as the binding energy.
Here, all nucleon masses are assumed as the proton mass, and the effective binding energy is taken for
a $^{16}\text{O}$ nucleus to be \SI{27}{\MeV}.
The subscript $\tit{l}$ represents the final-state charged lepton,
and thus $\tit{m}_\tit{l}$, $\tit{E}_\tit{l}$, $\tit{p}_\tit{l}$ and $\theta_\tit{l}$
respectively represent its mass, energy, momentum, and angle relative the neutrino beam axis.

\emph{The pion-like criteria} ---
The purpose of these criteria is to reject events that are likely to be caused by a neutral pion,
and utilises the {\nllnd} of the neutral pion hypothesis, {\nllpizero}, relative to other particle hypotheses ({\nlle}, {\nllmu}). Events including a $\pi^0$ often appear more electron-like than muon-like, so this criterion is only applied to events identified as electron-like, using Eq.~(\ref{eqn:detectors:nd_pid}).
\begin{align}
\nllratiomath &\geq 1.01 ~~ \text{({\eid} criterion)} \nonumber \\
\nllmupizeroratiomath &\leq 1.07 \nonumber \\
\nllepizeroratiomath  &\leq 1.02
\label{eqn:detectors:nd_nuvtx_cutsL6}
\end{align}

Event distributions over these variables are shown in Fig.~\ref{fig:detectors:nd_nuvtx_cutL6}
along with illustrations of the selection criteria.

\emph{The multi--sub-event criterion} ---
This criterion is used to reject events where the initial energy reconstruction is too low
due to a significant amount of energy being lost to non-primary final-state particles.
As the additional final-state particles must be noticeably energetic, they are assumed to be registered in the detector using a multi-particle final-state hypothesis.
This is quantified in the {\nse} variable, representing the number of reconstructed sub-events.
An electron-like event is accepted if it has no more than one registered sub-event,
and a muon-like event is accepted if the number of sub-events does not exceed two.
Muon events are expected to produce up to two sub-events: one containing the primary muon,
and one containing its decay product -- the Michel electron.
Event distributions over the {\nse} variable and the energy reconstruction performance can be found in Fig.~\ref{fig:detectors:nd_nuvtx_cutL7}.

% FIGURE
% L7 cuts
\begin{figure}[!thb]
\centering
\begin{subfigure}[b]{0.495\textwidth}
\centering
\includegraphics[width=\textwidth]{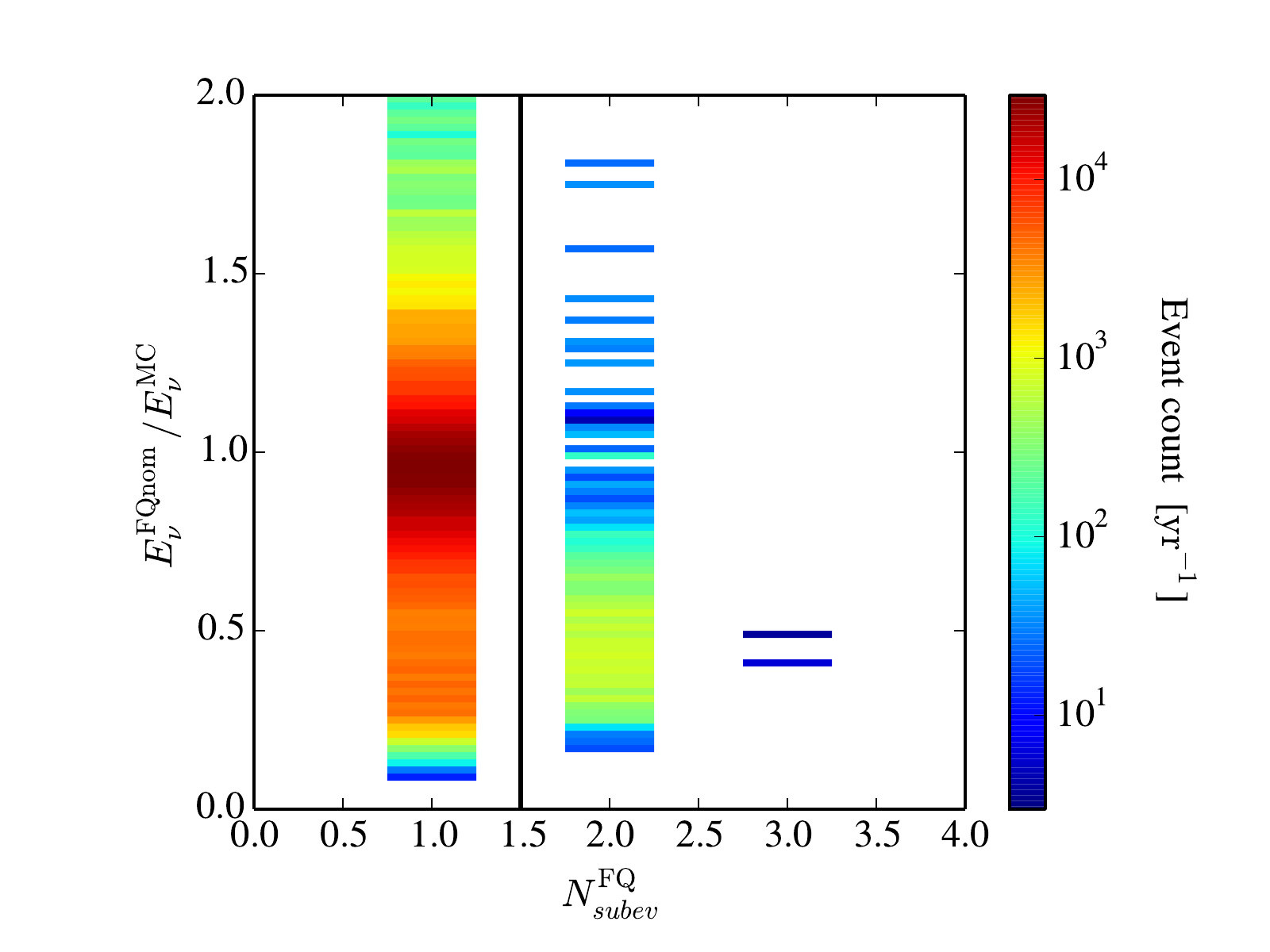}
\caption{Electron neutrino event distributions.}
\label{fig:detectors:nd_nuvtx_cutL7_nue}
\end{subfigure}
\hfill
\begin{subfigure}[b]{0.495\textwidth}  
\centering 
\includegraphics[width=\textwidth]{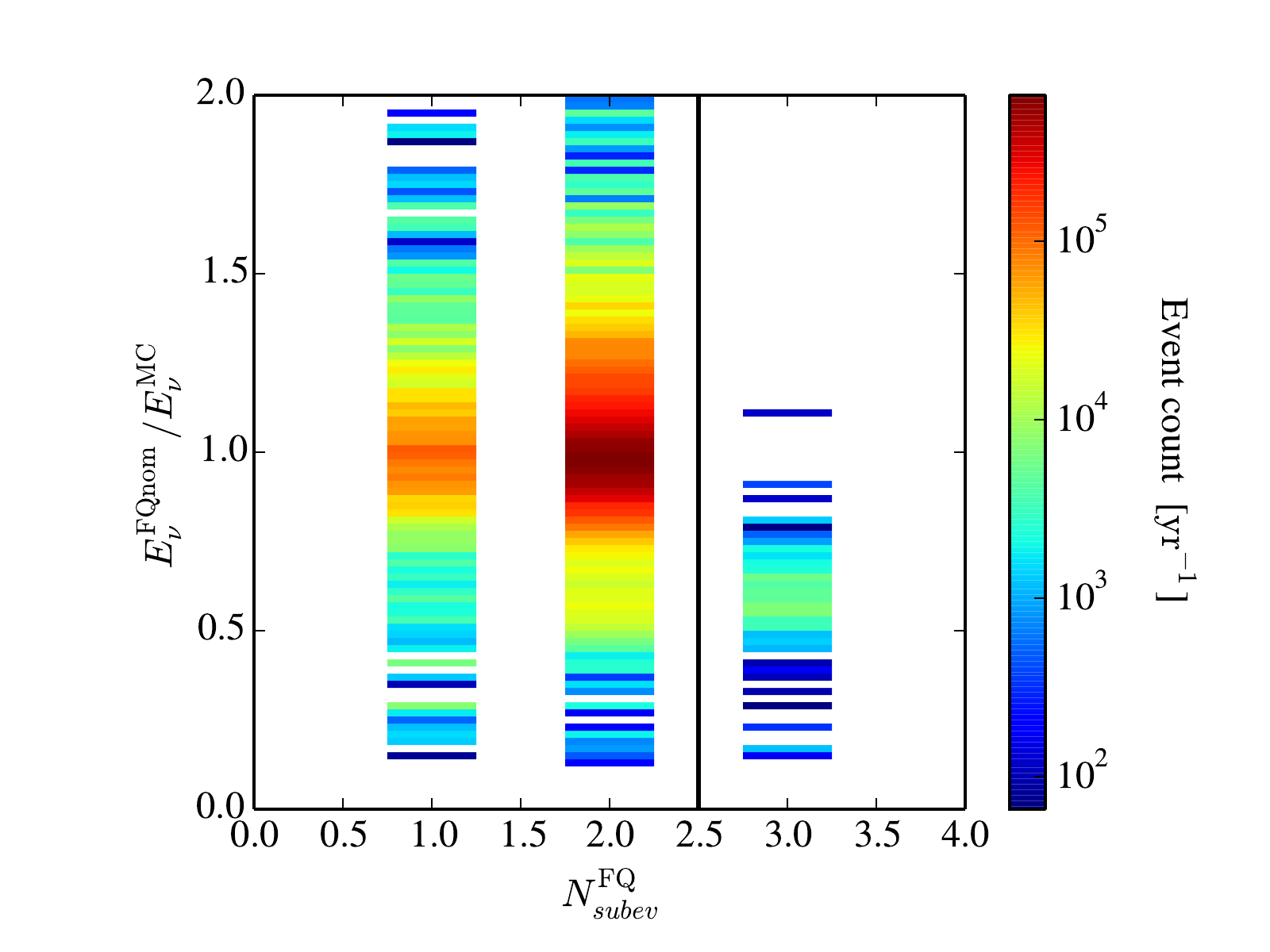}
\caption{Muon neutrino event distributions.}
\label{fig:detectors:nd_nuvtx_cutL7_numu}
\end{subfigure}
\caption{Event distributions for
electron and muon neutrinos over the
{\nse} and {\enunommcratio} variables.
The \emph{multi--sub-event criterion} is displayed as a black line.
All events to the right of this line are rejected.
\label{fig:detectors:nd_nuvtx_cutL7}}
\end{figure}

The selection efficiencies for electron-neutrino ($\nu_e+\bar\nu_e$) and muon-neutrino ($\nu_\mu+\bar\nu_\mu$)
interactions through charged-current with initial energy $<\SI{1.6}{\GeV}$ are \SI{54}{\percent} and \SI{47}{\percent}, respectively.
The selection efficiencies for each neutrino flavor and both interaction types are shown in Fig.~\ref{fig:detectors:nd_nuvtx_eff} as a function of the initial neutrino energy.

The present data selection shows an efficient rejection of NC events, while maintaining a high selection efficiency for CC events of all flavors. The CC selection efficiency after the final selection stage varies between \SI{40}{\percent} and \SI{60}{\percent} for neutrinos of all flavors.
The neutrino selection criteria (the final two criteria) have a higher rejection for high-energy CC events than for low-energy events, as can be seen when comparing Figs.~\ref{fig:detectors:nd_nuvtx_eff_L5cc} and \ref{fig:detectors:nd_nuvtx_eff_L7cc}.
This is attributed to the fact that both the pion production and sub-event multiplicity increase with energy, and these are the event properties which are used in the neutrino selection criteria.

\subsubsubsection{Neutrino Energy Migration Matrices}

The migration matrices for the ESS$\nu$SB near detector are shown in Fig.~\ref{fig:detectors:nd_nuvtx_migmat}.
The migration matrices show the transformation from true energy, on the $x$ axis, to reconstructed energy, on the $y$ axis, for a neutrino interaction with a given true energy.
The matrices are normalised such that, when projected to the $x$ axis, they will show the detection efficiency for a neutrino interaction within the detector volume as a function of the true neutrino energy. This allows for the multiplication of a spectrum containing the distribution of true neutrino energies for a neutrino source with the migration matrix in order to obtain the expected reconstructed spectrum. Conversely, the migration matrix can be inverted, for example with Bayesian methods, and thus used to deconvolute a spectrum of reconstructed neutrino energies in order to obtain an estimation of the true spectrum.

% FIGURE
% NUVTX efficiencies for L1, L5, L7
\begin{figure}[p]
\centering
\begin{subfigure}[b]{0.495\textwidth}
\centering
\includegraphics[width=\textwidth]{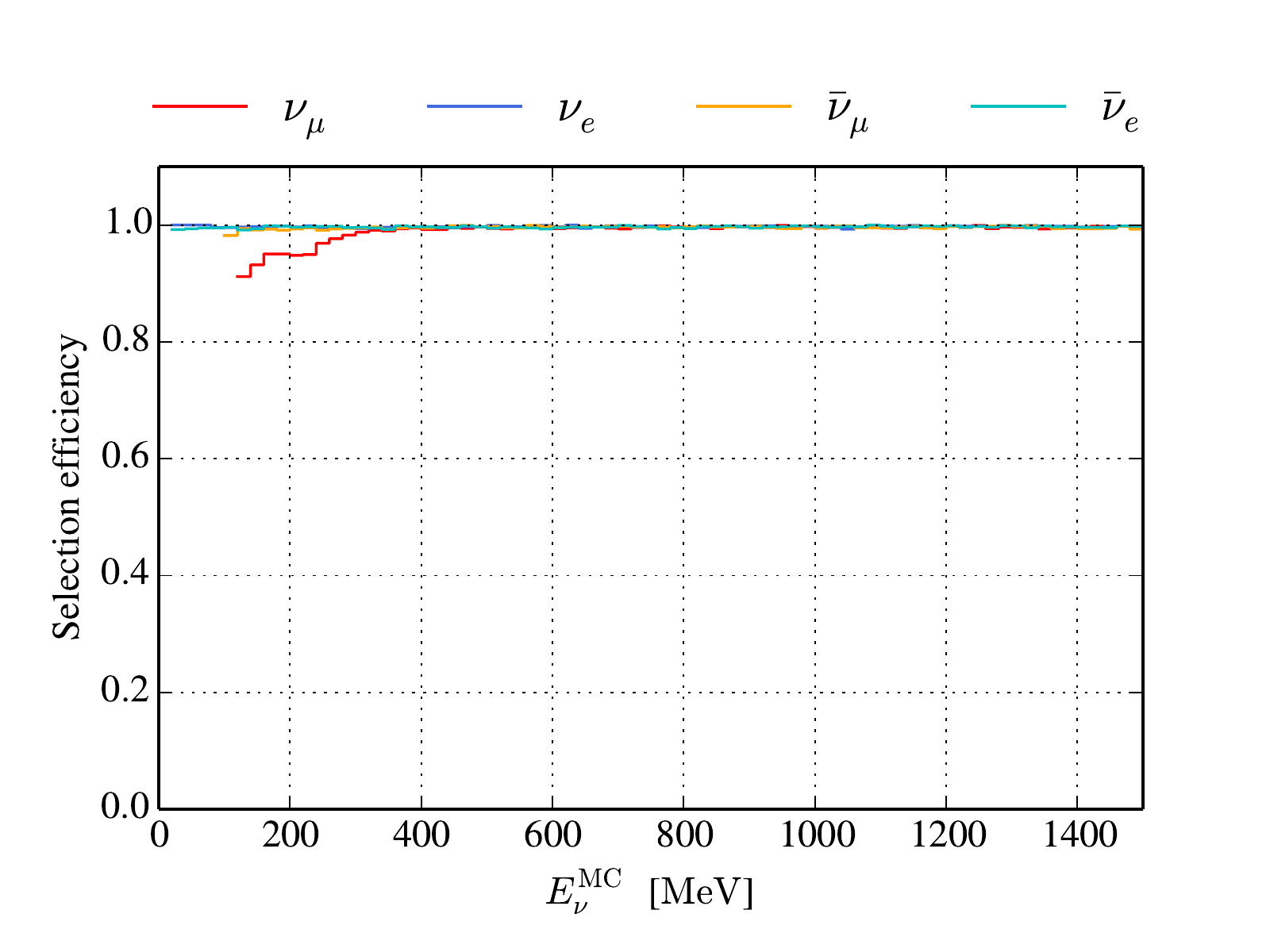}
\caption{}
\label{fig:detectors:nd_nuvtx_eff_L1cc}
\end{subfigure}
\hfill
\begin{subfigure}[b]{0.495\textwidth}  
\centering 
\includegraphics[width=\textwidth]{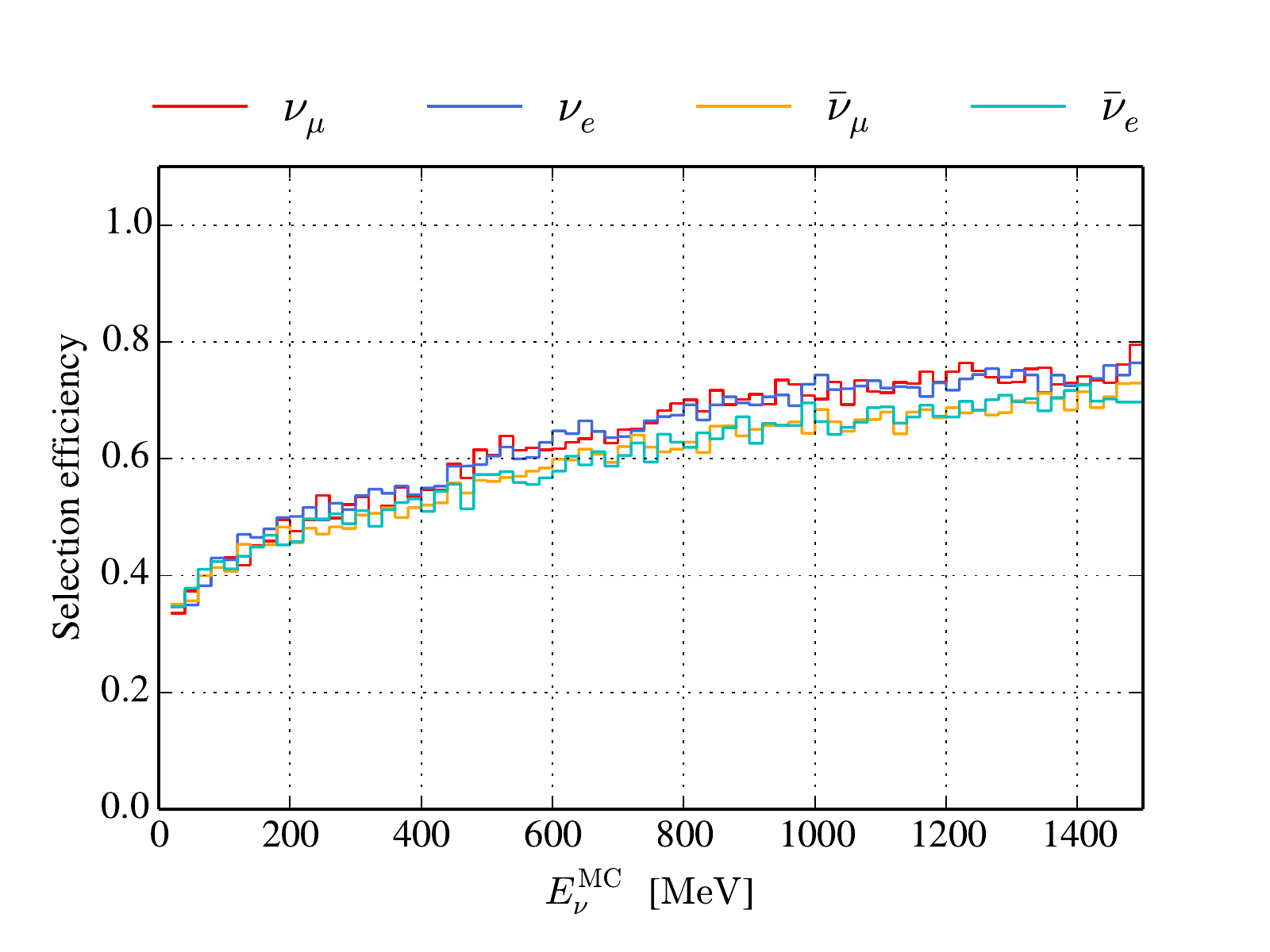}
\caption{}
\label{fig:detectors:nd_nuvtx_eff_L1nc}
\end{subfigure}
\hfill
\begin{subfigure}[b]{0.495\textwidth}  
\centering 
\includegraphics[width=\textwidth]{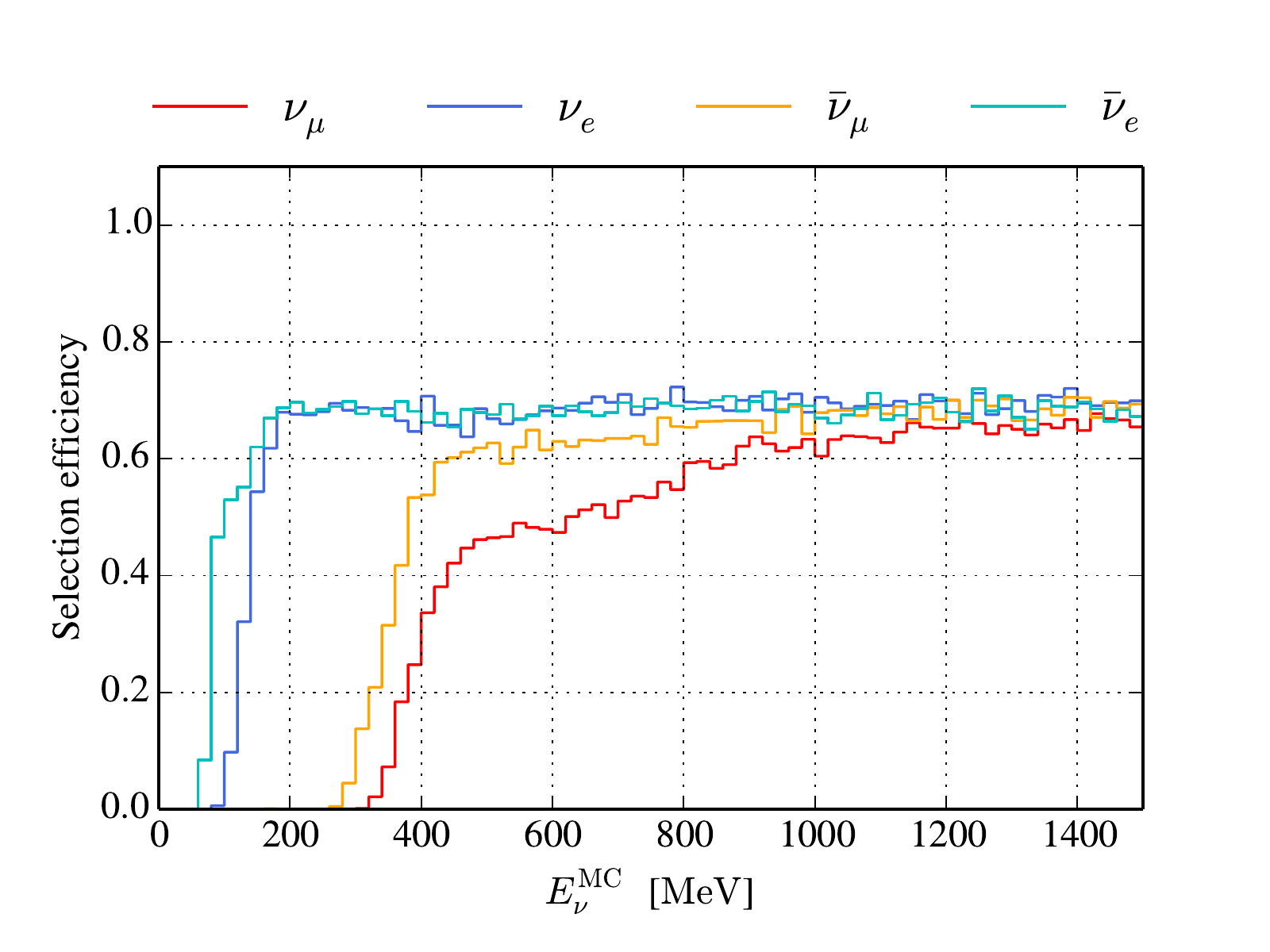}
\caption{}
\label{fig:detectors:nd_nuvtx_eff_L5cc}
\end{subfigure}
\hfill
\begin{subfigure}[b]{0.495\textwidth}  
\centering 
\includegraphics[width=\textwidth]{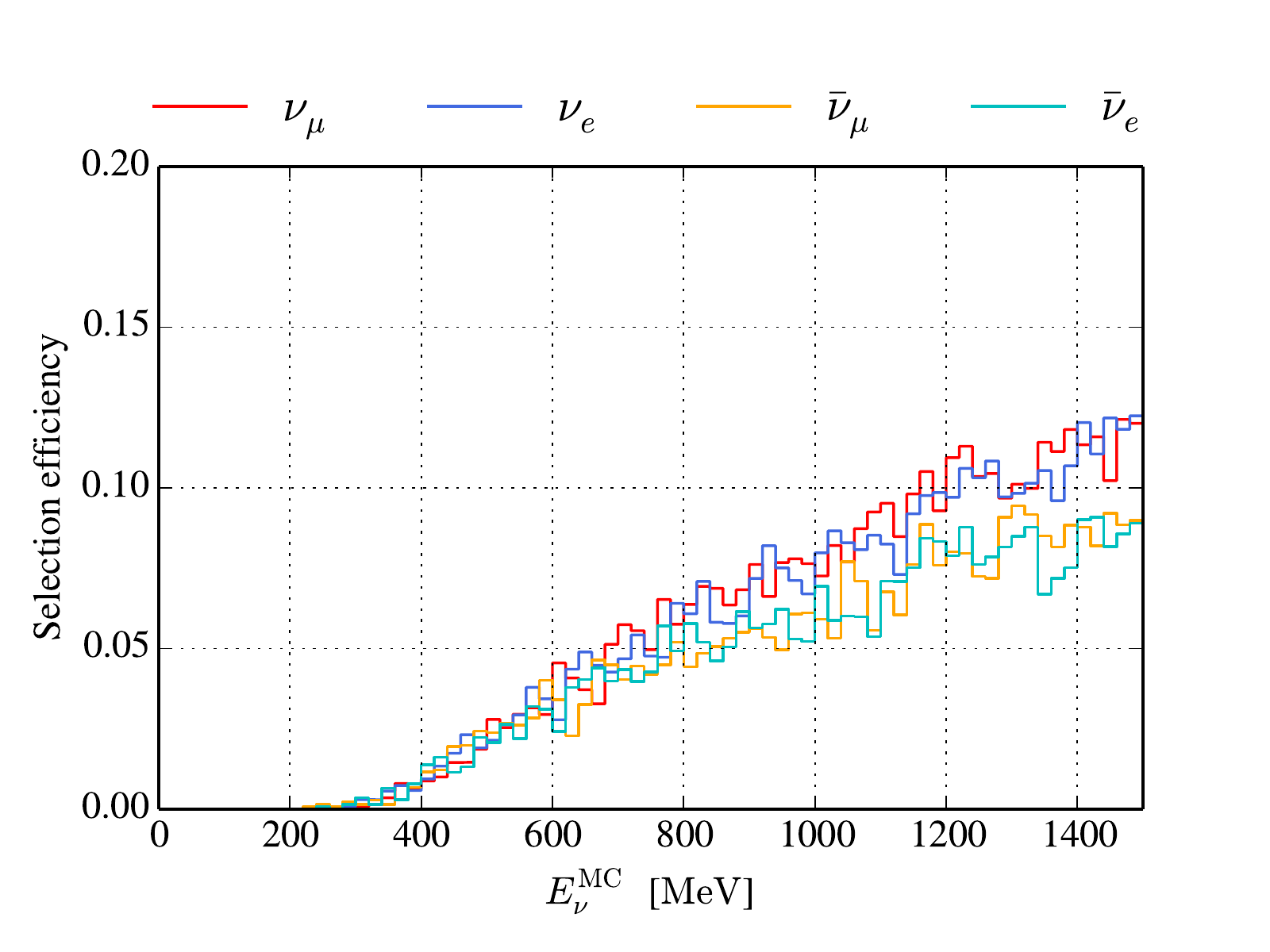}
\caption{}
\label{fig:detectors:nd_nuvtx_eff_L5nc}
\end{subfigure}
\hfill
\begin{subfigure}[b]{0.495\textwidth}  
\centering 
\includegraphics[width=\textwidth]{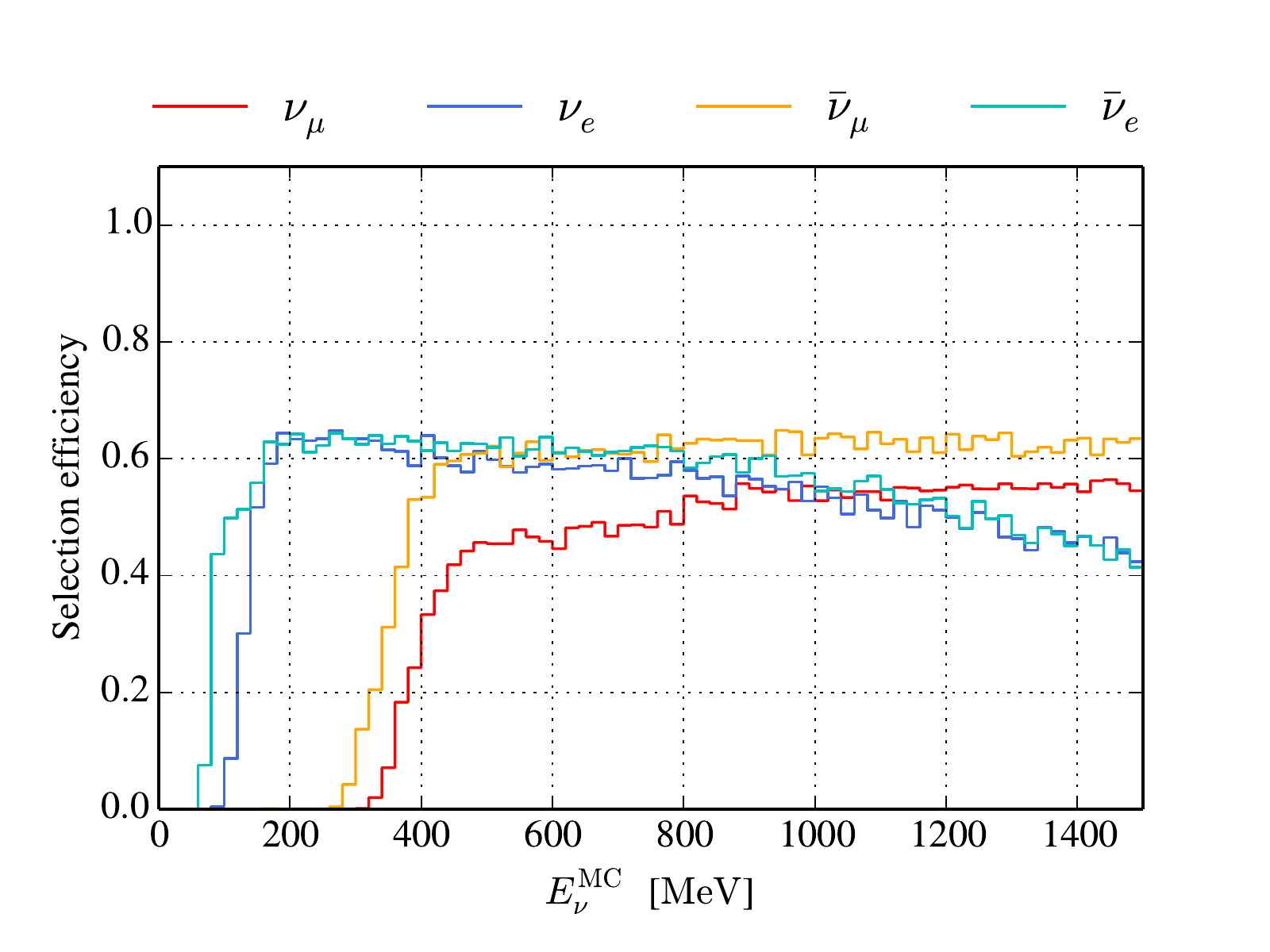}
\caption{}
\label{fig:detectors:nd_nuvtx_eff_L7cc}
\end{subfigure}
\hfill
\begin{subfigure}[b]{0.495\textwidth}  
\centering 
\includegraphics[width=\textwidth]{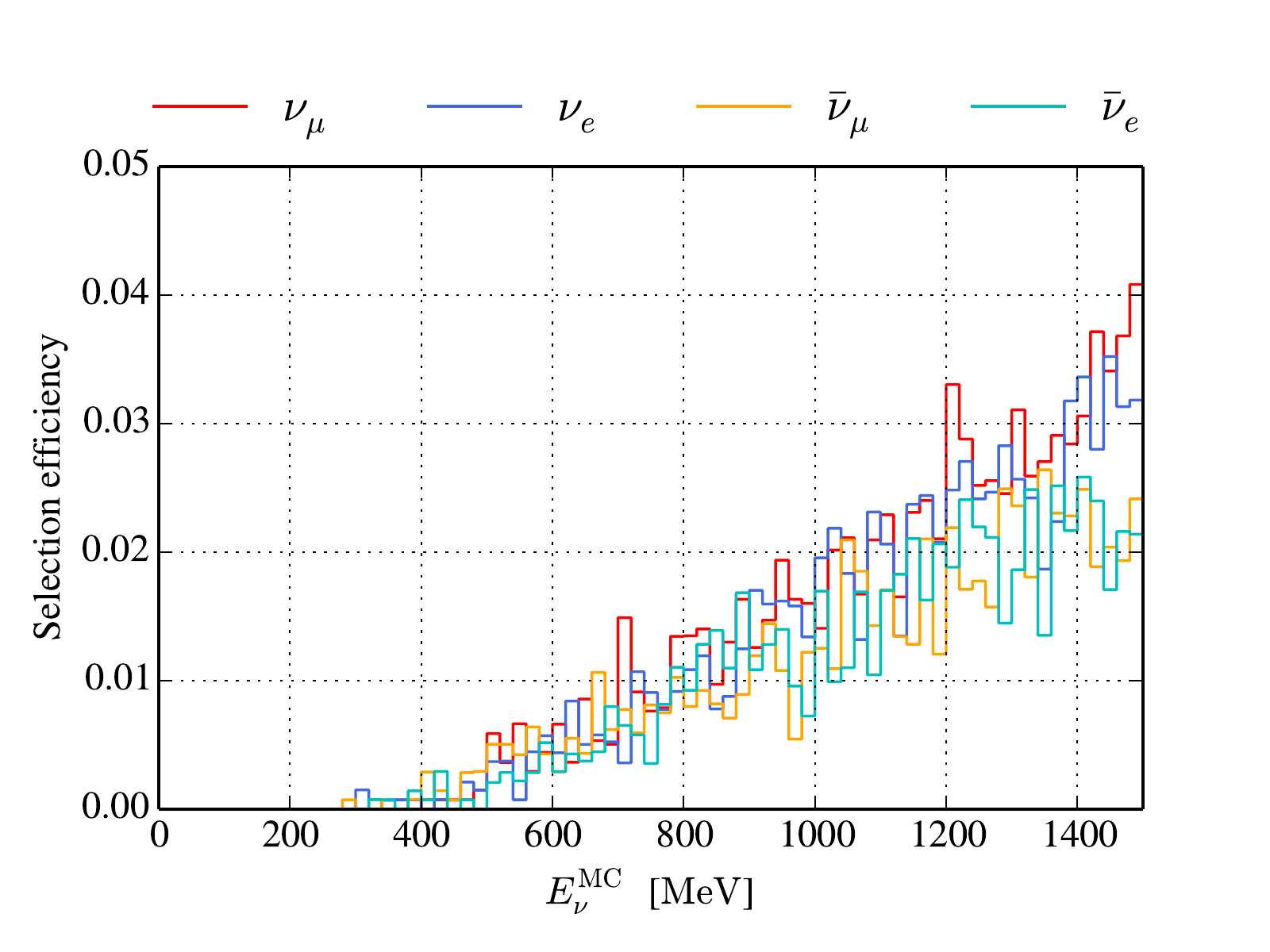}
\caption{}
\label{fig:detectors:nd_nuvtx_eff_L7nc}
\end{subfigure}
\caption{
The selection efficiency of the selection criteria for charged current events (\subref{fig:detectors:nd_nuvtx_eff_L1cc}, \subref{fig:detectors:nd_nuvtx_eff_L5cc}, \subref{fig:detectors:nd_nuvtx_eff_L7cc}) and neutral current events (\subref{fig:detectors:nd_nuvtx_eff_L1nc}, \subref{fig:detectors:nd_nuvtx_eff_L5nc}, \subref{fig:detectors:nd_nuvtx_eff_L7nc}) over the {\enumc} variable.
The events are shown after the \emph{trigger}
(\subref{fig:detectors:nd_nuvtx_eff_L1cc}, \subref{fig:detectors:nd_nuvtx_eff_L1nc}), \emph{charged lepton criteria}
(\subref{fig:detectors:nd_nuvtx_eff_L5cc}, \subref{fig:detectors:nd_nuvtx_eff_L5nc}),  and \emph{neutrino criteria} (\subref{fig:detectors:nd_nuvtx_eff_L7cc}, \subref{fig:detectors:nd_nuvtx_eff_L7nc})  have been applied.
\label{fig:detectors:nd_nuvtx_eff}}
\end{figure}

The reconstructed neutrino energy follows the true neutrino energy well for CC events, as can be seen in Fig.~\ref{fig:detectors:nd_nuvtx_migmat} where the majority of the events fall close to the diagonal line. A small population of events are also reconstructed with an energy which is lower than the true energy by a constant amount, shown in this figure as a band of events parallel with the diagonal band. These events have mainly been produced through resonant-scattering in the nucleus via the production of a heavier hadron (e.g.\ $\Delta^{++}$), as opposed to quasi-elastic scattering.

% FIGURE
\begin{figure}[p]
\centering
\begin{subfigure}[b]{0.495\textwidth}
\centering
\includegraphics[width=\textwidth]{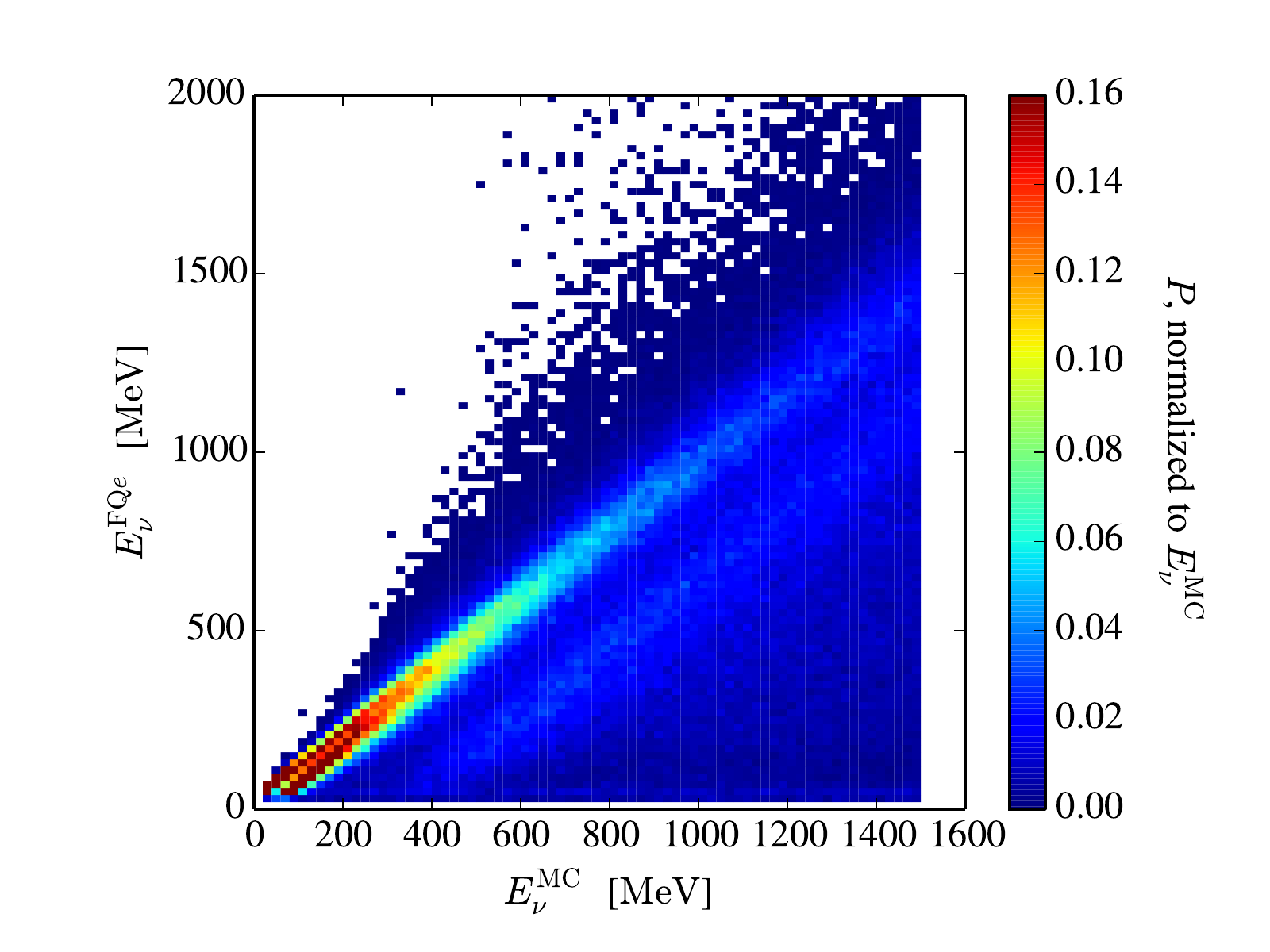}
\caption{}
\label{fig:detectors:nd_nuvtx_migmat_L1e}
\end{subfigure}
\hfill
\begin{subfigure}[b]{0.495\textwidth}  
\centering 
\includegraphics[width=\textwidth]{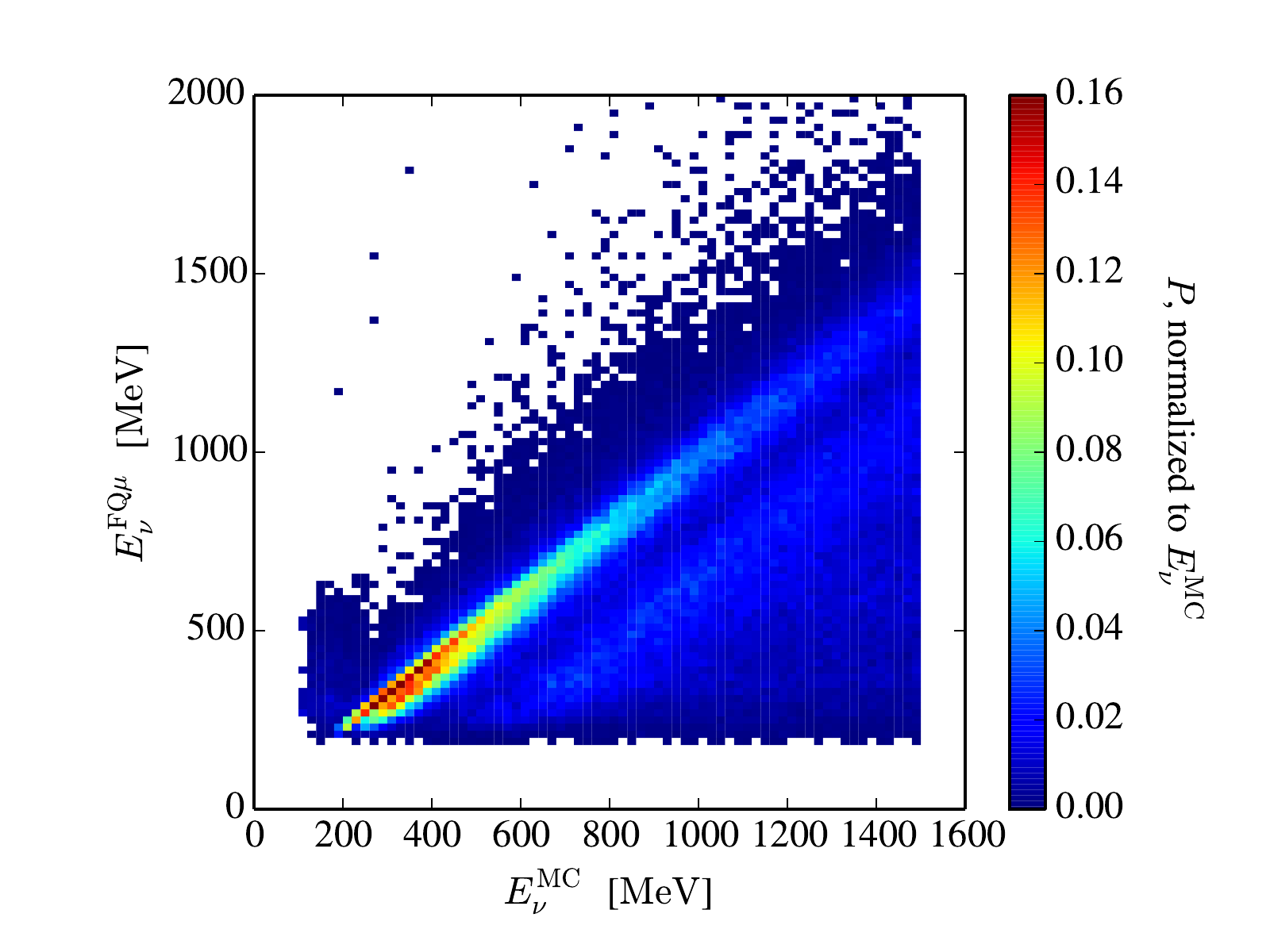}
\caption{}
\label{fig:detectors:nd_nuvtx_migmat_L1mu}
\end{subfigure}
\hfill
\begin{subfigure}[b]{0.495\textwidth}  
\centering 
\includegraphics[width=\textwidth]{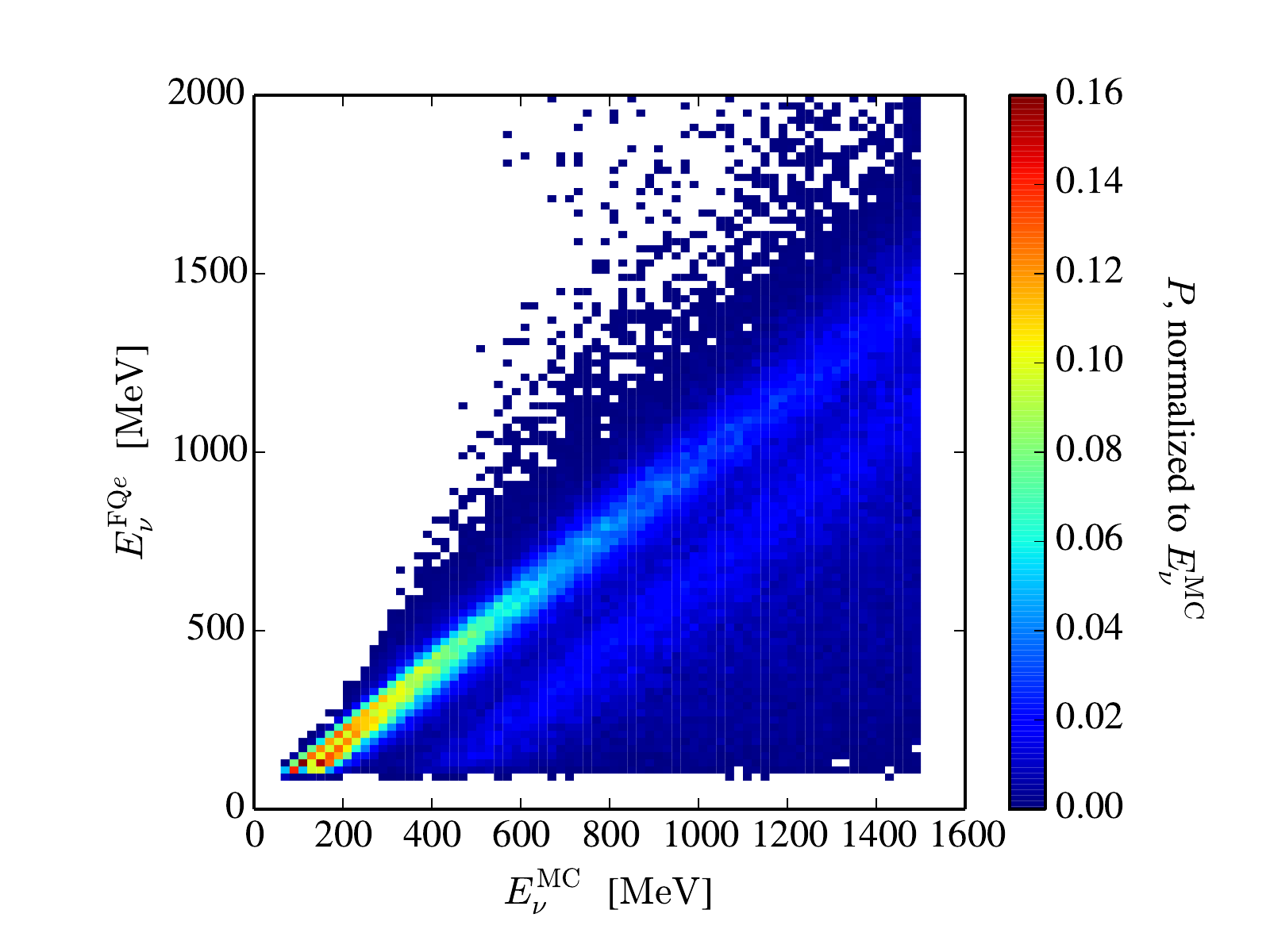}
\caption{}
\label{fig:detectors:nd_nuvtx_migmat_L5e}
\end{subfigure}
\hfill
\begin{subfigure}[b]{0.495\textwidth}  
\centering 
\includegraphics[width=\textwidth]{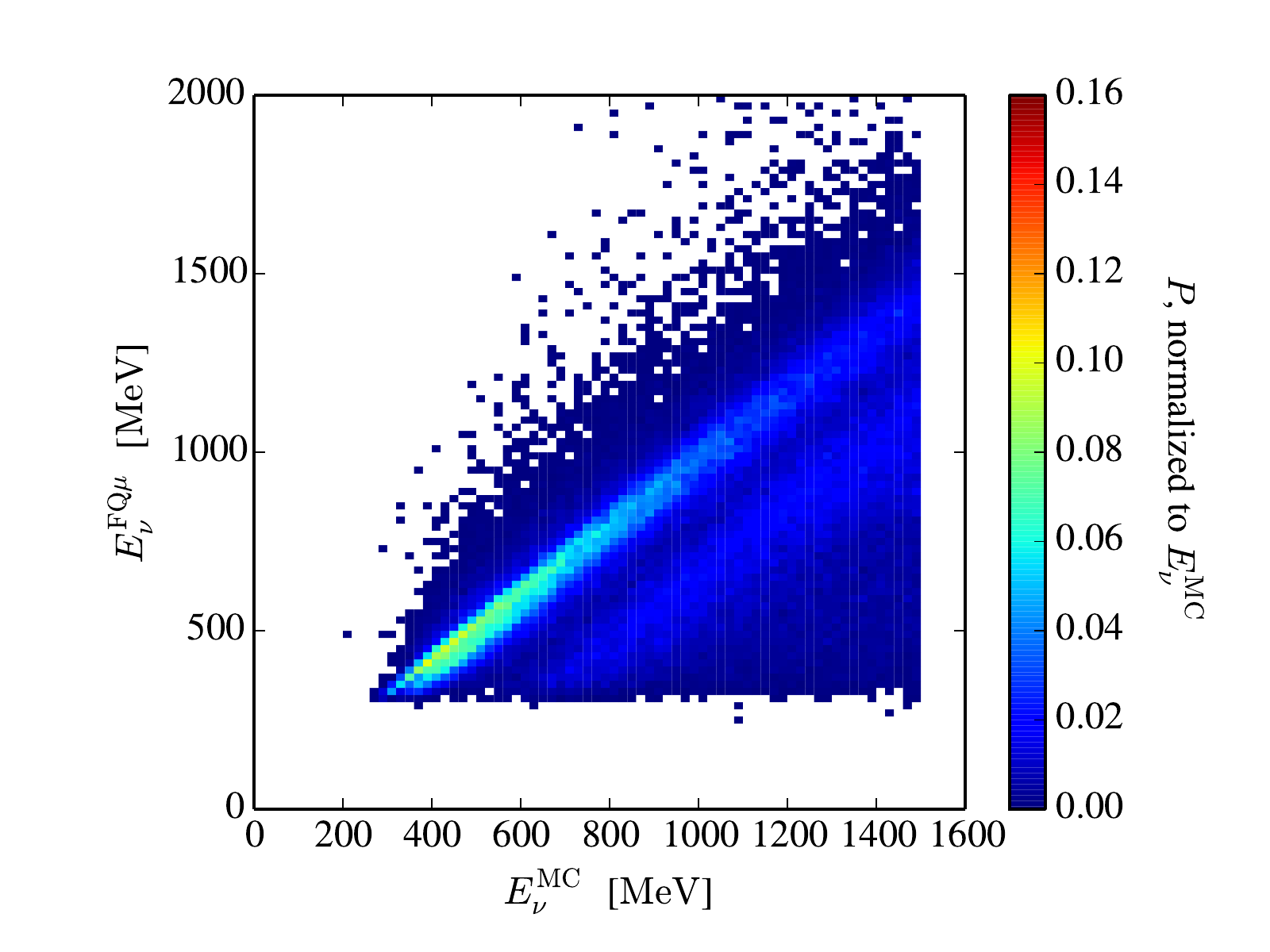}
\caption{}
\label{fig:detectors:nd_nuvtx_migmat_L5mu}
\end{subfigure}
\hfill
\begin{subfigure}[b]{0.495\textwidth}  
\centering 
\includegraphics[width=\textwidth]{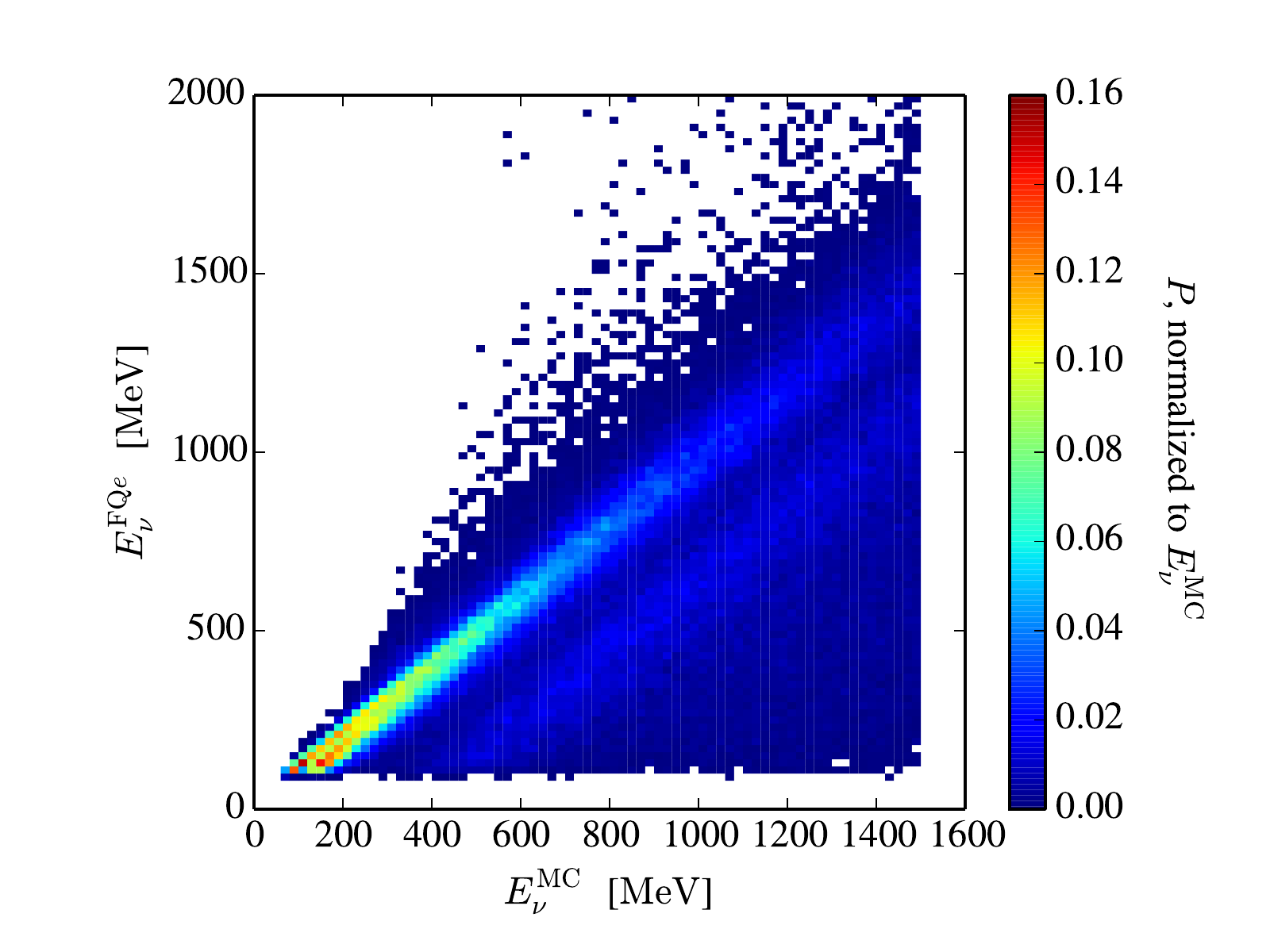}
\caption{}
\label{fig:detectors:nd_nuvtx_migmat_L7e}
\end{subfigure}
\hfill
\begin{subfigure}[b]{0.495\textwidth}  
\centering 
\includegraphics[width=\textwidth]{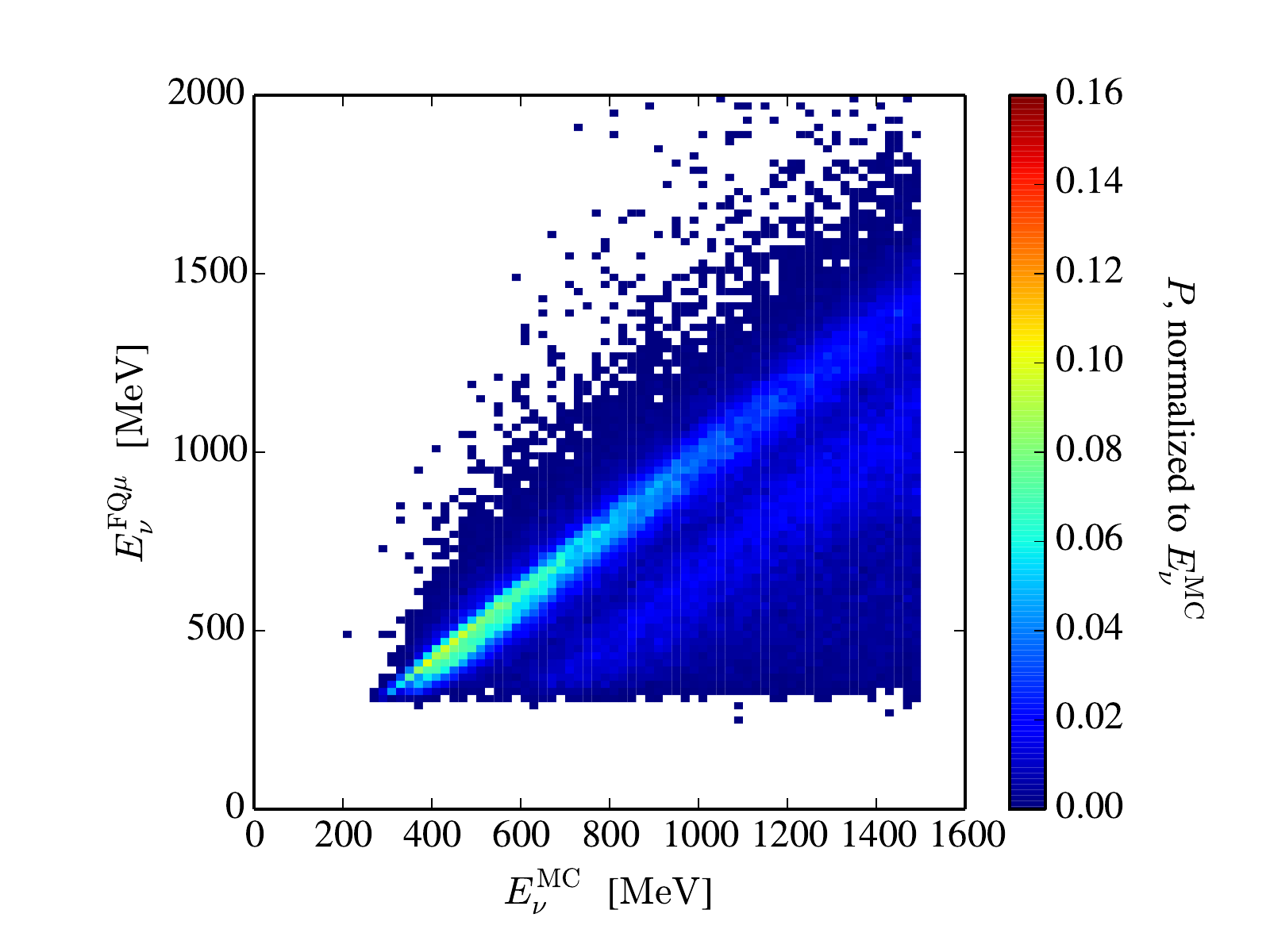}
\caption{}
\label{fig:detectors:nd_nuvtx_migmat_L7mu}
\end{subfigure}
\caption{Migration matrices between the true and reconstructed neutrino energies. The matrices are shown for $\nu_e$~{\eid}
(\subref{fig:detectors:nd_nuvtx_migmat_L1e}, \subref{fig:detectors:nd_nuvtx_migmat_L5e}, \subref{fig:detectors:nd_nuvtx_migmat_L7e})  and  $\nu_\mu$~{\muid} events
(\subref{fig:detectors:nd_nuvtx_migmat_L1mu}, \subref{fig:detectors:nd_nuvtx_migmat_L5mu}, \subref{fig:detectors:nd_nuvtx_migmat_L7mu}), and after the \emph{trigger}
(\subref{fig:detectors:nd_nuvtx_migmat_L1e}, \subref{fig:detectors:nd_nuvtx_migmat_L1mu}), \emph{charged lepton criteria}
(\subref{fig:detectors:nd_nuvtx_migmat_L5e}, \subref{fig:detectors:nd_nuvtx_migmat_L5mu}),  and \emph{neutrino criteria}
(\subref{fig:detectors:nd_nuvtx_migmat_L7e}, \subref{fig:detectors:nd_nuvtx_migmat_L7mu}) have been applied.
\label{fig:detectors:nd_nuvtx_migmat}}
\end{figure}

\subsubsubsection{Expected Number of Neutrino Events}
\label{sec:detectors:ndwc_exp_nev}

The expected number of neutrino interaction events in the near-detector water Cherenkov instrumented volume are described in Section~\ref{sec:detectors:NDbeamcomposition} per flavour, interaction type, and horn polarity.

The number of events that are expected per running year (\SI{200}{\d}, corresponding to \SI{2.16e23}{} protons-on-target) at each level of the data selection are found in Table~\ref{tbl:detectors:ndwc_nev}. The trigger efficiency is close to unity over this spectrum, as is reflected in Figs.~\ref{fig:detectors:nd_nuvtx_eff_L1cc} and \ref{fig:detectors:nd_nuvtx_eff_L1nc} (using a flat spectrum).
The corresponding event distributions are shown in Fig.~\ref{fig:detectors:ndwc_nev} as functions of {\enumc} for events that pass the \emph{trigger}, \emph{charged lepton criteria}, and \emph{neutrino criteria} levels of the data selection.

Tables~\ref{tbl:detectors:ndwc_nev_eid} and \ref{tbl:detectors:ndwc_nev_muid} report the number of expected events at each level of the analysis that are identified as electron-like and muon-like, according to Eq.~(\ref{eqn:detectors:nd_pid}), per flavour and interaction type.
The final purity of $\nu_e+\bar\nu_e$ in the {\eid} sample is \SI{74}{\percent} for positive horn polarity, and \SI{66}{\percent} for negative polarity.
The final-level purity is highly correlated with the $\nu_e$ content of the incident neutrino beam. If, for example, the $\nu_e$ contribution was a factor of two greater than estimated above (while maintaining the same spectral shape) the final purity would be \SI{85}{\percent} and \SI{80}{\percent} for positive and negative horn polarity; whereas if it was a factor of two smaller, the purity would be \SI{59}{\percent} and \SI{49}{\percent}, respectively.
An enhancement of the $\nu_e+\bar\nu_e$ contribution by a factor of five yields a final purity of \SI{93}{\percent} and \SI{91}{\percent} for positive and negative polarity, respectively. An enhancement of this magnitude of the $\nu_e+\bar\nu_e$ contribution in the initial neutrino beam would not affect the performance of the oscillation measurements with the far detector, as the contribution is still very small compared to the $\nu_\mu+\bar\nu_\mu$ bulk.

{
\begin{table}[!p]
\scriptsize
\centering
\caption{Number of expected events per running year, per level of the analysis, per flavour and interaction type, and per each horn polarity.
\label{tbl:detectors:ndwc_nev}}
\begin{tabular}{ r r r r r r r r r }
          \textbf{Positive polarity}  &                ~  &                ~  &                ~  &                ~  &                ~  &                ~  &                ~  &                ~  \\
                                   ~  &  \textbf{$\nu_\mu$ CC}  &  \textbf{$\nu_e$ CC}  &  \textbf{$\bar\nu_\mu$ CC}  &  \textbf{$\bar\nu_e$ CC}
                                      &  \textbf{$\nu_\mu$ NC}  &  \textbf{$\nu_e$ NC}  &  \textbf{$\bar\nu_\mu$ NC}  &  \textbf{$\bar\nu_e$ NC} \\
\hline
%                    All interactions  &  \SI{6.69e+07}{}  &  \SI{5.62e+05}{}  &  \SI{1.86e+05}{}  &  \SI{4.02e+02}{}  &  \SI{4.73e+07}{}  &  \SI{3.09e+05}{}  &  \SI{1.63e+05}{}  &  \SI{3.01e+02}{}  \\
%                             Trigger  &  \SI{6.63e+07}{}  &  \SI{5.60e+05}{}  &  \SI{1.85e+05}{}  &  \SI{4.00e+02}{}  &  \SI{2.64e+07}{}  &  \SI{1.79e+05}{}  &  \SI{8.49e+04}{}  &  \SI{1.52e+02}{}  \\
%             Sub-Cherenkov criterion  &  \SI{3.35e+07}{}  &  \SI{5.34e+05}{}  &  \SI{1.16e+05}{}  &  \SI{3.72e+02}{}  &  \SI{9.47e+05}{}  &  \SI{9.25e+03}{}  &  \SI{3.24e+03}{}  &  \SI{3.88e+00}{}  \\
%     Reconstruction quality criteria  &  \SI{2.80e+07}{}  &  \SI{4.83e+05}{}  &  \SI{1.01e+05}{}  &  \SI{3.36e+02}{}  &  \SI{7.89e+05}{}  &  \SI{7.77e+03}{}  &  \SI{2.74e+03}{}  &  \SI{3.25e+00}{}  \\
% Cherenkov-ring resolution criterion  &  \SI{2.18e+07}{}  &  \SI{3.81e+05}{}  &  \SI{7.91e+04}{}  &  \SI{2.63e+02}{}  &  \SI{6.76e+05}{}  &  \SI{6.72e+03}{}  &  \SI{2.30e+03}{}  &  \SI{2.80e+00}{}  \\
%                  Pion-like criteria  &  \SI{2.13e+07}{}  &  \SI{3.42e+05}{}  &  \SI{7.72e+04}{}  &  \SI{2.41e+02}{}  &  \SI{9.30e+04}{}  &  \SI{1.04e+03}{}  &  \SI{4.07e+02}{}  &  \SI{3.40e-01}{}  \\
%            multi--sub-event criterion  &  \SI{2.10e+07}{}  &  \SI{3.32e+05}{}  &  \SI{7.70e+04}{}  &  \SI{2.40e+02}{}  &  \SI{9.20e+04}{}  &  \SI{1.02e+03}{}  &  \SI{4.00e+02}{}  &  \SI{3.40e-01}{}  \\
                    All interactions  &  \SI{6.69e+07}{}  &  \SI{5.62e+05}{}  &  \SI{1.86e+05}{}  &       \SI{402}{}  &  \SI{4.73e+07}{}  &  \SI{3.09e+05}{}  &  \SI{1.63e+05}{}  &       \SI{301}{}  \\
                             Trigger  &  \SI{6.63e+07}{}  &  \SI{5.60e+05}{}  &  \SI{1.85e+05}{}  &       \SI{400}{}  &  \SI{2.64e+07}{}  &  \SI{1.79e+05}{}  &  \SI{8.49e+04}{}  &       \SI{152}{}  \\
             Sub-Cherenkov criterion  &  \SI{3.35e+07}{}  &  \SI{5.34e+05}{}  &  \SI{1.16e+05}{}  &       \SI{372}{}  &  \SI{9.47e+05}{}  &      \SI{9250}{}  &      \SI{3240}{}  &       \SI{3.9}{}  \\
     Reconstruction quality criteria  &  \SI{2.80e+07}{}  &  \SI{4.83e+05}{}  &  \SI{1.01e+05}{}  &       \SI{336}{}  &  \SI{7.89e+05}{}  &      \SI{7770}{}  &      \SI{2740}{}  &       \SI{3.3}{}  \\
 Cherenkov-ring resolution criterion  &  \SI{2.18e+07}{}  &  \SI{3.81e+05}{}  &  \SI{7.91e+04}{}  &       \SI{263}{}  &  \SI{6.76e+05}{}  &      \SI{6720}{}  &      \SI{2300}{}  &       \SI{2.8}{}  \\
                  Pion-like criteria  &  \SI{2.13e+07}{}  &  \SI{3.42e+05}{}  &  \SI{7.72e+04}{}  &       \SI{241}{}  &  \SI{9.30e+04}{}  &      \SI{1040}{}  &       \SI{407}{}  &       \SI{0.3}{}  \\
            multi--sub-event criterion  &  \SI{2.10e+07}{}  &  \SI{3.32e+05}{}  &  \SI{7.70e+04}{}  &       \SI{240}{}  &  \SI{9.20e+04}{}  &      \SI{1020}{}  &       \SI{400}{}  &       \SI{0.3}{}  \\
\hline
                                   ~  &                ~  &                ~  &                ~  &                ~  &                ~  &                ~  &                ~  &                ~  \\
          \textbf{Negative polarity}  &                ~  &                ~  &                ~  &                ~  &                ~  &                ~  &                ~  &                ~  \\
                                   ~  &  \textbf{$\nu_\mu$ CC}  &  \textbf{$\nu_e$ CC}  &  \textbf{$\bar\nu_\mu$ CC}  &  \textbf{$\bar\nu_e$ CC}
                                      &  \textbf{$\nu_\mu$ NC}  &  \textbf{$\nu_e$ NC}  &  \textbf{$\bar\nu_\mu$ NC}  &  \textbf{$\bar\nu_e$ NC} \\
\hline
%                    All interactions  &  \SI{6.82e+05}{}  &  \SI{3.44e+03}{}  &  \SI{1.09e+07}{}  &  \SI{5.55e+04}{}  &  \SI{5.30e+05}{}  &  \SI{2.05e+03}{}  &  \SI{9.65e+06}{}  &  \SI{4.02e+04}{}  \\
%                             Trigger  &  \SI{6.75e+05}{}  &  \SI{3.43e+03}{}  &  \SI{1.08e+07}{}  &  \SI{5.53e+04}{}  &  \SI{2.90e+05}{}  &  \SI{1.16e+03}{}  &  \SI{4.95e+06}{}  &  \SI{2.08e+04}{}  \\
%             Sub-Cherenkov criterion  &  \SI{3.41e+05}{}  &  \SI{3.26e+03}{}  &  \SI{5.98e+06}{}  &  \SI{5.26e+04}{}  &  \SI{1.07e+04}{}  &  \SI{5.81e+01}{}  &  \SI{1.27e+05}{}  &  \SI{6.07e+02}{}  \\
%     Reconstruction quality criteria  &  \SI{2.88e+05}{}  &  \SI{2.95e+03}{}  &  \SI{5.06e+06}{}  &  \SI{4.78e+04}{}  &  \SI{8.96e+03}{}  &  \SI{4.88e+01}{}  &  \SI{1.09e+05}{}  &  \SI{5.08e+02}{}  \\
% Cherenkov-ring resolution criterion  &  \SI{2.25e+05}{}  &  \SI{2.32e+03}{}  &  \SI{3.93e+06}{}  &  \SI{3.74e+04}{}  &  \SI{7.70e+03}{}  &  \SI{4.21e+01}{}  &  \SI{9.22e+04}{}  &  \SI{4.38e+02}{}  \\
%                  Pion-like criteria  &  \SI{2.18e+05}{}  &  \SI{2.08e+03}{}  &  \SI{3.88e+06}{}  &  \SI{3.42e+04}{}  &  \SI{1.31e+03}{}  &  \SI{6.92e+00}{}  &  \SI{1.42e+04}{}  &  \SI{5.65e+01}{}  \\
%            multi--sub-event criterion  &  \SI{2.13e+05}{}  &  \SI{2.02e+03}{}  &  \SI{3.87e+06}{}  &  \SI{3.41e+04}{}  &  \SI{1.29e+03}{}  &  \SI{6.75e+00}{}  &  \SI{1.41e+04}{}  &  \SI{5.62e+01}{}  \\
                    All interactions  &  \SI{6.82e+05}{}  &      \SI{3440}{}  &  \SI{1.09e+07}{}  &  \SI{5.55e+04}{}  &  \SI{5.30e+05}{}  &      \SI{2050}{}  &  \SI{9.65e+06}{}  &  \SI{4.02e+04}{}  \\
                             Trigger  &  \SI{6.75e+05}{}  &      \SI{3430}{}  &  \SI{1.08e+07}{}  &  \SI{5.53e+04}{}  &  \SI{2.90e+05}{}  &      \SI{1160}{}  &  \SI{4.95e+06}{}  &  \SI{2.08e+04}{}  \\
             Sub-Cherenkov criterion  &  \SI{3.41e+05}{}  &      \SI{3260}{}  &  \SI{5.98e+06}{}  &  \SI{5.26e+04}{}  &  \SI{1.07e+04}{}  &      \SI{58.1}{}  &  \SI{1.27e+05}{}  &       \SI{607}{}  \\
     Reconstruction quality criteria  &  \SI{2.88e+05}{}  &      \SI{2950}{}  &  \SI{5.06e+06}{}  &  \SI{4.78e+04}{}  &      \SI{8960}{}  &      \SI{48.8}{}  &  \SI{1.09e+05}{}  &       \SI{508}{}  \\
 Cherenkov-ring resolution criterion  &  \SI{2.25e+05}{}  &      \SI{2320}{}  &  \SI{3.93e+06}{}  &  \SI{3.74e+04}{}  &      \SI{7700}{}  &      \SI{42.1}{}  &  \SI{9.22e+04}{}  &       \SI{438}{}  \\
                  Pion-like criteria  &  \SI{2.18e+05}{}  &      \SI{2080}{}  &  \SI{3.88e+06}{}  &  \SI{3.42e+04}{}  &      \SI{1310}{}  &       \SI{6.9}{}  &  \SI{1.42e+04}{}  &      \SI{56.5}{}  \\
            multi--sub-event criterion  &  \SI{2.13e+05}{}  &      \SI{2020}{}  &  \SI{3.87e+06}{}  &  \SI{3.41e+04}{}  &      \SI{1290}{}  &       \SI{6.8}{}  &  \SI{1.41e+04}{}  &      \SI{56.2}{}  \\
\hline
\end{tabular}
\end{table}
}

{
\begin{table}[!p]
\scriptsize
\centering
\caption{Number of expected {\eid} events per running year, per level of the analysis, per flavour and interaction type, and per each horn polarity.
\label{tbl:detectors:ndwc_nev_eid}}
\begin{tabular}{ r r r r r r r r r }
          \textbf{Positive polarity}  &                ~  &                ~  &                ~  &                ~  &                ~  &                ~  &                ~  &                ~  \\
                                   ~  &  \textbf{$\nu_\mu$ CC {\eid}}  &  \textbf{$\nu_e$ CC {\eid}}  &  \textbf{$\bar\nu_\mu$ CC {\eid}}  &  \textbf{$\bar\nu_e$ CC {\eid}}
                                      &  \textbf{$\nu_\mu$ NC {\eid}}  &  \textbf{$\nu_e$ NC {\eid}}  &  \textbf{$\bar\nu_\mu$ NC {\eid}}  &  \textbf{$\bar\nu_e$ NC {\eid}} \\
\hline
%                    All interactions  &  \SI{1.50e+07}{}  &  \SI{5.33e+05}{}  &  \SI{4.28e+04}{}  &  \SI{3.82e+02}{}  &  \SI{2.44e+07}{}  &  \SI{1.65e+05}{}  &  \SI{7.87e+04}{}  &  \SI{1.42e+02}{}  \\
%                             Trigger  &  \SI{1.50e+07}{}  &  \SI{5.33e+05}{}  &  \SI{4.28e+04}{}  &  \SI{3.82e+02}{}  &  \SI{2.44e+07}{}  &  \SI{1.65e+05}{}  &  \SI{7.87e+04}{}  &  \SI{1.42e+02}{}  \\
%             Sub-Cherenkov criterion  &  \SI{2.57e+06}{}  &  \SI{5.14e+05}{}  &  \SI{1.00e+04}{}  &  \SI{3.59e+02}{}  &  \SI{8.93e+05}{}  &  \SI{8.57e+03}{}  &  \SI{3.06e+03}{}  &  \SI{3.68e+00}{}  \\
%     Reconstruction quality criteria  &  \SI{2.11e+06}{}  &  \SI{4.69e+05}{}  &  \SI{8.38e+03}{}  &  \SI{3.27e+02}{}  &  \SI{7.62e+05}{}  &  \SI{7.36e+03}{}  &  \SI{2.63e+03}{}  &  \SI{3.16e+00}{}  \\
% Cherenkov-ring resolution criterion  &  \SI{6.22e+05}{}  &  \SI{3.70e+05}{}  &  \SI{2.19e+03}{}  &  \SI{2.56e+02}{}  &  \SI{6.55e+05}{}  &  \SI{6.39e+03}{}  &  \SI{2.20e+03}{}  &  \SI{2.73e+00}{}  \\
%                  Pion-like criteria  &  \SI{9.63e+04}{}  &  \SI{3.32e+05}{}  &  \SI{2.09e+02}{}  &  \SI{2.34e+02}{}  &  \SI{7.19e+04}{}  &  \SI{7.18e+02}{}  &  \SI{3.13e+02}{}  &  \SI{2.69e-01}{}  \\
%            multi--sub-event criterion  &  \SI{3.95e+04}{}  &  \SI{3.22e+05}{}  &  \SI{8.09e+01}{}  &  \SI{2.34e+02}{}  &  \SI{7.09e+04}{}  &  \SI{6.91e+02}{}  &  \SI{3.07e+02}{}  &  \SI{2.69e-01}{}  \\
                    All interactions  &  \SI{1.50e+07}{}  &  \SI{5.33e+05}{}  &  \SI{4.28e+04}{}  &       \SI{382}{}  &  \SI{2.44e+07}{}  &  \SI{1.65e+05}{}  &  \SI{7.87e+04}{}  &       \SI{142}{}  \\
                             Trigger  &  \SI{1.50e+07}{}  &  \SI{5.33e+05}{}  &  \SI{4.28e+04}{}  &       \SI{382}{}  &  \SI{2.44e+07}{}  &  \SI{1.65e+05}{}  &  \SI{7.87e+04}{}  &       \SI{142}{}  \\
             Sub-Cherenkov criterion  &  \SI{2.57e+06}{}  &  \SI{5.14e+05}{}  &  \SI{1.00e+04}{}  &       \SI{359}{}  &  \SI{8.93e+05}{}  &      \SI{8570}{}  &      \SI{3060}{}  &       \SI{3.7}{}  \\
     Reconstruction quality criteria  &  \SI{2.11e+06}{}  &  \SI{4.69e+05}{}  &      \SI{8380}{}  &       \SI{327}{}  &  \SI{7.62e+05}{}  &      \SI{7360}{}  &      \SI{2630}{}  &       \SI{3.2}{}  \\
 Cherenkov-ring resolution criterion  &  \SI{6.22e+05}{}  &  \SI{3.70e+05}{}  &      \SI{2190}{}  &       \SI{256}{}  &  \SI{6.55e+05}{}  &      \SI{6390}{}  &      \SI{2200}{}  &       \SI{2.7}{}  \\
                  Pion-like criteria  &  \SI{9.63e+04}{}  &  \SI{3.32e+05}{}  &       \SI{209}{}  &       \SI{234}{}  &  \SI{7.19e+04}{}  &       \SI{718}{}  &       \SI{313}{}  &       \SI{0.3}{}  \\
            multi--sub-event criterion  &  \SI{3.95e+04}{}  &  \SI{3.22e+05}{}  &      \SI{80.9}{}  &       \SI{234}{}  &  \SI{7.09e+04}{}  &       \SI{691}{}  &       \SI{307}{}  &       \SI{0.3}{}  \\
\hline
                                   ~  &                ~  &                ~  &                ~  &                ~  &                ~  &                ~  &                ~  &                ~  \\
          \textbf{Negative polarity}  &                ~  &                ~  &                ~  &                ~  &                ~  &                ~  &                ~  &                ~  \\
                                   ~  &  \textbf{$\nu_\mu$ CC {\eid}}  &  \textbf{$\nu_e$ CC {\eid}}  &  \textbf{$\bar\nu_\mu$ CC {\eid}}  &  \textbf{$\bar\nu_e$ CC {\eid}}
                                      &  \textbf{$\nu_\mu$ NC {\eid}}  &  \textbf{$\nu_e$ NC {\eid}}  &  \textbf{$\bar\nu_\mu$ NC {\eid}}  &  \textbf{$\bar\nu_e$ NC {\eid}} \\
\hline
%                    All interactions  &  \SI{1.66e+05}{}  &  \SI{3.26e+03}{}  &  \SI{2.49e+06}{}  &  \SI{5.29e+04}{}  &  \SI{2.68e+05}{}  &  \SI{1.07e+03}{}  &  \SI{4.61e+06}{}  &  \SI{1.93e+04}{}  \\
%                             Trigger  &  \SI{1.66e+05}{}  &  \SI{3.26e+03}{}  &  \SI{2.49e+06}{}  &  \SI{5.29e+04}{}  &  \SI{2.68e+05}{}  &  \SI{1.07e+03}{}  &  \SI{4.61e+06}{}  &  \SI{1.93e+04}{}  \\
%             Sub-Cherenkov criterion  &  \SI{2.87e+04}{}  &  \SI{3.14e+03}{}  &  \SI{4.31e+05}{}  &  \SI{5.09e+04}{}  &  \SI{9.86e+03}{}  &  \SI{5.32e+01}{}  &  \SI{1.22e+05}{}  &  \SI{5.74e+02}{}  \\
%     Reconstruction quality criteria  &  \SI{2.39e+04}{}  &  \SI{2.86e+03}{}  &  \SI{3.49e+05}{}  &  \SI{4.66e+04}{}  &  \SI{8.50e+03}{}  &  \SI{4.58e+01}{}  &  \SI{1.06e+05}{}  &  \SI{4.92e+02}{}  \\
% Cherenkov-ring resolution criterion  &  \SI{8.00e+03}{}  &  \SI{2.26e+03}{}  &  \SI{6.89e+04}{}  &  \SI{3.66e+04}{}  &  \SI{7.33e+03}{}  &  \SI{3.97e+01}{}  &  \SI{8.95e+04}{}  &  \SI{4.26e+02}{}  \\
%                  Pion-like criteria  &  \SI{1.18e+03}{}  &  \SI{2.02e+03}{}  &  \SI{9.64e+03}{}  &  \SI{3.34e+04}{}  &  \SI{9.40e+02}{}  &  \SI{4.46e+00}{}  &  \SI{1.14e+04}{}  &  \SI{4.37e+01}{}  \\
%            multi--sub-event criterion  &  \SI{3.94e+02}{}  &  \SI{1.95e+03}{}  &  \SI{5.40e+03}{}  &  \SI{3.33e+04}{}  &  \SI{9.18e+02}{}  &  \SI{4.29e+00}{}  &  \SI{1.13e+04}{}  &  \SI{4.34e+01}{}  \\
                    All interactions  &  \SI{1.66e+05}{}  &      \SI{3260}{}  &  \SI{2.49e+06}{}  &  \SI{5.29e+04}{}  &  \SI{2.68e+05}{}  &      \SI{1070}{}  &  \SI{4.61e+06}{}  &  \SI{1.93e+04}{}  \\
                             Trigger  &  \SI{1.66e+05}{}  &      \SI{3260}{}  &  \SI{2.49e+06}{}  &  \SI{5.29e+04}{}  &  \SI{2.68e+05}{}  &      \SI{1070}{}  &  \SI{4.61e+06}{}  &  \SI{1.93e+04}{}  \\
             Sub-Cherenkov criterion  &  \SI{2.87e+04}{}  &      \SI{3140}{}  &  \SI{4.31e+05}{}  &  \SI{5.09e+04}{}  &      \SI{9860}{}  &      \SI{53.2}{}  &  \SI{1.22e+05}{}  &       \SI{574}{}  \\
     Reconstruction quality criteria  &  \SI{2.39e+04}{}  &      \SI{2860}{}  &  \SI{3.49e+05}{}  &  \SI{4.66e+04}{}  &      \SI{8500}{}  &      \SI{45.8}{}  &  \SI{1.06e+05}{}  &       \SI{492}{}  \\
 Cherenkov-ring resolution criterion  &      \SI{8000}{}  &      \SI{2260}{}  &  \SI{6.89e+04}{}  &  \SI{3.66e+04}{}  &      \SI{7330}{}  &      \SI{39.7}{}  &  \SI{8.95e+04}{}  &       \SI{426}{}  \\
                  Pion-like criteria  &      \SI{1180}{}  &      \SI{2020}{}  &      \SI{9640}{}  &  \SI{3.34e+04}{}  &       \SI{940}{}  &       \SI{4.5}{}  &  \SI{1.14e+04}{}  &      \SI{43.7}{}  \\
            multi--sub-event criterion  &       \SI{394}{}  &      \SI{1950}{}  &      \SI{5400}{}  &  \SI{3.33e+04}{}  &       \SI{918}{}  &       \SI{4.3}{}  &  \SI{1.13e+04}{}  &      \SI{43.4}{}  \\
\hline
\end{tabular}
\end{table}
}

{
\begin{table}[!p]
\scriptsize
\centering
\caption{Number of expected {\muid} events per running year, per level of the analysis, per flavour and interaction type, and per each horn polarity.
\label{tbl:detectors:ndwc_nev_muid}}
\begin{tabular}{ r r r r r r r r r }
          \textbf{Positive polarity}  &                ~  &                ~  &                ~  &                ~  &                ~  &                ~  &                ~  &                ~  \\
                                   ~  &  \textbf{$\nu_\mu$ CC {\muid}}  &  \textbf{$\nu_e$ CC {\muid}}  &  \textbf{$\bar\nu_\mu$ CC {\muid}}  &  \textbf{$\bar\nu_e$ CC {\muid}}
                                      &  \textbf{$\nu_\mu$ NC {\muid}}  &  \textbf{$\nu_e$ NC {\muid}}  &  \textbf{$\bar\nu_\mu$ NC {\muid}}  &  \textbf{$\bar\nu_e$ NC {\muid}} \\
\hline
%                    All interactions  &  \SI{5.19e+07}{}  &  \SI{2.88e+04}{}  &  \SI{1.43e+05}{}  &  \SI{1.97e+01}{}  &  \SI{2.29e+07}{}  &  \SI{1.44e+05}{}  &  \SI{8.44e+04}{}  &  \SI{1.59e+02}{}  \\
%                             Trigger  &  \SI{5.13e+07}{}  &  \SI{2.71e+04}{}  &  \SI{1.42e+05}{}  &  \SI{1.81e+01}{}  &  \SI{1.98e+06}{}  &  \SI{1.36e+04}{}  &  \SI{6.15e+03}{}  &  \SI{1.02e+01}{}  \\
%             Sub-Cherenkov criterion  &  \SI{3.10e+07}{}  &  \SI{2.00e+04}{}  &  \SI{1.06e+05}{}  &  \SI{1.26e+01}{}  &  \SI{5.40e+04}{}  &  \SI{6.78e+02}{}  &  \SI{1.79e+02}{}  &  \SI{2.05e-01}{}  \\
%     Reconstruction quality criteria  &  \SI{2.59e+07}{}  &  \SI{1.43e+04}{}  &  \SI{9.29e+04}{}  &  \SI{8.74e+00}{}  &  \SI{2.69e+04}{}  &  \SI{4.07e+02}{}  &  \SI{1.11e+02}{}  &  \SI{8.81e-02}{}  \\
% Cherenkov-ring resolution criterion  &  \SI{2.12e+07}{}  &  \SI{1.03e+04}{}  &  \SI{7.69e+04}{}  &  \SI{6.25e+00}{}  &  \SI{2.11e+04}{}  &  \SI{3.27e+02}{}  &  \SI{9.36e+01}{}  &  \SI{7.08e-02}{}  \\
%                  Pion-like criteria  &  \SI{2.12e+07}{}  &  \SI{1.03e+04}{}  &  \SI{7.69e+04}{}  &  \SI{6.25e+00}{}  &  \SI{2.11e+04}{}  &  \SI{3.27e+02}{}  &  \SI{9.36e+01}{}  &  \SI{7.08e-02}{}  \\
%            multi--sub-event criterion  &  \SI{2.10e+07}{}  &  \SI{1.03e+04}{}  &  \SI{7.69e+04}{}  &  \SI{6.25e+00}{}  &  \SI{2.11e+04}{}  &  \SI{3.26e+02}{}  &  \SI{9.34e+01}{}  &  \SI{7.08e-02}{}  \\
                    All interactions  &  \SI{5.19e+07}{}  &  \SI{2.88e+04}{}  &  \SI{1.43e+05}{}  &      \SI{19.7}{}  &  \SI{2.29e+07}{}  &  \SI{1.44e+05}{}  &  \SI{8.44e+04}{}  &       \SI{159}{}  \\
                             Trigger  &  \SI{5.13e+07}{}  &  \SI{2.71e+04}{}  &  \SI{1.42e+05}{}  &      \SI{18.1}{}  &  \SI{1.98e+06}{}  &  \SI{1.36e+04}{}  &      \SI{6150}{}  &      \SI{10.2}{}  \\
             Sub-Cherenkov criterion  &  \SI{3.10e+07}{}  &  \SI{2.00e+04}{}  &  \SI{1.06e+05}{}  &      \SI{12.6}{}  &  \SI{5.40e+04}{}  &       \SI{678}{}  &       \SI{179}{}  &       \SI{0.2}{}  \\
     Reconstruction quality criteria  &  \SI{2.59e+07}{}  &  \SI{1.43e+04}{}  &  \SI{9.29e+04}{}  &       \SI{8.7}{}  &  \SI{2.69e+04}{}  &       \SI{407}{}  &       \SI{111}{}  &       \SI{0.1}{}  \\
 Cherenkov-ring resolution criterion  &  \SI{2.12e+07}{}  &  \SI{1.03e+04}{}  &  \SI{7.69e+04}{}  &       \SI{6.3}{}  &  \SI{2.11e+04}{}  &       \SI{327}{}  &      \SI{93.6}{}  &       \SI{0.1}{}  \\
                  Pion-like criteria  &  \SI{2.12e+07}{}  &  \SI{1.03e+04}{}  &  \SI{7.69e+04}{}  &       \SI{6.3}{}  &  \SI{2.11e+04}{}  &       \SI{327}{}  &      \SI{93.6}{}  &       \SI{0.1}{}  \\
            multi--sub-event criterion  &  \SI{2.10e+07}{}  &  \SI{1.03e+04}{}  &  \SI{7.69e+04}{}  &       \SI{6.3}{}  &  \SI{2.11e+04}{}  &       \SI{326}{}  &      \SI{93.4}{}  &       \SI{0.1}{}  \\
\hline
                                   ~  &                ~  &                ~  &                ~  &                ~  &                ~  &                ~  &                ~  &                ~  \\
          \textbf{Negative polarity}  &                ~  &                ~  &                ~  &                ~  &                ~  &                ~  &                ~  &                ~  \\
                                   ~  &  \textbf{$\nu_\mu$ CC {\muid}}  &  \textbf{$\nu_e$ CC {\muid}}  &  \textbf{$\bar\nu_\mu$ CC {\muid}}  &  \textbf{$\bar\nu_e$ CC {\muid}}
                                      &  \textbf{$\nu_\mu$ NC {\muid}}  &  \textbf{$\nu_e$ NC {\muid}}  &  \textbf{$\bar\nu_\mu$ NC {\muid}}  &  \textbf{$\bar\nu_e$ NC {\muid}} \\
\hline
%                    All interactions  &  \SI{5.17e+05}{}  &  \SI{1.79e+02}{}  &  \SI{8.36e+06}{}  &  \SI{2.61e+03}{}  &  \SI{2.62e+05}{}  &  \SI{9.83e+02}{}  &  \SI{5.05e+06}{}  &  \SI{2.08e+04}{}  \\
%                             Trigger  &  \SI{5.10e+05}{}  &  \SI{1.68e+02}{}  &  \SI{8.31e+06}{}  &  \SI{2.40e+03}{}  &  \SI{2.20e+04}{}  &  \SI{8.69e+01}{}  &  \SI{3.46e+05}{}  &  \SI{1.41e+03}{}  \\
%             Sub-Cherenkov criterion  &  \SI{3.12e+05}{}  &  \SI{1.25e+02}{}  &  \SI{5.55e+06}{}  &  \SI{1.69e+03}{}  &  \SI{7.99e+02}{}  &  \SI{4.93e+00}{}  &  \SI{5.49e+03}{}  &  \SI{3.34e+01}{}  \\
%     Reconstruction quality criteria  &  \SI{2.65e+05}{}  &  \SI{8.90e+01}{}  &  \SI{4.71e+06}{}  &  \SI{1.17e+03}{}  &  \SI{4.56e+02}{}  &  \SI{3.05e+00}{}  &  \SI{3.05e+03}{}  &  \SI{1.57e+01}{}  \\
% Cherenkov-ring resolution criterion  &  \SI{2.17e+05}{}  &  \SI{6.55e+01}{}  &  \SI{3.87e+06}{}  &  \SI{8.06e+02}{}  &  \SI{3.72e+02}{}  &  \SI{2.47e+00}{}  &  \SI{2.72e+03}{}  &  \SI{1.28e+01}{}  \\
%                  Pion-like criteria  &  \SI{2.17e+05}{}  &  \SI{6.55e+01}{}  &  \SI{3.87e+06}{}  &  \SI{8.06e+02}{}  &  \SI{3.72e+02}{}  &  \SI{2.47e+00}{}  &  \SI{2.72e+03}{}  &  \SI{1.28e+01}{}  \\
%            multi--sub-event criterion  &  \SI{2.13e+05}{}  &  \SI{6.55e+01}{}  &  \SI{3.86e+06}{}  &  \SI{8.06e+02}{}  &  \SI{3.71e+02}{}  &  \SI{2.46e+00}{}  &  \SI{2.72e+03}{}  &  \SI{1.28e+01}{}  \\
                    All interactions  &  \SI{5.17e+05}{}  &       \SI{179}{}  &  \SI{8.36e+06}{}  &      \SI{2610}{}  &  \SI{2.62e+05}{}  &       \SI{983}{}  &  \SI{5.05e+06}{}  &  \SI{2.08e+04}{}  \\
                             Trigger  &  \SI{5.10e+05}{}  &       \SI{168}{}  &  \SI{8.31e+06}{}  &      \SI{2400}{}  &  \SI{2.20e+04}{}  &      \SI{86.9}{}  &  \SI{3.46e+05}{}  &      \SI{1410}{}  \\
             Sub-Cherenkov criterion  &  \SI{3.12e+05}{}  &       \SI{125}{}  &  \SI{5.55e+06}{}  &      \SI{1690}{}  &       \SI{799}{}  &       \SI{4.9}{}  &      \SI{5490}{}  &      \SI{33.4}{}  \\
     Reconstruction quality criteria  &  \SI{2.65e+05}{}  &      \SI{89.0}{}  &  \SI{4.71e+06}{}  &      \SI{1170}{}  &       \SI{456}{}  &       \SI{3.1}{}  &      \SI{3050}{}  &      \SI{15.7}{}  \\
 Cherenkov-ring resolution criterion  &  \SI{2.17e+05}{}  &      \SI{65.5}{}  &  \SI{3.87e+06}{}  &       \SI{806}{}  &       \SI{372}{}  &       \SI{2.5}{}  &      \SI{2720}{}  &      \SI{12.8}{}  \\
                  Pion-like criteria  &  \SI{2.17e+05}{}  &      \SI{65.5}{}  &  \SI{3.87e+06}{}  &       \SI{806}{}  &       \SI{372}{}  &       \SI{2.5}{}  &      \SI{2720}{}  &      \SI{12.8}{}  \\
            multi--sub-event criterion  &  \SI{2.13e+05}{}  &      \SI{65.5}{}  &  \SI{3.86e+06}{}  &       \SI{806}{}  &       \SI{371}{}  &       \SI{2.5}{}  &      \SI{2720}{}  &      \SI{12.8}{}  \\
\hline
\end{tabular}
\end{table}
}

% FIGURE
% Migration matrices
\begin{figure}[p]
\centering

\begin{subfigure}[b]{0.495\textwidth}
\centering
\includegraphics[width=\textwidth]{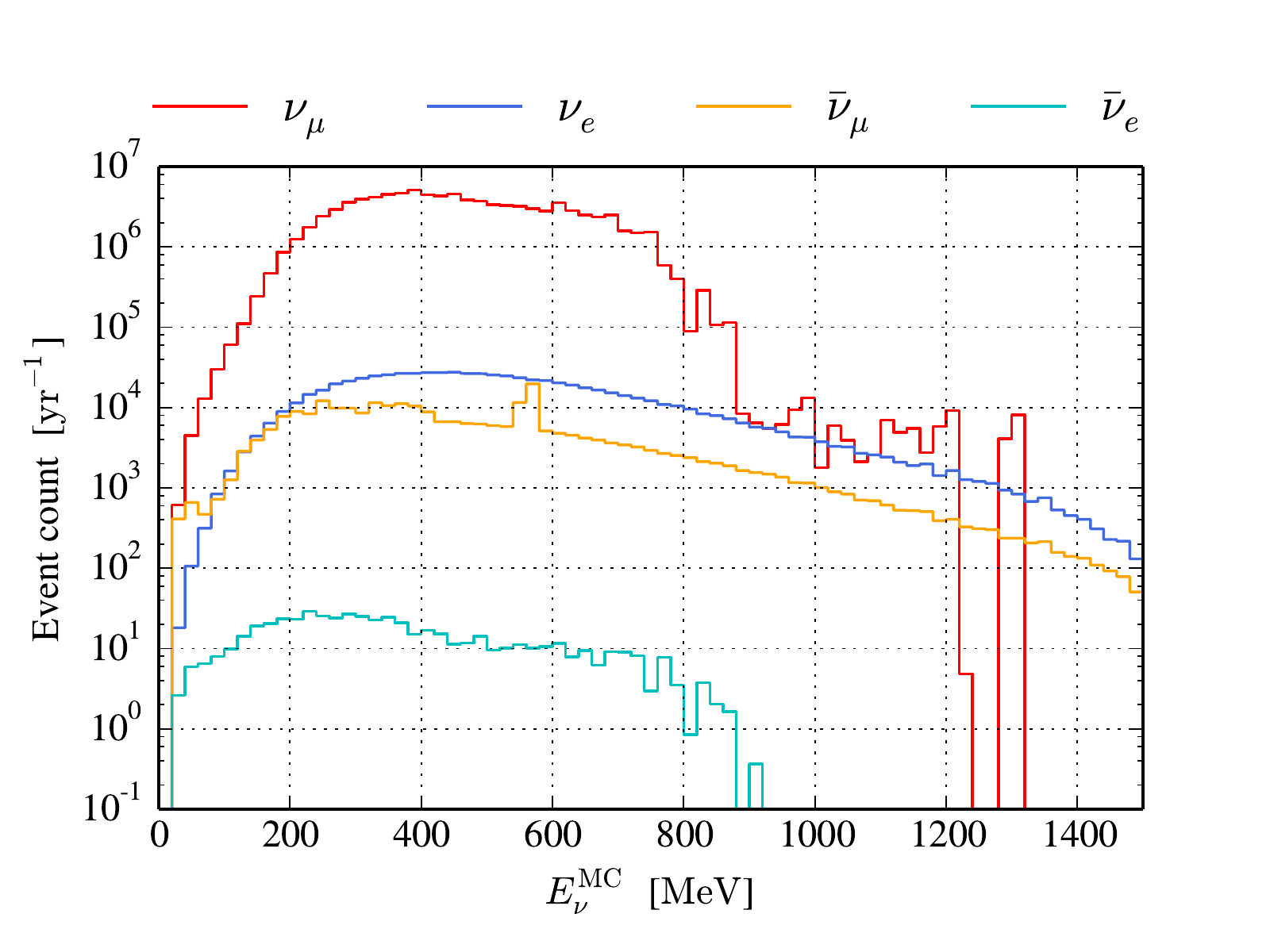}
\caption{}
\label{fig:detectors:ndwc_nev_L1pos}
\end{subfigure}
\hfill
\begin{subfigure}[b]{0.495\textwidth}  
\centering 
\includegraphics[width=\textwidth]{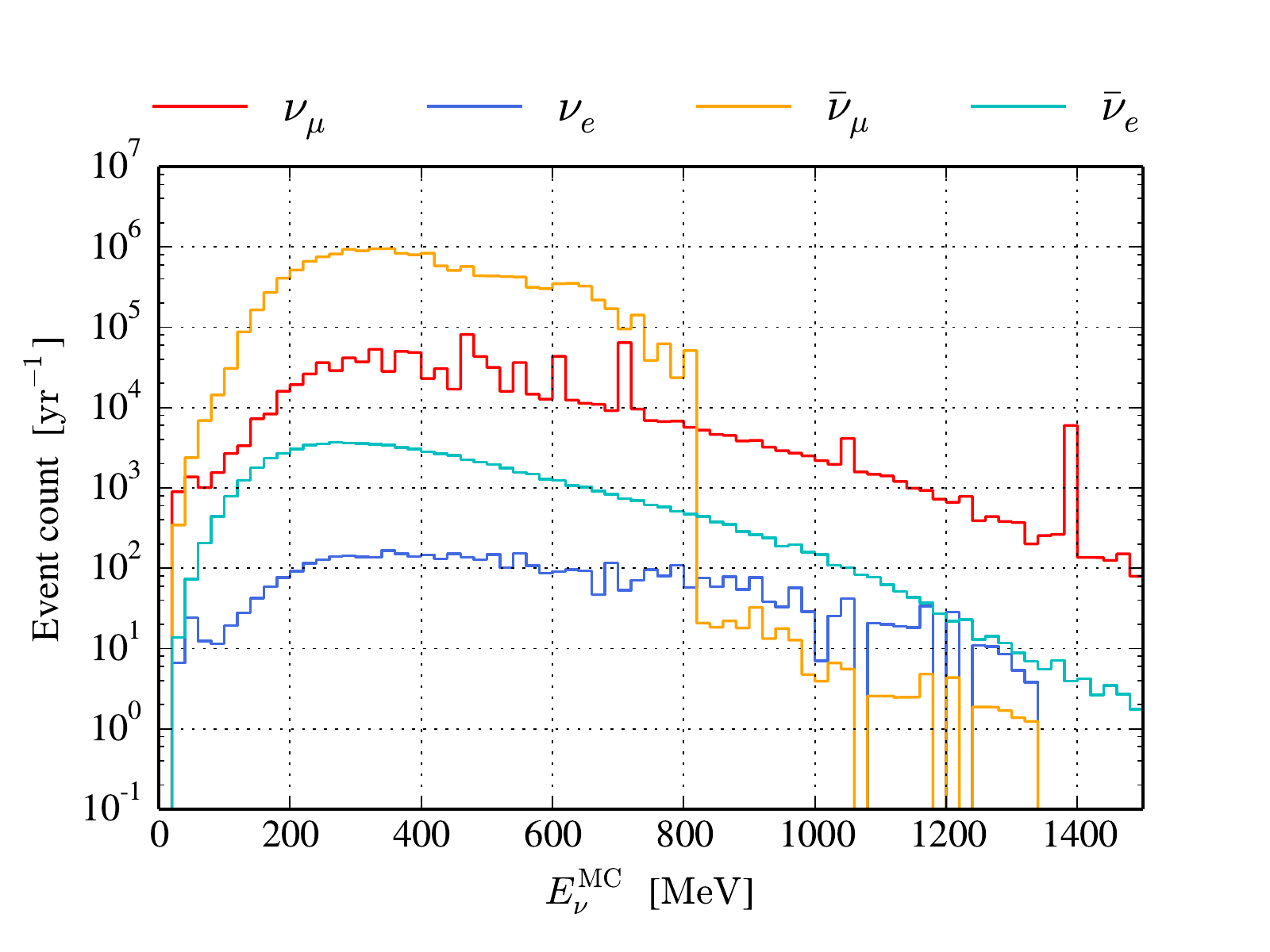}
\caption{}
\label{fig:detectors:ndwc_nev_L1neg}
\end{subfigure}
\hfill
\begin{subfigure}[b]{0.495\textwidth}  
\centering 
\includegraphics[width=\textwidth]{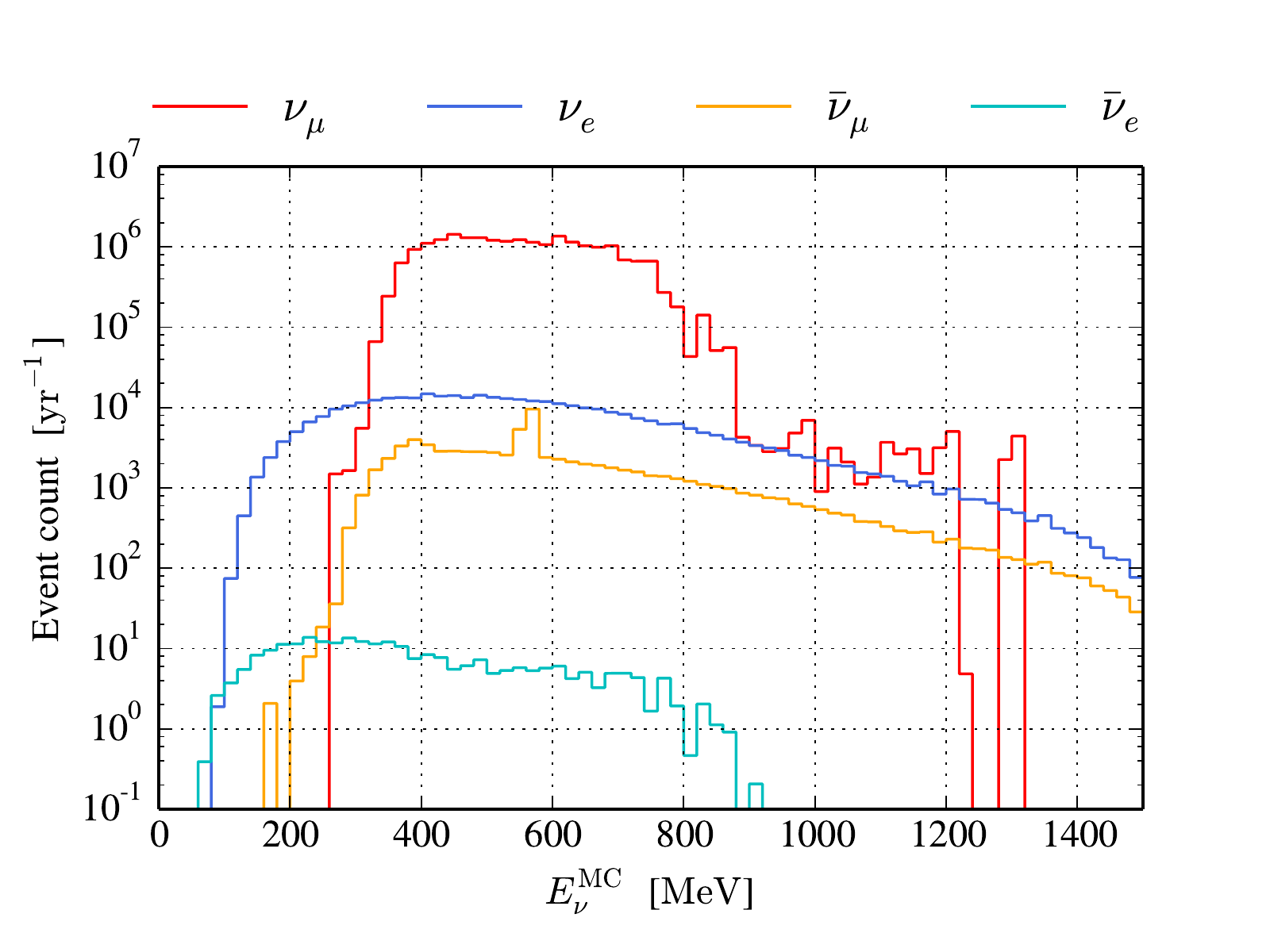}
\caption{}
\label{fig:detectors:ndwc_nev_L5pos}
\end{subfigure}
\hfill
\begin{subfigure}[b]{0.495\textwidth}  
\centering 
\includegraphics[width=\textwidth]{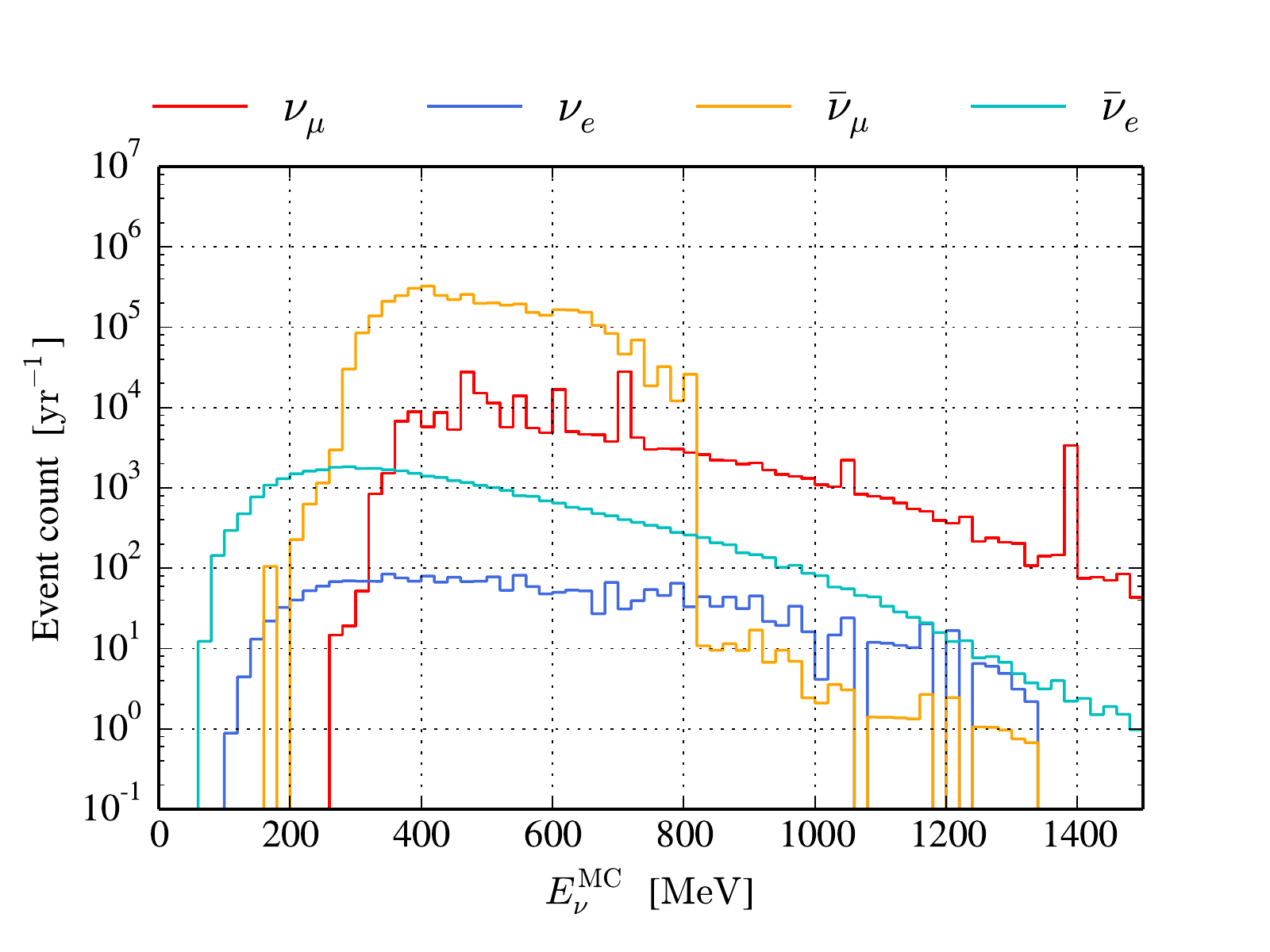}
\caption{}
\label{fig:detectors:ndwc_nev_L5neg}
\end{subfigure}
\hfill
\begin{subfigure}[b]{0.495\textwidth}  
\centering 
\includegraphics[width=\textwidth]{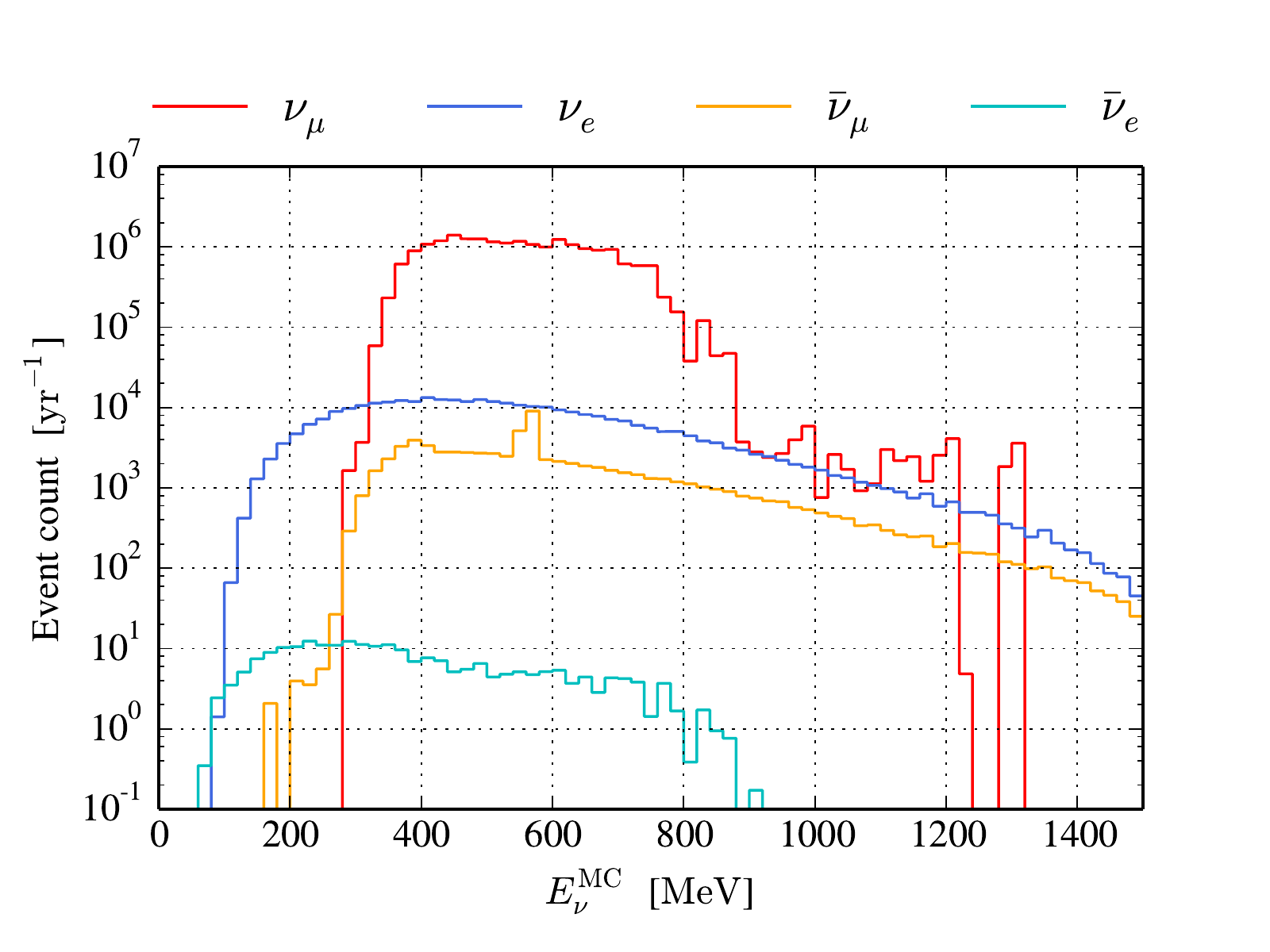}
\caption{}
\label{fig:detectors:ndwc_nev_L7pos}
\end{subfigure}
\hfill
\begin{subfigure}[b]{0.495\textwidth}  
\centering 
\includegraphics[width=\textwidth]{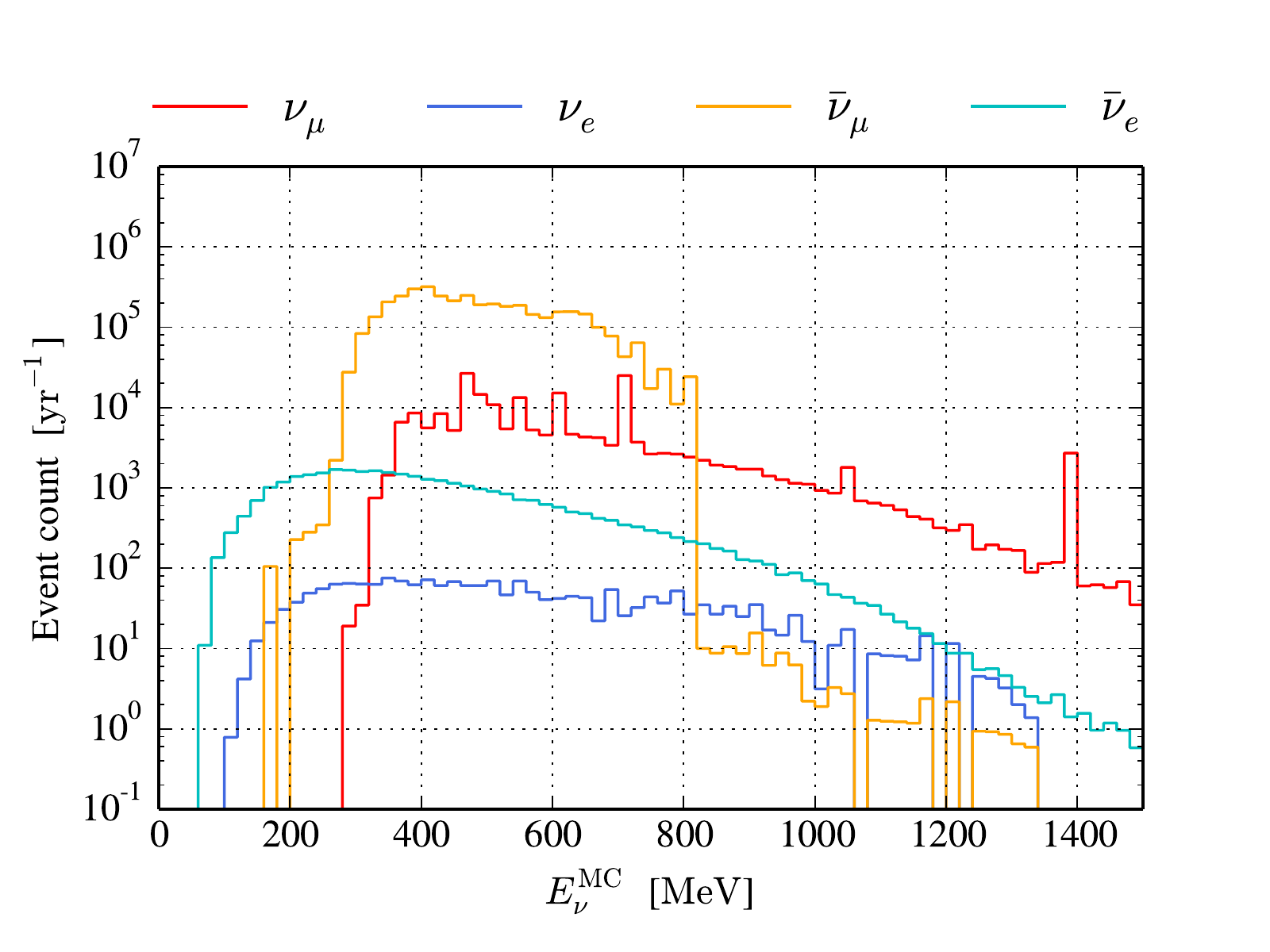}
\caption{}
\label{fig:detectors:ndwc_nev_L7neg}
\end{subfigure}
\caption{Number of expected events per running year, summed over interaction type for each neutrino flavour, as functions of the incident neutrino energy.
Given for
(\subref{fig:detectors:ndwc_nev_L1pos}, \subref{fig:detectors:ndwc_nev_L5pos}, \subref{fig:detectors:ndwc_nev_L7pos}) positive and
(\subref{fig:detectors:ndwc_nev_L1neg}, \subref{fig:detectors:ndwc_nev_L5neg}, \subref{fig:detectors:ndwc_nev_L7neg}) negative horn polarity, after the application of the
(\subref{fig:detectors:ndwc_nev_L1pos}, \subref{fig:detectors:ndwc_nev_L1neg}) \emph{trigger},
(\subref{fig:detectors:ndwc_nev_L5pos}, \subref{fig:detectors:ndwc_nev_L5neg}) \emph{charged lepton criteria} and
(\subref{fig:detectors:ndwc_nev_L7pos}, \subref{fig:detectors:ndwc_nev_L7neg}) \emph{neutrino criteria} have been applied.
\label{fig:detectors:ndwc_nev}}
\end{figure}

\subsubsubsection{Flux Measurement Using Neutrino Elastic Scattering on Atomic Electrons}

One of the main objectives for the water Cherenkov near detector is to measure the interaction cross-section between neutrinos and the ambient nuclei. This cross-section is inherently difficult to calculate, due to nuclear effects, and there is little experimental data in the relevant energy range. One large source of uncertainty for this measurement is the uncertainty in the incident neutrino flux.

An independent measurement of the incident neutrino flux can be achieved by observing elastic scattering of neutrinos on atomic electrons in the water. This process yields a free electron in the final state, and is well-suited for a flux measurement because its cross-section can be calculated theoretically with high precision. This cross-section is, however, three to four orders of magnitude smaller than that of neutrino-nucleus scattering, so events produced with this process must be efficiently selected in a large sample of $\nu_e$ CC background events.

The incident neutrino energy, $E_\nu$, can be calculated from the outgoing electron energy, $E_e$, and the angle between the neutrino and electron directions, $\theta_e$, through the following relation:
\begin{align}
E_\nu = \frac{m_e\left(E_e-m_e\right)}{m_e-E_e+p_e\cos\theta_e}
\label{eqn:detectors:ndwc_electronscattering_enureco}
\end{align}

The relation between the incident neutrino energy and the outgoing electron energy is shown in Fig.~\ref{fig:detectors:ndwc_kaja_enuee_comparison}, for an event sample produced with the ESS$\nu$SB energy spectrum using the \textsc{Genie} neutrino event generator.

\begin{figure}[!thbp]
\centering
\begin{subfigure}[b]{0.495\textwidth}
\centering
\includegraphics[width=\textwidth]{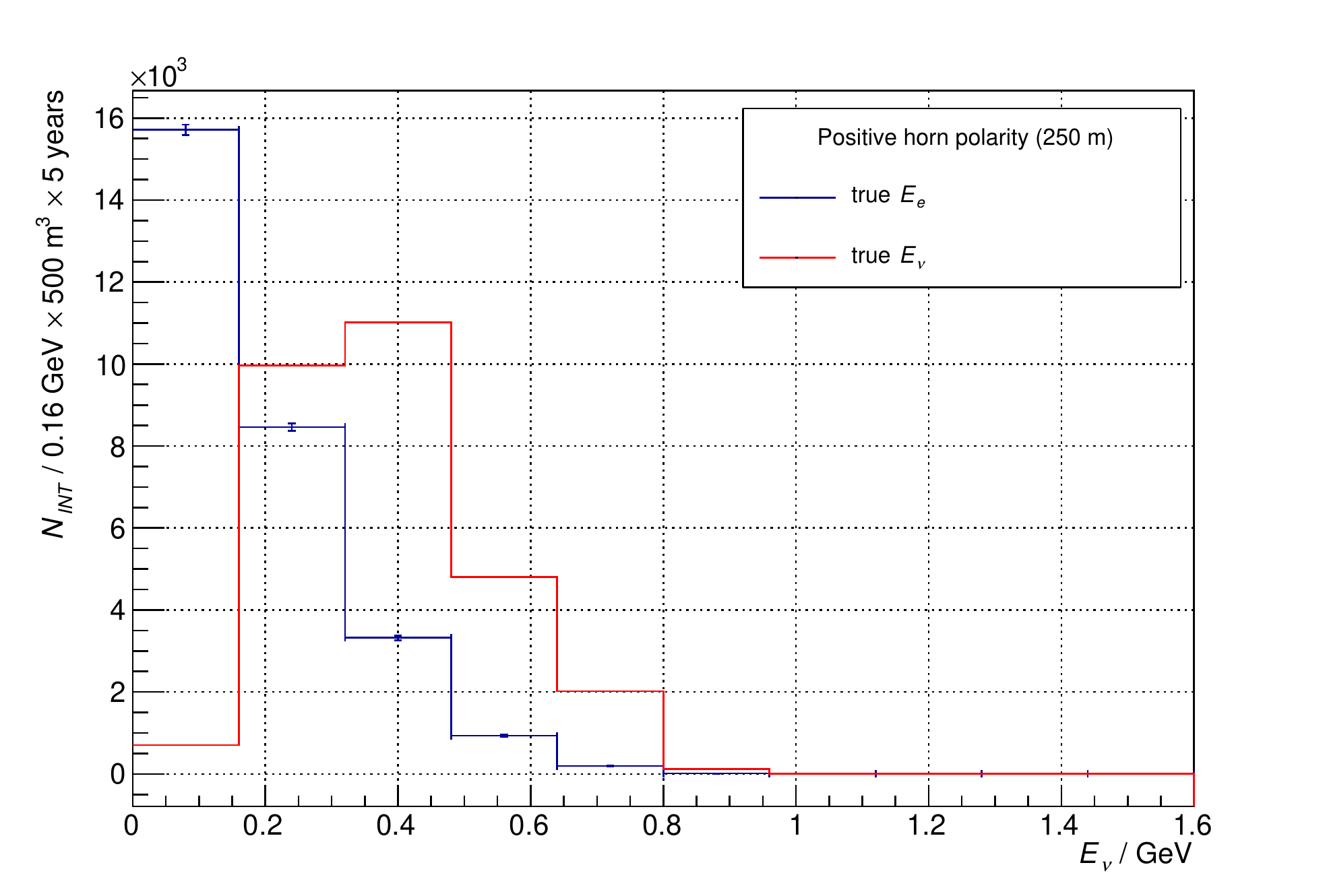}
\caption{Event distributions over the particle energy of the incident neutrino (red) and outgoing electron (blue).}
\label{fig:detectors:ndwc_kaja_enuee_comparison_1d}
\end{subfigure}
\hfill
\begin{subfigure}[b]{0.495\textwidth}  
\centering 
\includegraphics[width=\textwidth]{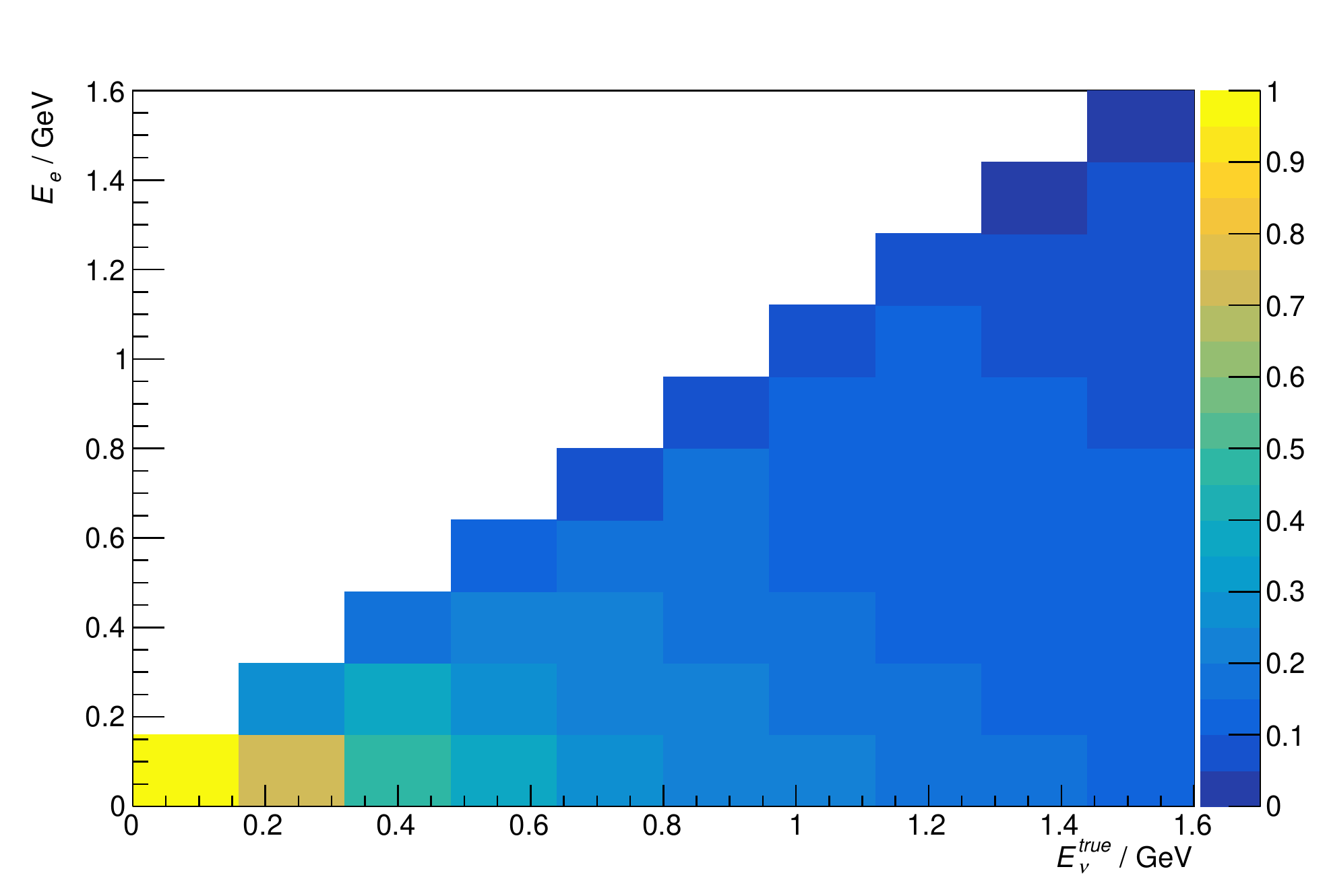}
\caption{2D event distributions over the incident neutrino energy, $E_\nu^{true}$, and outgoing electron energy, $E_e$.}
\label{fig:detectors:ndwc_kaja_enuee_comparison_2d}
\end{subfigure}
%\begin{tikzpicture}
%\node at (0.0\textwidth,0.0) { \includegraphics[width=0.5\textwidth]{figures/detectors/kaja_measur_pos_0} };
%\node at (0.5\textwidth,0.0) { \includegraphics[width=0.5\textwidth]{figures/detectors/kaja_pFLAT_resp_mtx_0_10_5} };
%\end{tikzpicture}
\caption{1D and 2D event distributions over the particle energy of the incident neutrino and outgoing electron for a sample of events simulated with the ESS$\nu$SB energy spectrum using \textsc{Genie}.}
\label{fig:detectors:ndwc_kaja_enuee_comparison}
\end{figure}

Following the analysis of the MINERvA experiment \cite{MINERvA:2019} the neutrino-electron scattering events can be identified using $E_e\theta_e^2$ as the discriminating variable.
%, where $E_e$ is the energy of the outgoing electron and $\theta_e$ is the angle between the outgoing electron direction and the incident neutrino direction.
The range of this variable is constrained by kinematics to the region $0 \leq E_e\theta_e^2 \leq 2m_e$, where $m_e$ is the mass of the electron.

A sample of events produced with the neutrino-electron scattering process were produced for each neutrino flavor ($\nu_e$, $\bar\nu_e$, $\nu_\mu$, $\bar\nu_\mu$) using the \textsc{Genie} event generator with the expected ESS$\nu$SB energy spectrum.
The discriminating power of $E_e\theta_e^2$ can be demonstrated using the MC-true values for $E_e$ and $\theta_e$ with this sample of simulated events.
The results are shown in Fig.~\ref{fig:detectors:Kaja-1},
and it is observed that events produced with neutrino-electron scattering dominate for values of $E_e\theta_e^2$ less than ${\sim}\SI{1}{\square\rad\MeV}$.

\begin{figure}[!thbp]
\centering
\begin{subfigure}[b]{0.495\textwidth}
\centering
\includegraphics[width=\textwidth]{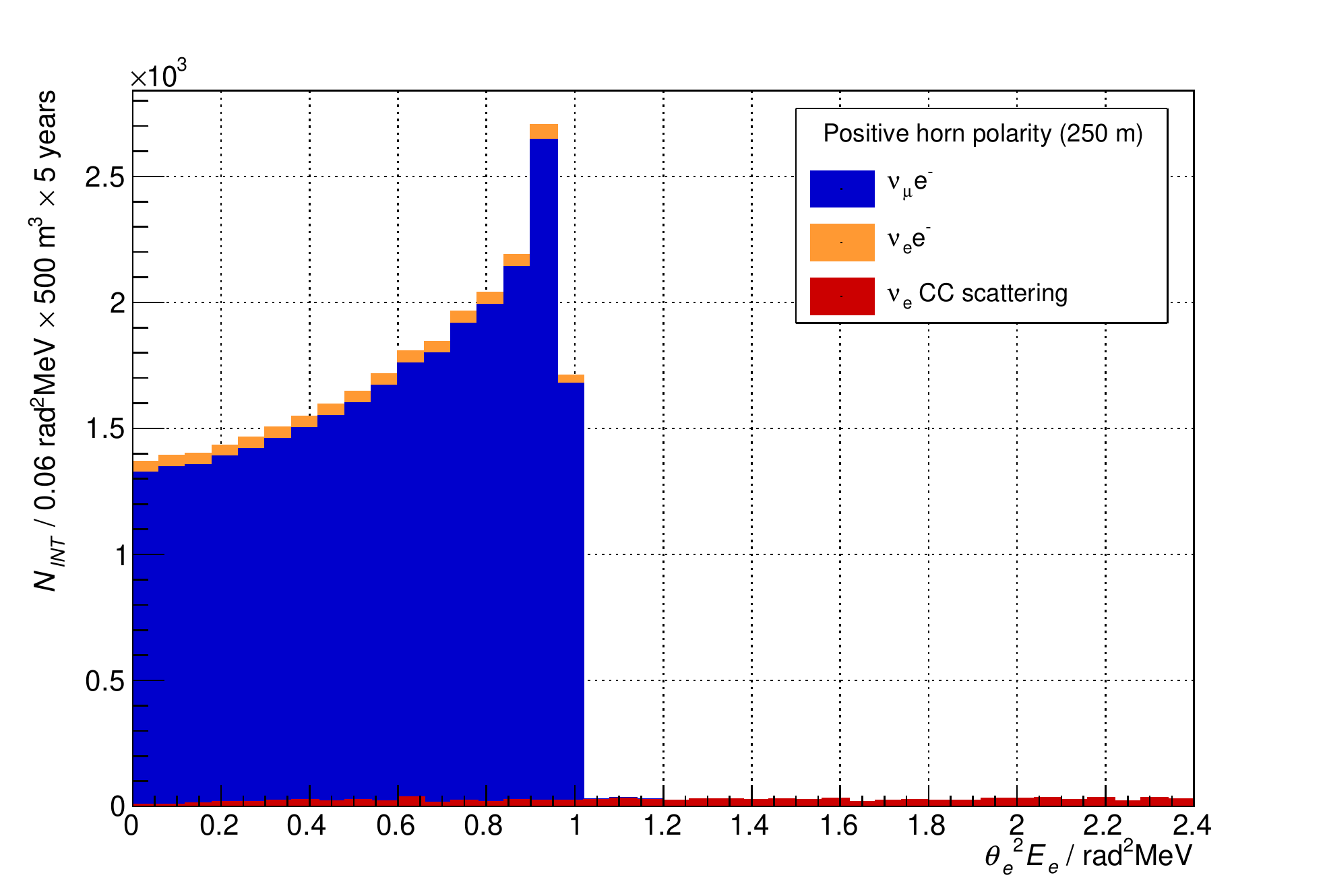}
\caption{Positive horn polarity.}
\label{fig:detectors:Kaja-1_pos}
\end{subfigure}
\hfill
\begin{subfigure}[b]{0.495\textwidth}  
\centering 
\includegraphics[width=\textwidth]{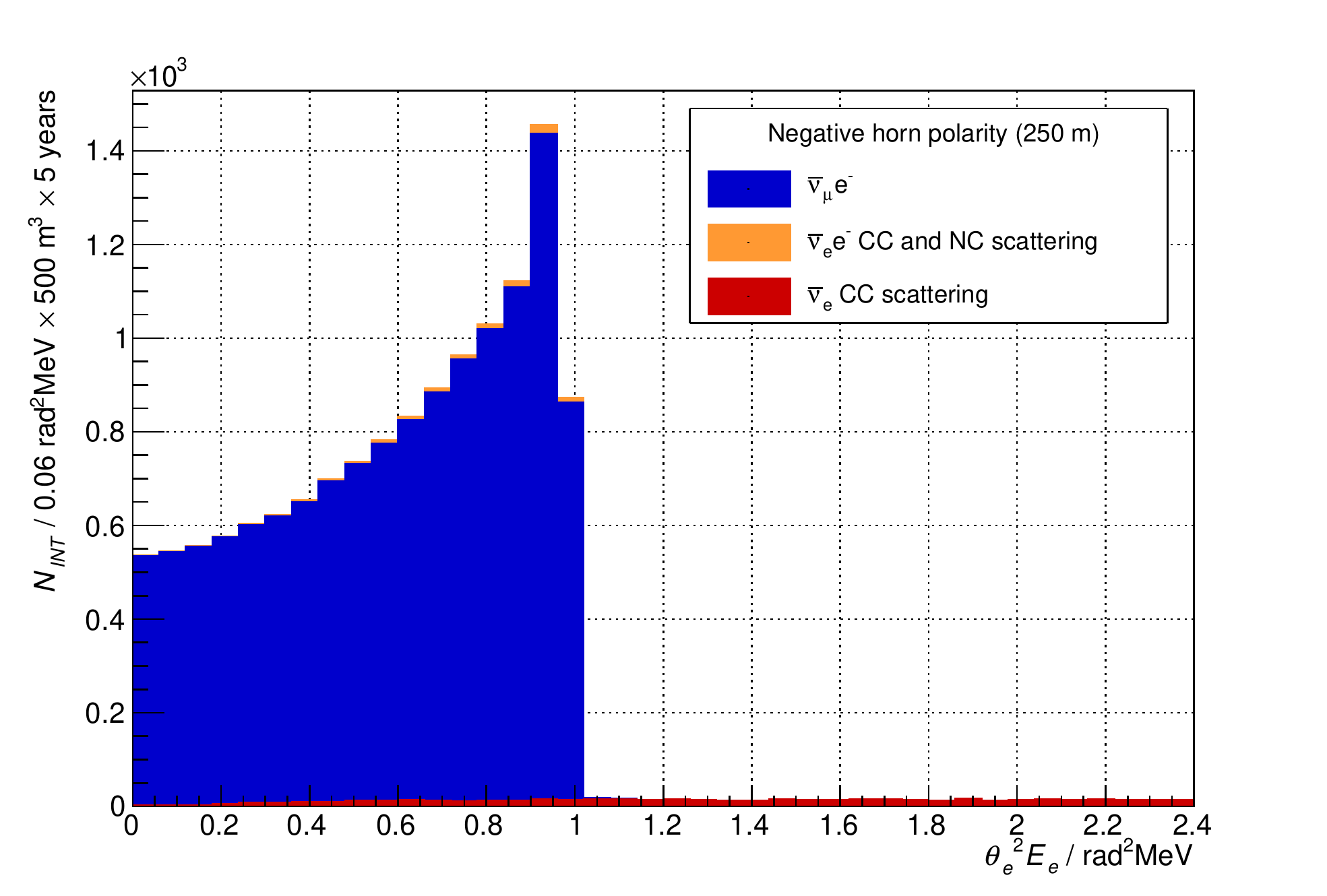}
\caption{Negative horn polarity.}
\label{fig:detectors:Kaja-1_neg}
\end{subfigure}
%\begin{tikzpicture}
%\node at (0.0\textwidth,0.0) { \includegraphics[width=0.5\textwidth]{figures/detectors/kaja_th2E_stacked_pos} };
%\node at (0.5\textwidth,0.0) { \includegraphics[width=0.5\textwidth]{figures/detectors/kaja_th2E_stacked_neg} };
%\end{tikzpicture}
\caption{Stacked event distributions of events produced by electron scattering with muon-neutrinos (blue) and electron-neutrinos (yellow) as well as events produced through electron-neutrino charged current nuclear scattering (red), over the $E_e\theta_e^2$ discriminating variable. Shown for each focusing horn polarity.
}
\label{fig:detectors:Kaja-1}
\end{figure}

These events were inserted into the \textsc{WCSim} model of the ESS$\nu$SB water Cherenkov near detector, with a homogeneous distribution over the volume, and an assumed direction for the incident neutrino along the $z$ axis. The full event simulation and reconstruction procedure was applied, as well as the \emph{charged lepton} and \emph{neutrino selection criteria}, which were discussed in Sections~\ref{sct:detectors:ndwc_chlep} and \ref{sct:detectors:ndwc_nuvtx}.

The number of expected events per running year can be found in Table~\ref{tbl:detectors:ndwc_nev_electronscattering}. This corresponds to Table~\ref{tbl:detectors:ndwc_nev_eid} for neutrino-nuclear CC and NC interactions with {\eid}. Comparing these tables, it is clear that electron-scattering events constitute a negligible background to the CC event sample.

{
\begin{table}[phtb]
\footnotesize
%\scriptsize
\centering
\caption{Number of expected neutrino-electron scattering events events per running year, per level of the analysis, per flavour, and per each horn polarity.
\label{tbl:detectors:ndwc_nev_electronscattering}}
\begin{tabular}{ r r r r r r r r r }
                        ~  &  \textbf{Positive polarity}  &                 ~  &                       ~  &                     ~
                           &  \textbf{Negative polarity}  &                 ~  &                       ~  &                     ~  \\
                        ~  &  \textbf{$\nu_\mu$}          &  \textbf{$\nu_e$}  &  \textbf{$\bar\nu_\mu$}  &  \textbf{$\bar\nu_e$}
                           &  \textbf{$\nu_\mu$}          &  \textbf{$\nu_e$}  &  \textbf{$\bar\nu_\mu$}  &  \textbf{$\bar\nu_e$}  \\
\hline
         All interactions  &  \SI{1.50e+04}{}  &  \SI{1580}{}  &  \SI{362}{}  &  \SI{23.0}{}  &  \SI{520}{}  &  \SI{347}{}  &  \SI{6950}{}  &   \SI{229}{}  \\
                  Trigger  &  \SI{1.47e+04}{}  &  \SI{1550}{}  &  \SI{355}{}  &  \SI{22.1}{}  &  \SI{510}{}  &  \SI{338}{}  &  \SI{6800}{}  &   \SI{222}{}  \\
 Charged lepton criterion  &      \SI{7470}{}  &   \SI{812}{}  &  \SI{158}{}  &   \SI{6.4}{}  &  \SI{240}{}  &  \SI{136}{}  &  \SI{3210}{}  &  \SI{77.6}{}  \\
       Neutrino criterion  &      \SI{7000}{}  &   \SI{753}{}  &  \SI{148}{}  &   \SI{6.1}{}  &  \SI{224}{}  &  \SI{127}{}  &  \SI{3000}{}  &  \SI{72.8}{}  \\
\hline
\end{tabular}
\end{table}
}

{
\begin{table}[phtb]
\footnotesize
%\scriptsize
\centering
\caption{Number of expected events per running year that are accepted by the selection criterion on $E_e\theta_e^2$,
at the final level of the analysis, per flavour and interaction type, and per each horn polarity.
\label{tbl:detectors:ndwc_nev_cutthetasquaredekin}}
\begin{tabular}{ r r r r r r r r r }
                        ~  &  \textbf{Positive polarity}  &                 ~  &                       ~  &                     ~
                           &  \textbf{Negative polarity}  &                 ~  &                       ~  &                     ~  \\
                        ~  &  \textbf{$\nu_\mu$}          &  \textbf{$\nu_e$}  &  \textbf{$\bar\nu_\mu$}  &  \textbf{$\bar\nu_e$}
                           &  \textbf{$\nu_\mu$}          &  \textbf{$\nu_e$}  &  \textbf{$\bar\nu_\mu$}  &  \textbf{$\bar\nu_e$}  \\
\hline
  Electron-scattering  &  \SI{6010}{}  &   \SI{653}{}  &   \SI{125}{}  &   \SI{5.0}{}  &  \SI{192}{}  &     \SI{108}{}  &    \SI{2560}{}  &  \SI{60.3}{}  \\
      Charged current  &     \SI{0}{}  &  \SI{1300}{}  &     \SI{0}{}  &   \SI{4.3}{}  &    \SI{0}{}  &    \SI{11.9}{}  &       \SI{0}{}  &  \SI{1480}{}  \\
      Neutral current  &  \SI{0.007}{}  &   \SI{3.8}{}  &  \SI{17.5}{}  &  \SI{0.01}{}  &  \SI{1.0}{}  &    \SI{0.02}{}  &     \SI{214}{}  &   \SI{4.2}{}  \\
\hline
\end{tabular}
\end{table}
}

% FIGURE
\begin{figure}[p]
\centering
\begin{subfigure}[b]{0.495\textwidth}
\centering
\includegraphics[width=\textwidth]{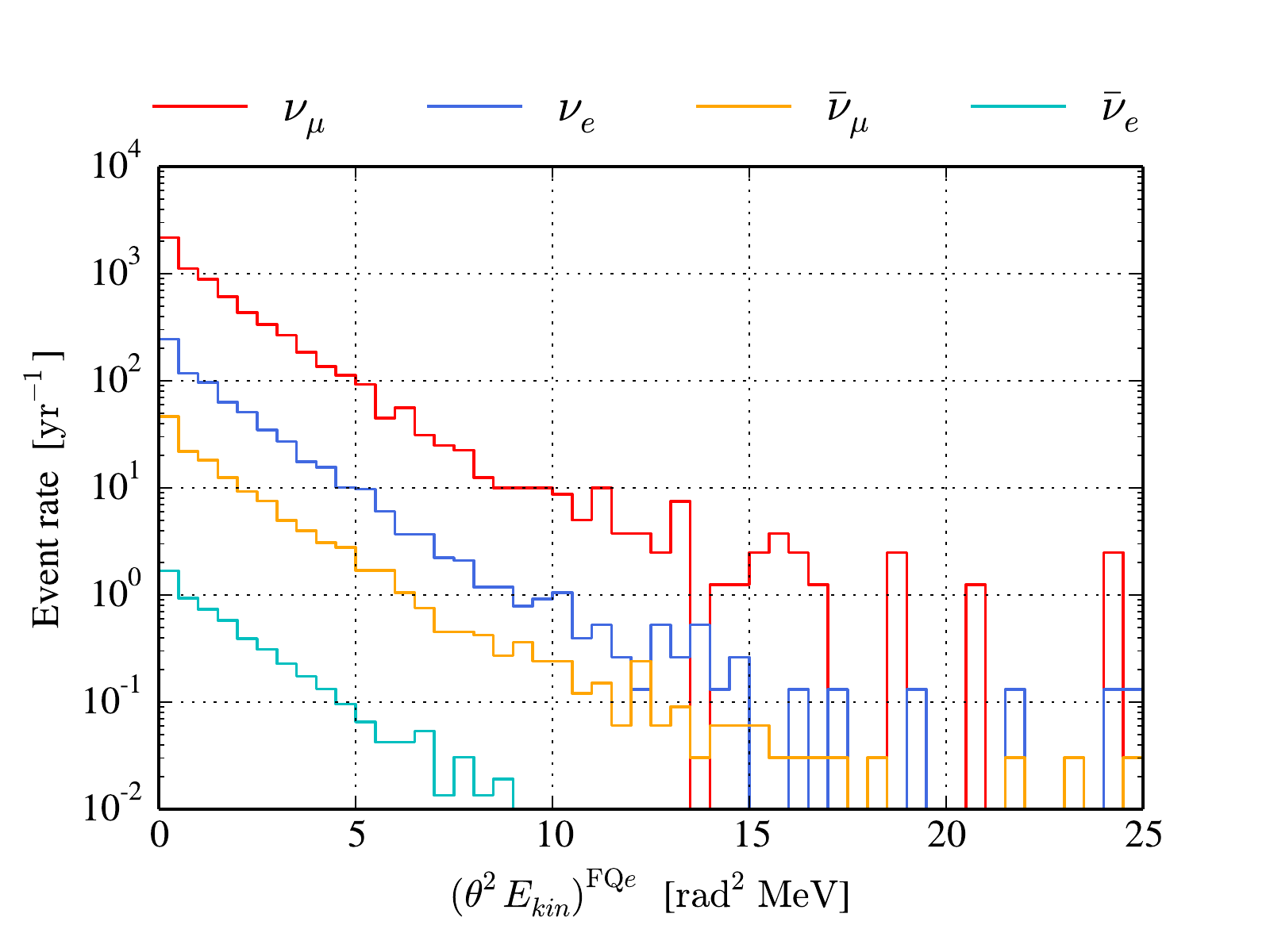}
\caption{}
\label{fig:detectors:ndwc_thetasquaredenergy_poses}
\end{subfigure}
\hfill
\begin{subfigure}[b]{0.495\textwidth}  
\centering 
\includegraphics[width=\textwidth]{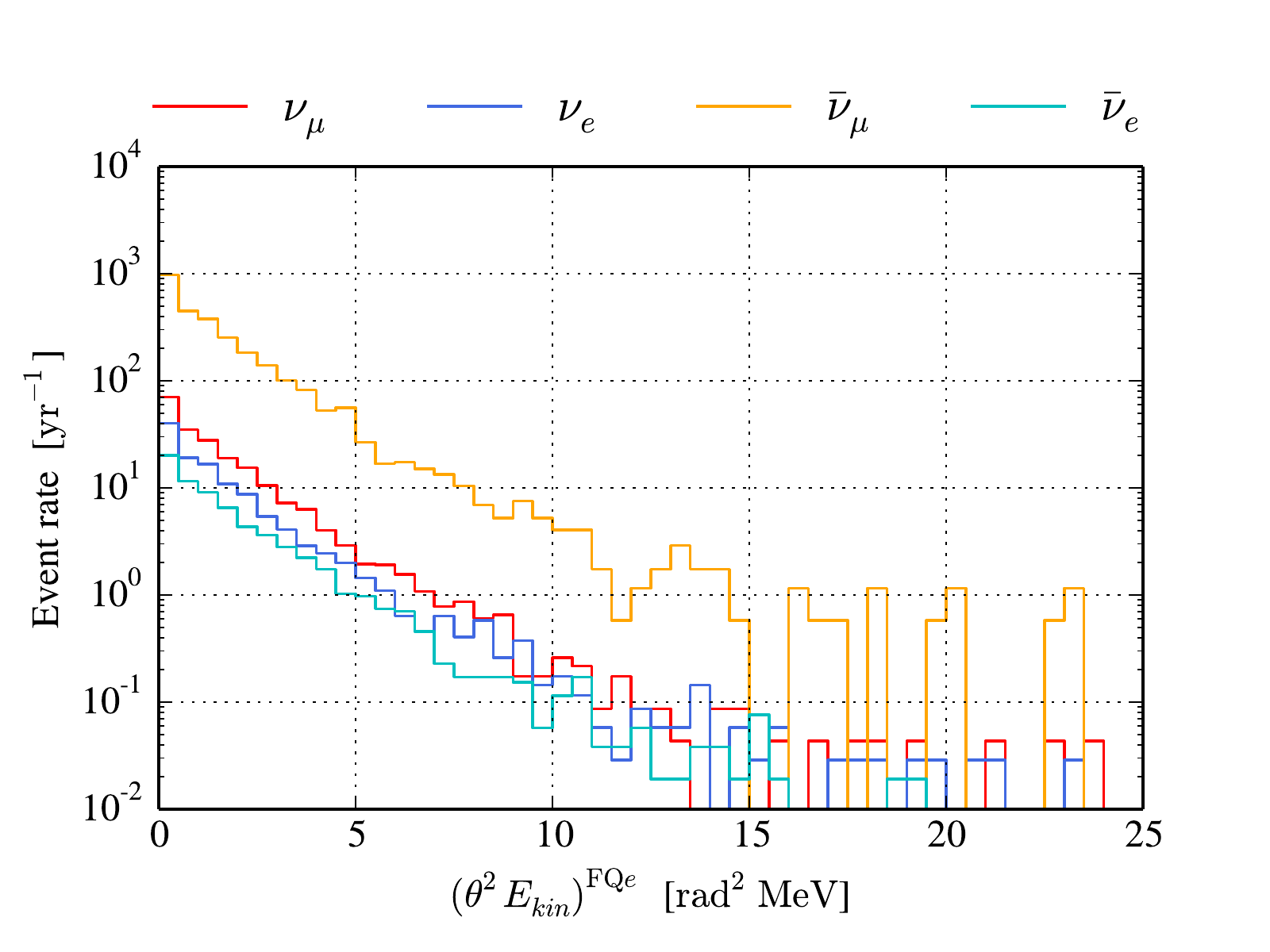}
\caption{}
\label{fig:detectors:ndwc_thetasquaredenergy_neges}
\end{subfigure}
\hfill
\begin{subfigure}[b]{0.495\textwidth}  
\centering 
\includegraphics[width=\textwidth]{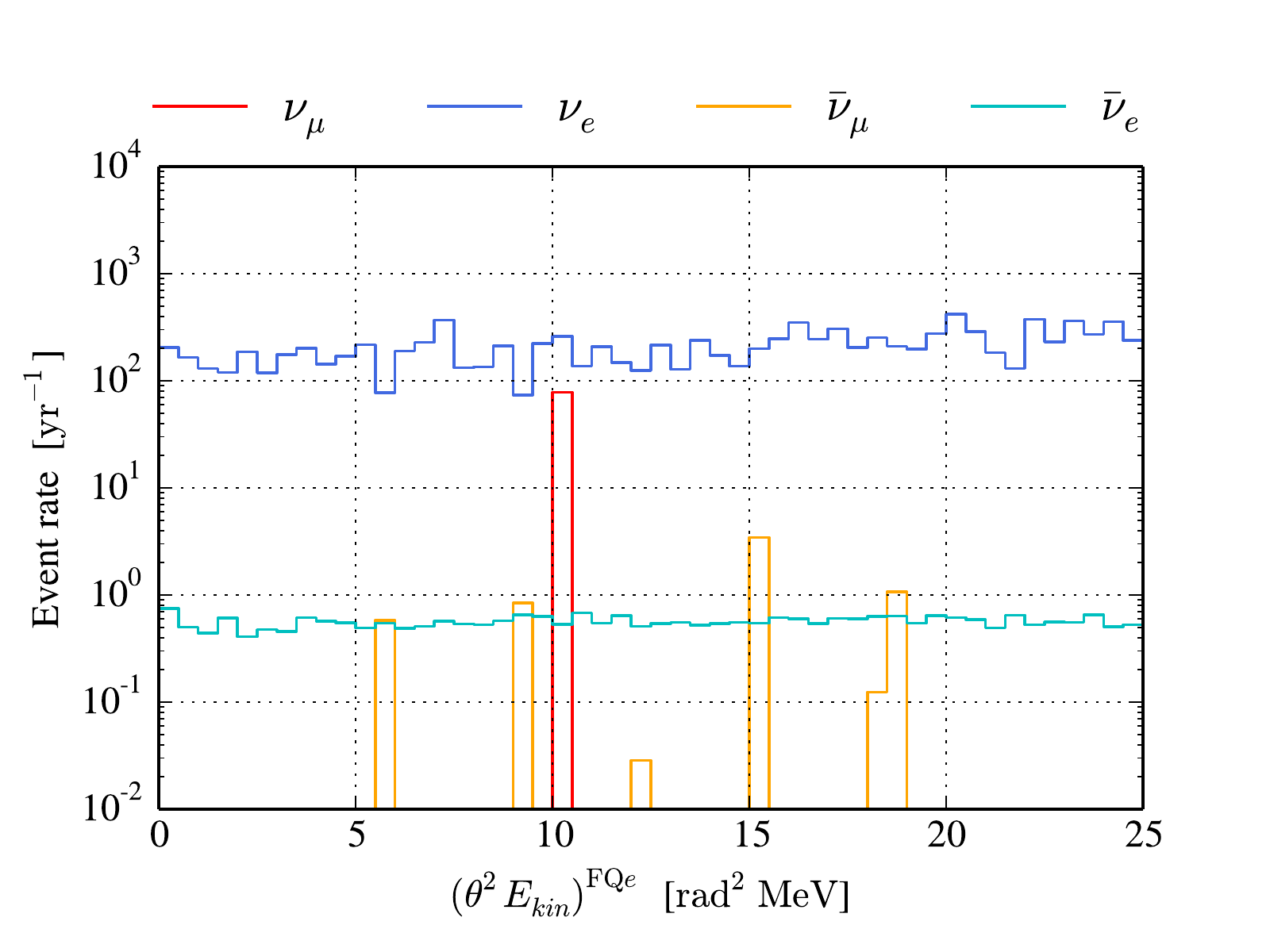}
\caption{}
\label{fig:detectors:ndwc_thetasquaredenergy_poscc}
\end{subfigure}
\hfill
\begin{subfigure}[b]{0.495\textwidth}  
\centering 
\includegraphics[width=\textwidth]{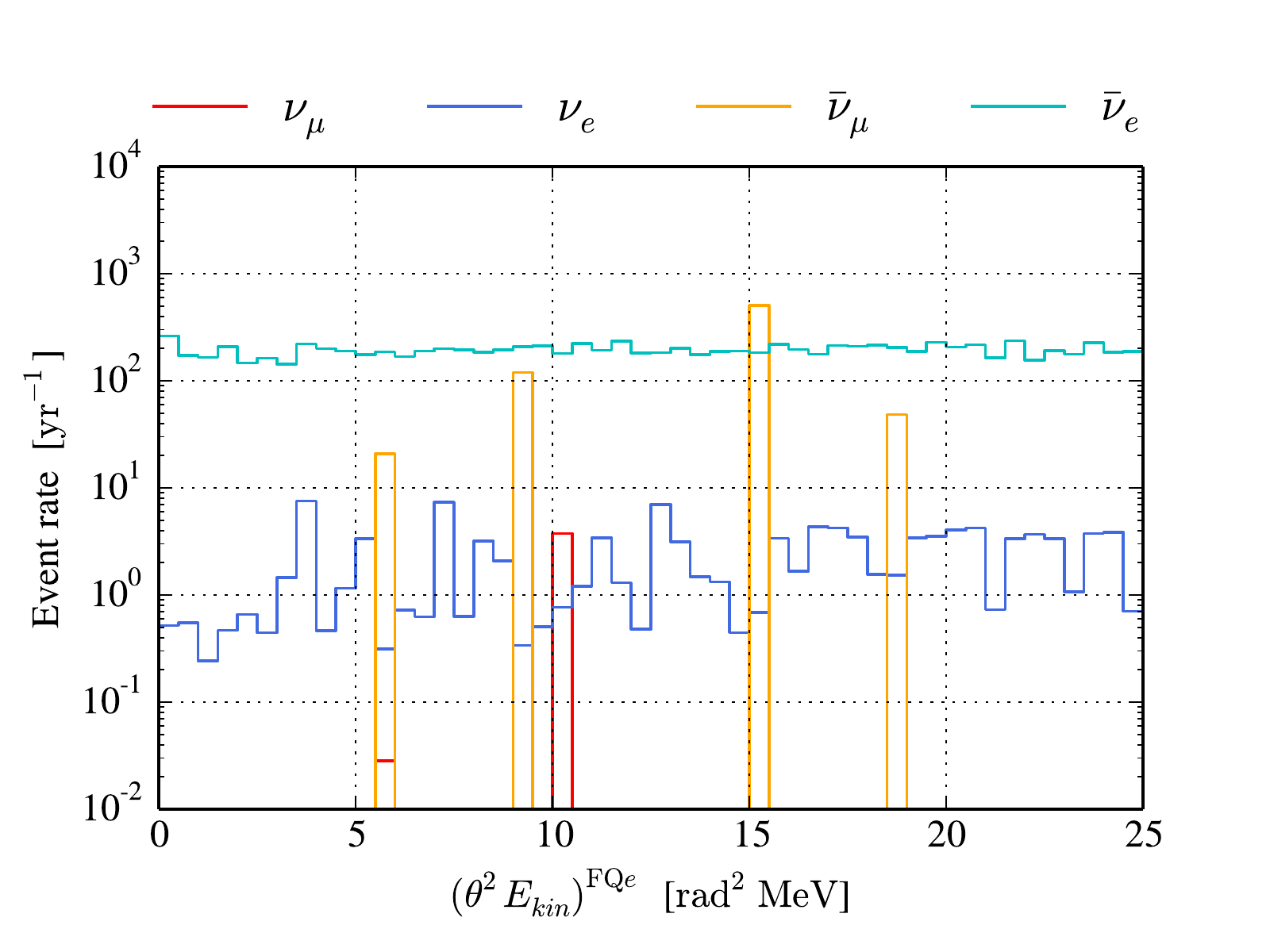}
\caption{}
\label{fig:detectors:ndwc_thetasquaredenergy_negcc}
\end{subfigure}
\caption{Expected event distributions over the reconstructed $E_e\theta_e^2$ variable for {\eid}
(\subref{fig:detectors:ndwc_thetasquaredenergy_poses}, \subref{fig:detectors:ndwc_thetasquaredenergy_neges}) neutrino-electron scattering events and %along with
%(center) CC and (bottom) NC events
CC events (\subref{fig:detectors:ndwc_thetasquaredenergy_poscc}, \subref{fig:detectors:ndwc_thetasquaredenergy_negcc})
after the application of the full event selection scheme, for
positive (\subref{fig:detectors:ndwc_thetasquaredenergy_poses}, \subref{fig:detectors:ndwc_thetasquaredenergy_poscc}),  and negative
(\subref{fig:detectors:ndwc_thetasquaredenergy_neges}, \subref{fig:detectors:ndwc_thetasquaredenergy_negcc}) polarity.
\label{fig:detectors:ndwc_thetasquaredenergy}}
\end{figure}

% FIGURE
\begin{figure}[p]
\centering
\begin{subfigure}[b]{0.495\textwidth}
\centering
\includegraphics[width=\textwidth]{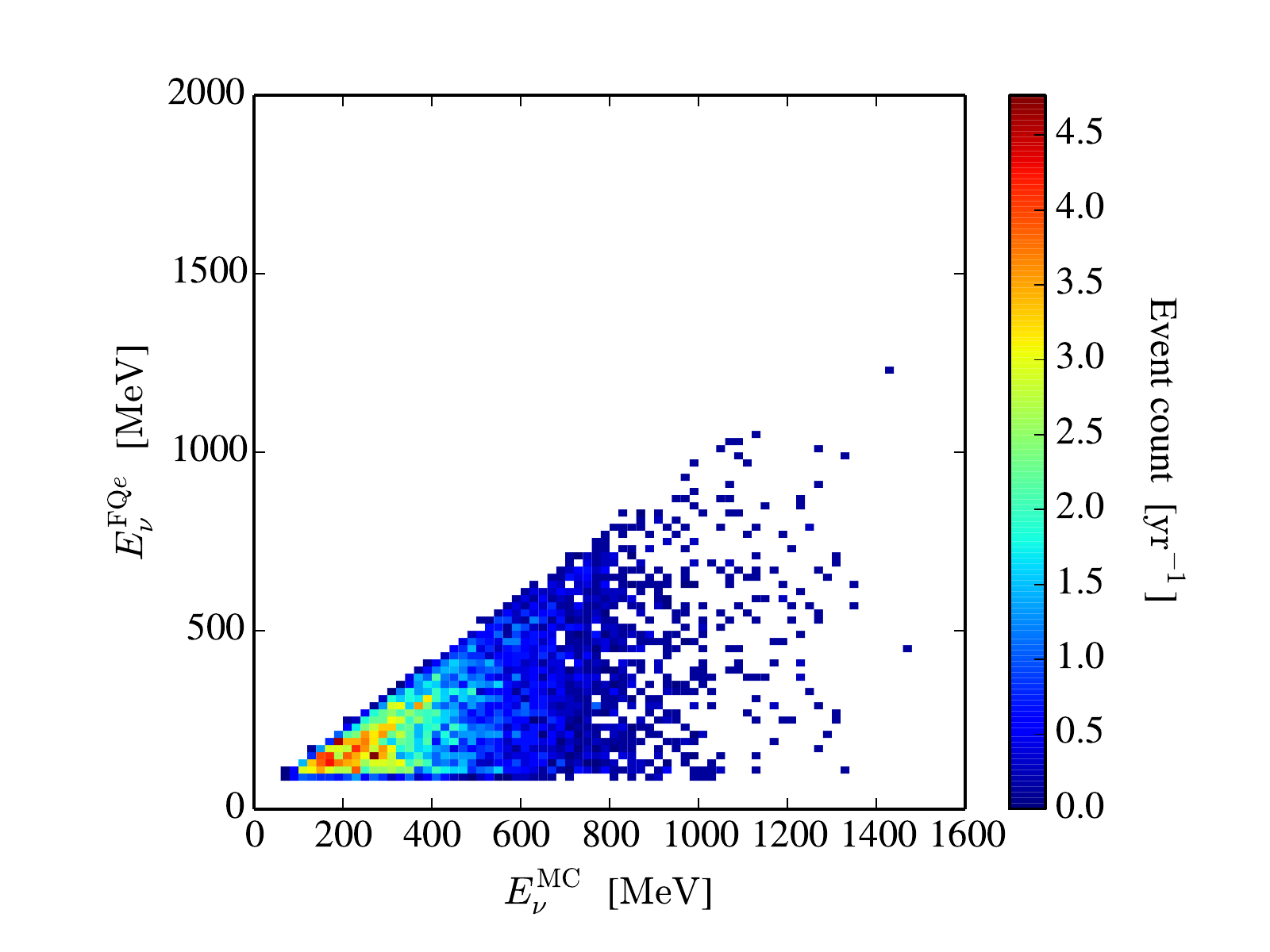}
\caption{}
\label{fig:detectors:ndwc_escatt_ereco_nuepos}
\end{subfigure}
\hfill
\begin{subfigure}[b]{0.495\textwidth}  
\centering 
\includegraphics[width=\textwidth]{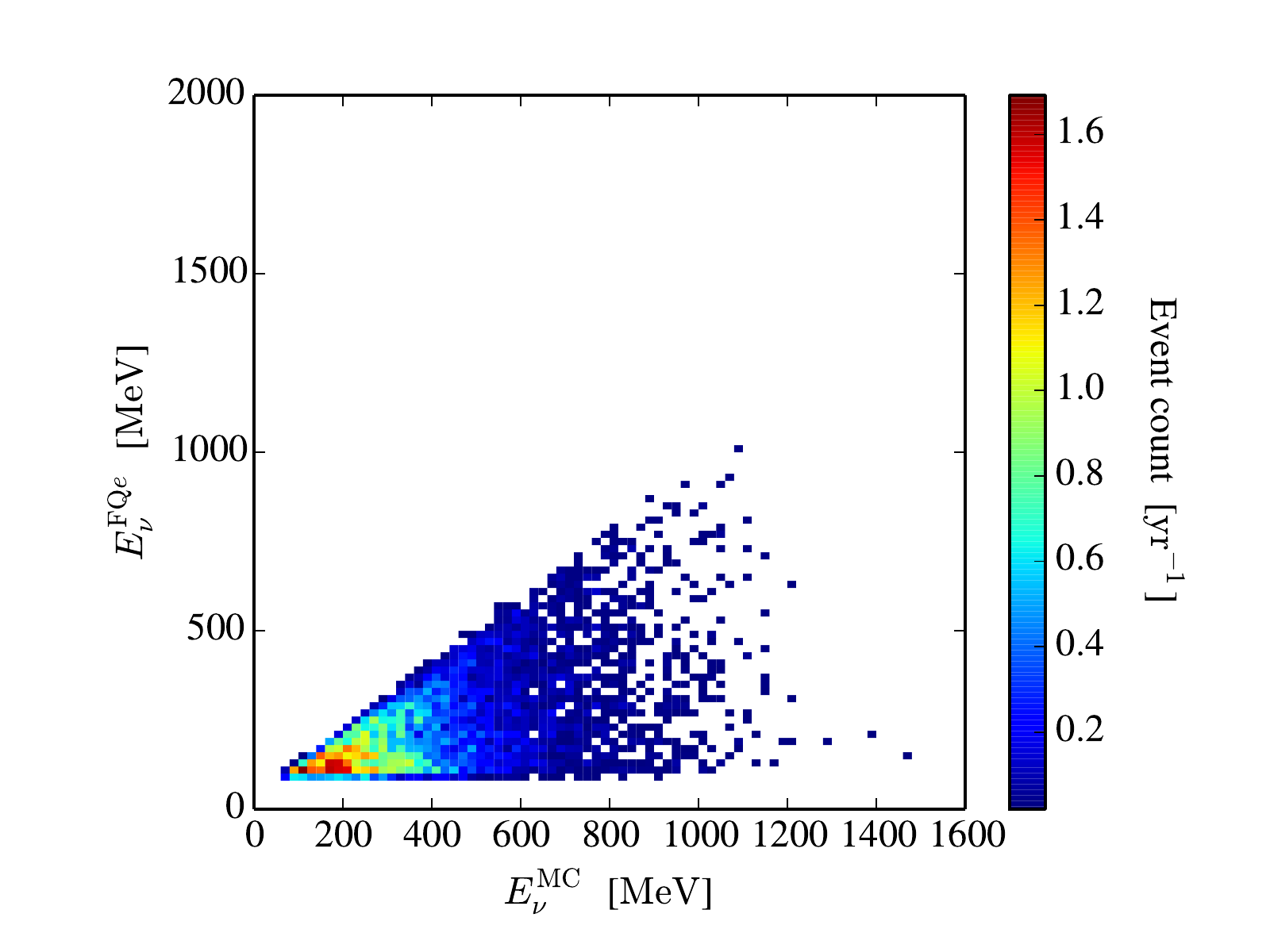}
\caption{}
\label{fig:detectors:ndwc_escatt_ereco_nueneg}
\end{subfigure}
\hfill
\begin{subfigure}[b]{0.495\textwidth}  
\centering 
\includegraphics[width=\textwidth]{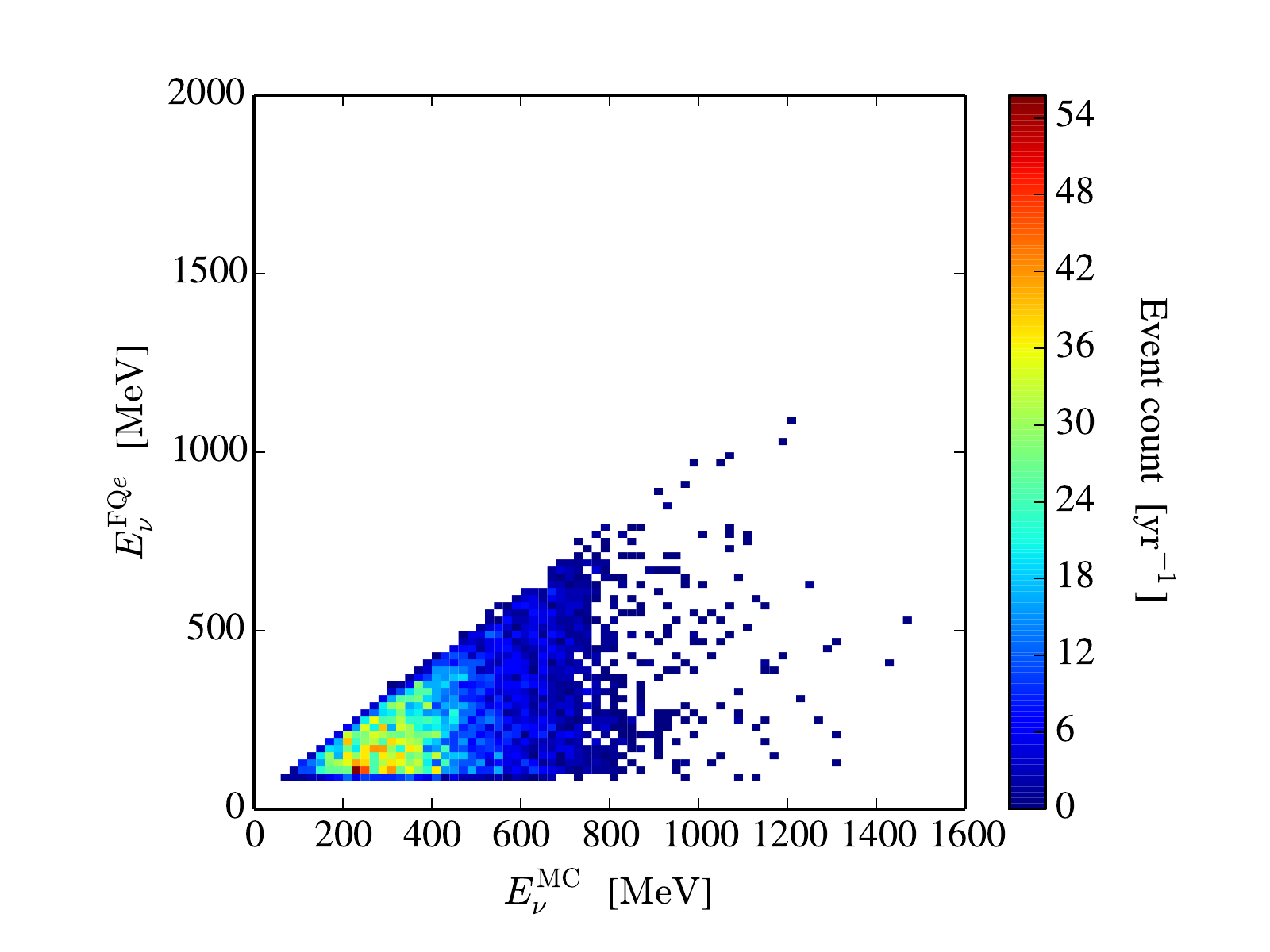}
\caption{}
\label{fig:detectors:ndwc_escatt_ereco_numupos}
\end{subfigure}
\hfill
\begin{subfigure}[b]{0.495\textwidth}  
\centering 
\includegraphics[width=\textwidth]{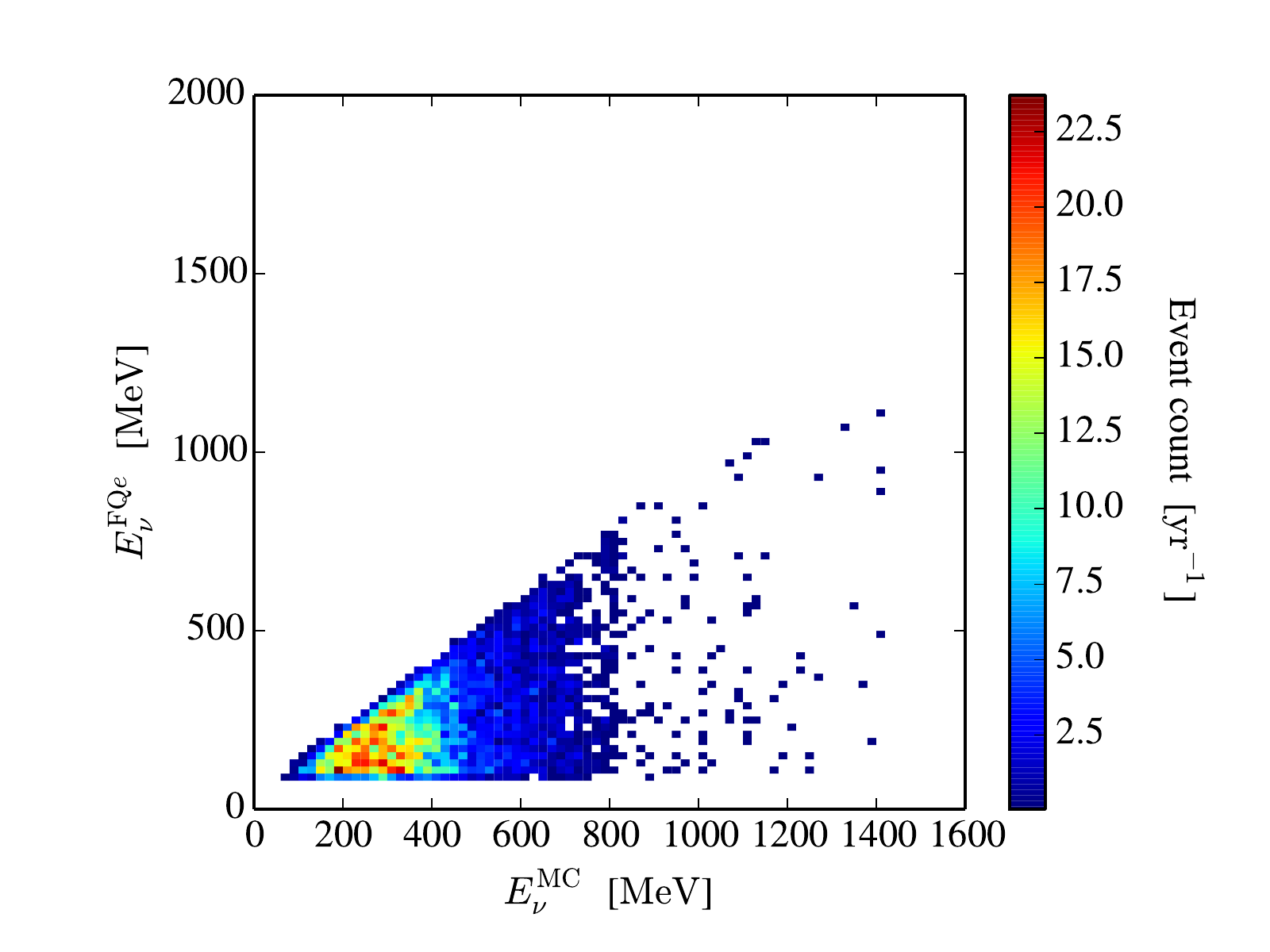}
\caption{}
\label{fig:detectors:ndwc_escatt_ereco_numuneg}
\end{subfigure}
\caption{
Event distributions over true and reconstructed neutrino energy (using Eq.~\ref{eqn:detectors:ndwc_electronscattering_enureco}) for neutrino-electron scattering events, including {\eid} events at the final level of the analysis that pass the $E_e\theta_e^2<\SI{4}{}~\text{rad}^2~\text{MeV}$
selection criterion.
The distributions are shown for positive
(\subref{fig:detectors:ndwc_escatt_ereco_nuepos}, \subref{fig:detectors:ndwc_escatt_ereco_numupos})  and negative
(\subref{fig:detectors:ndwc_escatt_ereco_nueneg}, \subref{fig:detectors:ndwc_escatt_ereco_numuneg}) horn polarity, and for
$\nu_e+\bar\nu_e$ (\subref{fig:detectors:ndwc_escatt_ereco_nuepos}, \subref{fig:detectors:ndwc_escatt_ereco_nueneg})  and  $\nu_\mu+\bar\nu_\mu$
(\subref{fig:detectors:ndwc_escatt_ereco_numupos}, \subref{fig:detectors:ndwc_escatt_ereco_numuneg}).
\label{fig:detectors:ndwc_escatt_ereco}}
\end{figure}

Figure~\ref{fig:detectors:ndwc_thetasquaredenergy} shows the expected event distributions over the reconstructed $E_e\theta_e^2$ variable for {\eid} neutrino-electron scattering events along with CC nuclear interaction events after the application of the full event selection scheme.
NC nuclear interaction events can be fully neglected here.
The neutrino-electron scattering events are observed to dominate the CC and NC events at $E_e\theta_e^2$ values below ${\sim}\SI{4}{\square\rad\MeV}$ (where a smearing has occurred in the reconstructed variables, compare with Fig.~\ref{fig:detectors:Kaja-1}).
Therefore, we place a selection criterion to only accept events with $E_e\theta_e^2<\SI{4}{\square\rad\MeV}$, accepting \SI{91}{\percent} of neutrino-electron scattering events.
Table~\ref{tbl:detectors:ndwc_nev_cutthetasquaredekin} contains the number of expected events per flavor and interaction type after the application of this selection criterion, at which point neutrino-electron scattering events dominate.

The neutrino energy-reconstruction performance for neutrino-electron scattering events is shown in Fig.~\ref{fig:detectors:ndwc_escatt_ereco}.
The poor energy reconstruction is largely attributed to the variability in the energy of the outgoing electron.
This is supported by the well-performing energy reconstruction of the \textsc{fiTQun} package, as is described in Section~\ref{sct:detectors:ndwc_chlep}, as well as the contents of Fig.~\ref{fig:detectors:ndwc_kaja_enuee_comparison}, where the incident neutrino energy for all simulated neutrino-electron scattering events is compared to the (MC-true) outgoing electron energy.

Using the detector response matrix, Fig.~\ref{fig:detectors:ndwc_escatt_ereco}, an unfolding procedure can be applied to transform the reconstructed energy to the incident neutrino energy.
A test of this concept has been performed using the true energy of the outgoing electron as a proxy for the reconstructed neutrino energy, and a comparison between the resulting unfolded neutrino spectrum and the true incident neutrino energy is made in Fig.~\ref{fig:detectors:ndwc_kaja_enu_unfold}.
This unfolding procedure reproduces the incident neutrino spectrum well, which indicates that a similar performance could be expected when applied to the proper reconstructed energy spectrum.

\begin{figure}[htb]
\centering
\begin{tikzpicture}
\node at (0.0\textwidth,0.0) { \includegraphics[width=0.495\textwidth]{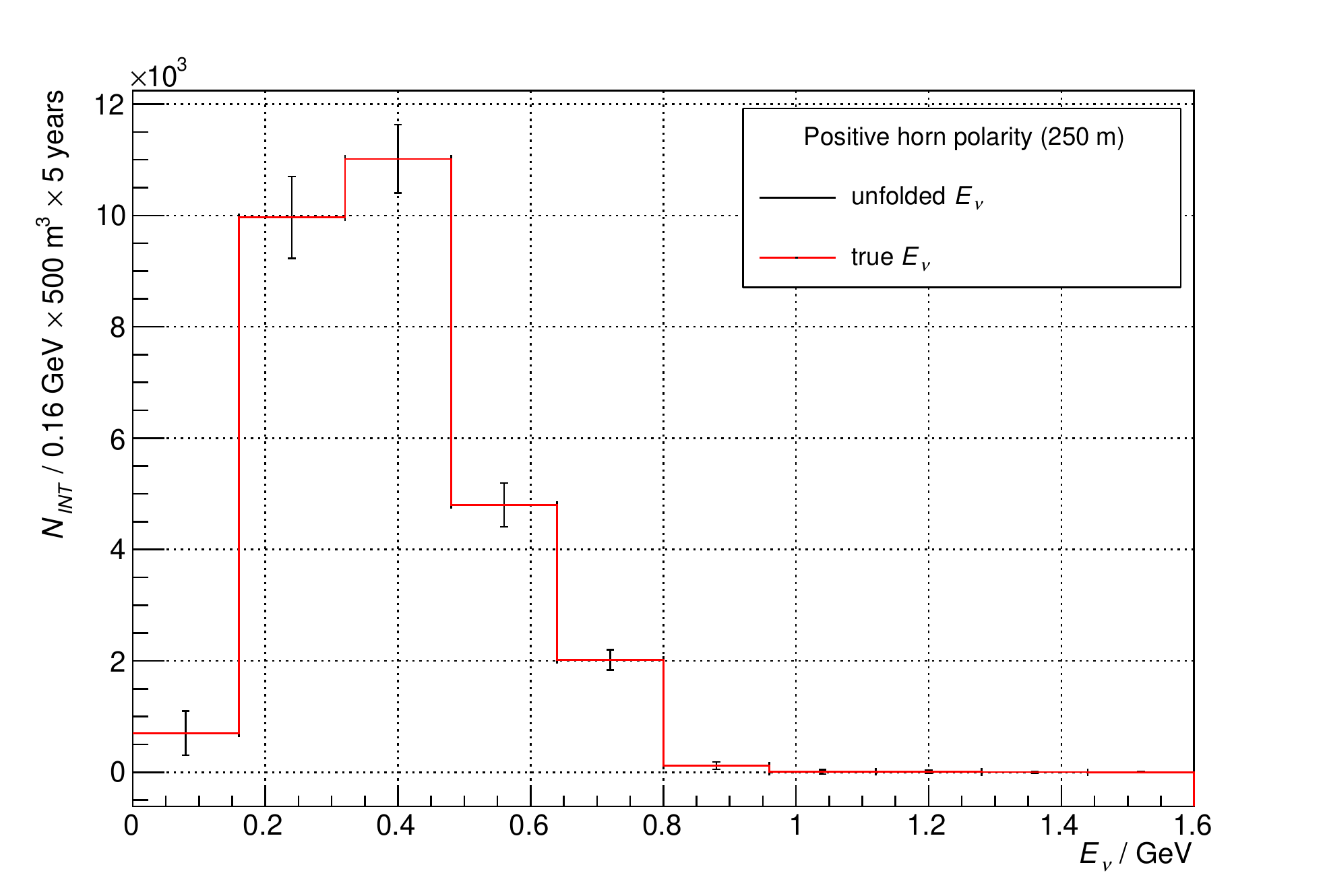} };
\end{tikzpicture}
\caption{Event distributions over the true energy of the incident neutrino sample (red) and the unfolded energy of the same sample (black). The displayed uncertainty interval represents the statistical uncertainty.}
\label{fig:detectors:ndwc_kaja_enu_unfold}
\end{figure}

\subsubsection {Super Fine Grained Detector (SFGD)}
%\bigskip

Several options have been considered for the fine-grained tracker. The baseline was chosen to match the design in development as an upgrade option for the ND280 near detector of the T2K long baseline neutrino oscillation experiment in Japan. It is called the Super Fine-Grained Detector (SFGD) \cite{Sgalaberna:2017}, and consists of a number of scintillator cubes with dimensions $1\times1\times\SI{1}{\cm\cubed}$ each, read out by a three-dimensional pattern of wave-length-shifting optical fibres (see Fig.~\ref{fig:detectors:SFGD-1}). MPPC (Multi-Pixel Photon Counters, equivalent to SiPMs) are connected at each end of the optical fibres. Our baseline is to use $\sim10^6$ cubes, forming a rectangular cuboid with dimensions $1.4\times1.4\times\SI{0.5}{\m\cubed}$. The thickness of 0.5 m along the neutrino beam is chosen in order to let more charged leptons penetrate into the water Cherenkov detector tank, thus allow for combined event analysis. A magnetic field of up to 1\,T oriented perpendicular to the beam is applied in the tracker by a dipole magnet (see Fig.~\ref{fig:detectors:nd_cavern_close}). 

\begin{figure}[p]
    \centering
     \includegraphics[width=0.8\linewidth]{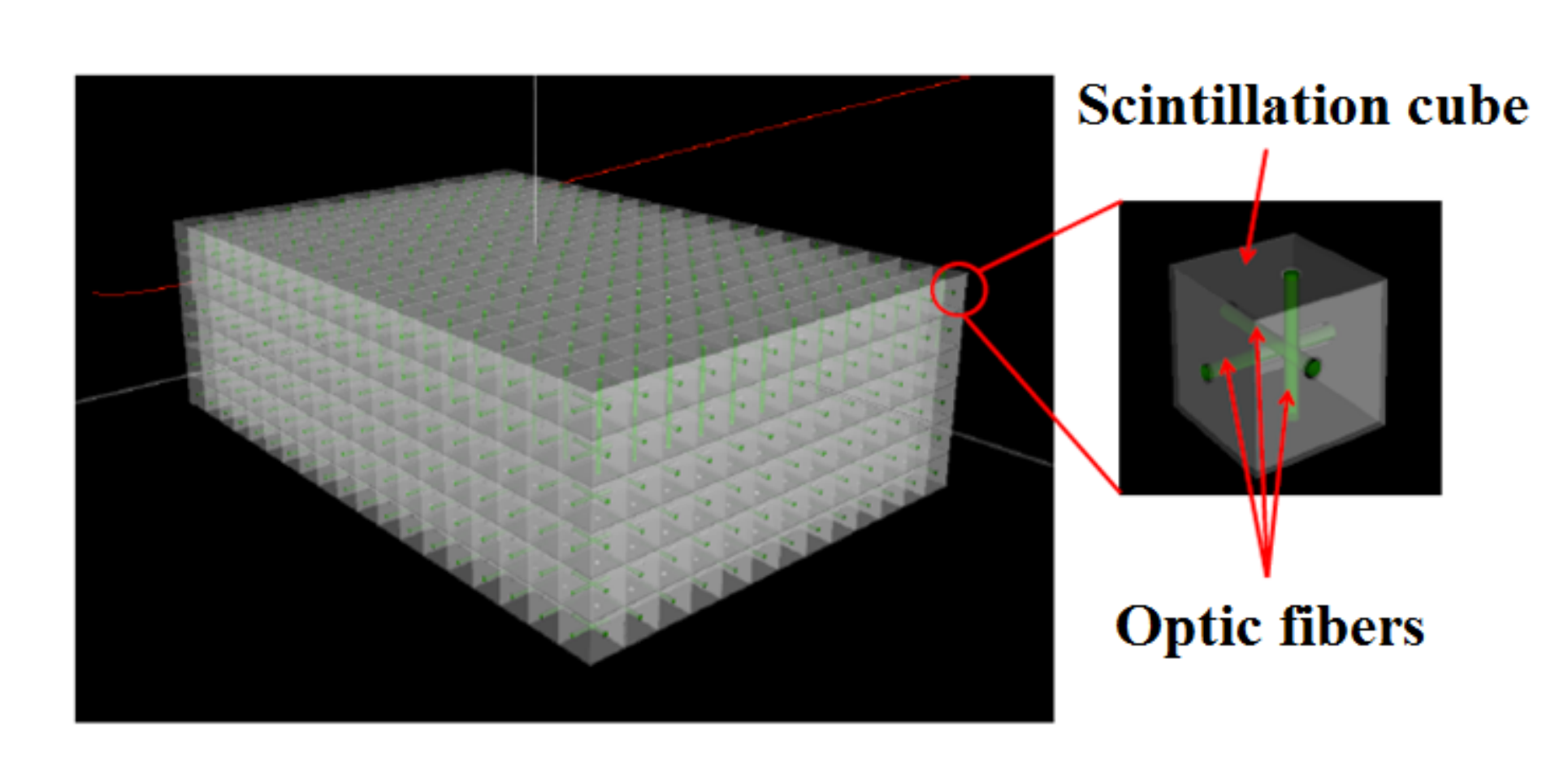} 
    \caption{ Design of the Super Fine-Grained Detector with three-dimensional read-out.\cite{Sgalaberna:2017}}
    \label{fig:detectors:SFGD-1}
\end{figure}

For the purpose of illustration, Figure~\ref{fig:detectors:SFGD_interaction} shows a simulated interaction of a 0.3\,GeV muon neutrino inside an artificially enlarged SFGD, using $200\times200\times200$ scintillator cubes.

\begin{figure}[p]
    \centering
     \includegraphics[width=0.6\linewidth]{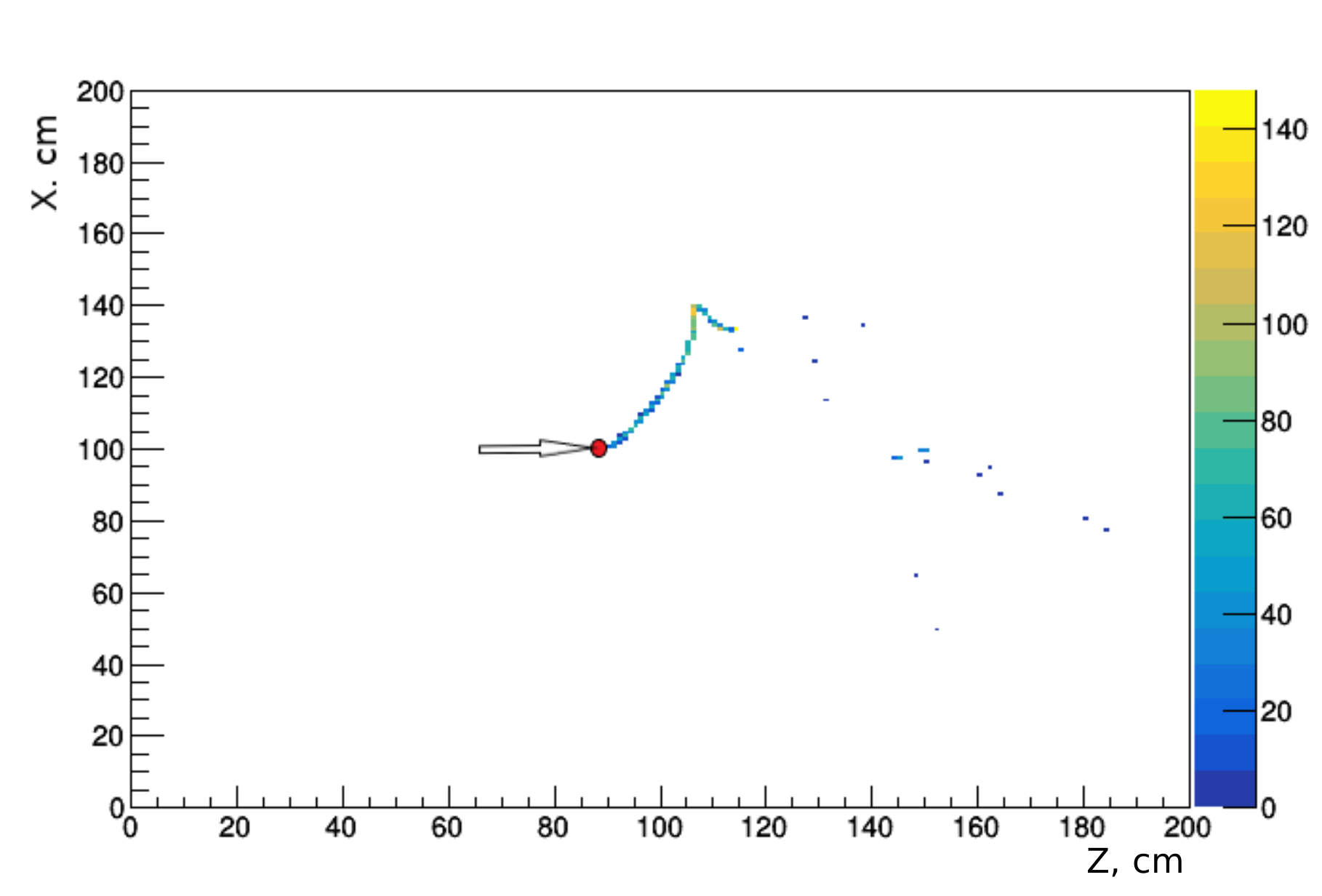} 
    \caption{Display of a quasi-elastic muon neutrino event with a neutrino energy of \SI{0.3}{GeV} in a granular detector made of $200\times200\times200$ cubes. The neutrino comes from the left and interacts at point $(\SI{100}{\cm},\SI{100}{\cm},\SI{91}{\cm})$. The event topology represents decay of the muon into electron and appearance of bremsstrahlung photons. The magnetic field of \SI{0.5}{T} is directed out of the page perpendicular to the image plane. The colour bar shows the number of photoelectrons in the corresponding cell.}
    \label{fig:detectors:SFGD_interaction}
\end{figure}

%\begin{figure}[!htbp]
%    \centering
%     \includegraphics[width=0.9\linewidth]{figures/detectors/SFGD-2.pdf} 
%    \caption{ The SFGD is shown, within its magnet yoke close to the window into the water Cherenkov detector volume, on the right. The emulsion detector in front of it, here with the cover removed, runs on the same set of rails that is used to position the SFGD and the magnet in place. The adjacent room needed for work with the emulsion detector can be seen in Fig.\ref{fig:detectors:nd_cavern_layout}. The SFGD, the magnet and the emulsion detector can be put in and lifted out of the detector room using a crane in the upper staging area.}
%    \label{fig:detectors:SFGD-2}
%\end{figure}

%\subsubsection {Detector performance}
%
%
%\subsubsubsection {Super Fine-Grained Detector}

\subsubsubsection{Neutrino Interactions in the SFGD}

% \begin{figure}[!htbp]
%     \centering
%      \includegraphics[width=0.6\linewidth]{figures/detectors/SFGD_interaction.pdf} 
%     \caption{Display of a quasi-elastic muon neutrino event with a neutrino energy of 0.3 GeV in a granular detector made of $200\times200\times200$ cubes. The neutrino comes from the left and interacts at point (100, 100, 91). The event topology represents decay of the muon into electron and appearance of bremsstrahlung photons.   The magnetic field of 0.5 T is perpendicular to the picture. The colour schema shows the number of photoelectrons in the corresponding cell.}
%     \label{fig:detectors:SFGD_interaction}
% \end{figure}

The number of expected neutrino interactions in the SFGD, obtained by the \textsc{GENIE} neutrino event generator \cite{GENIE:2010, GENIE:2015, GENIE:2021}, is given in Table~\ref{tab:detectors:SFGD_rates}.

{
\begin{table}[htb]
\footnotesize
\centering
\caption{Number of expected interactions per running year in the SFGD listed over the four charged lepton flavours and two horn polarities.
\label{tab:detectors:SFGD_rates}}
\begin{tabular}{ r r r r r }
             \textbf{Positive polarity}  &                 ~  &                ~  &                 ~  &                ~  \\
                                      ~  &  \textbf{$\nu_\mu$}  &   \textbf{$\nu_e$}  &  \textbf{$\bar{\nu}_\mu$}  &   \textbf{$\bar{\nu}_e$}  \\
\hline
                        Charged current  &   \SI{5.82e+04}{}  &      \SI{313.6}{}  &       \SI{122.3}{}  &       \SI{0.6}{}  \\
                        Neutral current  &   \SI{4.00e+04}{}  &      \SI{170.1}{}  &       \SI{118.8}{}  &       \SI{0.5}{}  \\
                                Total    &   \SI{9.82e+04}{}  &      \SI{483.7}{}  &       \SI{241.1}{}  &       \SI{1.1}{}  \\
\hline
                                      ~  &                 ~  &                ~  &                 ~  &                ~  \\
             \textbf{Negative polarity}  &                 ~  &                ~  &                 ~  &                ~  \\
                                      ~  &  \textbf{$\nu_\mu$}  &   \textbf{$\nu_e$}  &  \textbf{$\bar{\nu}_\mu$}  &   \textbf{$\bar{\nu}_e$}  \\
\hline
                        Charged current  &      \SI{531.9}{}  &       \SI{3.9}{}  &       \SI{9009}{}  &      \SI{29.1}{}  \\
                        Neutral current  &      \SI{396.7}{}  &       \SI{2.5}{}  &       \SI{8954}{}  &      \SI{22.8}{}  \\
                                Total    &      \SI{928.6}{}  &       \SI{6.4}{}  &   \SI{1.75e+04}{}  &       \SI{51.9}{}  \\
\hline
\end{tabular}
\end{table}
}

The main task of the SFGD is to reconstruct the topology of neutrino interactions in order to facilitate the cross-section measurement. To do so, we need identification of the outgoing lepton -- that could also be accomplished in the water Cherenkov detector if it penetrates there. Thus, we have to maximise the number of events in which the outgoing lepton enters the water Cherenkov detector. Figure~\ref{fig:detectors:SFGD_mu-tracklenght}, top-right plot, displays the track length of the outgoing negative muons produced by CC interactions of muon neutrinos. In a design where the length of the SFGD along the beam axis is 50 cm, around 12\,\% of those muons enter the water Cherenkov tank and can be identified there. For the muon antineutrino interactions (bottom plots in the same figure), the respective percentage is as high as 20\,\%. These numbers motivate a choice of 50 cm thickness of the SFGD. If its length along the beam axis were \SI{1}{m} and keeping the same total sensitive volume of the detector, the respective values would be halved. 

\begin{figure}[!htb]
    \centering
     \includegraphics[width=0.45\linewidth]{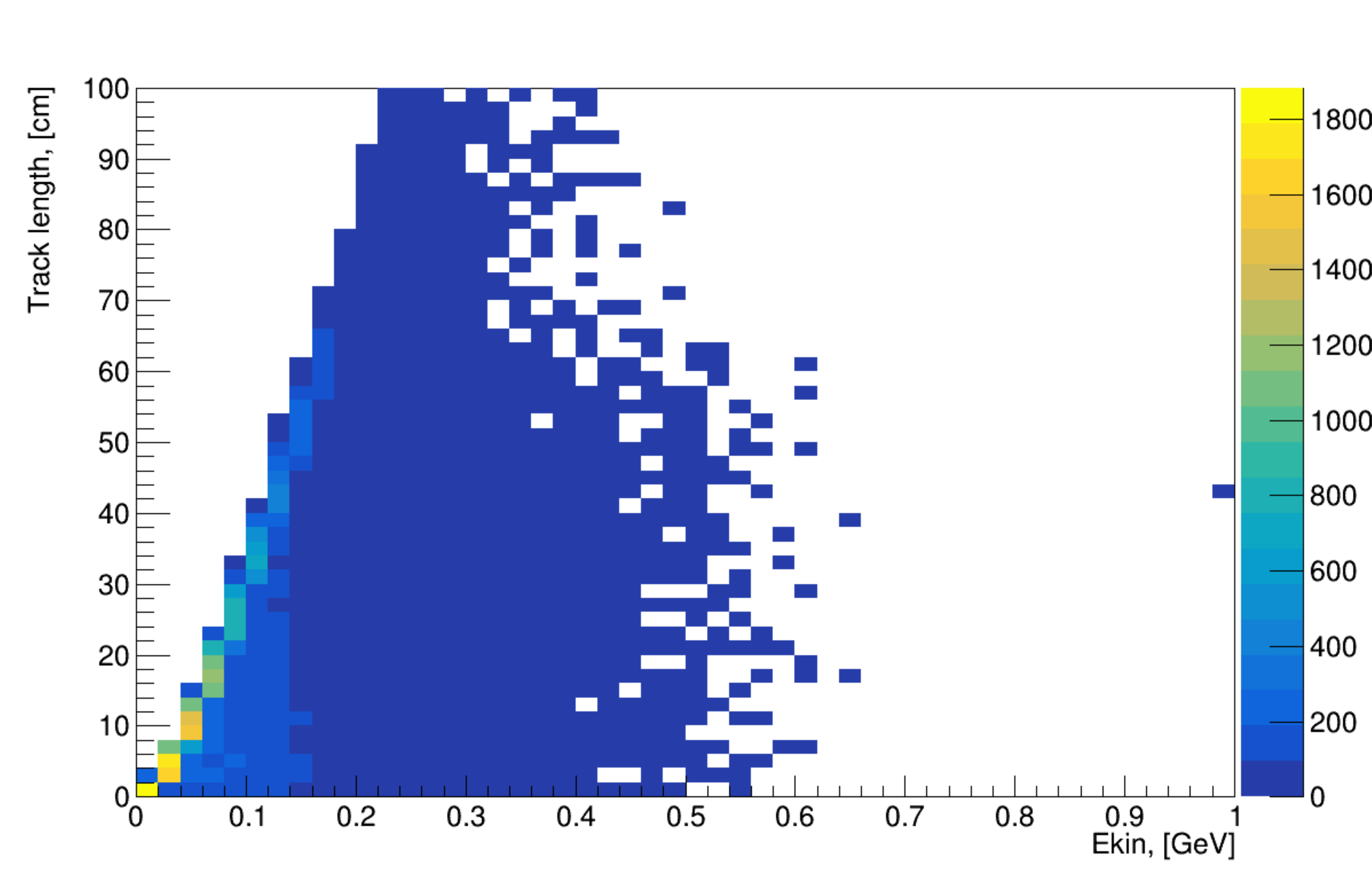} 
     \includegraphics[width=0.45\linewidth]{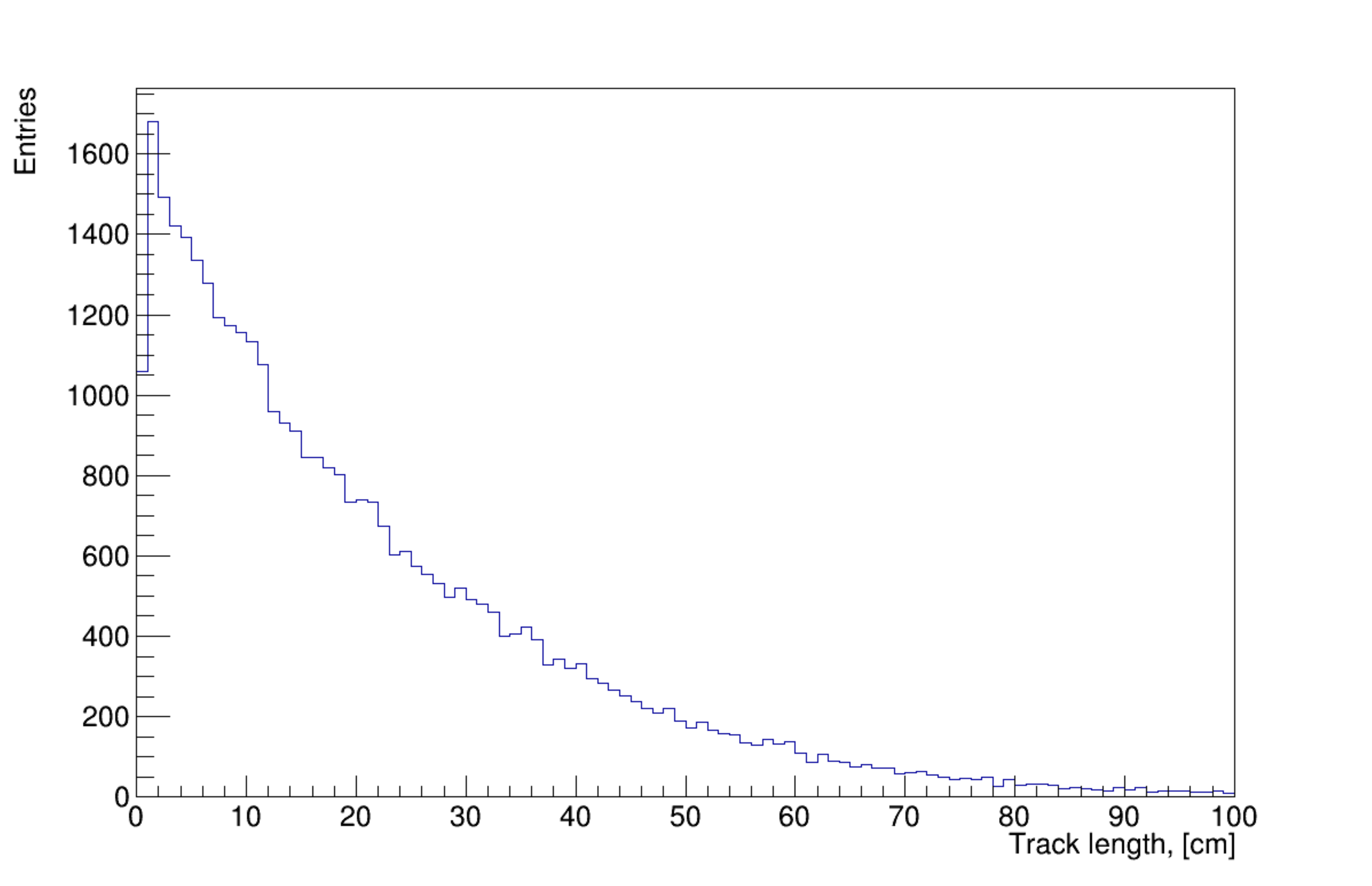} 
     \includegraphics[width=0.45\linewidth]{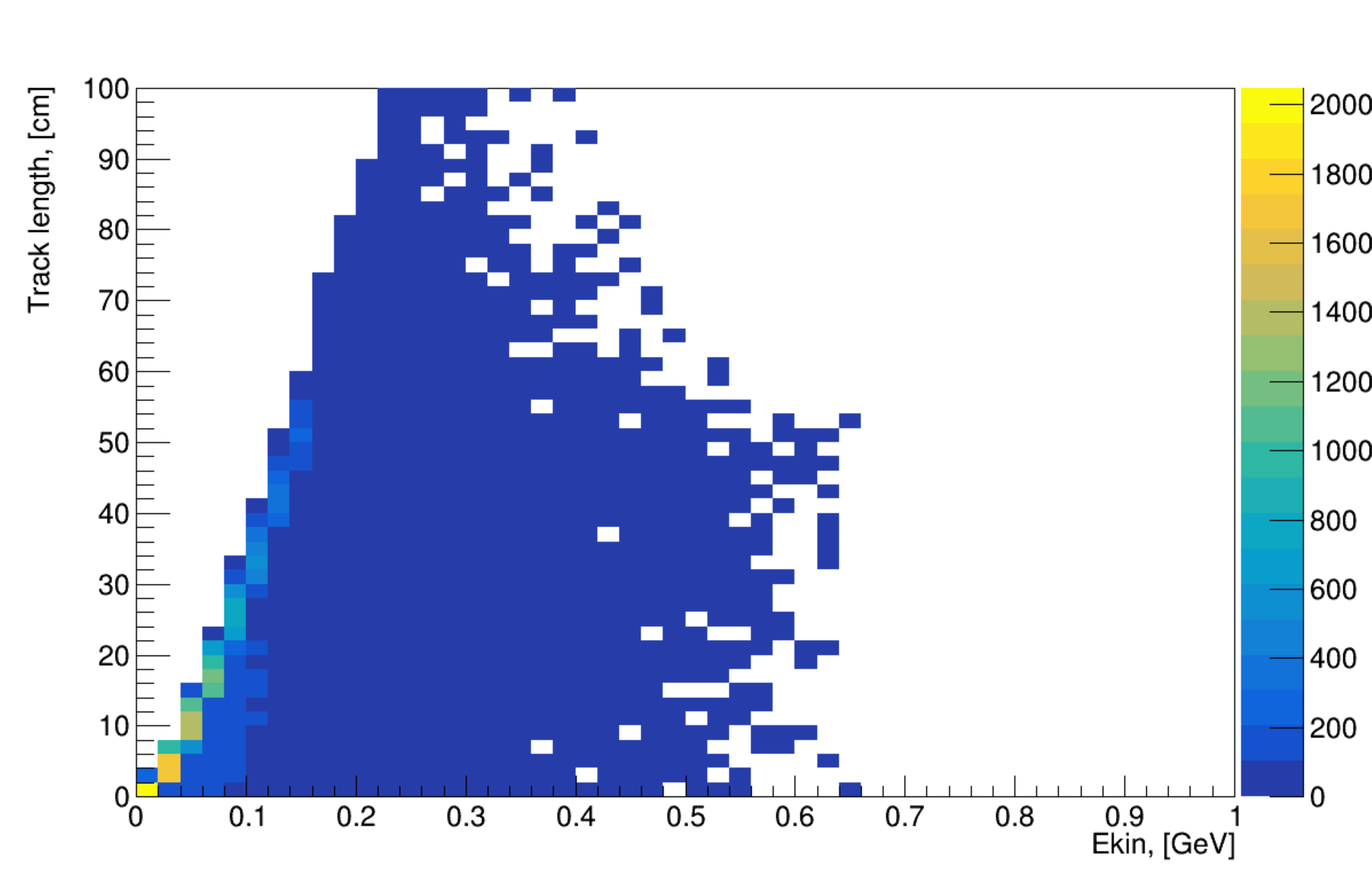} 
     \includegraphics[width=0.45\linewidth]{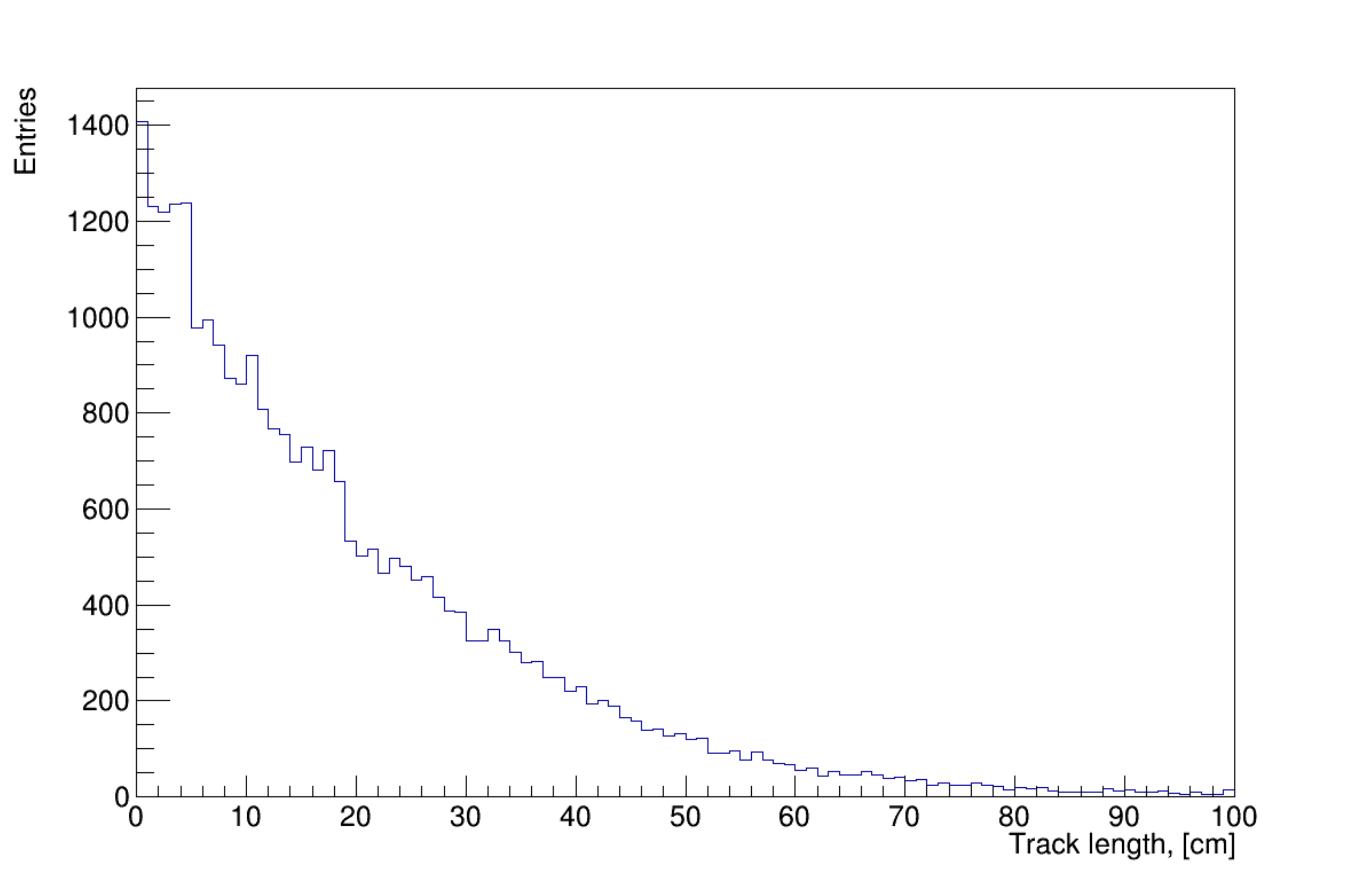} 
    \caption{ Left: Kinetic energy vs. track length of the outgoing negative (upper plots) and positive muons (lower plots) from CC interactions of muon (anti)neutrinos in the SFGD. Right: The respective projections on the track length.}
    \label{fig:detectors:SFGD_mu-tracklenght}
\end{figure}

\subsubsubsection {Magnetic field}
The SFGD detector should be magnetised in order to measure momenta of the outgoing leptons produced by neutrino interactions. Figure~\ref{fig:detectors:SFGD_magfield} shows the distributions of the outgoing charged leptons in muon neutrino and antineutrino interactions versus its scattering angle. While the distribution is flat for negative muons, a clear trend towards forward angles is seen for the positive muons. Thus, the magnetic field should be perpendicular to the neutrino beam. 

\begin{figure}[p]
    \centering
     \includegraphics[width=0.49\linewidth]{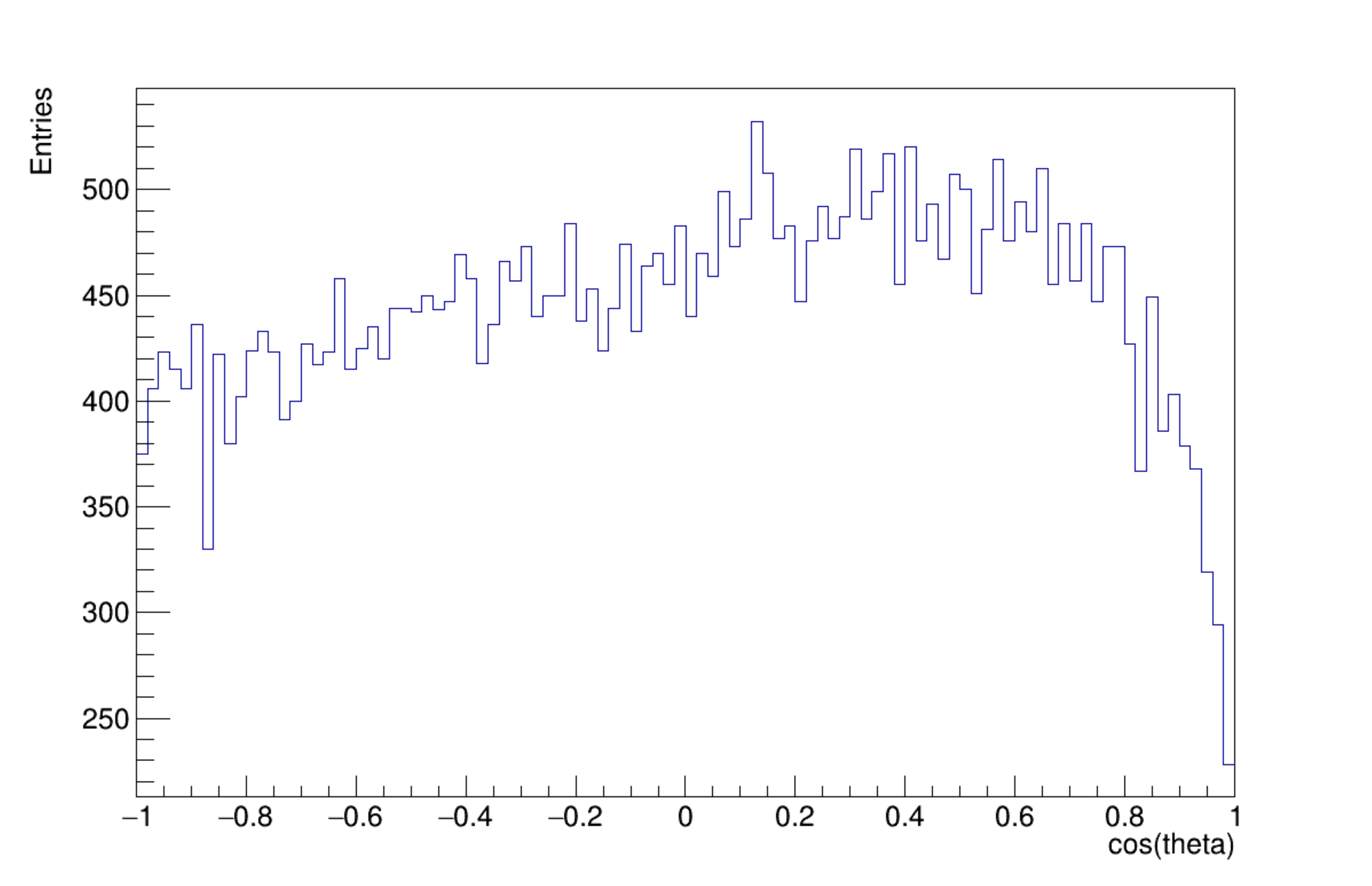} 
     \includegraphics[width=0.49\linewidth]{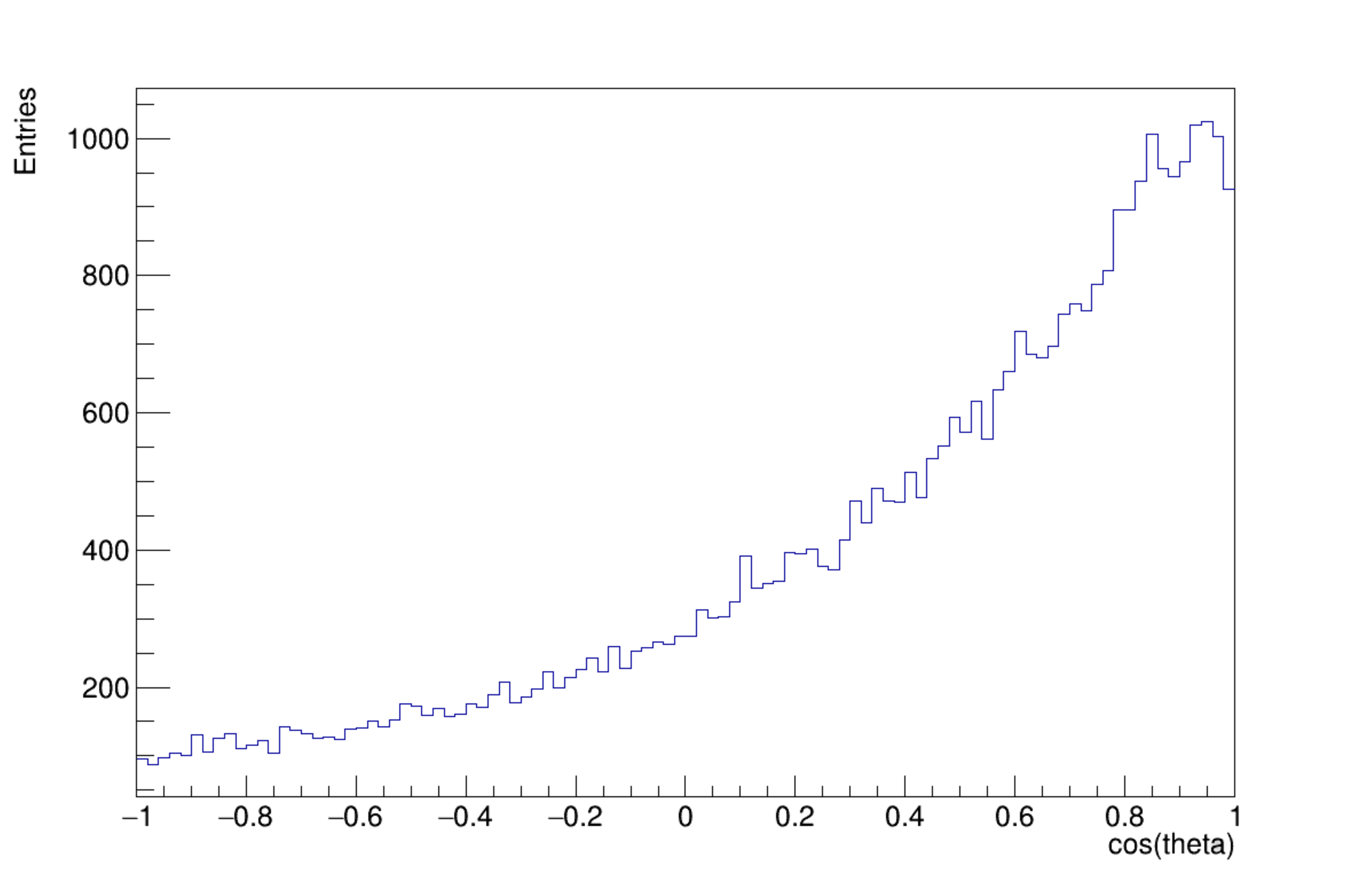} 
    \caption{Left: Distribution over $\cos\theta$ ($\theta$ is the polar angle) of the outgoing negative (left) and positive (right) muons from respective neutrino(antineutrino) CC interactions in the SFGD.}
    \label{fig:detectors:SFGD_magfield}
\end{figure}

\begin{figure}[p]
    \centering
     \includegraphics[width=0.5\linewidth]{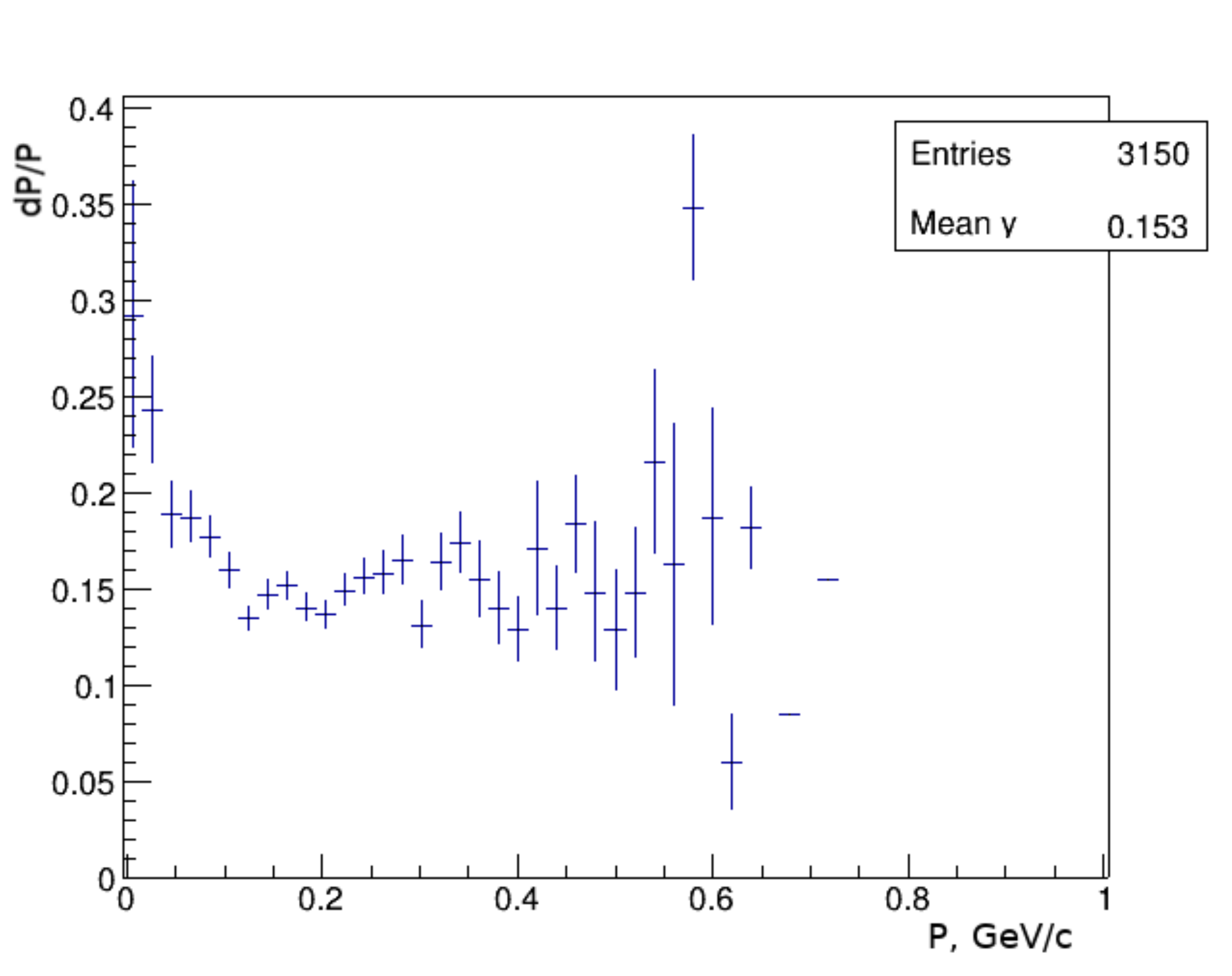} 
    \caption{Dependence of the full muon momentum resolution $\Delta p/p$ versus the true muon momentum over the neutrino beam spectrum with a magnetic field of \SI{1}{T}.}
    \label{fig:detectors:SFGD_mures}
\end{figure}

Theory predicts that the resolution of muon momentum improves with increasing the magnetic field. As expected, our simulations prove the well-known dependence between the resolution of the momentum measurement and the magnetic field. Another observation is that the track length is an essential factor for the quality of the momentum measurement (the longer the track, the better the measurement) while the angle of the track with respect to the beam axis is not particularly important.
 
Due to anticipated technical difficulties to build a relatively large conventional (``warm'') magnet with a field exceeding \SI{1}{T}, a magnetic field strength of \SI{1}{T} is retained. Figure~\ref{fig:detectors:SFGD_mures} shows the dependence of the $\mu^-$ full momentum resolution $\Delta p/p$ versus the true momentum over our neutrino spectrum for this field. The resolution is in the vicinity of 15--20\,\% with no strong momentum dependence. Theoretically expected growth with the momentum is not seen, presumably due to the strong influence of track length on the resolution.

\subsubsubsection {Neutrino Energy Reconstruction}
  
Figure~\ref{fig:detectors:SFGD_erec} shows the relative difference between the true and reconstructed muon neutrino energy. The calculation assumes CCQE interaction in the SFGD. The neutrino energy is calculated from its kinematic dependence on the outgoing muon momentum. In addition, the proton kinetic energy in the SFGD is estimated using the amount of light it produces in the scintillator cubes. Then, an attempt is made to obtain the total neutrino energy as a sum of the kinematically calculated energy and the proton kinetic energy; the result is shown in the figure. The obtained resolution averaged over the neutrino spectrum is \SI{69}{MeV}. The right tail is due to apparent double counting of the proton energy. 
 
\begin{figure}[thb]
    \centering
     \includegraphics[width=0.55\linewidth]{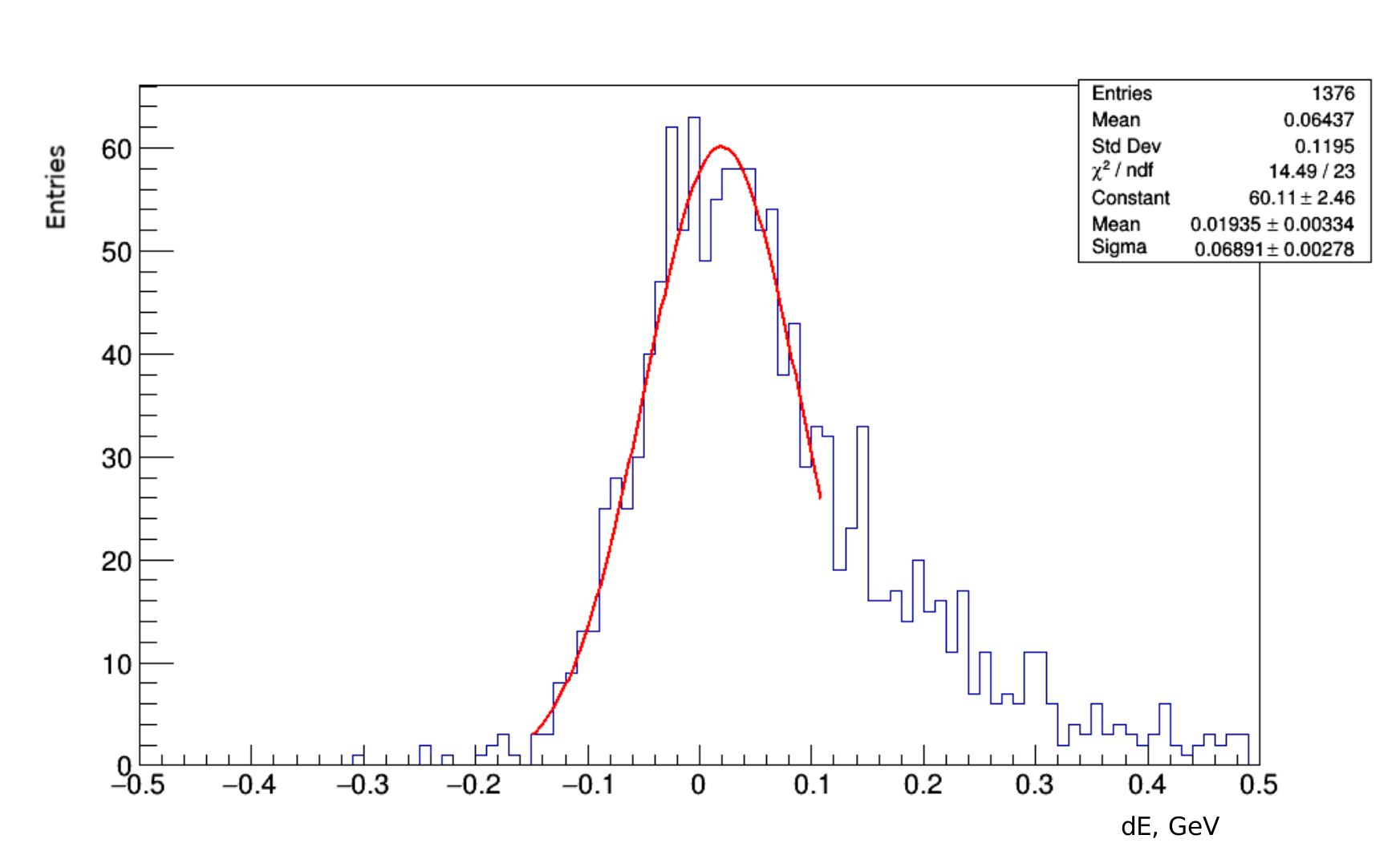} 
    \caption{ Relative difference between the reconstructed and true muon neutrino energy (see the text for details).}
    \label{fig:detectors:SFGD_erec}
\end{figure}

In parallel, reconstruction algorithms based on machine-learning methods for estimation of the SFGD performance have been developed. They are based on the software package Toolkit for Multivariate Data Analysis with ROOT (\textsc{TMVA)}) \cite{TMVA:2009}. In what follows, we use reconstructed values of the calorimetric variables (deposited energy in the scintillator cubes), Monte Carlo truth for the electron kinematic variables (momentum and angle), Monte Carlo truth for the muon kinematic variables (momentum and angle) and/or reconstructed with the \textsc{GenFit} \cite{GenFit:2010} package muon momentum,  where relevant. 
 
In Fig.\ref{fig:detectors:SFGD_muerecml} and Fig.\ref{fig:detectors:SFGD_eerecml}, we show the results of the application of the TMVA methods for reconstruction of muon and electron neutrino energies. 

\begin{figure}[!htbp]
    \centering
     \includegraphics[width=0.9\linewidth]{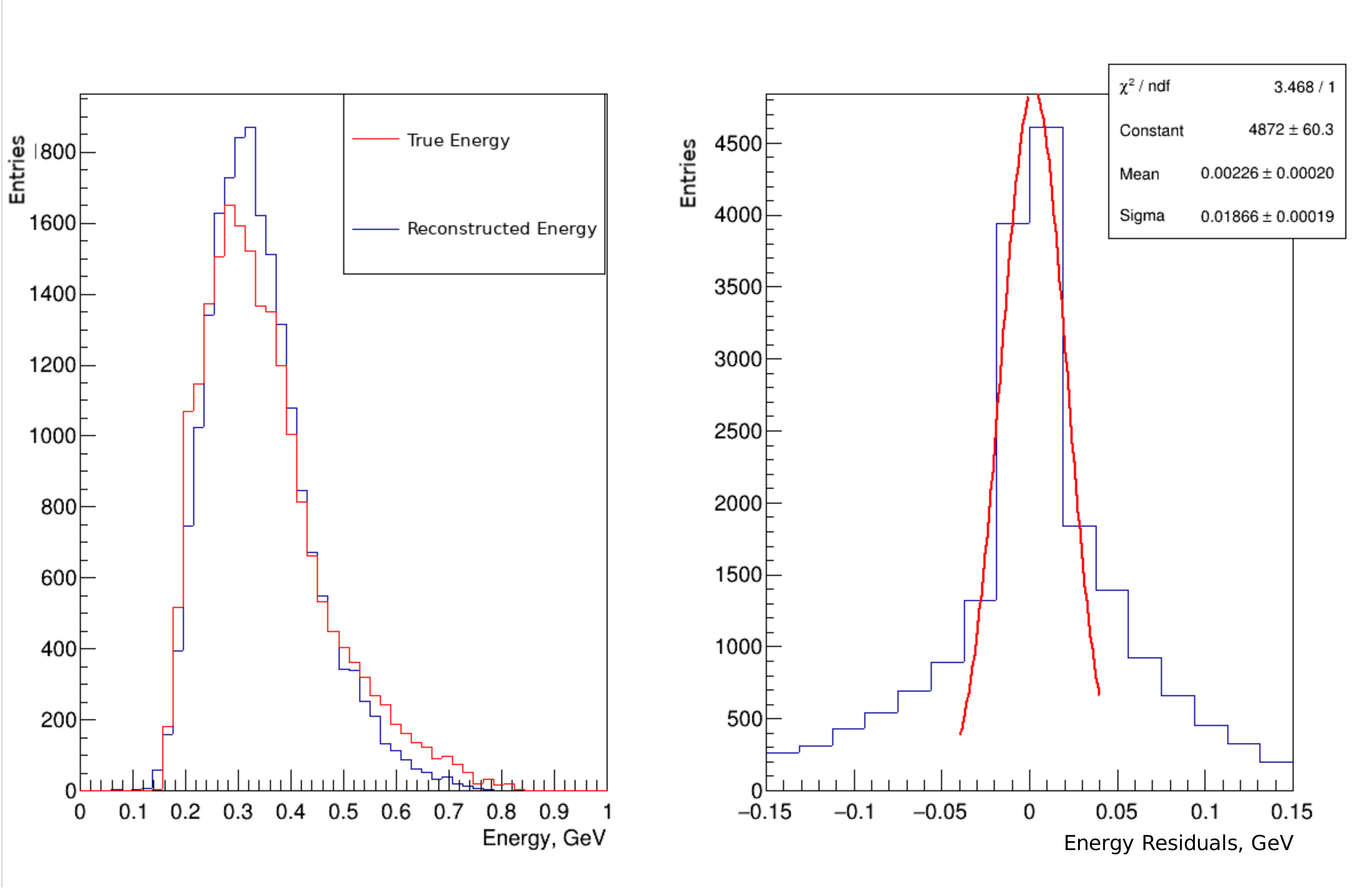} 
    \caption{Muon neutrino energy reconstruction using fitted muon momentum. The right hand plot shows the residuals, i.e., the difference between the true and reconstructed energy.}
    \label{fig:detectors:SFGD_muerecml}
\end{figure}

\begin{figure}[!htbp]
    \centering
     \includegraphics[width=0.9\linewidth]{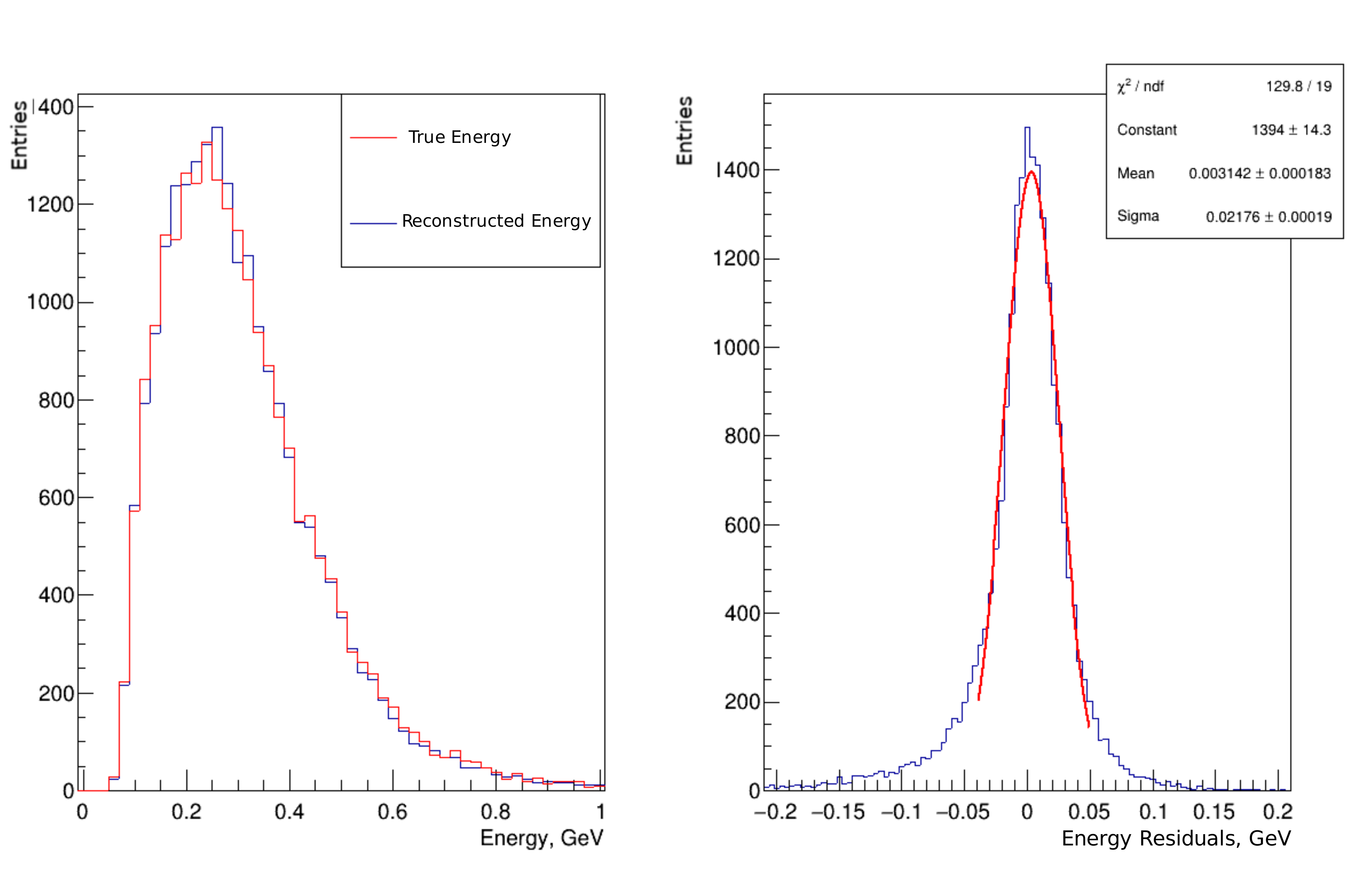} 
    \caption{Electron neutrino energy reconstruction. The right-hand plot shows the residuals, i.e., the difference between the true and reconstructed energy.}
    \label{fig:detectors:SFGD_eerecml}
\end{figure}

The results show that the achieved resolution, integrated over the spectrum, is {$\sim$}\SI{20}{MeV} and that the relative one, $(E_{true}-E_{rec})/E_{true}$, is of the order of 6\,\% at \SI{350}{MeV}. It is important to keep in mind that these are obtained with the fitted muon momentum for muon neutrino interactions and with Monte Carlo truth for the electron  momentum and angle in the case of electron neutrino interaction.

\subsubsubsection {CC and NC event separation}

Figure~\ref{fig:detectors:SFGD_muCCNC} and Fig.~\ref{fig:detectors:SFGD_eCCNC} demonstrate the performance of the SFGD in separation of CC from NC neutrino interactions applying TMVA methods. Our results show that the best training method is the Boosted Decision Tree with Gradient Boosting (BDTG). 

\begin{figure}[!htbp]
    \centering
     \includegraphics[width=0.5\linewidth]{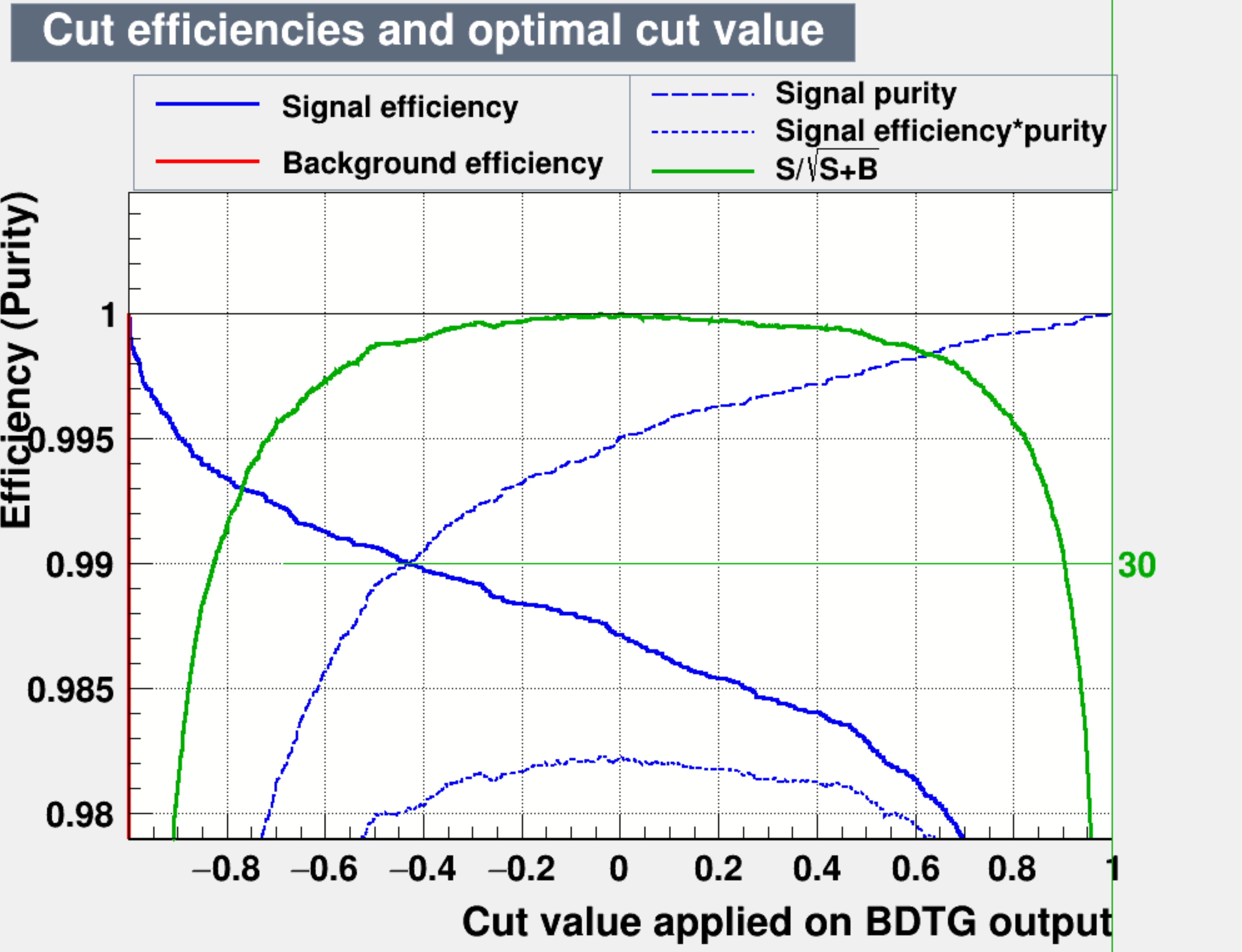} 
     \includegraphics[width=0.5\linewidth]{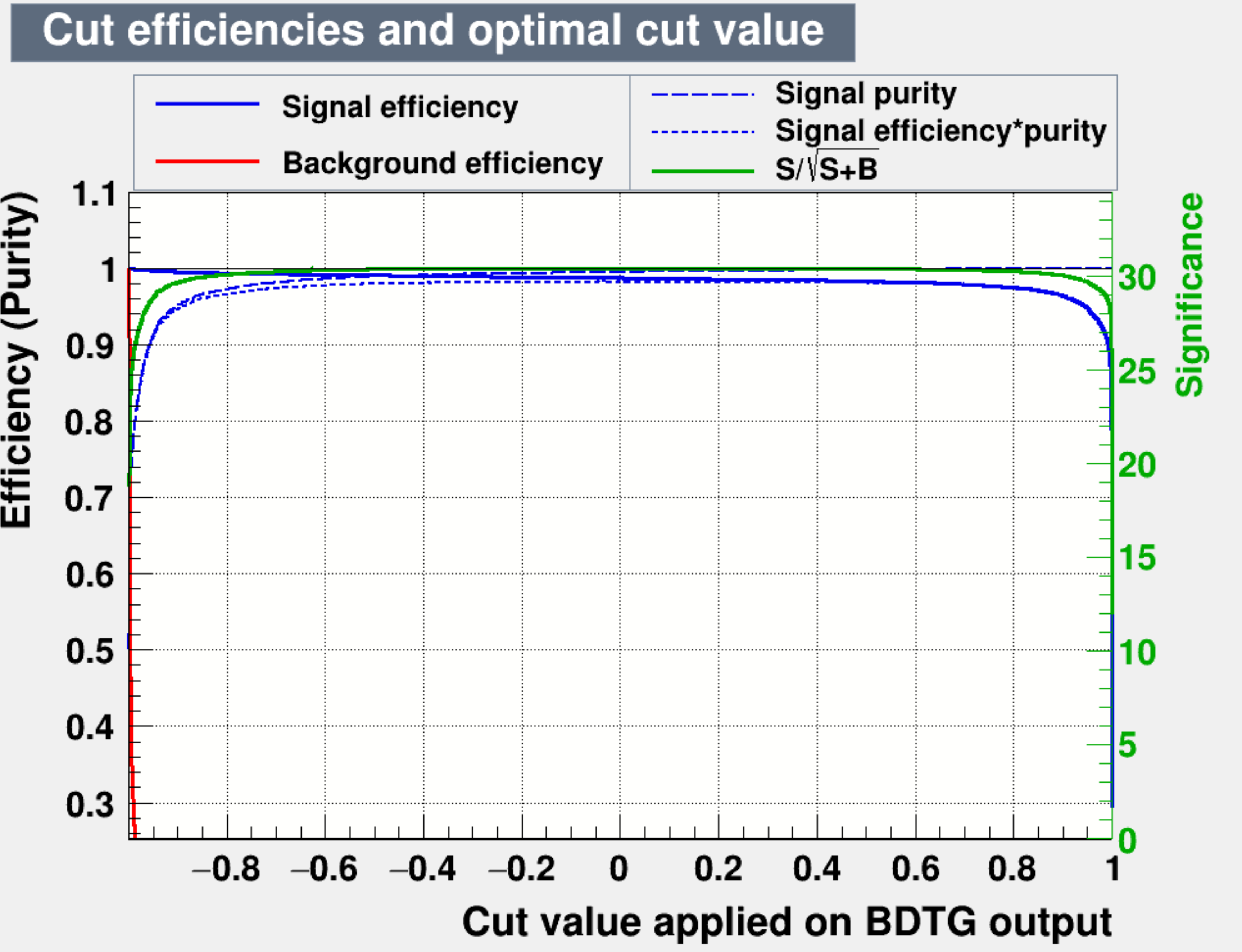} 
     \includegraphics[width=0.5\linewidth]{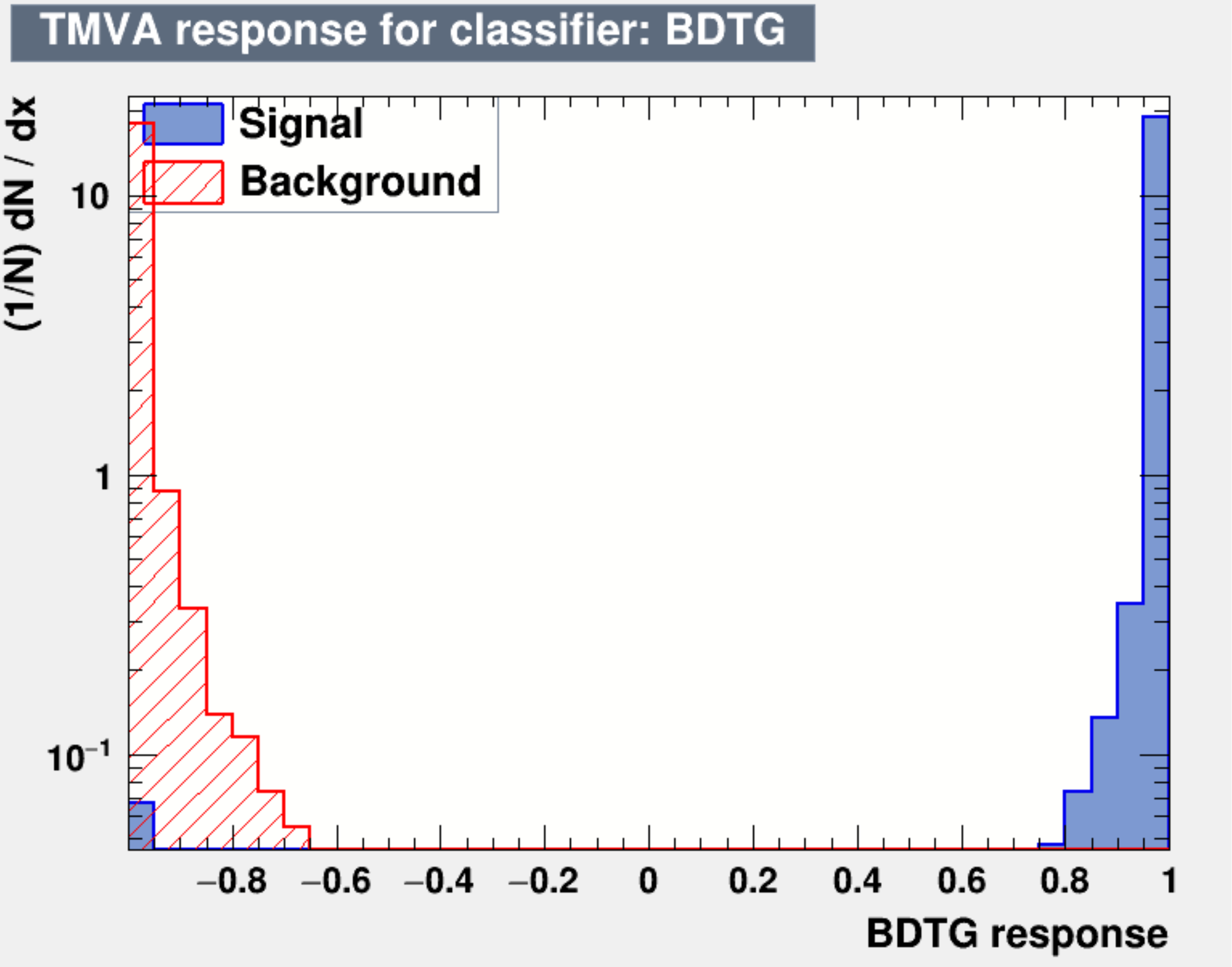} 
    \caption{The upper and central figures show the efficiency and purity for CC/NC separation of muon neutrino interactions in the SFGD. Signal efficiency (blue curve, CC interactions), background efficiency (red curve, NC interactions, almost invisible) and signal purity (dashed blue curve) are shown as a function of the method-specific cut. The upper figure shows the top range on the $y$ axis, while the central figure shows the full range in order to display the steep decline in the background efficiency. The lower plot shows the signal (blue) to background (red) separation capabilities when applied to the test sample.}
    \label{fig:detectors:SFGD_muCCNC}
\end{figure}

  \begin{figure}[!htbp]
    \centering
     \includegraphics[width=0.5\linewidth]{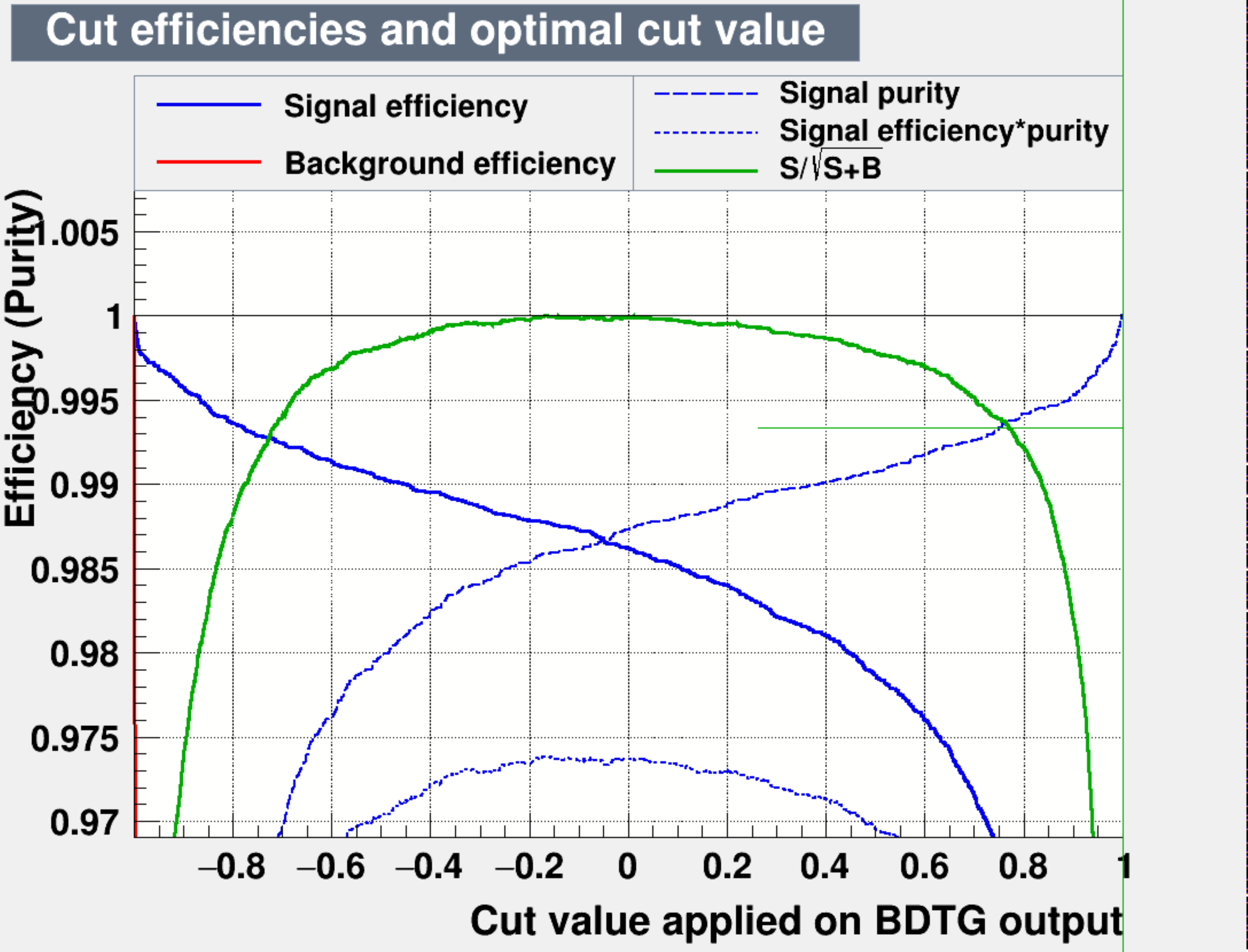} 
     \includegraphics[width=0.5\linewidth]{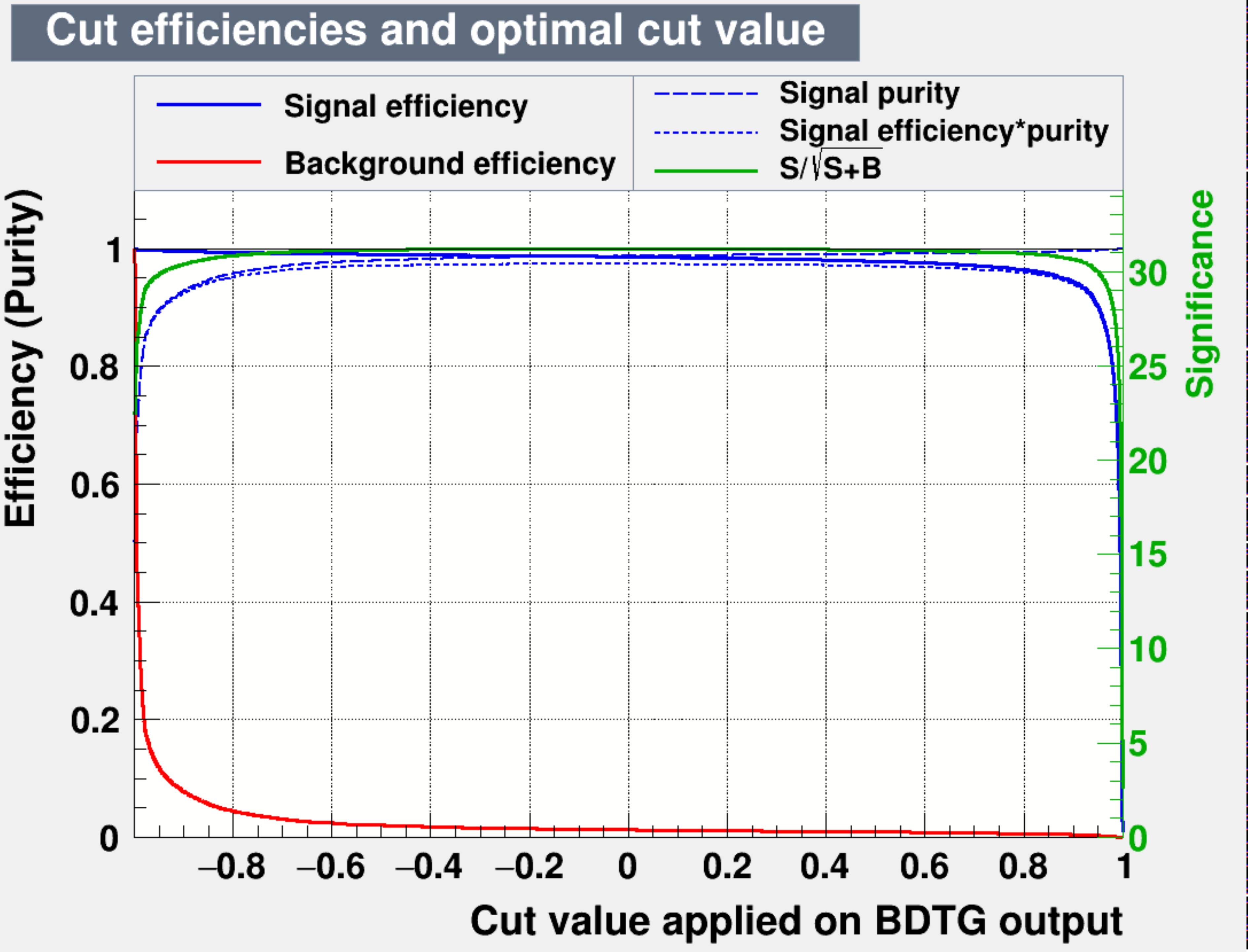} 
     \includegraphics[width=0.5\linewidth]{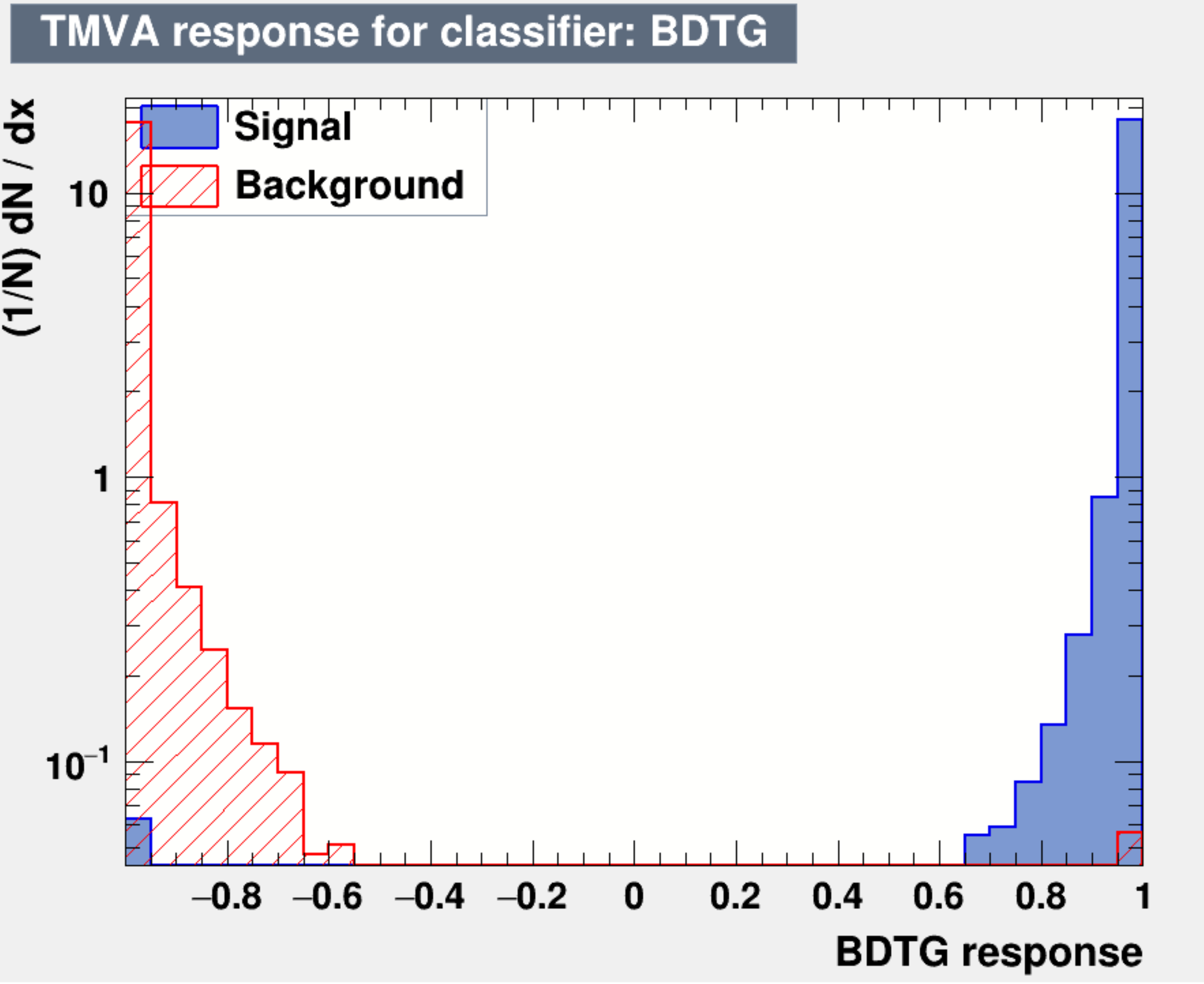} 
    \caption{The upper and central figures show the efficiency and purity performance of the SFGD for charge/neutral current interaction separation for electron neutrino interactions. Signal efficiency (blue curve, charge current interactions), background efficiency (red curve, neutral current interactions), and signal purity (dashed blue curve) are shown as a function of the method specific cut. The upper figure shows the top range on the $y$ axis, while the central figure shows the full range in order to display the steep decline in the background efficiency. The lower plot shows signal (blue) to background (red) separation capabilities when applied to the test sample.}
    \label{fig:detectors:SFGD_eCCNC}
\end{figure}

In Table~\ref{tab:detectors:SFGD_TableCCNC}, we show aggregated results for CC/NC separation performance of the SFGD in different energy bins with properly selected method-specific cuts. An excellent NC interaction rejection is demonstrated. A signal efficiency well above 95\% and signal purity against a NC admixture of 99.8\% can be achieved. 

\begin{table}[ht!] 
\begin{center}
\footnotesize
\caption{CC/NC separation performance (efficiency $\varepsilon$ and purity $p$) of the SFGD for all neutrino species and in various energy bins. }
\label{tab:detectors:SFGD_TableCCNC}
\begin{tabular}{ r r r r r r r r r } 
% \hline
 \multirow{2}{*}{$E_\nu$~~[GeV]} & \multicolumn{2}{c}{$\nu_\mu$} & \multicolumn{2}{c}{$\nu_e$} & \multicolumn{2}{c}{$\bar{\nu}_\mu$} & \multicolumn{2}{c}{$\bar{\nu}_e$} \\
% \cline{2-9}
  & $\varepsilon$~~[\%] & $p$~~[\%] & $\varepsilon$~~[\%] & $p$~~[\%] & $\varepsilon$~~[\%] & $p$~~[\%] & $\varepsilon$~~[\%] & $p$~~[\%] \\ 
 \hline
 0.0 -- 0.2 & 98.1 & 99.7 & 99.1 & 99.9 & 99.5 & 99.9 & 99.5 & 100 \\
% \hline
0.2 -- 0.3 & 99.3 & 99.9 & 99.6 & 99.9 & 99.5 & 99.8 & 99.5 & 99.9 \\
% \hline
0.3 -- 0.4 & 99.3 & 99.9 & 98.8 & 99.9 & 99.0 & 99.9 & 99.1 & 99.0 \\
% \hline
0.4 -- 0.5 & 98.3 & 99.5 & 97.8 & 98.0 & 98.5 & 99.5 & 98.5 & 98.2 \\
% \hline
0.5 -- 0.6 & 97.0 & 99.0 & 97.0 & 95.4 & 96.8 & 99.6 & 96.8 & 97.8 \\
% \hline
0.6 -- 0.8 & 95.8 & 98.4 & 95.2 & 95.1 & 98.0 & 99.8 & 95.9 & 98.5 \\
% \hline
0.8 -- 1.0 & 95.5 & 95.5 & n/a & n/a & n/a & n/a & n/a & n/a \\
 \hline
 \end{tabular}
 \end{center}
\end{table} 

\subsubsubsection {Neutrino Flavour Identification}

After selecting a CC interaction in the SFGD, a procedure for identification of the interacting neutrino flavour (i.e. $\nu_\mu$ or $\nu_e$) should be performed. The procedure is based again on the multi-variate analysis methods. In Fig.\ref{fig:detectors:SFGD_mu-e} and in Table~\ref{tab:detectors:SFGD_mu-e}, we show the performance of the SFGD to separate muon-neutrino from electron-neutrino interactions. In this case, the signal is an electron-neutrino interaction and the background is a muon-neutrino one. The same is shown for antineutrinos in Fig.\ref{fig:detectors:SFGD_antimu-e} and Table~\ref{tab:detectors:SFGD_antimu-e}. It is seen that by choosing an appropriate cut value, the admixture of the wrong flavour events could be constrained below {\it per mill} level, with a signal efficiency well above \SI{90}{\percent}. 

\begin{figure}[p]
    \centering
     \includegraphics[width=0.45\linewidth]{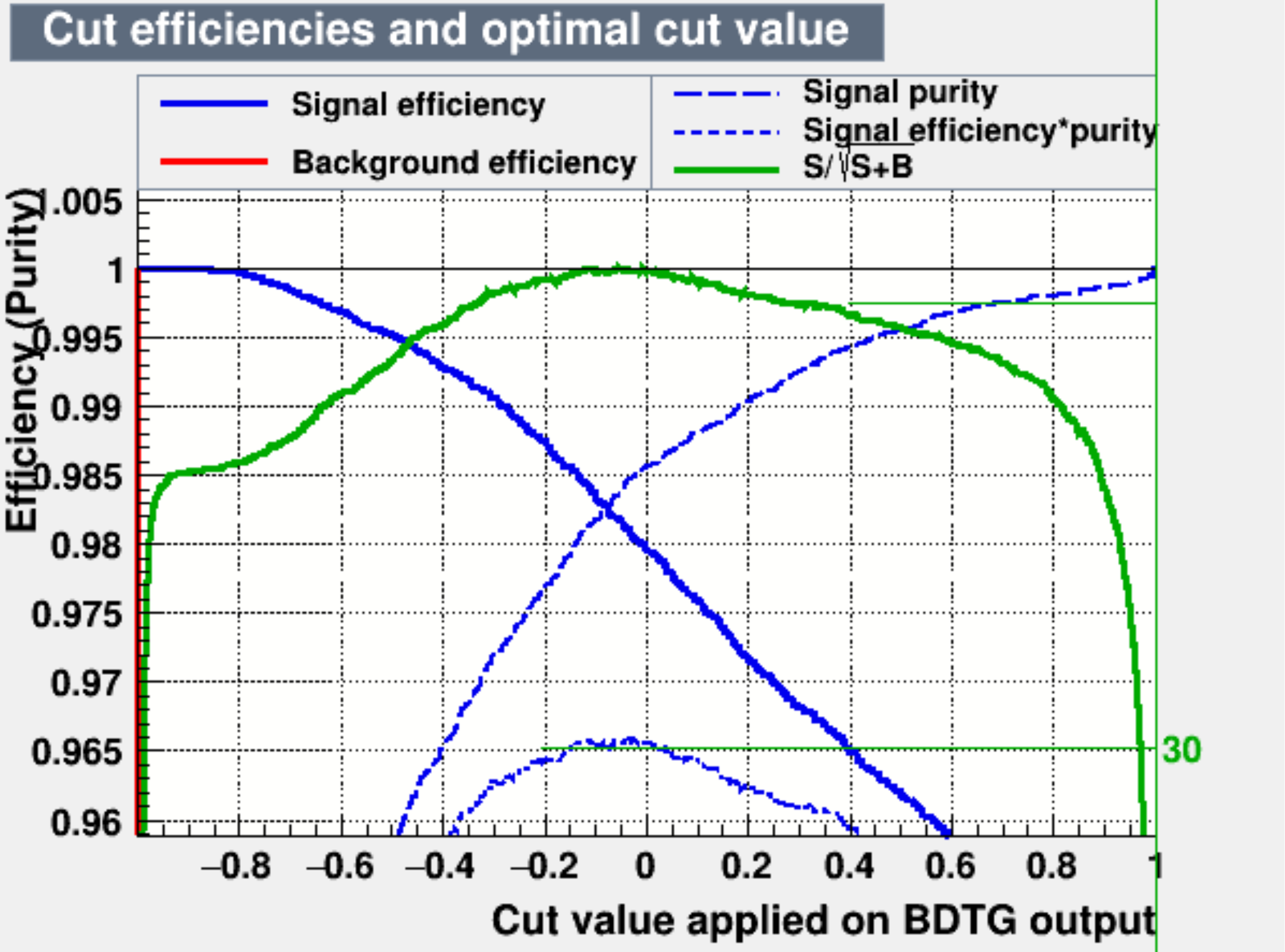} 
     \includegraphics[width=0.45\linewidth]{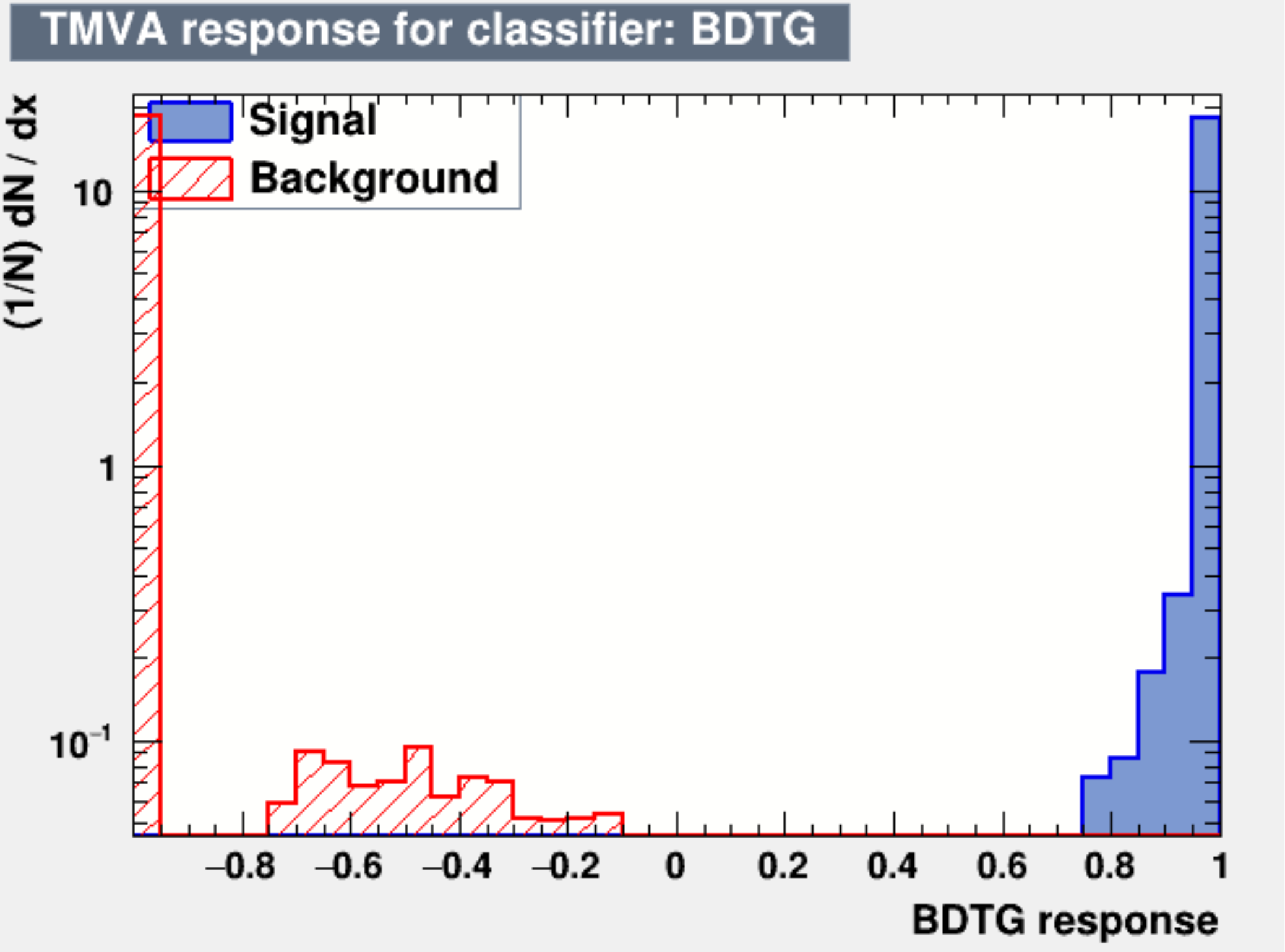} 
    \caption{Left: Efficiency and purity performance of SFGD for distinguishing between $\nu_\mu$ and $\nu_e$ CC events. Signal efficiency (blue curve, $\nu_e$ interactions), background efficiency (red curve, $\nu_\mu$ interactions), and signal purity (dashed blue curve) are shown as a function of the method-specific cut value. Right: signal/background separation capabilities as a function of the applied cut value.}
    \label{fig:detectors:SFGD_mu-e}
\end{figure}

\begin{figure}[p]
    \centering
     \includegraphics[width=0.45\linewidth]{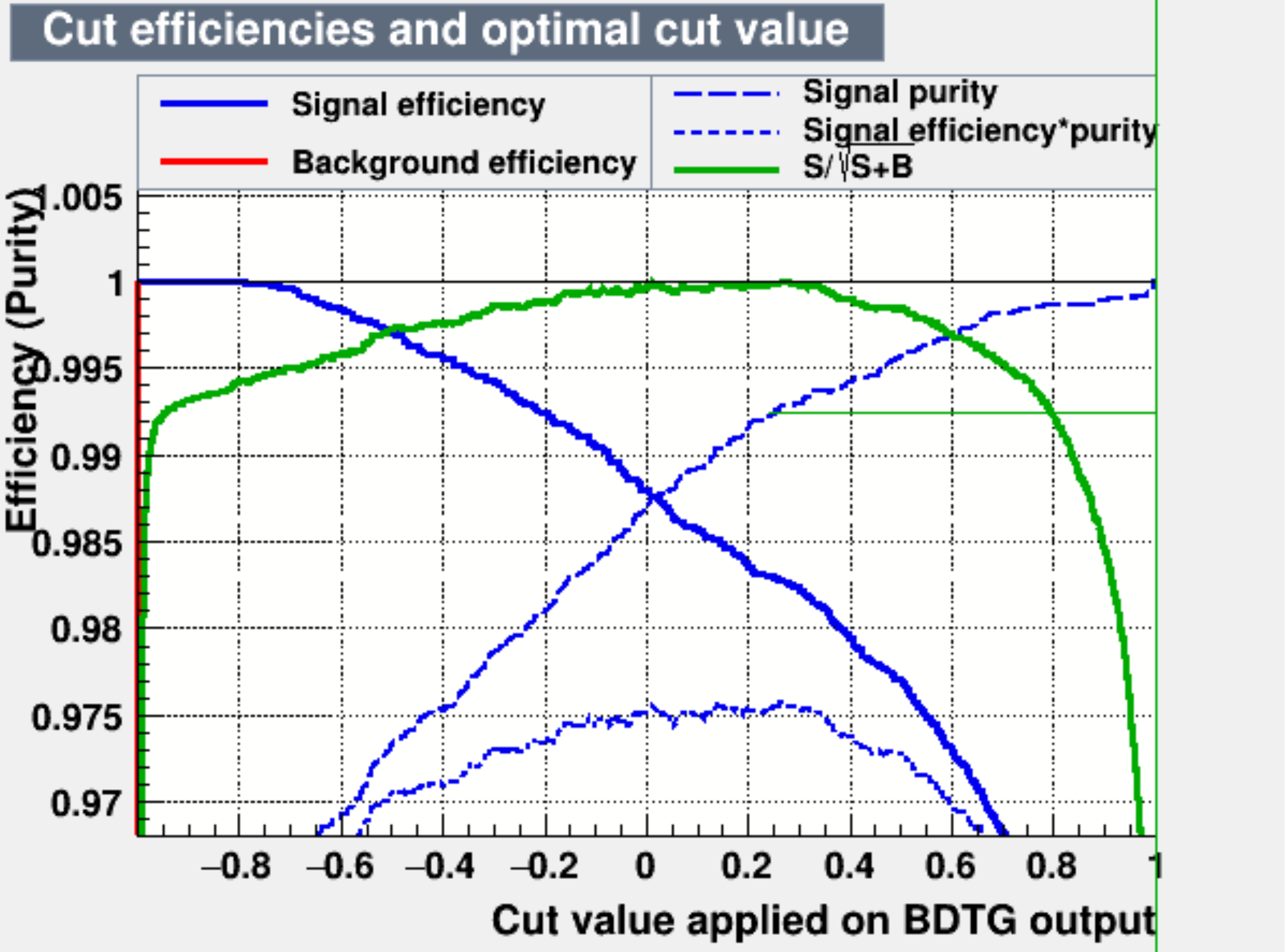} 
     \includegraphics[width=0.45\linewidth]{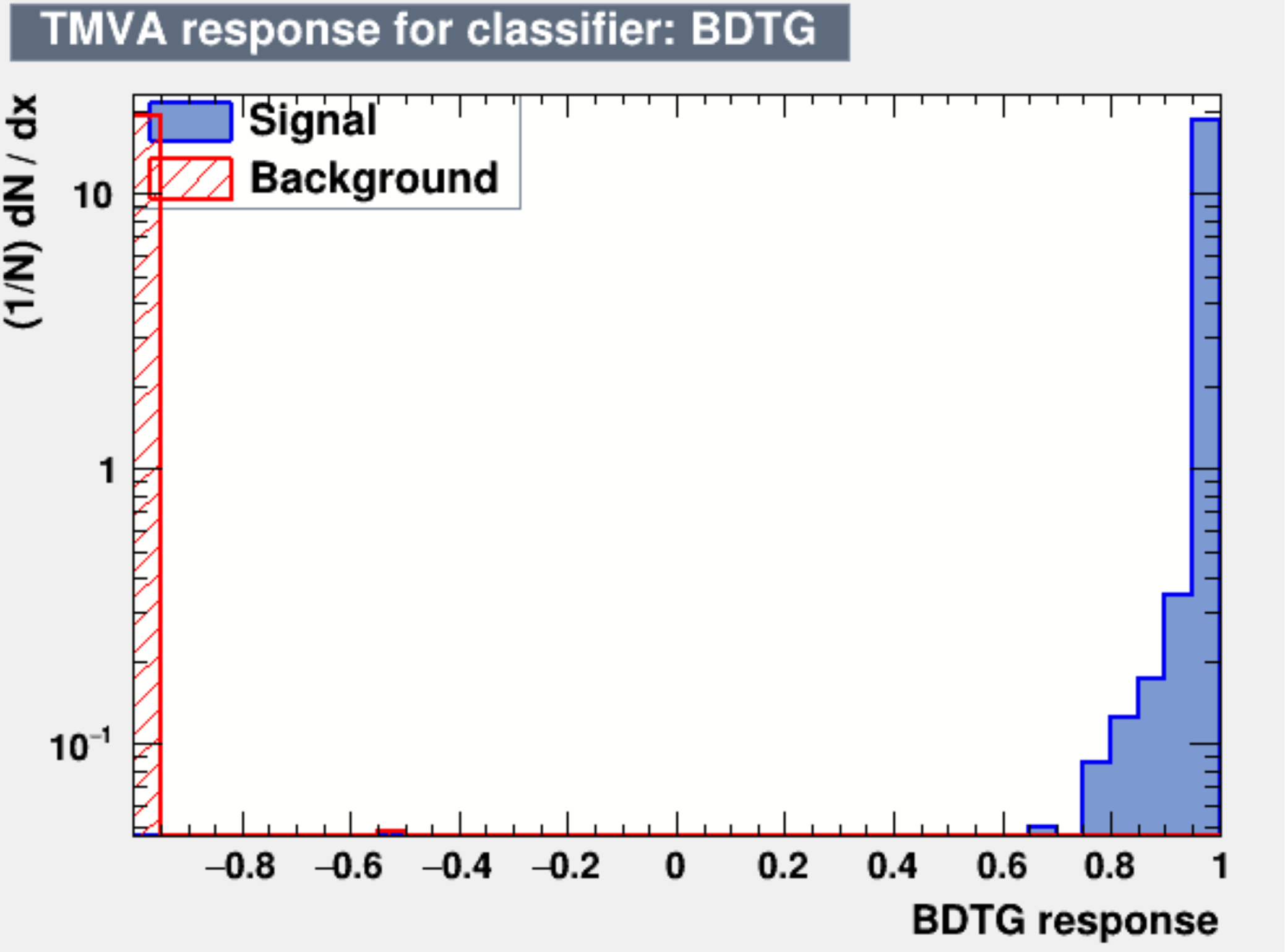} 
     \caption{Left: Efficiency and purity performance of the SFGD for distinguishing between anti-$\nu_\mu$ and anti-$\nu_e$ CC events. Signal efficiency (blue curve, anti-$\nu_e$ interactions), background efficiency (red curve, anti-$\nu_\mu$ interactions), and signal purity (dashed blue curve) are shown as a function of the method-specific cut value. Right: signal/background separation capabilities as a function of the applied cut value. }
    \label{fig:detectors:SFGD_antimu-e}
\end{figure}

 \begin{figure}[p]
    \centering
     \includegraphics[width=0.45\linewidth]{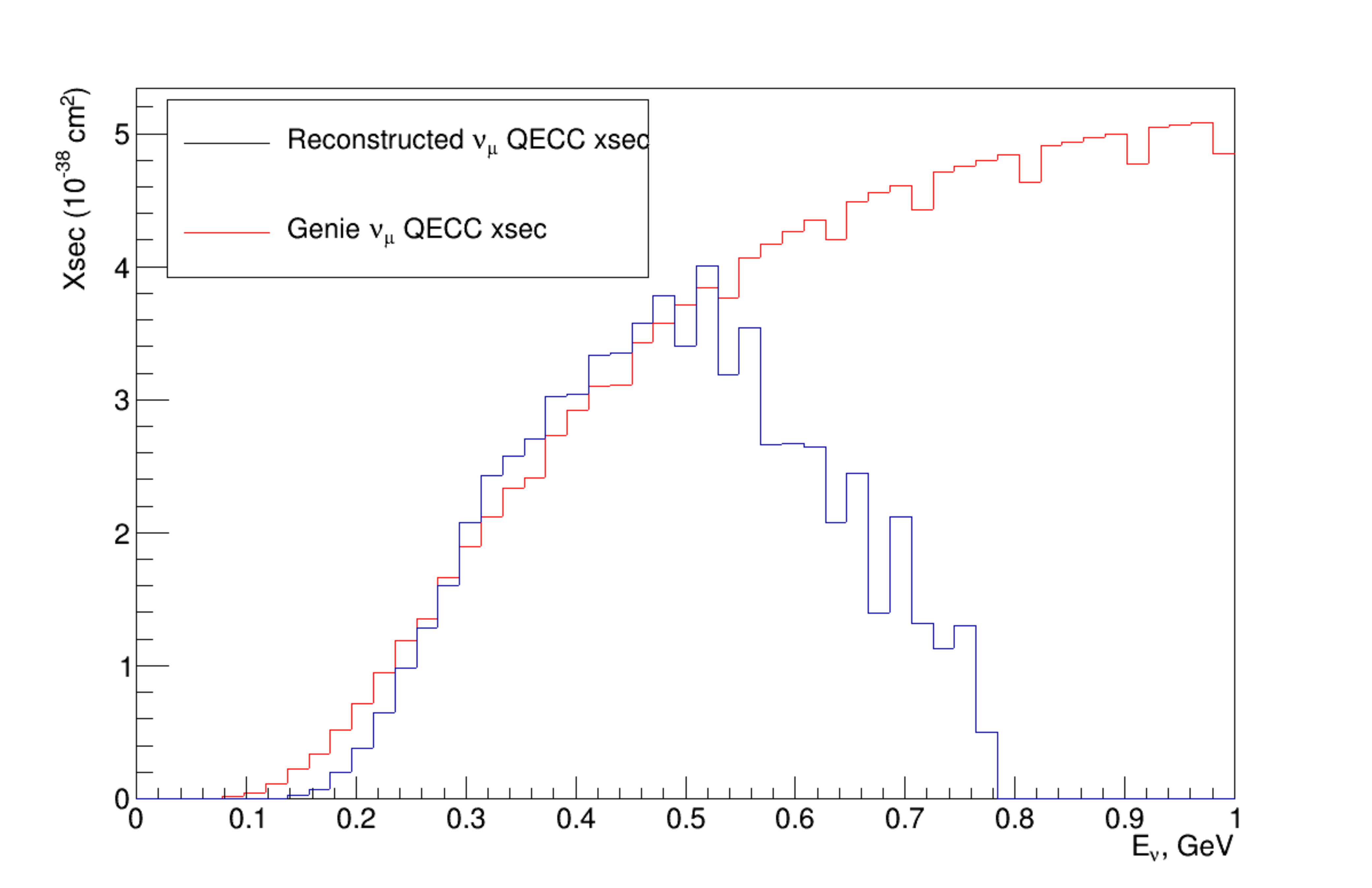} 
     \includegraphics[width=0.45\linewidth]{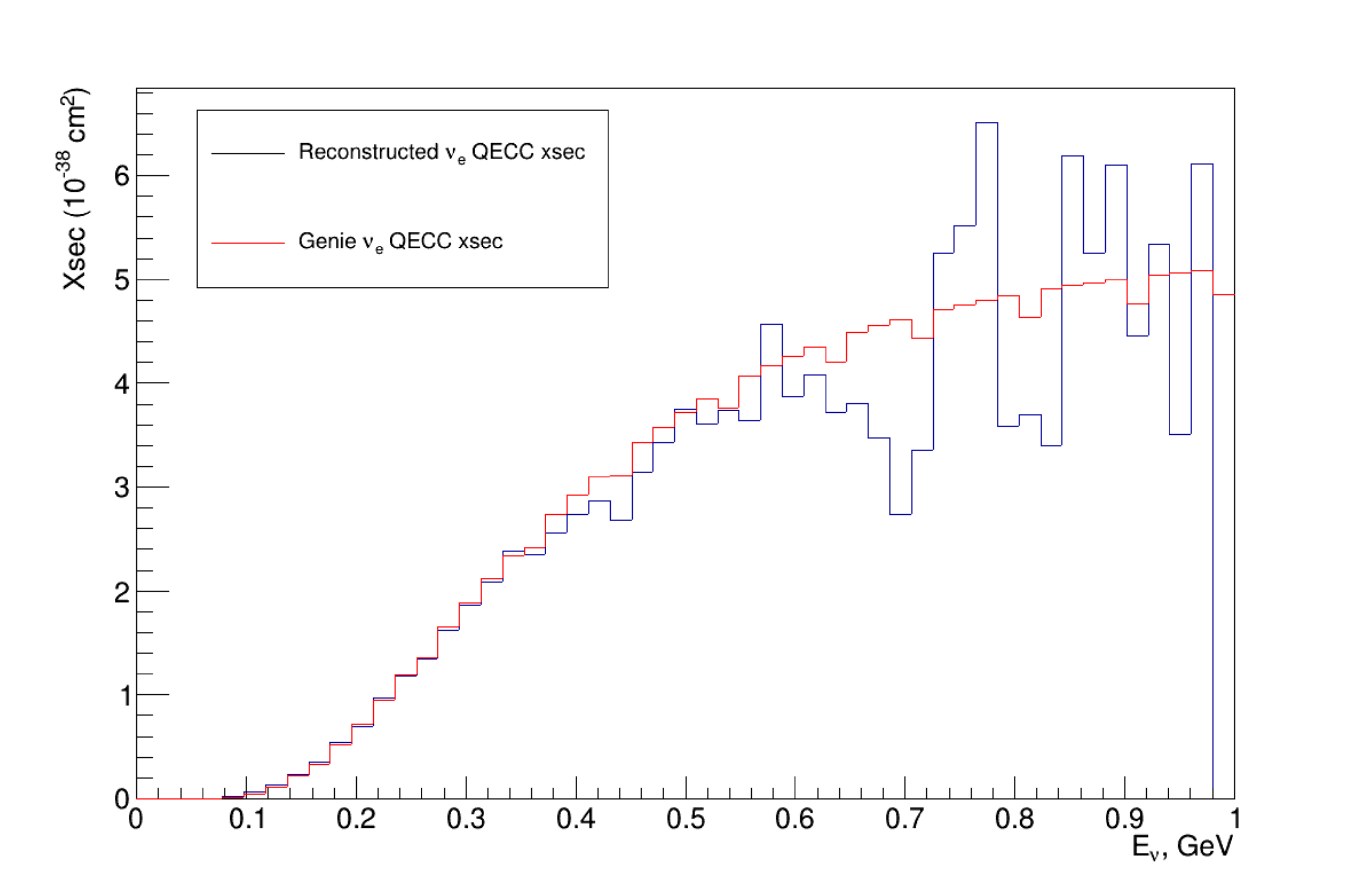} 
     \caption{Left: "Measured" $\nu_\mu$ cross-section (blue) compared to the cross-section used by \textsc{GENIE} (red) to simulate the interactions. Muon neutrino energy calculation is based on the fitted muon momentum. Right: "Measured" $\nu_e$ cross-section (blue) compared to the cross-section used by \textsc{GENIE} (red) to simulate the interactions. Electron neutrino energy calculation is based on the true electron momentum.}
    \label{fig:detectors:SFGD_xsec}
\end{figure}

\begin{table}[ht!] 
\begin{center}
\footnotesize
\caption{Efficiency and purity performance of the SFGD for distinguishing between $\nu_\mu$ and $\nu_e$ CC events as a function of the applied cut value.}
\label{tab:detectors:SFGD_mu-e}
\begin{tabular}{ r r r l } 
% \hline
 \textbf{BDTG cut} & \textbf{efficiency~~[\%]} & \textbf{purity~~[\%]} & ~ \\
 \hline
 $-0.80$ & 100 & 93.9 & ~ \\ 
% \hline
 $-0.60$ & 99.7 & 95.2 & ~ \\ 
% \hline
 $-0.40$ & 99.2 & 96.6 & ~ \\ 
% \hline
 $-0.20$ & 98.7 & 97.7 & ~ \\ 
% \hline
 $-0.03$  & 98.1 & 98.5 & \textit{best cut} \\ 
% \hline
 $0.20$ & 97.2 & 99.0 & ~ \\ 
% \hline
 $0.40$ & 96.5 & 99.4 & ~ \\ 
% \hline
 $0.60$ & 95.9 & 99.7 & ~ \\ 
% \hline
 $0.80$ & 95.0 & 99.8 & ~ \\ 
% \hline
 $0.95$ & 92.0 & 99.9 & ~ \\ 
 \hline
\end{tabular}
 \end{center}
\end{table}

\begin{table}[ht!] 
\begin{center}
\footnotesize
\caption{Efficiency and purity performance of SFGD for distinguishing between $\bar{\nu}_\mu$ and $\bar{\nu}_e$ CC events as a function of the applied cut value. }
\label{tab:detectors:SFGD_antimu-e}
\begin{tabular}{ r r r l } 
% \hline
 \textbf{BDTG cut} & \textbf{efficiency~~[\%]} & \textbf{purity~~[\%]} & ~ \\
 \hline
 $-0.80$ & 100 & 96.5 & ~ \\ 
% \hline
 $-0.60$ & 99.8 & 96.9 & ~ \\ 
% \hline
 $-0.40$ & 99.6 & 97.5 & ~ \\ 
% \hline
 $-0.20$ & 99.2 & 98.1 & ~ \\ 
% \hline
  $0.00$ & 98.8 & 98.7 & ~ \\ 
% \hline
  $0.26$ & 98.3 & 99.3 & \textit{best cut}\\ 
% \hline
 $0.40$ & 97.9 & 99.4 & ~ \\ 
% \hline
 $0.60$ & 97.3 & 99.7 & ~ \\ 
% \hline
 $0.80$ & 96.2 & 99.8 & ~ \\ 
% \hline
 $0.95$ & 93.0 & 99.9 & ~ \\ 
 \hline
\end{tabular}
 \end{center}
\end{table}

\subsubsubsection {Neutrino Cross-Section Measurement}

As mentioned above, one of the purposes of the SFGD is to measure neutrino cross-sections. In Fig.~\ref{fig:detectors:SFGD_xsec} the ``measured'' $\nu_\mu$ and $\nu_e$ cross-sections are compared with the ones used for simulation of neutrino events. Good agreement up to $E_{\nu} \sim \SI{500}{\MeV}$ is observed in both cases. For higher energies, that discrepancy becomes substantial, especially in the muon case. The higher momentum region suffers from low statistics, due to the low-energy nature of the neutrino beam, which reduces the accuracy of the neural network reconstruction in this region.

\subsubsubsection {SFGD and Water Cherenkov Detector Combined Analysis}

Figure~\ref{fig:detectors:event_crossover} shows a muon neutrino interaction in the SFGD cube where the secondary muon penetrates into the water Cherenkov volume, producing Cherenkov light. We call such multi-detector-spanning events crossover events.
As mentioned above, roughly \SI{12}{\percent} (\SI{20}{\percent}) of positive (negative) muons produced in SFGD will exit the detector in the direction of the water Cherenkov volume, and thus have a chance of being detected there. The corresponding proportion of electron neutrino events is about \SI{6}{\percent} of the sample.
The possible identification of exiting muon events in the water Cherenkov tank would serve as an additional veto for muon neutrinos in the electron neutrino event sample, and thus yield a better purity.
%For the crossover events we aim at a good purity of the electron neutrino event sample (by efficiently  rejecting events originated in the SFGD that have muons penetrated into the water Cherenkov detector by exploiting the SFGD and water Cherenkov ID capabilities together) and that the latter could be used for electron neutrino cross-section measurement.

\begin{figure}[!htbp]
\centering
\includegraphics[width=0.6\linewidth]{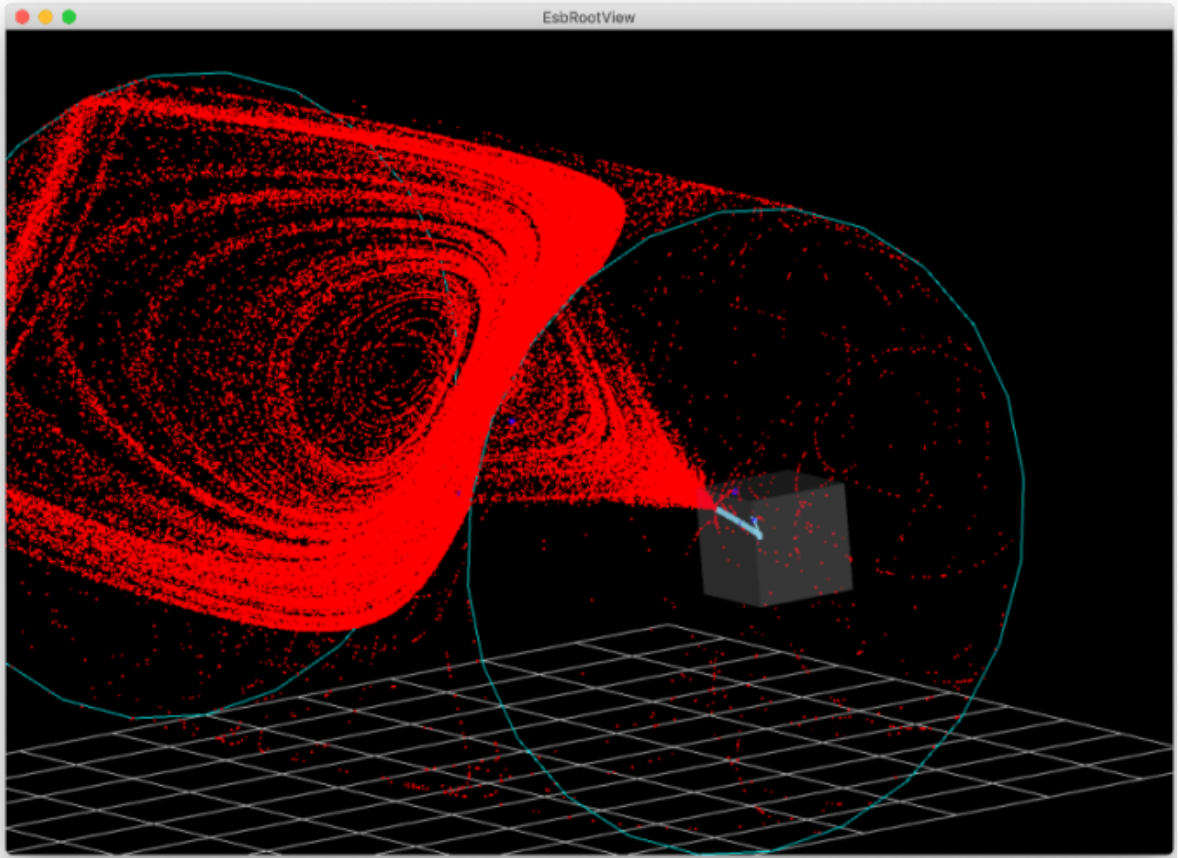}
\caption{Muon neutrino interaction in the SFGD cube with a secondary muon producing Cherenkov light in the near water Cherenkov detector.}
\label{fig:detectors:event_crossover}
\end{figure}

Figure~\ref{fig:detectors:spectra_crossover} contains momentum spectra of events in which the primary muon or electron potentially enters both the SFGD and the water Cherenkov detector. The total number of potential muons and electrons is shown, along with events with a positive $z$-component of the momentum, as well as those that would have chance to enter the water Cherenkov volume (assumed as  $p_z >\SI{120}{\MeV\per\speedoflight}$). It is seen that in ${\sim}\SI{12}{\percent}$ of the $\nu_\mu$ induced events, the outgoing muons may produce light in the water Cherenkov detector; whereas for the  $\nu_e$ induced events, this fraction is halved, ${\sim}\SI{6}{\percent}$, due to their softer spectrum.

\begin{figure}[p]
\centering
 \includegraphics[width=0.45\linewidth]{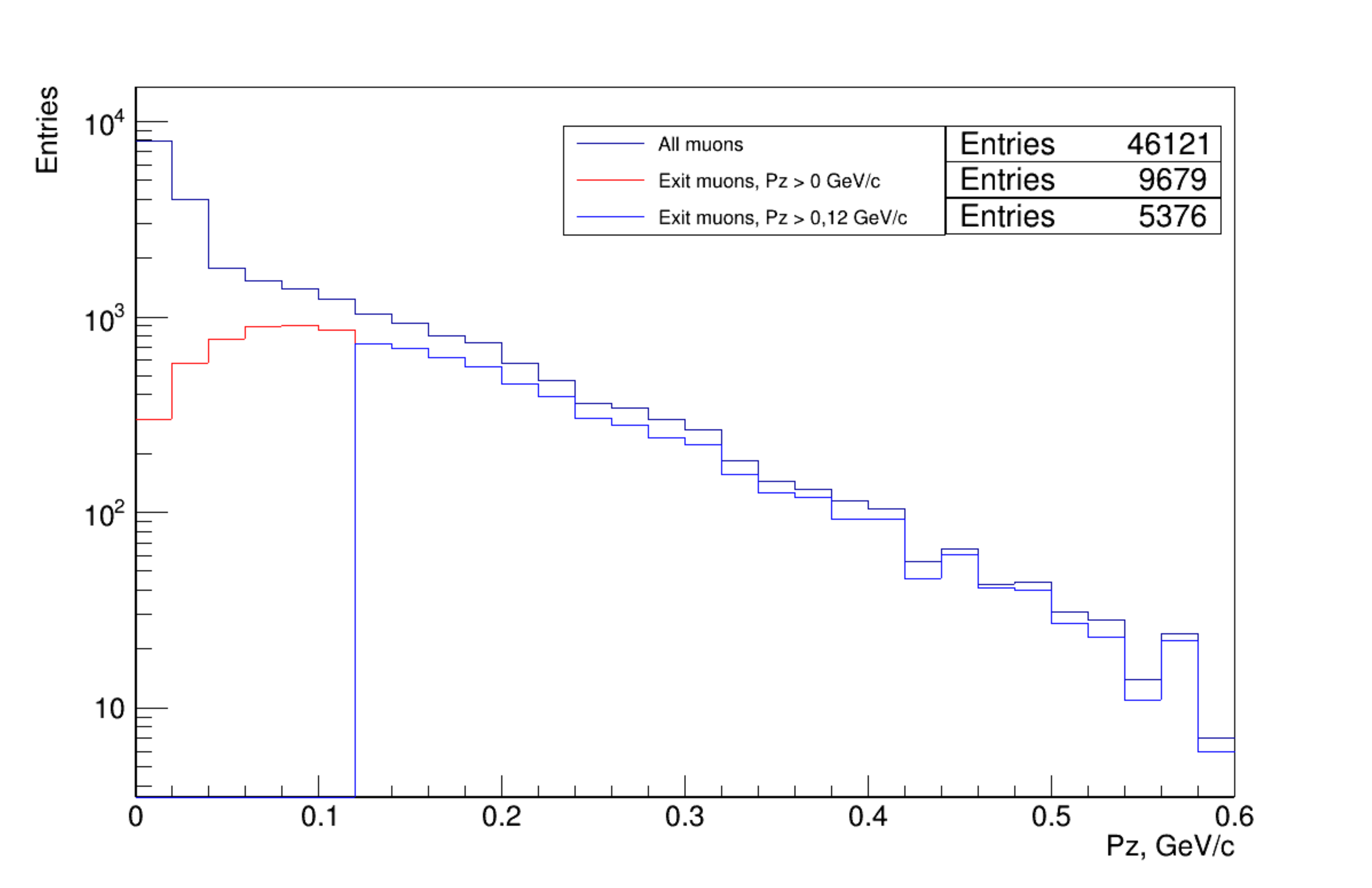} 
 \includegraphics[width=0.45\linewidth]{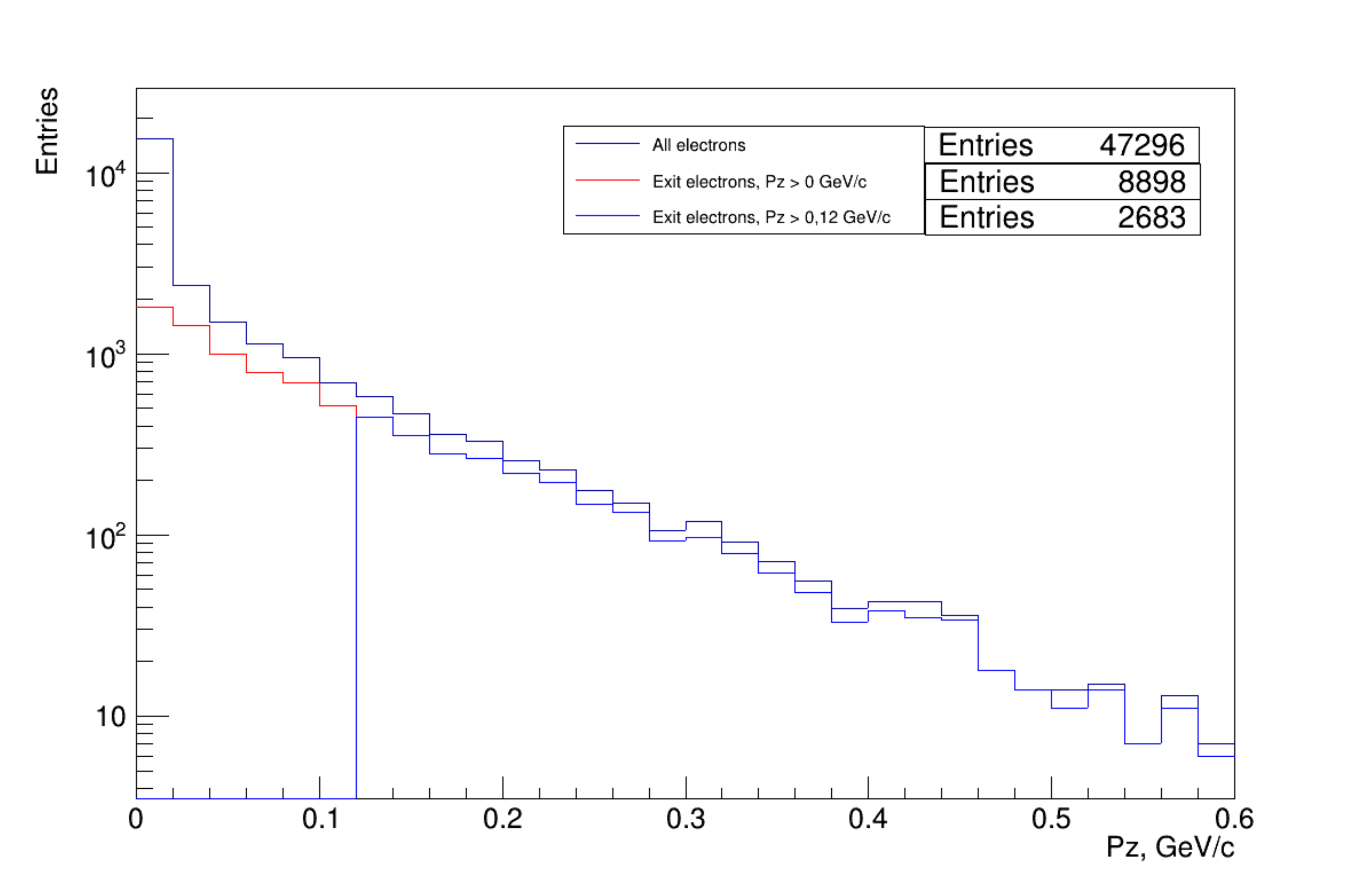} 
\caption{Momentum spectra of muons (left) and electrons (right) which potentially span the SFGD and the water Cherenkov detector: all candidates that reach the last plane of the SFGD (dark blue), candidates with positive $p_z$ (red), and candidates with high enough energy to potentially allow them to enter the water Cherenkov component (blue, $p_z>\SI{120}{\MeV}c^{-1}$).}
\label{fig:detectors:spectra_crossover}
\end{figure}

\begin{figure}[p]
\centering
\includegraphics[width=0.45\linewidth]{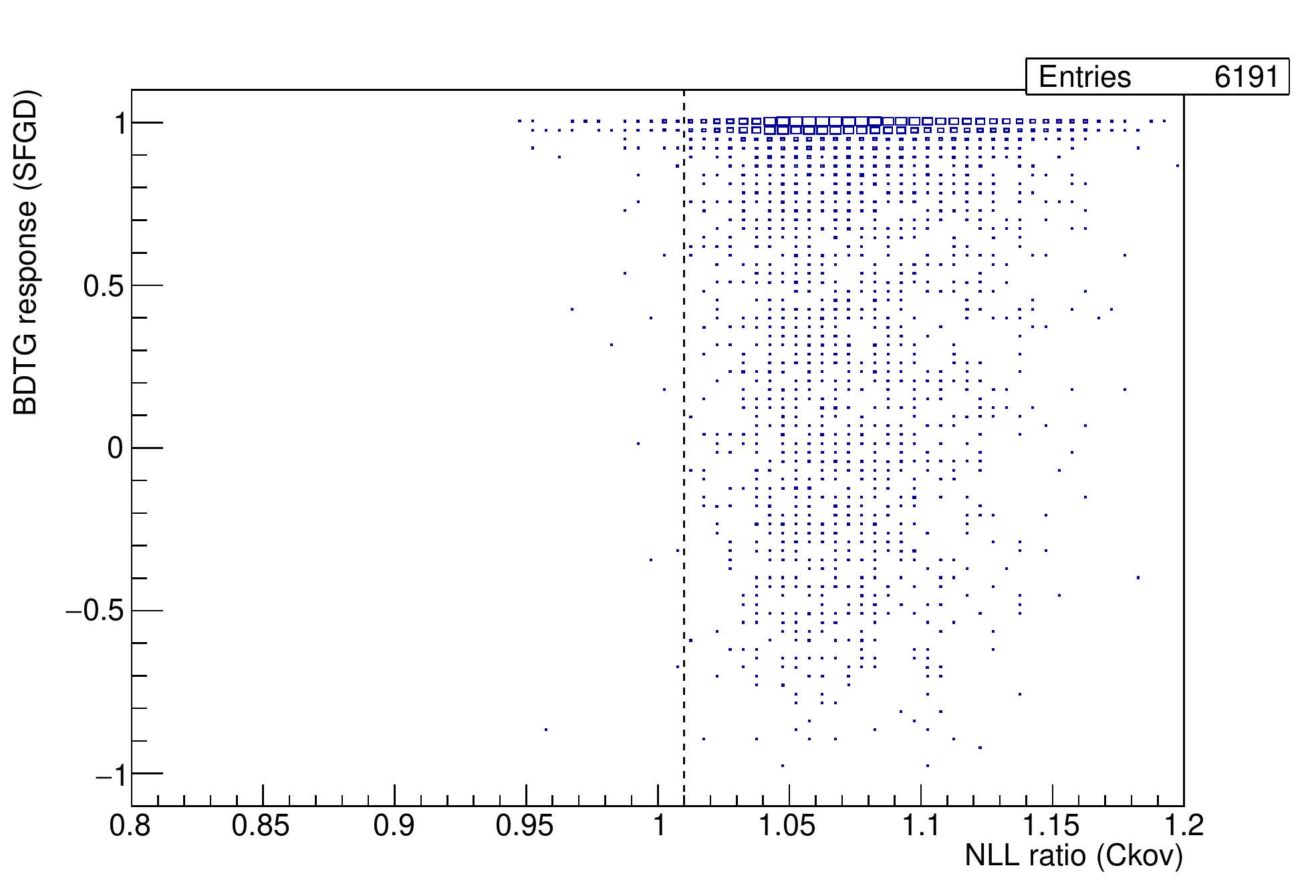}
\includegraphics[width=0.45\linewidth]{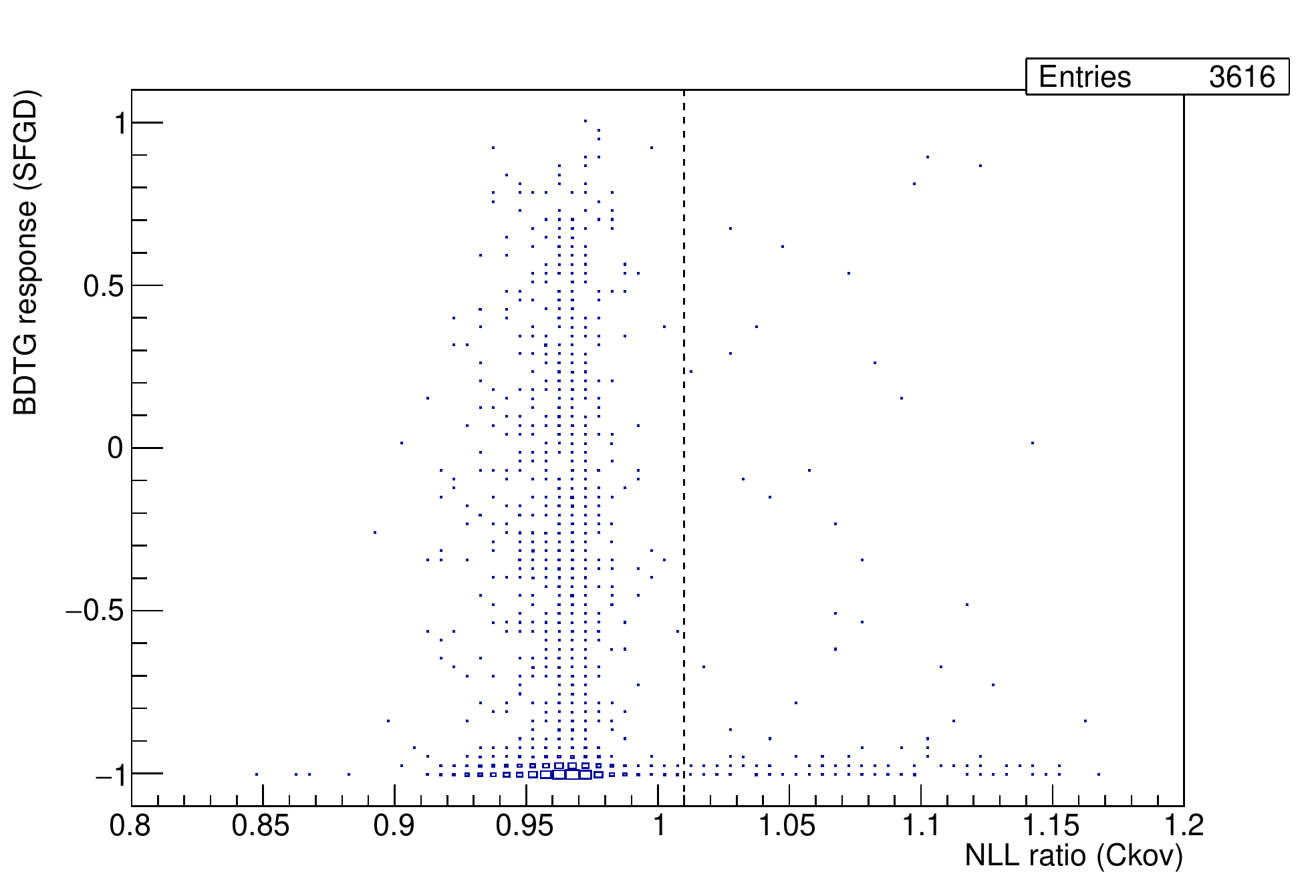}
\includegraphics[width=0.45\linewidth]{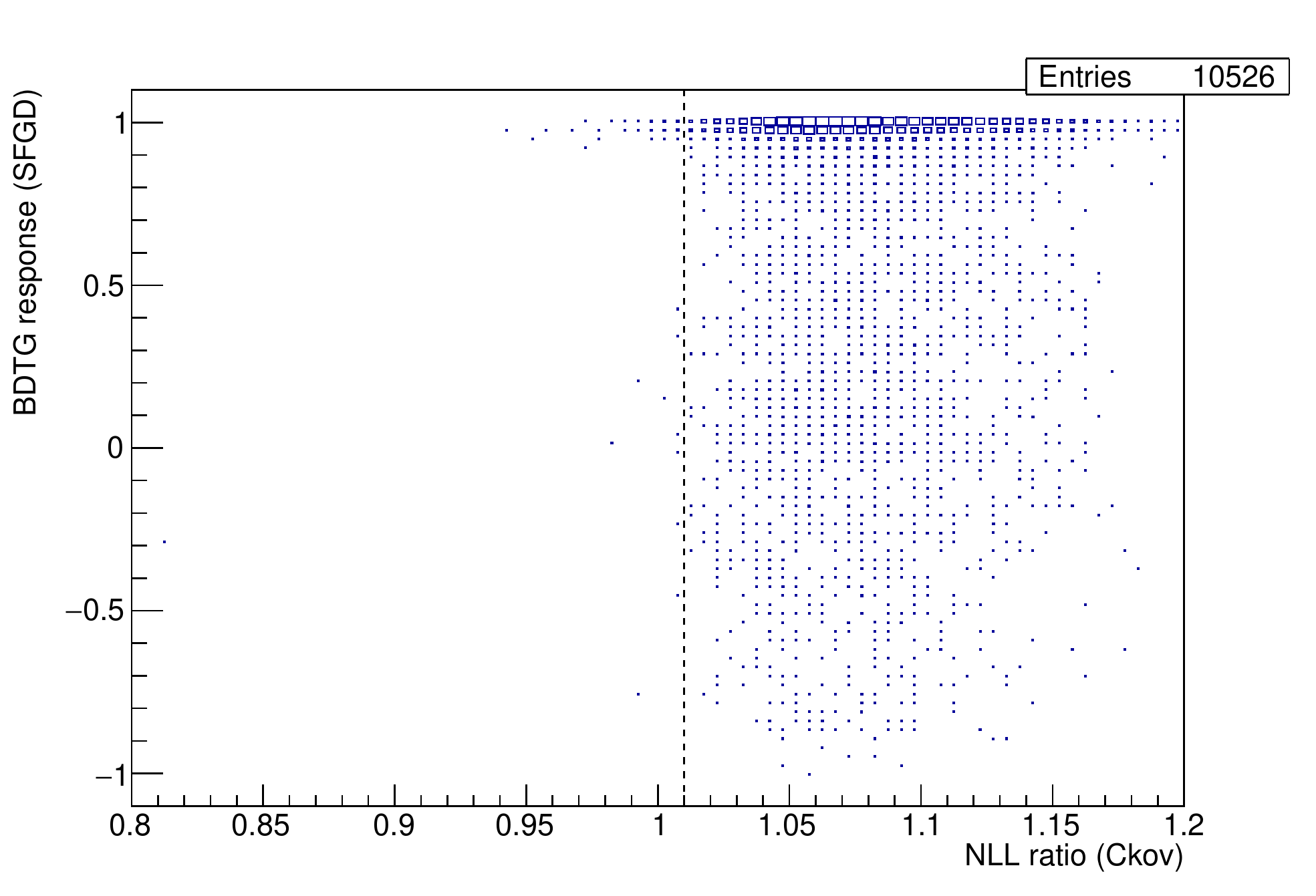}
\includegraphics[width=0.45\linewidth]{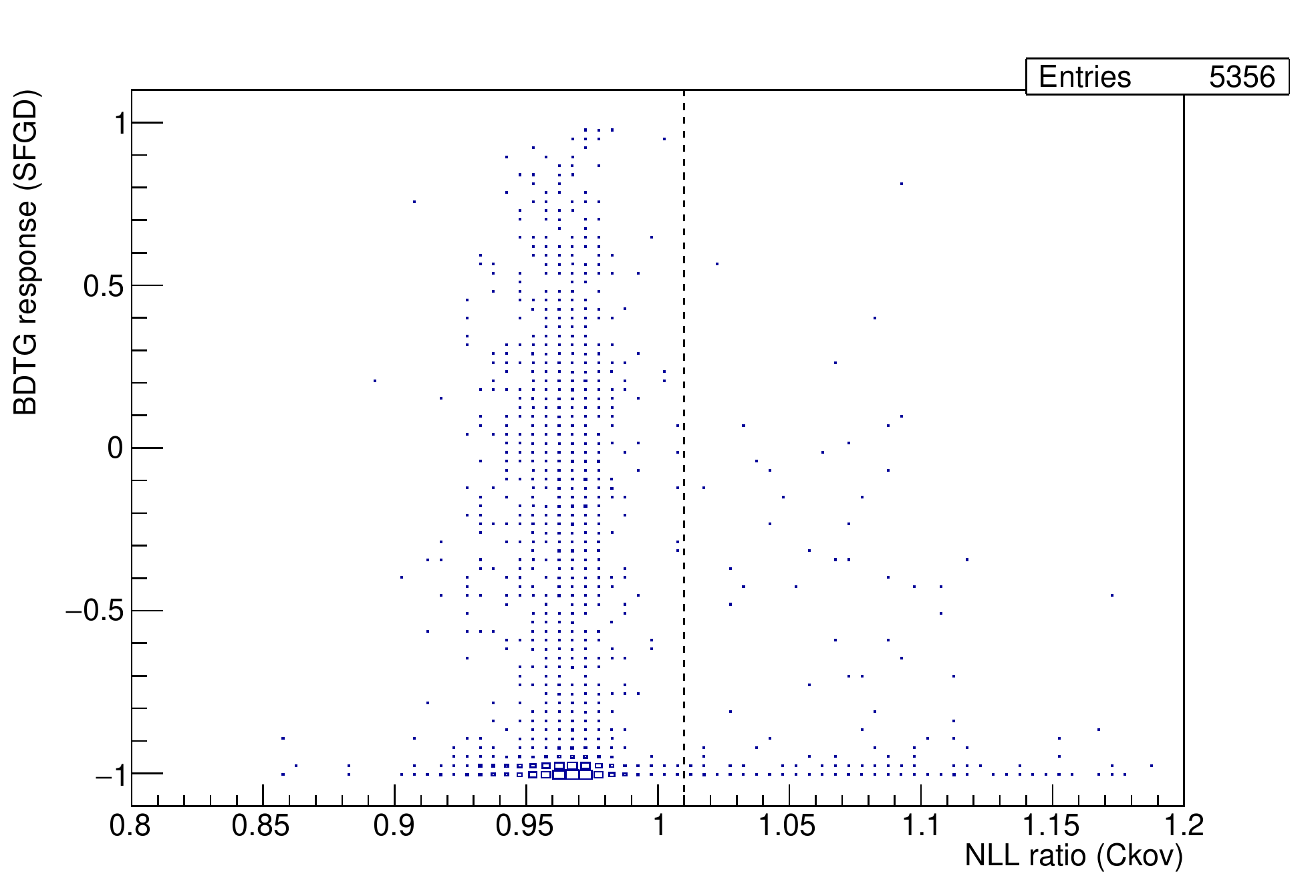}
\caption{Particle identification for charged lepton tracks which begin in the SFGD and end in the water Cherenkov detector.  The {\nllratio} from the water Cherenkov analysis is measured on the $x$ axis and  the BDTG value from the SFGD analysis is on the $y$ axis. Top left: $\nu_e$  beam, top right: $\nu_\mu$ beam, bottom left: anti-$\nu_e$ beam, bottom right: anti-$\nu_\mu$ beam. Note that events with $\nllratiomath < 1.01$ are classified as muon-like and events with $\nllratiomath \geq 1.01$ as electron-like in the water Cherenkov analysis (see Eq.~(\ref{eqn:detectors:nd_pid}).}
\label{fig:detectors:crossover_analysis}
\end{figure}

One sample of $10^5$ events interacting in the SFGD was simulated per flavour ($e^+$, $e^-$, $\mu^+$, $\mu^-$).
These were propagated to their end-point, and the events containing a charged lepton that exited through the backside of the SFGD (in the direction of the water Cherenkov component) were registered.
These particles were inserted into the water Cherenkov detector simulation with the best-case assumption that they immediately entered the instrumented volume of the water Cherenkov tank as they exited the SFGD.

Table~\ref{tbl:detectors:nd_xsfgd_nev} shows the number of events expected per running-year at each level of the water Cherenkov analysis, up to the \textit{sub-Cherenkov criterion}.
These events are efficiently rejected by the next selection level, the \textit{reconstruction quality criteria}, because they require a minimum distance between the event vertex and the wall of $\SI{30}{\cm}$ (see Eq.~\ref{eqn:detectors:cutsL3}).

{
\begin{table}[htbp]
\footnotesize
\centering
\caption{Number of expected events per running year in the SFGD, as well as those entering the water Cherenkov detector and passing the water Cherenkov selection criteria.
Listed over the four charged lepton flavours and two horn polarities.
\label{tbl:detectors:nd_xsfgd_nev}}
\begin{tabular}{ r r r r r }
             \textbf{Positive polarity}  &                 ~  &                ~  &                 ~  &                ~  \\
                                      ~  &  \textbf{$\mu^-$}  &   \textbf{$e^-$}  &  \textbf{$\mu^+$}  &   \textbf{$e^+$}  \\
\hline
                        All events SFGD  &   \SI{9.82e+04}{}  &       \SI{484}{}  &        \SI{241}{}  &       \SI{1.0}{}  \\
%Exiting SFGD (entering water Cherenkov)  &   \SI{1.82e+04}{}  &      \SI{90.4}{}  &       \SI{41.3}{}  &       \SI{0.2}{}  \\
%                                Trigger  &       \SI{8700}{}  &      \SI{50.4}{}  &       \SI{29.2}{}  &       \SI{0.1}{}  \\
%                Sub-Cherenkov criterion  &       \SI{3000}{}  &      \SI{32.1}{}  &       \SI{10.0}{}  &       \SI{0.1}{}  \\
Exiting SFGD (entering water Cherenkov)  &   \SI{1.78e+04}{}  &      \SI{87.9}{}  &       \SI{39.9}{}  &       \SI{0.2}{}  \\
                                Trigger  &   \SI{1.18e+04}{}  &      \SI{60.4}{}  &       \SI{33.6}{}  &       \SI{0.1}{}  \\
                Sub-Cherenkov criterion  &       \SI{4100}{}  &      \SI{40.0}{}  &       \SI{12.0}{}  &       \SI{0.1}{}  \\
\hline
                                      ~  &                 ~  &                ~  &                 ~  &                ~  \\
             \textbf{Negative polarity}  &                 ~  &                ~  &                 ~  &                ~  \\
                                      ~  &  \textbf{$\mu^-$}  &   \textbf{$e^-$}  &  \textbf{$\mu^+$}  &   \textbf{$e^+$}  \\
\hline
                        All events SFGD  &        \SI{929}{}  &       \SI{6.4}{}  &   \SI{1.75e+04}{}  &      \SI{51.9}{}  \\
%Exiting SFGD (entering water Cherenkov)  &        \SI{115}{}  &       \SI{0.6}{}  &       \SI{3930}{}  &      \SI{11.1}{}  \\
%                                Trigger  &         \SI{51}{}  &       \SI{0.3}{}  &       \SI{2810}{}  &       \SI{9.0}{}  \\
%                Sub-Cherenkov criterion  &         \SI{16}{}  &       \SI{0.2}{}  &        \SI{980}{}  &       \SI{6.6}{}  \\
Exiting SFGD (entering water Cherenkov)  &        \SI{120}{}  &       \SI{0.6}{}  &       \SI{3890}{}  &      \SI{11.1}{}  \\
                                Trigger  &         \SI{78}{}  &       \SI{0.4}{}  &       \SI{3350}{}  &       \SI{9.6}{}  \\
                Sub-Cherenkov criterion  &         \SI{26}{}  &       \SI{0.2}{}  &       \SI{1210}{}  &       \SI{6.8}{}  \\
\hline
\end{tabular}
\end{table}
}

The events that enter the instrumented water Cherenkov volume trigger the detector with an efficiency between \SI{44}{\percent} and \SI{82}{\percent}, depending on flavor, and pass the \textit{sub-Cherenkov criterion} with a \SI{14}{\percent} to \SI{60}{\percent} efficiency.

In Fig.\ref{fig:detectors:crossover_analysis} the capability of the two detectors for particle identification in the crossover events using a combined analysis is shown.
The identification criterion from the water Cherenkov analysis is shown, using the {\nllratio} (see Eq.~\ref{eqn:detectors:nd_pid}), along with the BDTG characterisation value of the SFGD.
The assumption is that the available information from the SFGD is the track momentum and length, number of tracks, and number of photoelectrons.
This figure includes events that pass the \textit{sub-Cherenkov criterion} in the water Cherenkov analysis (i.e.\ they have an energy above the Cherenkov threshold).
It is evident from the figures that the water Cherenkov detector can be used to identify the tracks exhibiting high efficiency. In addition, the water Cherenkov detector provides the particle energy, which could be used subsequently as a seed for momentum measurement of the charged lepton track at the vertex in the SFGD.

%\bigskip
\subsubsection {Emulsion detector}
%\bigskip

A NINJA-type emulsion detector will be situated immediately upstream of the SFGD tracker: it will be used for precise neutrino interaction cross-section measurements. The original NINJA detector \cite{Fukuda:2017clt} is an emulsion-based detector composed of modules using either iron or water as a target; it operates concurrently with the T2K near detectors at J-PARC (see Fig.~\ref{fig:detectors:nd_emulsion_ninja_placement}). Its primary purpose is to precisely measure the neutrino interaction topology and double differential cross-sections. There is a joint effort with the members of the NINJA collaboration to design a similar detector using water as a target, which will be included in the ESS$\nu$SB near-detector suite.
%\begin{figure}[!htp]
%    \centering
%     \includegraphics[width=0.9\linewidth]{figures/detectors/NINJAdetector.png} 
%    \caption{ Schematic view of the NINJA detector operation. Charged particle tracks are detected with a very high spatial resolution and %efficiency in theso called Emulsion Cloud Chamber (ECC). Tracks may exit ECC and enter the emulsion shifter and the scintillator tracker which %provide the information on the neutrino interaction time.}
%    \label{fig:detectors:NINJAdetector}
%\end{figure}

\begin{figure}[!htp]
    \centering
    \begin{minipage}[t]{0.45\textwidth}
        \centering
        \includegraphics[width=\textwidth]{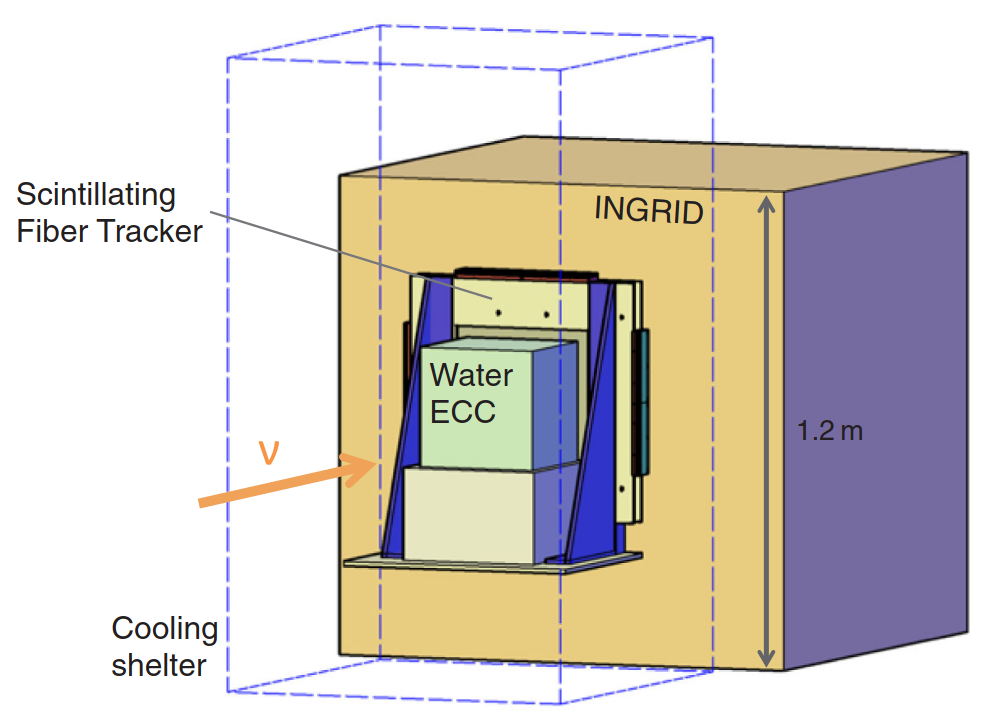} 
        \caption{A schematic view of the NINJA detector situated upstream of the INGRID detector at J-PARC. Taken from \cite{NINJA:2020gbg}.}
        \label{fig:detectors:nd_emulsion_ninja_placement}
    \end{minipage}
    \hfill
    \begin{minipage}[t]{0.45\textwidth}
        \centering
        \includegraphics[width=\textwidth]{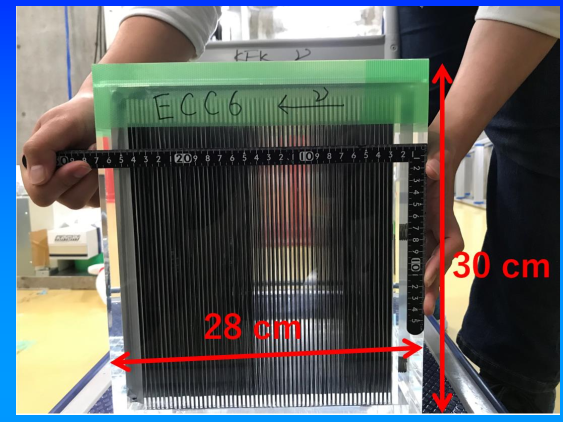} 
        \caption{A photograph of the NINJA ECC element using water as target.}
        \label{fig:detectors:ECC}
%        \includegraphics[width=0.9\linewidth]{figures/detectors/NINJAdetector.png} 
%        \caption{ Schematic view of the NINJA detector operation. Charged particle tracks are detected with a very high spatial resolution and efficiency in the Emulsion Cloud Chamber (ECC). Tracks may exit ECC and enter the emulsion shifter and the scintillator tracker which provide the information on the neutrino interaction time.}
%        \label{fig:detectors:NINJAdetector}    
    \end{minipage}
    
\end{figure}

A conceptual design for the NINJA-like detector adapted to the ESS$\nu$SB project has been developed and included in the technical drawings for the near detector. It would consist of 130 emulsion cloud chamber (ECC) elements like the one shown in Fig.~\ref{fig:detectors:ECC}, giving the water target a total mass of about \SI{1}{t}. The ECCs will be placed in a dedicated cooling shelter immediately upstream of the SFGD. The overall volume of the shelter would be on the order of $3.5\times3.5\times\SI{3}{\m\cubed}$, with a total mass including support structures of around \SI{8}{t}.

%\begin{figure}[!htbp]
%    \centering
%     \includegraphics[width=0.45\textwidth]{figures/detectors/ECC.png} 
%    \caption{A photograph of the NINJA ECC element.}
%    \label{fig:detectors:ECC}
%\end{figure}

\subsubsubsection {Detector Performance}

% Roumen suggests:
% arXiv:2008.03895,
% Phys. Rev. D 102, 072006 (2020)
The NINJA Collaboration has published \cite{NINJA:2020gbg} measurements of the antimuon neutrino interactions in water in the energy range of the T2K experiment, using a prototype of their emulsion detector. The main uncertainties in the measurement come from the low number of registered events. Currently, the analysis of the first physics run with 9 ECC elements in a \SI{75}{kg} water target is underway to increase the statistics of neutrino-water interactions, as shown in Fig.~\ref{fig:detectors:nd_emulsion_ninja_detector}. The event detection efficiency of the detector is estimated to be about 80\%. In one tonne of water there will be \SI{5.5e4}{} muon neutrino interactions %(see Section ZZZ)
and out of these, \SI{4.4e4}{} events would be detected in the emulsion. That would lead to less than 0.5\,\% statistical uncertainty for the total cross-section.

\begin{figure}[!htbp]
    \centering
     \includegraphics[width=0.9\linewidth]{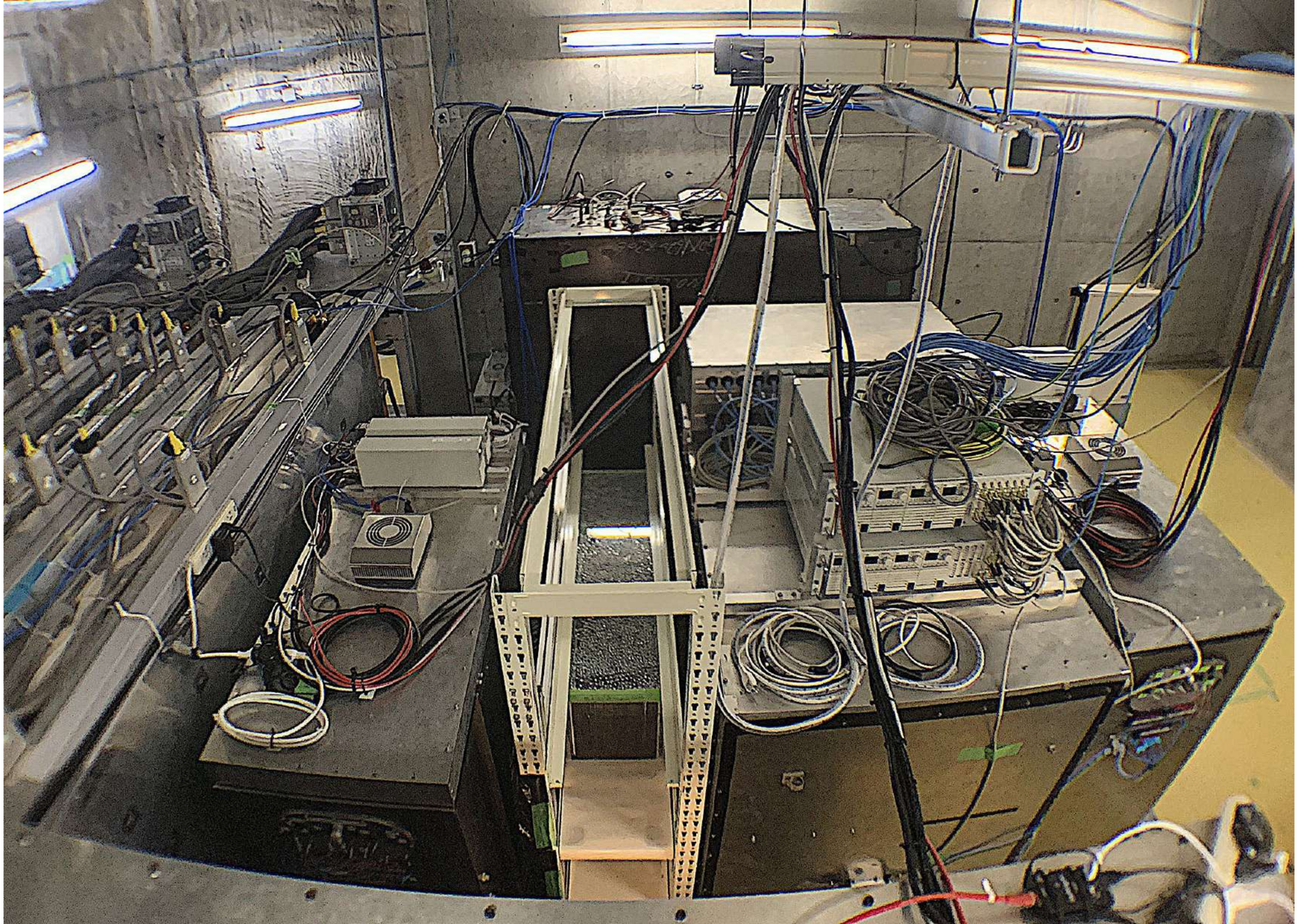} 
    \caption{A NINJA detector includes 9 water-target ECC elements. The NINJA detector is in the centre of the figure. The white rack consists of three tiers, with 3 ECC elements on each tier, for a total of 9 ECC elements in use. An emulsion shifter and a scintillation tracker, which add timing information to emulsion tracks, are placed downstream of ECC elements for $\mu$ identification (by connecting to the Baby-MIND -- one of the T2K near detectors).}
    \label{fig:detectors:nd_emulsion_ninja_detector}
\end{figure}

The scanning technique of nuclear emulsion has been developed throughout the NINJA experiment, and allows for fast automatic recognition of tracks emitted at large angles up to \SI{80}{\degree} with track-detection efficiency of \SI{98}{\percent} \cite{Suzuki:2021las}. This enables the measurement of charged particles emitted at large angles, which is important for the analysis of low energy neutrino-nucleus interactions. An even faster emulsion scanning system is currently under development -- it is expected to be 10 times faster.

The lepton energy resolution of the emulsion detector itself is based on evaluation of the multiple Coulomb scattering of the charged lepton in the final state and is about \SI{10}{}--\SI{20}{\percent}. This could be substantially improved by using information from the downstream SFGD (for leptons that reach it). Thanks to high granularity of the detector, low-momentum protons from neutrino interactions are also detected in the emulsion detector. In the case where \SI{500}{\micro m}-thick iron plates are used in the detector, the momentum threshold of protons is \SI{200}{MeV/c} and the energy measurement can be made with 5 \% accuracy by measuring the range. If thinner iron plates are used for the detector, the momentum threshold and energy resolution will be improved.

\subsubsubsection {Charged lepton identification}

In addition to the momentum and range measurement described above, the ECC technique can be used to measure the ionisation loss by measuring the blackness of tracks recorded in the emulsion layers. Blackness is a density of number of silver grains in an emulsion layer along the track. This is useful for discrimination between low energy $e$ and $\mu$. 

The lepton identification capability in water-target ECC was investigated using the \textsc{GEANT4} framework. The \SI{100}{MeV} momentum $e$ and $\mu$ are exposed to a NINJA type water target ECC; they are clearly separated by using the averaged blackness and range of each track, as shown in Fig.~\ref{fig:detectors:nd_emulsion_ninja_emu}. A detailed study of the capabilities for identification of electron neutrino events in the ECC will be necessary for further design work.

\begin{figure}[!htbp]
    \centering
    \includegraphics[width=0.45\linewidth]{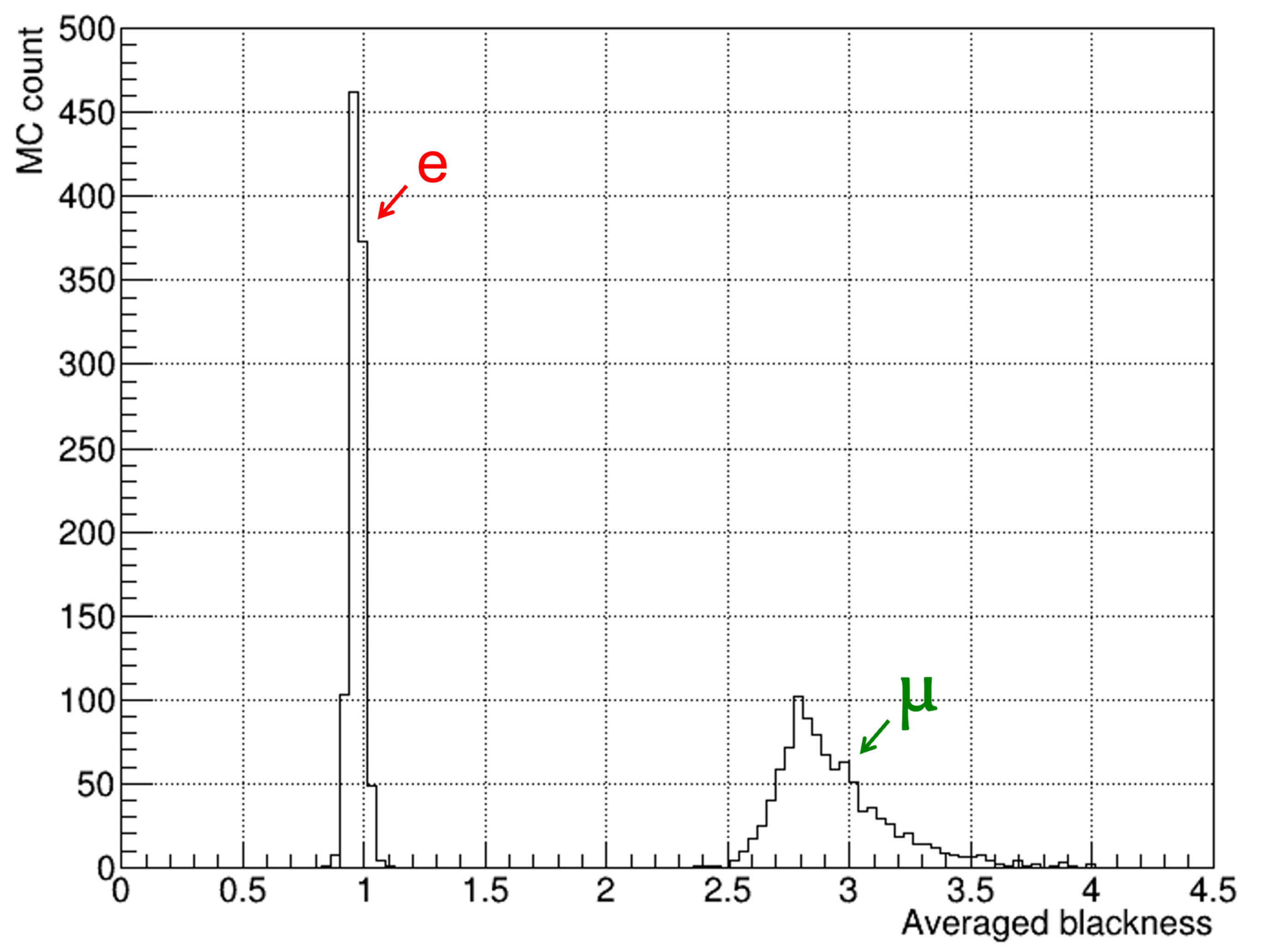} 
    \caption{$e$/$\mu$ separation in ECC. 1000 $e$ and 1000 $\mu$ are simulated perpendicular to a NINJA type ECC by using the GEANT4 framework. The figure shows their averaged blackness in 50 emulsion layers downstream of the stopping layer -- defined as the point at which a particle loses energy and stops or goes out of angular acceptance of \SI{80}{\degree}.}
    \label{fig:detectors:nd_emulsion_ninja_emu}
\end{figure}

\subsection{Far detector}
The primary purpose of the far detector (FD) is to measure the $\nu_e$ ($\overline{\nu}_e$ appearance signal in the $\nu_\mu \rightarrow \nu_e$ ($\overline{\nu}_\mu \rightarrow \overline{\nu}_e$) oscillation channel. A measurement of difference in oscillation probabilities for neutrinos and antineutrinos would imply a CP violation in lepton sector.

%\subsubsection{Mine evaluation}
\subsubsection{Expected neutrino energy spectra and charged lepton distributions}
%Neutrino flux plot: \ref{fig:NuFLux1}; Neutrino flux table: \ref{tab:NuFlux}.

The expected number of neutrino interactions and their energy spectrum in the \SI{538}{kt} fiducial volume of the far detector tanks at a distance of \SI{340}{km} (Zinkgruvan mine) has been calculated \cite{Halic:Thesis:2021} in order to fine-tune the detector response for maximal CP violation discovery potential. The neutrino flux used is shown in Fig.~\ref{fig:NuFLux1}. Oscillation probability was calculated using the PMNS matrix defined in Eq.~(\ref{eq:mixingmatrix}), using parameter values from Table~\ref{tab:ap_events}. The effects of matter on oscillations have been taken into account with an assumed constant matter density of \SI{2.8}{g/cm^3}. The expected spectra are shown in Fig.~\ref{fig:detectors:fd_exp_spec}, from which it is evident that interacting neutrino energies at the FD fall roughly in an energy interval of 150--700\,\si{MeV}. Hence, the neutrino flavour selection should be optimised to achieve high efficiency and purity of $\nu_e$ ($\overline{\nu}_e$) samples in that energy region. The total number of expected interactions is given in Table~\ref{tab:detectors:FD_expected_interactions}. 

%To better understand the results of the flat spectrum simulations, one needs to know how the expected (realistic) spectrum looks like. In this section, expected distribution of charged lepton momenta and scattering angle will be shown as well as the expected neutrino spectrum. All the realistic distributions are produced with flat spectrum simulations but each event was weighted to correspond to the realistic spectrum. The expected neutrino spectrum at FD is shown in Figure \ref{fig:detectors:fd_exp_spec}. Other channels are not shown as they have much smaller expected spectra. One can see that the expected spectrum is negligible above \SI{0.8}{GeV} which is an important note for plots which will be discussed later. Expected distributions of charged lepton momentum are shown in Figure \ref{fig:detectors:fd_exp_mom} while expected distributions of the cosine of charged lepton scattering angle are shown in Figure \ref{fig:detectors:fd_exp_theta}. The important thing to notice is that there are relatively few expected back-scattered antimuons and positrons. On the other hand, there is a noticeable number of back-scattered muons and electrons. 

%%%%%%%%%%%%%%%%%%%%%%%%%%%%%%%%%%%%%%%%%%%%%%%%%%% Expected neutrino spectrum
\begin{figure}[htp!]
        \centering
        \begin{subfigure}[b]{0.475\textwidth}
            \centering
            \includegraphics[width=\textwidth]{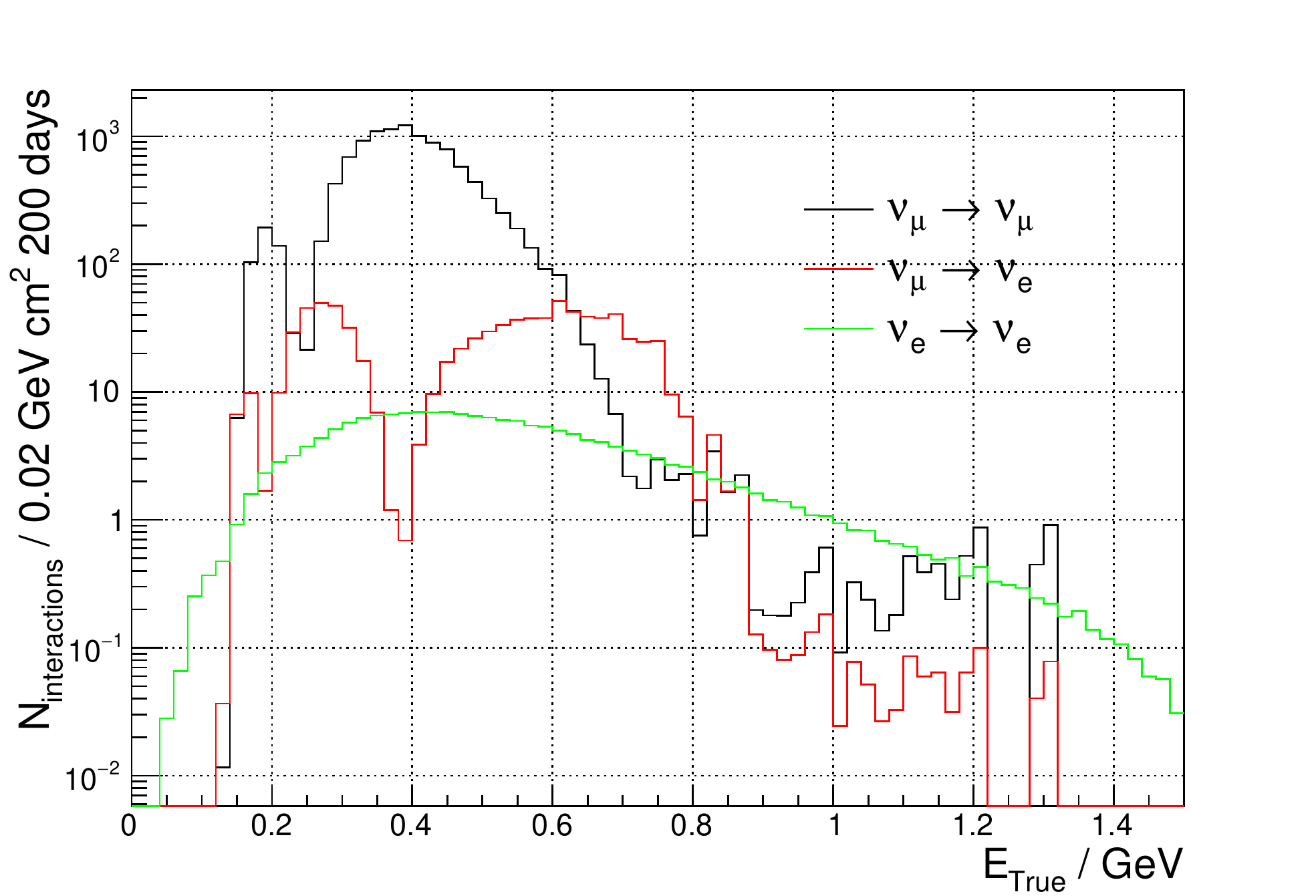}
            \caption{Positive polarity}
            \label{fig:detectors:fd_exp_spec_p}
        \end{subfigure}
        \hfill
        \begin{subfigure}[b]{0.475\textwidth}  
            \centering 
            \includegraphics[width=\textwidth]{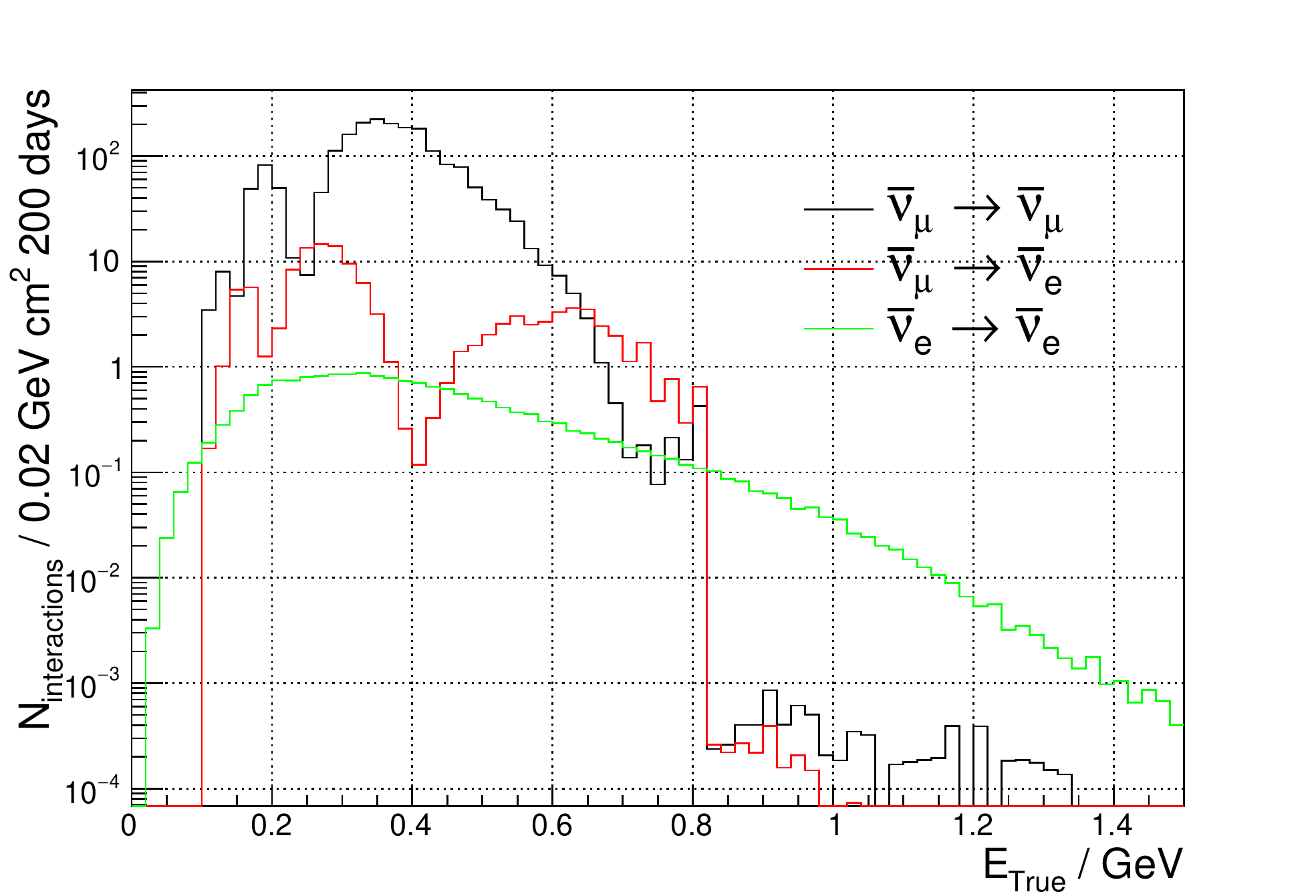}
            \caption{Negative polarity}
            \label{fig:detectors:fd_exp_spec_m}
        \end{subfigure}
        
        \caption{Expected neutrino spectrum in the FD using a realistic flux for positive (a) and negative (b) polarity. The black line represents the $\nu_{\mu} \to \nu_{\mu}$ channel, while the red line represents the $\nu_{\mu} \to \nu_e$ channel.}
        \label{fig:detectors:fd_exp_spec}
    \end{figure}

\begin{table}[htp!]
\setlength{\tabcolsep}{1pt}
\centering
\caption{Expected number of neutrino interactions in the \SI{538}{kt} FD fiducial volume at a distance of \SI{360}{km} (Zinkgruvan mine) in 200 days (one effective year). Shown for positive (negative) horn polarity.}
\label{tab:detectors:FD_expected_interactions}
\footnotesize

\begin{tabular}{p{0.03\textwidth}p{0.1\textwidth}p{0.08\textwidth}p{0.1\textwidth}p{0.08\textwidth}p{0.12\textwidth}p{0.08\textwidth}p{0.12\textwidth}p{0.08\textwidth}p{0.1\textwidth}}
\multicolumn{1}{c}{\multirow{2}{*}{}} & \multirow{2}{*}{\textbf{Channel}}  & \multicolumn{2}{c}{\multirow{2}{*}{\textbf{Non oscillated}}} & \multicolumn{6}{c}{\textbf{Oscillated}} \\ 
\multicolumn{1}{c}{} &  &  &  &  \multicolumn{2}{c}{$\delta_{CP}=0$} &  \multicolumn{2}{c}{$\delta_{CP}=\pi /2$} & \multicolumn{2}{c}{$\delta_{CP}=-\pi /2$}  \\ 
\hline

\multirow{8}{*}{CC} & $\nu_{\mu} \to \nu_{\mu}$ & 22 630.4 & (231.0) & 10 508.7 & (101.6) & 10 430.6 & (5.8) & 10 430.6 & (100.9) \\ 
 & $\nu_{\mu} \to \nu_e$ & \raggedright 0 & (0) & 768.3 & (8.6) & 543.8 & (5.8) & 1 159.9 & (12.8) \\ 
 & $\nu_e \to \nu_e$ & 190.2 & (1.2) & 177.9 & (1.1) & 177.9 & (1.1) & 177.9 & (1.1) \\
 & $\nu_e \to \nu_{\mu}$ & 0 & (0) & 5.3 & (\SI{3.3e-2}{}) & 7.3 & (\SI{4.5e-2}{}) & 3.9 & (\SI{2.4e-2}{}) \\ 
 & $\overline{\nu}_{\mu} \to \overline{\nu}_{\mu}$ & 62.4 & (3 640.3) & 26.0 & (1 896.8) & 26.0 & (1 898.9) & 26.0 & (1 898.9) \\ 
 & $\overline{\nu}_{\mu} \to \overline{\nu}_e$ & 0 & (0) & 2.6 & (116.1) & 3.5 & (164.0) & 1.4 & (56.8) \\ 
 & $\overline{\nu}_e \to \overline{\nu}_e$ & \SI{1.3e-1}{} & (18.5) & \SI{1.3e-1}{} & (17.5) & \SI{1.3e-1}{} & (17.5) & \SI{1.2e-1}{} & (17.5) \\ 
 & $\overline{\nu}_e \to \overline{\nu}_{\mu}$ & 0 & (0) & \SI{3.0e-3}{} & (\SI{4.0e-1}{}) & \SI{1.5e-3}{} & (\SI{2.1e-1}{}) & \SI{4.1e-3}{} & (\SI{5.6e-1}{}) \\ 
\hline

\multirow{4}{*}{NC}
 & $\nu_\mu$ & \multicolumn{7}{c}{\multirow{4}{*}{\begin{tabular}{rl} 16 015.1 & (179.3) \\ 103.7 & (0.7) \\ 55.2 & (3 265.5) \\ \SI{1e-1}{} & (13.6) \end{tabular}}} \\
 & $\nu_e$ & \\ 
 & $\overline{\nu}_\mu$ & \\  
 & $\overline{\nu}_e$ & \\ 
\hline
\end{tabular}
\end{table}   

Neutrino interactions will be observed by measuring charged particles in the final state of the scattering process -- most importantly charged leptons exiting a CC interaction vertex. Neutrino energy is determined using Eq.~(\ref{eqn:detectors:enu_qes}), for which knowledge of the final-state charged lepton momentum and cosine of scattering angle $\cos\theta$ is required. The expected distributions of these quantities for the appearance channel $\nu_\mu \rightarrow \nu_e$ ($\overline{\nu}_\mu \rightarrow \overline{\nu}_e$) and the disappearance channel $\nu_\mu \rightarrow \nu_\mu$ ($\overline{\nu}_\mu \rightarrow \overline{\nu}_\mu$) are shown in Figs.~\ref{fig:detectors:fd_exp_mom} and \ref{fig:detectors:fd_exp_theta}.

%%%%%%%%%%%%%%%%%%%%%%%%%%%%%%%%%%%%%%%%%%%%%%%%%%% Expected distribution of charged lepton momentum
\begin{figure}[htp!]
        \centering
        \begin{subfigure}[b]{0.475\textwidth}
            \centering
            \includegraphics[width=\textwidth]{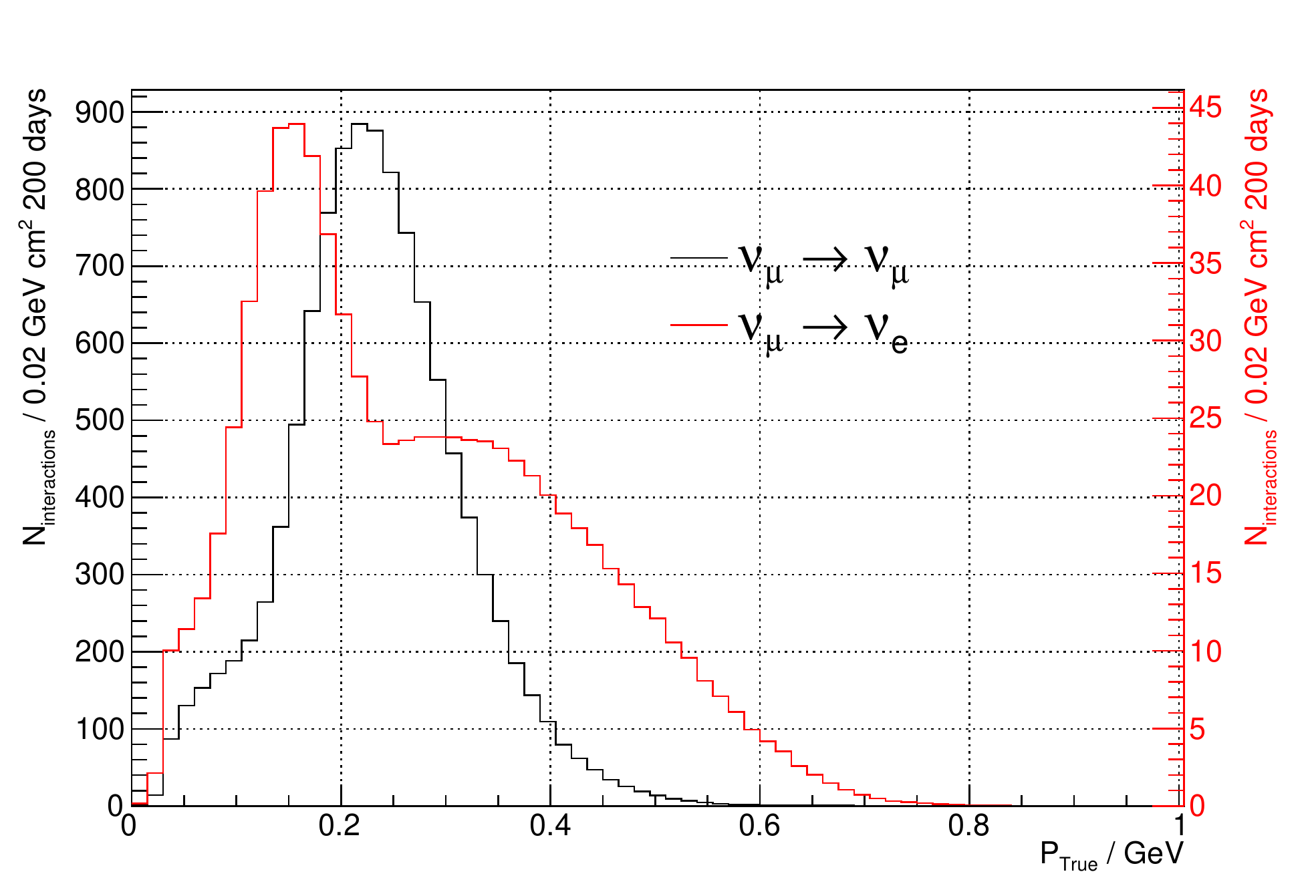}
            \caption{Positive polarity}
            \label{fig:detectors:fd_exp_mom_p}
        \end{subfigure}
        \hfill
        \begin{subfigure}[b]{0.475\textwidth}  
            \centering 
            \includegraphics[width=\textwidth]{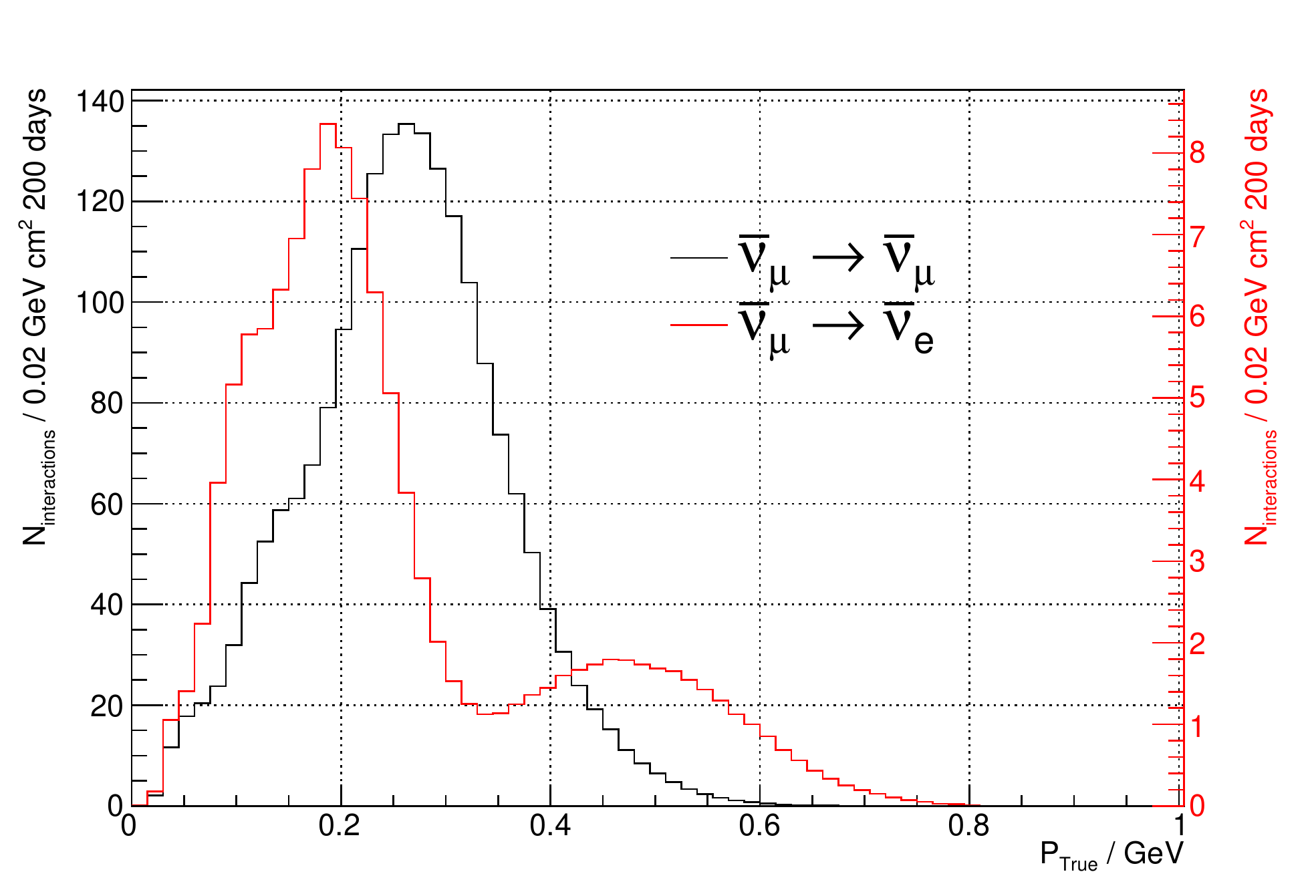}
            \caption{Negative polarity}
            \label{fig:detectors:fd_exp_mom_m}
        \end{subfigure}
        
        \caption{Expected distributions of charged lepton momentum in the FD using the realistic flux for positive (a) and negative (b) polarity. The black line represents muons originating from the $\nu_{\mu} \to \nu_{\mu}$ ($\overline{\nu}_{\mu} \to \overline{\nu}_{\mu}$) channel. The red line represents electrons originating from the $\nu_{\mu} \to \nu_e$ ($\overline{\nu}_{\mu} \to \overline{\nu}_e$) channel.}
        \label{fig:detectors:fd_exp_mom}
    \end{figure}

%%%%%%%%%%%%%%%%%%%%%%%%%%%%%%%%%%%%%%%%%%%%%%%%%%% Expected distribution of charged lepton Scattering angle
\begin{figure}[htp!]
        \centering
        \begin{subfigure}[b]{0.475\textwidth}
            \centering
            \includegraphics[width=\textwidth]{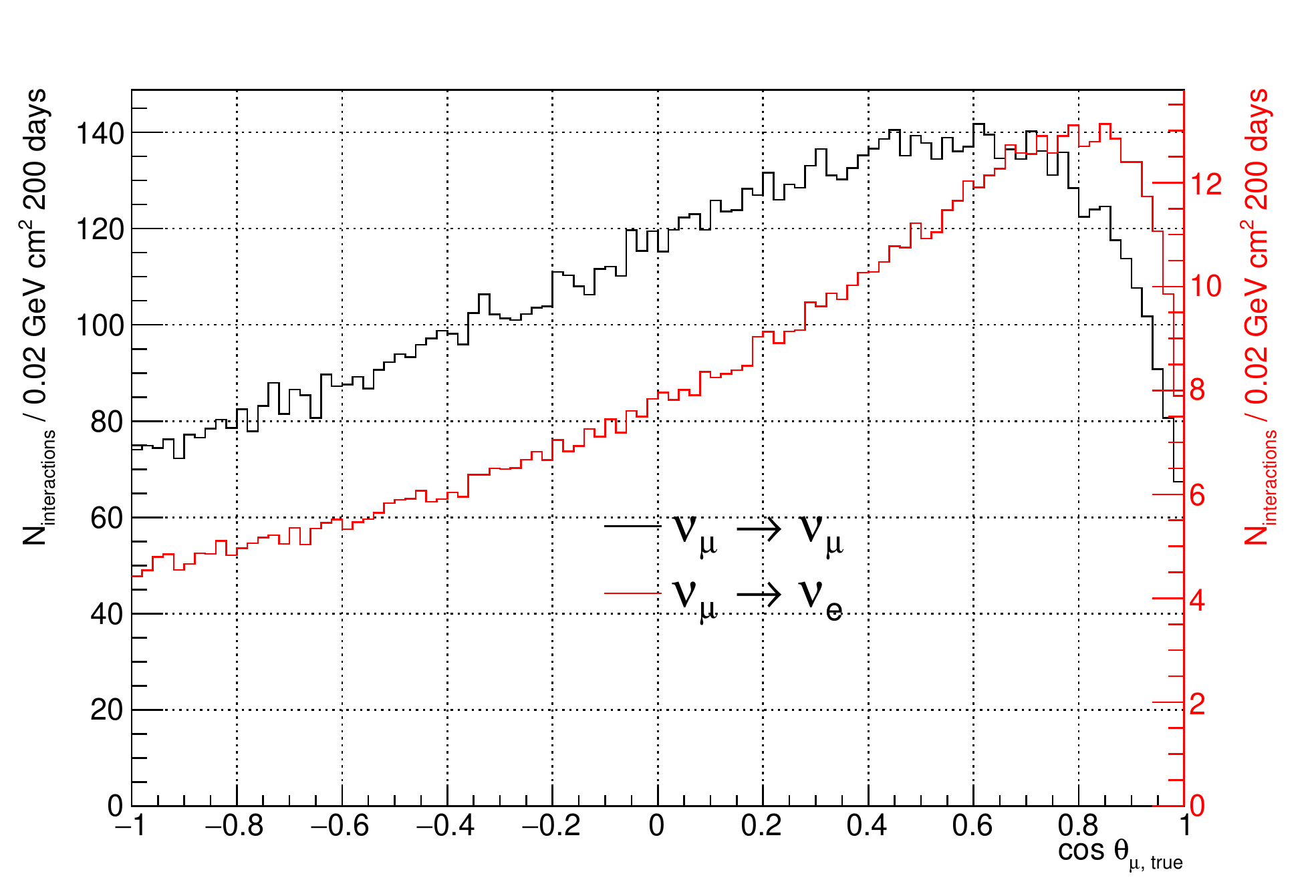}
            \caption{Positive polarity}
            \label{fig:detectors:fd_exp_theta_p}
        \end{subfigure}
        \hfill
        \begin{subfigure}[b]{0.475\textwidth}  
            \centering 
            \includegraphics[width=\textwidth]{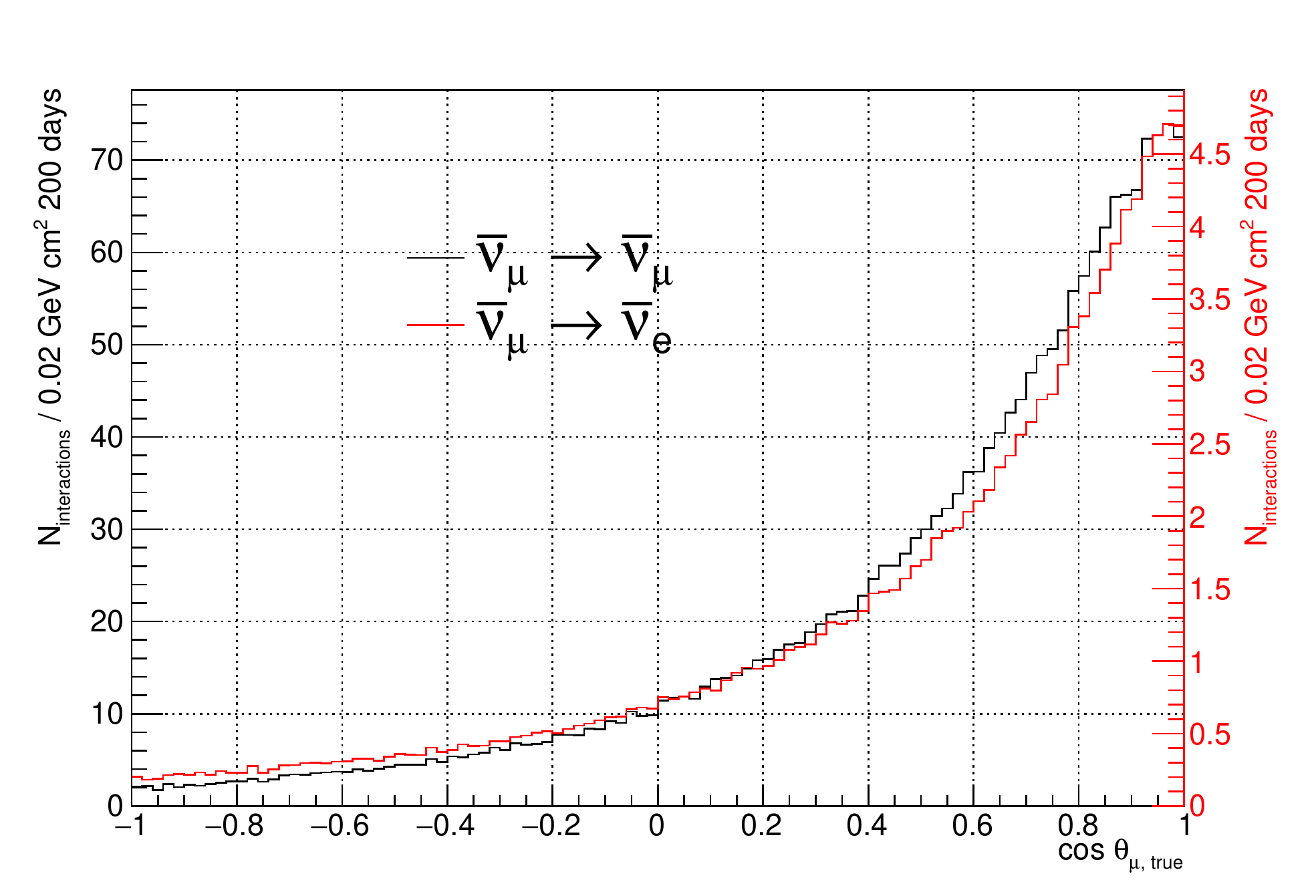}
            \caption{Negative polarity}
            \label{fig:detectors:fd_exp_theta_m}
        \end{subfigure}
        
        \caption{Expected distributions for the cosine of the charged lepton scattering angle in the FD using the realistic flux for positive (a) and negative (b) polarity. The black line represents muons originating from the $\nu_{\mu} \to \nu_{\mu}$ ($\overline{\nu}_{\mu} \to \overline{\nu}_{\mu}$) channel. The red line represents electrons originating from the $\nu_{\mu} \to \nu_e$ ($\overline{\nu}_{\mu} \to \overline{\nu}_e$) channel.}
        \label{fig:detectors:fd_exp_theta}
    \end{figure}
    
From the expected electron (positron) true-momentum distributions for the signal oscillation channel $\nu_{\mu} \to \nu_e$ ($\overline{\nu}_{\mu} \to \overline{\nu}_e$) shown in Fig.~\ref{fig:detectors:fd_exp_mom_m}, it can be seen that the relevant electron momentum falls roughly between 50 and \SI{600}{MeV}. There are virtually zero neutrino interactions in which electron (positron) momenta exceed \SI{800}{MeV}. The true momentum range of muons (anti-muons) exiting neutrino interaction vertices for the $\nu_{\mu} \to \nu_{\mu}$ ($\overline{\nu}_{\mu} \to \overline{\nu}_{\mu}$) disappearance channel is expected to be somewhat narrower, approximately covering the interval between 100 and \SI{400}{MeV}. The Cherenkov threshold momentum for muons in water is about $p_\mu = \SI{120}{MeV}$, so a fraction of low-momentum muons is expected to be undetectable. Muon (antimuon) momenta are not expected to exceed \SI{600}{MeV}.

The distributions of $\cos\theta$ for electrons and muons shown in Fig.~\ref{fig:detectors:fd_exp_theta_p} are somewhat flat, in contrast with $\cos \theta$ distributions of positrons and antimuons (Fig.~\ref{fig:detectors:fd_exp_theta_m}) which show a strong preference for forward scattering ($\cos\theta = 1$) due to the helicity dependence of weak forces.

\subsubsection{Detector Design}
The far detector of the ESS$\nu$SB project will be composed of two standing cylindrical tanks. This choice is mandated by immense technical difficulties and costs to excavate a single cavern large enough to hold the entire desired volume of target water. The photomultiplier tubes (PMT) used to detect the Cherenkov light will be placed on the walls of each cylinder, making the number of PMTs directly proportional to the surface area of the cylinders. Each cylindrical tank will have the diameter of its base equal to its height to minimise the surface-to-volume ratio, while minimising the number of PMTs for a given target mass. The price of PMTs is a major factor in the overall cost of the FD facility; thus, this design is the most cost-effective.

A single FD cavern will be a standing cylinder with base diameter $d_\mathrm{C} = \SI{78}{m}$ and height $h_\mathrm{C} = \SI{78}{m}$, with an additional volume atop the cavern for access and placement of infrastructure. A free-standing cylindrical detector with inner dimensions $d_\mathrm{I} = \SI{74}{m}$ and $h_\mathrm{I} = \SI{74}{m}$ will be placed in the cavern (see Fig.~\ref{fig:detectors:FD_main_technical_drawing}). The water in the inner volume of the detector will serve as the neutrino target, while the volume between the detector and cavern walls will be used as a veto detector. The purpose of the veto detector is to reject events coming from outside the inner-detector volume.

\begin{figure}[htp]
    \centering
    \includegraphics[width=\textwidth]{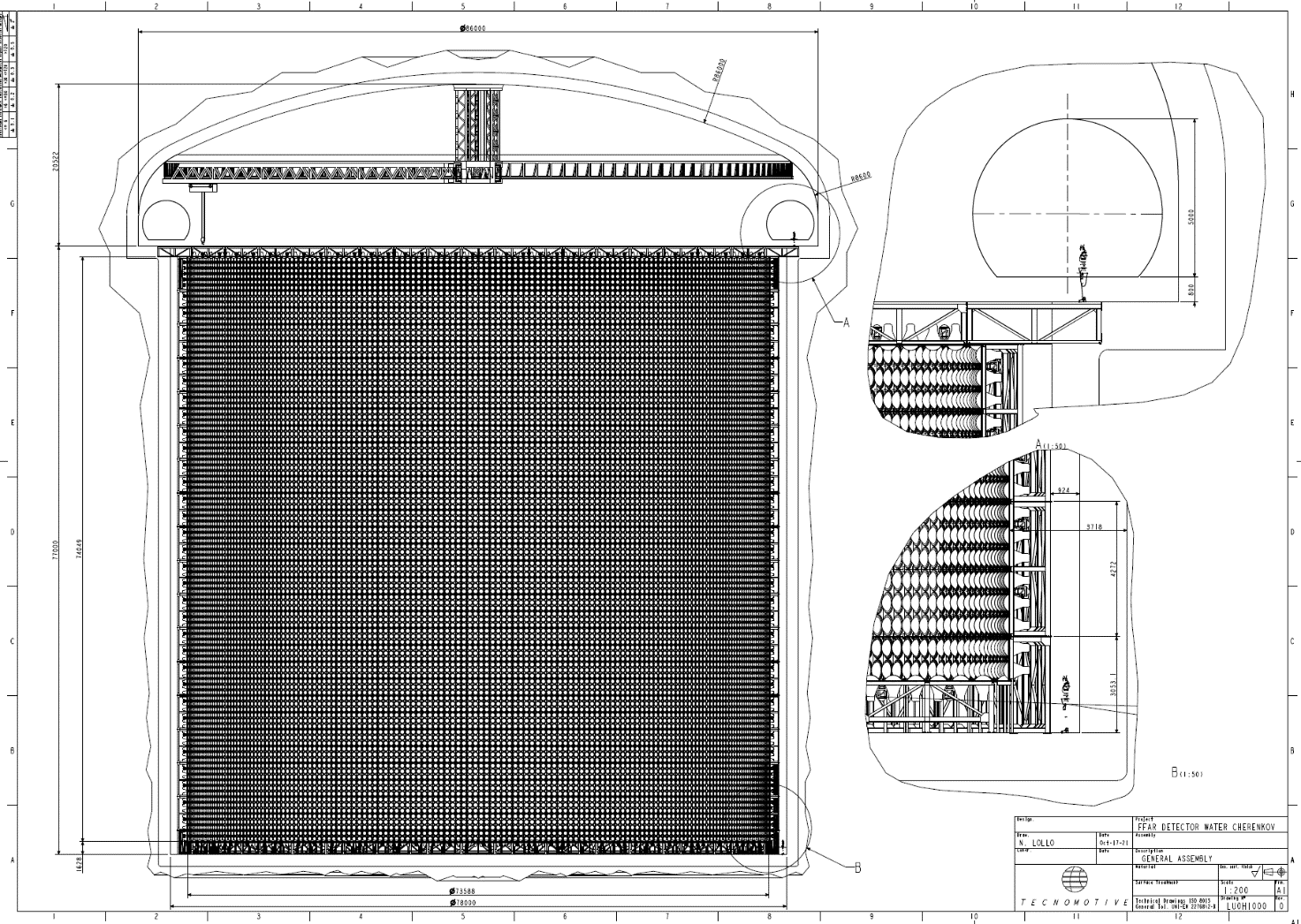}
    \caption{Overall view of a single far-detector tank with indicated dimensions.}
    \label{fig:detectors:FD_main_technical_drawing}
\end{figure}

The PMT coverage of the detector is defined as the fraction of its inner surface area covered by PMTs. Designs using 20, 30, and 40\% coverage have been studied. Large (\SI{20}{inch}) PMTs have been chosen for use in the inner detector to reduce the number of tubes required to achieve a given coverage. The numbers of required PMTs per single tank for different coverages are shown in Table~\ref{tab:detectors:fd_coverage_pmts}. It should be noted that the exact number of required PMTs may vary by a small amount, depending on the technical details of the tank construction.

\begin{table}[phtb]
\footnotesize
\centering
\caption{Number of PMTs required for various detector-coverage options.}
\label{tab:detectors:fd_coverage_pmts}
\begin{tabular}{cc}
\textbf{Coverage} & \textbf{Number of PMTs} \\
\hline
20\% & 24888 \\
30\% & 37830 \\
40\% & 49914 \\
\hline
\end{tabular}
\end{table}

A dedicated study has been performed in order to estimate the effect of coverage on the detector performance. Electrons and muons were simulated in the inner detector volume using \textsc{WCSim} software package and their tracks and momenta have been reconstructed by the \textsc{fiTQun} track reconstruction software. A dedicated optimisation procedure of \textsc{fiTQun}, also known as \emph{tuning}, has been performed for each of the three coverages. The distribution of difference between true and reconstructed charged lepton momentum is shown in Fig.~\ref{fig:detectors:FD_coverages_reconstruction}. 

\begin{figure}[ht!]
    \centering
    \begin{subfigure}[t]{0.49\textwidth}
        \centering
        \includegraphics[width=\textwidth]{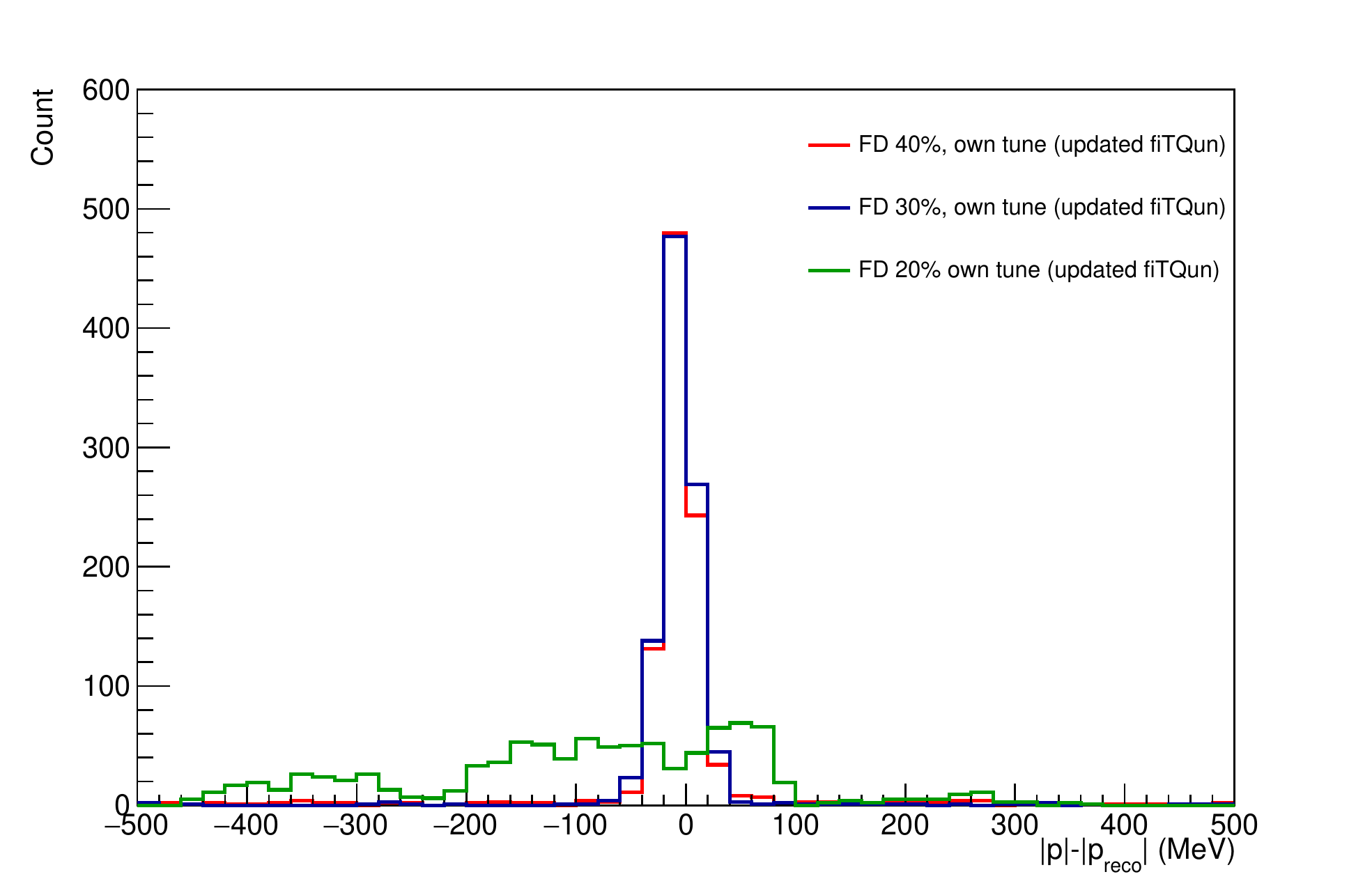}
        \caption{Electrons}
        \label{fig:detectors:FD_coverages_reconstruction_electrons}
    \end{subfigure}
    \hfill
    \begin{subfigure}[t]{0.49\textwidth}
        \centering
        \includegraphics[width=\textwidth]{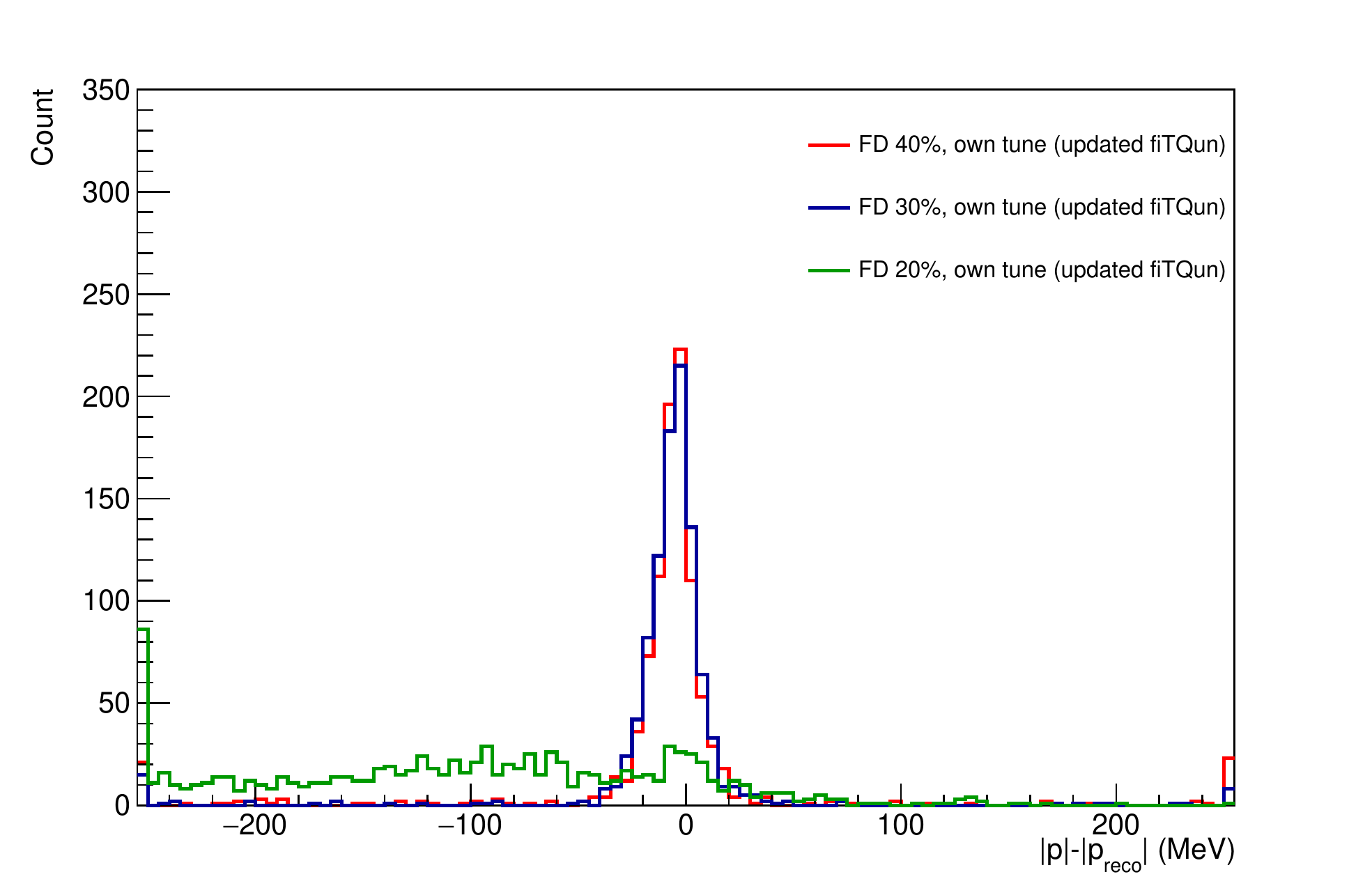}
        \caption{Muons}
        \label{fig:detectors:FD_coverages_reconstruction_muons}
    \end{subfigure}
    \caption{Distribution of difference between true and reconstructed momenta of electrons and muons for 20, 30, and 40\% coverage options.}
    \label{fig:detectors:FD_coverages_reconstruction}
\end{figure}
	 
As illustrated by Fig.~\ref{fig:detectors:FD_coverages_reconstruction}, the 30 and 40\% coverage options perform comparably well in the range of expected electron/muon momenta, while the 20\% option has essentially no momentum-reconstruction ability. The 30\% coverage option is chosen as the baseline for the far detector design because it provides a satisfactory performance with an optimal number of PMTs.

Further study is needed to evaluate the performance of the 30 and 40\% options for non-beam physics such as atmospheric, galactic, and supernova neutrino studies, proton decay studies, and more. To account for the possibility that these additional physics searches might benefit from higher PMT coverage, the frame of the detector has been designed to hold the number of PMTs corresponding to 40\% coverage. The 30\% coverage can be achieved in this design by evenly distributing PMTs across $3/4$ of the available mounts. The baseline far detector parameters are shown in Table \ref{tab:detectors:fd_parameters}.

\begin{table}[!htp]
\footnotesize
\centering
\caption{Parameters of the ESS$\nu$SB far detector}
\label{tab:detectors:fd_parameters}
\begin{tabular}{l l}
\textbf{Parameter} & \textbf{Value} \\
\hline
Type & Cylindrical water Cherenkov \\
Tank geometry & Standing cylinder \\
Number of tanks & 2 \\
Tank diameter & \SI{78}{m} \\
Tank height & \SI{78}{m} \\
Inner detector diameter & \SI{74}{m} \\
Inner detector height & \SI{74}{m} \\
Target water mass & \SI{318}{\kilo\tonne} per tank (\SI{636}{\kilo\tonne} total) \\
Inner detector PMT coverage & 30\% \\
Inner PMT diameter & \SI{20}{\inch} \\
Number of inner PMTs & 37,830 per tank (75,660 total) \\
Fiducial volume cut & \SI{2}{\meter} inwards from inner detector walls \\
Fiducial water mass & \SI{269}{\kilo\tonne} per tank (\SI{538}{\kilo\tonne} total) \\
Outer (veto) PMT size & \SI{8}{\inch} \\
Number of outer (veto) PMTs & 8226 per tank (16452) \\
\hline
\end{tabular}
\end{table}

\subsubsubsection{Preliminary technical design}
A preliminary technical design of the far detectors has been made by the company \textsc{Tecnomotive}.

Each PMT will be placed within a protective cover made of stainless steel bell and a transparent acrylic cover on top (see Fig. \ref{fig:detectors:FD_technical_Fig_1} for inward facing PMTs, and Fig.~\ref{fig:detectors:FD_technical_Fig_3} for outward facing PMTs). The bell will be interfaced to the acrylic window by a waterproof joint. The PMT will be locked to a bell using a plastic centering ring. The readout electronics and the high-voltage source will be placed in a dedicated space within the bell. Two cable feed-throughs are going to be implemented in the bell to be used for electric connections and pressurisation of the protective cover.

\begin{figure}[htp!]
    \centering
    \includegraphics[width=0.9\textwidth]{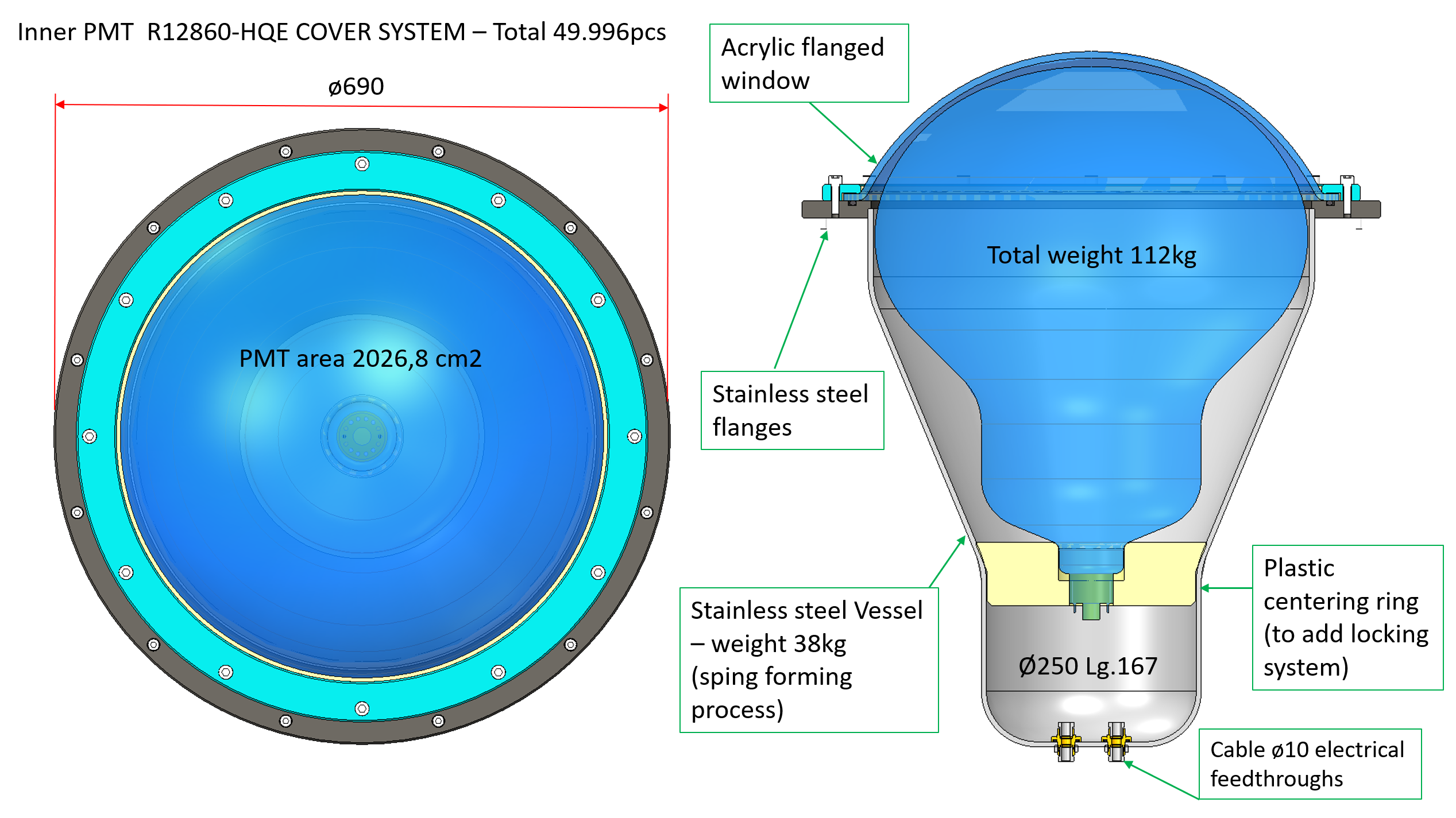}
    \caption{A schematic view of an inward-facing \SI{20}{inch} PMT embedded in a protective cover.}
    \label{fig:detectors:FD_technical_Fig_1}
\end{figure}

\begin{figure}[htp!]
    \centering
    \includegraphics[width=0.9\textwidth]{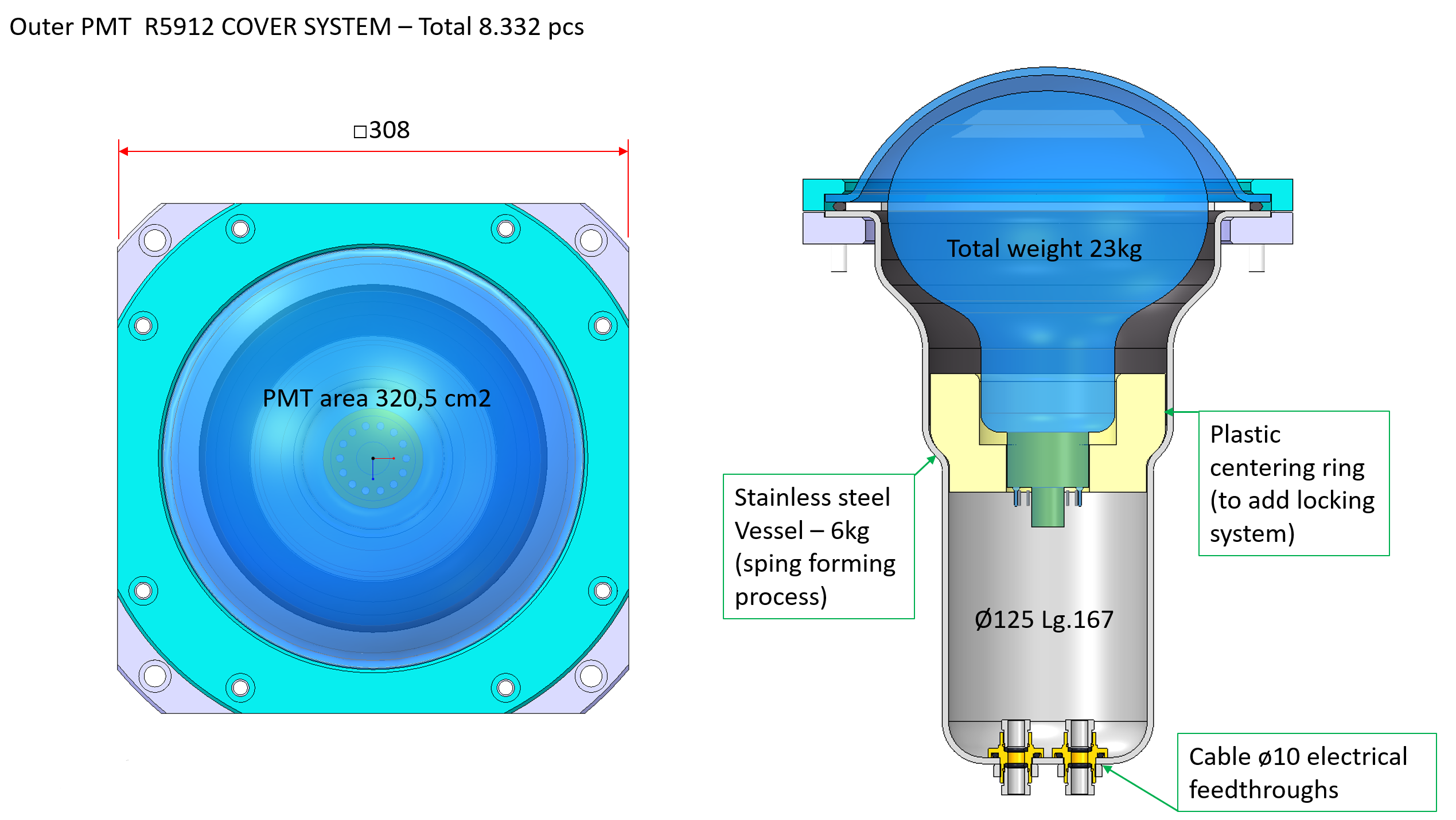}
    \caption{A schematic view of an outward-facing (veto) \SI{8}{inch} PMT embedded in a protective cover.}
    \label{fig:detectors:FD_technical_Fig_3}
\end{figure}

To simplify the construction phase, the inward-facing PMTs on the wall of the cylinder will be arranged in sub-modules consisting of three PMTs (see Fig.~\ref{fig:detectors:FD_technical_Fig_5.5.png}). A mechanical interface will be constructed for each outward-facing PMT (see Fig.~\ref{fig:detectors:FD_technical_Fig_7.png}). These will then be mounted on a frame as shown in Fig.~\ref{fig:detectors:FD_technical_Fig_8.png}, which will in turn be arranged to form the cylindrical structure of the detector as illustrated in Fig.~\ref{fig:detectors:FD_technical_Fig_9.png}.

\begin{figure}[htp!]
    \centering
    \includegraphics[width=0.8\textwidth]{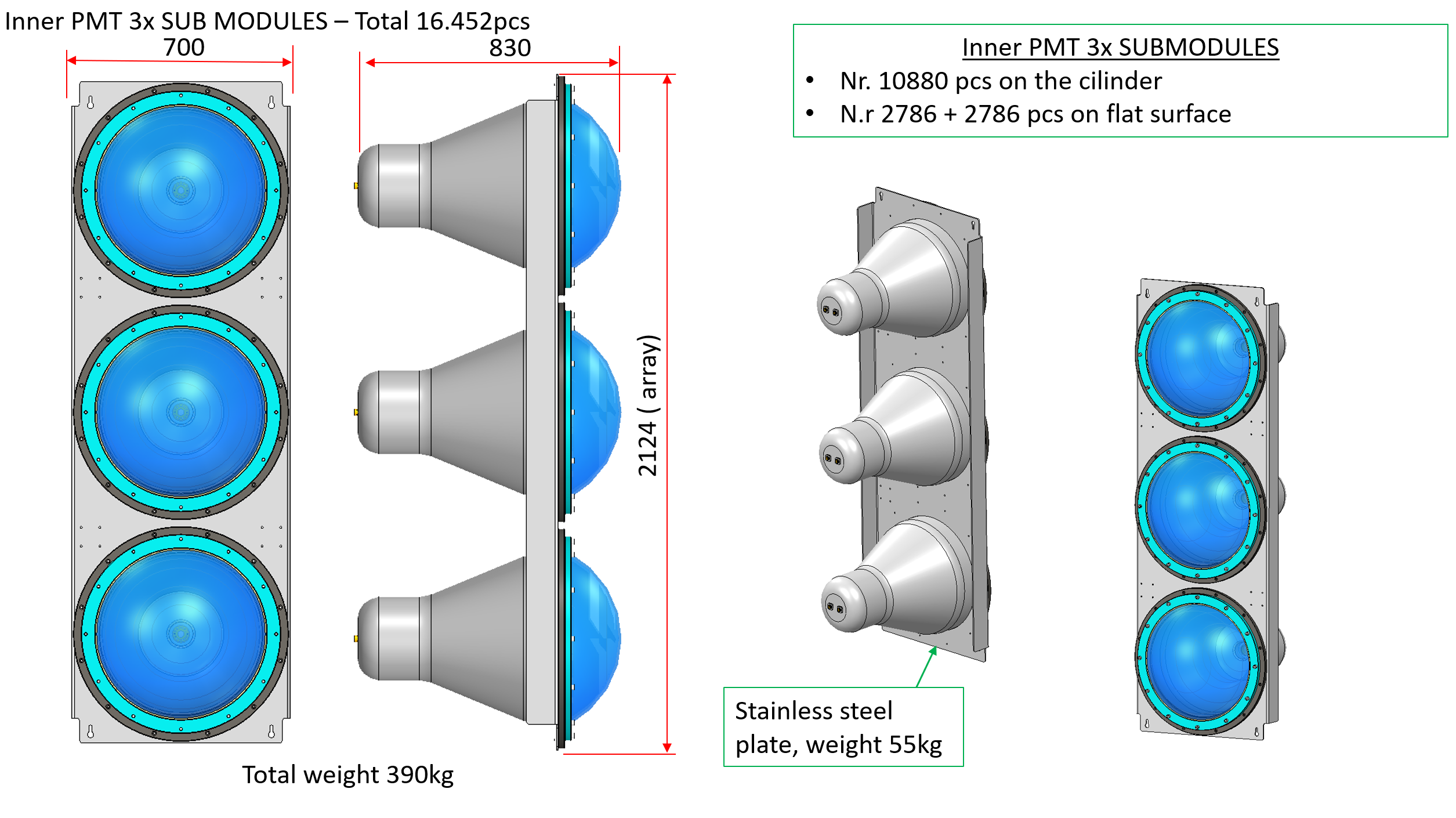}
    \caption{Basic inward-facing PMT module consisting of 3 PMTs within a protective cover.}
    \label{fig:detectors:FD_technical_Fig_5.5.png}
\end{figure}

\begin{figure}[htp!]
    \centering
    \includegraphics[width=0.8\textwidth]{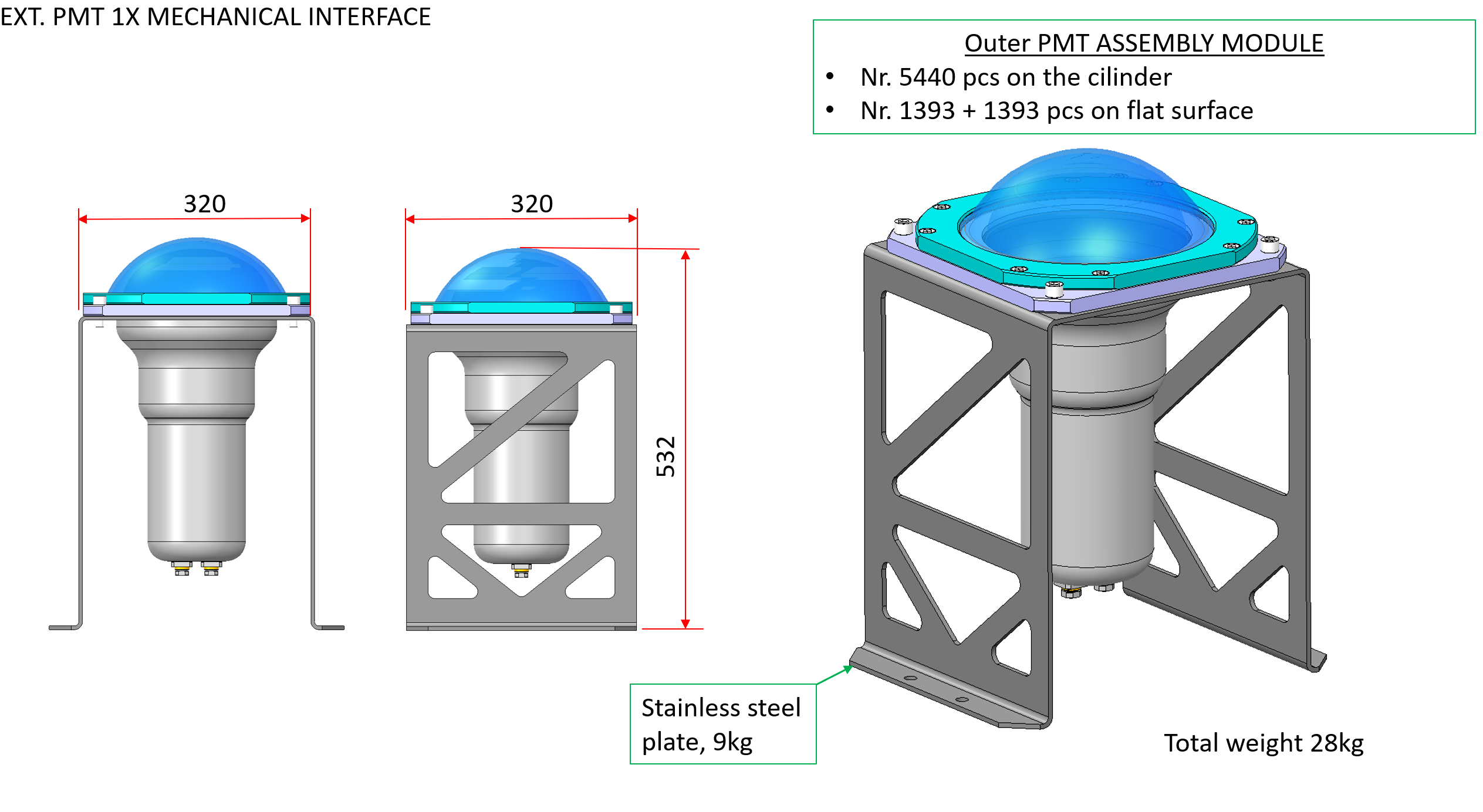}
    \caption{Mechanical interface for the outward-facing (veto) PMT within a protective cover.}
    \label{fig:detectors:FD_technical_Fig_7.png}
\end{figure}

\begin{figure}[htp!]
    \centering
    \includegraphics[width=0.8\textwidth]{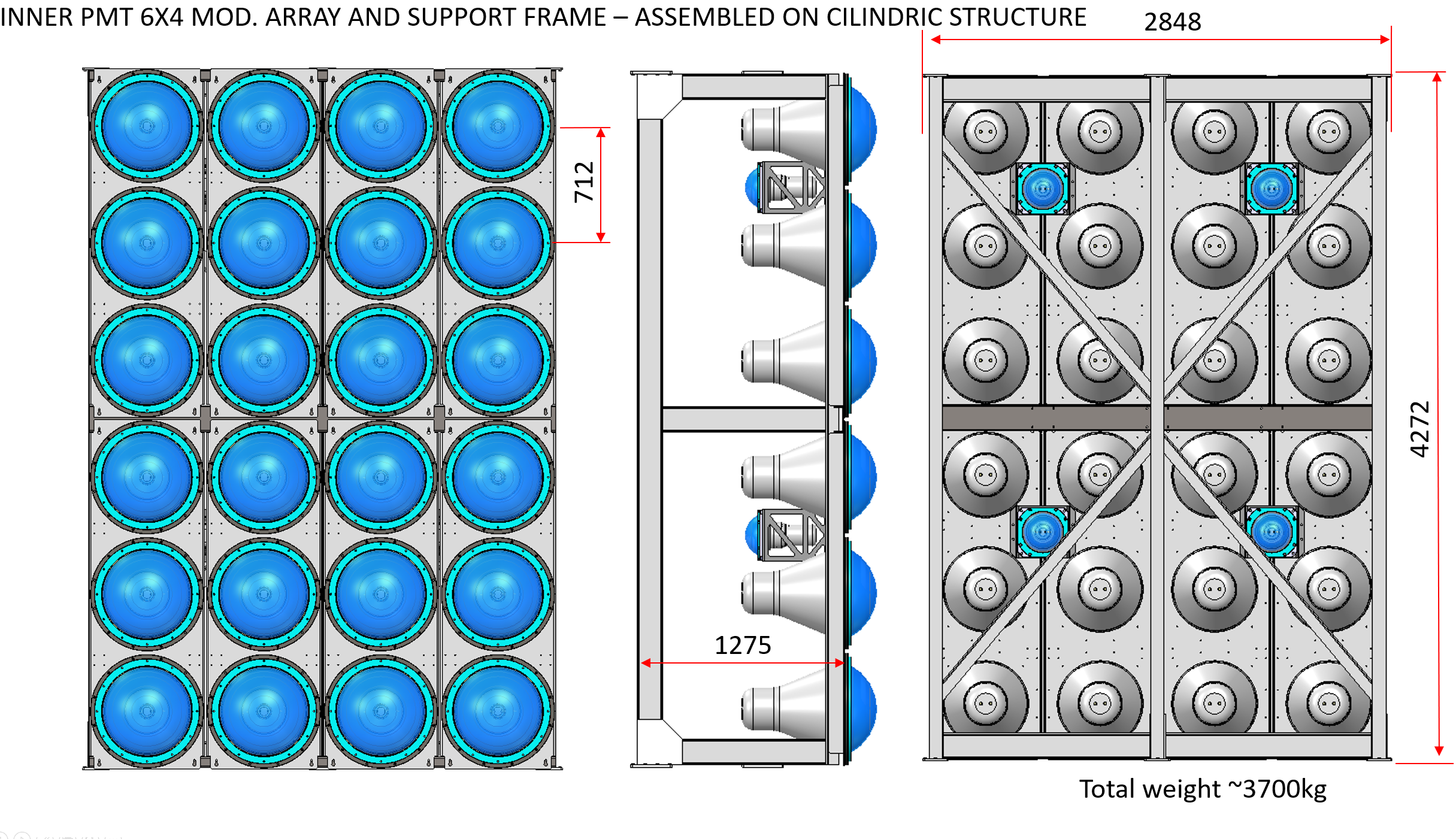}
    \caption{Basic frame used for the construction of the detector.}
    \label{fig:detectors:FD_technical_Fig_8.png}
\end{figure}

\begin{figure}[htp!]
    \centering
    \includegraphics[width=0.7\textwidth]{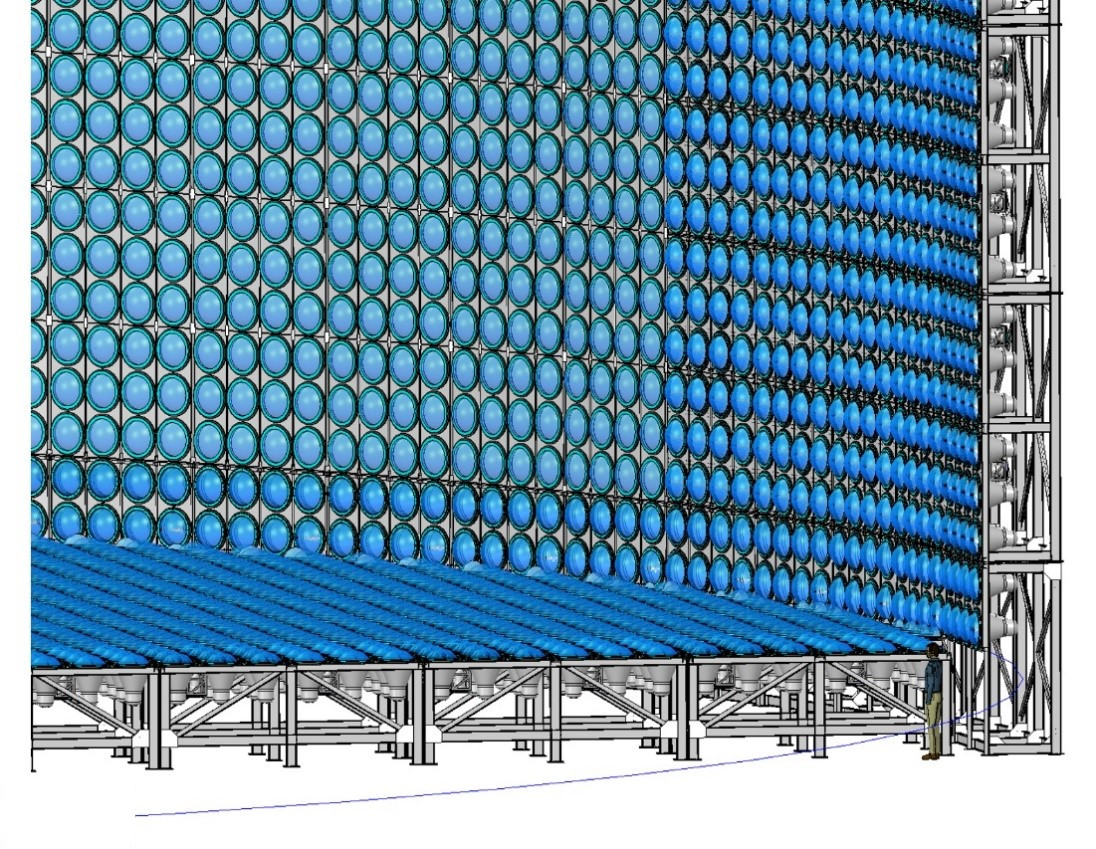}
    \caption{A detail of the lower part of the detector construction, using the frames described in Fig.~\ref{fig:detectors:FD_technical_Fig_8.png}.}
    \label{fig:detectors:FD_technical_Fig_9.png}
\end{figure}

The cylindrical frame will be constructed in an excavated cavern as a free-standing structure, with optional mechanical connections to the cavern walls and ceiling for additional stability. The entire cavern will be filled with ultra-pure water, which must be continuously purified to maintain the required purity. The top of the detector will be covered with a flooring made of steel sheets to minimise the introduction of impurities to the water from outside and to support required equipment.

The top part of the cavern will be dome-shaped. It will feature a crane with a fulcrum on the cylinder axis and with its arm extending through the entire width of the cavern for handling and assembly of components. All components will be transported by trucks to the top of the cavern through two service tunnels. This is illustrated in Fig.~\ref{fig:detectors:FD_technical_Fig_20.png}.

\begin{figure}[htp!]
    \centering
    \includegraphics[width=0.55\textwidth]{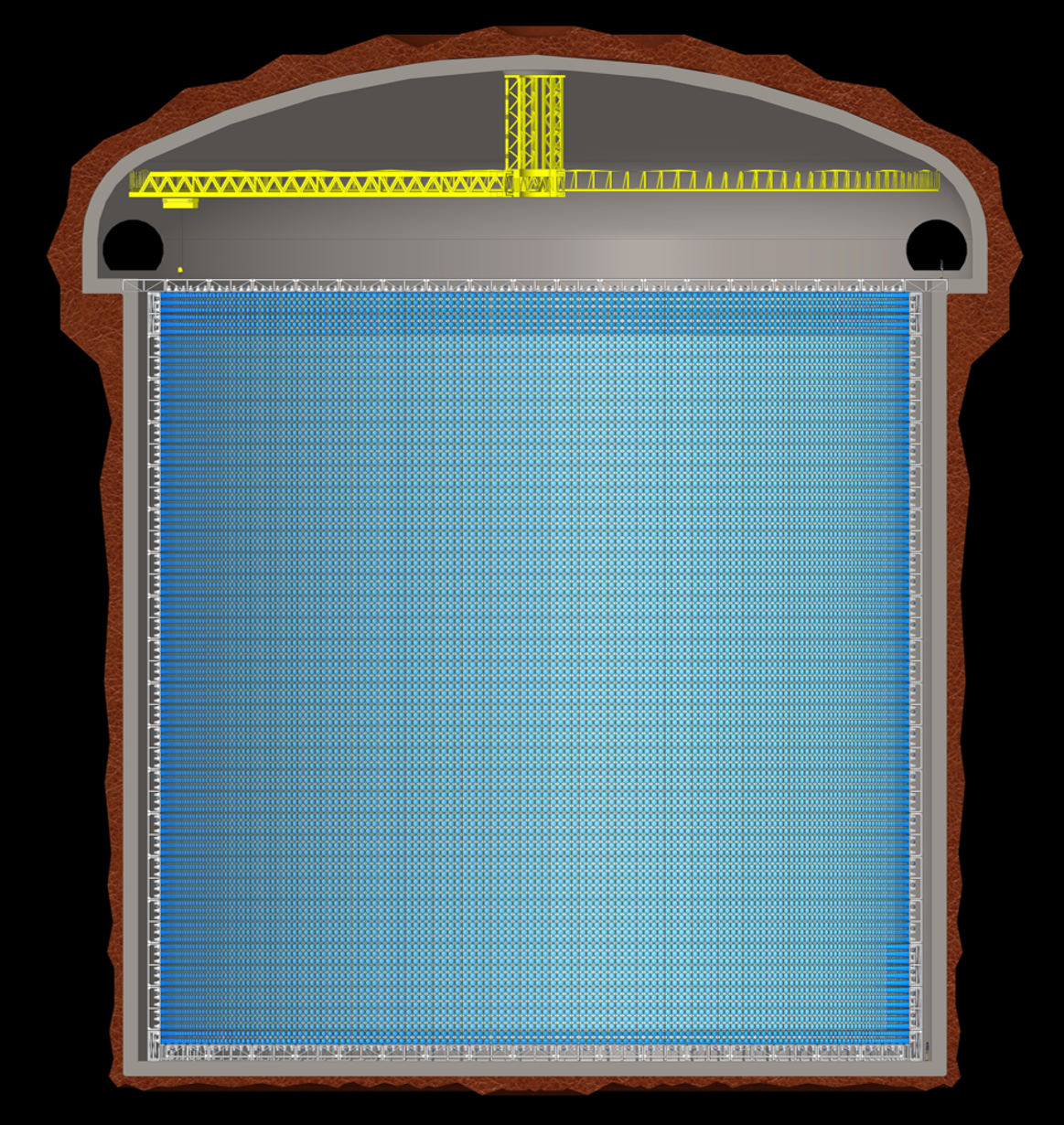}
    \caption{An illustration of the far detector cavern.}
    \label{fig:detectors:FD_technical_Fig_20.png}
\end{figure}

A more detailed technical design will be produced once the detailed geological properties of the mine site are determined.

\subsubsection{Detector Performance}

The CP violation will be measured in ESS$\nu$SB by observing CC interactions of electron neutrinos and antineutrinos coming from $\nu_\mu \rightarrow \nu_e$ and $\overline{\nu}_\mu \rightarrow \overline{\nu}_e$ oscillation channels, respectively. Apart from the parameters of neutrino mixing model (see equations \ref{eq:mixingmatrix} and \ref{Eq:Probability}), the oscillation probabilities depend on neutrino energy. Hence, the far detector should have a very good performance for discrimination between $\nu_\mu(\overline{\nu}_\mu)$, $\nu_e(\overline{\nu}_e)$ CC interactions, and $\nu(\overline{\nu})$ NC interactions, together with a reasonable neutrino energy resolution.

Determination of the incoming neutrino energy and flavour requires reconstruction of the momentum, scattering angle and flavour of the final state charged lepton (electron, muon, or their antiparticles); these three quantities are directly measured by the detector. The final state of neutrino interactions may contain additional charged particles, which degrade the performance of the detector for identification and momentum/angle reconstruction of charged leptons in the neutrino branch. The analysis of the charged lepton reconstruction performance in the FD is done in two parts: (\textit{i}) using a simulation of pure charged lepton flux to determine the intrinsic detector response, and (\textit{ii}) using full neutrino interaction simulation to determine a realistic detector response which includes effects of multi-particle final state of neutrino interactions.

Detector performance for the neutrino energy reconstruction is evaluated using full neutrino interaction simulation with a flavour selection algorithm which depends solely on observable quantities. The energy reconstruction performance (migration matrices) were determined not only for correctly determined neutrino flavour and interaction type, but also for all combinations of true neutrino flavour and reconstructed flavour.

Reconstruction performances in this section are shown mostly for the correctly identified flavour of charged leptons and neutrinos for clarity. The physics performance, however, has been determined using the full set of migration matrices for all combinations of identified/simulated neutrino flavours.

\subsubsubsection{Monte Carlo Simulation and Analysis}
\label{sec:detectors:FD_MC_simulation}
An extensive MC simulation of neutrino interactions in the far detector and its response has been performed to obtain the detector performance. Neutrino interactions with a water target have been simulated using the \textsc{GENIE} interaction generator, particle propagation through the detector and PMT response using \textsc{WCSim} software based on \textsc{Geant4}, and detector response using \textsc{fiTQun} reconstruction software. More details on the simulation software may be found in Section~\ref{sec:detectors:software}. The number of simulated events corresponds to 10--100 times expected number of neutrino interactions.

\paragraph{Pure charged lepton production} The study of the detector response to pure charged leptons is performed using a dedicated MC production that does not contain simulation of neutrino interactions: instead, propagation of charged leptons through the detector is simulated. Each simulated charged lepton originates from a point within the inner detector. Creation points of simulated leptons were uniformly distributed along the full inner water volume,\footnote{As opposed to fiducial volume only.} with an isotropic momentum-direction distribution. Their kinetic energies were uniformly distributed within predetermined ranges. Kinetic energy ranges and sizes of MC sample for each species are shown in Table~\ref{tab:detectors:fd_mc_events}.

\begin{table}[!htbp]
\footnotesize
\centering
\caption{Kinetic energy ($E_\text{kin}$) range and MC set size for each species of simulated charged leptons.}
\begin{tabular}{rcl}
\textbf{Charged Lepton} & \boldmath$E_\mathbf{kin} / \mathbf{MeV}$ & \textbf{Simulated events} \\
\hline
electron & 10--1200 & 500 000 \\
muon & 10--1100 & 500 000 \\
positron & 10--1200 & 500 000 \\
antimuon & 10--1100 & 500 000 \\
\hline
\end{tabular}
\label{tab:detectors:fd_mc_events}
\end{table}

\begin{figure}[htbp!]
    \centering
    \includegraphics[width=0.5\textwidth]{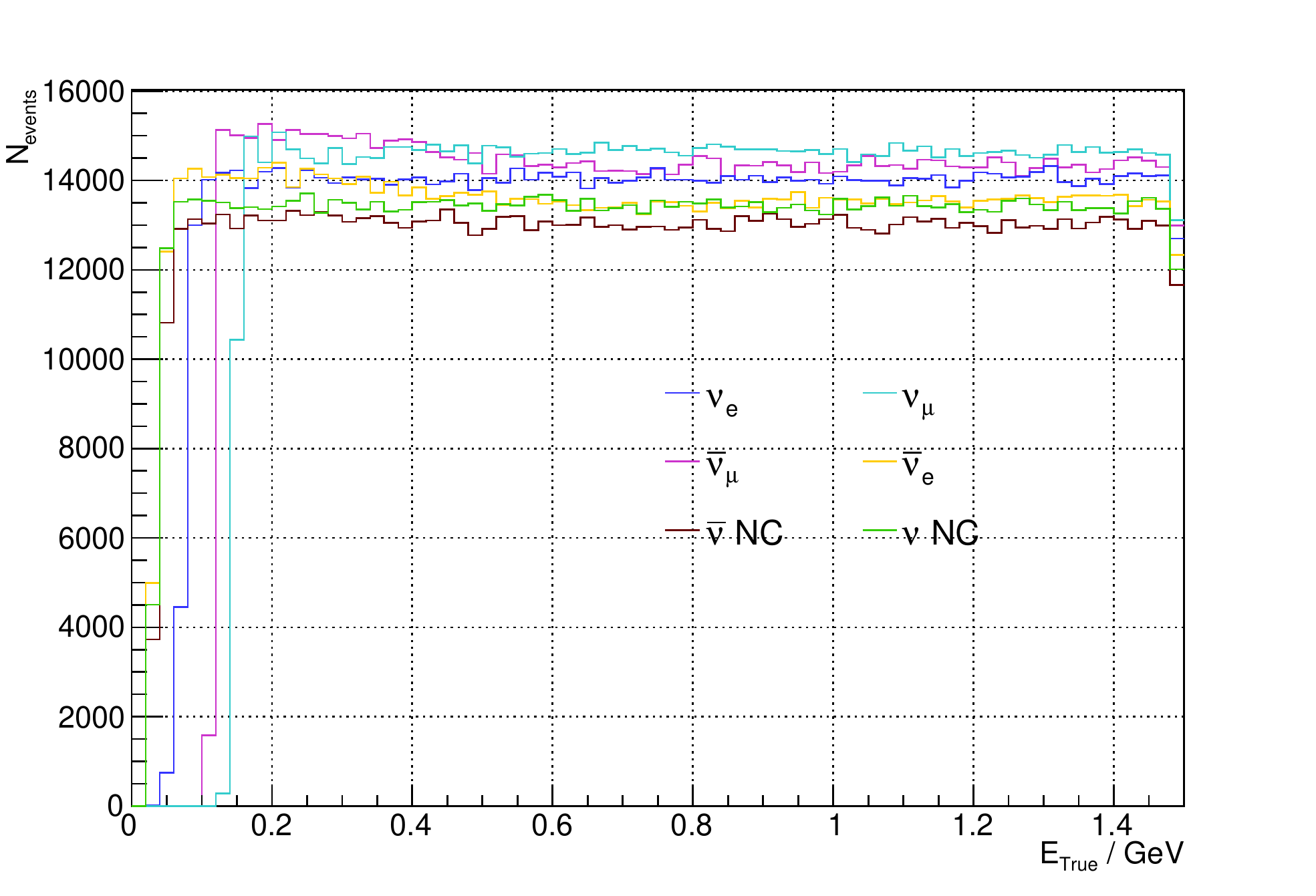}
    \caption{Number of simulated neutrino interactions as a function of neutrino energy for all simulated neutrino flavours and interaction types.}
    \label{fig:detectors:fd_flat_energy_spectrum}
\end{figure}

\paragraph{Full neutrino interaction production} Neutrino interaction vertices have been simulated uniformly across the inner water volume of the far detector, assuming a horizontal neutrino beam. A small expected beam tilt due to the curvature of the Earth is expected to have negligible contribution to the detector performance; it will be implemented in the future analysis once the exact position of the far detectors with respect to the beam source is established. The total number of simulated events for each neutrino flavour and interaction type are shown in Table \ref{tab:detectors:fd_mc_neutrino_events}. The interacting neutrino energy distribution (flux multiplied by cross-section) of simulated samples is shown in Fig.~\ref{fig:detectors:fd_flat_energy_spectrum}; since the distributions are approximately uniform, these samples will be referred to as ``flat energy'' MC samples.

\begin{table}[!htbp]
\footnotesize
\centering
\caption{Number of Monte Carlo simulated events per neutrino flavour}
\begin{tabular}{c c}
\textbf{Neutrino Flavour} & \textbf{Simulated events} \\
\hline
$\nu_e$ CC & 1 000 000 \\
$\nu_\mu$ CC & 991 000 \\
$\overline{\nu}_e$ CC & 1 000 000 \\
$\overline{\nu}_\mu$ CC & 1 000 000 \\
$\nu$ NC & 984 000\\
$\overline{\nu}$ NC & 953 000\\
\hline
\end{tabular}
\label{tab:detectors:fd_mc_neutrino_events}
\end{table}

\paragraph{Construction of uncertainty intervals} For each simulated MC event, the reconstructed value of a certain parameter --- like charged lepton momentum or netrino energy --- will be different than the true value used as an input to the simulation. In general, reconstructed values for a single true value follow a probability distribution. Symmetric uncertainty intervals have been calculated for the reconstructed parameter value as a function of the true value, for various parameters as shown below. For each bin in the true value, a cumulative probability distribution function (CDF) for reconstructed values has been constructed using subset of MC events that fall into the bin. The lower (upper) edge of the $1\,\sigma$ uncertainty interval is defined as the reconstructed value for which CDF has a value of 0.159 (0.841). The absolute resolution is defined as $1/2$ of the uncertainty interval width, while the relative resolution is defined as absolute resolution divided by mean energy of the bin.

\subsubsubsection{Neutrino Flavour Identification}
\label{section:detectors:fd_flavour_identification}

Neutrino flavour selection is based on the identification of the charged lepton in the final state of the neutrino scattering process. This procedure is complicated by the fact that there may be other particles which emit Cherenkov light in the final state. The discrimination between muons and electrons in the detector is based on the observation of the decay of the muon. At the relatively low neutrino energies of ESS$\nu$SB of (150--400)$\ \si{MeV}$ most neutrino interactions are of the QES type. In the case of CC QES interactions, the final state contains a detectable charged lepton (muon for $\nu_\mu$ and electron for $\nu_e$) and an undetectable proton or neutron for the neutrino and antineutrino, respectively. The produced proton is well below the Cherenkov threshold because its mass is large compared to ESS$\nu$SB neutrino energies. Neutral current elastic interactions do not have detectable particles in their final state, except for a possible gamma ray emitted by de-excitation of a target nucleus. Therefore, one can discriminate between $\nu_\mu$ and $\nu_e$ QES interactions by observing a single charged lepton in the final state and determining its flavour. Neutral current interactions can be rejected by discarding events which have only one gamma ray in the final state.

Complications arise with increasing neutrino energies because additional interaction modes become available. In particular, these are resonant (RES) interactions -- in which a short-lived hadron resonance is produced which then immediately decays into visible pions; and deep inelastic scattering (DIS) interactions -- in which a target nucleon is destroyed producing a number of particles in the final state. An additional complication is the existence of inelastic NC scattering modes which often contain a single $\pi^0$ in the final state. This particle promptly decays into a pair of gamma quanta, each of which produces an electromagnetic shower similar to that of an electron. Mis-tagging a $\pi^0$ as an electron can therefore pollute a selected $\nu_e$ sample with NC interactions.

The algorithm for selection of $\nu_e$-like and $\nu_\mu$-like samples is a sequential rejection algorithm. First, the decision is made whether a particular interaction is a $\nu_e$ interaction, and if it is not then the decision is made whether it is a $\nu_\mu$ interaction. Events not classified as either $\nu_\mu$ or $\nu_e$ are discarded. When running in neutrino mode, all events are assumed to be neutrino interactions, while in antineutrino mode all events are assumed to be antineutrino interactions. Due to the high purity of the beam, this results in very small pollution of the samples by wrong-sign\footnote{The sign here denotes whether neutrino is a particle or antiparticle.} neutrinos.

The algorithm for classifying an event as $\nu_e$ is a sequential rejection with following steps:
\begin{enumerate}
	\item Perform a fiducial cut assuming a particle is an electron: if the interaction vertex is produced outside of the fiducial volume, the event is not classified as $\nu_e$;
	\item If there are two global detector triggers within \SI{50}{ms}, with the first during the beam time-window, the event is not classified as $\nu_e$ --- this step effectively discards events containing muon decays since the muon will produce the first trigger, while the Michel electron will produce the second trigger;
	\item If \textsc{fiTQun} PID favours a muon over electron in the first trigger, event is not classified as $\nu_e$ --- this step rejects $\nu_\mu$ events in which a Michel electron is not detected;
	\item If the total PMT charge used for reconstruction is less than $1000\ \text{p.e.}$ the event is not classified as $\nu_e$ --- this step rejects most NC events;
	\item If \textsc{fiTQun} PID favours $\pi^0$ over electron with a negative log-likelihood difference of $\nllemath - \nllpizeromath > 150$ and fits the $\pi^0$ mass between \SI{55}{MeV} and \SI{215}{MeV}, the event is not classified as $\nu_e$ --- this step rejects $\pi^0$ produced in inelastic NC events;
	\item If the momentum of the particle in the first trigger, assuming it is an electron, is less than \SI{70}{MeV}, the event is not classified as $\nu_e$ --- this step rejects “dark muons”, i.e. muons with momenta below the Cherenkov threshold which decay to visible Michel electrons;
	\item The event is classified as $\nu_e$.
\end{enumerate}

Selection algorithm for $\nu_\mu$ events is as follows:
\begin{enumerate}
	\item If the event is classified as $\nu_e$ it is not classified as $\nu_\mu$;
	\item Perform a fiducial cut assuming a particle is a muon: if the interaction vertex is produced outside of the fiducial volume, the event is not classified as $\nu_\mu$;
	\item If there is more than one global detector trigger within 50 ms, and the first trigger is in the beam time window, the event is not classified as $\nu_\mu$;
	\item The event is classified as $\nu_\mu$.
\end{enumerate}

These algorithms are applied independently for each of the two FD tanks. The composition on $\nu_e$-like and $\nu_\mu$-like samples obtained by this selection are shown in Fig.~\ref{fig:rates}.

\subsubsubsection{Pure Charged Lepton Reconstruction}

The analysis in this section is based on the simulation of a pure charged lepton flux described in Section \ref{sec:detectors:FD_MC_simulation}.  The selection algorithm described in Section~\ref{section:detectors:fd_flavour_identification} is applied to each simulated event, keeping only those which are correctly identified (electrons and positrons pass $\nu_e$ selection, muons and antimuons pass $\nu_\mu$ selection). This significantly reduces the number of events events in which the muon is below the Cherenkov threshold and a Michel electron is erroneously reconstructed in its stead.

Figure~\ref{fig:detectors:fd_mom_2d_lep} shows the distribution of reconstructed charged lepton momenta as a function of true charged lepton momentum for each of the four species. Figure~\ref{fig:detectors:fd_mom_lep} shows absolute and relative $1\ \sigma$ uncertainty intervals for the reconstructed momentum as a function of the true charged lepton momentum.

%%%%%%%%%%%%%%%%%%%%%%%%%%%%%%%%%%%%%%%%%%%%%%%%%%% 2D momentum lepton production
\begin{figure}[htbp!]
        \centering
        \begin{subfigure}[b]{0.475\textwidth}
            \centering
            \includegraphics[width=\textwidth]{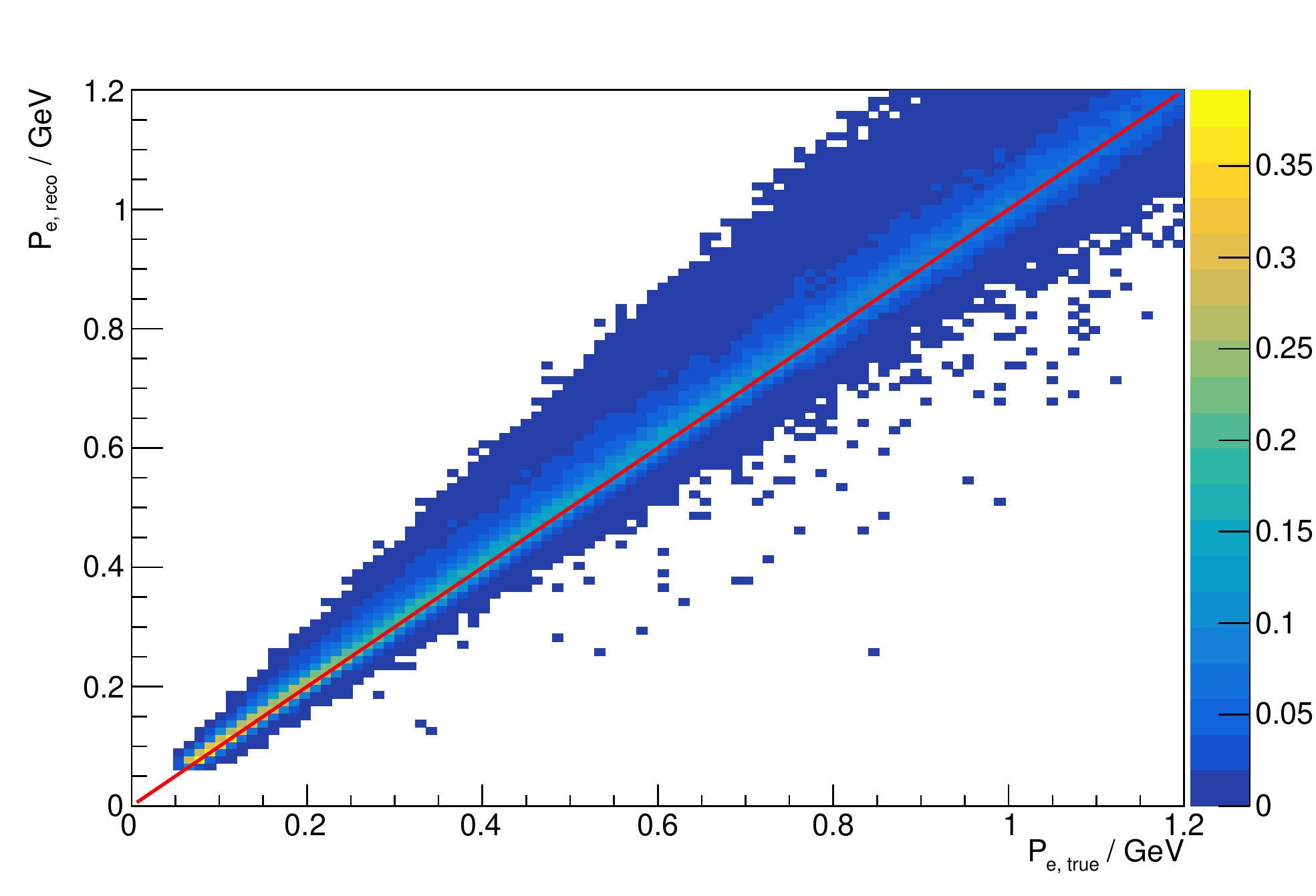}
%            \caption[]%
%            {{\small Reconstructed e$^-$ momentum vs True e$^-$ momentum.}}
%            \caption{Reconstructed e$^-$ momentum vs true e$^-$ momentum.}
            \caption{Electrons}
            \label{fig:detectors:fd_mom_2d_lep_e}
        \end{subfigure}
        \hfill
        \begin{subfigure}[b]{0.475\textwidth}  
            \centering 
            \includegraphics[width=\textwidth]{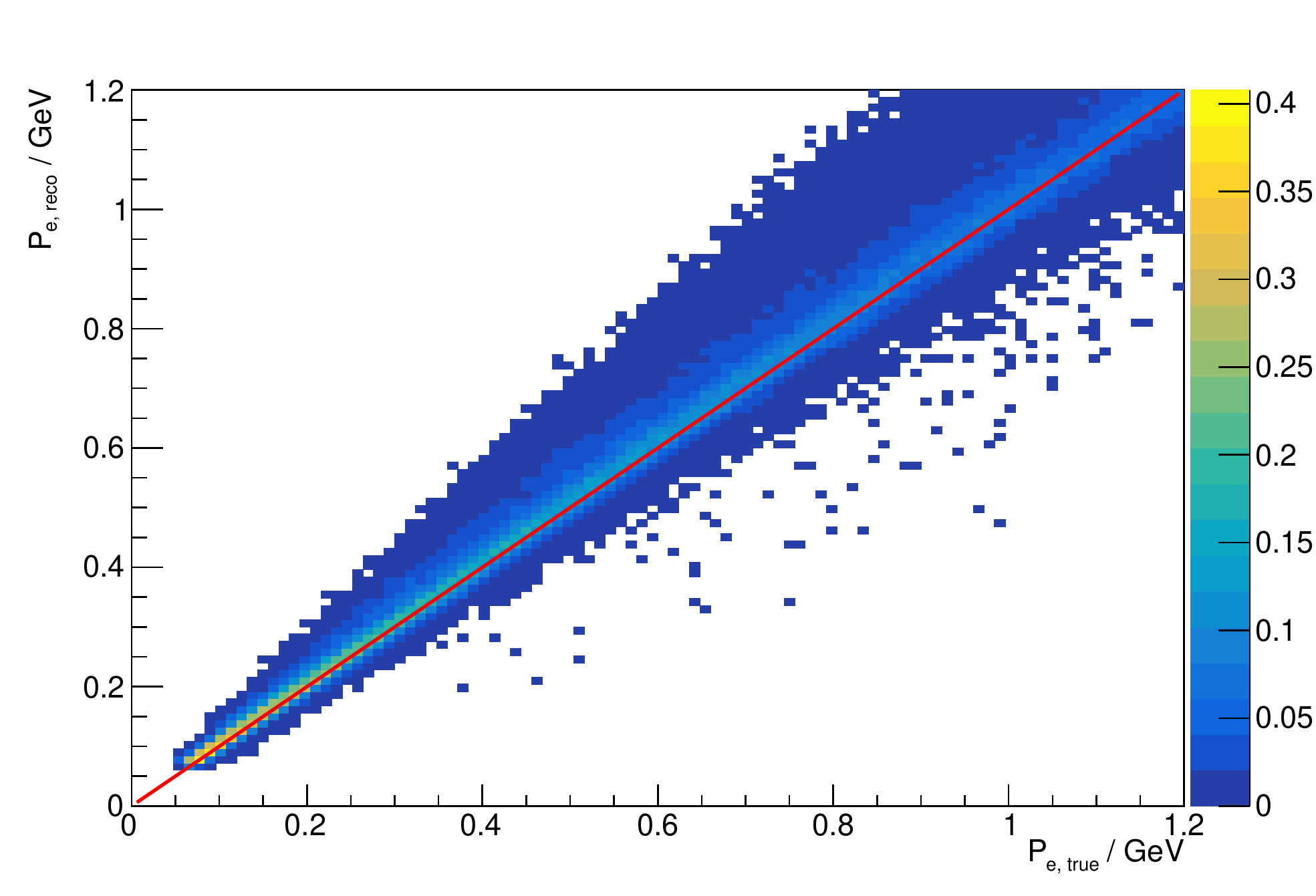}
%            \caption{Reconstructed e$^+$ momentum vs true e$^+$ momentum.}
            \caption{Positrons}
            \label{fig:detectors:fd_mom_2d_lep_ae}
        \end{subfigure}
        \begin{subfigure}[b]{0.475\textwidth}   
            \centering 
            \includegraphics[width=\textwidth]{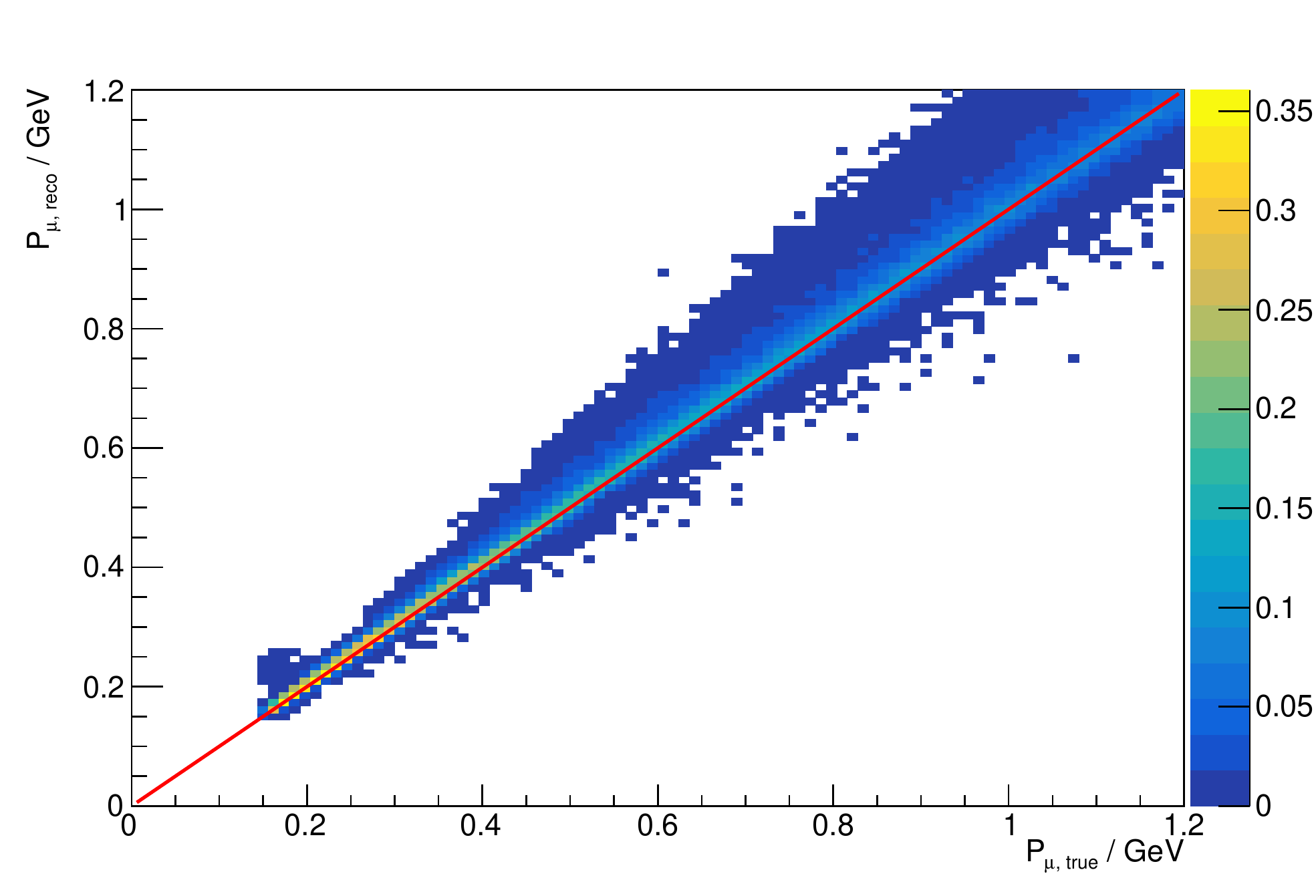}
%            \caption{Reconstructed $\mu^-$ momentum vs true $\mu^-$ momentum.}
            \caption{Muons}
            \label{fig:detectors:fd_mom_2d_lep_mu}
        \end{subfigure}
        \hfill
        \begin{subfigure}[b]{0.475\textwidth}   
            \centering 
            \includegraphics[width=\textwidth]{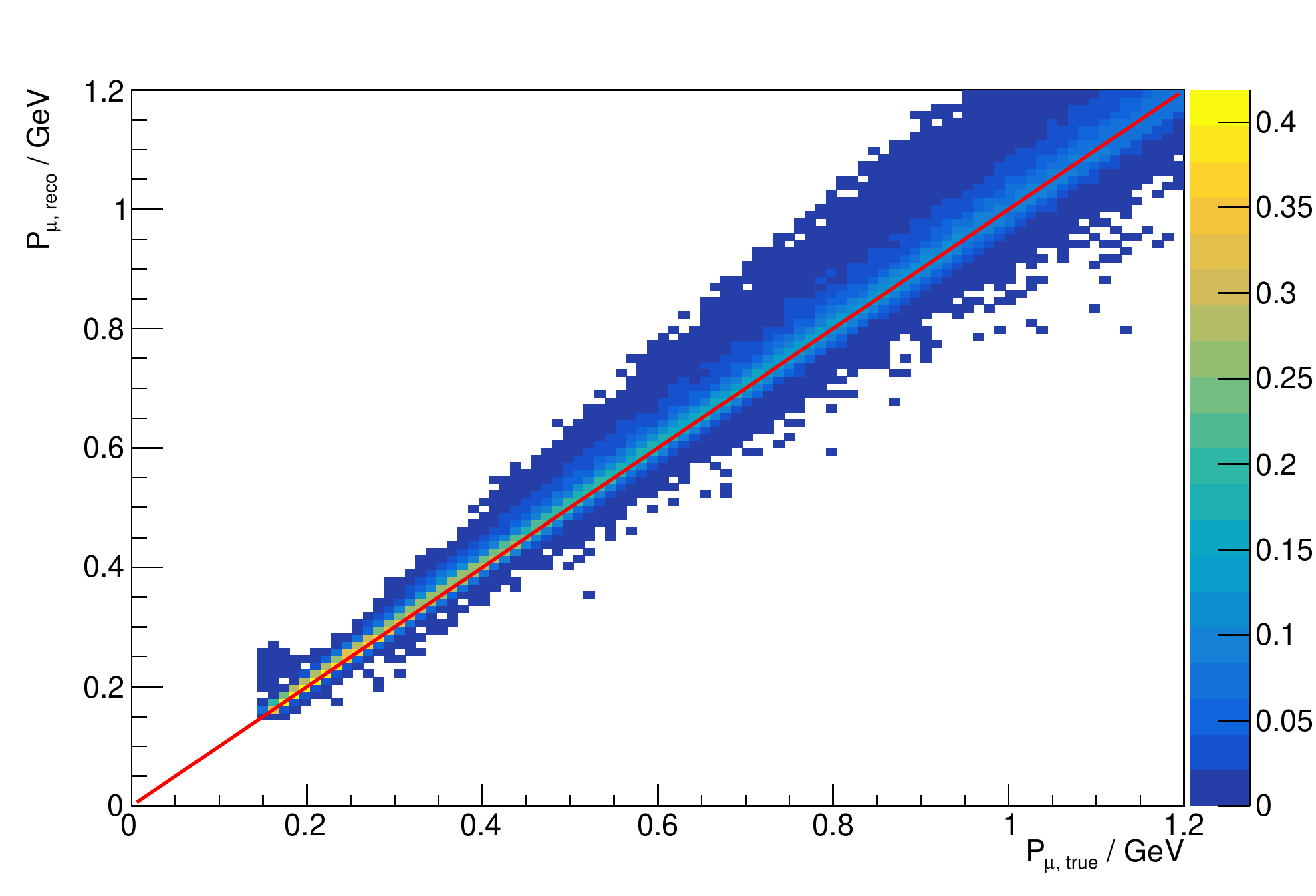}
%            \caption{Reconstructed $\mu^+$ momentum vs true $\mu^+$ momentum.}
            \caption{Antimuons}
            \label{fig:detectors:fd_mom_2d_lep_amu}
        \end{subfigure}
        \caption{Distribution of reconstructed momentum as a function of true momentum for different flavours of charged leptons. These plots were produced using the charged lepton production.}
        \label{fig:detectors:fd_mom_2d_lep}
    \end{figure}

\begin{figure}[hbtp!]
    \centering
    \begin{subfigure}[b]{0.475\textwidth}
        \centering
        \includegraphics[width=\textwidth]{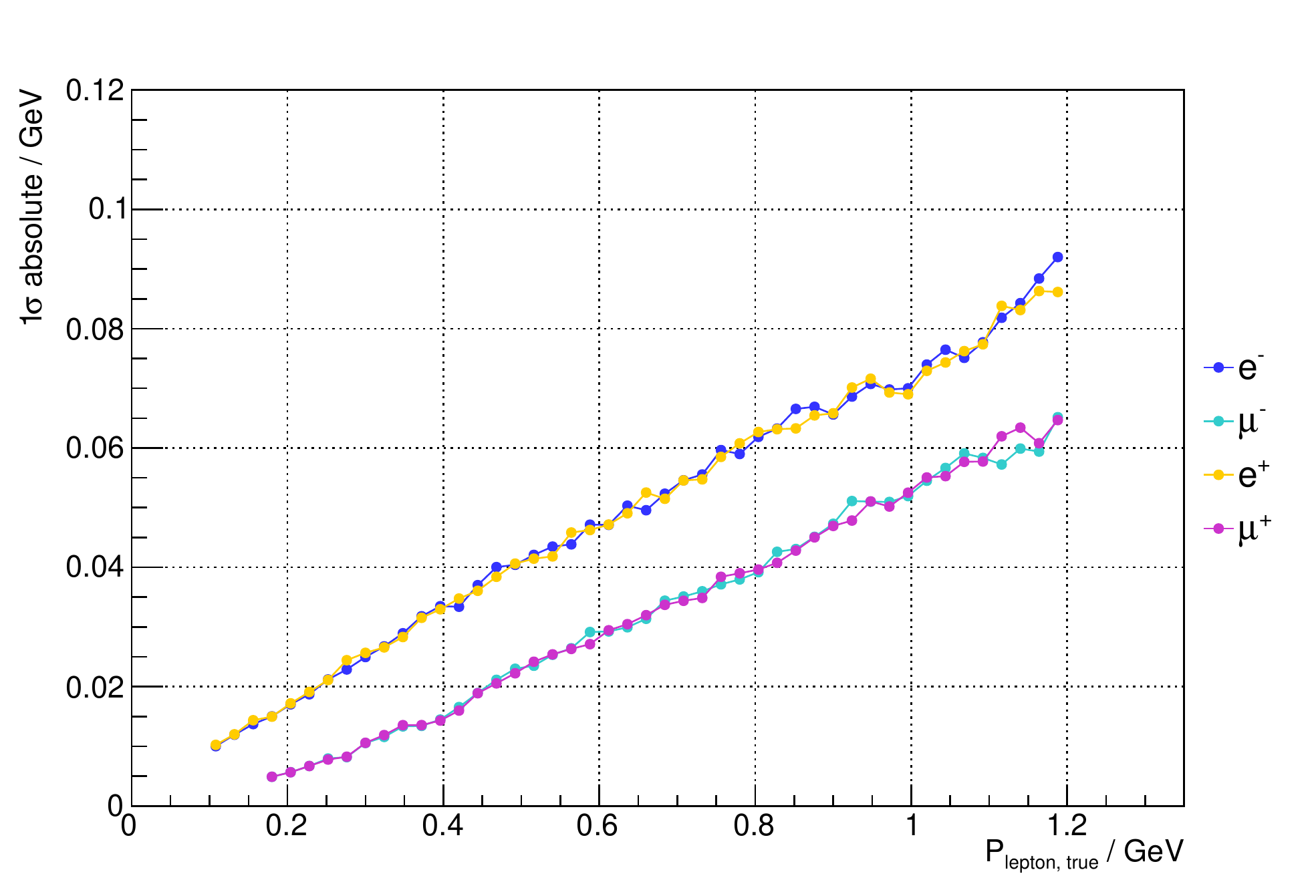}
        \caption{Absolute momentum resolution}
        \label{fig:detectors:fd_mom_abs_lep}
    \end{subfigure}
    \hfill
    \begin{subfigure}[b]{0.475\textwidth}  
        \centering 
        \includegraphics[width=\textwidth]{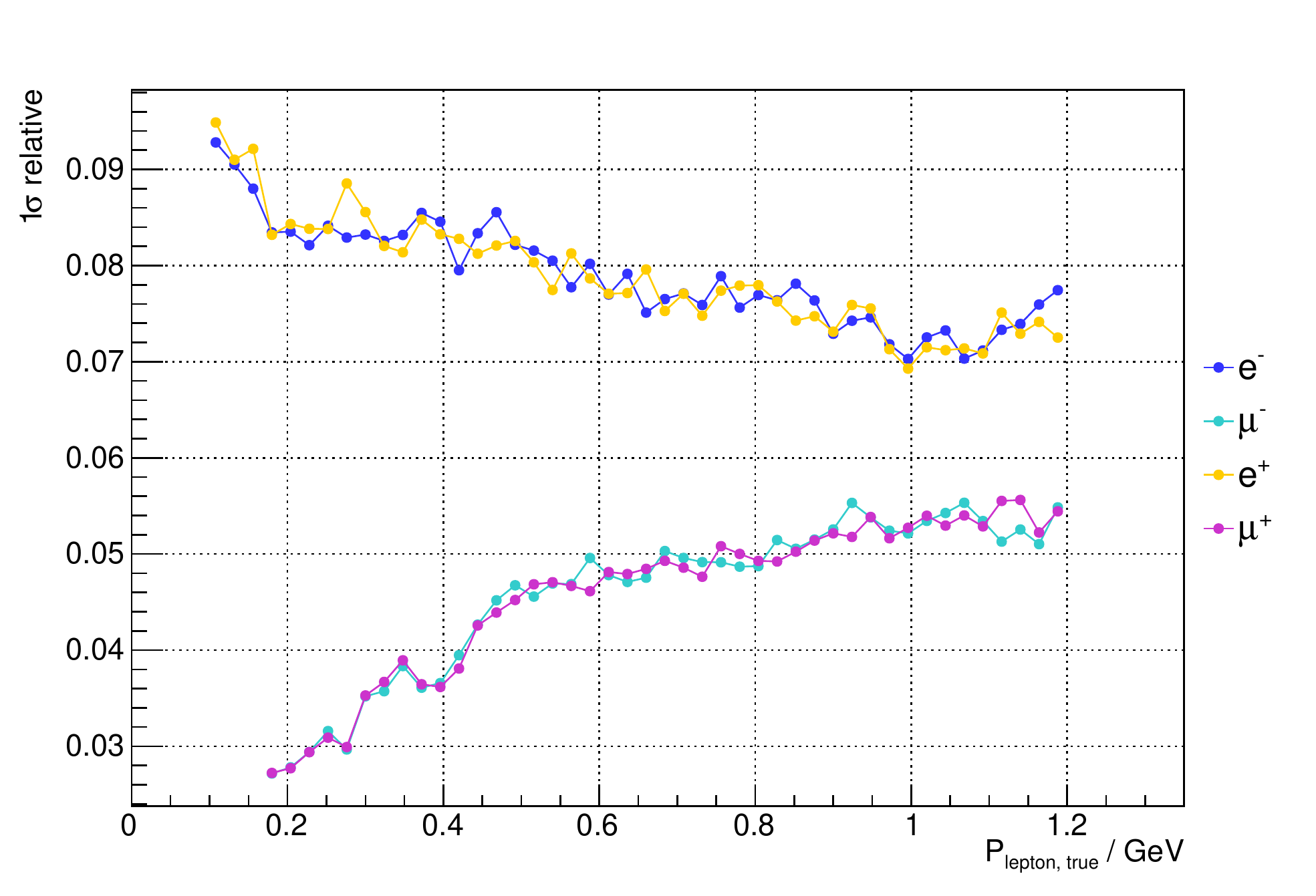}
        \caption{Relative momentum resolution}
        \label{fig:detectors:fd_mom_rel_lep}
    \end{subfigure}
    \caption{Momentum reconstruction performance for a pure charged lepton flux. 1\,$\sigma$ absolute resolution of reconstructed produced lepton momentum as a function of true lepton momentum is shown in Fig.~\ref{fig:detectors:fd_mom_abs_lep}. 1\,$\sigma$ relative resolution of reconstructed produced lepton momentum as a function of true lepton momentum is shown in Fig.~\ref{fig:detectors:fd_mom_rel_lep}.}
    \label{fig:detectors:fd_mom_lep}
\end{figure}
Absolute momentum reconstruction uncertainty for all four species increases with the true momentum value, as shown in Fig.~\ref{fig:detectors:fd_mom_abs_lep}. As can be seen in Fig.~\ref{fig:detectors:fd_mom_rel_lep}, the relative resolution of the (anti)electron momentum is below 10\,\% for momenta larger than \SI{100}{MeV} and dropping below 8\% for momenta greater than \SI{800}{MeV}. For (anti)muon momenta, relative resolution is best at lower momenta, beginning below 4\% for momenta up to \SI{400}{MeV} and reaching 6\% for energies above \SI{1}{GeV}.

Figure~\ref{fig:detectors:fd_cos_theta_2d_lep} shows the distribution of reconstructed $\cos \theta$ of the charged lepton as a function of true $\cos \theta$ of the charged lepton for each species. The absolute resolution of $\cos \theta$ shown in \ref{fig:detectors:fd_cos_abs_lep} peaks at $0.035$ for (anti)electron and at $0.03$ for (anti)muons, with both peaks at $\cos \theta =0$.
%Another important parameter in the neutrino energy reconstruction formula is the cosine of the charged lepton scattering angle.  
%%%%%%%%%%%%%%%%%%%%%%%%%%%%%%%%%%%%%%%%%%%%%%%%%%% 2D cos theta lepton production    
    \begin{figure*}[htp!]
        \centering
        \begin{subfigure}[b]{0.475\textwidth}
            \centering
            \includegraphics[width=\textwidth]{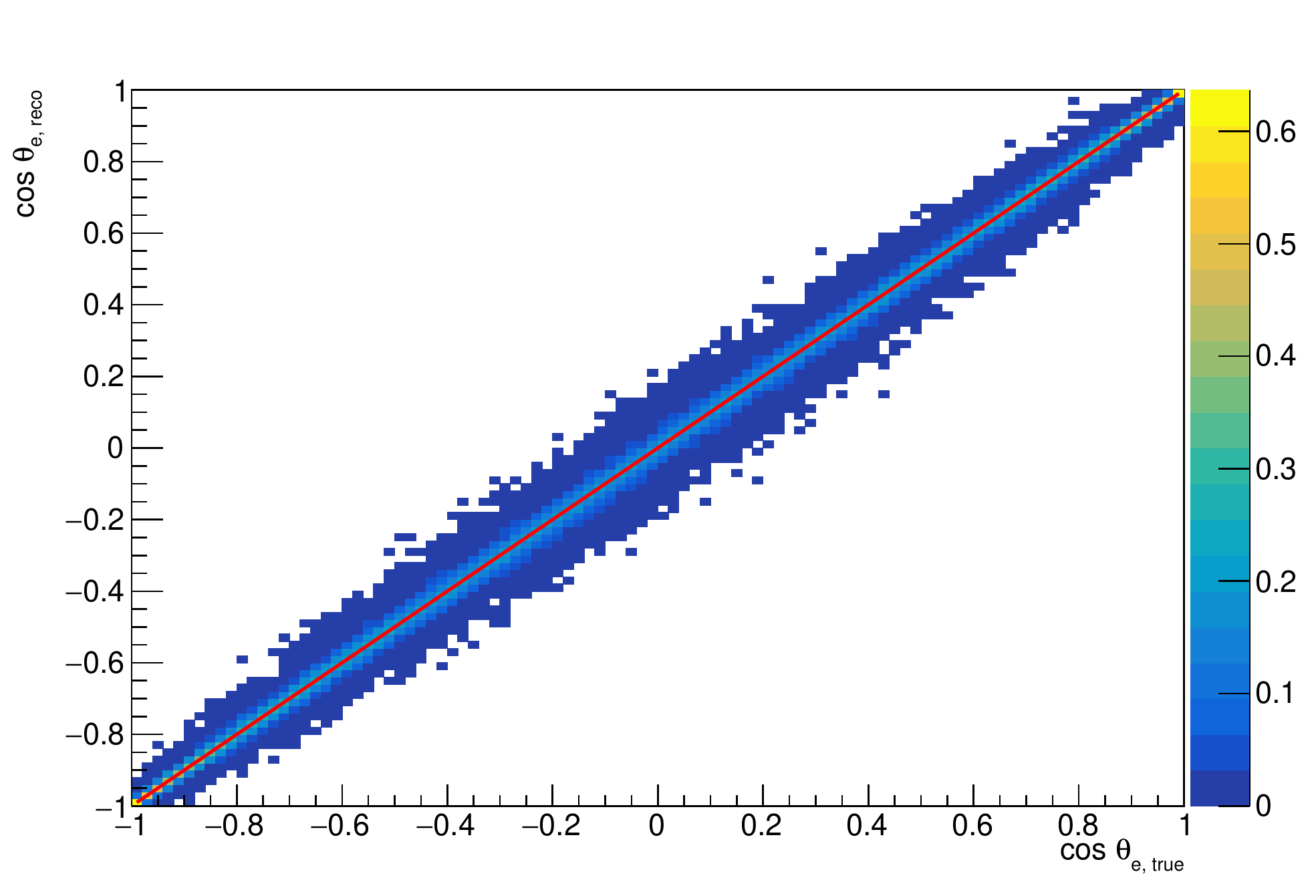}
            \caption{Electrons}
            \label{fig:detectors:fd_cos_theta_2d_lep_e}
        \end{subfigure}
        \hfill
        \begin{subfigure}[b]{0.475\textwidth}  
            \centering 
            \includegraphics[width=\textwidth]{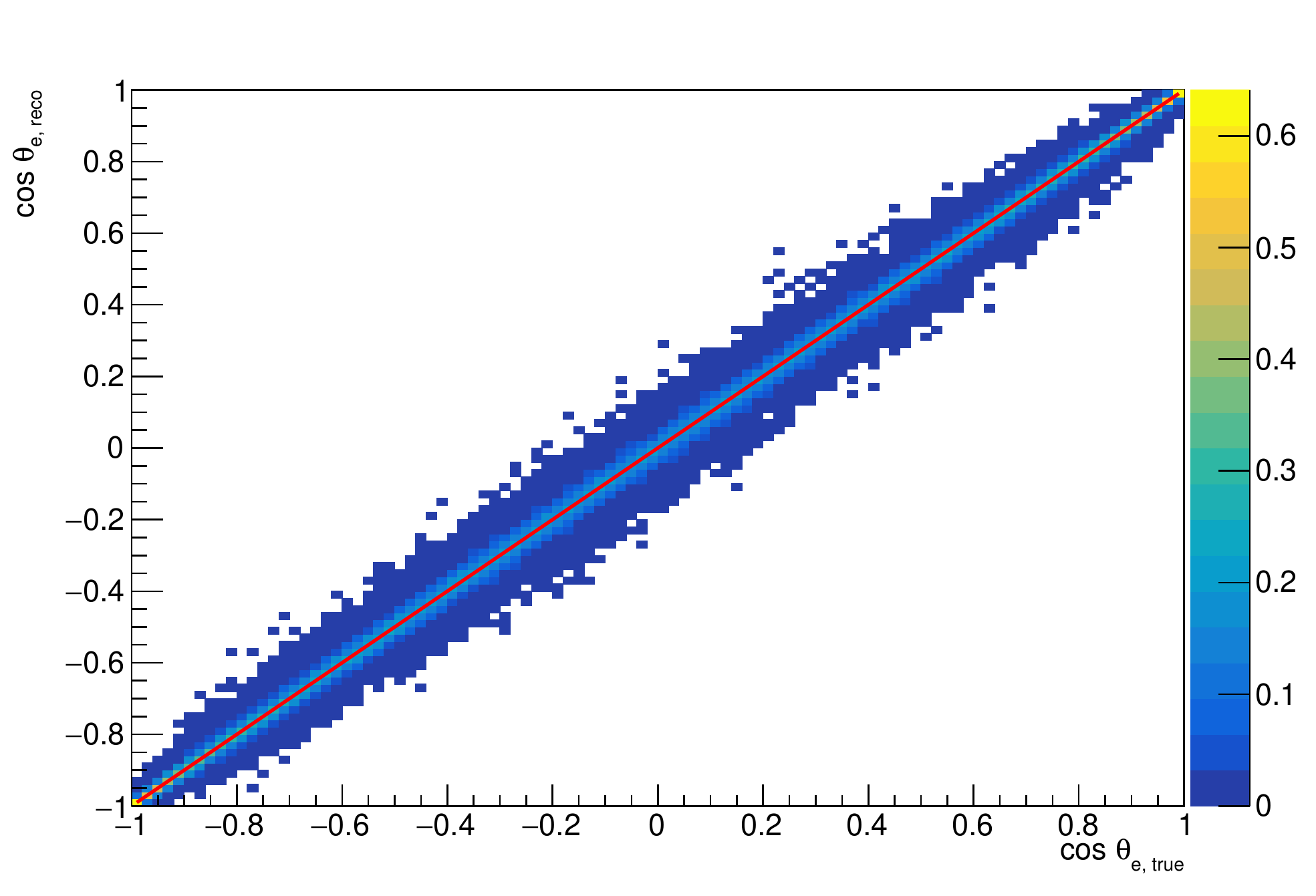}
            \caption{Positrons}
            \label{fig:detectors:fd_cos_tehta_2d_lep_ae}
        \end{subfigure}
        \begin{subfigure}[b]{0.475\textwidth}   
            \centering 
            \includegraphics[width=\textwidth]{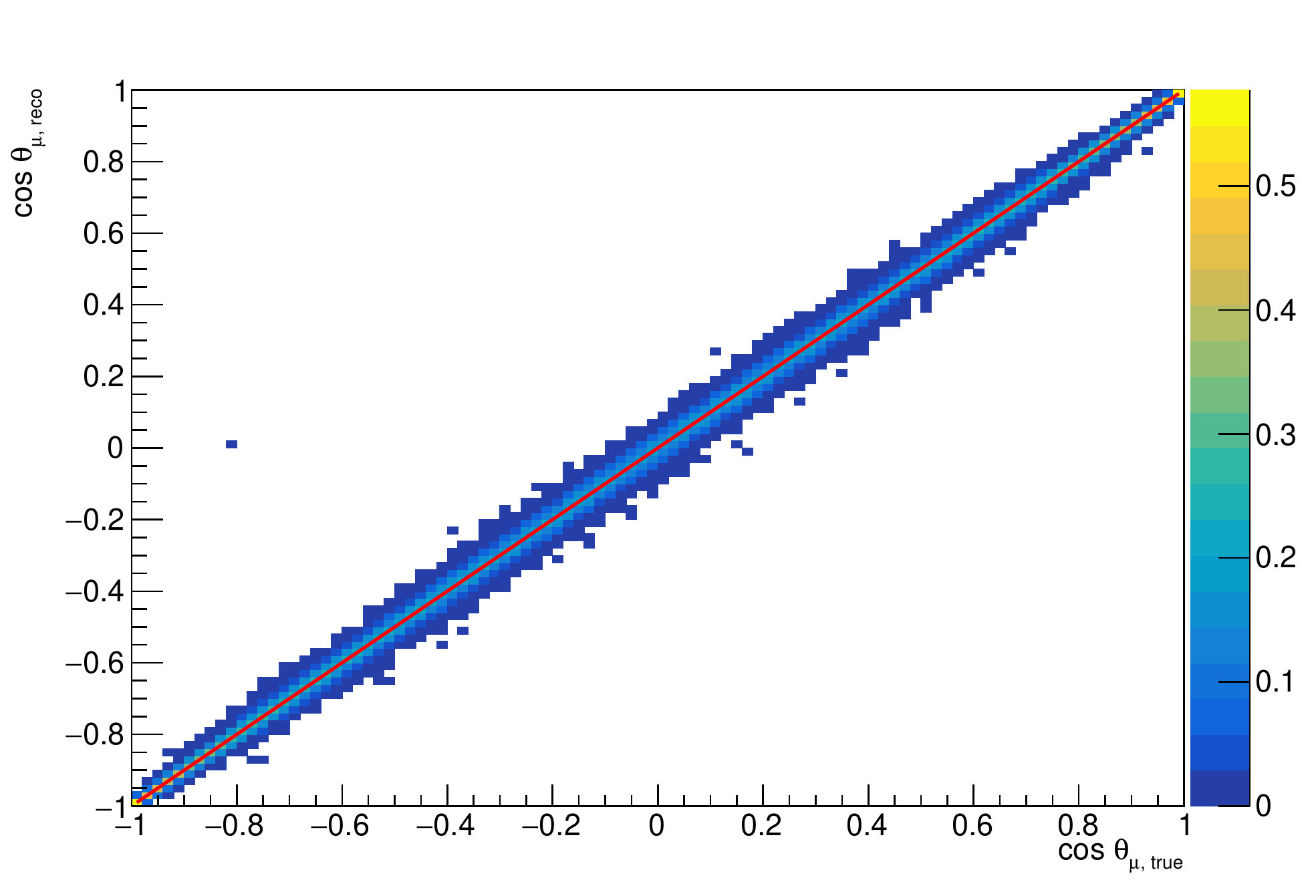}
            \caption{Muons}
            \label{fig:detectors:fd_cos_theta_2d_lep_mu}
        \end{subfigure}
        \hfill
        \begin{subfigure}[b]{0.475\textwidth}   
            \centering 
            \includegraphics[width=\textwidth]{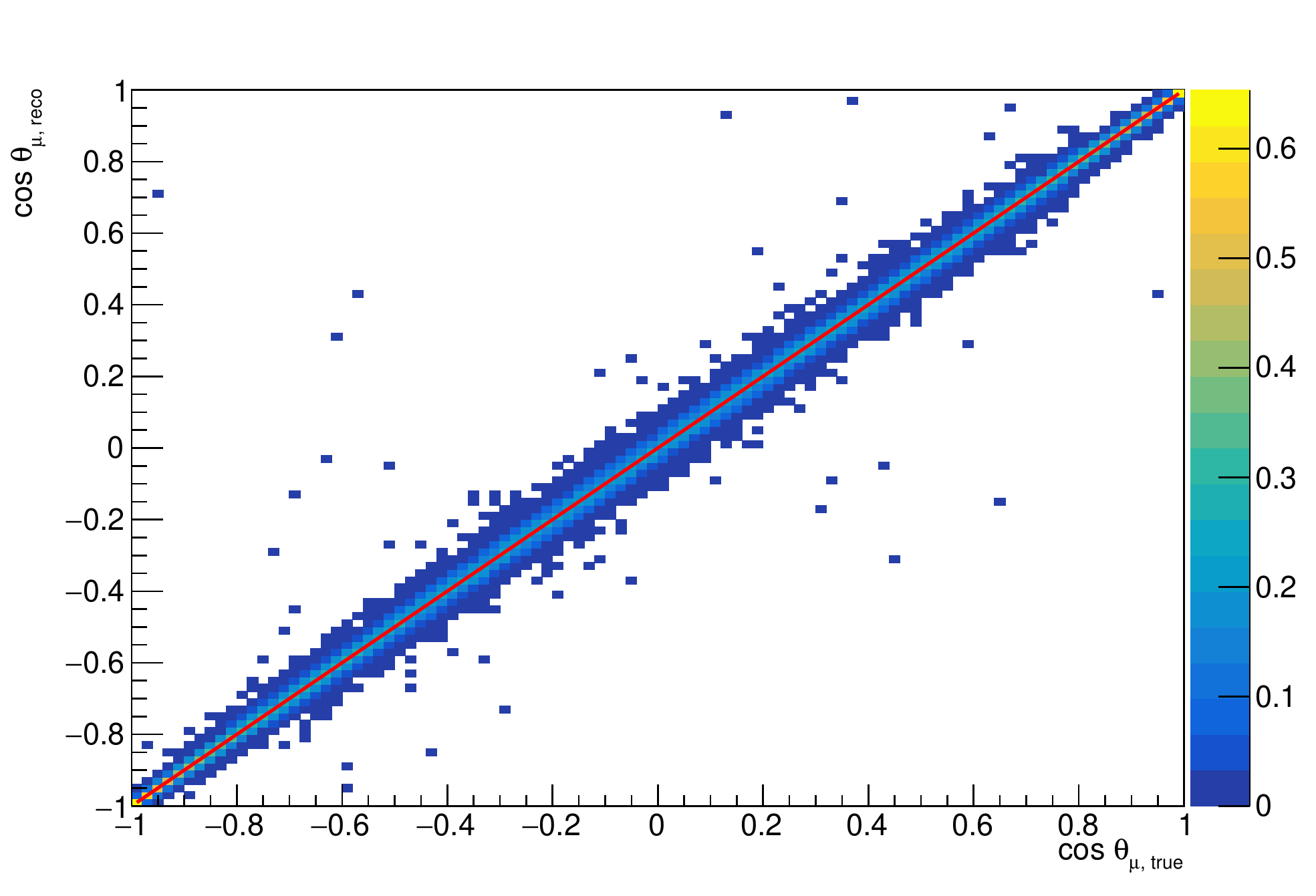}
            \caption{Antimuons}
            \label{fig:detectors:fd_cos_theta_2d_lep_amu}
        \end{subfigure}
        
        \caption{Distribution of reconstructed $\cos \theta$ as a function of true $\cos \theta$ for different flavours of charged leptons. These plots were produced using the charged lepton production.}
        \label{fig:detectors:fd_cos_theta_2d_lep}
    \end{figure*}

\begin{figure}[hbtp!]
    \centering 
    \includegraphics[width=0.475\textwidth]{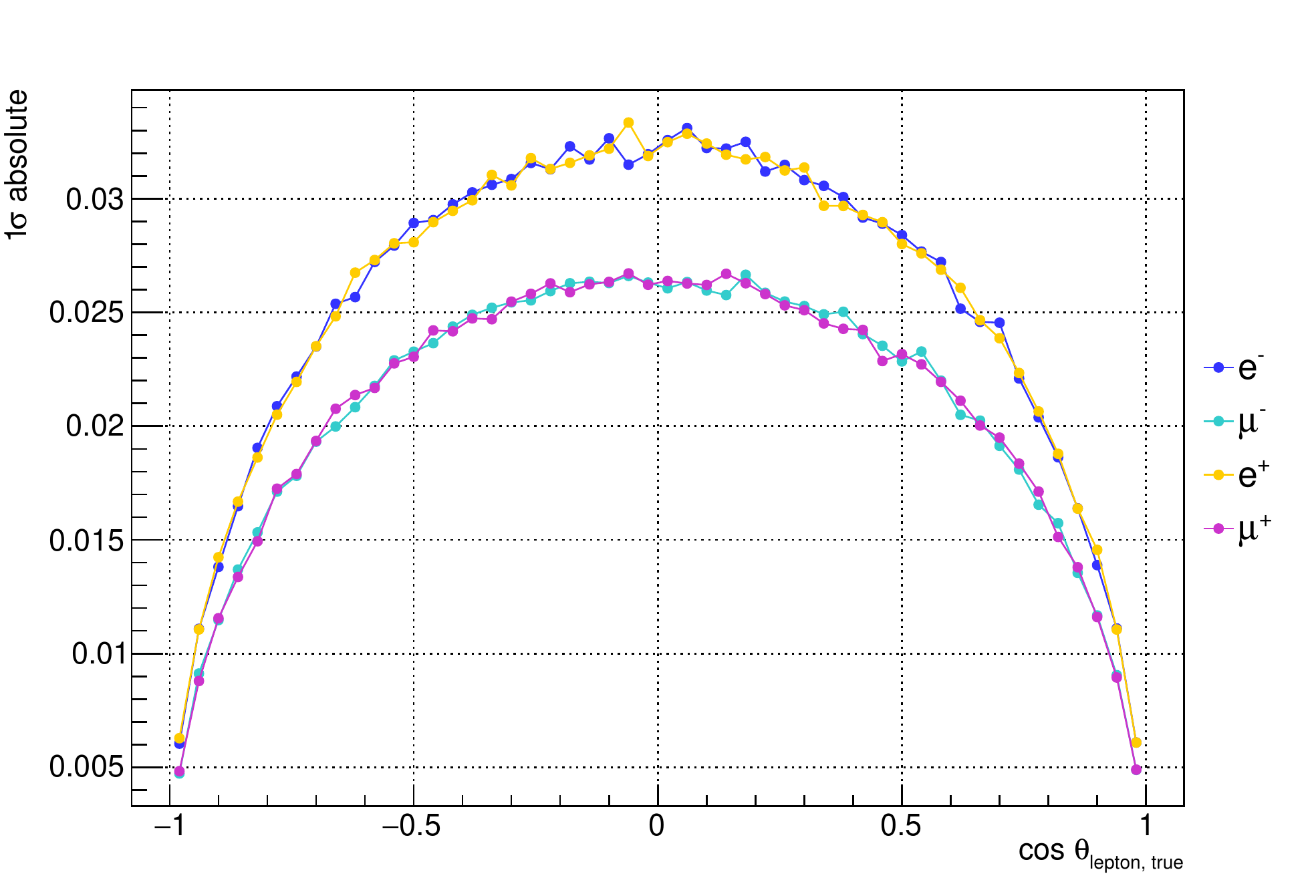}
%    \caption{Absolute $\cos \theta$ resolution}
    \caption{1\,$\sigma$ absolute resolution of reconstructed $\cos \theta$ as a function of true $\cos\theta$ for a pure lepton flux.} 
    \label{fig:detectors:fd_cos_abs_lep}
\end{figure}

The selection efficiency for identifying the correct charged lepton flavour by applying the neutrino flavour selection algorithm is shown in Fig.~\ref{fig:detectors:fd_diagonal_efficiency_lepton}. \emph{The selection efficiency for electrons and positrons} is almost at the level of the fiducial volume cut at momenta in the range 80--300\,\si{MeV}, which coincides with the main peak of the expected $\nu_\mu \rightarrow \nu_e$ and $\overline{\nu}_\mu \rightarrow \overline{\nu}_e$ signal shown in Fig.~\ref{fig:detectors:fd_exp_mom}. The sharp decrease of efficiency below \SI{70}{MeV} is due to the Michel electron cut. \emph{The selection efficiency for muons and antimuons} sharply rises after the muon Cherenkov threshold and reaches about 0.8 for antimuons and 0.7 for muons.

\begin{figure}[hbtp!]
    \centering 
    \includegraphics[width=0.475\textwidth]{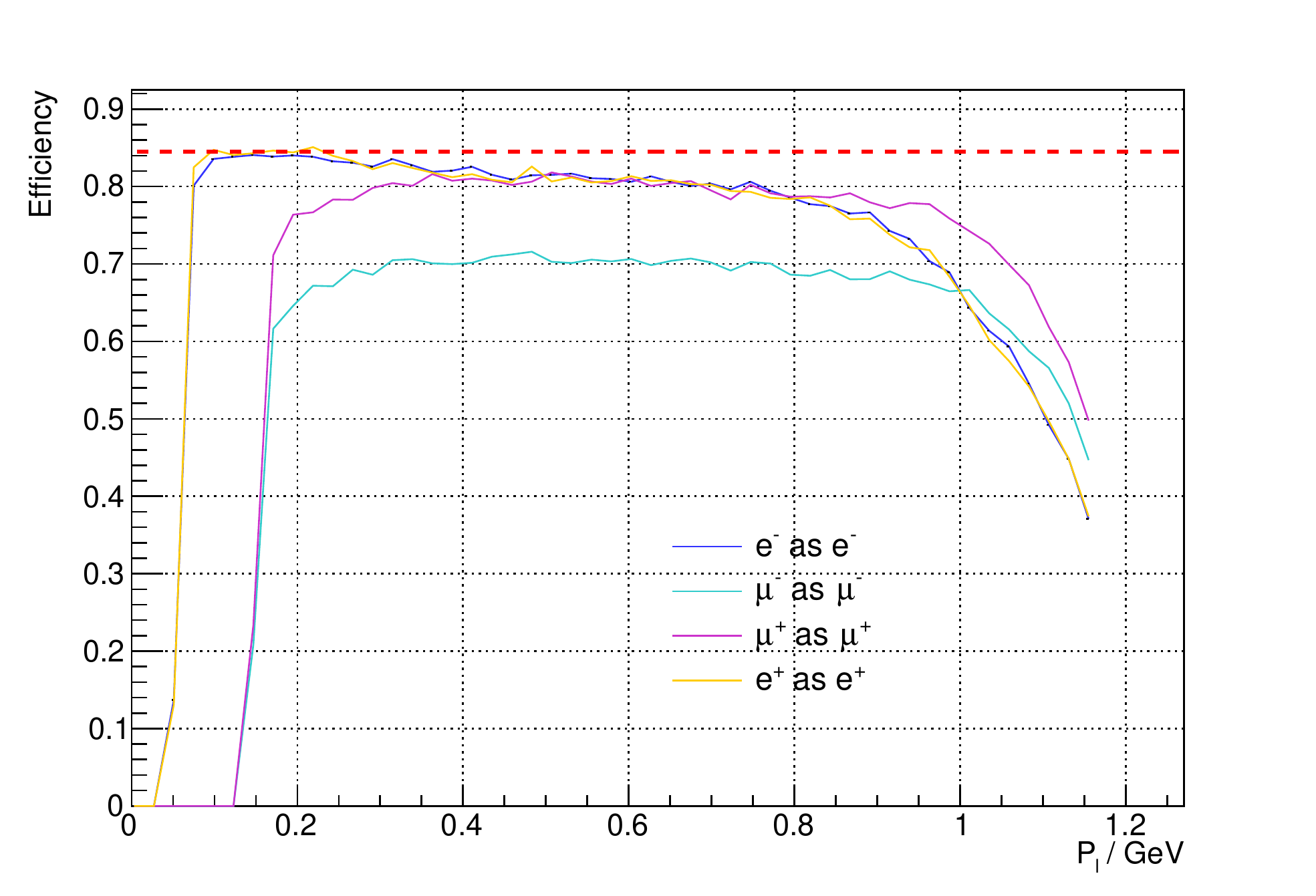}
%    \caption{Absolute $\cos \theta$ resolution}
    \caption{Efficiency for correct identification of charged lepton flavour by applying the neutrino flavour selection algorithm (electrons and positrons that pass $\nu_e$ selection, muons and antimuons that pass $\nu_\mu$ selection) as a function of true momentum. The horizontal dashed red line is the efficiency of a fiducial cut alone.} 
    \label{fig:detectors:fd_diagonal_efficiency_lepton}
\end{figure}

This is an intrinsic detector performance for reconstruction of charged lepton momentum and its direction, which is considered the best-case scenario. It is expected to worsen for full neutrino interaction events due to the presence of additional particles in the final state.

\subsubsubsection{Reconstruction of Charged Leptons in the Final State of Neutrino Interaction}

The analysis in this section is performed using a full simulation of neutrino interactions described in Section~\ref{sec:detectors:FD_MC_simulation}. The performance is shown separately for simulated samples containing (\textit{i}) only QES neutrino interactions, and (\textit{ii}) a full model of neutrino-nucleus scattering. The QES sample is expected to perform better because the final state consists only of one detectable charged particle -- the charged lepton in the neutrino branch. The neutrino flavour selection algorithm described (see Section~\ref{section:detectors:fd_flavour_identification}) is applied to each simulated event, keeping only those which are correctly identified.

The reconstructed momentum distributions of the final-state charged leptons as a function of true momentum for the four relevant neutrino species are presented in Fig.~\ref{fig:detectors:fd_mom_2d}. Similar plots for the reconstructed $\cos\theta$ as a function of true $\cos\theta$ are presented in Fig.~\ref{fig:detectors:fd_cos_theta_2d}.

%%%%%%%%%%%%%%%%%%%%%%%%%%%%%%%%%%%%%%%%%%%%%%%%%%% 2D momentum positive polarity neutrino production
\begin{figure}[hbtp!]
        \centering
        \begin{subfigure}[b]{0.4\textwidth}
            \centering
            \includegraphics[width=\textwidth]{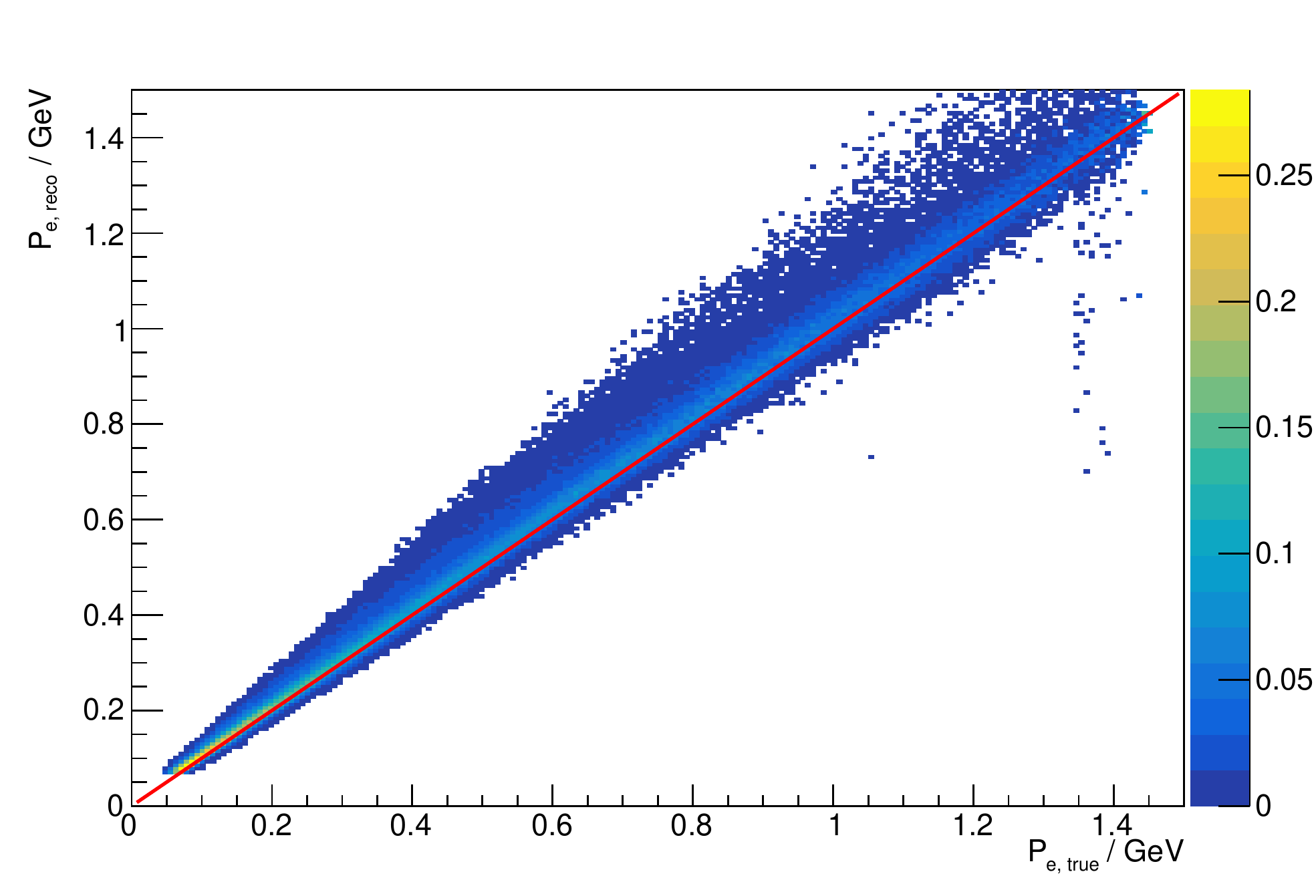}
            \caption{QES e$^-$ events}
            \label{fig:detectors:fd_mom_2d_qes_e}
        \end{subfigure}
        \hspace{1em}
        \begin{subfigure}[b]{0.4\textwidth}
            \centering 
            \includegraphics[width=\textwidth]{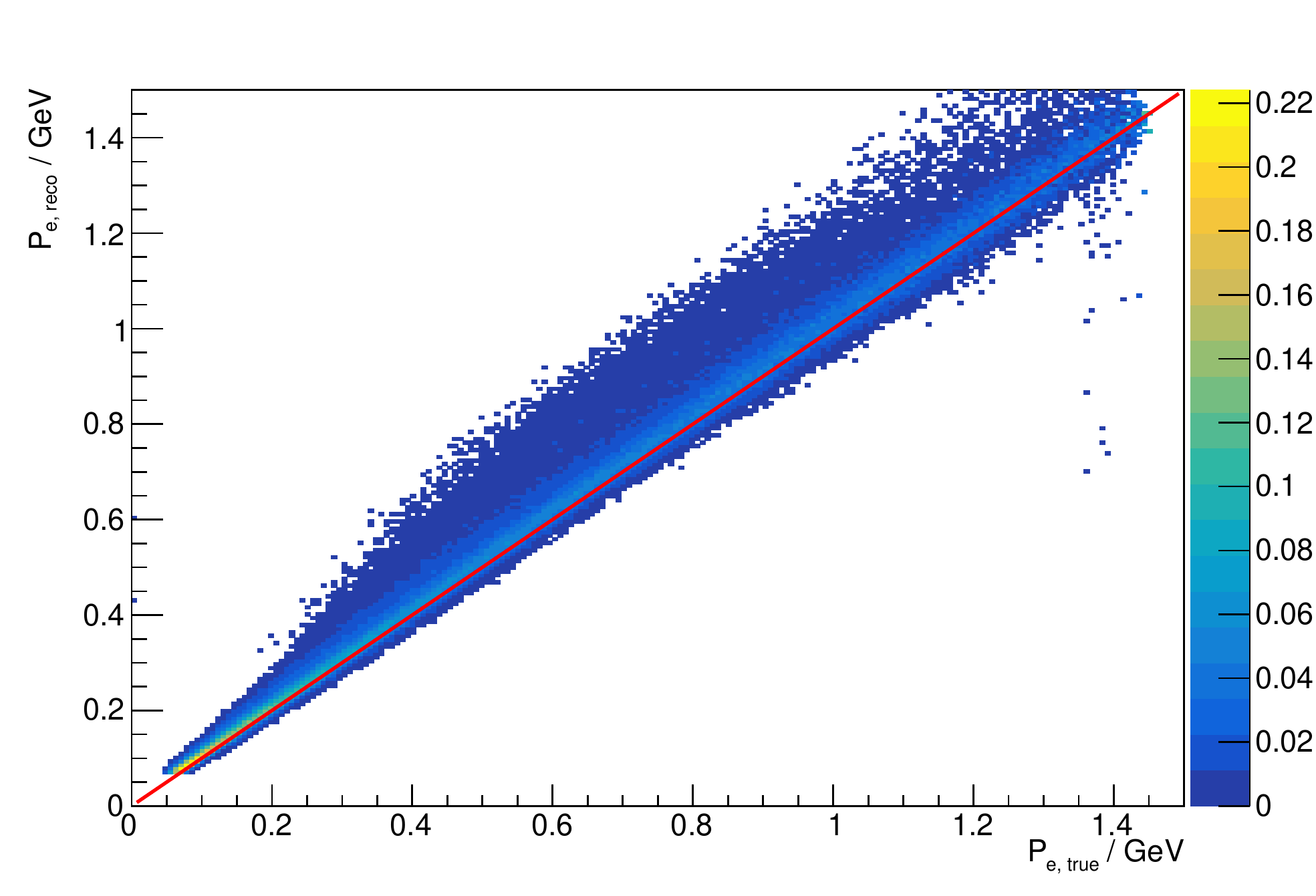}
            \caption{All e$^-$ events}
            \label{fig:detectors:fd_mom_2d_all_e}
        \end{subfigure}
        \smallskip
        
        \begin{subfigure}[b]{0.4\textwidth}
            \centering 
            \includegraphics[width=\textwidth]{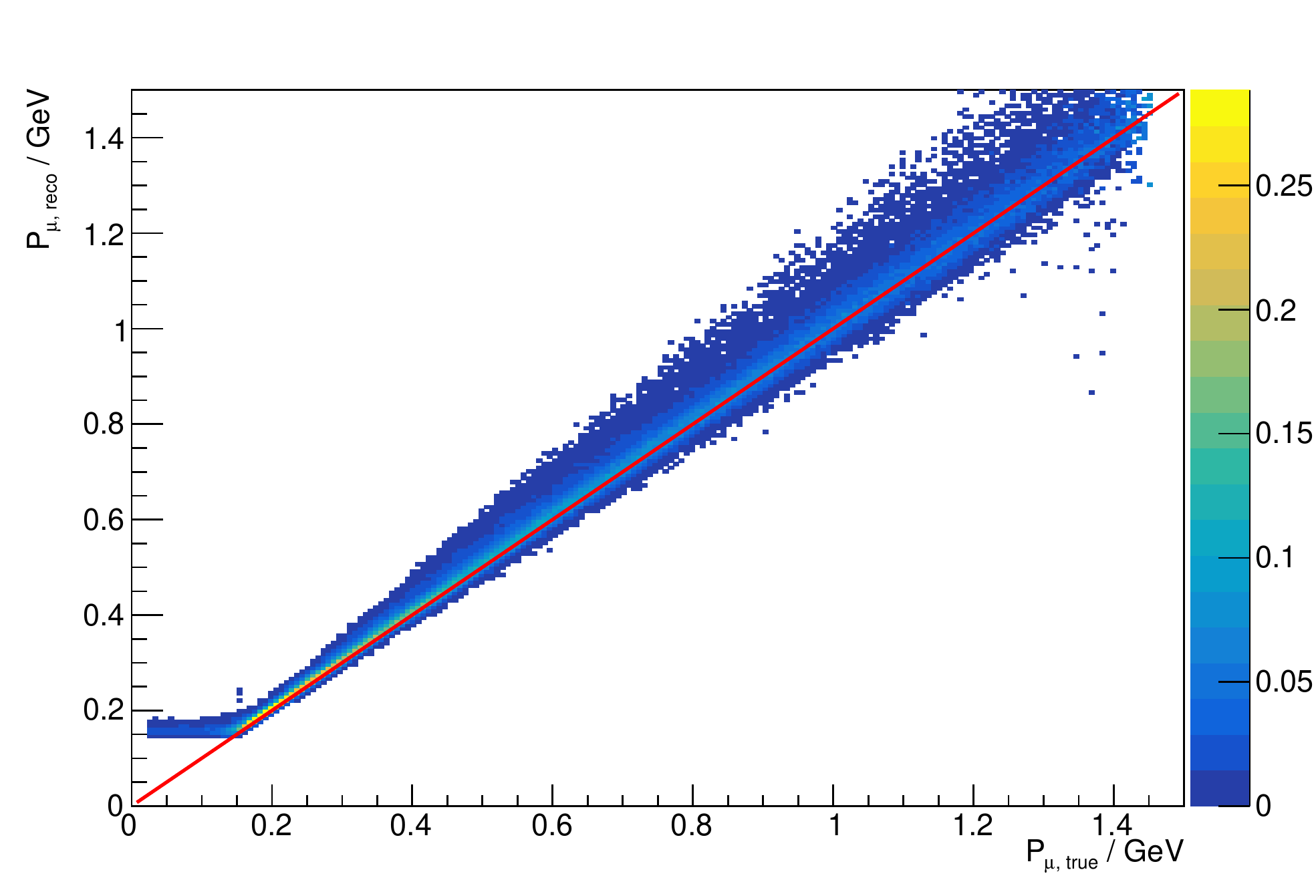}
            \caption{QES $\mu^-$ events}
            \label{fig:detectors:fd_mom_2d_qes_mu}
        \end{subfigure}
        \hspace{1em}
        \begin{subfigure}[b]{0.4\textwidth}
            \centering 
            \includegraphics[width=\textwidth]{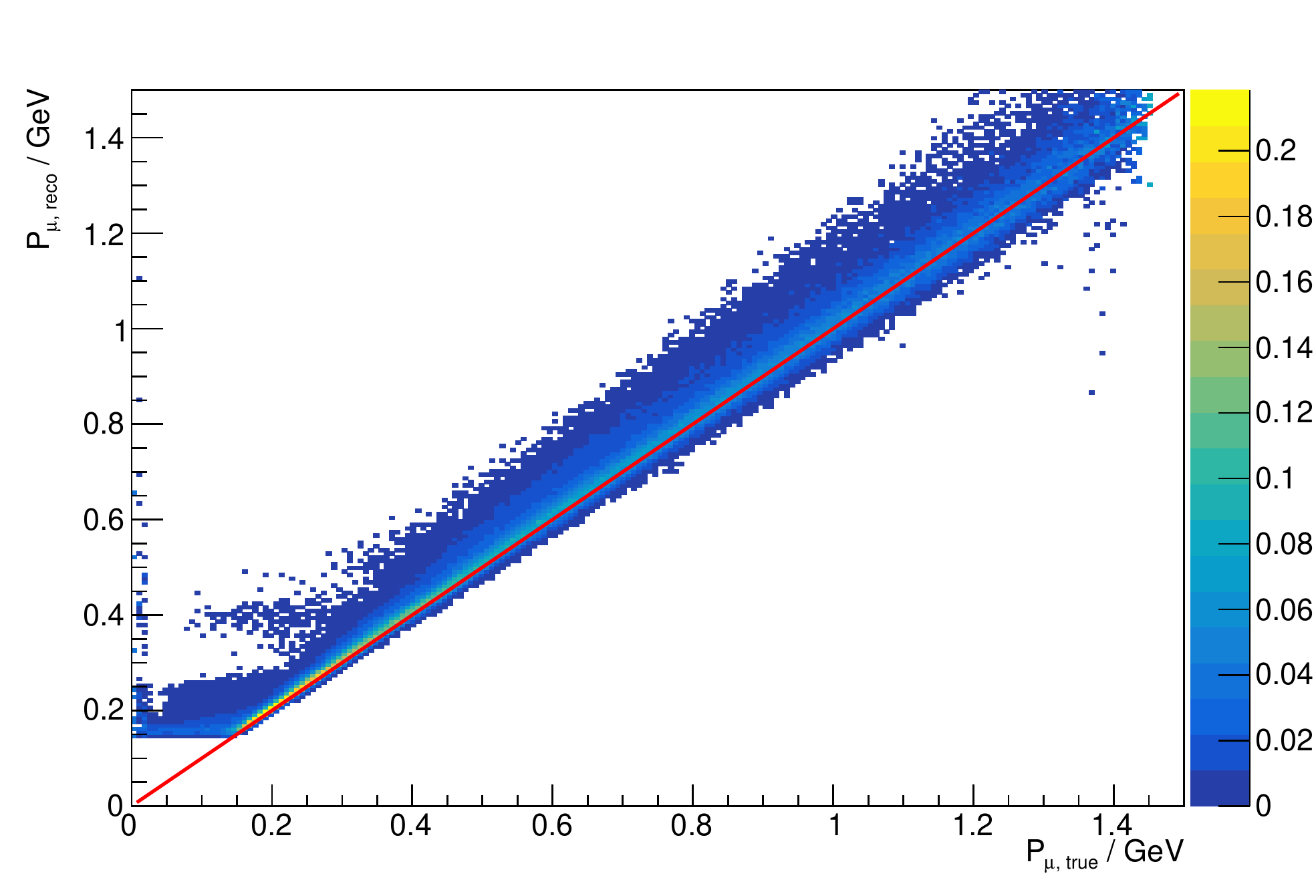}
            \caption{All $\mu^-$ events}
            \label{fig:detectors:fd_mom_2d_all_mu}
        \end{subfigure}
        \smallskip
        \begin{subfigure}[b]{0.4\textwidth}
            \centering 
            \includegraphics[width=\textwidth]{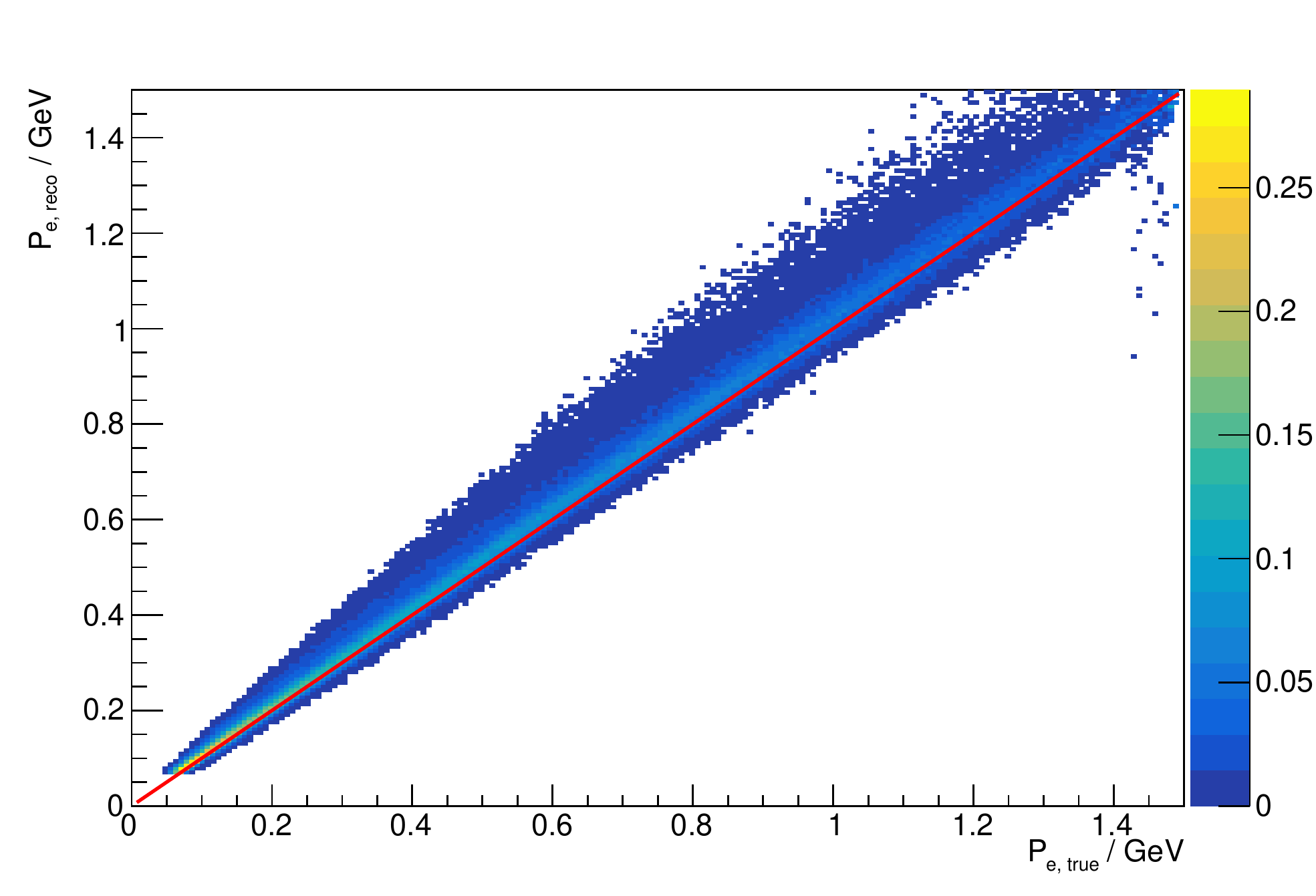}
            \caption{QES e$^+$ events}
            \label{fig:detectors:fd_mom_2d_qes_ae}
        \end{subfigure}
        \hspace{1em}
        \begin{subfigure}[b]{0.4\textwidth}
            \centering 
            \includegraphics[width=\textwidth]{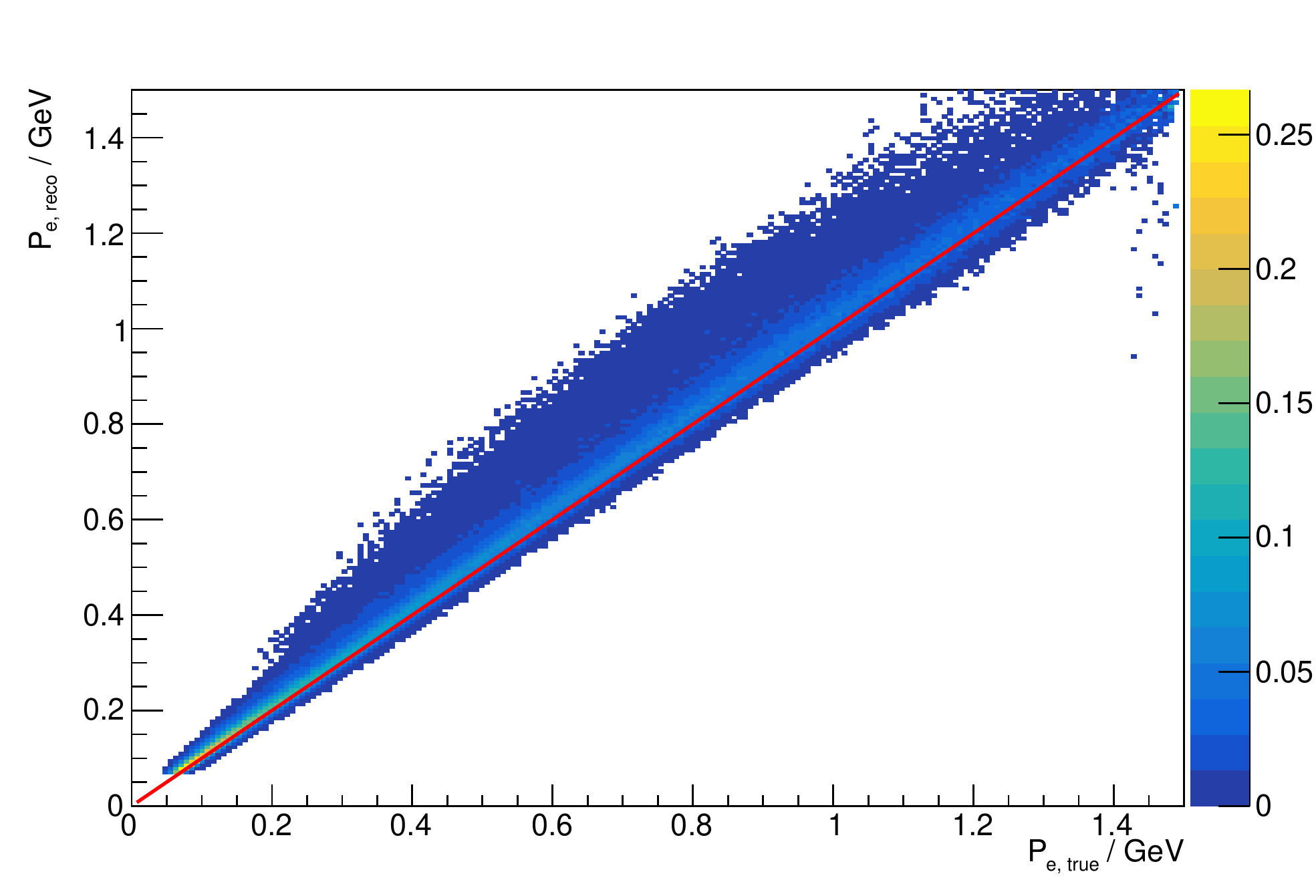}
            \caption{All e$^+$ events}
            \label{fig:detectors:fd_mom_2d_all_ae}
        \end{subfigure}
        \smallskip
        \begin{subfigure}[b]{0.4\textwidth}
            \centering 
            \includegraphics[width=\textwidth]{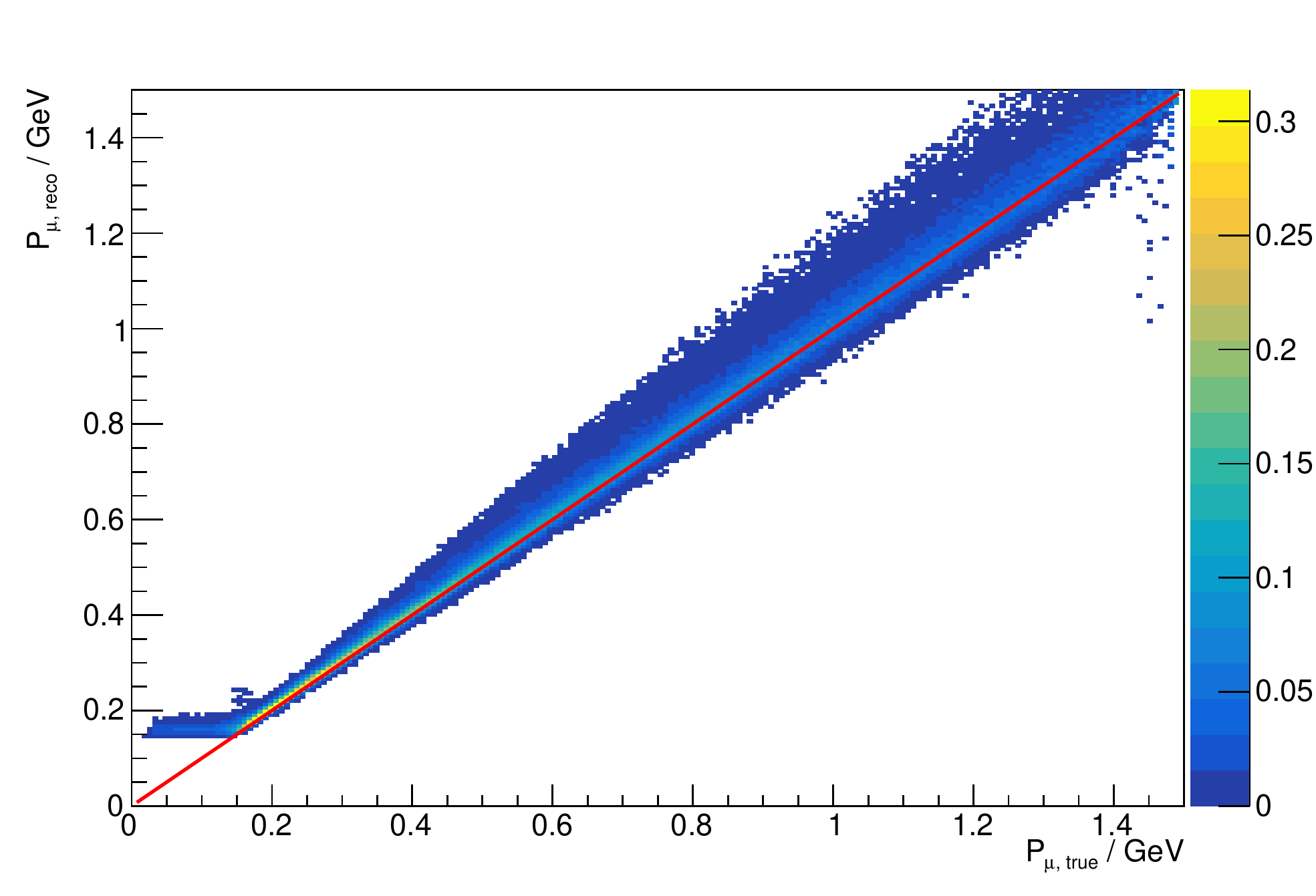}
            \caption{QES $\mu^+$ events}
            \label{fig:detectors:fd_mom_2d_qes_amu}
        \end{subfigure}
        \hspace{1em}
        \begin{subfigure}[b]{0.4\textwidth}
            \centering 
            \includegraphics[width=\textwidth]{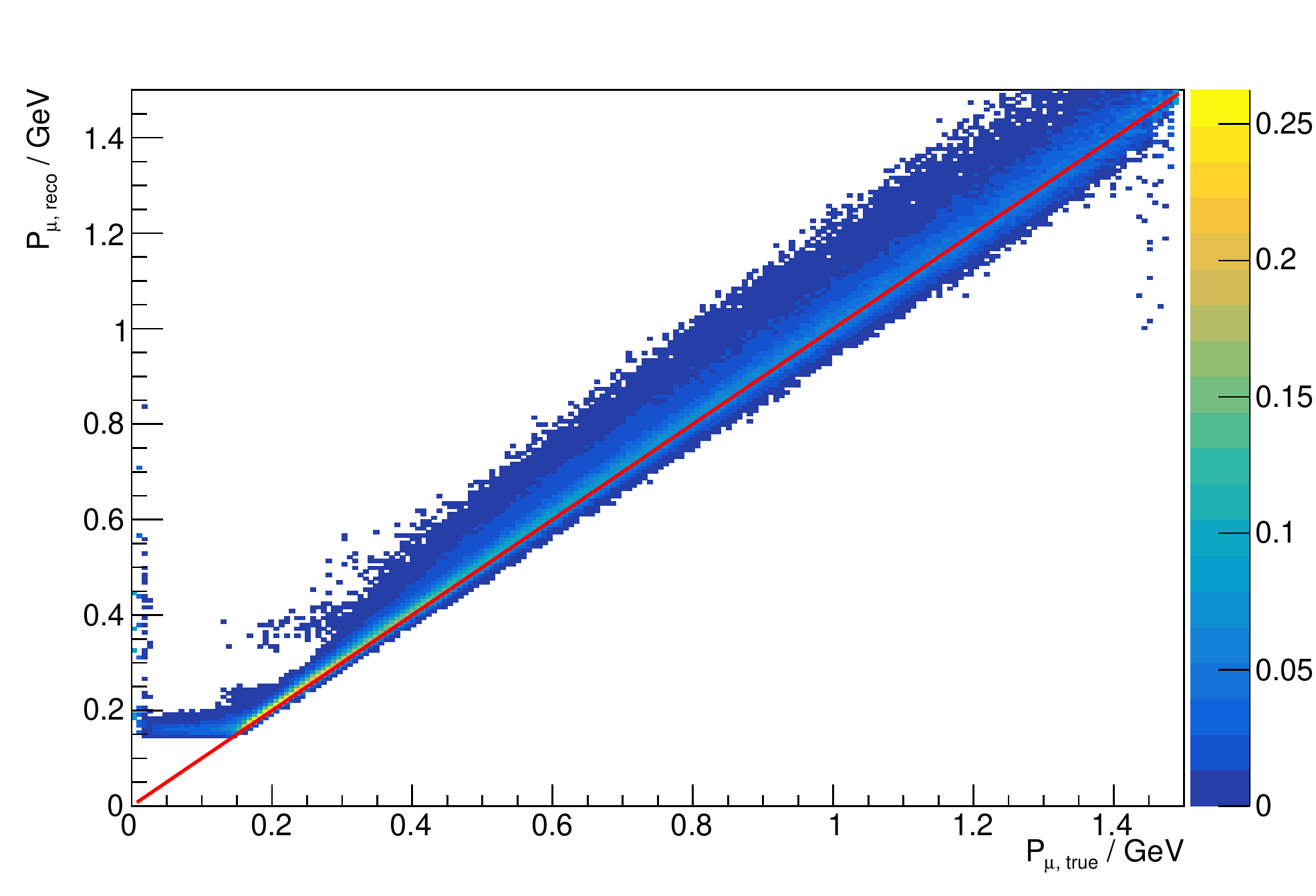}
            \caption{ALL $\mu^+$ events}
            \label{fig:detectors:fd_mom_2d_all_amu}
        \end{subfigure}
        \caption{Distribution of reconstructed momentum as a function of true momentum for different flavours of charged leptons in the final state of neutrino interactions. The left column shows plots using a MC subset composed only of QES events, while the right column shows plots using full event sample.}
        \label{fig:detectors:fd_mom_2d}
    \end{figure}

The absolute and relative resolutions for reconstructed lepton momentum and $\cos\theta$ as a function of their true values are shown in Fig.~\ref{fig:detectors:fd_lepton_comp_all_qes}, both for the QES and all event samples. The $\cos\theta$ resolution is similar to that of the pure lepton analysis, with the exception of the antimuons produced in $\overline{\nu}_\mu$ CC interactions. The reason for this exception is that $\overline{\nu}_\mu$ CC interactions strongly prefer forward scattering (see Fig.~\ref{fig:detectors:fd_exp_theta_m}), so there is a very small number of back-scattered antimuons; the signal in the $\cos\theta < 0$ region is therefore drowned in the noise coming from the random mis-reconstruction caused by photons in the final state of neutrino interactions. This effect is more pronounced for the full event sample, since additional final-state particles further increase the odds for mis-recontrsuction of muon tracks. Since the expected number of back-scattered antimuons in the ESS$\nu$SB beam is very small, this does not have a significant impact on the experiment's performance. Electrons and positrons produced in $\nu_e$ and $\overline{\nu}_e$ CC interactions do not exhibit these effects because the reconstruction and selection algorithms are tuned for maximal performance in $\nu_e$ and $\overline{\nu}_e$ detection -- their interactions are the signal for CP violation measurement.

%%%%%%%%%%%%%%%%%%%%%%%%%%%%%%%%%%%%%%%%%%%%%%%%%%% 2D cos theta neutrino production
    \begin{figure*}[htp!]
        \centering
        \begin{subfigure}[b]{0.4\textwidth}
            \centering
            \includegraphics[width=\textwidth]{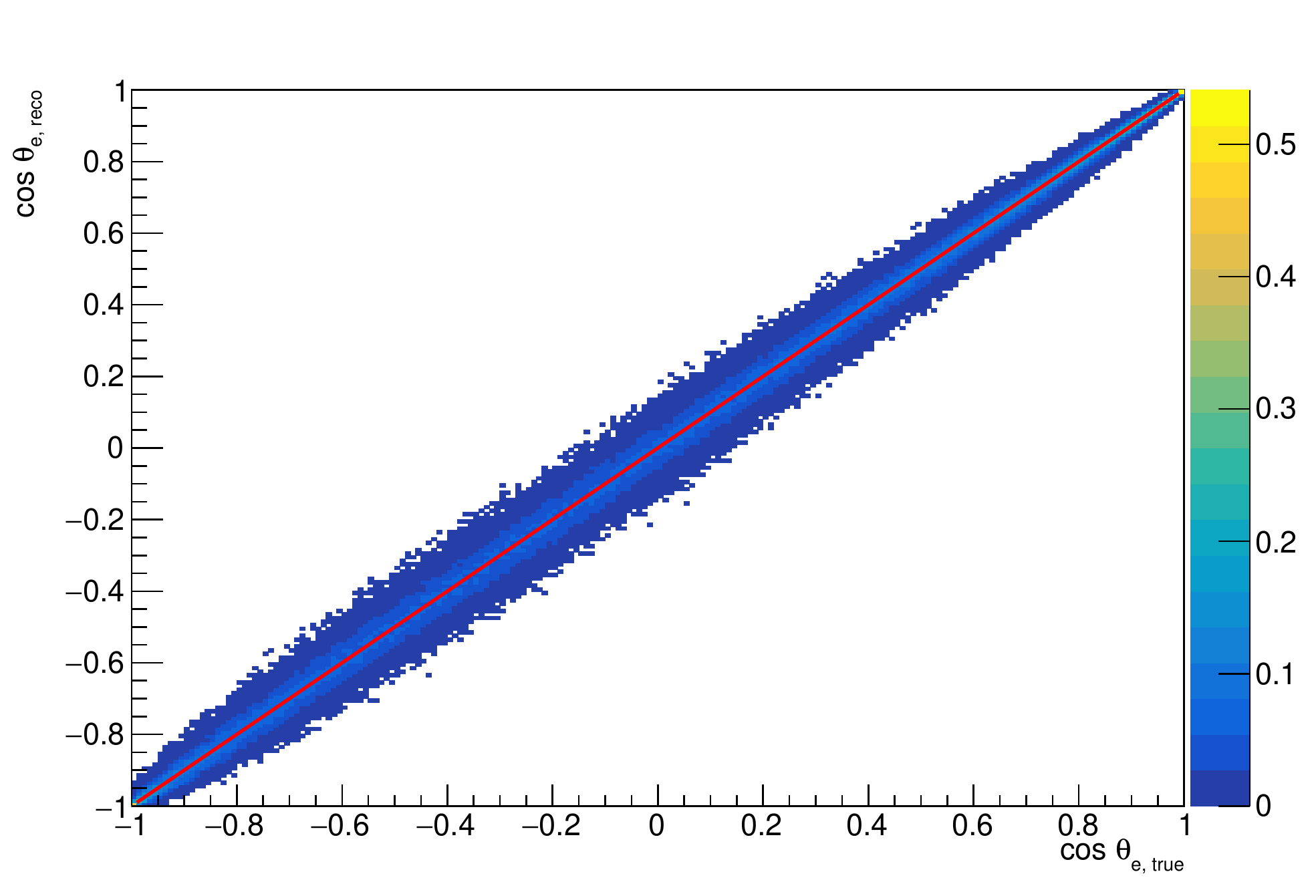}
            \caption{QES e$^-$ events}
            \label{fig:detectors:fd_cos_theta_2d_qes_e}
        \end{subfigure}
        \hspace{1em}
        \begin{subfigure}[b]{0.4\textwidth}  
            \centering 
            \includegraphics[width=\textwidth]{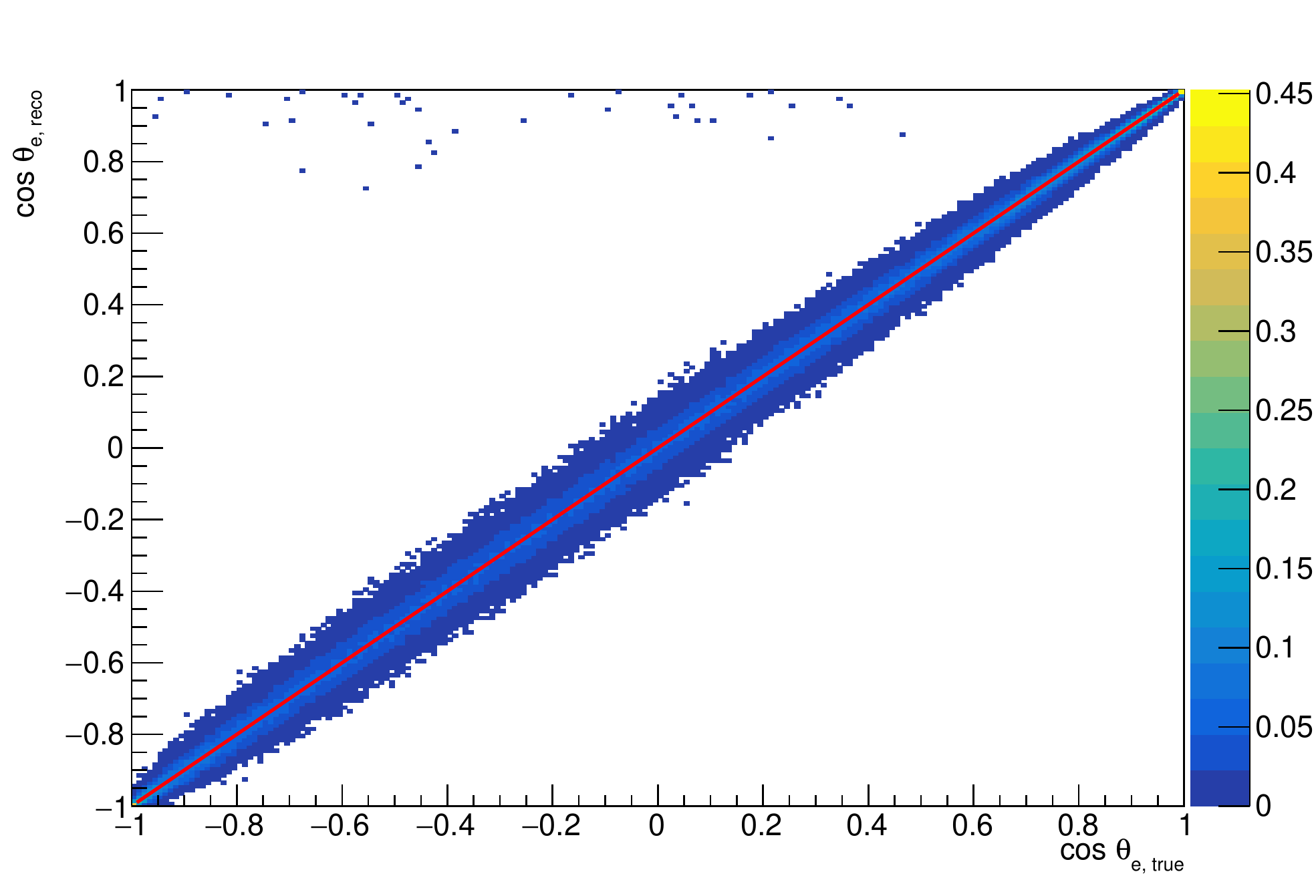}
            \caption{All e$^-$ events}
            \label{fig:detectors:fd_cos_theta_2d_all_e}
        \end{subfigure}
        \begin{subfigure}[b]{0.4\textwidth}   
            \centering 
            \includegraphics[width=\textwidth]{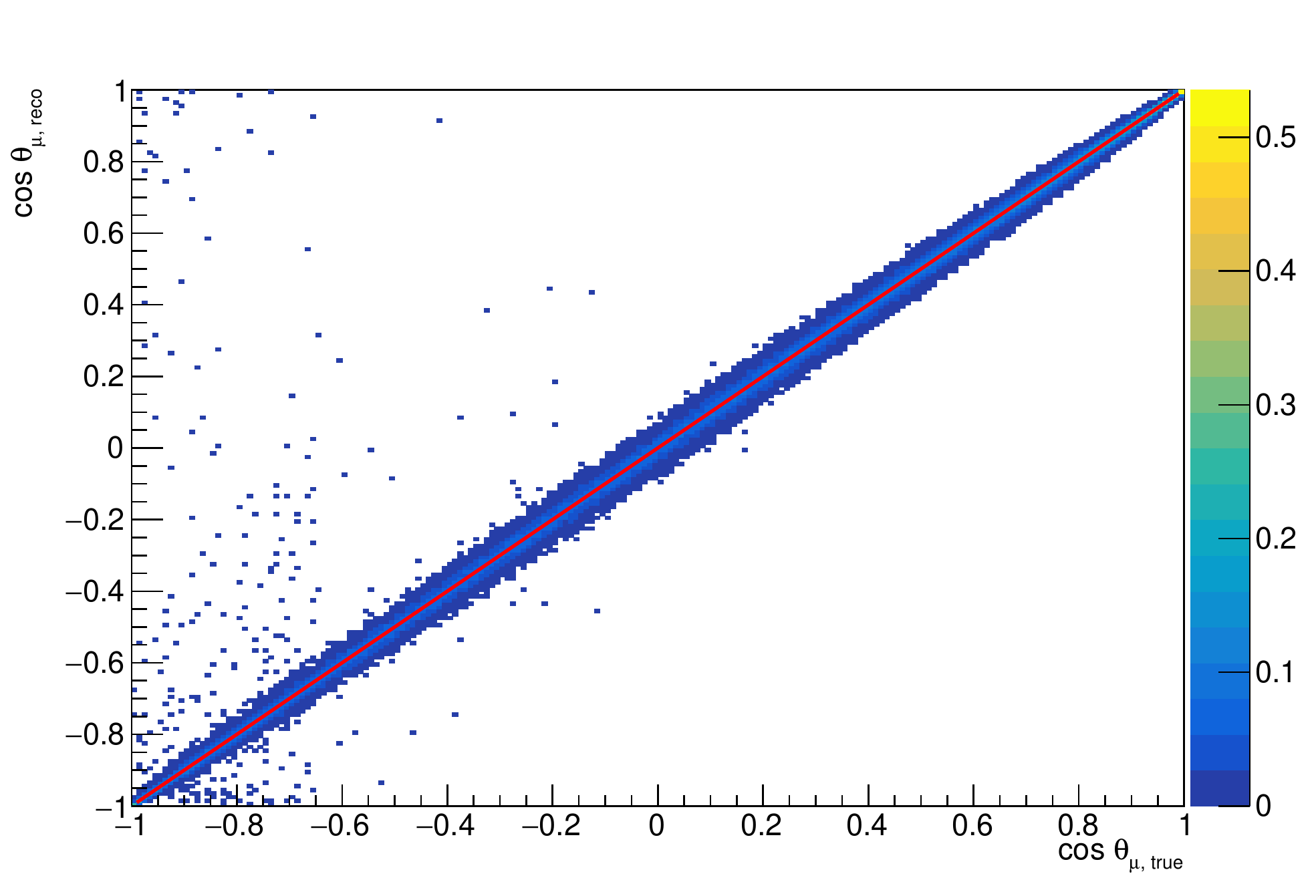}
            \caption{QES $\mu^-$ events}
            \label{fig:detectors:fd_cos_theta_2d_qes_mu}
        \end{subfigure}
        \hspace{1em}
        \begin{subfigure}[b]{0.4\textwidth}   
            \centering 
            \includegraphics[width=\textwidth]{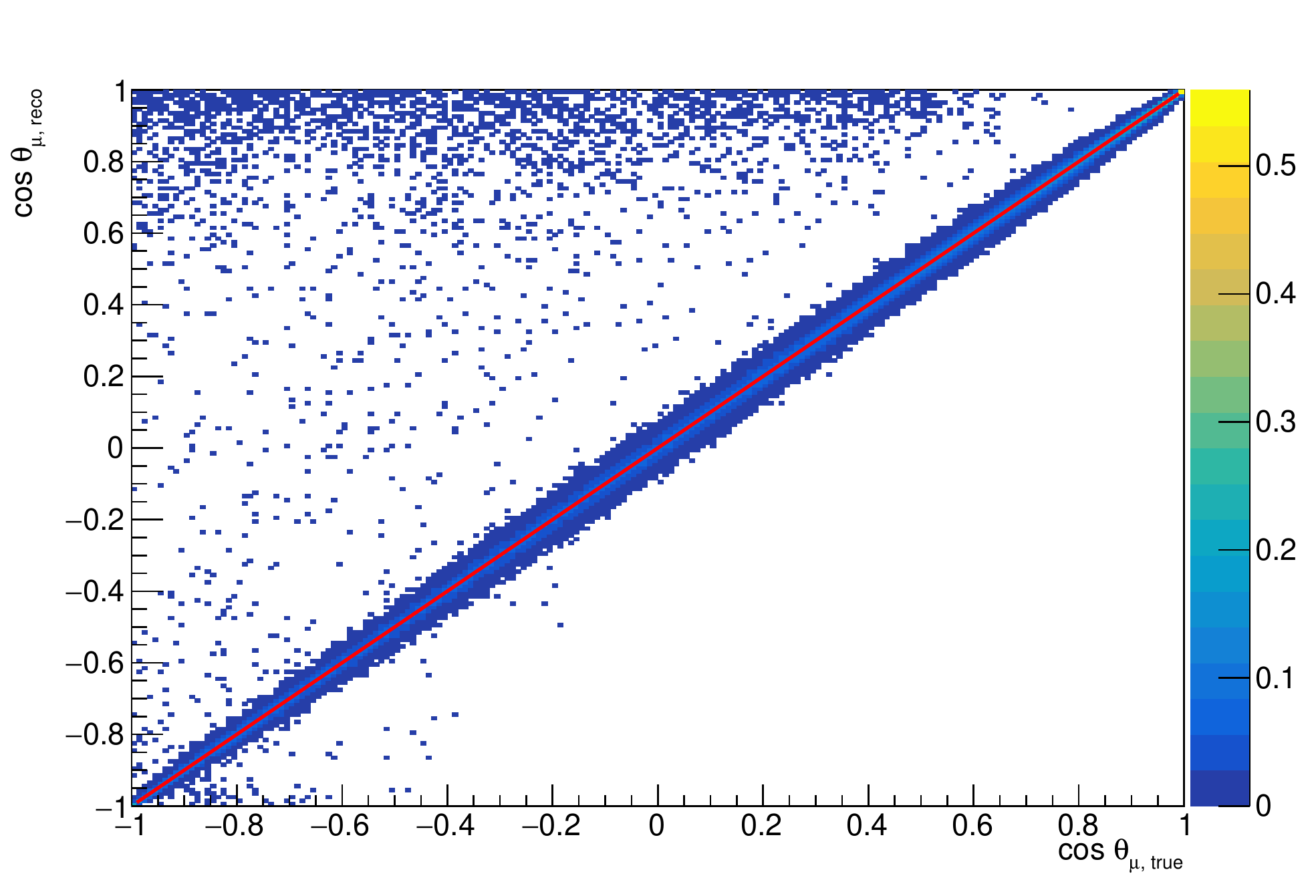}
            \caption{All $\mu^-$ events}
            \label{fig:detectors:fd_cos_theta_2d_all_mu}
        \end{subfigure}
        \begin{subfigure}[b]{0.4\textwidth}   
            \centering 
            \includegraphics[width=\textwidth]{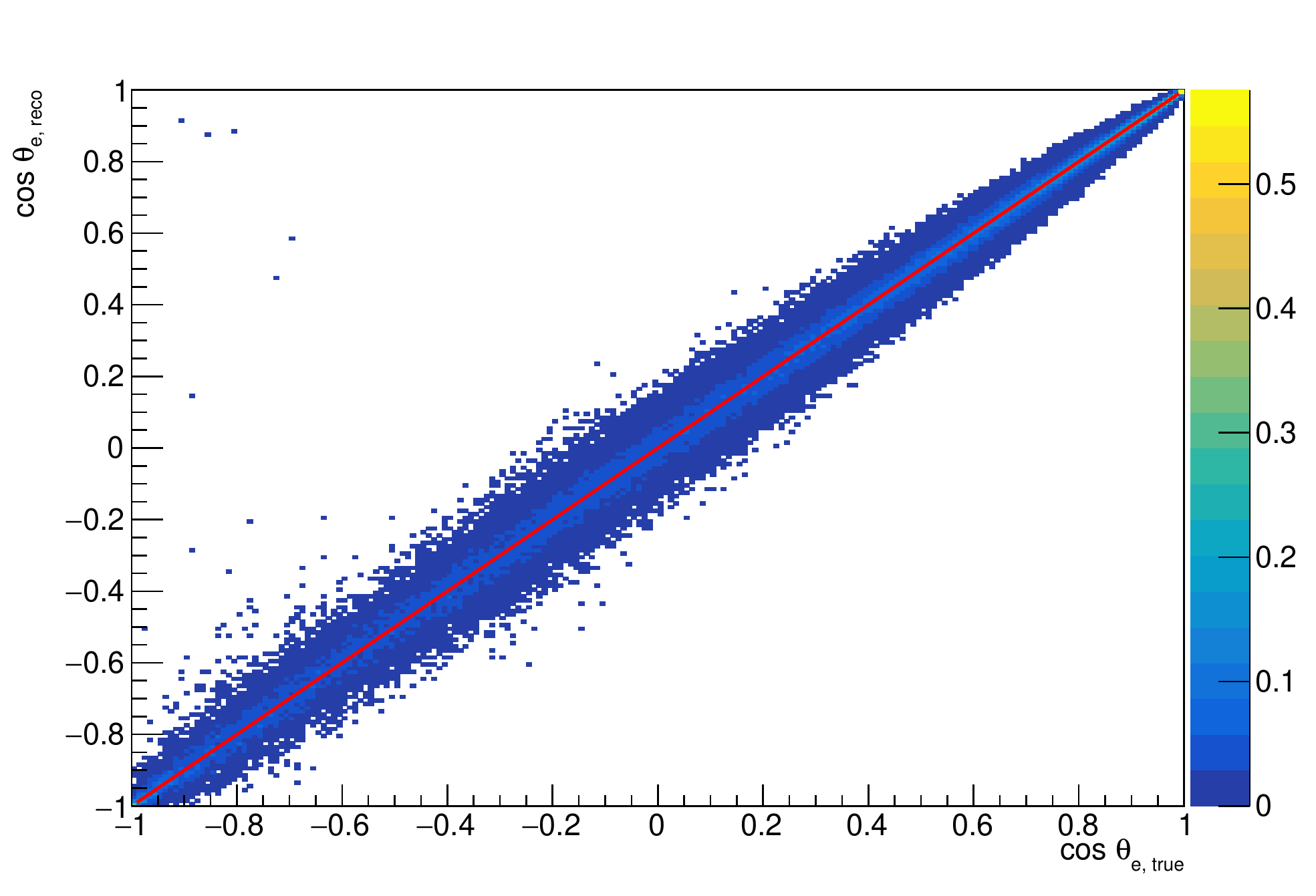}
            \caption{QES e$^+$ events}
            \label{fig:detectors:fd_cos_theta_2d_qes_ae}
        \end{subfigure}
        \hspace{1em}
        \begin{subfigure}[b]{0.4\textwidth}   
            \centering 
            \includegraphics[width=\textwidth]{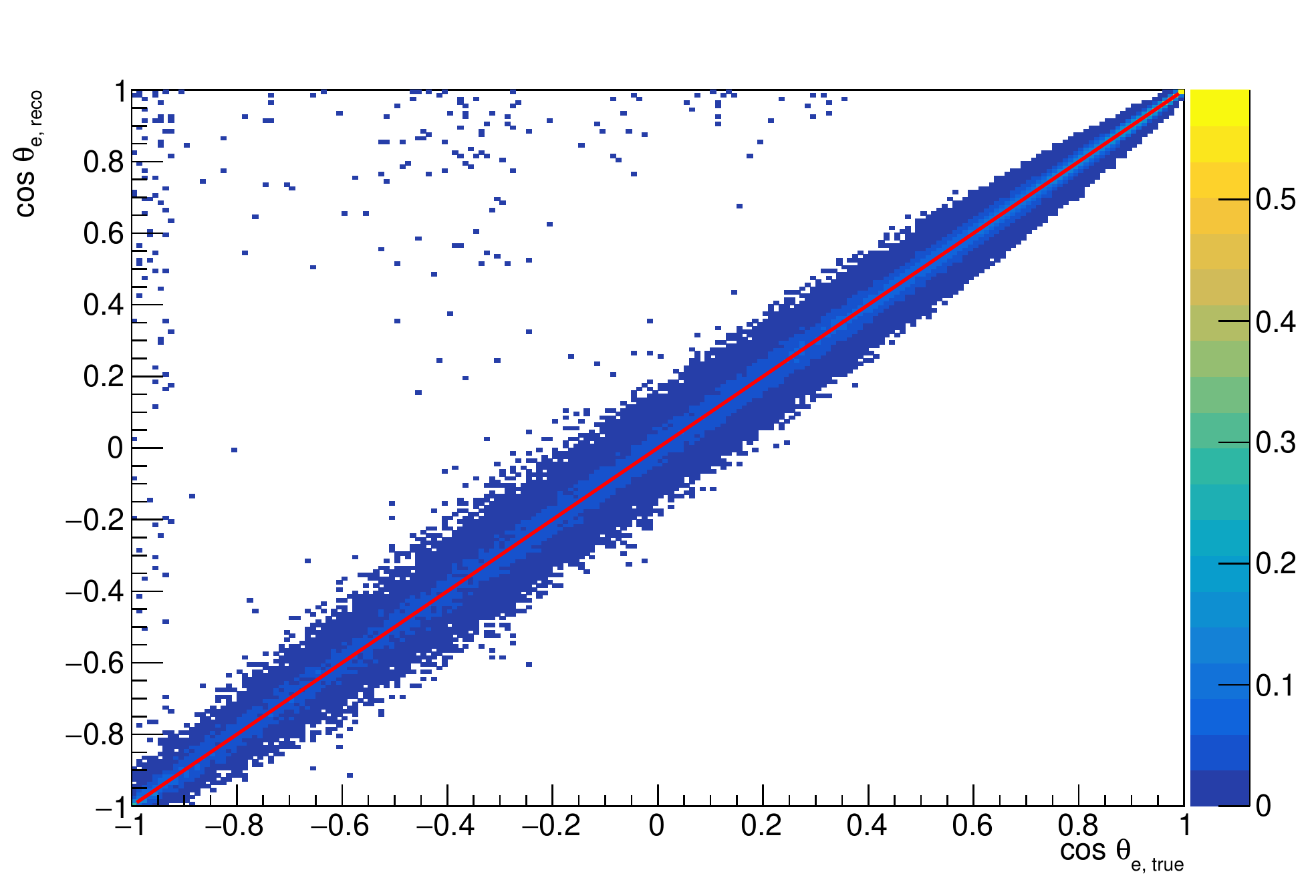}
            \caption{All e$^+$ events}
            \label{fig:detectors:fd_cos_theta_2d_all_ae}
        \end{subfigure}
        \begin{subfigure}[b]{0.4\textwidth}   
            \centering 
            \includegraphics[width=\textwidth]{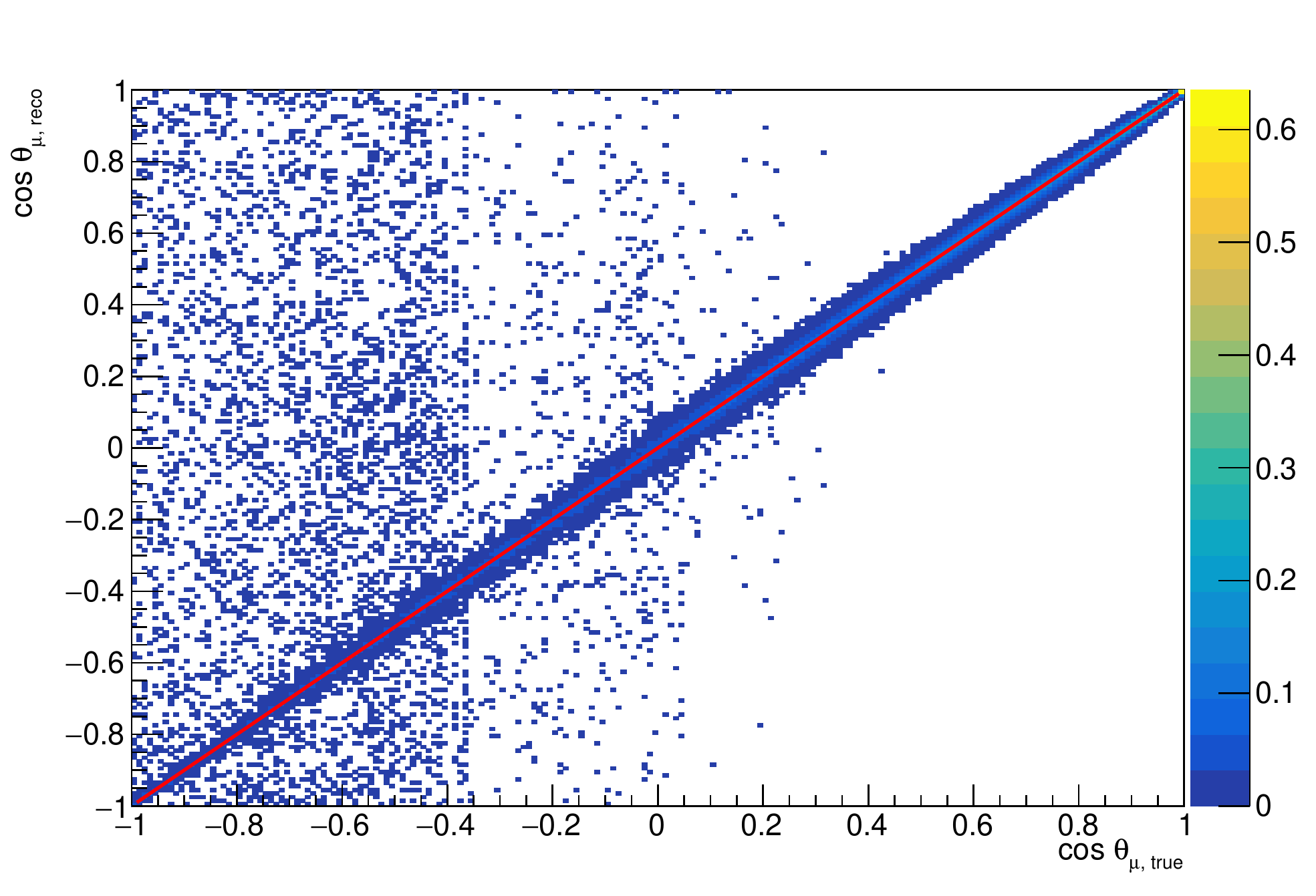}
            \caption{QES $\mu^+$ events}
            \label{fig:detectors:fd_cos_theta_2d_qes_amu}
        \end{subfigure}
        \hspace{1em}
        \begin{subfigure}[b]{0.4\textwidth}   
            \centering 
            \includegraphics[width=\textwidth]{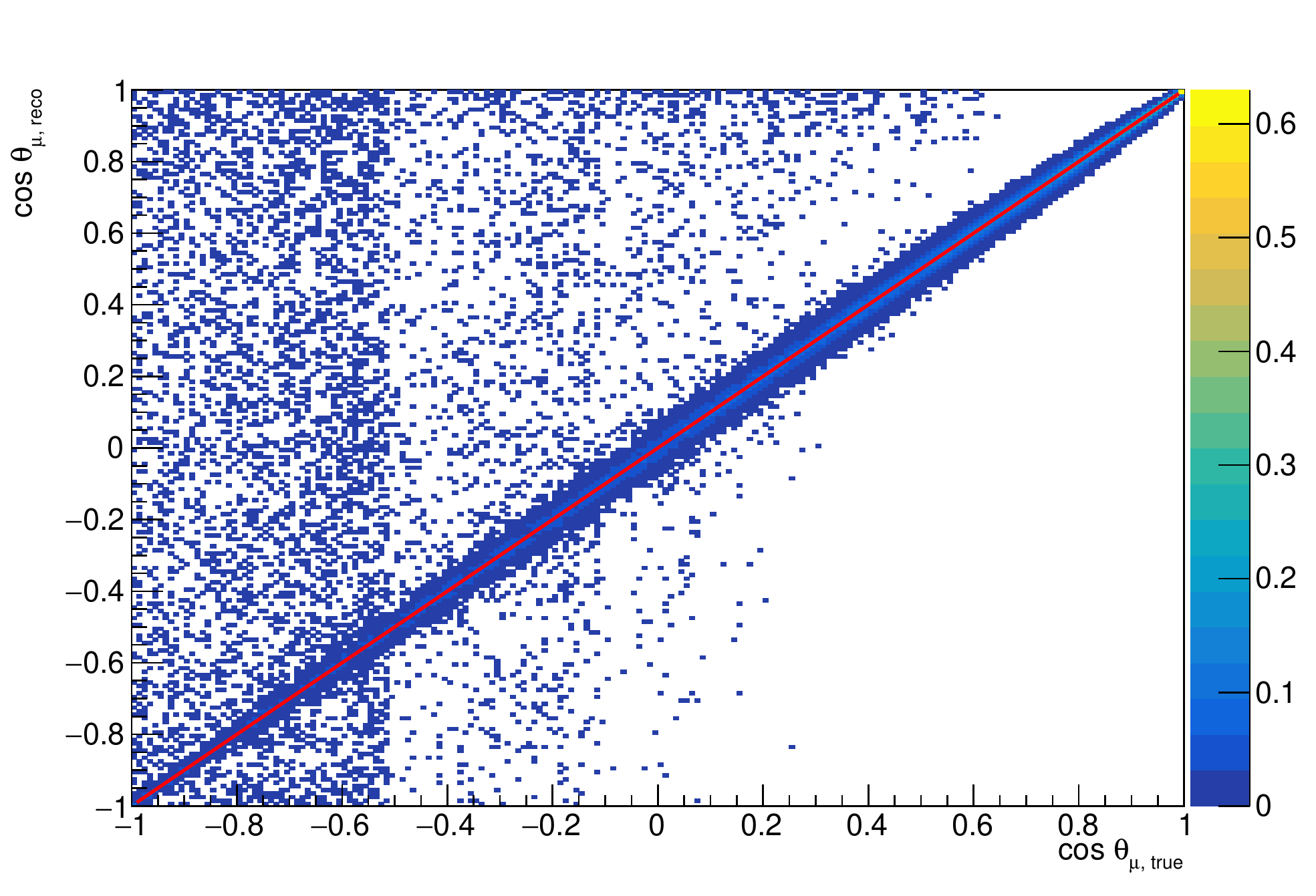}
            \caption{All $\mu^+$ events}
            \label{fig:detectors:fd_cos_theta_2d_all_amu}
        \end{subfigure}
        \caption{Distribution of reconstructed $\cos \theta$ as a function of true momentum for different flavours of final-state charged leptons. The left column shows plots using QES event sample, the right column shows plots using full event sample.}
        \label{fig:detectors:fd_cos_theta_2d}

    \end{figure*}

\begin{figure*}[htp!]
        \centering
        \begin{subfigure}[b]{0.475\textwidth}
            \centering
            \includegraphics[width=\textwidth]{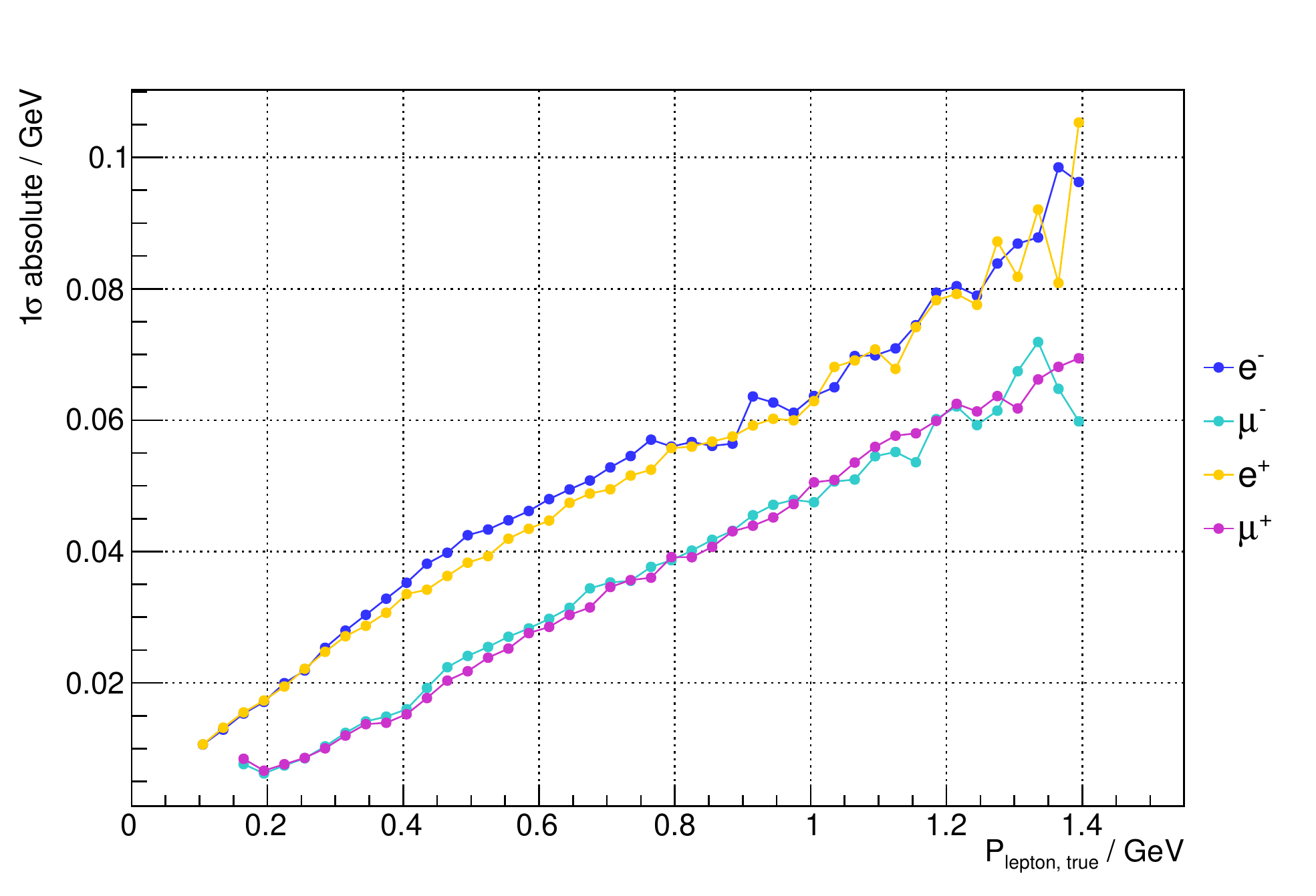}
            \caption{Absolute momentum resolution -- QES events}
            \label{fig:detectors:fd_mom_abs_qes}
        \end{subfigure}
        \hfill
        \begin{subfigure}[b]{0.475\textwidth}  
            \centering 
            \includegraphics[width=\textwidth]{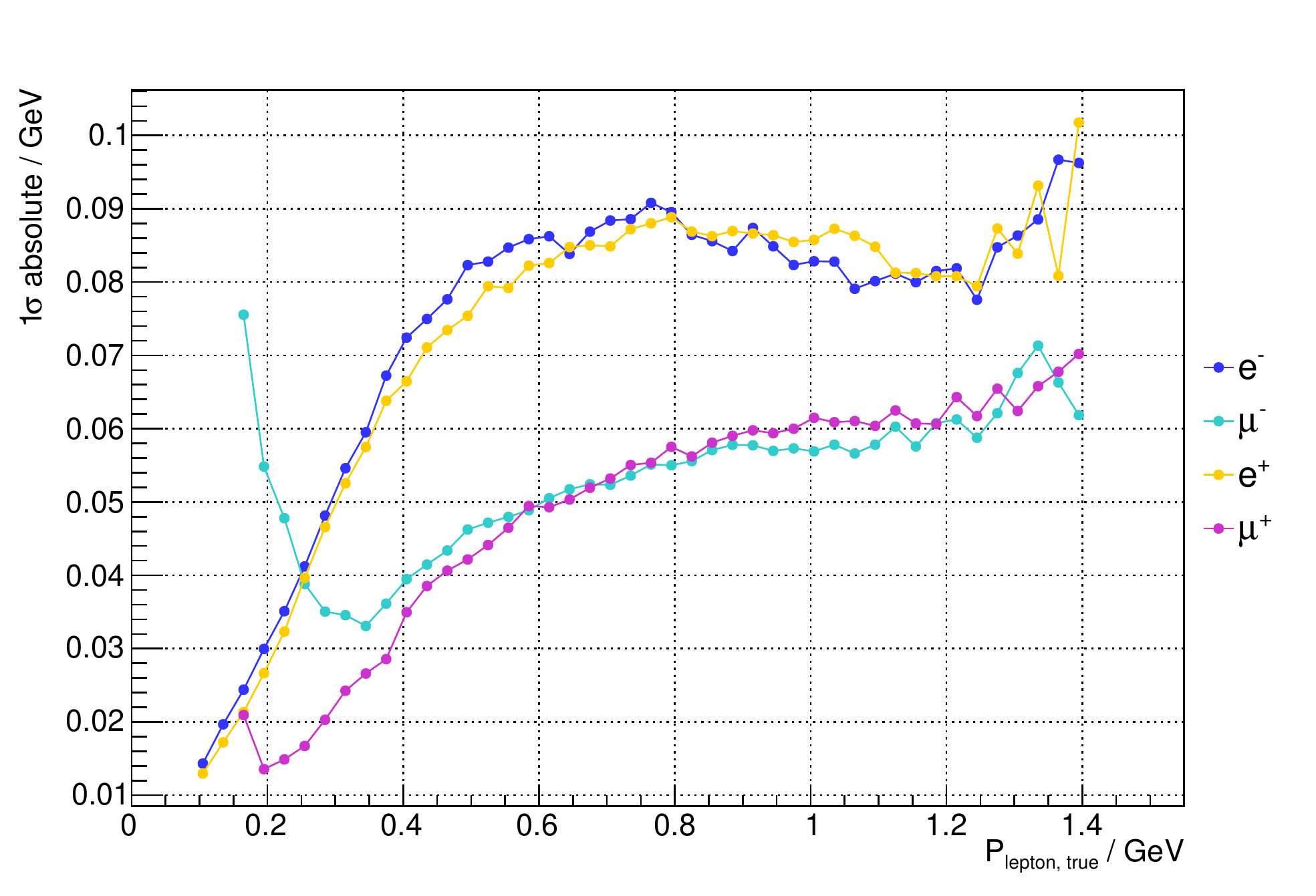}
            \caption{Absolute momentum resolution -- all events}
            \label{fig:detectors:fd_mom_abs_all}
        \end{subfigure}
        \begin{subfigure}[b]{0.475\textwidth}   
            \centering 
            \includegraphics[width=\textwidth]{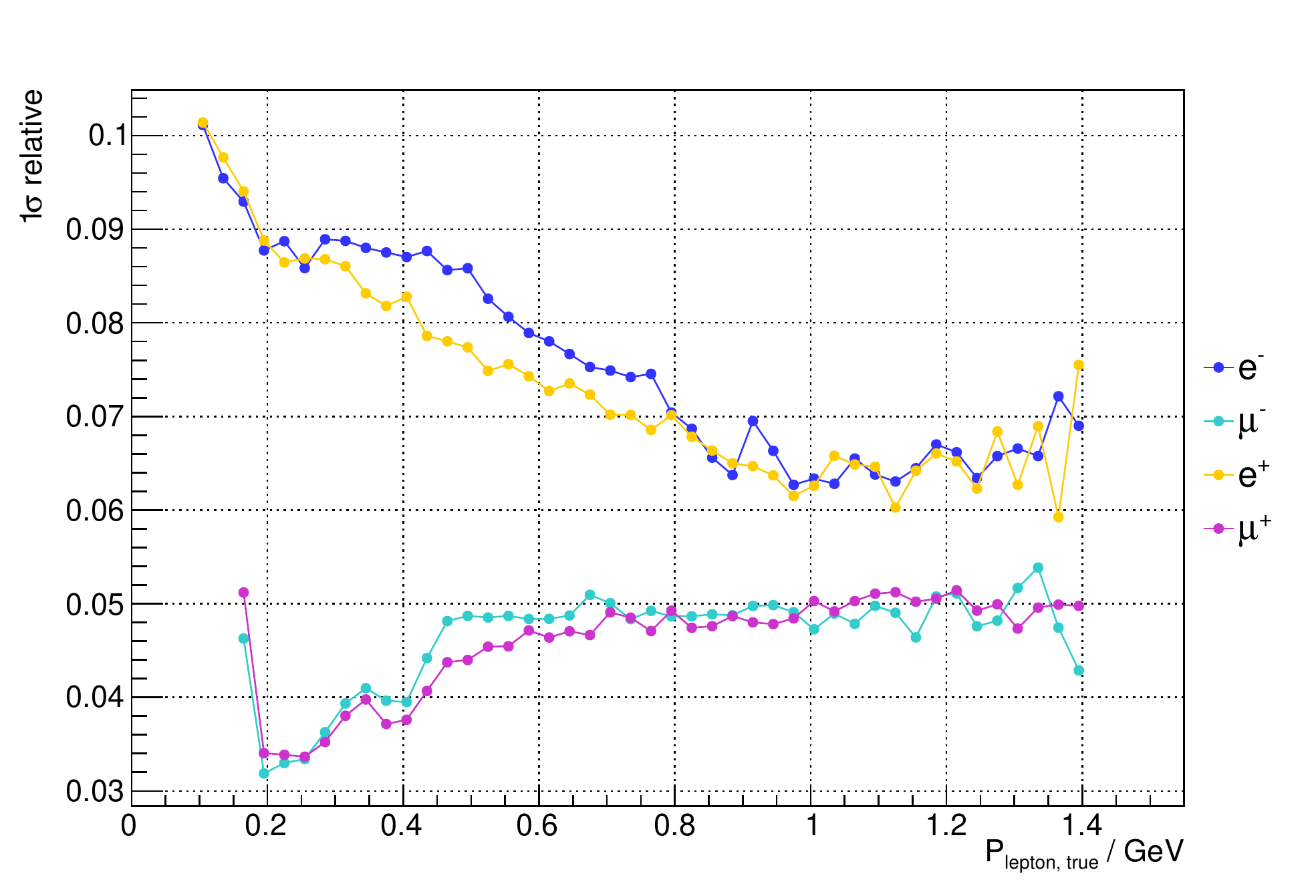}
            \caption{Relative momentum resolution -- QES events}
            \label{fig:detectors:fd_mom_rel_qes}
        \end{subfigure}
        \hfill
        \begin{subfigure}[b]{0.475\textwidth}   
            \centering 
            \includegraphics[width=\textwidth]{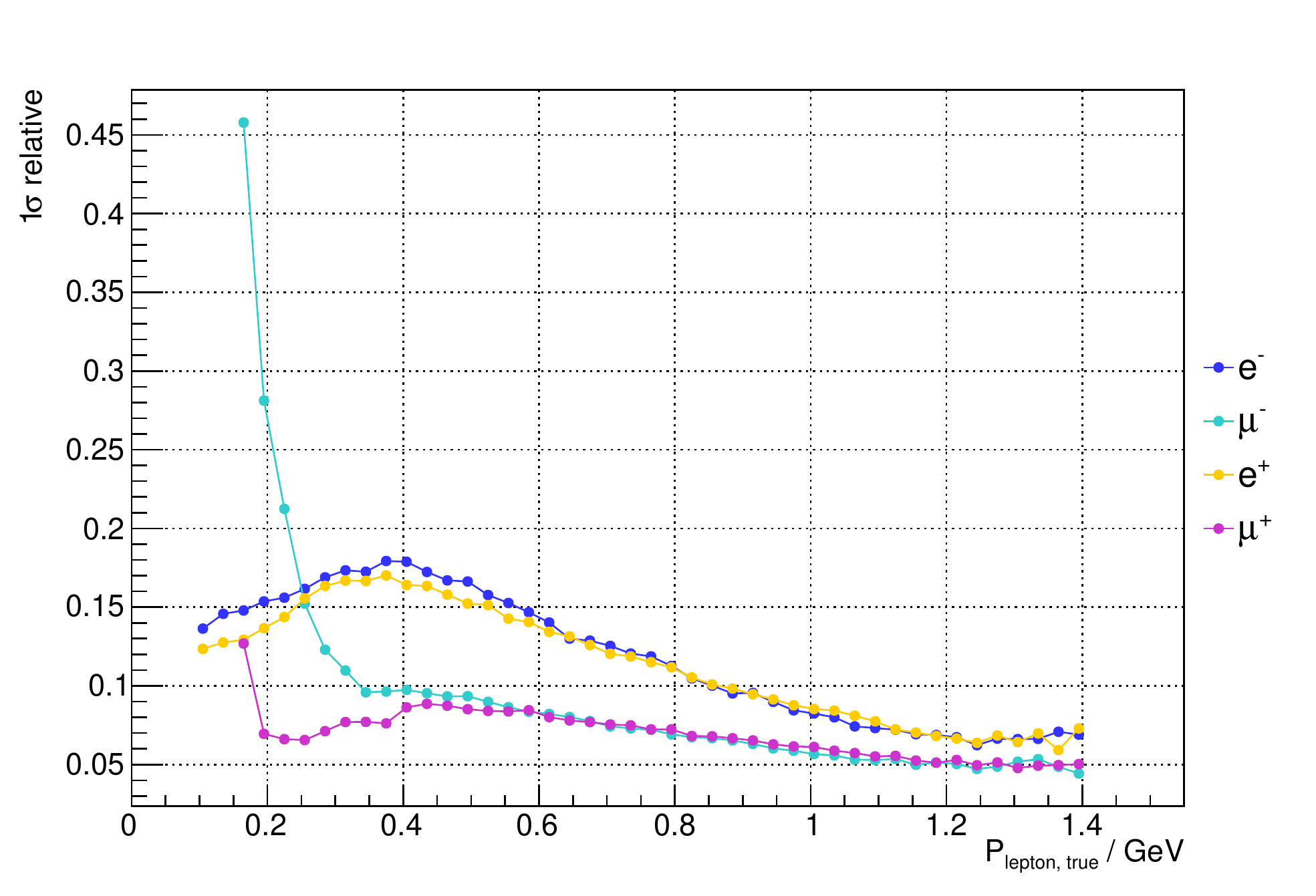}
            \caption{Relative momentum resolution -- all events}
            \label{fig:detectors:fd_mom_rel_all}
        \end{subfigure}
        \begin{subfigure}[b]{0.475\textwidth}   
            \centering 
            \includegraphics[width=\textwidth]{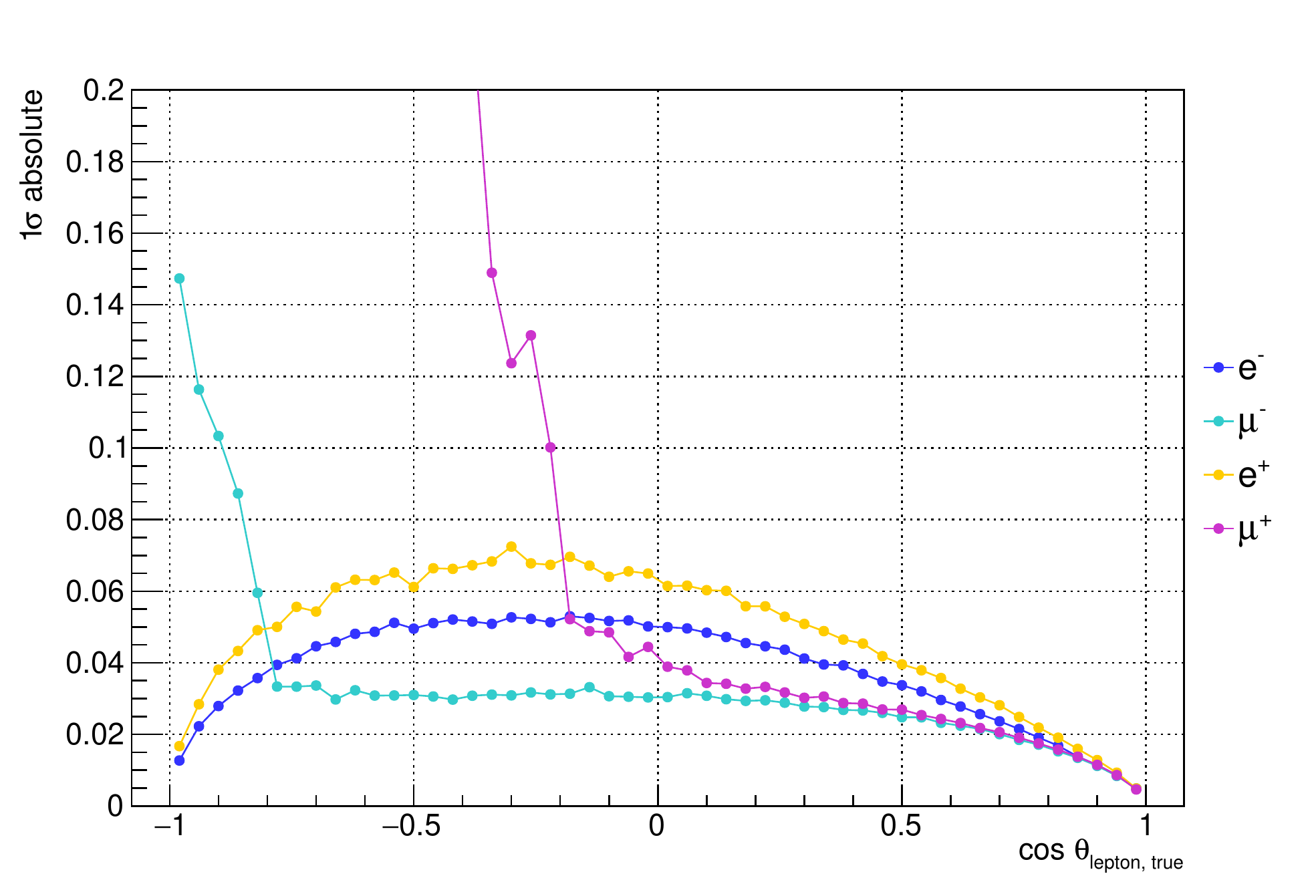}
            \caption{Absolute $\cos \theta$ resolution -- QES events}
            \label{fig:detectors:fd_cos_abs_qes}
        \end{subfigure}
        \hfill
        \begin{subfigure}[b]{0.475\textwidth}   
            \centering 
            \includegraphics[width=\textwidth]{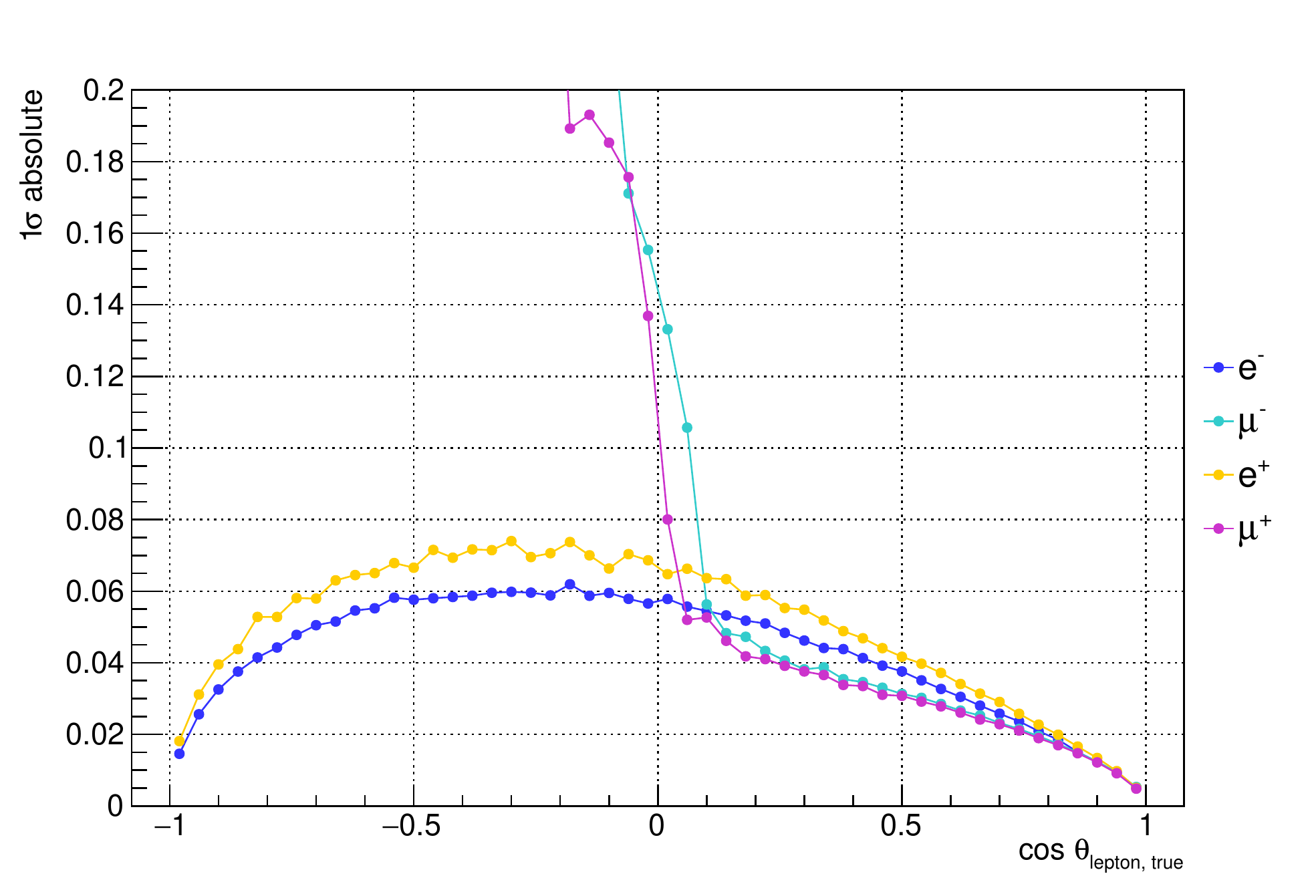}
            \caption{Absolute $\cos \theta$ resolution -- all events}
            \label{fig:detectors:fd_cos_abs_all}
        \end{subfigure}
        \caption{Resolution plots for charged leptons in the final state of neutrino interactions. Plots on the left are produced by selecting only QES events, while those on the right are produced using full event samples. Absolute 1 $\sigma$ resolutions of reconstructed charged lepton momentum as a function of true lepton momentum are shown in Figs.~\ref{fig:detectors:fd_mom_abs_qes} and \ref{fig:detectors:fd_mom_abs_all}. Relative 1 $\sigma$ resolutions of reconstructed lepton momentum as a function of true lepton momentum are shown in Figs.~\ref{fig:detectors:fd_mom_rel_qes} and \ref{fig:detectors:fd_mom_rel_all}, respectively. Absolute 1 $\sigma$ resolutions of reconstructed cos $\theta$ as a function of true cos $\theta$ are shown in Figs.~\ref{fig:detectors:fd_cos_abs_qes} and \ref{fig:detectors:fd_cos_abs_all}; in the QES sample, the resolution for back-scattered antimuons reaches the maximum of 0.75 in $\cos\theta$, in the full event sample, it reaches 0.81 for muons and 0.79 for antimuons -- these two plots are cropped at resolution of 0.2 for clarity.} 
        \label{fig:detectors:fd_lepton_comp_all_qes}
    \end{figure*}

%clearpage can be removed in the final version
%\clearpage

\subsubsubsection{Neutrino Energy Reconstruction and Selection Efficiency}

Neutrino energy is reconstructed using measurements of the final-state charged lepton momentum and its scattering angle ($\cos\theta$) as an input to the Eq.~(\ref{eqn:detectors:enu_qes}). The main source of the uncertainty comes from the fact that nucleons in the nucleus undergo a random Fermi motion: even if charged lepton momentum and scattering angle are perfectly measured, there will still be a residual uncertainty in reconstructed energy up to \SI{100}{MeV}, its exact value depending on the actual Fermi momentum of the struck nucleus and momentum transferred from neutrino to the nucleus. The contribution coming from the uncertainty of measurement of charged lepton momentum and $\cos\theta$ is lower than that, so further improvement of the detector response would yield only minor improvements to the neutrino energy resolution. The number of generated MC events is shown in Table~\ref{tab:detectors:fd_mc_neutrino_events}. 

Figure~\ref{fig:detectors:fd_mm} shows the distribution of the reconstructed neutrino energy as a function of true neutrino energy -- the migration matrices. The information on the selection efficiency is encoded in these matrices: the integral of a column for each true energy bin is less than one and represents the selection efficiency. The matrices are produced using the ``flat energy'' MC set, so they are valid for any flux within the neutrino energy region they cover. However, these do depend on the neutrino flavour-selection algorithm, which is optimised for the ESS$\nu$SB flux, making them most efficient in the ESS$\nu$SB neutrino energy region. The matrices are shown both for the QES and full simulated interaction sample. Some effort has been put into selecting only QES events to improve the physics reach of the project, but it has been shown that the optimal sensitivity is reached when keeping a full event sample; the loss of the sensitivity due to degraded energy resolution is compensated by the increased number of events. As seen in Fig.~\ref{fig:detectors:fd_mm}, only those matrices for the neutrino flavour are correctly selected for clarity; the full set of matrices containing the addition of all possible combinations of selected/true neutrino flavour have been produced and are used for physics reach determination.

In the matrices produced using the full event sample, there is a ridge in the area below the perfect reconstruction line (red diagonal line), which is absent from the matrices made using the QES subsample. This ridge is caused by the presence of additional particles in the final state of non-QES interactions, which carry away energy that is unaccounted for in the Eq.~(\ref{eqn:detectors:enu_qes}), resulting in an underestimation of the neutrino energy. This is the primary reason that the resolution for the full event sample is worse than that of the QES sample.

%%%%%%%%%%%%%%%%%%%%%%%%%%%%%%%%%%%%%%%%%%%%%%%%%%% 2D energy positive polarity neutrino production
    \begin{figure}[htp!]
        \centering
        \begin{subfigure}[b]{0.4\textwidth}
            \centering
            \includegraphics[width=\textwidth]{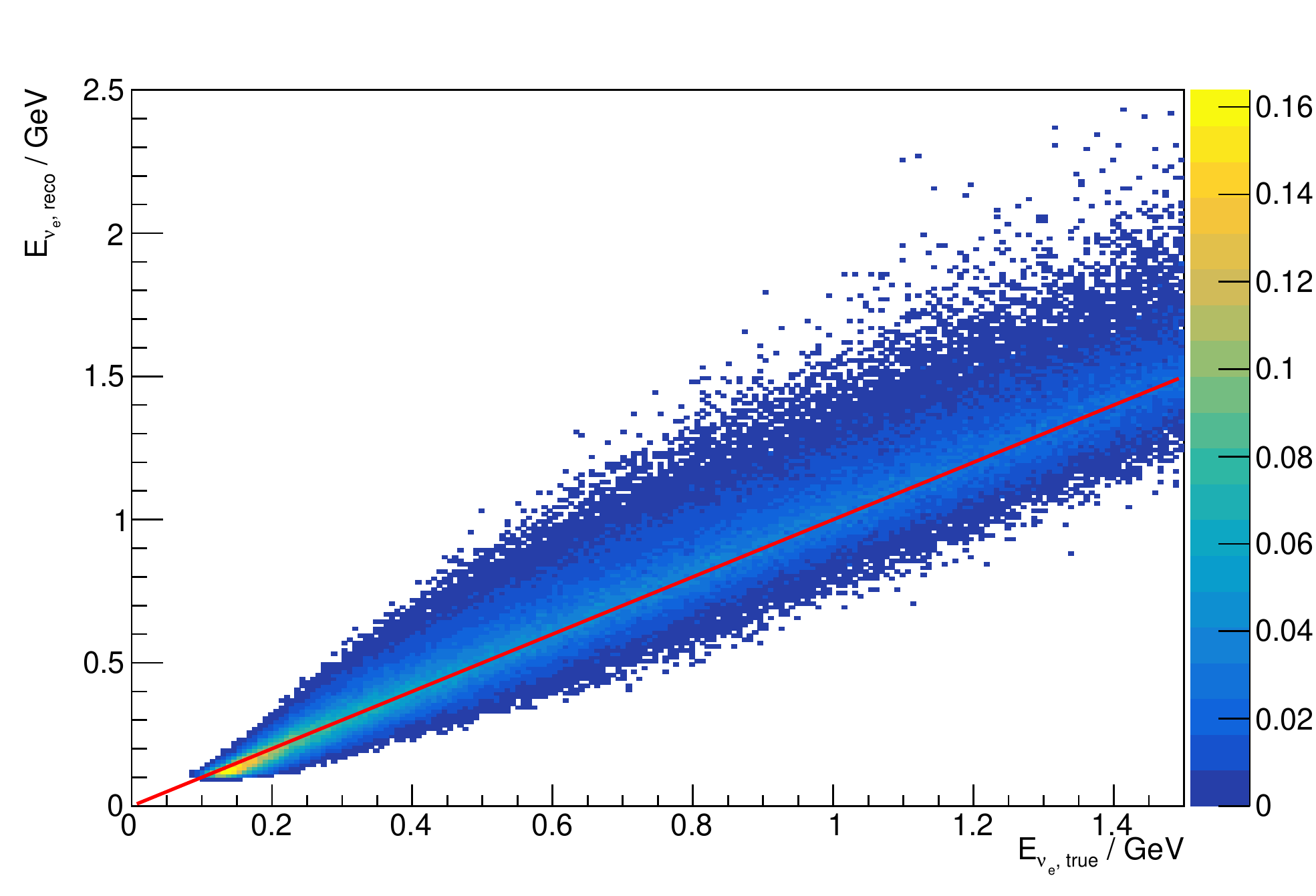}
            \caption{QES $\nu_e$ events.}
            \label{fig:detectors:fd_mm_qes_e}
        \end{subfigure}
        \hspace{1em}
        \begin{subfigure}[b]{0.4\textwidth}  
            \centering 
            \includegraphics[width=\textwidth]{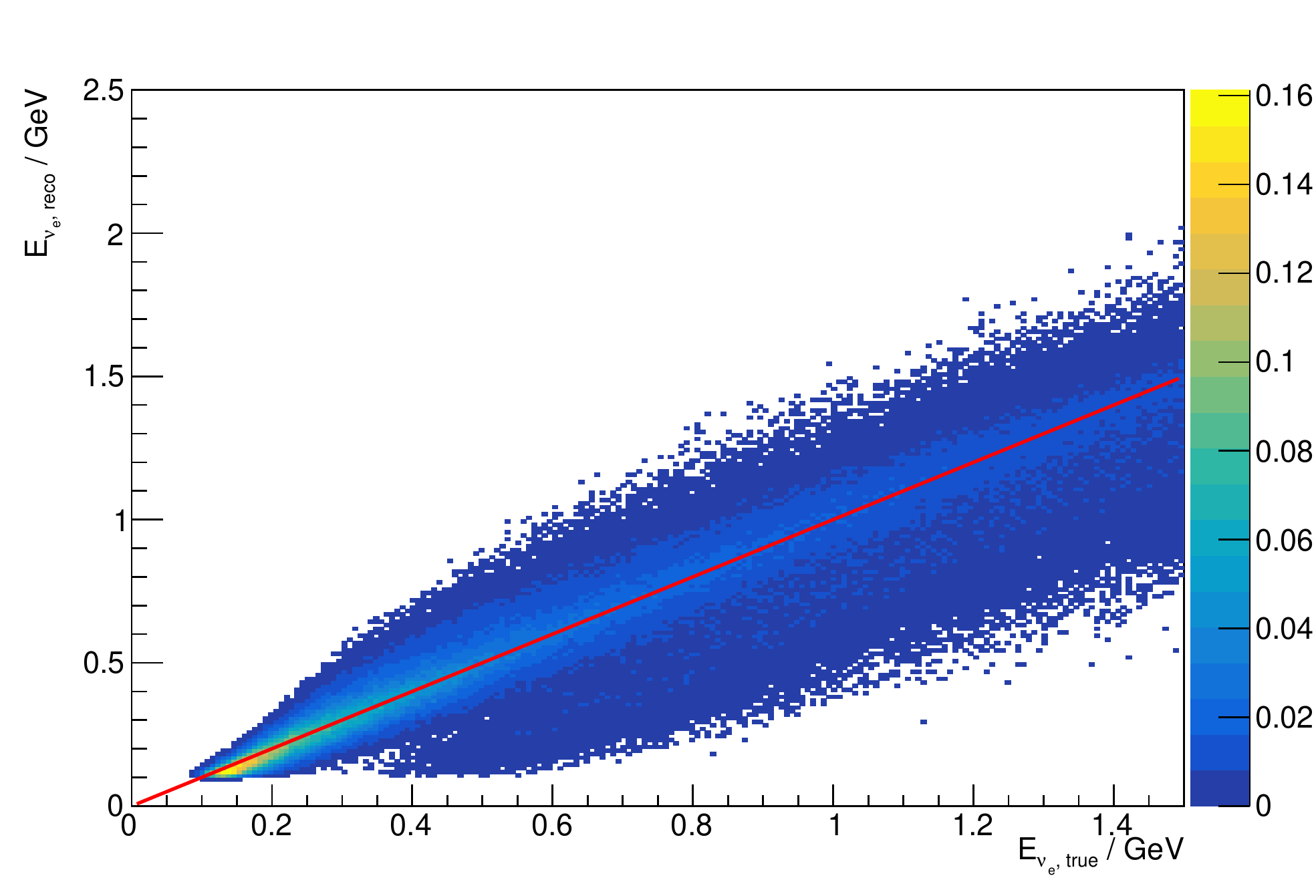}
            \caption{All $\nu_e$ events.}
            \label{fig:detectors:fd_mm_all_e}
        \end{subfigure}
        \begin{subfigure}[b]{0.4\textwidth}   
            \centering 
            \includegraphics[width=\textwidth]{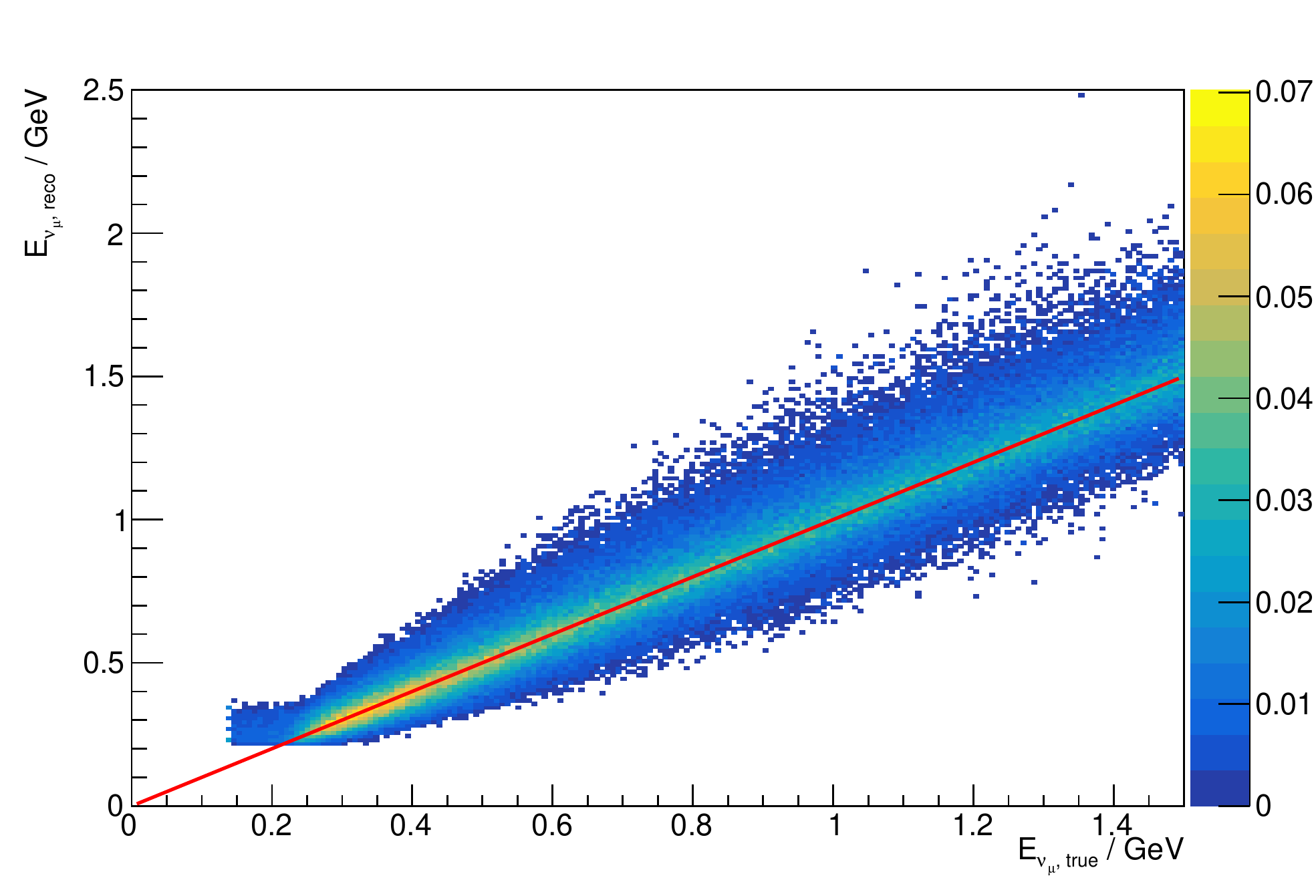}
            \caption{QES $\nu_\mu$ events.}
            \label{fig:detectors:fd_mm_qes_mu}
        \end{subfigure}
        \hspace{1em}
        \begin{subfigure}[b]{0.4\textwidth}   
            \centering 
            \includegraphics[width=\textwidth]{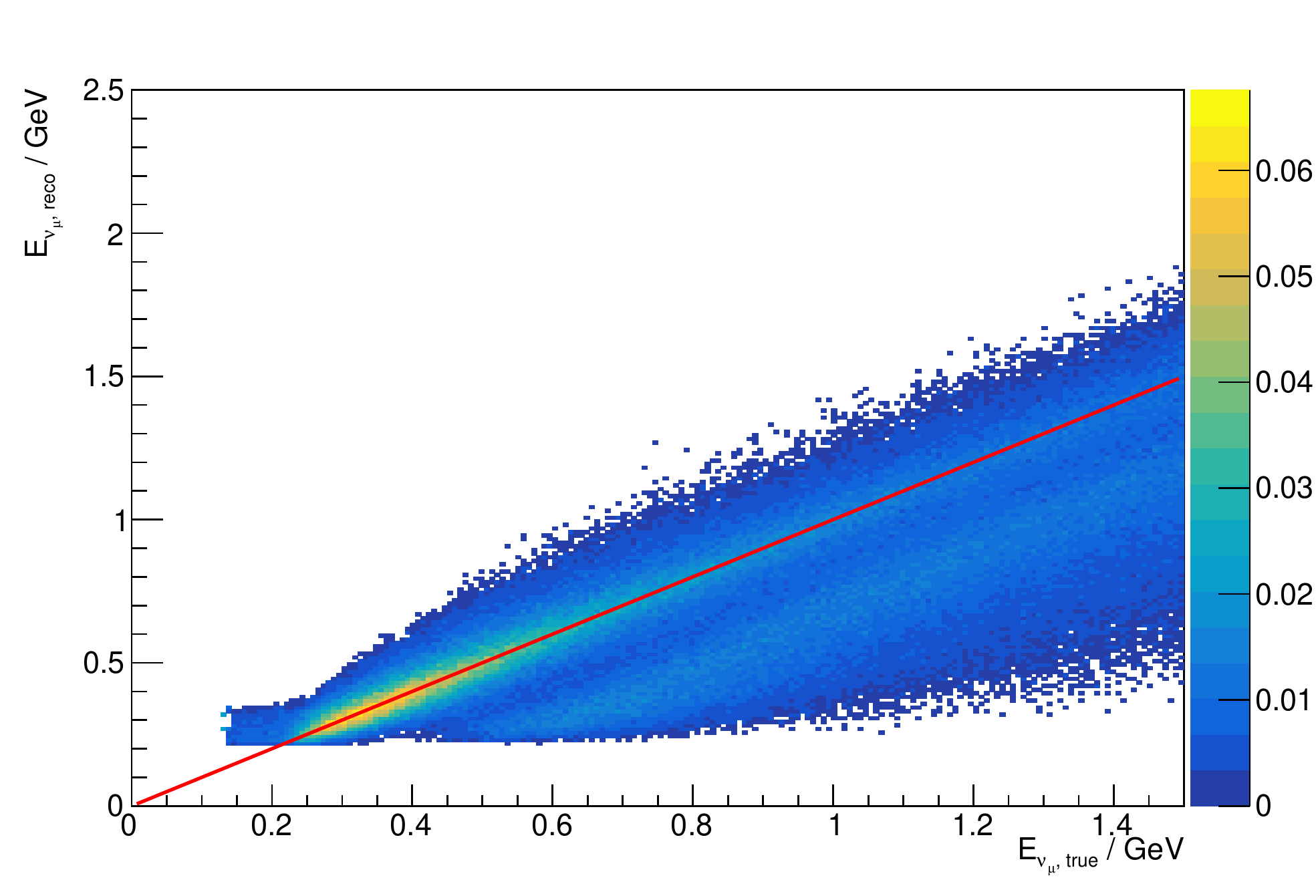}
            \caption{All $\nu_\mu$ events.}
            \label{fig:detectors:fd_mm_all_mu}
        \end{subfigure}
        \begin{subfigure}[b]{0.4\textwidth}   
            \centering 
            \includegraphics[width=\textwidth]{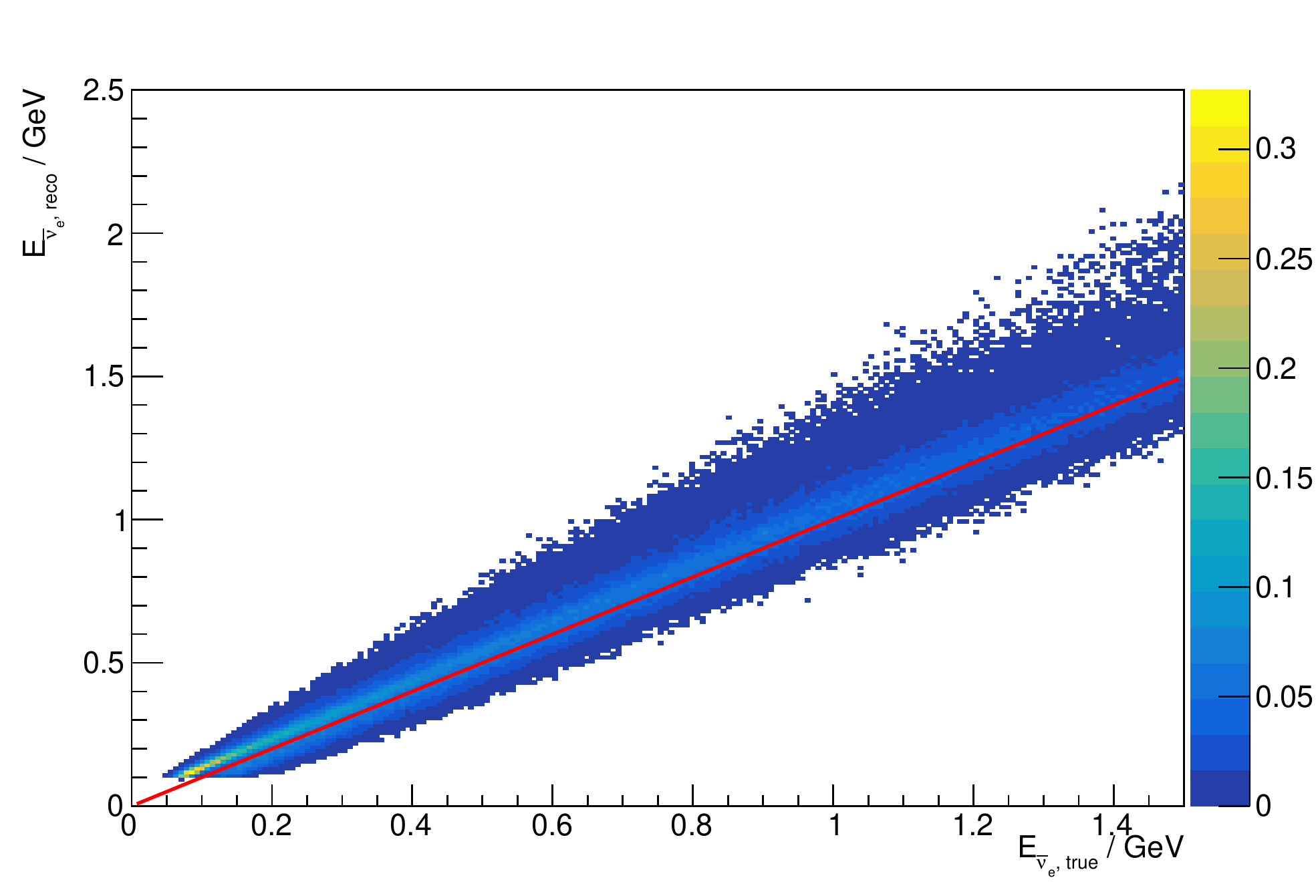}
            \caption{QES $\overline{\nu}_e$ events.}
            \label{fig:detectors:fd_mm_qes_ae}
        \end{subfigure}
        \hspace{1em}
        \begin{subfigure}[b]{0.4\textwidth}   
            \centering 
            \includegraphics[width=\textwidth]{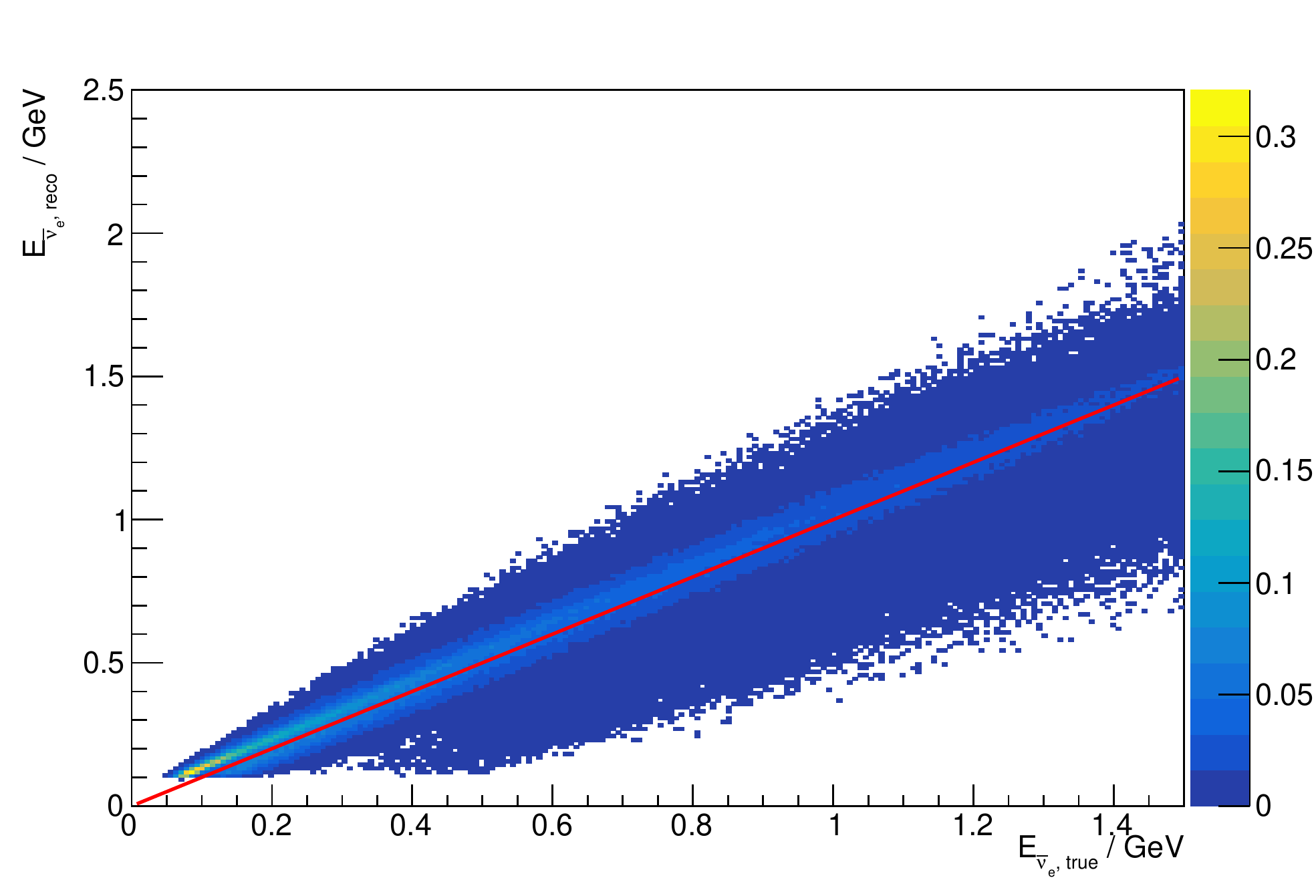}
            \caption{All $\overline{\nu}_e$ events.}
            \label{fig:detectors:fd_mm_all_ae}
        \end{subfigure}
        \begin{subfigure}[b]{0.4\textwidth}   
            \centering 
            \includegraphics[width=\textwidth]{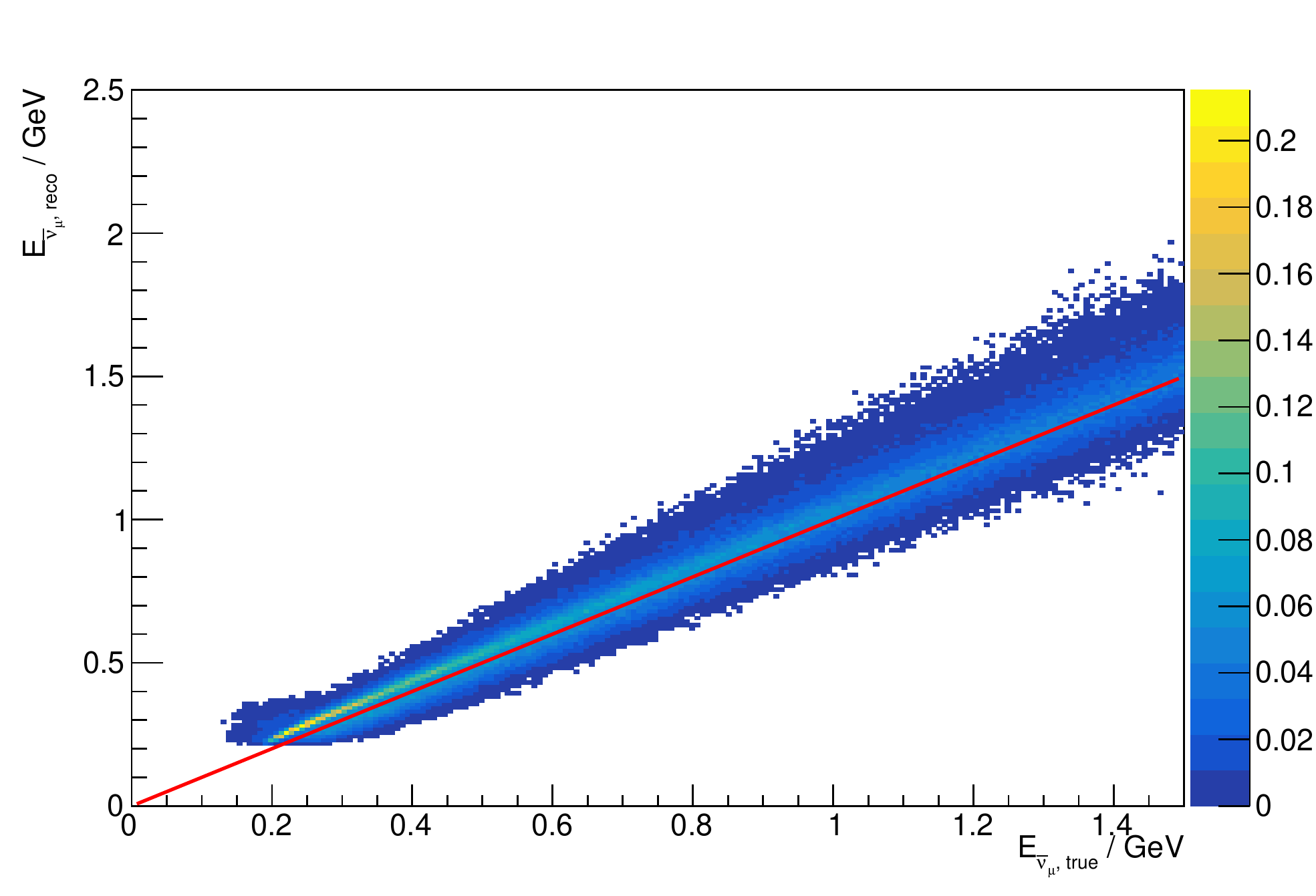}
            \caption{QES $\overline{\nu}_\mu$ events.}
            \label{fig:detectors:fd_mm_qes_amu}
        \end{subfigure}
        \hspace{1em}
        \begin{subfigure}[b]{0.4\textwidth}   
            \centering 
            \includegraphics[width=\textwidth]{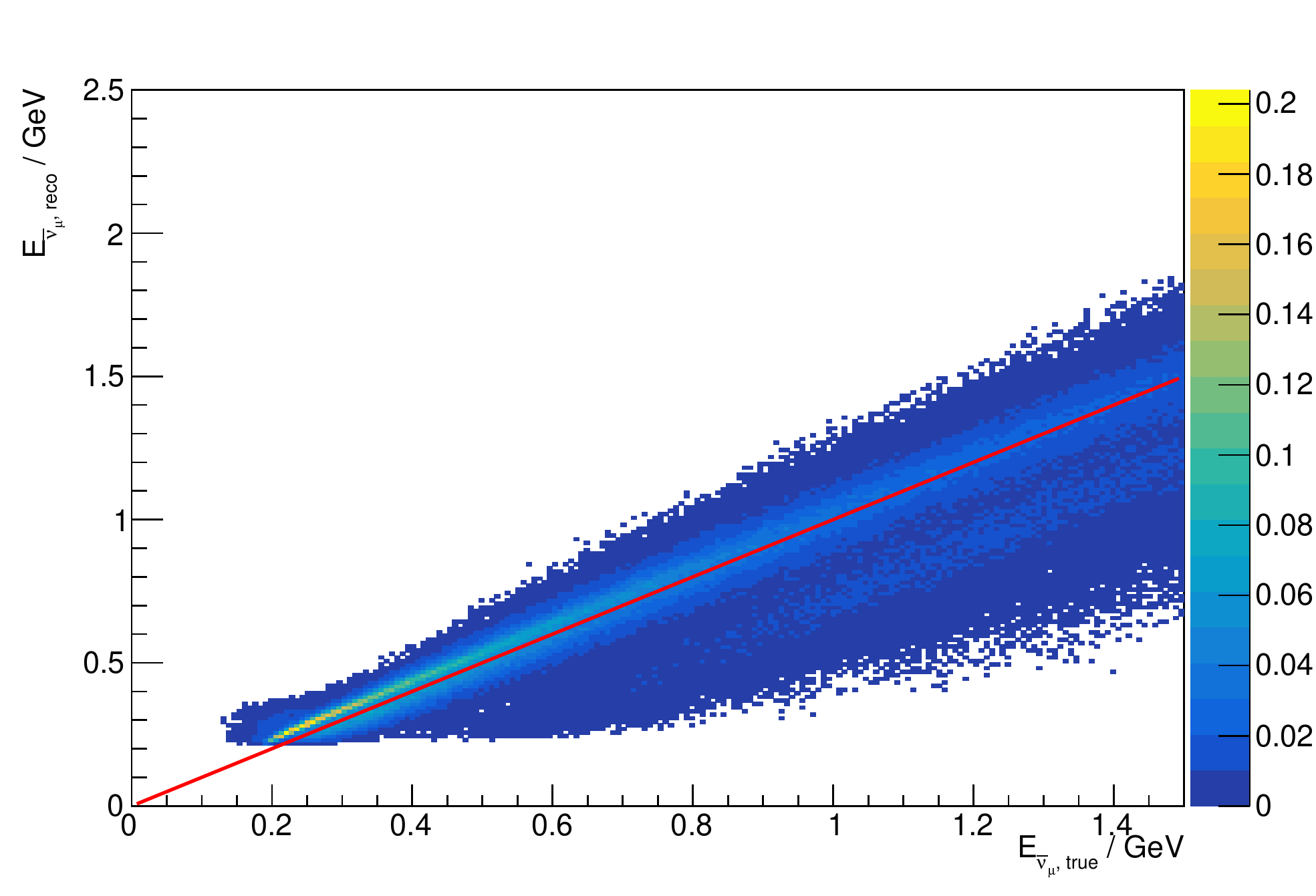}
            \caption{All $\overline{\nu}_\mu$ events.}
            \label{fig:detectors:fd_mm_all_amu}
        \end{subfigure}
        \caption{Distribution of reconstructed energy as a function of true energy (MM) for different flavours of charged leptons. The left column shows plots with only QES events, while the right column shows plots with full event sample. These plots were produced using the neutrino interaction production.}
        \label{fig:detectors:fd_mm}

    \end{figure}

The absolute and relative resolution of the reconstructed neutrino energy is shown in Fig.~\ref{fig:detectors:fd_neutrino_all_qes}. As expected, the resolution for the QES sample is better than the one for full event sample, which is mainly due to the ridge visible in the matrices in the left column of Fig.~\ref{fig:detectors:fd_mm}. The resolutions are comparable between QES and total samples in the energies up to roughly \SI{400}{MeV}, since non-QES interactions become prominent above that energy. The energy resolution for neutrinos is consistently worse than that for antineutrinos. The reason behind this fact is that, at ESS$\nu$SB energies, antineutrinos prefer forward scattering, unlike neutrinos which prefer isotropic scattering, which is due to the nature of the weak forces. The momentum transfer between neutrino and the target increases with the scattering angle because of the kinematics, which means that the forward-scattered particles transfer less of their momentum. Less momentum transfer implies less dependence of the final state on the initial target momentum, thereby reducing the uncertainty on the neutrino energy coming from the Fermi momentum of nucleons.

In the energy range above \SI{250}{MeV}, the resolution is better for muon (anti)neutrinos than for electron (anti)neutrinos because muons in the final state do not produce an electromagnetic shower (unlike electrons) which make their Cherenkov rings sharper and easier to reconstruct than those of electrons. In the lower energy region, muon (anti)neutrino reconstruction is worse because the resulting muons are very close to or below the Cherenkov threshold, which makes their observation more difficult or impossible; the produced electrons are well above the Cherenkov threshold for all ESS$\nu$SB neutrino energies.

The relative energy resolution for the QES sample is 12--22\% (15--28)\% for electron (muon) neutrinos, and 11--23\% (8--30)\% for electron (muon) antineutrinos. For the full event sample, the energy resolution is in the range of 12--33\% (16--33\%) for electron (muon) neutrinos, and 12--23\% (9--29\%) for electron (muon) antineutrinos. However, for energies corresponding to the second oscillation maximum at the FD site (150--400 \si{MeV}), the QES and full-sample resolutions are comparable.

\begin{figure}[htp!]
        \centering
        \begin{subfigure}[b]{0.475\textwidth}
            \centering
            \includegraphics[width=\textwidth]{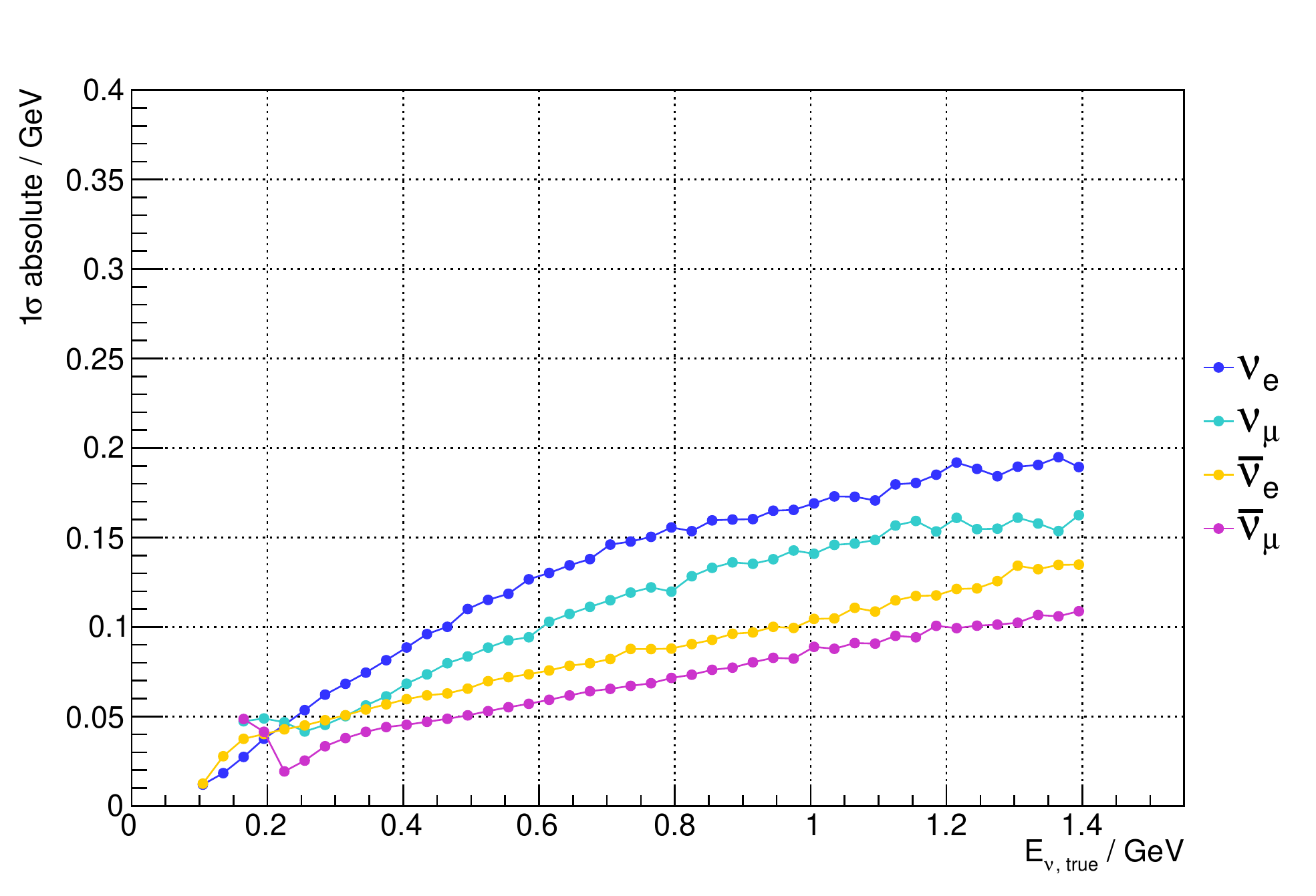}
            \caption{Absolute energy resolution -- QES events.}
            \label{fig:detectors:fd_nrg_abs_qes}
        \end{subfigure}
        \hfill
        \begin{subfigure}[b]{0.475\textwidth}  
            \centering 
            \includegraphics[width=\textwidth]{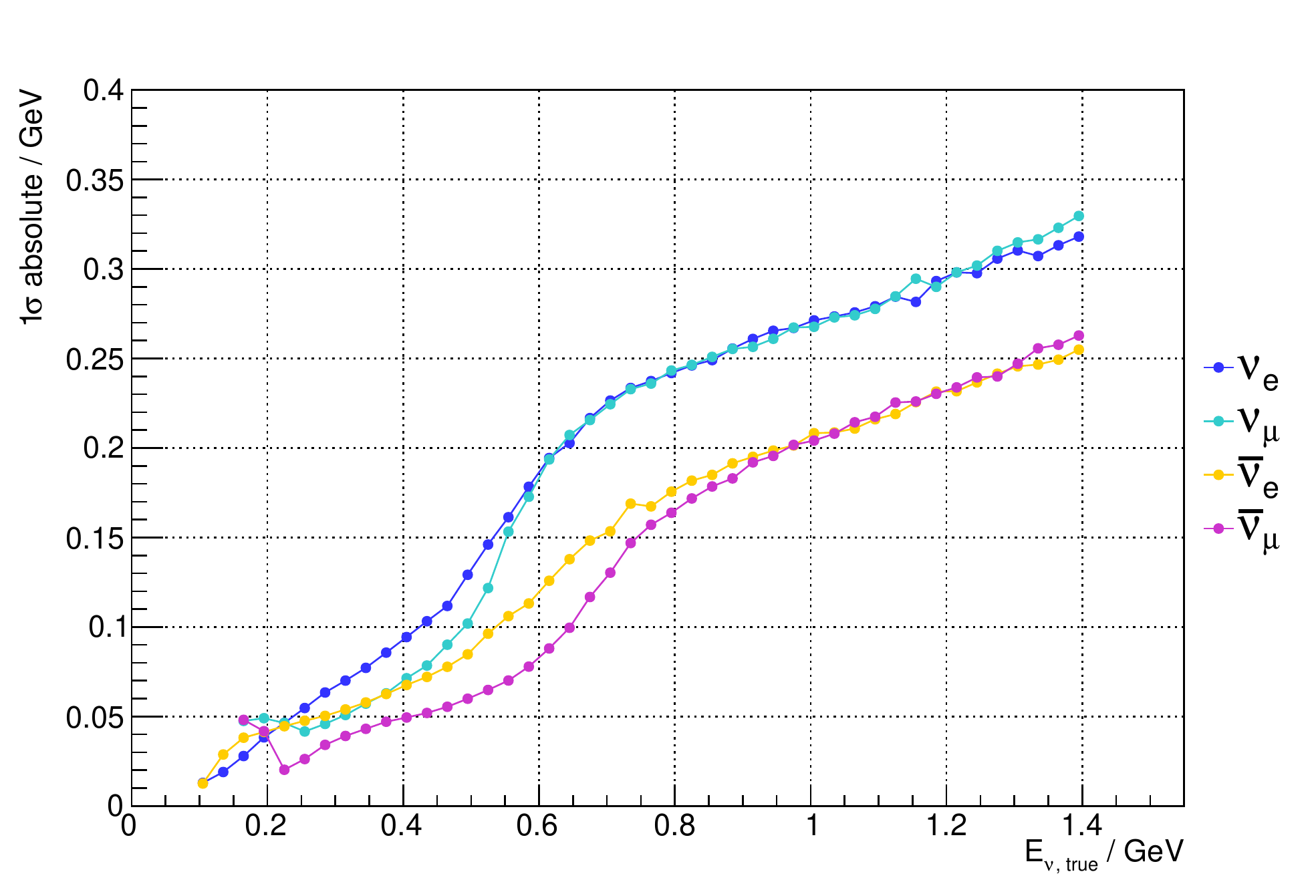}
            \caption{Absolute energy resolution -- all events.}
            \label{fig:detectors:fd_nrg_abs_all}
        \end{subfigure}
        \begin{subfigure}[b]{0.475\textwidth}   
            \centering 
            \includegraphics[width=\textwidth]{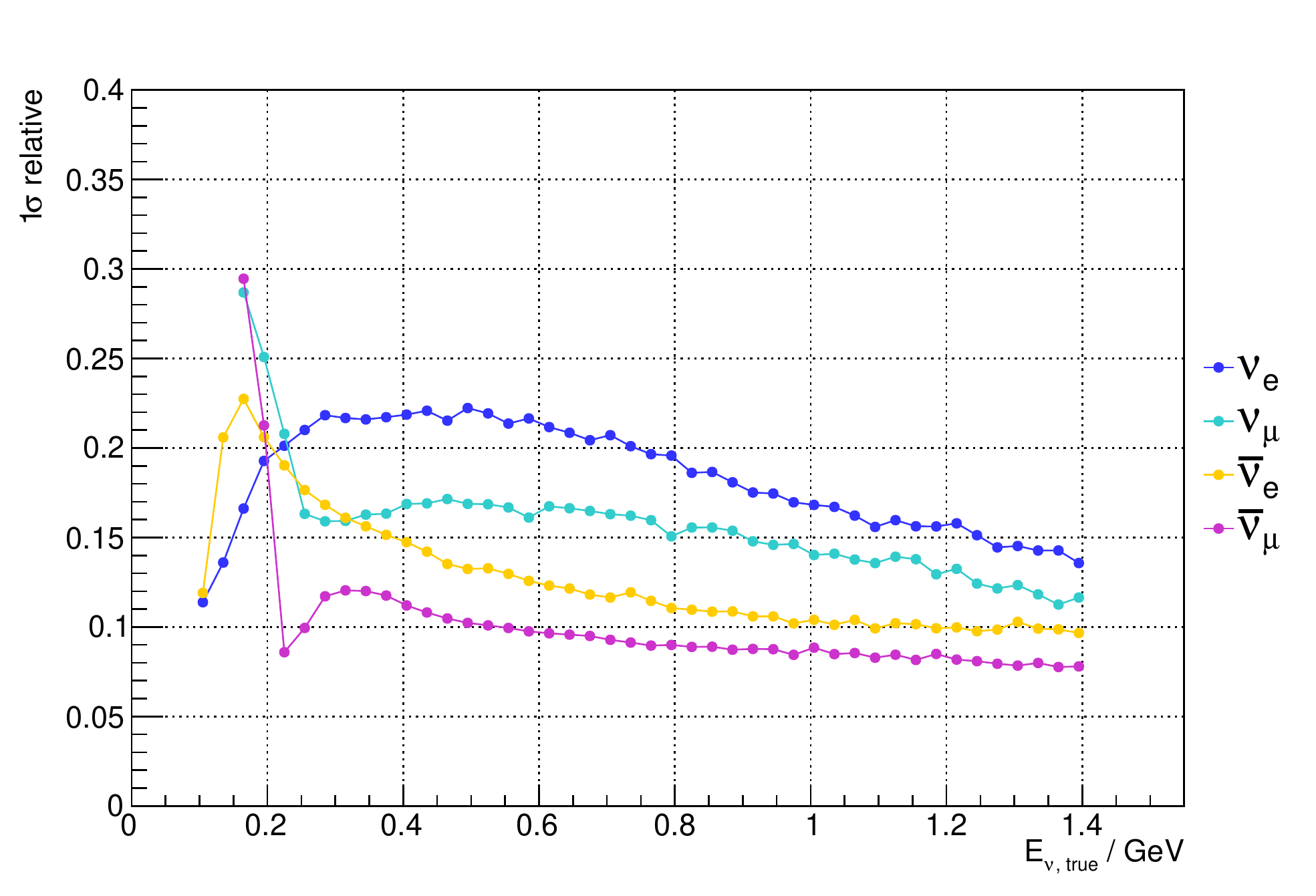}
            \caption{Relative resolution -- QES events.}
            \label{fig:detectors:fd_nrg_rel_qes}
        \end{subfigure}
        \hfill
        \begin{subfigure}[b]{0.475\textwidth}   
            \centering 
            \includegraphics[width=\textwidth]{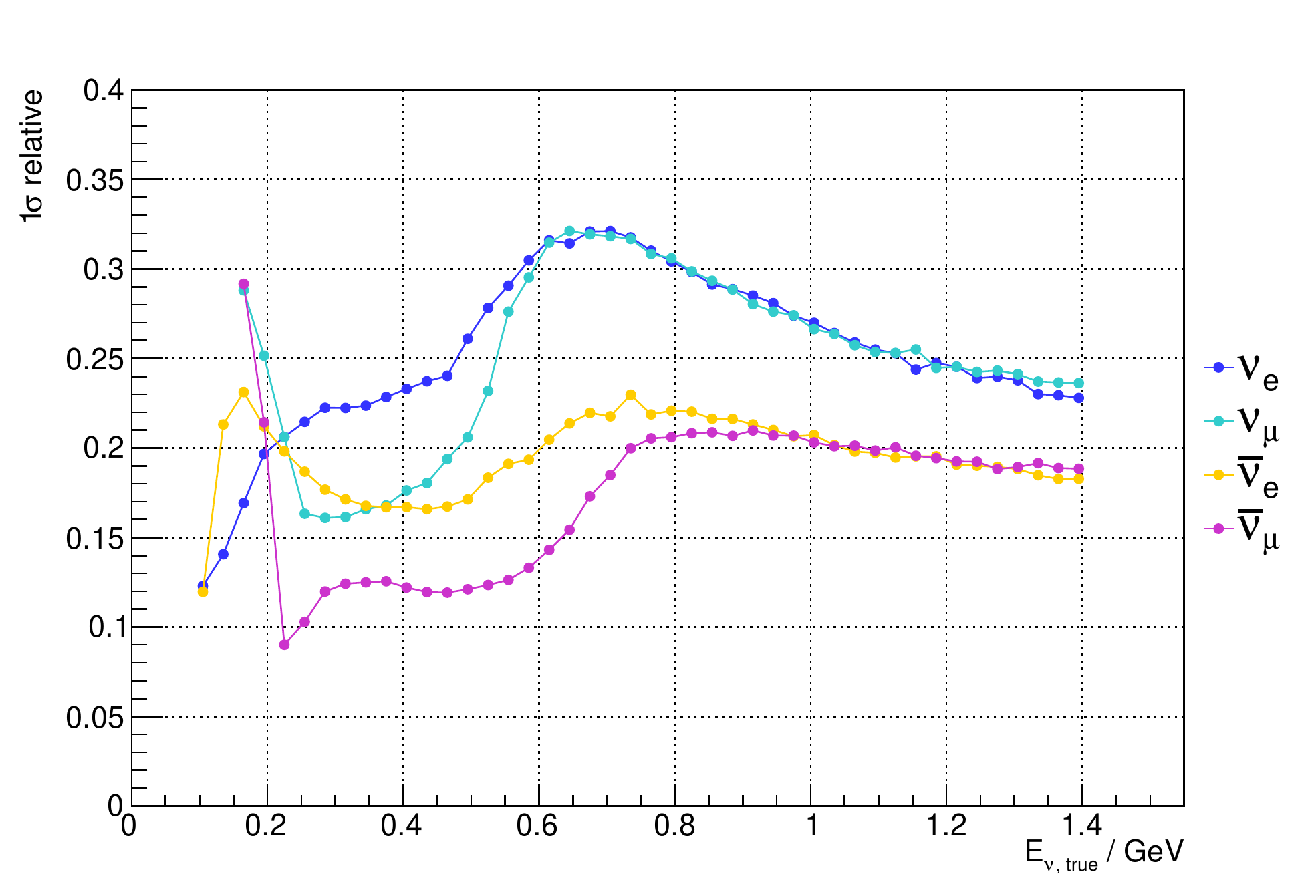}
            \caption{Relative resolution -- all events.}
            \label{fig:detectors:fd_nrg_rel_all}
        \end{subfigure}
        \caption{Neutrino energy reconstruction resolutions. Figures on the left are produced using only QES events, figures on the right are produced using the full event sample. Absolute 1 $\sigma$ resolutions of reconstructed neutrino energy as a function of true energy are shown in Figs.~\ref{fig:detectors:fd_nrg_abs_qes} and \ref{fig:detectors:fd_nrg_abs_all} for the QES and full sample, respectively. Relative 1 $\sigma$ resolutions of reconstructed neutrino energy as a function of true energy are shown in Figs.~\ref{fig:detectors:fd_nrg_rel_qes} and \ref{fig:detectors:fd_nrg_rel_all} for QES and full sample, respectively.} 
        \label{fig:detectors:fd_neutrino_all_qes}
\end{figure}

The neutrino energy reconstruction bias (difference between reconstructed and true neutrino energy) as a function of true neutrino energy for each neutrino flavor, using full event sample, is presented in Fig.~\ref{fig:detectors:fd_nrg_diff_all}. The bias is close to zero for energies corresponding to the second oscillation maximum at the FD site (150--400\,MeV), after which it drops below zero, meaning that the reconstructed energy is underestimated for energies above 600\,MeV where non-QES events prevail.

\begin{figure}[htpb!]
        \centering
        \begin{subfigure}[b]{0.475\textwidth}
            \centering
            \includegraphics[width=\textwidth]{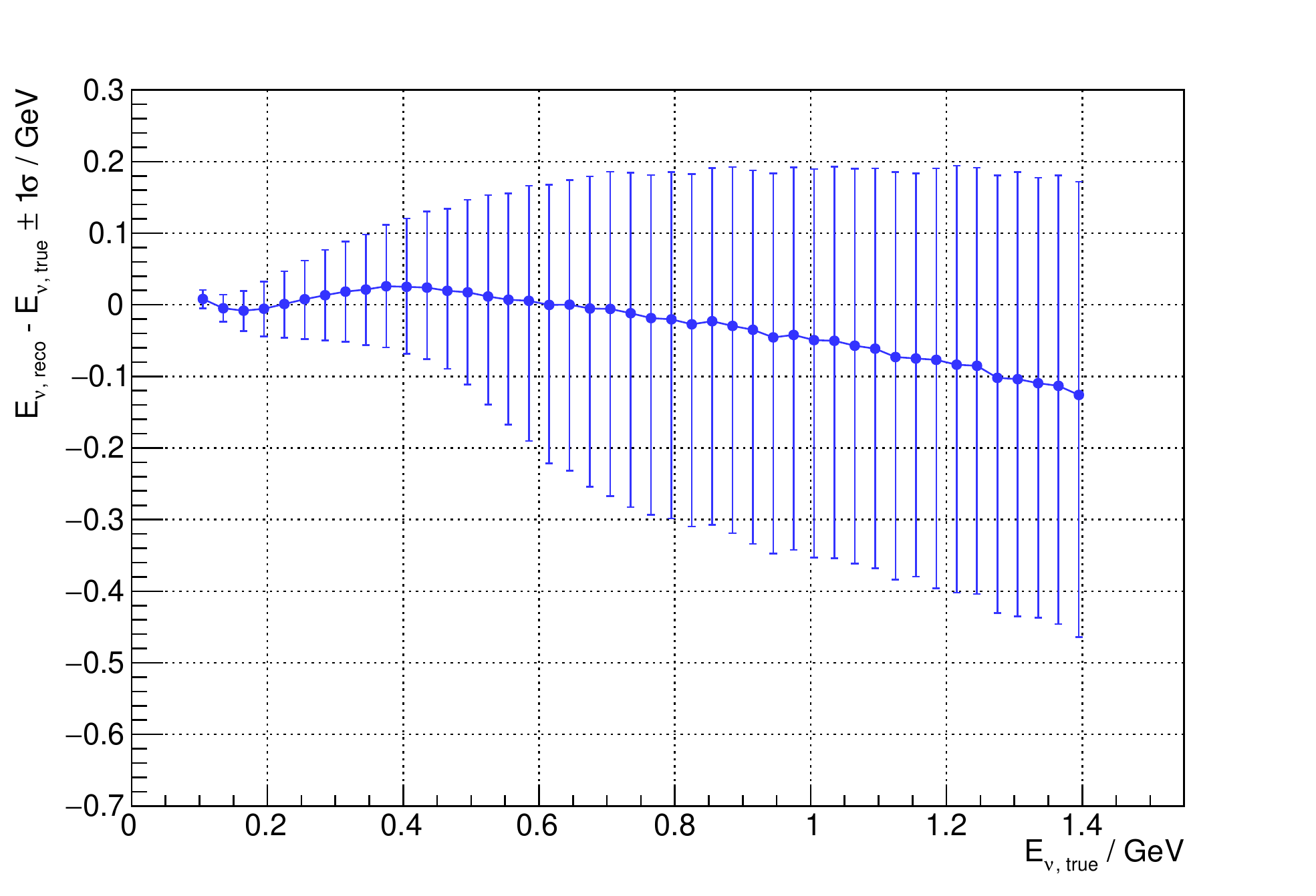}
            \caption{Electron neutrinos}
            \label{fig:detectors:fd_nrg_nue_all}
        \end{subfigure}
        \hfill
        \begin{subfigure}[b]{0.475\textwidth}  
            \centering 
            \includegraphics[width=\textwidth]{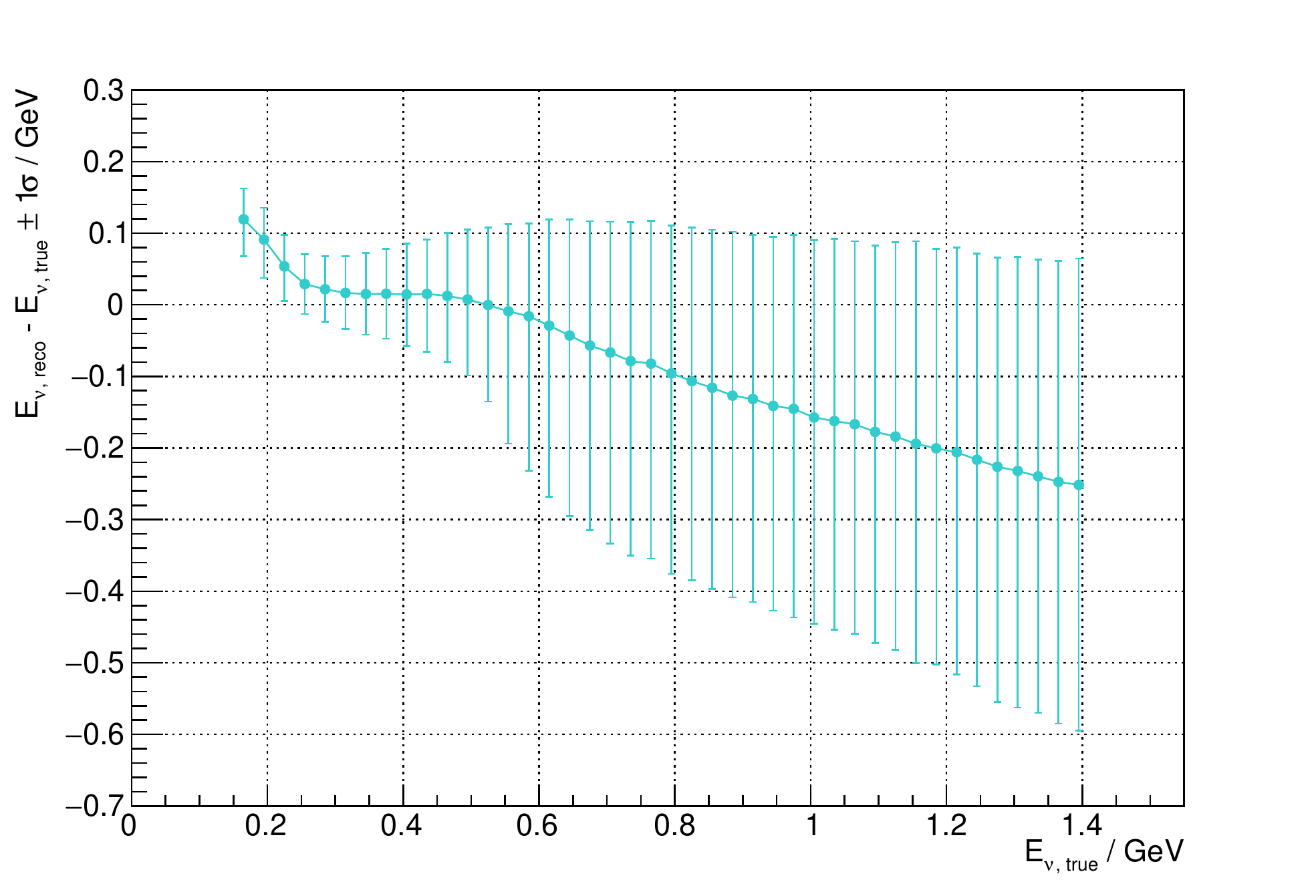}
            \caption{Muon neutrinos}
            \label{fig:detectors:fd_nrg_numu_all}
        \end{subfigure}
        \begin{subfigure}[b]{0.475\textwidth}   
            \centering 
            \includegraphics[width=\textwidth]{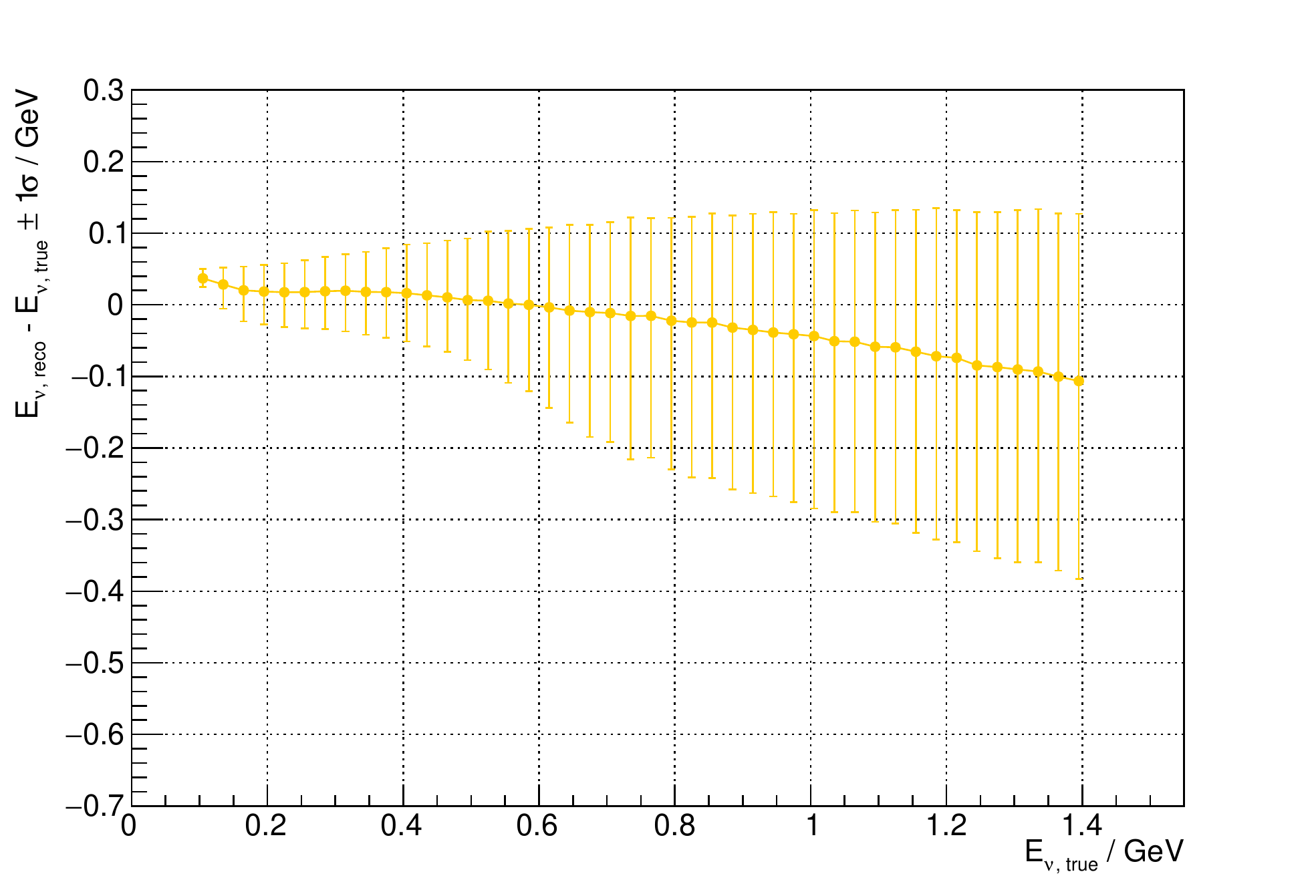}
            \caption{Electron antineutrinos}
            \label{fig:detectors:fd_nrg_anue_all}
        \end{subfigure}
        \hfill
        \begin{subfigure}[b]{0.475\textwidth}   
            \centering 
            \includegraphics[width=\textwidth]{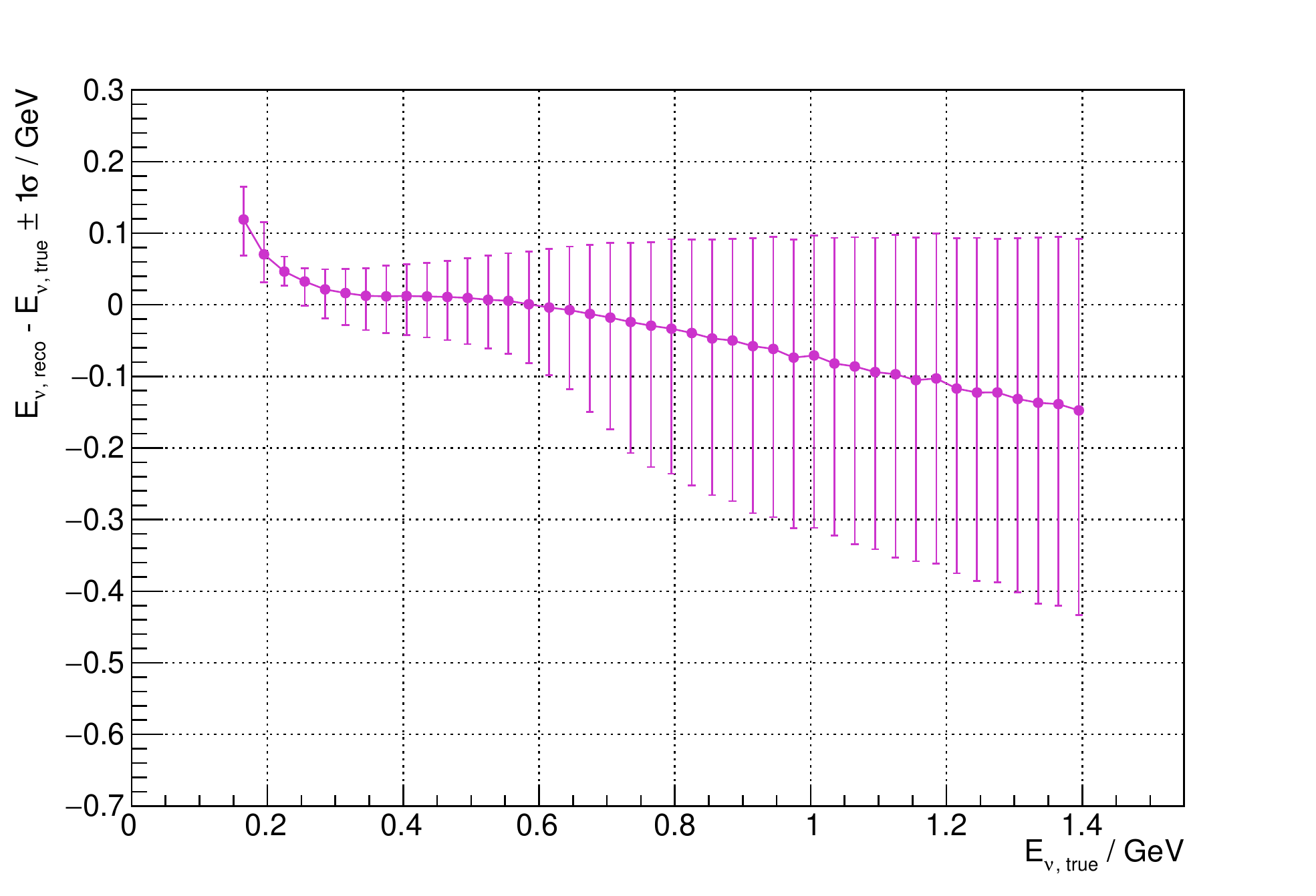}
            \caption{Muon antineutrinos}
            \label{fig:detectors:fd_nrg_anumu_all}
        \end{subfigure}
        \caption{Bias of the reconstructed neutrino energy with 1\,$\sigma$ uncertainty interval for the full MC event sample for the four relevant neutrino species.} 
        \label{fig:detectors:fd_nrg_diff_all}
\end{figure}

%\subsubsubsection{Selection efficiency}
%In this section efficiency of the selection process is presented
The selection efficiency is shown in Fig.~\ref{fig:detectors:fd_efficiency}. It is calculated using the migration matrices described above. Diagonal efficiency (Fig.~\ref{fig:detectors:fd_diagonal_efficiency}) refers to correct neutrino flavour selection, while off-diagonal (Fig.~\ref{fig:detectors:fd_off_diagonal_efficency}) refers to the misidentified neutrino flavour. The red dashed lines in Fig.~\ref{fig:detectors:fd_diagonal_efficiency} refer to fiducial selection efficiency.
\begin{figure}[htp!]
        \centering
        \begin{subfigure}[b]{0.475\textwidth}
            \centering
            \includegraphics[width=\textwidth]{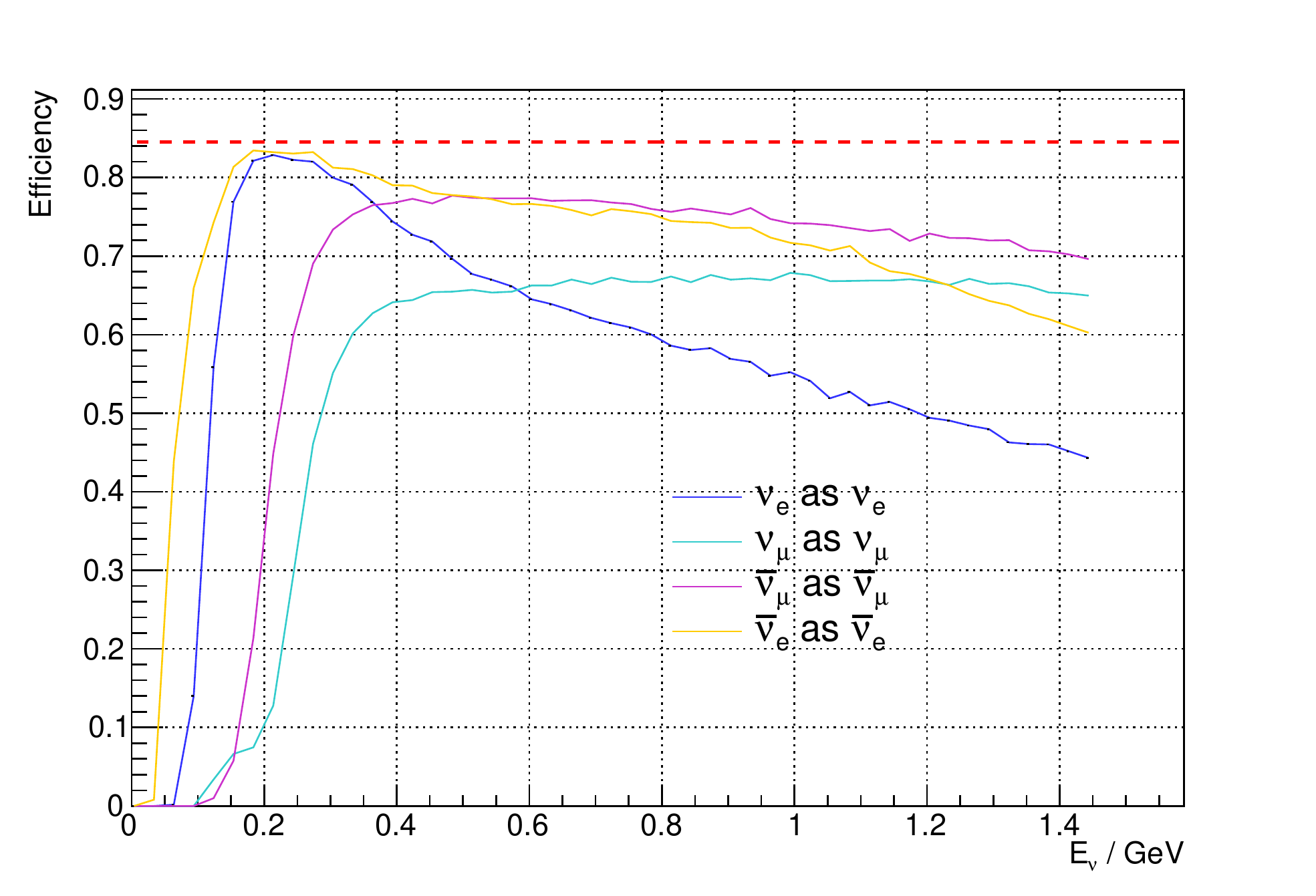}
            \caption{Diagonal}
            \label{fig:detectors:fd_diagonal_efficiency}
        \end{subfigure}
        \hfill
        \begin{subfigure}[b]{0.475\textwidth}  
            \centering 
            \includegraphics[width=\textwidth]{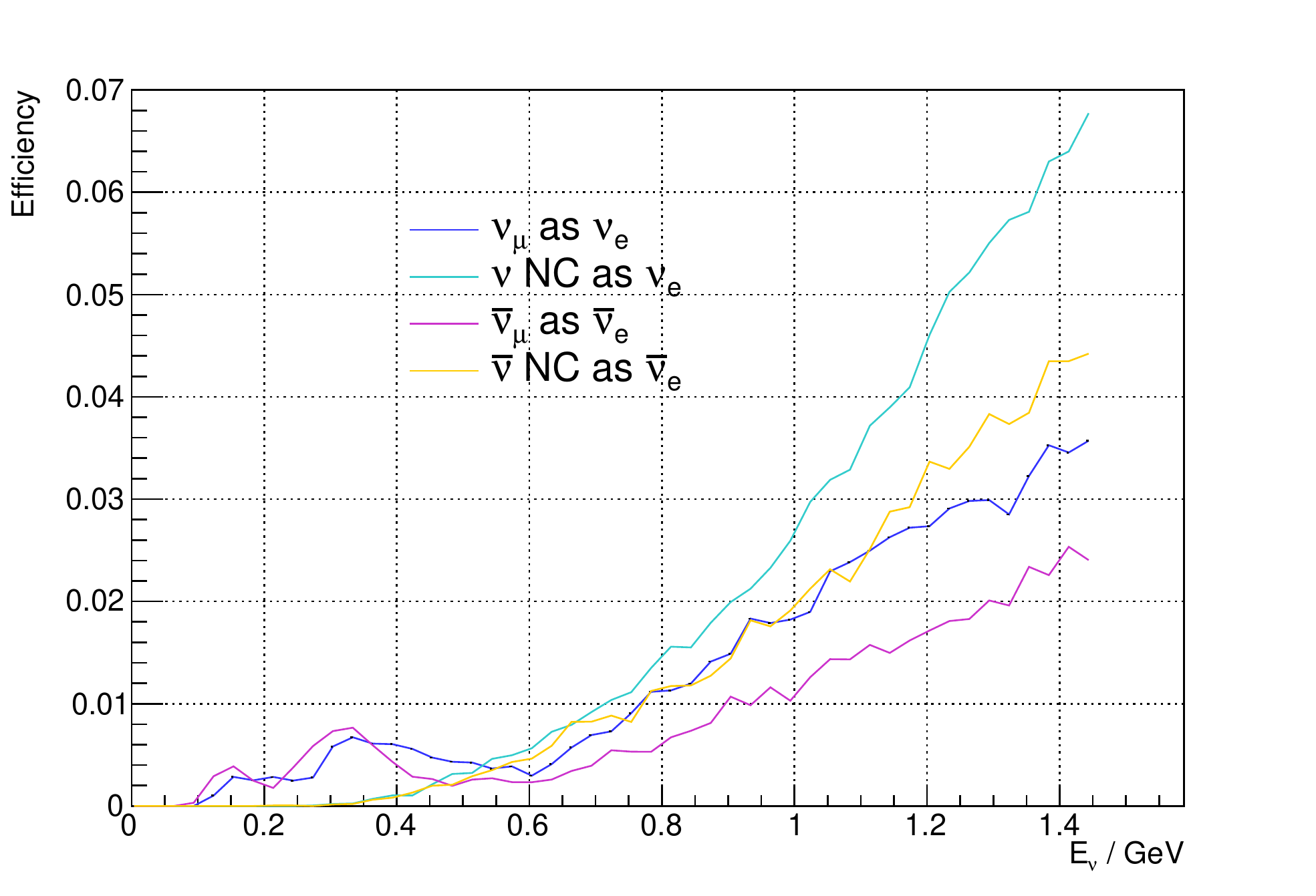}
            \caption{Off-diagonal}
            \label{fig:detectors:fd_off_diagonal_efficency}
        \end{subfigure}
        
        \caption{Efficiency of the selection algorithm. Figure~\ref{fig:detectors:fd_diagonal_efficiency} shows the diagonal selection -- the neutrinos that were selected correctly. Figure~ \ref{fig:detectors:fd_off_diagonal_efficency} shows off-diagonal efficiency -- the neutrinos that were selected incorrectly. The red dashed line represents the fiducial volume cut. Off-diagonal efficiency is shown only for the most important cases, the physics reach determination is made using all possible selected/true combinations.}
        \label{fig:detectors:fd_efficiency}
    \end{figure}

\subsubsection{External Background}

The external background events are events that are not a result of ESS$\nu$SB beam neutrino interactions and yet they trigger the detector. Most of the external background is rejected by discarding triggers which occur outside of the beam time window (BTW). In the current design, the duration of neutrino beam pulse will be \SI{1.3}{\micro \second}, which is by definition also the length of the BTW. The beam will operate in batches of four such pulses separated by \SI{0.75}{\milli\second}. The frequency of the batches themselves will be \SI{14}{Hz}, matching the native frequency of the ESS accelerator in neutrino mode. The duty cycle of such a design is \SI{7.28e-5}{}. This is the rejection factor for external background not related to the beam.

\emph{Atmospheric neutrinos} are neutrinos produced by cosmic rays colliding with the upper parts of the atmosphere. The interaction rate of these neutrinos in the energy range of 0--10\,\si{GeV} in the \SI{538}{kt} fiducial volume of FD is estimated to be \SI{1.5e-3}{} per second \cite{Honda:2015fha}. Therefore, the probability of their interaction in the BTW is \SI{2.0e-9} and the expected number of such interactions in a year is $1.89$, which is negligible compared to the expected number of beam neutrino interactions. Given that event selection in FD is based on the observation of muon decay, one must also take into account the probability to have an atmospheric neutrino interaction while waiting for the decay of the muon created in the beam neutrino interaction. This probability is \SI{7.5e-9} for a \SI{50}{ms} acceptance window between muon creation and its decay. Given the number of expected interactions shown in Table~\ref{tab:detectors:FD_expected_interactions}, one can put a conservative upper bound on the number of beam muons produced per year as $10000$. In this case, the expected number of events per year in which a muon is created and the atmospheric neutrino interacts before this muon decays is \SI{7.5e-3}. Therefore, the atmospheric neutrino background is completely negligible.

\emph{Atmospheric muons} are muons created by cosmic rays colliding with the upper part of the atmosphere. Since the detector will be placed \SI{1000}{m} underground, a very small number of the muons will reach it, and this small number will be further decreased by the BTW rejection. Therefore, this background is negligible.

\emph{Beam rock muons} are muons created by ESS$\nu$SB beam neutrino interactions in the rock and structures upstream of the FD and which enter the detector. This source of external background cannot be reduced by the BTW rejection. However, most muons produced by ESS$\nu$SB beam will have momenta of less than \SI{400}{MeV} due to the low energy of the beam. A muon passing through water will lose about \SI{200}{MeV} per meter travelled, which means that most of them will fully stop in the veto water volume of the detector and not reach the inner fiducial volume. Therefore, this background is negligible as well.

\subsection{Visualisation and Event Display}
An event display \textsc{EsbRoot} \cite{Barrand:2021azm} was developed based on the graphics technology developed at Orsay from 2010 (developments done at LAL up to 2020 and now continued in the new Orsay IN2P3 IJCLab laboratory that absorbed LAL). This technology allows programs to run natively on most modern interactive devices. It can run with the same core technology in most web browsers by using the \textsc{WebAssembly} and \textsc{WebGL} toolkits \cite{EsbRootView:inproceedings}. Current versions (3.x) can visualise simulated data for the near water Cherenkov detector, named in the event display ``neard'', the far Water Cherenkov (``fard'') and the SFGD (``fgd'').

\textsc{EsbRootView} is probably the first event display in high energy physics written in C++ capable of natively exploiting most of the modern devices: including desktops, laptops, smartphones, and tablets. Its flexible architecture also permits the use of a web version on the same C++ core as with other platforms. It permits to have animations, which can strengthen the intuition about what is occurring in the detectors. Its scripting mechanism, inspired by the \textsc{Bourne shell} makes in-depth customisation possible, with a short learning curve.

\textsc{EsbRootView} is fully written in C++ (98 flavour) with a large part of the code being pure header. The capability of running on most interactive devices is made possible by the fact that \textsc{C++98} and \textsc{GL-ES} (\textsc{OpenGL for Embedded Systems}, a graphics rendering standard) are available for all of them. A key point is that a light, portable code is available at Orsay (\textsc{inlib}/\textsc{rroot} classes) to read \textsc{ROOT} \cite{Brun:1997pa} files, which allows to read the files describing the detectors and the events issued from the ESS$\nu$SB detector simulations on all devices. \textsc{inlib} is strongly inspired by the \textsc{OpenInventor} developed by Silicon Graphics Inc. in the 1980's \cite{EsbRootView:OpenInventor}.

The web site for \textsc{EsbRootView} is at \url{https://gbarrand.github.io} -- the "softinex portalle" \cite{Barrand:2014rya} -- under the section "EsbRootView" at left. The source code of stable releases is on github at \url{https://github.com/gbarrand/EsbRootView.git}. The license is a customisation of the FreeType license. For each release binaries are provided at least for Linux, macOS and Windows. A \textsc{WebAssembly} version is available through the web pages.

%section{Geant4/g4tools}
It should be noted that \textsc{inlib}/\textsc{exlib} classes are delivered under the namespace ``tools'' in \textsc{Geant4} \cite{AGOSTINELLI2003250} as a base-layer to the analysis ``category''; these facilitate the creation of histograms and ntuples, I/O at various formats (ROOT, csv, HDF5), and offscreen plotting with our scene graph logic (see \cite{Dotti:2016ors} and \cite{Hrivnacova:2017hur}).

An example of the near and far detector visualisations are shown in Figs.~\ref{fig:detectors:layout-event-1} and \ref{fig:detectors:fard-setup}.

\begin{figure}[!htbp]
\centering
\includegraphics[width=10cm,clip]{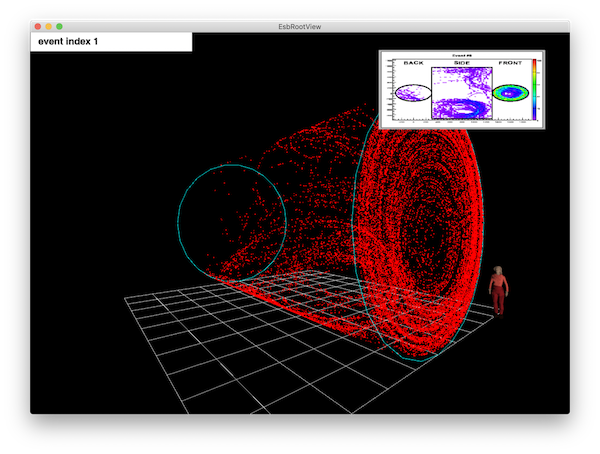}
\caption{Visualisation of a neutrino interaction in the near WC detector. Red dots depict positions where Cherenkov photons reach the surface of the detector.}
\label{fig:detectors:layout-event-1}
\end{figure}

\begin{figure}[!htbp]
\centering
\includegraphics[width=10cm,clip]{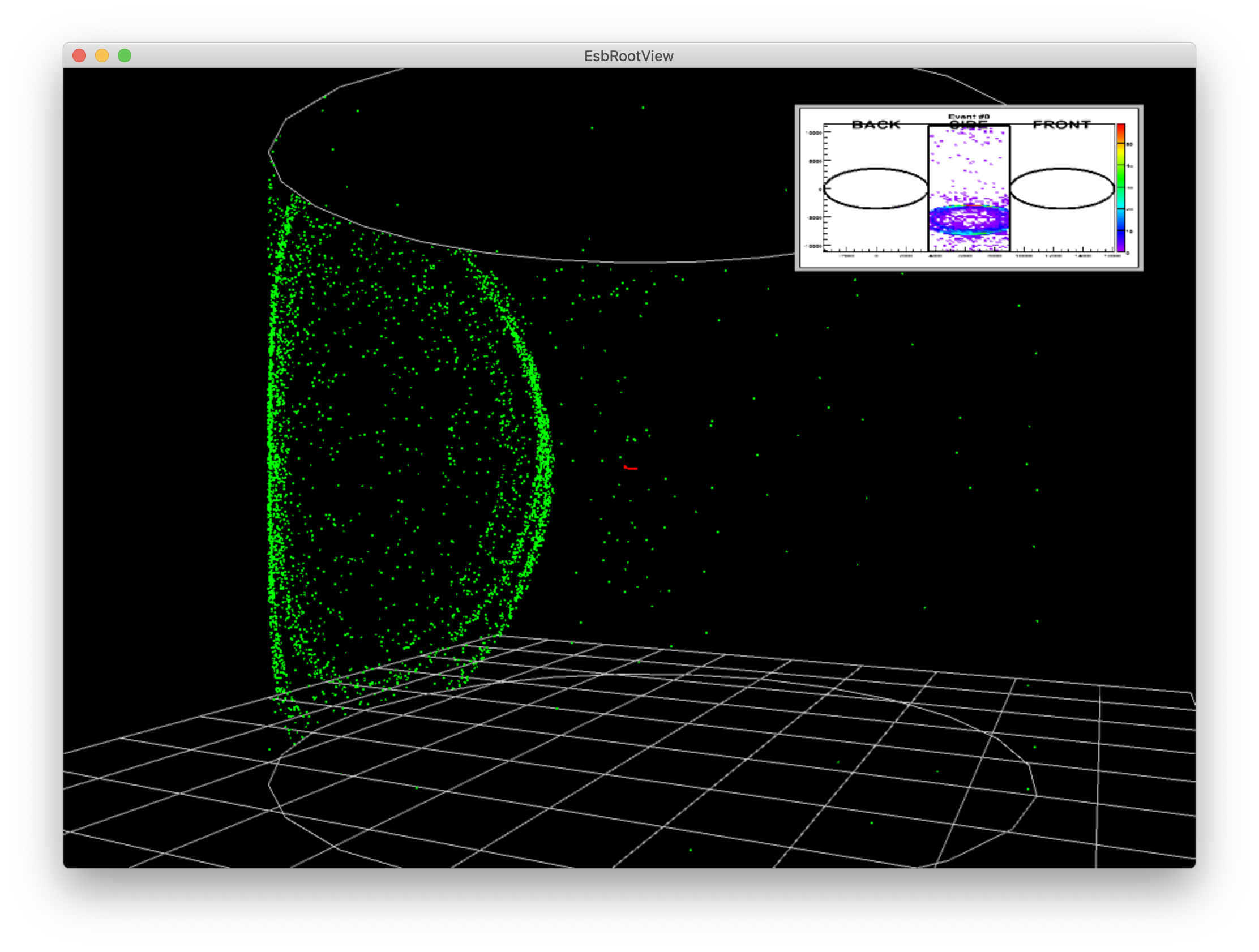}
\caption{Visualisation of a neutrino interaction in the far water Cherenkov detector. Green points depict positions where Cherenkov photons reach the surface of the detector.}
\label{fig:detectors:fard-setup}
\end{figure}

\subsection{Safety}

\subsubsection{Near Detector Site}
Safety aspects of the near detector complex at ESS are considered in Chapter~\ref{infrastructure} which is dedicated to the ESS$\nu$SB infrastructure. These concerns are adapted to the overarching safety and security strategy of ESS (General Safety Objectives for ESS 
%ESS-0000004
) and according to the ESS licensing strategy.

\subsubsection{Far Detector Site}

An active mine in Zinkgruvan, Sweden has been selected as a far detector site for the ESS$\nu$SB project. The far detector infrastructure will be located at an average depth of \SI{1000}{m} below ground level. Hence, the safety of the infrastructure will depend primarily on the geology and quality of the rock mass that will house the infrastructure at such a depth. Additionally, the ground stresses will be relatively high and will thus be critical for the design, construction, and operation of the infrastructure. The rock mass and ground stresses, as well as ground water concerns, will need to be thoroughly investigated, before engaging in detailed design studies of the far detector infrastructure. Furthermore, ground reinforcement will principally depend on the rock mass and ground stress conditions. Therefore, a proposal has been made to perform diamond core drilling to extract rock samples from the two proposed sites to investigate the rock mass quality and other geological parameters. Ground stress measurements will also be performed at the same time to determine stress conditions at the proposed sites. 

Other safety aspects relate to the interaction of the infrastructure with the operating mine. These concerns are not seen as highly significant since the mine presently operates at depths well beyond \SI{1000}{m}, and therefore the interaction will be minimal. However, mining-induced seismicity can not be totally ignored, but is considered manageable from the perspective of the ESS$\nu$SB project. 

As the ESS$\nu$SB project progresses, the mine and the ESS$\nu$SB project team will hold continuous discussions to understand the risks at play (both for ESS$\nu$SB and the Zinkgruvan mine) in order to adapt these as part of the design study and the development of risk-management strategy.

\subsection{Costing}

The costing of the detector complex is based on information gathered from manufacturers as well as from other long-baseline and/or large neutrino detector projects and proposals (i.e. EUROnu, JUNO, HyperK, KM3NeT, miniMIND and SuperFGD of T2K). Appropriate scaling was used to estimate the various costs. Assembly and installation costs (person-hours) assume a 5 year effort.

The total cost to construct all detectors is estimated to be 953\,M\texteuro. Details are given in Table~\ref{tab:detectors:costing}.

\begin{table}[!htp]
\footnotesize
\centering
\caption{Costing of the detector complex.}
\label{tab:detectors:costing}
\begin{tabular}{l l l}
\textbf{Item} & 
\multicolumn{2}{c}{\textbf{Cost [M\texteuro]}} \\
\hline
130 water-emulsion cloud chamber modules & 1.0 & \\
Detector Assembly, installation, air-conditioning, etc. & 1.0 & \\
\textbf{Emulsion Detector total} & & \textbf{2.0} \\
\hline
Scintillator cubes (980 000, $1\times 1 \times 1 \SI{}{cm^3}$ each) & 1.25 & \\
Wavelength shifting fibers ($\SI{29 400}{m}$) & 0.22 &\\
Multipixel photon counters + optical couplers (67 200 of them, double end readout) & 0.735 & \\
Mechanical structure & 0.5 & \\
Front-end readout and other electronics & 2.1 & \\
Data acquisition and calibration systems & 0.08 &\\
Superconducting magnet and power supply & 0.5 & \\
Assembly and installation effort (Personnel) & 0.1 & \\
\textbf{Super Fine Grained Detector total} & & \textbf{5.485} \\
\hline
Detector truss and carbon window & 0.5 & \\
Detector water storage & 0.5 & \\
Water pumping and purification plant & 10.0 & \\
PMT modules, support, power and signal transmission & 11.315 & \\
Calibration, monitoring and control systems & 0.4 & \\
Assembly and installation effort (Personnel) & 2.5 & \\
\textbf{Near water Cherenkov detector total} & & \textbf{25.215} \\
\hline
Two detector tanks and water storage & 0.8 & \\
PMT modules (PMTs, housing, power, readout) & 305.35 & \\
PMT module support & 15.0 & \\
Data logging, processing and transmission & 0.5 & \\
Calibration, monitoring and control systems & 1.5 & \\
Water pumping and purification plant & 40.0 & \\
Counting, testing, assembly storage rooms ($\times 3$) & 3.0 & \\
Veto detector (smaller PMTs looking outward) & 23.202 & \\
Assembly and installation effort (Personnel) & 10.0 & \\
\textbf{Far water Cherenkov detector total} & & \textbf{399.35} \\
\hline
Civil engineering - excavation of two caverns & 507.154 & \\
Cranes and other mechanical infrastructure & 0.5 & \\
Access systems & 3.0 & \\
Air conditioning and ventilation & 5.5 & \\
Monitoring systems & 2.5 & \\
Installation effort (Personnel) & 2.5 & \\
\textbf{Underground cavern excavations total} & & \textbf{521.154} \\
\hline
\textbf{Total} & & \textbf{953.21} \\

\hline
\end{tabular}

\end{table}

%\clearpage %Budimir: can be removed in the final version
\subsection{Summary}

The near and far detectors have been designed for optimal CP violation discovery potential and precision of $\delta_\text{CP}$ measurement. 

The near detector suite is composed of (in the downstream-to-upstream order) a water Cherenkov detector, an SFGD-like scintillator cube detector, and a water-target emulsion detector. The main purpose of near detectors is to measure the initial neutrino flux intensity and flavour composition, and to measure the neutrino--water interaction cross-section. The cross-section measurement is of particular importance since there is currently very little data on it in the ESS$\nu$SB neutrino energy region (approximately \SIrange{150}{450}{MeV}). The water Cherenkov detector will record the bulk of the neutrino interactions at the ND site due to its large \SI{0.75}{kt} fiducial mass. It will use the same technology and target material as the far detector, which will help reduce the systematic errors of the experiment. The scintillator detector will have a smaller target mass of \SI{1}{\tonne}, but will feature better tracking and event reconstruction capability. Since it will be magnetised, this detector will have the ability to determine particle charge and perform momentum spectroscopy. Charged particles originating in this detector and passing downstream into the water Cherenkov detector will provide an opportunity for additional calibration of the two detectors. The emulsion detector will have a target mass of \SI{1}{\tonne}, and will feature the best tracking and event-reconstruction performance at the expense of limited timing information. It will be used to study the detailed topology of neutrino interaction events and measure various double differential cross-sections.

The far detector site of the ESS$\nu$SB experiment will house two large water Cherenkov tanks, each containing a \SI{269}{kt} fiducial mass, for a total \SI{538}{kt} neutrino target mass. The decision to use two tanks instead of a single larger one is driven by the extreme technical difficulties in excavating a single cavern large enough to hold the desired volume of water. The far detector will be placed in the active Zinkgruvan mine in Sweden at a distance of \SI{360}{km} from the ESS site. The detectors will be placed about \SI{1000}{m} underground. At the \SI{360}{km} distance, a significant number of neutrino interactions is expected from both the first and second oscillation maxima. The signal detection efficiency of the far detectors is expected to be greater than 90\,\% at neutrino energies corresponding to the second oscillation maximum, and the relative neutrino energy resolution is expected to be less than 20\,\% (30\,\%) for electron (muon) neutrinos.

The total cost of all detectors and associated facilities is estimated to be 953~M\texteuro, the bulk of which will be the price of excavating and instrumenting the caverns for the large far detectors.
\clearpage

\setcounter{figure}{0}
\numberwithin{figure}{section}
\setcounter{equation}{0}
\numberwithin{equation}{section}
\setcounter{table}{0}
\numberwithin{table}{section}

\section{Infrastructure}
\label{infrastructure}
\setcounter{figure}{0}
\numberwithin{figure}{section}

%April 2022, 
%N. Gazis 
%M. Eshraqi 
%R. Johansson 
%B. Kideltoft 
%D. Patrzalek
%D. Saiang 
%E. Trachanas 

\subsection{Infrastructure and Conventional Facilities}
With the proposed upgrade of ESS to accommodate ESS$\nu$SB, the facility in Lund, Sweden will have to host the users of both a world-class neutron spallation source and the driver for the world's brightest neutrino beam. The working environment and atmosphere should be inviting, communication should be uninhibited, and obstacles should be minimised where possible -- whether among personnel and guests or between buildings and facilities. 

The upgrade would require several new buildings and auxiliary structures to be integrated into the ESS site. These new facilities should maintain an aesthetic balance with the architectural design of the ESS, respect the underlying safety needs (whether conventional, cryogenic, radiological or other) and comply with Swedish regulations on environmental stewardship and sustainability at all times.
The experience from ESS and other similar facilities indicates that the cost of the technical infrastructure, site facilities upgrades, and late-stage additions can reach the order of 30$\,\%$ of the total cost of the project. The impact of a new machine design, site layout upgrade, and new conventional facilities being delivered on-schedule will clearly be substantial. How the design and management of the machine, civil, site, and infrastructure works influence the safety of the staff and the general public gives paramount importance to the tasks defined within this sub-project. 
 
\subsubsection{Buildings $\&$ General Safety}
The ESS$\nu$SB project relies on the existing high-power linac of ESS, upgrading it to deliver a 5\,MW beam of $H^{-}$ ions to an accumulator ring, while continuing to provide the 5\,MW beam of protons for the nominal neutron-spallation program. The structural modifications to the linac tunnel are expected to be relatively minimal, with the primary task of creating access in the south tunnel wall to allow for extraction of the proton beam roughly downstream of the high-energy end. This modification will allow for the connection to a curved transfer line which leads to the accumulator ring. 

For radiological purposes, the target building will be underground, with the axis of the target modules at approximately $-20$\,m with respect to the ESS landscape. A geological study, together with licensing for excavation, must be performed before a formal design and site landscaping plan can be produced. 
The building design, construction, and shielding shall follow ESS safety rules, which in general terms tend to be more demanding and restrictive than the underlying national regulations \cite{esssafetey}. Definition and shielding-specification documents for ESS$\nu$SB must be derived from the ESS rules supervised and controlled radiation areas (\cite{essr}) and the ESS radiation protection strategy for employees (\cite{essrps}) frameworks that are currently valid for the existing ESS facility.

Several of the ESS buildings may also serve the needs of ESS$\nu$SB in terms of campus, lab space, radiological waste handling, logistics, utilities, fire-safety sprinkler systems, and, of course, the klystron gallery. However, practically all of these buildings have been sized according to the needs of ESS, and their feasibility under the expanded scope must be evaluated and modified accordingly. 
Beyond these, a utility building (or a pair of buildings) similar to the klystron gallery will be needed to serve the accumulator. Moreover, the transfer lines' power and control systems need hosting as well as heat exchangers. 

The target facility in itself must be designed to include several auxiliary buildings, for power, cooling, waste management, hot cells, the target cavern, and the decay tunnel at the downstream from the target. The target building will house a complex of systems for the hadronic target array, dedicated for ESS$\nu$SB, and therefore the studies and engineering design shall be based on new ESS$\nu$SB requirements, rather than benchmarking against existing target designs.

\subsubsection{Far Detector Site Assessment}

The Garpenberg and Zinkgruvan mines, located 360\,km and 540\,km respectively from ESS in Lund, have been assessed for potentially hosting the far detector (see Fig.~\ref{fig:0}). The Zinkgruvan mine appears to be the more ideal choice, and will be subject to further investigation in the next phase of study. This mine has all the appropriate access infrastructure and available baseline data for a detailed site-feasibility study. The far detector infrastructure, with unprecedented cavern dimensions (78\,m by 78\,m cylinders) will be located at an average depth of 1000\,m below ground level. Therefore, the safety of the installation will primarily depend on the geological capabilities to host the infrastructure. Moreover, at this depth the ground stresses will be relatively high and thus raise significant stability concerns.

A detailed site investigation is required to confirm its general feasibility for hosting the infrastructure as well as to provide assurance on its fitness in terms of safety. The site investigation is to include the following: (i) diamond drilling to investigate engineering rock properties and assessment of rock quality, (ii) ground stress measurements, (iii) ground water assessment, (iv) laboratory testing of rock specimens for rock strength parameters, and (v) extensive numerical simulations of the conceptual infrastructure located at the 1000\,m depth. The inputs for the simulations will be those derived from tasks (i) to (iv). The design and layout of the infrastructure will follow as a separate work package after the site’s capabilities have been established. The design studies will include methodology for rock re-enforcement, excavation procedures and associated rock excavation technology, and continuous monitoring and ground management planning. These studies are anticipated to precede any actual construction.

Initial discussions have been held with Zinkgruvan mine personnel, at the executive and operational levels, who have shown keen interest and support. However, the mine presently has a new management team. It is therefore essential to re-affirm commitment from the new team: without their support, progress on ESS$\nu$SB could be in jeopardy. Specifically, the ESS$\nu$SB project will require extensive use of the mine’s access infrastructure to reach the far detector site located at the back of the mine. Additionally, the interactions between the mining activities and the far-detector infrastructure must be as minimal as possible, which means appropriate distancing of the far detector infrastructure from active mining areas. There are also legal issues which must be investigated, particularly that the ESS$\nu$SB experiment is not a mining activity and may not be governed by mining legislation.

\begin{figure}[H]
  \centering
\includegraphics[width=0.55\textwidth]{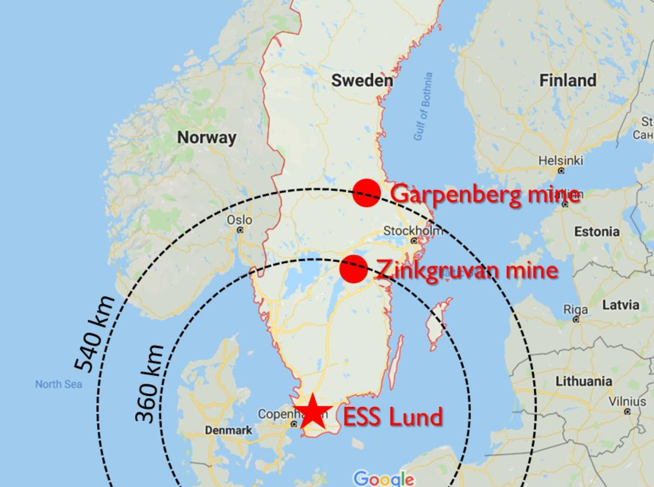}
  \caption{Mines investigated for hosting the far detector.}
    \label{fig:0}
\end{figure}

\subsubsection{Machine 3D Model Design }

The machine design imposes several constraints on the ESS site layout, building design, and infrastructure needed for upgrading the facility to accommodate ESS$\nu$SB. The entire landscape of the ESS site, and the dedicated layout as it is presented in Fig.~\ref{fig:layout}, needs to be thoroughly revised in order to accommodate for the new ESS$\nu$SB design for a new additional accelerator, second target building, near detector, and the associated buildings. The relevant draft layout is presented in Fig.~\ref{fig:layoutint}. 
%\vspace*{-0.5cm}

\begin{figure}[H]
  \centering
\includegraphics[width=0.6\textwidth]{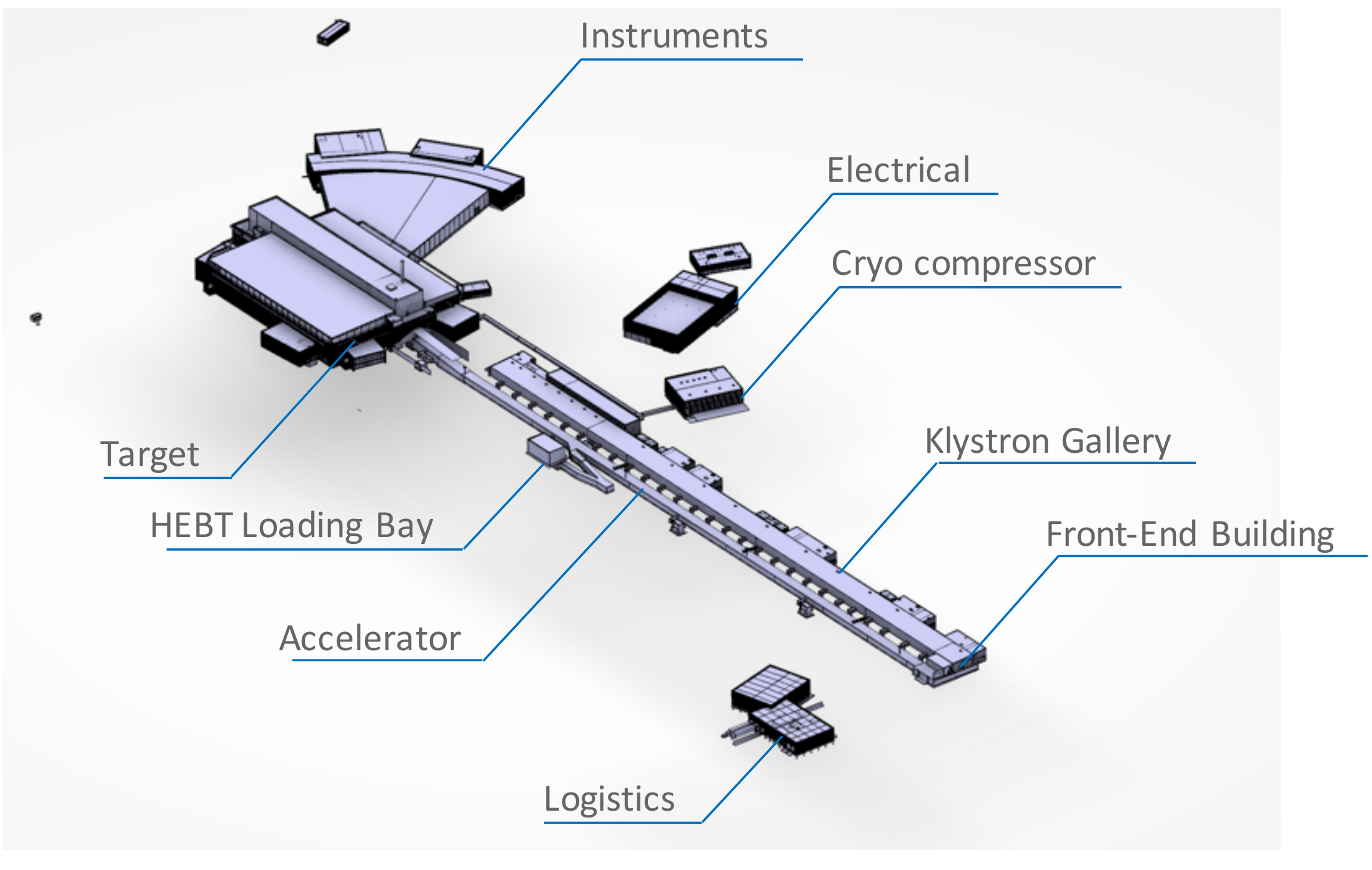}
  \caption{Current ESS facilities layout.}
    \label{fig:layout}
\end{figure}
The machine design of ESS$\nu$SB will follow the ESS procedure for mechanical engineering design \cite{esse}. Using CAD modeling, the conceptual design of machines and buildings will be integrated in the ESS master model in \textsc{CATIA V6}. The detailed design of magnets, accelerating structures, cryomodules, target, detectors, and other components will be used as basis for operations licensing; therefore, it is mandatory to follow the established ESS process for design under revision control and releases \cite{esse}. The detailed drawings issued for installation will follow the ISO GPS norms that are used and implemented by the ESS Mechanical Engineering \& Technology (MET) Section.

%\begin{figure}[H]
 % \centering
%\includegraphics[width=0.8\textwidth]{figures/infrastructure/2.pdf}
%  \caption{ESS facilities layout including the ESS$\nu$SB buildings}
%    \label{fig:layout_nu}
%\end{figure}

\subsubsection{Cooling $\&$ HVAC }
The baseline design for ESS$\nu$SB requires installation of new cooling systems, which rely mainly on three coolants: air, water and helium-saturated gas. The scaling and design of these systems will be derived from the needs of ESS$\nu$SB when consolidated into overall functional requirements. 
A heating, ventilation, and air conditioning (HVAC) system - operating independently of the HVAC system of ESS -- is required for ventilating the transfer lines, the ring, and the target, as well as ancillary buildings. The exhaust air and gas must pass through nuclear-grade filtering to avoid releasing any activated airborne aerosol or particulates to the atmosphere. 

\subsubsection{Electricity}
Delivering twice the power from the linac may require substantial increase of the nominal ESS input power, with some additional contingency factor also necessary (the efficiency of the two beams will not be the same due to heavy chopping of the $H^-$ beam). 
This calls for an upgrade of the entire electrical network (and grounding network) from the main high-voltage (HV) power station at the interface between ESS and the municipal grid; and the internal HV network from the HV station to the distribution sub-stations at the klystron gallery, the ancillary buildings of the ring, and those of the target. 
These electrical studies and design to be performed are an essential part of the licensing application required for constructing ESS$\nu$SB.

\subsubsection{Network}
A dedicated network for the ESS$\nu$SB control and interlocks of its systems will need to follow the pre-existing technical network of the ESS and designed to be fully integrated to it. It is anticipated that the number of control signals and complexity of ESS$\nu$SB control system might be analogous to those of the current ESS control system. A detailed study will need to be conducted in the next phase of the project to quantify the scale of the required update.

\subsubsection{Transportation and Handling}
The transportation and handling of magnets and equipment to be installed at the new tunnels and accumulator ring is planned to be expedited via overhead cranes and wheeled crane bridges. An investigation is on-going related to the lifting capacity and needs for the ESS$\nu$SB target station. The ESS rigging team, licensed to handle heavy loads, heavy lifts, and transportation of sensitive equipment on ESS site, is expected to handle all related cases as instructed by the ESS$\nu$SN infrastructure team. One of the typical large ESS overhead cranes is illustrated in Fig.~\ref{fig:3}. 

\begin{figure}[htbp]
  \centering
\includegraphics[width=0.6\textwidth]{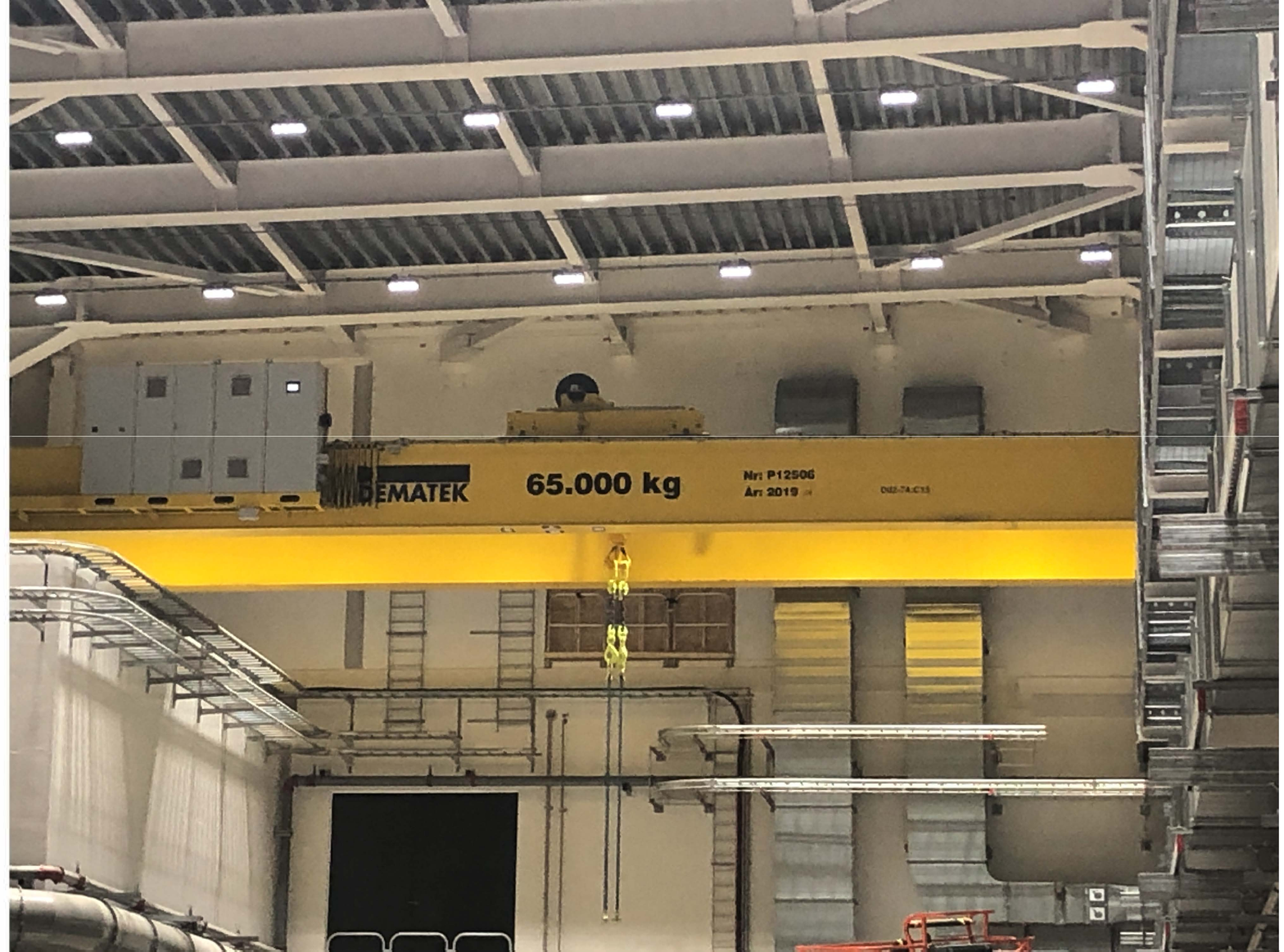}
  \caption{Largest existing ESS overhead crane currently in operation in the D02 Target building}
    \label{fig:3}
\end{figure}

%\vspace*{1cm}
\subsection{3D Model Design}
\subsubsection{Facility Layout 3D Model Design}
Structural requirements will need to be considered when sizing the cross-section of the new tunnels for ESS$\nu$SB. In Fig.~\ref{fig:4}, the new ESS$\nu$SB buildings are highlighted in color. Their dimensions follow a rough estimation and need to be further developed to match the ESS$\nu$SB machine design.

\begin{figure}[htbp]
  \centering
\includegraphics[width=0.9\textwidth]{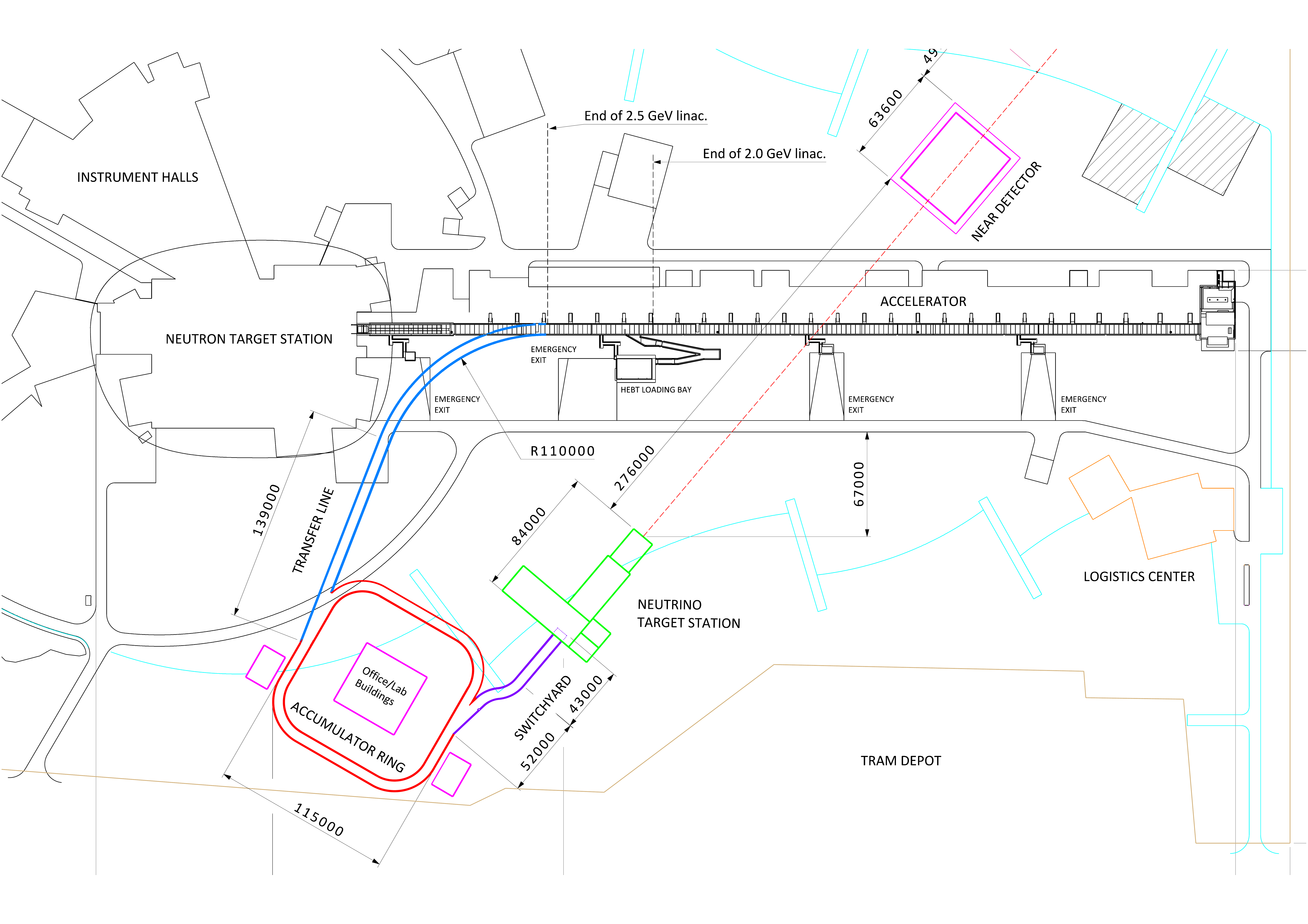}
  \caption{Overview of ESS$\nu$SB baseline layout (units in mm).}
    \label{fig:4}
\end{figure}

\subsubsection{Machine 3D Model Design }
The ESS$\nu$SB machine design includes the phases of 3D modelling, machine integration, conceptual design, detailed design, issuing of 2D drawings; and from that point, depending on the status of the components, the manufacturing cycle and installation. The ESS$\nu$SB buildings will be sized according to the machines' requirements and dimensions, once these are studied and baselined.

It is worth mentioning that, for the time being, start of the ESS$\nu$SB accelerator, the second ion source, will be housed in the existing ESS Front-End building, most likely without any modification of the civil infrastructure. Further studies are planned to confirm this design. 
\subsection{Costing and Budget Aspects}
Preliminary costing of ESS$\nu$SB can been benchmarked against similar facilities such as ORNL--SNS. Price comparison and adaption is necessary with respect to economic factors such as the year of construction start and the overall timeline. 
\subsection{Summary}
In conclusion, the preceding sections delineate the critical and detailed work which must take place in terms of infrastructure, engineering design, and landscaping in order to study and develop ESS$\nu$SB. As discussed, ESS$\nu$SB is a feasible project that can be constructed, licensed, hosted, and operated on the ESS premises. 
%\newpage

\clearpage

\setcounter{figure}{0}
\numberwithin{figure}{section}
\setcounter{equation}{0}
\numberwithin{equation}{section}
\setcounter{table}{0}
\numberwithin{table}{section}

\section{Physics Performance} \label{physics}
\subsection{Introduction}

The overwhelming evidence for neutrino flavour change from the observation of the neutrino oscillation phenomenon is one of the few precious pieces of evidence for physics beyond the Standard Model of particle physics, and therefore an excellent window to the underlying new physics to be discovered. The explanation of the neutrino flavour change observed in solar, atmospheric, reactor, and accelerator neutrinos requires, at least, three non-degenerate neutrino mass eigenstates $\nu_i$ as well as a unitary $3 \times 3$ mixing matrix $U$, the PMNS~\cite{Pontecorvo:1957cp,Pontecorvo:1957qd,Maki:1960ut,Maki:1962mu,Pontecorvo:1967fh}, relating them to the flavour eigenstates $\nu_\alpha = U_{\alpha i} \nu_i$, with $\alpha = e,\mu,\tau$. This lepton mixing matrix can be parameterised by 3 mixing angles
$\theta_{12}$, $\theta_{23}$ and $\theta_{13}$ as well as one CP violating phase
$\delta$:
\begin{eqnarray}
 U&=&
\left( \begin{array}{ccc} 1 & 0 & 0 \\ 0 & \phantom{-}c_{23} & s_{23} \\ 0 &
-s_{23} & c_{23}
            \end{array} \right)
            \left( \begin{array}{ccc}
    \phantom{-}c_{13} & 0 & s_{13}e^{-i\delta} \\ 0 & 1 & 0 \\  -s_{13}e^{i\delta} & 0 & c_{13}
            \end{array} \right)
            \left( \begin{array}{ccc}
    \phantom{-}c_{12} & s_{12} & 0 \\ -s_{12} &c_{12} & 0 \\ 0 & 0 & 1
            \end{array} \right) , ~~
\label{eq:mixingmatrix}
\end{eqnarray}
where, $c_{ij} \equiv \cos \theta_{ij}$ and $s_{ij} \equiv \sin
\theta_{ij}$.

The emerging picture from the last decades of neutrino oscillation measurements~\cite{Esteban:2020cvm,Capozzi:2021fjo} is that of two distinct mass splittings, the ``solar'' mass splitting $\Delta m^2_{21} = 7.4~\cdot~10^{-5}$~eV$^2$ and the larger ``atmospheric'' one, $|\Delta m^2_{31} |= 2.5~\cdot~10^{-3}$~eV$^2$, with $\Delta m^2_{ij} = m^2_j - m^2_i$. On the other hand, after the discovery of a non-zero $\theta_{13}$~\cite{An:2012eh,Ahn:2012nd,Abe:2012tg,Adamson:2011qu}, the structure of the PMNS matrix describing lepton flavour mixing is surprisingly different from its CKM counterpart in the quark sector, making the Standard Model flavour puzzle even more intriguing. Instead of the hierarchical structure characterised by very small mixing angles of the CKM, large mixing angles describe the PMNS. Indeed, the ``atmospheric'' mixing angle $\theta_{23}$ is large, with present data favouring a value near the maximal, in either the first or the second octant. Similarly, $\theta_{12}$, the ``solar'' mixing angle, is roughly $33^\circ$ and only $\theta_{13}~{\sim}~8.5^\circ$ is relatively small; although still similar to the Cabibbo angle, the largest in the CKM. This large mixing allows the present and next generation of neutrino oscillation experiments to explore new questions that may help to answer open problems in fundamental physics. 

Present experiments such as T2K~\cite{Abe:2017uxa, Abe:2019vii} and NO$\nu$A~\cite{Acero:2019ksn} have begun to provide the first hints on the phase $\delta$, whose value could imply the violation of the particle-antiparticle symmetry in the lepton sector if $\sin \delta \neq 0,\pi$. Given that CP violation is a necessary ingredient to explain the excess of matter over antimatter to which we owe our existence, and that the CKM contribution is insufficient~\cite{Gavela:1993ts,Gavela:1994dt},
such a discovery would be extremely suggestive. Furthermore, present data favour normal ordering (positive $\Delta m^2_{31}$) with respect to inverted ordering. This parameter, combined with the searches for the neutrinoless double beta decay, is necessary to probe the Majorana nature of neutrinos, which again could help to understand the origin of matter in the Universe. Despite these present hints for $\delta$ and the mass ordering, a new generation of experiments will be required to establish their discovery to a high degree of significance and precisely determine their values. 

The ``golden'' measurement sensitive to all the remaining unknown parameters is the $\nu_e$ appearance oscillation probability from a $\nu_\mu$ beam. In matter, it is given by the following expression~\cite{Cervera:2000kp} (see also~\cite{Freund:1999gy, Akhmedov:2004ny, Minakata:2015gra}) expanded up to second order in the two small parameters $\sin 2 \theta_{13}$ and $\Delta_{12} L/2$:

\begin{equation}
    \begin{split}
        P(\barparenb{\nu}_{\mu}\rightarrow&\barparenb{\nu}_e)  = s_{23}^2\sin^2{2\theta_{13}}\left(\frac{\Delta_{31}}{\tilde{B}_{\mp}}\right)^2\sin^2{\left(\frac{\tilde{B}_{\mp}L}{2}\right)}+c_{23}^2\sin^2{2\theta_{12}}\left(\frac{\Delta_{21}}{A}\right)^2\sin^2{\left(\frac{A L}{2}\right)}\\
        & + \tilde{J}\frac{\Delta_{21}}{A}\frac{\Delta_{31}}{\tilde{B}_{\mp}}\sin{\left(\frac{A L}{2}\right)}\sin{\left(\frac{\tilde{B}_{\mp}L}{2}\right)} \cos \left( \frac{\Delta_{31}L}{2} \pm \delta\right) ,
    \end{split}
    \label{Eq:Probability}
\end{equation}
where $\Delta_{i j} \equiv \Delta m^2_{i j}/ 2E$; $A = \sqrt{2} G_F n_e$ is the matter potential~\cite{Wolfenstein:1977ue} with electron density $n_e$ and the Fermi constant $G_F$; and $\tilde{B}_{\mp}\equiv |A\mp \Delta_{13}|$. The upper (lower) signs correspond to the (anti)neutrino oscillation probability. The first ``atmospheric'' term oscillates fast with a high frequency characterised by $\Delta m^2_{31}$, but with a small amplitude corresponding to $\sin^2 2 \theta_{13}$. Conversely, the second ``solar'' term has a slower frequency driven by $\Delta m^2_{21}$ but dominates once it develops, given its larger $\sin^2 2 \theta_{12}$ amplitude. The only dependence in the CP violating phase $\delta$ appears in the last term, which is the interference between the other two. Since $\sin 2 \theta_{13}~{\sim}~0.3$ while $\Delta_{12} L/2~{\sim}~0.05$ at the first oscillation peak, the ``atmospheric'' term tends to dominate the oscillation probability and the interesting CP interference is only subleading -- easily hidden behind any systematic uncertainty affecting the signal. Conversely, at the second oscillation maximum $\Delta_{12} L~{\sim}~0.14$, such that the dependence on $\delta$ of the oscillation probability is much more important and can allow for improvement on the sensitivity to this parameter~\cite{Coloma:2011pg}. This can be seen in Fig.~\ref{fig:probs} where the change in the probability upon varying the values of $\delta$ is much more significant at the second peak compared to the first.

%%%%% fig:probs; Probability %%%%%
\begin{figure}[ht]
\centering
\includegraphics[width=8cm]{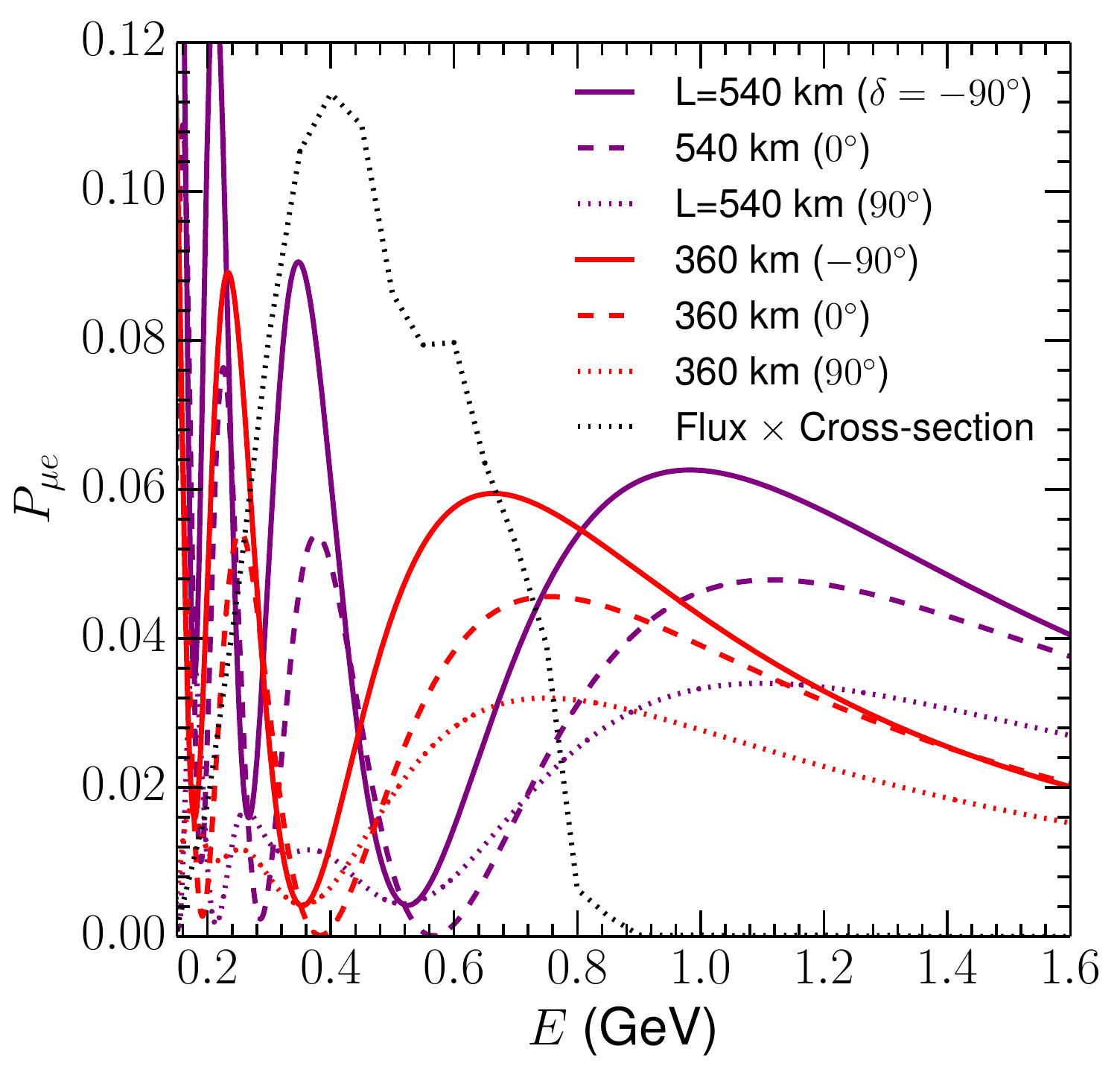}
\includegraphics[width=8cm]{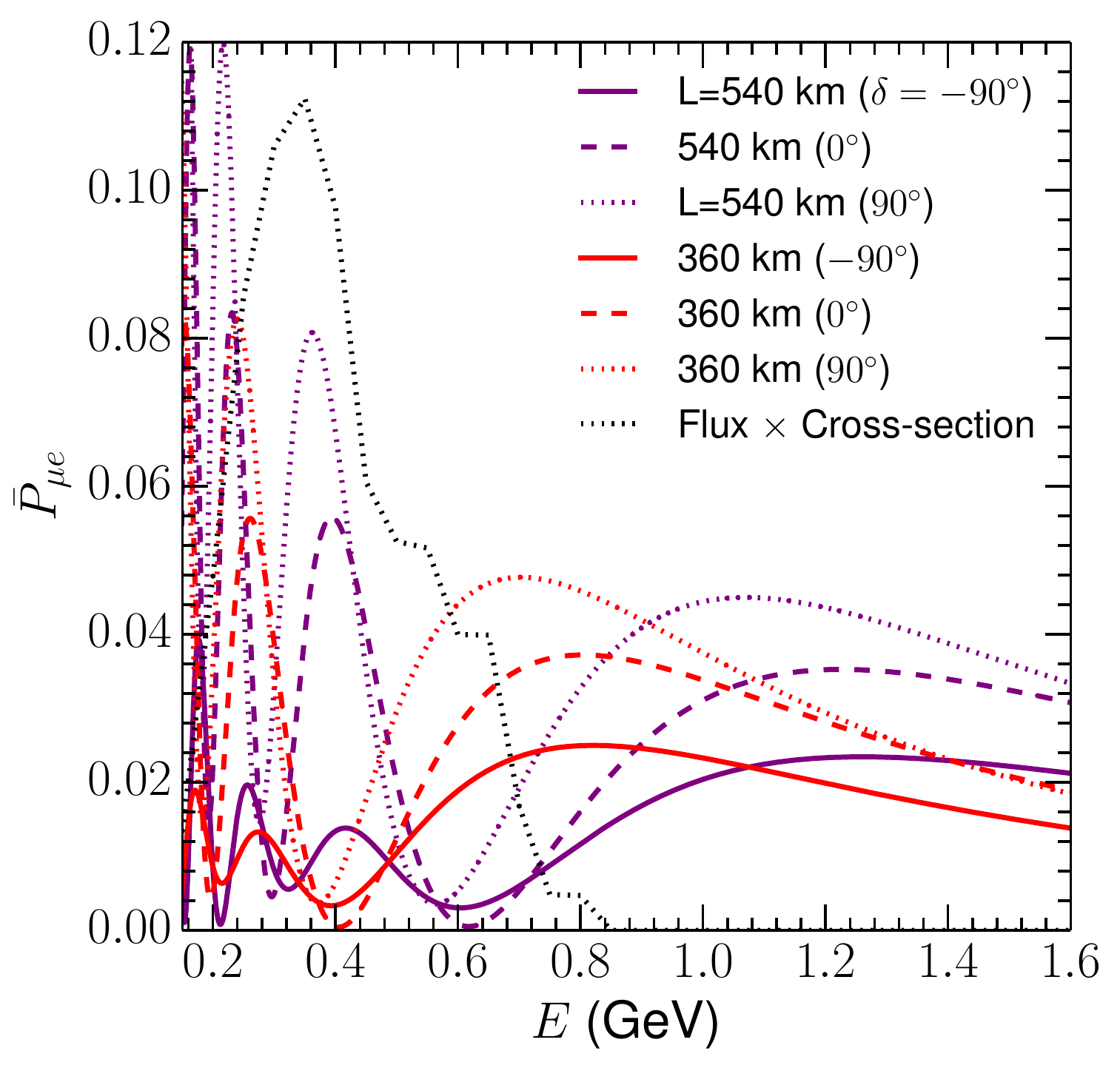}
\caption{Oscillation probabilities for the Zinkgruvan (red curves) and Garpenberg (purple curves) baselines as a function of the energy for neutrinos (left panel) and antineutrinos (right panel). The solid/dashed/dotted lines are for $\delta = -90^\circ$/$0^\circ$/$90^\circ$. The grey dotted lines show the convolution of the signal component of the neutrino flux with the detection cross section. Thus, these serve as a guide of what energies of the oscillation probability would be well-sampled by the ESS$\nu$SB setup.}
\label{fig:probs}
\end{figure}

In \cite{Baussan:2013zcy}, a greenfield study optimising the physics reach to leptonic CP-violation at the ESS$\nu$SB facility was performed. Interestingly, the outcome of this optimisation, as well as follow-up studies~\cite{Agarwalla:2014tpa,Chakraborty:2017ccm,Chakraborty:2019jlv,Ghosh:2019sfi,Blennow:2019bvl,ESSnuSB:2021azq}, was that the best baseline at which to study the neutrino beam would be between 300 and 600\,km. Two candidate mines that could host the detector were thus identified: Garpenberg, at 540\,km, and Zinkgruvan, at 360\,km from the ESS site. This choice makes the ESS$\nu$SB design unique, as the neutrino flux observed by the detector mainly corresponds to the second maximum of the $\nu_\mu \to \nu_e$ oscillation probability, with a more marginal contribution of events at the first oscillation peak. Conversely, other present and next generation oscillation experiments focus instead on the first maximum of the neutrino oscillation probability.

%%%%% fig:probs; Rates %%%%%
\begin{figure}[htp]
\centering
\includegraphics[width=8cm]{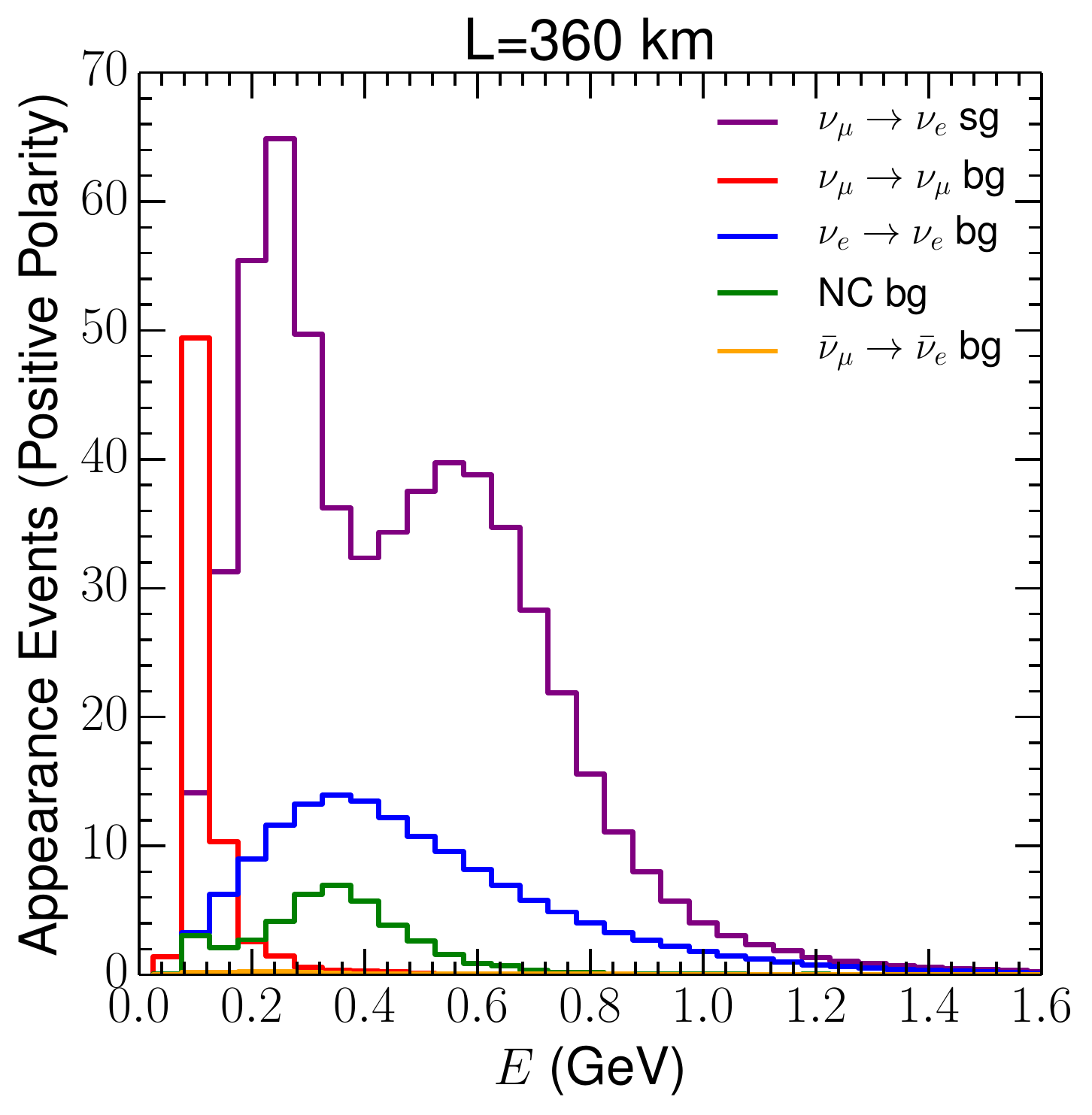}
\includegraphics[width=8cm]{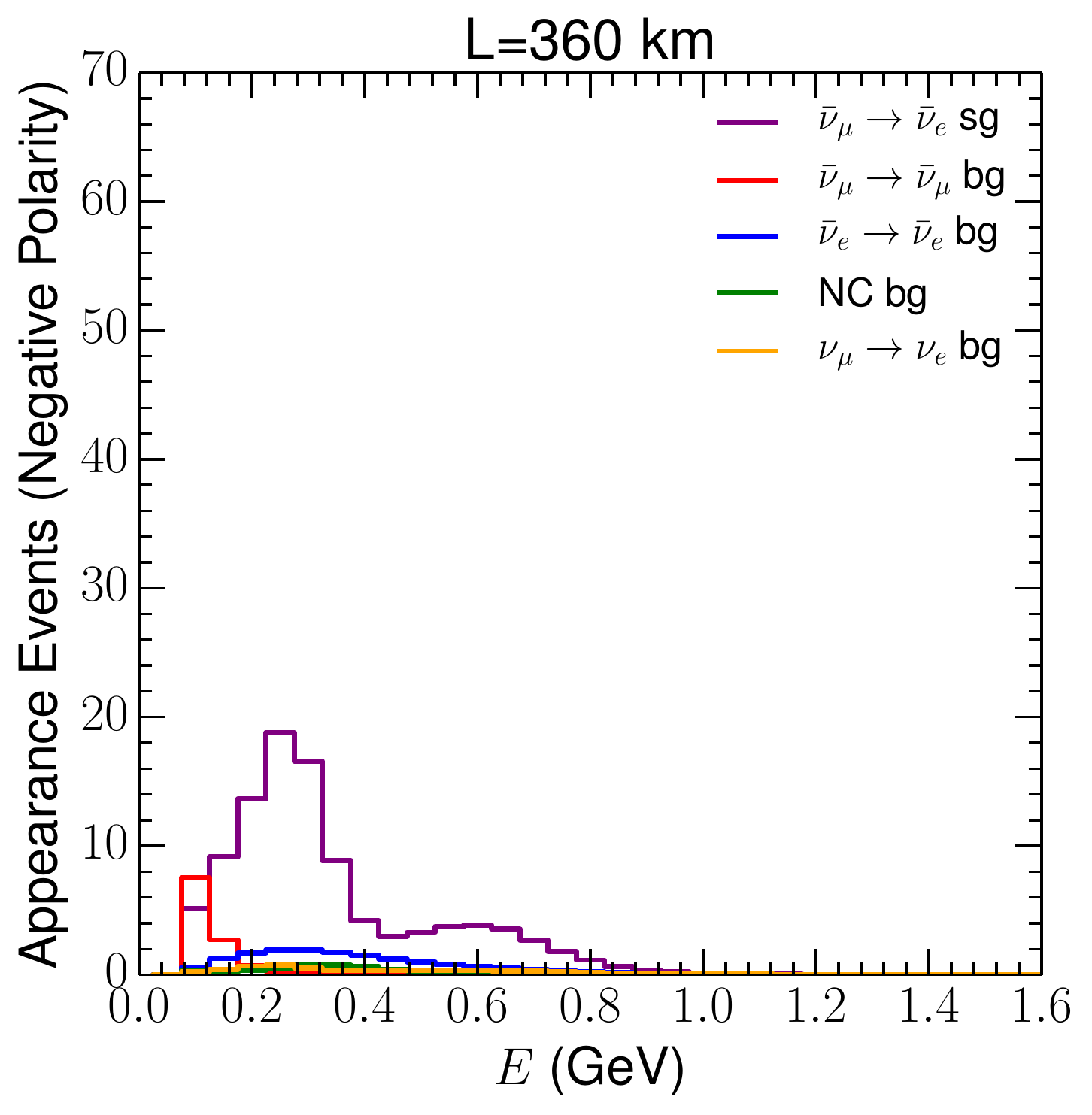}
\includegraphics[width=8cm]{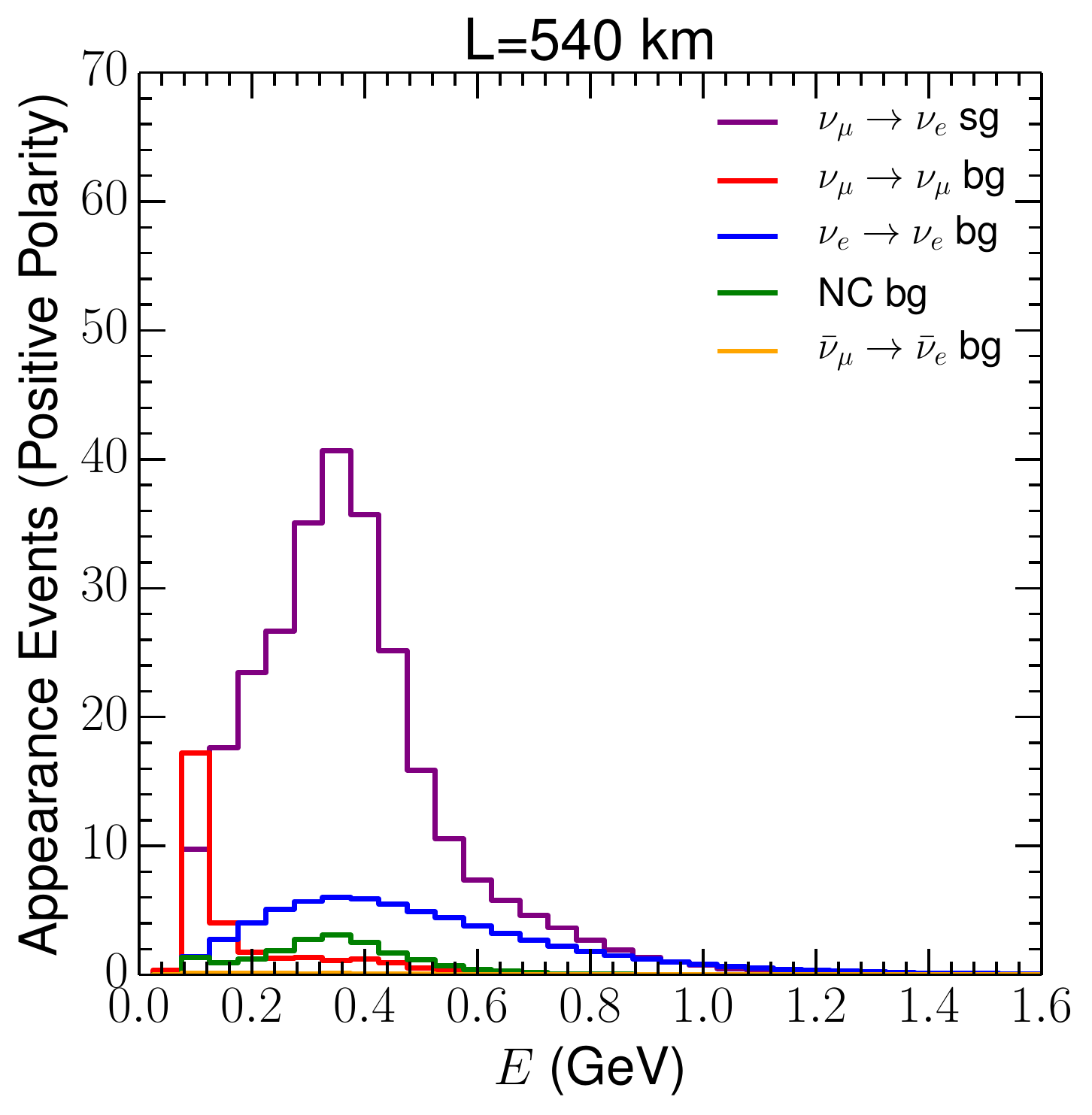}
\includegraphics[width=8cm]{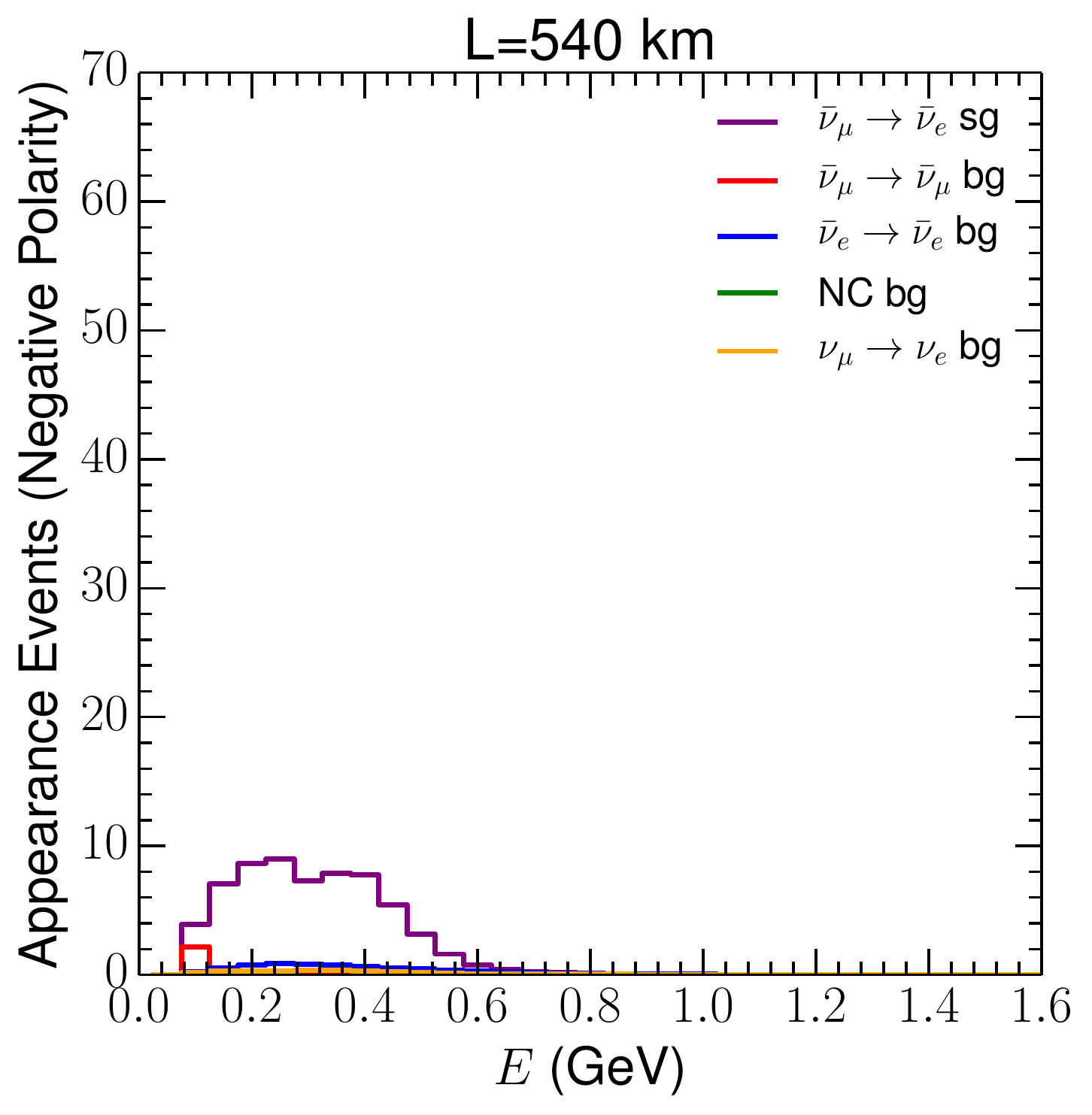}
\caption{Event rates for the different signal and background components of the $e$-like sample with positive (left panels) and negative focusing (right panels) and for the Zinkgruvan 360\,km (upper panels) and Garpenberg 540\,km (lower panels) baselines after one year of data taking in each neutrino and antineutrino mode. These events are calculated for $\delta = 0^\circ$.}
\label{fig:rates}
\end{figure}

\begin{table*}[ht]
	\centering
	\caption{Signal and major background events for the appearance channel corresponding to positive (negative) polarity per year for $\delta = 0^\circ$.}
	\label{tab:ap_events}
%\hspace{-2.01 cm}
	\setlength{\extrarowheight}{0.1cm}
	%\begin{tabular}{|p{0.12\textwidth}|p{0.21\textwidth}|p{0.15\textwidth}|p{0.19\textwidth}|}
	\begin{tabular}{lccc}
		 & \textbf{Channel} & \boldmath{$L = 540$} \textbf{km} & \boldmath{$L =  360$} \textbf{km} \\
		\hline
	Signal & $ \nu_\mu \rightarrow \nu_e$ ($ \bar{\nu}_\mu \rightarrow \bar{\nu}_e$)   & 272.22 (63.75) & 578.62 (101.18)\\
		 \hline
	    &	 $\nu_\mu \rightarrow \nu_\mu$ ($ \bar{\nu}_\mu \rightarrow \bar{\nu}_\mu$)  & 31.01 (3.73) & 67.23 (11.51) \\
Background	&	$\nu_e \rightarrow \nu_e$ ($\bar{\nu}_e \rightarrow \bar{\nu}_e$)  & 67.49  (7.31) & 151.12  (16.66) \\
	&	$\nu_\mu $ NC ($\bar{\nu}_\mu $ NC)   & 18.57 (2.10) & 41.78 (4.73)\\
	&	$\bar{\nu}_\mu \rightarrow \bar{\nu}_e$ ($\nu_\mu \rightarrow \nu_e$)  & 1.08 (3.08) & 1.94 (6.47)\\
		\hline
	\end{tabular}
	\end{table*}

There is a notable cost required in order to observe the oscillation probability at its second maximum. Although it is the optimal choice for maximising the dependence of the oscillation probability on $\delta$, the ratio of the oscillation baseline to the neutrino energy ($L/E$) must to be a factor of three greater when compared to the first maximum. This means that statistics will be about an order of magnitude smaller than if the detector were located at the first peak. Indeed, the neutrino flux decreases with $L^{-2}$ and the neutrino cross section and beam collimation decrease for smaller neutrino energies. The 360\,km Zinkgruvan baseline, corresponding to a point between the first and second oscillation maxima as seen in Fig.~\ref{fig:probs}, represents a compromise between the two choices. The neutrino flux would be 2.25 times larger than for the 540\,km of Garpenberg, and roughly the same number of events belonging to the second oscillation peak would be observed at either site. At the higher-energy end of the spectrum, events from the first oscillation maximum would also be observed for the shorter Zinkgruvan baseline. This can be seen in the event rates expected at each detector and depicted in Fig.~\ref{fig:rates}. The corresponding total number of events are shown in Table~\ref{tab:ap_events}.

In any case, choosing for ESS$\nu$SB to have its detector operate near the second oscillation maximum was shown to be optimal for its increased dependence on $\delta$ and despite the reduced event rate. In fact, as will be shown in the following sections, it could provide unprecedented discovery potential to leptonic CP-violation and the most precise measurement of $\delta$, which could prove an instrumental piece in understanding the flavour puzzle. Furthermore, the study also demonstrates that this design makes the physics performance extremely resilient against unexpected sources of systematic uncertainties~\cite{Coloma:2011pg}. However, statistics may become the bottleneck of the ESS$\nu$SB physics reach and therefore longer periods of data taking may be needed to significantly improve its capabilities.

Conversely, for measurements other than the CP violation search, the choice of the second oscillation maximum baseline may not be optimal. For instance, the sensitivity to the octant of $\theta_{23}$ mainly relies on the ``atmospheric'' term of the oscillation probability, which is leading at the first maximum. It also requires $\Delta m^2_{31}$ and $\sin^2 2\theta_{23}$ from $\nu_\mu$ disappearance, which is also more suitable at the first oscillation peak. 

Similarly, in Eq.~(\ref{Eq:Probability}) the ``atmospheric'' term already depends on the mass ordering, as it is inversely proportional to the square of $\tilde{B}_{\mp}$. For $E \sim |\Delta m_{31}^2|/(2A) $ the MSW resonance~\cite{Wolfenstein:1977ue,Mikheyev:1985zog,Mikheev:1986wj} will enhance the neutrino oscillation probability compared to the antineutrino one or vice versa, depending on the mass ordering. For an average matter density of $3.0~\text{g}/\text{cm}^3$, the energy for this resonance to occur is approximately $E \sim \mathcal{O}(\text{GeV})$. Since the peak of the flux for ESS$\nu$SB is at $E\sim \mathcal{O}(100)~\text{MeV}$ (see Fig.~\ref{fig:probs}), matter effects, and hence the sensitivity to the mass ordering for this facility, are not very significant. Conversely, since these matter effects are small, the confusion between genuine CP violation and the matter potential~\cite{Minakata:2001qm} is largely avoided and does not compromise the facility's sensitivity to CP violation~\cite{Baussan:2013zcy,Bernabeu:2018use,Ghosh:2019sfi}.

A pragmatic and very effective way of increasing both the octant and mass ordering sensitivity of a neutrino oscillation experiment is to combine the signal from the neutrino beam with the substantial atmospheric neutrino sample that will also be seen at the far detector~\cite{Huber:2005ep,Campagne:2006yx}. For ESS$\nu$SB, this combination was explored in \cite{Blennow:2019bvl}, where it was found that the combination of the ESS$\nu$SB beam data with the atmospheric neutrino sample could significantly improve the overall physics reach to the CP violating phase $\delta$ by solving intrinsic parametric degeneracies, and also providing a determination of the mass ordering and good sensitivity to the octant of $\theta_{23}$. After this analysis was performed, new developments have also come about, which increase the expected statistics from the optimised beam and detector simulations discussed in the previous sections. When these new results are considered, the studies show that the impact of the atmospheric neutrino sample on the CP violation searches is more marginal, although it is still very useful to determine the mass hierarchy and provide significant sensitivity to the octant. For this reason, and to facilitate the comparison with other proposals, this study will concentrate only on the physics reach of the beam data and focus on the CP-violation sensitivity. For the sensitivity of the combined beam and atmospheric samples to the mass hierarchy and the octant of $\theta_{23}$ see \cite{Blennow:2019bvl}.

The updated estimates of the ESS$\nu$SB neutrino flux and detector response, detailed in the previous sections, imply an important increase in statistics and, therefore, a significant improvement in the beam physics performance. Preliminary results in this vein were reported in \cite{ESSnuSB:2021lre}. In the following sections, that analysis will be updated and complemented with a dedicated discussion of the impact of the possible systematic uncertainties which may affect the far-detector event rates. 

\subsection{Simulation Details}
 
The estimation of the physics sensitivity arising from beam events makes use of the \textsc{GLoBES} software~\cite{Huber:2004ka,Huber:2007ji}. A water Cherenkov detector of fiducial volume 538\,kt located either at a distance of 540\,km or 360\,km from the neutrino source, as discussed in Chapter~\ref{Detectors}, has been considered. A value of $2.7 \times 10^{23}$ protons on target per year with a beam power of 5\,MW, and proton kinetic energy of 2.5\,GeV as discussed in Chapter~\ref{protondriver} has been assumed for the neutrino beam. The optimised fluxes from the genetic algorithm, as discussed in Chapter~\ref{targetstation}, have been implemented, together with the event selection obtained from the full Monte-Carlo simulations presented in Chapter~\ref{Detectors} in the form of migration matrices. The events are distributed in 50 bins between 0 to 2.5\,GeV of the reconstructed energy. Both the appearance channels ($\nu_\mu \rightarrow \nu_e$) and disappearance channels ($\nu_\mu \rightarrow \nu_\mu$) for signal events have been analysed. The relevant background channels as discussed in Chapter~\ref{Detectors} have also been implemented. Finally, a total run-time of 10 years (divided into 5 years of neutrino beam and 5 years of antineutrino beam) has been assumed, unless otherwise specified.

A detailed and exhaustive list of all the parameters encoding relevant systematic uncertainties that might impact the physics reach of the facility, together with an accurate estimation of their magnitude, energy dependence, and level of correlation propagated from the near detector measurements to those of the far detector is challenging and beyond the scope of this work. It is, however, the subject of dedicated ongoing studies and will be pursued and presented in future works.

Here, the focus will instead be on demonstrating how the ESS$\nu$SB setup, mainly exploiting the information from events at the second oscillation maximum, is relatively resilient against the effect of systematic uncertainties. To this end, all results will be presented, explicitly showing their dependence on the size of the systematic errors assumed. Furthermore, three different types of uncertainties will be taken into account. The effect of systematic uncertainties is incorporated by the pull method~\cite{Fogli:2002pt,Huber:2002mx}.

First, the impact on the physics reach of overall normalisation uncertainties will be studied, allowing them to be different between the different channels and signal and background components. Second, the impact of energy calibration systematics will be analysed, again allowing for a different effect in different channels and between signal and background. Finally, more general bin-to-bin uncorrelated uncertainties will be implemented (one independent nuisance parameter per bin) which may encode more complex situations such as the unknown shape of the neutrino cross sections at the energies of interest. These sets of nuisance parameters are meant to capture all sources of uncorrelated uncertainties affecting the far detector expectation after the measurements at the near detector and beam-monitoring information have been taken into account. Thus, the near-detector events are not simulated explicitly, unlike some previous works such as~\cite{Blennow:2019bvl}, to avoid double counting its impact.

For the estimation of the sensitivity, the Poisson log-likelihood will be used:
\begin{equation}
 \chi^2_{{\rm stat}} = 2 \sum_{i=1}^n \bigg[ N^{{\rm test}}_i - N^{{\rm true}}_i - N^{{\rm true}}_i \log\bigg(\frac{N^{{\rm test}}_i}{N^{{\rm true}}_i}\bigg) \bigg]\,,
 \label{eq:chi2}
\end{equation}
where $N^{{\rm test}}$ is the number of events expected for the values of the oscillation parameters tested for, $N^{{\rm true}}$ is the number of events expected for the parameter values assumed to be realised in nature (Asimov dataset) and $i$ is the number of energy bins. Unless otherwise specified, the best-fit values of the oscillation parameters are adopted from NuFIT \cite{Esteban:2020cvm} and summarised in Table~\ref{Tab:Param}. These results are updated with the latest parameters and are in good agreement with other global fits from other groups, see \cite{deSalas:2020pgw,Capozzi:2021fjo}. All results are shown for the normal ordering of neutrino masses, but similar results are obtained when an inverted hierarchy is instead assumed. 
%\begin{table}[ht]
%	\centering
%	\setlength{\extrarowheight}{0.1cm}
%	\begin{tabular}{|c|c|c|}
%		\hline
%		 Parameter & Best-fit value & $1\sigma$ range\\
%		\hline
%		$\sin^2 \theta_{12}$ & 0.304 & $\pm 0.012$ \\
%		$\sin^3 \theta_{13}$ & 0.02246 &  $\pm 0.00062$ \\
%		$\sin^2 2 \theta_{23}$ & 0.9898 & $\pm 0.0077$ \\
%		$\Delta m^2_{21}$ & $7.42\times10^{-5}$ eV$^2$ & $\pm 0.21 \times 10^{-5}$ eV$^2$ \\
%		$\Delta m^2_{31}$ &  $2.510\times10^{-3}$ eV$^2$ & $\pm 0.027 \times 10^{-3}$ eV$^2$\\
%		\hline
%	\end{tabular}
%	\caption{The best-fit values and $1 \sigma$ allowed regions of the oscillation parameters used in our calculation as given in Ref. \cite{Esteban:2020cvm}.}
%	\label{Tab:Param}
%\end{table}
\begin{table}[ht]
	\centering
	\caption{The best-fit values and $1 \sigma$ allowed regions of the oscillation parameters used in the calculation as given in Ref. \cite{Esteban:2020cvm}.}
	\label{Tab:Param}
	\setlength{\extrarowheight}{0.1cm}
	\begin{tabular}{lc}
		 \textbf{Parameter} & \textbf{Best-fit value \boldmath{$\pm1\sigma$} range}\\
		\hline
		$\sin^2 \theta_{12}$ & $0.304 \pm 0.012$ \\
		$\sin^2 \theta_{13}$ & $0.02246\pm 0.00062$ \\
		$\sin^2 2 \theta_{23}$ & $0.9898\pm 0.0077$ \\
		$\Delta m^2_{21}$ & $\left(7.42\pm 0.21\right) \times 10^{-5}$~eV$^2$ \\
		$\Delta m^2_{31}$ &  $\left(2.510\pm 0.027\right) \times 10^{-3}$~eV$^2$\\
		\hline
	\end{tabular}
\end{table}

\subsection{Physics Reach}

This study of the physics reach of the ESS$\nu$SB concentrates on its capability to probe for CP violation and perform a precision measurement of the CP-violating phase $\delta$. First the \emph{CP discovery potential} of ESS$\nu$SB is estimated. To that end, the significance with which ESS$\nu$SB would be able to disfavour CP-conservation, that is $\delta = 0$ or $\delta = \pi$, is computed for all possible ``true'' values of $\delta$ for the Asimov data. This is estimated through the $\Delta \chi^2$ as defined in Eq.~(\ref{eq:chi2}) with $N^{{\rm true}}$, the expected number of events for the chosen value of $\delta$ and $N^{{\rm test}}$ the expected number of events that minimises $\chi^2$ when fixing $\delta = 0$ or $\delta = \pi$ while letting the rest of the oscillation parameters in Table~\ref{Tab:Param} vary freely with their corresponding Gaussian priors added to $\chi^2_{{\rm stat}}$. Under the assumption that this $\Delta \chi^2$ is indeed $\chi^2$-distributed\footnote{See \cite{Blennow:2014sja} for a detailed discussion on the validity of this approximation.} the square root of $\Delta \chi^2$ provides the significance with which ESS$\nu$SB would be able to disfavor CP conservation if that value of $\delta$ is realised in nature. 

%%%%% Fig:cpsens; CP discovery potential %%%%%
\begin{figure}[htp]
\centering
\includegraphics[width=8cm]{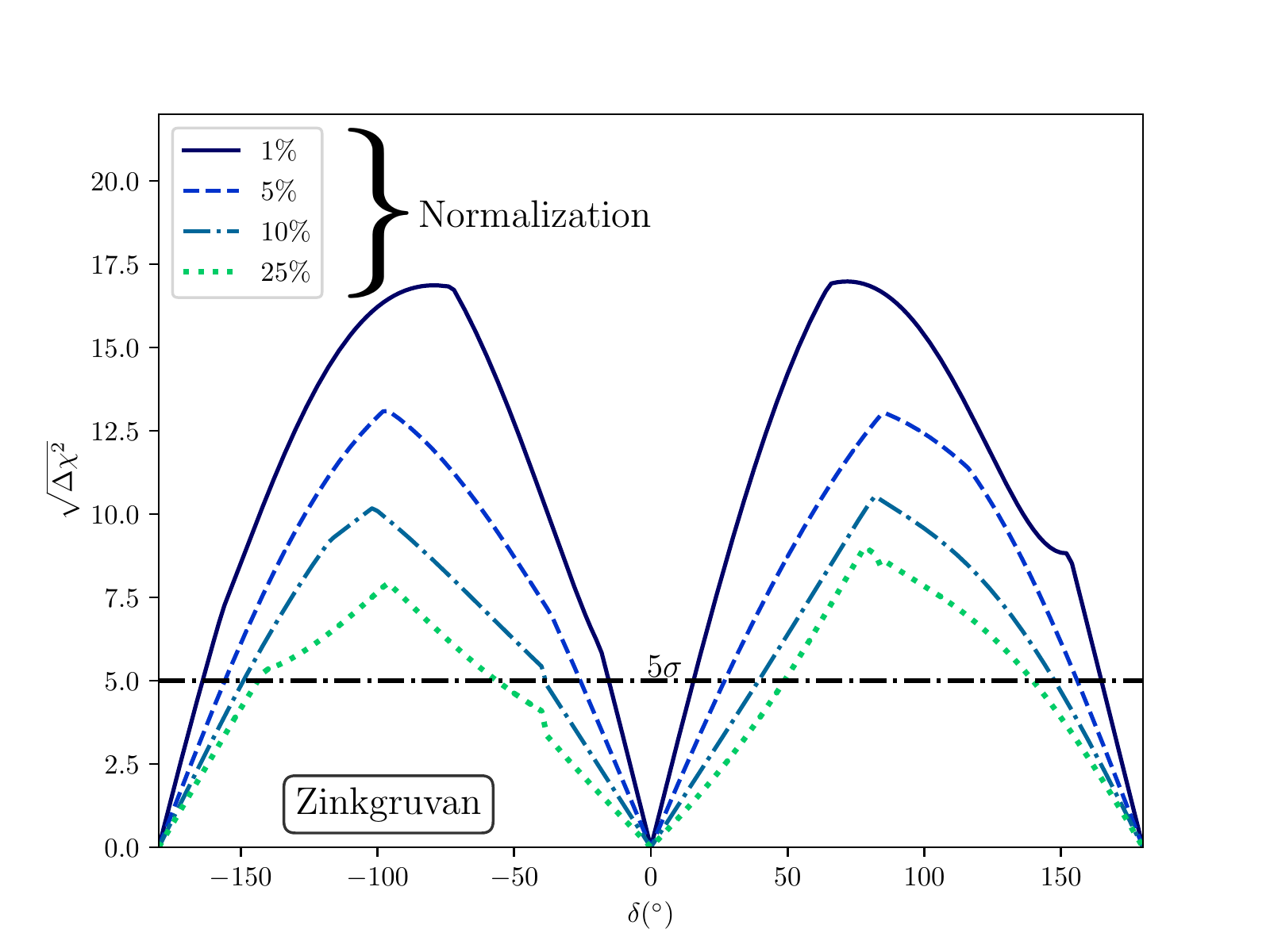}
\includegraphics[width=8cm]{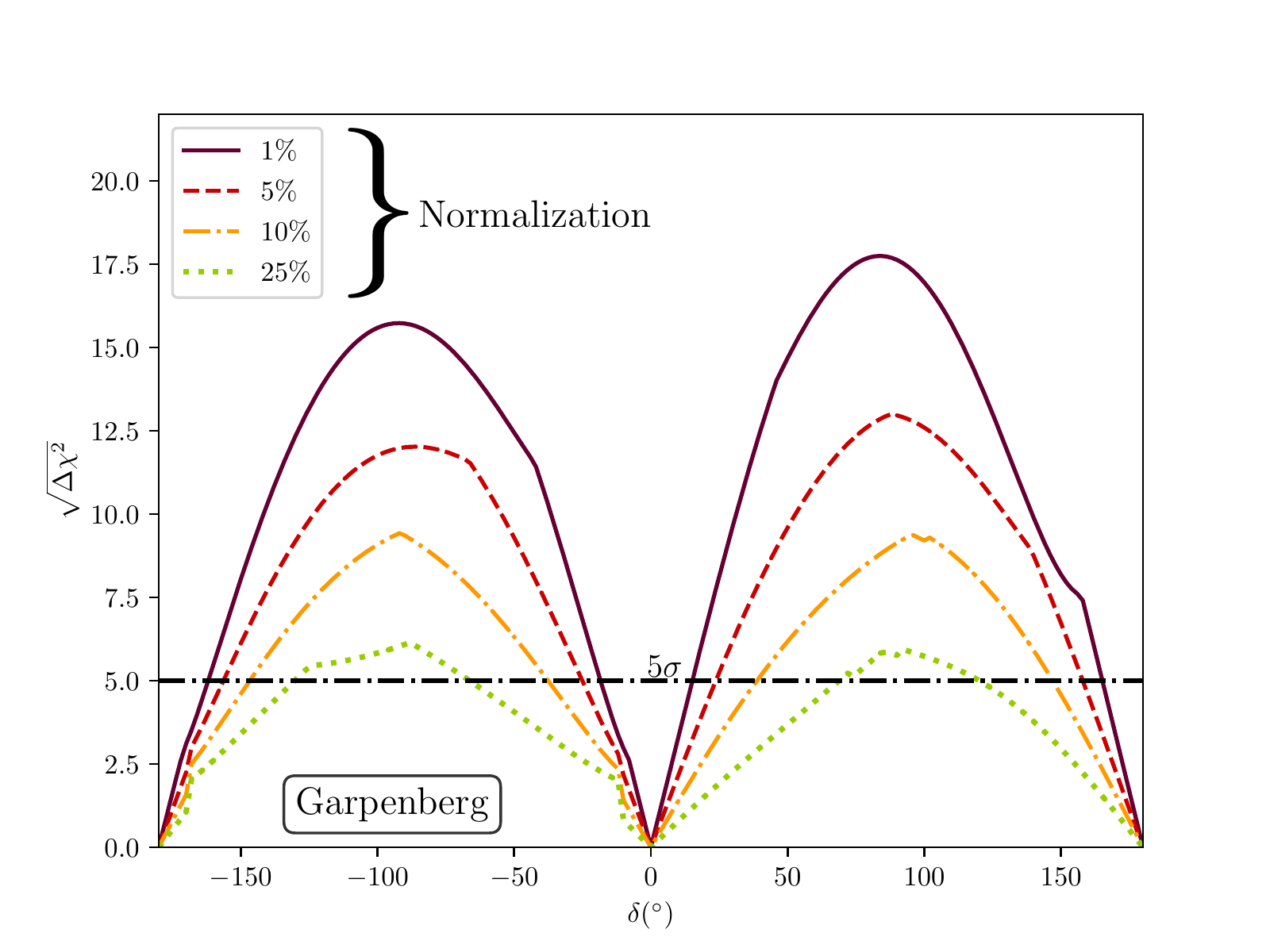}
\includegraphics[width=8cm]{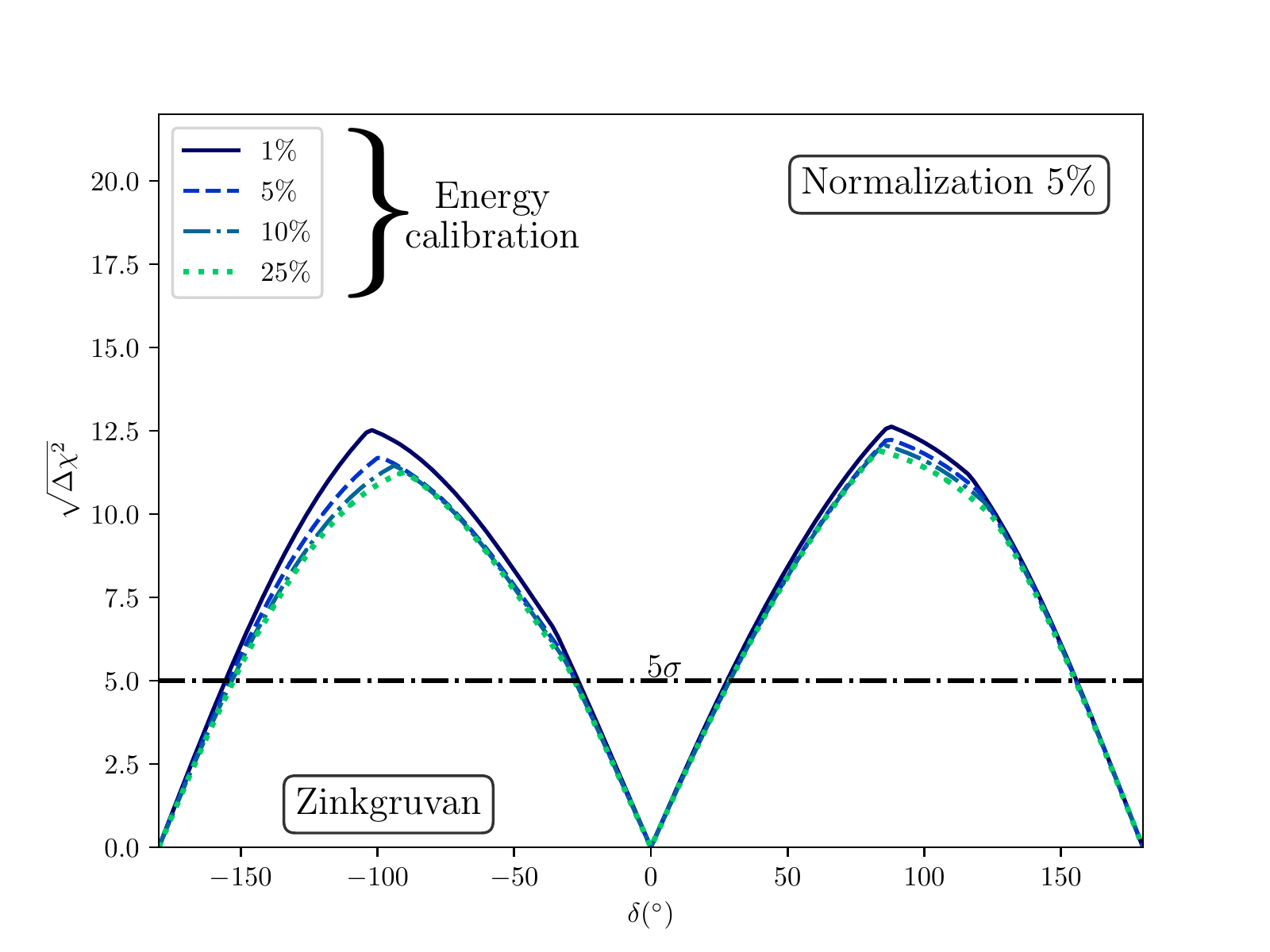}
\includegraphics[width=8cm]{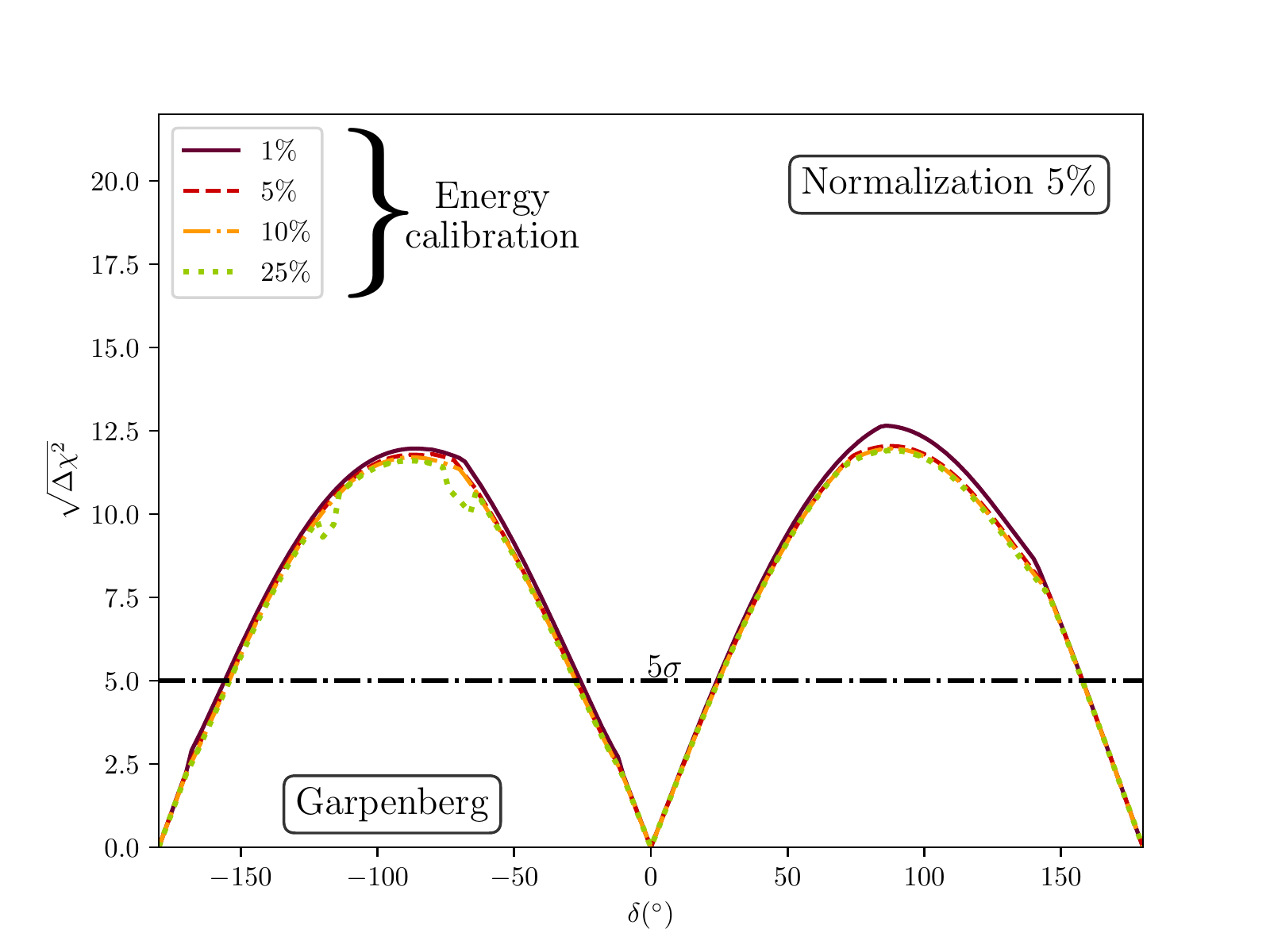}
\includegraphics[width=8cm]{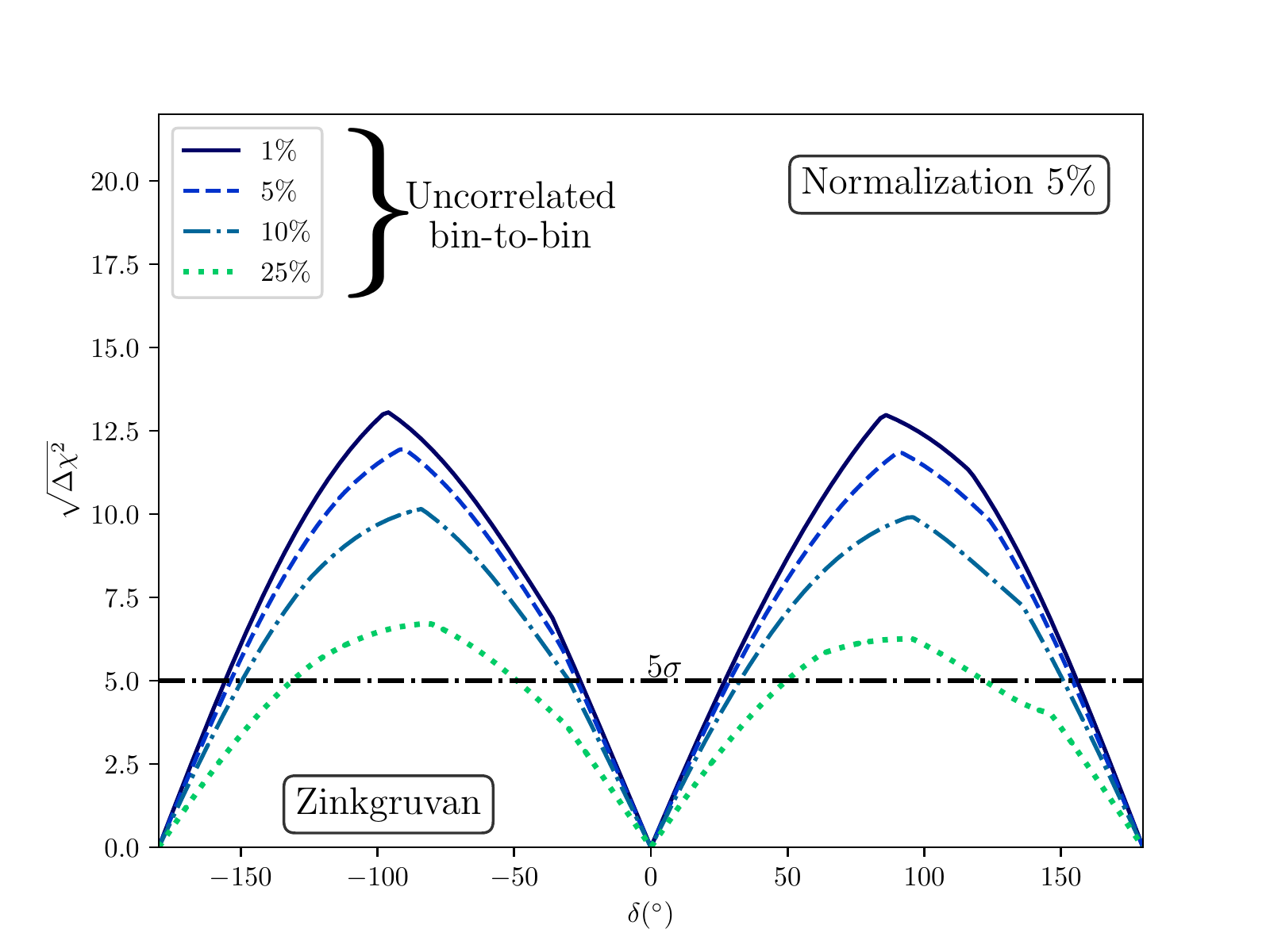}
\includegraphics[width=8cm]{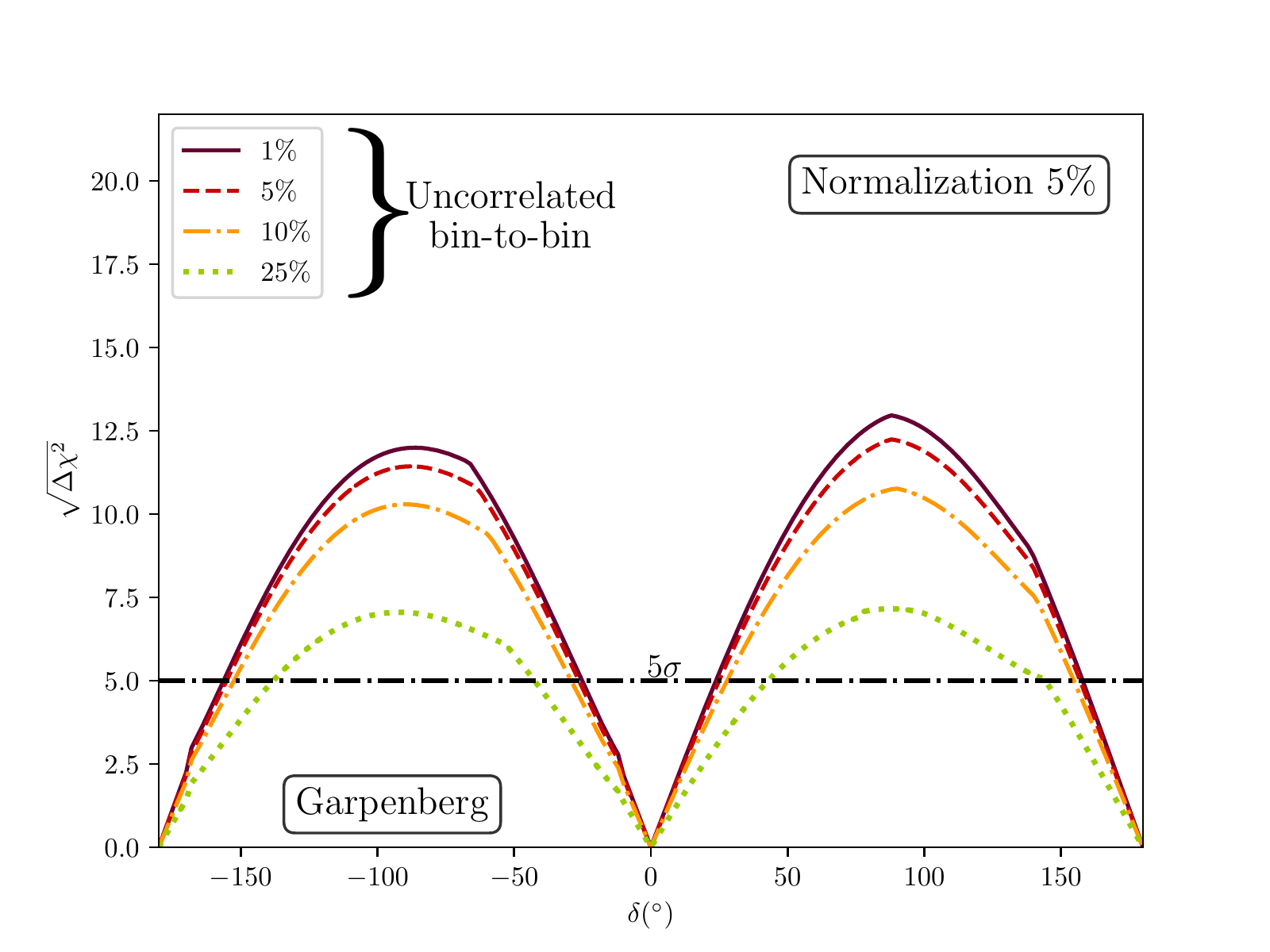} 
\caption{CP discovery potential of the ESS$\nu$SB facility. Left (right) panels for the Zinkgruvan 360\,km (Garpenberg 540\,km) baseline. The upper panels show the impact of an overall normalisation uncertainty with curves ranging from $1\%$ to $25\%$ uncertainties uncorrelated between the different signal and background samples for each channel. The middle panels show the impact of an energy-calibration uncertainty, again ranging from $1\%$ to $25\%$. The normalisation uncertainty has been fixed to $5\%$ to allow for the interplay between these two sources of systematics. The bottom panels show the impact of a more generic uncertainty, capable of affecting both the shape and normalisation parameterised as bin-to-bin uncorrelated nuisance parameters for all signal and background components. The lines again range from $1\%$ to $25\%$; and a $5\%$ normalisation uncertainty has also been included.}
\label{fig:cpsens}
\end{figure}

Figure~\ref{fig:cpsens} shows the CP discovery potential obtained for the Zinkgruvan and Garpenberg baselines and reports its dependence with the systematic uncertainties introduced as nuisance parameters. Generally, the performance of both baselines is very similar, with only slightly different areas covered beyond the $5 \sigma$ mark, depending on the systematic uncertainties under consideration. In the upper panels, the impact of an overall normalisation uncertainty, uncorrelated among the different signal and background samples, is analysed. This nuisance parameter has been sampled for $1\%$, $5\%$, $10\%$ and $25\%$ uncertainties. It is notable that, even for a highly conservative $25\%$ uncertainty, a significant portion of the values of $\delta$ would still allow a discovery of CP violation above the $5 \sigma$ level. Note that this uncertainty is uncorrelated between the neutrino and antineutrino samples, and therefore capable of mimicking the effect of CP violation. Nevertheless, a discovery would be possible even in this scenario. This can be understood from Fig.~\ref{fig:probs} where, close to the second oscillation peak, changes in $\delta$ can lead to very sizeable changes in the probability.

The dependence of the shape of the oscillation probability with $\delta$ may also play a relevant role in the sensitivity obtained. Thus, the middle and lower panels of Fig.~\ref{fig:cpsens} also explore the robustness of the results obtained against other systematic uncertainties which might also affect the shape of the recovered spectrum. In particular, in the middle panels, the impact of a $1\%$, $5\%$, $10\%$ and $25\%$ uncertainty in the energy calibration is analysed. Finally, in the lower panels, a more general bin-to-bin uncorrelated set of nuisance parameters has been considered with the same values, which allows for the modelling of more complex scenarios. In both cases, a $5\%$ normalisation uncertainty has been added to allow for the possible interplay between the different sets of systematic uncertainties that may be relevant. It can be observed that the energy calibration uncertainty has a relatively minor impact in the CP discovery potential for both baselines under study. Conversely, the more general implementation of uncorrelated uncertainties in each bin can have a more significant impact, but still less consequential than the overall normalisation considered in the upper panels.

These results demonstrate that the ESS$\nu$SB design has remarkable CP discovery potential, even for very conservative assumptions on the systematic uncertainties that could affect the far detector. Indeed, measurements at the near detector and beam monitoring keep these uncertainties at or below the level of a few percent for the present generation of neutrino oscillation experiments. In particular, assuming a $5\%$ normalisation uncertainty, in line with similar facilities, CP violation could be established for $71\%$ ($73\%$) of the values of $\delta$ for the Zinkgruvan (Garpenberg) baseline. This is further illustrated in the left panel of Fig.~\ref{fig:cpexpocov}, where the fraction of values of $\delta$ for which a given significance for CPV could be established for a $5\%$ normalisation uncertainty. The right panel of Fig.~\ref{fig:cpexpocov} shows instead the impact of statistics on the ESS$\nu$SB performance. The fraction of values of $\delta$ for which a $5~\sigma$ discovery of CPV violation are shown as a function of the total running time, assuming equal time spent in neutrino and antineutrino modes. The figure confirms that indeed the ESS$\nu$SB can significantly benefit from an extended running time, while the effect of the systematic uncertainties is less pronounced, as expected for a facility studying the second oscillation maximum.    

\begin{figure}[htp]
\centering
\includegraphics[width=8cm]{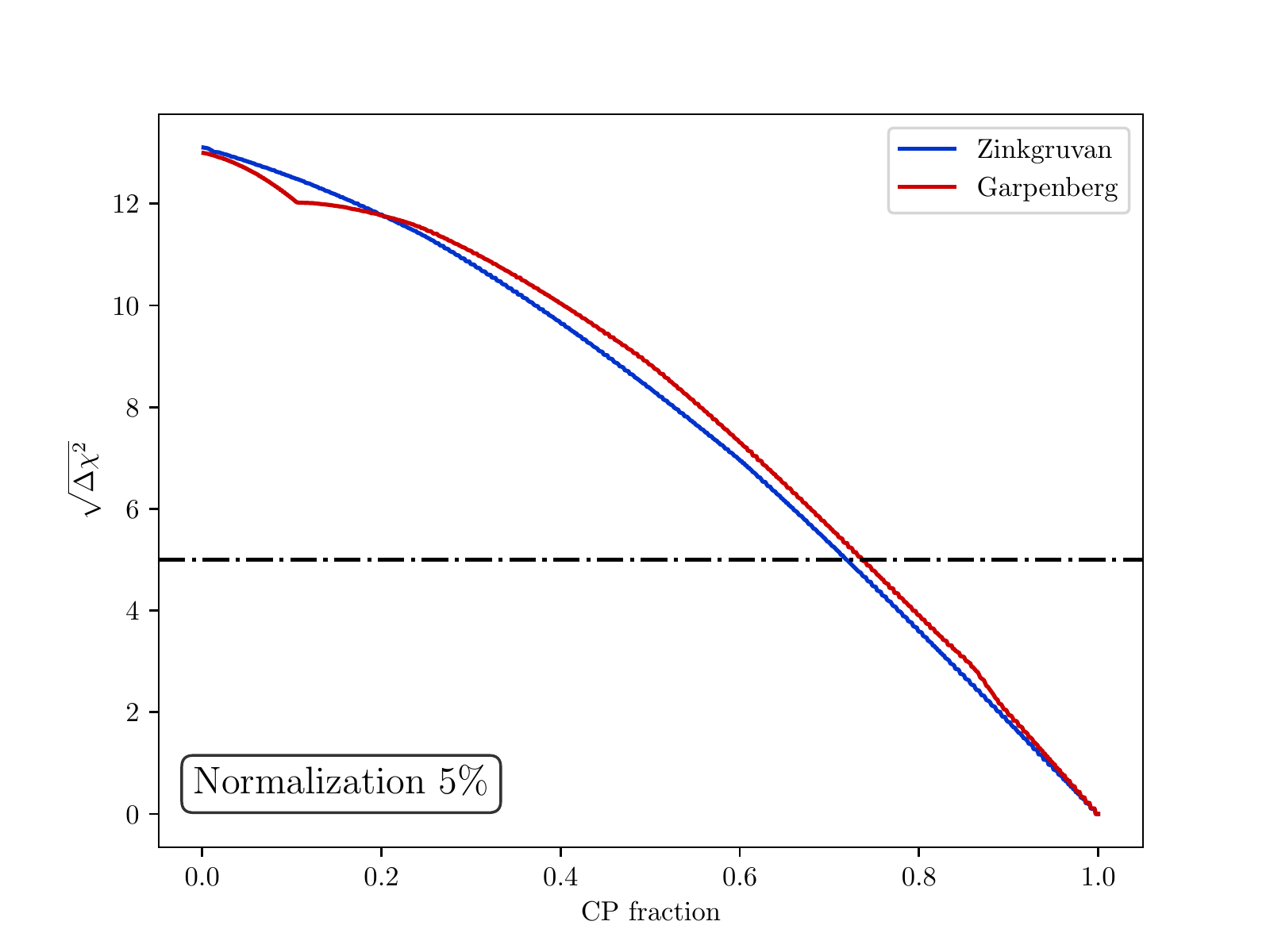}
\includegraphics[width=8cm]{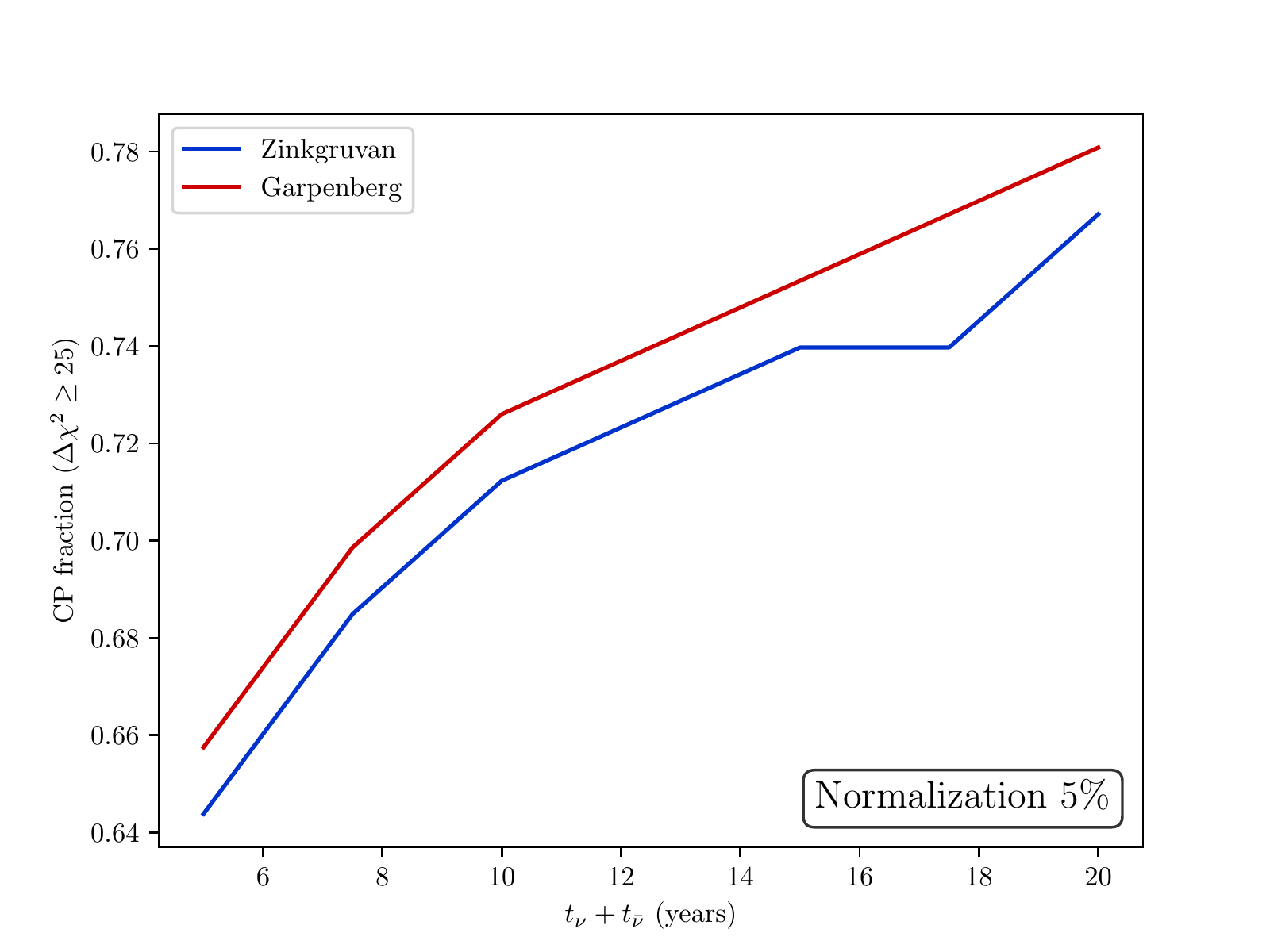}
\caption{Right panel: fraction of values of $\delta$ for which CP violation could be discovered at more than $5 \sigma$ as a function of experimental run-time, assuming equal time in neutrino and antineutrino modes. Left panel: Significance with which CPV violation could be established by ESS$\nu$SB as a function of the fraction of values of $\delta$ for which it would be possible. The horizontal black line corresponds to $5~\sigma$.  The blue (red) curves in both panels correspond to Garpenberg (Zinkgruvan) baseline. These plots represent a normalisation systematic uncertainty of 5\%.}
\label{fig:cpexpocov}
\end{figure}

%To further illustrate the capability of ESS$\nu$SB to dicscover CP violation, the left panel of Fig~\ref{fig:cpexpocov} shows the fraction of $\delta$ for which CP violation can be discovered at more than 5 $\sigma$ C.L. vs run-time and the right panel shows the CP discovery potential vs fraction of $\delta$ for the both the baselines. These panels are generated for normalization error of 5\%. Both the panels show the sensitivity of Zinkgruvan baseline is slightly better than the Garpenberg baseline. The left plot shows for a 10 years of run-time, ESS$\nu$SB can discover CP violation at $5 \sigma$ for 70\% of the total $\delta$ values and it can go upto 75\% if the run-time increased to 20 years. The right panel depicts the significance with which CP violation can be discovered for different $\delta$ fractions for a run-time of 10 years.

%%%%% Fig:cpsens; delta precision %%%%%
\begin{figure}[htp]
\centering
\includegraphics[width=8cm]{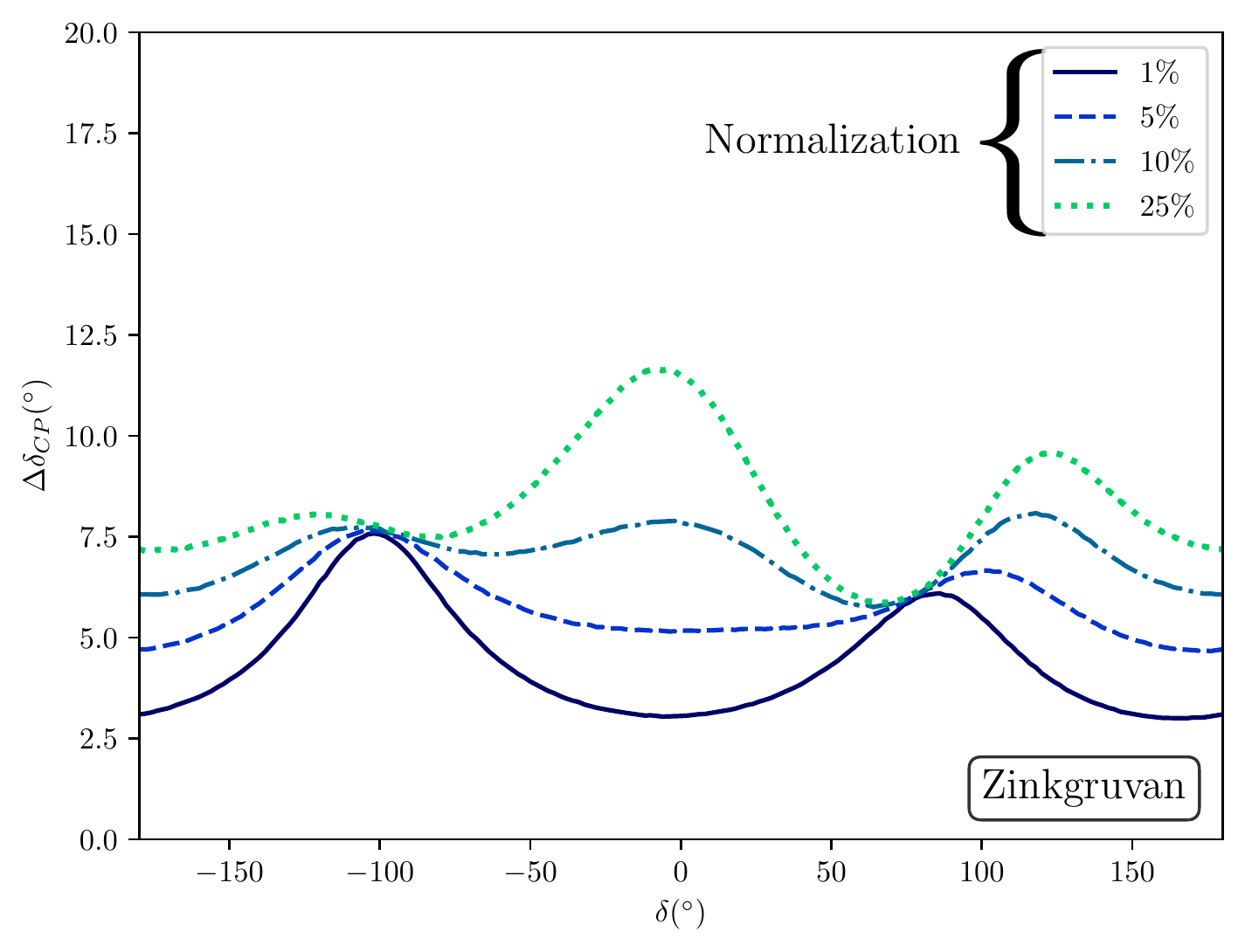}
\includegraphics[width=8cm]{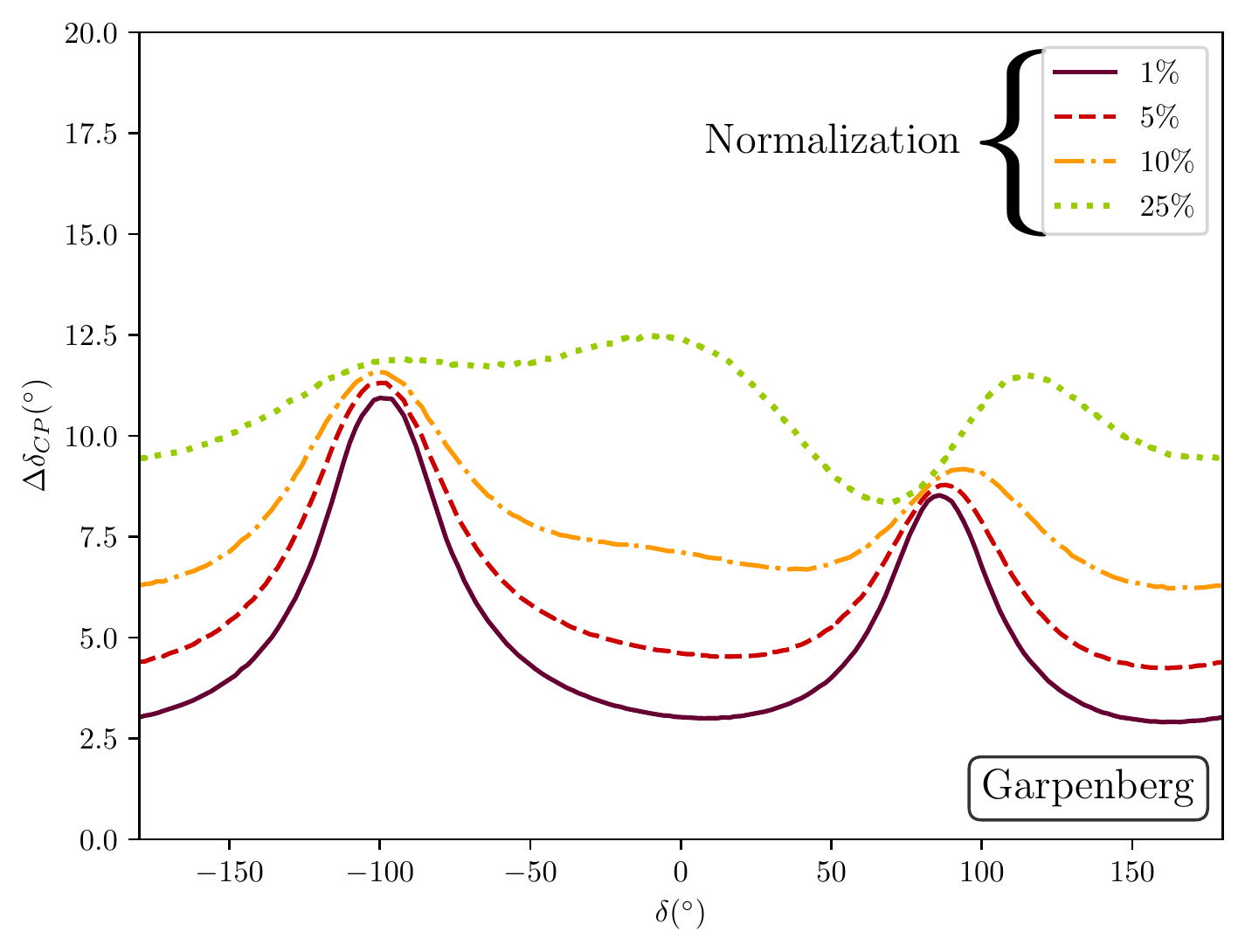}
\includegraphics[width=8cm]{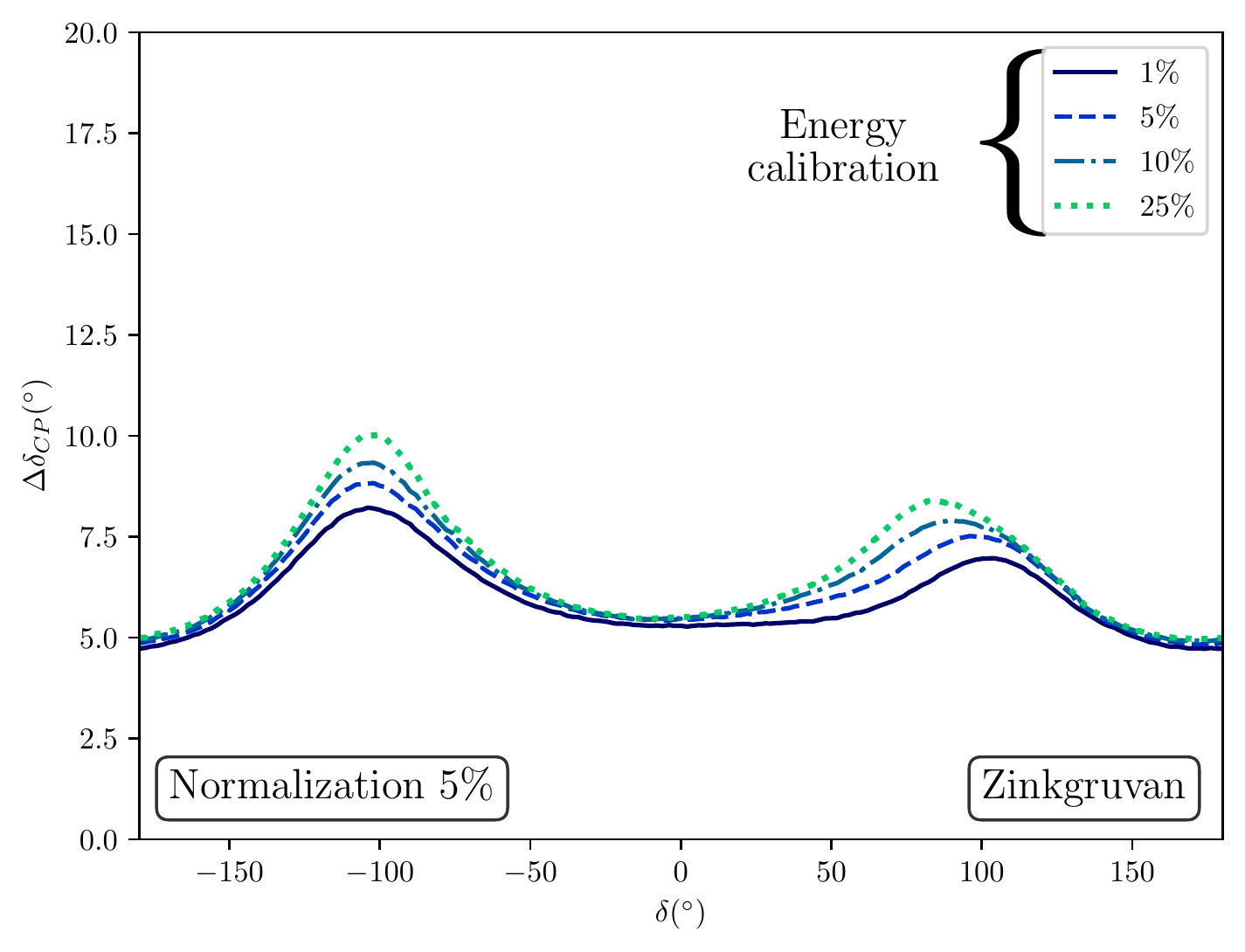}
\includegraphics[width=8cm]{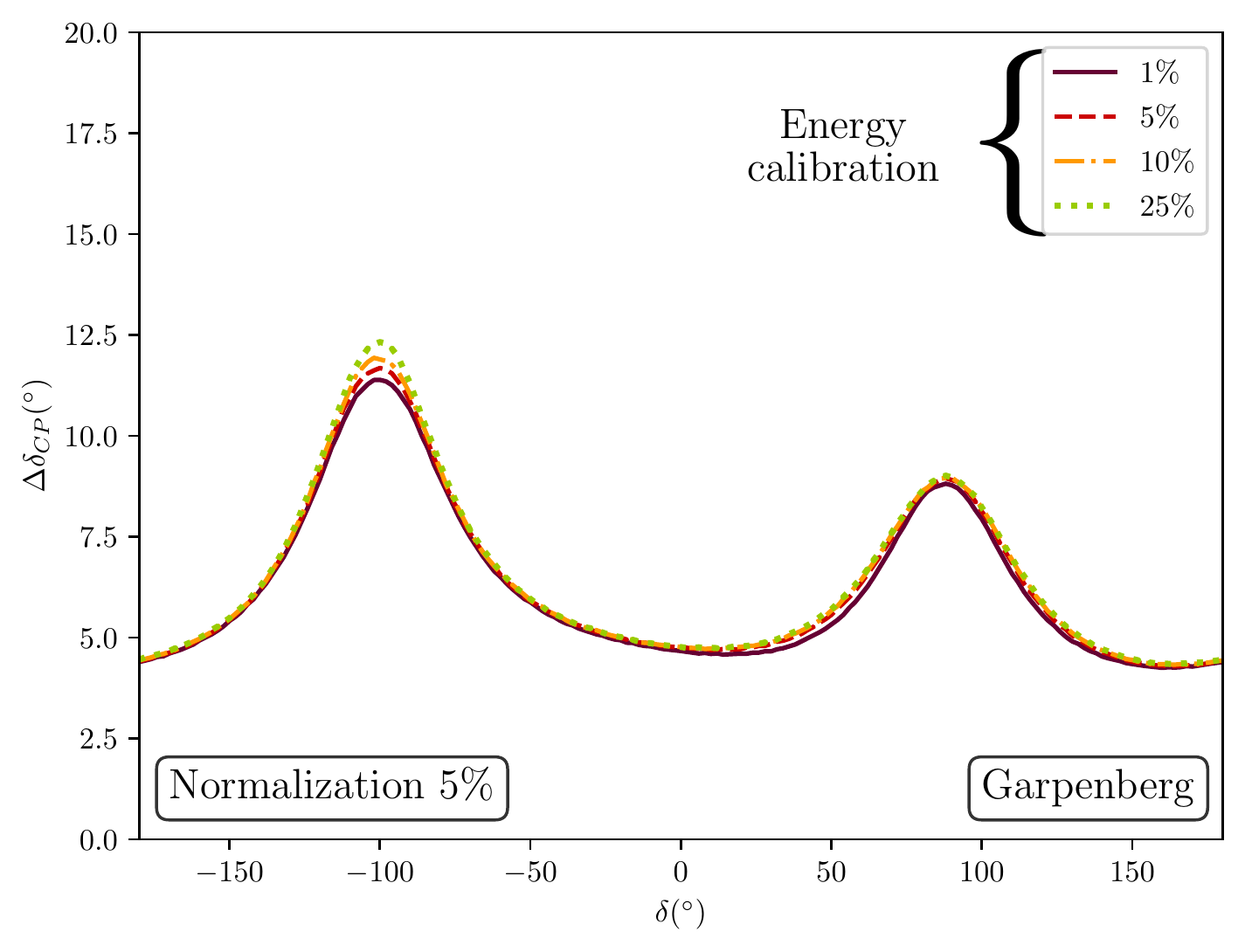}
\includegraphics[width=8cm]{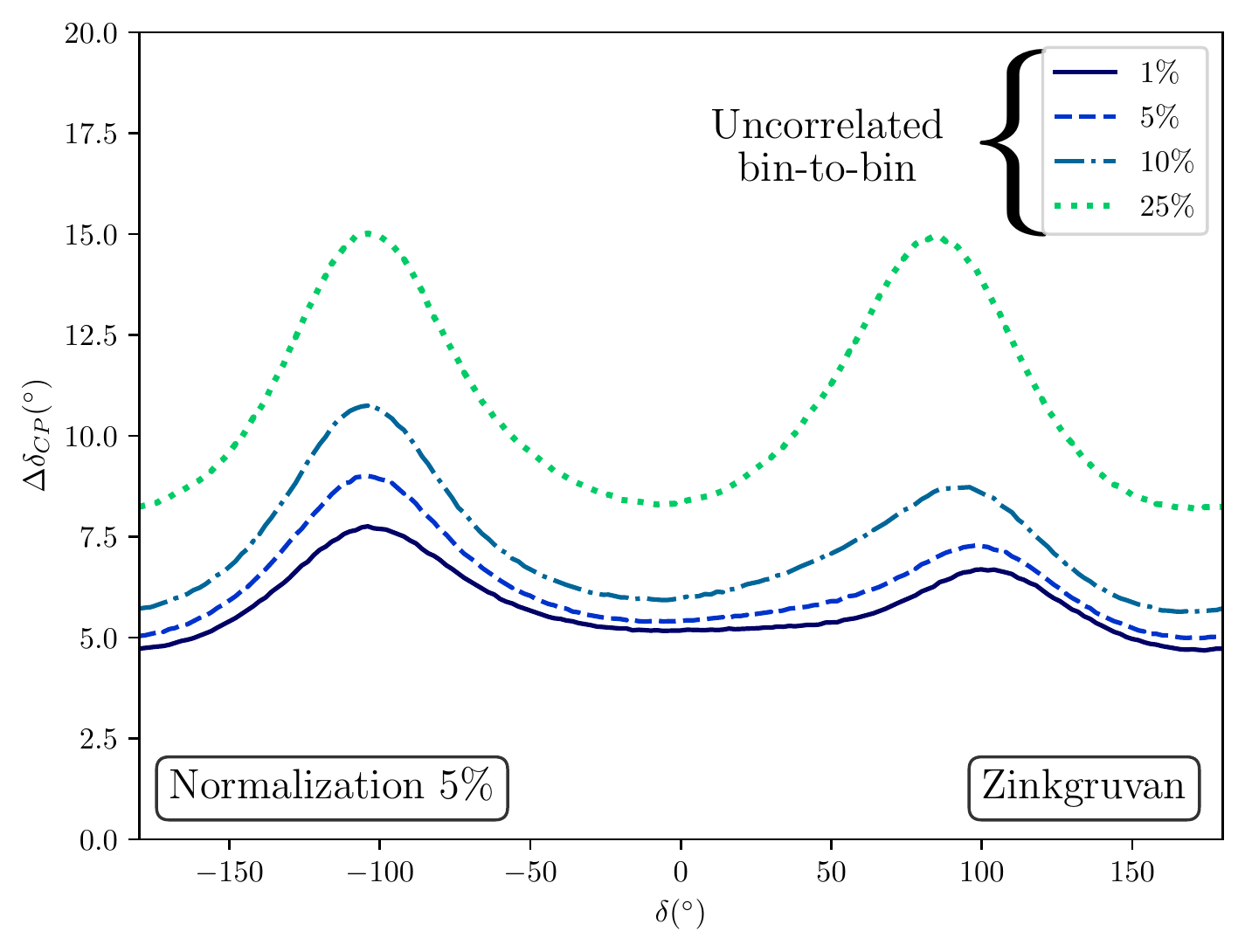}
\includegraphics[width=8cm]{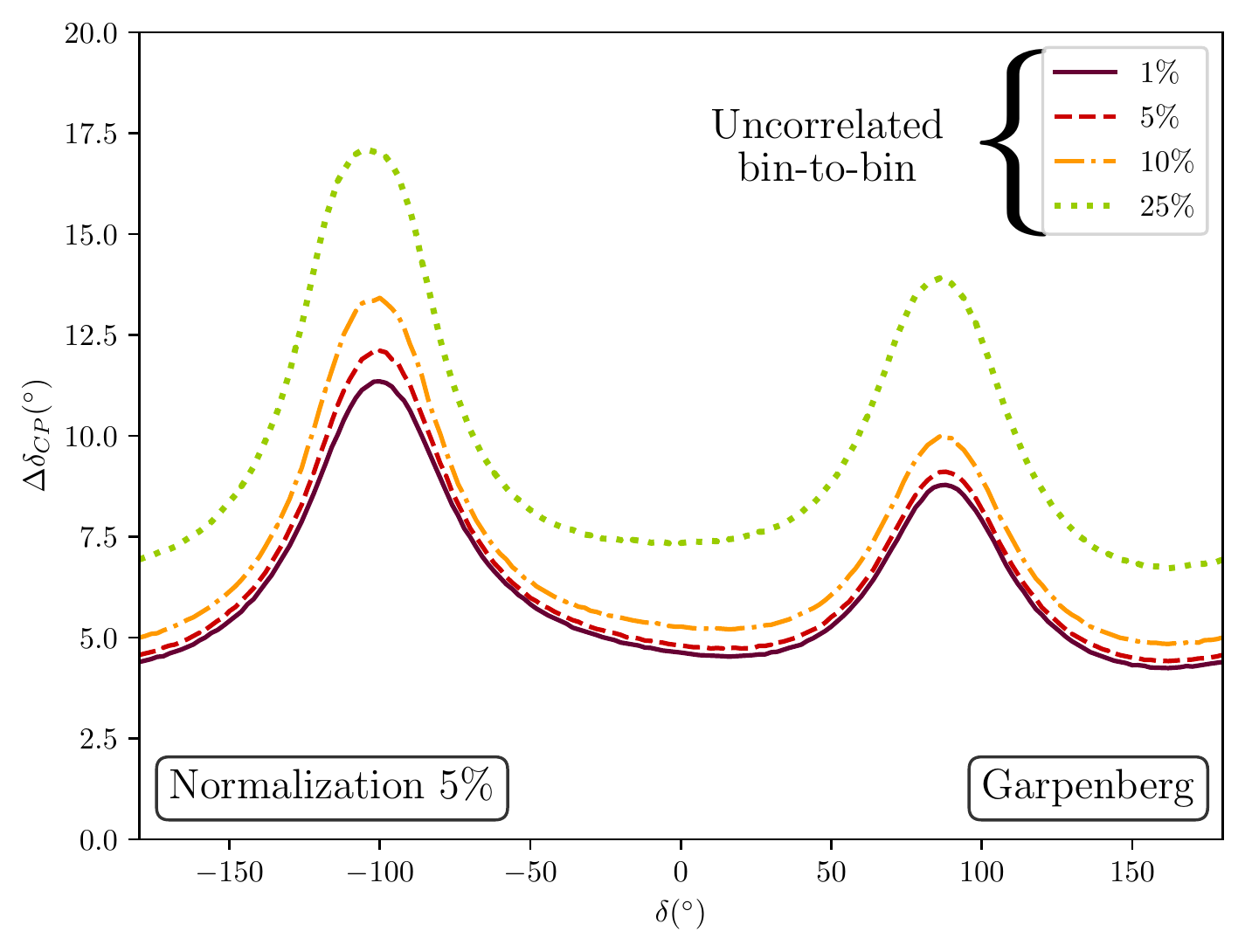}
\caption{Precision on the ESS$\nu$SB measurement of $\delta$. Left (right) panels for the Zinkgruvan 360~km (Garpenberg 540~km) baseline. The upper panels show the impact of an overall normalisation uncertainty with lines ranging from $1\%$ to $25\%$ uncertainties uncorrelated between the different signal and background samples for each channel. The middle panels show the impact of an energy calibration uncertainty again ranging from $1\%$ to $25\%$. The normalisation uncertainty has been fixed to $5\%$ to allow for the interplay between these two sources of systematics. The bottom panels show the impact of a more generic uncertainty capable of affecting both the shape and normalisation parameterised as bin-to-bin uncorrelated nuisance parameters for all signal and background components. The lines again range from $1\%$ to $25\%$; and a $5\%$ normalisation uncertainty has also been included.}
\label{fig:prec}
\end{figure}

Figure~\ref{fig:prec} shows an estimation of the precision with which the ESS$\nu$SB would be able to measure the CP-violating phase $\delta$. For each possible value of $\delta$ that could be realised in nature, Fig.~\ref{fig:prec} shows half of the region in $\delta$ for which $\Delta \chi^2 = 1$ -- as defined in Eq.~(\ref{eq:chi2}) -- and marginalising over all nuisance as well as oscillation parameters differing from $\delta$ within their priors summarised in Table~\ref{Tab:Param}. A similar behaviour to that of Fig.~\ref{fig:cpsens} is observed. In particular, the energy-calibration uncertainty has a very minor impact, while the two other sources of uncertainties studied have a more important effect.

Interestingly, the uncertainty on the overall normalisation is most important for values of $\delta \sim 0$. Conversely, the bin-to-bin uncorrelated systematics that can also affect the shape of the recovered spectrum are more relevant near the maximally CP violating values, that is $\delta \sim \pm \pi/2$. This varying dependence is related to the characteristic peak structure with poorest precision near $\delta \sim \pm \pi/2$, as shown in the plot. Following a discussion in Ref.~\cite{Coloma:2012wq}, this structure is a consequence of the dependence of the oscillation probability on $\delta$ shown in  Eq.~(\ref{Eq:Probability}). At an oscillation maximum $|\Delta m_{31}^{2}| L/(4E) = (2n-1)\pi/2$ and therefore mainly $\sin \delta$ is probed. Since the derivative of $\sin \delta$ vanishes at $\delta = \pm \pi/2$, the precision with which $\delta$ can be determined is poorest near these values.

In order to better constrain $\delta$ around $\delta = \pm \pi/2$, measurements further from the oscillation maxima are needed to determine $\cos \delta$. Thus, systematics affecting the shape of the spectrum affect these measurements and have a relevant impact around $\delta = \pm \pi/2$. These off-peak measurements are also easier to perform at Zinkgruvan, given its higher statistics and since events from the first peak down to the second maximum are observed. This also explains its better performance around $\delta = \pm \pi/2$ as compared to Garpenberg.

Conversely, nearer to $\delta \sim 0$, on-peak measurements are optimal for providing good precision on $\delta$, with energy information being less relevant but having an increased susceptibility to uncertainties on the overall normalisation. Since the precision achieved around CP-conserving values also controls the CP discovery potential (i.e. how small a value of $\delta$ can still allow for the exclusion CP-conservation) it also explains why the overall normalisation was also the most important systematic uncertainty in Fig.~\ref{fig:cpsens}. Finally, notice that for the largest values of the normalisation uncertainty considered ($10\%$ and $25\%$) a new peak develops around $\delta \sim 0$.

This worsening of the precision is due to the appearance of intrinsic~\cite{Burguet-Castell:2001ppm} and octant~\cite{Fogli:1996pv} degeneracies. These were found to be solved with the addition of atmospheric neutrino data in \cite{Blennow:2019bvl}. Moreover, such large values of the final systematic uncertainty affecting the far detector are overly conservative, so these degeneracies should not hinder the final physics reach of the facility. They are instead shown here to illustrate the resilience of the facility against even particularly large unexpected sources of systematic uncertainties, given the stronger dependence on $\delta$ of the oscillation probability at the second oscillation maximum.  

%%%%% Fig:cpsens; delta precision %%%%%
\begin{figure}[ht]
\centering
\includegraphics[width=8cm]{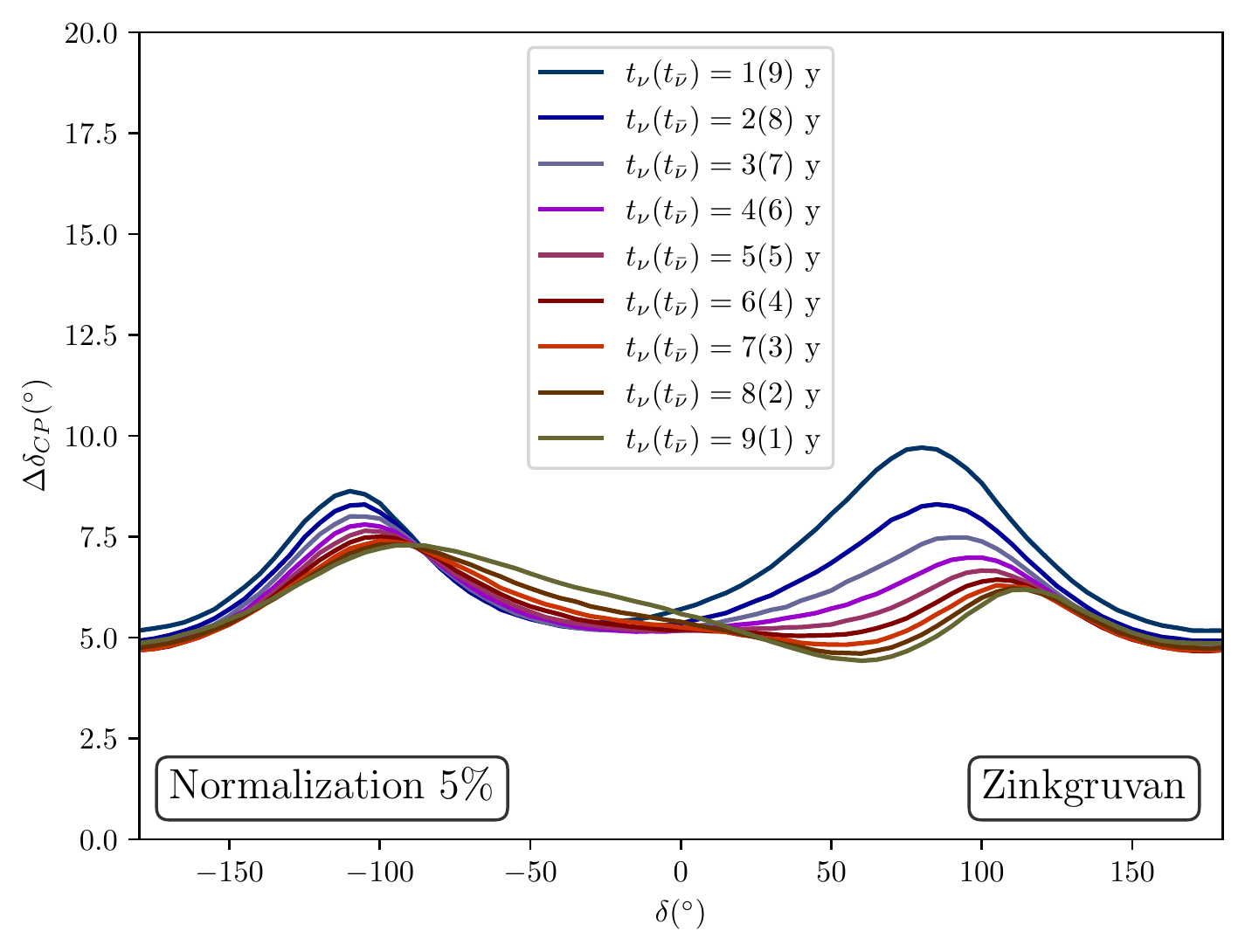}
\includegraphics[width=8cm]{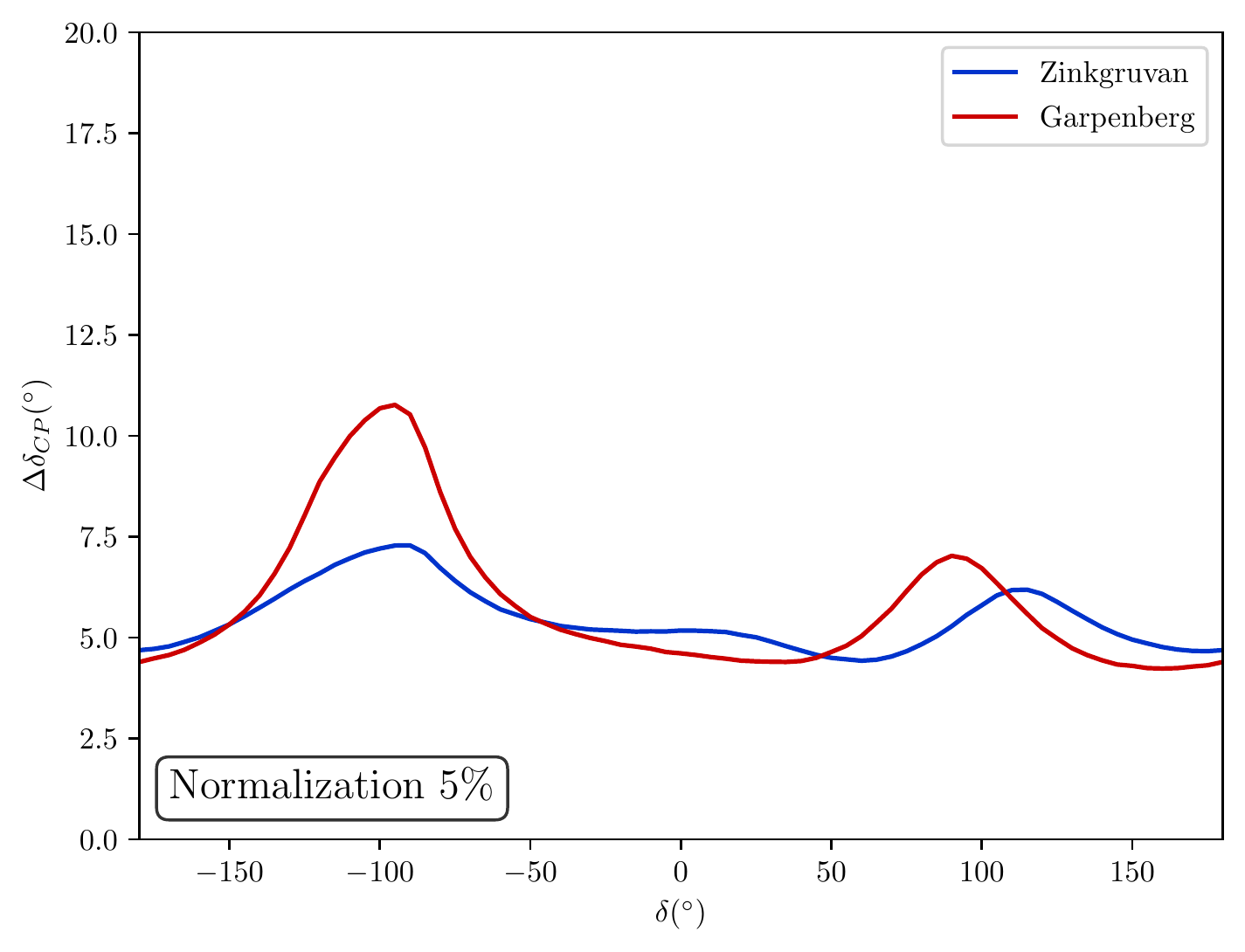}
\caption{Left panel, precision on the ESS$\nu$SB measurement of $\delta$ for different splittings of the running time between nenutrino and antineutrino modes at the Zinkgruvan baseline. The lines span from 9 (1) years to 1 (9) for neutrinos (antineutrinos); the total running time is always 10 years. Right panel, the precision on the measurement of $\delta$ when the running time is optimised as in the left panel, here comparing the Zinkgruvan and Garpenberg options.}
\label{fig:optprec}
\end{figure}

Finally, in the left panel of Fig.~\ref{fig:optprec} the dependence of the precision with which $\delta$ would be measured is studied as a function of the splitting of the total running time between positive focusing (neutrino mode) and negative focusing (antineutrino mode). As an example, the Zinkgruvan option is shown, but the behaviour is very similar for Garpenberg. As can be seen, the optimal splitting depends in the actual value of $\delta$.

As discussed above, for $\delta = \pm \pi/2$, off-peak measurements are required and statistics are of critical importance. Given the larger fluxes and cross sections, it is easier to accumulate statistics in neutrino mode and thus the best precision would be obtained by devoting longer periods of data-taking to positive focusing. Conversely, around $\delta=0$ or $\pi$ the complementarity between the neutrino and antineutrino samples is worthwhile, with more balanced partitioning of the running time providing better sensitivity. 

The measurement strategy of ESS$\nu$SB can follow a key lesson taken from preceding oscillation experiments and adapt the splitting between neutrino and antineutrino modes according to the left panel of Fig.~\ref{fig:optprec}, depending on the value of $\delta$ indicated by the data. The precision that could be obtained by following such a strategy for each value of $\delta$ is presented in the right panel of Fig.~\ref{fig:optprec}. Although the precision achievable near CP-conserving values in the measurement of $\delta$ is roughly $5^\circ$ for both the Zinkgruvan and Garpenberg options, Zinkgruvan outperforms Garpenberg near $\delta = \pm \pi/2$, with the former providing a sensitivity better than $7^\circ$ for any possible value of $\delta$. The conclusion of this study is thus that ESS$\nu$SB, with its far detector located at 360\,km in the Zinkgruvan site, can provide unprecedented precision on the measurement of $\delta$, ranging between $5^\circ$ and $7^\circ$ depending on its value. The same setup could deliver a $5\,\sigma$ discovery of CP violation for $71\,\%$ of all possible values of $\delta$.

\setcounter{figure}{0}
\numberwithin{figure}{section}
\setcounter{equation}{0}
\numberwithin{equation}{section}
\setcounter{table}{0}
\numberwithin{table}{section}

\section{Project Costing} \label{costing}
%{\bfseries TODO: Ilias}

%----------------------------------------------------
The estimated costs of the ESS$\nu$SB main components and their subcomponents are provided in Table~\ref{tab:summarycosting}. The total cost of the infrastructure is of the order of 1.38 B\texteuro, which does not include the cost of civil engineering on the ESS site. A cost estimate of the civil engineering will require a detailed study of the implementation of the components on the ESS site, that will be made only in the next phase of the study. None of these estimates include contingency. It is noteworthy that about 69\% of the total cost is for the far detector. Further details on the costs can be found in the preceding chapters.
            
\begin{table}[ht!]
\begin{center}
\caption{Cost summary for the ESS$\nu$SB experiment.}
\begin{tabular}{llcc}
\label{tab:summarycosting}
\textbf{Item} & \textbf{Sub-item} & \textbf{Cost (M€)} & \textbf{Cost (\%)} \\
\hline \textbf{Linac Upgrade} & Ion Source and Low-Energy Beam Transport (LEBT) & 5.00 & 0.36\%\\
 & Radio-Frequency Quadrupole & 6.90 & 0.50 \% \\
 & Medium Energy Beam Transport (MEBT) Upgrade & 3.00 & 0.22\%\\
 & Drift-Tube Linac with BPMs, BCMs & 13.40 & 0.97\%\\
 & High-Beta Linac (HBL) Upgrade & 10.40  & 0.75\% \\
 & 33 Modulator Upgrades & 3.50  & 0.25\% \\
 & 8 New Modulators  & 9.00  & 0.65\% \\
 & 15 Grid-Modulator Transformers & 5.60 & 0.41\% \\
 & 11 Grid-Modulator Transformers Retrofitted  & 0.50  & 0.04\% \\
 & 26 Solid-State Spoke Amplifiers & 26.00  & 1.88\% \\
 & New Klystrons for upgraded HBL & 12.10  & 0.88\% \\
 & Remaining Klystron Refurbishment/Replacement  & 25.20 & 1.82\% \\
  & Cryogenics, Water Cooling, Civil Eng. & 12.00  & 0.87\%\\
   & \textbf{Total}  & \textbf{132.60 }& \textbf{9.59}\textbf{\% }\\
 \hline\textbf{ Accumulator} & \textbf{Item}&\textbf{ Cost (M€)} & \textbf{Cost}\textbf{ (\%)}\\
 & DC Magnets and Power Supplies & 50.00 & 3.62\%\\
 & Injection system & 11.00 & 0.80\%\\
 & Extraction System & 7.00& 0.51\%\\
 & RF Systems & 16.00& 1.16\%\\
 & Collimation & 8.00& 0.58\%\\
 & Beam Instrumentation & 19.00& 1.37\%\\
 & Vacuum System& 24.00 & 1.74\%\\
 & Control System & 30.00 & 2.17\%\\
 & \textbf{Total }&\textbf{ 165.00}& \textbf{11.94}\%\\
  \hline \textbf{Target Station} &\textbf{ Item}& \textbf{Cost (M€)} & \textbf{Cost (\%)}\\
  &Target Station & 32.00& 2.32\%\\
  & Proton Beam Window System & 5.20&0.38\%\\
  & PSU + Striplines& 5.40& 0.39\%\\
  & Target and Horn Exchange System &40.42 &2.92\%\\
  & Facility Building Structure &26.60 &1.92\%\\
  & General System and Services & 21.80 &1.58\%\\
  & \textbf{Total}& \textbf{131.42}& \textbf{9.51}\%\\
  \hline \textbf{Detectors} & \textbf{Item}& \textbf{Cost (M€)} & \textbf{Cost (\%)}\\
  &Emulsion Detectors& 2.00& 0.14\%\\
  &Super Fine-Grained Detector & 5.49& 0.40\%\\
  &Near Water Cherenkov Detector &25.22& 1.82\%\\
  &Far Water Detector & 399.35& 28.89\%\\
  &Underground Cavern Excavations & 521.15&37.70\%\\
  &\textbf{Total}& \textbf{953.21}& \textbf{68.93}\textbf{\%}\\
  \hline   \hline\textbf{ Grand Total} &&\textbf{1382.23}&\textbf{100.00}\textbf{\%}\\
 
\end{tabular}
\end{center}
\end{table}

%\begin{figure}[hbt]
%\begin{center}
%\epsfig{file=figures/costing/costing.pdf,width=\linewidth}
%\caption{\small Costing figure.}
%\label{fig:costing}
%\end{center}
%\end{figure}

\clearpage

\section{Conclusion}
%{\bfseries TODO: Marcos}

Great efforts are currently being made to reduce the systematic errors in long baseline neutrino measurements. These errors have shown to be notoriously difficult to reduce further, in part because of their dependence on the modelling of the neutrino-nucleus interactions. A consequence of the unexpectedly high measured value of $\theta_{13}$, published in 2012, is that the CP signal is close to 3 times larger at the second oscillation maximum as compared to the first maximum and that the relative size of the systematic errors is close to 3 times smaller at the second maximum. This makes the sensitivity for CP violation discovery and, in particular, of the measurement of the value of CP violation phase $\delta_{CP}$, close to 3 times higher at the second maximum. As the second maximum is situated further away from the neutrino source and the neutrino beam is divergent, measurements at the second maximum require exceptionally intense neutrino beams. The ESS linac with its \SI{5}{\mega\watt} will be the most powerful proton linac in the world and will make possible the generation of a neutrino beam intense enough to measure neutrino oscillations with the detector placed at the second oscillation maximum. 

The present Conceptual Design Report describes the required modifications of the ESS linac in order to double its pulse frequency from \SI{14}{\hertz} to \SI{28}{\hertz} to provide a \SI{5}{\mega\watt} beam for neutron production through the spallation process concurrently with a \SI{5}{\mega\watt} beam for neutrino production from decay of the produced mesons. An accumulation ring has been designed for the purpose of compressing the primary \SI{2.86}{\milli\second} long beam pulses to \SI{1.2}{\micro\second}. This is required in order to reduce the neutrino cosmic ray background in the large far detector and to avoid excessive heating of the hadron collectors. Extensive studies have led to a design of a target station capable of handling the \SI{5}{\mega\watt} proton power and adequately focusing the produced mesons as well as to a design of the necessary radiation shielding for the target station and the decay tunnel. A near and a far detector have been designed, both of the water Cherenkov type. The near detector, which shall be used to monitor the neutrino flux as well as measure neutrino cross-sections, is complemented with a scintillator cube tracker and an emulsion detector immersed in water. The designs and computer simulations of the ESS$\nu$SB components resulting from this work have demonstrated their technical feasibility and that they satisfy the technical design specifications.

On the basis of extensive computer simulations it has been demonstrated that with the neutrino beam, generated with the use of the \SI{5}{\mega\watt} beam from the ESS linac, and the \SI{538}{\kilo\tonne} fiducial-mass Cherenkov far detector, ESS$\nu$SB will achieve, after 10 years of data collection, 70\% leptonic CP violation discovery coverage at $5\sigma$ significance and a measurement of the CP violation phase $\delta_{CP}$ with an error not exceeding $8^{\circ}$, irrespectively of the value of $\delta_{CP}$. Measurements of $\delta_{CP}$ with such accuracy are of prime interest for the verification of theories explaining the matter-antimatter asymmetry in the universe. 

Further studies have shown that this project has strong potential for future upgrades by using the muons produced concurrently with the neutrinos to feed a low energy nuSTORM race-track ring and an ENUBET-type Monitored Neutrino Beam for enhanced measurements of neutrino cross-sections. The high power of the proton beam and the pulse compression of the accumulator ring could, together with further compression in a second ring, be used for adequate tests of a high-power target station for the Muon Collider project.

The results of this EU H2020 Design INFRADEV-1 Design Study open the possibility to realise a facility capable of producing the most intense man-made neutrino beam to date, with very high physics performance and strong potential for other long-term projects involving neutrinos and muons. The study has also produced an estimate of the total cost of the ESS$\nu$SB research facility which amounts to around 1.38 B\texteuro~excluding civil engineering costs at the ESS site.

\section{Acknowledgments}
%{\bfseries TODO: Marcos}

This project has been supported by the COST Action EuroNuNet Combining forces for a novel European facility for neutrino-antineutrino symmetry-violation discovery.
It has also received funding from the European Union’s Horizon 2020 research and innovation programme under grant agreement No. 777419.

\noindent We acknowledge further support provided by the following research funding agencies:

\noindent Centre National de la Recherche Scientifique and Institut National de Physique Nucl\'eaire et de Physique des Particules, France,

\noindent Deutsche Forschungsgemeinschaft, Germany, Projektnummer 423761110,

\noindent Agencia Estatal de Investigacion through the grants IFT Centro de Excelencia Severo Ochoa, Spain, contract No. CEX2020-001007-S and PID2019-108892RB funded by MCIN/AEI/10.13039/501100011033, 

\noindent Polish Ministry of Science and Higher Education, grant No. W129/H2020/2018, with the science resources for the years 2018–2021 for the realisation of a co-funded project.

\noindent Ministry of Science and Education of Republic of Croatia grant No. KK.01.1.1.01.0001, Ramanujan Fellowship of SERB, Govt. of India, through grant no: RJF/2020/000082,

\noindent as well as support provided by the universities and laboratories to which the authors of this report are affiliated, see the author list on the first page.

\noindent The authors wish to express their deep gratitude to the members of the ESS$\nu$SB International Advisor Panel Chris Densham (RAL), Eligio Lisi (INFN) - IAP Chair, Shinji Machida (RAL), Alexander Olshevsky (JINR) and Jingyu Tang (IHEP) for all the very constructive and helpful advice they have provided throughout the four years of the design study.

\noindent We acknowledge the assistance from C. Vilela, E. O’Sullivan, H. Tanaka, B. Quilain and M. Wilking for providing support for the use of the WCSim and fiTQun software packages. 

\clearpage

\bibliographystyle{unsrt_modjp}
\bibliography{20_biblio}

\end{document}